# An Improved Parameterization Procedure for NDDO-Descendant Semiempirical Methods

Adrian Wee Wen Ong[1-2], Steve Yueran Cao[1-2], Leong Chuan Kwek[2-5]

[1] NUS High School of Mathematics and Science, 20 Clementi Avenue 1, Singapore 129957
[2] Centre for Quantum Technologies, National University of Singapore, Singapore 117543
[3] MajuLab, CNRS-UNS-NUS-NTU International Joint Research Unit, UMI 3654, Singapore 117543
[4] National Institute of Education, Nanyang Technological University, 1 Nanyang Walk, Singapore 637616
[5] School of Electrical and Electronic Engineering Block S2.1, 50 Nanyang Avenue, Singapore 639798

**Abstract**

MNDO-based semiempirical methods in quantum chemistry have found widespread application in the modelling of large and complex systems. A method for the analytic evaluation of first and second derivatives of molecular properties against semiempirical parameters in MNDO-based NDDO-descendant models is presented, and the resultant parameter Hessian is compared against the approximant currently used in parameterization for the PMx models. As a proof of concept, the exact parameter Hessian is employed in a limited reparameterization of MNDO for the elements C, H, N, O and F using 1206 molecules for reference data.

**Introduction**

In modern computational chemistry, semiempirical methods based on the Neglect of Diatomic Differential Overlap (NDDO) [1-12] have found widespread applications in studies where more computationally-intensive ab initio methods are unfeasible. Essential to the success of a developed semiempirical model is a robust parameterization procedure that allows the model to best reproduce experimental data; as such, parameter optimization is an important area of study for the development of effective and accurate semiempirical models.

While alternative formulations based on the NDDO approximation [13-17] have been proposed in recent years, most NDDO-descendant semiempirical models [1-2,5,9] use an identical formalism for the construction of the one-electron matrix **H** and two-electron matrix **G**; with the exception of minor corrections to the asymptotic behavior of two-electron integrals in PM7 [10], these methods also employ the same approximations and semiempirical expressions [18-19] for the evaluation of relevant molecular integrals. The difference between most NDDO-descendant models hence lie only in the parameter values chosen as well as the empirical expressions for core-core repulsion terms, which have been modified significantly between models [2,9,10-11]; as such, any parameterization scheme developed for one NDDO-descendant model can be readily applied with few modifications to other NDDO-descendant models.

As noted by Stewart when developing PM7 [10], the parameter Hessian can be used to determine the nature of stationary points detected during parameterization; likewise, accurate second-derivative information can be used to great effect in searching for minima on surfaces. Construction of the parameter Hessian via direct differentiation of the error function appears straightforward; in differentiating the error function $\mathcal{S}$ with respect to parameters ${}^{Z_A}p_i$ and ${}^{Z_B}p_j$, we find that:

$$\mathcal{S} = \sum_\alpha \mathcal{C}_\alpha^2 \left(\xi_\alpha^{\text{ref}} - \xi_\alpha\right)^2 \tag{1}$$

$$\frac{\partial^2 \mathcal{S}}{\partial {}^{Z_A}p_i \partial {}^{Z_B}p_j} = 2 \sum_\alpha \mathcal{C}_\alpha^2 \left[ \left(\xi_\alpha - \xi_\alpha^{\text{ref}}\right) \frac{\partial^2 \xi_\alpha}{\partial {}^{Z_A}p_i \partial {}^{Z_B}p_j} + \frac{\partial \xi_\alpha}{\partial {}^{Z_A}p_i} \frac{\partial \xi_\alpha}{\partial {}^{Z_B}p_j} \right] \tag{2}$$

It should be noted, however, that the parameter Hessian constructed in the development of PM7 appears to neglect second derivatives of the reference functions; as given in [10], the expression for $\frac{\partial^2 \mathcal{S}}{\partial {}^{Z_A}p_i \partial {}^{Z_B}p_j}$ appears to be:

$$\frac{\partial^2 \mathcal{S}^{\text{PM7}}}{\partial {}^{Z_A}p_i \partial {}^{Z_B}p_j} = 2 \sum_\alpha \mathcal{C}_\alpha^2 \frac{\partial \xi_\alpha}{\partial {}^{Z_A}p_i} \frac{\partial \xi_\alpha}{\partial {}^{Z_B}p_j} \tag{3}$$

The neglect of the second-derivative term is expected to significantly affect the nature of the constructed Hessian matrix and its eigenvalues, impacting the quality of parameter optimization.

Evaluation of parameter derivatives of molecular properties via finite difference, as in standard procedure, results in numerical instability and may lead to an undesirable irreproducibility in results; thus, an analytical method for evaluation of parameter first and second derivatives is sought. This requires the evaluation of second derivatives of the density matrix as the idempotency condition [20] does not apply for derivatives of the ionization energy or dipole moment; an efficient method for solution of the second order coupled-perturbed Hartree-Fock (CPHF) equations is hence necessary for analytical derivative evaluation. NDDO methods formally operate in the Lowdin basis, where the overlap matrix between basis functions is substituted for the identity matrix; thus, the relevant equations for CPHF equation solution are greatly simplified and may be easily implemented.

**The Second Order CPHF Equations Under the NDDO Approximation**

The form of the first order CPHF equations in the Lowdin basis for both closed- and open-shell species under the NDDO approximation has been well-documented [21] due to their implementation in geometry optimization routines [21-22]; the second-order CPHF equations have been presented only in the MO basis [23] and are hence presented in analogous form for the UHF case in the Lowdin basis, with a full derivation given in the Supplementary Information.

We first define the generalized Coulomb and exchange matrices **J(Δ)** and **K(Λ)** as well as their associated static derivatives as the contraction of the relevant 2-electron integrals with arbitrary matrices **Δ** and **Λ**:

$$J_{\mu\nu}(\mathbf{\Delta}) = \begin{cases} \sum\limits_{\lambda \in A} \Delta_{\lambda\lambda}(\mu\nu|\lambda\lambda) + \sum\limits_{\lambda,\sigma \in B \neq A} \Delta_{\lambda\sigma}(\mu\nu|\lambda\sigma), & \mu = \nu, \quad \mu,\nu \in A \\ 2\Delta_{\mu\nu}(\mu\nu|\mu\nu) + \sum\limits_{\lambda,\sigma \in B \neq A} \Delta_{\lambda\sigma}(\mu\nu|\lambda\sigma), & \mu,\nu \in A \\ 0, & \mu \in A, \quad \nu \in B \neq A \end{cases} \quad (4)$$

$$J_{\mu\nu}^{q_1}(\mathbf{\Delta}) = \begin{cases} \sum\limits_{\lambda \in A} \Delta_{\lambda\lambda} \frac{d(\mu\nu|\lambda\lambda)}{dq_1} + \sum\limits_{\lambda,\sigma \in B \neq A} \Delta_{\lambda\sigma} \frac{d(\mu\nu|\lambda\sigma)}{dq_1}, & \mu = \nu, \quad \mu,\nu \in A \\ 2\Delta_{\mu\nu} \frac{d(\mu\nu|\mu\nu)}{dq_1} + \sum\limits_{\lambda,\sigma \in B \neq A} \Delta_{\lambda\sigma} \frac{d(\mu\nu|\lambda\sigma)}{dq_1}, & \mu,\nu \in A \\ 0, & \mu \in A, \quad \nu \in B \neq A \end{cases} \quad (5)$$

$$J_{\mu\nu}^{q_1 q_2}(\mathbf{\Delta}) = \begin{cases} \sum\limits_{\lambda \in A} \Delta_{\lambda\lambda} \frac{d^2(\mu\nu|\lambda\lambda)}{dq_1 dq_2} + \sum\limits_{\lambda,\sigma \in B \neq A} \Delta_{\lambda\sigma} \frac{d^2(\mu\nu|\lambda\sigma)}{dq_1 dq_2}, & \mu = \nu, \quad \mu,\nu \in A \\ 2\Delta_{\mu\nu} \frac{d^2(\mu\nu|\mu\nu)}{dq_1 dq_2} + \sum\limits_{\lambda,\sigma \in B \neq A} \Delta_{\lambda\sigma} \frac{d^2(\mu\nu|\lambda\sigma)}{dq_1 dq_2}, & \mu,\nu \in A \\ 0, & \mu \in A, \quad \nu \in B \neq A \end{cases} \quad (6)$$

$$K_{\mu\nu}(\mathbf{\Lambda}) = \begin{cases} \sum\limits_{\lambda \in A} \Lambda_{\lambda\lambda}(\mu\lambda|\nu\lambda), & \mu = \nu, \quad \mu,\nu \in A \\ \Lambda_{\mu\nu}[(\mu\nu|\mu\nu) + (\mu\mu|\nu\nu)], & \mu,\nu \in A \\ \sum\limits_{\substack{\lambda \in A \\ \sigma \in B}} \Lambda_{\lambda\sigma}(\mu\lambda|\nu\sigma), & \mu \in A, \quad \nu \in B \neq A \end{cases} \quad (7)$$

$$K_{\mu\nu}^{q_1}(\mathbf{\Lambda}) = \begin{cases} \sum_{\lambda \in A} \Lambda_{\lambda\lambda} \frac{d(\mu\lambda|\nu\lambda)}{dq_1}, & \mu = \nu, \quad \mu,\nu \in A \\ \Lambda_{\mu\nu}\left[\frac{d(\mu\nu|\mu\nu)}{dq_1} + \frac{d(\mu\mu|\nu\nu)}{dq_1}\right], & \mu,\nu \in A \\ \sum_{\substack{\lambda \in A \\ \sigma \in B}} \Lambda_{\lambda\sigma} \frac{d(\mu\lambda|\nu\sigma)}{dq_1}, & \mu \in A, \quad \nu \in B \neq A \end{cases} \quad (8)$$

$$K_{\mu\nu}^{q_1 q_2}(\mathbf{\Lambda}) = \begin{cases} \sum_{\lambda \in A} \Lambda_{\lambda\lambda} \frac{d^2(\mu\lambda|\nu\lambda)}{dq_1 dq_2}, & \mu = \nu, \quad \mu,\nu \in A \\ \Lambda_{\mu\nu}\left[\frac{d^2(\mu\nu|\mu\nu)}{dq_1 dq_2} + \frac{d^2(\mu\mu|\nu\nu)}{dq_1 dq_2}\right], & \mu,\nu \in A \\ \sum_{\substack{\lambda \in A \\ \sigma \in B}} \Lambda_{\lambda\sigma} \frac{d^2(\mu\lambda|\nu\sigma)}{dq_1 dq_2}, & \mu \in A, \quad \nu \in B \neq A \end{cases} \quad (9)$$

The first-order CPHF equations for the UHF case are then expressed as:

$$^\sigma \mathfrak{F}^{q_1} = {^\sigma\mathbf{C}^\mathrm{T}} \left(\frac{d\mathbf{H}}{dq_1} + \mathbf{J}^{q_1}(\mathbf{P}) - \mathbf{K}^{q_1}(^\sigma\mathbf{P})\right) {^\sigma\mathbf{C}} \quad (10)$$

$$^\sigma \mathfrak{R}^{q_1} = {^\sigma\mathbf{C}^\mathrm{T}} \left(\mathbf{J}\left(\frac{d\mathbf{P}}{dq_1}\right) - \mathbf{K}\left(\frac{d^\sigma\mathbf{P}}{dq_1}\right)\right) {^\sigma\mathbf{C}} \quad (11)$$

$$\left(\frac{d^\sigma\mathbf{P}}{dq_1}\right)_{\mu\nu} = -\sum_{i \in occ.} \sum_{j \in virt.} \left({^\sigma c_{\mu j}}{^\sigma c_{\nu i}} + {^\sigma c_{\mu i}}{^\sigma c_{\nu j}}\right) {^\sigma x_{ij}^{q_1}} \quad (12)$$

$$\left(\frac{d\mathbf{H}}{dq_1}\right)_{\mu\nu} = \begin{cases} \frac{d^{Z_A}U_{\mu\mu}}{dq_1} + \sum_{B \neq A} \frac{dV_{\mu\mu,B}}{dq_1}, & \mu = \nu, \quad \mu,\nu \in A \\ \sum_{B \neq A} \frac{dV_{\mu\nu,B}}{dq_1}, & \mu,\nu \in A \\ \frac{d\beta_{\mu\nu}}{dq_1}, & \mu \in A, \quad \nu \in B \neq A \end{cases} \quad (13)$$

$$\left({^\sigma\epsilon_j} - {^\sigma\epsilon_i}\right) {^\sigma x_{ij}^{q_1}} - {^\sigma\mathfrak{R}_{ij}^{q_1}} = {^\sigma\mathfrak{F}_{ij}^{q_1}} \quad (14)$$

In the above, $\sigma \in \{\alpha, \beta\}$ denotes an arbitrary spin while $\mathbf{P} = {^\alpha\mathbf{P}} + {^\beta\mathbf{P}}$ indicates the total density matrix as a sum of the alpha- and beta-spin components.

Since only the occupied-virtual block of $^\sigma \mathbf{x}^{q_1}$ is necessary to solve for $\frac{d^\sigma \mathbf{P}}{dq_1}$, the remaining elements of $^\sigma \mathbf{x}^{q_1}$ are evaluated afterwards via direct substitution:

$$^\sigma x_{ij}^{q_1} = \begin{cases} 0, & i = j \\ \dfrac{^\sigma \mathfrak{F}^{q_1} + {}^\sigma \mathfrak{R}^{q_1}}{^\sigma \epsilon_j - {}^\sigma \epsilon_i}, & i \neq j \end{cases} \tag{15}$$

The full matrix $^\sigma \mathbf{x}^{q_1}$ is necessary for solution of the second-order CPHF equations, obtained via further direct differentiation:

$$^\sigma \rho_{\mu\nu}^{q_1 q_2} = -\sum_{i \in occ.} \sum_{j \in virt.} \left( {}^\sigma c_{\mu j} {}^\sigma c_{\nu i} + {}^\sigma c_{\mu i} {}^\sigma c_{\nu j} \right) {}^\sigma \gamma_{ij}^{q_1 q_2} \tag{16}$$

$$\begin{aligned}
^\sigma \varsigma_{\mu\nu}^{q_1 q_2} &= \sum_{i \in occ.} \sum_{j,k} \left( {}^\sigma c_{\mu k} {}^\sigma c_{\nu i} + {}^\sigma c_{\mu i} {}^\sigma c_{\nu k} \right) \left( {}^\sigma x_{kj}^{q_2} {}^\sigma x_{ji}^{q_1} + {}^\sigma x_{kj}^{q_1} {}^\sigma x_{ji}^{q_2} \right) \\
&+ \sum_{i \in occ.} \sum_{j \in occ.} \sum_k {}^\sigma c_{\mu i} {}^\sigma c_{\nu j} \left( {}^\sigma x_{ik}^{q_1} {}^\sigma x_{kj}^{q_2} + {}^\sigma x_{ik}^{q_2} {}^\sigma x_{kj}^{q_1} \right) \\
&+ \sum_{i \in occ.} \sum_{j,k} \left( {}^\sigma c_{\mu j} {}^\sigma c_{\nu k} + {}^\sigma c_{\mu k} {}^\sigma c_{\nu j} \right) {}^\sigma x_{ki}^{q_1} {}^\sigma x_{ji}^{q_2}
\end{aligned} \tag{17}$$

$$\begin{aligned}
^\sigma \mathbf{F}^{q_1 q_2} &= \frac{d^2 \mathbf{H}}{dq_1 dq_2} + \mathbf{J}^{q_1 q_2}(\mathbf{P}) + \mathbf{J}^{q_1}\left(\frac{d\mathbf{P}}{dq_2}\right) + \mathbf{J}^{q_2}\left(\frac{d\mathbf{P}}{dq_1}\right) \\
&- \mathbf{K}^{q_1 q_2}({}^\sigma \mathbf{P}) - \mathbf{K}^{q_1}\left(\frac{d^\sigma \mathbf{P}}{dq_2}\right) - \mathbf{K}^{q_2}\left(\frac{d^\sigma \mathbf{P}}{dq_1}\right) + \mathbf{J}(\varsigma^{q_1 q_2}) - \mathbf{K}({}^\sigma \varsigma^{q_1 q_2})
\end{aligned} \tag{18}$$

$$\begin{aligned}
^\sigma \mathfrak{F}^{q_1 q_2} &= {}^\sigma \mathbf{C}^{\mathrm{T}} {}^\sigma \mathbf{F}^{q_1 q_2} {}^\sigma \mathbf{C} + \left( {}^\sigma \mathbf{x}^{q_1} {}^\sigma \mathbf{x}^{q_2} + {}^\sigma \mathbf{x}^{q_2} {}^\sigma \mathbf{x}^{q_1} \right) {}^\sigma \boldsymbol{\epsilon} \\
&+ ({}^\sigma \mathbf{x}^{q_1})^{\mathrm{T}} ({}^\sigma \mathfrak{F}^{q_2} + {}^\sigma \mathfrak{R}^{q_2}) + ({}^\sigma \mathbf{x}^{q_1})^{\mathrm{T}} {}^\sigma \boldsymbol{\epsilon} {}^\sigma \mathbf{x}^{q_2} + ({}^\sigma \mathfrak{F}^{q_2} + {}^\sigma \mathfrak{R}^{q_2}) {}^\sigma \mathbf{x}^{q_1} \\
&+ ({}^\sigma \mathbf{x}^{q_2})^{\mathrm{T}} ({}^\sigma \mathfrak{F}^{q_1} + {}^\sigma \mathfrak{R}^{q_2}) + ({}^\sigma \mathbf{x}^{q_2})^{\mathrm{T}} {}^\sigma \boldsymbol{\epsilon} {}^\sigma \mathbf{x}^{q_1} + ({}^\sigma \mathfrak{F}^{q_1} + {}^\sigma \mathfrak{R}^{q_1}) {}^\sigma \mathbf{x}^{q_2}
\end{aligned} \tag{19}$$

$$^\sigma \mathfrak{R}^{q_1 q_2} = {}^\sigma \mathbf{C}^{\mathrm{T}} \left( \mathbf{J}(\rho^{q_1 q_2}) - \mathbf{K}({}^\sigma \rho^{q_1 q_2}) \right) {}^\sigma \mathbf{C} \tag{20}$$

$$\left(\frac{d^2\mathbf{H}}{dq_1 dq_2}\right)_{\mu\nu} = \begin{cases} \dfrac{d^{2\,Z_A}U_{\mu\mu}}{dq_1 dq_2} + \sum_{B \neq A} \dfrac{d^2 V_{\mu\mu,B}}{dq_1 dq_2}, & \mu = \nu, \quad \mu,\nu \in A \\ \sum_{B \neq A} \dfrac{d^2 V_{\mu\nu,B}}{dq_1 dq_2}, & \mu,\nu \in A \\ \dfrac{d^2 \beta_{\mu\nu}}{dq_1 dq_2}, & \mu \in A, \quad \nu \in B \neq A \end{cases} \quad (21)$$

$$\left(^\sigma\epsilon_j - {}^\sigma\epsilon_i\right){}^\sigma\gamma_{ij}^{q_1 q_2} - {}^\sigma\mathfrak{R}_{ij}^{q_1 q_2} = {}^\sigma\mathfrak{F}_{ij}^{q_1 q_2} \quad (22)$$

The form of the second-order CPHF equations have intentionally been cast into a form that resemble the first-order CPHF equations; as such, the same algorithms [21,24] employed to solve the first-order CPHF equations may be applied for solution of the second-order CPHF equations.

**Derivatives of Molecular Properties Under the MNDO Formalism**

The relevant expressions for the derivatives of heats of formation, ionization energies, dipole moments and energy gradients at a specified reference geometry are provided in the Supplementary Information.

## Nature of the Hessian Approximant in PM7

To illustrate the differences between the exact Hessian **H** and the approximant $^{PM7}\mathbf{H}$, a pictorial representation of the two matrices is given below. The matrices were computed using original MNDO parameters [1] for a chosen training set of 1206 molecules (see Supplementary Information):

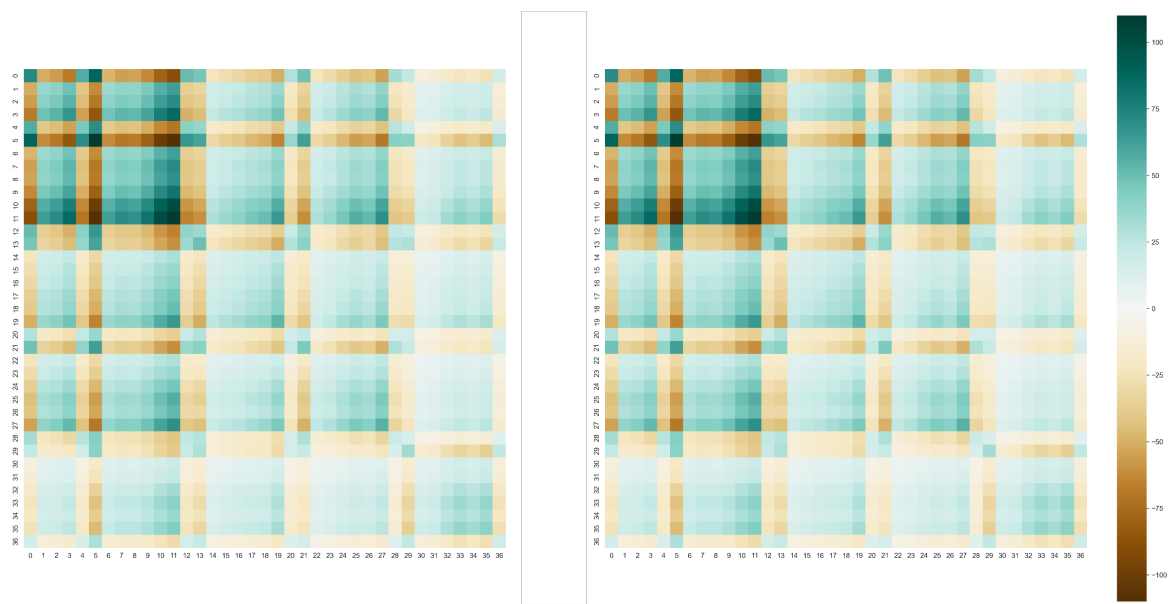

Fig. 1: A heatmap of the elements of the exact Hessian **H** (left) and Hessian approximant $^{PM7}\mathbf{H}$ (right), raised to the fifth root. The $37 \times 37$ Hessian contains second derivatives for the MNDO parameters $\alpha, \beta_s, \beta_p, U_{ss}, U_{pp}, \zeta_s, \zeta_p, E_{isol}$ for the elements H, C, N, O and F.

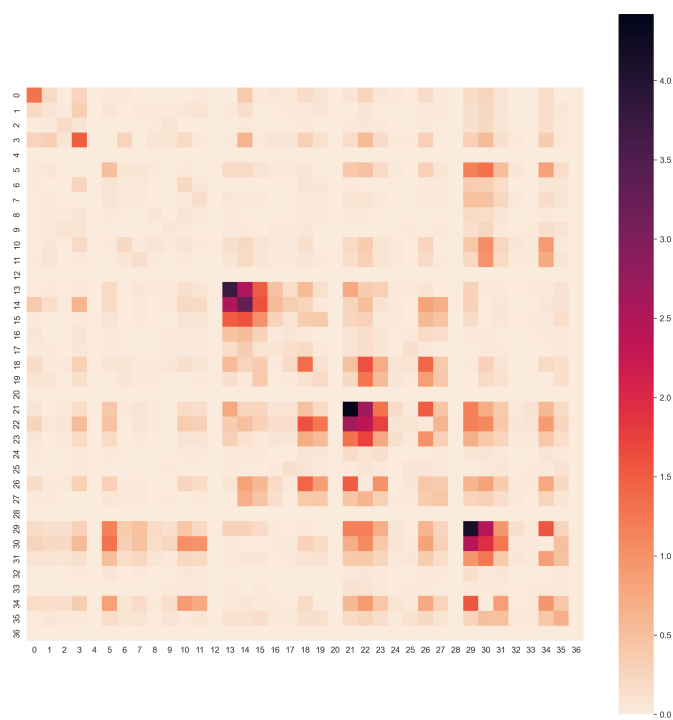

Fig. 2: A heatmap of percentage errors of each Hessian element, calculated as $\left|\frac{^{PM7}\mathbf{H}_{ij} - \mathbf{H}_{ij}}{\mathbf{H}_{ij}}\right| \times 100\%$.

Furthermore, the approximant $^{PM7}\mathbf{H}$ is observed to be positive definite while $\mathbf{H}$ reflects that the parameter surface is nonconvex:

| No. | Eigenvalue ($^{PM7}\mathbf{H}$) | Eigenvalue ($\mathbf{H}$) |
|---|---|---|
| 1 | 0.128368 | -512816 |
| 2 | 0.934416928 | -460096.7651 |
| 3 | 15.41202367 | -267302.7955 |
| 4 | 30.05635772 | -216687.9532 |
| 5 | 61.02869235 | -167602.8167 |
| 6 | 89.83798894 | -39421.75011 |
| 7 | 133.3580448 | -7503.560232 |
| 8 | 314.5683375 | -2616.97078 |
| 9 | 333.6432758 | -983.335745 |
| 10 | 1440.455065 | -533.6458032 |
| 11 | 1645.025654 | -211.742801 |
| 12 | 2550.646282 | -101.2912023 |
| 13 | 3835.543251 | -41.93718193 |
| 14 | 5680.70884 | 248.1694688 |
| 15 | 8639.457771 | 775.4405619 |
| 16 | 13231.619 | 2083.752434 |
| 17 | 20549.13091 | 3068.359634 |
| 18 | 38971.27902 | 11361.85088 |
| 19 | 103579.4856 | 19381.78711 |
| 20 | 121144.8051 | 68127.12467 |
| 21 | 182104.4115 | 119463.3803 |
| 22 | 247015.4439 | 298112.1443 |
| 23 | 313366.6596 | 894952.2348 |
| 24 | 536170.167 | 1204079.117 |
| 25 | 755156.5518 | 1925370.492 |
| 26 | 968678.9509 | 2107406.892 |
| 27 | 1153754.339 | 2604117.089 |
| 28 | 2096012.221 | 8556379.32 |
| 29 | 3620532.328 | 8932577.125 |
| 30 | 5794531.367 | 17840724.68 |
| 31 | 12084613.12 | 38505843.45 |
| 32 | 38639528.07 | 72107717.07 |
| 33 | 187190499.4 | 190046809.8 |
| 34 | 446920022 | 450868384.7 |
| 35 | 509186668.9 | 534566784.3 |
| 36 | 854723315.4 | 862489769.1 |
| 37 | 47330423728 | 47359707042 |

## Methods for Parameter Optimization

In PM7, parameter optimization is performed via an approximated line search, with the search direction obtained via a direct Hessian descent (HD) step on the approximated Hessian [5]:

$$\hat{\mathbf{d}} = -\frac{{}^{PM7}\mathbf{H}^{-1}\mathbf{g}}{|{}^{PM7}\mathbf{H}^{-1}\mathbf{g}|}, \qquad |\mathbf{d}| = \mathrm{argmax}_k\left(\tilde{\mathcal{S}}|_{\mathbf{p}=\mathbf{p}_0+k\hat{\mathbf{d}}}\right) \qquad (23)$$

$$\tilde{\mathcal{S}} = \sum_\alpha c_\alpha^2 \left( \xi_\alpha^{\mathrm{ref}} - \xi_\alpha|_{\mathbf{p}=\mathbf{p}_0} - \sum_{Z_A}\sum_{Z_A p_i} \frac{d\xi_\alpha}{d^{Z_A}p_i}\bigg|_{\mathbf{p}=\mathbf{p}_0} \left({}^{Z_A}p_i|_{\mathbf{p}} - {}^{Z_A}p_i|_{\mathbf{p}=\mathbf{p}_0}\right) \right)^2 \qquad (24)$$

This method of determining the step size shall be termed the approximated line search (ALS) method; an alternative would be a trust radius (TR), where the step size is constrained by a dynamic trust radius. The modification of the trust radius in our attempt is given by the following computation, with **B** representing an arbitrary Hessian or Hessian approximant:

$$Q_n = \mathbf{d}_n^T \mathbf{g}_n + \frac{1}{2}\mathbf{d}_n^T \mathbf{B}_n \mathbf{d}_n, \qquad \rho_n = \frac{\mathcal{S}|_{\mathbf{p}=\mathbf{p}_n+\mathbf{d}_n} - \mathcal{S}|_{\mathbf{p}=\mathbf{p}_n}}{Q_n} \qquad (25\mathrm{a,b})$$

$$|\mathbf{d}_n| = R_n, \qquad R_{n+1} = \begin{cases} \frac{5}{4}R_n, & \rho_n > 0.8 \text{ and } \mathcal{S}|_{\mathbf{p}=\mathbf{p}_n+\mathbf{d}_n} < \mathcal{S}|_{\mathbf{p}=\mathbf{p}_n} \\ \frac{1}{2}R_n, & \rho_n < 0.25 \text{ or } \mathcal{S}|_{\mathbf{p}=\mathbf{p}_n+\mathbf{d}_n} \geq \mathcal{S}|_{\mathbf{p}=\mathbf{p}_n} \\ R_n, & \text{otherwise} \end{cases} \qquad (26)$$

In addition to determination of $\hat{\mathbf{d}}$ via direct Hessian descent, a trust region optimization (TRO) was also attempted, where **d** is computed via:

$$\mathbf{d} = (\mathbf{B} - \lambda\mathbf{I})^{-1}\mathbf{g}, \qquad |\mathbf{d}| = R \qquad (27)$$

The shift parameter $\lambda$ is computed iteratively, with the chosen numerical method detailed in the Supplementary Information.

Lastly, three different choices for the second-derivative matrix **B** were investigated. In addition to the exact Hessian **H** and the approximant ${}^{PM7}\mathbf{H}$, a modified Hessian $\mathbb{H}$ was found to yield promising results:

$$\mathbb{H} = \mathbf{P}\mathbf{D}'\mathbf{P}^{-1}, \qquad \mathbf{D}'_{ii} = |\mathbf{D}_{ii}|, \qquad \mathbf{H} = \mathbf{P}\mathbf{D}\mathbf{P}^{-1} \qquad (28\mathrm{a,b,c})$$

The modified Hessian $\mathbb{H}$ preserves the eigenvectors of the exact Hessian while converting it to be positive (semi)definite, and ensures that a direct Hessian descent step will not traverse in the uphill direction given a surface of the wrong concavity.

**Results of Limited Parameterisation**

As a proof of concept, a limited parameterization of 1206 molecules consisting of the atom types C, H, N, O, F was performed using the MNDO formalism; the relevant routines for the semiempirical evaluation of molecular properties using MNDO were correspondingly implemented and compared against MOPAC [25] for accuracy (see the Supplementary Information). Geometrical data, e.g., bond angles or bond lengths, are accounted for in our parameterization procedure by using the norm of the gradient vector |**g**| at a reference geometry (either experimental or the result of high-level calculations), with the corresponding reference function set to zero to simulate a perfect correspondence between the semiempirical and reference geometries; this was chosen to facilitate the ease of preparing the relevant inputs. The weighting functions $C_i$ for the error function were chosen as:

| Property | Units | $C_i$ |
|---|---|---|
| $\Delta H_f$ | kcal/mol | 1 mol/kcal |
| $I.E.$ | eV | 10 eV$^{-1}$ |
| $\langle \mu \rangle$ | D | 20 D$^{-1}$ |
| \|**g**\| | kcal/(mol · bohr) | 0.5 mol · bohr/kcal |

For all optimization methods, the graphs shown below are terminated once the decrease in $S$ is no longer appreciable; the optimization runs discussed in this Section hence do not represent the identification of minima on the parameter surface.

First, application of the optimization method used in the PMx models (Hessian descent with step size determined by the approximate line search) leads to surprisingly similar behavior when both $\mathbb{H}$ and $^{PM7}\mathbf{H}$ are used; nonetheless, there is an appreciable difference in the final parameter values for the two methods. Direct employment of the Hessian matrix $\mathbf{H}$ results in a poor optimisation procedure as the evaluated step size shrinks significantly around $S = 575000$; this is expected as the line-search procedure seeks to minimise $S$ even as the Newton-Raphson step traverses uphill for nonconvex regions.

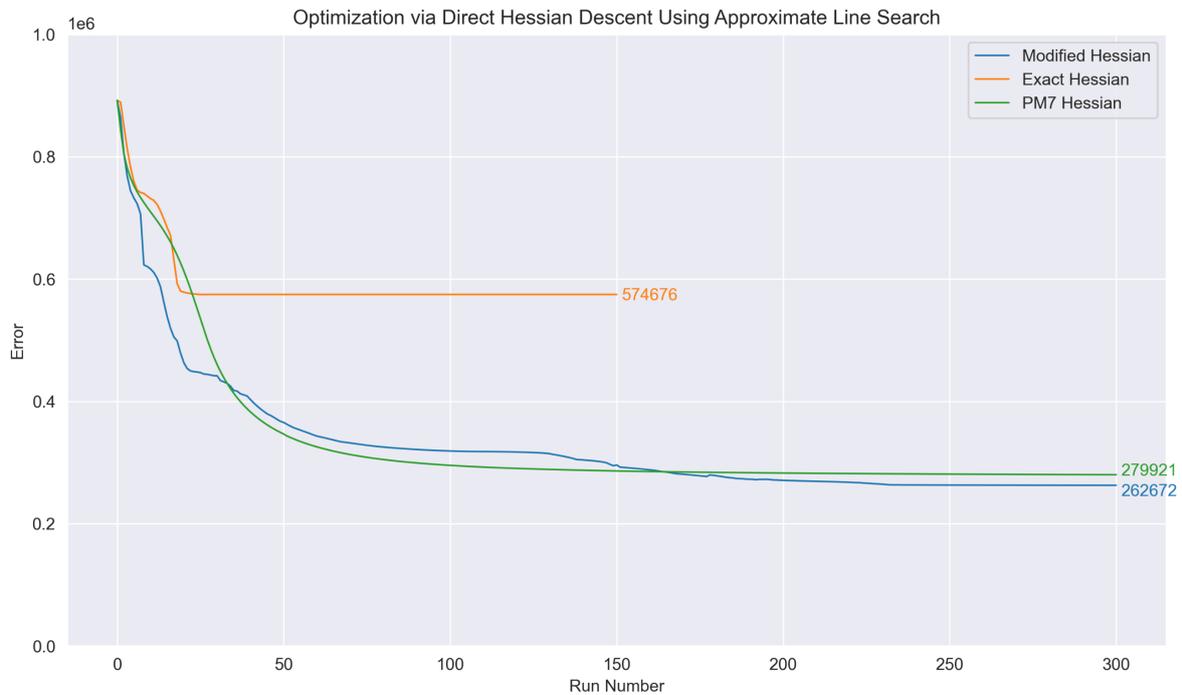

Fig. 3: Optimization curve obtained when using the parameterization procedure detailed in [5] and [10].

To compare the line-search method used in the PMx models with a trust radius method, optimization using direct Hessian descent but with the step size determined via a dynamic trust radius was attempted. We note in passing that the resultant optimization curve strongly resembles that of the according line-search method (cf. Fig. 3); however, once again, direct employment of **H** leads to very poor optimisation:

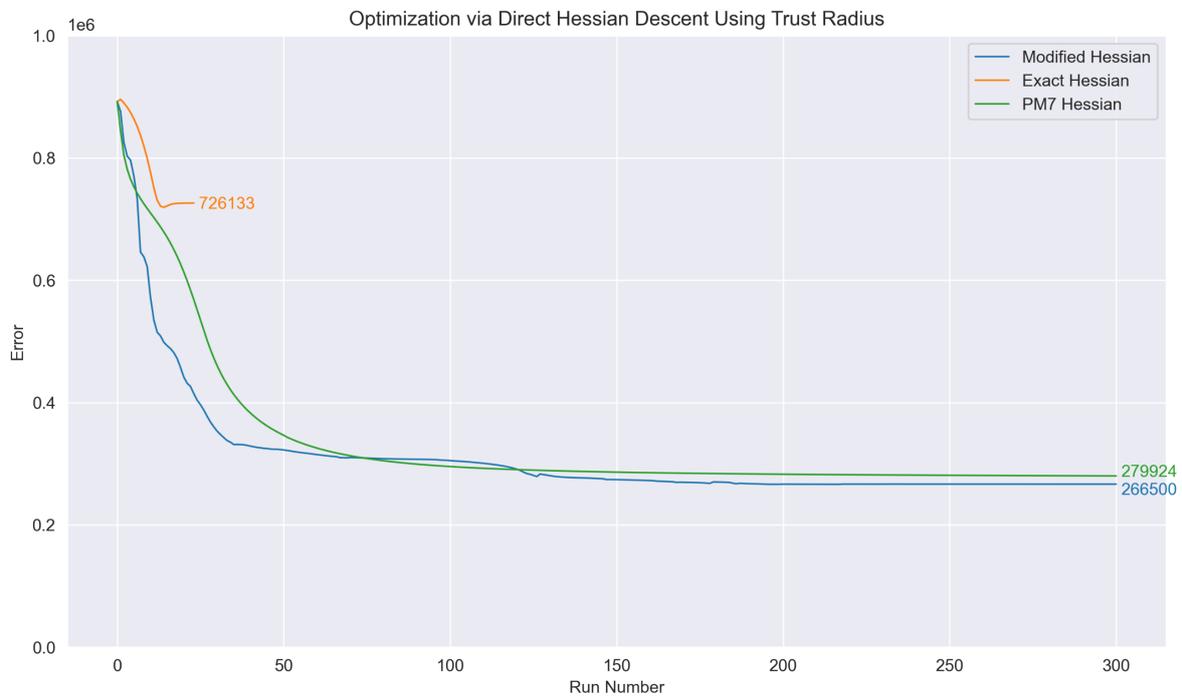

Fig. 4: Optimization curve obtained when using a trust radius with the direct Hessian descent (Newton-Raphson) step. The initial step size at run #0 is set at 2.0 for $\mathbb{H}$ and $^{PM7}\mathbf{H}$ and reduced to 0.5 for **H** (orange curve), and parameterization was halted for the attempt using **H** once the trust radius became sufficiently small.

Lastly, optimization was attempted with a trust region optimizer; this led to similar performance for $\mathbb{H}$ while $^{PM7}\mathbf{H}$ performed markedly poorer. Optimization with $\mathbf{H}$ led to suboptimal parameters that encountered problems with SCF convergence and geometry optimization and was hence abandoned.

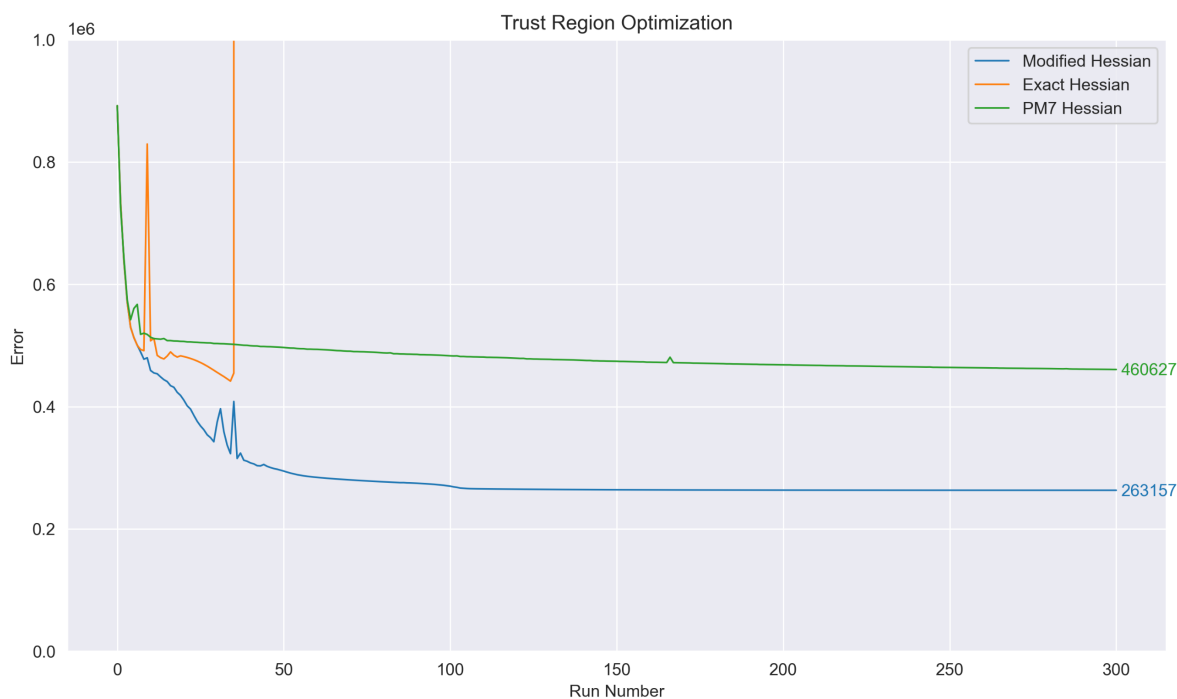

Fig. 5: Optimization curve obtained when using a trust region optimizer. The initial step size at run #0 is set at 0.1 for all three curves. Optimization using **H** (orange curve) was abandoned once suboptimal parameters led to issues with SCF convergence.

Optimization with $^{PM7}\mathbf{H}$ appears to result in the identification of a saddle point, which is incorrectly identified as a minimum due to the positive-definite nature of $^{PM7}\mathbf{H}$; the Hessian eigenvalues (from the exact Hessian **H**) for the resultant parameters obtained via optimization with $^{PM7}\mathbf{H}$ are given below:

| No. | Eigenvalue | No. | Eigenvalue | No. | Eigenvalue |
| --- | --- | --- | --- | --- | --- |
| 1 | -41.234 | 14 | 6818.55467 | 27 | 1512211.907 |
| 2 | -3.823761 | 15 | 8866.28568 | 28 | 2277598.426 |
| 3 | -2.9794242 | 16 | 19648.5226 | 29 | 4749192.867 |
| 4 | 7.18597596 | 17 | 34039.3551 | 30 | 11509664.69 |
| 5 | 42.2832925 | 18 | 42849.9713 | 31 | 20200473.71 |
| 6 | 53.6581437 | 19 | 122256.195 | 32 | 37260138.2 |
| 7 | 114.412963 | 20 | 166404.131 | 33 | 146715815.1 |
| 8 | 351.125533 | 21 | 219075.729 | 34 | 359303765.9 |
| 9 | 778.324332 | 22 | 233887.679 | 35 | 404915300.3 |
| 10 | 923.225124 | 23 | 328171.073 | 36 | 771116378.5 |
| 11 | 1804.53062 | 24 | 499506.976 | 37 | 39386113276 |
| 12 | 3531.45311 | 25 | 878679.923 | | |
| 13 | 4324.88892 | 26 | 1287823.88 | | |

In all methods, the step size is constrained to a maximum of $|\mathbf{d}|_{max} = 2$ to ensure that there are no significant and unexpected increases in the error function. We note in passing that the choice of $|\mathbf{d}|_{max} = 2$, while arbitrary, does not play a significant role in the optimization of parameters; choices of $|\mathbf{d}|_{max} = 1$ and $|\mathbf{d}|_{max} = 3$ lead to very similar optimization curves:

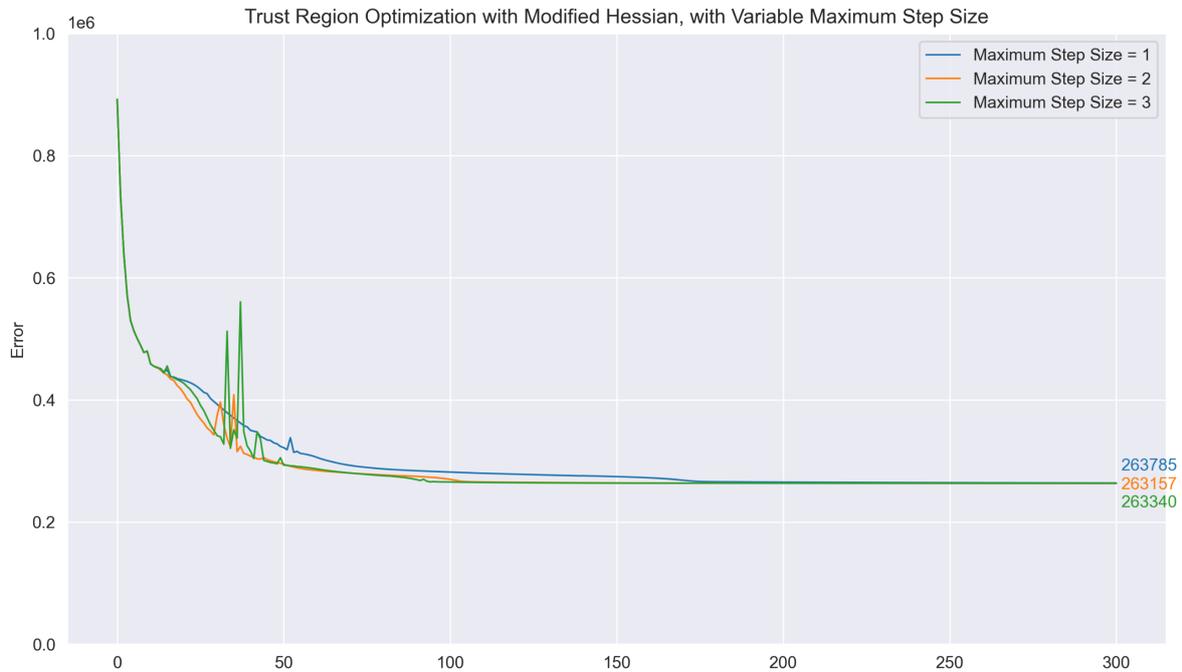

Fig. 6: Optimization curves obtained when using a trust region optimizer and the modified Hessian $\mathbb{H}$. The initial step size at run #0 is set at 0.1 for all three curves, but the maximum step size $|\mathbf{d}|_{max}$ was varied.

In the course of parameterization, it was realized that identification of a true local minimum would be difficult and require manual intervention in the parameterization procedure; thus, any set of parameters with a reasonably small gradient vector magnitude, a positive definite (exact) Hessian matrix and a marked resistance to further reduction in the error function should be accepted as a reasonable result from parameter optimization.

In this work, we report the identification of a local minimum from our limited parameterization, obtained with the imposition of no constraints on the parameters. The parameters, as well as the values of the associated gradient element, for an identified minimum with $S = 261996$ are given below:

| Parameter | Element | | | | |
|---|---|---|---|---|---|
| | C | H | N | O | F |
| $z_A \alpha$ | 2.660499 | 2.993408 | 3.021163 | 3.334554 | 3.590480 |
| $z_A \beta_s$ | -13.719865 | -7.334656 | -17.522551 | -73.616234 | -1753.340685 |
| $z_A \beta_p$ | -6.970837 | NIL | -15.550446 | -22.327739 | -22.382738 |
| $z_A U_{ss}$ | -47.292652 | -10.314472 | -64.905273 | -94.106196 | -107.380890 |
| $z_A U_{pp}$ | -40.221274 | NIL | -57.638287 | -78.565181 | -106.952003 |
| $z_A \zeta_s$ | 2.326858 | 1.124161 | 2.914488 | 5.594470 | 50.167806 |
| $z_A \zeta_p$ | 1.637021 | NIL | 2.126141 | 2.425014 | 2.463961 |
| $z_A E_{isol}$ | -112.533538 | -11.423363 | -189.262258 | -309.024988 | -434.411284 |

| Gradient | Element | | | | |
|---|---|---|---|---|---|
| | C | H | N | O | F |
| $z_A \alpha$ | -5631.2989 | -122.511277 | -265.68094 | -170.18825 | -1319.4913 |
| $z_A \beta_s$ | 122.967088 | 39.777392 | 4.0270245 | 0.92212293 | 1.65447933 |
| $z_A \beta_p$ | 284.128795 | NIL | 14.8445277 | 14.6824465 | 110.38728 |
| $z_A U_{ss}$ | 264.36961 | 51.0183803 | 29.0786636 | 40.8640734 | 655.636142 |
| $z_A U_{pp}$ | 655.367166 | NIL | 55.1677978 | 94.2737931 | 2154.95912 |
| $z_A \zeta_s$ | 1724.84129 | 97.0883125 | 57.8930721 | 36.7178843 | 86.2475113 |
| $z_A \zeta_p$ | 4822.82283 | NIL | 297.807752 | 334.865333 | 1932.25505 |
| $z_A E_{isol}$ | -245.1164 | -66.13715 | -17.695251 | -21.930875 | -391.28695 |

| No. | Eigenvalue | No. | Eigenvalue | No. | Eigenvalue |
|---|---|---|---|---|---|
| 1 | 0.000307191 | 14 | 7097.672134 | 27 | 1169583.719 |
| 2 | 2.110409103 | 15 | 11248.65204 | 28 | 4374828.93 |
| 3 | 9.894574497 | 16 | 13983.01 | 29 | 4455736.981 |
| 4 | 39.09862218 | 17 | 27423.51093 | 30 | 9638553.272 |
| 5 | 88.84495043 | 18 | 31607.2681 | 31 | 21918237.20 |
| 6 | 120.8695565 | 19 | 87137.37687 | 32 | 32920532.04 |
| 7 | 418.3575145 | 20 | 180143.5721 | 33 | 104814193.86 |
| 8 | 460.8171646 | 21 | 209812.6169 | 34 | 241208268.54 |
| 9 | 755.3529097 | 22 | 324571.6458 | 35 | 294810061.12 |
| 10 | 1256.25836 | 23 | 416778.2783 | 36 | 555086343.83 |
| 11 | 2304.566797 | 24 | 555533.5715 | 37 | 33782858461.66 |
| 12 | 3297.414171 | 25 | 688452.7501 | | |
| 13 | 3864.018903 | 26 | 1078765.379 | | |

While clearly unphysical, these parameters indicate a position close to a local minimum on the parameter surface and are an indication of the significant deficiencies of the MNDO model in modelling charged or radical species.

Notably, ammonia is predicted to be planar using these parameters; as the nonplanarity of ammonia was not a constraint imposed upon the system for parameterization, this is not an erroneous result. A detailed tabulation of the predicted molecular properties at the given parameters is provided in the Supplementary Information.

**Conclusion**

In this paper, we report a fully analytic differentiation routine for the evaluation of parameter derivatives for construction of the parameter gradient and Hessian in MNDO-based semiempirical methods and have applied these equations for a proof-of-concept optimization of the MNDO parameters for the elements C, H, N, O and F. While the PM7 Hessian $^{PM7}\mathbf{H}$ appears to work remarkably well in parameterization schemes, it does not guarantee that the identified stationary point on the parameter surface will be a local minimum; furthermore, $^{PM7}\mathbf{H}$ does appear to perform worse when the optimization nears a local minimum or stationary point. The full Hessian **H** may be necessary for optimization near the stationary point, and the accurate eigenvalue information may also be helpful in ensuring that a local minimum is attained at the termination of parameterization.

**References**

[1] M. J. S. Dewar, W. Thiel, "Ground States of Molecules. 38. The MNDO Method. Approximations and Parameters", *J. Am. Chem. Soc*. 1977, 99, 4899–4907.

[2] Dewar, M. J.; Zoebisch, E. G.; Healy, E. F.; Stewart, J. J. Development and Use of Quantum Mechanical Molecular Models. 76. AM1: A New General Purpose Quantum Mechanical Molecular Model. *Journal of the American Chemical Society* 1985, *107*(13), 3902–3909.

[3] Rocha, G. B.; Freire, R. O.; Simas, A. M.; Stewart, J. J. RM1: A Reparameterization of AM1 for H, C, N, O, P, S, F, Cl, Br, and I. *Journal of Computational Chemistry* 2006, *27*, 1101–1111.

[4] Thiel, W.; Voityuk, A. A. Extension of the MNDO Formalism To d Orbitals: Integral Approximations and Preliminary Numerical Results. *Theoretica Chimica Acta* 1992, *81*, 391–404.

[5] J. J. P. Stewart, "Optimization of Parameters for Semiempirical Methods I. Method", *J. Comput. Chem*. 1989, 10, 209–220.

[6] Stewart, J. J. Optimization of Parameters for Semiempirical Methods II. Applications. *Journal of Computational Chemistry* 1989, *10*, 221–264.

[7] Stewart, J. J. Optimization of Parameters for Semiempirical Methods. III Extension of PM3 to Be, Mg, Zn, Ga, Ge, as, SE, CD, in, Sn, Sb, Te, Hg, Tl, Pb, and Bi. *Journal of Computational Chemistry* 1991, *12*, 320–341.

[8] Stewart, J. J. Optimization of Parameters for Semiempirical Methods IV: Extension of MNDO, AM1, and PM3 to More Main Group Elements. *Journal of Molecular Modeling* 2004, *10*, 155–164.

[9] J. J. P. Stewart, "Optimization of Parameters for Semiempirical Methods V: Modification of NDDO Approximations and Application to 70 Elements", *J. Mol. Model.* 2007, 13, 1173–1213.

[10] Stewart, J. Optimization Of Parameters For Semiempirical Methods VI: More Modifications To The NDDO Approximations And Re-Optimization Of Parameters. *Journal of Molecular Modeling* 2012, *19* (1), 1-32

[11] Repasky, M. P.; Chandrasekhar, J.; Jorgensen, W. L. PDDG/PM3 and PDDG/MNDO: Improved Semiempirical Methods. *Journal of Computational Chemistry* 2002, *23*, 1601–1622.

[12] Husch, T.; Vaucher, A. C.; Reiher, M. Semiempirical Molecular Orbital Models Based on the Neglect of Diatomic Differential Overlap Approximation. *International Journal of Quantum Chemistry* 2018, *118*.

[13] Kolb, M., Thiel, W. "Beyond the MNDO Model: Methodical Considerations and Numerical Results", *J. Comput. Chem.* 1993, 14, 775–789.

[14] Weber, W., Thiel, W. "Orthogonalization Corrections for Semiempirical Methods", *Theor. Chem. Acc.* 2000, 103, 495–506.

[15] Scholten, M. *Semiemirische Verfahren mit Orthogonalisierungskorrek- turen: Die OM3 Methode*, Thesis, Heinrich-Heine-Universität Düsseldorf, 2003.

[16] Dral, P. O.; Wu, X.; Spörkel, L.; Koslowski, A.; Weber, W.; Steiger, R.; Scholten, M.; Thiel, W. Semiempirical Quantum-Chemical Orthogonalization-Corrected Methods: Theory, Implementation, and Parameters. *Journal of Chemical Theory and Computation* 2016, *12*, 1082–1096.

[17] Sattelmeyer, K. W.; Tubert-Brohman, I.; Jorgensen, W. L. No-MNDO: Reintroduction of the Overlap Matrix into MNDO. *Journal of Chemical Theory and Computation* 2006, *2*, 413–419.

[18] M. Goeppert-Mayer, A. L. Sklar, "Calculations of the Lower Excited Levels of Benzene", *J. Chem. Phys.* 1938, 6, 645–652.

[19] M. J. S., Dewar, W. Thiel, "A Semiempirical Model for the Two-Center Repulsion Integrals in the NDDO Approximation", *Theor. Chim. Acta* 1976, 46, 89–104.

[20] P. Pulay, "Ab initio calculation of force constants and equilibrium geometries in polyatomic molecules", *Mol. Physics*, 1969, 17

[21] S. Patchokovskii, W. Thiel, "Analytical Second Derivatives of the Energy in MNDO Methods", *Journal of Computational Chemistry*, 1996, 17, 11, 1318-1327

[22] Frisch, M.; Scalmani, G.; Vreven, T.; Zheng, G. Analytic Second Derivatives for Semiempirical Models Based on MNDO. *Molecular Physics* 2009, *107*, 881–887.

[23] Y. Osamura, Y. Yamaguchi, H.F. Schaefer, "Second-Order Coupled Perturbed Hartree-Fock Equations for Closed-Shell and Open-Shell Self-Consistent-Field Wavefunctions", *Chem. Phys.* 1986, 103, 227-243

[24] J. A. Pople, R. Krishnan, H. B. Schelegel, J. S. Binkley, "Derivative studies in Hartree-Fock and Møller-Plesset Theories", *Intl J. Quantum. Chem.*, 1979, 13, 225

[25] Stewart, J. J. Mopac: A Semiempirical Molecular Orbital Program. *Journal of Computer-Aided Molecular Design* 1990, *4*, 1–103.

# An Improved Parameterization Procedure for NDDO-Descendant Semiempirical Methods: Supplementary Information

**Derivation of the First- and Second-Order CPHF Equations**

All equations presented in this section is based on the Unrestricted Hartree-Fock (UHF); equations for the restricted case are obtained as a special case of these results. Matrices for a specific spin (e.g. the Fock matrices for alpha- and beta-spin) are denoted as $^\sigma\mathbf{M}$, where $\sigma \in \{\alpha, \beta\}$ is an arbitrary spin.

In matrix form, the first-order orbital coefficients $^\sigma\mathbf{x}^{q_1}$ are related to the direct derivative of the (AO) coefficient matrix $^\sigma\mathbf{C}$ as follows:

$$\frac{d\,^\sigma\mathbf{C}}{dq_1} = {}^\sigma\mathbf{C}\,^\sigma\mathbf{x}^{q_1} \tag{S1}$$

The second-order orbital coefficients can be obtained via differentiation of the first-order orbital coefficients:

$$\frac{d\,^\sigma\mathbf{x}^{q_1}}{dq_2} = {}^\sigma\boldsymbol{\gamma}^{q_1 q_2} - {}^\sigma\mathbf{x}^{q_2}\,{}^\sigma\mathbf{x}^{q_1} \tag{S2}$$

Application of the orthonormality condition $^\sigma\mathbf{C}^\mathrm{T}\,^\sigma\mathbf{C} = \mathbf{1}$ yields the relevant commutator relations for the first- and second-order orbital coefficients:

$$^\sigma x_{ij}^{q_1} + {}^\sigma x_{ji}^{q_1} = 0 \tag{S3}$$

$$^\sigma \gamma_{ij}^{q_1 q_2} + {}^\sigma \gamma_{ji}^{q_1 q_2} = \sum_k \left( {}^\sigma x_{ik}^{q_1}\,{}^\sigma x_{kj}^{q_2} + {}^\sigma x_{ik}^{q_2}\,{}^\sigma x_{kj}^{q_1} \right) \tag{S4}$$

The derivatives of the off-diagonal elements of the MO-basis Fock matrix must be zero due to the variational condition:

$$^\sigma\mathfrak{F} = {}^\sigma\mathbf{C}^\mathrm{T}\,^\sigma\mathbf{F}\,^\sigma\mathbf{C} \tag{S5}$$

$$\left(\frac{d\,^\sigma\mathfrak{F}}{dq_1}\right)_{ij} = 0, \qquad \left(\frac{d^2\,^\sigma\mathfrak{F}}{dq_1 dq_2}\right)_{ij} = 0, \qquad i \neq j \tag{S6}$$

The first-order CPHF equations are well-documented in the literature and are hence only presented for completeness.

Evaluation of the first derivative $\frac{d^\sigma\mathfrak{F}}{dq_1}$ yields the following:

$$\frac{d^\sigma\mathfrak{F}}{dq_1} = (^\sigma\mathbf{x}^{q_1})^{\mathrm{T}}{}^\sigma\mathfrak{F} + {}^\sigma\mathbf{C}^{\mathrm{T}}\frac{d^\sigma\mathbf{F}}{dq_1}{}^\sigma\mathbf{C} + {}^\sigma\mathfrak{F}{}^\sigma\mathbf{x}^{q_1} \tag{S7}$$

$$\frac{d^\sigma\mathbf{F}}{dq_1} = \frac{d\mathbf{H}}{dq_1} + \mathbf{J}^{q_1}(\mathbf{P}) - \mathbf{K}^{q_1}({}^\sigma\mathbf{P}) + \mathbf{J}\left(\frac{d\mathbf{P}}{dq_1}\right) - \mathbf{K}\left(\frac{d^\sigma\mathbf{P}}{dq_1}\right) \tag{S8}$$

The "static" derivative term $\frac{d\mathbf{H}}{dq_1} + \mathbf{J}^{q_1}(\mathbf{P}) - \mathbf{K}^{q_1}({}^\sigma\mathbf{P})$ is referred to as ${}^\sigma\mathbf{F}^{q_1}$ and the response term $\mathbf{J}\left(\frac{d\mathbf{P}}{dq_1}\right) - \mathbf{K}\left(\frac{d^\sigma\mathbf{P}}{dq_1}\right)$ is referred to as ${}^\sigma\mathbf{R}^{q_1}$; collection of terms yields:

$$\begin{aligned}\frac{d^\sigma\mathfrak{F}}{dq_1} &= (^\sigma\mathbf{x}^{q_1})^{\mathrm{T}}{}^\sigma\mathfrak{F} + {}^\sigma\mathbf{C}^{\mathrm{T}}({}^\sigma\mathbf{F}^{q_1} + {}^\sigma\mathbf{R}^{q_1}){}^\sigma\mathbf{C} + {}^\sigma\mathfrak{F}{}^\sigma\mathbf{x}^{q_1} \\ &= (^\sigma\mathbf{x}^{q_1})^{\mathrm{T}}{}^\sigma\mathfrak{F} + {}^\sigma\mathfrak{F}^{q_1} + {}^\sigma\mathfrak{R}^{q_1} + {}^\sigma\mathfrak{F}{}^\sigma\mathbf{x}^{q_1}\end{aligned} \tag{S9}$$

Thus, the first-order CPHF equations are given as:

$$\left(\frac{d^\sigma\mathfrak{F}}{dq_1}\right)_{ij} = {}^\sigma x_{ji}^{q_1}{}^\sigma\epsilon_j + {}^\sigma x_{ij}^{q_1}{}^\sigma\epsilon_i + {}^\sigma\mathfrak{F}_{ij}^{q_1} + {}^\sigma\mathfrak{R}_{ij}^{q_1} = 0, \qquad i \neq j \tag{S10}$$

$$\left({}^\sigma\epsilon_j - {}^\sigma\epsilon_i\right){}^\sigma x_{ij}^{q_1} - {}^\sigma\mathfrak{R}_{ij}^{q_1} = {}^\sigma\mathfrak{F}_{ij}^{q_1} \tag{S11}$$

To evaluate ${}^\sigma\mathfrak{R}_{ij}^{q_1}, \frac{\partial^\sigma\mathbf{P}}{\partial q_1}$ must be cast in terms of ${}^\sigma\mathbf{x}^{q_1}$; since ${}^\sigma\mathbf{P} = {}^\sigma\mathbf{C}_{occ}^{\mathrm{T}}{}^\sigma\mathbf{C}_{occ}$,

$$\begin{aligned}\left(\frac{d^\sigma\mathbf{P}}{dq_1}\right)_{\mu\nu} &= \sum_{i\in occ.}\sum_{j}\left({}^\sigma c_{\mu j}{}^\sigma c_{\nu i} + {}^\sigma c_{\mu i}{}^\sigma c_{\nu j}\right){}^\sigma x_{ji}^{q_1} \\ &= -\sum_{i\in occ.}\sum_{j\in virt.}\left({}^\sigma c_{\mu j}{}^\sigma c_{\nu i} + {}^\sigma c_{\mu i}{}^\sigma c_{\nu j}\right)x_{ij}^{q_1}\end{aligned} \tag{S12}$$

The first-order CPHF equations are hence linear in the first-order coefficients $x_{ij}^{q_1}$.

Direct differentiation of the first derivative and collection of terms yields

$$\frac{d^2{}^\sigma\mathfrak{F}}{dq_1 dq_2} = \left(\frac{d^\sigma\mathbf{x}^{q_1}}{dq_2}\right)^{\mathrm{T}} {}^\sigma\mathfrak{F} + ({}^\sigma\mathbf{x}^{q_1})^{\mathrm{T}} \frac{d^\sigma\mathfrak{F}}{dq_2} + ({}^\sigma\mathbf{x}^{q_2})^{\mathrm{T}}({}^\sigma\mathfrak{F}^{q_1} + {}^\sigma\mathfrak{R}^{q_1})$$
$$+ {}^\sigma\mathbf{C}^{\mathrm{T}} \frac{d^2{}^\sigma\mathbf{F}}{dq_1 dq_2} {}^\sigma\mathbf{C} + ({}^\sigma\mathfrak{F}^{q_1} + {}^\sigma\mathfrak{R}^{q_1}){}^\sigma\mathbf{x}^{q_2} + \frac{d^\sigma\mathfrak{F}}{dq_2}{}^\sigma\mathbf{x}^{q_1} + {}^\sigma\mathfrak{F}\frac{d^\sigma\mathbf{x}^{q_1}}{dq_2} \quad (S13)$$

Simplifying, this reduces to:

$$\left(\frac{d^2{}^\sigma\mathfrak{F}}{dq_1 dq_2}\right)_{ij} = {}^\sigma\mathbf{C}^{\mathrm{T}} \frac{d^2{}^\sigma\mathbf{F}}{dq_1 dq_2} {}^\sigma\mathbf{C} + {}^\sigma\gamma_{ji}^{q_1 q_2}{}^\sigma\epsilon_j$$
$$+ (({}^\sigma\mathbf{x}^{q_1})^{\mathrm{T}}({}^\sigma\mathfrak{F}^{q_2} + {}^\sigma\mathfrak{R}^{q_2}) + ({}^\sigma\mathbf{x}^{q_1})^{\mathrm{T}}{}^\sigma\boldsymbol{\epsilon}{}^\sigma\mathbf{x}^{q_2} + ({}^\sigma\mathfrak{F}^{q_2} + {}^\sigma\mathfrak{R}^{q_2}){}^\sigma\mathbf{x}^{q_1})_{ij} + {}^\sigma\gamma_{ij}^{q_1 q_2}{}^\sigma\epsilon_i$$
$$+ (({}^\sigma\mathbf{x}^{q_2})^{\mathrm{T}}({}^\sigma\mathfrak{F}^{q_1} + {}^\sigma\mathfrak{R}^{q_2}) + ({}^\sigma\mathbf{x}^{q_2})^{\mathrm{T}}{}^\sigma\boldsymbol{\epsilon}{}^\sigma\mathbf{x}^{q_1} + ({}^\sigma\mathfrak{F}^{q_1} + {}^\sigma\mathfrak{R}^{q_1}){}^\sigma\mathbf{x}^{q_2})_{ij} \quad (S14)$$

Application of the commutator relation then yields:

$$\left(\frac{d^2{}^\sigma\mathfrak{F}}{dq_1 dq_2}\right)_{ij} = {}^\sigma\mathbf{C}^{\mathrm{T}} \frac{d^2{}^\sigma\mathbf{F}}{dq_1 dq_2} {}^\sigma\mathbf{C} - ({}^\sigma\epsilon_j - {}^\sigma\epsilon_i){}^\sigma\gamma_{ij}^{q_1 q_2} + (({}^\sigma\mathbf{x}^{q_1}{}^\sigma\mathbf{x}^{q_2} + {}^\sigma\mathbf{x}^{q_2}{}^\sigma\mathbf{x}^{q_1}){}^\sigma\boldsymbol{\epsilon})_{ij}$$
$$+ (({}^\sigma\mathbf{x}^{q_1})^{\mathrm{T}}({}^\sigma\mathfrak{F}^{q_2} + {}^\sigma\mathfrak{R}^{q_2}) + ({}^\sigma\mathbf{x}^{q_1})^{\mathrm{T}}{}^\sigma\boldsymbol{\epsilon}{}^\sigma\mathbf{x}^{q_2} + ({}^\sigma\mathfrak{F}^{q_2} + {}^\sigma\mathfrak{R}^{q_2}){}^\sigma\mathbf{x}^{q_1})_{ij}$$
$$+ (({}^\sigma\mathbf{x}^{q_2})^{\mathrm{T}}({}^\sigma\mathfrak{F}^{q_1} + {}^\sigma\mathfrak{R}^{q_2}) + ({}^\sigma\mathbf{x}^{q_2})^{\mathrm{T}}{}^\sigma\boldsymbol{\epsilon}{}^\sigma\mathbf{x}^{q_1} + ({}^\sigma\mathfrak{F}^{q_1} + {}^\sigma\mathfrak{R}^{q_1}){}^\sigma\mathbf{x}^{q_2})_{ij} \quad (S15)$$

Evaluation of $\frac{d^2{}^\sigma\mathbf{F}}{dq_1 dq_2}$ is a bit more complicated:

$$\frac{d^2{}^\sigma\mathbf{F}}{dq_1 dq_2} = \frac{d^2\mathbf{H}}{dq_1 dq_2} + \mathbf{J}^{q_1 q_2}(\mathbf{P}) + \mathbf{J}^{q_1}\left(\frac{d\mathbf{P}}{dq_2}\right) + \mathbf{J}^{q_2}\left(\frac{d\mathbf{P}}{dq_1}\right) - \mathbf{K}^{q_1 q_2}({}^\sigma\mathbf{P})$$
$$- \mathbf{K}^{q_1}\left(\frac{d^\sigma\mathbf{P}}{dq_2}\right) - \mathbf{K}^{q_2}\left(\frac{d^\sigma\mathbf{P}}{dq_1}\right) + \mathbf{J}\left(\frac{d^2\mathbf{P}}{dq_1 dq_2}\right) - \mathbf{K}\left(\frac{d^2{}^\sigma\mathbf{P}}{dq_1 dq_2}\right) \quad (S16)$$

Lastly, direct differentiation of the density matrix yields:

$$\left(\frac{d^{2\,\sigma}\mathbf{P}}{dq_1 dq_2}\right)_{\mu\nu} = \sum_{i\in occ.}\sum_{j,k}\left(^\sigma c_{\mu k}{}^\sigma c_{\nu i} + {}^\sigma c_{\mu i}{}^\sigma c_{\nu k}\right)\left(^\sigma x_{kj}^{q_2}{}^\sigma x_{ji}^{q_1} + {}^\sigma x_{kj}^{q_1}{}^\sigma x_{ji}^{q_2}\right)$$
$$+ \sum_{i\in occ.}\sum_{j\in occ.}\sum_{k} {}^\sigma c_{\mu i}{}^\sigma c_{\nu j}\left(^\sigma x_{ik}^{q_1}{}^\sigma x_{kj}^{q_2} + {}^\sigma x_{ik}^{q_2}{}^\sigma x_{kj}^{q_1}\right)$$
$$+ \sum_{i\in occ.}\sum_{j,k}\left(^\sigma c_{\mu j}{}^\sigma c_{\nu k} + {}^\sigma c_{\mu k}{}^\sigma c_{\nu j}\right){}^\sigma x_{ki}^{q_1}{}^\sigma x_{ji}^{q_2}$$
$$- \sum_{i\in occ.}\sum_{j\in virt.}\left(^\sigma c_{\mu j}{}^\sigma c_{\nu i} + {}^\sigma c_{\mu i}{}^\sigma c_{\nu j}\right){}^\sigma \gamma_{ij}^{q_1 q_2} \tag{S17}$$

Since the first three terms in $\left(\frac{d^{2\,\sigma}\mathbf{P}}{dq_1 dq_2}\right)_{\mu\nu}$ are independent of the second-order coefficients, we define

$$^\sigma \rho_{\mu\nu}^{q_1 q_2} = -\sum_{i\in occ.}\sum_{j\in virt.}\left(^\sigma c_{\mu j}{}^\sigma c_{\nu i} + {}^\sigma c_{\mu i}{}^\sigma c_{\nu j}\right){}^\sigma \gamma_{ij}^{q_1 q_2} \tag{S18}$$

$$\frac{d^{2\,\sigma}\mathbf{P}}{dq_1 dq_2} = {}^\sigma\rho^{q_1 q_2} + {}^\sigma\varsigma^{q_1 q_2} \tag{S19}$$

The static derivative term in $\frac{d^{2\,\sigma}\mathbf{F}}{dq_1 dq_2}$ is hence termed $^\sigma\mathbf{F}^{q_1 q_2}$, with the response term correspondingly referred to as $^\sigma\mathbf{R}^{q_1 q_2}$:

$$^\sigma\mathbf{F}^{q_1 q_2} = \frac{d^2 \mathbf{H}}{dq_1 dq_2} + \mathbf{J}^{q_1 q_2}(\mathbf{P}) + \mathbf{J}^{q_1}\left(\frac{d\mathbf{P}}{dq_2}\right) + \mathbf{J}^{q_2}\left(\frac{d\mathbf{P}}{dq_1}\right)$$
$$-\mathbf{K}^{q_1 q_2}(^\sigma\mathbf{P}) - \mathbf{K}^{q_1}\left(\frac{d^\sigma\mathbf{P}}{dq_2}\right) - \mathbf{K}^{q_2}\left(\frac{d^\sigma\mathbf{P}}{dq_1}\right) + \mathbf{J}(\varsigma^{q_1 q_2}) - \mathbf{K}(^\sigma\varsigma^{q_1 q_2}) \tag{S20}$$

$$^\sigma\mathbf{R}^{q_1 q_2} = \mathbf{J}(\rho^{q_1 q_2}) - \mathbf{K}(^\sigma\rho^{q_1 q_2}) \tag{S21}$$

Thus, by defining $^\sigma\mathfrak{F}^{q_1 q_2}$ and $^\sigma\mathfrak{R}^{q_1 q_2}$, we obtain the second-order CPHF equations:

$$^\sigma\mathfrak{F}^{q_1 q_2} = {}^\sigma\mathbf{C}^{\mathrm{T}\,\sigma}\mathbf{F}^{q_1 q_2\,\sigma}\mathbf{C} + \left(^\sigma\mathbf{x}^{q_1\,\sigma}\mathbf{x}^{q_2} + {}^\sigma\mathbf{x}^{q_2\,\sigma}\mathbf{x}^{q_1}\right){}^\sigma\boldsymbol{\epsilon}$$
$$+(^\sigma\mathbf{x}^{q_1})^{\mathrm{T}}(^\sigma\mathfrak{F}^{q_2} + {}^\sigma\mathfrak{R}^{q_2}) + (^\sigma\mathbf{x}^{q_1})^{\mathrm{T}\,\sigma}\boldsymbol{\epsilon}{}^\sigma\mathbf{x}^{q_2} + (^\sigma\mathfrak{F}^{q_2} + {}^\sigma\mathfrak{R}^{q_2}){}^\sigma\mathbf{x}^{q_1}$$
$$+(^\sigma\mathbf{x}^{q_2})^{\mathrm{T}}(^\sigma\mathfrak{F}^{q_1} + {}^\sigma\mathfrak{R}^{q_2}) + (^\sigma\mathbf{x}^{q_2})^{\mathrm{T}\,\sigma}\boldsymbol{\epsilon}{}^\sigma\mathbf{x}^{q_1} + (^\sigma\mathfrak{F}^{q_1} + {}^\sigma\mathfrak{R}^{q_1}){}^\sigma\mathbf{x}^{q_2} \tag{S22}$$

$$^\sigma\mathfrak{R}^{q_1 q_2} = {}^\sigma\mathbf{C}^{\mathrm{T}\,\sigma}\mathbf{R}^{q_1 q_2\,\sigma}\mathbf{C} \tag{S23}$$

$$\left(^\sigma\epsilon_j - {}^\sigma\epsilon_i\right){}^\sigma\gamma_{ij}^{q_1 q_2} - {}^\sigma\mathfrak{R}_{ij}^{q_1 q_2} = {}^\sigma\mathfrak{F}_{ij}^{q_1 q_2} \tag{S24}$$

## First Derivatives of Heats of Formation

First derivatives of heats of formation do not require any first-order response terms; differentiation of the expression for $\Delta H_f$ yields the compact equation that

$$\frac{d(\Delta H_f)}{d^{Z_A}p} = k_{conv}\left(\frac{dE_{el}}{d^{Z_A}p} + \frac{dV_{core}}{d^{Z_A}p} - \sum_A \frac{d^{Z_A}E_{eisol}}{d^{Z_A}p}\right) \tag{S25}$$

The value $k_{conv}$ represents the conversion factor between the various terms (evaluated in eV) and the value $\Delta H_f$ (in kcal/mol).

Thus, the derivatives $\frac{d(\Delta H_f)}{d^{Z_A}\alpha}$ and $\frac{d(\Delta H_f)}{d^{Z_A}E_{eisol}}$ are easily evaluated:

$$\frac{d(\Delta H_f)}{d^{Z_A}\alpha} = k_{conv}\sum_{C>B}\frac{dV_{BC}^{CRF}}{d^{Z_A}\alpha}, \qquad \frac{d(\Delta H_f)}{d^{Z_A}E_{eisol}} = -k_{conv}n_{Z_A} \tag{S26}$$

In the above expression, $n_{Z_A}$ represents the stoichiometric ratio for the element in question (e.g. $n_{Z_6} = 6$ in cyclohexane, $C_6H_{12}$).

The remaining derivatives (against $^{Z_A}\beta_s$, $^{Z_A}\beta_p$, $^{Z_A}U_{ss}$, $^{Z_A}U_{pp}$, $^{Z_A}\zeta_s$, $^{Z_A}\zeta_p$) are significantly more complex, and requires particular attention:

$$\frac{d(\Delta H_f)}{d^{Z_A}\beta_x} = 2k_{conv}\sum_{\substack{B\\C>B}}\sum_{\substack{\mu\in B\\\nu\in C}}P_{\mu\nu}\frac{d\beta_{\mu\nu}}{d^{Z_A}\beta_x}, \qquad \frac{d\beta_{\mu\nu}}{d^{Z_A}\beta_x} = \frac{\left(\delta_{\beta_\mu}{}^{Z_A}\beta_x + \delta_{\beta_\nu}{}^{Z_A}\beta_x\right)S_{\mu\nu}}{2} \tag{S27a, b}$$

$$\frac{d(\Delta H_f)}{d^{Z_A}U_{xx}} = k_{conv}\sum_\mu P_{\mu\mu}\delta_{U_{\mu\mu}{}^{Z_A}U_{xx}} \tag{S28}$$

Alternatively, one may employ the matrices $\frac{\partial \mathbf{H}}{\partial^{Z_A}p}$, $\mathbf{J}^{Z_Ap}(\mathbf{P})$ and $\mathbf{K}^{Z_Ap}(^\sigma\mathbf{P})$ to evaluate $\frac{\partial E_{el}}{\partial^{Z_A}p}$ accordingly, with the final expression given by:

$$\frac{d(E_{el})}{d^{Z_A}p} = k_{conv}\left[\text{Tr}\left(\frac{d\mathbf{H}}{d^{Z_A}p}\mathbf{P}\right) + \frac{1}{2}\text{Tr}\left(\mathbf{J}^{Z_Ap}(\mathbf{P})\mathbf{P} - \mathbf{K}^{Z_Ap}(^\alpha\mathbf{P})^\alpha\mathbf{P} - \mathbf{K}^{Z_Ap}(^\beta\mathbf{P})^\beta\mathbf{P}\right)\right] \tag{S29}$$

The expressions for such an approach are given by:

$$\frac{dH_{\mu\nu}}{d^{Z_A}\beta_x} = \begin{cases} \frac{d\beta_{\mu\nu}}{d^{Z_A}\beta_x}, & \mu \in C, \nu \in B \neq C \\ 0, & \text{otherwise} \end{cases} \quad (S30)$$

$$\frac{dH_{\mu\nu}}{d^{Z_A}U_{xx}} = \begin{cases} \delta_{U_{\mu\mu}}{}^{Z_A}U_{xx}, & \mu = \nu \\ 0, & \text{otherwise} \end{cases} \quad (S31)$$

$$\mathbf{J}^{Z_A\beta_x}(\mathbf{P}) = \mathbf{J}^{Z_AU_{xx}}(\mathbf{P}) = \mathbf{K}^{Z_A\beta_x}({}^\sigma\mathbf{P}) = \mathbf{K}^{Z_AU_{xx}}({}^\sigma\mathbf{P}) = \mathbf{0} \quad (S32)$$

$$\frac{dH_{\mu\nu}}{d^{Z_A}\zeta_x} = \begin{cases} \sum_{C \neq B} \frac{dV_{\mu\nu,C}}{d^{Z_A}\zeta_x}, & \mu, \nu \in B \\ \frac{d\beta_{\mu\nu}}{d^{Z_A}\zeta_x}, & \mu \in C, \nu \in B \neq C \end{cases} \quad (S33)$$

In the above equations,

$$\frac{dV_{\mu\nu,C}}{d^{Z_A}\zeta_x} = -Q_A Q_B \frac{d(\mu\nu|s^C s^C)}{d^{Z_A}\zeta_x}, \quad \frac{d\beta_{\mu\nu}}{d^{Z_A}\zeta_x} = \frac{\beta_\mu + \beta_\nu}{2} \frac{dS_{\mu\nu}}{d^{Z_A}\zeta_x} \quad (S34a, b)$$

$$J_{\mu\nu}^{Z_A\zeta_x}(\mathbf{P}) = \begin{cases} \sum_{\lambda,\sigma \in C} P_{\lambda\sigma} \frac{d(\mu\nu|\lambda\sigma)}{d^{Z_A}\zeta_x}, & \mu, \nu \in B \\ 0, & \mu \in B, \nu \in C \neq B \end{cases} \quad (S35)$$

$$K_{\mu\nu}^{Z_A\zeta_x}({}^\sigma\mathbf{P}) = \begin{cases} 0, & \mu, \nu \in B \\ \sum_{\lambda \in B} \sum_{\sigma \in C} {}^\sigma P_{\lambda\sigma} \frac{d(\mu\lambda|\nu\sigma)}{d^{Z_A}\zeta_x}, & \mu \in B, \nu \in C \neq B \end{cases} \quad (S36)$$

## Second Derivatives of Heats of Formation

Further differentiation of the expression for $\Delta H_f$ yields

$$\frac{\partial^2}{\partial^{Z_A}p_i \partial^{Z_B}p_j}(\Delta H_f) = k_{conv}\left(\frac{\partial^2 E_{el}}{\partial^{Z_A}p_i \partial^{Z_B}p_j} + \frac{\partial^2 V_{core}}{\partial^{Z_A}p_i \partial^{Z_B}p_j} - \sum_A \frac{\partial^{2\,Z_A} E_{eisol}}{\partial^{Z_A}p_i \partial^{Z_B}p_j}\right) \quad (S37)$$

Thus, the derivative $\frac{\partial^2(\Delta H_f)}{\partial^{Z_A}\alpha \partial^{Z_B}\alpha}$ is easily evaluated:

$$\frac{\partial^2(\Delta H_f)}{\partial^{Z_A}\alpha \partial^{Z_B}\alpha} = k_{conv} \sum_{D>C} \frac{\partial^2 V_{CD}^{CRF}}{\partial^{Z_A}\alpha \partial^{Z_B}\alpha} \quad (S38)$$

Furthermore,

$$\frac{\partial^2(\Delta H_f)}{\partial^{Z_A}\alpha \partial^{Z_B}p_i} = 0 \;\forall\; {}^{Z_B}p_i \neq {}^{Z_B}\alpha \quad (S39)$$

$$\frac{\partial^2(\Delta H_f)}{\partial^{Z_A}E_{eisol} \partial^{Z_B}p_i} = 0 \;\forall\; {}^{Z_B}p_i \quad (S40)$$

Most static second derivatives of the heats of formation are zero. The non-zero static second derivative matrices are given below:

$$\frac{\partial^2 H_{\mu\nu}}{\partial^{Z_A}\beta_x \partial^{Z_B}\zeta_x} = \begin{cases} \dfrac{\partial^2 \beta_{\mu\nu}}{\partial^{Z_A}\beta_x \partial^{Z_B}\zeta_x}, & \mu \in C, \nu \in D \neq C \\ 0, & \text{otherwise} \end{cases} \quad (S41)$$

$$\frac{\partial^2 H_{\mu\nu}}{\partial^{Z_A}\zeta_{x_1} \partial^{Z_B}\zeta_{x_2}} = \begin{cases} \displaystyle\sum_{D \neq C} \dfrac{\partial^2 V_{\mu\nu,D}}{\partial^{Z_A}\zeta_x \partial^{Z_B}\zeta_x}, & \mu, \nu \in C \\ \dfrac{\partial^2 \beta_{\mu\nu}}{\partial^{Z_A}\zeta_{x_1} \partial^{Z_B}\zeta_{x_2}}, & \mu \in C, \nu \in D \neq C \end{cases} \quad (S42)$$

In the above,

$$\frac{\partial^2 \beta_{\mu\nu}}{\partial^{Z_A}\beta_x \partial^{Z_B}\zeta_x} = \frac{\left(\delta_{\beta_\mu}{}^{Z_A}\beta_x + \delta_{\beta_\nu}{}^{Z_A}\beta_x\right)}{2} \frac{\partial S_{\mu\nu}}{\partial^{Z_B}\zeta_x} \quad (S43a)$$

$$\frac{\partial^2 V_{\mu\nu,D}}{\partial^{Z_A}\zeta_x \partial^{Z_B}\zeta_x} = -Q_C Q_D \frac{\partial^2 (\mu\nu|s^D s^D)}{\partial^{Z_A}\zeta_x \partial^{Z_B}\zeta_x} \quad (S43b)$$

Lastly, the nonzero derivatives of the two-electron matrices are given by

$$J_{\mu\nu}^{Z_A\zeta_{x_1} Z_B\zeta_{x_2}}(\mathbf{P}) = \begin{cases} \sum\limits_{\lambda,\sigma \in D} P_{\lambda\sigma} \dfrac{d^2(\mu\nu|\lambda\sigma)}{d^{Z_A}\zeta_{x_1} \partial^{Z_B}\zeta_{x_2}}, & \mu,\nu \in C \\ \\ 0, & \mu \in C, \nu \in D \neq C \end{cases} \tag{S44}$$

$$K_{\mu\nu}^{Z_A\zeta_{x_1} Z_B\zeta_{x_2}}(^{\sigma}\mathbf{P}) = \begin{cases} 0, & \mu,\nu \in C \\ \\ \sum\limits_{\lambda \in C}\sum\limits_{\sigma \in D} {}^{\sigma}P_{\lambda\sigma} \dfrac{d^2(\mu\lambda|\nu\sigma)}{d^{Z_A}\zeta_{x_1} \partial^{Z_B}\zeta_{x_2}}, & \mu \in C, \nu \in D \neq C \end{cases} \tag{S45}$$

The second derivatives of the heat of formation are evaluated using the expression:

$$\frac{\partial^2(\Delta H_f)}{\partial^{Z_A}p_i \partial^{Z_B}p_j} = k_{conv}\mathrm{Tr}\left(\frac{d^2\mathbf{H}}{d^{Z_A}p_i d^{Z_B}p_j}\mathbf{P} + \frac{d\mathbf{H}}{d^{Z_A}p_i}\frac{d\mathbf{P}}{d^{Z_B}p_j}\right)$$
$$+ \frac{1}{2}k_{conv}\mathrm{Tr}\left(\mathbf{J}^{Z_Ap_i Z_Bp_j}(\mathbf{P})\mathbf{P} + \mathbf{J}^{Z_Ap}\left(\frac{d\mathbf{P}}{d^{Z_B}p_j}\right)\mathbf{P} + \mathbf{J}^{Z_Ap}(\mathbf{P})\frac{d\mathbf{P}}{d^{Z_B}p_j}\right)$$
$$- \frac{1}{2}k_{conv}\mathrm{Tr}\left(\mathbf{K}^{Z_Ap_i Z_Bp_j}(^{\alpha}\mathbf{P})^{\alpha}\mathbf{P} + \mathbf{K}^{Z_Ap}\left(\frac{d^{\alpha}\mathbf{P}}{d^{Z_B}p_j}\right)^{\alpha}\mathbf{P} + \mathbf{K}^{Z_Ap}(^{\alpha}\mathbf{P})\frac{d^{\alpha}\mathbf{P}}{d^{Z_B}p_j}\right)$$
$$- \frac{1}{2}k_{conv}\mathrm{Tr}\left(\mathbf{K}^{Z_Ap_i Z_Bp_j}(^{\beta}\mathbf{P})^{\beta}\mathbf{P} + \mathbf{K}^{Z_Ap}\left(\frac{d^{\beta}\mathbf{P}}{d^{Z_B}p_j}\right)^{\beta}\mathbf{P} + \mathbf{K}^{Z_Ap}(^{\beta}\mathbf{P})\frac{d^{\beta}\mathbf{P}}{d^{Z_B}p_j}\right) \tag{S46}$$

**First and Second Derivatives of Dipole Moments**

Derivatives of dipole moments are presented for the Restricted Hartree-Fock case, as no extension to UHF systems is necessary.

The semiempirical dipole moment is formally evaluated as the magnitude of a 3-element dipole moment vector:

$$\langle \mu \rangle = |\boldsymbol{\mu}| \tag{S47}$$

$$\mu_\tau = -2 \sum_B {}^{Z_B}D_1 P_{sp_\tau} + \sum_B \tau_{CM,B} \left( Q_A - \sum_{m \in B} P_{mm} \right) \tag{S48}$$

The distance $\tau_{CM,B}$ denotes the projection of the vector $\mathbf{r}_B - \mathbf{r}_{CM}$ onto the axis $\tau(= x, y, z)$; the centre-of-mass position vector $\mathbf{r}_{CM}$ is evaluated using standard empirical measurements for atomic masses.

Differentiation hence yields the first and second derivatives of the dipole moment:

$$\frac{d\langle \mu \rangle}{d^{Z_A}p} = \frac{\boldsymbol{\mu} \cdot \frac{d\boldsymbol{\mu}}{d^{Z_A}p}}{|\boldsymbol{\mu}|} \tag{S49}$$

$$\frac{d\mu_\tau}{d^{Z_A}p} = -2 \sum_B \left( \frac{d^{Z_B}D_1}{d^{Z_A}p} P_{sp_\tau} + {}^{Z_B}D_1 \frac{dP_{sp_\tau}}{d^{Z_A}p} \right) + \sum_B \tau_{CM,B} \left( Q_A - \sum_{m \in B} \frac{dP_{mm}}{d^{Z_A}p} \right) \tag{S50}$$

$$\frac{d^2 \langle \mu \rangle}{d^{Z_A}p_i d^{Z_B}p_j} = \frac{\boldsymbol{\mu} \cdot \frac{d^2 \boldsymbol{\mu}}{d^{Z_A}\zeta_{x_1} d^{Z_B}\zeta_{x_2}} + \frac{d\boldsymbol{\mu}}{d^{Z_B}p_j} \cdot \frac{d\boldsymbol{\mu}}{d^{Z_A}p_i} - \frac{d\langle \mu \rangle}{d^{Z_A}p_i} \frac{d\langle \mu \rangle}{d^{Z_B}p_j}}{|\boldsymbol{\mu}|} \tag{S51}$$

$$\frac{d^2 \mu_\tau}{d^{Z_A}p_i d^{Z_B}p_j} = \sum_B \tau_{CM,B} \left( Q_A - \sum_{m \in B} \frac{d^2 P_{mm}}{d^{Z_A}p_i d^{Z_B}p_j} \right)$$
$$-2 \sum_C \left( \frac{d^{Z_C}D_1}{d^{Z_A}p_i} \frac{dP_{sp_\tau}}{d^{Z_B}p_j} + \frac{d^{2Z_C}D_1}{d^{Z_A}p_i d^{Z_B}p_j} P_{sp_\tau} + \frac{d^{Z_C}D_1}{d^{Z_B}p_j} \frac{dP_{sp_\tau}}{d^{Z_A}p_i} + {}^{Z_C}D_1 \frac{d^2 P_{sp_\tau}}{d^{Z_A}p_i d^{Z_B}p_j} \right) \tag{S52}$$

# First and Second Derivatives of Ionization Energies

First derivatives of the ionization energy can be obtained directly via solution of the CPHF equations:

$$\frac{d^\sigma \boldsymbol{\epsilon}}{d^{Z_A}p} = \frac{d^\sigma \mathbf{C}^{\mathrm{T}}}{d^{Z_A}p}{}^\sigma\mathbf{F}^\sigma\mathbf{C} + {}^\sigma\mathbf{C}^{\mathrm{T}}\frac{d^\sigma \mathbf{F}}{d^{Z_A}p}{}^\sigma\mathbf{C} + {}^\sigma\mathbf{C}^{\mathrm{T}}{}^\sigma\mathbf{F}\frac{\partial^\sigma \mathbf{C}}{\partial^{Z_A}p} \tag{S53}$$

$$\begin{aligned}\frac{d^{2\,\sigma} \boldsymbol{\epsilon}}{d^{Z_A}p_i d^{Z_B}p_j} &= \frac{d^{2\,\sigma}\mathbf{C}^{\mathrm{T}}}{d^{Z_A}p_i d^{Z_B}p_j}{}^\sigma\mathbf{F}^\sigma\mathbf{C} + \frac{d^\sigma\mathbf{C}^{\mathrm{T}}}{d^{Z_A}p_i}\frac{d^\sigma\mathbf{F}}{d^{Z_B}p_j}{}^\sigma\mathbf{C} + \frac{d^\sigma\mathbf{C}^{\mathrm{T}}}{d^{Z_A}p_i}{}^\sigma\mathbf{F}\frac{d^\sigma\mathbf{C}}{d^{Z_B}p_j} \\ &+ \frac{d^\sigma\mathbf{C}^{\mathrm{T}}}{d^{Z_B}p_j}\frac{d^\sigma\mathbf{F}}{d^{Z_A}p_i}{}^\sigma\mathbf{C} + {}^\sigma\mathbf{C}^{\mathrm{T}}\frac{d^\sigma\mathbf{F}}{d^{Z_A}p_i d^{Z_B}p_j}{}^\sigma\mathbf{C} + {}^\sigma\mathbf{C}^{\mathrm{T}}\frac{d^\sigma\mathbf{F}}{d^{Z_A}p_i}\frac{d^\sigma\mathbf{C}}{d^{Z_B}p_j} \\ &+ \frac{d^\sigma\mathbf{C}^{\mathrm{T}}}{d^{Z_B}p_j}{}^\sigma\mathbf{F}\frac{d^\sigma\mathbf{C}}{d^{Z_A}p_i} + {}^\sigma\mathbf{C}^{\mathrm{T}}\frac{d^\sigma\mathbf{F}}{d^{Z_B}p_j}\frac{d^\sigma\mathbf{C}}{d^{Z_A}p_i} + {}^\sigma\mathbf{C}^{\mathrm{T}}{}^\sigma\mathbf{F}\frac{d^{2\,\sigma}\mathbf{C}}{d^{Z_A}p_i d^{Z_B}p_j}\end{aligned} \tag{S54}$$

# First and Second Derivatives of Reference Geometry Gradients

The derivative $\frac{d|\mathbf{g}|}{d^{Z_C}p}$ can be trivially cast in terms of the derivative vector $\frac{d\mathbf{g}}{d^{Z_C}p}$:

$$\frac{d|\mathbf{g}|}{d^{Z_C}p} = \frac{\mathbf{g}\cdot\frac{d\mathbf{g}}{d^{Z_C}p}}{|\mathbf{g}|} \tag{S55}$$

Likewise,

$$\frac{d^2|\mathbf{g}|}{d^{Z_C}p_i d^{Z_D}p_j} = \frac{\frac{d\mathbf{g}}{d^{Z_C}p_i}\cdot\frac{d\mathbf{g}}{d^{Z_D}p_j} + \mathbf{g}\cdot\frac{d^2\mathbf{g}}{d^{Z_C}p_i d^{Z_D}p_j} - \frac{d|\mathbf{g}|}{d^{Z_C}p_i}\frac{d|\mathbf{g}|}{d^{Z_D}p_j}}{|\mathbf{g}|} \tag{S56}$$

The elements of the reference geometry gradient $\mathbf{g}$ are evaluated as follows:

$$\frac{dE}{d^A\tau} = \sum_{B\neq A}\frac{\partial E_{AB}}{\partial^A\tau} \tag{S57}$$

$$\begin{aligned}\frac{\partial E_{AB}}{\partial^A\tau} &= \sum_{\mu,\nu\in A}P_{\mu\nu}\frac{dV_{\mu\nu,B}}{d^A\tau} + \sum_{\lambda,\sigma\in B}P_{\lambda\sigma}\frac{dV_{\lambda\sigma,A}}{d^A\tau} + 2\sum_{\substack{\mu\in A \\ \lambda\in B}}P_{\mu\lambda}\frac{d\beta_{\mu\lambda}}{d^A\tau} \\ &+ \sum_{\substack{\mu,\nu\in A \\ \lambda,\sigma\in B}}\left(P_{\mu\nu}P_{\lambda\sigma} - {}^\alpha P_{\mu\nu}{}^\alpha P_{\lambda\sigma} - {}^\beta P_{\mu\nu}{}^\beta P_{\lambda\sigma}\right)\frac{d(\mu\nu|\lambda\sigma)}{d^A\tau}\end{aligned} \tag{S58}$$

$$\frac{d\beta_{\mu\lambda}}{d^A\tau} = \frac{\beta_\mu + \beta_\lambda}{2}\frac{dS_{\mu\lambda}}{d^A\tau} \tag{S59}$$

Thus, elements of $\frac{d\mathbf{g}}{d^Z c_p}$ are given by:

$$\frac{d^2 E}{d^A \tau d^Z c_p} = \sum_{B \neq A} \frac{d}{d^Z c_p}\left(\frac{\partial E_{AB}}{\partial^A \tau}\right) \tag{S60}$$

$$\frac{d}{d^Z c_p}\left(\frac{\partial E_{AB}}{\partial^A \tau}\right) = \sum_{\mu,\nu \in A}\left(\frac{dP_{\mu\nu}}{d^Z c_p}\frac{dV_{\mu\nu,B}}{d^A \tau} + P_{\mu\nu}\frac{d^2 V_{\mu\nu,B}}{d^A \tau d^Z c_p}\right)$$
$$+ \sum_{\lambda,\sigma \in B}\left(\frac{dP_{\lambda\sigma}}{d^Z c_p}\frac{dV_{\lambda\sigma,A}}{d^A \tau} + P_{\lambda\sigma}\frac{d^2 V_{\lambda\sigma,A}}{d^A \tau d^Z c_p}\right) + 2\sum_{\substack{\mu \in A \\ \lambda \in B}}\left(\frac{dP_{\mu\lambda}}{d^Z c_p}\frac{d\beta_{\mu\lambda}}{d^A \tau} + P_{\mu\lambda}\frac{d^2 \beta_{\mu\lambda}}{d^A \tau d^Z c_p}\right)$$
$$+ \sum_{\substack{\mu,\nu \in A \\ \lambda,\sigma \in B}}\left(P_{\mu\nu}P_{\lambda\sigma} - {}^\alpha P_{\mu\nu}{}^\alpha P_{\lambda\sigma} - {}^\beta P_{\mu\nu}{}^\beta P_{\lambda\sigma}\right)\frac{d^2(\mu\nu|\lambda\sigma)}{d^A \tau d^Z c_p}$$
$$+ \sum_{\substack{\mu,\nu \in A \\ \lambda,\sigma \in B}}\left(\frac{dP_{\mu\nu}}{d^Z c_p}P_{\lambda\sigma} + P_{\mu\nu}\frac{dP_{\lambda\sigma}}{d^Z c_p}\right)\frac{d(\mu\nu|\lambda\sigma)}{d^A \tau}$$
$$- \sum_{\substack{\mu,\nu \in A \\ \lambda,\sigma \in B}}\frac{d(\mu\nu|\lambda\sigma)}{d^A \tau}\sum_{\sigma \in \{\alpha,\beta\}}\left(\frac{d^\alpha P_{\mu\nu}}{d^Z c_p}{}^\alpha P_{\lambda\sigma} + {}^\alpha P_{\mu\nu}\frac{d^\alpha P_{\lambda\sigma}}{d^Z c_p}\right) \tag{S61}$$

Elements of $\frac{d^2 \mathbf{g}}{d^Z c_{p_i} d^Z D_{p_j}}$ are likewise obtained by direct differentiation of the above expression and are omitted for brevity.

## The Trust Region Optimizer

In a quadratic trust region optimiser, we seek to optimise the Lagrangian function

$$\mathcal{L} = \mathbf{g}^T\mathbf{d} + \frac{1}{2}\mathbf{d}^T\mathbf{B}\mathbf{d} + \kappa(\mathbf{d}^T\mathbf{d} - R^2) \tag{S62}$$

Differentiation hence yields an expression for the Hessian shift parameter $\lambda$ as a function of the Lagrange multiplier $\kappa$:

$$\mathbf{d} = -(\mathbf{B} + 2\kappa\mathbf{I})^{-1}\mathbf{g} = -(\mathbf{B} + \lambda\mathbf{I})^{-1}\mathbf{g} \tag{S63}$$

By constraining the step length, we hence obtain an iterative algorithm for $\lambda$:

$$\Phi = 1 - \frac{\sqrt{\mathbf{d}^T\mathbf{d}}}{R} = 0 \tag{S64}$$

$$\lambda_{n+1} = \lambda_n + \frac{\Phi}{\frac{1}{2R\sqrt{\mathbf{d}^T\mathbf{d}}}\frac{d(\mathbf{d}^T\mathbf{d})}{d\lambda}} = \lambda_n + \frac{2(R\sqrt{\mathbf{d}^T\mathbf{d}} - \mathbf{d}^T\mathbf{d})}{\frac{d(\mathbf{d}^T\mathbf{d})}{d\lambda}} \tag{S65a}$$

$$\frac{d(\mathbf{d}^T\mathbf{d})}{d\lambda} = -2\sum_i \frac{(\mathbf{u}_i^T\mathbf{g})^2}{(B_{ii} + \lambda)^3} \tag{S65b}$$

In the above, $\mathbf{u}_i$ represents the eigenvectors of the Hessian $\mathbf{B}$.

# Comparison of Our Program to MOPAC

Calculated Values for $\Delta H_f$ Using the MNDO Formalism (given in kcal/mol)

| Molecular Formula | Molecule Name | MOPAC | Our Program |
|---|---|---|---|
| H | Hydrogen, cation | 326.7 | 326.7 |
| H | Hydrogen, atom | 52.1 | 52.1 |
| H2 | Hydrogen | 0.7 | 0.7 |
| C | Carbon, cation | 389.4 | 389.4 |
| C | Carbon, atom | 170.9 | 170.9 |
| CH | Methylidyne | 143.3 | 143.3 |
| CH2 | Methylene, singlet | 107.4 | 107.4 |
| CH2 | Methylene, triplet | 73.9 | 73.9 |
| CH3 | Methyl, cation | 243.9 | 243.9 |
| CH4 | Methane | -12 | -12.0 |
| C2H2 | Acetylene | 57.9 | 57.9 |
| C2H3 | Vinyl, cation | 265.7 | 265.7 |
| C2H3 | Vinyl | 59 | 59.0 |
| C2H4 | Ethylene, cation | 237.7 | 237.7 |
| C2H4 | Ethylene | 15.4 | 15.4 |
| C2H4 | Methylmethylene | 88.3 | 88.3 |
| C2H5 | Ethyl, cation | 219.6 | 219.6 |
| C2H5 | Ethyl radical | 10.5 | 10.5 |
| C2H6 | Ethane | -19.7 | -19.8 |
| C3 | Carbon, trimer | 220.3 | 220.3 |
| C3H3 | Cyclopropenyl, cation | 272.5 | 272.5 |
| C3H3 | Propynyl, cation | 265.4 | 265.4 |
| C3H4 | Allene | 43.9 | 43.9 |
| C3H4 | Cyclopropene | 68.3 | 68.3 |
| C3H4 | Propyne | 41.4 | 41.4 |
| C3H5 | Allyl, cation | 221.4 | 221.4 |
| C3H5 | Cyclopropyl, cation | 258.1 | 258.1 |
| C3H5 | Propenyl, cation | 240.1 | 240.1 |
| C3H5 | Allyl | 25.3 | 25.3 |
| C3H6 | Cyclopropane | 11.2 | 11.2 |
| C3H6 | Propene | 4.9 | 4.9 |
| C3H7 | i-Propyl radical | -1.6 | -1.6 |
| C3H8 | Propane | -25 | -25.0 |
| C4 | Carbon, tetramer | 271.3 | 271.3 |
| C4H2 | Diacetylene | 103.2 | 103.2 |
| C4H4 | Vinylacetylene | 65.5 | 65.5 |
| C4H4 | Butatriene | 71.2 | 71.2 |
| C4H6 | 1,2-Butadiene | 33.5 | 33.5 |
| C4H6 | 1,3-Butadiene | 28.9 | 28.9 |
| C4H6 | 1-Butyne | 36.1 | 36.1 |

| Formula | Name | Value 1 | Value 2 |
|---|---|---|---|
| C4H6 | 2-Butyne | 24.8 | 24.9 |
| C4H6 | Bicyclobutane | 64 | 64.0 |
| C4H6 | Cyclobutene | 31 | 31.0 |
| C4H6 | Methyl cyclopropene | 53.7 | 53.7 |
| C4H6 | Methylenecyclopropane | 37.8 | 37.8 |
| C4H7 | 2-Butenyl, cation | 206.9 | 206.9 |
| C4H7 | Cyclobutyl, cation | 221.4 | 221.3 |
| C4H8 | 1-Butene | 0.3 | -0.3 |
| C4H8 | cis-2-Butene | -4.4 | -4.4 |
| C4H8 | Cyclobutane | -11.9 | -12.0 |
| C4H8 | Isobutene | -2 | -2.0 |
| C4H8 | trans-2-Butene | -5.1 | -5.1 |
| C4H9 | Isobutyl, cation | 187.9 | 187.3 |
| C4H9 | Isobutyl | -10.1 | -10.1 |
| C4H10 | Isobutane | -26.8 | -26.8 |
| C4H10 | n-Butane, trans | -29.8 | -29.8 |
| C5H5 | Cyclopentadienyl, anion | 18.9 | 18.9 |
| C5H6 | Cyclopentadiene | 32 | 32.0 |
| C5H8 | 1,2-Dimethyl cyclopropene | 39.2 | 39.3 |
| C5H8 | 1,4-Pentadiene | 24.5 | 24.5 |
| C5H8 | 1,cis-3-Pentadiene | 20 | 20.0 |
| C5H8 | 1,trans-3-Pentadiene | 19.1 | 19.1 |
| C5H8 | Bicyclo(2.1.0)-pentane | 30.1 | 30.1 |
| C5H8 | Cyclopentene | -0.4 | -0.4 |
| C5H8 | Isoprene | 23.3 | 23.3 |
| C5H8 | Methylene cyclobutane | 10.8 | 10.8 |
| C5H8 | Spiropentane | 33.6 | 33.6 |
| C5H9 | Cyclopentyl, cation | 194.1 | 194.1 |
| C5H10 | 2-Methyl-2-butene | -10.2 | -10.2 |
| C5H10 | cis-2-Pentene | -8.9 | -8.9 |
| C5H10 | cis-Dimethylcyclopropane | -2.2 | -2.2 |
| C5H10 | Cyclopentane | -30.5 | -30.6 |
| C5H10 | trans-2-Pentene | -10.3 | -10.4 |
| C5H12 | n-Pentane | -34.5 | -34.5 |
| C5H12 | Neopentane | -24.7 | -24.7 |
| C6H6 | Benzene | 21.2 | 21.3 |
| C6H6 | Fulvene | 53.6 | 53.6 |
| C6H8 | (E)-1,3,5-Hexatriene | 42.5 | 42.5 |
| C6H8 | (Z)-1,3,5-Hexatriene | 43.8 | 43.8 |
| C6H8 | 1,3-Cyclohexadiene | 14.4 | 14.4 |
| C6H8 | 1,4-Cyclohexadiene | 14.3 | 14.3 |
| C6H10 | 1,2-Dimethylcyclobutene | 6.3 | 6.3 |
| C6H10 | 1,5-Hexadiene | 19.6 | 19.5 |
| C6H10 | 1-Methyl cyclopentene | -10.3 | -10.3 |
| C6H10 | 3-Methyl cyclopentene | -3.7 | -3.7 |
| C6H10 | 4-Methyl cyclopentene | -3.1 | -3.1 |

| Formula | Name | | |
|---|---|---|---|
| C6H10 | Bicyclopropyl | 28.5 | 28.5 |
| C6H10 | Cyclohexene | -10 | -10.0 |
| C6H11 | Cyclohexyl, cation | 186.8 | 186.8 |
| C6H12 | 1-Hexene | -9.1 | -9.1 |
| C6H12 | 2,3-Dimethyl-1-butene | -7.3 | -7.3 |
| C6H12 | 2,3-Dimethyl-2-butene | -13.3 | -13.2 |
| C6H12 | (Z)-3-Methyl-2-pentene | -12.5 | -12.4 |
| C6H12 | 4-Methyl-1-pentene | -5.8 | -5.9 |
| C6H12 | Cyclohexane | -34.8 | -34.8 |
| C6H14 | 2,2-Dimethyl butane | -26.1 | -26.1 |
| C6H14 | 2,3-Dimethyl butane | -27.8 | -27.8 |
| C6H14 | 2-Methyl pentane | -34.6 | -34.7 |
| C6H14 | 3-Methyl pentane | -33 | -33.1 |
| C6H14 | n-Hexane | -39.2 | -39.2 |
| C7H7 | Benzyl, cation | 218 | 218.0 |
| C7H7 | Tropylium cation | 207.6 | 207.6 |
| C7H8 | Cycloheptatriene | 33.7 | 33.8 |
| C7H8 | Norbornadiene | 62.8 | 62.8 |
| C7H8 | Toluene | 13.5 | 13.5 |
| C7H12 | 1,2-Dimethyl cyclopentene | -18.5 | -18.7 |
| C7H12 | 1-Ethyl cyclopentene | -14 | -14.0 |
| C7H12 | 1-Methyl cyclohexene | -17.1 | -17.1 |
| C7H12 | Norbornane | -10.5 | -10.6 |
| C7H14 | 1,1-Dimethyl cyclopentane | -30.1 | -30.1 |
| C7H14 | 1,2-cis-Dimethyl cyclopentane | -32.4 | -32.4 |
| C7H14 | 1,2-trans-Dimethyl cyclopentane | -34.6 | -35.3 |
| C7H14 | 1,3-cis-Dimethyl cyclopentane | -35.3 | -34.6 |
| C7H14 | 1-Heptene | -13.9 | -13.9 |
| C7H14 | Ethyl cyclopentane | -36.7 | -36.7 |
| C7H14 | Methyl-cyclohexane | -36.3 | -36.3 |
| C7H16 | 2,2,3-Trimethyl butane | -22.2 | -22.2 |
| C7H16 | 2,2-Dimethyl pentane | -30.4 | -30.4 |
| C7H16 | 2,3-Dimethyl pentane | -32 | -32.0 |
| C7H16 | 2,4-Dimethyl pentane | -34.6 | -34.6 |
| C7H16 | 2-Methyl hexane | -39.3 | -39.4 |
| C7H16 | 3,3-Dimethyl pentane | -27 | -27.0 |
| C7H16 | 3-Ethyl pentane | -35.9 | -36.0 |
| C7H16 | 3-Methyl hexane | -37.6 | -37.6 |
| C7H16 | n-Heptane | -43.9 | -43.9 |
| C8H8 | Cubane | 98.9 | 98.9 |
| C8H8 | Cyclooctatetraene | 56.1 | 56.1 |
| C8H8 | Styrene | 37.5 | 37.6 |
| C8H10 | Ethylbenzene | 8.7 | 8.7 |
| C8H10 | m-Xylene | 5.8 | 5.9 |
| C8H10 | o-Xylene | 8.2 | 8.3 |
| C8H10 | p-Xylene | 5.6 | 5.7 |
| C8H12 | 1,5-Cyclooctadiene | 11.3 | 11.3 |
| C8H12 | 4-Vinyl cyclohexene | 13 | 13.1 |
| C8H14 | 1-Octyne | 17 | 17.0 |
| C8H14 | 2,5-Dimethyl 2,4-hexadiene | -2.6 | -2.4 |
| C8H14 | 2-Octyne | 5.3 | 5.3 |
| C8H14 | 3,4-Dimethyl-(E,E)-2,4-hexadiene | -2.4 | -2.4 |
| C8H14 | 3,4-Dimethyl-(E,Z)-2,4-hexadiene | -2.2 | -1.9 |
| C8H14 | 3,4-Dimethyl-(Z,Z)-2,4-hexadiene | -2 | -2.0 |
| C8H14 | 3-Octyne | 4.7 | 4.7 |
| C8H14 | 4-Octyne | 4.6 | 4.6 |
| C8H14 | Bicyclo(2.2.2)-octane | -26.5 | -26.5 |
| C8H16 | 1-Octene | -19.3 | -19.3 |
| C8H16 | Ethylcyclohexane | -39.5 | -39.5 |
| C8H18 | 2,2,3,3-Tetramethyl butane | -12.8 | -12.8 |
| C8H18 | 2,2,3-Trimethyl pentane | -24.6 | -24.6 |
| C8H18 | 2,2,4-Trimethyl pentane | -29.3 | -29.2 |
| C8H18 | 2,2-Dimethyl hexane | -35.1 | -35.1 |
| C8H18 | 2,3,3-Trimethyl pentane | -24.5 | -24.5 |
| C8H18 | 2,3,4-Trimethyl pentane | -28.8 | -28.8 |
| C8H18 | 2,3-Dimethyl hexane | -36.6 | -36.6 |
| C8H18 | 2,4-Dimethyl hexane | -37.5 | -37.5 |
| C8H18 | 2,5-Dimethyl hexane | -39.4 | -39.5 |
| C8H18 | 2-Methyl heptane | -44.1 | -44.1 |
| C8H18 | 3,3-Dimethyl hexane | -31.2 | -31.2 |
| C8H18 | 3,4-Dimethyl hexane | -31.7 | -31.6 |
| C8H18 | 3-Ethyl hexane | -39.8 | -39.8 |
| C8H18 | 3-Ethyl-2-methyl pentane | -31.6 | -31.6 |
| C8H18 | 3-Ethyl-3-methyl pentane | -27.8 | -27.8 |
| C8H18 | 3-Methyl heptane | -42.3 | -42.3 |
| C8H18 | 4-Methyl heptane | -42.1 | -42.1 |
| C8H18 | n-Octane | -48.7 | -48.7 |
| C9H10 | alpha-Methyl styrene | 30.4 | 30.5 |
| C9H10 | Cyclopropyl benzene | 37.5 | 37.6 |
| C9H18 | 1,3,5-Trimethyl cyclohexane | -39 | -39.0 |
| C9H18 | cis-cis-trans-1,3,5-Trimethyl cyclohexane | -37.8 | -37.8 |
| C9H20 | 3,3-Diethylpentane | -27.9 | -27.9 |
| C9H20 | n-Nonane | -53.4 | -53.4 |
| C10H8 | Azulene | 72 | 72.1 |
| C10H8 | Naphthalene | 38.2 | 38.3 |
| C10H10 | 1,4-Dicyclopropylbuta-1,3-diyne | 117.7 | 117.7 |
| C10H10 | 1-Butynl benzene | 52 | 52.0 |
| C10H10 | 2a,4a,6a,6b-Tetrahydrocyclopentapentalene | 47.6 | 47.6 |
| C10H10 | Bulvalene | 63.4 | 63.4 |
| C10H10 | Diisopropenyldiacetylene | 103 | 103.0 |
| C10H10 | Tricyclo[6.2.0.0]deca-1(8),2,6-triene | 54.1 | 54.1 |
| C10H12 | 1,2,6,7-Cyclodecatetraene | 67.8 | 67.8 |

| Formula | Name | | |
|---|---|---|---|
| C10H12 | Dispiro[2.2.2.2]deca-4,9-diene | 65.1 | 65.1 |
| C10H12 | Tetralin | 1.3 | 1.3 |
| C10H14 | (1-Methylpropyl) benzene | 4.7 | 4.5 |
| C10H14 | (2-Methylpropyl) benzene | 3.6 | 3.7 |
| C10H14 | 1,2,3,4-Tetramethyl benzene | 0.9 | 1.0 |
| C10H14 | 1,2,3,4-Tetramethylfulvene | 20 | 20.1 |
| C10H14 | 1,2,3,5-Tetramethyl benzene | -3.1 | -3.0 |
| C10H14 | 1,2,4,5-Tetramethyl benzene | -4.7 | -4.6 |
| C10H14 | tert-Butyl benzene | 15.4 | 15.6 |
| C10H14 | Tetrahydrotriquinacene | -10.8 | -10.8 |
| C10H16 | 1,2,3,4,5-Pentamethyl-1,3-cyclopentadiene | -6.5 | -6.5 |
| C10H16 | Adamantane | -26.5 | -26.5 |
| C10H16 | Camphene | 14.6 | 14.6 |
| C10H16 | Perhydrotriquinacene | -38.9 | -38.9 |
| C10H18 | 1-Methyl-4-(1-methylethyl)-cyclohexene | -20.4 | -20.5 |
| C10H18 | 4-Methyl-1-(1-methylethyl)-cyclohexene | -23.8 | -23.8 |
| C10H18 | cis-Decalin | -37.2 | -37.3 |
| C10H18 | Spiro(4-5)decane | -37.3 | -37.3 |
| C10H18 | trans-Decalin | -41.9 | -41.9 |
| C10H20 | (E)-2,2,5,5-Tetramethyl-3-hexene | -10.9 | -10.9 |
| C10H20 | (Z)-2,2,5,5-Tetramethyl-3-hexene | -0.3 | -0.3 |
| C10H20 | 1-Decene | -28 | -28.0 |
| C10H20 | Butyl cyclohexane | -48.8 | -48.8 |
| C10H20 | Pentyl cyclopentane | -50.7 | -50.7 |
| C10H22 | 2,2,5,5-Tetramethylhexane | -30.7 | -11.8 |
| C10H22 | 3,3,4,4-Tetramethylhexane | -11.8 | -30.7 |
| C10H22 | n-Decane | -58.1 | -58.1 |
| C11H16 | Pentamethylbenzene | -2.6 | -2.5 |
| C11H22 | 1,1,4-Trimethylcycloheptane | -26.5 | -26.5 |
| C11H22 | Hexyl cyclopentane | -55.5 | -55.5 |
| C11H22 | Pentyl cyclohexane | -53.4 | -53.5 |
| C11H24 | Undecane | -62.8 | -62.8 |
| C12H8 | Acenaphthylene | 67 | 67.1 |
| C12H8 | Biphenylene | 94.6 | 94.7 |
| C12H10 | Acenaphthene | 33 | 33.1 |
| C12H10 | Biphenyl | 45.9 | 46.0 |
| C12H18 | Hexamethylbenzene | 0.1 | 0.1 |
| C12H24 | Hexylcyclohexane | -58.2 | -58.2 |
| C12H26 | n-Duodecane | -67.6 | -67.6 |
| C13H10 | Fluorene | 44.8 | 44.9 |
| C13H28 | Tri-t-butylmethane | 32.4 | 32.4 |
| C13H28 | Tridecane | -72.3 | -72.3 |
| C14H10 | Anthracene | 58.7 | 58.8 |
| C14H10 | Diphenylethyne | 89.4 | 89.5 |
| C14H10 | Phenanthrene | 55.5 | 55.6 |
| C14H12 | 9,10-Dihydro-phenanthrene | 38.3 | 38.3 |

| Formula | Name | Value 1 | Value 2 |
|---|---|---|---|
| C14H12 | 9-Methyl-9H-fluorene | 42.2 | 42.3 |
| C14H12 | Octalene | 106.1 | 106.2 |
| C14H12 | Stilbene | 63.3 | 63.4 |
| C14H14 | 1,2,3,4-Tetrahydrophenanthrene | 21.8 | 21.9 |
| C14H14 | 4,4'-Dimethylbiphenyl | 30.3 | 30.4 |
| C14H14 | Bibenzyl | 37.5 | 37.6 |
| C14H16 | 1,4,5,8-Tetramethynaphthalene | 31.3 | 31.4 |
| C14H18 | 1,2,3,4,5,6,7,8-Octahydro-anthracene | -18.6 | -18.6 |
| C14H20 | Diadamantane | -23.6 | -23.6 |
| C14H24 | 1,3,5,7-Tetramethyladamantane | -22.8 | -22.8 |
| C14H28 | (E)-3,4-Di-tert-butyl-3-hexene | 19.2 | 19.2 |
| C14H28 | Cyclotetradecane | -53.2 | -53.2 |
| C14H28 | n-Nonylcyclopentane | -69.6 | -69.6 |
| C14H28 | Octylcyclohexane | -67.7 | -67.7 |
| C14H30 | 3,3,4,4-Tetraethylhexane | 6.2 | 6.2 |
| C14H30 | Octamethylhexane | 46.7 | 46.7 |
| C14H30 | Tetradecane | -77 | -77.0 |
| C15H12 | 4-Methylphenanthrene | 56.8 | 56.9 |
| C15H22 | 1-Methyldiadamantane | -18.1 | -18.0 |
| C15H22 | 3-Methyladamantane | -22.7 | -22.7 |
| C15H22 | 4-Methyldiadamantane | -23 | -23.0 |
| C15H22 | 6-(1,1-dimethylethyl)-2,3-dihydro-1,1-dimethyl-1H-Indene | 0.5 | 0.6 |
| C15H30 | n-Nonylcyclohexane | -72.3 | -72.3 |
| C15H32 | Pentadecane | -81.7 | -81.7 |
| C16H10 | Fluoranthene | 72.6 | 72.7 |
| C16H10 | Pyrene | 60.5 | 60.7 |
| C16H14 | 2,7-Dimethylphenanthrene | 39.9 | 40.0 |
| C16H14 | 4,5,9,10-Tetrahydropyrene | 25.4 | 25.5 |
| C16H14 | 9,10-Dimethylphenanthrene | 52.7 | 52.9 |
| C16H16 | (2.2)Metaparacyclophane | 62.6 | 62.7 |
| C16H16 | [2.2]Metacyclophane | 58.6 | 58.7 |
| C16H16 | [2.2]Paracyclophane | 66.5 | 66.6 |
| C16H18 | 1,2,3,6,7,8-Hexahydropyrene | 9.2 | 9.3 |
| C16H28 | Tricyclo[8.2.2.2]-hexadecane | -15.8 | -15.8 |
| C16H32 | 1-Hexadecene | -56.4 | -56.4 |
| C16H32 | Decylcyclohexane | -77.1 | -77.1 |
| C16H32 | n-Undecylcyclopentane | -79 | -79.1 |
| C17H34 | n-Dodecylcyclopentane | -83.8 | -83.8 |
| C17H34 | Undecylcyclohexane | -81.8 | -81.8 |
| C17H36 | Heptadecane | -91.2 | -91.2 |
| C18H14 | p-Terphenyl | 70.6 | 70.7 |
| C18H18 | 2,5-Diphenyl-1,5-hexadiene | 72.5 | 72.6 |
| C18H18 | 3,4,5,6-Tetramethylphenanthrene | 48.9 | 49.0 |
| C18H20 | [3.3]Paracyclophane | 36.9 | 37.0 |
| C18H22 | 1,1'-(1,1,2,2-Tetramethyl-1,2-ethanediyl)bis-benzene | 70.2 | 70.3 |

| Formula | Name | Value 1 | Value 2 |
|---|---|---|---|
| C18H36 | Dodecylcyclohexane | -86.5 | -86.5 |
| C18H36 | n-Tridecylcyclopentane | -88.5 | -88.5 |
| C18H38 | 1,1,2,2-Tetra-t-butylethane | 70.2 | 70.2 |
| C18H38 | Octadecane | -95.9 | -95.9 |
| C19H20 | 2,6-Diphenyl-1,6-heptadiene | 67.8 | 67.9 |
| C19H38 | n-Tetradecylcyclopentane | -93.2 | -93.2 |
| C19H38 | n-Tridecylcyclohexane | -91.2 | -91.3 |
| C19H40 | Nonadecane | -100.6 | -100.6 |
| C20H14 | 9,10-Dihydro-9,10[1',2']benzanthracene | 87.1 | 87.3 |
| C20H16 | 3,9-Dimethylbenz[a]anthracene | 59.1 | 59.3 |
| C20H16 | 5,6-Dimethyl chrysene | 76.2 | 76.4 |
| C20H16 | 9,10-Dimethyl-1,2-benzanthracene | 79.6 | 79.8 |
| C20H30 | 1,3,5-Tri-tert-butyl pentalene | 54.6 | 54.7 |
| C20H36 | Tetra-tert-butyltetrahedrane | 80.9 | 80.9 |
| C20H38 | Meso-3,4-dicyclohexyl-2,5-dimethylhexane | -36 | -36.0 |
| C20H40 | Tetradecylcyclohexane | -96 | -96.0 |
| C20H42 | Eicosane | -105.3 | -105.4 |
| N | Nitrogen, cation | 417.5 | 417.5 |
| N | Nitrogen, atom | 113 | 113.0 |
| NH2 | Amidogen | 36.4 | 36.4 |
| NH3 | Ammonia | -6.4 | -6.4 |
| NH4 | Ammonium, cation | 164.6 | 164.6 |
| CN | Cyanide | 126.3 | 126.3 |
| CNH | Hydrogen cyanide | 35.3 | 35.3 |
| CNH4 | CH2-NH2, cation | 186.8 | 186.8 |
| CNH4 | CH3-NH. | 29.3 | 29.3 |
| CNH4 | CH3NH, anion | 23.5 | 23.5 |
| CNH5 | Methylamine | -7.6 | -7.6 |
| C2NH3 | Acetonitrile | 19.2 | 19.2 |
| C2NH3 | Methyl isocyanide | 60.3 | 60.3 |
| C2NH5 | Ethyleneimine (Azirane) | 25 | 25.0 |
| C2NH6 | Dimethyl nitrogen, anion | 8.5 | 8.5 |
| C2NH7 | Dimethylamine | -6.7 | -6.7 |
| C2NH7 | Ethylamine | -13.3 | -13.3 |
| C3NH3 | Acrylonitrile | 43.8 | 43.8 |
| C3NH5 | Ethyl cyanide | 13.7 | 13.7 |
| C3NH7 | Cyclopropylamine | 15.3 | 15.3 |
| C3NH9 | Isopropylamine | -16.4 | -16.4 |
| C3NH9 | n-Propylamine | -18.3 | -18.3 |
| C3NH9 | Trimethylamine | -2.8 | -2.9 |
| C4NH5 | (E)-2-Butenenitrile | 33.8 | 33.8 |
| C4NH5 | (Z)-2-Butenenitrile | 34.1 | 34.1 |
| C4NH5 | 3-Butenenitrile | 39.2 | 39.2 |
| C4NH5 | Pyrrole | 32.4 | 32.4 |
| C4NH7 | Butanenitrile | 8.9 | 8.9 |
| C4NH7 | Isobutane nitrile | 11.2 | 11.2 |

| Formula | Name | Value 1 | Value 2 |
|---|---|---|---|
| C4NH9 | Pyrrolidine | -15.9 | -15.9 |
| C4NH11 | 2-Butylamine | -20.4 | -20.4 |
| C4NH11 | 2-Methyl-1-propylamine | -19.3 | -19.3 |
| C4NH11 | N-Butylamine | -23.1 | -23.1 |
| C4NH11 | t-Butylamine | -15.5 | -15.5 |
| C5NH5 | Pyridine | 28.7 | 28.8 |
| C5NH7 | N-Methyl pyrrole | 32.2 | 32.2 |
| C5NH9 | 1,2,3,6-Tetrahydropyridine | 6.1 | 6.1 |
| C5NH9 | 2-Cyanobutane | 7.7 | 7.8 |
| C5NH9 | Butyl cyanide | 4.1 | 4.1 |
| C5NH9 | t-Butylnitrile | 12.5 | 12.5 |
| C5NH11 | Cyclopentylamine | -22.6 | -22.6 |
| C5NH11 | Piperidine | -18.6 | -18.6 |
| C5NH13 | N-Methyl-n-butylamine | -21.7 | -21.8 |
| C6NH7 | 1-Cyclopentenecarbonitrile | 28.5 | 28.6 |
| C6NH7 | 2-Cyclopentenecarbonitrile | 34.9 | 34.9 |
| C6NH7 | 2-Methyl pyridine | 19.7 | 19.7 |
| C6NH7 | 3-Methyl pyridine | 20.4 | 20.4 |
| C6NH7 | 4-Methyl pyridine | 20.8 | 20.8 |
| C6NH7 | Aniline | 21.6 | 21.7 |
| C6NH9 | 2,5-Dimethyl-1H-pyrrole | 10.1 | 10.1 |
| C6NH9 | Cyclopentanecarbonitrile | 5.4 | 5.4 |
| C6NH13 | 2-Methylpiperidine | -21.6 | -21.6 |
| C6NH13 | Cyclohexamethylenimine | -17.7 | -17.7 |
| C6NH13 | Cyclohexanamine | -25.8 | -25.8 |
| C6NH15 | Di-n-propylamine | -27.3 | -27.3 |
| C6NH15 | Diisopropylamine | -20.3 | -20.3 |
| C6NH15 | Triethylamine | -14.8 | -14.8 |
| C7NH5 | Phenyl cyanide | 51.9 | 51.9 |
| C7NH9 | 1-Cyclohexenecarbonitrile | 21 | 21.0 |
| C7NH9 | 2,6-Dimethylpyridine | 10.6 | 10.7 |
| C7NH9 | 2-Cyclohexenecarbonitrile | 26.5 | 26.6 |
| C7NH9 | Benzylamine | 19.4 | 19.5 |
| C7NH9 | m-Toluidine | 14.1 | 14.1 |
| C7NH9 | N-Methylaniline | 24.2 | 24.2 |
| C7NH9 | o-Toluidine | 16.1 | 16.1 |
| C7NH9 | p-Toluidine | 13.7 | 13.8 |
| C7NH11 | Cyclohexanecarbonitrile | 1.7 | 1.7 |
| C7NH13 | Hexahydro-1H-pyrrolizine | -20.1 | -20.1 |
| C7NH13 | n-Heptanenitrile | -5.3 | -5.3 |
| C7NH17 | Isopropylbutylamine | -28.9 | -29.0 |
| C8NH11 | 1-Norbornylcyanide | 24.2 | 24.3 |
| C8NH11 | 1-Norbornylisocyanide | 63.1 | 63.0 |
| C8NH11 | 5-Ethyl-2-methyl-pyridine | 6.4 | 6.4 |
| C8NH11 | N,N-Dimethyl aniline | 33.4 | 28.8 |
| C8NH11 | N-Ethyl aniline | 19.2 | 19.3 |

| Formula | Name | Value 1 | Value 2 |
|---|---|---|---|
| C8NH15 | 3-Azabicyclo[3.2.2]nonane | -13.9 | -13.9 |
| C8NH15 | n-Heptyl cyanide | -10 | -10.0 |
| C8NH17 | N-(2-Methylpropylidene)-butylamine | -15.2 | -15.2 |
| C8NH19 | 2-Methyl-N-(2-methylpropyl)-1-propanamine | -28.5 | -28.5 |
| C8NH19 | Di-sec-butylamine | -26.6 | -26.7 |
| C8NH19 | Dibutylamine | -36.6 | -36.7 |
| C8NH19 | N-(2-Methylpropyl)-1-butanamine | -32.9 | -32.9 |
| C8NH19 | n-Octylamine | -42 | -42.0 |
| C9NH7 | Isoquinoline | 45.2 | 45.2 |
| C9NH7 | Quinoline | 44.6 | 44.7 |
| C9NH9 | 2,6-Dimethylbenzonitrile | 39.9 | 40.0 |
| C9NH11 | (1a,2a,4a)-Bicyclo[2.2.2]oct-5-ene-2-carbonitrile | 39.7 | 39.7 |
| C9NH11 | (1a,2b,4a)-Bicyclo[2.2.2]oct-5-ene-2-carbonitrile | 40.1 | 40.1 |
| C9NH11 | 1,2,3,4-Tetrahydroquinoline | 14 | 14.1 |
| C9NH11 | 5,6,7,8-Tetrahydroquinoline | 6.9 | 7.0 |
| C9NH13 | N,N-Dimethyl m-toluidine | 26.1 | 26.2 |
| C9NH13 | N,N-Dimethyl p-toluidine | 25.3 | 25.4 |
| C9NH13 | N-Ethyl m-toluidine | 11.7 | 11.7 |
| C9NH17 | cis-3,7a-H-cis-5,8-H-3,5-Dimethylpyrrolizidine | -25.9 | -25.9 |
| C9NH17 | Decahydro trans-quinoline | -26.9 | -26.9 |
| C9NH19 | 2,2,6,6-Tetramethyl piperidine | -14 | -14.0 |
| N2 | Nitrogen | 8.3 | 8.3 |
| N2H2 | Diazene | 31.8 | 31.8 |
| N2H4 | Hydrazine | 14.1 | 14.1 |
| CN2H2 | Diazomethane | 67.2 | 67.2 |
| CN2H2 | N=N-CH2- | 72.4 | 72.4 |
| CN2H6 | Methylhydrazine | 14.3 | 14.3 |
| C2N2 | Cyanogen | 66.6 | 66.6 |
| C2N2H8 | 1,1-Dimethylhydrazine | 18.1 | 18.0 |
| C2N2H8 | 1,2-Dimethylhydrazine | 15 | 14.9 |
| C3N2H4 | 1H-Pyrazole | 45.3 | 45.3 |
| C3N2H4 | Imidazole | 33.2 | 33.2 |
| C3N2H10 | 1,2-Propanediamine | -10.6 | -10.7 |
| C4N2 | Dicyanoacetylene | 111.3 | 111.4 |
| C4N2H2 | Fumaronitrile | 74.7 | 74.7 |
| C4N2H4 | 1,3-Diazine | 34.9 | 34.9 |
| C4N2H4 | Pyrazine | 37.7 | 37.7 |
| C4N2H4 | Pyridazine | 43.5 | 43.5 |
| C4N2H4 | Succinonitrile | 48.8 | 48.8 |
| C4N2H6 | 2-Methyl-1H-imidazole | 21.5 | 21.5 |
| C4N2H8 | (Dimethylamino) acetonitrile | 31.5 | 31.5 |
| C4N2H8 | 1,4,5,6-Tetrahydropyrimidine | 8.9 | 8.9 |
| C4N2H10 | Piperazine | -2.9 | -2.9 |
| C5N2H6 | 2-Aminopyridine | 25.7 | 25.8 |
| C5N2H6 | 3-Aminopyridine | 29.2 | 29.3 |

| Formula | Name | Value 1 | Value 2 |
|---|---|---|---|
| C5N2H6 | 4-Aminopyridine | 28.3 | 28.3 |
| C5N2H6 | Dimethyl propanedinitrile | 52.8 | 52.8 |
| C5N2H8 | 2-Ethyl-1H-imidazole | 16.5 | 16.5 |
| C5N2H10 | Diethylcyanamide | 29.3 | 29.4 |
| C5N2H12 | Butylmethyldiazene | 5.7 | 5.6 |
| C5N2H14 | N,N-Dimethyl-1,3-propanediamine | -8.8 | -8.8 |
| C6N2H4 | 2-Cyanopyridine | 60.2 | 60.3 |
| C6N2H4 | 3-Cyanopyridine | 59.6 | 59.7 |
| C6N2H4 | 4-Cyanopyridine | 60.8 | 60.9 |
| C6N2H8 | 2,3-Dimethyl pyrazine | 20.2 | 20.2 |
| C6N2H8 | Hexanedinitrile | 38.5 | 38.5 |
| C6N2H8 | Phenylhydrazine | 45.2 | 45.2 |
| C6N2H12 | 3(Dimethylamino) propanenitrile | 22.1 | 22.1 |
| C6N2H12 | Tetramethyldiazetine | 33.6 | 33.6 |
| C6N2H12 | Triethylenediamine | 19.3 | 19.3 |
| C6N2H14 | 1,2-Diisopropyldiazene | 3.7 | 3.7 |
| C6N2H14 | Dipropyldiazene | -0.9 | -0.9 |
| C7N2H6 | 1H-Benzimidazole | 45.7 | 45.8 |
| C7N2H6 | 1H-Indazole | 58 | 58.0 |
| C7N2H10 | 1-Methyl-1-phenylhydrazine | 24.9 | 24.9 |
| C7N2H10 | t-Butylmalononitrile | 52.5 | 52.5 |
| C7N2H10 | Trimethyl pyrazine | 10.5 | 10.6 |
| C7N2H12 | 1-Piperidineacetonitrile | 19.6 | 19.6 |
| C7N2H14 | 3,3,5,5-Tetramethyl-1-pyrazoline | 5.2 | 5.2 |
| C8N2H4 | m-Dicyanobenzene | 83.8 | 83.8 |
| C8N2H4 | o-Dicyanobenzene | 85.3 | 85.3 |
| C8N2H4 | p-Dicyanobenzene | 83.9 | 83.9 |
| C8N2H6 | Phthalazine | 59 | 59.0 |
| C8N2H6 | Quinazoline | 50.2 | 50.3 |
| C8N2H6 | Quinoxaline | 52.6 | 52.7 |
| C8N2H12 | n-Pentylmalonodinitrile | 34.2 | 34.2 |
| C8N2H12 | Tetramethylbutanedinitrile | 56.8 | 56.8 |
| C8N2H12 | Tetramethylpyrazine | 2.6 | 2.7 |
| C8N2H14 | 1,4-Dimethyl-2,3-diaza-bicyclo[2.2.2]oct-2-ene | 13.7 | 13.7 |
| C8N2H16 | 3,4,5,6-Tetrahydro-3,3,6,6-tetramethylpyridazine | 6.7 | 6.7 |
| C8N2H18 | Di-n-butyldiazene | -10.4 | -10.4 |
| C8N2H18 | Di-tert-butyldiazene | 4.9 | 4.9 |
| C8N2H20 | 1,2-Dibutylhydrazine | -14.4 | -14.4 |
| N3 | Azide radical | 97.1 | 97.1 |
| N3H | Hydrazoic acid | 73 | 73.1 |
| C3N3H3 | 1,3,5-Triazine | 39.9 | 40.0 |
| C5N3H | Ethylenetricarbonitrile | 110.4 | 110.4 |
| C5N3H3 | 1,1,1-Ethanetricarbonitrile | 96.1 | 96.2 |
| CN4H2 | 1-H Tetrazole | 53.8 | 53.7 |
| CN4H2 | 2-H-Tetrazole | 58 | 58.0 |
| C6N4 | Tetracyanoethylene | 147.9 | 148.0 |

| Formula | Name | Value 1 | Value 2 |
|---|---|---|---|
| C6N4H12 | 1,3,5,7-Tetraazaadamantane | 51.4 | 51.5 |
| C3N6H6 | Melamine | 21.4 | 21.7 |
| C10NH9 | 2-Methyl-quinoline | 35.7 | 35.8 |
| C10NH9 | 4-Methyl-quinoline | 38.9 | 39.0 |
| C10NH9 | 6-Methyl-quinoline | 36.9 | 37.0 |
| C10NH9 | 8-Methyl-quinoline | 38.1 | 38.1 |
| C10NH11 | 2,4,6-Trimethyl-benzonitrile | 32.4 | 32.5 |
| C10NH11 | 2,4,6-Trimethylphenyl isocyanide | 70.8 | 70.9 |
| C10NH15 | N,N-Diethyl aniline | 26.9 | 26.9 |
| C10NH19 | n-Nonyl cyanide | -19.5 | -19.5 |
| C11NH11 | 2,6-Dimethyl quinoline | 28 | 28.1 |
| C11NH11 | 2,7-Dimethyl quinoline | 28 | 28.1 |
| C11NH15 | 1-Adamantyl cyanide | 11.4 | 11.4 |
| C11NH15 | 1-Adamantyl isocyanide | 50 | 50.0 |
| C11NH17 | 2-Methyl-6-t-butylaniline | 19.7 | 19.7 |
| C11NH21 | n-Undecanenitrile | -24.2 | -24.2 |
| C12NH9 | Carbazole | 53.4 | 53.5 |
| C12NH11 | 2-Biphenylamine | 47.8 | 47.9 |
| C12NH11 | Biphenylamine | 56 | 55.7 |
| C12NH23 | 2-n-Butyl-2-methylhexanenitrile | -7.3 | -7.3 |
| C13NH9 | 6,7-Benzoquinoline | 60.5 | 60.6 |
| C13NH9 | a-Benzoquinoline | 59.8 | 59.9 |
| C13NH9 | Acridine | 64.2 | 64.3 |
| C13NH9 | Benzo[f]quinoline | 61.4 | 61.5 |
| C13NH9 | Phenanthridine | 60.5 | 60.6 |
| C13NH11 | N-Methylcarbazole | 56 | 56.1 |
| C13NH15 | 1,2,3,4-Tetrahydro-N-methylcarbazole | 22.8 | 22.9 |
| C14NH13 | 9-Ethyl-9H-carbazole | 51 | 51.1 |
| C14NH27 | Tetradecanenitrile | -38.4 | -38.4 |
| C9N2H8 | 3-Quinolinamine | 45.1 | 45.2 |
| C9N2H8 | 5-Quinolinamine | 47.2 | 47.2 |
| C9N2H8 | 6-Quinolinamine | 45.2 | 45.2 |
| C9N2H8 | 8-Quinolinamine | 45.2 | 45.3 |
| C9N2H18 | 2-(Diethylamino)-pentanenitrile | 14 | 14.0 |
| C10N2H8 | 2,2-Bipyridyl | 58.7 | 58.8 |
| C10N2H8 | 2,4-Bipyridyl | 59.9 | 60.0 |
| C10N2H8 | 4,4'-Bipyridine | 61.1 | 61.1 |
| C10N2H10 | 2,3-Dimethyl quinoxaline | 36.3 | 36.4 |
| C10N2H12 | alpha N,N-dimethylamino phenylacetonitrile | 58.8 | 58.9 |
| C10N2H16 | Ethyl(1,1-dimethylpropyl)malonodinitrile | 56.5 | 56.6 |
| C10N2H16 | Meso-2,3-diethyl-2,3-dimethylsuccinodinitrile | 55 | 55.0 |
| C10N2H16 | Methyl(1,1,2-trimethylpropyl)malonodinitrile | 66.2 | 66.3 |
| C12N2H8 | Phenazine | 71.6 | 71.7 |
| C12N2H8 | Phenazone | 72.5 | 72.6 |
| C12N2H10 | cis-Azobenzene | 83.4 | 83.5 |
| C12N2H12 | 4,4'-Dimethyl-2,2'-bipyridine | 42.9 | 43.0 |

| Formula | Name | Value 1 | Value 2 |
|---|---|---|---|
| C12N2H20 | 1,(1-Piperidinyl) cyclohexanecarbonitrile | 17.6 | 17.6 |
| C13N2H16 | a-Phenyl-1-piperidineacetonitrile | 53.5 | 53.5 |
| C9N3H3 | 1,3,5-Tricyanobenzene | 116.8 | 116.8 |
| C11N3H7 | 1,1,1-Tricyano-2-phenyl ethane | 126.7 | 126.7 |
| C16NH35 | Dioctylamine | -74.5 | -74.5 |
| C18NH15 | Triphenylamine | 93.7 | 93.9 |
| C18N2H12 | 2,2'-Biquinoline | 90.9 | 91.1 |
| C15N3H11 | 2,2',6',2'-Terpyridine | 88.5 | 88.6 |
| O | Oxygen, atom | 59.6 | 59.6 |
| HO | Hydroxyl radical | 0.2 | 0.2 |
| HO | Hydroxide, anion | -5.8 | -5.8 |
| H2O | Water | -60.9 | -61.0 |
| H3O | Hydronium, cation | 134.1 | 134.2 |
| CO | Carbon monoxide | -5.9 | -6.0 |
| CHO | HCO, cation | 184.9 | 184.8 |
| CHO | HCO | -1.4 | -1.4 |
| CH2O | Formaldehyde | -32.9 | -32.9 |
| CH3O | CH2OH, cation | 155.5 | 155.5 |
| CH3O | Methoxy, anion | -39.8 | -39.8 |
| CH4O | Methanol | -57.4 | -57.4 |
| C2H2O | Ketene | -6.8 | -6.9 |
| C2H4O | Acetaldehyde | -42.3 | -42.3 |
| C2H4O | Ethylene oxide | -15.6 | -15.5 |
| C2H5O | Ethoxy, anion | -45.3 | -45.3 |
| C2H6O | Dimethyl ether | -51.3 | -51.2 |
| C2H6O | Ethanol | -63 | -63.0 |
| C3H6O | Acetone | -49.4 | -49.5 |
| C3H6O | Propanal | -47.4 | -47.4 |
| C3H6O | Trimethylene oxide | -37.2 | -37.2 |
| C3H8O | Isopropanol | -65.1 | -65.2 |
| C3H8O | Methyl ethyl ether | -56.7 | -56.7 |
| C3H8O | Propanol | -67.6 | -67.6 |
| C4H4O | Acetyl acetylene | 12.3 | 12.3 |
| C4H4O | Furan | -8.7 | -8.7 |
| C4H6O | 2,3-Dihydrofuran | -29.9 | -29.9 |
| C4H6O | Crotonaldehyde | -27.9 | -27.9 |
| C4H6O | Divinyl ether | -2.1 | -2.0 |
| C4H8O | Butanal | -52.2 | -52.9 |
| C4H8O | Isobutanal | -50.6 | -50.6 |
| C4H8O | Methyl ethyl ketone | -54.1 | -54.1 |
| C4H8O | Tetrahydrofuran | -59.3 | -59.3 |
| C4H10O | Diethyl ether | -62.1 | -62.0 |
| C4H10O | t-Butanol | -64.3 | -64.4 |
| C5H8O | 2,3-Dihydro-5-methyl-furan | -39.9 | -39.9 |
| C5H8O | 2-Ethylacrolein | -28 | -30.7 |
| C5H8O | 3,4-Dihydro-2H-pyran | -39.4 | -39.4 |
| C5H8O | 3-Penten-2-one | -34.9 | -34.9 |
| C5H8O | Cyclopentanone | -57.1 | -57.1 |
| C5H10O | Diethyl ketone | -59.5 | -59.5 |
| C5H10O | Tetrahydropyran | -62.1 | -62.1 |
| C5H12O | t-Butyl methyl ether | -54.6 | -54.6 |
| C6H5O | Phenoxy, anion | -42.3 | -42.3 |
| C6H6O | Phenol | -26.8 | -26.7 |
| C6H10O | 4-Methyl-3-penten-2-one | -40.9 | -40.9 |
| C6H10O | Cyclohexanone | -60.2 | -60.2 |
| C6H12O | Methyl neopentyl ketone | -50.9 | -50.9 |
| C6H14O | Di-isopropyl ether | -62.5 | -62.5 |
| C7H6O | Benzaldehyde | -9.7 | -9.6 |
| C7H8O | Anisole | -17.8 | -17.7 |
| C7H8O | m-Cresol | -34.3 | -34.2 |
| C7H8O | o-Cresol | -33.3 | -33.3 |
| C7H8O | p-Cresol | -34.7 | -34.6 |
| C7H10O | 2-Methyl-5-hexen-3-yn-2-ol | 5.2 | 5.3 |
| C7H10O | 2-Norbornanone | -37.4 | -37.4 |
| C7H10O | cis-2,3-Epoxybicyclo[2.2.1]heptane | 2.8 | 2.8 |
| C7H10O | Norbornan-7-one | -37.7 | -37.7 |
| C7H12O | 1-Methoxy cyclohexene | -49.1 | -49.1 |
| C7H12O | Bicyclo[2.2.1]heptan-7-ol | -51.2 | -51.2 |
| C7H12O | cis-1,2-Epoxycycloheptane | -35.7 | -35.6 |
| C7H12O | Cycloheptanone | -59.5 | -59.6 |
| C7H14O | 2,4-Dimethyl 3-pentanone | -63.1 | -63.1 |
| C7H14O | 2-Methyl cis-cyclohexanol | -73.8 | -73.8 |
| C7H14O | 3,3-Dimethyl-2-pentanone | -52.3 | -52.3 |
| C7H14O | 4-Heptanone | -67.8 | -67.8 |
| C7H14O | Heptanal | -66.4 | -67.1 |
| C7H14O | t-Butyl ethyl ketone | -55.4 | -55.4 |
| C7H16O | n-Heptanol | -86.7 | -86.7 |
| C7H16O | t-Butyl isopropyl ether | -60.5 | -60.5 |
| C8H6O | Benzofuran | 3.8 | 3.8 |
| C8H8O | 1,3-Dihydro isobenzofuran | -20.5 | -20.4 |
| C8H8O | 1-Phenylethenol | -9.5 | -9.5 |
| C8H8O | 2,3-Dihydro-benzofuran | -22.1 | -22.1 |
| C8H8O | Acetophenone | -17.1 | -17.1 |
| C8H10O | 2,3-Dimethyl phenol | -36.1 | -36.0 |
| C8H10O | 2,4-Dimethyl phenol | -40.3 | -40.2 |
| C8H10O | 2,5-Dimethyl phenol | -40.1 | -40.1 |
| C8H10O | 2,6-Dimethyl phenol | -38.6 | -38.5 |
| C8H10O | 2-Ethyl phenol | -35.3 | -35.2 |
| C8H10O | 3,4-Dimethyl phenol | -39.6 | -39.5 |
| C8H10O | 3,5-Dimethyl phenol | -41.7 | -41.6 |
| C8H10O | 3-Ethyl phenol | -39.1 | -39.1 |
| C8H10O | 4-Ethyl phenol | -39.5 | -39.4 |
| C8H10O | 2-phenylethanol | -34.1 | -34.0 |
| C8H10O | Ethoxybenzene | -23 | -23.0 |
| C8H10O | Phenetole | -21.9 | -22.6 |
| C8H12O | 1-Methylnorcamphor | -40.8 | -40.9 |
| C8H12O | Bicyclo[2.2.2]octanone | -53.4 | -53.4 |
| C8H12O | Bicyclo[3.2.1]octan-2-one | -52.1 | -52.1 |
| C8H12O | Bicyclo[3.2.1]octan-3-one | -52 | -52.1 |
| C8H12O | Bicyclo[3.2.1]octan-8-one | -52.6 | -52.6 |
| C8H12O | cis-Bicyclo[3.3.0]-octan-2-one | -64 | -64.0 |
| C8H12O | trans-Bicyclo[3.3.0]-octan-2-one | -48.1 | -48.2 |
| C8H14O | 3-Oxabicyclo[3,2,2]nonane | -58.1 | -58.0 |
| C8H14O | 6-Methyl-5-hepten-2-one | -48.3 | -48.3 |
| C8H14O | 8-Oxatricyclo[3,2,1,0(1,5)]octane | 50.3 | 50.3 |
| C8H14O | Bicyclo(2.2.2)octan-2-ol | -66 | -66.0 |
| C8H14O | cis-1,2-Epoxycyclooctane | -34.3 | -34.3 |
| C8H14O | Cyclooctanone | -61.5 | -61.5 |
| C8H16O | 2,2,4-Trimethyl-3-pentanone | -56.6 | -56.6 |
| C8H16O | 2-Octanone | -72.9 | -72.9 |
| C8H16O | 3,3,4-Trimethyl pentan-2-one | -48.5 | -48.5 |
| C8H16O | 3-Octanone | -72.9 | -74.0 |
| C8H16O | 4-Octanone | -72.5 | -74.0 |
| C8H16O | Octanal | -71.1 | -71.1 |
| C8H18O | 1-Octanol | -91.4 | -91.4 |
| C8H18O | 1-Tert-butoxybutane | -69.2 | -69.2 |
| C8H18O | 2-(1,1-Dimethylethoxy)-butane | -63.2 | -63.2 |
| C8H18O | Di-n-butyl ether | -81 | -81.0 |
| C8H18O | Di-sec-butyl ether | -69 | -69.1 |
| C8H18O | tert-Butyl ether | -51.8 | -51.8 |
| C8H18O | tert-Butyl isobutyl ether | -65.3 | -65.3 |
| C9H10O | 3,4-Dihydro-1H-2-benzopyran | -26.8 | -26.7 |
| C9H10O | 3,4-Dihydro-2H-1-benzopyran | -28.8 | -28.8 |
| C9H10O | Benzyl methyl ketone | -20.8 | -20.8 |
| C9H12O | 2(-1-Methylethyl)-phenol | -37.2 | -37.2 |
| C9H12O | 2,4,6-Trimethyl phenol | -46.3 | -46.2 |
| C9H12O | 3(-1-Methylethyl)-phenol | -39.6 | -39.6 |
| C9H12O | 4-(-1-Methylethyl)-phenol | -40 | -40.0 |
| C9H14O | 2,6,6-Trimethyl-2-cyclohexen-1-one | -39.9 | -39.9 |
| C9H14O | Bicycle[3.3.1]nonan-9-one -check -3-one | -58.4 | -58.4 |
| C9H14O | cis Octahydro-2H-inden-2-one | -64.2 | -64.3 |
| C9H14O | trans Octahydro-2H-inden-2-one | -62.1 | -62.1 |
| C9H16O | Cyclononanone | -61.7 | -61.7 |
| C9H18O | 2,6-Dimethyl-4-heptanone | -69.7 | -69.7 |
| C9H18O | 2-Nonanone | -78.3 | -78.3 |
| C9H18O | 3,3,4,4-Tetramethyl-2-pentanone | -37.6 | -37.6 |
| C9H18O | 5-Nonanone | -77.3 | -77.3 |
| C9H18O | Di-tert-butyl ketone | -43.6 | -43.6 |

| Formula | Name | | |
|---|---|---|---|
| C9H20O | 1-Nonanol | -96.1 | -96.1 |
| C9H20O | Amyl-t-butyl ether | -73.9 | -73.9 |
| C9H20O | Butyl 1,1-dimethylpropyl ether | -71.5 | -71.5 |
| NO | Nitric oxide, cation | 230.6 | 230.6 |
| NO | Nitric oxide | -0.5 | -0.5 |
| CNO | NCO | 31.6 | 31.6 |
| CNHO | Hydrogen isocyanate | -10.8 | -10.8 |
| CNH3O | Formamide | -39.4 | -40.2 |
| C2NH5O | Acetaldoxime | -18.3 | -18.3 |
| C2NH5O | Acetamide | -47.2 | -48.2 |
| C3NH3O | Isoxazole | 19.2 | 19.2 |
| C3NH3O | Oxalone (oxazole) | -8.3 | -8.3 |
| C3NH5O | Acrylamine | -22.4 | -24.4 |
| C3NH5O | Methoxyacetonitrile | -15.7 | -15.6 |
| C3NH7O | Dimethylformamide | -37 | -37.0 |
| C3NH7O | N-Methyl acetamide | -46.9 | -47.0 |
| C3NH7O | Propanamide | -51.7 | -53.5 |
| C3NH9O | Dimethylaminomethanol | -50.2 | -50.2 |
| C4NH5O | 3-Methyl isoxazole | 7.4 | 7.4 |
| C4NH5O | 5-Methyl isoxazole | 8.1 | 8.1 |
| C4NH7O | 2-Pyrrolidinone | -51.8 | -51.9 |
| C4NH7O | 4,5-Dihydro-2-methyl oxazole | -39.7 | -39.6 |
| C4NH7O | Methacrylamide | -30.9 | -32.0 |
| C4NH9O | 2-Methyl propanamide | -54.1 | -55.2 |
| C4NH9O | Butanamide | -56.2 | -58.2 |
| C4NH9O | Isobutylamide | -54.1 | -55.2 |
| C4NH11O | N,N-Diethyl-hydroxylamine | -28.2 | -28.3 |
| C5NH5O | 2-Pyridinol | -25.1 | -25.0 |
| C5NH5O | 3-Pyridinol | -18.7 | -18.7 |
| C5NH5O | 4-Pyridinol | -20 | -20.0 |
| C5NH5O | Pyridine 1 oxide | 44.4 | 44.4 |
| C5NH7O | 3,5-Dimethyl isoxazole | -3.6 | -3.6 |
| C5NH9O | 1-Methyl-2-pyrrolidinone | -51.1 | -51.0 |
| C5NH9O | 2-Ethyl-4,5-dihydro-oxazole | -44.1 | -44.1 |
| C5NH9O | N,N-Dimethylamino-2-propen-3-al | -14.6 | -14.6 |
| C5NH11O | 1-(Dimethylamino)-2-propanone | -37.3 | -37.3 |
| C5NH11O | 2,2-Dimethyl-propanamide | -49 | -50.1 |
| C5NH11O | N,N-Dimethyl propanamide | -45.6 | -45.6 |
| C6NH7O | 2-Hydroxy-6-methylpyridine | -34 | -33.9 |
| C6NH7O | 3-Hydroxy-2-methylpyridine | -26.7 | -26.6 |
| C6NH7O | 3-Hydroxy-6-methylpyridine | -27.8 | -27.8 |
| C6NH7O | 4-Hydroxy-2-methylpyridine | -28.9 | -28.9 |
| C6NH7O | 6-Methyl-2(1H)-pyridinone | -23.1 | -23.0 |
| C6NH7O | m-Amino phenol | -26.5 | -26.4 |
| C6NH7O | o-Amino phenol | -26.2 | -26.2 |
| C6NH7O | p-Amino phenol | -25.9 | -25.9 |

| Formula | Name | Value 1 | Value 2 |
|---|---|---|---|
| C6NH9O | Trimethyl isoxazole | -13.3 | -13.3 |
| C6NH11O | Caprolactam | -52.9 | -52.9 |
| C6NH11O | Cyclohexanone oxime | -35.3 | -35.3 |
| C6NH13O | N,N-Diethyl acetamide | -48.9 | -48.9 |
| C6NH13O | N,N-Dimethylbutyramine | -50.1 | -50.1 |
| C7NH5O | Benzoxazole | 5.1 | 5.2 |
| C7NH5O | Isocyanatobenzene | 14 | 14.0 |
| C7NH7O | Benzamide | -15.2 | -16.2 |
| C7NH11O | N,N-Dimethylamino-2,4-pentadiene-5-al | -0.1 | -0.1 |
| C7NH13O | 2-Methoxy-3,3-dimethylbutanenitrile | -18.9 | -18.9 |
| C7NH15O | N,N-Diethylaminoacetone | -43.9 | -44.0 |
| C7NH15O | N,N-Dimethyl-tert-butylcarboxamide | -35 | -41.3 |
| C8NH5O | alpha-oxo Benzeneacetonitrile | 25.4 | 25.4 |
| C8NH9O | 1,3-dimethyl-2-nitroso-benzene | 16.7 | 16.8 |
| C8NH9O | N-methyl-N-phenyl formamide | -6.8 | -6.7 |
| C8NH17O | Octanone-1-oxime | -46.4 | -46.5 |
| C8NH17O | Octanone-2-oxime | -47.5 | -47.5 |
| C8NH17O | Octanone-3-oxime | -49 | -49.0 |
| C8NH17O | Octanone-4-oxime | -48.3 | -48.3 |
| N2O | Nitrous oxide | 31 | 31.0 |
| CN2H4O | Urea | -42.8 | -44.8 |
| C2N2H6O | N-Methyl urea | -41.7 | -43.9 |
| C4N2H6O | Dimethyl furazan | 20.9 | 20.9 |
| C4N2H10O | Isopropylurea | -50.7 | -51.3 |
| C5N2H8O | 5-Amino-3,4-dimethylisoxazole | -4.8 | -4.7 |
| C5N2H12O | (1-Methylpropyl) urea | -54.7 | -55.3 |
| C5N2H12O | N,N-diethylurea | -43.6 | -45.5 |
| C5N2H12O | Tetramethylurea | -30.8 | -30.7 |
| C4N3H5O | 4-Amino-2(1H)-pyrimidinone | -23 | -22.9 |
| O2 | Oxygen (Singlet) | 12.1 | 12.1 |
| O2 | Oxygen (Triplet) | -16 | -16.1 |
| H2O2 | Hydrogen peroxide | -38.3 | -38.2 |
| CO2 | Carbon dioxide | -75.1 | -75.1 |
| CHO2 | Formate, anion | -112.9 | -101.6 |
| CH2O2 | Formic acid | -92.6 | -92.6 |
| C2H2O2 | trans Glyoxal | -61.4 | -61.4 |
| C2H3O2 | Acetate, anion | -126.7 | -110.0 |
| C2H4O2 | Acetic acid | -101.2 | -101.1 |
| C2H4O2 | Methyl formate | -85.6 | -85.5 |
| C2H6O2 | Dimethyl peroxide | -28.4 | -28.3 |
| C2H6O2 | Ethylene glycol | -106 | -106.0 |
| C3O2 | Carbon suboxide | -23.5 | -23.6 |
| C3H4O2 | 2-Oxo-propanal | -70.9 | -70.9 |
| C3H4O2 | 2-Propenoic acid | -76.2 | -76.2 |
| C3H4O2 | beta-Propiolactone | -68.9 | -68.9 |
| C3H6O2 | 1,3-Dioxalane | -93.1 | -93.0 |
| C3H6O2 | Ethyl formate | -90.2 | -90.2 |
| C3H6O2 | Methyl acetate | -93.7 | -93.7 |
| C3H6O2 | Propionic acid | -105.7 | -105.7 |
| C3H8O2 | 1,3-Propanediol | -110.2 | -110.2 |
| C3H8O2 | 2-Methoxyethanol | -99.8 | -99.8 |
| C3H8O2 | Dimethoxymethane | -94.4 | -94.4 |
| C3H8O2 | Propylene glycol | -107.6 | -107.6 |
| C4H6O2 | 2-Butenoic acid | -85.9 | -85.9 |
| C4H6O2 | 2-Methyl-2-propenic acid | -83.8 | -83.7 |
| C4H6O2 | Diacetyl | -78.8 | -78.9 |
| C4H6O2 | gamma Butyrolactone | -94 | -94.0 |
| C4H6O2 | Methyl 2-propenoate | -68.6 | -68.6 |
| C4H8O2 | 1,1 Dimethoxy ethene | -71 | -70.9 |
| C4H8O2 | 1,3 Dioxan | -94.3 | -94.3 |
| C4H8O2 | 1,4-Dioxane | -89.3 | -89.3 |
| C4H8O2 | Ethyl acetate | -99 | -99.0 |
| C4H10O2 | 1,2-Dimethoxyethane | -93.4 | -93.4 |
| C4H10O2 | 1,4 Butandiol | -115.3 | -115.3 |
| C4H10O2 | Diethyl peroxide | -38.3 | -38.5 |
| C4H10O2 | Dimethyl acetal | -95.2 | -95.2 |
| C5H8O2 | Acetylacetone | -84.3 | -84.4 |
| C5H10O2 | Ethyl propionate | -103.5 | -103.5 |
| C5H10O2 | Isopropyl acetate | -100.5 | -100.6 |
| C5H12O2 | 1,5 Pentandiol | -119.9 | -119.9 |
| C6H4O2 | p-Benzoquinone | -32.9 | -33.0 |
| C6H6O2 | 1,2-Benzenediol | -72.8 | -72.7 |
| C6H6O2 | Hydroquinone | -74 | -74.0 |
| C6H6O2 | Resorcinol | -75 | -75.0 |
| C6H8O2 | 1,3-Cyclohexanedione | -83.6 | -83.7 |
| C6H8O2 | 1,4-Cyclohexanedione | -84.6 | -84.7 |
| C6H10O2 | 2,4-Hexanedione | -88.9 | -89.0 |
| C6H10O2 | 2-Oxepanone | -94.4 | -94.4 |
| C6H10O2 | 3-Methyl-2,4-pentandione | -84.4 | -84.5 |
| C6H10O2 | Ethyl-(E)-2-butenoate | -83.8 | -83.7 |
| C6H12O2 | 1,1-Dimethoxy-2-butene | -79.8 | -79.7 |
| C6H12O2 | 4-Hydroxy-4-methylpentan-2-one | -96.4 | -96.4 |
| C6H12O2 | 5,5-Dimethyl-1,3-dioxane | -93.7 | -93.6 |
| C6H12O2 | cis-2,4-Dimethyl-1,3-dioxane | -100.8 | -100.7 |
| C6H12O2 | Ethyl butanoate | -108.1 | -108.1 |
| C6H12O2 | Hexanoic acid | -119.8 | -119.7 |
| C6H12O2 | Methyl 2-methylbutanoate | -104 | -103.9 |
| C6H12O2 | Methyl 2,2-dimethyl-propanoate | -96.3 | -96.2 |
| C6H12O2 | Methyl 3-methylbutanoate | -104 | -103.9 |
| C6H12O2 | Methyl pentanoate | -107.6 | -107.5 |
| C6H12O2 | t-Butyl acetate | -96.6 | -96.5 |
| C6H12O2 | trans 4,5-Dimethyl-1,3-dioxane | -98 | -97.9 |

| Formula | Name | | |
|---|---|---|---|
| C6H14O2 | 1,1-Diethoxy ethane | -105.8 | -106.5 |
| C6H14O2 | 1,1-Dimethoxy-butane | -104 | -104.0 |
| C6H14O2 | 1,2-Diethoxy ethane | -104.2 | -104.1 |
| C6H14O2 | 1,6-Hexanediol | -124.7 | -124.7 |
| C6H14O2 | 2,3-Dimethyl-2,3-butanediol | -99.2 | -99.2 |
| C7H6O2 | 3-(2-Furanyl)-2-propenal | -27.8 | -27.8 |
| C7H6O2 | Benzoic acid | -65.9 | -67.7 |
| C7H6O2 | Phenyl formate | -50.1 | -50.0 |
| C7H6O2 | Tropolone | -40.1 | -40.1 |
| C7H8O2 | 3-Methyl-1,2-benzenediol | -80.3 | -80.3 |
| C7H8O2 | 4-Methyl 1,2-Benzenediol | -80.4 | -80.3 |
| C7H10O2 | Ethyl 2-methylene-3-butenoate | -57.6 | -57.6 |
| C7H10O2 | Ethyl 2-pentynoate | -56.2 | -56.2 |
| C7H10O2 | Ethyl 3-pentynoate | -58.3 | -58.2 |
| C7H10O2 | Ethyl 4-pentynoate | -47.1 | -47.0 |
| C7H12O2 | 3,5-Heptanedione | -93.5 | -93.6 |
| C7H12O2 | 3-Ethyl-2,4-pentanedione | -87.8 | -87.8 |
| C7H12O2 | 5-Methyl-2,4-hexanedione | -90.9 | -91.0 |
| C7H12O2 | Butyl 2-propenoate | -83.4 | -83.3 |
| C7H12O2 | Ethyl (Z)-2-pentenoate | -88.4 | -88.4 |
| C7H12O2 | Ethyl (Z)-3-pentenoate | -89.6 | -89.5 |
| C7H12O2 | Ethyl 4-pentenoate | -82.8 | -82.7 |
| C7H12O2 | Ethyl trans-2-pentenoate | -88.4 | -88.4 |
| C7H12O2 | Heptanolactone | -92.4 | -92.4 |
| C7H12O2 | Isopropyl 2-butenoate | -85.4 | -85.3 |
| C7H12O2 | Propyl (E)-2-butenoate | -88.6 | -88.5 |
| C7H14O2 | (2a,4a,6b)-2,4,6-Trimethyl-1,3-dioxane | -103.4 | -103.4 |
| C7H14O2 | 1,1-Dimethoxycyclopentane | -97.2 | -97.3 |
| C7H14O2 | 1,1-Dimethylpropyl acetate | -97.4 | -97.3 |
| C7H14O2 | Ethyl 2-methylbutanoate | -108.5 | -108.5 |
| C7H14O2 | Ethyl 3-methylbutanoate | -108.5 | -108.4 |
| C7H14O2 | Ethyl pentanoate | -112.8 | -112.8 |
| C7H14O2 | Methyl 3,3-dimethylbutanoate | -99.7 | -99.6 |
| C7H14O2 | Methyl hexanoate | -112.3 | -112.2 |
| C7H16O2 | 1,3-Diethoxypropane | -108.3 | -108.2 |
| C7H16O2 | 1,7-Heptanediol | -129.4 | -129.4 |
| C8H8O2 | m-Methylbenzoic acid | -75.5 | -75.3 |
| C8H8O2 | Methyl benzoate | -60.4 | -60.3 |
| C8H8O2 | o-Methylbenzoic acid | -73.4 | -73.3 |
| C8H8O2 | p-Methylbenzoic acid | -75.5 | -75.4 |
| NO2 | Nitrogen dioxide, cation | 240.6 | 240.7 |
| NO2 | Nitrogen dioxide | -6.1 | -6.0 |
| NHO2 | Nitrous acid, trans | -40.7 | -40.7 |
| CNH3O2 | Methyl nitrite | -36.7 | -36.7 |
| CNH3O2 | Nitromethane | 3.3 | 3.3 |
| C2NH5O2 | Ethyl nitrite | -42 | -42.0 |

| Formula | Name | Value 1 | Value 2 |
|---|---|---|---|
| C2NH5O2 | Glycine | -95.7 | -95.7 |
| C2NH5O2 | Methyl carbamate | -87.3 | -89.0 |
| C2NH5O2 | Nitroethane | -2.1 | -2.1 |
| C3NH7O2 | Alanine | -98.8 | -98.7 |
| C3NH7O2 | beta-Alanine | -99.1 | -99.0 |
| C3NH7O2 | Isopropylnitrite | -44.8 | -44.7 |
| C3NH7O2 | N-Methyl glycine | -94.1 | -94.0 |
| C3NH7O2 | Propyl nitrite | -47 | -47.0 |
| C3NH7O2 | Urethane | -92.6 | -94.3 |
| C4NH5O2 | Methyl cyanoacetate | -57.1 | -57.1 |
| C4NH5O2 | Succinimide | -87.8 | -87.7 |
| C4NH9O2 | 2-Nitrobutane | -10.2 | -10.2 |
| C4NH9O2 | 2-Nitroisobutane | -3.7 | -3.7 |
| C4NH9O2 | 4-Aminobutanoic acid | -104.3 | -104.3 |
| C4NH9O2 | Isobutyl nitrite | -48 | -47.9 |
| C4NH9O2 | n-Butyl nitrite | -51.6 | -51.6 |
| C4NH9O2 | Sec-butyl nitrite | -48.3 | -48.2 |
| C4NH9O2 | t-Butyl nitrite | -42.6 | -42.6 |
| C4NH11O2 | Diethanolamine | -103.4 | -103.4 |
| C5NH5O2 | N-Methylmaleimide | -51.2 | -51.1 |
| C5NH7O2 | Glutarimide | -92.1 | -92.0 |
| C5NH7O2 | N-Methylsuccinimide | -86.1 | -86.1 |
| C5NH9O2 | Proline | -101 | -101.0 |
| C5NH11O2 | 5-Aminovaleric acid | -108.4 | -109.1 |
| C5NH11O2 | N,N-Dimethylglycine methyl ester | -81.9 | -81.9 |
| C5NH11O2 | tert-Pentyl nitrite | -44.2 | -44.2 |
| C5NH11O2 | Valine | -101.1 | -101.1 |
| C5NH13O2 | 1,1-Dimethoxy-trimethylamine | -84.5 | -84.4 |
| C6NH5O2 | Niacin | -60.1 | -60.1 |
| C6NH5O2 | Nitrobenzene | 37.5 | 35.8 |
| C6NH9O2 | Ethyl 2-cyanopropionate | -64.7 | -64.6 |
| C6NH13O2 | Ethyl N,N-dimethylglycinate | -86.5 | -86.4 |
| C6NH13O2 | Hexanoic acid, 6-amino- | -112.9 | -112.9 |
| C6NH13O2 | Isoleucine | -103.3 | -103.2 |
| C6NH13O2 | Leucine | -106.8 | -106.7 |
| C6NH13O2 | Methyl N,N-dimethylalaninate | -81.7 | -81.6 |
| C6NH15O2 | N,N-Dimethylacetamide dimethyl acetal | -78.5 | -78.4 |
| C7NH7O2 | m-Aminobenzoic acid | -67 | -67.0 |
| C7NH7O2 | o-Aminobenzoic acid | -66.4 | -66.3 |
| C7NH7O2 | p-Aminobenzoic acid | -67.8 | -67.7 |
| C7NH7O2 | p-Nitrotoluene | 28 | 28.1 |
| C7NH7O2 | Phenylnitromethane | 29 | 29.0 |
| C7NH15O2 | Methyl N,N-,a,a-tetramethylglycinate | -73.1 | -73.0 |
| C2N2H4O2 | Oxalamide | -73.3 | -76.3 |
| C2N2H6O2 | N-Nitrodimethylamine | 22.3 | 22.3 |
| C3N2H6O2 | Acetyl-urea | -80.1 | -82.4 |

| Formula | Name | Value 1 | Value 2 |
|---|---|---|---|
| C3N2H6O2 | Propanediamide | -80 | -82.2 |
| C4N2H4O2 | Pyrazine-1,4-dioxide | 71.7 | 71.7 |
| C4N2H4O2 | Uracil | -64.8 | -64.7 |
| C5N2H6O2 | Thymine | -72.7 | -72.5 |
| C6N2H6O2 | m-Nitroaniline | 36.4 | 36.5 |
| C6N2H6O2 | p-Nitroaniline | 35.5 | 35.6 |
| C6N2H14O2 | Lysine | -104.4 | -104.4 |
| C2N3H5O2 | Imidodicarbonic diamide | -76.3 | -77.8 |
| O3 | Ozone | 48.5 | 48.5 |
| C3H6O3 | 1,3,5-Trioxane | -130.3 | -130.2 |
| C3H6O3 | Methyl hydroxyacetate | -137.1 | -137.0 |
| C3H8O3 | Glycerol | -150.8 | -150.8 |
| C4H2O3 | Malaic anhydride | -88.6 | -88.5 |
| C4H6O3 | Acetic anhydride | -132.7 | -132.6 |
| C4H10O3 | Trimethoxymethane | -136 | -135.9 |
| C6H14O3 | 2,5,8-Trioxanonane | -135.5 | -135.4 |
| C7H6O3 | m-Salicylic acid | -115.2 | -115.1 |
| C7H6O3 | o-Salicylic acid | -114.2 | -114.1 |
| C7H6O3 | p-Salicylic acid | -114.6 | -116.0 |
| C7H14O3 | 2,3-Butanediol, 2,3-dimethyl-, monoformate | -121.8 | -121.8 |
| NO3 | Nitrate anion | -67.1 | -66.9 |
| NHO3 | Nitric acid | -17.6 | -17.4 |
| CNH3O3 | Methyl nitrate | -12.5 | -12.3 |
| C2NH3O3 | Oxamic acid | -126.5 | -127.3 |
| C2NH5O3 | Ethyl nitrate | -18 | -17.8 |
| C3NH7O3 | Serine | -141.3 | -141.3 |
| C4NH3O3 | 2-Nitrofuran | 7.8 | 7.9 |
| C4NH9O3 | Threonine | -142.6 | -142.5 |
| C6NH5O3 | m-Nitrophenol | -11.5 | -11.4 |
| C6NH5O3 | o-Nitrophenol | -10.3 | -10.2 |
| C6NH5O3 | p-Nitrophenol | -12.6 | -12.5 |
| N2O3 | Dinitrogen trioxide | 13.6 | 12.5 |
| C2H2O4 | Oxalic acid | -170.6 | -170.5 |
| C2H6O4 | Dioxybismethanol | -129.2 | -129.1 |
| C4H4O4 | 1,4-Dioxan-2,5-dione | -154.4 | -154.3 |
| C4H6O4 | Dimethyl oxalate | -164.1 | -164.0 |
| C4H8O4 | 1,3,5,7-Tetroxane | -168.5 | -168.3 |
| C5H8O4 | Dimethyl malonate | -170.4 | -170.4 |
| C5H8O4 | Ethylmalonic acid | -190.3 | -190.2 |
| C5H8O4 | Methylene diacetate | -178 | -178.2 |
| C5H12O4 | Tetramethoxymethane | -181.3 | -181.1 |
| C6H10O4 | 1,1-Diacetoxyethane | -180.6 | -180.5 |
| C6H10O4 | Dimethyl methylmalonate | -172.2 | -172.1 |
| C4NH7O4 | Aspartic acid | -182.8 | -182.7 |
| N2O4 | Dinitrogen tetroxide | 29.9 | 30.2 |
| CN2H2O4 | Dinitromethane | 27.8 | 27.9 |

| Formula | Name | Value 1 | Value 2 |
|---|---|---|---|
| CN3H3O4 | Methyldinitramine | 53 | 53.2 |
| C3H4O5 | Tartronic acid | -226.5 | -226.4 |
| C5H10O5 | 1,3,5,7,9-pentaoxecane | -209.1 | -208.8 |
| N2O5 | Dinitrogen pentoxide | 34.1 | 34.4 |
| C10H10O | 4-Phenyl-3-buten-2-one | -2.9 | -2.8 |
| C10H14O | 2-Adamantone | -53.7 | -53.7 |
| C10H14O | 2-Isopropyl-4-methylphenol | -42.6 | -42.5 |
| C10H14O | 2-Isopropyl-5-methylphenol | -44.8 | -44.8 |
| C10H14O | 2-Isopropyl-6-methylphenol | -40.6 | -40.5 |
| C10H14O | 2-Methyl-5-isopropylphenol | -45.5 | -45.5 |
| C10H14O | 3-Isopropyl-2-methylphenol | -39.1 | -39.1 |
| C10H14O | 3-Methyl-2-isopropylphenol | -36.5 | -38.8 |
| C10H14O | 3-Methyl-5-isopropylphenol | -47 | -46.9 |
| C10H14O | 4-Isopropyl-2-methylphenol | -45.6 | -45.6 |
| C10H14O | 4-Isopropyl-3-methylphenol | -40.4 | -40.3 |
| C10H14O | 4-Methyl-3-isopropylphenol | -40.3 | -40.3 |
| C10H14O | m-tert-Butylphenol | -32.1 | -32.1 |
| C10H14O | o-sec-Butylphenol | -38.2 | -38.1 |
| C10H14O | o-tert-Butylphenol | -26.8 | -26.7 |
| C10H14O | p-sec-Butylphenol | -43.4 | -43.4 |
| C10H14O | p-tert-Butyl phenol | -32.7 | -32.6 |
| C10H16O | 1-Adamantol | -65.9 | -65.9 |
| C10H16O | 2-Adamantol | -65.2 | -65.2 |
| C10H16O | Camphor | -33.9 | -33.9 |
| C10H16O | Octahydro-3a-methyl-cis-2H-inden-2-one | -60.7 | -60.7 |
| C10H16O | Octahydro-3a-methyl-trans-2H-inden-2-one | -54.2 | -54.2 |
| C10H18O | Beta-caran-3-ol | -45.3 | -46.8 |
| C10H18O | Cyclodecanone | -63.2 | -63.3 |
| C10H20O | 2,2,5,5-Tetramethyl-3-hexanone | -54.6 | -54.6 |
| C10H22O | Decanol | -100.8 | -100.8 |
| C10H22O | Dipentyl ether | -89.4 | -89.4 |
| C11H14O | 2,4,5-Trimethyl-acetophenone | -35.4 | -35.4 |
| C11H14O | 2,4,6-Trimethyl-acetophenone | -34.7 | -34.7 |
| C11H16O | 2-tert-Butyl-p-cresol | -34.3 | -34.3 |
| C11H16O | 3-Methyl-2-phenylbutane-2-ol | -25.7 | -25.6 |
| C11H20O | Cycloundecanone | -66.3 | -66.4 |
| C11H22O | 2,2,6,6-Tetramethyl-4-heptanone | -60.4 | -60.4 |
| C11H22O | Dipentyl ketone | -86.7 | -87.8 |
| C11H24O | Decyl methyl ether | -94.4 | -94.5 |
| C12H8O | Dibenzofuran | 14.2 | 14.4 |
| C12H10O | m-Hydroxybiphenyl | -1.8 | -1.7 |
| C12H10O | o-Hydroxybiphenyl | -0.8 | -0.7 |
| C12H10O | p-Hydroxybiphenyl | -2.1 | -2.1 |
| C12H16O | Isobutyl phenyl ketone | -31.9 | -31.9 |
| C12H18O | 2,6-Diisopropylphenol | -44 | -44.7 |
| C13H10O | Benzophenone | 15.9 | 16.0 |

| Formula | Name | Value 1 | Value 2 |
|---|---|---|---|
| C14H10O | Anthone | 12.2 | 12.3 |
| C9NH7O | 2(1H)-Quinolinone | -3.4 | -3.4 |
| C9NH7O | 3-Phenyl isoxazole | 40.5 | 40.6 |
| C9NH7O | 4-Quinolinol | -3.1 | -3.1 |
| C9NH7O | 5-Phenyl isoxazole | 41.5 | 41.6 |
| C9NH7O | 8-Quinolinol | -4.7 | -4.6 |
| C9NH7O | a-Cyanoacetophenone | 17.8 | 17.9 |
| C9NH11O | 2,4,6-Trimethylnitrosobenzene | 9.2 | 9.3 |
| C9NH11O | N,N-Dimethyl benzamide | -9.3 | -9.3 |
| C9NH13O | N,N-Dimethyamino-2,4,6-heptatriene-7-al | 13.7 | 13.8 |
| C9NH17O | 2,2,6,6-Tetramethyl-4-piperidinone | -39.6 | -39.7 |
| C10NH9O | 2-Methyl-4-hydroxyquinoline | -11.8 | -11.8 |
| C10NH9O | 2-Methyl-8-quinolinol | -13.6 | -13.6 |
| C10NH9O | 3-Methyl-5-phenyl isoxazole | 29.7 | 29.9 |
| C10NH9O | 4-Methyl-2-hydroxyquinoline | -14.8 | -14.7 |
| C10NH9O | 5-Methyl-3-phenyl isoxazole | 29.5 | 29.6 |
| C10NH9O | beta-Cyanopropiophenone | 12.3 | 12.4 |
| C10NH11O | 2,4,6-Trimethylbenzonitrile, N-oxide | 48.5 | 48.6 |
| C10NH13O | 2-(Dimethylamino)-acetophenone | -4.7 | -4.7 |
| C10NH13O | N,N,4-Trimethyl benzamide | -17.1 | -17.1 |
| C11NH13O | (E)-3-(Methylamino)-1-phenyl-but-2-enone | 2.8 | 2.9 |
| C11NH15O | 1-Propanone, 2-(dimethylamino)-1-phenyl- | -4.8 | -4.9 |
| C11NH15O | 2-Propanamine, 2-methyl-N-(phenylmethylene)-, N-oxide | 50.3 | 50.4 |
| C11NH17O | 1-Adamantanecarboxamide | -51.2 | -52.4 |
| C11NH23O | N,N-Dimethylnonamide | -74 | -74.1 |
| C12NH9O | Phenoxazine | 16.6 | 16.8 |
| C12NH17O | 2-(Diethylamino)-1-phenylethanone | -12 | -12.0 |
| C13NH9O | 9,10-Dihydro-9-oxoacridine | 17.6 | 17.7 |
| C13NH11O | Benzenamine, N-(phenylmethylene)-, N-oxide | 85.9 | 86.1 |
| C13NH19O | 1-Propanone, 2-(diethylamino)-1-phenyl- | -11.7 | -11.8 |
| C13NH21O | N,N-Dimethyl-1-adamantylcarboxamide | -43.8 | -43.8 |
| C9H10O2 | 1,3-Dioxolane-2-phenyl | -61.3 | -61.2 |
| C9H10O2 | 2,3-Dimethylbenzoic acid | -77.4 | -77.4 |
| C9H10O2 | 2,4-Dimethylbenzoic acid | -81 | -80.9 |
| C9H10O2 | 2,5-Dimethylbenzoic acid | -81.1 | -81.0 |
| C9H10O2 | 2,6-Dimethylbenzoic acid | -78.2 | -78.1 |
| C9H10O2 | 3,4-Dimethylbenzoic acid | -80.6 | -80.6 |
| C9H10O2 | 3,5-Dimethylbenzoic acid | -83 | -82.9 |
| C9H10O2 | 3-Ethylbenzoic acid | -80.2 | -80.2 |
| C9H10O2 | 4-Ethylbenzoic acid | -80.3 | -80.3 |
| C9H10O2 | Methyl 4-methylbenzoate | -65.9 | -65.8 |
| C10H8O2 | 1,2-Naphthalenediol | -54.7 | -54.7 |
| C10H8O2 | 1,3-Naphthalenediol | -56.5 | -56.5 |
| C10H8O2 | 1,4-Naphthalenediol | -53.9 | -53.9 |
| C10H8O2 | 2,3-Naphthalenediol | -55.8 | -55.7 |
| C10H10O2 | 1-Phenyl-1,3-butanedione | -51.2 | -51.9 |

| Formula | Name | | |
|---|---|---|---|
| C10H12O2 | 2,3,4-Trimethylbenzoic acid | -80.9 | -80.9 |
| C10H12O2 | 2,3,5-Trimethylbenzoic acid | -85 | -85.0 |
| C10H12O2 | 2,3,6-Trimethylbenzoic acid | -82 | -81.9 |
| C10H12O2 | 2,4,5-Trimethylbenzoic acid | -86.2 | -86.1 |
| C10H12O2 | 2,4,6-Trimethylbenzoic acid | -85.7 | -85.6 |
| C10H12O2 | 2-Isopropyl benzoic acid | -77.5 | -77.5 |
| C10H12O2 | 2-Methyl-2-phenyl-1,3-dioxolane | -62.2 | -62.2 |
| C10H12O2 | 3,4,5-Trimethylbenzoic acid | -82.1 | -82.1 |
| C10H12O2 | 3-Isopropyl benzoic acid | -80.8 | -80.7 |
| C10H12O2 | 4-Isopropyl benzoic acid | -80.9 | -80.9 |
| C10H14O2 | 2-Isopropyl-6-methyl-pyrocatechol | -89.7 | -89.7 |
| C10H18O2 | Cyclohexyl butanoate | -119.1 | -119.1 |
| C10H20O2 | Ethyl octanoate | -127 | -127.0 |
| C10H22O2 | 1,10-Decanediol | -143.5 | -143.6 |
| C11H8O2 | 1-Naphthoic acid | -48.4 | -48.3 |
| C11H8O2 | Isonaphthoic acid | -50.6 | -50.5 |
| C11H14O2 | 2,3,4,5-Tetramethylbenzoic acid | -84.5 | -84.4 |
| C11H14O2 | 2,3,4,6-Tetramethylbenzoic acid | -85.5 | -85.4 |
| C11H14O2 | 2,3,5,6-Tetramethylbenzoic acid | -85.8 | -85.8 |
| C11H14O2 | p-tert-Butyl benzoic acid | -73.6 | -73.6 |
| C11H20O2 | Oxacyclododecan-2-one | -111.8 | -111.8 |
| C11H22O2 | Ethyl nonanoate | -131.7 | -131.7 |
| C11H24O2 | 1,1-Dibutoxypropane | -128.5 | -128.5 |
| C12H16O2 | 5,5-Dimethyl-2-phenyl-1,3-dioxane | -62 | -61.9 |
| C12H16O2 | Pentamethylbenzoic acid | -83.2 | -83.2 |
| C12H22O2 | 2,2,6,6-Tetramethyl-3,5-heptanedione | -89.9 | -90.0 |
| C12H24O2 | Ethyl decanoate | -136.4 | -136.4 |
| C13H8O2 | Xanthone | -22.3 | -22.3 |
| C13H10O2 | Phenyl benzoate | -27 | -26.9 |
| C8NH9O2 | 2,6-Dimethylnitrobenzene | 24.6 | 24.8 |
| C8NH9O2 | 2-Amino-2-phenylacetic acid | -63.8 | -63.7 |
| C8NH9O2 | 2-Nitro-m-xylene | 24.6 | 24.8 |
| C8NH9O2 | N-Phenylglycine | -62.6 | -62.6 |
| C8NH17O2 | 8-Aminocaprylic acid | -122.5 | -123.3 |
| C9NH7O2 | 2-Methyl-1H-isoindole-1,3(2H)-dione | -45.9 | -45.9 |
| C9NH11O2 | Nitromesitylene | 17.2 | 17.3 |
| C9NH11O2 | Phenylalanine | -68.8 | -68.8 |
| C10NH7O2 | 1-Nitroso-2-naphthalenol | -0.4 | -0.4 |
| C10NH13O2 | N,N-Dimethyl 4-methoxybenzamide | -49 | -49.0 |
| C10NH15O2 | 1-Nitroadamantane | -4.6 | -4.6 |
| C10NH15O2 | 2-Nitroadamantane | -3.8 | -3.8 |
| C8N2H8O2 | Isophthalamide | -51.4 | -53.3 |
| C8N2H8O2 | Teraphthalamide | -51.2 | -53.0 |
| C8N2H10O2 | N,N-Dimethyl-m-nitroaniline | 43.4 | 43.6 |
| C8N2H10O2 | N,N-Dimethyl-p-nitroaniline | 43.3 | 43.4 |
| C11N2H12O2 | Tryptophan | -48.3 | -48.2 |

| Formula | Name | | |
|---|---|---|---|
| C6N3H9O2 | Histidine | -60.4 | -61.3 |
| C6N4H14O2 | Argenine | -75.7 | -75.6 |
| C10H14O3 | Trimethoxymethyl benzene | -98.2 | -98.1 |
| C10H22O3 | 1-tert-Butoxy-3-propoxy-2-propanol | -151.7 | -151.7 |
| C11H24O3 | 1-Butoxy-3-tert-butyl-2-propanol | -157.2 | -157.2 |
| C9NH11O3 | Tyrosine | -117 | -117.0 |
| C4N2H4O3 | Barbituric acid | -118.8 | -118.7 |
| C4N2H8O3 | Asparagine | -129 | -130.4 |
| C4N2H8O3 | GLY-GLY | -128.9 | -129.2 |
| C5N2H10O3 | Glutamine | -132.8 | -134.4 |
| C6N2H12O3 | ALA-ALA | -133.9 | -134.2 |
| C9N2H6O3 | 8-Hydroxy-5-nitroquinoline | 11.6 | 11.8 |
| C10N2H16O3 | PRO-PRO | -136.1 | -136.1 |
| C10N2H20O3 | VAL-VAL | -140.1 | -140.2 |
| C3N3H3O3 | 1,3,5-Triazine-2,4,6(1H,3H,5H)-trione | -119 | -118.8 |
| C6N3H9O3 | 1,3,5-Trimethyl-s-triazine-2,4,6-trione | -103.7 | -103.5 |
| C6N3H9O3 | 2,4,6-Trimethoxy-s-triazine | -100 | -99.7 |
| C7H12O4 | 2,2-Diacetoxypropane | -171 | -171.0 |
| C7H12O4 | Diethyl malonate | -181.1 | -181.0 |
| C7H12O4 | Dimethyl dimethylmalonate | -166.6 | -166.5 |
| C7H16O4 | 3,5,7,9-Tetraoxyundecane | -184.3 | -189.8 |
| C8H16O4 | 12-Crown-4 | -163.9 | -164.0 |
| C10H6O4 | 5,8-Dihydroxy-1,4-naphthalenedione | -113.6 | -113.6 |
| C10H10O4 | Dimethyl isophthalate | -141.1 | -141.0 |
| C10H10O4 | Dimethyl phthalate | -138.3 | -138.2 |
| C10H22O4 | 1-(tert-Butyldioxy)-3-propoxy-2-propanol | -130 | -130.0 |
| C11H12O4 | Benzal diacetate | -146 | -145.8 |
| C11H24O4 | 1-Butoxy-1-tert-butyldioxy-2-propanol | -139 | -138.9 |
| C6NH5O4 | 4-Nitrocatechol | -59.3 | -59.2 |
| C7NH5O4 | p-Nitrobenzoic acid | -51.2 | -51.2 |
| C6N2H4O4 | m-Dinitrobenzene | 54.6 | 54.8 |
| C6N2H4O4 | o-Dinitrobenzene | 54.6 | 54.8 |
| C6N2H4O4 | p-Dinitrobenzene | 54.4 | 54.6 |
| C7N2H6O4 | 2,4-Dinitrotoluene | 48.7 | 48.9 |
| C7N2H6O4 | Dinitrophenylmethane | 56 | 56.2 |
| C8H18O5 | 3,5,7,9,11-Pentaoxa-tridecane | -230.1 | -235.4 |
| C10H20O5 | 15-Crown-5 | -208.6 | -208.6 |
| C6N2H12O5 | SER-SER | -220.2 | -220.9 |
| C8N2H16O5 | THR-THR | -217.4 | -217.6 |
| C7H10O6 | Trimethyl methanetricarboxylate | -242.6 | -242.5 |
| C2N3H3O6 | 1,1,1-Trinitroethane | 56.2 | 56.4 |
| CN4O8 | Tetranitromethane | 94.8 | 95.0 |
| C3N3H5O9 | Glycerol trinitrate | -6.7 | -6.3 |
| C14NH21O | 4-Isopropylbenzylidene t-butylamine N-oxide | 37.1 | 37.2 |
| C14H8O2 | 9,10-Anthroquinone | -10.7 | -13.2 |
| C14H8O2 | 9,10-Phenanthroquinone | -10.7 | -10.7 |

| Formula | Name | Value 1 | Value 2 |
|---|---|---|---|
| C13NH11O2 | Phenol, 2-[(phenylimino)methyl]-, N-oxide | 38.1 | 38.2 |
| C14NH13O2 | Benzenamine, N-[(4-methoxyphenyl)methylene]-, N-oxide | 45.7 | 45.8 |
| C15NH17O2 | N-(3-Phenoxy-2-hydroxypropyl)aniline | -28.4 | -28.4 |
| C14H10O3 | Benzoic acid, anhydride | -66.3 | -66.1 |
| C12N2H24O3 | ILE-ILE | -144.5 | -144.8 |
| C12N2H24O3 | LEU-LEU | -150.2 | -150.6 |
| C12H14O4 | 1,1-Ethanediol, 2-phenyl-, diacetate | -150.8 | -151.0 |
| C12H14O4 | Benzyl diacetate | -151.1 | -151.0 |
| C14H6O4 | 1,4,9,10-Anthracenetetrone | -62.1 | -62.1 |
| C14H8O4 | 1,4-Dihydroxy-9,10-anthracenedione | -105 | -104.9 |
| C14H10O4 | Diphenyl oxalate | -97.2 | -97.0 |
| C14H12O4 | Dimethyl naphthalene-2,6-dicarboxylate | -124.1 | -124.0 |
| C10H16O6 | Triethyl methanetricarboxylate | -258.6 | -258.5 |
| C11H18O6 | Triethyl 1,1,1-ethanetricarboxylate | -253 | -252.9 |
| C7N3H5O6 | 2,4,6-Trinitrotoluene | 74.4 | 74.6 |
| C9NH9O7 | 2-(Diacetoxymethyl)-5-nitrofuran | -161.4 | -161.3 |
| C7N3H5O7 | 2,4,6-Trinitroanisole | 39.7 | 39.9 |
| C8N3H7O7 | 2,4,6-Trinitrophenetole | 36.6 | 36.7 |
| F | Fluorine, atom | 18.9 | 18.9 |
| F | Fluoride, anion | -17.1 | -17.1 |
| HF | Hydrogen fluoride | -59.7 | -59.7 |
| CF | Fluoromethylidyne radical | 38.2 | 38.2 |
| CH2F | Fluoromethyl, cation | 182.8 | 182.8 |
| CH3F | Fluoromethane | -60.9 | -60.9 |
| C2HF | Fluoroacetylene | 15.6 | 15.6 |
| C2H3F | Fluoroethylene | -34.5 | -34.6 |
| C2H4F | CH3CHF, cation | 164.7 | 164.7 |
| C2H5F | Fluoroethane | -65.1 | -65.1 |
| C3H7F | 2-Fluoropropane | -66.6 | -66.7 |
| C6H5F | Fluorobenzene | -25.3 | -25.3 |
| C6H11F | Fluorocyclohexane | -76.5 | -76.5 |
| C9H19F | 1-Fluorononane | -98.5 | -98.5 |
| CNF | Cyanogen fluoride | -2.4 | -2.3 |
| OF | Fluorine oxide | 20.4 | 20.4 |
| HOF | Hypofluorous acid | -18.6 | -18.6 |
| CHOF | HCOF | -88.8 | -88.8 |
| C2H3OF | Acetyl fluoride | -96.5 | -96.5 |
| NOF | Nitrosyl fluoride | -24.8 | -24.7 |
| O2F | Fluorine dioxide | 22.7 | 22.8 |
| C7H5O2F | m-fluorobenzoic acid | -113.3 | -113.2 |
| NO2F | Fluorine nitrite | 0.6 | 0.8 |
| NO3F | Fluorine nitrate | 27.9 | 28.2 |
| CN2HO4F | Fluorodinitromethane | -18.3 | -18.2 |
| F2 | Fluorine molecule | 7.3 | 7.4 |
| CF2 | Difluoromethylene | -65.2 | -65.2 |
| CHF2 | Difluoromethyl, cation | 132.4 | 132.4 |

| Formula | Name | Value 1 | Value 2 |
|---|---|---|---|
| CH2F2 | Difluoromethane, cation | 177.6 | 177.6 |
| CH2F2 | Difluoromethane | -111.8 | -111.7 |
| C2F2 | Difluoroacetylene | -21 | -21.0 |
| C2H2F2 | gem-Difluoroethylene | -83.7 | -83.7 |
| C2H3F2 | CH3CF2, cation | 116.5 | 116.6 |
| C2H4F2 | 1,1-Difluoroethane | -113.5 | -113.4 |
| C6H4F2 | 1,2-Difluorobenzene | -70.7 | -70.6 |
| C6H4F2 | 1,3-Difluorobenzene | -71 | -71.0 |
| C6H4F2 | 1,4-Difluorobenzene | -71.1 | -71.1 |
| C4NH9F2 | t-Butyldifluoroamine | -21.2 | -21.1 |
| C7NH7F2 | N,N'-Difluorobenzylamine | 11.3 | 11.3 |
| N2F2 | cis-Difluorodiazene | -2.3 | -2.3 |
| N2F2 | trans-Difluorodiazene | 2.3 | 2.3 |
| OF2 | Difluorine oxide | 18.2 | 18.3 |
| COF2 | Carbonyl fluoride | -138.6 | -138.5 |
| CF3 | Trifluoromethyl, cation | 100.9 | 101.0 |
| CF3 | Trifluoromethyl | -138.7 | -138.6 |
| CF3 | Trifluoromethyl, anion | -178.8 | -178.7 |
| CHF3 | Trifluoromethane | -163.8 | -163.7 |
| C2HF3 | Trifluoroethylene | -131.1 | -131.1 |
| C2H2F3 | CF3CH2, cation | 121.2 | 121.3 |
| C2H2F3 | CH2F.CF2, radical cation | 82.4 | 82.4 |
| C2H2F3 | CF3CH2 radical | -131.2 | -131.1 |
| C2H3F3 | 1,1,1-Trifluoroethane | -164.4 | -164.3 |
| C7H5F3 | Trifluoromethylbenzene | -127.7 | -127.5 |
| NF3 | Nitrogen trifluoride | -34.3 | -34.1 |
| C2NF3 | Trifluoroacetonitrile | -113.3 | -113.1 |
| C2NF3 | Trifluoromethylisocyanide | -90.5 | -90.3 |
| NOF3 | F3NO | 22.7 | 23.0 |
| C2HO2F3 | Trifluoroacetic acid | -238.2 | -238.0 |
| CF4 | Carbon tetrafluoride | -214.2 | -214.0 |
| C2F4 | Tetrafluoroethylene | -175.7 | -175.6 |
| C6H2F4 | 1,2,4,5-Tetrafluorobenzene | -159.3 | -159.2 |
| N2F4 | Tetrafluorohydrazine | -16.2 | -19.6 |
| COF4 | Perfluoromethanol | -163.4 | -163.1 |
| CO2F4 | Bis(fluoroxy)perfluoromethane | -112.5 | -112.1 |
| CNF5 | Pentafluoromethylamine | -163.4 | -163.1 |
| CN3F5 | Pentafluoroguanidine | 6.1 | 6.2 |
| C2F6 | Hexafluoroethane | -299.7 | -299.4 |
| C4F6 | Perfluorobutadiene | -250.1 | -250.0 |
| CN2F6 | Hexafluorodimethylamine | -111.7 | -111.9 |
| C2OF6 | Dimethyl perfluoroether | -357.5 | -357.3 |
| C3OF6 | Perfluoroacetone | -322 | -322.2 |
| C7N2H5O4F | Fluorodinitrophenylmethane | 14.5 | 14.6 |
| C7H4F4 | 1-Fluoro-3-(trifluoro-methyl)benzene | -172.5 | -172.3 |
| C6N2H10F4 | N,N,N',N'-tetrafluoro-1,1-cyclohexanediamine | -24.8 | -24.7 |
| C6N2H12F4 | N,N,N',N'-Tetrafluoro-4-methyl-1,2-pentane | -26.6 | -26.5 |
| C7N2H14F4 | 1,1-Bis(difluoroamine)heptane | -42.5 | -42.5 |
| C6HF5 | Pentafluorobenzene | -201.8 | -201.7 |
| C7H3F5 | 2,3,4,5,6-Pentafluorotoluene | -206.6 | -206.6 |
| C6HOF5 | Pentafluorophenol | -247.9 | -247.9 |
| C6F6 | Hexafluorobenzene | -243.6 | -243.5 |
| C3F8 | Perfluoropropane | -384.6 | -384.3 |
| C4F8 | Perfluorobut-2-ene | -366.2 | -365.9 |
| C4F8 | Perfluorocyclobutane | -363.8 | -363.7 |
| C7F8 | Octafluorotoluene | -340.6 | -340.4 |
| CN4F8 | Octafluoromethanetetramine | -4 | -3.8 |
| C3O2F8 | Perfluorodimethoxymethane | -502.1 | -501.3 |
| C4F10 | n-Perfluorobutane | -469.4 | -469.0 |
| C7H6O4F6 | Hexafluoropentanedioic acid, dimethyl ester | -420 | -419.7 |
| C6F10 | Decafluorocyclohexene | -423.3 | -423.1 |
| C5NF11 | Undecafluoropiperidine | -437.2 | -436.9 |
| C2N5F11 | Tetrakis(difluoroamine)-N-1,1-trifluorodimethylamine | -85.7 | -85.4 |
| C6F12 | Dodecafluorocyclohexane | -522.8 | -522.5 |

Calculated Values for $I.E.$ Using the MNDO Formalism (given in eV)

| Molecular Formula | Molecule Name | MOPAC | Our Program |
|---|---|---|---|
| H2 | Hydrogen | 15.75 | 15.75 |
| CH4 | Methane | 13.86 | 13.87 |
| C2H2 | Acetylene | 11.01 | 11.01 |
| C2H4 | Ethylene | 10.18 | 10.18 |
| C2H6 | Ethane | 12.7 | 12.70 |
| C3 | Carbon, trimer | 11.04 | 11.05 |
| C3H4 | Allene | 10.02 | 10.02 |
| C3H4 | Cyclopropene | 9.88 | 9.88 |
| C3H4 | Propyne | 10.72 | 10.72 |
| C3H6 | Cyclopropane | 11.43 | 11.43 |
| C3H6 | Propene | 9.96 | 9.96 |
| C3H8 | Propane | 12.34 | 12.34 |
| C4H2 | Diacetylene | 9.99 | 9.99 |
| C4H4 | Vinylacetylene | 9.5 | 9.50 |
| C4H4 | Butatriene | 9.01 | 9.01 |
| C4H6 | 1,2-Butadiene | 9.84 | 9.84 |
| C4H6 | 1,3-Butadiene | 9.14 | 9.14 |
| C4H6 | 1-Butyne | 10.68 | 10.68 |
| C4H6 | 2-Butyne | 10.47 | 10.47 |
| C4H6 | Cyclobutene | 9.77 | 9.77 |
| C4H8 | 1-Butene | 9.95 | 9.97 |
| C4H8 | Cyclobutane | 11.8 | 11.81 |
| C4H10 | Isobutane | 12.12 | 12.12 |
| C4H10 | n-Butane, trans | 12.21 | 12.21 |
| C5H6 | Cyclopentadiene | 9.04 | 9.04 |
| C5H8 | Cyclopentene | 9.72 | 9.72 |
| C5H10 | Cyclopentane | 12.06 | 12.06 |
| C5H12 | n-Pentane | 12.16 | 12.16 |
| C5H12 | Neopentane | 12.11 | 12.12 |
| C6H6 | Benzene | 9.39 | 9.39 |
| C6H10 | Cyclohexene | 9.75 | 9.75 |
| C6H12 | Cyclohexane | 11.74 | 11.74 |
| C7H8 | Cycloheptatriene | 8.58 | 8.72 |
| C7H8 | Toluene | 9.28 | 9.28 |
| C8H10 | Ethylbenzene | 9.28 | 9.28 |
| C8H14 | Bicyclo(2.2.2)-octane | 11.4 | 11.41 |
| C10H8 | Naphthalene | 8.58 | 8.57 |
| C10H16 | Adamantane | 11.27 | 11.27 |
| C14H10 | Anthracene | 8.05 | 8.05 |
| NH3 | Ammonia | 11.19 | 11.19 |
| CNH | Hydrogen cyanide | 13.41 | 13.41 |
| CNH5 | Methylamine | 10.56 | 10.56 |
| C2NH3 | Acetonitrile | 12.79 | 12.79 |

| Formula | Name | Value 1 | Value 2 |
|---|---|---|---|
| C2NH3 | Methyl isocyanide | 12.24 | 12.24 |
| C2NH5 | Ethyleneimine (Azirane) | 10.68 | 10.68 |
| C2NH7 | Dimethylamine | 10.04 | 10.04 |
| C2NH7 | Ethylamine | 10.5 | 10.50 |
| C3NH3 | Acrylonitrile | 10.61 | 10.61 |
| C3NH5 | Ethyl cyanide | 12.59 | 12.59 |
| C3NH9 | Trimethylamine | 9.59 | 9.59 |
| C4NH5 | Pyrrole | 8.56 | 8.56 |
| C5NH5 | Pyridine | 9.69 | 9.69 |
| C6NH7 | Aniline | 8.75 | 8.75 |
| C7NH5 | Phenyl cyanide | 9.81 | 9.81 |
| N2 | Nitrogen | 14.87 | 14.87 |
| CN2H2 | Diazomethane | 8.66 | 8.67 |
| CN2H6 | Methylhydrazine | 9.66 | 9.66 |
| C2N2 | Cyanogen | 13.2 | 13.20 |
| H2O | Water | 12.19 | 12.19 |
| CO | Carbon monoxide | 13.43 | 13.43 |
| CH2O | Formaldehyde | 11.04 | 11.04 |
| CH4O | Methanol | 11.41 | 11.41 |
| C2H2O | Ketene | 9.29 | 9.29 |
| C2H4O | Acetaldehyde | 10.88 | 10.88 |
| C2H4O | Ethylene oxide | 11.49 | 11.49 |
| C2H6O | Dimethyl ether | 11.04 | 11.04 |
| C2H6O | Ethanol | 11.3 | 11.30 |
| C3H6O | Acetone | 10.75 | 10.75 |
| C3H6O | Propanal | 10.82 | 10.82 |
| C4H4O | Furan | 9.14 | 9.14 |
| C4H8O | Butanal | 10.81 | 10.80 |
| C4H10O | Diethyl ether | 10.91 | 10.91 |
| C7H6O | Benzaldehyde | 9.74 | 9.73 |
| C7H8O | Anisole | 8.84 | 8.84 |
| CNHO | Hydrogen isocyanate | 11.1 | 11.10 |
| CH2O2 | Formic acid | 11.74 | 11.74 |
| C2H2O2 | trans Glyoxal | 10.75 | 10.75 |
| C2H4O2 | Acetic acid | 11.57 | 11.57 |
| C2H4O2 | Methyl formate | 11.61 | 11.61 |
| C2H6O2 | Dimethyl peroxide | 10.57 | 10.57 |
| C3O2 | Carbon suboxide | 10.07 | 10.07 |
| C3H4O2 | beta-Propiolactone | 11.4 | 11.40 |
| C3H6O2 | Methyl acetate | 11.46 | 11.46 |
| C3H6O2 | Propionic acid | 11.52 | 11.51 |
| C3H8O2 | 2-Methoxyethanol | 11.02 | 11.02 |
| C5H8O2 | Acetylacetone | 10.79 | 10.78 |
| C7H6O2 | Benzoic acid | 9.76 | 9.77 |
| CNH3O2 | Methyl nitrite | 11.42 | 11.41 |
| C2NH5O2 | Ethyl nitrite | 11.36 | 11.36 |

| Formula | Name | | |
|---|---|---|---|
| C3NH7O2 | Alanine | 10.81 | 10.81 |
| C6NH5O2 | Nitrobenzene | 10.31 | 10.31 |
| O3 | Ozone | 12.71 | 12.71 |
| C4H2O3 | Malaic anhydride | 11.7 | 11.70 |
| C7H6O3 | o-Salicylic acid | 9.26 | 9.26 |
| C2H2O4 | Oxalic acid | 11.67 | 11.66 |
| N2O4 | Dinitrogen tetroxide | 12.05 | 12.05 |
| N2O5 | Dinitrogen pentoxide | 13.18 | 13.18 |
| HF | Hydrogen fluoride | 14.82 | 14.82 |
| CH3F | Fluoromethane | 13.05 | 13.05 |
| C2HF | Fluoroacetylene | 11.06 | 11.06 |
| C2H3F | Fluoroethylene | 10.17 | 10.17 |
| C2H5F | Fluoroethane | 12.61 | 12.61 |
| C3H7F | 2-Fluoropropane | 12.33 | 12.33 |
| C6H5F | Fluorobenzene | 9.47 | 9.47 |
| NOF | Nitrosyl fluoride | 12.93 | 12.93 |
| C7H5O2F | m-fluorobenzoic acid | 9.83 | 9.83 |
| NO2F | Fluorine nitrite | 12.99 | 12.99 |
| CH2F2 | Difluoromethane | 13.09 | 13.09 |
| C2F2 | Difluoroacetylene | 11.17 | 11.17 |
| C2H2F2 | gem-Difluoroethylene | 10.18 | 10.18 |
| C2H4F2 | 1,1-Difluoroethane | 12.73 | 12.73 |
| N2F2 | trans-Difluorodiazene | 13 | 13.00 |
| OF2 | Difluorine oxide | 13.52 | 13.52 |
| CHF3 | Trifluoromethane | 14.57 | 14.57 |
| C2HF3 | Trifluoroethylene | 10.46 | 10.46 |
| C2H3F3 | 1,1,1-Trifluoroethane | 14.01 | 14.01 |
| C7H5F3 | Trifluoromethylbenzene | 10.07 | 10.07 |
| NF3 | Nitrogen trifluoride | 13.93 | 13.93 |
| C2HO2F3 | Trifluoroacetic acid | 12.73 | 12.72 |
| CF4 | Carbon tetrafluoride | 16.81 | 16.81 |
| C2F4 | Tetrafluoroethylene | 10.74 | 10.74 |
| N2F4 | Tetrafluorohydrazine | 13.19 | 13.04 |
| C2F6 | Hexafluoroethane | 14.5 | 14.49 |
| C3OF6 | Perfluoroacetone | 13 | 13.00 |
| C6HF5 | Pentafluorobenzene | 10.4 | 10.40 |
| C6F6 | Hexafluorobenzene | 10.78 | 10.77 |

Calculated Values for $\langle \mu \rangle$ Using the MNDO Formalism (given in Debyes)

| Molecular Formula | Molecule Name | MOPAC | Our Program |
|---|---|---|---|
| C3H4 | Cyclopropene | 0.48 | 0.48 |
| C3H4 | Propyne | 0.12 | 0.12 |
| C3H6 | Propene | 0.04 | 0.04 |
| C3H8 | Propane | 0.00 | 0.00 |
| C4H6 | Bicyclobutane | 0.41 | 0.41 |
| C4H6 | Cyclobutene | 0.08 | 0.08 |
| C5H6 | Cyclopentadiene | 0.18 | 0.18 |
| C5H8 | Cyclopentene | 0.06 | 0.05 |
| C6H6 | Fulvene | 0.70 | 0.69 |
| C7H8 | Toluene | 0.05 | 0.05 |
| NH3 | Ammonia | 1.75 | 1.75 |
| CNH | Hydrogen cyanide | 2.50 | 2.50 |
| CNH5 | Methylamine | 1.48 | 1.48 |
| C2NH3 | Acetonitrile | 2.63 | 2.63 |
| C2NH3 | Methyl isocyanide | 2.17 | 2.17 |
| C2NH5 | Ethyleneimine (Azirane) | 1.75 | 1.75 |
| C2NH7 | Dimethylamine | 1.17 | 1.17 |
| C2NH7 | Ethylamine | 1.52 | 1.52 |
| C3NH3 | Acrylonitrile | 2.97 | 2.97 |
| C3NH9 | Trimethylamine | 0.75 | 0.75 |
| C4NH5 | Pyrrole | 1.81 | 1.81 |
| C5NH5 | Pyridine | 1.96 | 1.96 |
| C6NH7 | Aniline | 1.46 | 1.46 |
| CN2H2 | Diazomethane | 1.25 | 1.25 |
| CN2H2 | N=N-CH2- | 1.55 | 1.55 |
| CN2H6 | Methylhydrazine | 0.24 | 0.24 |
| H2O | Water | 1.78 | 1.78 |
| CO | Carbon monoxide | 0.20 | 0.20 |
| CH2O | Formaldehyde | 2.17 | 2.16 |
| CH4O | Methanol | 1.48 | 1.48 |
| C2H2O | Ketene | 1.04 | 1.04 |
| C2H4O | Acetaldehyde | 2.38 | 2.38 |
| C2H4O | Ethylene oxide | 1.92 | 1.92 |
| C2H6O | Dimethyl ether | 1.27 | 1.27 |
| C2H6O | Ethanol | 1.40 | 1.40 |
| C3H6O | Acetone | 2.50 | 2.51 |
| C4H4O | Furan | 0.42 | 0.42 |
| C4H10O | Diethyl ether | 1.09 | 1.10 |
| C6H6O | Phenol | 1.17 | 1.16 |
| C7H8O | Anisole | 1.07 | 1.07 |
| C3NH7O | Dimethylformamide | 3.06 | 3.06 |
| N2O | Nitrous oxide | 0.77 | 0.76 |
| CH2O2 | Formic acid | 1.49 | 1.49 |

| Formula | Name | | |
|---|---|---|---|
| C2H4O2 | Acetic acid | 1.68 | 1.68 |
| C2H4O2 | Methyl formate | 1.63 | 1.63 |
| C3H6O2 | Methyl acetate | 1.75 | 1.75 |
| C3H6O2 | Propionic acid | 1.64 | 1.64 |
| NHO2 | Nitrous acid, trans | 2.28 | 2.28 |
| O3 | Ozone | 1.18 | 1.18 |
| NHO3 | Nitric acid | 2.78 | 2.78 |
| HF | Hydrogen fluoride | 1.99 | 1.99 |
| CH3F | Fluoromethane | 1.76 | 1.76 |
| C2HF | Fluoroacetylene | 1.57 | 1.58 |
| C2H3F | Fluoroethylene | 1.70 | 1.70 |
| C2H5F | Fluoroethane | 1.87 | 1.87 |
| C6H5F | Fluorobenzene | 1.96 | 1.96 |
| CNF | Cyanogen fluoride | 0.89 | 0.89 |
| HOF | Hypofluorous acid | 1.81 | 1.81 |
| CHOF | HCOF | 2.50 | 2.50 |
| NOF | Nitrosyl fluoride | 0.51 | 0.51 |
| NO2F | Fluorine nitrite | 0.66 | 0.66 |
| CH2F2 | Difluoromethane | 2.21 | 2.22 |
| C2H4F2 | 1,1-Difluoroethane | 2.50 | 2.50 |
| N2F2 | cis-Difluorodiazene | 0.02 | 0.02 |
| OF2 | Difluorine oxide | 0.32 | 0.32 |
| COF2 | Carbonyl fluoride | 0.81 | 0.81 |
| CHF3 | Trifluoromethane | 2.23 | 2.23 |
| C2HF3 | Trifluoroethylene | 1.82 | 1.82 |
| C2H3F3 | 1,1,1-Trifluoroethane | 2.87 | 2.87 |
| NF3 | Nitrogen trifluoride | 0.20 | 0.20 |
| C2NF3 | Trifluoroacetonitrile | 0.36 | 0.36 |
| C2HO2F3 | Trifluoroacetic acid | 2.45 | 2.45 |

# Computed Molecular Properties Using Optimized Parameters

Values for $\Delta H_f$ (given in kcal/mol)

| Molecular Formula | Molecule Name | Ref. | Initial | Final |
|---|---|---|---|---|
| H | Hydrogen, cation | 365.7 | 326.7 | 315.5 |
| H | Hydrogen, atom | 52.1 | 52.1 | 77.7 |
| H2 | Hydrogen | 0 | 0.7 | -7.9 |
| C | Carbon, cation | 430.6 | 389.4 | 412.3 |
| C | Carbon, atom | 170.9 | 170.9 | 170.3 |
| CH | Methylidyne | 142.4 | 143.3 | 150.1 |
| CH2 | Methylene, singlet | 99.8 | 107.4 | 115.0 |
| CH2 | Methylene, triplet | 99.8 | 73.9 | 65.0 |
| CH3 | Methyl, cation | 261 | 243.9 | 262.8 |
| CH4 | Methane | -17.9 | -12.0 | -7.6 |
| C2H2 | Acetylene | 54.3 | 57.9 | 43.0 |
| C2H3 | Vinyl, cation | 266 | 265.7 | 249.8 |
| C2H3 | Vinyl | 59.6 | 59.0 | 49.6 |
| C2H4 | Ethylene, cation | 257 | 237.7 | 242.9 |
| C2H4 | Ethylene | 12.5 | 15.4 | 17.2 |
| C2H4 | Methylmethylene | 90.3 | 88.3 | 84.7 |
| C2H5 | Ethyl, cation | 216 | 219.6 | 211.1 |
| C2H5 | Ethyl radical | 25 | 10.5 | 13.6 |
| C2H6 | Ethane | -20 | -19.8 | -14.0 |
| C3 | Carbon, trimer | 196 | 220.3 | 215.9 |
| C3H3 | Cyclopropenyl, cation | 257 | 272.5 | 273.7 |
| C3H3 | Propynyl, cation | 281 | 265.4 | 274.5 |
| C3H4 | Allene | 45.6 | 43.9 | 38.4 |
| C3H4 | Cyclopropene | 66.2 | 68.3 | 62.2 |
| C3H4 | Propyne | 44.4 | 41.4 | 31.0 |
| C3H5 | Allyl, cation | 226 | 221.4 | 234.8 |
| C3H5 | Cyclopropyl, cation | 235 | 258.1 | 249.1 |
| C3H5 | Propenyl, cation | 237 | 240.1 | 218.9 |
| C3H5 | Allyl | 40 | 25.3 | 28.5 |
| C3H6 | Cyclopropane | 12.7 | 11.2 | 10.9 |
| C3H6 | Propene | 4.9 | 4.9 | 6.5 |
| C3H7 | i-Propyl radical | 16.8 | -1.6 | 0.7 |
| C3H8 | Propane | -24.8 | -25.0 | -19.8 |
| C4 | Carbon, tetramer | 232 | 271.3 | 208.2 |
| C4H2 | Diacetylene | 113 | 103.2 | 85.0 |
| C4H4 | Vinylacetylene | 72.8 | 65.5 | 56.3 |
| C4H4 | Butatriene | 83 | 71.2 | 65.8 |
| C4H6 | 1,2-Butadiene | 38.8 | 33.5 | 30.6 |
| C4H6 | 1,3-Butadiene | 26 | 28.9 | 29.6 |
| C4H6 | 1-Butyne | 39.5 | 36.1 | 26.8 |
| C4H6 | 2-Butyne | 34.7 | 24.9 | 20.1 |
| C4H6 | Bicyclobutane | 51.9 | 64.0 | 54.2 |
| C4H6 | Cyclobutene | 37.5 | 31.0 | 25.7 |
| C4H6 | Methyl cyclopropene | 58.2 | 53.7 | 50.9 |
| C4H6 | Methylenecyclopropane | 47.9 | 37.8 | 35.4 |
| C4H7 | 2-Butenyl, cation | 200 | 206.9 | 206.9 |
| C4H7 | Cyclobutyl, cation | 213 | 221.3 | 203.6 |
| C4H8 | 1-Butene | -0.2 | -0.3 | 1.5 |
| C4H8 | cis-2-Butene | -1.9 | -4.4 | -3.5 |
| C4H8 | Cyclobutane | 6.8 | -12.0 | -8.7 |
| C4H8 | Isobutene | -4.3 | -2.0 | -1.4 |
| C4H8 | trans-2-Butene | -3 | -5.1 | -3.2 |
| C4H9 | Isobutyl, cation | 176 | 187.3 | 161.3 |
| C4H9 | Isobutyl | 4.5 | -10.1 | -8.3 |
| C4H10 | Isobutane | -32.4 | -26.8 | -24.3 |
| C4H10 | n-Butane, trans | -30.4 | -29.8 | -25.4 |
| C5H5 | Cyclopentadienyl, anion | 21.3 | 18.9 | 4.1 |
| C5H6 | Cyclopentadiene | 32.1 | 32.0 | 23.9 |
| C5H8 | 1,2-Dimethyl cyclopropene | 46.4 | 39.3 | 40.3 |
| C5H8 | 1,4-Pentadiene | 25.3 | 24.5 | 23.3 |
| C5H8 | 1,cis-3-Pentadiene | 19.1 | 20.0 | 19.2 |
| C5H8 | 1,trans-3-Pentadiene | 18.1 | 19.1 | 19.9 |
| C5H8 | Bicyclo(2.1.0)-pentane | 37.3 | 30.1 | 23.3 |
| C5H8 | Cyclopentene | 8.3 | -0.4 | -1.1 |
| C5H8 | Isoprene | 18 | 23.3 | 22.6 |
| C5H8 | Methylene cyclobutane | 29.1 | 10.8 | 10.6 |
| C5H8 | Spiropentane | 44.3 | 33.6 | 32.1 |
| C5H9 | Cyclopentyl, cation | 188 | 194.1 | 174.7 |
| C5H10 | 2-Methyl-2-butene | -9.9 | -10.2 | -10.7 |
| C5H10 | cis-2-Pentene | -6.1 | -8.9 | -8.4 |
| C5H10 | cis-Dimethylcyclopropane | 1.3 | -2.2 | -3.1 |
| C5H10 | Cyclopentane | -18.3 | -30.6 | -25.4 |
| C5H10 | trans-2-Pentene | -7.9 | -10.4 | -8.0 |
| C5H12 | n-Pentane | -35.1 | -34.5 | -30.9 |
| C5H12 | Neopentane | -40.3 | -24.7 | -26.7 |
| C6H6 | Benzene | 19.8 | 21.3 | 21.3 |
| C6H6 | Fulvene | 47.5 | 53.6 | 50.4 |
| C6H8 | (E)-1,3,5-Hexatriene | 40.1 | 42.5 | 42.3 |
| C6H8 | (Z)-1,3,5-Hexatriene | 41.1 | 43.8 | 42.3 |
| C6H8 | 1,3-Cyclohexadiene | 25.4 | 14.4 | 15.3 |
| C6H8 | 1,4-Cyclohexadiene | 25 | 14.3 | 13.5 |
| C6H10 | 1,2-Dimethylcyclobutene | 19.8 | 6.3 | 6.8 |
| C6H10 | 1,5-Hexadiene | 20.1 | 19.5 | 17.6 |
| C6H10 | 1-Methyl cyclopentene | -1.1 | -10.3 | -8.8 |
| C6H10 | 3-Methyl cyclopentene | 2.3 | -3.7 | -5.4 |
| C6H10 | 4-Methyl cyclopentene | 3.5 | -3.1 | -6.3 |
| C6H10 | Bicyclopropyl | 30.9 | 28.5 | 23.9 |

| Formula | Name | | | |
|---|---|---|---|---|
| C6H10 | Cyclohexene | -1.1 | -10.0 | -8.7 |
| C6H11 | Cyclohexyl, cation | 177 | 186.8 | 165.0 |
| C6H12 | 1-Hexene | -10.1 | -9.1 | -9.5 |
| C6H12 | 2,3-Dimethyl-1-butene | -15.7 | -7.3 | -8.5 |
| C6H12 | 2,3-Dimethyl-2-butene | -16.8 | -13.2 | -17.6 |
| C6H12 | (Z)-3-Methyl-2-pentene | -14.8 | -12.4 | -15.5 |
| C6H12 | 4-Methyl-1-pentene | -11.8 | -5.9 | -8.3 |
| C6H12 | Cyclohexane | -29.5 | -34.8 | -31.8 |
| C6H14 | 2,2-Dimethyl butane | -44.4 | -26.1 | -31.7 |
| C6H14 | 2,3-Dimethyl butane | -42.5 | -27.8 | -34.2 |
| C6H14 | 2-Methyl pentane | -41.7 | -34.7 | -35.5 |
| C6H14 | 3-Methyl pentane | -41.1 | -33.1 | -34.9 |
| C6H14 | n-Hexane | -39.9 | -39.2 | -36.5 |
| C7H7 | Benzyl, cation | 212 | 218.0 | 227.9 |
| C7H7 | Tropylium cation | 209 | 207.6 | 213.9 |
| C7H8 | Cycloheptatriene | 43.2 | 33.8 | 35.0 |
| C7H8 | Norbornadiene | 59.7 | 62.8 | 45.6 |
| C7H8 | Toluene | 12 | 13.5 | 13.9 |
| C7H12 | 1,2-Dimethyl cyclopentene | -9.9 | -18.7 | -16.2 |
| C7H12 | 1-Ethyl cyclopentene | -6 | -14.0 | -13.1 |
| C7H12 | 1-Methyl cyclohexene | -19.4 | -17.1 | -15.8 |
| C7H12 | Norbornane | -12.4 | -10.6 | -19.1 |
| C7H14 | 1,1-Dimethyl cyclopentane | -33.1 | -30.1 | -33.0 |
| C7H14 | 1,2-cis-Dimethyl cyclopentane | -31 | -32.4 | -34.4 |
| C7H14 | 1,2-trans-Dimethyl cyclopentane | -32.7 | -35.3 | -35.3 |
| C7H14 | 1,3-cis-Dimethyl cyclopentane | -31.9 | -34.6 | -35.1 |
| C7H14 | 1-Heptene | -14.9 | -13.9 | -15.0 |
| C7H14 | Ethyl cyclopentane | -30.4 | -36.7 | -36.1 |
| C7H14 | Methyl-cyclohexane | -37 | -36.3 | -36.3 |
| C7H16 | 2,2,3-Trimethyl butane | -48.7 | -22.2 | -35.2 |
| C7H16 | 2,2-Dimethyl pentane | -49.3 | -30.4 | -37.5 |
| C7H16 | 2,3-Dimethyl pentane | -47.6 | -32.0 | -38.6 |
| C7H16 | 2,4-Dimethyl pentane | -48.3 | -34.6 | -39.9 |
| C7H16 | 2-Methyl hexane | -46.6 | -39.4 | -40.9 |
| C7H16 | 3,3-Dimethyl pentane | -48.2 | -27.0 | -36.1 |
| C7H16 | 3-Ethyl pentane | -45.3 | -36.0 | -40.9 |
| C7H16 | 3-Methyl hexane | -46 | -37.6 | -40.5 |
| C7H16 | n-Heptane | -44.9 | -43.9 | -42.0 |
| C8H8 | Cubane | 148.7 | 98.9 | 74.5 |
| C8H8 | Cyclooctatetraene | 70.7 | 56.1 | 57.6 |
| C8H8 | Styrene | 35.3 | 37.6 | 37.4 |
| C8H10 | Ethylbenzene | 7.2 | 8.7 | 9.6 |
| C8H10 | m-Xylene | 4.1 | 5.9 | 6.7 |
| C8H10 | o-Xylene | 4.6 | 8.3 | 6.6 |
| C8H10 | p-Xylene | 4.3 | 5.7 | 6.8 |
| C8H12 | 1,5-Cyclooctadiene | 13.7 | 11.3 | 9.1 |

| Formula | Name | | | |
|---|---|---|---|---|
| C8H12 | 4-Vinyl cyclohexene | 16.6 | 13.1 | 9.3 |
| C8H14 | 1-Octyne | 19.3 | 17.0 | 4.7 |
| C8H14 | 2,5-Dimethyl 2,4-hexadiene | -4.6 | -2.4 | -6.0 |
| C8H14 | 2-Octyne | 15.2 | 5.3 | -0.9 |
| C8H14 | 3,4-Dimethyl-(E,E)-2,4-hexadiene | -0.5 | -2.4 | -3.8 |
| C8H14 | 3,4-Dimethyl-(E,Z)-2,4-hexadiene | 0.7 | -1.9 | -2.1 |
| C8H14 | 3,4-Dimethyl-(Z,Z)-2,4-hexadiene | -0.9 | -2.0 | -4.1 |
| C8H14 | 3-Octyne | 14.9 | 4.7 | 0.3 |
| C8H14 | 4-Octyne | 14.4 | 4.6 | -0.1 |
| C8H14 | Bicyclo(2.2.2)-octane | -24.1 | -26.5 | -30.0 |
| C8H16 | 1-Octene | -19.8 | -19.3 | -20.5 |
| C8H16 | Ethylcyclohexane | -41.1 | -39.5 | -41.5 |
| C8H18 | 2,2,3,3-Tetramethyl butane | -54 | -12.8 | -35.8 |
| C8H18 | 2,2,3-Trimethyl pentane | -52.6 | -24.6 | -40.7 |
| C8H18 | 2,2,4-Trimethyl pentane | -53.6 | -29.2 | -43.4 |
| C8H18 | 2,2-Dimethyl hexane | -53.7 | -35.1 | -43.0 |
| C8H18 | 2,3,3-Trimethyl pentane | -51.7 | -24.5 | -40.7 |
| C8H18 | 2,3,4-Trimethyl pentane | -52 | -28.8 | -42.3 |
| C8H18 | 2,3-Dimethyl hexane | -51.1 | -36.6 | -44.3 |
| C8H18 | 2,4-Dimethyl hexane | -52.4 | -37.5 | -45.2 |
| C8H18 | 2,5-Dimethyl hexane | -53.2 | -39.5 | -45.6 |
| C8H18 | 2-Methyl heptane | -51.5 | -44.1 | -46.5 |
| C8H18 | 3,3-Dimethyl hexane | -52.6 | -31.2 | -41.9 |
| C8H18 | 3,4-Dimethyl hexane | -50.9 | -31.6 | -44.9 |
| C8H18 | 3-Ethyl hexane | -50.4 | -39.8 | -47.5 |
| C8H18 | 3-Ethyl-2-methyl pentane | -50.5 | -31.6 | -46.3 |
| C8H18 | 3-Ethyl-3-methyl pentane | -51.4 | -27.8 | -42.3 |
| C8H18 | 3-Methyl heptane | -50.8 | -42.3 | -46.0 |
| C8H18 | 4-Methyl heptane | -50.7 | -42.1 | -46.1 |
| C8H18 | n-Octane | -49.9 | -48.7 | -47.5 |
| C9H10 | alpha-Methyl styrene | 28.3 | 30.5 | 31.0 |
| C9H10 | Cyclopropyl benzene | 36 | 37.6 | 35.1 |
| C9H18 | 1,3,5-Trimethyl cyclohexane | -51.5 | -39.0 | -44.8 |
| C9H18 | cis-cis-trans-1,3,5-Trimethyl cyclohexane | -49.4 | -37.8 | -44.3 |
| C9H20 | 3,3-Diethylpentane | -55.4 | -27.9 | -46.2 |
| C9H20 | n-Nonane | -54.7 | -53.4 | -53.0 |
| C10H8 | Azulene | 73.5 | 72.1 | 72.0 |
| C10H8 | Naphthalene | 36.1 | 38.3 | 39.3 |
| C10H10 | 1,4-Dicyclopropylbuta-1,3-diyne | 134.3 | 117.7 | 104.8 |
| C10H10 | 1-Butynl benzene | 59.4 | 52.0 | 48.5 |
| C10H10 | 2a,4a,6a,6b-Tetrahydrocyclopentapentalene | 53.1 | 47.6 | 37.7 |
| C10H10 | Bulvalene | 79.9 | 63.4 | 59.2 |
| C10H10 | Diisopropenyldiacetylene | 118.1 | 103.0 | 96.7 |
| C10H10 | Tricyclo[6.2.0.0]deca-1(8),2,6-triene | 74 | 54.1 | 48.6 |
| C10H12 | 1,2,6,7-Cyclodecatetraene | 85.1 | 67.8 | 59.4 |
| C10H12 | Dispiro[2.2.2.2]deca-4,9-diene | 72.3 | 65.1 | 58.3 |

| Formula | Name | | | |
|---|---|---|---|---|
| C10H12 | Tetralin | 6.2 | 1.3 | 1.5 |
| C10H14 | (1-Methylpropyl) benzene | -4.2 | 4.5 | 0.7 |
| C10H14 | (2-Methylpropyl) benzene | -5.2 | 3.7 | -0.2 |
| C10H14 | 1,2,3,4-Tetramethyl benzene | -8.6 | 1.0 | -6.6 |
| C10H14 | 1,2,3,4-Tetramethylfulvene | 19.9 | 20.1 | 19.6 |
| C10H14 | 1,2,3,5-Tetramethyl benzene | -10.3 | -3.0 | -6.8 |
| C10H14 | 1,2,4,5-Tetramethyl benzene | -11.3 | -4.6 | -7.5 |
| C10H14 | tert-Butyl benzene | -5.4 | 15.6 | 4.5 |
| C10H14 | Tetrahydrotriquinacene | 3 | -10.8 | -15.9 |
| C10H16 | 1,2,3,4,5-Pentamethyl-1,3-cyclopentadiene | -5.9 | -6.5 | -8.3 |
| C10H16 | Adamantane | -31.9 | -26.5 | -36.4 |
| C10H16 | Camphene | -6.8 | 14.6 | -4.3 |
| C10H16 | Perhydrotriquinacene | -24.5 | -38.9 | -42.2 |
| C10H18 | 1-Methyl-4-(1-methylethyl)-cyclohexene | -26.5 | -20.5 | -29.4 |
| C10H18 | 4-Methyl-1-(1-methylethyl)-cyclohexene | -26.6 | -23.8 | -27.7 |
| C10H18 | cis-Decalin | -40.5 | -37.3 | -41.3 |
| C10H18 | Spiro(4-5)decane | -34.7 | -37.3 | -44.1 |
| C10H18 | trans-Decalin | -43.5 | -41.9 | -46.8 |
| C10H20 | (E)-2,2,5,5-Tetramethyl-3-hexene | -39.9 | -10.9 | -23.1 |
| C10H20 | (Z)-2,2,5,5-Tetramethyl-3-hexene | -30.3 | -0.3 | -22.6 |
| C10H20 | 1-Decene | -29.8 | -28.0 | -31.6 |
| C10H20 | Butyl cyclohexane | -51 | -48.8 | -52.6 |
| C10H20 | Pentyl cyclopentane | -45.2 | -50.7 | -52.3 |
| C10H22 | 2,2,5,5-Tetramethylhexane | -68 | -11.8 | -45.6 |
| C10H22 | 3,3,4,4-Tetramethylhexane | -63.5 | -30.7 | -49.4 |
| C10H22 | n-Decane | -59.7 | -58.1 | -58.5 |
| C11H16 | Pentamethylbenzene | -16.1 | -2.5 | -13.1 |
| C11H22 | 1,1,4-Trimethylcycloheptane | -50.3 | -26.5 | -50.8 |
| C11H22 | Hexyl cyclopentane | -50.1 | -55.5 | -57.8 |
| C11H22 | Pentyl cyclohexane | -55.9 | -53.5 | -58.1 |
| C11H24 | Undecane | -64.6 | -62.8 | -64.0 |
| C12H8 | Acenaphthylene | 61.6 | 67.1 | 68.2 |
| C12H8 | Biphenylene | 100.5 | 94.7 | 89.4 |
| C12H10 | Acenaphthene | 37.4 | 33.1 | 35.2 |
| C12H10 | Biphenyl | 43.5 | 46.0 | 46.2 |
| C12H18 | Hexamethylbenzene | -18.5 | 0.1 | -18.9 |
| C12H24 | Hexylcyclohexane | -60.8 | -58.2 | -63.6 |
| C12H26 | n-Duodecane | -69.2 | -67.6 | -69.5 |
| C13H10 | Fluorene | 41.8 | 44.9 | 45.1 |
| C13H28 | Tri-t-butylmethane | -56.2 | 32.4 | -48.7 |
| C13H28 | Tridecane | -74.5 | -72.3 | -75.1 |
| C14H10 | Anthracene | 55.2 | 58.8 | 61.2 |
| C14H10 | Diphenylethyne | 92 | 89.5 | 86.3 |
| C14H10 | Phenanthrene | 49.5 | 55.6 | 54.9 |
| C14H12 | 9,10-Dihydro-phenanthrene | 37.1 | 38.3 | 36.3 |
| C14H12 | 9-Methyl-9H-fluorene | 35.4 | 42.3 | 42.5 |

| Formula | Name | | | |
|---|---|---|---|---|
| C14H12 | Octalene | 131.8 | 106.2 | 106.8 |
| C14H12 | Stilbene | 53.4 | 63.4 | 58.1 |
| C14H14 | 1,2,3,4-Tetrahydrophenanthrene | 22.1 | 21.9 | 20.4 |
| C14H14 | 4,4'-Dimethylbiphenyl | 26.6 | 30.4 | 31.6 |
| C14H14 | Bibenzyl | 32.4 | 37.6 | 33.4 |
| C14H16 | 1,4,5,8-Tetramethynaphthalene | 19.5 | 31.4 | 18.1 |
| C14H18 | 1,2,3,4,5,6,7,8-Octahydro-anthracene | -8.9 | -18.6 | -17.4 |
| C14H20 | Diadamantane | -34.9 | -23.6 | -42.2 |
| C14H24 | 1,3,5,7-Tetramethyladamantane | -68 | -22.8 | -47.4 |
| C14H28 | (E)-3,4-Di-tert-butyl-3-hexene | -40.2 | 19.2 | -42.9 |
| C14H28 | Cyclotetradecane | -57.2 | -53.2 | -80.5 |
| C14H28 | n-Nonylcyclopentane | -64.9 | -69.6 | -74.4 |
| C14H28 | Octylcyclohexane | -70.7 | -67.7 | -74.7 |
| C14H30 | 3,3,4,4-Tetraethylhexane | -63.5 | 6.2 | -64.7 |
| C14H30 | Octamethylhexane | -59.4 | 46.7 | -44.1 |
| C14H30 | Tetradecane | -79.4 | -77.0 | -80.6 |
| C15H12 | 4-Methylphenanthrene | 46.8 | 56.9 | 49.9 |
| C15H22 | 1-Methyldiadamantane | -39.9 | -18.0 | -43.8 |
| C15H22 | 3-Methyladamantane | -37.6 | -22.7 | -45.6 |
| C15H22 | 4-Methyldiadamantane | -43.5 | -23.0 | -45.1 |
| C15H22 | 6-(1,1-dimethylethyl)-2,3-dihydro-1,1-dimethyl-1H-Indene | -24.9 | 0.6 | -14.0 |
| C15H30 | n-Nonylcyclohexane | -75.6 | -72.3 | -80.2 |
| C15H32 | Pentadecane | -84.8 | -81.7 | -86.1 |
| C16H10 | Fluoranthene | 69.8 | 72.7 | 76.9 |
| C16H10 | Pyrene | 53.9 | 60.7 | 65.9 |
| C16H14 | 2,7-Dimethylphenanthrene | 34.2 | 40.0 | 40.3 |
| C16H14 | 4,5,9,10-Tetrahydropyrene | 21.6 | 25.5 | 27.1 |
| C16H14 | 9,10-Dimethylphenanthrene | 40 | 52.9 | 41.9 |
| C16H16 | (2.2)Metaparacyclophane | 52.2 | 62.7 | 52.0 |
| C16H16 | [2.2]Metacyclophane | 40.7 | 58.7 | 47.4 |
| C16H16 | [2.2]Paracyclophane | 58.5 | 66.6 | 57.7 |
| C16H18 | 1,2,3,6,7,8-Hexahydropyrene | 9.7 | 9.3 | 4.4 |
| C16H28 | Tricyclo[8.2.2.2]-hexadecane | -36.4 | -15.8 | -63.7 |
| C16H32 | 1-Hexadecene | -59.4 | -56.4 | -64.6 |
| C16H32 | Decylcyclohexane | -80.5 | -77.1 | -85.7 |
| C16H32 | n-Undecylcyclopentane | -74.7 | -79.1 | -85.4 |
| C17H34 | n-Dodecylcyclopentane | -79.6 | -83.8 | -90.9 |
| C17H34 | Undecylcyclohexane | -85.4 | -81.8 | -91.2 |
| C17H36 | Heptadecane | -94.2 | -91.2 | -97.1 |
| C18H14 | p-Terphenyl | 66.6 | 70.7 | 71.3 |
| C18H18 | 2,5-Diphenyl-1,5-hexadiene | 68 | 72.6 | 69.0 |
| C18H18 | 3,4,5,6-Tetramethylphenanthrene | 38.3 | 49.0 | 37.7 |
| C18H20 | [3.3]Paracyclophane | 30.9 | 37.0 | 26.5 |
| C18H22 | 1,1'-(1,1,2,2-Tetramethyl-1,2-ethanediyl)bis-benzene | 13.7 | 70.3 | 31.2 |
| C18H36 | Dodecylcyclohexane | -90.4 | -86.5 | -96.7 |

| Formula | Name | | | |
|---|---|---|---|---|
| C18H36 | n-Tridecylcyclopentane | -84.6 | -88.5 | -96.5 |
| C18H38 | 1,1,2,2-Tetra-t-butylethane | -59.9 | 70.2 | -63.3 |
| C18H38 | Octadecane | -99.1 | -95.9 | -102.6 |
| C19H20 | 2,6-Diphenyl-1,6-heptadiene | 61.9 | 67.9 | 59.1 |
| C19H38 | n-Tetradecylcyclopentane | -89.5 | -93.2 | -102.0 |
| C19H38 | n-Tridecylcyclohexane | -95.3 | -91.3 | -102.2 |
| C19H40 | Nonadecane | -104 | -100.6 | -108.1 |
| C20H14 | 9,10-Dihydro-9,10[1',2']benzanthracene | 76.9 | 87.3 | 81.7 |
| C20H16 | 3,9-Dimethylbenz[a]anthracene | 45.1 | 59.3 | 60.7 |
| C20H16 | 5,6-Dimethyl chrysene | 62.7 | 76.4 | 61.8 |
| C20H16 | 9,10-Dimethyl-1,2-benzanthracene | 66.3 | 79.8 | 65.0 |
| C20H30 | 1,3,5-Tri-tert-butyl pentalene | 3.4 | 54.7 | 30.0 |
| C20H36 | Tetra-tert-butyltetrahedrane | 6.2 | 80.9 | 31.7 |
| C20H38 | Meso-3,4-dicyclohexyl-2,5-dimethylhexane | -71.6 | -36.0 | -97.3 |
| C20H40 | Tetradecylcyclohexane | -100.2 | -96.0 | -107.7 |
| C20H42 | Eicosane | -108.9 | -105.4 | -113.6 |
| N | Nitrogen, cation | 448.3 | 417.5 | 413.3 |
| N | Nitrogen, atom | 113 | 113.0 | 98.0 |
| NH2 | Amidogen | 45.5 | 36.4 | 51.6 |
| NH3 | Ammonia | -11 | -6.4 | 9.7 |
| NH4 | Ammonium, cation | 155 | 164.6 | 170.4 |
| CN | Cyanide | 104 | 126.3 | 86.4 |
| CNH | Hydrogen cyanide | 32.3 | 35.3 | 35.1 |
| CNH4 | CH2-NH2, cation | 178 | 186.8 | 193.9 |
| CNH4 | CH3-NH. | 37 | 29.3 | 35.1 |
| CNH4 | CH3NH, anion | 30.5 | 23.5 | 49.0 |
| CNH5 | Methylamine | -5.5 | -7.6 | 5.7 |
| C2NH3 | Acetonitrile | 17.7 | 19.2 | 20.0 |
| C2NH3 | Methyl isocyanide | 39.1 | 60.3 | 59.8 |
| C2NH5 | Ethyleneimine (Azirane) | 30.2 | 25.0 | 33.6 |
| C2NH6 | Dimethyl nitrogen, anion | 24.7 | 8.5 | 35.8 |
| C2NH7 | Dimethylamine | -4.4 | -6.7 | 2.2 |
| C2NH7 | Ethylamine | -11.4 | -13.3 | -3.9 |
| C3NH3 | Acrylonitrile | 44.1 | 43.8 | 46.8 |
| C3NH5 | Ethyl cyanide | 12.1 | 13.7 | 15.7 |
| C3NH7 | Cyclopropylamine | 18.4 | 15.3 | 18.4 |
| C3NH9 | Isopropylamine | -20 | -16.4 | -9.3 |
| C3NH9 | n-Propylamine | -16.8 | -18.3 | -9.2 |
| C3NH9 | Trimethylamine | -5.7 | -2.9 | 0.5 |
| C4NH5 | (E)-2-Butenenitrile | 33.6 | 33.8 | 35.6 |
| C4NH5 | (Z)-2-Butenenitrile | 32 | 34.1 | 35.9 |
| C4NH5 | 3-Butenenitrile | 37.7 | 39.2 | 38.2 |
| C4NH5 | Pyrrole | 25.9 | 32.4 | 21.0 |
| C4NH7 | Butanenitrile | 7.5 | 8.9 | 10.1 |
| C4NH7 | Isobutane nitrile | 5.6 | 11.2 | 12.3 |
| C4NH9 | Pyrrolidine | -0.8 | -15.9 | -9.6 |

| Formula | Name | | | |
|---|---|---|---|---|
| C4NH11 | 2-Butylamine | -25.4 | -20.4 | -14.2 |
| C4NH11 | 2-Methyl-1-propylamine | -23.6 | -19.3 | -13.4 |
| C4NH11 | N-Butylamine | -22.7 | -23.1 | -14.7 |
| C4NH11 | t-Butylamine | -28.9 | -15.5 | -12.5 |
| C5NH5 | Pyridine | 34.6 | 28.8 | 30.4 |
| C5NH7 | N-Methyl pyrrole | 24.6 | 32.2 | 21.1 |
| C5NH9 | 1,2,3,6-Tetrahydropyridine | 7.1 | 6.1 | 6.5 |
| C5NH9 | 2-Cyanobutane | 0.6 | 7.8 | 7.0 |
| C5NH9 | Butyl cyanide | 2.7 | 4.1 | 4.7 |
| C5NH9 | t-Butylnitrile | -0.8 | 12.5 | 10.4 |
| C5NH11 | Cyclopentylamine | -13.1 | -22.6 | -15.9 |
| C5NH11 | Piperidine | -11.3 | -18.6 | -16.9 |
| C5NH13 | N-Methyl-n-butylamine | -25.9 | -21.8 | -17.7 |
| C6NH7 | 1-Cyclopentenecarbonitrile | 37.4 | 28.6 | 31.2 |
| C6NH7 | 2-Cyclopentenecarbonitrile | 33.9 | 34.9 | 32.1 |
| C6NH7 | 2-Methyl pyridine | 23.7 | 19.7 | 22.7 |
| C6NH7 | 3-Methyl pyridine | 24.8 | 20.4 | 22.8 |
| C6NH7 | 4-Methyl pyridine | 24.8 | 20.8 | 22.4 |
| C6NH7 | Aniline | 20.8 | 21.7 | 23.5 |
| C6NH9 | 2,5-Dimethyl-1H-pyrrole | 9.5 | 10.1 | 5.2 |
| C6NH9 | Cyclopentanecarbonitrile | 11.7 | 5.4 | 6.3 |
| C6NH13 | 2-Methylpiperidine | -20.2 | -21.6 | -22.1 |
| C6NH13 | Cyclohexamethylenimine | -10.8 | -17.7 | -20.7 |
| C6NH13 | Cyclohexanamine | -25.1 | -25.8 | -21.0 |
| C6NH15 | Di-n-propylamine | -27.8 | -27.3 | -28.2 |
| C6NH15 | Diisopropylamine | -32.6 | -20.3 | -27.1 |
| C6NH15 | Triethylamine | -22.1 | -14.8 | -26.1 |
| C7NH5 | Phenyl cyanide | 51.5 | 51.9 | 55.0 |
| C7NH9 | 1-Cyclohexenecarbonitrile | 24.3 | 21.0 | 23.2 |
| C7NH9 | 2,6-Dimethylpyridine | 13.4 | 10.7 | 15.2 |
| C7NH9 | 2-Cyclohexenecarbonitrile | 26.2 | 26.6 | 24.8 |
| C7NH9 | Benzylamine | 21 | 19.5 | 25.7 |
| C7NH9 | m-Toluidine | 14.6 | 14.1 | 16.3 |
| C7NH9 | N-Methylaniline | 20.1 | 24.2 | 22.2 |
| C7NH9 | o-Toluidine | 12.7 | 16.1 | 17.4 |
| C7NH9 | p-Toluidine | 10 | 13.8 | 16.9 |
| C7NH11 | Cyclohexanecarbonitrile | -0.9 | 1.7 | 0.3 |
| C7NH13 | Hexahydro-1H-pyrrolizine | -0.9 | -20.1 | -22.5 |
| C7NH13 | n-Heptanenitrile | -7.4 | -5.3 | -6.3 |
| C7NH17 | Isopropylbutylamine | -39.4 | -29.0 | -33.2 |
| C8NH11 | 1-Norbornylcyanide | 18 | 24.3 | 15.1 |
| C8NH11 | 1-Norbornylisocyanide | 39.6 | 63.0 | 48.0 |
| C8NH11 | 5-Ethyl-2-methyl-pyridine | 8.3 | 6.4 | 11.0 |
| C8NH11 | N,N-Dimethyl aniline | 24 | 28.8 | 23.5 |
| C8NH11 | N-Ethyl aniline | 13.4 | 19.3 | 13.8 |
| C8NH15 | 3-Azabicyclo[3.2.2]nonane | -10.4 | -13.9 | -20.2 |

| Formula | Name | | | |
|---|---|---|---|---|
| C8NH15 | n-Heptyl cyanide | -12.1 | -10.0 | -11.8 |
| C8NH17 | N-(2-Methylpropylidene)-butylamine | -21.8 | -15.2 | -13.2 |
| C8NH19 | 2-Methyl-N-(2-methylpropyl)-1-propanamine | -43.2 | -28.5 | -37.8 |
| C8NH19 | Di-sec-butylamine | -43.9 | -26.7 | -37.8 |
| C8NH19 | Dibutylamine | -40.9 | -36.7 | -39.2 |
| C8NH19 | N-(2-Methylpropyl)-1-butanamine | -41.8 | -32.9 | -39.0 |
| C8NH19 | n-Octylamine | -41.5 | -42.0 | -36.7 |
| C9NH7 | Isoquinoline | 48.9 | 45.2 | 48.5 |
| C9NH7 | Quinoline | 47.9 | 44.7 | 49.1 |
| C9NH9 | 2,6-Dimethylbenzonitrile | 34.9 | 40.0 | 41.0 |
| C9NH11 | (1a,2a,4a)-Bicyclo[2.2.2]oct-5-ene-2-carbonitrile | 36.2 | 39.7 | 29.6 |
| C9NH11 | (1a,2b,4a)-Bicyclo[2.2.2]oct-5-ene-2-carbonitrile | 36.6 | 40.1 | 29.9 |
| C9NH11 | 1,2,3,4-Tetrahydroquinoline | 19.6 | 14.1 | 13.1 |
| C9NH11 | 5,6,7,8-Tetrahydroquinoline | 17 | 7.0 | 9.7 |
| C9NH13 | N,N-Dimethyl m-toluidine | 17.4 | 26.2 | 16.3 |
| C9NH13 | N,N-Dimethyl p-toluidine | 16.5 | 25.4 | 16.9 |
| C9NH13 | N-Ethyl m-toluidine | 7.3 | 11.7 | 6.6 |
| C9NH17 | cis-3,7a-H-cis-5,8-H-3,5-Dimethylpyrrolizidine | -15.9 | -25.9 | -31.7 |
| C9NH17 | Decahydro trans-quinoline | -27 | -26.9 | -32.4 |
| C9NH19 | 2,2,6,6-Tetramethyl piperidine | -38.2 | -14.0 | -32.0 |
| N2 | Nitrogen | 0 | 8.3 | 34.5 |
| N2H2 | Diazene | 36 | 31.8 | 51.1 |
| N2H4 | Hydrazine | 22.8 | 14.1 | 30.2 |
| CN2H2 | Diazomethane | 71 | 67.2 | 64.1 |
| CN2H2 | N=N-CH2- | 79 | 72.4 | 86.1 |
| CN2H6 | Methylhydrazine | 22.6 | 14.3 | 24.4 |
| C2N2 | Cyanogen | 73.8 | 66.6 | 71.4 |
| C2N2H8 | 1,1-Dimethylhydrazine | 20 | 18.0 | 22.4 |
| C2N2H8 | 1,2-Dimethylhydrazine | 22 | 14.9 | 20.7 |
| C3N2H4 | 1H-Pyrazole | 42.9 | 45.3 | 34.7 |
| C3N2H4 | Imidazole | 31.8 | 33.2 | 25.3 |
| C3N2H10 | 1,2-Propanediamine | -12.8 | -10.7 | 1.4 |
| C4N2 | Dicyanoacetylene | 126.5 | 111.4 | 114.1 |
| C4N2H2 | Fumaronitrile | 81.3 | 74.7 | 79.6 |
| C4N2H4 | 1,3-Diazine | 47 | 34.9 | 39.2 |
| C4N2H4 | Pyrazine | 46.9 | 37.7 | 41.8 |
| C4N2H4 | Pyridazine | 66.5 | 43.5 | 47.2 |
| C4N2H4 | Succinonitrile | 50.1 | 48.8 | 48.2 |
| C4N2H6 | 2-Methyl-1H-imidazole | 21.5 | 21.5 | 17.1 |
| C4N2H8 | (Dimethylamino) acetonitrile | 27.3 | 31.5 | 31.6 |
| C4N2H8 | 1,4,5,6-Tetrahydropyrimidine | 13.2 | 8.9 | 8.5 |
| C4N2H10 | Piperazine | 6 | -2.9 | -1.0 |
| C5N2H6 | 2-Aminopyridine | 28.2 | 25.8 | 29.9 |
| C5N2H6 | 3-Aminopyridine | 34.5 | 29.3 | 32.8 |
| C5N2H6 | 4-Aminopyridine | 31.1 | 28.3 | 31.1 |

| Formula | Name | | | |
|---|---|---|---|---|
| C5N2H6 | Dimethyl propanedinitrile | 47.1 | 52.8 | 51.6 |
| C5N2H8 | 2-Ethyl-1H-imidazole | 16.3 | 16.5 | 12.4 |
| C5N2H10 | Diethylcyanamide | 15.2 | 29.4 | 14.3 |
| C5N2H12 | Butylmethyldiazene | 18.9 | 5.6 | 13.4 |
| C5N2H14 | N,N-Dimethyl-1,3-propanediamine | -8.3 | -8.8 | -9.3 |
| C6N2H4 | 2-Cyanopyridine | 67.1 | 60.3 | 66.8 |
| C6N2H4 | 3-Cyanopyridine | 66.4 | 59.7 | 65.1 |
| C6N2H4 | 4-Cyanopyridine | 67.8 | 60.9 | 65.7 |
| C6N2H8 | 2,3-Dimethyl pyrazine | 30.1 | 20.2 | 25.7 |
| C6N2H8 | Hexanedinitrile | 35.7 | 38.5 | 35.9 |
| C6N2H8 | Phenylhydrazine | 48.5 | 45.2 | 47.7 |
| C6N2H12 | 3(Dimethylamino) propanenitrile | 21.6 | 22.1 | 19.7 |
| C6N2H12 | Tetramethyldiazetine | 35.9 | 33.6 | 30.0 |
| C6N2H12 | Triethylenediamine | 21.6 | 19.3 | 1.0 |
| C6N2H14 | 1,2-Diisopropyldiazene | 8.6 | 3.7 | 9.1 |
| C6N2H14 | Dipropyldiazene | 12.4 | -0.9 | 7.7 |
| C7N2H6 | 1H-Benzimidazole | 43.4 | 45.8 | 43.4 |
| C7N2H6 | 1H-Indazole | 58.1 | 58.0 | 52.9 |
| C7N2H10 | 1-Methyl-1-phenylhydrazine | 50.4 | 24.9 | 14.3 |
| C7N2H10 | t-Butylmalononitrile | 30.3 | 52.5 | 43.1 |
| C7N2H10 | Trimethyl pyrazine | 17.8 | 10.6 | 18.1 |
| C7N2H12 | 1-Piperidineacetonitrile | 19.8 | 19.6 | 12.1 |
| C7N2H14 | 3,3,5,5-Tetramethyl-1-pyrazoline | 9.4 | 5.2 | 2.8 |
| C8N2H4 | m-Dicyanobenzene | 86.7 | 83.8 | 90.3 |
| C8N2H4 | o-Dicyanobenzene | 87.8 | 85.3 | 91.6 |
| C8N2H4 | p-Dicyanobenzene | 85.6 | 83.9 | 90.3 |
| C8N2H6 | Phthalazine | 78.9 | 59.0 | 65.0 |
| C8N2H6 | Quinazoline | 58.1 | 50.3 | 57.6 |
| C8N2H6 | Quinoxaline | 57.4 | 52.7 | 60.9 |
| C8N2H12 | n-Pentylmalonodinitrile | 32.5 | 34.2 | 31.3 |
| C8N2H12 | Tetramethylbutanedinitrile | 24.1 | 56.8 | 39.9 |
| C8N2H12 | Tetramethylpyrazine | 13.1 | 2.7 | 10.5 |
| C8N2H14 | 1,4-Dimethyl-2,3-diaza-bicyclo[2.2.2]oct-2-ene | 22.1 | 13.7 | 8.8 |
| C8N2H16 | 3,4,5,6-Tetrahydro-3,3,6,6-tetramethylpyridazine | 10 | 6.7 | 0.2 |
| C8N2H18 | Di-n-butyldiazene | 2.2 | -10.4 | -3.1 |
| C8N2H18 | Di-tert-butyldiazene | -8.7 | 4.9 | 5.0 |
| C8N2H20 | 1,2-Dibutylhydrazine | -14.2 | -14.4 | -18.6 |
| N3 | Azide radical | 99 | 97.1 | 98.7 |
| N3H | Hydrazoic acid | 70.3 | 73.1 | 77.7 |
| C3N3H3 | 1,3,5-Triazine | 54 | 40.0 | 48.3 |
| C5N3H | Ethylenetricarbonitrile | 124.4 | 110.4 | 118.5 |
| C5N3H3 | 1,1,1-Ethanetricarbonitrile | 101 | 96.2 | 96.4 |
| CN4H2 | 1-H Tetrazole | 76.6 | 53.7 | 54.6 |
| CN4H2 | 2-H-Tetrazole | 80 | 58.0 | 56.2 |
| C6N4 | Tetracyanoethylene | 168.5 | 148.0 | 159.4 |
| C6N4H12 | 1,3,5,7-Tetraazaadamantane | 47.6 | 51.5 | 18.7 |

| Formula | Name | | | |
|---|---|---|---|---|
| C3N6H6 | Melamine | 12.4 | 21.7 | 34.8 |
| C10NH9 | 2-Methyl-quinoline | 38 | 35.8 | 41.3 |
| C10NH9 | 4-Methyl-quinoline | 38.7 | 39.0 | 41.3 |
| C10NH9 | 6-Methyl-quinoline | 38.5 | 37.0 | 41.6 |
| C10NH9 | 8-Methyl-quinoline | 40.1 | 38.1 | 42.7 |
| C10NH11 | 2,4,6-Trimethyl-benzonitrile | 25.4 | 32.5 | 33.4 |
| C10NH11 | 2,4,6-Trimethylphenyl isocyanide | 56.6 | 70.9 | 67.4 |
| C10NH15 | N,N-Diethyl aniline | 14.8 | 26.9 | 7.0 |
| C10NH19 | n-Nonyl cyanide | -21.9 | -19.5 | -22.8 |
| C11NH11 | 2,6-Dimethyl quinoline | 28.9 | 28.1 | 33.9 |
| C11NH11 | 2,7-Dimethyl quinoline | 29.1 | 28.1 | 33.9 |
| C11NH15 | 1-Adamantyl cyanide | -1.8 | 11.4 | -2.1 |
| C11NH15 | 1-Adamantyl isocyanide | 17.5 | 50.0 | 30.4 |
| C11NH17 | 2-Methyl-6-t-butylaniline | -10.7 | 19.7 | 3.1 |
| C11NH21 | n-Undecanenitrile | -27.1 | -24.2 | -28.3 |
| C12NH9 | Carbazole | 50.1 | 53.5 | 51.7 |
| C12NH11 | 2-Biphenylamine | 44.1 | 47.9 | 48.9 |
| C12NH11 | Biphenylamine | 48.2 | 55.7 | 45.8 |
| C12NH23 | 2-n-Butyl-2-methylhexanenitrile | -31.8 | -7.3 | -27.7 |
| C13NH9 | 6,7-Benzoquinoline | 58.2 | 60.6 | 64.5 |
| C13NH9 | a-Benzoquinoline | 59.9 | 59.9 | 65.3 |
| C13NH9 | Acridine | 65.5 | 64.3 | 71.6 |
| C13NH9 | Benzo[f]quinoline | 55.9 | 61.5 | 64.7 |
| C13NH9 | Phenanthridine | 57.5 | 60.6 | 64.5 |
| C13NH11 | N-Methylcarbazole | 47.6 | 56.1 | 52.7 |
| C13NH15 | 1,2,3,4-Tetrahydro-N-methylcarbazole | 22.3 | 22.9 | 20.1 |
| C14NH13 | 9-Ethyl-9H-carbazole | 40.6 | 51.1 | 44.8 |
| C14NH27 | Tetradecanenitrile | -41.8 | -38.4 | -44.8 |
| C9N2H8 | 3-Quinolinamine | 49.8 | 45.2 | 51.7 |
| C9N2H8 | 5-Quinolinamine | 50.3 | 47.2 | 51.8 |
| C9N2H8 | 6-Quinolinamine | 49.3 | 45.2 | 51.3 |
| C9N2H8 | 8-Quinolinamine | 44.8 | 45.3 | 50.5 |
| C9N2H18 | 2-(Diethylamino)-pentanenitrile | -1.1 | 14.0 | -0.6 |
| C10N2H8 | 2,2-Bipyridyl | 69.1 | 58.8 | 66.2 |
| C10N2H8 | 2,4-Bipyridyl | 67.9 | 60.0 | 65.9 |
| C10N2H8 | 4,4'-Bipyridine | 70.1 | 61.1 | 65.0 |
| C10N2H10 | 2,3-Dimethyl quinoxaline | 41.3 | 36.4 | 44.6 |
| C10N2H12 | alpha N,N-dimethylamino phenylacetonitrile | 52.7 | 58.9 | 48.4 |
| C10N2H16 | Ethyl(1,1-dimethylpropyl)malonodinitrile | 14.6 | 56.6 | 33.1 |
| C10N2H16 | Meso-2,3-diethyl-2,3-dimethylsuccinodinitrile | 15.3 | 55.0 | 30.4 |
| C10N2H16 | Methyl(1,1,2-trimethylpropyl)malonodinitrile | 19 | 66.3 | 36.2 |
| C12N2H8 | Phenazine | 80.9 | 71.7 | 84.0 |
| C12N2H8 | Phenazone | 89.9 | 72.6 | 80.7 |
| C12N2H10 | cis-Azobenzene | 107.7 | 83.5 | 86.8 |
| C12N2H12 | 4,4'-Dimethyl-2,2'-bipyridine | 50 | 43.0 | 50.6 |
| C12N2H20 | 1,(1-Piperidinyl) cyclohexanecarbonitrile | 0.8 | 17.6 | -3.2 |

| Formula | Name | | | |
|---|---|---|---|---|
| C13N2H16 | a-Phenyl-1-piperidineacetonitrile | 45.4 | 53.5 | 41.9 |
| C9N3H3 | 1,3,5-Tricyanobenzene | 121.8 | 116.8 | 126.9 |
| C11N3H7 | 1,1,1-Tricyano-2-phenyl ethane | 129.2 | 126.7 | 123.5 |
| C16NH35 | Dioctylamine | -76.5 | -74.5 | -83.3 |
| C18NH15 | Triphenylamine | 78.1 | 93.9 | 81.2 |
| C18N2H12 | 2,2'-Biquinoline | 101.8 | 91.1 | 104.3 |
| C15N3H11 | 2,2',6',2'-Terpyridine | 94.3 | 88.6 | 102.2 |
| O | Oxygen, atom | 59.6 | 59.6 | 40.2 |
| HO | Hydroxyl radical | 9.5 | 0.2 | -1.5 |
| HO | Hydroxide, anion | -33.2 | -5.8 | -14.2 |
| H2O | Water | -57.8 | -61.0 | -48.4 |
| H3O | Hydronium, cation | 138.9 | 134.2 | 144.3 |
| CO | Carbon monoxide | -26.4 | -6.0 | 5.5 |
| CHO | HCO, cation | 199 | 184.8 | 193.7 |
| CHO | HCO | 10.4 | -1.4 | -11.2 |
| CH2O | Formaldehyde | -26 | -32.9 | -26.5 |
| CH3O | CH2OH, cation | 168 | 155.5 | 170.7 |
| CH3O | Methoxy, anion | -36 | -39.8 | -26.5 |
| CH4O | Methanol | -48.1 | -57.4 | -46.0 |
| C2H2O | Ketene | -11.4 | -6.9 | -21.9 |
| C2H4O | Acetaldehyde | -39.7 | -42.3 | -43.2 |
| C2H4O | Ethylene oxide | -12.6 | -15.5 | -15.1 |
| C2H5O | Ethoxy, anion | -47.5 | -45.3 | -39.9 |
| C2H6O | Dimethyl ether | -44 | -51.2 | -42.4 |
| C2H6O | Ethanol | -56.2 | -63.0 | -55.1 |
| C3H6O | Acetone | -52 | -49.5 | -55.8 |
| C3H6O | Propanal | -45.5 | -47.4 | -48.7 |
| C3H6O | Trimethylene oxide | -19.3 | -37.2 | -32.2 |
| C3H8O | Isopropanol | -65.1 | -65.2 | -61.8 |
| C3H8O | Methyl ethyl ether | -51.7 | -56.7 | -50.0 |
| C3H8O | Propanol | -61.2 | -67.6 | -60.9 |
| C4H4O | Acetyl acetylene | 15.6 | 12.3 | -1.0 |
| C4H4O | Furan | -8.3 | -8.7 | -13.2 |
| C4H6O | 2,3-Dihydrofuran | -17.3 | -29.9 | -30.8 |
| C4H6O | Crotonaldehyde | -24 | -27.9 | -28.2 |
| C4H6O | Divinyl ether | -3.3 | -2.0 | -6.6 |
| C4H8O | Butanal | -48.9 | -52.9 | -54.1 |
| C4H8O | Isobutanal | -51.6 | -50.6 | -51.4 |
| C4H8O | Methyl ethyl ketone | -57.1 | -54.1 | -60.8 |
| C4H8O | Tetrahydrofuran | -44 | -59.3 | -53.4 |
| C4H10O | Diethyl ether | -60.3 | -62.0 | -58.5 |
| C4H10O | t-Butanol | -74.7 | -64.4 | -63.2 |
| C5H8O | 2,3-Dihydro-5-methyl-furan | -31.1 | -39.9 | -37.5 |
| C5H8O | 2-Ethylacrolein | -31.4 | -30.7 | -29.9 |
| C5H8O | 3,4-Dihydro-2H-pyran | -27 | -39.4 | -38.5 |
| C5H8O | 3-Penten-2-one | -32.6 | -34.9 | -40.6 |
| C5H8O | Cyclopentanone | -46 | -57.1 | -63.2 |
| C5H10O | Diethyl ketone | -61.6 | -59.5 | -64.5 |
| C5H10O | Tetrahydropyran | -53.4 | -62.1 | -58.7 |
| C5H12O | t-Butyl methyl ether | -67.8 | -54.6 | -56.9 |
| C6H5O | Phenoxy, anion | -40.5 | -42.3 | -54.1 |
| C6H6O | Phenol | -23 | -26.7 | -22.0 |
| C6H10O | 4-Methyl-3-penten-2-one | -42.6 | -40.9 | -47.5 |
| C6H10O | Cyclohexanone | -54 | -60.2 | -67.0 |
| C6H12O | Methyl neopentyl ketone | -76.6 | -50.9 | -64.1 |
| C6H14O | Di-isopropyl ether | -76.3 | -62.5 | -68.2 |
| C7H6O | Benzaldehyde | -8.8 | -9.6 | -9.7 |
| C7H8O | Anisole | -17.3 | -17.7 | -15.0 |
| C7H8O | m-Cresol | -31.9 | -34.2 | -29.4 |
| C7H8O | o-Cresol | -30.7 | -33.3 | -27.7 |
| C7H8O | p-Cresol | -29.9 | -34.6 | -28.9 |
| C7H10O | 2-Methyl-5-hexen-3-yn-2-ol | 11 | 5.3 | 0.7 |
| C7H10O | 2-Norbornanone | -40.8 | -37.4 | -54.5 |
| C7H10O | cis-2,3-Epoxybicyclo[2.2.1]heptane | -12.9 | 2.8 | -14.8 |
| C7H10O | Norbornan-7-one | -32 | -37.7 | -52.1 |
| C7H12O | 1-Methoxy cyclohexene | -38.5 | -49.1 | -45.0 |
| C7H12O | Bicyclo[2.2.1]heptan-7-ol | -52 | -51.2 | -60.8 |
| C7H12O | cis-1,2-Epoxycycloheptane | -36.4 | -35.6 | -41.5 |
| C7H12O | Cycloheptanone | -59.3 | -59.6 | -68.4 |
| C7H14O | 2,4-Dimethyl 3-pentanone | -74.4 | -63.1 | -68.7 |
| C7H14O | 2-Methyl cis-cyclohexanol | -71.9 | -73.8 | -76.8 |
| C7H14O | 3,3-Dimethyl-2-pentanone | -72.6 | -52.3 | -69.1 |
| C7H14O | 4-Heptanone | -71.3 | -67.8 | -74.7 |
| C7H14O | Heptanal | -63.1 | -67.1 | -70.6 |
| C7H14O | t-Butyl ethyl ketone | -75 | -55.4 | -68.3 |
| C7H16O | n-Heptanol | -81.2 | -86.7 | -82.3 |
| C7H16O | t-Butyl isopropyl ether | -85.5 | -60.5 | -69.4 |
| C8H6O | Benzofuran | 3.3 | 3.8 | 4.4 |
| C8H8O | 1,3-Dihydro isobenzofuran | -7.2 | -20.4 | -15.7 |
| C8H8O | 1-Phenylethenol | -11 | -9.5 | -4.9 |
| C8H8O | 2,3-Dihydro-benzofuran | -11.1 | -22.1 | -18.2 |
| C8H8O | Acetophenone | -20.7 | -17.1 | -21.7 |
| C8H10O | 2,3-Dimethyl phenol | -37.6 | -36.0 | -36.0 |
| C8H10O | 2,4-Dimethyl phenol | -39 | -40.2 | -35.9 |
| C8H10O | 2,5-Dimethyl phenol | -38.7 | -40.1 | -36.3 |
| C8H10O | 2,6-Dimethyl phenol | -38.7 | -38.5 | -34.7 |
| C8H10O | 2-Ethyl phenol | -34.7 | -35.2 | -32.3 |
| C8H10O | 3,4-Dimethyl phenol | -37.4 | -39.5 | -36.3 |
| C8H10O | 3,5-Dimethyl phenol | -38.6 | -41.6 | -36.6 |
| C8H10O | 3-Ethyl phenol | -34.9 | -39.1 | -33.6 |
| C8H10O | 4-Ethyl phenol | -34.5 | -39.4 | -33.3 |
| C8H10O | 2-phenylethanol | -22.6 | -34.0 | -30.6 |
| C8H10O | Ethoxybenzene | -24.3 | -23.0 | -22.5 |
| C8H10O | Phenetole | -26.3 | -22.6 | -22.5 |
| C8H12O | 1-Methylnorcamphor | -48.9 | -40.9 | -57.7 |
| C8H12O | Bicyclo[2.2.2]octanone | -52.2 | -53.4 | -66.2 |
| C8H12O | Bicyclo[3.2.1]octan-2-one | -52 | -52.1 | -65.0 |
| C8H12O | Bicyclo[3.2.1]octan-3-one | -52.9 | -52.1 | -67.7 |
| C8H12O | Bicyclo[3.2.1]octan-8-one | -46.2 | -52.6 | -63.9 |
| C8H12O | cis-Bicyclo[3.3.0]-octan-2-one | -55 | -64.0 | -72.2 |
| C8H12O | trans-Bicyclo[3.3.0]-octan-2-one | -49.4 | -48.2 | -60.6 |
| C8H14O | 3-Oxabicyclo[3,2,2]nonane | -53.2 | -58.0 | -61.1 |
| C8H14O | 6-Methyl-5-hepten-2-one | -60.1 | -48.3 | -61.3 |
| C8H14O | 8-Oxatricyclo[3,2,1,0(1,5)]octane | 6.4 | 50.3 | 22.7 |
| C8H14O | Bicyclo(2.2.2)octan-2-ol | -68 | -66.0 | -72.6 |
| C8H14O | cis-1,2-Epoxycyclooctane | -39.5 | -34.3 | -45.8 |
| C8H14O | Cyclooctanone | -65.1 | -61.5 | -75.0 |
| C8H16O | 2,2,4-Trimethyl-3-pentanone | -80.9 | -56.6 | -72.1 |
| C8H16O | 2-Octanone | -82.5 | -72.9 | -81.6 |
| C8H16O | 3,3,4-Trimethyl pentan-2-one | -78.5 | -48.5 | -71.9 |
| C8H16O | 3-Octanone | -80.9 | -74.0 | -80.9 |
| C8H16O | 4-Octanone | -83.5 | -74.0 | -81.3 |
| C8H16O | Octanal | -69.8 | -71.1 | -76.1 |
| C8H18O | 1-Octanol | -85.3 | -91.4 | -87.8 |
| C8H18O | 1-Tert-butoxybutane | -86.3 | -69.2 | -76.6 |
| C8H18O | 2-(1,1-Dimethylethoxy)-butane | -90.8 | -63.2 | -75.0 |
| C8H18O | Di-n-butyl ether | -79.8 | -81.0 | -80.5 |
| C8H18O | Di-sec-butyl ether | -86.3 | -69.1 | -78.7 |
| C8H18O | tert-Butyl ether | -86.3 | -51.8 | -70.1 |
| C8H18O | tert-Butyl isobutyl ether | -88 | -65.3 | -75.8 |
| C9H10O | 3,4-Dihydro-1H-2-benzopyran | -15.1 | -26.7 | -24.1 |
| C9H10O | 3,4-Dihydro-2H-1-benzopyran | -19.7 | -28.8 | -25.4 |
| C9H10O | Benzyl methyl ketone | -22.6 | -20.8 | -29.8 |
| C9H12O | 2(-1-Methylethyl)-phenol | -41.9 | -37.2 | -34.9 |
| C9H12O | 2,4,6-Trimethyl phenol | -42.3 | -46.2 | -41.4 |
| C9H12O | 3(-1-Methylethyl)-phenol | -41.9 | -39.6 | -37.0 |
| C9H12O | 4-(-1-Methylethyl)-phenol | -41.9 | -40.0 | -36.8 |
| C9H14O | 2,6,6-Trimethyl-2-cyclohexen-1-one | -55.6 | -39.9 | -49.5 |
| C9H14O | Bicycle[3.3.1]nonan-9-one -check -3-one | -57.3 | -58.4 | -70.2 |
| C9H14O | cis Octahydro-2H-inden-2-one | -59.7 | -64.3 | -77.0 |
| C9H14O | trans Octahydro-2H-inden-2-one | -59.6 | -62.1 | -76.0 |
| C9H16O | Cyclononanone | -66.9 | -61.7 | -80.8 |
| C9H18O | 2,6-Dimethyl-4-heptanone | -85.5 | -69.7 | -83.0 |
| C9H18O | 2-Nonanone | -81.5 | -78.3 | -87.1 |
| C9H18O | 3,3,4,4-Tetramethyl-2-pentanone | -83.1 | -37.6 | -72.9 |
| C9H18O | 5-Nonanone | -82.4 | -77.3 | -85.4 |
| C9H18O | Di-tert-butyl ketone | -82.7 | -43.6 | -70.9 |
| C9H20O | 1-Nonanol | -89.8 | -96.1 | -93.3 |

| Formula | Name | | | |
|---|---|---|---|---|
| C9H20O | Amyl-t-butyl ether | -91 | -73.9 | -82.2 |
| C9H20O | Butyl 1,1-dimethylpropyl ether | -91.4 | -71.5 | -81.4 |
| NO | Nitric oxide, cation | 237 | 230.6 | 268.1 |
| NO | Nitric oxide | 21.6 | -0.5 | 17.1 |
| CNO | NCO | 38.1 | 31.6 | 19.0 |
| CNHO | Hydrogen isocyanate | -24.3 | -10.8 | -29.6 |
| CNH3O | Formamide | -44.5 | -40.2 | -43.2 |
| C2NH5O | Acetaldoxime | -5.4 | -18.3 | -9.2 |
| C2NH5O | Acetamide | -57 | -48.2 | -55.2 |
| C3NH3O | Isoxazole | 19.6 | 19.2 | 17.6 |
| C3NH3O | Oxalone (oxazole) | -3.7 | -8.3 | -9.8 |
| C3NH5O | Acrylamine | -31.1 | -24.4 | -28.7 |
| C3NH5O | Methoxyacetonitrile | -8.5 | -15.6 | -11.0 |
| C3NH7O | Dimethylformamide | -45.8 | -37.0 | -43.8 |
| C3NH7O | N-Methyl acetamide | -59.3 | -47.0 | -57.1 |
| C3NH7O | Propanamide | -61.9 | -53.5 | -59.7 |
| C3NH9O | Dimethylaminomethanol | -48.6 | -50.2 | -46.1 |
| C4NH5O | 3-Methyl isoxazole | 8.5 | 7.4 | 9.4 |
| C4NH5O | 5-Methyl isoxazole | 8.1 | 8.1 | 9.6 |
| C4NH7O | 2-Pyrrolidinone | -47.2 | -51.9 | -60.0 |
| C4NH7O | 4,5-Dihydro-2-methyl oxazole | -31.2 | -39.6 | -35.1 |
| C4NH7O | Methacrylamide | -37.7 | -32.0 | -36.3 |
| C4NH9O | 2-Methyl propanamide | -67.5 | -55.2 | -61.7 |
| C4NH9O | Butanamide | -66.7 | -58.2 | -64.5 |
| C4NH9O | Isobutylamide | -67.5 | -55.2 | -61.7 |
| C4NH11O | N,N-Diethyl-hydroxylamine | -29.1 | -28.3 | -29.1 |
| C5NH5O | 2-Pyridinol | -19 | -25.0 | -16.9 |
| C5NH5O | 3-Pyridinol | -10.4 | -18.7 | -11.9 |
| C5NH5O | 4-Pyridinol | -7.2 | -20.0 | -13.3 |
| C5NH5O | Pyridine 1 oxide | 21 | 44.4 | 25.4 |
| C5NH7O | 3,5-Dimethyl isoxazole | -4.3 | -3.6 | 1.5 |
| C5NH9O | 1-Methyl-2-pyrrolidinone | -50.4 | -51.0 | -58.6 |
| C5NH9O | 2-Ethyl-4,5-dihydro-oxazole | -35.6 | -44.1 | -39.1 |
| C5NH9O | N,N-Dimethylamino-2-propen-3-al | -24.9 | -14.6 | -22.2 |
| C5NH11O | 1-(Dimethylamino)-2-propanone | -43 | -37.3 | -44.8 |
| C5NH11O | 2,2-Dimethyl-propanamide | -74.8 | -50.1 | -63.0 |
| C5NH11O | N,N-Dimethyl propanamide | -59.8 | -45.6 | -57.9 |
| C6NH7O | 2-Hydroxy-6-methylpyridine | -28.8 | -33.9 | -24.7 |
| C6NH7O | 3-Hydroxy-2-methylpyridine | -20.2 | -26.6 | -18.3 |
| C6NH7O | 3-Hydroxy-6-methylpyridine | -16.7 | -27.8 | -19.1 |
| C6NH7O | 4-Hydroxy-2-methylpyridine | -17.1 | -28.9 | -21.1 |
| C6NH7O | 6-Methyl-2(1H)-pyridinone | -28.8 | -23.0 | -34.2 |
| C6NH7O | m-Amino phenol | -23.6 | -26.4 | -20.1 |
| C6NH7O | o-Amino phenol | -25 | -26.2 | -17.8 |
| C6NH7O | p-Amino phenol | -21.6 | -25.9 | -18.2 |
| C6NH9O | Trimethyl isoxazole | -4.8 | -13.3 | -5.2 |
| C6NH11O | Caprolactam | -57.3 | -52.9 | -65.9 |
| C6NH11O | Cyclohexanone oxime | -17.9 | -35.3 | -29.2 |
| C6NH13O | N,N-Diethyl acetamide | -68.6 | -48.9 | -70.4 |
| C6NH13O | N,N-Dimethylbutyramine | -64.7 | -50.1 | -63.6 |
| C7NH5O | Benzoxazole | 10.8 | 5.2 | 9.2 |
| C7NH5O | Isocyanatobenzene | -3.5 | 14.0 | -4.1 |
| C7NH7O | Benzamide | -24.1 | -16.2 | -20.4 |
| C7NH11O | N,N-Dimethylamino-2,4-pentadiene-5-al | -7 | -0.1 | -8.1 |
| C7NH13O | 2-Methoxy-3,3-dimethylbutanenitrile | -39.4 | -18.9 | -22.2 |
| C7NH15O | N,N-Diethylaminoacetone | -55.8 | -44.0 | -64.1 |
| C7NH15O | N,N-Dimethyl-tert-butylcarboxamide | -68.4 | -41.3 | -61.0 |
| C8NH5O | alpha-oxo Benzeneacetonitrile | 28.1 | 25.4 | 27.6 |
| C8NH9O | 1,3-dimethyl-2-nitroso-benzene | 33.4 | 16.8 | 22.7 |
| C8NH9O | N-methyl-N-phenyl formamide | -18.1 | -6.7 | -18.2 |
| C8NH17O | Octanone-1-oxime | -35.7 | -46.5 | -41.7 |
| C8NH17O | Octanone-2-oxime | -43.2 | -47.5 | -45.5 |
| C8NH17O | Octanone-3-oxime | -41.3 | -49.0 | -43.5 |
| C8NH17O | Octanone-4-oxime | -43.3 | -48.3 | -44.6 |
| N2O | Nitrous oxide | 19.6 | 31.0 | 26.2 |
| CN2H4O | Urea | -58.7 | -44.8 | -52.0 |
| C2N2H6O | N-Methyl urea | -56.3 | -43.9 | -52.0 |
| C4N2H6O | Dimethyl furazan | 25.6 | 20.9 | 29.1 |
| C4N2H10O | Isopropylurea | -69.3 | -51.3 | -66.4 |
| C5N2H8O | 5-Amino-3,4-dimethylisoxazole | 1.2 | -4.7 | 4.6 |
| C5N2H12O | (1-Methylpropyl) urea | -73.4 | -55.3 | -71.3 |
| C5N2H12O | N,N-diethylurea | -65.1 | -45.5 | -66.9 |
| C5N2H12O | Tetramethylurea | -49.1 | -30.7 | -45.2 |
| C4N3H5O | 4-Amino-2(1H)-pyrimidinone | -14.2 | -22.9 | -9.9 |
| O2 | Oxygen (Singlet) | 22 | 12.1 | 19.7 |
| O2 | Oxygen (Triplet) | 0 | -16.1 | -8.2 |
| H2O2 | Hydrogen peroxide | -32.5 | -38.2 | -43.4 |
| CO2 | Carbon dioxide | -94.1 | -75.1 | -101.9 |
| CHO2 | Formate, anion | -106.6 | -101.6 | -117.6 |
| CH2O2 | Formic acid | -90.5 | -92.6 | -94.6 |
| C2H2O2 | trans Glyoxal | -50.7 | -61.4 | -60.8 |
| C2H3O2 | Acetate, anion | -122.5 | -110.0 | -131.0 |
| C2H4O2 | Acetic acid | -103.3 | -101.1 | -104.5 |
| C2H4O2 | Methyl formate | -83.6 | -85.5 | -86.7 |
| C2H6O2 | Dimethyl peroxide | -30.1 | -28.3 | -35.5 |
| C2H6O2 | Ethylene glycol | -93.9 | -106.0 | -95.6 |
| C3O2 | Carbon suboxide | -22.4 | -23.6 | -41.5 |
| C3H4O2 | 2-Oxo-propanal | -64.8 | -70.9 | -74.3 |
| C3H4O2 | 2-Propenoic acid | -79 | -76.2 | -76.5 |
| C3H4O2 | beta-Propiolactone | -67.6 | -68.9 | -77.2 |
| C3H6O2 | 1,3-Dioxalane | -72.1 | -93.0 | -85.5 |
| C3H6O2 | Ethyl formate | -95.2 | -90.2 | -94.0 |

| Formula | Name | | | |
|---|---|---|---|---|
| C3H6O2 | Methyl acetate | -97.9 | -93.7 | -96.1 |
| C3H6O2 | Propionic acid | -108.4 | -105.7 | -108.5 |
| C3H8O2 | 1,3-Propanediol | -97.6 | -110.2 | -100.7 |
| C3H8O2 | 2-Methoxyethanol | -90.1 | -99.8 | -91.2 |
| C3H8O2 | Dimethoxymethane | -83.2 | -94.4 | -85.3 |
| C3H8O2 | Propylene glycol | -102.7 | -107.6 | -102.1 |
| C4H6O2 | 2-Butenoic acid | -88.1 | -85.9 | -87.8 |
| C4H6O2 | 2-Methyl-2-propenic acid | -87.8 | -83.7 | -84.0 |
| C4H6O2 | Diacetyl | -78.2 | -78.9 | -87.5 |
| C4H6O2 | gamma Butyrolactone | -87 | -94.0 | -99.6 |
| C4H6O2 | Methyl 2-propenoate | -79.6 | -68.6 | -68.5 |
| C4H8O2 | 1,1 Dimethoxy ethene | -67.1 | -70.9 | -62.7 |
| C4H8O2 | 1,3 Dioxan | -80.9 | -94.3 | -89.8 |
| C4H8O2 | 1,4-Dioxane | -75.5 | -89.3 | -84.9 |
| C4H8O2 | Ethyl acetate | -106.5 | -99.0 | -103.4 |
| C4H10O2 | 1,2-Dimethoxyethane | -81.9 | -93.4 | -84.8 |
| C4H10O2 | 1,4 Butandiol | -101.8 | -115.3 | -105.6 |
| C4H10O2 | Diethyl peroxide | -46.1 | -38.5 | -49.6 |
| C4H10O2 | Dimethyl acetal | -93.1 | -95.2 | -87.8 |
| C5H8O2 | Acetylacetone | -91.9 | -84.4 | -98.9 |
| C5H10O2 | Ethyl propionate | -111.5 | -103.5 | -107.2 |
| C5H10O2 | Isopropyl acetate | -115.1 | -100.6 | -107.7 |
| C5H12O2 | 1,5 Pentandiol | -105.6 | -119.9 | -111.3 |
| C6H4O2 | p-Benzoquinone | -29.3 | -33.0 | -37.4 |
| C6H6O2 | 1,2-Benzenediol | -65.7 | -72.7 | -61.2 |
| C6H6O2 | Hydroquinone | -66.2 | -74.0 | -63.8 |
| C6H6O2 | Resorcinol | -68 | -75.0 | -65.4 |
| C6H8O2 | 1,3-Cyclohexanedione | -80.2 | -83.7 | -100.9 |
| C6H8O2 | 1,4-Cyclohexanedione | -79.5 | -84.7 | -99.6 |
| C6H10O2 | 2,4-Hexanedione | -105.1 | -89.0 | -102.9 |
| C6H10O2 | 2-Oxepanone | -94.7 | -94.4 | -107.2 |
| C6H10O2 | 3-Methyl-2,4-pentandione | -102.5 | -84.5 | -101.2 |
| C6H10O2 | Ethyl-(E)-2-butenoate | -89.8 | -83.7 | -86.5 |
| C6H12O2 | 1,1-Dimethoxy-2-butene | -72.4 | -79.7 | -72.3 |
| C6H12O2 | 4-Hydroxy-4-methylpentan-2-one | -116.2 | -96.4 | -107.9 |
| C6H12O2 | 5,5-Dimethyl-1,3-dioxane | -100.7 | -93.6 | -96.2 |
| C6H12O2 | cis-2,4-Dimethyl-1,3-dioxane | -102.3 | -100.7 | -97.0 |
| C6H12O2 | Ethyl butanoate | -115.9 | -108.1 | -112.8 |
| C6H12O2 | Hexanoic acid | -122.5 | -119.7 | -124.9 |
| C6H12O2 | Methyl 2-methylbutanoate | -117.7 | -103.9 | -108.2 |
| C6H12O2 | Methyl 2,2-dimethyl-propanoate | -118.2 | -96.2 | -102.4 |
| C6H12O2 | Methyl 3-methylbutanoate | -119 | -103.9 | -109.7 |
| C6H12O2 | Methyl pentanoate | -112.7 | -107.5 | -110.9 |
| C6H12O2 | t-Butyl acetate | -123.4 | -96.5 | -109.2 |
| C6H12O2 | trans 4,5-Dimethyl-1,3-dioxane | -98.2 | -97.9 | -98.3 |
| C6H14O2 | 1,1-Diethoxy ethane | -108.4 | -106.5 | -104.5 |

| Formula | Name | | | |
|---|---|---|---|---|
| C6H14O2 | 1,1-Dimethoxy-butane | -101.7 | -104.0 | -98.3 |
| C6H14O2 | 1,2-Diethoxy ethane | -97.6 | -104.1 | -102.5 |
| C6H14O2 | 1,6-Hexanediol | -109.8 | -124.7 | -116.9 |
| C6H14O2 | 2,3-Dimethyl-2,3-butanediol | -129.2 | -99.2 | -108.2 |
| C7H6O2 | 3-(2-Furanyl)-2-propenal | -25.3 | -27.8 | -28.8 |
| C7H6O2 | Benzoic acid | -70.1 | -67.7 | -67.9 |
| C7H6O2 | Phenyl formate | -51.6 | -50.0 | -52.3 |
| C7H6O2 | Tropolone | -37.2 | -40.1 | -36.4 |
| C7H8O2 | 3-Methyl-1,2-benzenediol | -71.5 | -80.3 | -69.8 |
| C7H8O2 | 4-Methyl 1,2-Benzenediol | -71.3 | -80.3 | -68.2 |
| C7H10O2 | Ethyl 2-methylene-3-butenoate | -69.2 | -57.6 | -58.4 |
| C7H10O2 | Ethyl 2-pentynoate | -59.8 | -56.2 | -62.9 |
| C7H10O2 | Ethyl 3-pentynoate | -56.8 | -58.2 | -69.7 |
| C7H10O2 | Ethyl 4-pentynoate | -55.7 | -47.0 | -65.4 |
| C7H12O2 | 3,5-Heptanedione | -104.9 | -93.6 | -106.9 |
| C7H12O2 | 3-Ethyl-2,4-pentanedione | -105.1 | -87.8 | -106.9 |
| C7H12O2 | 5-Methyl-2,4-hexanedione | -108.2 | -91.0 | -105.5 |
| C7H12O2 | Butyl 2-propenoate | -89.7 | -83.3 | -86.2 |
| C7H12O2 | Ethyl (Z)-2-pentenoate | -94.3 | -88.4 | -91.4 |
| C7H12O2 | Ethyl (Z)-3-pentenoate | -92.6 | -89.5 | -94.7 |
| C7H12O2 | Ethyl 4-pentenoate | -92.1 | -82.7 | -90.9 |
| C7H12O2 | Ethyl trans-2-pentenoate | -94.2 | -88.4 | -91.4 |
| C7H12O2 | Heptanolactone | -98.3 | -92.4 | -108.0 |
| C7H12O2 | Isopropyl 2-butenoate | -98.2 | -85.3 | -90.9 |
| C7H12O2 | Propyl (E)-2-butenoate | -94.4 | -88.5 | -91.6 |
| C7H14O2 | (2a,4a,6b)-2,4,6-Trimethyl-1,3-dioxane | -106.8 | -103.4 | -100.9 |
| C7H14O2 | 1,1-Dimethoxycyclopentane | -95 | -97.3 | -96.3 |
| C7H14O2 | 1,1-Dimethylpropyl acetate | -128.8 | -97.3 | -114.1 |
| C7H14O2 | Ethyl 2-methylbutanoate | -123.4 | -108.5 | -115.4 |
| C7H14O2 | Ethyl 3-methylbutanoate | -126 | -108.4 | -116.9 |
| C7H14O2 | Ethyl pentanoate | -121.2 | -112.8 | -118.1 |
| C7H14O2 | Methyl 3,3-dimethylbutanoate | -122.3 | -99.6 | -110.4 |
| C7H14O2 | Methyl hexanoate | -118 | -112.2 | -116.4 |
| C7H16O2 | 1,3-Diethoxypropane | -104.3 | -108.2 | -107.9 |
| C7H16O2 | 1,7-Heptanediol | -114.1 | -129.4 | -122.5 |
| C8H8O2 | m-Methylbenzoic acid | -78.4 | -75.3 | -75.2 |
| C8H8O2 | Methyl benzoate | -66.8 | -60.3 | -59.5 |
| C8H8O2 | o-Methylbenzoic acid | -76.6 | -73.3 | -74.1 |
| C8H8O2 | p-Methylbenzoic acid | -79 | -75.4 | -75.8 |
| NO2 | Nitrogen dioxide, cation | 233 | 240.7 | 238.7 |
| NO2 | Nitrogen dioxide | 7.9 | -6.0 | -14.4 |
| NHO2 | Nitrous acid, trans | -18.8 | -40.7 | -27.3 |
| CNH3O2 | Methyl nitrite | -15.8 | -36.7 | -23.1 |
| CNH3O2 | Nitromethane | -17.9 | 3.3 | -14.5 |
| C2NH5O2 | Ethyl nitrite | -25.9 | -42.0 | -30.9 |
| C2NH5O2 | Glycine | -93.7 | -95.7 | -92.9 |

| Formula | Name | | | |
|---|---|---|---|---|
| C2NH5O2 | Methyl carbamate | -101.6 | -89.0 | -94.3 |
| C2NH5O2 | Nitroethane | -23.5 | -2.1 | -21.2 |
| C3NH7O2 | Alanine | -99.1 | -98.7 | -96.9 |
| C3NH7O2 | beta-Alanine | -101 | -99.0 | -97.7 |
| C3NH7O2 | Isopropylnitrite | -31.9 | -44.7 | -35.7 |
| C3NH7O2 | N-Methyl glycine | -87.8 | -94.0 | -95.9 |
| C3NH7O2 | Propyl nitrite | -28.4 | -47.0 | -36.2 |
| C3NH7O2 | Urethane | -106.7 | -94.3 | -101.7 |
| C4NH5O2 | Methyl cyanoacetate | -58.2 | -57.1 | -60.8 |
| C4NH5O2 | Succinimide | -89.8 | -87.7 | -105.9 |
| C4NH9O2 | 2-Nitrobutane | -39.1 | -10.2 | -32.4 |
| C4NH9O2 | 2-Nitroisobutane | -42.2 | -3.7 | -28.7 |
| C4NH9O2 | 4-Aminobutanoic acid | -105 | -104.3 | -103.0 |
| C4NH9O2 | Isobutyl nitrite | -36.1 | -47.9 | -40.5 |
| C4NH9O2 | n-Butyl nitrite | -34.8 | -51.6 | -41.4 |
| C4NH9O2 | Sec-butyl nitrite | -36.5 | -48.2 | -40.8 |
| C4NH9O2 | t-Butyl nitrite | -41 | -42.6 | -37.9 |
| C4NH11O2 | Diethanolamine | -94.9 | -103.4 | -96.8 |
| C5NH5O2 | N-Methylmaleimide | -61.2 | -51.1 | -64.7 |
| C5NH7O2 | Glutarimide | -94.1 | -92.0 | -111.0 |
| C5NH7O2 | N-Methylsuccinimide | -93.1 | -86.1 | -103.5 |
| C5NH9O2 | Proline | -87.5 | -101.0 | -100.8 |
| C5NH11O2 | 5-Aminovaleric acid | -110 | -109.1 | -108.5 |
| C5NH11O2 | N,N-Dimethylglycine methyl ester | -88.5 | -81.9 | -84.7 |
| C5NH11O2 | tert-Pentyl nitrite | -45.8 | -44.2 | -42.7 |
| C5NH11O2 | Valine | -108.2 | -101.1 | -103.5 |
| C5NH13O2 | 1,1-Dimethoxy-trimethylamine | -85 | -84.4 | -76.5 |
| C6NH5O2 | Niacin | -52.9 | -60.1 | -58.4 |
| C6NH5O2 | Nitrobenzene | 15.4 | 35.8 | 19.3 |
| C6NH9O2 | Ethyl 2-cyanopropionate | -74.6 | -64.6 | -70.4 |
| C6NH13O2 | Ethyl N,N-dimethylglycinate | -97.3 | -86.4 | -91.9 |
| C6NH13O2 | Hexanoic acid, 6-amino- | -115.3 | -112.9 | -114.1 |
| C6NH13O2 | Isoleucine | -110 | -103.2 | -109.0 |
| C6NH13O2 | Leucine | -113.1 | -106.7 | -111.7 |
| C6NH13O2 | Methyl N,N-dimethylalaninate | -94.3 | -81.6 | -89.4 |
| C6NH15O2 | N,N-Dimethylacetamide dimethyl acetal | -92.5 | -78.4 | -79.8 |
| C7NH7O2 | m-Aminobenzoic acid | -69.2 | -67.0 | -65.2 |
| C7NH7O2 | o-Aminobenzoic acid | -70.8 | -66.3 | -66.9 |
| C7NH7O2 | p-Aminobenzoic acid | -70.2 | -67.7 | -67.2 |
| C7NH7O2 | p-Nitrotoluene | 7.4 | 28.1 | 11.0 |
| C7NH7O2 | Phenylnitromethane | 7.3 | 29.0 | 10.6 |
| C7NH15O2 | Methyl N,N-,a,a-tetramethylglycinate | -108.1 | -73.0 | -89.1 |
| C2N2H4O2 | Oxalamide | -92.5 | -76.3 | -87.2 |
| C2N2H6O2 | N-Nitrodimethylamine | -3.2 | 22.3 | -4.8 |
| C3N2H6O2 | Acetyl-urea | -105.6 | -82.4 | -104.4 |
| C3N2H6O2 | Propanediamide | -99.5 | -82.2 | -98.7 |

| Formula | Name | | | |
|---|---|---|---|---|
| C4N2H4O2 | Pyrazine-1,4-dioxide | 36.3 | 71.7 | 34.8 |
| C4N2H4O2 | Uracil | -72.4 | -64.7 | -87.0 |
| C5N2H6O2 | Thymine | -78.6 | -72.5 | -92.8 |
| C6N2H6O2 | m-Nitroaniline | 14.9 | 36.5 | 21.6 |
| C6N2H6O2 | p-Nitroaniline | 13.2 | 35.6 | 18.9 |
| C6N2H14O2 | Lysine | -102.5 | -104.4 | -101.8 |
| C2N3H5O2 | Imidodicarbonic diamide | -104.4 | -77.8 | -104.1 |
| O3 | Ozone | 34.1 | 48.5 | 45.7 |
| C3H6O3 | 1,3,5-Trioxane | -111.3 | -130.2 | -123.1 |
| C3H6O3 | Methyl hydroxyacetate | -133.1 | -137.0 | -133.7 |
| C3H8O3 | Glycerol | -138.1 | -150.8 | -141.4 |
| C4H2O3 | Malaic anhydride | -95.2 | -88.5 | -100.9 |
| C4H6O3 | Acetic anhydride | -137.1 | -132.6 | -145.0 |
| C4H10O3 | Trimethoxymethane | -127.1 | -135.9 | -128.3 |
| C6H14O3 | 2,5,8-Trioxanonane | -124.6 | -135.4 | -128.2 |
| C7H6O3 | m-Salicylic acid | -112.1 | -115.1 | -110.5 |
| C7H6O3 | o-Salicylic acid | -118.5 | -114.1 | -112.7 |
| C7H6O3 | p-Salicylic acid | -117.7 | -116.0 | -111.6 |
| C7H14O3 | 2,3-Butanediol, 2,3-dimethyl-, monoformate | -161.7 | -121.8 | -141.6 |
| NO3 | Nitrate anion | -74.7 | -66.9 | -98.2 |
| NHO3 | Nitric acid | -32.1 | -17.4 | -37.0 |
| CNH3O3 | Methyl nitrate | -29.1 | -12.3 | -33.1 |
| C2NH3O3 | Oxamic acid | -132 | -127.3 | -132.8 |
| C2NH5O3 | Ethyl nitrate | -36.8 | -17.8 | -40.9 |
| C3NH7O3 | Serine | -133.5 | -141.3 | -135.9 |
| C4NH3O3 | 2-Nitrofuran | -6.9 | 7.9 | -9.8 |
| C4NH9O3 | Threonine | -140.5 | -142.5 | -138.5 |
| C6NH5O3 | m-Nitrophenol | -25.2 | -11.4 | -22.2 |
| C6NH5O3 | o-Nitrophenol | -30.8 | -10.2 | -21.5 |
| C6NH5O3 | p-Nitrophenol | -27.4 | -12.5 | -24.4 |
| N2O3 | Dinitrogen trioxide | 19.8 | 12.5 | 11.7 |
| C2H2O4 | Oxalic acid | -175 | -170.5 | -174.3 |
| C2H6O4 | Dioxybismethanol | -124.5 | -129.1 | -133.3 |
| C4H4O4 | 1,4-Dioxan-2,5-dione | -146.3 | -154.3 | -164.9 |
| C4H6O4 | Dimethyl oxalate | -169.5 | -164.0 | -164.4 |
| C4H8O4 | 1,3,5,7-Tetroxane | -148.2 | -168.3 | -166.1 |
| C5H8O4 | Dimethyl malonate | -176.4 | -170.4 | -177.2 |
| C5H8O4 | Ethylmalonic acid | -203.1 | -190.2 | -199.6 |
| C5H8O4 | Methylene diacetate | -186.6 | -178.2 | -190.4 |
| C5H12O4 | Tetramethoxymethane | -173.8 | -181.1 | -161.4 |
| C6H10O4 | 1,1-Diacetoxyethane | -194.2 | -180.5 | -193.0 |
| C6H10O4 | Dimethyl methylmalonate | -183.7 | -172.1 | -177.9 |
| C4NH7O4 | Aspartic acid | -185.9 | -182.7 | -189.1 |
| N2O4 | Dinitrogen tetroxide | 2.2 | 30.2 | 2.5 |
| CN2H2O4 | Dinitromethane | -13.3 | 27.9 | -8.6 |
| CN3H3O4 | Methyldinitramine | 10.3 | 53.2 | 8.8 |

| Formula | Name | | | |
|---|---|---|---|---|
| C3H4O5 | Tartronic acid | -227.5 | -226.4 | -228.9 |
| C5H10O5 | 1,3,5,7,9-pentaoxecane | -186.4 | -208.8 | -206.1 |
| N2O5 | Dinitrogen pentoxide | 2.7 | 34.4 | -8.8 |
| C10H10O | 4-Phenyl-3-buten-2-one | -11.5 | -2.8 | -8.8 |
| C10H14O | 2-Adamantone | -55.1 | -53.7 | -70.4 |
| C10H14O | 2-Isopropyl-4-methylphenol | -47.4 | -42.5 | -40.2 |
| C10H14O | 2-Isopropyl-5-methylphenol | -44.1 | -44.8 | -42.2 |
| C10H14O | 2-Isopropyl-6-methylphenol | -45.4 | -40.5 | -40.3 |
| C10H14O | 2-Methyl-5-isopropylphenol | -46.4 | -45.5 | -44.0 |
| C10H14O | 3-Isopropyl-2-methylphenol | -43.6 | -39.1 | -43.7 |
| C10H14O | 3-Methyl-2-isopropylphenol | -41 | -38.8 | -40.4 |
| C10H14O | 3-Methyl-5-isopropylphenol | -50.2 | -46.9 | -44.2 |
| C10H14O | 4-Isopropyl-2-methylphenol | -49.4 | -45.6 | -43.8 |
| C10H14O | 4-Isopropyl-3-methylphenol | -44 | -40.3 | -45.1 |
| C10H14O | 4-Methyl-3-isopropylphenol | -44.2 | -40.3 | -44.8 |
| C10H14O | m-tert-Butylphenol | -44.3 | -32.1 | -38.7 |
| C10H14O | o-sec-Butylphenol | -45.8 | -38.1 | -38.9 |
| C10H14O | o-tert-Butylphenol | -47.6 | -26.7 | -33.7 |
| C10H14O | p-sec-Butylphenol | -45.6 | -43.4 | -42.5 |
| C10H14O | p-tert-Butyl phenol | -44.5 | -32.6 | -38.4 |
| C10H16O | 1-Adamantol | -74.3 | -65.9 | -75.8 |
| C10H16O | 2-Adamantol | -71.5 | -65.2 | -78.0 |
| C10H16O | Camphor | -63.9 | -33.9 | -61.6 |
| C10H16O | Octahydro-3a-methyl-cis-2H-inden-2-one | -68.6 | -60.7 | -80.0 |
| C10H16O | Octahydro-3a-methyl-trans-2H-inden-2-one | -65.8 | -54.2 | -75.9 |
| C10H18O | Beta-caran-3-ol | -61.9 | -46.8 | -62.1 |
| C10H18O | Cyclodecanone | -72.9 | -63.3 | -85.1 |
| C10H20O | 2,2,5,5-Tetramethyl-3-hexanone | -94.1 | -54.6 | -78.1 |
| C10H22O | Decanol | -94.5 | -100.8 | -98.8 |
| C10H22O | Dipentyl ether | -93.1 | -89.4 | -88.0 |
| C11H14O | 2,4,5-Trimethyl-acetophenone | -45.2 | -35.4 | -43.7 |
| C11H14O | 2,4,6-Trimethyl-acetophenone | -49 | -34.7 | -42.3 |
| C11H16O | 2-tert-Butyl-p-cresol | -49.5 | -34.3 | -40.4 |
| C11H16O | 3-Methyl-2-phenylbutane-2-ol | -50.5 | -25.6 | -37.1 |
| C11H20O | Cycloundecanone | -77 | -66.4 | -93.1 |
| C11H22O | 2,2,6,6-Tetramethyl-4-heptanone | -100.7 | -60.4 | -88.8 |
| C11H22O | Dipentyl ketone | -92.6 | -87.8 | -96.4 |
| C11H24O | Decyl methyl ether | -91.1 | -94.5 | -94.1 |
| C12H8O | Dibenzofuran | 11.3 | 14.4 | 19.8 |
| C12H10O | m-Hydroxybiphenyl | 5.1 | -1.7 | 3.2 |
| C12H10O | o-Hydroxybiphenyl | 4 | -0.7 | 5.7 |
| C12H10O | p-Hydroxybiphenyl | 0 | -2.1 | 3.2 |
| C12H16O | Isobutyl phenyl ketone | -38.4 | -31.9 | -43.0 |
| C12H18O | 2,6-Diisopropylphenol | -60.7 | -44.7 | -48.2 |
| C13H10O | Benzophenone | 11.9 | 16.0 | 14.8 |
| C14H10O | Anthone | 5.6 | 12.3 | 8.5 |

| Formula | Name | | | |
|---|---|---|---|---|
| C9NH7O | 2(1H)-Quinolinone | -6.1 | -3.4 | -11.6 |
| C9NH7O | 3-Phenyl isoxazole | 33.3 | 40.6 | 43.3 |
| C9NH7O | 4-Quinolinol | 5 | -3.1 | 6.5 |
| C9NH7O | 5-Phenyl isoxazole | 38.3 | 41.6 | 44.4 |
| C9NH7O | 8-Quinolinol | 1.6 | -4.6 | 4.5 |
| C9NH7O | a-Cyanoacetophenone | 16.8 | 17.9 | 13.6 |
| C9NH11O | 2,4,6-Trimethylnitrosobenzene | 25.7 | 9.3 | 14.8 |
| C9NH11O | N,N-Dimethyl benzamide | -20.6 | -9.3 | -17.8 |
| C9NH13O | N,N-Dimethyamino-2,4,6-heptatriene-7-al | 5.2 | 13.8 | 5.2 |
| C9NH17O | 2,2,6,6-Tetramethyl-4-piperidinone | -65.4 | -39.7 | -68.0 |
| C10NH9O | 2-Methyl-4-hydroxyquinoline | -5.6 | -11.8 | -1.2 |
| C10NH9O | 2-Methyl-8-quinolinol | -9.4 | -13.6 | -3.4 |
| C10NH9O | 3-Methyl-5-phenyl isoxazole | 24.5 | 29.9 | 36.2 |
| C10NH9O | 4-Methyl-2-hydroxyquinoline | -14.6 | -14.7 | -6.1 |
| C10NH9O | 5-Methyl-3-phenyl isoxazole | 23.6 | 29.6 | 35.3 |
| C10NH9O | beta-Cyanopropiophenone | 7.2 | 12.4 | 6.2 |
| C10NH11O | 2,4,6-Trimethylbenzonitrile, N-oxide | 32.7 | 48.6 | 30.4 |
| C10NH13O | 2-(Dimethylamino)-acetophenone | -11.1 | -4.7 | -9.0 |
| C10NH13O | N,N,4-Trimethyl benzamide | -26.9 | -17.1 | -25.3 |
| C11NH13O | (E)-3-(Methylamino)-1-phenyl-but-2-enone | -12.8 | 2.9 | -8.4 |
| C11NH15O | 1-Propanone, 2-(dimethylamino)-1-phenyl- | -16.7 | -4.9 | -13.8 |
| C11NH15O | 2-Propanamine, 2-methyl-N-(phenylmethylene)-, N-oxide | 7.4 | 50.4 | 21.5 |
| C11NH17O | 1-Adamantanecarboxamide | -76.2 | -52.4 | -77.4 |
| C11NH23O | N,N-Dimethylnonamide | -89.4 | -74.1 | -91.0 |
| C12NH9O | Phenoxazine | 22.5 | 16.8 | 18.9 |
| C12NH17O | 2-(Diethylamino)-1-phenylethanone | -23 | -12.0 | -26.0 |
| C13NH9O | 9,10-Dihydro-9-oxoacridine | 8.3 | 17.7 | 11.8 |
| C13NH11O | Benzenamine, N-(phenylmethylene)-, N-oxide | 62.9 | 86.1 | 68.8 |
| C13NH19O | 1-Propanone, 2-(diethylamino)-1-phenyl- | -29.7 | -11.8 | -32.1 |
| C13NH21O | N,N-Dimethyl-1-adamantylcarboxamide | -68.4 | -43.8 | -74.4 |
| C9H10O2 | 1,3-Dioxolane-2-phenyl | -49.1 | -61.2 | -58.5 |
| C9H10O2 | 2,3-Dimethylbenzoic acid | -82.7 | -77.4 | -81.4 |
| C9H10O2 | 2,4-Dimethylbenzoic acid | -84.9 | -80.9 | -81.7 |
| C9H10O2 | 2,5-Dimethylbenzoic acid | -83.9 | -81.0 | -81.3 |
| C9H10O2 | 2,6-Dimethylbenzoic acid | -81.6 | -78.1 | -80.5 |
| C9H10O2 | 3,4-Dimethylbenzoic acid | -86.6 | -80.6 | -83.0 |
| C9H10O2 | 3,5-Dimethylbenzoic acid | -87.1 | -82.9 | -82.4 |
| C9H10O2 | 3-Ethylbenzoic acid | -82.9 | -80.2 | -79.7 |
| C9H10O2 | 4-Ethylbenzoic acid | -85 | -80.3 | -80.4 |
| C9H10O2 | Methyl 4-methylbenzoate | -74.7 | -65.8 | -67.2 |
| C10H8O2 | 1,2-Naphthalenediol | -47.9 | -54.7 | -45.2 |
| C10H8O2 | 1,3-Naphthalenediol | -50.5 | -56.5 | -45.5 |
| C10H8O2 | 1,4-Naphthalenediol | -47.1 | -53.9 | -42.7 |
| C10H8O2 | 2,3-Naphthalenediol | -46.1 | -55.7 | -43.0 |
| C10H10O2 | 1-Phenyl-1,3-butanedione | -58.3 | -51.9 | -64.4 |
| C10H12O2 | 2,3,4-Trimethylbenzoic acid | -90.2 | -80.9 | -88.1 |

| Formula | Name | | | |
|---|---|---|---|---|
| C10H12O2 | 2,3,5-Trimethylbenzoic acid | -91.4 | -85.0 | -88.4 |
| C10H12O2 | 2,3,6-Trimethylbenzoic acid | -88.7 | -81.9 | -87.5 |
| C10H12O2 | 2,4,5-Trimethylbenzoic acid | -92.3 | -86.1 | -88.8 |
| C10H12O2 | 2,4,6-Trimethylbenzoic acid | -89.4 | -85.6 | -87.9 |
| C10H12O2 | 2-Isopropyl benzoic acid | -85.9 | -77.5 | -81.2 |
| C10H12O2 | 2-Methyl-2-phenyl-1,3-dioxolane | -62.6 | -62.2 | -59.9 |
| C10H12O2 | 3,4,5-Trimethylbenzoic acid | -93.2 | -82.1 | -89.4 |
| C10H12O2 | 3-Isopropyl benzoic acid | -89.8 | -80.7 | -83.3 |
| C10H12O2 | 4-Isopropyl benzoic acid | -91.5 | -80.9 | -83.5 |
| C10H14O2 | 2-Isopropyl-6-methyl-pyrocatechol | -90.6 | -89.7 | -84.8 |
| C10H18O2 | Cyclohexyl butanoate | -130.4 | -119.1 | -128.7 |
| C10H20O2 | Ethyl octanoate | -136.2 | -127.0 | -134.7 |
| C10H22O2 | 1,10-Decanediol | -125 | -143.6 | -139.1 |
| C11H8O2 | 1-Naphthoic acid | -53.3 | -48.3 | -47.4 |
| C11H8O2 | Isonaphthoic acid | -55.6 | -50.5 | -49.5 |
| C11H14O2 | 2,3,4,5-Tetramethylbenzoic acid | -95.3 | -84.4 | -94.7 |
| C11H14O2 | 2,3,4,6-Tetramethylbenzoic acid | -95.2 | -85.4 | -94.1 |
| C11H14O2 | 2,3,5,6-Tetramethylbenzoic acid | -95.6 | -85.8 | -94.4 |
| C11H14O2 | p-tert-Butyl benzoic acid | -95.2 | -73.6 | -85.3 |
| C11H20O2 | Oxacyclododecan-2-one | -123.6 | -111.8 | -136.8 |
| C11H22O2 | Ethyl nonanoate | -141.1 | -131.7 | -140.2 |
| C11H24O2 | 1,1-Dibutoxypropane | -132.1 | -128.5 | -130.0 |
| C12H16O2 | 5,5-Dimethyl-2-phenyl-1,3-dioxane | -74.4 | -61.9 | -65.0 |
| C12H16O2 | Pentamethylbenzoic acid | -101.1 | -83.2 | -100.1 |
| C12H22O2 | 2,2,6,6-Tetramethyl-3,5-heptanedione | -126.3 | -90.0 | -123.5 |
| C12H24O2 | Ethyl decanoate | -146.3 | -136.4 | -145.7 |
| C13H8O2 | Xanthone | -23.5 | -22.3 | -18.7 |
| C13H10O2 | Phenyl benzoate | -34 | -26.9 | -28.3 |
| C8NH9O2 | 2,6-Dimethylnitrobenzene | 2.1 | 24.8 | 5.7 |
| C8NH9O2 | 2-Amino-2-phenylacetic acid | -67 | -63.7 | -64.3 |
| C8NH9O2 | 2-Nitro-m-xylene | 2.1 | 24.8 | 5.7 |
| C8NH9O2 | N-Phenylglycine | -64.2 | -62.6 | -73.1 |
| C8NH17O2 | 8-Aminocaprylic acid | -125 | -123.3 | -125.1 |
| C9NH7O2 | 2-Methyl-1H-isoindole-1,3(2H)-dione | -55.9 | -45.9 | -55.8 |
| C9NH11O2 | Nitromesitylene | -6.4 | 17.3 | -2.0 |
| C9NH11O2 | Phenylalanine | -69.3 | -68.8 | -71.9 |
| C10NH7O2 | 1-Nitroso-2-naphthalenol | 8.6 | -0.4 | 9.6 |
| C10NH13O2 | N,N-Dimethyl 4-methoxybenzamide | -53.7 | -49.0 | -54.2 |
| C10NH15O2 | 1-Nitroadamantane | -45.7 | -4.6 | -41.5 |
| C10NH15O2 | 2-Nitroadamantane | -43 | -3.8 | -41.6 |
| C8N2H8O2 | Isophthalamide | -70.3 | -53.3 | -62.1 |
| C8N2H8O2 | Teraphthalamide | -69.7 | -53.0 | -62.0 |
| C8N2H10O2 | N,N-Dimethyl-m-nitroaniline | 17.4 | 43.6 | 20.8 |
| C8N2H10O2 | N,N-Dimethyl-p-nitroaniline | 16.1 | 43.4 | 17.9 |
| C11N2H12O2 | Tryptophan | -51.6 | -48.2 | -54.8 |
| C6N3H9O2 | Histidine | -63.5 | -61.3 | -66.8 |

| Formula | Name | | | |
|---|---|---|---|---|
| C6N4H14O2 | Argenine | -84.3 | -75.6 | -78.9 |
| C10H14O3 | Trimethoxymethyl benzene | -103.6 | -98.1 | -88.1 |
| C10H22O3 | 1-tert-Butoxy-3-propoxy-2-propanol | -169.2 | -151.7 | -160.2 |
| C11H24O3 | 1-Butoxy-3-tert-butyl-2-propanol | -172.8 | -157.2 | -169.3 |
| C9NH11O3 | Tyrosine | -111.6 | -117.0 | -114.7 |
| C4N2H4O3 | Barbituric acid | -121.8 | -118.7 | -150.4 |
| C4N2H8O3 | Asparagine | -137.8 | -130.4 | -139.6 |
| C4N2H8O3 | GLY-GLY | -136.4 | -129.2 | -137.8 |
| C5N2H10O3 | Glutamine | -142.4 | -134.4 | -145.1 |
| C6N2H12O3 | ALA-ALA | -152.4 | -134.2 | -146.8 |
| C9N2H6O3 | 8-Hydroxy-5-nitroquinoline | -0.4 | 11.8 | 5.5 |
| C10N2H16O3 | PRO-PRO | -131.4 | -136.1 | -153.4 |
| C10N2H20O3 | VAL-VAL | -160.6 | -140.2 | -162.2 |
| C3N3H3O3 | 1,3,5-Triazine-2,4,6(1H,3H,5H)-trione | -134.8 | -118.8 | -157.0 |
| C6N3H9O3 | 1,3,5-Trimethyl-s-triazine-2,4,6-trione | -141.1 | -103.5 | -141.8 |
| C6N3H9O3 | 2,4,6-Trimethoxy-s-triazine | -70.1 | -99.7 | -68.6 |
| C7H12O4 | 2,2-Diacetoxypropane | -199.3 | -171.0 | -191.0 |
| C7H12O4 | Diethyl malonate | -190.1 | -181.0 | -191.9 |
| C7H12O4 | Dimethyl dimethylmalonate | -191.9 | -166.5 | -175.3 |
| C7H16O4 | 3,5,7,9-Tetraoxyundecane | -177.1 | -189.8 | -187.3 |
| C8H16O4 | 12-Crown-4 | -150.8 | -164.0 | -169.8 |
| C10H6O4 | 5,8-Dihydroxy-1,4-naphthalenedione | -119.3 | -113.6 | -113.4 |
| C10H10O4 | Dimethyl isophthalate | -150.4 | -141.0 | -139.6 |
| C10H10O4 | Dimethyl phthalate | -144.9 | -138.2 | -137.0 |
| C10H22O4 | 1-(tert-Butyldioxy)-3-propoxy-2-propanol | -157 | -130.0 | -148.4 |
| C11H12O4 | Benzal diacetate | -167.3 | -145.8 | -158.3 |
| C11H24O4 | 1-Butoxy-1-tert-butyldioxy-2-propanol | -161.3 | -138.9 | -159.1 |
| C6NH5O4 | 4-Nitrocatechol | -69.3 | -59.2 | -65.9 |
| C7NH5O4 | p-Nitrobenzoic acid | -68.7 | -51.2 | -67.6 |
| C6N2H4O4 | m-Dinitrobenzene | 11.3 | 54.8 | 22.3 |
| C6N2H4O4 | o-Dinitrobenzene | 20.2 | 54.8 | 22.3 |
| C6N2H4O4 | p-Dinitrobenzene | 13.3 | 54.6 | 22.3 |
| C7N2H6O4 | 2,4-Dinitrotoluene | 7.9 | 48.9 | 14.6 |
| C7N2H6O4 | Dinitrophenylmethane | 8.3 | 56.2 | 19.2 |
| C8H18O5 | 3,5,7,9,11-Pentaoxa-tridecane | -216.5 | -235.4 | -230.3 |
| C10H20O5 | 15-Crown-5 | -191.1 | -208.6 | -214.9 |
| C6N2H12O5 | SER-SER | -211.6 | -220.9 | -225.2 |
| C8N2H16O5 | THR-THR | -230.2 | -217.6 | -233.7 |
| C7H10O6 | Trimethyl methanetricarboxylate | -252.1 | -242.5 | -251.6 |
| C2N3H3O6 | 1,1,1-Trinitroethane | -12.4 | 56.4 | -0.3 |
| CN4O8 | Tetranitromethane | 18.5 | 95.0 | 25.9 |
| C3N3H5O9 | Glycerol trinitrate | -64.7 | -6.3 | -84.0 |
| C14NH21O | 4-Isopropylbenzylidene t-butylamine N-oxide | -12.2 | 37.2 | 6.4 |
| C14H8O2 | 9,10-Anthroquinone | -18.1 | -13.2 | -14.7 |
| C14H8O2 | 9,10-Phenanthroquinone | -11.1 | -10.7 | -12.7 |
| C13NH11O2 | Phenol, 2-[(phenylimino)methyl]-, N-oxide | 12.9 | 38.2 | 20.2 |
| C14NH13O2 | Benzenamine, N-[(4-methoxyphenyl)methylene]-, N-oxide | 26.2 | 45.8 | 31.2 |
| C15NH17O2 | N-(3-Phenoxy-2-hydroxypropyl)aniline | -44.1 | -28.4 | -39.1 |
| C14H10O3 | Benzoic acid, anhydride | -76.3 | -66.1 | -71.0 |
| C12N2H24O3 | ILE-ILE | -176.5 | -144.8 | -173.1 |
| C12N2H24O3 | LEU-LEU | -177.1 | -150.6 | -177.2 |
| C12H14O4 | 1,1-Ethanediol, 2-phenyl-, diacetate | -167.3 | -151.0 | -167.6 |
| C12H14O4 | Benzyl diacetate | -161.8 | -151.0 | -167.6 |
| C14H6O4 | 1,4,9,10-Anthracenetetrone | -50 | -62.1 | -66.4 |
| C14H8O4 | 1,4-Dihydroxy-9,10-anthracenedione | -112.6 | -104.9 | -101.0 |
| C14H10O4 | Diphenyl oxalate | -104.5 | -97.0 | -101.2 |
| C14H12O4 | Dimethyl naphthalene-2,6-dicarboxylate | -132.7 | -124.0 | -121.3 |
| C10H16O6 | Triethyl methanetricarboxylate | -280.4 | -258.5 | -274.0 |
| C11H18O6 | Triethyl 1,1,1-ethanetricarboxylate | -286.2 | -252.9 | -269.2 |
| C7N3H5O6 | 2,4,6-Trinitrotoluene | 12.9 | 74.6 | 23.3 |
| C9NH9O7 | 2-(Diacetoxymethyl)-5-nitrofuran | -184.4 | -161.3 | -186.5 |
| C7N3H5O7 | 2,4,6-Trinitroanisole | -5.8 | 39.9 | -3.4 |
| C8N3H7O7 | 2,4,6-Trinitrophenetole | -20.1 | 36.7 | -10.9 |
| F | Fluorine, atom | 18.9 | 18.9 | 1.8 |
| F | Fluoride, anion | -61 | -17.1 | -61.2 |
| HF | Hydrogen fluoride | -65.1 | -59.7 | -77.9 |
| CF | Fluoromethylidyne radical | 61 | 38.2 | 46.0 |
| CH2F | Fluoromethyl, cation | 200.3 | 182.8 | 198.1 |
| CH3F | Fluoromethane | -56.8 | -60.9 | -60.5 |
| C2HF | Fluoroacetylene | 30 | 15.6 | 5.6 |
| C2H3F | Fluoroethylene | -32.5 | -34.6 | -37.2 |
| C2H4F | CH3CHF, cation | 166 | 164.7 | 159.4 |
| C2H5F | Fluoroethane | -62.9 | -65.1 | -66.9 |
| C3H7F | 2-Fluoropropane | -69.4 | -66.7 | -71.5 |
| C6H5F | Fluorobenzene | -27.8 | -25.3 | -25.7 |
| C6H11F | Fluorocyclohexane | -80.5 | -76.5 | -82.9 |
| C9H19F | 1-Fluorononane | -101.2 | -98.5 | -104.7 |
| CNF | Cyanogen fluoride | 8.6 | -2.3 | -8.4 |
| OF | Fluorine oxide | 26.1 | 20.4 | 10.1 |
| HOF | Hypofluorous acid | -20.9 | -18.6 | -30.7 |
| CHOF | HCOF | -90 | -88.8 | -96.3 |
| C2H3OF | Acetyl fluoride | -106.4 | -96.5 | -107.1 |
| NOF | Nitrosyl fluoride | -15.7 | -24.7 | -15.5 |
| O2F | Fluorine dioxide | 3 | 22.8 | -6.3 |
| C7H5O2F | m-fluorobenzoic acid | -118.4 | -113.2 | -113.5 |
| NO2F | Fluorine nitrite | -26 | 0.8 | -22.5 |
| NO3F | Fluorine nitrate | 2.5 | 28.2 | -7.0 |
| CN2HO4F | Fluorodinitromethane | -56.1 | -18.2 | -53.6 |
| F2 | Fluorine molecule | 0 | 7.4 | -9.1 |
| CF2 | Difluoromethylene | -45 | -65.2 | -61.8 |
| CHF2 | Difluoromethyl, cation | 142.4 | 132.4 | 141.7 |
| CH2F2 | Difluoromethane, cation | 185.2 | 177.6 | 142.3 |

| Formula | Name | | | |
|---|---|---|---|---|
| CH2F2 | Difluoromethane | -108.1 | -111.7 | -118.0 |
| C2F2 | Difluoroacetylene | 5 | -21.0 | -23.0 |
| C2H2F2 | gem-Difluoroethylene | -80.5 | -83.7 | -86.6 |
| C2H3F2 | CH3CF2, cation | 107 | 116.6 | 112.0 |
| C2H4F2 | 1,1-Difluoroethane | -118.8 | -113.4 | -122.9 |
| C6H4F2 | 1,2-Difluorobenzene | -67.7 | -70.6 | -69.7 |
| C6H4F2 | 1,3-Difluorobenzene | -73.9 | -71.0 | -71.2 |
| C6H4F2 | 1,4-Difluorobenzene | -73.3 | -71.1 | -71.3 |
| C4NH9F2 | t-Butyldifluoroamine | -46 | -21.1 | -31.8 |
| C7NH7F2 | N,N'-Difluorobenzylamine | 1.8 | 11.3 | 5.8 |
| N2F2 | cis-Difluorodiazene | 16.4 | -2.3 | 12.6 |
| N2F2 | trans-Difluorodiazene | 19.4 | 2.3 | 25.0 |
| OF2 | Difluorine oxide | 5.9 | 18.3 | -9.9 |
| COF2 | Carbonyl fluoride | -152.7 | -138.5 | -152.0 |
| CF3 | Trifluoromethyl, cation | 99.3 | 101.0 | 106.5 |
| CF3 | Trifluoromethyl | -112.4 | -138.6 | -151.4 |
| CF3 | Trifluoromethyl, anion | -163.4 | -178.7 | -194.7 |
| CHF3 | Trifluoromethane | -166.3 | -163.7 | -175.8 |
| C2HF3 | Trifluoroethylene | -117.3 | -131.1 | -131.6 |
| C2H2F3 | CF3CH2, cation | 114 | 121.3 | 117.4 |
| C2H2F3 | CH2F.CF2, radical cation | 81 | 82.4 | 79.2 |
| C2H2F3 | CF3CH2 radical | -123.6 | -131.1 | -139.7 |
| C2H3F3 | 1,1,1-Trifluoroethane | -178.9 | -164.3 | -176.7 |
| C7H5F3 | Trifluoromethylbenzene | -143.2 | -127.5 | -137.4 |
| NF3 | Nitrogen trifluoride | -31.6 | -34.1 | -32.8 |
| C2NF3 | Trifluoroacetonitrile | -118.4 | -113.1 | -120.9 |
| C2NF3 | Trifluoromethylisocyanide | -99.7 | -90.3 | -98.5 |
| NOF3 | F3NO | -39 | 23.0 | -17.0 |
| C2HO2F3 | Trifluoroacetic acid | -255 | -238.0 | -249.4 |
| CF4 | Carbon tetrafluoride | -223.3 | -214.0 | -223.5 |
| C2F4 | Tetrafluoroethylene | -157.9 | -175.6 | -173.5 |
| C6H2F4 | 1,2,4,5-Tetrafluorobenzene | -154.6 | -159.2 | -155.5 |
| N2F4 | Tetrafluorohydrazine | -2 | -19.6 | -14.7 |
| COF4 | Perfluoromethanol | -182.8 | -163.1 | -184.5 |
| CO2F4 | Bis(fluoroxy)perfluoromethane | -134.9 | -112.1 | -144.6 |
| CNF5 | Pentafluoromethylamine | -169 | -163.1 | -167.6 |
| CN3F5 | Pentafluoroguanidine | 22.9 | 6.2 | 19.1 |
| C2F6 | Hexafluoroethane | -321.2 | -299.4 | -319.0 |
| C4F6 | Perfluorobutadiene | -253.4 | -250.0 | -244.6 |
| CN2F6 | Hexafluorodimethylamine | -108.8 | -111.9 | -110.3 |
| C2OF6 | Dimethyl perfluoroether | -369 | -357.3 | -370.4 |
| C3OF6 | Perfluoroacetone | -342.6 | -322.2 | -343.7 |
| C7N2H5O4F | Fluorodinitrophenylmethane | -16.9 | 14.6 | -19.5 |
| C7H4F4 | 1-Fluoro-3-(trifluoro-methyl)benzene | -189.4 | -172.3 | -182.3 |
| C6N2H10F4 | N,N,N',N'-tetrafluoro-1,1-cyclohexanediamine | -41.6 | -24.7 | -38.4 |
| C6N2H12F4 | N,N,N',N'-Tetrafluoro-4-methyl-1,2-pentane | -52.8 | -26.5 | -53.3 |

| Formula | Name | | | |
|---|---|---|---|---|
| C7N2H14F4 | 1,1-Bis(difluoroamine)heptane | -52.8 | -42.5 | -54.6 |
| C6HF5 | Pentafluorobenzene | -192.5 | -201.7 | -194.8 |
| C7H3F5 | 2,3,4,5,6-Pentafluorotoluene | -201.6 | -206.6 | -199.9 |
| C6HOF5 | Pentafluorophenol | -228.8 | -247.9 | -233.3 |
| C6F6 | Hexafluorobenzene | -242.5 | -243.5 | -233.2 |
| C3F8 | Perfluoropropane | -426.2 | -384.3 | -413.5 |
| C4F8 | Perfluorobut-2-ene | -389.9 | -365.9 | -383.5 |
| C4F8 | Perfluorocyclobutane | -369.5 | -363.7 | -385.8 |
| C7F8 | Octafluorotoluene | -356.8 | -340.4 | -341.9 |
| CN4F8 | Octafluoromethanetetramine | 0.4 | -3.8 | 3.4 |
| C3O2F8 | Perfluorodimethoxymethane | -520.6 | -501.3 | -517.0 |
| C4F10 | n-Perfluorobutane | -515.3 | -469.0 | -507.7 |
| C7H6O4F6 | Hexafluoropentanedioic acid, dimethyl ester | -460.5 | -419.7 | -448.9 |
| C6F10 | Decafluorocyclohexene | -461.9 | -423.1 | -453.6 |
| C5NF11 | Undecafluoropiperidine | -478.9 | -436.9 | -475.2 |
| C2N5F11 | Tetrakis(difluoroamine)-N-1,1-trifluorodimethylamine | -84.4 | -85.4 | -84.9 |
| C6F12 | Dodecafluorocyclohexane | -590.5 | -522.5 | -574.3 |

Values for $I.E.$ (given in eV)

| Molecular Formula | Molecule Name | Ref. | Initial | Final |
|---|---|---|---|---|
| H2 | Hydrogen | 15.4 | 15.75 | 15.01 |
| CH4 | Methane | 13.6 | 13.86 | 12.94 |
| C2H2 | Acetylene | 11.4 | 11.01 | 12.06 |
| C2H4 | Ethylene | 10.51 | 10.18 | 10.87 |
| C2H6 | Ethane | 12 | 12.7 | 11.10 |
| C3 | Carbon, trimer | 11.1 | 11.04 | 10.54 |
| C3H4 | Allene | 10.07 | 10.02 | 10.30 |
| C3H4 | Cyclopropene | 9.86 | 9.88 | 10.01 |
| C3H4 | Propyne | 10.37 | 10.72 | 10.86 |
| C3H6 | Cyclopropane | 11 | 11.43 | 11.38 |
| C3H6 | Propene | 9.88 | 9.96 | 10.02 |
| C3H8 | Propane | 11.5 | 12.34 | 10.53 |
| C4H2 | Diacetylene | 10.17 | 9.99 | 10.99 |
| C4H4 | Vinylacetylene | 9.1 | 9.5 | 10.27 |
| C4H4 | Butatriene | 9.15 | 9.01 | 9.74 |
| C4H6 | 1,2-Butadiene | 9.15 | 9.84 | 9.56 |
| C4H6 | 1,3-Butadiene | 9.08 | 9.14 | 9.70 |
| C4H6 | 1-Butyne | 10.2 | 10.68 | 10.56 |
| C4H6 | 2-Butyne | 9.6 | 10.47 | 9.99 |
| C4H6 | Cyclobutene | 9.43 | 9.77 | 9.76 |
| C4H8 | 1-Butene | 9.7 | 9.95 | 9.92 |
| C4H8 | Cyclobutane | 10.7 | 11.8 | 10.11 |
| C4H10 | Isobutane | 11.4 | 12.12 | 10.53 |
| C4H10 | n-Butane, trans | 11.2 | 12.21 | 10.31 |
| C5H6 | Cyclopentadiene | 8.57 | 9.04 | 9.29 |
| C5H8 | Cyclopentene | 9.18 | 9.72 | 9.15 |
| C5H10 | Cyclopentane | 10.5 | 12.06 | 10.14 |
| C5H12 | n-Pentane | 10.3 | 12.16 | 10.23 |
| C5H12 | Neopentane | 11.3 | 12.11 | 10.96 |
| C6H6 | Benzene | 9.25 | 9.39 | 9.83 |
| C6H10 | Cyclohexene | 10.3 | 9.75 | 9.20 |
| C6H12 | Cyclohexane | 10.3 | 11.74 | 10.17 |
| C7H8 | Cycloheptatriene | 8.5 | 8.58 | 8.81 |
| C7H8 | Toluene | 8.82 | 9.28 | 9.31 |
| C8H10 | Ethylbenzene | 8.8 | 9.28 | 9.40 |
| C8H14 | Bicyclo(2.2.2)-octane | 9.45 | 11.4 | 9.80 |
| C10H8 | Naphthalene | 8.15 | 8.58 | 9.01 |
| C10H16 | Adamantane | 9.6 | 11.27 | 9.64 |
| C14H10 | Anthracene | 8.16 | 8.05 | 8.46 |
| NH3 | Ammonia | 10.85 | 11.19 | 9.77 |
| CNH | Hydrogen cyanide | 13.6 | 13.41 | 12.63 |
| CNH5 | Methylamine | 9.6 | 10.56 | 8.93 |
| C2NH3 | Acetonitrile | 12.21 | 12.79 | 12.25 |

| Formula | Name | | | |
|---|---|---|---|---|
| C2NH3 | Methyl isocyanide | 11.32 | 12.24 | 10.61 |
| C2NH5 | Ethyleneimine (Azirane) | 9.9 | 10.68 | 9.79 |
| C2NH7 | Dimethylamine | 8.93 | 10.04 | 8.63 |
| C2NH7 | Ethylamine | 9.5 | 10.5 | 9.17 |
| C3NH3 | Acrylonitrile | 10.91 | 10.61 | 11.19 |
| C3NH5 | Ethyl cyanide | 11.9 | 12.59 | 11.50 |
| C3NH9 | Trimethylamine | 8.54 | 9.59 | 8.53 |
| C4NH5 | Pyrrole | 8.21 | 8.56 | 9.09 |
| C5NH5 | Pyridine | 9.67 | 9.69 | 10.11 |
| C6NH7 | Aniline | 7.7 | 8.75 | 8.50 |
| C7NH5 | Phenyl cyanide | 9.7 | 9.81 | 10.20 |
| N2 | Nitrogen | 15.6 | 14.87 | 12.54 |
| CN2H2 | Diazomethane | 9 | 8.66 | 9.36 |
| CN2H6 | Methylhydrazine | 9.3 | 9.66 | 8.88 |
| C2N2 | Cyanogen | 13.36 | 13.2 | 13.05 |
| H2O | Water | 12.62 | 12.19 | 12.03 |
| CO | Carbon monoxide | 14.01 | 13.43 | 11.12 |
| CH2O | Formaldehyde | 10.1 | 11.04 | 10.74 |
| CH4O | Methanol | 10.96 | 11.41 | 10.72 |
| C2H2O | Ketene | 9.64 | 9.29 | 10.09 |
| C2H4O | Acetaldehyde | 10.21 | 10.88 | 10.70 |
| C2H4O | Ethylene oxide | 10.57 | 11.49 | 11.23 |
| C2H6O | Dimethyl ether | 10.04 | 11.04 | 10.18 |
| C2H6O | Ethanol | 10.6 | 11.3 | 10.36 |
| C3H6O | Acetone | 9.72 | 10.75 | 10.64 |
| C3H6O | Propanal | 10 | 10.82 | 10.49 |
| C4H4O | Furan | 8.88 | 9.14 | 9.64 |
| C4H8O | Butanal | 9.83 | 10.81 | 10.52 |
| C4H10O | Diethyl ether | 9.6 | 10.91 | 10.28 |
| C7H6O | Benzaldehyde | 9.7 | 9.74 | 10.15 |
| C7H8O | Anisole | 8.4 | 8.84 | 9.07 |
| CNHO | Hydrogen isocyanate | 11.6 | 11.1 | 11.50 |
| CH2O2 | Formic acid | 11.51 | 11.74 | 11.78 |
| C2H2O2 | trans Glyoxal | 10.59 | 10.75 | 10.67 |
| C2H4O2 | Acetic acid | 10.8 | 11.57 | 11.57 |
| C2H4O2 | Methyl formate | 11.02 | 11.61 | 11.37 |
| C2H6O2 | Dimethyl peroxide | 10.6 | 10.57 | 10.40 |
| C3O2 | Carbon suboxide | 10.6 | 10.07 | 10.85 |
| C3H4O2 | beta-Propiolactone | 10.6 | 11.4 | 11.41 |
| C3H6O2 | Methyl acetate | 10.6 | 11.46 | 11.22 |
| C3H6O2 | Propionic acid | 10.5 | 11.52 | 11.26 |
| C3H8O2 | 2-Methoxyethanol | 9.8 | 11.02 | 10.11 |
| C5H8O2 | Acetylacetone | 9.15 | 10.79 | 10.69 |
| C7H6O2 | Benzoic acid | 9.8 | 9.76 | 10.17 |
| CNH3O2 | Methyl nitrite | 11 | 11.42 | 11.05 |
| C2NH5O2 | Ethyl nitrite | 11.3 | 11.36 | 10.87 |

| Formula | Name | | | |
|---|---|---|---|---|
| C3NH7O2 | Alanine | 8.9 | 10.81 | 9.55 |
| C6NH5O2 | Nitrobenzene | 9.9 | 10.31 | 10.76 |
| O3 | Ozone | 12.75 | 12.71 | 12.69 |
| C4H2O3 | Malaic anhydride | 10.84 | 11.7 | 12.21 |
| C7H6O3 | o-Salicylic acid | 9.8 | 9.26 | 9.47 |
| C2H2O4 | Oxalic acid | 11.2 | 11.67 | 11.32 |
| N2O4 | Dinitrogen tetroxide | 11.4 | 12.05 | 12.60 |
| N2O5 | Dinitrogen pentoxide | 12.3 | 13.18 | 13.63 |
| HF | Hydrogen fluoride | 16.06 | 14.82 | 13.82 |
| CH3F | Fluoromethane | 13.31 | 13.05 | 12.59 |
| C2HF | Fluoroacetylene | 11.3 | 11.06 | 12.01 |
| C2H3F | Fluoroethylene | 10.58 | 10.17 | 10.93 |
| C2H5F | Fluoroethane | 12.43 | 12.61 | 11.40 |
| C3H7F | 2-Fluoropropane | 11.08 | 12.33 | 11.35 |
| C6H5F | Fluorobenzene | 9.19 | 9.47 | 9.95 |
| NOF | Nitrosyl fluoride | 12.94 | 12.93 | 13.01 |
| C7H5O2F | m-fluorobenzoic acid | 9.9 | 9.83 | 10.25 |
| NO2F | Fluorine nitrite | 13.51 | 12.99 | 13.96 |
| CH2F2 | Difluoromethane | 13.17 | 13.09 | 12.76 |
| C2F2 | Difluoroacetylene | 11.2 | 11.17 | 11.92 |
| C2H2F2 | gem-Difluoroethylene | 10.72 | 10.18 | 10.88 |
| C2H4F2 | 1,1-Difluoroethane | 12.8 | 12.73 | 12.45 |
| N2F2 | trans-Difluorodiazene | 13.4 | 13 | 12.74 |
| OF2 | Difluorine oxide | 13.26 | 13.52 | 14.33 |
| CHF3 | Trifluoromethane | 14.8 | 14.57 | 14.12 |
| C2HF3 | Trifluoroethylene | 10.54 | 10.46 | 11.02 |
| C2H3F3 | 1,1,1-Trifluoroethane | 13.8 | 14.01 | 13.71 |
| C7H5F3 | Trifluoromethylbenzene | 9.68 | 10.07 | 10.47 |
| NF3 | Nitrogen trifluoride | 13.73 | 13.93 | 14.19 |
| C2HO2F3 | Trifluoroacetic acid | 12 | 12.73 | 12.61 |
| CF4 | Carbon tetrafluoride | 16.23 | 16.81 | 15.73 |
| C2F4 | Tetrafluoroethylene | 10.5 | 10.74 | 11.11 |
| N2F4 | Tetrafluorohydrazine | 12 | 13.19 | 13.62 |
| C2F6 | Hexafluoroethane | 14.6 | 14.5 | 14.11 |
| C3OF6 | Perfluoroacetone | 12.1 | 13 | 12.85 |
| C6HF5 | Pentafluorobenzene | 9.75 | 10.4 | 10.81 |
| C6F6 | Hexafluorobenzene | 10.9 | 10.78 | 11.12 |

Values for ⟨$\mu$⟩ (given in Debyes)

| Molecular Formula | Molecule Name | Expt. | Initial | Final |
|---|---|---|---|---|
| C3H4 | Cyclopropene | 0.45 | 0.48 | 0.02 |
| C3H4 | Propyne | 0.78 | 0.12 | 1.15 |
| C3H6 | Propene | 0.37 | 0.04 | 0.59 |
| C3H8 | Propane | 0.08 | 0.00 | 0.02 |
| C4H6 | Bicyclobutane | 0.68 | 0.41 | 0.20 |
| C4H6 | Cyclobutene | 0.13 | 0.08 | 0.57 |
| C5H6 | Cyclopentadiene | 0.42 | 0.18 | 0.96 |
| C5H8 | Cyclopentene | 0.2 | 0.06 | 0.51 |
| C6H6 | Fulvene | 0.42 | 0.70 | 0.63 |
| C7H8 | Toluene | 0.36 | 0.05 | 0.74 |
| NH3 | Ammonia | 1.47 | 1.75 | 0.00 |
| CNH | Hydrogen cyanide | 2.98 | 2.50 | 2.11 |
| CNH5 | Methylamine | 1.31 | 1.48 | 0.65 |
| C2NH3 | Acetonitrile | 3.92 | 2.63 | 3.25 |
| C2NH3 | Methyl isocyanide | 3.85 | 2.17 | 3.26 |
| C2NH5 | Ethyleneimine (Azirane) | 1.9 | 1.75 | 1.94 |
| C2NH7 | Dimethylamine | 1.03 | 1.17 | 1.30 |
| C2NH7 | Ethylamine | 1.22 | 1.52 | 1.10 |
| C3NH3 | Acrylonitrile | 3.87 | 2.97 | 3.04 |
| C3NH9 | Trimethylamine | 0.61 | 0.75 | 1.37 |
| C4NH5 | Pyrrole | 1.74 | 1.81 | 1.83 |
| C5NH5 | Pyridine | 2.22 | 1.96 | 2.07 |
| C6NH7 | Aniline | 1.53 | 1.46 | 1.41 |
| CN2H2 | Diazomethane | 1.5 | 1.25 | 1.65 |
| CN2H2 | N=N-CH2- | 1.59 | 1.55 | 1.94 |
| CN2H6 | Methylhydrazine | 1.66 | 0.24 | 2.05 |
| H2O | Water | 1.85 | 1.78 | 2.11 |
| CO | Carbon monoxide | 0.11 | 0.20 | 0.23 |
| CH2O | Formaldehyde | 2.33 | 2.17 | 2.65 |
| CH4O | Methanol | 1.7 | 1.48 | 2.05 |
| C2H2O | Ketene | 1.42 | 1.04 | 1.67 |
| C2H4O | Acetaldehyde | 2.69 | 2.38 | 3.08 |
| C2H4O | Ethylene oxide | 1.89 | 1.92 | 2.26 |
| C2H6O | Dimethyl ether | 1.3 | 1.27 | 1.87 |
| C2H6O | Ethanol | 1.69 | 1.40 | 2.14 |
| C3H6O | Acetone | 2.88 | 2.50 | 3.15 |
| C4H4O | Furan | 0.66 | 0.42 | 0.63 |
| C4H10O | Diethyl ether | 1.15 | 1.09 | 1.87 |
| C6H6O | Phenol | 1.45 | 1.17 | 1.41 |
| C7H8O | Anisole | 1.38 | 1.07 | 1.70 |
| C3NH7O | Dimethylformamide | 3.82 | 3.06 | 4.05 |
| N2O | Nitrous oxide | 0.17 | 0.77 | 0.83 |
| CH2O2 | Formic acid | 1.41 | 1.49 | 1.43 |

| Formula | Name | | | |
|---|---|---|---|---|
| C2H4O2 | Acetic acid | 1.74 | 1.68 | 1.78 |
| C2H4O2 | Methyl formate | 1.77 | 1.63 | 1.72 |
| C3H6O2 | Methyl acetate | 1.72 | 1.75 | 1.73 |
| C3H6O2 | Propionic acid | 1.75 | 1.64 | 1.79 |
| NHO2 | Nitrous acid, trans | 1.86 | 2.28 | 2.64 |
| O3 | Ozone | 0.53 | 1.18 | 1.29 |
| NHO3 | Nitric acid | 2.17 | 2.78 | 2.76 |
| HF | Hydrogen fluoride | 1.83 | 1.99 | 1.62 |
| CH3F | Fluoromethane | 1.86 | 1.76 | 2.10 |
| C2HF | Fluoroacetylene | 0.7 | 1.57 | 1.25 |
| C2H3F | Fluoroethylene | 1.43 | 1.70 | 1.87 |
| C2H5F | Fluoroethane | 1.96 | 1.87 | 2.28 |
| C6H5F | Fluorobenzene | 1.66 | 1.96 | 2.07 |
| CNF | Cyanogen fluoride | 2.17 | 0.89 | 0.75 |
| HOF | Hypofluorous acid | 2.23 | 1.81 | 2.32 |
| CHOF | HCOF | 2.02 | 2.50 | 2.51 |
| NOF | Nitrosyl fluoride | 1.81 | 0.51 | 1.36 |
| NO2F | Fluorine nitrite | 0.47 | 0.66 | 0.77 |
| CH2F2 | Difluoromethane | 1.96 | 2.21 | 2.37 |
| C2H4F2 | 1,1-Difluoroethane | 2.3 | 2.50 | 2.82 |
| N2F2 | cis-Difluorodiazene | 0.16 | 0.02 | 0.62 |
| OF2 | Difluorine oxide | 0.3 | 0.32 | 0.86 |
| COF2 | Carbonyl fluoride | 0.95 | 0.81 | 0.44 |
| CHF3 | Trifluoromethane | 1.65 | 2.23 | 2.18 |
| C2HF3 | Trifluoroethylene | 1.3 | 1.82 | 1.65 |
| C2H3F3 | 1,1,1-Trifluoroethane | 2.32 | 2.87 | 3.07 |
| NF3 | Nitrogen trifluoride | 0.24 | 0.20 | 0.76 |
| C2NF3 | Trifluoroacetonitrile | 1.26 | 0.36 | 0.32 |
| C2HO2F3 | Trifluoroacetic acid | 2.28 | 2.45 | 2.85 |

Values for |**g**| (given in kcal/(mol · bohr))

| Molecular Formula | Molecule Name | Initial | Final |
|---|---|---|---|
| H2 | Hydrogen | 40.249 | 52.187 |
| CH2 | Methylene, singlet | 19.276 | 11.754 |
| CH2 | Methylene, triplet | 26.903 | 15.460 |
| CH3 | Methyl, cation | 3.512 | 5.395 |
| CH4 | Methane | 13.162 | 3.430 |
| C2H2 | Acetylene | 9.043 | 15.677 |
| C2H3 | Vinyl, cation | 44.877 | 29.066 |
| C2H3 | Vinyl | 0.391 | 17.207 |
| C2H4 | Ethylene, cation | 8.075 | 11.443 |
| C2H4 | Ethylene | 7.216 | 7.344 |
| C2H4 | Methylmethylene | 13.604 | 24.882 |
| C2H5 | Ethyl, cation | 47.440 | 20.017 |
| C2H5 | Ethyl radical | 16.467 | 15.797 |
| C2H6 | Ethane | 21.822 | 10.730 |
| C3 | Carbon, trimer | 17.716 | 14.392 |
| C3H3 | Cyclopropenyl, cation | 123.215 | 92.829 |
| C3H3 | Propynyl, cation | 18.987 | 11.366 |
| C3H4 | Allene | 6.419 | 8.729 |
| C3H4 | Propyne | 21.322 | 15.551 |
| C3H5 | Allyl, cation | 16.256 | 15.540 |
| C3H5 | Allyl | 15.268 | 9.568 |
| C3H6 | Cyclopropane | 24.308 | 16.632 |
| C3H6 | Propene | 19.881 | 9.559 |
| C3H7 | i-Propyl radical | 26.969 | 14.921 |
| C3H8 | Propane | 23.422 | 9.306 |
| C4 | Carbon, tetramer | 0.054 | 8.193 |
| C4H2 | Diacetylene | 16.398 | 37.058 |
| C4H4 | Vinylacetylene | 16.331 | 20.697 |
| C4H4 | Butatriene | 10.203 | 24.536 |
| C4H6 | 1,2-Butadiene | 20.758 | 15.980 |
| C4H6 | 1,3-Butadiene | 10.563 | 13.962 |
| C4H6 | 1-Butyne | 26.050 | 17.390 |
| C4H6 | 2-Butyne | 28.664 | 14.399 |
| C4H6 | Bicyclobutane | 71.719 | 63.762 |
| C4H6 | Cyclobutene | 27.983 | 16.017 |
| C4H6 | Methyl cyclopropene | 67.177 | 60.092 |
| C4H6 | Methylenecyclopropane | 26.722 | 30.665 |
| C4H8 | 1-Butene | 24.330 | 11.238 |
| C4H8 | cis-2-Butene | 27.688 | 16.122 |
| C4H8 | Cyclobutane | 21.936 | 17.444 |
| C4H8 | Isobutene | 26.057 | 11.277 |
| C4H8 | trans-2-Butene | 28.551 | 15.413 |
| C4H9 | Isobutyl | 23.183 | 12.969 |
| C4H10 | Isobutane | 24.439 | 7.759 |
| C4H10 | n-Butane, trans | 26.302 | 9.570 |
| C5H5 | Cyclopentadienyl, anion | 16.858 | 23.711 |
| C5H6 | Cyclopentadiene | 30.305 | 14.418 |
| C5H8 | 1,4-Pentadiene | 24.387 | 9.954 |
| C5H8 | 1,cis-3-Pentadiene | 29.799 | 12.709 |
| C5H8 | 1,trans-3-Pentadiene | 20.673 | 15.786 |
| C5H8 | Bicyclo(2.1.0)-pentane | 37.177 | 30.853 |
| C5H8 | Cyclopentene | 30.052 | 16.229 |
| C5H8 | Isoprene | 21.221 | 15.038 |
| C5H8 | Methylene cyclobutane | 23.437 | 18.069 |
| C5H8 | Spiropentane | 28.996 | 24.211 |
| C5H10 | 2-Methyl-2-butene | 32.403 | 18.438 |
| C5H10 | cis-2-Pentene | 30.831 | 16.176 |
| C5H10 | Cyclopentane | 29.826 | 16.038 |
| C5H10 | trans-2-Pentene | 31.253 | 14.941 |
| C5H12 | n-Pentane | 29.531 | 10.259 |
| C5H12 | Neopentane | 24.782 | 8.028 |
| C6H6 | Benzene | 20.254 | 6.026 |
| C6H6 | Fulvene | 28.377 | 17.945 |
| C6H8 | (E)-1,3,5-Hexatriene | 11.884 | 20.958 |
| C6H8 | (Z)-1,3,5-Hexatriene | 12.152 | 18.315 |
| C6H8 | 1,3-Cyclohexadiene | 20.353 | 13.487 |
| C6H8 | 1,4-Cyclohexadiene | 29.609 | 14.660 |
| C6H10 | 1,5-Hexadiene | 26.448 | 10.864 |
| C6H10 | 1-Methyl cyclopentene | 36.114 | 19.957 |
| C6H10 | 3-Methyl cyclopentene | 32.239 | 15.151 |
| C6H10 | 4-Methyl cyclopentene | 31.351 | 14.494 |
| C6H10 | Cyclohexene | 30.643 | 13.430 |
| C6H12 | 1-Hexene | 28.993 | 9.829 |
| C6H12 | 2,3-Dimethyl-1-butene | 33.167 | 16.006 |
| C6H12 | 2,3-Dimethyl-2-butene | 28.397 | 17.338 |
| C6H12 | (Z)-3-Methyl-2-pentene | 36.289 | 19.672 |
| C6H12 | Cyclohexane | 29.769 | 10.258 |
| C6H14 | 2,2-Dimethyl butane | 30.004 | 12.138 |
| C6H14 | 2,3-Dimethyl butane | 30.809 | 11.911 |
| C6H14 | 2-Methyl pentane | 31.084 | 10.973 |
| C6H14 | 3-Methyl pentane | 32.647 | 13.664 |
| C6H14 | n-Hexane | 31.879 | 10.562 |
| C7H7 | Benzyl, cation | 43.009 | 43.100 |
| C7H7 | Tropylium cation | 70.279 | 61.893 |
| C7H8 | Norbornadiene | 45.765 | 20.212 |
| C7H8 | Toluene | 28.952 | 15.622 |
| C7H12 | Norbornane | 28.769 | 15.425 |
| C7H16 | 2,4-Dimethyl pentane | 33.228 | 11.057 |
| C7H16 | 3-Ethyl pentane | 34.650 | 13.695 |
| C7H16 | n-Heptane | 33.964 | 10.744 |

| Formula | Name | Value 1 | Value 2 |
|---|---|---|---|
| C8H8 | Cubane | 10.043 | 20.375 |
| C8H8 | Cyclooctatetraene | 32.305 | 11.589 |
| C8H8 | Styrene | 24.666 | 10.270 |
| C8H10 | Ethylbenzene | 31.377 | 14.900 |
| C8H10 | m-Xylene | 35.794 | 21.375 |
| C8H10 | p-Xylene | 42.984 | 24.998 |
| C8H14 | Bicyclo(2.2.2)-octane | 38.031 | 19.785 |
| C8H18 | 2,2,3,3-Tetramethyl butane | 40.281 | 15.966 |
| C8H18 | n-Octane | 35.933 | 10.962 |
| C10H8 | Azulene | 30.682 | 17.700 |
| C10H8 | Naphthalene | 23.013 | 13.449 |
| C10H16 | Camphene | 31.978 | 9.212 |
| C10H22 | n-Decane | 39.694 | 11.480 |
| C12H8 | Acenaphthylene | 41.638 | 28.347 |
| C12H8 | Biphenylene | 36.731 | 30.415 |
| C12H10 | Biphenyl | 31.376 | 12.517 |
| C14H10 | Anthracene | 22.757 | 16.905 |
| C14H10 | Phenanthrene | 22.757 | 16.905 |
| C16H10 | Pyrene | 29.386 | 13.055 |
| NH2 | Amidogen | 22.160 | 13.559 |
| NH3 | Ammonia | 7.739 | 20.484 |
| NH4 | Ammonium, cation | 4.007 | 12.680 |
| CN | Cyanide | 52.121 | 62.391 |
| CNH | Hydrogen cyanide | 14.852 | 7.393 |
| CNH4 | CH2-NH2, cation | 49.919 | 23.295 |
| CNH5 | Methylamine | 24.489 | 17.299 |
| C2NH3 | Acetonitrile | 28.120 | 20.165 |
| C2NH3 | Methyl isocyanide | 29.041 | 10.348 |
| C2NH5 | Ethyleneimine (Azirane) | 25.641 | 17.491 |
| C2NH7 | Dimethylamine | 26.785 | 10.852 |
| C2NH7 | Ethylamine | 22.568 | 23.890 |
| C3NH3 | Acrylonitrile | 28.833 | 34.467 |
| C3NH5 | Ethyl cyanide | 24.752 | 16.005 |
| C3NH7 | Cyclopropylamine | 47.982 | 37.357 |
| C3NH9 | Isopropylamine | 35.110 | 17.257 |
| C3NH9 | n-Propylamine | 20.692 | 28.640 |
| C3NH9 | Trimethylamine | 43.326 | 26.499 |
| C4NH5 | (E)-2-Butenenitrile | 22.960 | 14.547 |
| C4NH5 | (Z)-2-Butenenitrile | 24.208 | 17.381 |
| C4NH5 | Pyrrole | 41.911 | 16.167 |
| C4NH7 | Butanenitrile | 37.375 | 17.034 |
| C4NH7 | Isobutane nitrile | 39.260 | 37.022 |
| C4NH9 | Pyrrolidine | 24.169 | 21.793 |
| C4NH11 | 2-Butylamine | 29.298 | 22.550 |
| C4NH11 | 2-Methyl-1-propylamine | 28.162 | 22.519 |
| C4NH11 | N-Butylamine | 19.993 | 29.097 |

| Formula | Name | Value 1 | Value 2 |
|---|---|---|---|
| C4NH11 | t-Butylamine | 34.573 | 19.032 |
| C5NH5 | Pyridine | 28.533 | 13.794 |
| C5NH7 | N-Methyl pyrrole | 48.235 | 22.594 |
| C5NH9 | 1,2,3,6-Tetrahydropyridine | 29.857 | 13.086 |
| C5NH9 | 2-Cyanobutane | 32.884 | 16.028 |
| C5NH9 | Butyl cyanide | 29.213 | 15.217 |
| C5NH9 | t-Butylnitrile | 25.843 | 12.721 |
| C5NH11 | Cyclopentylamine | 34.233 | 22.549 |
| C5NH11 | Piperidine | 27.905 | 13.458 |
| C6NH7 | 2-Methyl pyridine | 30.916 | 18.164 |
| C6NH7 | Aniline | 27.972 | 23.368 |
| C6NH13 | 2-Methylpiperidine | 47.118 | 16.748 |
| C6NH13 | Cyclohexanamine | 30.709 | 25.733 |
| C6NH15 | Triethylamine | 48.195 | 22.261 |
| C7NH5 | Phenyl cyanide | 34.374 | 32.433 |
| N2 | Nitrogen | 12.911 | 53.256 |
| N2H2 | Diazene | 62.752 | 45.406 |
| N2H4 | Hydrazine | 51.307 | 47.591 |
| CN2H2 | Diazomethane | 16.022 | 15.972 |
| CN2H2 | N=N-CH2- | 25.646 | 47.502 |
| C2N2 | Cyanogen | 13.439 | 17.804 |
| C2N2H8 | 1,2-Dimethylhydrazine | 47.844 | 46.011 |
| C3N2H4 | 1H-Pyrazole | 64.670 | 41.402 |
| C3N2H4 | Imidazole | 38.444 | 21.186 |
| C3N2H10 | 1,2-Propanediamine | 32.068 | 30.070 |
| C4N2 | Dicyanoacetylene | 16.131 | 40.175 |
| C4N2H2 | Fumaronitrile | 21.894 | 13.354 |
| C4N2H4 | 1,3-Diazine | 36.420 | 19.129 |
| C4N2H4 | Pyrazine | 27.650 | 17.958 |
| C4N2H4 | Pyridazine | 68.081 | 51.188 |
| C4N2H4 | Succinonitrile | 33.855 | 22.737 |
| C4N2H6 | 2-Methyl-1H-imidazole | 68.113 | 42.656 |
| C6N2H12 | Triethylenediamine | 40.188 | 16.402 |
| N3 | Azide radical | 6.325 | 2.801 |
| N3H | Hydrazoic acid | 14.904 | 22.552 |
| C3N3H3 | 1,3,5-Triazine | 52.462 | 30.879 |
| CN4H2 | 1-H Tetrazole | 90.922 | 62.378 |
| C6N4 | Tetracyanoethylene | 66.329 | 83.534 |
| HO | Hydroxyl radical | 31.279 | 8.823 |
| H2O | Water | 22.001 | 8.813 |
| H3O | Hydronium, cation | 14.123 | 12.051 |
| CHO | HCO | 35.416 | 38.024 |
| CH2O | Formaldehyde | 20.263 | 12.600 |
| CH4O | Methanol | 40.995 | 12.185 |
| C2H2O | Ketene | 47.680 | 26.278 |
| C2H4O | Acetaldehyde | 31.664 | 13.002 |

| Formula | Name | Value 1 | Value 2 |
|---|---|---|---|
| C2H6O | Dimethyl ether | 41.952 | 9.502 |
| C2H6O | Ethanol | 44.052 | 10.309 |
| C3H6O | Acetone | 29.050 | 13.580 |
| C3H6O | Propanal | 35.135 | 12.467 |
| C3H8O | Isopropanol | 42.592 | 12.235 |
| C3H8O | Propanol | 40.301 | 13.695 |
| C4H4O | Acetyl acetylene | 50.842 | 60.869 |
| C4H4O | Furan | 46.322 | 21.539 |
| C4H6O | 2,3-Dihydrofuran | 53.967 | 24.513 |
| C4H6O | Crotonaldehyde | 19.547 | 18.366 |
| C4H6O | Divinyl ether | 26.252 | 27.973 |
| C4H8O | Butanal | 26.539 | 13.585 |
| C4H8O | Isobutanal | 26.579 | 19.180 |
| C4H8O | Methyl ethyl ketone | 27.976 | 16.756 |
| C4H10O | Diethyl ether | 122.746 | 65.731 |
| C4H10O | t-Butanol | 40.595 | 15.797 |
| C5H8O | 2,3-Dihydro-5-methyl-furan | 44.539 | 20.200 |
| C5H8O | 3,4-Dihydro-2H-pyran | 38.999 | 10.955 |
| C5H8O | Cyclopentanone | 14.852 | 21.854 |
| C5H10O | Diethyl ketone | 15.856 | 30.893 |
| C5H10O | Tetrahydropyran | 41.446 | 11.000 |
| C6H6O | Phenol | 42.904 | 21.436 |
| C6H10O | Cyclohexanone | 40.023 | 13.202 |
| C7H6O | Benzaldehyde | 23.121 | 18.518 |
| C7H8O | Anisole | 40.437 | 14.445 |
| C8H8O | Acetophenone | 27.322 | 19.921 |
| NO | Nitric oxide, cation | 63.858 | 105.225 |
| NO | Nitric oxide | 76.737 | 0.867 |
| CNO | NCO | 52.221 | 22.084 |
| CNHO | Hydrogen isocyanate | 31.070 | 21.519 |
| CNH3O | Formamide | 44.898 | 21.631 |
| C2NH5O | Acetaldoxime | 113.114 | 49.098 |
| C2NH5O | Acetamide | 59.735 | 23.494 |
| C3NH3O | Isoxazole | 100.540 | 46.398 |
| C3NH3O | Oxalone (oxazole) | 74.534 | 39.594 |
| C3NH5O | Acrylamine | 44.379 | 19.382 |
| C3NH5O | Methoxyacetonitrile | 52.272 | 36.381 |
| C3NH7O | Dimethylformamide | 46.462 | 29.996 |
| C3NH7O | Propanamide | 44.520 | 44.804 |
| C4NH5O | 3-Methyl isoxazole | 102.120 | 50.462 |
| C4NH5O | 5-Methyl isoxazole | 101.461 | 49.339 |
| C4NH7O | 2-Pyrrolidinone | 44.389 | 23.334 |
| C4NH7O | 4,5-Dihydro-2-methyl oxazole | 101.095 | 62.684 |
| C4NH7O | Methacrylamide | 42.331 | 22.758 |
| C4NH9O | 2-Methyl propanamide | 41.477 | 20.699 |
| C4NH9O | Butanamide | 43.063 | 21.544 |

| Formula | Name | Value 1 | Value 2 |
|---|---|---|---|
| C4NH11O | N,N-Diethyl-hydroxylamine | 96.665 | 44.518 |
| C5NH5O | 2-Pyridinol | 51.392 | 24.891 |
| C5NH5O | 3-Pyridinol | 39.319 | 19.221 |
| C5NH5O | 4-Pyridinol | 38.034 | 15.619 |
| N2O | Nitrous oxide | 10.055 | 21.044 |
| CN2H4O | Urea | 48.383 | 19.329 |
| C2N2H6O | N-Methyl urea | 44.261 | 22.193 |
| O2 | Oxygen (Singlet) | 177.061 | 9.171 |
| O2 | Oxygen (Triplet) | 179.698 | 11.883 |
| H2O2 | Hydrogen peroxide | 131.677 | 51.097 |
| CO2 | Carbon dioxide | 60.709 | 4.126 |
| CHO2 | Formate, anion | 38.670 | 30.917 |
| CH2O2 | Formic acid | 60.991 | 22.717 |
| C2H2O2 | trans Glyoxal | 9.206 | 29.323 |
| C2H3O2 | Acetate, anion | 24.275 | 24.544 |
| C2H4O2 | Acetic acid | 70.252 | 26.769 |
| C2H4O2 | Methyl formate | 84.411 | 35.178 |
| C2H6O2 | Dimethyl peroxide | 128.050 | 39.281 |
| C2H6O2 | Ethylene glycol | 59.646 | 20.064 |
| C3O2 | Carbon suboxide | 69.482 | 2.243 |
| C3H4O2 | 2-Oxo-propanal | 40.003 | 17.963 |
| C3H4O2 | 2-Propenoic acid | 72.339 | 20.764 |
| C3H4O2 | beta-Propiolactone | 45.404 | 13.018 |
| C3H6O2 | 1,3-Dioxalane | 35.024 | 19.398 |
| C3H6O2 | Ethyl formate | 55.845 | 19.728 |
| C3H6O2 | Methyl acetate | 58.786 | 17.588 |
| C3H6O2 | Propionic acid | 49.264 | 23.786 |
| C3H8O2 | 1,3-Propanediol | 69.176 | 22.604 |
| C3H8O2 | Dimethoxymethane | 51.576 | 12.255 |
| C3H8O2 | Propylene glycol | 62.492 | 23.854 |
| C4H6O2 | Diacetyl | 33.111 | 20.406 |
| C4H6O2 | Methyl 2-propenoate | 59.743 | 20.999 |
| C4H8O2 | 1,1 Dimethoxy ethene | 46.555 | 28.790 |
| C4H8O2 | 1,3 Dioxan | 49.970 | 14.450 |
| C4H8O2 | 1,4-Dioxane | 51.484 | 10.918 |
| C4H8O2 | Ethyl acetate | 55.013 | 17.263 |
| C4H10O2 | 1,2-Dimethoxyethane | 109.426 | 46.306 |
| C4H10O2 | 1,4 Butandiol | 105.161 | 51.670 |
| C4H10O2 | Dimethyl acetal | 40.931 | 35.803 |
| C5H8O2 | Acetylacetone | 23.312 | 34.343 |
| C6H12O2 | Hexanoic acid | 66.356 | 27.592 |
| C7H6O2 | Benzoic acid | 62.148 | 16.946 |
| NO2 | Nitrogen dioxide | 51.862 | 20.159 |
| NHO2 | Nitrous acid, trans | 119.630 | 48.545 |
| CNH3O2 | Methyl nitrite | 124.706 | 48.513 |
| CNH3O2 | Nitromethane | 75.774 | 24.700 |

| Formula | Name | Value1 | Value2 |
|---|---|---|---|
| C2NH5O2 | Glycine | 68.591 | 29.796 |
| C3NH7O2 | Alanine | 57.661 | 26.591 |
| C3NH7O2 | beta-Alanine | 60.677 | 24.817 |
| C3NH7O2 | Isopropylnitrite | 104.117 | 42.688 |
| C3NH7O2 | Propyl nitrite | 114.786 | 47.167 |
| C3NH7O2 | Urethane | 65.416 | 20.773 |
| C5NH9O2 | Proline | 87.871 | 33.473 |
| C6NH5O2 | Nitrobenzene | 48.802 | 12.490 |
| C2N2H4O2 | Oxalamide | 60.700 | 25.436 |
| C2N2H6O2 | N-Nitrodimethylamine | 62.493 | 25.710 |
| C4N2H4O2 | Uracil | 50.352 | 35.848 |
| O3 | Ozone | 180.153 | 20.589 |
| C3H6O3 | 1,3,5-Trioxane | 55.828 | 13.778 |
| C3H8O3 | Glycerol | 70.627 | 24.147 |
| C4H2O3 | Malaic anhydride | 42.613 | 30.476 |
| NHO3 | Nitric acid | 75.722 | 33.525 |
| CNH3O3 | Methyl nitrate | 45.529 | 29.312 |
| C2NH3O3 | Oxamic acid | 66.760 | 16.641 |
| C2NH5O3 | Ethyl nitrate | 69.365 | 32.820 |
| C3NH7O3 | Serine | 62.759 | 29.878 |
| N2O3 | Dinitrogen trioxide | 90.632 | 50.463 |
| C2H2O4 | Oxalic acid | 76.342 | 24.934 |
| N2O4 | Dinitrogen tetroxide | 58.535 | 27.883 |
| N2O5 | Dinitrogen pentoxide | 81.852 | 48.352 |
| HF | Hydrogen fluoride | 42.440 | 42.534 |
| CF | Fluoromethylidyne radical | 31.773 | 3.386 |
| CH3F | Fluoromethane | 56.439 | 15.925 |
| C2HF | Fluoroacetylene | 12.129 | 35.863 |
| C2H3F | Fluoroethylene | 49.943 | 7.548 |
| C2H5F | Fluoroethane | 61.873 | 14.238 |
| C3H7F | 2-Fluoropropane | 65.938 | 12.780 |
| C6H5F | Fluorobenzene | 59.242 | 14.933 |
| CNF | Cyanogen fluoride | 4.332 | 47.266 |
| OF | Fluorine oxide | 146.425 | 10.348 |
| HOF | Hypofluorous acid | 139.415 | 26.606 |
| CHOF | HCOF | 89.492 | 25.404 |
| C2H3OF | Acetyl fluoride | 89.311 | 17.028 |
| NOF | Nitrosyl fluoride | 101.644 | 80.951 |
| O2F | Fluorine dioxide | 137.358 | 38.408 |
| NO3F | Fluorine nitrate | 154.290 | 59.658 |
| F2 | Fluorine molecule | 144.153 | 0.208 |
| CF2 | Difluoromethylene | 29.545 | 7.532 |
| CH2F2 | Difluoromethane | 41.805 | 10.761 |
| C2H2F2 | gem-Difluoroethylene | 43.293 | 10.860 |
| C2H4F2 | 1,1-Difluoroethane | 55.299 | 10.281 |
| C6H4F2 | 1,2-Difluorobenzene | 73.597 | 21.793 |
| C6H4F2 | 1,3-Difluorobenzene | 79.806 | 19.928 |
| C6H4F2 | 1,4-Difluorobenzene | 83.772 | 23.537 |
| N2F2 | cis-Difluorodiazene | 136.754 | 28.861 |
| N2F2 | trans-Difluorodiazene | 162.092 | 36.801 |
| OF2 | Difluorine oxide | 172.425 | 12.186 |
| CF3 | Trifluoromethyl | 42.882 | 7.120 |
| CHF3 | Trifluoromethane | 28.935 | 23.916 |
| C2HF3 | Trifluoroethylene | 58.367 | 18.977 |
| C2H3F3 | 1,1,1-Trifluoroethane | 44.162 | 23.641 |
| NF3 | Nitrogen trifluoride | 140.046 | 23.111 |
| C2NF3 | Trifluoroacetonitrile | 181.028 | 148.417 |
| NOF3 | F3NO | 126.938 | 38.928 |
| C2HO2F3 | Trifluoroacetic acid | 57.566 | 32.190 |
| CF4 | Carbon tetrafluoride | 39.994 | 24.187 |
| C2F6 | Hexafluoroethane | 64.367 | 30.206 |
| C4F6 | Perfluorobutadiene | 89.146 | 30.442 |
| C6F6 | Hexafluorobenzene | 149.749 | 78.901 |
| C3F8 | Perfluoropropane | 83.956 | 44.568 |

**Training Set for Limited Parameterization**

The training set employed for our limited parameterization is provided on the next page, with all atomic coordinates in Angstroms. EXPGEOM refers to the reference geometry (either experimental or from high-level calculation).

```
H1, RHF, CHARGE=1, MULT=1
HF=365.7
1    H    0    0    0
---
H1, UHF, CHARGE=0, MULT=2
HF=52.1
1    H    0    0    0
---
H2, RHF, CHARGE=0, MULT=1
HF=0, IE=15.4
1    H     0.331664257    -0.010000000    -0.010000000
2    H    -0.331664257    -0.010000000    -0.010000000
EXPGEOM
1    H     0.371300000     0.000000000     0.000000000
2    H    -0.371300000     0.000000000     0.000000000
---
C1, UHF, CHARGE=1, MULT=2
HF=430.6
1    C    0    0    0
---
C1, UHF, CHARGE=0, MULT=3
HF=170.9
1    C    0    0    0
---
H1C1, UHF, CHARGE=0, MULT=2
HF=142.4
1    C     0.000000000     2.000000000     0.000000000
2    H     0.000000000     2.000000000     1.098011582
---
H2C1, RHF, CHARGE=0, MULT=1
HF=99.8
1    C     0.330140612    -0.010000000    -0.022014279
2    H    -0.348909250    -0.010000000     0.832051876
3    H    -0.221442589    -0.010000000    -0.963534882
EXPGEOM
1    C     0.106700000     0.000000000     0.000000000
2    H    -0.320200000     0.000000000     0.993400000
3    H    -0.320200000     0.000000000    -0.993400000
---
H2C1, UHF, CHARGE=0, MULT=3
HF=99.8
1    C     0.028272289     0.002210737    -0.002505502
2    H    -0.267372093    -0.018024304     1.006803715
3    H    -0.168921821     0.014944742    -1.035651495
EXPGEOM
1    C     0.106700000     0.000000000     0.000000000
2    H    -0.320200000     0.000000000     0.993400000
3    H    -0.320200000     0.000000000    -0.993400000
---
H3C1, RHF, CHARGE=1, MULT=1
HF=261
1    C     0.000000000     0.000000000     0.000000000
2    H    -0.000000000     0.000000000    -1.098159290
3    H    -0.951033843     0.000000000     0.549079645
4    H     0.951033843     0.000000000     0.549079645
EXPGEOM
1    C     0.000000000     0.000000000     0.000000000
2    H     0.000000000     0.000000000     1.095400000
3    H     0.000000000     0.948700000    -0.547700000
```

```
4       H       0.000000000     -0.948700000    -0.547700000
---
H4C1, RHF, CHARGE=0, MULT=1
HF=-17.9, IE=13.6
1       C       -0.002109905    0.002109805     -0.002109843
2       H       0.635848274     0.639091331     0.635846910
3       H       0.635846971     -0.635848277    -0.639091242
4       H       -0.639091099    -0.635846998    0.635848354
5       H       -0.639693041    0.639692700     -0.639692831
EXPGEOM
1       C       0.000000000     0.000000000     0.000000000
2       H       0.630400000     0.630400000     0.630400000
3       H       0.630400000     -0.630400000    -0.630400000
4       H       -0.630400000    -0.630400000    0.630400000
5       H       -0.630400000    0.630400000     -0.630400000
---
H2C2, RHF, CHARGE=0, MULT=1
HF=54.3, IE=11.4
1       C       0.597385389     0.000000000     0.000000000
2       C       -0.597385389    0.000000000     0.000000000
3       H       1.648326422     0.000000000     0.000000000
4       H       -1.648326422    0.000000000     0.000000000
EXPGEOM
1       C       0.599700000     0.000000000     0.000000000
2       C       -0.599700000    0.000000000     0.000000000
3       H       1.662000000     0.000000000     0.000000000
4       H       -1.662000000    0.000000000     0.000000000
---
H3C2, RHF, CHARGE=1, MULT=1
HF=266
1       C       0.000000000     0.000000000     0.000000000
2       C       0.000000000     1.285301930     0.000000000
3       H       0.000000000     2.352436000     -0.000020717
4       H       -0.352748393    -0.621995606    0.842358180
5       H       0.352585636     -0.621580107    -0.842761049
EXPGEOM
1       H       -1.146200000    0.000000000     0.948300000
2       C       -0.584700000    0.000000000     0.000000000
3       C       0.673900000     0.000000000     0.000000000
4       H       -1.146200000    0.000000000     -0.948300000
5       H       1.757500000     0.000000000     0.000000000
---
H3C2, UHF, CHARGE=0, MULT=2
HF=59.6
1       C       0.000000000     0.000000000     0.000000000
2       C       0.000000000     1.306649284     0.000000000
3       H       0.000000000     2.355370242     0.000000000
4       H       -0.057066192    -0.594074326    0.913852202
5       H       0.057045813     -0.593816349    -0.914014247
EXPGEOM
1       C       0.000000071     -0.108746090    0.693904737
2       C       -0.000000096    0.047111511     -0.603724177
3       H       -0.000000091    -0.269457622    1.730411538
4       H       0.000000092     -0.791229815    -1.302477509
5       H       0.000000086     1.028345042     -1.081839773
---
H4C2, UHF, CHARGE=1, MULT=2
HF=257
1       C       0.000000000     0.000000000     0.000000000
```

```
2    C     0.000000000     1.421626560     0.000000000
3    H     0.000000000     1.997810054     0.929353699
4    H    -0.391490177    -0.577373943     0.841931576
5    H     0.388820100    -0.575971552    -0.844211745
6    H     0.001696423     1.998561295    -0.928809372
EXPGEOM
1    C     0.710200000     0.000000000     0.000000000
2    C    -0.710200000     0.000000000     0.000000000
3    H     1.268700000     0.000000000     0.935200000
4    H     1.268700000     0.000000000    -0.935200000
5    H    -1.268700000     0.000000000     0.935200000
6    H    -1.268700000     0.000000000    -0.935200000
---
H4C2, RHF, CHARGE=0, MULT=1
HF=12.5, IE=10.51
1    C     0.667272761     0.000000000    -0.004308789
2    C    -0.667272761     0.000000000     0.004308790
3    H     1.269732555     0.000000000     0.902979043
4    H     1.257380268     0.000000000    -0.919470263
5    H    -1.269732555     0.000000000    -0.902979043
6    H    -1.257380268     0.000000000     0.919470263
EXPGEOM
1    C     0.664200000     0.000000000     0.000000000
2    C    -0.664200000     0.000000000     0.000000000
3    H     1.239000000     0.000000000     0.921500000
4    H     1.239000000     0.000000000    -0.921500000
5    H    -1.239000000     0.000000000    -0.921500000
6    H    -1.239000000     0.000000000     0.921500000
---
H4C2, RHF, CHARGE=0, MULT=1
HF=90.3
1    C     0.000000000     0.000000000     0.000000000
2    C     0.000000000     1.462361850     0.000000000
3    H     0.000000000    -0.361918691     1.051274014
4    H    -0.854068281    -0.483215857    -0.522901646
5    H     0.939432225    -0.350862946    -0.480630665
6    H    -0.778250487     2.003839255    -0.527855116
EXPGEOM
1    C     0.000000000     0.065900000     0.871500000
2    C     0.000000000     0.065900000    -0.594700000
3    H     0.000000000    -0.929000000     1.322000000
4    H     0.000000000     1.079900000    -1.003300000
5    H     0.876300000    -0.470600000    -0.989900000
6    H    -0.876300000    -0.470600000    -0.989900000
---
H5C2, RHF, CHARGE=1, MULT=1
HF=216
1    C     0.000000000     0.000000000     0.000000000
2    C     0.000000000     1.457600590     0.000000000
3    H     0.000000000     1.891190744     1.024872616
4    H     0.834467992     1.892670589    -0.595354604
5    H    -0.953395536     1.799735274    -0.495366068
6    H    -0.400269573    -0.578208317     0.840216909
7    H     0.389068117    -0.577335651    -0.846124208
EXPGEOM
1    H     1.055200000     0.000000000     0.000000000
2    C    -0.064000000     0.000000000     0.691200000
3    C    -0.064000000     0.000000000    -0.691200000
4    H    -0.071700000     0.935400000     1.249600000
```

```
5      H     -0.071700000    -0.935400000     1.249600000
6      H     -0.071700000    -0.935400000    -1.249600000
7      H     -0.071700000     0.935400000    -1.249600000
---
H5C2, UHF, CHARGE=0, MULT=2
HF=25
1      C      0.000000000     0.000000000     0.000000000
2      C      0.000000000     1.475310220     0.000000000
3      H      0.000000000     1.885519649     1.031539202
4      H      0.882972630     1.887502416    -0.531152626
5      H     -0.910553587     1.858627039    -0.512524873
6      H     -0.487005000    -0.564063100     0.785071839
7      H      0.474946731    -0.565192771    -0.791544562
EXPGEOM
1      C      0.000000000    -0.010900000    -0.693100000
2      C      0.000000000    -0.010900000     0.794600000
3      H      0.000000000     1.012100000    -1.108300000
4      H      0.886800000    -0.508000000    -1.101800000
5      H     -0.886800000    -0.508000000    -1.101800000
6      H     -0.926500000     0.067300000     1.351600000
7      H      0.926500000     0.067300000     1.351600000
---
H6C2, RHF, CHARGE=0, MULT=1
HF=-20, IE=12
1      C      0.760473919     0.003148967    -0.003015508
2      C     -0.760473919     0.003148967     0.003015508
3      H      1.165664482     0.004200547     1.029446173
4      H      1.160845896    -0.892018392    -0.521281685
5      H      1.157579713     0.899126580    -0.522377155
6      H     -1.165664482     0.004200547    -1.029446173
7      H     -1.160845896    -0.892018392     0.521281685
8      H     -1.157579713     0.899126580     0.522377155
EXPGEOM
1      C      0.765300000     0.000000000     0.000000000
2      C     -0.765300000     0.000000000     0.000000000
3      H      1.163700000     0.000000000     1.019700000
4      H      1.163700000    -0.883100000    -0.509900000
5      H      1.163700000     0.883100000    -0.509900000
6      H     -1.163700000     0.000000000    -1.019700000
7      H     -1.163700000    -0.883100000     0.509900000
8      H     -1.163700000     0.883100000     0.509900000
---
C3, RHF, CHARGE=0, MULT=1
HF=196, IE=11.1
1      C      0.000000000     0.000000000     0.000000000
2      C      0.000000000     1.289765220     0.000000000
3      C      0.000000000     2.579530430     0.000006380
EXPGEOM
1      C      0.114900000     0.000000000     0.000000000
2      C     -0.057400000     0.000000000     1.281100000
3      C     -0.057400000     0.000000000    -1.281100000
---
H3C3, RHF, CHARGE=1, MULT=1
HF=257
1      C      0.000000000     0.000000000     0.000000000
2      C      0.000000000     1.412108010     0.000000000
3      C      0.000000000     0.705531538     1.223120579
4      H     -0.000000022    -0.928515500    -0.536903044
5      H     -0.000001702     2.341346238    -0.535631834
```

```
6    H      0.000002175      0.706150183      2.295675910
EXPGEOM
1    C      0.000000000      0.000000000      0.785700000
2    C      0.000000000      0.680400000     -0.392800000
3    C      0.000000000     -0.680400000     -0.392800000
4    H      0.000000000      0.000000000      1.867600000
5    H      0.000000000      1.617400000     -0.933800000
6    H      0.000000000     -1.617400000     -0.933800000
---
H3C3, RHF, CHARGE=1, MULT=1
HF=281
1    C      0.000000000      2.000000000      0.000000000
2    C      0.000000000      2.000000000      1.224478500
3    H      0.000000000      2.000000000     -1.062084880
4    C     -0.000000000      2.000000000      2.585089830
5    H      0.466967001      1.193494521      3.160805627
6    H     -0.467202589      2.806242843      3.160967278
EXPGEOM
1    C     -1.239900000      0.000000000      0.000000000
2    C      0.105200000      0.000000000      0.000000000
3    C      1.332500000      0.000000000      0.000000000
4    H     -1.797000000      0.000000000      0.938800000
5    H     -1.797000000      0.000000000     -0.938800000
6    H      2.407100000      0.000000000      0.000000000
---
H4C3, RHF, CHARGE=0, MULT=1
HF=45.6, IE=10.07
1    C     -0.000051684     -0.000582416     -0.000318291
2    C      1.305823994     -0.002271376      0.002549445
3    C     -1.305930131      0.002447688     -0.002193376
4    H      1.894592338     -0.001257067      0.920441859
5    H      1.899827143     -0.004804246     -0.911943660
6    H     -1.895304236      0.919846097     -0.004768947
7    H     -1.901298345     -0.910994362     -0.001404238
EXPGEOM
1    C      0.000000000      0.000000000      0.000000000
2    C      1.302300000      0.000000000      0.000000000
3    C     -1.302300000      0.000000000      0.000000000
4    H      1.868200000      0.000000000      0.927100000
5    H      1.868200000      0.000000000     -0.927100000
6    H     -1.868200000      0.927100000      0.000000000
7    H     -1.868200000     -0.927100000      0.000000000
---
H4C3, RHF, CHARGE=0, MULT=1
HF=66.2, DIP=0.45, IE=9.86
1    C      0.000000000      0.000000000      0.000000000
2    C      1.512740660      0.000000000      0.000000000
3    C      0.582832117      1.193198627      0.000000000
4    H      0.536942049      2.254131693     -0.000079732
5    H     -0.864841754     -0.616204590     -0.000110466
6    H      2.066301671     -0.270521255     -0.907623677
7    H      2.066261446     -0.271460823      0.893790386
---
H4C3, RHF, CHARGE=0, MULT=1
HF=44.4, DIP=0.78, IE=10.37
1    H     -1.626758579     -0.894740519     -0.520370529
2    H     -1.620217456      0.902275533     -0.528924856
3    H     -1.627775005      0.011123061      1.031396237
4    H      2.466268102     -0.001056138      0.000985901
```

```
5    C     1.415594963   -0.000225867   -0.000369904
6    C     0.218450578    0.001114350   -0.001778276
7    C    -1.226413570    0.004765339   -0.004874411
EXPGEOM
1    H    -1.629100000   -0.885600000   -0.511300000
2    H    -1.629100000    0.885600000   -0.511300000
3    H    -1.629100000    0.000000000    1.022600000
4    H     2.482600000    0.000000000    0.000000000
5    C     1.421000000    0.000000000    0.000000000
6    C     0.219100000    0.000000000    0.000000000
7    C    -1.239300000    0.000000000    0.000000000
---
H5C3, RHF, CHARGE=1, MULT=1
HF=226
1    H     0.000000000    0.000000000    0.000000000
2    C     0.000000000    1.100794050    0.000000000
3    C     0.000000000    1.731487829    1.240502867
4    C     0.000000000    1.730812611   -1.240847622
5    H     0.000004012    1.157649745    2.171176047
6    H    -0.000004839    2.817177654    1.366186547
7    H    -0.000005029    1.156701047   -2.171325908
8    H     0.000005750    2.816441749   -1.367054603
EXPGEOM
1    C     0.494900000    0.000000000    0.000000000
2    H     1.580300000    0.000000000    0.000000000
3    C    -0.211100000    0.000000000    1.188600000
4    C    -0.211100000    0.000000000   -1.188600000
5    H     0.291500000    0.000000000    2.154300000
6    H     0.291500000    0.000000000   -2.154300000
7    H    -1.299600000    0.000000000    1.198800000
8    H    -1.299600000    0.000000000   -1.198800000
---
H5C3, RHF, CHARGE=1, MULT=1
HF=235
1    C     0.000000000    1.662325418    0.000000000
2    C    -0.265303210    0.293800103    0.611706849
3    C     0.316396991    0.431787686   -0.729910654
4    H     0.807921069    2.300298421    0.373553986
5    H    -0.827376219    2.298569863   -0.334268187
6    H    -1.286283760   -0.082804777    0.728065336
7    H     0.348501939   -0.082156200    1.437333167
8    H     0.696654043   -0.043155936   -1.607532297
---
H5C3, RHF, CHARGE=1, MULT=1
HF=237
1    C     0.000000000    0.000000000    0.000000000
2    C     0.000000000    1.293897860    0.000000000
3    C     0.000000000    2.722034380    0.000000000
4    H    -0.000000019    3.077631591   -1.064812405
5    H     0.917927915    3.133858479    0.479871644
6    H    -0.906413660    3.157062801    0.506910176
7    H    -0.000000016   -0.610018536   -0.916518188
8    H     0.000000016   -0.604750373    0.920660084
---
H5C3, UHF, CHARGE=0, MULT=2
HF=40
1    C     0.000000000    0.000000000    0.000000000
2    C     0.000000000    1.397184480    0.000000000
3    C     0.000000000    2.222081422    1.127880711
```

```
4     H      0.000000000     1.841004704     2.144576784
5     H      0.000932118     3.304470070     1.039864537
6     H      0.000000000     1.893318688    -0.977996355
7     H     -0.000000000    -0.568011803    -0.925569889
8     H      0.000000000    -0.595985216     0.907604478
EXPGEOM
1     C      0.443200000     0.000000000     0.000000000
2     H      1.531600000     0.000000000     0.000000000
3     C     -0.196100000     0.000000000     1.225900000
4     C     -0.196100000     0.000000000    -1.225900000
5     H      0.360500000     0.000000000     2.155500000
6     H      0.360500000     0.000000000    -2.155500000
7     H     -1.279300000     0.000000000     1.291900000
8     H     -1.279300000     0.000000000    -1.291900000
---
H6C3, RHF, CHARGE=0, MULT=1
HF=12.7, IE=11
1     C      0.003825017     0.001574296     0.875978202
2     C      0.004244677     0.763631384    -0.445619672
3     C     -0.005121029    -0.761920016    -0.444624635
4     H      0.908243187    -0.004757679     1.495258815
5     H      0.908476855     1.297680263    -0.759758343
6     H      0.892949712    -1.306615617    -0.758371968
7     H     -0.895942915     0.007946506     1.501910287
8     H     -0.895527695     1.308251817    -0.754433611
9     H     -0.910971637    -1.296500066    -0.752835270
EXPGEOM
1     C      0.000000000     0.000000000     0.869500000
2     C      0.000000000     0.753000000    -0.434700000
3     C      0.000000000    -0.753000000    -0.434700000
4     H      0.909300000     0.000000000     1.460600000
5     H      0.909300000     1.264900000    -0.730300000
6     H      0.909300000    -1.264900000    -0.730300000
7     H     -0.909300000     0.000000000     1.460600000
8     H     -0.909300000     1.264900000    -0.730300000
9     H     -0.909300000    -1.264900000    -0.730300000
---
H6C3, RHF, CHARGE=0, MULT=1
HF=4.9, DIP=0.37, IE=9.88
1     C     -0.005082703    -1.140548940    -0.500049733
2     C     -0.001768944    -0.000806215     0.468295917
3     C     -0.006606847     1.306462072     0.172886393
4     H     -0.013940618     1.706729460    -0.839716214
5     H     -0.003372550     2.077957505     0.942124000
6     H      0.005716497    -0.313309393     1.518892851
7     H     -0.007005996    -0.811857706    -1.559207090
8     H      0.892000503    -1.777675983    -0.348665414
9     H     -0.902175618    -1.776698027    -0.345503506
EXPGEOM
1     C      0.000000000    -1.137600000    -0.507500000
2     C      0.000000000     0.000000000     0.473400000
3     C      0.000000000     1.291300000     0.154200000
4     H      0.000000000     1.625400000    -0.880100000
5     H      0.000000000     2.068700000     0.911600000
6     H      0.000000000    -0.285800000     1.524800000
7     H      0.000000000    -0.774900000    -1.539300000
8     H      0.879700000    -1.777800000    -0.368900000
9     H     -0.879700000    -1.777800000    -0.368900000
---
```

```
H7C3, UHF, CHARGE=0, MULT=2
HF=16.8
1    C     0.000000000    0.000000000    0.000000000
2    C     0.000000000    1.482947490    0.000000000
3    C     0.000000000   -0.852355770    1.214403624
4    H     0.000592575   -0.505542350   -0.964086603
5    H     0.492114762    1.880338276   -0.912254366
6    H    -1.043175877    1.868543149    0.014209045
7    H     0.528302665    1.908258298    0.878268593
8    H    -0.490907022   -1.827719815    1.015121737
9    H    -0.529721054   -0.378623304    2.066597176
10   H     1.043056846   -1.060896790    1.539077214
EXPGEOM
1    C     0.000000000   -0.013300000    0.537800000
2    C     1.295700000   -0.013300000   -0.198900000
3    C    -1.295700000   -0.013300000   -0.198900000
4    H     0.000000000    0.277400000    1.581600000
5    H     1.285900000   -0.733400000   -1.021600000
6    H    -1.285900000   -0.733400000   -1.021600000
7    H     1.507100000    0.969300000   -0.645100000
8    H     2.133500000   -0.255200000    0.456200000
9    H    -1.507100000    0.969300000   -0.645100000
10   H    -2.133500000   -0.255200000    0.456200000
---
H8C3, RHF, CHARGE=0, MULT=1
HF=-24.8, DIP=0.08, IE=11.5
1    C     0.562179883   -0.001959122   -0.000816915
2    C    -0.256387052    0.001996076    1.291789362
3    C    -0.254657618   -0.001918317   -1.294458535
4    H     1.237508957    0.885264389   -0.001779757
5    H     1.230615926   -0.894467493    0.000881338
6    H     0.416688896    0.000123955    2.173876673
7    H     0.419595741    0.000937646   -2.175651103
8    H    -0.903275477    0.899645995    1.367579872
9    H    -0.907054665   -0.892706973    1.367572143
10   H    -0.901293318   -0.899461678   -1.372311654
11   H    -0.905754348    0.892529537   -1.369884737
EXPGEOM
1    C     0.586800000    0.000000000    0.000000000
2    C    -0.259900000    0.000000000    1.276100000
3    C    -0.259900000    0.000000000   -1.276100000
4    H     1.246800000    0.876500000    0.000000000
5    H     1.246800000   -0.876500000    0.000000000
6    H     0.366000000    0.000000000    2.174300000
7    H     0.366000000    0.000000000   -2.174300000
8    H    -0.906900000    0.883500000    1.320400000
9    H    -0.906900000   -0.883500000    1.320400000
10   H    -0.906900000   -0.883500000   -1.320400000
11   H    -0.906900000    0.883500000   -1.320400000
---
C4, UHF, CHARGE=0, MULT=3
HF=232
1    C     0.000000000    0.000000000    0.000000000
2    C     0.000000000    1.314467467    0.000000000
3    C     0.000000000    2.593808313    0.000003901
4    C     0.000000777    3.908275782    0.000012317
EXPGEOM
1    C     0.639676057   -0.000000000   -0.000000000
2    C    -0.639676058   -0.000000000    0.000000000
```

```
3    C     1.954121073    0.000000000     0.000000000
4    C    -1.954121072    0.000000000    -0.000000000
---
H2C4, RHF, CHARGE=0, MULT=1
HF=113, IE=10.17
1    C     0.683875887    0.000000000     0.000000000
2    C    -0.683875887    0.000000000     0.000000000
3    C     1.883403311    0.000000000     0.000000000
4    C    -1.883403311    0.000000000     0.000000000
5    H     2.933892771    0.000000000     0.000000000
6    H    -2.933892774    0.000000000     0.000000000
EXPGEOM
1    C     0.682800000    0.000000000     0.000000000
2    C    -0.682800000    0.000000000     0.000000000
3    C     1.888600000    0.000000000     0.000000000
4    C    -1.888600000    0.000000000     0.000000000
5    H     2.950500000    0.000000000     0.000000000
6    H    -2.950500000    0.000000000     0.000000000
---
H4C4, RHF, CHARGE=0, MULT=1
HF=72.8, IE=9.1
1    C    -0.010000000    0.571704665    -0.548834154
2    C    -0.010000000    0.003252367     0.749086013
3    C    -0.010000000   -0.119325624    -1.702544937
4    C    -0.010000000   -0.457660234     1.855340764
5    H    -0.010000000    1.668539089    -0.565086428
6    H    -0.010000000    0.373011634    -2.673920967
7    H    -0.010000000   -1.206187108    -1.761706992
8    H    -0.010000000   -0.861553273     2.825138877
EXPGEOM
1    C     0.000000000    0.581400000    -0.556500000
2    C     0.000000000    0.000000000     0.742900000
3    C     0.000000000   -0.120000000    -1.693800000
4    C     0.000000000   -0.455500000     1.858300000
5    H     0.000000000    1.668500000    -0.584500000
6    H     0.000000000    0.377600000    -2.657400000
7    H     0.000000000   -1.204100000    -1.696700000
8    H     0.000000000   -0.877400000     2.832900000
---
H4C4, RHF, CHARGE=0, MULT=1
HF=83, IE=9.15
1    H     0.000000000    0.000000000     0.000000000
2    C     0.000000000    1.089797590     0.000000000
3    C     0.000000000    1.802892321     1.099565331
4    C     0.000000000    2.494064439     2.165326815
5    C     0.000000000    3.207275708     3.265071846
6    H    -0.000029600    2.757573626     4.257823067
7    H     0.000015975    1.538926102    -0.993351679
8    H     0.000025817    4.297114078     3.259673511
EXPGEOM
1    C     0.632800000    0.000000000     0.000000000
2    C    -0.632800000    0.000000000     0.000000000
3    C     1.945400000    0.000000000     0.000000000
4    C    -1.945400000    0.000000000     0.000000000
5    H     2.514200000    0.000000000     0.926100000
6    H     2.514200000    0.000000000    -0.926100000
7    H    -2.514200000    0.000000000     0.926100000
8    H    -2.514200000    0.000000000    -0.926100000
---
```

```
H6C4, RHF, CHARGE=0, MULT=1
HF=38.8, IE=9.15
1    C     0.000000000     0.000000000     0.000000000
2    C     0.000000000     1.305704480     0.000000000
3    C    -0.000000000     2.616561460     0.000000000
4    C    -0.000000000     3.496546442     1.213049906
5    H    -0.923029239    -0.582128055     0.000000000
6    H     0.901170679    -0.598345767     0.000000000
7    H    -0.000000000     3.154665916    -0.955978387
8    H    -0.000000000     2.918578606     2.159473013
9    H     0.902233033     4.144512886     1.214896943
10   H    -0.886742277     4.147770537     1.214906229
EXPGEOM
1    C     0.000000000    -0.677100000     1.827100000
2    C     0.000000000     0.000000000     0.713500000
3    C     0.000000000     0.678100000    -0.400000000
4    C     0.000000000     0.069800000    -1.782100000
5    H     0.926200000    -0.975900000     2.311100000
6    H    -0.926200000    -0.975900000     2.311100000
7    H     0.000000000     1.766400000    -0.343400000
8    H     0.000000000    -1.021000000    -1.732400000
9    H    -0.881800000     0.390800000    -2.348600000
10   H     0.881800000     0.390800000    -2.348600000
---
H6C4, RHF, CHARGE=0, MULT=1
HF=26, IE=9.08
1    C     0.000000000    -0.316965541     0.660592339
2    C     0.000000000     0.316962422    -0.660590730
3    C     0.000000000     0.328379265     1.840062904
4    C     0.000000000    -0.328370821    -1.840060424
5    H     0.000000000    -1.413261403     0.661795069
6    H     0.000000000     1.413265588    -0.661796230
7    H     0.000000000    -0.201597245     2.791724574
8    H     0.000000000     1.411653920     1.947519812
9    H     0.000000000     0.201546923    -2.791771761
10   H     0.000000000    -1.411712794    -1.947561627
EXPGEOM
1    C     0.000000000    -0.325300000     0.652500000
2    C     0.000000000     0.325300000    -0.652500000
3    C     0.000000000     0.325300000     1.819900000
4    C     0.000000000    -0.325300000    -1.819900000
5    H     0.000000000    -1.414100000     0.639200000
6    H     0.000000000     1.414100000    -0.639200000
7    H     0.000000000    -0.199700000     2.768900000
8    H     0.000000000     1.410700000     1.867200000
9    H     0.000000000     0.199700000    -2.768900000
10   H     0.000000000    -1.410700000    -1.867200000
---
H6C4, RHF, CHARGE=0, MULT=1
HF=39.5, IE=10.2
1    C     0.000000000     0.000000000     0.000000000
2    C     0.000000000     1.197098590     0.000000000
3    C    -0.000000000     2.647341100     0.000000000
4    C    -0.000000000     3.275753998     1.400151542
5    H     0.000000000    -1.050718350     0.000000000
6    H     0.897707488     3.008153287    -0.559543385
7    H    -0.885621947     3.015584667    -0.571067883
8    H    -0.000000000     4.382556393     1.320847558
9    H    -0.902084499     2.981026039     1.974197573
```

```
10      H       0.883532587     2.979032162     1.978081077
EXPGEOM
1       C       0.000000000     -0.660400000    -1.856200000
2       C       0.000000000     0.000000000     -0.851300000
3       C       0.000000000     0.750600000     0.403400000
4       C       0.000000000     -0.160600000    1.643800000
5       H       0.000000000     -1.227600000    -2.753800000
6       H       0.877500000     1.408300000     0.425700000
7       H       -0.877500000    1.408300000     0.425700000
8       H       0.000000000     0.440700000     2.557900000
9       H       -0.883900000    -0.803700000    1.653200000
10      H       0.883900000     -0.803700000    1.653200000
---
H6C4, RHF, CHARGE=0, MULT=1
HF=34.7, IE=9.6
1       C       0.599732216     0.000827445     -0.002213842
2       C       -0.599815819    0.000845394     -0.002205076
3       C       2.043923202     -0.002331621    0.000712324
4       C       -2.043975029    0.002207142     0.000791584
5       H       2.439747840     -0.003270088    1.039251134
6       H       2.440931401     -0.901967985    -0.517254644
7       H       2.446431972     0.894887905     -0.516924784
8       H       -2.439604589    0.002200855     1.039408102
9       H       -2.441692404    0.901577401     -0.516891315
10      H       -2.445731477    -0.895268565    -0.517240977
EXPGEOM
1       C       0.602000000     0.000000000     0.000000000
2       C       -0.602000000    0.000000000     0.000000000
3       C       2.061800000     0.000000000     0.000000000
4       C       -2.061800000    0.000000000     0.000000000
5       H       2.455500000     0.000000000     1.021700000
6       H       2.455500000     -0.884800000    -0.510800000
7       H       2.455500000     0.884800000     -0.510800000
8       H       -2.455500000    0.000000000     1.021700000
9       H       -2.455500000    0.884800000     -0.510800000
10      H       -2.455500000    -0.884800000    -0.510800000
---
H6C4, RHF, CHARGE=0, MULT=1
HF=51.9, DIP=0.68
1       C       -0.342633297    0.766569133     -0.000251899
2       C       -0.345218959    -0.769806456    0.000452001
3       C       0.289007576     -0.002313860    1.157638815
4       C       0.289316490     -0.003338387    -1.157390151
5       H       -1.014099260    1.604902539     -0.000673601
6       H       -1.019809166    -1.605595051    0.000633113
7       H       -0.240796656    -0.000918761    2.119138816
8       H       -0.240202850    -0.002737149    -2.119087993
9       H       1.374253376     -0.004301295    1.319637172
10      H       1.374521396     -0.005719353    -1.319161909
EXPGEOM
1       C       -0.323100000    0.739300000     0.000000000
2       C       -0.323100000    -0.739300000    0.000000000
3       C       0.314900000     0.000000000     1.135300000
4       C       0.314900000     0.000000000     -1.135300000
5       H       -1.135900000    1.447600000     0.000000000
6       H       -1.135900000    -1.447600000    0.000000000
7       H       -0.217500000    0.000000000     2.084300000
8       H       -0.217500000    0.000000000     -2.084300000
9       H       1.402500000     0.000000000     1.236900000
```

```
10      H      1.402500000     0.000000000    -1.236900000
---
H6C4, RHF, CHARGE=0, MULT=1
HF=37.5, DIP=0.13, IE=9.43
1       C     -0.000000000     0.000000000    -1.353349390
2       C      0.000000000     0.000000000     0.001015250
3       C     -0.000000000     1.523714192    -1.449596916
4       C      0.000000000     1.521260985     0.118724492
5       H     -0.897452164     1.966224832    -1.916624462
6       H      0.888354414     1.973857981    -1.924224663
7       H     -0.888586261     1.964602604     0.595893484
8       H      0.888585244     1.964588679     0.596036936
9       H     -0.000153396    -0.747615178     0.773169715
10      H     -0.000030569    -0.723515032    -2.123789333
EXPGEOM
1       C      0.814900000     0.000000000     0.669000000
2       C      0.814900000     0.000000000    -0.669000000
3       C     -0.700000000     0.000000000     0.786500000
4       C     -0.700000000     0.000000000    -0.786500000
5       H      1.599000000     0.000000000     1.418300000
6       H      1.599000000     0.000000000    -1.418300000
7       H     -1.144400000     0.888500000     1.247200000
8       H     -1.144400000    -0.888500000    -1.247200000
9       H     -1.144400000    -0.888500000     1.247200000
10      H     -1.144400000     0.888500000    -1.247200000
---
H6C4, RHF, CHARGE=0, MULT=1
HF=58.2
1       C      0.000000000     0.000000000     0.000000000
2       C      0.000000000     1.507095520     0.000000000
3       C      0.000000000     0.575787298     1.200946959
4       C      0.000047627     0.505845519     2.660653731
5       H     -0.000007221    -0.864353799    -0.617446850
6       H     -0.905091920     2.065386927    -0.267298607
7       H      0.904860546     2.065689325    -0.267466456
8       H      0.002775164     1.519443467     3.112314346
9       H     -0.899246381    -0.031913868     3.030291310
10      H      0.896378260    -0.036867174     3.030266419
EXPGEOM
1       H      0.00020        -1.38470         1.78070
2       C      0.00030        -0.91200         0.81410
3       C     -0.00070         0.15510         0.08640
4       H      0.91130        -1.48510        -1.17350
5       H     -0.91090        -1.48600        -1.17330
6       C      0.00010        -1.14580        -0.67200
7       H      0.88170         1.94790        -0.63580
8       H     -0.00090         2.13130         0.89750
9       H     -0.87960         1.94900        -0.63830
10      C      0.00010         1.62400        -0.07140
---
H6C4, RHF, CHARGE=0, MULT=1
HF=47.9
1       C      0.000000000     0.000000000     0.000000000
2       C      0.000000000     1.320491520     0.000000000
3       C      0.000000000     2.595917451     0.775099121
4       C      0.000000000     2.576017949    -0.763921292
5       H     -0.000000000    -0.588313575    -0.917883436
6       H      0.000000000    -0.593272256     0.902745062
7       H     -0.909125297     2.900142402     1.308589439
```

```
8      H     0.890246952    2.905977840    1.317194018
9      H     0.890177655    2.885739311   -1.306323704
10     H    -0.890185136    2.885947489   -1.306192499
EXPGEOM
1      C     1.63420        0.00000        0.00000
2      C     0.31720        0.00000        0.00000
3      H     2.20270        0.00000        0.92540
4      H     2.20270        0.00000       -0.92540
5      C    -0.93120        0.00000        0.76840
6      C    -0.93120        0.00000       -0.76840
7      H    -1.23460        0.91170        1.27640
8      H    -1.23460       -0.91170        1.27640
9      H    -1.23460       -0.91170       -1.27640
10     H    -1.23460        0.91170       -1.27640
---
H7C4, RHF, CHARGE=1, MULT=1
HF=200
1      C     0.000000000    0.000000000    0.000000000
2      C     0.000000000    1.385704800    0.000000000
3      C     0.000000000    2.196541675    1.148478803
4      C     0.000000000    3.686116202    1.137133772
5      H     0.000000000   -0.607469081    0.907814827
6      H    -0.000000000   -0.567774733   -0.933976950
7      H     0.000000000    1.885260010   -0.979514098
8      H    -0.000043393    1.724810313    2.141613658
9      H    -0.000000000    4.102670120    2.164839077
10     H     0.898744453    4.073058838    0.606057597
11     H    -0.899325267    4.072590548    0.606672589
---
H7C4, RHF, CHARGE=1, MULT=1
HF=213
1      C     0.000000000    0.000000000    0.000000000
2      C     0.000000000    2.183632450    0.000000000
3      C     0.000000000    1.091924233    1.116893795
4      C     0.001180432    1.091466748   -1.024489642
5      H     0.889058970   -0.666535829   -0.034026662
6      H    -0.889957315   -0.665341489   -0.034555957
7      H     0.889365640    2.849704750   -0.034269348
8      H    -0.890251564    2.848523016   -0.034800751
9      H    -0.894537935    1.092963935    1.763376306
10     H     0.893966087    1.092463810    1.764193295
11     H     0.002195686    1.093113124   -2.109676078
---
H8C4, RHF, CHARGE=0, MULT=1
HF=-0.2, IE=9.7
1      C    -0.292666052    1.739941214   -0.250134083
2      C     0.300399294    0.549234688    0.510552008
3      C     0.344397906   -0.728438240   -0.281875773
4      C    -0.282345413   -1.876067438    0.013118506
5      H     0.979597214   -0.682265177   -1.173860881
6      H    -1.344593885    1.551917742   -0.546060561
7      H    -0.280511245    2.649139310    0.385772200
8      H     0.284093343    1.968883133   -1.169103895
9      H     1.342179635    0.800104600    0.823391551
10     H    -0.270889944    0.408920261    1.457417920
11     H    -0.929605948   -2.020193622    0.876716861
12     H    -0.185089921   -2.767172126   -0.606328769
EXPGEOM
1      C    -0.290800000    1.728400000   -0.247900000
```

```
2    C     0.302700000    0.538100000    0.523400000
3    C     0.337400000   -0.722100000   -0.293600000
4    C    -0.276400000   -1.860000000    0.014900000
5    H     0.912800000   -0.664200000   -1.218400000
6    H    -1.327900000    1.529800000   -0.535700000
7    H    -0.273400000    2.638900000    0.359100000
8    H     0.275700000    1.926700000   -1.164300000
9    H     1.324800000    0.795800000    0.835300000
10    H    -0.269300000    0.365000000    1.442300000
11    H    -0.862700000   -1.962500000    0.924200000
12    H    -0.217300000   -2.736000000   -0.622600000
---
H8C4, RHF, CHARGE=0, MULT=1
HF=-1.9
1    C     0.000000000    0.000000000    0.000000000
2    C     0.000000000    1.349380530    0.000000000
3    C    -0.000364291    2.267519842    1.186109976
4    C    -0.000004114   -0.829356024    1.236815833
5    H     0.000100023   -0.545044962   -0.947241793
6    H     0.000006972    1.909529749   -0.933321888
7    H     0.897541340    2.116208284    1.821578855
8    H    -0.000797653    3.329855742    0.866362979
9    H    -0.898278911    2.115515979    1.821369865
10    H    -0.884088778   -0.621960257    1.857934699
11    H     0.000607074   -1.901179582    1.003299656
12    H     0.892012999   -0.621307156    1.867965138
EXPGEOM
1    C     0.66320    0.00000    0.66710
2    C     0.66320    0.00000   -0.66710
3    C    -0.52160    0.00000    1.59230
4    C    -0.52160    0.00000   -1.59230
5    H     1.62910    0.00000    1.17110
6    H     1.62910    0.00000   -1.17110
7    H    -1.47530    0.00000    1.05990
8    H    -1.47530    0.00000   -1.05990
9    H    -0.50180   -0.87940    2.24760
10    H    -0.50180    0.87940    2.24760
11    H    -0.50180    0.87940   -2.24760
12    H    -0.50180   -0.87940   -2.24760
---
H8C4, RHF, CHARGE=0, MULT=1
HF=6.8, IE=10.7
1    C    -0.000187808   -0.000021240    1.095387023
2    C    -0.000187806    0.000021405   -1.095386966
3    C     0.000187808   -1.095387004   -0.000019150
4    C     0.000187807    1.095386986    0.000018637
5    H     0.890891057   -0.000024791    1.748379904
6    H    -0.892042061   -0.000027058    1.747302998
7    H     0.890891058    0.000024749   -1.748379897
8    H    -0.892042064    0.000027046   -1.747302996
9    H    -0.890891057   -1.748379901   -0.000025505
10    H     0.892042062   -1.747302997   -0.000027296
11    H    -0.890891057    1.748379899    0.000025635
12    H     0.892042063    1.747302997    0.000027335
EXPGEOM
1    C     0.123100000    0.000000000    1.085200000
2    C     0.123100000    0.000000000   -1.085200000
3    C    -0.123100000   -1.085200000    0.000000000
4    C    -0.123100000    1.085200000    0.000000000
```

```
5     H    1.162600000    0.000000000    1.426500000
6     H   -0.527100000    0.000000000    1.963600000
7     H    1.162600000    0.000000000   -1.426500000
8     H   -0.527100000    0.000000000   -1.963600000
9     H   -1.162600000   -1.426500000    0.000000000
10    H    0.527100000   -1.963600000    0.000000000
11    H   -1.162600000    1.426500000    0.000000000
12    H    0.527100000    1.963600000    0.000000000
---
H8C4, RHF, CHARGE=0, MULT=1
HF=-4.3
1     C    0.116952433    0.002720080   -0.001284250
2     C    1.464999026    0.001249435   -0.003593640
3     H    2.070830614    0.000787880    0.901809107
4     H    2.066212478    0.000403303   -0.912245621
5     C   -0.679910739    0.003009694    1.279948604
6     C   -0.682691402    0.003104207   -1.280849628
7     H   -0.034679162   -0.000974574    2.182245193
8     H   -1.327727582    0.902542770    1.344781714
9     H   -1.333418075   -0.892608832    1.341199173
10    H   -0.039538695    0.000133418   -2.184715012
11    H   -1.335904371   -0.892764339   -1.341235169
12    H   -1.331313755    0.902203653   -1.343788234
EXPGEOM
1     C    0.126000000    0.000000000    0.000000000
2     C    1.458900000    0.000000000    0.000000000
3     H    2.031100000    0.000000000    0.922800000
4     H    2.031100000    0.000000000   -0.922800000
5     C   -0.679900000    0.000000000    1.275100000
6     C   -0.679900000    0.000000000   -1.275100000
7     H   -0.037700000    0.000000000    2.159400000
8     H   -1.334400000    0.879300000    1.325200000
9     H   -1.334400000   -0.879300000    1.325200000
10    H   -0.037700000    0.000000000   -2.159400000
11    H   -1.334400000   -0.879300000   -1.325200000
12    H   -1.334400000    0.879300000   -1.325200000
---
H8C4, RHF, CHARGE=0, MULT=1
HF=-3
1     C   -0.004952672    0.320611497    0.591900329
2     C    0.001394006   -0.320690107   -0.591768196
3     C   -0.006003303   -0.318172070    1.945170128
4     C   -0.004872947    0.317832719   -1.945134523
5     H   -0.010036015    1.415825173    0.628960045
6     H    0.010622052   -1.415890261   -0.628756980
7     H   -0.009439162   -1.426296574    1.905712054
8     H   -0.012434371    1.425926694   -1.906010352
9     H    0.892392126   -0.006715697    2.519107774
10    H   -0.901560681   -0.001329367    2.520828811
11    H    0.893207591    0.009559781   -2.521307663
12    H   -0.900959683   -0.002663554   -2.518134496
EXPGEOM
1     C    0.000000000    0.324100000    0.581600000
2     C    0.000000000   -0.324100000   -0.581600000
3     C    0.000000000   -0.324100000    1.937400000
4     C    0.000000000    0.324100000   -1.937400000
5     H    0.000000000    1.414600000    0.577800000
6     H    0.000000000   -1.414600000   -0.577800000
7     H    0.000000000   -1.415200000    1.857900000
```

```
  8    H     0.000000000     1.415200000    -1.857900000
  9    H     0.879700000    -0.025300000     2.520700000
 10    H    -0.879700000    -0.025300000     2.520700000
 11    H     0.879700000     0.025300000    -2.520700000
 12    H    -0.879700000     0.025300000    -2.520700000
---
H9C4, RHF, CHARGE=1, MULT=1
HF=176
  1    C     0.000000000     1.127201000     0.000000000
  2    C     0.000000000     1.145875017     1.501804755
  3    C     1.257096103     1.140818530    -0.725784773
  4    C    -1.257096103     1.140818530    -0.725784773
  5    H     0.000000000     2.213109577     1.833830650
  6    H    -0.903445226     0.666405723     1.933514474
  7    H     0.890948343     0.650973315     1.947409672
  8    H     1.613386228     2.184978600    -0.931488973
  9    H     2.115267097     0.675630524    -0.192471496
 10    H     1.224318754     0.675630524    -1.735639293
 11    H    -1.613386228     2.184978600    -0.931488973
 12    H    -1.224318754     0.675630524    -1.735639293
 13    H    -2.115267097     0.675630524    -0.192471496
---
H9C4, UHF, CHARGE=0, MULT=2
HF=4.5
  1    C     0.511974407     0.005627583    -0.004262508
  2    C     0.539481576     1.500248963    -0.003352616
  3    C     0.540988820    -0.741636035     1.289745339
  4    C     0.540812550    -0.740523190    -1.299011523
  5    H     1.591315959     1.863308648    -0.002343676
  6    H     0.038295241     1.931112170    -0.894480245
  7    H     0.036981506     1.929502627     0.888319909
  8    H     1.592415115    -0.923863714     1.603427995
  9    H     0.039814929    -0.185035710     2.108743896
 10    H     0.038839701    -1.728619149     1.216494265
 11    H     1.592463208    -0.920598776    -1.614234425
 12    H     0.040118584    -1.728216395    -1.226165075
 13    H     0.037439964    -0.183407720    -2.116041364
EXPGEOM
  1    C     0.00000    -0.26890     0.23220
  2    C     0.00000     0.86650     1.20810
  3    H     0.00000    -1.21140     0.80910
  4    H     0.92900     1.22760     1.63830
  5    H    -0.92900     1.22760     1.63830
  6    C    -1.27210    -0.26890    -0.63320
  7    C     1.27210    -0.26890    -0.63320
  8    H    -1.32030     0.63480    -1.25010
  9    H    -1.29390    -1.13630    -1.30130
 10    H    -2.17410    -0.29960    -0.01320
 11    H     1.32030     0.63480    -1.25010
 12    H     1.29390    -1.13630    -1.30130
 13    H     2.17410    -0.29960    -0.01320
---
H10C4, RHF, CHARGE=0, MULT=1
HF=-32.4, IE=11.4
  1    C     0.346092775    -0.000214552    -0.000994168
  2    H     1.467470736    -0.000045968    -0.001155683
  3    C    -0.091440593    -0.002696923     1.476733377
  4    C    -0.091942513     1.280358441    -0.738542667
  5    C    -0.091922348    -1.279694772    -0.740717411
```

```
 6      H      -1.194635286     -0.013686317      1.587008000
 7      H      -1.195188947      1.374178464     -0.796582359
 8      H      -1.195261168     -1.372018316     -0.800681492
 9      H       0.289992653      0.893020236      2.008349519
10      H       0.308634898     -0.890449745      2.008205641
11      H       0.301713601      1.296425656     -1.775512144
12      H       0.295627832      2.186774710     -0.229774365
13      H       0.293702128     -2.186966477     -0.232036351
14      H       0.303389459     -1.295302388     -1.776996745
EXPGEOM
 1      C       0.372700000      0.000000000      0.000000000
 2      H       1.471500000      0.000000000      0.000000000
 3      C      -0.095800000      0.000000000      1.462100000
 4      C      -0.095800000      1.266200000     -0.731000000
 5      C      -0.095800000     -1.266200000     -0.731000000
 6      H      -1.190900000      0.000000000      1.520300000
 7      H      -1.190900000      1.316700000     -0.760200000
 8      H      -1.190900000     -1.316700000     -0.760200000
 9      H       0.265000000      0.885200000      1.996700000
10      H       0.265000000     -0.885200000      1.996700000
11      H       0.265000000      1.286600000     -1.764900000
12      H       0.265000000      2.171800000     -0.231700000
13      H       0.265000000     -2.171800000     -0.231700000
14      H       0.265000000     -1.286600000     -1.764900000
---
H10C4, RHF, CHARGE=0, MULT=1
HF=-30.4, IE=11.2
 1      C      -0.003525682     -0.409702870      0.652186980
 2      C      -0.001715555      0.409398497     -0.651483700
 3      C      -0.003391991      0.426312222      1.934821852
 4      C       0.003802306     -0.426304791     -1.934252807
 5      H       0.885963024     -1.081424625      0.668593644
 6      H      -0.895693234     -1.077847767      0.667470668
 7      H       0.886816605      1.082352300     -0.664820344
 8      H      -0.894870745      1.076255594     -0.669546024
 9      H      -0.004577515     -0.237478290      2.824027891
10      H       0.892894525      1.075444949      2.004979511
11      H      -0.898815498      1.076728099      2.003706388
12      H       0.008449048      0.237798085     -2.823213075
13      H       0.898925610     -1.077419575     -2.000193134
14      H      -0.892900789     -1.074413052     -2.007867336
EXPGEOM
 1      C       0.000000000     -0.421600000      0.640000000
 2      C       0.000000000      0.421600000     -0.640000000
 3      C       0.000000000      0.421600000      1.918100000
 4      C       0.000000000     -0.421600000     -1.918100000
 5      H       0.876900000     -1.082900000      0.636200000
 6      H      -0.876900000     -1.082900000      0.636200000
 7      H       0.876900000      1.082900000     -0.636200000
 8      H      -0.876900000      1.082900000     -0.636200000
 9      H       0.000000000     -0.207300000      2.814000000
10      H       0.883600000      1.068100000      1.965000000
11      H      -0.883600000      1.068100000      1.965000000
12      H       0.000000000      0.207300000     -2.814000000
13      H       0.883600000     -1.068100000     -1.965000000
14      H      -0.883600000     -1.068100000     -1.965000000
---
H5C5, RHF, CHARGE=-1, MULT=1
HF=21.3
```

```
 1    C     0.000000000    0.000000000    0.000000000
 2    C     0.000000000    1.416281070    0.000000000
 3    C     0.000000000    1.852440737    1.342361425
 4    C     0.000000000    0.710559905    2.171986411
 5    C     0.000000000   -0.431320927    1.342361425
 6    H    -0.000000000   -0.636768013   -0.876435981
 7    H    -0.000000000    2.039806511   -0.858209143
 8    H    -0.000000000    2.861326092    1.670168149
 9    H    -0.000000000    0.710559905    3.232791251
10    H    -0.000000000   -1.439130988    1.673459278
EXPGEOM
 1    C     0.00000    0.00000    1.20170
 2    C     0.00000    1.14280    0.37130
 3    C     0.00000    0.70630   -0.97220
 4    C     0.00000   -1.14280    0.37130
 5    C     0.00000   -0.70630   -0.97220
 6    H     0.00000    0.00000    2.28930
 7    H     0.00000    2.17720    0.70740
 8    H     0.00000    1.34560   -1.85210
 9    H     0.00000   -2.17720    0.70740
10    H     0.00000   -1.34560   -1.85210
---
H6C5, RHF, CHARGE=0, MULT=1
HF=32.1, DIP=0.42, IE=8.57
 1    C     0.286425232    0.003635216    1.183687128
 2    C     0.286809782    0.002957203   -1.183185981
 3    C     1.241899290    0.004274116    0.000434534
 4    C    -1.000602967    0.000634654    0.738971413
 5    C    -1.000338026    0.000297919   -0.738811143
 6    H     1.898449506    0.899298420    0.000229797
 7    H     1.899310602   -0.890181419    0.000968839
 8    H     0.636933281    0.005163132    2.206958250
 9    H     0.637880354    0.003571042   -2.206233656
10    H    -1.906087728   -0.001769795    1.331513175
11    H    -1.905578173   -0.002311737   -1.331669727
EXPGEOM
 1    C     0.279400000    0.000000000    1.179100000
 2    C     0.279400000    0.000000000   -1.179100000
 3    C     1.219700000    0.000000000    0.000000000
 4    C    -0.990100000    0.000000000    0.736800000
 5    C    -0.990100000    0.000000000   -0.736800000
 6    H     1.880400000    0.878100000    0.000000000
 7    H     1.880400000   -0.878100000    0.000000000
 8    H     0.611000000    0.000000000    2.209300000
 9    H     0.611000000    0.000000000   -2.209300000
10    H    -1.886100000    0.000000000    1.345700000
11    H    -1.886100000    0.000000000   -1.345700000
---
H8C5, RHF, CHARGE=0, MULT=1
HF=46.4
 1    C     0.000000000    0.000000000    0.000000000
 2    C     0.000000000    1.513860040    0.000000000
 3    C     0.000000000    0.589565422    1.199042302
 4    C     0.000083364    0.523121760    2.659051487
 5    C     0.000074275   -1.196589941   -0.839144050
 6    H    -0.905439804    2.069029731   -0.272951793
 7    H     0.905214163    2.069332730   -0.273102024
 8    H     0.003245265    1.537788268    3.108398236
 9    H    -0.899256273   -0.013359323    3.030662053
```

```
10      H        0.896047132     -0.019046928      3.030575122
11      H        0.003072342     -0.932619722     -1.916986053
12      H        0.896091183     -1.821901642     -0.636992147
13      H       -0.899157965     -1.818650834     -0.641313107
---
H8C5, RHF, CHARGE=0, MULT=1
HF=25.3
1       C       -0.191583330      0.004215960     -0.065246070
2       C       -0.084586610      1.340053510     -0.055614150
3       C       -0.088995880      2.232841200      1.156900730
4       C       -1.277063100      3.156978360      1.197726450
5       C       -1.245602490      4.490769990      1.325206920
6       H       -0.310739830     -0.610123430      0.826172260
7       H       -0.166758300     -0.573736690     -0.988390120
8       H        0.033279300      1.875558780     -1.004426310
9       H        0.868053750      2.806614350      1.165242450
10      H       -0.086055090      1.634297130      2.098802380
11      H       -2.245693430      2.648836460      1.132387300
12      H       -2.154212550      5.090953810      1.358749040
13      H       -0.333576360      5.080385400      1.405216240
EXPGEOM
1       C        0.66460          0.00000          0.00000
2       C       -0.17830          0.00000          1.25230
3       C       -0.17830          0.00000         -1.25230
4       C       -0.17600          0.95530          2.17770
5       C       -0.17600         -0.95530         -2.17770
6       H        1.31590         -0.88340         -0.01010
7       H        1.31590          0.88340          0.01010
8       H       -0.83810         -0.85920          1.36610
9       H       -0.83810          0.85920         -1.36610
10      H       -0.80980          0.90210          3.05710
11      H       -0.80980         -0.90210         -3.05710
12      H        0.46430          1.82970          2.09630
13      H        0.46430         -1.82970         -2.09630
---
H8C5, RHF, CHARGE=0, MULT=1
HF=19.1
1       C        0.000000000      0.000000000      0.000000000
2       C        0.000000000      1.345514600      0.000000000
3       C        0.000000000      2.189769774      1.196231915
4       C        0.003819445      3.537027121      1.284894268
5       C        0.009499040      4.577143542      0.211765562
6       H       -0.000149366     -0.580853670     -0.921318303
7       H        0.000254301     -0.614048937      0.898962235
8       H       -0.000378088      1.859944780     -0.967001457
9       H       -0.003754804      1.632676919      2.143170450
10      H        0.002977403      3.973643523      2.292599143
11      H        0.005738198      4.167091377     -0.816898695
12      H       -0.883575289      5.230805677      0.311340807
13      H        0.911012576      5.219429959      0.309567703
EXPGEOM
1       C       -0.11030          2.17330         -0.59440
2       C        0.10550          1.45990          0.50970
3       C       -0.09040          0.01030          0.64820
4       C        0.17060         -0.89740         -0.29330
5       C       -0.04920         -2.37210         -0.15120
6       H       -0.49080          1.72000         -1.50140
7       H        0.07790          3.23830         -0.62290
8       H        0.43870          1.97430          1.40670
```

```
9      H    -0.45460    -0.32630     1.61560
10     H     0.58590    -0.56440    -1.24010
11     H    -0.45210    -2.62400     0.82980
12     H    -0.74660    -2.73720    -0.90980
13     H     0.88390    -2.92380    -0.29140
---
H8C5, RHF, CHARGE=0, MULT=1
HF=18.1
1    C     0.052434270    -0.055345700    -0.094126220
2    C     0.027991170     1.284863780    -0.014244280
3    C     0.011640550     2.106698030     1.201792720
4    C    -0.752534950     1.917105070     2.296740210
5    C    -0.756051780     2.770202450     3.526019310
6    H     0.089690030    -0.580995980    -1.047807580
7    H     0.045876110    -0.721954910     0.766365970
8    H     0.050281730     1.875700240    -0.938391740
9    H     0.698670320     2.960095590     1.149995160
10   H    -1.456195710     1.079201030     2.345767730
11   H    -0.053085500     3.625878160     3.465319960
12   H    -1.770529540     3.183711070     3.709653440
13   H    -0.471531350     2.165125220     4.413386680
EXPGEOM
1    C     0.00000    -2.41900     0.77540
2    C     0.00000    -1.41300    -0.10520
3    C     0.00000     0.00000     0.25020
4    C     0.00000     1.00140    -0.63810
5    C     0.00000     2.46220    -0.29830
6    H     0.00000    -2.23910     1.84670
7    H     0.00000    -3.45640     0.45920
8    H     0.00000    -1.63230    -1.17200
9    H     0.00000     0.22540     1.31630
10   H     0.00000     0.75940    -1.70090
11   H     0.00000     2.62150     0.78370
12   H     0.87940     2.96590    -0.71840
13   H    -0.87940     2.96590    -0.71840
---
H8C5, RHF, CHARGE=0, MULT=1
HF=37.3
1    C     0.000000000     1.540948950     0.781790650
2    C     0.000000000     0.000000000     0.780967810
3    C    -0.000000000     0.000000000    -0.785081410
4    C     0.000000000     1.540948950    -0.769539130
5    C     1.172695444     2.145857595     0.000000000
6    H     2.163547666     1.677016157     0.000000000
7    H     1.267589125     3.239039711    -0.000000000
8    H    -0.587946573     2.112101798     1.492772808
9    H    -0.582077113     2.103852061    -1.466764413
10   H    -0.910923226    -0.435373639     1.229707438
11   H     0.872315759    -0.483639832     1.254751539
12   H    -0.902003044    -0.428059632    -1.226933381
13   H     0.866782965    -0.475619137    -1.246550823
EXPGEOM
1    C     0.76360    -0.24590    -0.53920
2    C    -0.76360    -0.24590    -0.53920
3    C     0.78360    -0.24590     0.99570
4    C    -0.78360    -0.24590     0.99570
5    C     0.00000     0.94270    -1.04850
6    H     1.44770    -0.79560    -1.17500
7    H    -1.44770    -0.79560    -1.17500
```

```
8       H      1.28130     0.60780     1.46780
9       H     -1.28130     0.60780     1.46780
10      H      1.19320    -1.16440     1.42190
11      H     -1.19320    -1.16440     1.42190
12      H      0.00000     1.87760    -0.49070
13      H      0.00000     1.07160    -2.12640
---
H8C5, RHF, CHARGE=0, MULT=1
HF=8.3, DIP=0.2, IE=9.18
1    C    0.000000000    0.000000000    0.000000000
2    C    0.000000000    1.552102240    0.000000000
3    C    0.000000000    1.990465767    1.488745707
4    C    0.003557293    0.696246745    2.270997536
5    C    0.003671536   -0.384073830    1.462358558
6    H   -0.891769620   -0.408528386   -0.520993280
7    H    0.888827301   -0.408798244   -0.525809049
8    H   -0.888484493    1.950184766   -0.532387174
9    H    0.888201351    1.950497825   -0.532643501
10   H   -0.891865749    2.605635973    1.733615337
11   H    0.888885167    2.610338312    1.732528459
12   H    0.005730521    0.700093644    3.354809600
13   H    0.005867382   -1.425007011    1.764668013
EXPGEOM
1    C    0.00000    0.25620   -1.20710
2    C    1.23780   -0.06630   -0.32650
3    C   -1.23780   -0.06630   -0.32650
4    C    0.66610   -0.06630    1.07460
5    C   -0.66610   -0.06630    1.07460
6    H    0.00000    1.32290   -1.45170
7    H    0.00000   -0.29080   -2.15320
8    H    1.66690   -1.04930   -0.56600
9    H   -1.66690   -1.04930   -0.56600
10   H    1.29270   -0.10070    1.95970
11   H   -1.29270   -0.10070    1.95970
12   H   -2.04630    0.66130   -0.45850
13   H    2.04630    0.66130   -0.45850
---
H8C5, RHF, CHARGE=0, MULT=1
HF=18
1    C    0.000000000    0.000000000    0.000000000
2    C    0.000000000    1.352495710    0.000000000
3    C    0.000000000    2.179004055    1.221964684
4    C    0.003105723    1.782501784    2.507007137
5    C   -0.000165757    2.118139026   -1.304846300
6    H   -0.000049092   -0.615717293    0.897305597
7    H    0.000001112   -0.597248555   -0.911316811
8    H   -0.003042069    3.261376831    1.036635955
9    H    0.006349265    0.748795936    2.845713590
10   H    0.002615549    2.499116207    3.328053168
11   H   -0.898496348    2.765705764   -1.383962649
12   H    0.898741981    2.764811207   -1.385083075
13   H   -0.000783651    1.449925812   -2.190053694
EXPGEOM
1    C    0.00000   -0.58480    1.73360
2    C    0.00000    0.00000    0.52840
3    C    0.00000   -0.82590   -0.68580
4    C    0.00000   -0.37630   -1.94390
5    C    0.00000    1.50020    0.37030
6    H    0.00000   -1.66520    1.84130
```

```
7    H    0.00000    -0.00750    2.65250
8    H    0.00000    -1.89980   -0.50700
9    H    0.00000     0.68140   -2.18560
10   H    0.00000    -1.05990   -2.78620
11   H    0.00000     1.99640    1.34340
12   H    0.88090     1.83780   -0.18720
13   H   -0.88090     1.83780   -0.18720
---
H8C5, RHF, CHARGE=0, MULT=1
HF=29.1
1    C     0.000000000    0.000000000    0.000000000
2    C     0.000000000    1.333938740    0.000000000
3    C     0.000000000    2.385086406    1.096701821
4    C     0.000000000    2.376634718   -1.094337906
5    C    -0.000000000    3.485969460    0.000000000
6    H    -0.000000000   -0.591098999   -0.915765112
7    H     0.000000000   -0.595003401    0.898729862
8    H     0.895437768    2.387590359    1.745833926
9    H    -0.885376930    2.387566135    1.739553874
10   H     0.885376930    2.374154989   -1.737189959
11   H    -0.885376930    2.374154989   -1.737189959
12   H     0.893088068    4.135838499    0.000000000
13   H    -0.886122026    4.121851465    0.000000000
EXPGEOM
1    C     1.09470     0.22040   -0.39510
2    C    -1.09470     0.22040   -0.39510
3    C     0.00000    -0.09970    0.61630
4    C     0.00000     0.22040   -1.50690
5    C     0.00000    -0.59010    1.84670
6    H     1.90900    -0.50320   -0.49020
7    H    -1.90900    -0.50320   -0.49020
8    H     1.53450     1.21180   -0.24020
9    H    -1.53450     1.21180   -0.24020
10   H     0.92450    -0.80380    2.37520
11   H    -0.92450    -0.80380    2.37520
12   H     0.00000     1.06650   -2.19710
13   H     0.00000    -0.70460   -2.08790
---
H8C5, RHF, CHARGE=0, MULT=1
HF=44.3
1    C     0.000000000    0.000000000    0.000000000
2    C     0.000000000    1.304824790    0.766053980
3    C    -0.000000000    1.304824790   -0.760182190
4    C     0.760182190   -1.276606230   -0.000000000
5    C    -0.760182190   -1.276606230   -0.000000000
6    H    -0.901099670    1.616236456    1.310752567
7    H     0.892278550    1.616236456    1.310752567
8    H    -0.892278550    1.614735983   -1.297797867
9    H     0.892278550    1.614735983   -1.297797867
10   H     1.297797867   -1.586517423   -0.892278550
11   H     1.297797867   -1.586517423    0.892278550
12   H    -1.297797867   -1.586517423   -0.892278550
13   H    -1.297797867   -1.586517423    0.892278550
EXPGEOM
1    C     0.00000     0.00000    0.00000
2    C     1.26900     0.00000    0.76310
3    C     1.26900     0.00000   -0.76310
4    C    -1.26900    -0.76310    0.00000
5    C    -1.26900     0.76310    0.00000
```

```
6      H     1.57320      0.91210     1.26820
7      H     1.57320     -0.91210     1.26820
8      H     1.57320     -0.91210    -1.26820
9      H     1.57320      0.91210    -1.26820
10     H    -1.57320     -1.26820    -0.91210
11     H    -1.57320     -1.26820     0.91210
12     H    -1.57320      1.26820     0.91210
13     H    -1.57320      1.26820    -0.91210
---
H9C5, RHF, CHARGE=1, MULT=1
HF=188
1      C     0.000000000     0.000000000     0.000000000
2      C     0.000000000     1.549286780     0.000000000
3      C     0.000000000     2.008936244     1.480855373
4      C    -0.001381923    -0.460093519     1.480292256
5      C    -0.000404785     0.774316692     2.305177856
6      H    -0.888795598    -0.398960562    -0.531549527
7      H     0.889575610    -0.399200913    -0.530079448
8      H    -0.889416225     1.948584917    -0.530216856
9      H     0.889065204     1.948849390    -0.530611766
10     H    -0.885842841     2.643959834     1.732465521
11     H     0.885727888     2.644037125     1.732682282
12     H    -0.888311045    -1.093740096     1.731511627
13     H     0.883000114    -1.096813061     1.732632754
14     H    -0.000018581     0.773400777     3.398032524
---
H10C5, RHF, CHARGE=0, MULT=1
HF=-9.9
1      C     0.000000000     0.000000000     0.000000000
2      C     0.000000000     1.508515170     0.000000000
3      C     0.000000000     2.140408184     1.373959460
4      C    -0.002170543     2.279926959    -1.113620688
5      C    -0.004395994     1.856747193    -2.549746954
6      H    -0.910094098    -0.391821678     0.502039264
7      H     0.885532615    -0.393860777     0.542644427
8      H     0.023703877    -0.442644431    -1.015353900
9      H     0.893808740     1.827865814     1.953865142
10     H    -0.902137449     1.840056296     1.947438996
11     H     0.007611343     3.248822838     1.337568752
12     H    -0.003000765     3.371506855    -1.000725968
13     H     0.888621214     1.246661466    -2.800668161
14     H    -0.908306883     1.263443529    -2.801985231
15     H     0.004613501     2.741786692    -3.219361794
EXPGEOM
1      C     0.00000     -1.50860     0.48690
2      H     0.00000     -1.96510    -0.50420
3      H     0.87900     -1.87940     1.02930
4      H    -0.87900     -1.87940     1.02930
5      C     0.00000      0.27880    -2.09660
6      H    -0.87920      0.66430    -2.62770
7      H     0.87920      0.66430    -2.62770
8      H     0.00000     -0.80850    -2.19760
9      C     0.00000      0.73930    -0.66550
10     H     0.00000      1.82290    -0.54720
11     C     0.00000      0.00000     0.44820
12     C     0.00000      0.64920     1.81270
13     H     0.00000      1.74010     1.74290
14     H     0.87920      0.34400     2.39470
15     H    -0.87920      0.34400     2.39470
```

```
---
H10C5, RHF, CHARGE=0, MULT=1
HF=-6.1
1    C     0.000000000    0.000000000    0.000000000
2    C     0.000000000    1.531732860    0.000000000
3    C     0.000000000    2.172132914    1.360721535
4    C    -0.005986860    3.485389970    1.661206716
5    C    -0.014155626    4.674829591    0.755027057
6    H    -0.000117407   -0.383253165   -1.041146283
7    H    -0.896923432   -0.410884438    0.506225277
8    H     0.896498771   -0.411218011    0.506763534
9    H    -0.892520346    1.881261487   -0.572377834
10   H     0.892154577    1.881738664   -0.572759337
11   H     0.005617658    1.467940087    2.201594337
12   H    -0.005145737    3.764096671    2.723156181
13   H     0.874843553    5.311218205    0.952865607
14   H    -0.005074280    4.419549921   -0.322898410
15   H    -0.919540720    5.290864663    0.942828219
EXPGEOM
1    C    -0.41390    2.19050   -0.29170
2    H    -0.56590    2.67320    0.67970
3    H    -0.04820    2.95080   -0.98930
4    H    -1.39000    1.84430   -0.64560
5    C     0.57680    1.01840   -0.17910
6    H     0.72740    0.56790   -1.16470
7    H     1.55400    1.41220    0.13260
8    C    -0.07860   -2.07010   -0.67730
9    H    -1.05300   -2.51220   -0.91850
10   H     0.22860   -1.45180   -1.52350
11   H     0.63080   -2.90250   -0.59340
12   C     0.12920   -0.01360    0.82120
13   H     0.01400    0.36960    1.83560
14   C    -0.15110   -1.30160    0.61300
15   H    -0.47200   -1.89250    1.47040
---
H10C5, RHF, CHARGE=0, MULT=1
HF=1.3
1    C     0.060301110   -0.020621210    0.002804470
2    C    -0.015120930    1.506770170    0.007515730
3    C     0.055489610    0.755429660    1.337822790
4    C     1.222335200   -0.807103080   -0.558005100
5    C     1.214400610    0.847748180    2.303449080
6    H    -0.887878590   -0.540966130   -0.223195260
7    H    -0.957260560    1.988035300   -0.282755600
8    H     0.838048290    2.104728540   -0.333041980
9    H    -0.894261290    0.692371640    1.898976190
10   H     2.194639080   -0.282948100   -0.466272740
11   H     1.317497010   -1.786244870   -0.044365680
12   H     1.057470330   -1.008432800   -1.637596960
13   H     2.192461310    1.003164540    1.805826950
14   H     1.061497940    1.696380900    3.003042240
15   H     1.289586530   -0.079451240    2.908458840
---
H10C5, RHF, CHARGE=0, MULT=1
HF=-18.3, IE=10.5
1    C     0.000000000    0.000000000    0.000000000
2    C    -0.000000000    1.537659990    0.000000000
3    C     0.000000000    2.013267720    1.469495000
4    C     0.000000000   -0.458804360    1.469495000
```

```
 5    C     0.111661292     0.766906660     2.377247045
 6    H    -0.885834733    -0.398574618    -0.539088619
 7    H     0.888293240    -0.395328560    -0.537973130
 8    H    -0.875839016     1.935861108    -0.535054243
 9    H     0.884345483     1.924567727    -0.531822002
10    H    -0.929200968     2.578642500     1.697441857
11    H     0.839745467     2.715569784     1.656652127
12    H    -0.923443538    -1.016746582     1.699135822
13    H     0.820519794    -1.162224521     1.675020823
14    H    -0.685990727     0.766877385     3.150363802
15    H     1.074642839     0.766874792     2.931595366
EXPGEOM
 1    C     0.25380     1.02710    -0.73040
 2    C     0.12570    -0.37250     1.23820
 3    C     0.01460    -1.30800    -0.00210
 4    C    -0.15350    -0.37160    -1.23190
 5    C    -0.24050     1.03550     0.72480
 6    H     1.15450    -0.36940     1.61430
 7    H    -0.51100    -0.69540     2.06650
 8    H    -0.83180    -1.99490     0.08630
 9    H     0.90940    -1.92990    -0.09840
10    H    -1.20280    -0.35240    -1.54740
11    H     0.43090    -0.70090    -2.09570
12    H    -0.16290     1.83710    -1.33640
13    H     1.34590     1.13310    -0.75060
14    H    -1.32890     1.17320     0.74120
15    H     0.19650     1.83620     1.32890
---
H10C5, RHF, CHARGE=0, MULT=1
HF=-7.9
 1    C    -0.013478030    -0.047059430    -0.035126030
 2    C    -0.041387470     1.448980990    -0.005839640
 3    C    -0.081637660     2.226070030     1.093126870
 4    C    -0.078165690     3.730804170     1.105374400
 5    C    -1.253085970     4.360017110     1.860764700
 6    H    -1.203483370     5.466445630     1.794389450
 7    H    -1.244276400     4.090877730     2.936263230
 8    H    -2.227712070     4.043181270     1.438248450
 9    H    -0.067970090     4.135809270     0.067168660
10    H     0.878256060     4.069660850     1.571003580
11    H    -0.097146910     1.771316250     2.089793580
12    H    -0.021181980     1.900887070    -1.003915210
13    H     0.951725390    -0.406742420    -0.451394620
14    H    -0.826194300    -0.435245170    -0.684955110
15    H    -0.138427000    -0.508656760     0.965202230
EXPGEOM
 1    H     0.00000    -0.73670    -1.23050
 2    C     0.00000    -0.98040    -0.16940
 3    H     0.00000    -0.26650     1.79000
 4    C     0.00000     0.00000     0.73190
 5    H     0.87960    -2.94920    -0.26610
 6    H    -0.87960    -2.94920    -0.26610
 7    H     0.00000    -2.61710     1.23760
 8    C     0.00000    -2.44820     0.15670
 9    H    -0.87150     1.92710     0.96230
10    H     0.87150     1.92710     0.96230
11    C     0.00000     1.48270     0.46080
12    H     0.00000     2.98030    -1.11220
13    H     0.88450     1.50820    -1.53420
```

```
14      H     -0.88450      1.50820     -1.53420
15      C      0.00000      1.89050     -1.01480
---
H12C5, RHF, CHARGE=0, MULT=1
HF=-35.1, IE=10.3
1    C     0.000000000     0.000000000     0.000000000
2    C     0.000000000     1.531268210     0.000000000
3    C     0.000000000     2.175263905     1.399812647
4    C     0.001836653     3.715853486     1.386803666
5    C     0.000458632     4.367765344     2.772333604
6    H     0.004126047     5.473063118     2.676225664
7    H     0.894102562     4.082265341     3.363473841
8    H    -0.898101958     4.087844423     3.358647241
9    H    -0.887726899     4.083242997     0.824366966
10   H     0.893718877     4.081243671     0.826781668
11   H     0.890293571     1.815071770     1.965382233
12   H    -0.892375622     1.817912607     1.963781602
13   H     0.890371376     1.883139646    -0.571017189
14   H    -0.890908628     1.882416974    -0.570490799
15   H     0.001647154    -0.383464502    -1.041086531
16   H    -0.897023623    -0.412300933     0.504847410
17   H     0.895166590    -0.412592998     0.507900906
EXPGEOM
1    C     0.31580     0.00000     0.00000
2    C    -0.52400     0.00000     1.28330
3    C    -0.52400     0.00000    -1.28330
4    C     0.32460     0.00000     2.55990
5    C     0.32460     0.00000    -2.55990
6    H     0.97740     0.87740     0.00000
7    H     0.97740    -0.87740     0.00000
8    H    -1.18430     0.87710     1.28280
9    H    -1.18430    -0.87710     1.28280
10   H    -1.18430    -0.87710    -1.28280
11   H    -1.18430     0.87710    -1.28280
12   H    -0.30180     0.00000     3.45770
13   H    -0.30180     0.00000    -3.45770
14   H     0.97090    -0.88360     2.60460
15   H     0.97090     0.88360     2.60460
16   H     0.97090     0.88360    -2.60460
17   H     0.97090    -0.88360    -2.60460
---
H12C5, RHF, CHARGE=0, MULT=1
HF=-40.3, IE=11.3
1    C    -0.000093793    -0.000093770     0.000093795
2    C     0.897429120     0.897429214     0.897727694
3    C     0.897429241    -0.897727609    -0.897429151
4    C    -0.897560824    -0.897560841     0.897560842
5    C    -0.897727709     0.897429135    -0.897429221
6    H     1.554412851     1.554414850     0.291978007
7    H     0.291839373     1.554497525     1.554835492
8    H     1.554498392     0.291838408     1.554832460
9    H     0.291838572    -1.554832463    -1.554498451
10   H     1.554414826    -0.291977811    -1.554412831
11   H     1.554497596    -1.554835393    -0.291839428
12   H    -1.554891868    -1.554895553     0.292527132
13   H    -0.292527092    -1.554891893     1.554895545
14   H    -1.554895572    -0.292527266     1.554891956
15   H    -0.291978008     1.554412830    -1.554414882
16   H    -1.554835485     0.291839343    -1.554497506
```

```
17      H     -1.554832497     1.554498451    -0.291838501
EXPGEOM
1    C      0.000000000     0.000000000     0.000000000
2    C      0.889000000     0.889000000     0.889000000
3    C      0.889000000    -0.889000000    -0.889000000
4    C     -0.889000000    -0.889000000     0.889000000
5    C     -0.889000000     0.889000000    -0.889000000
6    H      1.534500000     1.534500000     0.283000000
7    H      0.283000000     1.534500000     1.534500000
8    H      1.534500000     0.283000000     1.534500000
9    H      0.283000000    -1.534500000    -1.534500000
10     H      1.534500000    -0.283000000    -1.534500000
11     H      1.534500000    -1.534500000    -0.283000000
12     H     -1.534500000    -1.534500000     0.283000000
13     H     -0.283000000    -1.534500000     1.534500000
14     H     -1.534500000    -0.283000000     1.534500000
15     H     -0.283000000     1.534500000    -1.534500000
16     H     -1.534500000     0.283000000    -1.534500000
17     H     -1.534500000     1.534500000    -0.283000000
---
H6C6, RHF, CHARGE=0, MULT=1
HF=19.8, IE=9.25
1    C     -0.010000000    -0.002684144     1.406966068
2    C     -0.010000000     1.216466186     0.705416779
3    C     -0.010000000     1.219401661    -0.701255473
4    C     -0.010000000     0.002684142    -1.406967792
5    C     -0.010000000    -1.216466184    -0.705416826
6    C     -0.010000000    -1.219400590     0.701255369
7    H     -0.010000000    -0.004584588     2.497306245
8    H     -0.010000000     2.159773094     1.252267640
9    H     -0.010000000     2.164878035    -1.244287122
10     H     -0.010000000     0.004584396    -2.497267559
11     H     -0.010000000    -2.159773075    -1.252267688
12     H     -0.010000000    -2.164887851     1.244288887
EXPGEOM
1    C      0.000000000     0.000000000     1.394400000
2    C      0.000000000     1.207600000     0.697200000
3    C      0.000000000     1.207600000    -0.697200000
4    C      0.000000000     0.000000000    -1.394400000
5    C      0.000000000    -1.207600000    -0.697200000
6    C      0.000000000    -1.207600000     0.697200000
7    H      0.000000000     0.000000000     2.479800000
8    H      0.000000000     2.147600000     1.239900000
9    H      0.000000000     2.147600000    -1.239900000
10     H      0.000000000     0.000000000    -2.479800000
11     H      0.000000000    -2.147600000    -1.239900000
12     H      0.000000000    -2.147600000     1.239900000
---
H6C6, RHF, CHARGE=0, MULT=1
HF=47.5, DIP=0.42
1    C      0.790651442    -0.010000000    -0.001587137
2    C      2.135538131    -0.010000000    -0.003440916
3    C     -0.122040359    -0.010000000     1.177733955
4    C     -0.123882197    -0.010000000    -1.179527219
5    C     -1.415458071    -0.010000000     0.738159190
6    C     -1.416714390    -0.010000000    -0.738378434
7    H      2.733211617    -0.010000000     0.907234095
8    H      2.730202666    -0.010000000    -0.916182081
9    H      0.228156367    -0.010000000     2.201065592
```

```
10      H      0.224969040    -0.010000000    -2.203361464
11      H     -2.317340209    -0.010000000     1.335497153
12      H     -2.319453051    -0.010000000    -1.334372582
EXPGEOM
1       C      0.763800000     0.000000000     0.000000000
2       C      2.103100000     0.000000000     0.000000000
3       C     -0.128300000     0.000000000     1.176700000
4       C     -0.128300000     0.000000000    -1.176700000
5       C     -1.404200000     0.000000000     0.739200000
6       C     -1.404200000     0.000000000    -0.739200000
7       H      2.673100000     0.000000000     0.923900000
8       H      2.673100000     0.000000000    -0.923900000
9       H      0.220600000     0.000000000     2.200400000
10      H      0.220600000     0.000000000    -2.200400000
11      H     -2.299200000     0.000000000     1.348500000
12      H     -2.299200000     0.000000000    -1.348500000
---
H8C6, RHF, CHARGE=0, MULT=1
HF=40.1
1       H      0.000000000     0.000000000     0.000000000
2       C      0.000000000     1.088419340     0.000000000
3       C      0.000000000     1.848228065     1.109991754
4       C     -0.000365962     1.346977448     2.485871602
5       C     -0.000324352     2.123457225     3.597197375
6       C     -0.000935310     1.621598861     4.972667757
7       C     -0.001001824     2.378865146     6.084545397
8       H      0.000032325     1.522359512    -0.999080147
9       H      0.000319458     2.938812402     0.997744225
10      H     -0.000735386     0.255582981     2.587010782
11      H      0.000238662     3.214923415     3.497330908
12      H     -0.001388904     0.530801180     5.085292238
13      H     -0.000509467     3.467422473     6.085408278
14      H     -0.001609153     1.943806194     7.082956171
EXPGEOM
1       C      0.00000    -0.00180     0.67310
2       C      0.00000     0.00180    -0.67310
3       C      0.00000     1.19480     1.49650
4       C      0.00000    -1.19480    -1.49650
5       C      0.00000     1.19480     2.83490
6       C      0.00000    -1.19480    -2.83490
7       H      0.00000     0.95190    -1.20590
8       H      0.00000    -0.95190     1.20590
9       H      0.00000     2.14140     0.95840
10      H      0.00000    -2.14140    -0.95840
11      H      0.00000     2.11600     3.40700
12      H      0.00000    -2.11600    -3.40700
13      H      0.00000    -0.27030    -3.40510
14      H      0.00000     0.27030     3.40510
---
H8C6, RHF, CHARGE=0, MULT=1
HF=41.1
1       H      0.000000000     0.000000000     0.000000000
2       C      0.000000000     1.088442430     0.000000000
3       C      0.000000000     1.851016148     1.108754857
4       C      0.000056681     1.345109160     2.481608715
5       C     -0.001645838     2.034799488     3.648022906
6       C     -0.004271362     3.481258371     3.864397145
7       C     -0.005975552     4.086864941     5.066353631
8       H      0.000020916     1.521473648    -0.999594635
```

```
  9     H      0.000020661     2.939601133     0.988140313
 10     H      0.001772449     0.249320654     2.560383298
 11     H     -0.001081417     1.435422895     4.568588313
 12     H     -0.004920405     4.113987769     2.970269570
 13     H     -0.005557406     3.560491344     6.019240428
 14     H     -0.007971202     5.171069247     5.170011638
EXPGEOM
 1    C      0.00000       0.67440       0.75590
 2    C      0.00000      -0.67440       0.75590
 3    C      0.00000       1.55480      -0.40220
 4    C      0.00000      -1.55480      -0.40220
 5    C      0.00000       2.89050      -0.31640
 6    C      0.00000      -2.89050      -0.31640
 7    H      0.00000       1.18620       1.71650
 8    H      0.00000      -1.18620       1.71650
 9    H      0.00000       1.08930      -1.38430
 10    H      0.00000      -1.08930      -1.38430
 11    H      0.00000       3.40030       0.64300
 12    H      0.00000      -3.40030       0.64300
 13    H      0.00000       3.52100      -1.19880
 14    H      0.00000      -3.52100      -1.19880
---
H8C6, RHF, CHARGE=0, MULT=1
HF=25.4
 1    C      0.113922770    -0.002972070    -0.038442410
 2    C     -0.063670280     1.449860830    -0.038216620
 3    C     -0.168158830     2.135846760     1.122139310
 4    C      0.219514870    -0.689299400     1.121617370
 5    C     -0.128187060     1.480892560     2.476459180
 6    C      0.180915050    -0.034466640     2.475966410
 7    H      0.156575110    -0.502822590    -1.006874840
 8    H     -0.107379610     1.949930080    -1.006793150
 9    H     -0.299589740     3.219802680     1.129277020
 10    H     -1.111302680     1.653259180     2.976024240
 11    H      0.628120670     2.005558750     3.105647240
 12    H     -0.574766760    -0.559220090     3.105934720
 13    H      1.164166010    -0.207078670     2.974677330
 14    H      0.350059480    -1.773025870     1.128539700
EXPGEOM
 1    C      0.10930      -0.05050       1.42450
 2    C      1.25770       0.05050       0.72950
 3    C      1.25770      -0.05050      -0.72950
 4    C      0.10930       0.05050      -1.42450
 5    C     -1.19110       0.30980      -0.70350
 6    C     -1.19110      -0.30980       0.70350
 7    H      0.11100      -0.01900       2.51240
 8    H      2.20740       0.18470       1.24400
 9    H      2.20740      -0.18470      -1.24400
 10    H      0.11100       0.01900      -2.51240
 11    H     -2.04250      -0.06660      -1.28220
 12    H     -1.33130       1.40280      -0.62420
 13    H     -2.04250       0.06660       1.28220
 14    H     -1.33130      -1.40280       0.62420
---
H8C6, RHF, CHARGE=0, MULT=1
HF=25
 1    C      0.000000000     0.000000000     0.000000000
 2    C      0.000000000     1.506109110     0.000000000
 3    C      0.000000000     2.248173536     1.124053403
```

```
4    C    0.001356596    1.653879180    2.508232763
5    C    0.004318143    0.147914864    2.508324505
6    C    0.004400388   -0.593668292    1.384053849
7    H    0.890597712   -0.368319573   -0.562213839
8    H   -0.894863071   -0.367715164   -0.555958127
9    H   -0.000320263    1.975122768   -0.986236598
10   H   -0.000801570    3.339212604    1.081356671
11   H   -0.891805849    2.020917797    3.067382282
12   H    0.893504700    2.024790620    3.066597127
13   H    0.006933693   -0.321700488    3.494215184
14   H    0.007354846   -1.684745168    1.427062444
EXPGEOM
1    C    0.00000    1.50140    0.00000
2    C    0.00000   -1.50140    0.00000
3    C    0.00000   -0.66570    1.25320
4    C    0.00000    0.66570    1.25320
5    C    0.00000   -0.66570   -1.25320
6    C    0.00000    0.66570   -1.25320
7    H    0.00000    1.20750   -2.19640
8    H    0.00000   -1.20750   -2.19640
9    H    0.00000    1.20750    2.19640
10   H    0.00000   -1.20750    2.19640
11   H    0.87030    2.17560    0.00000
12   H    0.87030   -2.17560    0.00000
13   H   -0.87030    2.17560    0.00000
14   H   -0.87030   -2.17560    0.00000
---
H10C6, RHF, CHARGE=0, MULT=1
HF=19.8
1    C    0.000000000    0.000000000    0.000000000
2    C    0.000000000    1.365606730    0.000000000
3    C    0.000000000   -0.097921358    1.526995469
4    C   -0.000338394    1.463659792    1.526970566
5    H    0.894337822   -0.569201427    1.971374908
6    H   -0.894362929   -0.569394942    1.971097201
7    H    0.893663969    1.935400386    1.971584465
8    H   -0.895135236    1.934589859    1.970861408
9    C   -0.000243922    2.444439408   -1.012374481
10   C   -0.000205776   -1.079223832   -1.011886254
11   H    0.012945494    2.045471257   -2.047364753
12   H   -0.904168269    3.082132923   -0.909730075
13   H    0.889610962    3.098785359   -0.893278114
14   H    0.009606168   -0.680745061   -2.047126790
15   H    0.891341879   -1.731555695   -0.894613123
16   H   -0.902316406   -1.719037623   -0.906845501
---
H10C6, RHF, CHARGE=0, MULT=1
HF=20.1
1    C   -0.112264270   -0.002553700    0.004872200
2    C   -0.062957500    1.336631440    0.030141890
3    C   -0.100678580    2.214321290    1.251889030
4    C   -1.420364220    3.002766710    1.399600890
5    C   -1.456782430    3.883568520    2.619222960
6    C   -1.406197590    5.222783490    2.640645890
7    H   -1.454011420    5.793058330    3.567862600
8    H   -1.314773880    5.848356930    1.754051090
9    H   -1.550653140    3.336185810    3.564025310
10   H   -2.274744620    2.288316490    1.462984570
11   H   -1.595662310    3.603893780    0.477945120
```

```
12      H       0.754069880     2.928690170     1.191331690
13      H       0.073056490     1.610694000     2.172162990
14      H       0.030543110     1.886289600    -0.913372670
15      H      -0.203114100    -0.631312790     0.889253310
16      H      -0.063549000    -0.569839560    -0.924136360
EXPGEOM
1       C      -0.29540        -0.42280         0.64690
2       C      -0.29540         0.42280        -0.64690
3       C      -0.33540         0.42280         1.88980
4       C      -0.33540        -0.42280        -1.88980
5       C       0.59610         0.44060         2.83920
6       C       0.59610        -0.44060        -2.83920
7       H      -1.17040        -1.08780         0.63020
8       H       0.59020        -1.06750         0.66240
9       H      -1.17040         1.08780        -0.63020
10      H       0.59020         1.06750        -0.66240
11      H      -1.20570         1.07290         1.98550
12      H      -1.20570        -1.07290        -1.98550
13      H       0.51290         1.07980         3.71240
14      H       1.48090        -0.18820         2.78410
15      H       0.51290        -1.07980        -3.71240
16      H       1.48090         0.18820        -2.78410
---
H10C6, RHF, CHARGE=0, MULT=1
HF=-1.1
1       C       0.000000000     0.000000000     0.000000000
2       C       0.000000000     1.510080140     0.000000000
3       C       0.000000000     1.903170997     1.498423366
4       C      -0.001954556     0.574759997     2.298410574
5       C      -0.000398693    -0.528771842     1.248867452
6       C      -0.000360463    -1.973699081     1.627325785
7       H       0.000554122    -0.546137862    -0.937411971
8       H      -0.890504151     1.911597091    -0.529192542
9       H       0.890028582     1.911865687    -0.529747738
10      H      -0.888329389     2.519187825     1.749559283
11      H       0.888758708     2.518123775     1.750360366
12      H      -0.894403522     0.503097476     2.956047898
13      H       0.887190327     0.502541214     2.960349224
14      H       0.892787157    -2.225336802     2.237679199
15      H      -0.901211249    -2.228597645     2.224776481
16      H       0.007583087    -2.637924596     0.738729960
EXPGEOM
1       C      -0.00050        -0.75690         0.07600
2       H      -0.06930        -0.33550         2.17980
3       C      -0.02190         0.02590         1.15700
4       H       0.85800        -2.65860        -0.45270
5       H      -0.89730        -2.61110        -0.55040
6       H      -0.10600        -2.68460         1.03610
7       C      -0.03930        -2.25440         0.03300
8       H       1.13350        -0.03960        -1.61540
9       H      -0.57010        -0.25740        -1.97680
10      C       0.12400         0.06800        -1.19260
11      H      -1.20090         1.76690        -0.91270
12      H       0.45830         2.25920        -1.23000
13      C      -0.15440         1.51550        -0.71390
14      H       1.07860         1.88370         1.08380
15      H      -0.63910         2.11430         1.36290
16      C       0.08450         1.49570         0.81980
---
```

```
H10C6, RHF, CHARGE=0, MULT=1
HF=2.3
1    C      0.000000000      0.000000000      0.000000000
2    C      0.000000000      1.509894660      0.000000000
3    C      0.000000000      1.903759897      1.500471703
4    C      0.062978168      0.572605849      2.317992074
5    C      0.027298107     -0.508442547      1.248898845
6    C     -1.013194313      0.422072537      3.401248876
7    H     -0.013807516     -0.558551071     -0.928885596
8    H     -0.888801942      1.914957012     -0.528691586
9    H      0.891508719      1.911072174     -0.527644375
10   H     -0.902109422      2.502051206      1.744043816
11   H      0.871396201      2.547007394      1.743006311
12   H      1.054675191      0.518616844      2.832813111
13   H     -2.038498166      0.442279047      2.979862722
14   H     -0.893227160     -0.537673346      3.944527046
15   H     -0.938848381      1.237984288      4.149096154
16   H      0.045382255     -1.559056239      1.516839645
EXPGEOM
1    H      0.11410    -0.29810     2.21410
2    C      0.09890     0.10210     1.20490
3    H     -0.38710     2.16200     1.58600
4    C     -0.16010     1.36810     0.88200
5    H      0.87410     2.16590    -0.83640
6    H     -0.87260     2.20030    -1.01270
7    C     -0.04460     1.60790    -0.60760
8    H      0.65000     0.07210    -2.05220
9    H     -1.01210    -0.12050    -1.50160
10   C     -0.00450     0.16770    -1.18140
11   H      1.51370    -0.91330    -0.03790
12   C      0.42510    -0.75250    -0.00500
13   H      0.07720    -2.73900     0.83770
14   H     -0.01030    -2.68150    -0.92770
15   H     -1.33870    -2.02840     0.04600
16   C     -0.24950    -2.12980    -0.01210
---
H10C6, RHF, CHARGE=0, MULT=1
HF=3.5
1    C      0.000000000      0.000000000      0.000000000
2    C      0.000000000      1.510899870      0.000000000
3    C      0.000000000      1.918878363      1.509045863
4    C     -0.121155596      0.575415623      2.299730983
5    C     -0.066041639     -0.507231396      1.247487135
6    C     -1.044097896      2.977174973      1.885139771
7    H      0.046748584     -0.556644733     -0.928844767
8    H     -0.882076261      1.900992956     -0.551131202
9    H      0.895590216      1.913523197     -0.519688460
10   H      1.000170440      2.360625135      1.741032444
11   H      0.708138473      0.470804884      3.031502445
12   H     -2.080330336      2.644856148      1.672929611
13   H     -0.984559094      3.224572875      2.965007212
14   H     -0.871940959      3.917292099      1.321483196
15   H     -0.082970369     -1.554846569      1.525364923
16   H     -1.063530133      0.504194484      2.883456964
EXPGEOM
1    H     -0.14210     2.43840     1.29370
2    C     -0.06650     1.55640     0.66630
3    H     -0.14120     2.43850    -1.29380
4    C     -0.06620     1.55650    -0.66630
```

```
5    H    -0.98440    -0.14360     1.63030
6    H     0.70860     0.05190     2.05790
7    C    -0.00360     0.15480     1.23110
8    H     0.70780     0.05150    -2.05830
9    H    -0.98500    -0.14320    -1.62960
10   C    -0.00370     0.15470    -1.23110
11   H     1.48660    -0.78320     0.00010
12   C     0.39240    -0.70790    -0.00000
13   H     0.12910    -2.68510     0.88400
14   H     0.12890    -2.68500    -0.88420
15   H    -1.28530    -2.09360     0.00020
16   C    -0.18960    -2.12220    -0.00000
---
H10C6, RHF, CHARGE=0, MULT=1
HF=30.9
1    C     0.000000000     0.000000000     0.000000000
2    C     0.000000000     1.506412730     0.000000000
3    H     0.000000000     1.953413846     1.005744919
4    H    -0.000000000    -0.437279165    -1.005837626
5    C    -0.773346210     2.243172874    -1.090754080
6    C     0.754174250     2.243172874    -1.090754080
7    C     0.754174250    -0.815833357     1.062876881
8    C    -0.754174250    -0.815833357     1.062876881
9    H    -1.320097195     1.675588510    -1.854378569
10   H    -1.320097195     3.156612336    -0.854007039
11   H     1.294563428     3.153040669    -0.848719279
12   H     1.294563428     1.678903903    -1.844438879
13   H     1.294563428    -1.709376584     0.766193969
14   H     1.294563428    -0.298235084     1.849345111
15   H    -1.294563428    -0.298235084     1.849345111
16   H    -1.294563428    -1.709376584     0.766193969
---
H10C6, RHF, CHARGE=0, MULT=1
HF=-1.1, IE=10.3
1    C     1.311867126    -0.094424137    -0.666319987
2    C     1.311867466     0.094425231     0.666319699
3    C     0.062974368     0.227150170     1.492860790
4    C     0.062975622    -0.227158189    -1.492861134
5    C    -1.225676412    -0.175321857     0.749429006
6    C    -1.225663187     0.175301004    -0.749426464
7    H     2.249459917    -0.175050204    -1.221345798
8    H     2.249460252     0.175051906     1.221345568
9    H    -0.019426680     1.282889979     1.844181393
10   H    -0.019422384    -1.282889594    -1.844178717
11   H     0.161333221    -0.394111011     2.412730695
12   H     0.161331576     0.394111111    -2.412732724
13   H    -1.390266222    -1.270790755     0.869926134
14   H    -1.390719798     1.270815367    -0.869974394
15   H    -2.096273171     0.317857727     1.237767002
16   H    -2.096272210    -0.317818562    -1.237764388
EXPGEOM
1    C     1.305100000    -0.118700000    -0.656000000
2    C     1.305100000     0.118700000     0.656000000
3    C     0.048000000     0.251700000     1.481700000
4    C     0.048000000    -0.251700000    -1.481700000
5    C    -1.191900000    -0.251700000     0.725100000
6    C    -1.191900000     0.251700000    -0.725100000
7    H     2.250400000    -0.224100000    -1.184400000
8    H     2.250400000     0.224100000     1.184400000
```

```
 9    H    -0.087800000     1.304400000     1.770100000
10    H    -0.087800000    -1.304400000    -1.770100000
11    H     0.166200000    -0.297100000     2.424300000
12    H     0.166200000     0.297100000    -2.424300000
13    H    -1.190600000    -1.349200000     0.722700000
14    H    -1.190600000     1.349200000    -0.722700000
15    H    -2.104700000     0.061400000     1.243500000
16    H    -2.104700000    -0.061400000    -1.243500000
---
H11C6, RHF, CHARGE=1, MULT=1
HF=177
 1    C     0.000000000     0.000000000     0.000000000
 2    C     0.000000000     1.544051710     0.000000000
 3    C     0.000000000     2.176596183     1.402647804
 4    C     1.036950275     1.572078692     2.365146093
 5    C     1.037543577     0.028556666     2.404129619
 6    C     0.687774268    -0.650713260     1.136941321
 7    H     0.424370668    -0.386697040    -0.960756665
 8    H    -1.057555160    -0.386840575     0.016237608
 9    H    -0.893326958     1.896135385    -0.562112483
10    H     0.886233808     1.897235820    -0.574550486
11    H     0.201993012     3.266774759     1.301258970
12    H    -1.017190185     2.094881774     1.849237566
13    H     0.848761861     1.950463222     3.394415097
14    H     2.058392294     1.919461890     2.089264076
15    H     2.025770976    -0.349208098     2.768845901
16    H     0.295746068    -0.339657552     3.167354370
17    H     0.931254082    -1.719135034     1.044050771
---
H12C6, RHF, CHARGE=0, MULT=1
HF=-10.1
 1    C     0.000000000     0.000000000     0.000000000
 2    C     0.000000000     1.531249540     0.000000000
 3    C     0.000000000     2.176453434     1.399335227
 4    C    -0.000603931     3.718531249     1.383801386
 5    C     0.001882486     4.359733374     2.746460976
 6    C     0.005234912     5.680967843     2.980680184
 7    H    -0.893996302    -0.412244460     0.510016382
 8    H     0.897521577    -0.412810566     0.503367237
 9    H    -0.004076377    -0.383661233    -1.040886619
10    H    -0.892536755     1.881529791    -0.568768321
11    H     0.889680440     1.883275796    -0.571625757
12    H    -0.893563991     1.819812706     1.961875879
13    H     0.891687917     1.819003596     1.963836745
14    H    -0.895538665     4.077298593     0.821458151
15    H     0.889606934     4.078701655     0.815397046
16    H     0.000526996     3.675310428     3.601240381
17    H     0.006876901     6.441216313     2.201268552
18    H     0.006484403     6.094277008     3.988401630
EXPGEOM
 1    C    -0.43700    3.08550   -0.16870
 2    C     0.38420    2.04130   -0.21030
 3    C     0.22580    0.77650    0.58300
 4    C     0.08440   -0.47380   -0.30260
 5    C    -0.02670   -1.77310    0.50170
 6    C    -0.16860   -3.01580   -0.38140
 7    H    -1.31020    3.09700    0.47770
 8    H    -0.26810    3.96990   -0.77380
 9    H     1.24650    2.07490   -0.87660
```

```
10      H        1.10200      0.64440      1.23510
11      H       -0.64520      0.86270      1.24430
12      H       -0.79880     -0.36010     -0.94450
13      H        0.94670     -0.53770     -0.98050
14      H        0.85820     -1.87780      1.14340
15      H       -0.88700     -1.70600      1.18070
16      H       -0.24710     -3.92650      0.22080
17      H       -1.06350     -2.95450     -1.01060
18      H        0.69460     -3.12980     -1.04650
---
H12C6, RHF, CHARGE=0, MULT=1
HF=-15.7
1       C        0.000000000     0.000000000     0.000000000
2       C        0.000000000     1.543183200     0.000000000
3       C        0.000000000     2.130014523     1.411911082
4       C       -1.002249727     1.930607415     2.292240289
5       C       -1.115343234     2.141846718    -0.882941656
6       C        1.206059331     2.957791312     1.787157867
7       H        0.804440012    -0.396602285     0.652366652
8       H       -0.959755205    -0.427130688     0.352560315
9       H        0.183174167    -0.390408508    -1.022356753
10      H        0.962280464     1.841128298    -0.493485425
11      H       -1.896729401     1.346394138     2.081888038
12      H       -1.007233312     2.347029930     3.299292168
13      H       -2.129767400     1.851936242    -0.543811699
14      H       -1.066278137     3.249816738    -0.886401994
15      H       -1.006680243     1.805305401    -1.934618592
16      H        1.383662127     3.770636378     1.052257747
17      H        2.119813042     2.327754029     1.826854887
18      H        1.097191337     3.438945428     2.780895774
EXPGEOM
1       C        0.00000     -0.95040     -1.81860
2       C        0.00000     -0.06490     -0.82140
3       C        0.00000     -0.51950      0.63030
4       C        0.00000      1.42160     -1.09440
5       C        1.26760     -0.06490      1.37860
6       C       -1.26760     -0.06490      1.37860
7       H        0.00000     -0.64150     -2.85990
8       H        0.00000     -2.02080     -1.63470
9       H        0.00000     -1.61600      0.61700
10      H        0.00000      1.62020     -2.16900
11      H       -0.87950      1.91140     -0.66090
12      H        0.87950      1.91140     -0.66090
13      H        1.28830     -0.49000      2.38760
14      H        2.17310     -0.38790      0.85570
15      H        1.30630      1.02470      1.48140
16      H       -1.28830     -0.49000      2.38760
17      H       -2.17310     -0.38790      0.85570
18      H       -1.30630      1.02470      1.48140
---
H12C6, RHF, CHARGE=0, MULT=1
HF=-16.8
1       C       -0.000032134    -0.000056672    -0.000084610
2       C        1.511708436     0.000044664     0.000151724
3       C        2.248310443     1.152191844     0.000170954
4       C        3.759516817     1.151860821     0.044759797
5       C        2.146754309    -1.372090895     0.000160463
6       C        1.614850822     2.524255078    -0.043370493
7       H       -0.393706958     0.265865020    -1.004357377
```

```
 8     H    -0.423951976   -0.990798115    0.264184852
 9     H    -0.418128249    0.718162938    0.735611745
10     H     4.183158625    0.862529290   -0.940468860
11     H     4.176547040    2.148308322    0.298051665
12     H     4.154166499    0.450970951    0.809591828
13     H     2.555314998   -1.617386657    1.003698653
14     H     1.425259750   -2.173093059   -0.262370299
15     H     2.972806270   -1.450118901   -0.737104434
16     H     1.184749211    2.790737196    0.945691269
17     H     2.342800401    3.319102169   -0.306926980
18     H     0.805229351    2.587361302   -0.800028541
EXPGEOM
 1    C     0.67490    0.00000    0.00000
 2    C    -0.67490    0.00000    0.00000
 3    C     1.52460   -0.01190    1.24960
 4    C     1.52460    0.01190   -1.24960
 5    C    -1.52460   -0.01190   -1.24960
 6    C    -1.52460    0.01190    1.24960
 7    H     0.95720   -0.11430    2.17540
 8    H     0.95720    0.11430   -2.17540
 9    H    -0.95720   -0.11430   -2.17540
10    H    -0.95720    0.11430    2.17540
11    H     2.24490   -0.84150    1.20810
12    H     2.12040    0.91000    1.31920
13    H     2.24490    0.84150   -1.20810
14    H     2.12040   -0.91000   -1.31920
15    H    -2.24490   -0.84150   -1.20810
16    H    -2.12040    0.91000   -1.31920
17    H    -2.24490    0.84150    1.20810
18    H    -2.12040   -0.91000    1.31920
 ---
H12C6, RHF, CHARGE=0, MULT=1
HF=-14.8
 1    C     0.000000000    0.000000000    0.000000000
 2    C     0.000000000    1.496991620    0.000000000
 3    C     0.000000000    2.349039683    1.058301670
 4    C    -0.000179272    1.838974046    2.488021657
 5    C    -0.000835558    2.837154685    3.650040725
 6    C    -0.000016174    3.837908281    0.802599321
 7    H     0.892381625   -0.412489164    0.516226308
 8    H    -0.903424242   -0.412111918    0.496984751
 9    H     0.011192624   -0.392411878   -1.038092245
10    H    -0.000028470    1.912928277   -1.015243039
11    H    -0.895489801    1.183415589    2.621983471
12    H     0.893815584    1.181927993    2.621626530
13    H     0.891615139    3.494413159    3.640399943
14    H    -0.903914225    3.479762979    3.651635790
15    H     0.010014766    2.285296919    4.613438271
16    H    -0.897872340    4.320970088    1.241482608
17    H     0.902651805    4.320164408    1.232572624
18    H    -0.004782520    4.084198785   -0.279593477
EXPGEOM
 1    C    -0.44350    1.64270   -1.32660
 2    H    -0.98170    2.30420   -0.64020
 3    H    -0.01860    2.26350   -2.12210
 4    H    -1.17610    0.96640   -1.77750
 5    C     0.65860    0.85050   -0.59880
 6    H     1.40350    1.55650   -0.20990
 7    H     1.18450    0.21970   -1.32150
```

```
 8    C    -0.07550    0.77500    1.84390
 9    H    -0.51370    0.14130    2.61950
10    H     0.87670    1.16910    2.22200
11    H    -0.73550    1.63970    1.70110
12    C    -0.02480   -2.19130   -0.72550
13    H     0.65270   -3.02630   -0.50690
14    H    -0.99370   -2.63590   -0.98460
15    H     0.35110   -1.67680   -1.61250
16    C     0.12950    0.01730    0.55210
17    C    -0.16100   -1.28510    0.46690
18    H    -0.54860   -1.77630    1.35950
---
H12C6, RHF, CHARGE=0, MULT=1
HF=-11.8
 1    C     0.041197480   -0.054114120   -0.011789050
 2    C     0.003193560    1.483835090   -0.106920260
 3    C    -0.122974890    2.186246620    1.273150210
 4    C    -1.380998020    1.888473540    2.044388860
 5    C    -1.456087800    1.604790300    3.352881570
 6    C     1.205569010    2.031213580   -0.905860220
 7    H     0.918442070   -0.416831800    0.561140510
 8    H    -0.868877300   -0.454492710    0.478744820
 9    H     0.088593570   -0.510373750   -1.022375170
10    H    -0.912247880    1.747673310   -0.696785840
11    H     0.775026630    1.948827440    1.888785200
12    H    -0.098437200    3.291692020    1.112587270
13    H    -2.307270130    1.956311660    1.463206750
14    H    -0.595327070    1.524076410    4.014803440
15    H    -2.406506250    1.435087340    3.857961490
16    H     1.126386240    3.128053740   -1.051070720
17    H     2.171635170    1.827855050   -0.400649310
18    H     1.253212860    1.574064640   -1.915395330
---
H12C6, RHF, CHARGE=0, MULT=1
HF=-29.5, IE=10.3
 1    C     0.185055423   -0.000392903   -1.491770504
 2    C     0.193447289   -1.290345087    0.745201393
 3    C     0.190888591    1.291460661    0.744513000
 4    C    -0.185054508    0.000392122    1.491771402
 5    C    -0.190887279   -1.291460903   -0.744514276
 6    C    -0.193448098    1.290343858   -0.745200522
 7    H     1.279401085    0.000666664   -1.701896375
 8    H     1.289123707   -1.466934560    0.844710732
 9    H     1.286261542    1.470573096    0.843755137
10    H    -1.279402302   -0.000620419    1.701898354
11    H    -1.286260164   -1.470574214   -0.843756741
12    H    -1.289124413    1.466930262   -0.844707223
13    H    -0.313590024   -0.001429006   -2.487758174
14    H    -0.298403457   -2.155400614    1.245750272
15    H    -0.302700107    2.155536801    1.244967386
16    H     0.313586322    0.001435910    2.487757079
17    H     0.302704432   -2.155535951   -1.244965785
18    H     0.298399197    2.155402230   -1.245751747
EXPGEOM
 1    C     0.229400000    0.000000000   -1.467000000
 2    C     0.229400000   -1.270400000    0.733500000
 3    C     0.229400000    1.270400000    0.733500000
 4    C    -0.229400000    0.000000000    1.467000000
 5    C    -0.229400000   -1.270400000   -0.733500000
```

```
6     C      -0.229400000    1.270400000   -0.733500000
7     H       1.326200000   -0.000100000   -1.535100000
8     H       1.326200000   -1.329400000    0.767600000
9     H       1.326200000    1.329400000    0.767500000
10    H      -1.326200000    0.000100000    1.535100000
11    H      -1.326200000   -1.329400000   -0.767500000
12    H      -1.326200000    1.329400000   -0.767600000
13    H      -0.145900000    0.000400000   -2.496700000
14    H      -0.145900000   -2.162400000    1.248000000
15    H      -0.145900000    2.162100000    1.248700000
16    H       0.145900000   -0.000400000    2.496700000
17    H       0.145900000   -2.162100000   -1.248700000
18    H       0.145900000    2.162400000   -1.248000000
 ---
H14C6, RHF, CHARGE=0, MULT=1
HF=-44.4
1     C      -0.000779805    1.953550185   -1.064108721
2     C       0.000687182    1.174268018    0.254711114
3     C       0.000330953   -0.392500158    0.221354682
4     C      -0.000654529   -0.899657092    1.695972649
5     C       1.269410938   -0.940074146   -0.489977779
6     C      -1.268680898   -0.939218437   -0.490550051
7     H      -0.002598834   -2.007695320    1.748319703
8     H       0.004604607    3.042983947   -0.847167526
9     H       2.199424645   -0.538405169   -0.038192313
10    H      -2.198859855   -0.537716169   -0.038908117
11    H       1.288247807   -0.682945671   -1.568130900
12    H      -1.287079971   -0.681159396   -1.568419043
13    H       1.325664979   -2.046446259   -0.423907725
14    H      -1.325215183   -2.045577833   -0.425307523
15    H       0.891875201    1.742234128   -1.685569768
16    H      -0.899908030    1.749717254   -1.678604470
17    H       0.890224515    1.514372976    0.836483694
18    H      -0.887442147    1.513581618    0.839000407
19    H       0.892972408   -0.548497639    2.251100185
20    H      -0.892888801   -0.545730080    2.251851306
EXPGEOM
1     C       0.000000000    1.905800000   -1.075700000
2     C       0.000000000    1.171300000    0.271400000
3     C       0.000000000   -0.376600000    0.220100000
4     C       0.000000000   -0.898900000    1.670300000
5     C       1.258500000   -0.898900000   -0.499200000
6     C      -1.258500000   -0.898900000   -0.499200000
7     H       0.000000000   -1.994200000    1.695000000
8     H       0.000000000    2.988900000   -0.915500000
9     H       2.171000000   -0.524500000   -0.021600000
10    H      -2.171000000   -0.524500000   -0.021600000
11    H       1.282100000   -0.596800000   -1.550800000
12    H      -1.282100000   -0.596800000   -1.550800000
13    H       1.293900000   -1.993600000   -0.471400000
14    H      -1.293900000   -1.993600000   -0.471400000
15    H       0.884100000    1.665000000   -1.674000000
16    H      -0.884100000    1.665000000   -1.674000000
17    H       0.876800000    1.493600000    0.848600000
18    H      -0.876800000    1.493600000    0.848600000
19    H       0.884800000   -0.552400000    2.215900000
20    H      -0.884800000   -0.552400000    2.215900000
 ---
H14C6, RHF, CHARGE=0, MULT=1
```

```
HF=-42.5
1    C    -0.000114706   -0.218016928    0.753487452
2    C     0.000225838    0.217979805   -0.753451199
3    H    -0.000284052   -1.338224979    0.773157620
4    H     0.000259930    1.338187152   -0.773020985
5    C     1.263653459    0.228642253    1.523187013
6    C    -1.264050487    0.229259679    1.522558444
7    C     1.264249771   -0.228986885   -1.522599925
8    C    -1.263476160   -0.228863775   -1.523168417
9    H     1.435239103    1.321473884    1.450266795
10   H    -1.434932679    1.322218155    1.449883161
11   H     1.435404636   -1.321898643   -1.449739976
12   H    -1.434768750   -1.321765725   -1.450552334
13   H     1.183388469   -0.023460263    2.601185485
14   H     2.174268325   -0.285193901    1.154857302
15   H    -1.184706825   -0.023258474    2.600517642
16   H    -2.174531660   -0.284042536    1.153293891
17   H     1.184733185    0.023137356   -2.600692630
18   H     2.174658084    0.284511661   -1.153343865
19   H    -1.183250473    0.023657216   -2.601052556
20   H    -2.174192074    0.284702951   -1.154748972
EXPGEOM
1    C     0.000000000   -0.232400000    0.739200000
2    C     0.000000000    0.232400000   -0.739200000
3    H     0.000000000   -1.332700000    0.732100000
4    H     0.000000000    1.332700000   -0.732100000
5    C     1.256200000    0.232400000    1.496600000
6    C    -1.256200000    0.232400000    1.496600000
7    C     1.256200000   -0.232400000   -1.496600000
8    C    -1.256200000   -0.232400000   -1.496600000
9    H     1.343800000    1.325500000    1.475700000
10   H    -1.343800000    1.325500000    1.475700000
11   H     1.343800000   -1.325500000   -1.475700000
12   H    -1.343800000   -1.325500000   -1.475700000
13   H     1.207600000   -0.073500000    2.546800000
14   H     2.174800000   -0.185000000    1.076500000
15   H    -1.207600000   -0.073500000    2.546800000
16   H    -2.174800000   -0.185000000    1.076500000
17   H     1.207600000    0.073500000   -2.546800000
18   H     2.174800000    0.185000000   -1.076500000
19   H    -1.207600000    0.073500000   -2.546800000
20   H    -2.174800000    0.185000000   -1.076500000
---
H14C6, RHF, CHARGE=0, MULT=1
HF=-41.7
1    C     0.000000000    0.000000000    0.000000000
2    C     0.000000000    1.531965450    0.000000000
3    C     0.000000000    2.170706979    1.403507479
4    C     0.081206934    3.722501079    1.441518098
5    C    -1.235494052    4.416860974    1.039257171
6    C     0.586607147    4.224853405    2.811546935
7    H     0.895028438   -0.413208848    0.507703305
8    H    -0.896826298   -0.412421788    0.505129215
9    H     0.001782931   -0.383411018   -1.041107572
10   H     0.893440125    1.882386145   -0.566920616
11   H    -0.888402142    1.877636917   -0.575896079
12   H     0.870768858    1.757994239    1.964997257
13   H    -0.906366618    1.835583790    1.957761154
14   H     0.852705274    4.039863672    0.693295587
```

```
15      H       -1.531818938     4.159968912     0.002437953
16      H       -2.076421497     4.140551430     1.706989736
17      H       -1.128972877     5.520825114     1.075661764
18      H        1.595277378     3.823042722     3.038511097
19      H       -0.088054376     3.931170849     3.640850761
20      H        0.671152428     5.330909262     2.824404382
EXPGEOM
1       C        0.14720     2.84390    -0.21560
2       H       -0.18990     3.67810     0.40800
3       H        1.23910     2.90210    -0.28710
4       H       -0.25740     2.99550    -1.22250
5       C        0.19420    -1.44610     1.38040
6       H       -0.12670    -0.73450     2.14680
7       H       -0.16130    -2.43720     1.68120
8       H        1.29080    -1.47180     1.38600
9       C       -0.29780     1.49540     0.36350
10      H        0.08740     1.39940     1.38520
11      H       -1.39260     1.47670     0.44560
12      C        0.16490     0.30400    -0.48560
13      H        1.26390     0.29080    -0.52340
14      C       -0.33480    -1.07410    -0.01360
15      H       -1.43190    -1.02690     0.04400
16      C        0.03450    -2.15930    -1.03660
17      H       -0.37030    -1.92840    -2.02780
18      H        1.12290    -2.24940    -1.13490
19      H       -0.35280    -3.13850    -0.73550
20      H       -0.17090     0.46130    -1.51970
---
H14C6, RHF, CHARGE=0, MULT=1
HF=-41.1
1       C       -0.120856680     0.010995120    -0.111878730
2       C        0.146993830     1.252863630     0.744447020
3       C        1.471423320     2.018556550     0.460644010
4       C        1.412947210     3.453861710     1.058650560
5       C        2.427607810     4.466252210     0.518066140
6       C        2.718220800     1.236299830     0.923251200
7       H       -0.111502490     0.246750120    -1.195325380
8       H        0.617333490    -0.795846090     0.066933360
9       H       -1.121933570    -0.404163030     0.129295190
10      H       -0.713098300     1.945706600     0.587677330
11      H        0.106204210     0.960754160     1.818908710
12      H        1.553045740     2.128373900    -0.650564110
13      H        0.399602770     3.882810750     0.875135710
14      H        1.521435830     3.399711990     2.166243710
15      H        3.474860960     4.180769410     0.740852720
16      H        2.337176210     4.596213890    -0.579573110
17      H        2.253030850     5.457409720     0.987384430
18      H        3.653455180     1.771200560     0.660837980
19      H        2.725171060     1.069199590     2.019315310
20      H        2.779850930     0.243404620     0.433514790
EXPGEOM
1       C        0.00000    -0.09170    -0.18320
2       C        0.00000     1.29490    -0.84510
3       H        0.00000    -0.84850    -0.98260
4       C        1.26280    -0.32100     0.67070
5       C       -1.26280    -0.32100     0.67070
6       C        2.58430    -0.32100    -0.10850
7       C       -2.58430    -0.32100    -0.10850
8       H        0.00000     2.08640    -0.08550
```

```
9    H     0.88020    1.44070   -1.47780
10   H    -0.88020    1.44070   -1.47780
11   H     1.30360    0.44400    1.45830
12   H    -1.30360    0.44400    1.45830
13   H     1.16090   -1.28320    1.18850
14   H    -1.16090   -1.28320    1.18850
15   H     3.42210   -0.57650    0.54810
16   H    -3.42210   -0.57650    0.54810
17   H     2.56370   -1.05700   -0.92030
18   H     2.80290    0.65550   -0.55070
19   H    -2.56370   -1.05700   -0.92030
20   H    -2.80290    0.65550   -0.55070
---
H14C6, RHF, CHARGE=0, MULT=1
HF=-39.9
1    C     0.000000000    0.000000000    0.000000000
2    C     0.000000000    1.531265290    0.000000000
3    C     0.000000000    2.175011755    1.400136184
4    C    -0.000863171    3.717002891    1.384564571
5    C     0.001395589    4.361069243    2.784373423
6    C    -0.000091378    5.892313447    2.783809665
7    H     0.003605206    6.276308265    3.824719688
8    H    -0.898636008    6.303497825    2.280783648
9    H     0.893461717    6.305524697    2.273580187
10   H     0.893498784    4.010181863    3.353300887
11   H    -0.888062291    4.008894340    3.356563397
12   H    -0.893673481    4.075151183    0.821917045
13   H     0.889193799    4.076381619    0.818375784
14   H     0.891455517    1.815879256    1.964292373
15   H    -0.891355985    1.815694401    1.964261058
16   H     0.890411983    1.883064155   -0.570972384
17   H    -0.890922530    1.882373294   -0.570522175
18   H     0.001882890   -0.383542623   -1.041037095
19   H    -0.897119331   -0.412242310    0.504700752
20   H     0.895061164   -0.412519126    0.508131442
EXPGEOM
1    C     0.00000    1.41490    2.90370
2    C     0.00000   -1.41490   -2.90370
3    C     0.00000    0.00490    0.76680
4    C     0.00000   -0.00490   -0.76680
5    C     0.00000   -1.41490   -1.37070
6    C     0.00000    1.41490    1.37070
7    H     0.00000   -2.43350   -3.30470
8    H     0.00000    2.43350    3.30470
9    H     0.88360    0.90140    3.29860
10   H    -0.88360    0.90140    3.29860
11   H    -0.88360   -0.90140   -3.29860
12   H     0.88360   -0.90140   -3.29860
13   H    -0.87700    1.96460    1.00480
14   H     0.87700    1.96460    1.00480
15   H    -0.87700   -1.96460   -1.00480
16   H     0.87700   -1.96460   -1.00480
17   H     0.87750   -0.54540    1.13350
18   H    -0.87750   -0.54540    1.13350
19   H    -0.87750    0.54540   -1.13350
20   H     0.87750    0.54540   -1.13350
---
H7C7, RHF, CHARGE=1, MULT=1
HF=212
```

```
  1    C    0.000000000     0.000000000     0.000000000
  2    C    0.000000000     2.860992740     0.000000000
  3    C    0.000000000     0.707841965     1.230158830
  4    C    0.003494615     0.707472346    -1.230238915
  5    C   -0.002597057     2.097951581     1.242417836
  6    C    0.006233938     2.097641951    -1.242286208
  7    C   -0.005186202     4.243845379    -0.000427432
  8    H   -0.003018113    -1.093477674     0.001379882
  9    H    0.000139101     0.150079004     2.169732073
 10    H    0.004407347     0.149497382    -2.169652215
 11    H   -0.007948579     2.632114419     2.196213069
 12    H    0.013158428     2.631599072    -2.196165424
 13    H   -0.013684509     4.837771943     0.916009590
 14    H   -0.001783502     4.836807269    -0.917520295
EXPGEOM
  1    C    0.99240    0.00000    0.00000
  2    C    0.25170    0.00000    1.21570
  3    C   -1.13080    0.00000    1.20920
  4    C   -1.83460    0.00000    0.00000
  5    C   -1.13080    0.00000   -1.20920
  6    C    0.25170    0.00000   -1.21570
  7    C    2.39500    0.00000    0.00000
  8    H    0.79240    0.00000    2.15690
  9    H   -1.67290    0.00000    2.14930
 10    H   -2.91910    0.00000    0.00000
 11    H   -1.67290    0.00000   -2.14930
 12    H    0.79240    0.00000   -2.15690
 13    H    2.95590    0.00000    0.92740
 14    H    2.95590    0.00000   -0.92740
---
H7C7, RHF, CHARGE=1, MULT=1
HF=209
  1    C    0.000000000     0.000000000     0.000000000
  2    C    0.000000000     2.533250940     0.000000000
  3    C    0.000000000     2.846193078     1.370784429
  4    C   -0.000008435     1.266647410    -0.610169852
  5    C   -0.000278397    -0.312897986     1.370619645
  6    C   -0.000610583     0.563615645     2.469938608
  7    C   -0.000032799     1.969720744     2.470066168
  8    H   -0.000152797     2.447111604     3.461179502
  9    H    0.000020470     3.393318151    -0.685885424
 10    H   -0.000047539     1.266844598    -1.710226954
 11    H   -0.000751370     0.086240046     3.461013649
 12    H    0.000060768    -0.860196032    -0.685689058
 13    H   -0.000320342    -1.385455391     1.615167826
 14    H    0.000043128     3.918711795     1.615483841
EXPGEOM
  1    C    1.61810    0.00000    0.00000
  2    C    0.99380    0.00000    1.26370
  3    C    0.99380    0.00000   -1.26370
  4    C   -0.33990    0.00000    1.58550
  5    C   -0.33990    0.00000   -1.58550
  6    C   -1.46340    0.00000    0.67730
  7    C   -1.46340    0.00000   -0.67730
  8    H    2.69980    0.00000    0.00000
  9    H    1.67240    0.00000    2.10960
 10    H    1.67240    0.00000   -2.10960
 11    H   -0.58390    0.00000    2.63930
 12    H   -0.58390    0.00000   -2.63930
```

```
13    H    -2.43550     0.00000      1.15670
14    H    -2.43550     0.00000     -1.15670
---
H8C7, RHF, CHARGE=0, MULT=1
HF=43, IE=8.5
1     C     0.000000000    0.000000000    0.000000000
2     C    -0.417676952   -0.596323690    2.415710082
3     C    -1.042930977    0.677664860    2.911229474
4     C    -0.000000000   -0.881889730    1.162816580
5     C     0.000000000    1.355470110    0.000000000
6     C     0.000000000    2.237355860    1.162816580
7     C    -0.417676952    1.951683170    2.415710082
8     H    -0.333533075    2.727382063    3.184592859
9     H    -0.333558470   -1.372213973    3.184360806
10    H     0.389961034   -1.886500031    0.960501157
11    H     0.390336111    3.241828906    0.960512986
12    H     0.030570264   -0.523861245   -0.961946809
13    H     0.030574778    1.879240322   -0.962033812
14    H    -1.010476359    0.677664860    4.026773482
15    H    -2.128986799    0.677664860    2.643237706
---
H8C7, RHF, CHARGE=0, MULT=1
HF=59.7
1     C     0.000000000    0.000000000    0.000000000
2     C     0.000000000    1.579287260    0.000000000
3     C     0.000000000   -0.046959694    1.578726119
4     C     1.229906066    1.816022274    0.895617595
5     C     1.229668155    0.841098860    1.842074979
6     C    -1.230834858    1.815136994    0.894361762
7     C    -1.230855405    0.839866345    1.840367380
8     H     0.885941942   -0.460818396   -0.474077482
9     H    -0.886126475   -0.460633990   -0.473884815
10    H     0.000108400    2.088535660   -0.972241426
11    H     0.000090714   -1.034328762    2.057993867
12    H     1.920760074    2.631582940    0.744251484
13    H     1.920314887    0.664903697    2.652814574
14    H    -1.922369832    2.629941758    0.742280589
15    H    -1.922310638    0.662885693    2.650181586
EXPGEOM
1     C     1.35410      0.00000      0.00000
2     C     0.27280      0.00000      1.12370
3     C     0.27280      0.00000     -1.12370
4     C    -0.52120      1.24350      0.66670
5     C    -0.52120      1.24350     -0.66670
6     C    -0.52120     -1.24350      0.66670
7     C    -0.52120     -1.24350     -0.66670
8     H     1.97710     -0.89890      0.00000
9     H     1.97710      0.89890      0.00000
10    H     0.61000      0.00000      2.16000
11    H     0.61000      0.00000     -2.16000
12    H    -1.01620      1.93320      1.33740
13    H    -1.01620      1.93320     -1.33740
14    H    -1.01620     -1.93320      1.33740
15    H    -1.01620     -1.93320     -1.33740
---
H8C7, RHF, CHARGE=0, MULT=1
HF=12, DIP=0.36, IE=8.82
1     C    -0.000134600   -0.026479450   -0.001441860
2     C    -0.001191820    2.815042920   -0.009523020
```

```
  3    C       0.041327290     0.684567130     1.210657370
  4    C      -0.041823950     0.678033860    -1.216064650
  5    C       0.040567630     2.089236850     1.205807480
  6    C      -0.042503350     2.084243070    -1.219330940
  7    H       0.000343360    -1.116516800     0.001609960
  8    H       0.073863520     0.144687890     2.157688790
  9    H      -0.074035620     0.134129170    -2.160882110
 10    H       0.073245520     2.619714640     2.159347420
 11    H      -0.075405550     2.606113110    -2.177316920
 12    C      -0.001604740     4.319879820     0.001716210
 13    H       0.911392910     4.712880180     0.497029790
 14    H      -0.032973050     4.752305990    -1.018891410
 15    H      -0.883129340     4.711807380     0.551847910
EXPGEOM
 1    C      0.00000     0.00360     0.91130
 2    C      1.20060     0.00680     0.19420
 3    C      1.20340     0.00680    -1.19910
 4    C      0.00000     0.00610    -1.90200
 5    C     -1.20340     0.00680    -1.19910
 6    C     -1.20060     0.00680     0.19420
 7    C      0.00000    -0.02640     2.42250
 8    H      2.14390     0.01140     0.73340
 9    H      2.14720     0.01160    -1.73560
 10   H      0.00000     0.00950    -2.98710
 11   H     -2.14720     0.01160    -1.73560
 12   H     -2.14390     0.01140     0.73340
 13   H      0.00000    -1.05710     2.79720
 14   H      0.88560     0.46880     2.83100
 15   H     -0.88560     0.46880     2.83100
---
H12C7, RHF, CHARGE=0, MULT=1
HF=-9.9
 1    C      -0.003188466    -0.732666210    -0.085647203
 2    C      -0.005897809    -1.971248700    -0.968908574
 3    C      -0.006804666    -1.444355746    -2.422938111
 4    C      -0.007171319     0.098649753    -2.320491522
 5    C      -0.003748905     0.410162793    -0.831371248
 6    H      -0.897871597    -2.603484925    -0.769748979
 7    H       0.884151163    -2.606831058    -0.771724479
 8    H      -0.895113374    -1.808245830    -2.980164071
 9    H       0.880904687    -1.807952604    -2.981264306
 10   H      -0.900025994     0.533326245    -2.819128622
 11   H       0.882275276     0.534536732    -2.824079842
 12   C      -0.001021622    -0.854590676     1.405230842
 13   H       0.000579540    -1.912634174     1.738237877
 14   H      -0.898689619    -0.372273730     1.847367689
 15   H       0.896480933    -0.370214784     1.845523821
 16   C      -0.002355629     1.824244246    -0.342627783
 17   H       0.000725599     2.554628340    -1.177380515
 18   H       0.893812212     2.033788258     0.279465028
 19   H      -0.901086912     2.036328286     0.275121990
---
H12C7, RHF, CHARGE=0, MULT=1
HF=-6
 1    C       0.000000000     0.000000000     0.000000000
 2    C       0.000000000     1.510367420     0.000000000
 3    C       0.000000000     1.903895371     1.497220999
 4    C       0.000229959     0.574865592     2.294798839
 5    C      -0.000241456    -0.533896245     1.246310277
```

```
  6    H    -0.000234300    -0.531769512    -0.945454110
  7    H    -0.890327217     1.910878299    -0.530121935
  8    H     0.890092005     1.911333320    -0.530220949
  9    H    -0.888438094     2.519459494     1.749047619
 10    H     0.891327801     0.505540506     2.954588028
 11    C    -0.000790141    -1.971197371     1.682732498
 12    H    -0.890336290     0.505578555     2.955252862
 13    H     0.887891940     2.520059511     1.749423450
 14    H    -0.892695252    -2.139879937     2.333654704
 15    H     0.892289508    -2.139868251     2.332508756
 16    C    -0.000233859    -3.045764762     0.591992015
 17    H     0.893637593    -2.972714804    -0.059894305
 18    H    -0.900779905    -2.982593052    -0.051622556
 19    H     0.007865388    -4.055221366     1.052786076
---
H12C7, RHF, CHARGE=0, MULT=1
HF=-19.4
 1    C     0.000000000     0.000000000     0.000000000
 2    C     0.000000000     1.353670040     0.000000000
 3    C     0.000000000     2.216124274     1.231553212
 4    C     0.332167222     1.460996004     2.530337041
 5    C    -0.231995804     0.031359658     2.565012027
 6    C     0.053866089    -0.789389352     1.293057776
 7    C    -0.037366413    -0.800127253    -1.276669734
 8    H    -0.022013085     1.922109870    -0.934272886
 9    H    -1.002248601     2.698516147     1.322735243
 10   H     0.728918348     3.048706212     1.098144115
 11   H    -0.061965196     2.030786745     3.401878220
 12   H     1.437784700     1.427901107     2.664709716
 13   H    -1.332163973     0.074081421     2.737046314
 14   H     0.192771973    -0.504518756     3.444046806
 15   H    -0.674245093    -1.632435913     1.253238984
 16   H     1.062793797    -1.258219971     1.380572487
 17   H     0.757978025    -1.574584881    -1.292244731
 18   H    -1.015652519    -1.314416397    -1.387971609
 19   H     0.107188633    -0.168899550    -2.177473347
---
H12C7, RHF, CHARGE=0, MULT=1
HF=-12.4
 1    C    -0.000151580    -0.006579710     0.000774120
 2    C     1.134170400     1.067600570     0.000040410
 3    C    -1.134765220     1.067555030     0.000055910
 4    C     0.777983340     1.895970280     1.274544400
 5    C    -0.778656460     1.895823640     1.274583200
 6    C     0.778117380     1.893564450    -1.275917840
 7    C    -0.778707740     1.893762000    -1.275846430
 8    H    -0.000331750    -0.673818060     0.883934710
 9    H    -0.000309830    -0.674781610    -0.881418010
 10   H     2.165522850     0.679470970     0.000469520
 11   H    -2.166155340     0.679557370     0.000337850
 12   H     1.187986530     1.417473160     2.187659350
 13   H     1.198145640     2.921082070     1.250380510
 14   H    -1.188361940     1.417039770     2.187663190
 15   H    -1.198917820     2.920730350     1.250661780
 16   H     1.198490200     2.918677560    -1.253809190
 17   H     1.187714610     1.413383370    -2.188308100
 18   H    -1.198621900     2.918858790    -1.253363230
 19   H    -1.188417810     1.413910000    -2.188220190
EXPGEOM
```

```
    1    C     1.38960    0.00000    0.00000
    2    C     0.34030    0.00000    1.13280
    3    C     0.34030    0.00000   -1.13280
    4    C    -0.49380    1.25500    0.78260
    5    C    -0.49380   -1.25500    0.78260
    6    C    -0.49380   -1.25500   -0.78260
    7    C    -0.49380    1.25500   -0.78260
    8    H     2.02780   -0.88980    0.00000
    9    H     2.02780    0.88980    0.00000
   10    H     0.72920    0.00000    2.15320
   11    H     0.72920    0.00000   -2.15320
   12    H    -0.01960    2.15940    1.17640
   13    H    -0.01960   -2.15940   -1.17640
   14    H    -0.01960    2.15940   -1.17640
   15    H    -0.01960   -2.15940    1.17640
   16    H    -1.50170    1.20870   -1.20580
   17    H    -1.50170   -1.20870    1.20580
   18    H    -1.50170    1.20870    1.20580
   19    H    -1.50170   -1.20870   -1.20580
---
H14C7, RHF, CHARGE=0, MULT=1
HF=-33.1
    1    C    0.000000000    0.000000000    0.000000000
    2    C    0.000000000    1.567716800    0.000000000
    3    C    0.000000000    2.056658262    1.461515294
    4    C   -0.305465083    0.845388408    2.359533795
    5    C   -0.455747010   -0.389137012    1.449174812
    6    C   -0.980055309   -0.558611368   -1.062945211
    7    C    1.423551434   -0.546789207   -0.291168719
    8    H   -0.895480597    1.966361277   -0.524968645
    9    H    0.870749121    1.980518845   -0.551833230
   10    H   -0.752757818    2.860804402    1.606609534
   11    H    0.981057482    2.506317168    1.727128048
   12    H   -1.231861317    1.011927305    2.950018587
   13    H    0.506090582    0.690956964    3.102882784
   14    H    0.129060522   -1.244299677    1.849029213
   15    H   -1.514369453   -0.729103802    1.456692147
   16    H   -2.016335469   -0.196281612   -0.903740516
   17    H   -1.014019287   -1.667325855   -1.040691597
   18    H   -0.682164646   -0.256742258   -2.088177657
   19    H    1.449757879   -1.655219520   -0.250666023
   20    H    2.169033173   -0.174745805    0.440928942
   21    H    1.776481075   -0.247176385   -1.299417639
---
H14C7, RHF, CHARGE=0, MULT=1
HF=-31
    1    C    0.000000000    0.000000000    0.000000000
    2    C    0.000000000    1.552397060    0.000000000
    3    C    0.000000000    2.015260324    1.468733051
    4    C   -0.203092965    0.767086725    2.348022307
    5    C   -0.513495793   -0.437917575    1.414327693
    6    C   -0.664726627   -0.625671056   -1.233651607
    7    H    1.071535461   -0.328159607   -0.046071664
    8    H   -0.885704159    1.963382196   -0.530525197
    9    H    0.888782349    1.940822265   -0.541610846
   10    H   -0.804781071    2.760484472    1.646519152
   11    H    0.952486835    2.526891554    1.724639746
   12    H   -1.005581521    0.939709020    3.095254784
   13    H    0.715747322    0.556605843    2.938245033
```

```
14         H        0.111687656      -1.306079500     1.739683439
15         C       -1.972376180      -0.912014475     1.520885558
16         H       -1.723974049      -0.324948483    -1.358036123
17         H       -0.626672652      -1.733407363    -1.191798445
18         H       -0.126862697      -0.317958505    -2.155219616
19         H       -2.697864673      -0.145554179     1.182363717
20         H       -2.145344215      -1.831770154     0.926259710
21         H       -2.221817153      -1.165107754     2.572802247
---
H14C7, RHF, CHARGE=0, MULT=1
HF=-31.9
1     C     0.000000000      0.000000000      0.000000000
2     C     0.000000000      1.552400810      0.000000000
3     C     0.000000000      2.031172002      1.465928716
4     C     0.053813196      0.777708064      2.378467840
5     C    -0.282434292     -0.430089231      1.464265833
6     C    -0.928480040     -0.640485888     -1.041249024
7     H     1.037044318     -0.335245101     -0.255003526
8     H    -0.880030612      1.967750624     -0.535999683
9     H     0.892301433      1.935930212     -0.540063157
10        H    -0.902540286      2.648402508     1.664259216
11        H     0.870278256      2.692445861     1.665936068
12        C    -0.799962907      0.872404573     3.650649830
13        H     1.113252840      0.657126040     2.719690035
14        H     0.335783152     -1.311617818     1.737859916
15        H    -1.340133557     -0.745917686     1.592025352
16        H    -1.991123175     -0.360353156     -0.894507401
17        H    -0.863499835     -1.747385462     -0.999460933
18        H    -0.638938286     -0.332461073     -2.067312495
19        H    -1.877294918      1.015762061     3.431616600
20        H    -0.700526763     -0.048458424     4.261533739
21        H    -0.469893246      1.723485621     4.281747597
---
H14C7, RHF, CHARGE=0, MULT=1
HF=-32.7
1     C    -0.621576781     -6.449314353     1.005135944
2     C    -0.036798180     -5.166614424     1.624809542
3     H    -0.166974639     -7.354835650     1.461986478
4     C    -0.348067843     -6.406250633    -0.509318922
5     H    -1.712080541     -6.520306816     1.207953378
6     C     0.717699720     -4.392947859     0.508381559
7     H     0.654328118     -5.409353910     2.460703139
8     H    -0.851495444     -4.556833297     2.071261956
9     C     0.246031102     -5.012074954    -0.849351928
10        H    -1.281991616     -6.577600378    -1.086064233
11        H     0.341559703     -7.229686808    -0.794727668
12        H    -0.592762508     -4.388480559    -1.248783605
13        C     1.320365912     -5.047063958    -1.947409652
14        H     1.807168703     -4.611536649     0.638097899
15        C     0.568757954     -2.866135469     0.602830808
16        H     2.224989047     -5.607841834    -1.637246045
17        H     1.639921105     -4.022354489    -2.226898688
18        H     0.927233786     -5.527681039    -2.867141600
19        H    -0.482820701     -2.535237172     0.486894695
20        H     1.168302971     -2.353238215    -0.176506083
21        H     0.929733730     -2.497565791     1.585166559
---
H14C7, RHF, CHARGE=0, MULT=1
HF=-14.9
```

```
  1    C     0.000000000    0.000000000    0.000000000
  2    C     0.000000000    1.531298690    0.000000000
  3    C     0.000000000    2.175831700    1.399901063
  4    C     0.002523818    3.717857643    1.383823506
  5    C     0.002626261    4.362450774    2.785079676
  6    C     0.007701010    5.868439993    2.781970034
  7    C     0.006881603    6.645065547    3.876191108
  8    H    -0.894160880   -0.412228746    0.509741824
  9    H     0.897266216   -0.412752876    0.503844527
 10    H    -0.003613171   -0.383822303   -1.040807225
 11    H     0.892016607    1.882118741   -0.569154065
 12    H    -0.890173145    1.883274964   -0.570813424
 13    H     0.891865897    1.815876998    1.962993569
 14    H    -0.892039059    1.817581469    1.963268539
 15    H     0.896722945    4.074025034    0.822123465
 16    H    -0.888422554    4.077921629    0.819927281
 17    H     0.893814983    4.002664235    3.352696497
 18    H    -0.891380514    4.007259196    3.350652723
 19    H     0.012407759    6.348556521    1.797841045
 20    H     0.002345353    6.265565847    4.896672663
 21    H     0.010784723    7.732712225    3.819013852
---
H14C7, RHF, CHARGE=0, MULT=1
HF=-30.4
  1    C     0.000000000    0.000000000    0.000000000
  2    C     0.000000000    1.553675560    0.000000000
  3    C     0.000000000    2.025004719    1.468289969
  4    C    -0.263529538    0.794679385    2.357113892
  5    C    -0.443149975   -0.422921462    1.429078664
  6    C    -0.819701959   -0.669090411   -1.126457595
  7    C    -0.258462870   -0.518574295   -2.543808767
  8    H     1.060832087   -0.333897095   -0.124253231
  9    H    -0.885512504    1.966821561   -0.528789541
 10    H     0.889422522    1.943442820   -0.539419409
 11    H    -0.774106607    2.805126191    1.630786993
 12    H     0.970908691    2.498033652    1.729991578
 13    H    -1.164912507    0.946018126    2.988410476
 14    H     0.580154230    0.631124993    3.061623778
 15    H     0.159832989   -1.285432545    1.784390088
 16    H    -1.501897850   -0.759986525    1.449903718
 17    H    -0.895672910   -1.760690510   -0.908050442
 18    H    -1.865680592   -0.284396403   -1.116207669
 19    H    -0.213258347    0.539422041   -2.870459187
 20    H     0.761345823   -0.945105812   -2.631717930
 21    H    -0.906217672   -1.059204383   -3.265208734
---
H14C7, RHF, CHARGE=0, MULT=1
HF=-37
  1    C     0.000000000    0.000000000    0.000000000
  2    C     0.000000000    1.539133920    0.000000000
  3    C     0.000000000    2.167210907    1.403113463
  4    C    -1.027566879    1.539190641    2.358429311
  5    C    -1.029358704    0.000168425    2.355997263
  6    C    -1.058831621   -0.632074936    0.940677047
  7    C    -0.948769348   -2.168914813    0.989356326
  8    H    -0.184425631   -0.350500457   -1.041216796
  9    H     1.016931860   -0.364664926    0.271260780
 10    H    -0.880079513    1.910446979   -0.573668975
 11    H     0.895962093    1.896387051   -0.557436826
```

```
12      H       -0.205637347     3.258141658     1.312747869
13      H        1.018918985     2.087855634     1.846972664
14      H       -0.824509093     1.895868592     3.394149976
15      H       -2.046371497     1.913005766     2.105478821
16      H       -1.916794119    -0.350603669     2.930870974
17      H       -0.139135341    -0.363007836     2.918965474
18      H       -2.063980652    -0.406899344     0.499375279
19      H       -1.019022371    -2.605102744    -0.028315207
20      H        0.007336514    -2.510713745     1.434643133
21      H       -1.773489789    -2.606427911     1.588818502
---
H16C7, RHF, CHARGE=0, MULT=1
HF=-48.7
1       C        0.000000000     0.000000000     0.000000000
2       C        0.000000000     1.547194510     0.000000000
3       C        0.000000000     2.227181485     1.431812587
4       C        1.299654998     1.929360156     2.232936992
5       C       -0.150959098     3.773999658     1.294128058
6       C       -1.218698862     1.723792408     2.266342105
7       C        1.090882889     2.051590531    -0.973302302
8       H        0.911470468    -0.432762256     0.457208480
9       H       -0.877153970    -0.417842584     0.532509020
10      H       -0.059539253    -0.385814713    -1.039844737
11      H       -0.976290831     1.838878706    -0.470152598
12      H        1.440349304     0.845978585     2.418977480
13      H        2.206610233     2.296380420     1.712452868
14      H        1.279634351     2.423962951     3.226828339
15      H        0.749190490     4.251527467     0.858349386
16      H       -1.016293495     4.047566701     0.656706226
17      H       -0.310963991     4.254395615     2.281992795
18      H       -2.172892722     1.840770153     1.713484760
19      H       -1.125738960     0.657096561     2.552093773
20      H       -1.321513823     2.292377574     3.214238996
21      H        1.061383093     3.150226409    -1.110940511
22      H        2.114436410     1.779181989    -0.647882263
23      H        0.939355927     1.611152357    -1.981751795
---
H16C7, RHF, CHARGE=0, MULT=1
HF=-49.3
1       C        0.000000000     0.000000000     0.000000000
2       C        0.000000000     1.533034410     0.000000000
3       C        0.000000000     2.163713834     1.407638447
4       C        0.002605645     3.727712801     1.540186878
5       C        1.273440807     4.348731741     0.895842912
6       C       -1.263649578     4.353152658     0.891028740
7       C       -0.000107773     4.073195432     3.061335071
8       H        0.891726286    -0.413335704     0.513418114
9       H       -0.900127759    -0.413290101     0.498513967
10      H        0.008828849    -0.382840702    -1.041287385
11      H        0.892579071     1.874542052    -0.571967876
12      H       -0.889770775     1.875928494    -0.574860576
13      H       -0.892693807     1.774998339     1.952400186
14      H        0.887894943     1.770355494     1.956356599
15      H        2.201525881     3.890617493     1.294685745
16      H        1.290117509     4.225269749    -0.205583750
17      H        1.337145435     5.438912659     1.094329457
18      H       -1.278103357     4.225027933    -0.209955976
19      H       -2.194945652     3.899044275     1.287137083
20      H       -1.323646542     5.443983009     1.086038371
```

```
21      H     -0.895726163    3.667036002    3.574306834
22      H      0.890151416    3.661350692    3.579079809
23      H      0.002808572    5.169302641    3.232153690
---
H16C7, RHF, CHARGE=0, MULT=1
HF=-47.6
1    C      0.000000000    0.000000000    0.000000000
2    C      0.000000000    1.532747760    0.000000000
3    C      0.000000000    2.219260635    1.397480222
4    C      0.597770867    3.669592876    1.341108396
5    C     -0.308122727    4.731212522    0.684118150
6    C      1.120399282    4.154516922    2.712002709
7    C     -1.371778750    2.115603307    2.095894307
8    H      0.859689904   -0.416126164    0.563147694
9    H     -0.928092042   -0.423667737    0.432542407
10   H      0.075248987   -0.373079172   -1.043044940
11   H      0.900572107    1.857443716   -0.571863001
12   H     -0.874628732    1.881377367   -0.595723355
13   H      0.712860214    1.630869321    2.032075616
14   H      1.510420749    3.604276210    0.692097366
15   H     -0.653052429    4.413006722   -0.319987587
16   H     -1.206043850    4.963171091    1.290914647
17   H      0.247950797    5.681988169    0.543543780
18   H      0.309352604    4.331785516    3.445655021
19   H      1.822636882    3.420453952    3.157672854
20   H      1.679316465    5.107077321    2.600746013
21   H     -1.381631316    2.648836693    3.067646718
22   H     -2.196422262    2.527569197    1.480437357
23   H     -1.621812261    1.057387409    2.317494021
---
H16C7, RHF, CHARGE=0, MULT=1
HF=-48.3
1    C     -0.204632254   -2.062283114   -1.048784299
2    C      0.335548385   -0.958754975   -0.113665506
3    C     -0.587281751    0.293830384   -0.120339297
4    C      0.090053721    1.658558575    0.194448392
5    C      0.811281866    2.270877973   -1.024511093
6    C      0.604477586   -1.504498412    1.304128971
7    C     -0.909586609    2.659160917    0.813424572
8    H      0.473347173   -2.940548948   -1.061671973
9    H     -0.280328730   -1.697284563   -2.093829458
10   H     -1.208814869   -2.419532033   -0.742715067
11   H     -1.416131724    0.118599050    0.603901087
12   H     -1.086138857    0.375239401   -1.113664259
13   H      1.548398743    1.567799809   -1.461845791
14   H      1.372349885    3.182399160   -0.731398477
15   H      0.104891937    2.556939556   -1.829738160
16   H     -0.413121439    3.622592160    1.050876340
17   H     -1.328578593    2.269080078    1.763443761
18   H     -1.758013197    2.878715414    0.134517501
19   H      0.876064422    1.488780267    0.973308393
20   H      1.331003031   -0.664096512   -0.532267452
21   H      1.322852360   -2.349383800    1.272340491
22   H     -0.320180265   -1.869769602    1.794952977
23   H      1.049431127   -0.731849666    1.962900081
EXPGEOM
1    C     -0.04140   -2.82750   -0.90280
2    C     -0.34450   -1.64530    0.03050
3    C      0.18950   -0.33190   -0.57130
```

```
4    C     -0.20060    0.94300    0.18880
5    C      0.25920    2.22690   -0.51480
6    C     -0.12200    3.50200    0.24590
7    C      0.20940   -1.91430    1.43820
8    H     -0.45280   -3.76220   -0.50720
9    H     -0.46620   -2.66880   -1.89970
10   H      1.04010   -2.96490   -1.02050
11   H     -1.43730   -1.55270    0.11110
12   H      1.28520   -0.39250   -0.64050
13   H     -0.17450   -0.24900   -1.60440
14   H     -1.29200    0.97180    0.31420
15   H      0.22240    0.92300    1.20080
16   H      1.34810    2.19400   -0.65070
17   H     -0.17170    2.26000   -1.52390
18   H      0.21760    4.39950   -0.28090
19   H     -1.20800    3.57900    0.36890
20   H      0.32480    3.51400    1.24630
21   H      1.30400   -1.98180    1.41950
22   H     -0.06420   -1.12590    2.14540
23   H     -0.17350   -2.86070    1.83480
---
H16C7, RHF, CHARGE=0, MULT=1
HF=-46.6
1    C      0.000000000    0.000000000    0.000000000
2    C      0.000000000    1.542312470    0.000000000
3    C      0.000000000    2.161683706    1.425953008
4    C     -1.337499085    2.081715698    2.191079437
5    C     -1.314670806    2.780516362    3.565164891
6    C     -2.630622982    2.702993282    4.344581901
7    C      1.155741379    2.102076797   -0.857795336
8    H      0.878983050   -0.419233280    0.530052307
9    H     -0.909890271   -0.408453458    0.483813356
10   H      0.014264683   -0.397026229   -1.036125976
11   H     -0.944876202    1.863480511   -0.509657492
12   H      0.800681849    1.681413073    2.033827668
13   H      0.284399043    3.236588052    1.342386004
14   H     -2.141584915    2.536996080    1.568097072
15   H     -1.620788992    1.015066775    2.338166781
16   H     -0.508049528    2.336488503    4.193619980
17   H     -1.048209158    3.854499571    3.432051844
18   H     -3.467887677    3.171429287    3.788688510
19   H     -2.916717668    1.655254184    4.567814419
20   H     -2.534242664    3.236103037    5.312805323
21   H      1.117663425    3.209213515   -0.912436939
22   H      2.150177777    1.821633738   -0.455494210
23   H      1.095826743    1.724730117   -1.899141694
---
H16C7, RHF, CHARGE=0, MULT=1
HF=-48.2
1    C      0.000000000    0.000000000    0.000000000
2    C      0.000000000    1.532692700    0.000000000
3    C      0.000000000    2.313195546    1.365808676
4    C      0.004006948    3.842243331    0.995355660
5    C      0.003702909    4.896549316    2.107819327
6    C     -1.270923401    1.960452975    2.187656783
7    C      1.265626739    1.955914850    2.193517375
8    H      0.890916926   -0.428095987    0.500600322
9    H     -0.899893175   -0.428301801    0.484182254
10   H      0.009905132   -0.367812656   -1.048092818
```

```
11      H       0.890994457     1.854863571    -0.589733580
12      H      -0.888405078     1.857007233    -0.591942460
13      H       0.895858903     4.047079100     0.356314191
14      H      -0.882988484     4.051422113     0.351385696
15      H      -0.890974718     4.831294397     2.758176059
16      H       0.899830006     4.834726319     2.756522088
17      H       0.000748224     5.910064899     1.653389371
18      H      -1.309626763     2.509108448     3.150573419
19      H      -2.200180481     2.203560410     1.632977136
20      H      -1.314316257     0.882637029     2.444864331
21      H       2.198205468     2.191998779     1.641462596
22      H       1.303531621     2.507697025     3.154711422
23      H       1.302066509     0.878773277     2.454787891
---
H16C7, RHF, CHARGE=0, MULT=1
HF=-45.3
1       C      -0.664632323    -8.539869443    -0.176220379
2       C      -0.165312138    -7.092294627    -0.231608383
3       H      -1.506522603    -8.712304041    -0.877160200
4       H      -1.030136448    -8.773591160     0.845642614
5       H       0.130836399    -9.273472317    -0.414119634
6       H      -0.995012483    -6.451096341     0.146615211
7       C       0.297843363    -6.576151783    -1.625543369
8       H       0.660980110    -6.970157506     0.506265234
9       H      -0.393084806    -7.015463708    -2.387588929
10      C       1.734114328    -7.064068634    -1.975072635
11      C       0.173333974    -5.028023815    -1.731852306
12      H       0.530847161    -4.555881298    -0.787629238
13      C      -1.223598144    -4.486575231    -2.055611539
14      H       0.858749745    -4.651487428    -2.526052139
15      H      -1.982530820    -4.785982680    -1.305492996
16      H      -1.577177434    -4.830017172    -3.049299753
17      H      -1.201785501    -3.376888585    -2.077150213
18      H       1.904421062    -8.071702726    -1.530012863
19      H       2.485401341    -6.398462524    -1.490909933
20      C       2.065035172    -7.174130471    -3.467477842
21      H       1.402616891    -7.899158572    -3.982806222
22      H       3.109346541    -7.528639949    -3.595577481
23      H       1.981956544    -6.204052883    -3.997074286
EXPGEOM
1       C      -0.21180   -0.57180    0.13730
2       H      -1.31200   -0.58840    0.11000
3       C       0.22830   -0.77050    1.59610
4       C       0.28990   -1.71050   -0.77310
5       C       0.24570    0.78990   -0.42200
6       C      -0.26460   -3.10150   -0.44010
7       C      -0.35770    2.02240    0.26590
8       C       0.04040    3.33520   -0.42000
9       H       1.32240   -0.75890    1.67570
10      H      -0.16100    0.01760    2.24690
11      H      -0.12440   -1.72470    1.99790
12      H       1.38770   -1.73660   -0.73810
13      H       0.02580   -1.46650   -1.81000
14      H      -0.00910    0.83070   -1.48970
15      H       1.34260    0.84780   -0.37070
16      H       0.06270   -3.83750   -1.18150
17      H       0.06890   -3.45590    0.53990
18      H      -1.36050   -3.09890   -0.43760
19      H      -0.04660    2.05930    1.31610
```

```
20    H    -1.45200    1.93180    0.27320
21    H    -0.39690    4.20060    0.08810
22    H     1.12830    3.46550   -0.42260
23    H    -0.29640    3.35480   -1.46240
---
H16C7, RHF, CHARGE=0, MULT=1
HF=-46
1   C    0.000000000    0.000000000    0.000000000
2   C    0.000000000    1.532320220    0.000000000
3   C    0.000000000    2.170062754    1.404257483
4   C   -0.083448462    3.724436154    1.448937955
5   C   -0.515782695    4.206269558    2.864699361
6   C   -1.026208561    5.646486878    2.972877837
7   C    1.214463933    4.403288579    0.964504608
8   H    0.887330348   -0.413198120    0.521167110
9   H   -0.904464715   -0.413742092    0.490346079
10  H    0.017725761   -0.381724494   -1.041771806
11  H    0.889169517    1.872926597   -0.577751832
12  H   -0.891716215    1.883458793   -0.568880431
13  H   -0.869495006    1.749738012    1.961608437
14  H    0.907199671    1.834337126    1.956997774
15  H   -0.893868186    4.029628238    0.739602813
16  H   -1.331264350    3.542818315    3.237414761
17  H    0.330829686    4.073927553    3.577253741
18  H   -0.243252813    6.395000788    2.739308705
19  H   -1.884166494    5.831775910    2.295020046
20  H   -1.371720288    5.845993857    4.008842933
21  H    1.124968026    5.508404309    0.977333328
22  H    2.087808657    4.132071347    1.591742620
23  H    1.452601515    4.123091949   -0.081560314
---
H16C7, RHF, CHARGE=0, MULT=1
HF=-44.9
1   C    0.000000000    0.000000000    0.000000000
2   C    0.000000000    1.531318860    0.000000000
3   C    0.000000000    2.174517127    1.400324727
4   C    0.001264978    3.716659771    1.385731383
5   C    0.000272316    4.358033135    2.788241266
6   C    0.004002815    5.898805828    2.776311855
7   C    0.002035757    6.548970513    4.162732960
8   H    0.006867821    7.654382216    4.068243575
9   H    0.894882960    6.261750408    4.754228787
10  H   -0.897312116    6.269111958    4.747856915
11  H   -0.884536298    6.268249995    2.213649537
12  H    0.896830240    6.264112368    2.217724645
13  H    0.889682984    3.997517196    3.354861476
14  H   -0.893240556    4.002044127    3.351173118
15  H   -0.889521721    4.075793961    0.820701600
16  H    0.893437842    4.074465445    0.822058873
17  H    0.890582175    1.813656669    1.964839053
18  H   -0.892222767    1.815828288    1.963543745
19  H    0.890314885    1.883053624   -0.571136617
20  H   -0.890939985    1.882271872   -0.570546965
21  H    0.001658617   -0.383643524   -1.040997984
22  H   -0.897042135   -0.412190728    0.504881116
23  H    0.895162789   -0.412512739    0.507954915
EXPGEOM
1   C    0.49600    0.00000    0.00000
2   C   -0.34480    0.00000    1.28290
```

```
3    C     -0.34480    0.00000   -1.28290
4    C      0.49510    0.00000    2.56630
5    C      0.49510    0.00000   -2.56630
6    C     -0.35330    0.00000    3.84300
7    C     -0.35330    0.00000   -3.84300
8    H      1.15720   -0.87750    0.00000
9    H      1.15720    0.87750    0.00000
10   H     -1.00620   -0.87740    1.28380
11   H     -1.00620    0.87740    1.28380
12   H     -1.00620    0.87740   -1.28380
13   H     -1.00620   -0.87740   -1.28380
14   H      1.15540    0.87700    2.56560
15   H      1.15540   -0.87700    2.56560
16   H      1.15540   -0.87700   -2.56560
17   H      1.15540    0.87700   -2.56560
18   H      0.27340    0.00000    4.74060
19   H     -0.99950    0.88360    3.88810
20   H     -0.99950   -0.88360    3.88810
21   H      0.27340    0.00000   -4.74060
22   H     -0.99950   -0.88360   -3.88810
23   H     -0.99950    0.88360   -3.88810
---
H8C8, RHF, CHARGE=0, MULT=1
HF=148.7
1    C     0.000000000    0.000000000    0.000000000
2    C     0.000000000    1.575927890    0.000000000
3    C     0.000000000    1.575927890    1.575928740
4    C     0.000000000   -0.000000850    1.575928740
5    C    -1.575928740    0.000000000    0.000000000
6    C    -1.575928740    1.575928740    0.000000000
7    C    -1.575928740    1.575928740    1.575928740
8    C    -1.575929590    0.000000000    1.575928740
9    H     0.625710113   -0.625719140   -0.625719140
10   H     0.613646356    2.189582934   -0.613655375
11   H     0.613646356    2.189582934    2.189584115
12   H     0.613646356   -0.613656225    2.189583784
13   H    -2.189575096   -0.613655044   -0.613655375
14   H    -2.189574765    2.189584281   -0.613655210
15   H    -2.189575096    2.189583950    2.189583950
16   H    -2.189576277   -0.613654879    2.189583950
EXPGEOM
1    C      0.78540    0.78540    0.78540
2    C      0.78540   -0.78540    0.78540
3    C     -0.78540    0.78540    0.78540
4    C     -0.78540   -0.78540    0.78540
5    C      0.78540    0.78540   -0.78540
6    C      0.78540   -0.78540   -0.78540
7    C     -0.78540    0.78540   -0.78540
8    C     -0.78540   -0.78540   -0.78540
9    H      1.41420    1.41420    1.41420
10   H      1.41420   -1.41420    1.41420
11   H     -1.41420    1.41420    1.41420
12   H     -1.41420   -1.41420    1.41420
13   H      1.41420    1.41420   -1.41420
14   H      1.41420   -1.41420   -1.41420
15   H     -1.41420    1.41420   -1.41420
16   H     -1.41420   -1.41420   -1.41420
---
H8C8, RHF, CHARGE=0, MULT=1
```

```
       HF=70.7
 1     C      0.087091160     -0.001348280     0.100221950
 2     C      0.092173530      1.343446200    -0.003488100
 3     C      0.811150270      2.174827320    -0.977442540
 4     C      0.812150940      2.070934290    -2.322101010
 5     C      0.094852230      1.099574140    -3.157279970
 6     C      0.091039530     -0.245280270    -3.053829270
 7     C      0.801868910     -1.081487670    -2.078004390
 8     C      0.800614620     -0.977298110    -0.733313090
 9     H     -0.485227710     -0.465602960     0.913193750
10     H     -0.476792710      1.929941040     0.728310690
11     H      1.385013330      2.986562770    -0.513391920
12     H      1.386760710      2.801747960    -2.904626820
13     H     -0.471840050      1.566815220    -3.972037100
14     H     -0.478587790     -0.828233640    -3.787963270
15     H      1.371138350     -1.897262470    -2.540266490
16     H      1.368927370     -1.711990340    -0.149282850
EXPGEOM
 1     C      0.37710      0.66870      1.56420
 2     C      0.37710     -0.66870      1.56420
 3     C      0.37710      0.66870     -1.56420
 4     C      0.37710     -0.66870     -1.56420
 5     C     -0.37710      1.56420      0.66870
 6     C     -0.37710     -1.56420      0.66870
 7     C     -0.37710      1.56420     -0.66870
 8     C     -0.37710     -1.56420     -0.66870
 9     H      0.93740      1.17710      2.34880
10     H      0.93740     -1.17710      2.34880
11     H      0.93740      1.17710     -2.34880
12     H      0.93740     -1.17710     -2.34880
13     H     -0.93740      2.34880      1.17710
14     H     -0.93740     -2.34880      1.17710
15     H     -0.93740      2.34880     -1.17710
16     H     -0.93740     -2.34880     -1.17710
---
H8C8, RHF, CHARGE=0, MULT=1
       HF=35.3
 1     C     -0.005046669      0.006264190     0.000420674
 2     C      2.829366307      0.006633720    -0.000162165
 3     C      0.700796896      1.221775701     0.000078428
 4     C      0.700927237     -1.209071730     0.000130182
 5     C      2.106346296      1.222809931    -0.000645198
 6     C      2.106616974     -1.209566969    -0.000606281
 7     H     -1.095272832      0.006309154     0.000647582
 8     H      0.157734112      2.167375227     0.000042612
 9     H      0.158601075     -2.155204953     0.000137987
10     H      2.637619891      2.175836372    -0.001953420
11     H      2.638059028     -2.162533327    -0.001773605
12     C      4.308108556      0.008011717    -0.031015160
13     C      5.118551482      0.017667339     1.036724896
14     H      4.731453206      0.001312521    -1.042221349
15     H      6.203986787      0.019258737     0.944336979
16     H      4.769152365      0.024406409     2.068069457
EXPGEOM
 1     C     -0.06210      0.51310     -0.22130
 2     C     -0.08670      0.01010      1.08840
 3     C     -0.02830     -1.35830      1.32590
 4     C      0.05050     -2.25650      0.26030
 5     C      0.06380     -1.77220     -1.04550
```

```
6    C      0.00290    -0.40110    -1.28160
7    C     -0.10500     1.95510    -0.51790
8    C      0.13980     2.95520     0.33110
9    H     -0.16840     0.69510     1.92590
10   H     -0.05240    -1.72760     2.34630
11   H      0.09250    -3.32430     0.44850
12   H      0.11880    -2.46090    -1.88260
13   H      0.01350    -0.02960    -2.30230
14   H     -0.34870     2.19600    -1.55150
15   H      0.08100     3.98980     0.01100
16   H      0.41460     2.78920     1.36790
---
H10C8, RHF, CHARGE=0, MULT=1
HF=7.2, IE=8.8
1    C      0.000000000     0.000000000     0.000000000
2    C      0.000000000     2.843331540     0.000000000
3    C      0.000000000     0.708441796     1.213570731
4    C     -0.000000000     0.708365681    -1.213321908
5    C      0.000000000     2.113982663     1.213015713
6    C      0.000000000     2.114080596    -1.212750681
7    H      0.000000000    -1.090120211     0.000205869
8    H     -0.000000000     0.167949679     2.160755727
9    H     -0.000000000     0.168426049    -2.160843110
10   H      0.000000000     2.637059371     2.171023889
11   H      0.000000000     2.636617422    -2.171082607
12   C     -0.000000000     4.356453391     0.000165333
13   C     -1.396153862     4.989814035     0.000234538
14   H      0.563892172     4.733589476    -0.885102215
15   H      0.564277428     4.733579507     0.885197499
16   H     -1.979046423     4.698712773    -0.896721517
17   H     -1.315495528     6.096356646     0.002211513
18   H     -1.980786762     4.695589722     0.895039443
EXPGEOM
1    C      0.00000    -0.23240     0.49000
2    C      1.20080    -0.23430    -0.22720
3    C      1.20380    -0.23430    -1.62050
4    C      0.00000    -0.23370    -2.32280
5    C     -1.20380    -0.23430    -1.62050
6    C     -1.20080    -0.23430    -0.22720
7    C      0.00000    -0.19500     2.00330
8    C      0.00000     1.23890     2.56280
9    H      2.14400    -0.24030     0.31250
10   H      2.14740    -0.23970    -2.15750
11   H      0.00000    -0.23750    -3.40800
12   H     -2.14740    -0.23970    -2.15750
13   H     -2.14400    -0.24030     0.31250
14   H     -0.87870    -0.73000     2.38110
15   H      0.87870    -0.73000     2.38110
16   H      0.88330     1.79050     2.22580
17   H     -0.88330     1.79050     2.22580
18   H      0.00000     1.23230     3.65760
---
H10C8, RHF, CHARGE=0, MULT=1
HF=4.1
1    C      0.000000000     0.000000000     0.000000000
2    C      0.000000000     2.833346480     0.000000000
3    C      0.000000000     0.707689060     1.213774734
4    C      0.000147593     0.688688745    -1.232993195
5    C      0.000173018     2.110570485     1.216782118
```

```
  6     C      0.000530448     2.104462698    -1.209835889
  7     H     -0.004809505    -1.091007038     0.025091012
  8     H     -0.000037846     0.163400808     2.159029438
  9     C     -0.000066642    -0.047853894    -2.545855726
 10     H      0.000318898     2.638963946     2.171736604
 11     H      0.000923000     2.647630639    -2.157815531
 12     C      0.003788274     4.338714566     0.010701456
 13     H      0.911497252     4.728184323     0.518372684
 14     H     -0.012507115     4.771088757    -1.010325835
 15     H     -0.883862537     4.733617539     0.548586207
 16     H     -0.899783878     0.208972738    -3.144028646
 17     H      0.895803748     0.213941704    -3.147631710
 18     H      0.003281750    -1.148951335    -2.414512460
EXPGEOM
 1     C      0.00000      1.21900      0.27200
 2     C      0.00000      0.00000      0.94720
 3     C      0.00000     -1.22570      0.26670
 4     C      0.00000     -1.21090     -1.12700
 5     C      0.00000      0.00100     -1.82130
 6     C      0.00000      1.20540     -1.13000
 7     C      0.00000      2.53100      1.02420
 8     C      0.00000     -2.52550      1.04000
 9     H      0.00000     -0.00220      2.03490
10     H      0.00000     -2.14770     -1.67650
11     H      0.00000     -0.00070     -2.90710
12     H      0.00000      2.14480     -1.67590
13     H      0.88120      3.13170      0.77360
14     H      0.00000      2.37040      2.10550
15     H     -0.88120      3.13170      0.77360
16     H      0.88130     -2.60310      1.68620
17     H      0.00000     -3.38770      0.36810
18     H     -0.88130     -2.60310      1.68620
---
H10C8, RHF, CHARGE=0, MULT=1
HF=4.6
 1     C      0.000000000     0.000000000     0.000000000
 2     C      0.000000000     2.839208400     0.000000000
 3     C      0.000000000     0.712703820     1.208657319
 4     C     -0.001173207     0.699539787    -1.216389993
 5     C      0.000087994     2.128350630     1.236246646
 6     C     -0.000690543     2.102746568    -1.209025445
 7     H      0.002756453    -1.090303327     0.006646580
 8     H      0.000487583     0.147708404     2.143112337
 9     H     -0.001746029     0.158028916    -2.162774781
10     C     -0.001325596     2.830839916     2.570674204
11     H     -0.001303810     2.624508270    -2.168332043
12     C      0.003661904     4.345805336    -0.064261365
13     H      0.903280374     4.768284063     0.430798325
14     H      0.005660629     4.722704017    -1.107685538
15     H     -0.895135459     4.772301284     0.428855650
16     H      0.897165604     3.472190390     2.689905248
17     H     -0.901240309     3.470508636     2.688216199
18     H     -0.001417353     2.118521228     3.421205469
---
H10C8, RHF, CHARGE=0, MULT=1
HF=4.3
 1     C      0.000006960    -0.038857980     0.001134050
 2     C      0.002604780     2.830635610     0.006088140
 3     C      0.003101040     0.692288880     1.212985530
```

```
  4     C        -0.000971170     0.694994320    -1.205411760
  5     C         0.003994960     2.095308630     1.215205030
  6     C        -0.000119340     2.101012290    -1.202708940
  7     C        -0.001772920    -1.543144260     0.013841490
  8     H         0.004078660     0.166442020     2.169717750
  9     H        -0.003216610     0.176718390    -2.166064410
 10     H         0.006241510     2.617777840     2.174121630
 11     H        -0.001082580     2.622525990    -2.161954870
 12     C         0.005064770     4.334914110     0.024047060
 13     H         0.918316210     4.722649400     0.522906680
 14     H        -0.023883050     4.771728180    -0.994672680
 15     H        -0.876300500     4.726678820     0.574587120
 16     H        -0.900340920    -1.933330850     0.536507630
 17     H        -0.001401640    -1.976421280    -1.006803570
 18     H         0.894859930    -1.936206960     0.538197560
EXPGEOM
 1      C        -0.00690      0.00010     1.41540
 2      C        -0.00790     -1.19230     0.69590
 3      C        -0.00800     -1.19230    -0.69440
 4      C        -0.00690     -0.00010    -1.41540
 5      C        -0.00790      1.19230    -0.69590
 6      C        -0.00800      1.19230     0.69440
 7      H        -0.01350     -2.13360     1.23150
 8      H        -0.01360     -2.13380    -1.22920
 9      H        -0.01350      2.13360    -1.23150
 10     H        -0.01360      2.13380     1.22920
 11     C         0.01540      0.00330     2.92490
 12     C         0.01540     -0.00330    -2.92490
 13     H         1.03960      0.00500     3.30500
 14     H        -0.48440      0.88550     3.32520
 15     H        -0.48290     -0.87760     3.33080
 16     H         1.03960     -0.00500    -3.30500
 17     H        -0.48440     -0.88550    -3.32520
 18     H        -0.48290      0.87760    -3.33080
---
H12C8, RHF, CHARGE=0, MULT=1
HF=13.7
 1      C         0.000000000     0.000000000     0.000000000
 2      C         0.000000000     1.345977795     0.000000000
 3      C         0.000000000     2.302089968     1.160047542
 4      C         1.279139082     2.296002805     2.025186770
 5      C         1.267839616     1.342378887     3.187170368
 6      C         1.269111430    -0.003581521     3.187113056
 7      C         1.276459512    -0.956430259     2.024413160
 8      C        -0.003261911    -0.955973778     1.160162563
 9      H        -0.013567961    -0.506157263    -0.973434700
 10     H        -0.014179431     1.852053826    -0.973473706
 11     H        -0.132317099     3.334646368     0.757842179
 12     H        -0.896335787     2.120687245     1.797826903
 13     H         2.173553637     2.103873618     1.387913092
 14     H         1.422670627     3.327857709     2.425421019
 15     H         1.274216813     1.849511893     4.160076200
 16     H         1.277105422    -0.511057483     4.159783360
 17     H         2.171526637    -0.767755206     1.387007403
 18     H         1.415918839    -1.989244170     2.423616845
 19     H        -0.140860897    -1.987813232     0.757760322
 20     H        -0.898279762    -0.770807967     1.798692292
---
H12C8, RHF, CHARGE=0, MULT=1
```

```
HF=16.6
1    C     0.000000000     0.000000000     0.000000000
2    C     0.000000000     1.345620390     0.000000000
3    C     0.000000000     2.183817369     1.248919883
4    C    -0.413874314     1.396646034     2.523878426
5    C     0.184219646    -0.036283131     2.548520310
6    C    -0.047656583    -0.833837821     1.249420156
7    C    -0.067440596     2.162017983     3.782317357
8    C    -0.939608587     2.657551033     4.671963435
9    H     0.017644369    -0.561944490    -0.936694465
10   H     0.021767368     1.907480979    -0.936779985
11   H     1.016232868     2.626169047     1.371392840
12   H    -0.696559251     3.043095559     1.115015750
13   H    -1.527195675     1.278165865     2.478519502
14   H     1.278333941     0.011327607     2.751267648
15   H    -0.260732543    -0.600463970     3.398976891
16   H     0.714848112    -1.643931217     1.182061343
17   H    -1.036617201    -1.348149786     1.295414846
18   H     1.004749055     2.309117939     3.954466922
19   H    -2.021602819     2.565742715     4.593302214
20   H    -0.616847343     3.201172802     5.559523248
---
H14C8, RHF, CHARGE=0, MULT=1
HF=19.3
1    C     0.000000000     0.000000000     0.000000000
2    C     0.000000000     1.531265150     0.000000000
3    C     0.000000000     2.177934143     1.398814286
4    C    -0.001290016     3.719984658     1.381361131
5    C     0.002858352     4.365349893     2.781648462
6    C     0.000407829     5.910495779     2.763225008
7    C     0.008543596     6.497359450     4.090799660
8    C     0.014831631     6.992981456     5.180320047
9    H     0.892242577    -0.412920007     0.512837292
10   H    -0.899600572    -0.412772929     0.499914066
11   H     0.007748808    -0.382907593    -1.041316052
12   H     0.891815608     1.881143181    -0.570148744
13   H    -0.890204236     1.882280451    -0.571582569
14   H     0.892976628     1.820024018     1.961429862
15   H    -0.890896200     1.818334939     1.963271687
16   H     0.888800814     4.077132969     0.813815457
17   H    -0.894827224     4.076235027     0.818917491
18   H     0.898207897     4.014494382     3.344239776
19   H    -0.886502949     4.011968126     3.351706659
20   H     0.889660852     6.286166548     2.201892877
21   H    -0.897381938     6.282311609     2.213148742
22   H     0.020376513     7.431931696     6.134908033
---
H14C8, RHF, CHARGE=0, MULT=1
HF=-4.6
1    C     0.000000000     0.000000000     0.000000000
2    C     0.000000000     1.508929723     0.000000000
3    C     0.000000000     2.245031920     1.138600127
4    C     0.054481614     1.748445711     2.523346238
5    C    -0.945553753     1.716057914     3.437904672
6    C    -2.360013981     2.156161918     3.157657355
7    C    -0.005840333     2.175604852    -1.354325557
8    C    -0.671592366     1.215036142     4.837648275
9    H     0.033042549    -0.425286758    -1.023948154
10   H     0.881602895    -0.400988938     0.543458998
```

```
11   H    -0.913882827   -0.399421906    0.488102817
12   H    -0.003479440    3.341188302    1.081164654
13   H     1.058155102    1.403248554    2.806323246
14   H    -3.045470800    1.281993477    3.159600145
15   H    -2.475400905    2.661380451    2.178066454
16   H    -2.714089492    2.870812976    3.930579709
17   H    -0.901834839    1.875913659   -1.938027710
18   H     0.894775007    1.890331617   -1.938120832
19   H    -0.015150568    3.282755562   -1.289186202
20   H    -1.601548912    1.037101024    5.414886688
21   H    -0.112293254    0.256164494    4.824526390
22   H    -0.066659643    1.953573811    5.405706356
---
H14C8, RHF, CHARGE=0, MULT=1
HF=15.2
1    C    -0.463990761   -8.826498711    1.692735392
2    C    -0.301345370   -7.323704946    1.451126127
3    H    -1.062806177   -9.001728227    2.610223081
4    H     0.513445635   -9.331122879    1.831551104
5    H    -0.987861201   -9.325225196    0.852455397
6    H     0.179095715   -6.873876067    2.350759293
7    H    -1.316190697   -6.867939907    1.381545107
8    C     0.513763858   -6.955424289    0.195969864
9    H     1.527018191   -7.412567234    0.273001088
10   C     0.658614648   -5.437250704   -0.028470807
11   H     0.032278492   -7.413016976   -0.698715122
12   H     1.143142276   -4.979645823    0.864414210
13   H    -0.354366393   -4.980048228   -0.106126677
14   C     1.472945153   -5.055812500   -1.285110984
15   H     2.498076447   -5.494456652   -1.222550530
16   C     1.591304293   -3.622200483   -1.472120808
17   H     0.999126095   -5.497064631   -2.194895779
18   C     1.697528179   -2.439401004   -1.638756543
19   C     1.821131907   -1.013749828   -1.831910879
20   H     2.331857388   -0.539186009   -0.966363368
21   H     0.821768233   -0.539911397   -1.941186914
22   H     2.410526476   -0.783610978   -2.745340968
---
H14C8, RHF, CHARGE=0, MULT=1
HF=-0.5
1    C     0.000000000    0.000000000    0.000000000
2    C     0.000000000    1.496717650    0.000000000
3    C     0.000000000    2.333481745    1.066917788
4    C     0.003873393    1.872998739    2.503786533
5    C    -0.005875099    3.815669576    0.853072149
6    C     1.167436084    4.486579452    0.745673643
7    C     1.376035825    5.954146055    0.538063060
8    C    -1.364450892    4.467167093    0.772488777
9    H     0.906848124   -0.409743514    0.493060340
10   H    -0.888478118   -0.412790681    0.522320194
11   H    -0.016092348   -0.393085532   -1.037769274
12   H     0.000483010    1.925607003   -1.009285878
13   H     0.004822003    2.720871134    3.218549691
14   H     0.904289675    1.261290619    2.724236737
15   H    -0.891479093    1.256747657    2.731568214
16   H     2.106874684    3.925227670    0.815174561
17   H     0.980918166    6.547425987    1.389715757
18   H     0.881975796    6.315889826   -0.388036425
19   H     2.456988593    6.188455487    0.446785267
```

```
20      H    -1.512256692     4.959005923    -0.212323093
21      H    -2.191006580     3.738768648     0.898608612
22      H    -1.487856765     5.238307462     1.562032017
---
H14C8, RHF, CHARGE=0, MULT=1
HF=0.7
1       C     0.000000000     0.000000000     0.000000000
2       C     0.000000000     1.496383490     0.000000000
3       C     0.000000000     2.350105245     1.051580183
4       C     0.017735173     3.848288164     0.847580962
5       C    -0.013166924     1.905522025     2.480504588
6       C    -1.196981484     1.710955217     3.111447351
7       C    -1.468042668     1.303237565     4.524919770
8       C     1.338100175     1.723570246     3.127816389
9       H     1.020759005    -0.380882270    -0.219965935
10      H    -0.326201232    -0.450487609     0.958000092
11      H    -0.679048256    -0.383469502    -0.790939835
12      H     0.015802515     1.907906553    -1.017407017
13      H     0.956574936     4.290234945     1.242849620
14      H    -0.831917952     4.333148336     1.372170728
15      H    -0.053320909     4.133779557    -0.221853680
16      H    -2.125521969     1.869706402     2.547152965
17      H    -0.558907387     1.078959359     5.116660025
18      H    -2.105991196     0.393613222     4.543038558
19      H    -2.017341202     2.110358860     5.055659323
20      H     2.172036370     1.853568726     2.408324913
21      H     1.491668055     2.464009633     3.941437493
22      H     1.444548831     0.707191792     3.561537653
---
H14C8, RHF, CHARGE=0, MULT=1
HF=-0.9
1       C     0.000000000     0.000000000     0.000000000
2       C     0.000000000     1.496174733     0.000000000
3       C     0.000000000     2.351547937     1.051294428
4       C     0.024270546     3.847451518     0.825473024
5       C     0.010111568     1.912512405     2.477252508
6       C    -1.111691949     1.722974688     3.213991512
7       C    -2.543856409     1.886025763     2.812601227
8       C     1.376752224     1.709940454     3.094039359
9       H     1.005053379    -0.382077596    -0.281121230
10      H    -0.268426556    -0.449945929     0.975968241
11      H    -0.726150222    -0.382463718    -0.748795069
12      H     0.009641438     1.910179578    -1.016627472
13      H     1.024070137     4.172927953     0.467199208
14      H    -0.197202981     4.418715544     1.749494577
15      H    -0.724713084     4.155325756     0.066111169
16      H    -1.007679639     1.416124589     4.262874169
17      H    -2.708188721     1.863950181     1.717211317
18      H    -2.939987944     2.852338438     3.192699734
19      H    -3.160520462     1.073906525     3.253171242
20      H     2.194971677     1.798971474     2.351307695
21      H     1.568017061     2.465669987     3.885212959
22      H     1.463935926     0.704395062     3.556520230
---
H14C8, RHF, CHARGE=0, MULT=1
HF=14.9
1       C     0.000000000     0.000000000     0.000000000
2       C     0.000000000     1.534255350     0.000000000
3       C     0.000000000     2.124062666     1.324414507
```

```
 4    C    -0.000971965     2.623410305     2.414714370
 5    C    -0.011040197     3.252726223     3.721895331
 6    C     0.551861232     2.350579726     4.842710753
 7    C     0.546731589     3.002717759     6.238281400
 8    C     1.101081106     2.119354452     7.359199595
 9    H     0.892129115    -0.407979858     0.516172244
10    H    -0.901000752    -0.407576388     0.500876885
11    H     0.008690067    -0.381887548    -1.041436971
12    H     0.892832038     1.901732243    -0.561877825
13    H    -0.892525589     1.902048988    -0.561997773
14    H    -0.040802872     1.407796017     4.875060215
15    H     1.593454882     2.057094770     4.578097912
16    H     1.062350324     2.661336671     8.326771089
17    H    -0.494576571     3.298830547     6.504137998
18    H     1.138417950     3.946991098     6.208472262
19    H     2.157730649     1.836302223     7.177866979
20    H     0.514800400     1.185554240     7.476977446
21    H    -1.059854066     3.550430328     3.964680833
22    H     0.577457683     4.200302910     3.665844372
---
H14C8, RHF, CHARGE=0, MULT=1
HF=14.4
 1    C     0.000000000     0.000000000     0.000000000
 2    C     0.000000000     1.530855510     0.000000000
 3    C     0.000000000     2.177231501     1.402026788
 4    C     0.000482028     3.627511132     1.365034687
 5    C    -0.000571314     4.826655143     1.347729164
 6    C     0.000223403     6.276857348     1.309084087
 7    C     0.044091235     6.931514451     2.706474877
 8    C     0.044114220     8.461985076     2.696247068
 9    H     0.893064353    -0.413071306     0.511467364
10    H    -0.899082361    -0.412771611     0.501053228
11    H     0.005980943    -0.381865447    -1.041778925
12    H     0.891326430     1.884999704    -0.567459790
13    H    -0.890909587     1.886142043    -0.567251739
14    H     0.893176200     1.832997315     1.977280889
15    H    -0.893459241     1.833655915     1.977146525
16    H     0.875930871     6.616348846     0.704448345
17    H    -0.909666763     6.619655515     0.759489715
18    H     0.081003174     8.850619037     3.734943493
19    H     0.952493383     6.577380681     3.246183299
20    H    -0.828761352     6.578636150     3.302608077
21    H    -0.869375709     8.871534226     2.219250586
22    H     0.921759366     8.871267604     2.155843091
---
H14C8, RHF, CHARGE=0, MULT=1
HF=-24.1, IE=9.45
 1    C     0.000000000     0.000000000     0.000000000
 2    C     0.000000000     2.593839960     0.000000000
 3    C     0.000000000     2.069397029     1.464825726
 4    C     0.000000000     0.520920431     1.450349132
 5    C    -1.256039193     0.520920431    -0.725174566
 6    C     1.256039193     0.520920431    -0.725174566
 7    C     1.256039193     2.072919529    -0.725174566
 8    C    -1.256039193     2.072919529    -0.725174566
 9    H     0.000000000     3.706890840    -0.000000000
10    H    -0.000000000    -1.111038190    -0.000000000
11    H    -0.884751830     2.458202781     2.015438738
12    H     0.882192390     2.458202781     2.015438738
```

```
13      H        0.882192390     0.135612304     1.989150219
14      H       -0.882192390     0.135612304     1.989150219
15      H       -2.163750817     0.135612304    -0.230574089
16      H       -1.281558427     0.135612304    -1.758576130
17      H        1.281558427     0.135612304    -1.758576130
18      H        2.163750817     0.135612304    -0.230574089
19      H        2.163750817     2.458227656    -0.230574089
20      H        1.281558427     2.458227656    -1.758576130
21      H       -1.281558427     2.458227656    -1.758576130
22      H       -2.163750817     2.458227656    -0.230574089
EXPGEOM
1       C        1.29810     0.00000     0.00000
2       C       -1.29810     0.00000     0.00000
3       C        0.77940     0.00000     1.45130
4       C        0.77940     1.25680    -0.72560
5       C        0.77940    -1.25680    -0.72560
6       C       -0.77940     0.00000     1.45130
7       C       -0.77940    -1.25680    -0.72560
8       C       -0.77940     1.25680    -0.72560
9       H        2.39340     0.00000     0.00000
10      H       -2.39340     0.00000     0.00000
11      H        1.16670     0.87750     1.98060
12      H        1.16670    -0.87750     1.98060
13      H        1.16670     1.27650    -1.75020
14      H        1.16670     2.15400    -0.23040
15      H        1.16670    -2.15400    -0.23040
16      H        1.16670    -1.27650    -1.75020
17      H       -1.16670    -0.87750     1.98060
18      H       -1.16670     0.87750     1.98060
19      H       -1.16670    -1.27650    -1.75020
20      H       -1.16670    -2.15400    -0.23040
21      H       -1.16670     2.15400    -0.23040
22      H       -1.16670     1.27650    -1.75020
---
H16C8, RHF, CHARGE=0, MULT=1
HF=-19.8
1       C       -0.549376174    -5.473707169    -0.922926314
2       C        0.330400625    -4.475078538    -0.765594511
3       C        0.347051336    -3.167977363    -1.511325285
4       C        0.110343154    -1.943283479    -0.602323057
5       C        0.087049030    -0.597269068    -1.354375465
6       C        0.004854805     0.638740506    -0.435477840
7       C       -0.104727028     1.976300621    -1.193280538
8       C       -0.139797156     3.217936244    -0.297864275
9       H       -1.389535246    -5.451310644    -1.614789129
10      H       -0.482703649    -6.401658749    -0.356052127
11      H        1.146466920    -4.592056894    -0.043427456
12      H       -0.410330055    -3.174781833    -2.328332849
13      H        1.338251682    -3.075346234    -2.015514052
14      H        0.903462296    -1.917651735     0.179685385
15      H       -0.856117394    -2.075245096    -0.063753697
16      H       -0.779934857    -0.591145970    -2.054294888
17      H        1.000153029    -0.514006844    -1.987509457
18      H        0.906358740     0.664628586     0.219159981
19      H       -0.871965270     0.536372750     0.244317743
20      H       -1.024422141     1.970367271    -1.823075381
21      H        0.752375428     2.075960324    -1.898944239
22      H       -1.000266202     3.199392812     0.401220920
23      H       -0.234167841     4.133863507    -0.917153741
```

```
 24    H     0.785706104    3.318095845    0.304733295
---
H16C8, RHF, CHARGE=0, MULT=1
HF=-41.1
 1    C     0.000000000    0.000000000    0.000000000
 2    C     0.000000000    1.550736700    0.000000000
 3    C     0.000000000    2.180841048    1.404670652
 4    C    -1.058372360    1.598122447    2.354768063
 5    C    -1.113822273    0.062424224    2.348010028
 6    C    -1.098210583   -0.564047292    0.941866650
 7    C    -0.112195933   -0.613196708   -1.422795583
 8    C     1.135784187   -0.513895609   -2.305814075
 9    H     0.983782722   -0.331061699    0.420962707
 10   H    -0.878364969    1.933824686   -0.568160281
 11   H     0.899443466    1.923314549   -0.540493424
 12   H    -0.167100896    3.277697737    1.305285040
 13   H     1.010260798    2.070448182    1.861756104
 14   H    -0.852487691    1.951339431    3.390691847
 15   H    -2.060287001    2.007929082    2.090232433
 16   H    -2.037632035   -0.268544863    2.875575581
 17   H    -0.264483132   -0.342787862    2.944745944
 18   H    -0.951645478   -1.662628421    1.057002982
 19   H    -2.102933921   -0.435596620    0.478219876
 20   H    -0.362310831   -1.696452817   -1.329115189
 21   H    -0.968915308   -0.152784918   -1.966660551
 22   H     1.409248703    0.534549484   -2.538033947
 23   H     2.015115523   -0.996464725   -1.832730450
 24   H     0.955491760   -1.028205705   -3.272841809
---
H18C8, RHF, CHARGE=0, MULT=1
HF=-54
 1    C     0.000000000    0.000000000    0.000000000
 2    C     0.000000000    1.561588258    0.000000000
 3    C     0.000000000    2.164809176    1.489636592
 4    C     0.249902430    3.706048733    1.496380289
 5    C    -1.359750363    1.924717309    2.218681207
 6    C     1.110060110    1.522848763    2.380455383
 7    C     1.254322586    2.011321110   -0.814030589
 8    C    -1.252714819    2.013637314   -0.815142649
 9    H     0.945743695   -0.424508106    0.392397027
 10   H    -0.828894658   -0.429339122    0.596343818
 11   H    -0.115786043   -0.399513532   -1.030001618
 12   H     1.274986819    3.970928198    1.168084556
 13   H    -0.459080913    4.259108087    0.849523847
 14   H     0.134080220    4.125379164    2.518477963
 15   H    -2.190990365    2.503009355    1.767663913
 16   H    -1.661918960    0.859174726    2.224437365
 17   H    -1.303609488    2.240615179    3.282166904
 18   H     0.915929778    0.452829478    2.594939609
 19   H     2.118029667    1.593566951    1.926649071
 20   H     1.174063424    2.030420529    3.366497762
 21   H     1.250956120    3.097909455   -1.033030513
 22   H     2.206687727    1.779537606   -0.298135107
 23   H     1.297871151    1.496749787   -1.797530055
 24   H    -2.195147049    1.587994337   -0.416083280
 25   H    -1.372269151    3.114549558   -0.842948155
 26   H    -1.184542370    1.679367610   -1.872231582
EXPGEOM
 1    C     0.80660      0.00000     0.00000
```

```
  2    C      -0.80660     0.00000     0.00000
  3    C       1.40390    -0.71960     1.23620
  4    C       1.40390    -0.71080    -1.24130
  5    C       1.40390     1.43040     0.00510
  6    C      -1.40390     0.71080    -1.24130
  7    C      -1.40390    -1.43040     0.00510
  8    C      -1.40390     0.71960     1.23620
  9    H       2.48610    -0.56260     1.26370
 10    H       1.00620    -0.35580     2.18230
 11    H       1.23600    -1.80110     1.19810
 12    H       2.48610    -0.81300    -1.11910
 13    H       1.00620    -1.71210    -1.39930
 14    H       1.23600    -0.13700    -2.15880
 15    H       2.48610     1.37570    -0.14460
 16    H       1.00620     2.06790    -0.78300
 17    H       1.23600     1.93810     0.96080
 18    H      -2.48610     0.81300    -1.11910
 19    H      -1.00620     1.71210    -1.39930
 20    H      -1.23600     0.13700    -2.15880
 21    H      -2.48610    -1.37570    -0.14460
 22    H      -1.00620    -2.06790    -0.78300
 23    H      -1.23600    -1.93810     0.96080
 24    H      -2.48610     0.56260     1.26370
 25    H      -1.00620     0.35580     2.18230
 26    H      -1.23600     1.80110     1.19810
---
H18C8, RHF, CHARGE=0, MULT=1
HF=-52.6
  1    C      -0.000056559    0.000165560    0.000067918
  2    C       1.533896281   -0.000061022   -0.000081205
  3    C       2.206462199    1.406986090    0.000016673
  4    C       3.558426695    1.500623329    0.828173347
  5    C       3.275165210    1.240732364    2.341038976
  6    C       4.630512129    0.486944360    0.337440618
  7    C       4.161548904    2.936522088    0.727645727
  8    C       2.296434035    1.942573840   -1.449170902
  9    H      -0.411849960    0.501269764    0.899549859
 10    H      -0.426777169    0.501051437   -0.891420835
 11    H      -0.376178488   -1.044360039    0.001035415
 12    H       1.850193767   -0.595701420    0.884687290
 13    H       1.885014946   -0.584883773   -0.881831099
 14    H       1.500060117    2.101805288    0.524531956
 15    H       2.471635525    1.900032151    2.728040861
 16    H       2.972409447    0.195469614    2.549037319
 17    H       4.177226679    1.430511876    2.959626407
 18    H       4.301670351   -0.565448965    0.450232914
 19    H       4.891125899    0.635987138   -0.729485014
 20    H       5.573245770    0.588674103    0.914589104
 21    H       4.555216916    3.161985802   -0.283712356
 22    H       3.414617488    3.716827855    0.978886906
 23    H       5.013292445    3.067615779    1.427518190
 24    H       2.515070338    3.029110055   -1.474915630
 25    H       3.068736188    1.427491494   -2.053604093
 26    H       1.330687883    1.816864974   -1.981507905
---
H18C8, RHF, CHARGE=0, MULT=1
HF=-53.6
  1    C       0.000000000    0.000000000    0.000000000
  2    C       0.000000000    1.544163640    0.000000000
```

```
 3    C     0.000000000    2.139954550    1.439706454
 4    C    -1.264893731    2.915443808    1.957003116
 5    C    -1.538239730    4.185501393    1.102989952
 6    C    -2.529862929    2.013429639    1.966892550
 7    C    -0.973156849    3.365175712    3.423136646
 8    C     1.155061010    2.100865857   -0.861043300
 9    H     0.920507039   -0.421988655    0.451676629
10    H    -0.865829830   -0.404153167    0.562426155
11    H    -0.075180913   -0.394068142   -1.034567944
12    H    -0.939395746    1.847163678   -0.525629300
13    H     0.208723133    1.317612650    2.163539562
14    H     0.872176907    2.827898949    1.539118845
15    H    -0.646023778    4.841903894    1.046463612
16    H    -1.831431686    3.938384211    0.063131220
17    H    -2.363450889    4.791713247    1.531046502
18    H    -2.851360380    1.726641135    0.945803896
19    H    -2.360932454    1.077031641    2.536700018
20    H    -3.391601925    2.530644716    2.437932119
21    H    -0.785228475    2.498901720    4.089795537
22    H    -0.084449456    4.026039040    3.482326306
23    H    -1.826365621    3.927560999    3.854830778
24    H     1.101753216    3.205703451   -0.940868064
25    H     2.150490805    1.839197989   -0.448598263
26    H     1.106911049    1.701239038   -1.895075341
---
H18C8, RHF, CHARGE=0, MULT=1
HF=-53.7
 1    C     0.000000000    0.000000000    0.000000000
 2    C     0.000000000    1.531430390    0.000000000
 3    C     0.000000000    2.174433490    1.402958851
 4    C    -0.005597802    3.717828189    1.374751950
 5    C     0.002118384    4.501068565    2.734605183
 6    C    -1.254972294    4.170024236    3.586926331
 7    C     1.282134509    4.190275209    3.560203824
 8    C    -0.013745165    6.025163327    2.402738473
 9    H     0.891723169   -0.412984576    0.513739465
10    H    -0.900156647   -0.412780453    0.498947804
11    H     0.008702261   -0.383127451   -1.041271380
12    H     0.891816220    1.881685446   -0.569647767
13    H    -0.890304234    1.882737404   -0.570973057
14    H    -0.890422195    1.803677893    1.958977174
15    H     0.892849321    1.808818943    1.958029034
16    H     0.878606929    4.050334402    0.781539622
17    H    -0.900994303    4.044397523    0.795361909
18    H    -2.192109776    4.330514604    3.015772086
19    H    -1.258241905    3.119613720    3.940838762
20    H    -1.311124017    4.810752427    4.491507945
21    H     1.312229837    3.138215634    3.908047058
22    H     2.204093971    4.368363097    2.969687292
23    H     1.345878828    4.828221409    4.466044605
24    H     0.869498657    6.325289149    1.802703815
25    H    -0.916404648    6.311391492    1.825256786
26    H    -0.006958611    6.644093381    3.323402169
---
H18C8, RHF, CHARGE=0, MULT=1
HF=-51.7
 1    C     0.000000000    0.000000000    0.000000000
 2    C     0.000000000    1.533843070    0.000000000
 3    C     0.000000000    2.212112151    1.404207535
```

```
  4    C      0.764702190    3.603232642    1.470501487
  5    C      2.283039840    3.392872203    1.176091494
  6    C      0.198076707    4.642575809    0.462510948
  7    C      0.669575639    4.211401045    2.904719847
  8    C     -1.438450103    2.236210165    1.974149742
  9    H      0.870463892   -0.415403375    0.546944443
 10    H     -0.918237557   -0.425630488    0.451296410
 11    H      0.054690927   -0.374429796   -1.043740025
 12    H      0.885908504    1.850399463   -0.593784250
 13    H     -0.882014445    1.882833657   -0.585851915
 14    H      0.574624530    1.535765163    2.089222164
 15    H      2.726960969    2.621248587    1.837715783
 16    H      2.480711235    3.084055220    0.130678208
 17    H      2.859273084    4.328257524    1.335911530
 18    H      0.326371211    4.323496510   -0.591132187
 19    H     -0.882373412    4.833401492    0.619153079
 20    H      0.713823596    5.620928229    0.561899703
 21    H     -0.349471035    4.575628757    3.144770427
 22    H      0.958587617    3.478664729    3.685312420
 23    H      1.344162891    5.085936867    3.018453204
 24    H     -1.456720691    2.510744892    3.047819134
 25    H     -2.104960042    2.937976773    1.434748334
 26    H     -1.906701191    1.231839409    1.912450708
---
H18C8, RHF, CHARGE=0, MULT=1
HF=-52
  1    C      0.000000000    0.000000000    0.000000000
  2    C      0.000000000    1.541884910    0.000000000
  3    C      0.000000000    2.247495440    1.406304274
  4    C     -1.441742490    2.466415256    1.996763944
  5    C     -2.000139161    1.341016667    2.891613041
  6    C     -1.569019124    3.834702079    2.707413701
  7    C      1.121693081    2.084503361   -0.916784706
  8    C      1.020922937    1.671785115    2.408844114
  9    H      0.945914480   -0.432112249    0.382901091
 10    H     -0.830162857   -0.413712338    0.604773663
 11    H     -0.140594015   -0.384480564   -1.032510701
 12    H     -0.954759708    1.839144623   -0.507223091
 13    H      0.375154993    3.284576312    1.194765827
 14    H     -2.142813646    2.527378660    1.124347377
 15    H     -1.943034466    0.350082020    2.401203791
 16    H     -1.476930665    1.266051317    3.865866928
 17    H     -3.073833166    1.522599008    3.110961679
 18    H     -0.913031743    3.918948131    3.596483122
 19    H     -1.315537629    4.668401168    2.020648436
 20    H     -2.611744065    4.006861638    3.046216549
 21    H      1.067567030    3.188114992   -1.014278946
 22    H      2.135379505    1.828295654   -0.549620071
 23    H      1.025459535    1.668768890   -1.941405408
 24    H      1.059794981    2.284926980    3.332933192
 25    H      0.801473995    0.630012080    2.714148149
 26    H      2.045145124    1.681435695    1.982132618
---
H18C8, RHF, CHARGE=0, MULT=1
HF=-51.1
  1    C      0.000000000    0.000000000    0.000000000
  2    C      0.000000000    1.531944250    0.000000000
  3    C      0.000000000    2.170375627    1.403920098
  4    C     -0.082521359    3.726517317    1.437592474
```

```
5    C    -0.678941735    4.251492517    2.791517086
6    C     0.264149431    4.159060478    4.008494448
7    C    -1.292390480    5.664299698    2.667030277
8    C     1.245573468    4.392700703    1.022013310
9    H     0.888849154   -0.414172630    0.517560351
10   H    -0.903498214   -0.413911976    0.491767479
11   H     0.014358763   -0.380325871   -1.042445760
12   H     0.888301143    1.877395047   -0.576368170
13   H    -0.892802522    1.881019102   -0.568833841
14   H    -0.866952704    1.751592435    1.965591861
15   H     0.909059423    1.833019698    1.952096553
16   H    -0.829822140    4.018720991    0.654677088
17   H    -1.549013650    3.584088014    3.029515672
18   H     0.676348489    3.137774012    4.133361407
19   H     1.119276039    4.860958675    3.942330367
20   H    -0.285028974    4.397715871    4.943523035
21   H    -0.532912842    6.451466859    2.490726046
22   H    -2.026170165    5.713435671    1.836570857
23   H    -1.838702883    5.935458868    3.594386742
24   H     1.202644998    5.496205338    1.117745435
25   H     2.107079463    4.039806099    1.623797079
26   H     1.477866137    4.181244768   -0.042217115
---
H18C8, RHF, CHARGE=0, MULT=1
HF=-52.4
1    C     0.000000000    0.000000000    0.000000000
2    C     0.000000000    1.532209640    0.000000000
3    C     0.000000000    2.227734882    1.391644547
4    C     0.378453786    3.734218709    1.256741594
5    C     1.044156612    4.406506072    2.492082767
6    C     2.538639483    4.055685990    2.644001815
7    C     0.822652772    5.935076395    2.486663209
8    C    -1.325335846    2.026714402    2.156657299
9    H     0.819929920   -0.417782196    0.618228608
10   H    -0.955295065   -0.422565358    0.370365692
11   H     0.141688882   -0.375799408   -1.035042417
12   H     0.906944295    1.856023060   -0.563165249
13   H    -0.871689066    1.887077921   -0.596565659
14   H     0.794159876    1.726263887    1.999597953
15   H     1.062231472    3.868122166    0.387060816
16   H    -0.544100878    4.299421679    0.989003699
17   H     0.536187440    4.030562709    3.416310979
18   H     2.693031123    2.964190467    2.760483029
19   H     3.138691951    4.387656336    1.772661551
20   H     2.968068391    4.536353960    3.547393017
21   H     1.274138567    6.422337630    1.598967278
22   H    -0.258207173    6.184000311    2.495042089
23   H     1.267728873    6.404624892    3.388004985
24   H    -1.300094049    2.517082096    3.150844330
25   H    -2.196010787    2.435769472    1.605074383
26   H    -1.524235328    0.951967050    2.344204286
---
H18C8, RHF, CHARGE=0, MULT=1
HF=-53.2
1    C    -0.756400416   -4.746689202   -1.720631517
2    C    -0.347695912   -3.507955803   -0.893887724
3    C    -0.177347576   -2.258845875   -1.804990933
4    C    -0.145248310   -0.898166783   -1.077240690
5    C    -0.228010169    0.357294221   -1.991543928
```

```
6    C     -0.618020801    1.612615739    -1.181229878
7    C      0.879640634   -3.801397046    -0.007692592
8    C      1.046447612    0.601722733    -2.824726737
9    H     -1.686925462   -4.555878562    -2.294160720
10   H     -0.956554763   -5.617132710    -1.062280743
11   H      0.028762250   -5.047039564    -2.443760335
12   H     -1.022803085   -2.244148236    -2.531602912
13   H      0.747837098   -2.378588383    -2.412514107
14   H      0.775069246   -0.830373098    -0.454301374
15   H     -1.002291018   -0.865497910    -0.365027705
16   H     -1.600206266    1.479739575    -0.683406857
17   H     -0.709954593    2.499250370    -1.841172201
18   H      0.127310798    1.853827256    -0.396671839
19   H     -1.061362971    0.190747688    -2.721742221
20   H      0.947268432    1.518255279    -3.442785338
21   H      1.241300553   -0.234795963    -3.526325886
22   H      1.945816172    0.725055133    -2.188306543
23   H      1.807898199   -3.918612470    -0.602534880
24   H      0.736755494   -4.734720126     0.575448476
25   H      1.051474011   -2.990407672     0.728770706
26   H     -1.203694227   -3.304745884    -0.199954574
---
H18C8, RHF, CHARGE=0, MULT=1
HF=-51.5
1    C      0.000000000    0.000000000     0.000000000
2    C      0.000000000    1.531319670     0.000000000
3    C      0.000000000    2.175058971     1.400430369
4    C      0.006482348    3.718124999     1.384079585
5    C     -0.009791989    4.356851948     2.789010436
6    C      0.082980526    5.908236364     2.826355857
7    C     -1.218957531    6.613359596     2.395946316
8    C      0.564043511    6.407878924     4.206113578
9    H      0.890618345   -0.412691715     0.515798522
10   H     -0.901002430   -0.412310356     0.497756867
11   H      0.010545115   -0.383848043    -1.040882997
12   H      0.891983792    1.882115536    -0.569185026
13   H     -0.890146345    1.883383397    -0.570765962
14   H     -0.895581281    1.820037175     1.960606483
15   H      0.888345204    1.811343565     1.966031507
16   H      0.909094807    4.070179623     0.833720205
17   H     -0.874108454    4.073167022     0.802720806
18   H     -0.928622235    4.030735409     3.328176059
19   H      0.847503979    3.938665026     3.366380594
20   H      0.872799302    6.217907842     2.094183782
21   H     -1.488648436    6.367439017     1.349239072
22   H     -2.078725462    6.336353973     3.038862068
23   H     -1.107528831    7.716374752     2.444723420
24   H     -0.129822654    6.119509557     5.021355347
25   H      1.563936340    5.996829928     4.454481487
26   H      0.657731316    7.513087602     4.220509902
---
H18C8, RHF, CHARGE=0, MULT=1
HF=-52.6
1    C      0.000000000    0.000000000     0.000000000
2    C      0.000000000    1.533262210     0.000000000
3    C      0.000000000    2.161601472     1.409268736
4    C     -0.002192468    3.729767007     1.556750434
5    C     -0.000678372    4.019084438     3.103661461
6    C     -0.003007377    5.465892458     3.609658228
```

```
  7      C      1.264656009    4.340450256    0.895701434
  8      C     -1.271059770    4.337478124    0.897152144
  9      H      0.891864744   -0.413425241    0.513094514
 10      H     -0.900049125   -0.413449531    0.498507083
 11      H      0.008687489   -0.382574317   -1.041396872
 12      H      0.892530653    1.873578341   -0.572441781
 13      H     -0.889860664    1.874918301   -0.575174631
 14      H     -0.891484022    1.765066002    1.950205556
 15      H      0.890635332    1.765909746    1.951678039
 16      H     -0.889461709    3.518149710    3.556298034
 17      H      0.889658482    3.520416259    3.555242549
 18      H      0.885743643    6.037254640    3.276047341
 19      H     -0.904895759    6.027454431    3.295236980
 20      H      0.009554751    5.466801880    4.720313626
 21      H      1.301718720    5.443149995    1.007350494
 22      H      2.196632222    3.933105173    1.338102404
 23      H      1.303616411    4.141061223   -0.194393314
 24      H     -2.201603504    3.927904223    1.340466866
 25      H     -1.310955322    5.439929210    1.010788969
 26      H     -1.310756438    4.139232641   -0.193088201
---
H18C8, RHF, CHARGE=0, MULT=1
HF=-50.9
  1      C      0.000000000    0.000000000    0.000000000
  2      C      0.000000000    1.534143280    0.000000000
  3      C      0.000000000    2.223232079    1.398211520
  4      C      0.509920767    3.713919890    1.357908762
  5      C      0.533724805    4.395465847    2.760208718
  6      C      0.456911455    5.927448580    2.764047043
  7      C      1.892079739    3.866160892    0.678492754
  8      C     -1.394814279    2.082788690    2.054529566
  9      H      0.809145031   -0.417678593    0.632465812
 10      H     -0.960151924   -0.426657154    0.352073655
 11      H      0.160489894   -0.373303078   -1.033356741
 12      H      0.895320986    1.840216761   -0.585459106
 13      H     -0.875384343    1.890999702   -0.590484574
 14      H      0.717994313    1.659568463    2.045442455
 15      H     -0.219965758    4.278719757    0.724948540
 16      H     -0.316856392    4.041880311    3.383778962
 17      H      1.450239637    4.080411101    3.310801417
 18      H      1.368324478    6.402859896    2.350590965
 19      H     -0.413791822    6.301318768    2.188404371
 20      H      0.345304707    6.292710879    3.806783290
 21      H      2.624143076    3.116362993    1.040957760
 22      H      1.817504504    3.771309083   -0.423870206
 23      H      2.335775588    4.865912344    0.861769929
 24      H     -1.342848830    2.198092640    3.156111897
 25      H     -2.119977062    2.825564920    1.664764793
 26      H     -1.834785757    1.079773293    1.881584275
---
H18C8, RHF, CHARGE=0, MULT=1
HF=-50.4
  1      C      0.000000000    0.000000000    0.000000000
  2      C      0.000000000    1.532325406    0.000000000
  3      C      0.000000000    2.205992815    1.404638413
  4      C      1.246434694    3.112417722    1.637814493
  5      C      2.583867819    2.375782845    1.854214005
  6      C      3.770311500    3.295947918    2.159788874
  7      C     -1.305391594    3.006726099    1.687796238
```

```
8    C     -2.587065148    2.188965585    1.876840609
9    H      0.897857633   -0.415818118    0.500141933
10   H     -0.893088234   -0.418096555    0.506365168
11   H     -0.003308474   -0.379234058   -1.043180431
12   H      0.880782863    1.872131982   -0.592035755
13   H     -0.878210090    1.876228756   -0.593403217
14   H      0.045877444    1.401938828    2.181164901
15   H      1.350241334    3.824505707    0.787486768
16   H      1.062882818    3.743077804    2.539135954
17   H      2.477557991    1.651457352    2.694696352
18   H      2.840005595    1.770612298    0.955137525
19   H      3.959561099    4.013413288    1.335809778
20   H      3.612831498    3.879567006    3.089370304
21   H      4.693496116    2.695663237    2.296865604
22   H     -1.164694837    3.604108480    2.619680032
23   H     -1.472966394    3.751369508    0.875845021
24   H     -2.902503702    1.671635855    0.949116320
25   H     -2.477184110    1.424242141    2.672221699
26   H     -3.419697014    2.859991804    2.175345616
---
H18C8, RHF, CHARGE=0, MULT=1
HF=-50.5
1    C      0.000000000    0.000000000    0.000000000
2    C      0.000000000    1.533981560    0.000000000
3    C      0.000000000    2.230070242    1.394416995
4    C      0.374304745    3.757913841    1.330075762
5    C      0.550044701    4.394142715    2.729633368
6    C      1.620581823    4.081762721    0.472908909
7    C     -1.352957595    1.982886570    2.134413014
8    C     -1.258581619    1.219188485    3.460191254
9    H      0.834123387   -0.413458287    0.602400145
10   H     -0.944012691   -0.430717086    0.388165343
11   H      0.124457325   -0.372687379   -1.038415894
12   H      0.905909903    1.835091533   -0.573476180
13   H     -0.867672131    1.894817527   -0.598878188
14   H      0.815690926    1.740083197    1.983418816
15   H     -0.481725328    4.288910565    0.840568842
16   H     -0.385700202    4.372270257    3.322311826
17   H      1.338946724    3.888462815    3.322232149
18   H      0.831834993    5.464640435    2.645074969
19   H      2.479224205    3.420078943    0.704669535
20   H      1.401463801    3.995841627   -0.610864774
21   H      1.959301371    5.126616943    0.632535166
22   H     -1.881464402    2.944511206    2.325746136
23   H     -2.057006394    1.423777487    1.475436503
24   H     -0.831886138    0.204816304    3.325058503
25   H     -0.636845067    1.748827077    4.209377734
26   H     -2.270564868    1.096618800    3.899593655
---
H18C8, RHF, CHARGE=0, MULT=1
HF=-51.4
1    C      1.211664303   -5.700319635    0.084225776
2    C      0.395980885   -4.408249714    0.210624380
3    C      0.046894476   -3.841067229    1.635388055
4    C     -0.846929272   -2.558793986    1.463417209
5    C     -0.315372246   -1.364485712    0.662074278
6    C     -0.767915971   -4.928559179    2.426955120
7    C      1.350095671   -3.485636825    2.405351040
8    H      2.201075063   -5.630277088    0.577645954
```

```
 9      H     1.397381578    -5.910392888    -0.990490697
10      H     0.686444679    -6.583961140     0.496995720
11      H    -0.555081475    -4.562624812    -0.351750693
12      H     0.958144444    -3.629826923    -0.356673382
13      H    -1.105833980    -2.164893450     2.474031365
14      H    -1.815371813    -2.861417087     0.999665868
15      H     0.641764000    -0.973796478     1.061010220
16      H    -1.052091425    -0.534320024     0.711158048
17      H    -0.166791982    -1.598594498    -0.410707728
18      C    -1.328673016    -4.602627071     3.816163754
19      H    -0.125471320    -5.830833132     2.555837496
20      H    -1.627323427    -5.255817903     1.795639701
21      H    -0.545489098    -4.277499094     4.529515343
22      H    -1.799087771    -5.514832175     4.241811366
23      H    -2.113106048    -3.820088098     3.789745767
24      H     2.071524468    -2.934136409     1.768644958
25      H     1.874763196    -4.390379793     2.774554696
26      H     1.148956401    -2.847839764     3.290164443
---
H18C8, RHF, CHARGE=0, MULT=1
HF=-50.8
 1      C     0.000000000     0.000000000     0.000000000
 2      C     0.000000000     1.531454150     0.000000000
 3      C     0.000000000     2.175337462     1.401389047
 4      C    -0.006356706     3.718545556     1.379802787
 5      C     0.072303253     4.420789629     2.767746148
 6      C     0.501304466     5.907862877     2.601634706
 7      C     1.008515777     6.620994280     3.859020125
 8      C    -1.226682906     4.268691373     3.586167202
 9      H     0.889639746    -0.413260861     0.517066168
10      H    -0.902171739    -0.412903030     0.495182343
11      H     0.012638531    -0.382442373    -1.041522102
12      H     0.890878523     1.881041259    -0.571509693
13      H    -0.890605707     1.882916546    -0.570571840
14      H    -0.887721847     1.800812302     1.959449476
15      H     0.894862558     1.820312231     1.962241111
16      H     0.862567359     4.048784255     0.764021653
17      H    -0.914620084     4.072881564     0.840746887
18      H     0.882818935     3.910808815     3.347807946
19      H     1.317724864     5.964620110     1.843657195
20      H    -0.346216017     6.494057157     2.177374337
21      H     0.223007268     6.730619117     4.632773760
22      H     1.864244895     6.086604794     4.319446371
23      H     1.355923644     7.642335488     3.597414371
24      H    -1.140114426     4.751460996     4.580586239
25      H    -2.100404105     4.718430814     3.072367535
26      H    -1.462861157     3.202506464     3.778083372
---
H18C8, RHF, CHARGE=0, MULT=1
HF=-50.7
 1      C     0.000000000     0.000000000     0.000000000
 2      C     0.000000000     1.532098990     0.000000000
 3      C     0.000000000     2.169611507     1.403664506
 4      C    -0.117381609     3.722563500     1.448074879
 5      C    -0.566263341     4.197577603     2.861886251
 6      C    -1.165117132     5.616613751     2.937882724
 7      C    -1.702933366     6.003937557     4.318990954
 8      C     1.170016801     4.425234554     0.969376485
 9      H     0.887443244    -0.414372219     0.519902352
```

```
10      H      -0.905032088    -0.414282341     0.488724889
11      H       0.017414284    -0.379484884    -1.042829117
12      H       0.887814662     1.876194570    -0.577648599
13      H      -0.892738322     1.881453874    -0.568827863
14      H      -0.852780043     1.732119260     1.973286326
15      H       0.920309893     1.855267355     1.947529965
16      H      -0.928871357     4.009335379     0.732856154
17      H      -1.331988753     3.487458135     3.252005913
18      H       0.296369527     4.127551367     3.563783619
19      H      -0.401675224     6.371140958     2.640976883
20      H      -1.995276415     5.710815856     2.199956163
21      H      -2.521764554     5.332737211     4.648400886
22      H      -0.909651462     5.979284041     5.093501805
23      H      -2.109563864     7.036344259     4.293186886
24      H       1.057179612     5.528322914     0.971525888
25      H       2.043490641     4.177606822     1.606043100
26      H       1.422990838     4.141086264    -0.072289441
 ---
H18C8, RHF, CHARGE=0, MULT=1
HF=-49.9
1       C       0.000000000     0.000000000     0.000000000
2       C       0.000000000     1.531313370     0.000000000
3       C       0.000000000     2.174919986     1.400122998
4       C      -0.000658191     3.717083934     1.384327671
5       C       0.001935445     4.359694642     2.786475177
6       C       0.000412280     5.901818803     2.770461660
7       C       0.005424415     6.545839814     4.170297197
8       C       0.003088050     8.077108307     4.169686692
9       H       0.008803155     8.461288252     5.210500905
10      H      -0.896774468     8.487679467     3.668518372
11      H       0.895296298     8.490768384     3.657467935
12      H       0.898927753     6.195599682     4.737388129
13      H      -0.882536044     6.193295592     4.744638420
14      H      -0.893647830     6.259905854     2.209780128
15      H       0.889157623     6.261930958     2.202639615
16      H       0.894834044     4.002063856     3.349139996
17      H      -0.888261364     4.000929803     3.352755305
18      H      -0.893447140     4.074382524     0.821303140
19      H       0.889532929     4.075445910     0.817823720
20      H       0.891466054     1.815821300     1.964224959
21      H      -0.891324571     1.815301685     1.964208473
22      H       0.890462002     1.882648941    -0.571092668
23      H      -0.890791667     1.882497342    -0.570757114
24      H       0.001818144    -0.383676332    -1.040977000
25      H      -0.897103687    -0.412134330     0.504798013
26      H       0.895081365    -0.412514792     0.508106047
EXPGEOM
1       C       0.00000     0.00040     0.76710
2       C       0.00000    -0.00040    -0.76710
3       C       0.00000    -1.40570     1.37980
4       C       0.00000     1.40570    -1.37980
5       C       0.00000    -1.40570     2.91370
6       C       0.00000     1.40570    -2.91370
7       C       0.00000    -2.81440     3.51800
8       C       0.00000     2.81440    -3.51800
9       H       0.87750     0.55350     1.12920
10      H      -0.87750     0.55350     1.12920
11      H       0.87750    -0.55350    -1.12920
12      H      -0.87750    -0.55350    -1.12920
```

```
13      H       -0.87740        -1.95960        1.01840
14      H        0.87740        -1.95960        1.01840
15      H       -0.87740         1.95960       -1.01840
16      H        0.87740         1.95960       -1.01840
17      H        0.87700        -0.85270        3.27460
18      H       -0.87700        -0.85270        3.27460
19      H        0.87700         0.85270       -3.27460
20      H       -0.87700         0.85270       -3.27460
21      H        0.00000        -2.78080        4.61220
22      H       -0.88360        -3.38020        3.20230
23      H        0.88360        -3.38020        3.20230
24      H        0.00000         2.78080       -4.61220
25      H       -0.88360         3.38020       -3.20230
26      H        0.88360         3.38020       -3.20230
---
H10C9, RHF, CHARGE=0, MULT=1
HF=28.3
1       C       0.000000000     0.000000000     0.000000000
2       C       0.000000000     1.414999320     0.000000000
3       C       0.000000000     2.133181655     1.208555442
4       C       0.000338965     1.450817916     2.437104679
5       C       0.000031049     0.045421579     2.451431220
6       C      -0.000061700    -0.674419507     1.244154257
7       C      -0.003758228    -0.763018506    -1.280649894
8       C      -1.358906011    -1.056684023    -1.875322530
9       C       1.148186584    -1.161644013    -1.856466608
10      H       0.001704476     1.963940856    -0.943025786
11      H       0.000448071     3.223594012     1.190764245
12      H       0.000988841     2.008084770     3.374126642
13      H      -0.000682211    -0.489523749     3.401765850
14      H      -0.000048022    -1.765088944     1.277426002
15      H      -1.288627111    -1.639177900    -2.816793647
16      H      -1.902834892    -0.117714433    -2.109342257
17      H      -1.983199731    -1.644751392    -1.170516299
18      H       1.192290447    -1.722302082    -2.789335691
19      H       2.132083454    -0.955537190    -1.436926294
---
H10C9, RHF, CHARGE=0, MULT=1
HF=36
1       C       0.000000000     0.000000000     0.000000000
2       C       0.000000000     1.415341570     0.000000000
3       C       0.000000000     2.137703551     1.205884350
4       C      -0.005072110     1.460134645     2.436771808
5       C      -0.014557526     0.054840059     2.453795993
6       C      -0.015206601    -0.667568100     1.248265803
7       C      -0.046499441    -0.768439242    -1.279645004
8       C       0.945500030    -0.570915082    -2.435696728
9       C       0.952930156    -1.873546814    -1.652953408
10      H      -0.004945974     1.965039692    -0.942655838
11      H       0.002647006     3.228061441     1.183789104
12      H      -0.004053028     2.019659481     3.372410018
13      H      -0.023138094    -0.478104337     3.405253021
14      H      -0.031942812    -1.757947980     1.288654313
15      H      -1.095318915    -0.951690451    -1.575091337
16      H       1.786846174     0.123960314    -2.332336373
17      H       0.557587786    -0.557763684    -3.461472026
18      H       0.570362966    -2.788654942    -2.120973093
19      H       1.799144635    -2.101391962    -0.994627232
---
```

```
H18C9, RHF, CHARGE=0, MULT=1
HF=-51.5
1    C     0.000000000     0.000000000     0.000000000
2    C     0.000000000     1.549713390     0.000000000
3    C     0.000000000     2.128354942     1.437747038
4    C    -1.098199284     1.528524840     2.352131212
5    C    -1.091609284    -0.020993612     2.326718461
6    C    -1.095580923    -0.625929682     0.899994008
7    C     1.143399844     2.128394648    -0.857995700
8    C    -1.023313262     2.084976380     3.788564188
9    C    -1.014082659    -2.165877487     0.924142693
10   H    -0.153509721    -0.355225111    -1.044937247
11   H     1.001091709    -0.377089170     0.309785189
12   H    -0.950638504     1.882328795    -0.491768400
13   H    -0.149883704     3.230824114     1.378364989
14   H     1.000664077     1.978402954     1.903234784
15   H    -2.088001899     1.862415148     1.945763751
16   H    -1.991158865    -0.390015201     2.870986324
17   H    -0.211421670    -0.401232307     2.893685339
18   H    -2.085781081    -0.378819993     0.436620245
19   H     1.098531295     3.236514012    -0.885568543
20   H     2.144722140     1.844172088    -0.476060298
21   H     1.071262435     1.773196074    -1.906470462
22   H    -0.068653199     1.832073402     4.292272588
23   H    -1.848882026     1.686369668     4.413189593
24   H    -1.119802096     3.190295320     3.789981656
25   H    -1.066130112    -2.583206558    -0.102387367
26   H    -0.075833552    -2.533248395     1.386766341
27   H    -1.861126570    -2.598688871     1.495025657
---
H18C9, RHF, CHARGE=0, MULT=1
HF=-49.4
1    C     0.000000000     0.000000000     0.000000000
2    C     0.000000000     1.550033600     0.000000000
3    C     0.000000000     2.126393179     1.439223789
4    C    -1.032482486     1.492536430     2.404293497
5    C    -0.976046042    -0.055886009     2.376184631
6    C    -1.027105489    -0.663155649     0.951368882
7    H     0.980549272     1.857685784    -0.451395702
8    C    -0.903772989     2.050785080     3.836493317
9    C    -0.886372787    -2.199270645     0.972237917
10   H    -0.177536136    -0.367490030    -1.036385299
11   H     1.027060309    -0.344173705     0.264538772
12   C    -1.075540169     2.168044244    -0.915602294
13   H    -0.169886192     3.226150907     1.395840500
14   H     1.026264624     2.001751167     1.857275178
15   H    -2.052103501     1.793867069     2.051287935
16   H    -1.836396927    -0.454117947     2.961769988
17   H    -0.058241966    -0.406783890     2.900899614
18   H    -2.048670405    -0.458668661     0.539115650
19   H    -0.996856900     3.274827258    -0.928364363
20   H    -0.947521024     1.822415109    -1.962294197
21   H    -2.108075243     1.913112077    -0.605166318
22   H     0.082433844     1.829213534     4.291818758
23   H    -1.683899575     1.625229684     4.500585449
24   H    -1.036512453     3.152347599     3.844607696
25   H    -0.971052576    -2.620440174    -0.050597055
26   H     0.085680400    -2.530290700     1.390347462
27   H    -1.688940263    -2.662380703     1.582346473
```

```
---
H20C9, RHF, CHARGE=0, MULT=1
HF=-55.4
1    C    -0.005621680    -0.128218790    -0.063001760
2    C    -0.008014280     1.402822510     0.022883260
3    C     0.078197120     2.119230760     1.420689400
4    C     0.195248080     3.668550790     1.173763730
5    C    -0.953415680     4.426809530     0.496880080
6    C     1.335940600     1.647942430     2.238908000
7    C     2.740498670     1.794674300     1.640865700
8    C    -1.211757540     1.756908580     2.244866680
9    C    -1.457895170     2.410274940     3.610607830
10   H    -0.696884310     5.506524480     0.446982440
11   H    -1.911885120     4.348767430     1.046289940
12   H    -1.131996320     4.089480760    -0.543590130
13   H     1.103029050     3.862603220     0.554564330
14   H     0.392861260     4.174720600     2.147729210
15   H     0.829629020     1.766308710    -0.617212630
16   H    -0.938605610     1.740593940    -0.491617310
17   H    -0.026355730    -0.431524860    -1.131910830
18   H    -0.893729500    -0.583845140     0.418491340
19   H     0.896342730    -0.588994070     0.384880960
20   H     3.480247480     1.333927640     2.330114220
21   H     3.037206640     2.854604750     1.511817120
22   H     2.854178900     1.290588490     0.661518540
23   H     1.204366930     0.573930550     2.509800400
24   H     1.353535080     2.194695950     3.211288420
25   H    -2.455631580     2.101350370     3.989566900
26   H    -1.457489050     3.516997970     3.573407060
27   H    -0.713328650     2.097444110     4.369687370
28   H    -1.224245700     0.655018860     2.419595580
29   H    -2.108607450     1.962659930     1.615082230
---
H20C9, RHF, CHARGE=0, MULT=1
HF=-54.7
1    C     0.000000000     0.000000000     0.000000000
2    C     0.000000000     1.531328790     0.000000000
3    C     0.000000000     2.174792627     1.400301520
4    C     0.000719640     3.716934813     1.385245358
5    C    -0.000300298     4.358734447     2.787790637
6    C     0.002598447     5.901009287     2.772865548
7    C     0.000306574     6.542524736     4.175353824
8    C     0.005643500     8.083294698     4.163081291
9    C     0.002084163     8.733844589     5.549306020
10   H     0.008155918     9.839181086     5.454337883
11   H     0.893654589     8.445990683     6.142344207
12   H    -0.898452040     8.455123056     6.133074511
13   H    -0.881586183     8.453313870     3.598778130
14   H     0.899821897     8.447335944     3.605808351
15   H     0.888595567     6.181181539     4.743120550
16   H    -0.894325783     6.187317391     4.737046293
17   H    -0.886965505     6.261096696     2.206610179
18   H     0.896000316     6.257958230     2.210475952
19   H     0.889607563     3.998052295     3.353264685
20   H    -0.893552871     4.001256741     3.350158769
21   H    -0.890247843     4.075373380     0.820208212
22   H     0.892717492     4.074610074     0.821068118
23   H     0.890705116     1.814015059     1.964668090
24   H    -0.892122249     1.815814540     1.963507669
```

```
25      H        0.890230009     1.883156301    -0.571180290
26      H       -0.891057861     1.882200020    -0.570409844
27      H        0.001582363    -0.383517004    -1.041013016
28      H       -0.896956845    -0.412260501     0.504896880
29      H        0.895177385    -0.412549985     0.507833834
---
H8C10, RHF, CHARGE=0, MULT=1
HF=73.5
1    C     0.000000000    0.000000000    0.000000000
2    C     1.498407191    0.000000000    0.000000000
3    C     1.913514500    1.325183082    0.000000000
4    C     0.747830511    2.197467173    0.000119115
5    C    -0.387299369    1.421911697    0.000497027
6    C     2.347938378   -1.163969276   -0.000928219
7    C     1.994964264   -2.482006771   -0.001591405
8    C     0.681286812   -3.074526628   -0.001632228
9    C    -0.542095927   -2.468230051   -0.001246891
10   C    -0.858297674   -1.061815274   -0.000527708
11   H     2.932324848    1.690288750    0.000234342
12   H     0.802467842    3.278314684   -0.000042130
13   H    -1.412874311    1.766320248    0.000608768
14   H     3.424482610   -0.945180987   -0.001226417
15   H     2.812425588   -3.214442735   -0.002141027
16   H     0.689053669   -4.172828012   -0.001910176
17   H    -1.422077072   -3.124030218   -0.001336910
18   H    -1.935635821   -0.848733460   -0.000544494
EXPGEOM
1    C     0.00000    0.00000   -2.49860
2    C     0.00000    0.00000    2.70470
3    C     0.00000    1.26350   -1.90780
4    C     0.00000   -1.26350   -1.90780
5    C     0.00000    1.59070   -0.55200
6    C     0.00000   -1.59070   -0.55200
7    C     0.00000    0.74870    0.55320
8    C     0.00000   -0.74870    0.55320
9    C     0.00000    1.14860    1.89900
10   C     0.00000   -1.14860    1.89900
11   H     0.00000    0.00000   -3.58680
12   H     0.00000    0.00000    3.78830
13   H     0.00000    2.10410   -2.59630
14   H     0.00000   -2.10410   -2.59630
15   H     0.00000    2.65540   -0.32240
16   H     0.00000   -2.65540   -0.32240
17   H     0.00000    2.17350    2.24460
18   H     0.00000   -2.17350    2.24460
---
H8C10, RHF, CHARGE=0, MULT=1
HF=36.1, IE=8.15
1    C     1.409641675   -0.010000000    1.262643722
2    C     0.712444037   -0.010000000    2.455875883
3    C    -0.716568030   -0.010000000    2.454416252
4    C    -1.411669496   -0.010000000    1.260016438
5    C    -1.409641698   -0.010000000   -1.262643622
6    C    -0.712444129   -0.010000000   -2.455876946
7    C     0.716567847   -0.010000000   -2.454415152
8    C    -0.717480139   -0.010000000   -0.000702203
9    C     0.717480141   -0.010000000    0.000702202
10   C     1.411669478   -0.010000000   -1.260016440
11   H     2.500902294   -0.010000000    1.270172431
```

```
12    H     1.240398859   -0.010000000    3.410156874
13    H    -1.246887599   -0.010000000    3.407402698
14    H    -2.502956578   -0.010000000    1.265987876
15    H    -2.500902284   -0.010000000   -1.270172432
16    H    -1.240397699   -0.010000000   -3.410150673
17    H     1.246889955   -0.010000000   -3.407414845
18    H     2.502956592   -0.010000000   -1.265987868
EXPGEOM
1     C     0.000000000    1.244400000    1.400000000
2     C     0.000000000    2.429300000    0.709000000
3     C     0.000000000    2.429300000   -0.709000000
4     C     0.000000000    1.244400000   -1.400000000
5     C     0.000000000   -1.244400000   -1.400000000
6     C     0.000000000   -2.429300000   -0.709000000
7     C     0.000000000   -2.429300000    0.709000000
8     C     0.000000000    0.000000000   -0.712700000
9     C     0.000000000    0.000000000    0.712700000
10    C     0.000000000   -1.244400000    1.400000000
11    H     0.000000000    1.237500000    2.486100000
12    H     0.000000000    3.374600000    1.242200000
13    H     0.000000000    3.374600000   -1.242200000
14    H     0.000000000    1.237500000   -2.486100000
15    H     0.000000000   -1.237500000   -2.486100000
16    H     0.000000000   -3.374600000   -1.242200000
17    H     0.000000000   -3.374600000    1.242200000
18    H     0.000000000   -1.237500000    2.486100000
---
H10C10, RHF, CHARGE=0, MULT=1
HF=134.3
1     C     0.000000000    0.000000000    0.000000000
2     C     0.000000000    1.539879940    0.000000000
3     C     0.000000000    0.790312364    1.321545982
4     C     5.161544596   -3.612957421   -2.056078395
5     C     5.165574730   -4.337735107   -3.414655601
6     C     5.114552343   -5.150207308   -2.131776488
7     C     4.049822226   -2.834431225   -1.608083914
8     C     3.116160786   -2.173998086   -1.236912578
9     C     1.126363768   -0.764892847   -0.434600589
10    C     2.057775979   -1.425656252   -0.810655449
11    H    -0.950757161   -0.489658592   -0.277366018
12    H    -0.907495862    2.061114490   -0.326917540
13    H     0.896901220    2.085846755   -0.313897549
14    H     0.896892460    0.801305634    1.951260400
15    H    -0.907546941    0.777227946    1.936625760
16    H     6.120918824   -3.168157719   -1.735857260
17    H     6.085656922   -4.323265836   -4.010745375
18    H     4.282231899   -4.290065759   -4.061528078
19    H     4.194764300   -5.682165210   -1.863377611
20    H     5.998231618   -5.715355090   -1.812686262
---
H10C10, RHF, CHARGE=0, MULT=1
HF=59.4
1     C     0.000000000    0.000000000    0.000000000
2     C     0.000000000    1.418194990    0.000000000
3     C     0.000000000    2.132558082    1.209740901
4     C    -0.000622383    1.446824915    2.437286662
5     C    -0.001104578    0.041028818    2.451498014
6     C    -0.000444407   -0.678978709    1.245098615
7     C     0.001112919   -0.725832863   -1.221225163
```

```
8    C     0.002090165   -1.339088006   -2.253290584
9    C     0.000435303   -2.094180496   -3.490098579
10   C    -0.007912196   -1.228239323   -4.756176039
11   H     0.000295886    1.965272307   -0.943836548
12   H     0.000270984    3.223027615    1.195534626
13   H    -0.000711304    2.003803712    3.374625296
14   H    -0.001739561   -0.494244542    3.401677680
15   H    -0.000267096   -1.769497509    1.273926825
16   H    -0.889491516   -2.769702988   -3.501726119
17   H     0.895311369   -2.762975608   -3.508566967
18   H    -0.007043797   -1.871593137   -5.659656939
19   H     0.884977008   -0.573308331   -4.807511539
20   H    -0.908322797   -0.583311425   -4.802226550
---
H10C10, RHF, CHARGE=0, MULT=1
HF=53.1
1    C     0.000000000    0.000000000    0.000000000
2    C     0.000000000    1.350545210    0.000000000
3    C     0.000000000   -0.584473229    1.399861452
4    C     1.280355506   -1.283928499    1.816260733
5    C     1.956086572   -0.608877721    2.771222960
6    C     1.261072409    0.675279137    3.183156167
7    C     1.958048120    1.958289872    2.771036521
8    C     1.284409296    2.633061672    1.814464180
9    C     0.002189612    1.936616024    1.399330143
10   C    -0.066736790    0.676328820    2.338806606
11   H    -0.007554753   -0.645642972   -0.870926053
12   H    -0.008792495    1.995141641   -0.871818739
13   H    -0.872852439   -1.258451471    1.540218012
14   H     1.567847710   -2.227653161    1.366483079
15   H     2.889609756   -0.906990944    3.235197058
16   H     1.063188591    0.675786682    4.277095378
17   H     2.891101420    2.256842480    3.235766811
18   H     1.574544762    3.575631635    1.363902049
19   H    -0.869162711    2.612893352    1.537770051
20   H    -0.969093316    0.677317029    2.976067271
---
H10C10, RHF, CHARGE=0, MULT=1
HF=79.9
1    C     0.000000000    0.000000000    0.000000000
2    C     0.000000000    1.518311160    0.000000000
3    C     0.000000000    2.281265515    1.114205059
4    C    -0.000792050    1.769824026    2.507319772
5    C    -0.775392382    0.497562250    2.927185404
6    C    -1.574132492   -0.305159420    1.968114308
7    C    -1.248991294   -0.534983959    0.677713212
8    C     0.772288842    0.496919717    2.928163683
9    C     1.571374072   -0.306726878    1.970091077
10   C     1.247764991   -0.535832067    0.679246724
11   H     0.000461866   -0.350362405   -1.060235611
12   H     0.000945542    1.983629992   -0.987916014
13   H     0.000604013    3.371403162    1.026581322
14   H    -0.000824063    2.588784357    3.244499003
15   H    -1.249228446    0.538498506    3.921101257
16   H    -2.494616037   -0.728302989    2.380127068
17   H    -1.899413356   -1.141418249    0.044069470
18   H     1.244944228    0.537529829    3.922687726
19   H     2.490656448   -0.731167063    2.383496025
20   H     1.898362106   -1.142965025    0.046396166
```

```
---
H10C10, RHF, CHARGE=0, MULT=1
HF=118.1
1    C     0.000000000     0.000000000     0.000000000
2    C     0.000000000     1.510649590     0.000000000
3    C     0.000000000    -2.892186815     5.951637748
4    C    -0.009449445    -4.402861163     5.952156827
5    C     0.001510816    -0.739789645    -1.132749051
6    C     0.009311972    -2.152504861     7.084527155
7    C    -0.000762284    -1.148363704     2.362648042
8    C    -0.001040561    -1.743294722     3.589408154
9    C    -0.000521327    -0.623604238     1.279260964
10   C    -0.000749259    -2.268899468     4.672365877
11   H    -0.900365080     1.904369544     0.516202043
12   H     0.004438998     1.930455468    -1.026401461
13   H     0.895576562     1.904776711     0.524160300
14   H    -0.912187370    -4.791108463     5.436072441
15   H    -0.007456801    -4.822786014     6.978477728
16   H     0.883531044    -4.802524916     5.427854164
17   H     0.001550626    -1.828832502    -1.135246008
18   H     0.002701797    -0.307149468    -2.131991538
19   H     0.016255159    -1.063498090     7.087387220
20   H     0.010621806    -2.585698240     8.083520331
---
H10C10, RHF, CHARGE=0, MULT=1
HF=74
1    C     0.000000000     0.000000000     0.000000000
2    C     0.000000000     1.398203950     0.000000000
3    C     0.000000000     2.190772978     1.193276068
4    C     0.000102879     1.647809110     2.481800384
5    C    -0.000005272     0.249559270     2.482016952
6    C     0.000461898    -0.542753217     1.288734822
7    C     0.000476410    -1.843525665     2.067191728
8    C    -0.000073662    -0.973027414     3.378469993
9    C     0.000005453     3.491272333     0.414282539
10   C     0.000081882     2.620550738    -0.896682952
11   H    -0.000134022    -0.600885094    -0.904884876
12   H     0.001268004     2.248575721     3.386797220
13   H     0.895731112    -2.475079036     1.928757181
14   H    -0.894331228    -2.475607273     1.927975964
15   H    -0.895468310    -1.090438668     4.014336715
16   H     0.894709318    -1.090176175     4.015067394
17   H     0.895185901     4.122989965     0.552848388
18   H    -0.894965401     4.123245193     0.553106764
19   H    -0.894971987     2.737948071    -1.532810683
20   H     0.894945916     2.737470930    -1.533187974
---
H12C10, RHF, CHARGE=0, MULT=1
HF=85.1
1    C     0.028489530    -0.012763320    -0.029423520
2    C     0.002718030     1.297269410    -0.067317230
3    C    -0.023803940     2.607702250    -0.086196530
4    C     0.795394000     3.512002240     0.798501890
5    C     0.240926050     3.710808940     2.227586180
6    C     0.456427730     2.551563540     3.166420160
7    C    -0.206447400     1.420698800     3.150629190
8    C    -0.868397040     0.289822240     3.119904110
9    C    -0.483229030    -0.932093730     2.326188330
10   C    -0.855391520    -0.883000950     0.826694900
```

```
11     H     0.735636880    -0.554960950    -0.667870660
12     H    -0.681978720     3.122617000    -0.795523300
13     H     0.855292890     4.514383270     0.312942770
14     H     1.844697570     3.140966480     0.853355720
15     H    -0.843487100     3.962313980     2.176881560
16     H     0.737014250     4.609630340     2.663789700
17     H     1.240395520     2.703082440     3.917638970
18     H    -1.785069000     0.188118160     3.712772720
19     H    -0.994628470    -1.815165990     2.776680190
20     H     0.607794100    -1.130884110     2.433959860
21     H    -0.808937720    -1.924498200     0.430353840
22     H    -1.918036670    -0.566146800     0.715352480
---
H12C10, RHF, CHARGE=0, MULT=1
HF=72.3
1      C     0.000604880    -0.005972460    -0.003449410
2      C     0.010199120     1.343352170    -0.002900740
3      C    -0.061072100     2.172641630     1.239360340
4      C    -0.143232910     1.343505220     2.481171530
5      C    -0.152598970    -0.005802370     2.480703760
6      C    -0.082272630    -0.834916240     1.238279350
7      C    -0.849375700    -2.181669720     1.190891690
8      C     0.662141690    -2.194282050     1.284462990
9      C    -0.808033530     3.530612990     1.192726770
10     C     0.703685700     3.520560620     1.286346170
11     H     0.055963090    -0.544606490    -0.952870020
12     H     0.072753710     1.881818790    -0.952222050
13     H    -0.197691000     1.882309250     3.430703730
14     H    -0.214323790    -0.544487120     3.429712250
15     H    -1.339640970    -2.486792030     0.258935320
16     H    -1.451166370    -2.486903940     2.055030850
17     H     1.147448550    -2.508588800     2.215889430
18     H     1.258307960    -2.508610590     0.419723080
19     H    -1.405324460     3.844312610     2.057266010
20     H    -1.293534150     3.843703030     0.261100450
21     H     1.304471960     3.826417060     0.421827710
22     H     1.193774040     3.826866760     2.218205160
---
H12C10, RHF, CHARGE=0, MULT=1
HF=6.2
1      C     0.000000000     0.000000000     0.000000000
2      C     0.000000000     1.537618260     0.000000000
3      C     0.000000000     2.145868644     1.408947518
4      C    -1.097386110     1.576149573     2.323864247
5      C    -1.377256919     0.097839919     2.159205995
6      C    -0.883560012    -0.639662334     1.048870273
7      C    -1.204778612    -2.013765336     0.938059532
8      C    -1.997131094    -2.655232995     1.901837231
9      C    -2.483897446    -1.927203076     3.000069953
10     C    -2.175128108    -0.564204987     3.123122472
11     H    -0.304800567    -0.347451697    -1.014489276
12     H     1.043799857    -0.366664294     0.149518055
13     H    -0.883383244     1.910631338    -0.567445366
14     H     0.892946319     1.900756455    -0.557770420
15     H    -0.128112476     3.249517596     1.332992637
16     H     0.996579047     1.989079671     1.882185030
17     H    -2.047259546     2.134360240     2.143008752
18     H    -0.813017791     1.788328085     3.380872876
19     H    -0.834329640    -2.595563345     0.091806891
```

```
20     H    -2.233246868   -3.714689997    1.798261151
21     H    -3.100140005   -2.418186495    3.753768836
22     H    -2.564423968   -0.014319210    3.982240590
---
H14C10, RHF, CHARGE=0, MULT=1
HF=-4.2
1      C    -0.778496154    1.149479452    1.156352084
2      C    -1.096746968    0.109185111    0.265859058
3      C    -0.389923784   -0.032909477   -0.939578209
4      C     0.642479952    0.869807214   -1.243264331
5      C     0.962259584    1.905893066   -0.348915698
6      C     0.255564459    2.072803668    0.867624913
7      C     0.625283896    3.210771373    1.814355175
8      C    -0.522488759    4.232385502    2.049083433
9      C    -1.032552657    4.971825736    0.809078314
10     C     1.209288384    2.677205824    3.142300141
11     H     0.459139625    2.136935421    3.753235703
12     H     1.607282917    3.507940687    3.760906406
13     H     2.050887085    1.980572170    2.950625418
14     H     1.459230794    3.793171383    1.341622805
15     H    -0.162241332    4.999798326    2.774577568
16     H    -1.385904641    3.732972475    2.544132986
17     H    -0.222354591    5.527390732    0.295404252
18     H    -1.806489404    5.711851418    1.101720799
19     H    -1.495936302    4.286517710    0.072076535
20     H    -0.640252847   -0.835821011   -1.632712966
21     H     1.196915793    0.769488090   -2.177172645
22     H    -1.353192510    1.228243243    2.080677623
23     H    -1.897042260   -0.589179755    0.512778371
24     H     1.772510930    2.587252237   -0.615213595
---
H14C10, RHF, CHARGE=0, MULT=1
HF=-5.2
1      C     0.058844480   -0.085532290   -0.069049920
2      C     0.147414840    1.312252750    0.054197640
3      C     0.993156650    1.908547490    1.019550580
4      C     1.757294380    1.052114260    1.848962250
5      C     1.669602300   -0.345181060    1.727231140
6      C     0.819133690   -0.919703330    0.767441170
7      H    -0.602825960   -0.521564540   -0.818442060
8      H     2.199609330    3.664829900    1.200140180
9      H     0.767374140    3.890494180    0.184921610
10     H     0.651432550    3.492353990    3.252861070
11     H    -0.449955970    1.933606280   -0.615650230
12     C     1.113865300    3.414955760    1.131310290
13     H     2.434741560    1.468371420    2.596735100
14     H     2.265462080   -0.984254260    2.379658490
15     H     0.752726790   -2.003331640    0.670105660
16     H    -1.544227060    4.561048460    1.343937300
17     H    -1.522237480    2.960011230    2.144756470
18     H    -1.639948790    4.440134560    3.126918490
19     H     0.639354320    6.176560760    1.708493050
20     H     1.957929270    5.543184300    2.741953290
21     H     0.381640380    5.953323110    3.466901950
22     C     0.373752690    4.068602650    2.333491800
23     C    -1.162546770    4.005187450    2.224199380
24     C     0.862611710    5.513872480    2.568891340
---
H14C10, RHF, CHARGE=0, MULT=1
```

```
HF=-8.6
1    C     0.000000000    0.000000000    0.000000000
2    C     0.000000000    1.397616980    0.000000000
3    C     0.000000000    2.127528966    1.208163581
4    C    -0.000311315    1.412290845    2.441207654
5    C    -0.079296282   -0.016380556    2.439347717
6    C    -0.036023890   -0.730823320    1.207124431
7    H     0.023693881   -0.519314444   -0.960498919
8    H    -0.178425987    3.993559606    0.089821131
9    H     0.969577300    4.057248050    1.451577389
10   H    -0.805642048    4.079047006    1.750944585
11   H     0.005396256    1.917638439   -0.960427432
12   C    -0.003953895    3.634683739    1.125301116
13   C     0.079304213    2.173866191    3.746299586
14   C    -0.212441813   -0.772823929    3.743064649
15   C    -0.021787772   -2.237909464    1.121028444
16   H     0.627604477    1.617192963    4.533644025
17   H    -0.938213629    2.398692802    4.131764303
18   H     0.620099683    3.137035301    3.647585959
19   H    -0.872078717   -0.249262045    4.465887031
20   H     0.781258526   -0.914881116    4.219179263
21   H    -0.664416389   -1.777332103    3.618566945
22   H    -1.008795058   -2.664108064    1.399650220
23   H     0.202541826   -2.595238128    0.094441471
24   H     0.751983538   -2.681102556    1.781765652
---
H14C10, RHF, CHARGE=0, MULT=1
HF=19.9
1    C     0.000000000    0.000000000    0.000000000
2    C     0.000000000    1.377731030    0.000000000
3    C     0.000000000    1.859838848    1.414685700
4    C    -0.000128861    0.768686404    2.256097912
5    C    -0.000158645   -0.472942782    1.420110621
6    C     0.000012045   -0.941886940   -1.160463803
7    C    -0.000219194    2.307045957   -1.172390339
8    C    -0.000859435    3.311702822    1.775716992
9    C    -0.000260614    0.730979410    3.750289811
10   C    -0.000774864   -1.749028071    1.854396647
11   H     0.897370929   -1.596905936   -1.140487558
12   H     0.002700809   -0.425247620   -2.140765368
13   H    -0.900263109   -1.592965939   -1.143257278
14   H    -0.900350159    2.957962602   -1.163581438
15   H     0.001724865    1.778064861   -2.146269471
16   H     0.897519909    2.961235452   -1.161787862
17   H     0.001582855    3.487476724    2.869981809
18   H    -0.901498758    3.821188520    1.371659309
19   H     0.896310241    3.823839824    1.367322972
20   H     0.897190974    0.200107950    4.134328899
21   H    -0.900509025    0.204574915    4.133924835
22   H     0.002185299    1.738653651    4.211303232
23   H    -0.001034556   -2.031259302    2.906299934
24   H    -0.001124204   -2.614231814    1.192938182
---
H14C10, RHF, CHARGE=0, MULT=1
HF=-10.3
1    C     0.000000000    0.000000000    0.000000000
2    C     0.000000000    1.425832960    0.000000000
3    C     0.000000000    2.121815602    1.230070768
4    C     0.010187590    1.455171916    2.470481296
```

```
5    C     0.030846200    0.045467550    2.451910657
6    C     0.033942412   -0.691778123    1.247523418
7    C     0.082846709   -2.198091168    1.343123130
8    C    -0.007557399   -0.769266367   -1.300756078
9    C     0.005489546    2.241771080   -1.270628212
10   H    -0.007956845    3.214808009    1.221713272
11   C     0.000416473    2.205633362    3.775239697
12   H     0.047018622   -0.489407596    3.405065152
13   H    -0.850164967   -2.657042200    0.953709250
14   H     0.940353844   -2.613724364    0.773603165
15   H     0.199121414   -2.550441330    2.388843168
16   H     1.033179982   -0.981697943   -1.627284046
17   H    -0.542387269   -1.737521773   -1.222789461
18   H    -0.512900355   -0.224095625   -2.123223910
19   H     0.850432135    1.959973326   -1.933041389
20   H     0.112125550    3.328282446   -1.071450836
21   H    -0.941590746    2.112851647   -1.835664120
22   H     0.878110284    1.931907789    4.397896076
23   H    -0.916553032    1.973270714    4.357039574
24   H     0.029730976    3.304894455    3.633022289
---
H14C10, RHF, CHARGE=0, MULT=1
HF=-11.3
1    C    -0.000057182    0.000588699    0.002699727
2    C     1.412418741    0.000840191    0.014440639
3    C     2.187725394    1.181558584    0.005553601
4    C     1.512201026    2.432743079   -0.021863943
5    C     0.099777833    2.432649429   -0.027963695
6    C    -0.675775176    1.252103010   -0.015075496
7    C    -2.179947641    1.361268987   -0.019473360
8    C    -0.733695541   -1.316963461    0.009654031
9    H     1.932048314   -0.961199871    0.030506064
10   C     3.691753462    1.072373796    0.019103769
11   C     2.247271400    3.749534161   -0.043145668
12   H    -0.419596135    3.394807290   -0.043474732
13   H    -2.617667861    0.866031760   -0.911594846
14   H    -2.619317833    0.893486111    0.886630904
15   H    -2.526834758    2.414972569   -0.036018055
16   H    -1.375632143   -1.414602391    0.910263316
17   H    -1.378069826   -1.424406138   -0.887983656
18   H    -0.043015405   -2.185207734    0.012794249
19   H     4.126547447    1.444534623   -0.932679967
20   H     4.037948821    0.025926481    0.144392286
21   H     4.135328108    1.655431601    0.853150603
22   H     2.839987813    3.892235206    0.884856979
23   H     1.559302720    4.616194073   -0.121042812
24   H     2.939253994    3.812504202   -0.908845990
---
H14C10, RHF, CHARGE=0, MULT=1
HF=-5.4
1    C     0.000000000    0.000000000    0.000000000
2    C     0.000000000    1.404927133    0.000000000
3    C     0.000000000    2.115297956    1.211159499
4    C    -0.006345010    1.398234352    2.416458318
5    C    -0.003445903   -0.008642118    2.409936407
6    C     0.015290241   -0.753221553    1.204111077
7    C     0.007757900   -2.296089542    1.177767735
8    C    -1.448450526   -2.765969114    0.883649664
9    C     0.974334886   -2.838574158    0.080491819
```

```
10      C       0.471417125     -2.924873513     2.526826845
11      H      -0.019037559     -0.493214016    -0.973268511
12      H      -0.003099954      1.944152305    -0.948132278
13      H       0.001481409      3.205123087     1.213827861
14      H      -0.015292943      1.931043636     3.368254161
15      H      -0.021083615     -0.500525808     3.383327589
16      H      -2.155984662     -2.397699874     1.654091861
17      H      -1.812308856     -2.400566609    -0.097684084
18      H      -1.523473014     -3.872705597     0.867038914
19      H       0.614462509     -2.630958300    -0.946939843
20      H       1.985725770     -2.394654116     0.178364298
21      H       1.089425831     -3.940343146     0.149259804
22      H       1.469354165     -2.549859878     2.831957430
23      H      -0.236791469     -2.722611237     3.355514401
24      H       0.548780710     -4.029746197     2.450700698
---
H14C10, RHF, CHARGE=0, MULT=1
HF=3
1       C       0.041978150      0.041054700     0.031466540
2       C       1.310373940     -0.209055150    -0.355876590
3       C      -0.882364610     -1.138704920    -0.202835150
4       C       0.052721140     -2.235823710    -0.811983810
5       C       1.480112940     -1.603265610    -0.927820580
6       C       1.894834720     -1.680288970    -2.415375130
7       C       0.728295010     -2.304180320    -3.212136440
8       C      -0.451403240     -2.571406840    -2.243721920
9       C      -2.033603850     -0.911599810    -1.209207600
10      C      -1.746670500     -1.751476320    -2.472753710
11      H      -0.327727900      0.955458880     0.481830970
12      H       2.155865820      0.465127360    -0.273989780
13      H      -1.302295540     -1.471484870     0.772943160
14      H       0.069942390     -3.143040770    -0.174676710
15      H       2.213777030     -2.165460580    -0.307431320
16      H       2.812772770     -2.297657880    -2.523519730
17      H       2.153184250     -0.680297190    -2.824235520
18      H       1.050339830     -3.260867200    -3.679375540
19      H       0.440116630     -1.641989690    -4.055821720
20      H      -0.712155350     -3.653347530    -2.304267350
21      H      -3.000250210     -1.217447620    -0.752518960
22      H      -2.147879060      0.161789500    -1.470438680
23      H      -2.595604440     -2.439640670    -2.678323700
24      H      -1.674222780     -1.091492480    -3.362932100
---
H16C10, RHF, CHARGE=0, MULT=1
HF=-5.9
1       C      -0.029668780     -0.008620170    -0.017548790
2       C      -0.002446030      1.364573960    -0.018774350
3       C      -0.039427020      1.880527190     1.425394040
4       C       0.003235400      0.582790510     2.241951590
5       C      -0.022087600     -0.496846830     1.393523730
6       C      -0.009890120     -1.947853240     1.760885620
7       C      -0.040374530     -0.921169790    -1.204058870
8       C       0.085975690      2.289254130    -1.189965420
9       C      -1.249260920      2.777143480     1.739923820
10      C       0.089687340      0.589686560     3.734119360
11      H       0.892976950      2.463209790     1.624430560
12      H       0.959697910     -2.415135960     1.485658600
13      H      -0.818964550     -2.504328300     1.243124890
14      H      -0.157901570     -2.115566880     2.846984980
```

```
15    H    -0.855964860    -1.670758410    -1.129522090
16    H     0.922291020    -1.468511200    -1.291923110
17    H    -0.197271150    -0.377615760    -2.158136760
18    H    -0.893105900     2.371461640    -1.708731040
19    H     0.392207400     3.312914690    -0.891145130
20    H     0.834055410     1.939672840    -1.932302690
21    H    -2.210435310     2.251325340     1.571037950
22    H    -1.230375600     3.116334920     2.795676780
23    H    -1.246827410     3.689070990     1.109117670
24    H     0.446840570     1.564118290     4.127322110
25    H    -0.903761180     0.389481860     4.190379040
26    H     0.796223630    -0.180228540     4.109146190
---
H16C10, RHF, CHARGE=0, MULT=1
HF=-31.9, IE=9.6
1     C     0.000000000     0.000000000     0.000000000
2     C     0.000000000     2.535872510     0.000000000
3     C     0.000000000     1.268304737     2.196240742
4     C     2.070844349     1.268070233     0.732527660
5     C     1.554659580    -0.001710427    -0.001079038
6     C     1.554680940     2.537853383    -0.001206082
7     C    -0.519386350     2.538232995     1.465218395
8     C    -0.519373038    -0.001655193     1.465361241
9     C     1.554718870     1.268077604     2.198960905
10    C    -0.519393109     1.268021664    -0.734601767
11    H     1.945335543     2.150443530     2.751586285
12    H     1.944868178     0.385312710     2.751254623
13    H    -0.189474883     1.268028695    -1.796506569
14    H    -1.630542850     1.267985235    -0.777422929
15    H     1.945146246    -0.038460397    -1.041593843
16    H     1.944971244    -0.921622221     0.486668711
17    H     1.945348699     3.457537033     0.486747096
18    H     1.945105231     2.574701207    -1.041739977
19    H    -0.189388286     3.457866545     1.996180575
20    H    -1.630530662     2.575188063     1.486797353
21    H    -1.630530309    -0.038410904     1.486738702
22    H    -0.189658106    -0.921372463     1.996313925
23    H    -0.370608479    -0.908621074    -0.524145411
24    H    -0.370564499     3.444108509    -0.524943715
25    H    -0.371216181     1.268141298     3.245030422
26    H     3.183391592     1.268127862     0.732339798
---
H16C10, RHF, CHARGE=0, MULT=1
HF=-6.8
1     C     0.000000000     0.000000000     0.000000000
2     C     0.000000000     1.533536030     0.000000000
3     C     0.000000000     1.978634598     1.485049939
4     C    -0.033152923     0.553753719     2.194105100
5     C    -1.402446808    -0.173235593     1.986158070
6     C    -1.402808397    -0.506416171     0.466551898
7     C     0.875024467    -0.305644402     1.254612206
8     C    -1.209366100     2.884513259     1.833441220
9     C     1.292255659     2.755411647     1.862949980
10    C     0.012429552     2.307864309    -1.094786024
11    H     0.342022951    -0.460623302    -0.941814983
12    H     0.271267546     0.586563185     3.253450075
13    H    -1.445750906    -1.095506450     2.604253954
14    H    -2.282709719     0.422919958     2.292124172
15    H    -1.514085254    -1.595237712     0.285465206
```

```
16      H       -2.235559677     -0.015615199    -0.074513257
17      H        1.926959555      0.022736410     1.170685024
18      H        0.923636985     -1.377779712     1.528163481
19      H       -1.370102374      2.934678331     2.929964321
20      H       -2.154657504      2.538302427     1.370736188
21      H       -1.054513618      3.924053383     1.476583645
22      H        2.218745441      2.209334678     1.597959467
23      H        1.331197831      2.956428507     2.953563956
24      H        1.344406756      3.734597517     1.343981715
25      H        0.022726272      3.396339150    -1.072391145
26      H        0.014637462      1.903830260    -2.106761540
EXPGEOM
1       C        0.89190          0.89190         0.89190
2       C        0.89190         -0.89190        -0.89190
3       C       -0.89190         -0.89190         0.89190
4       C       -0.89190          0.89190        -0.89190
5       C        1.78050          0.00000         0.00000
6       C       -1.78050          0.00000         0.00000
7       C        0.00000          0.00000         1.78050
8       C        0.00000          0.00000        -1.78050
9       C        0.00000          1.78050         0.00000
10      C        0.00000         -1.78050         0.00000
11      H        1.52460          1.52460         1.52460
12      H        1.52460         -1.52460        -1.52460
13      H       -1.52460         -1.52460         1.52460
14      H       -1.52460          1.52460        -1.52460
15      H        0.62300          2.43460        -0.62300
16      H       -0.62300          2.43460         0.62300
17      H       -0.62300         -2.43460        -0.62300
18      H        0.62300         -2.43460         0.62300
19      H        0.62300         -0.62300         2.43460
20      H       -0.62300          0.62300         2.43460
21      H        0.62300          0.62300        -2.43460
22      H       -0.62300         -0.62300        -2.43460
23      H        2.43460         -0.62300         0.62300
24      H        2.43460          0.62300        -0.62300
25      H       -2.43460          0.62300         0.62300
26      H       -2.43460         -0.62300        -0.62300
---
H16C10, RHF, CHARGE=0, MULT=1
HF=-24.5
1       C        0.000000000      0.000000000     0.000000000
2       C        0.000000000      1.556098870     0.000000000
3       C        0.000000000     -0.497011569     1.474711048
4       C        1.302575792     -0.496909204    -0.691603942
5       C        2.081482452      0.761102240    -1.149200748
6       C        1.222082316      2.004444825    -0.836666591
7       C       -0.040184866      2.009749999     1.480066486
8       C       -0.166128295      0.754728453     2.369580325
9       C        1.288703027     -1.332850733     1.671617797
10      C        2.014784374     -1.433553412     0.313343805
11      H       -0.892653369     -0.386074253    -0.536451995
12      H       -0.922566527      1.930377991    -0.500965096
13      H       -0.875170987     -1.165783532     1.644553334
14      H        1.041819912     -1.090232572    -1.598231786
15      H        3.070655921      0.846443549    -0.651263551
16      H        2.297600875      0.700532813    -2.237814549
17      H        1.828356404      2.778529549    -0.321208540
18      H        0.875093429      2.478302440    -1.781290932
```

```
19   H    0.862455704    2.590439250    1.766025598
20   H   -0.901368529    2.692412258    1.646816272
21   H   -1.163030855    0.732074632    2.862393366
22   H    0.573951878    0.785828611    3.196428867
23   H    1.965030098   -0.888704175    2.432579080
24   H    1.032369072   -2.345228549    2.052162580
25   H    3.092183974   -1.192050837    0.428626793
26   H    1.983015145   -2.479890084   -0.062677808
---
H18C10, RHF, CHARGE=0, MULT=1
HF=-26.5
1    C   -0.512741853   -0.329206641   -1.183927344
2    C   -0.650502800    1.161216614   -1.031297861
3    C    0.037783320   -0.959188048   -2.248265101
4    C    0.545121311   -0.168329684   -3.435895853
5    C    0.719284264    1.341912785   -3.173518189
6    C   -0.436742272    1.959750747   -2.347375305
7    C    0.159675299   -2.460213718   -2.305770089
8    C   -0.357922805    3.506778590   -2.113602824
9    C    0.973327288    4.036557686   -1.542793417
10   C   -0.785517371    4.316747773   -3.358395511
11   H    0.938329501   -2.817669053   -1.598438850
12   H    0.438143568   -2.824688739   -3.315397137
13   H   -0.796046241   -2.956688703   -2.035861000
14   H   -0.890736184   -0.895116002   -0.327272688
15   H   -0.160180912   -0.314200871   -4.288171290
16   H    1.525735782   -0.576700989   -3.773324477
17   H    0.801163728    1.854207347   -4.158157094
18   H    1.695315231    1.502292540   -2.662446713
19   H    0.065047503    1.485080025   -0.240511695
20   H   -1.668567087    1.384468023   -0.634258195
21   H   -1.371684265    1.797223016   -2.946177904
22   H    0.852084387    5.082337493   -1.190127852
23   H    1.789445373    4.035737525   -2.292991729
24   H    1.319182415    3.445278583   -0.671768255
25   H   -1.135067767    3.734701001   -1.336248613
26   H   -0.886604833    5.393332396   -3.107393996
27   H   -1.771451211    3.982414703   -3.741271530
28   H   -0.058548039    4.237943940   -4.191159408
---
H18C10, RHF, CHARGE=0, MULT=1
HF=-26.6
1    C   -0.023987190   -0.036240140   -0.016962720
2    C    0.596860100    1.346776260   -0.072452470
3    C    1.180915080    1.846204490    1.262842460
4    C    0.294175530    1.528418680    2.490972370
5    C   -0.109179580    0.032457400    2.523967450
6    C   -0.355106880   -0.585916010    1.175128810
7    C    0.931390530    1.994399870    3.813998600
8    C   -0.250680220   -0.732947300   -1.357209500
9    C   -1.729896410   -1.089513750   -1.616361780
10   C    0.698168420   -1.936647180   -1.544243860
11   H   -0.652084080    2.117761690    2.375210690
12   H   -0.179397500    2.065824750   -0.427093520
13   H    1.412764390    1.374680440   -0.831112310
14   H    1.330798630    2.947377170    1.188110930
15   H    2.195816950    1.407929720    1.401145770
16   H   -1.029970900   -0.079246640    3.143059850
17   H   -0.823998200   -1.572140910    1.237496230
```

```
18      H        0.242844650     1.822673060     4.666681730
19      H        1.879722430     1.464482310     4.035912660
20      H        1.153668980     3.080746950     3.787524640
21      H        0.019105990    -0.007304980    -2.169455650
22      H       -2.074203150    -1.955491530    -1.016467540
23      H       -2.394325950    -0.232284980    -1.385721090
24      H       -1.885271800    -1.344123360    -2.685386550
25      H        1.754301220    -1.634152710    -1.391148390
26      H        0.481164720    -2.764851200    -0.840398060
27      H        0.619121210    -2.344877660    -2.572733290
28      H        0.678173360    -0.572914560     3.032837190
---
H18C10, RHF, CHARGE=0, MULT=1
HF=-40.5
1       C       -0.051543510    -0.026317810    -0.001206482
2       C        0.484356838     1.439388143    -0.124322654
3       C       -0.675703170    -0.352959680     1.380741152
4       C       -1.494746911     0.792848148     1.995934356
5       C       -0.759758847     2.143372299     2.043580266
6       C        0.541895241     2.198427786     1.225472218
7       C        1.831883626     1.555238720    -0.883831706
8       C        1.967947677     0.612445675    -2.089756383
9       C        1.637728274    -0.859104804    -1.786260986
10      C        0.982466109    -1.103300854    -0.415859577
11      H       -0.892069512    -0.103435139    -0.740180963
12      H       -0.267664241     1.983728479    -0.754073412
13      H       -1.342109790    -1.239675346     1.270887127
14      H        0.114303382    -0.658922373     2.104463601
15      H       -2.447575051     0.911027430     1.430325572
16      H       -1.795452414     0.500430201     3.027990127
17      H       -0.519845579     2.397927485     3.101573219
18      H       -1.450549353     2.944357422     1.693474648
19      H        1.378251036     1.809390911     1.850563459
20      H        0.784712586     3.268418085     1.029888517
21      H        2.687242547     1.378357926    -0.191908335
22      H        1.949095197     2.603630125    -1.244004779
23      H        1.314737568     0.974403995    -2.916747293
24      H        3.009017158     0.683442658    -2.479647374
25      H        0.973819750    -1.254256348    -2.588939389
26      H        2.569269255    -1.467569407    -1.843288625
27      H        0.480730451    -2.098040577    -0.442456268
28      H        1.780525611    -1.189318804     0.356889939
---
H18C10, RHF, CHARGE=0, MULT=1
HF=-34.7
1       C        0.000000000     0.000000000     0.000000000
2       C        0.000000000     1.536246290     0.000000000
3       C        0.000000000     2.162716825     1.406016687
4       C       -1.077866628     1.608504270     2.388558389
5       C       -1.018688750     0.046450092     2.375863561
6       C       -1.004289653    -0.616388313     0.985826243
7       C       -0.817702147     2.147663752     3.841389756
8       C       -2.176003157     2.461555237     4.496803582
9       C       -3.226825006     2.504764903     3.374940702
10      C       -2.516430456     2.139399938     2.055765801
11      H       -0.226705468    -0.365179280    -1.027505035
12      H        1.024532819    -0.371837814     0.232311151
13      H        0.899796797     1.897974931    -0.549015076
14      H       -0.873450221     1.908378003    -0.582390076
```

```
15    H     1.010372204    2.013649527    1.854483696
16    H    -0.114581536    3.265149505    1.294983584
17    H    -0.105177510   -0.287092502    2.922021192
18    H    -1.876381713   -0.372091106    2.949774795
19    H    -0.764851580   -1.697395987    1.114706931
20    H    -2.024471577   -0.591207847    0.539965954
21    H    -0.241416491    1.426955717    4.458503411
22    H    -0.200321322    3.072551039    3.815460429
23    H    -2.437739893    1.689337241    5.252182255
24    H    -2.135747765    3.425300367    5.047751733
25    H    -4.058842555    1.798800535    3.585120517
26    H    -3.692581672    3.510981554    3.302978587
27    H    -3.117900481    1.400197418    1.487467131
28    H    -2.467391151    3.040236258    1.405666901
---
H18C10, RHF, CHARGE=0, MULT=1
HF=-43.5
1    C     0.023975590   -0.020324670    0.000334250
2    C     0.296771630    1.516627240    0.021465690
3    C     0.425446980   -0.690111270    1.339162600
4    C    -0.179004190   -0.005177780    2.577906360
5    C    -0.030874570    1.524554480    2.586515090
6    C    -0.367907370    2.198295530    1.244455370
7    C    -0.104597650    2.186432390   -1.317297870
8    C     0.498472710    1.501277290   -2.556320110
9    C     0.349483230   -0.028491670   -2.564509770
10   C     0.687911460   -0.702096870   -1.222800280
11   H    -1.080489760   -0.171999690   -0.112213630
12   H     1.401191520    1.668429320    0.134415800
13   H     0.094666240   -1.754013170    1.330468660
14   H     1.534262120   -0.717605150    1.441885940
15   H    -1.257522390   -0.272978840    2.660885820
16   H     0.305654620   -0.421389450    3.490460480
17   H    -0.692123480    1.947938900    3.376665930
18   H     1.007813760    1.794221060    2.886724950
19   H    -1.474234560    2.226951260    1.118549790
20   H    -0.040139170    3.261877520    1.293983970
21   H    -1.213393790    2.214648690   -1.419486450
22   H     0.226912130    3.250127460   -1.308795750
23   H     0.013352470    1.917277620   -3.468707570
24   H     1.576996840    1.768575080   -2.640363060
25   H    -0.689917590   -0.297319350   -2.862898670
26   H     1.008919890   -0.452435740   -3.355808630
27   H     0.360559610   -1.765736960   -1.271966280
28   H     1.794301840   -0.730504380   -1.097751120
---
H20C10, RHF, CHARGE=0, MULT=1
HF=-39.9
1    C     0.000000000    0.000000000    0.000000000
2    C     0.000000000    1.557001930    0.000000000
3    C     0.000000000    2.024811993    1.455697266
4    C    -1.045601907    2.397310424    2.218993752
5    C    -1.044544989    2.865509332    3.674854300
6    C    -1.050502620    4.422643787    3.671204454
7    C    -1.218879453    2.076935211   -0.812328722
8    C     1.298786363    2.082040921   -0.683742264
9    C     0.177911992    2.353645122    4.486459462
10   C    -2.339540179    2.334179362    4.360633148
11   H    -0.913677232   -0.405865571    0.479910185
```

```
12      H       0.870510230     -0.409315351     0.551978181
13      H       0.041771719     -0.408955051    -1.030161378
14      H       1.006823928      2.018893172     1.889137932
15      H      -2.052379481      2.403146152     1.785787112
16      H      -0.142095639      4.830888222     3.183369317
17      H      -1.926756165      4.827166216     3.124799430
18      H      -1.086464001      4.834166789     4.700592362
19      H      -2.177753710      1.658608778    -0.445006372
20      H      -1.297369864      3.182337474    -0.767603670
21      H      -1.140374331      1.797110622    -1.883391705
22      H       1.354062297      3.189260697    -0.648349122
23      H       2.212977211      1.690284804    -0.192771305
24      H       1.350584592      1.781321395    -1.750103879
25      H       1.134504417      2.773595184     4.115008743
26      H       0.260274966      1.248302775     4.446104442
27      H       0.100704368      2.637782023     5.556480914
28      H      -2.386761079      1.226453408     4.329864178
29      H      -3.256833232      2.717122805     3.868566434
30      H      -2.393395799      2.639060839     5.425774247
---
H20C10, RHF, CHARGE=0, MULT=1
HF=-30.3
1       C       0.003478440     -0.003790670    -0.238739530
2       C       0.045399130      1.545692950    -0.169693060
3       C       0.018326730      2.096450170     1.253415850
4       C      -0.626451530      1.887339750     2.419606430
5       C      -1.704793010      0.975547040     2.998587950
6       C      -1.092305210     -0.414432590     3.336204590
7       C      -1.126278230      2.165738210    -0.984059990
8       C       1.392570420      1.993156480    -0.829896710
9       C      -2.942128880      0.819759420     2.074488360
10      C      -2.208975240      1.620481590     4.333185780
11      H      -0.987531250     -0.416610060     0.026985070
12      H       0.755475440     -0.454886440     0.439690830
13      H       0.224670680     -0.370208460    -1.263398360
14      H       0.738567690      2.930844270     1.338016830
15      H      -0.261856190      2.573982040     3.205542540
16      H      -0.790525940     -0.973215050     2.430010890
17      H      -0.191560320     -0.315317090     3.976001370
18      H      -1.816774580     -1.054504440     3.881277710
19      H      -2.115929920      1.846087090    -0.605687700
20      H      -1.103564030      3.274055690    -0.945834320
21      H      -1.078786500      1.869637480    -2.052424900
22      H       1.490903720      3.097298110    -0.867096970
23      H       2.266519660      1.599461360    -0.271745120
24      H       1.477492340      1.628906190    -1.874122130
25      H      -2.729841930      0.211796710     1.175361860
26      H      -3.319252610      1.805972040     1.735883420
27      H      -3.778199130      0.311585170     2.599541560
28      H      -2.641965410      2.626935840     4.160949740
29      H      -1.390915680      1.734090200     5.073428560
30      H      -2.995085540      1.002588360     4.813448140
---
H20C10, RHF, CHARGE=0, MULT=1
HF=-29.8
1       C       0.000000000      0.000000000     0.000000000
2       C       0.000000000      1.531330560     0.000000000
3       C       0.000000000      2.174990910     1.400444558
4       C       0.001190295      3.717274134     1.384891824
```

```
5    C     0.002280529    4.359539746    2.787449466
6    C     0.004305108    5.902002181    2.771632237
7    C     0.006631859    6.543960424    4.174161952
8    C     0.008836678    8.086322005    4.159785197
9    C     0.013071247    8.725231439    5.523669769
10   C     0.015409121   10.045977017    5.759418912
11   H    -0.894283163   -0.412211472    0.509596220
12   H     0.897178862   -0.412697542    0.504093476
13   H    -0.003454411   -0.383978470   -1.040765861
14   H     0.891945637    1.882334338   -0.569075268
15   H    -0.890115300    1.883518473   -0.570716811
16   H     0.892172859    1.815129668    1.963008116
17   H    -0.891677557    1.816046831    1.963868998
18   H     0.893762343    4.073696759    0.820929716
19   H    -0.889994289    4.076022841    0.820654298
20   H     0.894502341    4.000276558    3.350198310
21   H    -0.889372385    4.002170722    3.351838539
22   H     0.896820612    6.258067774    2.207305216
23   H    -0.886907251    6.261416565    2.207833720
24   H     0.899847347    6.186159759    4.736331053
25   H    -0.885300761    6.188494373    4.739388380
26   H     0.900952733    8.445326786    3.593228735
27   H    -0.884011432    8.447993667    3.596552884
28   H     0.014159558    8.039343514    6.377154100
29   H     0.014527811   10.807110860    4.980849895
30   H     0.018450694   10.458308493    6.767549665
---
H20C10, RHF, CHARGE=0, MULT=1
HF=-51
1    C     0.018388762    0.097252118   -0.047066206
2    C    -0.063822570    1.638241307    0.109999854
3    C    -0.205057713    2.122111332    1.564532755
4    C    -1.294666873    1.394286616    2.366980415
5    C    -1.252261673   -0.133944859    2.212294194
6    C    -1.101496059   -0.616452962    0.758811598
7    C     0.014244798   -0.371251106   -1.530376200
8    C     1.311469537   -0.108552008   -2.322840696
9    C     1.278640293   -0.646845656   -3.767294774
10   C     2.560537158   -0.403012770   -4.568447459
11   H     0.992752191   -0.227738830    0.398861302
12   H    -0.914470594    2.038472227   -0.487679694
13   H     0.853583544    2.104997213   -0.314749395
14   H    -0.427955148    3.213614954    1.557913983
15   H     0.773323100    2.017705257    2.087685161
16   H    -1.189168338    1.654542704    3.444686068
17   H    -2.297609608    1.771741233    2.061634188
18   H    -2.186753663   -0.566997227    2.637041454
19   H    -0.421646875   -0.546556768    2.829976056
20   H    -0.894887741   -1.711184839    0.779286042
21   H    -2.079377345   -0.499114219    0.237917575
22   H    -0.181127727   -1.469121534   -1.549492259
23   H    -0.842696509    0.099732972   -2.064104589
24   H     2.169323774   -0.571588993   -1.783251376
25   H     1.516425152    0.985328738   -2.356384583
26   H     1.077651617   -1.743052536   -3.752384289
27   H     0.426763968   -0.183145359   -4.316768563
28   H     2.778760012    0.678476331   -4.679327701
29   H     3.441528388   -0.880218181   -4.093539482
30   H     2.460745719   -0.829141773   -5.587982523
```

```
---
H20C10, RHF, CHARGE=0, MULT=1
HF=-45.2
1    C     0.000000000     0.000000000     0.000000000
2    C     0.000000000     1.553452130     0.000000000
3    C     0.000000000     2.025935641     1.468112295
4    C    -0.257622668     0.795017578     2.357666860
5    C    -0.444774969    -0.421572458     1.429344297
6    C    -0.820227318    -0.673596320    -1.125737145
7    C    -0.227860652    -0.538934099    -2.544229704
8    C    -1.048778145    -1.257525605    -3.634980802
9    C    -0.456249127    -1.133435575    -5.052149478
10   C    -1.260850036    -1.840502858    -6.146360708
11   H     1.061267082    -0.333863773    -0.120314187
12   H    -0.885674194     1.966698516    -0.528546761
13   H     0.889551800     1.943457739    -0.539040043
14   H    -0.776863053     2.803211291     1.630800156
15   H     0.969288962     2.503051441     1.728257085
16   H    -1.153773824     0.945954121     2.996427617
17   H     0.591308186     0.630108590     3.055614162
18   H     0.154225245    -1.287210937     1.783794071
19   H    -1.505363941    -0.752602722     1.451397933
20   H    -1.859914314    -0.273479828    -1.122340463
21   H    -0.908050581    -1.759282411    -0.885137699
22   H    -0.141352648     0.538653193    -2.811604190
23   H     0.809810684    -0.945110539    -2.544869769
24   H    -2.086353015    -0.851167630    -3.639865222
25   H    -1.137688549    -2.337229133    -3.373838564
26   H    -0.364116076    -0.056567457    -5.324802580
27   H     0.580320937    -1.543423550    -5.058245824
28   H    -2.291329023    -1.439347831    -6.227575800
29   H    -1.334468060    -2.931936332    -5.964623038
30   H    -0.770390426    -1.700345179    -7.131929724
---
H22C10, RHF, CHARGE=0, MULT=1
HF=-63.5
1    C     0.000000000     0.000000000     0.000000000
2    C     0.000000000     1.535890240     0.000000000
3    C     0.000000000     2.324156927     1.365873653
4    C    -0.293642046     3.902986366     1.119323987
5    C     0.039179392     4.741776873     2.413106791
6    C    -0.418434571     6.204031117     2.517345759
7    C    -1.071492926     1.698526404     2.312031127
8    C     1.391499505     2.100260455     2.037985622
9    C     0.578464353     4.472059464    -0.044411013
10   C    -1.785946587     4.137702308     0.729246994
11   H     0.865136828    -0.434438962     0.538370334
12   H    -0.923826988    -0.429917759     0.435042517
13   H     0.059956299    -0.361450942    -1.049042244
14   H    -0.885568613     1.835400229    -0.605356700
15   H     0.894770538     1.841885991    -0.590829907
16   H    -0.381828618     4.229620727     3.307904690
17   H     1.144643891     4.754991712     2.558575395
18   H    -0.025946101     6.840109552     1.699670492
19   H    -1.521252517     6.309227768     2.532181700
20   H    -0.040449984     6.634568651     3.469409953
21   H    -2.055375290     1.576796346     1.815634212
22   H    -1.241852891     2.298072624     3.228080647
23   H    -0.768088532     0.690648204     2.665298577
```

```
24    H     2.221375888    2.562464871    1.467708179
25    H     1.434839723    2.500357618    3.071116723
26    H     1.636148240    1.021527655    2.127427449
27    H     1.662230754    4.303035674    0.111757599
28    H     0.312235331    4.038934834   -1.029535962
29    H     0.446264179    5.568239647   -0.156937481
30    H    -2.134715460    3.459812727   -0.074633591
31    H    -2.475191079    4.005472027    1.587483629
32    H    -1.950302166    5.166587478    0.345528291
---
H22C10, RHF, CHARGE=0, MULT=1
HF=-68
1     C     0.000000000    0.000000000    0.000000000
2     C     0.000000000    1.554450830    0.000000000
3     C     0.000000000    2.133264920    1.461344364
4     C     1.195687338    1.806899631    2.382826769
5     C     1.189994879    2.376829117    3.847820573
6     C     1.156892469    3.930729927    3.855361522
7     C     1.218016005    2.069549824   -0.816966537
8     C    -1.304832282    2.044836153   -0.701167307
9     C    -0.013189467    1.831142880    4.666794052
10    C     2.508419937    1.909868233    4.538941051
11    H     0.940930162   -0.421239815    0.407260498
12    H    -0.833279686   -0.409851971    0.606729111
13    H    -0.112976181   -0.405714336   -1.026707178
14    H    -0.090725634    3.240693757    1.377681698
15    H    -0.935549036    1.784651658    1.955623220
16    H     1.292092988    0.699473834    2.457408197
17    H     2.128569550    2.164892619    1.890117522
18    H     1.968910134    4.361919971    3.234937449
19    H     0.197591857    4.333549623    3.473210698
20    H     1.283310056    4.333229134    4.882065210
21    H     1.293361411    3.175778820   -0.789280891
22    H     2.178383223    1.662612451   -0.441671342
23    H     1.143929696    1.774561932   -1.884377237
24    H    -2.214095025    1.692995276   -0.172287731
25    H    -1.357727932    3.151839660   -0.745633520
26    H    -1.372483949    1.674324737   -1.744615264
27    H    -0.985733866    2.197079102    4.280909191
28    H    -0.050037767    0.722561372    4.652301994
29    H     0.045317294    2.142185032    5.730497427
30    H     2.584255429    0.803953085    4.577401783
31    H     3.407239984    2.282798131    4.006694344
32    H     2.574263154    2.276511552    5.583930993
---
H22C10, RHF, CHARGE=0, MULT=1
HF=-59.7
1     C     0.000000000    0.000000000    0.000000000
2     C     0.000000000    1.531232890    0.000000000
3     C     0.000000000    2.174937918    1.400382795
4     C     0.001148188    3.717168445    1.384747339
5     C     0.001257157    4.359542707    2.787223557
6     C     0.004073721    5.901961388    2.771216429
7     C     0.004413025    6.544409901    4.173626008
8     C     0.008391801    8.086668641    4.157771681
9     C     0.009930755    8.730587786    5.557775527
10    C     0.013628093   10.261895161    5.557825575
11    H     0.891766699   -0.412593504    0.513837027
12    H    -0.899862519   -0.412423487    0.499632934
```

```
13      H      0.008289446    -0.383773487    -1.040911689
14      H      0.891973036     1.882072757    -0.569137863
15      H     -0.890060383     1.883358481    -0.570852282
16      H     -0.893275177     1.816722237     1.962195976
17      H      0.890478911     1.814425359     1.964685369
18      H      0.893977447     4.073672165     0.821258371
19      H     -0.889715015     4.075696528     0.819888632
20      H     -0.892610714     4.003351541     3.349262744
21      H      0.891182880     3.999229627     3.352422746
22      H      0.897486058     6.257542594     2.208057560
23      H     -0.886241109     6.261518369     2.206154225
24      H     -0.889840833     6.189191574     4.735658473
25      H      0.893930466     6.183615358     4.739166873
26      H      0.901756689     8.442627060     3.594631130
27      H     -0.881841264     8.447925811     3.593529632
28      H     -0.882638931     8.381826440     6.127242377
29      H      0.899308923     8.376518960     6.128596890
30      H      0.906656793    10.672367050     5.044466565
31      H     -0.884842754    10.676361806     5.057443938
32      H      0.022322032    10.645198265     6.598923756
EXPGEOM
1       C      0.00000    -0.00000     0.76710
2       C      0.00000     0.00000    -0.76710
3       C      0.00000    -1.40670     1.37930
4       C      0.00000     1.40670    -1.37930
5       C      0.00000    -1.40670     2.91320
6       C      0.00000     1.40670    -2.91320
7       C      0.00000    -2.81260     3.52630
8       C      0.00000     2.81260    -3.52630
9       C      0.00000    -2.80260     5.05920
10      C      0.00000     2.80260    -5.05920
11      H      0.87750     0.55290     1.12940
12      H     -0.87750     0.55290     1.12940
13      H      0.87750    -0.55290    -1.12940
14      H     -0.87750    -0.55290    -1.12940
15      H     -0.87750    -1.95970     1.01690
16      H      0.87750    -1.95970     1.01690
17      H     -0.87750     1.95970    -1.01690
18      H      0.87750     1.95970    -1.01690
19      H      0.87740    -0.85400     3.27640
20      H     -0.87740    -0.85400     3.27640
21      H      0.87740     0.85400    -3.27640
22      H     -0.87740     0.85400    -3.27640
23      H     -0.87700    -3.36460     3.16410
24      H      0.87700    -3.36460     3.16410
25      H     -0.87700     3.36460    -3.16410
26      H      0.87700     3.36460    -3.16410
27      H      0.00000    -3.81860     5.46690
28      H      0.88360    -2.28650     5.45080
29      H     -0.88360    -2.28650     5.45080
30      H      0.00000     3.81860    -5.46690
31      H      0.88360     2.28650    -5.45080
32      H     -0.88360     2.28650    -5.45080
---
H16C11, RHF, CHARGE=0, MULT=1
HF=-16.1
1       C      0.000541213     0.000162693     0.000458797
2       C      1.428712290    -0.001106676    -0.000198211
3       C      2.156323191     1.227833875     0.000029352
```

```
4    C     1.454583137    2.453802201   -0.169065877
5    C     0.049179335    2.421649245   -0.263992781
6    C    -0.695559547    1.229110144   -0.170243195
7    C    -2.199770612    1.322382521   -0.265223160
8    C    -0.772655233   -1.287749342    0.180666024
9    C     2.177932999   -1.317197635   -0.015409182
10   C     3.658760852    1.234538706    0.178450077
11   C     2.141414023    3.795506803   -0.261599412
12   H    -0.487182455    3.363694818   -0.409307416
13   H    -2.607049884    0.622642871   -1.024487637
14   H    -2.678272874    1.097634727    0.711389242
15   H    -2.538680961    2.336838120   -0.561398089
16   H    -1.780568302   -1.127145310    0.613928582
17   H    -0.908476066   -1.805638894   -0.792766959
18   H    -0.269084569   -1.988586598    0.878166741
19   H     2.380645400   -1.671397112    1.017755744
20   H     3.148549296   -1.248376812   -0.548204046
21   H     1.622327657   -2.116923574   -0.546916354
22   H     4.171985291    1.067337917   -0.792598974
23   H     4.001530726    0.455461523    0.890463246
24   H     4.039442543    2.189758933    0.592976462
25   H     1.442195607    4.604387398   -0.559230038
26   H     2.952992865    3.789972274   -1.018731650
27   H     2.575779664    4.091741064    0.716591586
---
H22C11, RHF, CHARGE=0, MULT=1
HF=-50.3
1    C    -0.000089054   -0.000037178   -0.000066833
2    C     1.565769361    0.000030760    0.000015721
3    C     2.364130099    1.318761889    0.000053146
4    C     2.733205049    1.924638057   -1.367725112
5    C     1.851494042    3.067006347   -1.903984842
6    C     0.493586712    2.683374583   -2.547627882
7    C    -0.561478838    2.162604630   -1.532211284
8    C    -0.609036610    0.636723417   -1.295087304
9    C    -0.447499430   -1.502165885    0.013384921
10   C    -0.566605413    0.655732458    1.290604976
11   C    -0.068384459    3.865019482   -3.373023450
12   H     1.916784383   -0.619659525   -0.858144758
13   H     1.896693720   -0.555305523    0.911816255
14   H     3.328696626    1.096019753    0.520587757
15   H     1.873837180    2.084270291    0.639216138
16   H     3.765427866    2.342220464   -1.269705218
17   H     2.819757805    1.125715181   -2.138147551
18   H     2.462948254    3.590626873   -2.678474656
19   H     1.689155027    3.820635676   -1.099872748
20   H     0.696009164    1.873625334   -3.293452380
21   H    -1.576033436    2.440732645   -1.908411543
22   H    -0.462099024    2.721901099   -0.577543219
23   H    -1.692758078    0.363606077   -1.318613922
24   H    -0.166414637    0.117051832   -2.176024169
25   H    -1.551425528   -1.601659718    0.057741139
26   H    -0.107164765   -2.044260411   -0.892355356
27   H    -0.039534004   -2.044947505    0.890563655
28   H    -0.254944533    1.710741619    1.416465026
29   H    -1.676278237    0.644848497    1.296968771
30   H    -0.229552505    0.116283505    2.200244484
31   H    -0.320923532    4.738880613   -2.739170006
32   H    -0.986624673    3.567821362   -3.920235093
```

```
33         H        0.661814780     4.203744568    -4.136476492
---
H22C11, RHF, CHARGE=0, MULT=1
HF=-50.1
1     C       -0.050126115    -0.097699077    -0.181963564
2     C        1.445685040    -0.152563866     0.181602471
3     C        1.962687185     1.298003610     0.265702273
4     C        0.732326224     2.245483675     0.223217475
5     C       -0.443405665     1.386048562    -0.323109660
6     C        0.930528063     3.575223301    -0.542787222
7     C        1.868815790     4.594143238     0.137283726
8     C        2.006901422     5.925501507    -0.630301211
9     C        2.934024149     6.949862426     0.055037666
10    C        3.085070867     8.275985535    -0.715131608
11    C        4.033225365     9.287520837    -0.065331877
12    H       -0.663120904    -0.594632918     0.600544878
13    H       -0.251216962    -0.650120894    -1.124527529
14    H        1.595682358    -0.680123684     1.148095626
15    H        2.017911987    -0.731353059    -0.574407620
16    H        2.543974490     1.454627535     1.199375500
17    H        2.666044318     1.496099361    -0.571387512
18    H        0.477718494     2.508614735     1.280507833
19    H       -1.376748362     1.604622105     0.238206182
20    H       -0.664640678     1.618020155    -1.387212014
21    H       -0.070079582     4.051445926    -0.669624892
22    H        1.302125385     3.362096427    -1.571023892
23    H        1.493459429     4.804878863     1.164989459
24    H        2.879889548     4.143786947     0.260240900
25    H        0.997655084     6.379475069    -0.760527263
26    H        2.390003710     5.716167255    -1.655429308
27    H        2.546073394     7.167332745     1.076753089
28    H        3.941410529     6.494328864     0.194196916
29    H        2.084654389     8.753953365    -0.830197938
30    H        3.448590347     8.067462769    -1.747944923
31    H        3.693236580     9.580333851     0.948715335
32    H        5.063957460     8.888423654     0.024613300
33    H        4.085900014    10.210583668    -0.678717787
---
H22C11, RHF, CHARGE=0, MULT=1
HF=-55.9
1     C       -0.154744788    -0.083045207    -0.034370558
2     C        1.361399226    -0.105607132     0.214791147
3     C        2.029389175     1.280606095     0.167347580
4     C        1.647998212     2.132885163    -1.071550833
5     C        0.110293567     2.157536573    -1.290717557
6     C       -0.565226795     0.775810763    -1.240809790
7     C        2.223494294     3.576655071    -1.024007032
8     C        3.740095448     3.708414322    -1.276652537
9     C        4.251490211     5.163174639    -1.213407300
10    C        5.776033187     5.306134490    -1.387054453
11    C        6.299517528     6.739897407    -1.263354756
12    H       -0.514241266    -1.125583639    -0.189992859
13    H       -0.676982831     0.282897271     0.879128069
14    H        1.557646060    -0.562801143     1.211574582
15    H        1.849437455    -0.779176179    -0.526927932
16    H        3.131966786     1.126091265     0.187949730
17    H        1.780424830     1.832753586     1.102473991
18    H        2.091590882     1.629390257    -1.968113162
19    H       -0.372520094     2.821419247    -0.537382695
```

```
 20      H        -0.109004729     2.611609509    -2.284636555
 21      H        -1.670121570     0.918440976    -1.224473691
 22      H        -0.347012035     0.220854535    -2.182204913
 23      H         1.973548725     4.039355513    -0.042055291
 24      H         1.696556200     4.189690974    -1.792207245
 25      H         4.300173407     3.098921608    -0.532575262
 26      H         3.982537559     3.280752029    -2.276426760
 27      H         3.955140503     5.612355655    -0.237816098
 28      H         3.740609311     5.763099377    -2.001054906
 29      H         6.298470346     4.675807501    -0.630393063
 30      H         6.074902743     4.905423825    -2.383133795
 31      H         6.093263973     7.168579884    -0.261401943
 32      H         5.846283592     7.412373725    -2.019149423
 33      H         7.398979113     6.760683015    -1.414349253
---
H24C11, RHF, CHARGE=0, MULT=1
HF=-64.6
 1    C     0.000141604    -0.000006592     0.000014581
 2    C     1.531322901     0.000130134     0.000054948
 3    C     2.174532439     1.400506818    -0.000015452
 4    C     3.716788144     1.385098317    -0.005445710
 5    C     4.358609003     2.787769549    -0.004016957
 6    C     5.901075479     2.772143201    -0.010307349
 7    C     6.542829316     4.174879292    -0.008953835
 8    C     8.085276177     4.159255941    -0.014984790
 9    C     8.727308949     5.561561324    -0.014031129
 10   C    10.268343880     5.548279589    -0.020570586
 11   C    10.919075076     6.934350293    -0.019385395
 12   H    -0.412922048     0.507132148     0.895503066
 13   H    -0.413018836     0.504860307    -0.896677817
 14   H    -0.382493089    -1.041672444     0.001340027
 15   H     1.883081355    -0.569594210     0.891349243
 16   H     1.883102232    -0.570547853    -0.890457628
 17   H     1.811911773     1.965740016    -0.889201249
 18   H     1.818298325     1.962482740     0.893803633
 19   H     4.071741223     0.823202489    -0.899746993
 20   H     4.077978788     0.818435720     0.883408316
 21   H     3.996662164     3.354490663    -0.892490547
 22   H     4.004070211     3.349328661     0.890699282
 23   H     6.262862929     2.205524056     0.878359837
 24   H     6.255766557     2.210516390    -0.904826465
 25   H     6.181330789     4.741394971    -0.897757853
 26   H     6.187829228     4.736597183     0.885366924
 27   H     8.447261420     3.593127982     0.873811359
 28   H     8.440129933     3.597182773    -0.909300242
 29   H     8.366028566     6.129013296    -0.902407684
 30   H     8.374033222     6.124002856     0.880722642
 31   H    10.638858079     4.985202064     0.867202368
 32   H    10.631488225     4.989951520    -0.914376298
 33   H    10.632856204     7.525650634    -0.912720144
 34   H    10.640202909     7.520773068     0.879495889
 35   H    12.024511975     6.837700980    -0.024189400
---
H8C12, RHF, CHARGE=0, MULT=1
HF=61.6
 1    C    -0.000298364    -0.000282632    -0.000019178
 2    C     1.380429199     0.000962267     0.000007280
 3    C     2.067513382     1.265382235    -0.000008440
 4    C     1.402289230     2.484197967    -0.000029416
```

```
  5    C    -0.037420144    2.531704428   -0.000076219
  6    C    -0.679777032    1.280131882   -0.000070933
  7    C    -2.116146981    1.086113147   -0.000164814
  8    C    -2.919743226    2.208926822   -0.000344958
  9    C    -2.293181895    3.504348740   -0.000383136
 10    C    -0.915093449    3.674079315   -0.000212043
 11    C    -1.083553607   -1.009489494   -0.000034285
 12    C    -2.304831628   -0.382390134   -0.000113346
 13    H     1.963706703   -0.919532990    0.000030462
 14    H     3.159041916    1.249069254   -0.000001362
 15    H     1.969253615    3.415950218   -0.000018071
 16    H    -4.007539793    2.146048788   -0.000450052
 17    H    -2.942934137    4.381648254   -0.000572990
 18    H    -0.488152365    4.677739253   -0.000180719
 19    H    -0.900927813   -2.076207438   -0.000005325
 20    H    -3.278062792   -0.855517672   -0.000125849
EXPGEOM
 1    C     0.00000    0.00000   -0.14000
 2    C     0.00000    0.00000    1.24960
 3    C     0.00000    1.27980    1.87410
 4    C     0.00000   -1.27980    1.87410
 5    C     0.00000    1.16090   -0.94630
 6    C     0.00000   -1.16090   -0.94630
 7    C     0.00000    2.42200    1.10210
 8    C     0.00000   -2.42200    1.10210
 9    C     0.00000    2.38510   -0.32260
 10    C    0.00000   -2.38510   -0.32260
 11    C    0.00000    0.67900   -2.33810
 12    C    0.00000   -0.67900   -2.33810
 13    H    0.00000    3.38740    1.58890
 14    H    0.00000   -3.38740    1.58890
 15    H    0.00000    1.35480    2.95340
 16    H    0.00000   -1.35480    2.95340
 17    H    0.00000    3.31470   -0.87590
 18    H    0.00000   -3.31470   -0.87590
 19    H    0.00000    1.31390   -3.21020
 20    H    0.00000   -1.31390   -3.21020
---
H8C12, RHF, CHARGE=0, MULT=1
HF=100.5
 1    C     0.000173566    0.000400143    0.000024546
 2    C     1.439946742   -0.000167637    0.000035212
 3    C     2.178881731    1.176913129   -0.000069106
 4    C     1.552318216    2.473190926    0.000061727
 5    C     0.183474572    2.477387917    0.000137572
 6    C    -0.598561729    1.231194670    0.000082044
 7    C    -1.854151172    2.019332366    0.000158144
 8    C    -3.222951668    2.023363304    0.000175894
 9    C    -3.849827118    3.319446891    0.000323204
 10    C   -3.111211652    4.496705200    0.000377270
 11    C   -1.671425324    4.496175752    0.000285700
 12    C   -1.072197581    3.265429795    0.000211602
 13    H   -0.550222070   -0.937683560   -0.000068378
 14    H    1.953272119   -0.962996031    0.000081925
 15    H    3.269076882    1.133279384   -0.000263245
 16    H    2.157835259    3.376778667    0.000097218
 17    H   -3.828378869    1.119762144    0.000089840
 18    H   -4.940094025    3.362768562    0.000416139
 19    H   -3.624795273    5.459382245    0.000500791
```

```
  20    H      -1.121230874     5.434401234      0.000329353
EXPGEOM
1     C     0.00000      0.71910      0.76170
2     C     0.00000      0.71910     -0.76170
3     C     0.00000     -0.71910      0.76170
4     C     0.00000     -0.71910     -0.76170
5     C     0.00000      1.46190      1.93210
6     C     0.00000      0.70190      3.14730
7     C     0.00000     -0.70190      3.14730
8     C     0.00000     -1.46190      1.93210
9     C     0.00000      1.46190     -1.93210
10    C     0.00000      0.70190     -3.14730
11    C     0.00000     -0.70190     -3.14730
12    C     0.00000     -1.46190     -1.93210
13    H     0.00000      2.55790      1.95190
14    H     0.00000      1.23250      4.10740
15    H     0.00000     -1.23250      4.10740
16    H     0.00000     -2.55790      1.95190
17    H     0.00000      2.55790     -1.95190
18    H     0.00000      1.23250     -4.10740
19    H     0.00000     -1.23250     -4.10740
20    H     0.00000     -2.55790     -1.95190
---
H10C12, RHF, CHARGE=0, MULT=1
HF=37.4
1     C     -0.000356715    -0.000938739    -0.000149349
2     C      1.379370140     0.001098257    -0.000001427
3     C      2.071268709     1.261537141    -0.000076118
4     C      1.402374243     2.476157895    -0.000262146
5     C     -0.036857156     2.518253268    -0.000233944
6     C     -0.689821343     1.265110713    -0.000186715
7     C     -2.122484672     1.104904676    -0.000223509
8     C     -2.911121903     2.237128709    -0.000090848
9     C     -2.274380873     3.526387599     0.000013462
10    C     -0.895700975     3.673938355    -0.000127236
11    C     -1.040844399    -1.101923513    -0.000342760
12    C     -2.428691980    -0.378676337    -0.000556570
13    H      1.960368065    -0.920743966     0.000162914
14    H      3.162712872     1.244464347    -0.000033409
15    H      1.964587631     3.410692874    -0.000425564
16    H     -3.999648693     2.186032071    -0.000187743
17    H     -2.913844818     4.411015961     0.000147453
18    H     -0.451832500     4.670119316    -0.000313812
19    H     -0.935168239    -1.755893536     0.890949596
20    H     -0.935023953    -1.755561733    -0.891834820
21    H     -3.024959961    -0.666634590    -0.892095318
22    H     -3.025333089    -0.666956882     0.890637338
---
H10C12, RHF, CHARGE=0, MULT=1
HF=43.5
1     C     -0.000166285    -0.000196443     0.000420093
2     C      1.414816395     0.000318750    -0.000506567
3     C     -0.677214030     1.242362834    -0.000330606
4     C      2.130998234     1.209831470    -0.002892711
5     C      0.041302094     2.450500151    -0.003091274
6     C      1.446678362     2.437421554    -0.004529275
7     C     -0.759087376    -1.278174098     0.002010124
8     C     -1.139041784    -1.895199233    -1.213379484
9     C     -1.856654394    -3.103871925    -1.210130773
```

```
10      C       -2.206419743    -3.715872985     0.005843998
11      C       -1.834933453    -3.113168011     1.219855774
12      C       -1.117343443    -1.904350441     1.219114654
13      H        1.963912449    -0.942616931     0.000668493
14      H       -1.767962766     1.272984777     0.001340955
15      H        3.221414338     1.193992099    -0.003056250
16      H       -0.494826751     3.400118969    -0.004111573
17      H        2.003213139     3.375001913    -0.006815604
18      H       -0.876770822    -1.435794929    -2.167715727
19      H       -2.142261667    -3.566998845    -2.155192918
20      H       -2.763138025    -4.653257161     0.007592380
21      H       -2.103461393    -3.583337859     2.166502782
22      H       -0.838346395    -1.451807972     2.171897358
EXPGEOM
1      C      0.00000     0.00000      0.74370
2      C      0.00000     0.00000     -0.74370
3      C     -0.46470     1.11290      1.45810
4      C      0.46470    -1.11290      1.45810
5      C     -0.46470    -1.11290     -1.45810
6      C      0.46470     1.11290     -1.45810
7      C     -0.46330     1.11350      2.85250
8      C      0.46330    -1.11350      2.85250
9      C     -0.46330    -1.11350     -2.85250
10     C      0.46330     1.11350     -2.85250
11     C      0.00000     0.00000      3.55440
12     C      0.00000     0.00000     -3.55440
13     H     -0.85040     1.97470      0.91560
14     H      0.85040    -1.97470      0.91560
15     H     -0.85040    -1.97470     -0.91560
16     H      0.85040     1.97470     -0.91560
17     H     -0.83380     1.98340      3.39250
18     H      0.83380    -1.98340      3.39250
19     H     -0.83380    -1.98340     -3.39250
20     H      0.83380     1.98340     -3.39250
21     H      0.00000     0.00000      4.64300
22     H      0.00000     0.00000     -4.64300
---
H18C12, RHF, CHARGE=0, MULT=1
HF=-18.5
1      C       0.000066105     0.000152058    -0.000199273
2      C       1.424272001     0.000018847     0.000492715
3      C       2.132939341     1.235203513    -0.000447812
4      C       1.424201507     2.459159637     0.166664938
5      C       0.000158821     2.459191473     0.162914016
6      C      -0.708162859     1.223690025     0.171090130
7      C      -0.758741172    -1.293551948    -0.212993598
8      C      -2.206397751     1.207977889     0.393370531
9      C      -0.758038202     3.769696311     0.117716124
10     C       2.183473980     3.752871047     0.377897686
11     C       3.633152238     1.250005845    -0.209439865
12     C       2.183067663    -1.310548652     0.032879407
13     H      -0.955235108    -1.806511270     0.752376020
14     H      -0.212261390    -2.000452160    -0.870970526
15     H      -1.733349147    -1.136579781    -0.719237570
16     H      -2.550137850     2.050756400     1.027744421
17     H      -2.544319789     0.297492032     0.930040840
18     H      -2.755779984     1.260523933    -0.570474142
19     H      -0.207310042     4.557434940    -0.436714360
20     H      -0.962223032     4.150643548     1.140816699
```

```
21         H        -1.728518681      3.680565384      -0.412495677
22         H         2.409744513      4.247390126      -0.590528349
23         H         3.142179424      3.597630710       0.914453080
24         H         1.624589969      4.476050742       1.006695381
25         H         3.981235317      0.410995813      -0.846452694
26         H         4.173135008      1.189725013       0.759250776
27         H         3.978811096      2.163313102      -0.736412723
28         H         2.401916438     -1.672209722      -0.994267261
29         H         1.627219378     -2.110211937       0.564236297
30         H         3.145605052     -1.228923668       0.578723635
---
H24C12, RHF, CHARGE=0, MULT=1
HF=-60.8
1     C     0.000195258     -0.000361259      0.000010620
2     C     1.539180108      0.000298116     -0.000074838
3     C     2.167236970      1.402916620      0.000055066
4     C     1.525927645      2.364444398      1.012833243
5     C    -0.013767507      2.352810889      1.011934821
6     C    -0.639325369      0.934555121      1.063034266
7     C    -2.189915488      0.936976963      0.942804046
8     C    -2.958567996      1.451773705      2.178169286
9     C    -4.489454626      1.287905646      2.070912959
10    C    -5.265475075      1.830099995      3.288430427
11    C    -6.794119323      1.663957934      3.184909160
12    C    -7.578670386      2.199801946      4.385739393
13    H    -0.361851779      0.273907841     -1.017301813
14    H    -0.344041060     -1.045286083      0.176180286
15    H     1.911395759     -0.572997106      0.880049098
16    H     1.897235204     -0.557222540     -0.895622912
17    H     2.101002360      1.840537874     -1.022494899
18    H     3.255353544      1.314758781      0.220268582
19    H     1.875018649      3.400593222      0.799258535
20    H     1.897663416      2.126142324      2.035969074
21    H    -0.358000408      2.944108879      1.890478882
22    H    -0.379275172      2.899940058      0.113028697
23    H    -0.389260955      0.498919365      2.063925878
24    H    -2.524029801     -0.107208585      0.738999691
25    H    -2.492179664      1.533030952      0.051525307
26    H    -2.723743266      2.528266041      2.339184094
27    H    -2.600448645      0.913391419      3.085316017
28    H    -4.731728655      0.208546083      1.937112668
29    H    -4.847904378      1.805268802      1.151494869
30    H    -5.026870937      2.910358549      3.422025816
31    H    -4.906604975      1.314947746      4.209181554
32    H    -7.042589056      0.585068362      3.055820653
33    H    -7.162194009      2.178016444      2.266892297
34    H    -8.666386370      2.041728408      4.233192623
35    H    -7.417970537      3.287145768      4.532267042
36    H    -7.296526708      1.685053281      5.326501120
---
H26C12, RHF, CHARGE=0, MULT=1
HF=-69.2
1     C     0.000013169     -0.000064473     -0.000033922
2     C     1.531190218      0.000058215     -0.000022660
3     C     2.173769132      1.400831317      0.000038930
4     C     3.715991925      1.386366795     -0.004138004
5     C     4.356929833      2.789436310     -0.002194704
6     C     5.899377999      2.774586943     -0.011220928
7     C     6.540399063      4.177645949     -0.007376679
```

```
 8    C     8.082814834    4.162812559   -0.020991550
 9    C     8.723877531    5.565820581   -0.015672290
10    C    10.265995969    5.551414311   -0.033427810
11    C    10.908720208    6.952112919   -0.027070803
12    C    12.439769403    6.952256935   -0.047149758
13    H    12.822401808    7.993813353   -0.037056107
14    H    12.840807309    6.459514436   -0.956020778
15    H    12.864486561    6.433313649    0.835978734
16    H    10.568803714    7.512808368    0.874193013
17    H    10.545524108    7.531946669   -0.907085757
18    H    10.615412134    4.996639133   -0.934458055
19    H    10.635793106    4.979197278    0.848436673
20    H     8.375726433    6.121661386    0.885066467
21    H     8.355367714    6.137910713   -0.897932997
22    H     8.433321946    3.605676254   -0.920037655
23    H     8.449204331    3.592027105    0.863056081
24    H     6.189660329    4.735396405    0.891193765
25    H     6.174026526    4.747616072   -0.891927193
26    H     6.252509662    2.215278838   -0.907884194
27    H     6.263163707    2.205950072    0.875282500
28    H     4.003519684    3.349588660    0.893833508
29    H     3.993116641    3.357006254   -0.889341073
30    H     4.071722941    0.824478496   -0.898115225
31    H     4.076885158    0.820150925    0.885039946
32    H     1.816400198    1.963080421    0.893274835
33    H     1.811687152    1.965448134   -0.889774254
34    H     1.882921868   -0.570254438    0.890812011
35    H     1.882932461   -0.570304686   -0.890735808
36    H    -0.382462655   -1.041730915    0.003978555
37    H    -0.412891269    0.502386043   -0.898211843
38    H    -0.413081230    0.509338194    0.894061762
---
H10C13, RHF, CHARGE=0, MULT=1
HF=41.8
 1    C     0.000156545    0.000356037    0.000009544
 2    C     1.414633948   -0.000259153   -0.000011578
 3    C     2.130217831    1.205022347   -0.000011065
 4    C     1.455766983    2.447384678   -0.000014497
 5    C     0.056824186    2.459968573   -0.000022171
 6    C    -0.676442143    1.221117413    0.000007193
 7    C    -0.912984694    3.571958911   -0.000027442
 8    C    -2.239966148    3.013879677    0.000007196
 9    C    -3.357142886    3.850418497    0.000088037
10    C    -3.165368667    5.251907778    0.000197836
11    C    -1.874162089    5.797031321    0.000170043
12    C    -0.735194003    4.959684646   -0.000020108
13    C    -2.169352356    1.497179384    0.000000478
14    H    -0.540867097   -0.946022194    0.000025330
15    H     1.950094007   -0.950340875   -0.000026701
16    H     3.220913723    1.188398208   -0.000016169
17    H     2.032429548    3.372032887    0.000003452
18    H    -4.368322138    3.443214356    0.000067904
19    H    -4.034124017    5.911365743    0.000312551
20    H    -1.741839726    6.879818279    0.000312724
21    H     0.259340691    5.405162937   -0.000155227
22    H    -2.666342281    1.063535447    0.894364783
23    H    -2.666621793    1.063688201   -0.894364870
---
H28C13, RHF, CHARGE=0, MULT=1
```

```
HF=-56.2
1    C    0.000657033    -0.000490900   -0.000620631
2    C    1.625111378     0.000886452    0.000402774
3    C   -0.715453334     1.457572672   -0.000026977
4    C   -0.712344209    -1.146781115   -0.903917084
5    H   -0.223921144    -0.360043225    1.036106898
6    C    2.276247626     0.838416065   -1.138676998
7    C    2.300972098    -1.409349318   -0.099900314
8    C    2.173787502     0.511497408    1.382880894
9    C   -2.092493883     1.394058802    0.757136418
10   C   -0.949023107     2.071856718   -1.410983290
11   C    0.032129086     2.573639813    0.806622452
12   C   -2.276270371    -1.075704658   -0.971184231
13   C   -0.251171766    -1.188187409   -2.390058444
14   C   -0.491100648    -2.566578006   -0.264890691
15   H    1.938910561     0.523763159   -2.144416465
16   H    2.088896712     1.925266892   -1.059843957
17   H    3.382822271     0.730640804   -1.130470207
18   H    1.981744962    -2.086382273    0.716885826
19   H    2.146417815    -1.918127246   -1.068531147
20   H    3.407268430    -1.326096479   -0.008057847
21   H    2.350266420     1.603150404    1.415037802
22   H    1.508294994     0.241944109    2.227027728
23   H    3.165262566     0.071087492    1.625115115
24   H   -2.953961421     1.188078743    0.094089918
25   H   -2.093231870     0.643191692    1.572050290
26   H   -2.348065880     2.362901892    1.238090551
27   H   -1.647530182     1.493195618   -2.043005216
28   H   -0.010987031     2.185959952   -1.986803013
29   H   -1.395296104     3.087397797   -1.333582801
30   H    0.192695253     2.288905930    1.865307686
31   H    0.998677084     2.879422085    0.366841412
32   H   -0.565582954     3.512384874    0.832421014
33   H   -2.741619860    -1.122573163    0.033212228
34   H   -2.666380289    -0.191090780   -1.506268815
35   H   -2.688717331    -1.941537350   -1.536350876
36   H   -0.403289776    -0.223312135   -2.909971790
37   H    0.811141067    -1.463104079   -2.523989028
38   H   -0.825338302    -1.948247160   -2.963911105
39   H    0.407668910    -3.088672721   -0.643835678
40   H   -0.430259641    -2.527573264    0.840874953
41   H   -1.325915819    -3.262443693   -0.498376872
---
H28C13, RHF, CHARGE=0, MULT=1
HF=-74.5
1    C    0.000084730    -0.000021505    0.000030100
2    C    1.531317961     0.000131160    0.000069450
3    C    2.174522910     1.400700480   -0.000035887
4    C    3.716778534     1.385074079   -0.005158409
5    C    4.358567004     2.787766528   -0.003578414
6    C    5.901032529     2.772179918   -0.008842337
7    C    6.542788227     4.174913616   -0.007531972
8    C    8.085273028     4.159213538   -0.012454467
9    C    8.727018588     5.561952347   -0.011262924
10   C   10.269472359     5.546171903   -0.015913880
11   C   10.911417480     6.948624368   -0.014849915
12   H   -0.412892685     0.507122361    0.895499721
13   H   -0.412974862     0.504873514   -0.896644764
14   H   -0.382485953    -1.041646416    0.001327779
```

```
15      H        1.883144575      -0.569534877     0.891344417
16      H        1.883124300      -0.570580340    -0.890411535
17      H        1.812190454       1.965668511    -0.889387679
18      H        1.818105939       1.962596229     0.893727620
19      H        4.071891449       0.823304964    -0.899526966
20      H        4.077806290       0.818274771     0.883616522
21      H        3.997296840       3.354287398    -0.892456112
22      H        4.003368387       3.349549801     0.890717858
23      H        6.262342654       2.205778800     0.880053941
24      H        6.256282165       2.210459284    -0.903161003
25      H        6.181773792       4.741148409    -0.896689330
26      H        6.187311801       4.736872560     0.886514938
27      H        8.446346163       3.593078504     0.876735094
28      H        8.440682229       3.597105115    -0.906470031
29      H        8.366101855       6.128091071    -0.900526080
30      H        8.371457176       6.124082843     0.882675123
31      H       10.630541511       4.980383190     0.873386458
32      H       10.625174145       4.984175428    -0.909794219
33      H       10.551368606       7.515579417    -0.904086690
34      H       10.556850429       7.511673674     0.879026054
35      C       12.452458882       6.935473319    -0.019767552
36      C       13.103100490       8.321580250    -0.019057815
37      H       12.816472130       6.376665542    -0.912931989
38      H       12.822169347       6.372929921     0.868650421
39      H       12.819102987       8.912264947    -0.913395913
40      H       12.822972050       8.908884373     0.878825563
41      H       14.208589629       8.225290870    -0.022274585
 ---
H10C14, RHF, CHARGE=0, MULT=1
HF=55.2, IE=8.16
1       C       -0.000195362     -0.000564639     0.000000000
2       C        1.444941243      0.000426645     0.000000000
3       C        2.129307111      1.236562628     0.000000000
4       C       -0.687676934      1.233537007     0.000000000
5       C        2.136895444     -1.276179653     0.000000000
6       C       -0.691127127     -1.277457262     0.000000000
7       C       -0.003366677      2.469321903     0.000000000
8       C        0.003270409     -2.461604316     0.000000000
9       C        1.441845063      2.470909349     0.000000000
10      C        1.443370620     -2.461005741     0.000000000
11      H       -1.780040982      1.232017489     0.000000000
12      H       -1.782485622     -1.281406623     0.000000000
13      H        3.221636660      1.237830545     0.000000000
14      H        3.228232937     -1.279401485     0.000000000
15      H       -0.518449310     -3.419287890     0.000000000
16      H        1.965447772     -3.418574055     0.000000000
17      C       -0.697753233      3.744425368     0.000000000
18      C        2.130419158      3.749277368     0.000000000
19      C       -0.006160064      4.930246475     0.000000000
20      C        1.433978144      4.932431993     0.000000000
21      H       -1.789094454      3.746396975     0.000000000
22      H        3.221700050      3.755344452     0.000000000
23      H        1.954230021      5.890970025     0.000000000
24      H       -0.529629777      5.887008422     0.000000000
EXPGEOM
1       C        0.00000          2.48080         1.40780
2       C        0.00000          3.65710         0.71670
3       C        0.00000          3.65710        -0.71670
4       C        0.00000          2.48080        -1.40780
```

```
 5     C      0.00000     -2.48080    -1.40780
 6     C      0.00000     -3.65710    -0.71670
 7     C      0.00000     -3.65710     0.71670
 8     C      0.00000     -2.48080     1.40780
 9     C      0.00000      0.00000     1.40040
10     C      0.00000      0.00000    -1.40040
11     C      0.00000      1.22070     0.71660
12     C      0.00000      1.22070    -0.71660
13     C      0.00000     -1.22070    -0.71660
14     C      0.00000     -1.22070     0.71660
15     H      0.00000      2.47680     2.49760
16     H      0.00000      4.60660     1.24930
17     H      0.00000      4.60660    -1.24930
18     H      0.00000      2.47680    -2.49760
19     H      0.00000     -2.47680    -2.49760
20     H      0.00000     -4.60660    -1.24930
21     H      0.00000     -4.60660     1.24930
22     H      0.00000     -2.47680     2.49760
23     H      0.00000      0.00000     2.49140
24     H      0.00000      0.00000    -2.49140
---
H10C14, RHF, CHARGE=0, MULT=1
HF=92
 1     C      0.000725443    -0.000159104     0.000007234
 2     C      1.405281418    -0.000348328    -0.000116429
 3     C      2.113726843     1.214151633    -0.000161566
 4     C      1.411344037     2.432111470    -0.000168267
 5     C      0.006520564     2.439540618    -0.000131088
 6     C     -0.720521764     1.221638493     0.000030794
 7     C     -4.761206697     1.200556810     0.000667916
 8     C     -5.474894937    -0.025649929    -0.002263814
 9     C     -6.879671461    -0.034688111    -0.002110040
10     C     -7.596304747     1.175000879     0.000959692
11     C     -6.901341638     2.397280153     0.003882763
12     C     -5.496627248     2.413614065     0.003755447
13     C     -2.139783919     1.220556260     0.000276577
14     C     -3.341831184     1.212663141     0.000468143
15     H     -0.534562823    -0.950790151     0.000090516
16     H      1.947082941    -0.946781851    -0.000134267
17     H      3.203966351     1.211358531    -0.000162316
18     H      1.958297049     3.375699773    -0.000099039
19     H     -0.523337342     3.393019394    -0.000238154
20     H     -4.934238009    -0.973187439    -0.004693709
21     H     -7.415487127    -0.984616592    -0.004401820
22     H     -8.686619057     1.165257769     0.001081684
23     H     -7.453837190     3.337491088     0.006233293
24     H     -4.972668504     3.370535324     0.006085885
---
H10C14, RHF, CHARGE=0, MULT=1
HF=49.5
 1     C     -0.000100726     0.000325647    -0.000000002
 2     C      1.432319197    -0.000601950    -0.000000002
 3     C      2.146820456     1.263146828    -0.000000002
 4     C     -0.718264128     1.282584690     0.000000005
 5     C      2.143976729    -1.240128658    -0.000000001
 6     C     -0.659014287    -1.270669332    -0.000000003
 7     C      0.029031639     2.504821584     0.000000002
 8     C      0.052568206    -2.466109562    -0.000000002
 9     C      1.479718552     2.454099539     0.000000000
```

```
  10   C      1.469448040   -2.454751177   -0.000000003
  11   C     -2.146173697    1.384284100    0.000000020
  12   H     -1.747881629   -1.331788218    0.000000001
  13   H      3.238389132    1.246121594   -0.000000001
  14   H      3.235388561   -1.232868178    0.000000003
  15   H     -0.476619851   -3.419945221   -0.000000001
  16   H      2.018749828   -3.396518005    0.000000003
  17   C     -0.656318007    3.759085701    0.000000012
  18   H      2.035448609    3.393833537   -0.000000012
  19   C     -2.044305788    3.817913330    0.000000042
  20   C     -2.793943562    2.615483630    0.000000055
  21   H     -0.080132546    4.686086141   -0.000000008
  22   H     -2.766838269    0.487604870    0.000000003
  23   H     -3.883695180    2.662441333    0.000000100
  24   H     -2.560925211    4.778026149    0.000000054
EXPGEOM
  1    C     0.00000    2.48080    1.40780
  2    C     0.00000    3.65710    0.71670
  3    C     0.00000    3.65710   -0.71670
  4    C     0.00000    2.48080   -1.40780
  5    C     0.00000   -2.48080   -1.40780
  6    C     0.00000   -3.65710   -0.71670
  7    C     0.00000   -3.65710    0.71670
  8    C     0.00000   -2.48080    1.40780
  9    C     0.00000    0.00000    1.40040
 10    C     0.00000    0.00000   -1.40040
 11    C     0.00000    1.22070    0.71660
 12    C     0.00000    1.22070   -0.71660
 13    C     0.00000   -1.22070   -0.71660
 14    C     0.00000   -1.22070    0.71660
 15    H     0.00000    2.47680    2.49760
 16    H     0.00000    4.60660    1.24930
 17    H     0.00000    4.60660   -1.24930
 18    H     0.00000    2.47680   -2.49760
 19    H     0.00000   -2.47680   -2.49760
 20    H     0.00000   -4.60660   -1.24930
 21    H     0.00000   -4.60660    1.24930
 22    H     0.00000   -2.47680    2.49760
 23    H     0.00000    0.00000    2.49140
 24    H     0.00000    0.00000   -2.49140
---
H12C14, RHF, CHARGE=0, MULT=1
HF=37.1
  1    C    -0.000184397    0.000196451    0.000463246
  2    C     1.404933333   -0.000030704   -0.000949928
  3    C     2.107258952    1.214011877    0.000254226
  4    C     1.396267376    2.425291556    0.028684249
  5    C    -0.014414092    2.439794234    0.045204424
  6    C    -0.738039221    1.208333528   -0.006335927
  7    C    -2.222610030    1.239574277   -0.043058179
  8    C    -2.907867331    2.413687548    0.398053499
  9    C    -2.119848612    3.626395467    0.837016247
 10    C    -0.762370085    3.750911185    0.122851195
 11    C    -4.318453312    2.436875866    0.416780896
 12    C    -5.066291117    1.333787564   -0.027508082
 13    C    -4.401483751    0.196036393   -0.507673279
 14    C    -2.997057851    0.154235221   -0.518910352
 15    H    -0.506417306   -0.966022745    0.017760465
 16    H     1.947074007   -0.946308176   -0.001211536
```

```
17         H        3.197335578     1.219725589    -0.013235802
18         H        1.956046976     3.362306285     0.044104172
19         H       -1.962217488     3.575181592     1.940462503
20         H       -2.699561886     4.559742975     0.653866618
21         H       -0.155612019     4.519165380     0.654014180
22         H       -0.914550547     4.142170450    -0.910971781
23         H       -4.849221987     3.321298408     0.774036108
24         H       -6.155878238     1.366950915    -0.005462351
25         H       -4.972363183    -0.657816131    -0.874312226
26         H       -2.520454992    -0.742135580    -0.918351525
---
H12C14, RHF, CHARGE=0, MULT=1
HF=35.4
1      C       -0.000510227     0.000042205    -0.000091302
2      C        1.412205743     0.000597474     0.000136697
3      C        2.126416047     1.206675531     0.000198695
4      C        1.448292301     2.447414759     0.009120207
5      C        0.051075928     2.461232571     0.021956826
6      C       -0.681025784     1.223260792     0.005048837
7      C       -2.121089606     1.535023888    -0.027099159
8      C       -2.275725882     2.964988851    -0.029494444
9      C       -0.910057993     3.650089264     0.021724482
10        C       -0.727978703     4.614075711     1.205542810
11        C       -3.551795690     3.529959474    -0.100342451
12        C       -4.681226640     2.680798291    -0.148779039
13        C       -4.530462016     1.287358972    -0.131144115
14        C       -3.245676254     0.702773208    -0.072786234
15        H       -0.537574943    -0.948105866    -0.006927920
16        H        1.949158319    -0.948841446    -0.002916194
17        H        3.217019260     1.192756871    -0.007066212
18        H        2.025846668     3.371947729     0.003315170
19        H       -0.763191372     4.231588547    -0.922043708
20        H       -0.863644190     4.106257199     2.181244836
21        H        0.284682507     5.066057301     1.196521861
22        H       -1.457066193     5.448131883     1.153163422
23        H       -3.694704539     4.610572495    -0.121409928
24        H       -5.678440241     3.119468190    -0.200766768
25        H       -5.411300781     0.644956020    -0.165566942
26        H       -3.148978228    -0.382615214    -0.066854895
---
H12C14, RHF, CHARGE=0, MULT=1
HF=131.8
1      C       -0.000155337     0.000272128    -0.001069980
2      C        1.491724414    -0.000468757    -0.000360892
3      C       -0.716745749     1.160177902     0.001941341
4      C       -0.226942522     2.542513772    -0.083936771
5      C        0.530179273     3.076450962    -1.065676078
6      C        1.088263647     2.369098857    -2.225997823
7      C        1.826256339     1.239708058    -2.192165444
8      C        2.263029557     0.524950538    -0.984312050
9      C       -0.726690171    -1.284850903     0.013564265
10        C       -0.328379089    -2.570275872    -0.123578688
11        C        0.977911856    -3.205145434    -0.290903299
12        C        2.137487488    -2.968778476     0.360375502
13        C        2.407399752    -1.920789859     1.347622359
14        C        2.123837043    -0.611648927     1.195710306
15        H       -1.811658230     1.123039696     0.073319892
16        H       -0.586354429     3.189110616     0.724935983
17        H        0.763014251     4.146827635    -1.040747759
```

```
18         H        0.878326617     2.848168719    -3.188495348
19         H        2.199528662     0.809298344    -3.127775544
20         H        3.355160258     0.440516229    -0.917615593
21         H       -1.813987152    -1.156974864     0.129721389
22         H       -1.133700524    -3.321858944    -0.123439685
23         H        0.956872176    -4.033795635    -1.011396451
24         H        2.996138922    -3.620537770     0.159600602
25         H        2.902607386    -2.267888030     2.260563752
26         H        2.371273254     0.100860636     1.989571387
---
H12C14, RHF, CHARGE=0, MULT=1
HF=53.4
1     C        0.000146120    -0.000083237    -0.000166371
2     C        1.403264780     0.000169125    -0.000074811
3     C        2.110868576     1.215160955     0.000104630
4     C        1.398998749     2.424860150     0.000502173
5     C       -0.007365930     2.423611140     0.000702016
6     C       -0.743765579     1.212573811     0.000042846
7     C       -2.215733370     1.143302259     0.000450446
8     C       -3.100460196     2.170420079    -0.016174887
9     C       -4.572481362     2.101029843    -0.016397559
10        C       -5.308441923     0.889820701    -0.006887970
11        C       -6.714844115     0.887807191    -0.008293766
12        C       -7.427405578     2.096967429    -0.019449803
13        C       -6.720289261     3.312200800    -0.029150481
14        C       -5.317126464     3.313144104    -0.027568104
15        H       -0.517627589    -0.961223029    -0.000410196
16        H        1.945933704    -0.945904462    -0.000166501
17        H        3.200877423     1.216857258     0.000116038
18        H        1.937147013     3.373628376     0.000768292
19        H       -0.515327712     3.388809886     0.001916381
20        H       -2.595987819     0.115256314     0.015750224
21        H       -2.720204741     3.198482982    -0.031768532
22        H       -4.800157474    -0.075177839     0.001710000
23        H       -7.252586312    -0.061113756    -0.000760150
24        H       -8.517498496     2.094625044    -0.020658403
25        H       -7.263469137     4.257972312    -0.037981999
26        H       -4.800019279     4.274625782    -0.035248161
---
H14C14, RHF, CHARGE=0, MULT=1
HF=22.1
1     C        0.043908661    -0.021504737    -0.050468822
2     C        1.546860008     0.106222048    -0.355075313
3     C        2.190291976     1.355522843     0.256249798
4     C        1.416390165     2.641495830    -0.070658902
5     C       -0.092367081     2.505803391    -0.074503336
6     C       -0.750253426     1.271888471    -0.060083473
7     C       -2.205150246     1.228052657    -0.037830041
8     C       -2.946892724     2.455575423    -0.088668151
9     C       -2.232953008     3.701349180    -0.131818661
10        C       -0.856921837     3.722604785    -0.115797256
11        C       -4.385714941     2.425775965    -0.087846025
12        C       -5.074570665     1.230724782    -0.021960778
13        C       -4.350094725     0.004262716     0.049810493
14        C       -2.967194852     0.005476552     0.042697501
15        H       -0.076599935    -0.502424937     0.949652408
16        H       -0.379499333    -0.730755348    -0.799192882
17        H        1.704093835     0.107755708    -1.458283337
18        H        2.069995222    -0.801382722     0.022958648
```

```
19         H         3.236330453     1.452884637    -0.113066590
20         H         2.269378055     1.235728907     1.361036929
21         H         1.738431383     3.017594249    -1.071195072
22         H         1.717541402     3.424602298     0.663969723
23         H        -2.792110001     4.637788408    -0.173121648
24         H        -0.334547244     4.681332295    -0.139181967
25         H        -4.935610761     3.367181650    -0.137710995
26         H        -6.164593873     1.209631186    -0.021307400
27         H        -4.902265008    -0.934348906     0.112068516
28         H        -2.456776860    -0.956389626     0.107513851
---
H14C14, RHF, CHARGE=0, MULT=1
HF=26.6
1     C     0.000184077    -0.000112379     0.000036073
2     C     1.413127761    -0.000588127     0.000055559
3     C     2.142219884     1.200763955    -0.000059651
4     C     1.481413777     2.449615078    -0.000197401
5     C     0.067377982     2.455117153     0.000097351
6     C    -0.657330412     1.252564850     0.000218181
7     H    -1.423592334    -1.353487232     0.910562115
8     H    -1.457790333    -1.335899151    -0.884355377
9     C     2.249574433     3.721320473    -0.000598409
10        C     2.621486216     4.346323952     1.210801903
11        C     3.347588048     5.549540372     1.209187036
12        C     3.728827049     6.172096249    -0.000606135
13        C     3.354526063     5.542822478    -1.211230019
14        C     2.628990749     4.340893426    -1.213598470
15        C    -0.790208903    -1.280476340     0.001291051
16        H     1.963528819    -0.943232195     0.000128148
17        H     3.232722324     1.156498108     0.000034136
18        H    -0.477190929     3.400858986     0.000241960
19        H    -1.747949412     1.298674065     0.000547820
20        H    -0.141205787    -2.179391323    -0.020294109
21        H     5.482324407     7.327271722    -0.529784298
22        H     3.945853022     8.255487060    -0.551017465
23        H     4.720961923     7.838544571     1.003782610
24        H     2.346605024     3.897442813     2.166813993
25        H     3.614041658     5.997948103     2.168173281
26        C     4.506650120     7.460055086    -0.016215790
27        H     3.629096446     5.990431856    -2.168324805
28        H     2.359559574     3.887510747    -2.169206219
---
H14C14, RHF, CHARGE=0, MULT=1
HF=32.4
1     C     0.000335616    -0.000006815    -0.000081254
2     C     1.405489741    -0.001334163     0.000023871
3     C     2.114522653     1.212164284     0.000002932
4     C     1.435485782     2.453964892     0.000593167
5     C     0.020307620     2.436389537    -0.001074802
6     C    -0.690021372     1.223624720    -0.001132850
7     C     2.199566278     3.761756374    -0.015812824
8     C     2.521526840     4.311413860     1.393089606
9     C     3.287665312     5.617892000     1.376452217
10        C     4.702971807     5.633523445     1.376681500
11        C     5.414933370     6.845462786     1.375982166
12        C     4.726152725     8.070117443     1.375475303
13        C     3.321079498     8.073393537     1.376841335
14        C     2.610395632     6.860650566     1.377753815
15        H    -0.549529677    -0.941351209     0.000107663
```

```
16      H       1.949846963     -0.946290215    -0.000222500
17      H       3.205567607      1.180403536    -0.001677396
18      H      -0.542127915      3.371818939    -0.003556486
19      H      -1.780570807      1.234572471    -0.002330998
20      H       3.147524026      3.629448471    -0.586832742
21      H       1.618020461      4.523102112    -0.585363737
22      H       1.573194238      4.445185686     1.963072524
23      H       3.101221483      3.549558410     1.963638349
24      H       5.264364310      4.697428930     1.378615409
25      H       6.505289344      6.833174967     1.376181438
26      H       5.277557822      9.010514084     1.374584562
27      H       2.778201318      9.019183953     1.377509070
28      H       1.519379484      6.893864611     1.380741447
---
H16C14, RHF, CHARGE=0, MULT=1
HF=19.5
1       C       0.000507261      0.000068489     0.000802398
2       H       1.108707869     -0.001759757     0.001111817
3       H      -0.354312589      1.051803982    -0.003958793
4       H      -0.304471533     -0.437211222     0.975942873
5       C      -1.763796532     -1.471223292    -0.929350158
6       C      -0.575260696     -0.787704699    -1.150342630
7       C       0.088319191     -0.945117616    -2.436492899
8       C      -0.282925731     -2.081211880    -3.251605186
9       C      -1.620509230     -2.630560021    -3.081474731
10      C      -2.322268389     -2.320140795    -1.924214191
11      C       0.706933728     -2.597655019    -4.186172677
12      C       1.775289730     -1.783312580    -4.537464640
13      C       1.920503691     -0.479724262    -3.987818207
14      C       1.099726686     -0.037155748    -2.958803679
15      H      -2.306250683     -1.373133207     0.013508733
16      H      -2.046882963     -3.138714588    -5.162343768
17      H      -2.143616313     -4.528473747    -4.016934557
18      H      -3.434399032     -3.300753416    -4.072513779
19      H       0.282304498     -4.730694494    -3.990414852
20      H       0.118497387     -4.085657919    -5.666772226
21      H       1.726170361     -4.358891516    -4.947977924
22      H       0.280523395      1.844662516    -2.209481687
23      C      -2.333799701     -3.441576191    -4.135404304
24      H      -3.321524678     -2.721989348    -1.742585587
25      C       0.696104261     -4.008181256    -4.722274222
26      H       2.533259005     -2.118898357    -5.248671380
27      H       2.704178957      0.158559568    -4.401786900
28      C       1.247533704      1.397580515    -2.515344427
29      H       1.967934443      1.496654953    -1.676902276
30      H       1.629044342      2.034523269    -3.342307421
---
H18C14, RHF, CHARGE=0, MULT=1
HF=-8.9
1       C       0.000079595      0.000050918     0.000305399
2       C       1.534737986     -0.000150450    -0.000086945
3       H      -0.372445782      1.050161966     0.001345555
4       H      -0.366901569     -0.450199502    -0.949982452
5       C       0.267992976     -1.613875610     7.437487255
6       C      -0.608800617     -1.419977421     6.189073244
7       C       0.132154707     -0.963388791     4.950842465
8       C       1.445542123     -0.429083899     5.019962942
9       C       2.204338921     -0.340967787     6.326595576
10      C       1.354559941     -0.541321777     7.592710313
```

```
11   C    2.066911668    0.004729232    3.828635470
12   C    1.430624842   -0.075568793    2.570809007
13   C    0.119201316   -0.614184600    2.500852879
14   C   -0.502890741   -1.045619938    3.692690685
15   H   -0.382083589   -1.611711131    8.341668131
16   H    1.901674597    0.676845579   -0.804609881
17   H    1.907711970   -1.015968858   -0.265819323
18   H    0.740610646   -2.622412576    7.406873143
19   H   -1.135792459   -2.380236775    5.981611570
20   H   -1.407678298   -0.673630170    6.415147566
21   H    2.711891844    0.648873809    6.401390575
22   H    3.019200886   -1.104018333    6.309996881
23   H    0.884374284    0.426617755    7.881784506
24   H    2.024212126   -0.816914272    8.438895979
25   H    3.077973148    0.417161719    3.882014798
26   C    2.147975941    0.437142735    1.340700639
27   C   -0.614712713   -0.763384940    1.185496525
28   H   -1.513243804   -1.459655315    3.639129947
29   H    3.212981103    0.108801194    1.359487807
30   H    2.169481790    1.552621041    1.386846134
31   H   -0.663482421   -1.849845266    0.933101179
32   H   -1.671370226   -0.427832444    1.304028946
---
H20C14, RHF, CHARGE=0, MULT=1
HF=-34.9
1    C    0.000186537    0.000684448   -0.000406365
2    C    1.568473168   -0.000410366    0.000939541
3    C    2.052823150    1.491300055   -0.000623087
4    C    1.571900848    2.153464530   -1.338434830
5    C    0.003577790    2.154883336   -1.339476235
6    C   -0.481082874    0.663378879   -1.337823902
7    H   -0.375335096   -1.046165547    0.054937212
8    C   -0.553500813    0.768360168    1.228293733
9    C    2.122765628   -0.763406128   -1.230688220
10   H    1.941546898   -0.514349403    0.915583706
11   H    3.164399816    1.523934625    0.054696953
12   C    1.494671911    2.256105736    1.227992244
13   H    1.947445113    3.200187024   -1.394270319
14   C    2.125586384    1.385758477   -2.567201512
15   H   -0.369392344    2.668882139   -2.254059827
16   C   -0.550115388    2.918102095   -0.107940393
17   H   -1.592814455    0.630807588   -1.392960248
18   C    0.077175717   -0.101385602   -2.566216000
19   C   -0.055801670    2.236914389    1.194873886
20   H   -1.665180587    0.732016406    1.239478847
21   H   -0.238303895    0.269653014    2.171329571
22   C    1.627674166   -0.082696198   -2.533387592
23   H    1.806121465   -1.829323289   -1.201244442
24   H    3.234363246   -0.791065523   -1.201371175
25   H    1.873389577    3.301867852    1.238422431
26   H    1.871793282    1.802721203    2.171136708
27   H    1.810239225    1.884047095   -3.510348785
28   H    3.237277570    1.422356875   -2.578041642
29   H   -0.232420978    3.983682187   -0.137331629
30   H   -1.661726203    2.946686938   -0.137213859
31   H   -0.299662445    0.351804155   -3.509595138
32   H   -0.301150938   -1.147337024   -2.576503626
33   H   -0.454441978    2.785477264    2.077024651
34   H    2.026029514   -0.631593498   -3.415502426
```

```
---
H24C14, RHF, CHARGE=0, MULT=1
HF=-68
1    C     0.000108604    -0.001655869    -0.000704928
2    C     1.563933691     0.000145955     0.000404957
3    C     2.056015860     1.484420611    -0.000411297
4    C     1.521550855     2.262774576    -1.247283252
5    C    -0.041669629     2.218369863    -1.228164226
6    C    -0.571691106     0.747242773    -1.248854482
7    H    -0.379569978    -1.047653197     0.022708487
8    C     2.062387660    -0.688204754    -1.312328115
9    C     2.105577055    -0.749897714     1.243233165
10   H     3.168345558     1.516933476     0.020371906
11   C     2.021301218     3.729058402    -1.224656848
12   C     2.021583799     1.534024612    -2.537526055
13   H    -0.448847048     2.781822476    -2.097240332
14   C    -2.120787724     0.727685203    -1.229591540
15   C    -0.033994444     0.045381112    -2.538715632
16   C     1.529076513     0.050431291    -2.583244478
17   H     1.741539046    -1.753699722    -1.328686339
18   H     3.174912572    -0.717259402    -1.327843001
19   H     1.671481440     2.079274604    -3.442377506
20   H     3.133084983     1.567756400    -2.587073209
21   H    -0.442186867     0.548341505    -3.443843484
22   H    -0.412512660    -1.000059066    -2.589024562
23   C     2.035999767    -0.649183106    -3.869341216
24   H    -0.381041264     0.469701280     0.932636117
25   H     1.730940039     1.993663969     0.934305266
26   H    -0.426785985     2.749245247    -0.328958603
27   H     3.214626038    -0.759957667     1.264599377
28   H     1.763442463    -0.278237565     2.187276292
29   H     1.767536510    -1.806305878     1.262994161
30   H     3.129112196     3.784308355    -1.238380714
31   H     1.677207819     4.264512561    -0.316096875
32   H     1.653358758     4.300445033    -2.101421743
33   H    -2.547652470     1.254605428    -2.107516909
34   H    -2.525644335     1.220588508    -0.322142747
35   H    -2.516785778    -0.308388672    -1.244373353
36   H     3.144005730    -0.656743686    -3.921582305
37   H     1.667812574    -0.140792288    -4.783955976
38   H     1.697827421    -1.704371908    -3.921734010
---
H28C14, RHF, CHARGE=0, MULT=1
HF=-40.2
1    C    -0.106760725    -0.087491710     0.123227740
2    C     1.416194337     0.056515447    -0.049619568
3    C     1.990092966     1.477004762    -0.043576755
4    C     2.598510921     2.015680864     1.079096614
5    C     3.171655850     3.436351714     1.085162572
6    C     2.786827141     1.287696257     2.460894748
7    C     3.258599401     2.201511181     3.648496630
8    C     1.451614556     0.691827653     3.012301091
9    C     3.895640158     0.197781332     2.321846420
10   C     1.803178278     2.206599070    -1.424817516
11   C     1.331002539     1.294572220    -2.613455547
12   C     0.695843873     3.298166557    -1.285596354
13   C     3.139471213     2.802012179    -1.974530409
14   H    -0.623883023    -0.246053650    -0.844285696
15   H    -0.340683572    -0.975712209     0.747134704
```

```
16      H       -0.576180073     0.791115507     0.606640979
17      H        1.898881120    -0.581364295     0.720636849
18      H        1.722567749    -0.475626977    -0.980346053
19      H        2.863372785     3.968844151     2.014995345
20      C        4.694899303     3.580805271     0.914428427
21      H        2.689735455     4.074098586     0.314295718
22      H        4.246803848     2.671890698     3.488324884
23      H        2.527160218     3.002028419     3.880613848
24      H        3.374760113     1.603071625     4.578599645
25      H        0.589495393     1.341850730     2.764163093
26      H        1.241374134    -0.333174787     2.654123742
27      H        1.463031886     0.597084054     4.119650096
28      H        3.774369310    -0.458264432     1.439718857
29      H        4.901525132     0.656204361     2.227878793
30      H        3.920818500    -0.466344808     3.210981853
31      H        2.061595250     0.493410816    -2.845744190
32      H        0.342200046     0.825060202    -2.453944134
33      H        1.216020239     1.893812066    -3.543293503
34      H       -0.311323489     2.841391485    -1.197073395
35      H        0.814933842     3.950016292    -0.400177760
36      H        0.674862585     3.965804781    -2.172274983
37      H        3.349379845     3.827039885    -1.616161628
38      H        4.001366223     2.151925265    -1.725236960
39      H        3.129637640     2.895857344    -3.081840155
40      H        4.929324039     4.464392508     0.284343444
41      H        5.209371295     3.747391322     1.881936930
42      H        5.166440618     2.698867049     0.439169808
---
H28C14, RHF, CHARGE=0, MULT=1
HF=-57.2
1       C       -0.104180293    -0.023810795     0.152657327
2       C        1.439435793     0.004964292     0.219584797
3       C        2.079230820     1.393794704     0.020804300
4       C        3.592434782     1.408793112    -0.293399585
5       C        3.998154099     1.408685378    -1.784908740
6       C        4.259164119     0.027035005    -2.417851317
7       C        4.285002382     0.006451469    -3.962858330
8       C        3.678005081    -1.254360614    -4.616063137
9       C        2.134261657    -1.285720777    -4.681204115
10      C        1.497048067    -2.675381731    -4.480276322
11      C       -0.016226041    -2.692218702    -4.165991872
12      C       -0.422033341    -2.690395717    -2.674618064
13      C       -0.686031335    -1.308031339    -2.044049305
14      C       -0.712525039    -1.284957210    -0.498962541
15      H       -0.499153933     0.046989621     1.193190397
16      H       -0.497963620     0.876794344    -0.369916206
17      H        1.858609094    -0.707298764    -0.523649784
18      H        1.753085751    -0.391081829     1.214164570
19      H        1.923481140     1.981594536     0.956477393
20      H        1.536052422     1.954300546    -0.773370930
21      H        4.104069700     0.578764925     0.243505781
22      H        4.005905495     2.343944759     0.155001997
23      H        3.229170379     1.958578270    -2.372981455
24      H        4.931537229     2.012973916    -1.880723117
25      H        3.504856132    -0.700301840    -2.046854576
26      H        5.239851830    -0.353584670    -2.046977811
27      H        5.347127205     0.084600963    -4.294243197
28      H        3.782470630     0.906778445    -4.382856280
29      H        4.073836715    -2.155789127    -4.096212480
```

```
30   H     4.071439994   -1.322227209   -5.657239409
31   H     1.715098216   -0.573624382   -3.937892819
32   H     1.818153406   -0.890922069   -5.675270786
33   H     2.041381964   -3.233580900   -3.685408429
34   H     1.653483485   -3.264589210   -5.415241496
35   H    -0.529060925   -1.863805996   -4.704326448
36   H    -0.429180995   -3.628348851   -4.612755765
37   H     0.347862351   -3.237775227   -2.085555644
38   H    -1.354357837   -3.295966540   -2.578033452
39   H    -1.667466352   -0.930244445   -2.415787209
40   H     0.066736954   -0.580000573   -2.415932937
41   H    -1.775000826   -1.361090539   -0.168146768
42   H    -0.211402736   -2.185290599   -0.077002105
---
H28C14, RHF, CHARGE=0, MULT=1
HF=-64.9
1    C     0.109456129   -0.109828584   -0.143139034
2    C     1.587601990   -0.170106804    0.290801049
3    C     2.110734405    1.277135674    0.362579006
4    C     0.967921694    2.212187020   -0.082865611
5    C    -0.169719389    1.328122842   -0.664997554
6    C    -1.611075385    1.820884387   -0.391997893
7    C    -2.025677151    3.082223669   -1.178314030
8    C    -3.478611248    3.537145944   -0.927282377
9    C    -3.856595582    4.832612825   -1.674422109
10   C    -5.300243952    5.314829300   -1.423745705
11   C    -5.656360035    6.616305651   -2.171373503
12   C    -7.101626235    7.104461733   -1.945118172
13   C    -7.449519233    8.396000590   -2.710964581
14   C    -8.881656441    8.897968966   -2.508739017
15   H    -0.546788630   -0.371912717    0.714739905
16   H    -0.104643245   -0.859977304   -0.933643915
17   H     2.184738853   -0.767864084   -0.430928939
18   H     1.691432812   -0.681394296    1.271367572
19   H     3.001702971    1.405875678   -0.288793384
20   H     2.445364121    1.527983269    1.391695221
21   H     1.326904022    2.937886606   -0.843338710
22   H     0.619334293    2.818915941    0.780378181
23   H    -0.035081634    1.299704354   -1.775432348
24   H    -2.315393444    0.996468969   -0.652551478
25   H    -1.744084497    2.004777846    0.698526687
26   H    -1.337048966    3.919044013   -0.921627870
27   H    -1.892483610    2.890797609   -2.268171162
28   H    -4.174285763    2.722239482   -1.232219075
29   H    -3.632931717    3.687272520    0.165589523
30   H    -3.149035046    5.641236008   -1.378583373
31   H    -3.711813267    4.674912569   -2.768039491
32   H    -6.011028084    4.512279154   -1.726838109
33   H    -5.450102840    5.468872687   -0.330568230
34   H    -4.949927184    7.419378781   -1.858163140
35   H    -5.492068216    6.464148469   -3.263036933
36   H    -7.811774095    6.300533969   -2.246106132
37   H    -7.264763144    7.272580401   -0.855938429
38   H    -6.747116721    9.206713122   -2.407904407
39   H    -7.284890558    8.236247685   -3.801808716
40   H    -9.631051792    8.161767866   -2.864145263
41   H    -9.097765169    9.114221959   -1.443043969
42   H    -9.041691351    9.837240851   -3.077762664
---
```

```
H28C14, RHF, CHARGE=0, MULT=1
HF=-70.7
1    C    0.000133851    0.000062983   -0.000079035
2    C    1.539070632    0.000006547    0.000083816
3    C    2.168471878    1.402093368   -0.000021783
4    C    1.528896110    2.363085870   -1.014512322
5    C   -0.010722006    2.352775825   -1.014251902
6    C   -0.638984424    0.935501416   -1.062826774
7    C   -2.189278356    0.941419768   -0.938926226
8    C   -2.960558751    1.458074875   -2.171807549
9    C   -4.493260737    1.325619271   -2.047114893
10   C   -5.269772032    1.855088781   -3.270382070
11   C   -6.801608355    1.722023857   -3.146545002
12   C   -7.578438277    2.251661690   -4.369162879
13   C   -9.108998147    2.119165758   -4.247776883
14   C   -9.893611005    2.642703143   -5.453999260
15   H   -0.344628672   -1.044652582   -0.176523792
16   H   -0.361743013    0.274117435    1.017380830
17   H    1.910978518   -0.573586908   -0.879977229
18   H    1.896643148   -0.557624267    0.895749970
19   H    3.256953092    1.313003336   -0.218797229
20   H    2.101009634    1.840657511    1.022077119
21   H    1.901168179    2.122941181   -2.037097018
22   H    1.878950305    3.398970758   -0.802164819
23   H   -0.353596908    2.942917482   -1.893835820
24   H   -0.375644026    2.902193295   -0.116485453
25   H   -0.391698022    0.498520363   -2.063869199
26   H   -2.525844389   -0.102103281   -0.735326718
27   H   -2.487311087    1.537552499   -0.046306050
28   H   -2.621593633    0.902086303   -3.075825478
29   H   -2.707104879    2.527793026   -2.348394295
30   H   -4.754855882    0.254506764   -1.886498441
31   H   -4.833015116    1.870733643   -1.136666926
32   H   -4.929536002    1.310833389   -4.181172930
33   H   -5.009490190    2.926553038   -3.430912842
34   H   -7.062238313    0.650635938   -2.986200552
35   H   -7.141893685    2.266282579   -2.235742016
36   H   -7.239052846    1.708230319   -5.280861547
37   H   -7.319171153    3.323440642   -4.529788968
38   H   -9.378106868    1.048926724   -4.090903477
39   H   -9.458011707    2.662048720   -3.338960965
40   H   -9.630626022    2.100435107   -6.384797867
41   H   -9.712020381    3.722769793   -5.627507067
42   H  -10.982856683    2.510709627   -5.288285845
---
H30C14, RHF, CHARGE=0, MULT=1
HF=-63.5
1    C    0.000162978   -0.000065447    0.000030224
2    C    1.535814700    0.000028135   -0.000015809
3    C    2.403797590    1.316350621   -0.000086323
4    C    2.341689086    2.154583849   -1.419788304
5    C    3.507887452    3.226508516   -1.560356768
6    C    3.654988169    4.453319530   -0.650817287
7    C    0.975078832    2.912883889   -1.593497431
8    C    0.757931499    3.869014718   -2.780090790
9    C    2.594025874    1.191618802   -2.660889628
10   C    1.472633111    0.477765001   -3.429975040
11   C    3.913994213    0.902390188    0.260139375
12   C    4.308432077   -0.353807819    1.053532501
```

```
13   C    1.916432016    2.211784642    1.203969515
14   C    2.041977903    1.732573366    2.658563957
15   H   -0.442918129    0.655434369    0.774528147
16   H   -0.446045516    0.280993400   -0.972226741
17   H   -0.352730487   -1.032271238    0.217695707
18   H    1.796542266   -0.576064676    0.920882156
19   H    1.864787953   -0.657532559   -0.835396286
20   H    3.446764921    3.646538011   -2.594926350
21   H    4.494674984    2.708743299   -1.543116464
22   H    4.170658278    4.217215724    0.301205188
23   H    2.698533085    4.954834162   -0.409325594
24   H    4.287960858    5.207993873   -1.167321565
25   H    0.147428862    2.172608654   -1.636945132
26   H    0.766040449    3.530448842   -0.690900082
27   H    1.149178364    3.495025293   -3.745161292
28   H    1.213240285    4.864339102   -2.602627039
29   H   -0.332113519    4.036835295   -2.915219691
30   H    3.316679293    0.396140481   -2.369642036
31   H    3.133026465    1.772358695   -3.449645797
32   H    0.736236576    1.169930381   -3.881336340
33   H    0.910922512   -0.258922238   -2.827068067
34   H    1.930500228   -0.088901219   -4.270825501
35   H    4.443417669    0.769860736   -0.710754853
36   H    4.426229170    1.755923419    0.762626903
37   H    3.842107301   -0.425525063    2.054112314
38   H    4.080529387   -1.288651049    0.502286748
39   H    5.408249428   -0.339855101    1.215881465
40   H    2.442147069    3.190275986    1.180259454
41   H    0.837803937    2.454263204    1.066481693
42   H    1.592668011    0.737387262    2.840165838
43   H    3.092957474    1.705990794    3.008256062
44   H    1.506551414    2.449540081    3.318257405
---
H30C14, RHF, CHARGE=0, MULT=1
HF=-59.4
1    C   -0.000023799   -0.000213118   -0.000096423
2    C    1.575841935    0.000171878    0.000397068
3    C    2.178158988    1.537587703   -0.000288841
4    C    3.780923849    1.773585048   -0.353512891
5    C    4.491202546    3.204347442    0.064362631
6    C    5.881268294    3.388367583   -0.654851075
7    C    1.942292258   -0.874192933   -1.238886809
8    C    1.952761733   -0.842879420    1.263225014
9    C    1.846902874    2.120694216    1.408531538
10   C    1.308730464    2.343487299   -1.024197645
11   C    4.647880180    0.662072774    0.315183147
12   C    3.987741833    1.621532064   -1.898865906
13   C    3.704394948    4.493499547   -0.326986278
14   C    4.832383202    3.338695120    1.585355843
15   H   -0.395557495   -1.028471196    0.147858172
16   H   -0.433216441    0.348364557   -0.958969461
17   H   -0.447950280    0.610830205    0.806879425
18   H    6.417768920    4.277637934   -0.257754117
19   H    5.785014250    3.563361427   -1.745109985
20   H    6.573285042    2.535756427   -0.516354828
21   H    3.015633112   -1.132422690   -1.302532987
22   H    1.644511286   -0.405597095   -2.197766675
23   H    1.410576844   -1.850975290   -1.204430810
24   H    1.598709116   -0.388677325    2.209710875
```

```
25      H        3.034946039    -1.031221288     1.375658866
26      H        1.484701687    -1.851032737     1.218585770
27      H        1.902760882     3.225910893     1.439095362
28      H        2.509698095     1.726321366     2.202880365
29      H        0.810315493     1.901523983     1.738145119
30      H        1.709985789     3.343506718    -1.264823920
31      H        0.287009095     2.543496955    -0.637136900
32      H        1.188626486     1.827532306    -1.997448369
33      H        5.738303148     0.829408966     0.196584366
34      H        4.486215487    -0.336619503    -0.134476038
35      H        4.468931602     0.575757205     1.404368207
36      H        3.707513301     2.529259304    -2.469479364
37      H        3.409859234     0.795967418    -2.350005678
38      H        5.043107847     1.391400098    -2.157802780
39      H        4.303745746     5.404623849    -0.106401371
40      H        2.759609129     4.628066426     0.231655879
41      H        3.473234823     4.542759679    -1.409481893
42      H        5.522779288     2.551133209     1.946706907
43      H        3.950406619     3.330295748     2.249724486
44      H        5.342905318     4.305315080     1.791125366
---
H30C14, RHF, CHARGE=0, MULT=1
HF=-79.4
1       C        0.000124931    -0.000032722     0.000030125
2       C        1.531366931     0.000180619     0.000158863
3       C        2.174505266     1.400562720    -0.000044461
4       C        3.716788862     1.385162434    -0.001334145
5       C        4.358575788     2.787868708    -0.001513968
6       C        5.901081407     2.772168977    -0.002979869
7       C        6.542821553     4.174916456    -0.004412762
8       C        8.085334003     4.159210035    -0.005862603
9       C        8.727068095     5.561955347    -0.008095928
10      C       10.269576332     5.546215563    -0.009618700
11      C       10.911260882     6.948971380    -0.012466855
12      H       -0.412907887     0.506445946     0.895853985
13      H       -0.412986843     0.505339657    -0.896362672
14      H       -0.382102660    -1.041671221     0.000602892
15      H        1.882864458    -0.569918486     0.891210160
16      H        1.882992320    -0.570475058    -0.890462803
17      H        1.814387504     1.964559573    -0.891024370
18      H        1.815846794     1.963811845     0.892038913
19      H        4.074082132     0.821226503    -0.893466281
20      H        4.075658546     0.820620642     0.889772927
21      H        3.999425196     3.352270437    -0.892547756
22      H        4.001313515     3.351696983     0.890715440
23      H        6.260135014     2.208426515     0.888497925
24      H        6.258546434     2.207684116    -0.894816658
25      H        6.183662044     4.738654184    -0.895856961
26      H        6.185626965     4.739385036     0.887426245
27      H        8.444420600     3.595964408     0.885903183
28      H        8.442785907     3.594249505    -0.897419174
29      H        8.367905278     6.125204902    -0.899867033
30      H        8.369838690     6.126898862     0.883416727
31      H       10.628794522     4.983433313     0.882436507
32      H       10.626806908     4.980860383    -0.900881230
33      H       10.551641541     7.512117247    -0.904118814
34      H       10.554633253     7.514259632     0.879074273
35      C       12.453457017     6.933667613    -0.014716692
36      C       13.096897414     8.334052047    -0.018568577
```

```
37         H          12.811598147       6.368251857      -0.905762861
38         H          12.814189069       6.371857855       0.877337318
39         H          12.743807927       8.903010736      -0.909644553
40         H          12.746869108       8.906035890       0.871790500
41         C          14.628074574       8.333533467      -0.021119033
42         H          15.042376120       7.829302101       0.875377335
43         H          15.039289836       7.825442559      -0.916873054
44         H          15.011003683       9.375025033      -0.024074178
---
H12C15, RHF, CHARGE=0, MULT=1
HF=46.8
1     C     0.000065523    -0.000024064     0.000086569
2     C     1.404216601     0.000121644     0.000068012
3     C     2.152969102     1.199030531    -0.000202096
4     C     1.487531033     2.414432336     0.078175447
5     C     0.061976109     2.458379869     0.066698369
6     C    -0.700239826     1.256509260    -0.121881878
7     C    -2.138943015     1.410168467    -0.409183595
8     C    -2.790377828     2.649000382    -0.097261381
9     C    -1.987149913     3.793674232     0.293327400
10    C    -0.625575894     3.717432686     0.300414320
11    C    -4.204907661     2.772231808    -0.250232158
12    C    -4.959404458     1.738377591    -0.792810437
13    C    -4.306698389     0.562415828    -1.234335905
14    C    -2.932879859     0.414680584    -1.060526815
15    C    -0.672312317    -1.333479889     0.221790473
16    H     1.955779679    -0.942657243     0.024771975
17    H     3.242085665     1.157932663    -0.028608803
18    H     2.059161096     3.341045592     0.152828280
19    H    -2.490257162     4.729525808     0.542582227
20    H    -0.026903618     4.604795099     0.515850650
21    H    -4.699598045     3.697897274     0.049256835
22    H    -6.039978805     1.833901188    -0.900927647
23    H    -4.884907634    -0.223594919    -1.721805802
24    H    -2.467591157    -0.477829692    -1.478754919
25    H    -1.645028426    -1.238419556     0.744657183
26    H    -0.832561740    -1.874143491    -0.734610028
27    H    -0.047675461    -1.992624747     0.862791390
---
H22C15, RHF, CHARGE=0, MULT=1
HF=-39.9
1     C    -0.000045536     0.000856731    -0.000625574
2     C     1.571078882    -0.000243275    -0.000072932
3     C     2.064837227     1.487272969    -0.000423546
4     C     1.568181386     2.148960435    -1.331421039
5     C    -0.002972547     2.144691717    -1.327130334
6     C    -0.518310276     0.647798874    -1.349484966
7     H    -0.361866936    -1.049836610     0.071418758
8     C    -0.510942967     0.764764301     1.252462906
9     C     2.144649192    -0.761494085    -1.220822675
10    H     1.936117015    -0.515119398     0.918073763
11    H     3.177457688     1.514012769     0.040049702
12    C     1.530981731     2.252401241     1.236637539
13    H     1.930944624     3.201035441    -1.384340506
14    C     2.141171734     1.396258303    -2.557840002
15    H    -0.367793029     2.677184504    -2.234504423
16    C    -0.514213547     2.923841086    -0.083363959
17    C    -2.057020369     0.542394503    -1.518396642
18    C     0.116652775    -0.107617668    -2.569834853
```

```
19      C      -0.018558045     2.234873195     1.214534747
20      H      -1.618706611     0.724604656     1.326268506
21      H      -0.153151502     0.259745384     2.177881989
22      C       1.662488697    -0.075854834    -2.522752060
23      H       1.830046626    -1.828259679    -1.198702590
24      H       3.255715859    -0.787056653    -1.176937364
25      H       1.913778645     3.296790432     1.244666056
26      H       1.917596900     1.796362257     2.174611242
27      H       1.824148206     1.890237265    -3.502745170
28      H       3.252308209     1.449241501    -2.562923299
29      H      -0.159099069     3.978081695    -0.122974516
30      H      -1.622348730     3.005204019    -0.085118682
31      H      -0.235357708     0.339034065    -3.527058247
32      H      -0.232686646    -1.164312562    -2.594969273
33      H      -0.411926251     2.781555331     2.100340619
34      H       2.070954058    -0.620860997    -3.403471761
35      H      -2.387850090     1.004359867    -2.472026953
36      H      -2.386957948    -0.517376194    -1.532190932
37      H      -2.630362661     1.039548987    -0.712535554
---
H22C15, RHF, CHARGE=0, MULT=1
HF=-37.6
1       C       0.004743788     0.015782883     0.084361235
2       C       1.570898344    -0.030047081     0.078685130
3       C       2.098516377     1.446051732     0.045299712
4       C       1.628775244     2.091764320    -1.303902190
5       C       0.061795473     2.137888707    -1.297741963
6       C      -0.468621433     0.659353138    -1.265068888
7       H      -0.398692100    -1.017320683     0.177019056
8       C      -0.520907236     0.826122827     1.299187061
9       C       2.092813073    -0.832080181    -1.140371785
10      H       1.934910575    -0.536692076     1.001099819
11      H       3.210868828     1.447966148     0.095067034
12      C       1.568911110     2.252296469     1.259736192
13      H       2.033959094     3.125697887    -1.385565390
14      C       2.152352077     1.279943213    -2.516698197
15      H      -0.300947001     2.646863158    -2.219420358
16      C      -0.463750791     2.943739557    -0.081027915
17      H      -1.581743924     0.672940152    -1.296876108
18      C       0.041536251    -0.120981055    -2.520899085
19      C       0.018453249     2.278316537     1.233604275
20      H      -1.633099570     0.821926897     1.316296857
21      H      -0.215263892     0.337880310     2.250790205
22      C       1.602794564    -0.171940072    -2.455115181
23      H       1.764657095    -1.892186399    -1.076613991
24      H       3.204734408    -0.879527228    -1.123869415
25      H       1.978220899     3.286468740     1.246563892
26      H       1.936592679     1.807375532     2.210620122
27      H       1.858599713     1.772892382    -3.469539802
28      H       3.264823195     1.274603907    -2.528257635
29      H      -0.116555957     3.999025538    -0.136513615
30      H      -1.574355907     3.003284480    -0.105599026
31      H      -0.233390584     0.488895209    -3.421058391
32      C      -0.649277350    -1.473480933    -2.773001687
33      H      -0.359504500     2.857105130     2.105426777
34      H       1.997749890    -0.745081243    -3.323606915
35      H      -0.275487930    -1.931312722    -3.712441007
36      H      -0.498015861    -2.213918482    -1.964054255
37      H      -1.744089774    -1.334248427    -2.891048758
```

```
---
H22C15, RHF, CHARGE=0, MULT=1
HF=-43.5
1    C    -0.000061370    0.000418739   -0.000693529
2    C     1.566515038   -0.000354594    0.000418185
3    C     2.052856941    1.488866783   -0.000506114
4    C     1.575114425    2.141895481   -1.341829693
5    C     0.008439081    2.149321512   -1.346898458
6    C    -0.472501169    0.658422581   -1.341377856
7    H    -0.376353699   -1.045958137    0.057890787
8    C    -0.557361463    0.772444972    1.223683272
9    C     2.123801951   -0.770476184   -1.224164109
10   H     1.939226770   -0.513863711    0.916031334
11   H     3.164395253    1.521261110    0.058261935
12   C     1.491558743    2.258475412    1.223718227
13   H     1.953643740    3.187791708   -1.403503724
14   C     2.132392710    1.372042762   -2.566653346
15   H    -0.361790173    2.660322241   -2.264303824
16   C    -0.549192598    2.917362816   -0.120133963
17   H    -1.584434331    0.624728011   -1.402375224
18   C     0.084578634   -0.111192449   -2.566216875
19   C    -0.058841857    2.240678357    1.186515703
20   H    -1.669108119    0.736478423    1.231332752
21   H    -0.245360581    0.276924334    2.169454149
22   C     1.648762441   -0.115536156   -2.562820789
23   H     1.803636916   -1.834950340   -1.173104117
24   H     3.235176770   -0.798067066   -1.173805633
25   H     1.870793145    3.304078139    1.231150062
26   H     1.865667542    1.808044414    2.169492367
27   H     1.818826492    1.884014826   -3.503530677
28   H     3.244017882    1.422744946   -2.565232593
29   H    -0.231376565    3.982818890   -0.152447791
30   H    -1.660691828    2.945707275   -0.153131019
31   H    -0.304270571    0.346698704   -3.502978534
32   H    -0.310522920   -1.151559166   -2.564801418
33   H    -0.459465698    2.792824082    2.065491532
34   C     2.204579390   -0.883600041   -3.784682923
35   H     3.313836018   -0.892096663   -3.792211645
36   H     1.867955301   -1.940686866   -3.790941771
37   H     1.872926001   -0.427481321   -4.739873823
---
H22C15, RHF, CHARGE=0, MULT=1
HF=-24.9
1    C    -0.000693127    0.001004030    0.000102498
2    C     1.423647038   -0.001130004   -0.000022729
3    C     1.977456623    1.399129168    0.000018141
4    C     0.739612630    2.327461071    0.000188157
5    C    -0.559322436    1.430755412   -0.000075649
6    C    -0.699027970   -1.213130771    0.000156369
7    C    -0.019035835   -2.467817256   -0.000041716
8    C     1.392509073   -2.428903321   -0.000230294
9    C     2.112904817   -1.214909147   -0.000172718
10   C    -0.826778449   -3.783780296    0.000700204
11   C    -1.718906000   -3.840099319   -1.276882521
12   C    -1.709982486   -3.843098390    1.284268096
13   C     0.068806871   -5.059634251   -0.004328228
14   C    -1.407944793    1.696546660    1.270757799
15   C    -1.407264795    1.696103816   -1.271668066
16   H     2.617774391    1.582373857    0.890230760
```

```
17   H    2.617612332    1.582625277   -0.890258176
18   H    0.768266729    2.999563282    0.883618972
19   H    0.768282891    2.999896174   -0.882932551
20   H   -1.789960706   -1.186199963    0.000390947
21   H    3.203244393   -1.241282494   -0.000140554
22   H   -1.110886010   -3.718259870   -2.196307088
23   H   -2.496121899   -3.050356877   -1.287885932
24   H   -2.252731257   -4.809557226   -1.357482155
25   H   -1.095249700   -3.726000620    2.199875602
26   H   -2.485279797   -3.051613372    1.303659931
27   H   -2.245262529   -4.811729768    1.364987931
28   H    0.720659284   -5.115918463    0.890930742
29   H    0.716225472   -5.111735154   -0.903101959
30   H   -0.548700909   -5.982393824   -0.004822718
31   H   -0.841155113    1.467854452    2.196137041
32   H   -2.327774457    1.077361757    1.288559567
33   H   -1.726689313    2.757661420    1.328672661
34   H   -0.841168686    1.463028739   -2.196370776
35   H   -2.329072533    1.079682932   -1.287820922
36   H   -1.722965162    2.757946289   -1.331971768
37   H    1.990604372   -3.341031103   -0.000414155
---
H30C15, RHF, CHARGE=0, MULT=1
HF=-75.6
1    C    0.150193054   -0.044213718   -0.026667811
2    C    1.684194450   -0.025561641    0.099151277
3    C    2.307817875    1.377505495    0.031149976
4    C    1.743476993    2.249963806   -1.100907240
5    C    0.208235970    2.225263946   -1.211082420
6    C   -0.405100392    0.800326404   -1.205615133
7    C   -1.960659078    0.801859999   -1.215331436
8    C   -2.613867210    1.148851003   -2.569632307
9    C   -4.156883052    1.119888622   -2.551235036
10   C   -4.797538760    1.498446743   -3.902433826
11   C   -6.339978179    1.507598862   -3.895680201
12   C   -6.961407347    1.887845699   -5.255376888
13   C   -8.503404507    1.897595624   -5.267991375
14   C   -9.113728692    2.278014818   -6.631280622
15   C  -10.644207780    2.286003060   -6.669831784
16   H   -0.172144621   -1.103547255   -0.151027349
17   H   -0.296255966    0.300573602    0.934036273
18   H    2.129716270   -0.667400432   -0.695420152
19   H    1.969143612   -0.501504185    1.065280895
20   H    3.409708999    1.280291737   -0.098908184
21   H    2.164046822    1.896215504    1.006793057
22   H    2.073869947    3.302657351   -0.946310175
23   H    2.190525281    1.933758556   -2.071376262
24   H   -0.073434468    2.750327842   -2.151481202
25   H   -0.223637573    2.830754905   -0.381419350
26   H   -0.071082236    0.292928773   -2.146316138
27   H   -2.317199060   -0.210194288   -0.912157925
28   H   -2.335473166    1.502267404   -0.434634965
29   H   -2.248301561    0.433579919   -3.342039816
30   H   -2.280599203    2.159608207   -2.896728034
31   H   -4.501057735    0.102172792   -2.256265078
32   H   -4.526450699    1.815486927   -1.763405556
33   H   -4.442105047    0.786783413   -4.682970293
34   H   -4.431935384    2.505899474   -4.207895108
35   H   -6.710398364    0.501679810   -3.591994378
```

```
36      H      -6.697962551     2.220321976    -3.117830363
37      H      -6.597255711     1.176343396    -6.031904532
38      H      -6.589956286     2.894840598    -5.554725652
39      H      -8.878066205     0.891413674    -4.970259318
40      H      -8.871539961     2.609490142    -4.493908676
41      H      -8.747101321     1.572122182    -7.412194467
42      H      -8.749338304     3.287860862    -6.931756299
43      H     -11.071790542     3.015503204    -5.952568604
44      H     -11.068391380     1.288997459    -6.433933394
45      H     -10.999322897     2.567641676    -7.683002448
---
H32C15, RHF, CHARGE=0, MULT=1
HF=-84.8
1       C       0.000002074    -0.000021546    -0.000042931
2       C       1.531243873     0.000089627     0.000145757
3       C       2.174251312     1.400636584     0.000004899
4       C       3.716487859     1.385913338     0.004315944
5       C       4.357295478     2.789034255    -0.000859760
6       C       5.899724772     2.774350454     0.009468269
7       C       6.540530573     4.177452876    -0.001587498
8       C       8.082939486     4.162872409     0.014029694
9       C       8.723803533     5.565917606    -0.001912810
10      C      10.266167039     5.551426754     0.018552681
11      C      10.907119423     6.954362289    -0.002024966
12      H      -0.413056492     0.506049114     0.896003241
13      H      -0.412952489     0.505789387    -0.896266652
14      H      -0.381939587    -1.041730690     0.000071782
15      H       1.882791300    -0.570198097     0.891033363
16      H       1.883129711    -0.570403700    -0.890508135
17      H       1.817093193     1.962955733    -0.893291263
18      H       1.812247326     1.965496738     0.889723309
19      H       4.077430749     0.817829127    -0.883647054
20      H       4.071968397     0.826085639     0.899559549
21      H       4.004435802     3.346679939    -0.898688243
22      H       3.992859115     3.359077466     0.884462364
23      H       6.252281811     2.220093109     0.909498555
24      H       6.264311199     2.200905175    -0.873597586
25      H       6.190707564     4.730061591    -0.903700845
26      H       6.173188557     4.752431617     0.879325799
27      H       8.432465994     3.613186973     0.918036783
28      H       8.450503401     3.584974696    -0.864887596
29      H       8.376851059     6.114039142    -0.907870729
30      H       8.353642408     6.145242178     0.874961891
31      H      10.612907826     5.006024591     0.926214176
32      H      10.636400842     4.969391617    -0.856504480
33      H      10.562103845     7.498584076    -0.911098949
34      H      10.534962301     7.537539222     0.871459942
35      C      12.449424614     6.940224014     0.021688287
36      C      13.090237879     8.342910554    -0.002758670
37      H      12.821527466     6.355490079    -0.850741039
38      H      12.794858212     6.398105334     0.931768120
39      H      12.748917250     8.883889949    -0.915285853
40      H      12.716207740     8.930709597     0.866953910
41      C      14.631000370     8.331352012     0.026187789
42      H      14.981128572     7.801106719     0.942000431
43      H      15.014601726     7.741554459    -0.838341712
44      C      15.281369811     9.717126426    -0.008610924
45      H      14.980971279    10.336507743     0.860706261
46      H      15.016698506    10.274206039    -0.930149086
```

```
 47      H       16.386643315     9.621683228     0.016726827
---
H10C16, RHF, CHARGE=0, MULT=1
HF=69.8
1     C      -0.000019129     0.000258706     0.000255286
2     C       1.381566191    -0.000597331     0.000042121
3     C      -2.356588296     0.350596953    -0.000033371
4     C      -1.066213110     1.024811401    -0.003441593
5     C      -3.534383558     1.098309975    -0.002203133
6     C      -3.453309228     2.512460656    -0.007908187
7     C       2.072728336    -1.261643720     0.004895280
8     C       1.407965748    -2.479290031     0.009824256
9     C      -0.031597963    -2.527054284     0.010356066
10    C      -0.684240421    -1.277799440     0.005599608
11    C      -2.213003045     3.160387949    -0.011383899
12    C      -1.006424836     2.418628255    -0.009176138
13    C      -2.124119598    -1.109791331     0.005593816
14    C      -2.912073248    -2.244726903     0.010552809
15    C      -2.271780651    -3.532346307     0.015380703
16    C      -0.892588357    -3.681771896     0.015353963
17    H       1.965195921     0.919398735    -0.003634388
18    H      -4.512179100     0.616908818     0.000430024
19    H       3.164237678    -1.243039483     0.004668920
20    H       1.974015395    -3.411456683     0.013423596
21    H      -4.373235597     3.098637328    -0.009614893
22    H      -2.168550385     4.250176248    -0.015805825
23    H      -4.000612288    -2.198729580     0.010982322
24    H      -2.910540259    -4.417618146     0.019211439
25    H      -0.450401511    -4.678758191     0.019199813
26    H      -0.052368569     2.945475748    -0.011935041
---
H10C16, RHF, CHARGE=0, MULT=1
HF=53.9
1     C      -0.000295534     0.000134309     0.000001893
2     C       1.435774251    -0.000279822    -0.000000748
3     C      -0.716490943     1.244856315    -0.000002736
4     C       0.012503019     2.456494366    -0.000008139
5     C       1.415805629     2.448060255    -0.000009182
6     C       2.122774299     1.235859686    -0.000006930
7     C       2.125846545    -1.283528283    -0.000005946
8     C       1.442270676    -2.464621462    -0.000015684
9     C      -0.014132188    -2.505711592    -0.000002958
10    C      -0.730084505    -1.260861225     0.000009392
11    C      -2.173012644     1.204105508     0.000000858
12    C      -2.856492160     0.023001652     0.000017151
13    C      -2.166153781    -1.260178649     0.000017415
14    C      -2.853286075    -2.496078128     0.000015014
15    C      -2.146635174    -3.708492364     0.000004468
16    C      -0.743413434    -3.717362726     0.000002676
17    H      -0.512963503     3.412895512    -0.000009584
18    H       1.961786196     3.392645749    -0.000013604
19    H       3.213715895     1.257523757    -0.000011771
20    H       3.217623889    -1.282060226    -0.000007871
21    H       1.984869524    -3.412037671    -0.000034908
22    H      -2.715418445     2.151642927    -0.000011236
23    H      -3.948210426     0.021350515     0.000024788
24    H      -3.944335088    -2.517441113     0.000017772
25    H      -2.693152952    -4.652782359    -0.000003026
26    H      -0.218336076    -4.674011098     0.000008565
```

```
EXPGEOM
1    C     0.00000      0.00000      0.71620
2    C     0.00000      0.00000     -0.71620
3    C     0.00000      1.22770      1.42550
4    C     0.00000     -1.22770      1.42550
5    C     0.00000     -1.22770     -1.42550
6    C     0.00000      1.22770     -1.42550
7    C     0.00000      1.20700      2.82360
8    C     0.00000     -1.20700      2.82360
9    C     0.00000     -1.20700     -2.82360
10   C     0.00000      1.20700     -2.82360
11   C     0.00000      2.46270      0.67600
12   C     0.00000     -2.46270      0.67600
13   C     0.00000     -2.46270     -0.67600
14   C     0.00000      2.46270     -0.67600
15   C     0.00000      0.00000      3.51370
16   C     0.00000      0.00000     -3.51370
17   H     0.00000      3.39650      1.22640
18   H     0.00000     -3.39650      1.22640
19   H     0.00000     -3.39650     -1.22640
20   H     0.00000      3.39650     -1.22640
21   H     0.00000      2.14420      3.36780
22   H     0.00000     -2.14420      3.36780
23   H     0.00000     -2.14420     -3.36780
24   H     0.00000      2.14420     -3.36780
25   H     0.00000      0.00000      4.59650
26   H     0.00000      0.00000     -4.59650
---
H14C16, RHF, CHARGE=0, MULT=1
HF=34.2
1    C     0.061028971    -0.028681005     0.010437309
2    C     1.449922586     0.010491033     0.007643076
3    C     2.156705736     1.249577529    -0.022566550
4    C     1.401660778     2.424200349    -0.048729242
5    C    -0.027404781     2.408279907    -0.046446275
6    C    -0.730152051     1.162719280    -0.017061496
7    C    -2.198627872     1.162953952    -0.016603549
8    C    -2.900821314     2.408903684    -0.046287549
9    C    -2.146575977     3.649275192    -0.074945711
10   C    -0.781243692     3.648946589    -0.074998911
11   C    -4.329831518     2.425858442    -0.048098057
12   C    -5.085380062     1.251675303    -0.021325869
13   C    -4.379322146     0.012178435     0.008764256
14   C    -2.990466304    -0.027956298     0.010980554
15   H    -0.415154271    -1.009939307     0.034052713
16   H     2.007172961    -0.928285082     0.028790355
17   C     3.661409396     1.262348974    -0.025788062
18   H     1.906997368     3.392500453    -0.072223548
19   H    -2.693647775     4.593815191    -0.096865048
20   H    -0.233964018     4.593378147    -0.097119771
21   H    -4.834498101     3.394452843    -0.071664875
22   C    -6.590079938     1.265456738    -0.023672578
23   H    -4.937231890    -0.926212557     0.030324962
24   H    -2.514940333    -1.009402718     0.034422351
25   H     4.077219343     2.289944394    -0.051222680
26   H     4.059920907     0.724172491    -0.911765953
27   H     4.063343756     0.767090840     0.883316305
28   H    -7.004891178     2.293305715    -0.057226818
29   H    -6.991758893     0.778215953     0.889743461
```

```
30      H       -6.989479929     0.720297229    -0.904835953
---
H14C16, RHF, CHARGE=0, MULT=1
HF=21.6
1       C       -0.000014592    -0.000244104    -0.001428398
2       C        1.423882544     0.000073221     0.001722287
3       C       -2.861032681     0.601337225     0.000609230
4       C       -2.223481216    -0.068262481     1.230902630
5       C       -0.712269406    -0.046412224     1.230499654
6       C        2.113185034    -0.044398233     1.231872806
7       C        2.184437066     0.075774114    -1.302075340
8       C        1.423230568    -0.508767222    -2.504832971
9       C       -0.053727285    -0.188752185    -2.511227414
10      C       -0.738483555     0.046711861    -1.285323200
11      C        1.409395153    -0.095823615     2.444866719
12      C        0.006178549    -0.098376482     2.444118853
13      C       -2.134238981     0.327179755    -1.295728739
14      C       -2.822212896     0.372376201    -2.526856660
15      C       -2.146054924     0.135480279    -3.733524776
16      C       -0.771296747    -0.145382838    -3.725339818
17      H       -3.921172385     0.265748631    -0.067947252
18      H       -2.898933446     1.706210495     0.156470455
19      H       -2.560592870    -1.130699915     1.292043508
20      H        3.160511930    -0.454377323    -1.216238706
21      H        1.906890829    -0.137772548    -3.437429953
22      H       -2.620791038     0.431742543     2.143825893
23      H        3.204381898    -0.040330582     1.254238859
24      H       -3.890974525     0.591975332    -2.554612224
25      H       -2.689417459     0.169964301    -4.678523278
26      H       -0.263594068    -0.329022810    -4.673983373
27      H        2.433561926     1.145126931    -1.503405188
28      H        1.543751989    -1.618539525    -2.517897518
29      H        1.954118155    -0.133778313     3.388981290
30      H       -0.522441830    -0.140398843     3.398093919
---
H14C16, RHF, CHARGE=0, MULT=1
HF=40
1       C       -0.001070096     0.003824971     0.028037891
2       C        1.384662148    -0.004764208    -0.077367973
3       C        2.083805890     1.218218925    -0.204195597
4       C        1.377564306     2.414866657    -0.232618050
5       C       -0.050824219     2.458386200    -0.123941857
6       C       -0.754161950     1.221574346     0.026129515
7       C       -2.214590607     1.238788673     0.155093518
8       C       -2.922442808     2.458762233    -0.083102945
9       C       -2.178675949     3.689734697    -0.379088697
10      C       -0.799510437     3.716994855    -0.230601036
11      C       -4.350474230     2.447300753     0.038109175
12      C       -5.053027108     1.296457151     0.374810146
13      C       -4.350135100     0.094865243     0.625739691
14      C       -2.963975550     0.076318396     0.525571629
15      H       -0.505042990    -0.960957410     0.098873004
16      H        1.935083062    -0.945967416    -0.069807182
17      H        3.171001865     1.215013718    -0.288732852
18      H        1.954152234     3.332233613    -0.359689618
19      C       -2.950791564     4.907033143    -0.838565116
20      C       -0.033634907     5.021164835    -0.184655049
21      H       -4.929258572     3.359183346    -0.116535832
22      H       -6.140285273     1.316042894     0.456824343
```

```
23   H     -4.897904288   -0.806223056    0.903169036
24   H     -2.456953190   -0.861928738    0.754519484
25   H     -3.394511872    5.452663492    0.020900453
26   H     -3.768687294    4.630979767   -1.536953603
27   H     -2.323989067    5.626136405   -1.404353923
28   H      0.789085010    4.985979364    0.559701435
29   H      0.403227319    5.268422532   -1.175379729
30   H     -0.663985914    5.879055750    0.127276050
---
H16C16, RHF, CHARGE=0, MULT=1
HF=52.2
1    C     -0.000302796   -0.000998645   -0.001173125
2    C      1.413430400   -0.000068159    0.000331094
3    C      2.062169169    1.261118368   -0.000962139
4    C      1.377663919    2.459514636    0.326765765
5    C     -0.035400392    2.421489880    0.317842989
6    C     -0.711666412    1.207158963    0.105406466
7    C      1.908339696    0.117092037    3.222557806
8    C      2.938933472   -0.467512701    2.441686037
9    C      4.137548582    0.277934574    2.302183125
10   C      4.114094350    1.670374816    2.490462760
11   C      2.891796389    2.312137739    2.815937799
12   C      1.884561268    1.507971151    3.409606895
13   C      2.657469490   -1.673056098    1.573819976
14   C      2.221366196   -1.280934687    0.122333971
15   C      2.149307403    3.684629647    0.786642197
16   C      2.567436583    3.694013581    2.295263287
17   H      5.011480693    2.255133567    2.285936684
18   H      3.135480303    1.303334480   -0.200054346
19   H     -0.617144394    3.328053430    0.492131029
20   H     -1.801642509    1.197180024    0.062915403
21   H      5.052576187   -0.201311928    1.952554861
22   H      1.039006079    1.965662259    3.924158145
23   H      1.862476225   -2.304893894    2.029572873
24   H      3.554401142   -2.328463851    1.501536869
25   H      1.644553116   -2.136739093   -0.296477815
26   H      3.134583617   -1.203890775   -0.511613649
27   H      3.067157253    3.799723546    0.165398848
28   H      1.550876663    4.604973012    0.599109615
29   H      3.439529225    4.376378804    2.409471556
30   H      1.748313922    4.153861900    2.892386498
31   H     -0.555550935   -0.937748404   -0.068005174
32   H      1.081227123   -0.488534840    3.594600810
---
H16C16, RHF, CHARGE=0, MULT=1
HF=40.7
1    C     -0.000464007   -0.000671701   -0.003056283
2    C      1.412924768    0.000199849    0.001394323
3    C      2.064416438    1.258890753    0.000743442
4    C      1.384538023    2.436955730    0.399883880
5    C     -0.028476194    2.409762731    0.385948387
6    C     -0.709380270    1.207482322    0.123652210
7    C      1.545697905   -0.597186865    3.633936815
8    C      2.526654588   -0.420011790    2.632124871
9    C      3.108358823    0.866707461    2.509086958
10   C      2.472912474    2.015154780    3.043665461
11   C      1.493479335    1.809971427    4.042141651
12   C      1.083123060    0.506163406    4.373379254
13   C      2.844853355   -1.485134329    1.601942780
```

```
   14    C      2.228889056    -1.264361881     0.178548555
   15    C      2.173976226     3.592688792     0.980424502
   16    C      2.737248003     3.373750857     2.426027331
   17    H      3.130965834     1.309267075    -0.232182854
   18    H     -0.604722143     3.310692358     0.601006274
   19    H     -1.798628691     1.203388068     0.069784924
   20    H      4.022884486     0.986299773     1.922938351
   21    H      1.030646830     2.657036655     4.550253979
   22    H      0.361860459     0.354263174     5.177283548
   23    H      2.505179064    -2.480307585     1.968506410
   24    H      3.950268832    -1.575274093     1.492351387
   25    H      1.605033708    -2.154632216    -0.063335793
   26    H      3.058873114    -1.277503595    -0.565187633
   27    H      3.028770574     3.833734109     0.306996602
   28    H      1.545959209     4.512055706     0.995292215
   29    H      2.317548025     4.172986903     3.078028657
   30    H      3.835593481     3.561226403     2.402309898
   31    H     -0.556019768    -0.936210278    -0.080422395
   32    H      1.121896001    -1.583095937     3.830018655
---
H16C16, RHF, CHARGE=0, MULT=1
HF=58.5
1     C      0.000509260     0.000341892     0.000462391
2     C      1.418379586    -0.000424465    -0.002813934
3     C      2.126791978     1.213384358     0.001239029
4     C      1.427831735     2.447109133     0.012209631
5     C      0.074750800     2.412144499    -0.410369476
6     C     -0.633136324     1.197936767    -0.417167352
7     C     -0.420743403     0.232210236     2.793449152
8     C      0.932964776     0.267002978     3.213628530
9     C      1.640226631     1.481493281     3.221833169
10    C      1.005615818     2.679497133     2.805991526
11    C     -0.412188334     2.679755231     2.812053312
12    C     -1.120152843     1.465535007     2.807150384
13    C     -1.020703388    -1.007959261     2.164074262
14    C     -0.787900747    -1.136158604     0.617488967
15    C      2.026609088     3.687690862     0.641580893
16    C      1.792862682     3.816305696     2.188039950
17    H      1.971549213    -0.937602086     0.076173099
18    H      3.214701088     1.192997644     0.079640093
19    H     -0.452682679     3.335572145    -0.653831653
20    H     -1.695021558     1.204185914    -0.667851782
21    H      1.461842425    -0.656611011     3.453127928
22    H      2.702333239     1.475692774     3.471385501
23    H     -0.965647450     3.616919869     2.736955907
24    H     -2.208206794     1.485806428     2.730148260
25    H     -2.117444189    -1.039607169     2.353707217
26    H     -0.611031872    -1.920340331     2.653733413
27    H     -1.781828608    -1.223436085     0.122827640
28    H     -0.276014133    -2.105627521     0.422504840
29    H      1.616922327     4.599757020     0.151531929
30    H      3.123510224     3.719576914     0.452478413
31    H      1.279488431     4.785273564     2.381979406
32    H      2.786440827     3.905276771     2.682735063
---
H18C16, RHF, CHARGE=0, MULT=1
HF=9.7
1     C     -0.007405345     0.016277627     0.075787677
2     C      1.504092631     0.049543114     0.296202701
```

```
 3    C     -0.724787002    1.227413277   -0.100859120
 4    C     -0.051587437    2.580909402   -0.186613321
 5    C      1.461661943    2.601486599    0.079464471
 6    C      2.174368295    1.301995268   -0.317651587
 7    C      2.168104176   -1.233471048   -0.149205713
 8    C      1.517708824   -2.410632250   -0.147497711
 9    C      0.077617169   -2.530893671    0.296206350
10    C     -0.698023569   -1.233088277    0.077046899
11    C     -2.130866833    1.172616768   -0.237653040
12    C     -2.807003053   -0.050486919   -0.236110705
13    C     -2.105415216   -1.270079942   -0.098203538
14    C     -2.894082620   -2.559916401   -0.180608519
15    C     -2.104831227   -3.853859038    0.072720700
16    C     -0.625441413   -3.762073740   -0.323714449
17    H     -0.541894465    3.282226768    0.528830032
18    H      1.917684960    3.452544364   -0.475806803
19    H      2.209014822    1.217588881   -1.427445503
20    H      3.217121152   -1.175016325   -0.445213731
21    H      2.027014700   -3.330518113   -0.440394894
22    H     -2.708668549    2.091710285   -0.354842992
23    H     -3.892800181   -0.050388994   -0.352234661
24    H     -3.358553462   -2.618228804   -1.194416273
25    H     -2.582308933   -4.687871756   -0.490604761
26    H     -0.107781349   -4.690333617    0.009658590
27    H      1.659000808    0.108154132    1.409284670
28    H     -0.240788921    2.998229525   -1.204908250
29    H      1.646767375    2.813514057    1.157833311
30    H      3.234195544    1.357200365    0.020436035
31    H      0.108485663   -2.696075178    1.408702027
32    H     -3.741443954   -2.519904901    0.543603588
33    H     -2.185436732   -4.133822058    1.148333572
34    H     -0.535462872   -3.740442175   -1.433373191
---
H28C16, RHF, CHARGE=0, MULT=1
HF=-36.4
 1    C      0.000074129    0.000153084    0.000164566
 2    C      1.545283506   -0.000418531    0.000127783
 3    C      2.255658983    1.366272014   -0.000821517
 4    C      1.417077608    2.595679742   -0.423549011
 5    C      0.347958325    2.170762592   -1.468471083
 6    C     -0.521207627    0.941915450   -1.125744414
 7    C      0.219951815    0.498291894    3.836393852
 8    C      1.733747816    0.823433177    3.795347831
 9    C      2.149743198    2.290789123    3.977796268
10    C      1.237381733    3.391130552    3.381475583
11    C     -0.257946637    3.122088382    3.689211817
12    C     -0.653443304    1.723408323    4.205344362
13    C     -0.284933269   -0.393057284    2.659295220
14    C     -0.761915134    0.242025536    1.336147057
15    C      0.803655922    3.497395357    0.687379121
16    C      1.620862533    3.872434367    1.943844469
17    H      1.937133441   -0.613487718    0.840180136
18    H      2.728631916    1.534299197    0.987660188
19    H     -0.325420546    3.029571060   -1.690987430
20    H     -1.552724283    1.288287651   -0.890623698
21    H      2.195626660    0.435066979    2.864864044
22    H      3.188807095    2.406044450    3.592294232
23    H     -0.903776082    3.379800134    2.823252690
24    H     -0.676049734    1.776234258    5.321562496
```

```
25    H    -1.159402846   -0.961974019    3.063245124
26    H     0.481048633   -1.176385197    2.464529024
27    H    -1.794565597   -0.153545121    1.160840375
28    H    -0.906144952    1.329164506    1.485888789
29    H    -0.186172797    3.095400546    0.977533921
30    H     0.564620010    4.467537648    0.181569424
31    H     1.583971465    4.990179456    1.994841117
32    H     2.696634228    3.646253477    1.772102800
33    H    -0.288216994   -1.041437487   -0.303353264
34    H     1.869577451   -0.555233743   -0.914738752
35    H     3.118269921    1.292101057   -0.707156830
36    H     2.131189118    3.267294985   -0.970226653
37    H     0.878260690    1.968043544   -2.429024591
38    H    -0.630829491    0.333943255   -2.054669944
39    H     0.103411309   -0.189140967    4.720778305
40    H     2.227828014    0.239648871    4.610106020
41    H     2.231361133    2.490412706    5.074386484
42    H     1.488953711    4.297154060    4.002236847
43    H    -0.568345756    3.852215295    4.476759619
44    H    -1.710939117    1.533584725    3.911100643
---
H32C16, RHF, CHARGE=0, MULT=1
HF=-59.4
1     C    -0.000028310    0.000063630    0.000068302
2     C     1.542558041   -0.000061017    0.000106853
3     C     2.169941230    1.409000238    0.000016119
4     C     3.712144452    1.410167733   -0.001134185
5     C     4.337656697    2.780389714    0.001132653
6     C     5.655821640    3.028918157    0.017039568
7     C    -0.627710997   -1.409035504   -0.000593582
8     C    -2.170278581   -1.408640865    0.000005590
9     C    -2.797922885   -2.817760920   -0.000659118
10    C    -4.340474877   -2.817488213    0.000132356
11    C    -4.968108273   -4.226583963   -0.000473676
12    C    -6.510660139   -4.226311787    0.000619851
13    C    -7.138279810   -5.635447057   -0.000143051
14    C    -8.680538038   -5.635594063    0.001165899
15    C    -9.309938476   -7.042353905   -0.000100437
16    C   -10.841075786   -7.056161336    0.002305956
17    H    -0.363975516    0.560079674    0.891971077
18    H    -0.363822631    0.560974385   -0.891246644
19    H     1.906572636   -0.560371666    0.891788960
20    H     1.906500150   -0.560512931   -0.891435386
21    H     1.808440020    1.969307087    0.892734028
22    H    -0.263544654   -1.969956287    0.890689038
23    H    -0.264167655   -1.969107073   -0.892629420
24    H    -2.533834360   -0.848799272    0.892153198
25    H    -2.534401645   -0.847529336   -0.891066041
26    H    -2.433573221   -3.378781361    0.890428319
27    H    -2.434480080   -3.377702667   -0.892800365
28    H    -4.703836844   -2.257462171    0.892271694
29    H    -4.704817906   -2.256424257   -0.890923214
30    H    -4.603481470   -4.787734010    0.890424199
31    H    -4.604886023   -4.786395753   -0.892825464
32    H    -6.873861059   -3.666460970    0.892934051
33    H    -6.875182449   -3.665076387   -0.890308168
34    H    -6.773726804   -6.196756961    0.890623453
35    H    -6.775281819   -6.195159970   -0.892594291
36    H    -9.044814000   -5.076725319    0.893769779
```

```
37    H     -9.046355775    -5.074371193    -0.889308121
38    H     -8.951702376    -7.610773812     0.889344843
39    H     -8.954341327    -7.608012860    -0.892274419
40    H    -11.256973056    -6.555671575     0.900150223
41    H    -11.259994897    -6.551895232    -0.892081478
42    H    -11.214437393    -8.101137764     0.000701333
43    H     1.807158789     1.969893157    -0.891802157
44    H     4.079425049     0.846332881     0.889062740
45    H     4.078158537     0.851238350    -0.895099798
46    H     3.643678466     3.627204453    -0.012559939
47    H     6.425812493     2.259085576     0.030966502
48    H     6.057196145     4.041778001     0.017175300
---
H32C16, RHF, CHARGE=0, MULT=1
HF=-80.5
1     C     -0.000066885     0.000104314    -0.000018882
2     C      1.538972909    -0.000317031     0.000120642
3     C      2.168827089     1.401447425    -0.000174284
4     C      1.528787803     2.362878062    -1.013820993
5     C     -0.010892353     2.353730917    -1.011228652
6     C     -0.639068404     0.936540927    -1.061899897
7     C     -2.189495423     0.942510975    -0.939004549
8     C     -2.959375064     1.451301217    -2.176002604
9     C     -4.492449604     1.324030485    -2.050623663
10    C     -5.267110616     1.840249146    -3.280720873
11    C     -6.799504497     1.716421314    -3.154174571
12    C     -7.574288459     2.227489021    -4.386267779
13    C     -9.106889263     2.110039825    -4.256742265
14    C     -9.881837953     2.613548961    -5.491542855
15    C    -11.413443450     2.503146899    -5.361085034
16    C    -12.196082283     2.996358110    -6.581260633
17    H     -0.344652830    -1.044487270    -0.177645951
18    H      1.910598758    -0.574067112    -0.879857736
19    H      3.257135607     1.312340539    -0.220012895
20    H      1.879816844     3.398569661    -0.801671252
21    H     -0.374250991     2.901343572    -0.111636044
22    H     -2.488502668     1.543660420    -0.050094757
23    H     -2.526079776    -0.099732295    -0.729232221
24    H     -2.703183336     2.518942321    -2.360963667
25    H     -2.621620467     0.887198677    -3.075465447
26    H     -4.831187933     1.880872349    -1.146891485
27    H     -4.756823847     0.255597216    -1.877247490
28    H     -5.001814737     2.908067001    -3.456581887
29    H     -4.929937736     1.281412109    -4.183798099
30    H     -7.136561474     2.278527395    -2.253084941
31    H     -7.064999240     0.649266323    -2.974537183
32    H     -7.305489712     3.293257636    -4.569190050
33    H     -7.240299238     1.661841052    -5.286271739
34    H     -9.440752408     2.680800564    -3.359885943
35    H     -9.376201129     1.045445143    -4.067933625
36    H     -9.610577370     3.676971626    -5.684291223
37    H     -9.552272158     2.039636289    -6.388085174
38    H    -11.752826676     3.079173495    -4.469179457
39    H    -11.694433404     1.441767188    -5.168757136
40    H    -12.001967060     4.067782557    -6.791058800
41    H    -11.943269502     2.420260550    -7.494394633
42    H    -13.286246507     2.883430576    -6.407615536
43    H     -0.362073980     0.272910085     1.017656080
44    H      1.896294716    -0.558128359     0.895801597
```

```
45      H        2.101978366      1.839733785      1.022170309
46      H        1.899377763      2.122764568     -2.036976976
47      H       -0.355204726      2.945990261     -1.889081389
48      H       -0.391196272      0.500721355     -2.063216181
---
H32C16, RHF, CHARGE=0, MULT=1
HF=-74.7
1       C       -0.079587732     -0.149005967     -0.089662950
2       C        1.458381257     -0.196451255     -0.018344814
3       C        1.969192229      1.255057385      0.083456681
4       C        0.742660447      2.175738476      0.332355387
5       C       -0.503130827      1.333058181     -0.063331637
6       C        0.794348229      3.564630175     -0.348802974
7       C        1.820902949      4.547468597      0.251817764
8       C        1.836609197      5.935777996     -0.421844093
9       C        2.886374621      6.898927454      0.169557821
10      C        2.940964850      8.277443146     -0.520254624
11      C        4.019844176      9.218267566      0.054268944
12      C        4.089228315     10.596391973     -0.635313030
13      C        5.162680621     11.532975136     -0.044020990
14      C        5.243513179     12.913781049     -0.725809626
15      C        6.315951216     13.844452543     -0.126576732
16      C        6.410661430     15.221800415     -0.788264584
17      H       -0.451530565     -0.649286157     -1.008955109
18      H       -0.529112838     -0.702168388      0.763090516
19      H        1.882494141     -0.700565894     -0.912990400
20      H        1.793219812     -0.793962818      0.856801392
21      H        2.509728076      1.529933823     -0.847054432
22      H        2.706408656      1.355267239      0.908493888
23      H        0.676560911      2.351735716      1.435441949
24      H       -1.328674525      1.497450795      0.661588857
25      H       -0.903284382      1.629284673     -1.056612906
26      H       -0.218503217      4.025226901     -0.273000774
27      H        0.994578026      3.440339897     -1.437493103
28      H        1.607186118      4.676395670      1.338131061
29      H        2.840747686      4.104945058      0.184124089
30      H        0.826523592      6.397875479     -0.332784132
31      H        2.028748469      5.810093389     -1.512073841
32      H        2.675064323      7.044333857      1.254132612
33      H        3.891782058      6.422304064      0.108085285
34      H        1.944281505      8.767895310     -0.434383434
35      H        3.125641417      8.133758650     -1.609531502
36      H        3.830590923      9.364955347      1.142578294
37      H        5.014812629      8.722577359     -0.025792663
38      H        3.093582412     11.091138219     -0.564101967
39      H        4.289194959     10.451134484     -1.721733713
40      H        4.960282046     11.676473321      1.042432990
41      H        6.157238222     11.035422341     -0.113786672
42      H        4.250029786     13.413630997     -0.658072373
43      H        5.449308976     12.773985232     -1.811898380
44      H        6.116228619     13.990059698      0.960316897
45      H        7.313832392     13.352698888     -0.194423797
46      H        5.459735575     15.786133588     -0.705714657
47      H        6.668701532     15.146348440     -1.863992763
48      H        7.200937141     15.827664285     -0.298567614
---
H34C17, RHF, CHARGE=0, MULT=1
HF=-79.6
1       C       -0.082515139     -0.158502241     -0.089814823
```

```
  2    C     1.455469364    -0.205066938    -0.017642415
  3    C     1.965742139     1.246594125     0.086328121
  4    C     0.738684590     2.167240112     0.333636999
  5    C    -0.507753284     1.322841458    -0.056494434
  6    C     0.787793608     3.553443176    -0.352941877
  7    C     1.817610938     4.538466474     0.239367086
  8    C     1.828992701     5.925559590    -0.436553744
  9    C     2.879907936     6.889721309     0.152056708
 10    C     2.934921457     8.268608263    -0.536937871
 11    C     4.014670453     9.207332946     0.040106674
 12    C     4.084988487    10.587404289    -0.644788684
 13    C     5.163444379    11.520699165    -0.056839133
 14    C     5.244065129    12.899743221    -0.742937007
 15    C     6.327824197    13.827474953    -0.156897641
 16    C     6.409690779    15.208297212    -0.836089686
 17    H    -0.453090226    -0.654468138    -1.012137916
 18    H    -0.531974998    -0.715989104     0.760000674
 19    H     1.880085240    -0.708143011    -0.912305215
 20    H     1.789575578    -0.803186132     0.857195290
 21    H     2.508558139     1.522012329    -0.843091093
 22    H     2.700869600     1.346007502     0.913151271
 23    H     0.674156247     2.347373468     1.436110805
 24    H    -1.328914888     1.483896804     0.674003603
 25    H    -0.915296476     1.621233813    -1.046163909
 26    H    -0.224910721     4.014468096    -0.275222092
 27    H     0.983915325     3.424715381    -1.441767662
 28    H     1.609900963     4.668749957     1.326226260
 29    H     2.837124948     4.095414776     0.166407150
 30    H     0.818728114     6.386899945    -0.346108595
 31    H     2.018057263     5.799198810    -1.526778165
 32    H     2.670664254     7.034931262     1.236979799
 33    H     3.884980143     6.412762352     0.088511357
 34    H     1.938742674     8.759882943    -0.450402943
 35    H     3.119980208     8.126297008    -1.626469821
 36    H     3.825970906     9.351139561     1.128928783
 37    H     5.009120402     8.710995288    -0.041958057
 38    H     3.090992965    11.084611913    -0.567212647
 39    H     4.279181176    10.444832142    -1.732588485
 40    H     4.964941601    11.666355099     1.029978478
 41    H     6.156342713    11.020152619    -0.128981996
 42    H     4.253117551    13.403221457    -0.666154877
 43    H     5.437920773    12.755185776    -1.830577645
 44    H     6.136515987    13.970056889     0.931601661
 45    H     7.319508574    13.325589427    -0.236505274
 46    H     5.421518763    15.718480946    -0.762318958
 47    H     6.608916248    15.075447441    -1.924918325
 48    C     7.479084326    16.139877250    -0.259262990
 49    H     8.495018110    15.702782909    -0.339362377
 50    H     7.292451680    16.370784685     0.809127223
 51    H     7.486323452    17.102288082    -0.811428759
---
H34C17, RHF, CHARGE=0, MULT=1
HF=-85.4
  1    C    -0.000017761    -0.000876058    -0.000277112
  2    C     1.536750170     0.000340235    -0.000068570
  3    C     2.168437203     1.404539269     0.000129919
  4    C     1.574987735     2.371135494    -1.057828867
  5    C     0.022484825     2.376018624    -1.008833884
  6    C     2.140503458     3.816745660    -0.959029911
```

```
7    C     3.590127986    4.004804048   -1.453911090
8    C     4.068876254    5.471970561   -1.435262194
9    C     5.518931696    5.664475546   -1.925076891
10   C     5.999779558    7.129985545   -1.901120143
11   C     7.448133742    7.323235808   -2.395604484
12   C     7.932456422    8.787398939   -2.361783095
13   C     9.378371494    8.981643948   -2.862782538
14   C     9.866388560   10.444002035   -2.819446917
15   C    11.308931787   10.641271594   -3.324606891
16   C    11.809009972   12.087916683   -3.282241534
17   H    -0.363448686   -1.030509853   -0.219538787
18   H    -0.372520087    0.239287399    1.022090934
19   H     1.907891294   -0.573659673   -0.880179345
20   H     1.899179137   -0.554406511    0.895480780
21   H     3.261391906    1.284854182   -0.175892531
22   H     2.069275152    1.844788711    1.018631460
23   H     1.861570189    1.974584236   -2.065119392
24   H    -0.369813175    2.936998715   -1.887899243
25   H    -0.329822853    2.931170912   -0.109452223
26   H     2.064117815    4.173640968    0.093366197
27   H     1.483186011    4.490112452   -1.557478333
28   H     4.280023675    3.395557080   -0.827384445
29   H     3.676684505    3.608716254   -2.491687874
30   H     3.978555683    5.871027934   -0.398881660
31   H     3.387420354    6.086227018   -2.067572954
32   H     6.200938149    5.047708031   -1.295791641
33   H     5.607880938    5.268910296   -2.962935273
34   H     5.913941956    7.524011994   -0.862420813
35   H     5.316201238    7.747929127   -2.527533122
36   H     8.131466418    6.699853992   -1.774289171
37   H     7.531629321    6.935777312   -3.436976528
38   H     7.853522251    9.172478302   -1.319178311
39   H     7.246583549    9.412382517   -2.978695094
40   H    10.063779166    8.351491911   -2.250709535
41   H     9.455244514    8.602998367   -3.907908163
42   H     9.794018871   10.821887059   -1.773619279
43   H     9.179901279   11.076473833   -3.428410662
44   H    12.003242673   10.011511438   -2.721438373
45   H    11.388516994   10.272170000   -4.373316231
46   H    11.807683131   12.495292152   -2.250980161
47   H    11.189574490   12.757598525   -3.912656957
48   H    12.850621048   12.143790166   -3.660694589
49   C    -0.616378913    0.975968193   -1.013972035
50   H    -1.705171621    1.078702392   -0.802166373
51   H    -0.548608159    0.540061702   -2.037246816
---
H36C17, RHF, CHARGE=0, MULT=1
HF=-94.2
1    C    -0.203892899   -0.137688518    0.012089952
2    C     1.326623026   -0.093977896   -0.002875385
3    C     1.929336570    1.324256960   -0.002052216
4    C     3.471308943    1.354830653   -0.001998183
5    C     4.067216533    2.777421736   -0.000919164
6    C     5.609112656    2.814524992    0.000394844
7    C     6.201300964    4.239096034   -0.001507328
8    C     7.743253736    4.276364102   -0.000011413
9    C     8.339357609    5.699070599   -0.004021714
10   C     9.881667737    5.729923113   -0.001511188
11   C    10.483575825    7.150125797   -0.007528305
```

```
12      H       -0.623564178     0.373136552     0.902293357
13      H       -0.638891639     0.339096059    -0.889632463
14      H       -0.556436775    -1.189660678     0.035349491
15      H        1.703180258    -0.658135583     0.881614929
16      H        1.685780529    -0.649700577    -0.899987224
17      H        1.553255634     1.877398424    -0.893320893
18      H        1.554013565     1.876924866     0.889722003
19      H        3.847144283     0.802248002    -0.893471834
20      H        3.846497186     0.801650796     0.889514621
21      H        3.691087044     3.329693709    -0.892624140
22      H        3.689632770     3.328898809     0.890777698
23      H        5.985822366     2.264593348     0.893252011
24      H        5.987644194     2.261683967    -0.889902051
25      H        5.824322724     4.788700362    -0.894450569
26      H        5.822622843     4.792014821     0.888672551
27      H        8.118855572     3.727280755     0.893901855
28      H        8.121102426     3.721572046    -0.889438541
29      H        7.965208207     6.248480226    -0.898227337
30      H        7.962167920     6.254980898     0.884867100
31      H       10.254611053     5.181023386     0.893604453
32      H       10.257896640     5.172173037    -0.889771269
33      H       10.112276657     7.699276952    -0.903100679
34      H       10.107470904     7.708929052     0.880025012
35      C       12.026145719     7.176949023    -0.003381145
36      C       12.626830388     8.597616530    -0.011496262
37      H       12.402859868     6.617063935    -0.889989862
38      H       12.397733788     6.629389081     0.893023740
39      H       12.255451629     9.144020752    -0.908784981
40      H       12.248258922     9.157178017     0.874513269
41      C       14.169113920     8.630115436    -0.005525505
42      H       14.541640800     8.085786775     0.892548983
43      H       14.548832024     8.070075527    -0.890609772
44      C       14.764466885    10.052794745    -0.015362114
45      H       14.384637583    10.613419579     0.869594894
46      H       14.391940149    10.597531657    -0.913449237
47      C       16.304913255    10.092360161    -0.009715409
48      C       16.908095347    11.499850380    -0.024272566
49      H       16.687151194     9.555430127     0.889188426
50      H       16.694373711     9.533845465    -0.892364393
51      H       18.016118245    11.440808250    -0.036878280
52      H       16.616037471    12.083073072     0.872484845
53      H       16.595366540    12.074474887    -0.919532058
---
H14C18, RHF, CHARGE=0, MULT=1
HF=66.6
1       C       -0.000772614    -0.000959934    -0.001081463
2       C        1.405003746     0.001085240     0.000353169
3       C        2.110415590     1.216671534     0.000157494
4       C        1.403228120     2.431385217    -0.000773152
5       C       -0.002459084     2.431749301    -0.003199125
6       C       -0.724434302     1.214843631    -0.004764921
7       C       -2.931382477     1.184875191    -1.226750312
8       C       -2.210566632     1.214275687    -0.010854162
9       C       -2.941360435     1.243242894     1.198989744
10      C       -4.346250875     1.242778125     1.193032698
11      C       -5.067104165     1.213115671    -0.022728404
12      C       -4.336224244     1.184188165    -1.232600516
13      C       -8.682843695     0.000288412     0.000098110
14      C       -7.277112618    -0.001910620     0.009983443
```

```
15      C       -6.553231930     1.213149984    -0.030002605
16      C       -7.275007623     2.429252521    -0.078979651
17      C       -8.680580877     2.428884849    -0.088763265
18      C       -9.387922387     1.215046710    -0.049496748
19      H       -0.530317441    -0.955004550     0.000349477
20      H        1.948338398    -0.944458416     0.001644818
21      H        3.200703584     1.217544028     0.000584133
22      H        1.945280956     3.377692494     0.000194473
23      H       -0.533413128     3.385002410    -0.004585615
24      H       -2.399672762     1.162433353    -2.179503622
25      H       -2.417343921     1.266194729     2.156006489
26      H       -4.878281728     1.266077122     2.145591587
27      H       -4.860378537     1.161134989    -2.189560762
28      H       -9.226538144    -0.944560296     0.030903258
29      H       -6.747752666    -0.955281836     0.048610839
30      H       -6.743805208     3.381964351    -0.109928307
31      H       -9.222486967     3.374490037    -0.127200581
32      H      -10.478119624     1.215546814    -0.057681802
---
H18C18, RHF, CHARGE=0, MULT=1
HF=68
1       C       -0.000132527     0.000084248    -0.000138930
2       C        1.405236138     0.000183203     0.000461964
3       C        2.112364902     1.214825112    -0.000060325
4       C        1.406089206     2.429566635     0.001622467
5       C        0.000187743     2.431414608     0.001765070
6       C       -0.724686825     1.216343987    -0.003172855
7       C       -2.215404131     1.209772627     0.000131554
8       C       -2.909260301     1.061146265    -1.343655524
9       C       -3.383190130     2.395951857    -1.961832648
10      C       -4.161111878     2.240323004    -3.257912532
11      C       -3.410841887     2.451284148    -4.528659109
12      C       -3.231514998     3.755419335    -5.049473231
13      C       -2.872141960     1.358870554    -5.248534516
14      C       -2.169268657     1.565392669    -6.448454659
15      C       -1.996473403     2.865542360    -6.952578340
16      C       -2.530453808     3.959644903    -6.250292958
17      C       -2.898635874     1.298866786     1.159910872
18      C       -5.480281597     1.958747418    -3.279948239
19      H       -0.529338868    -0.954177792     0.003818670
20      H        1.947206986    -0.946169081     0.001885516
21      H        3.202573438     1.214400737    -0.000646921
22      H        1.948980081     3.375564951     0.003712577
23      H       -0.526616219     3.386909060     0.008197471
24      H       -3.780308290     0.374994331    -1.224340137
25      H       -2.227310956     0.550716408    -2.060814666
26      H       -4.017578696     2.943340146    -1.226018267
27      H       -2.498847215     3.050330012    -2.134133112
28      H       -3.640330443     4.618460621    -4.521515517
29      H       -2.999370532     0.339956871    -4.879618752
30      H       -1.758200563     0.712036749    -6.988945622
31      H       -1.452452518     3.024797865    -7.883790220
32      H       -2.401840396     4.970962788    -6.637514262
33      H       -2.417709602     1.386983786     2.133312876
34      H       -3.986139500     1.287791666     1.222888143
35      H       -6.057041847     1.862789375    -4.198833915
36      H       -6.080302841     1.806133540    -2.383856946
---
H18C18, RHF, CHARGE=0, MULT=1
```

```
HF=38.3
1    C    -0.000293133   -0.000289834   -0.000842638
2    C     1.422497599    0.000498283    0.002752460
3    C     2.122290666    1.249887469   -0.000022712
4    C     1.395097012    2.427442026    0.412112572
5    C    -0.019980351    2.417378707    0.377447300
6    C    -0.700163548    1.191956509    0.118580154
7    C     2.180294091   -1.235642248    0.110392840
8    C     3.540755372   -1.209869622    0.214713633
9    C     4.277986498    0.021976325   -0.016406461
10   C     3.558035144    1.213165810   -0.353006722
11   C     4.267988421    2.246703823   -1.069415570
12   C     5.682740056    2.271775070   -1.030170235
13   C     6.381125066    1.175239428   -0.445490755
14   C     5.700496369    0.047403401   -0.009559863
15   H     7.470597700    1.196102062   -0.376316768
16   C     3.533579754    3.212057612   -1.971863841
17   C     2.113165308    3.614443725    1.012808443
18   H    -1.789990195    1.174917241    0.055475263
19   H    -0.544512504   -0.941626050   -0.091449825
20   C    -0.855755682    3.647450358    0.634788697
21   H     1.636810418   -2.180465567    0.161967065
22   H     4.100056002   -2.124426547    0.418087670
23   C     6.499942267    3.400067773   -1.610400264
24   H     6.259691061   -0.823991118    0.334840933
25   H     3.602799573    4.260296507   -1.614423487
26   H     2.459662168    2.968425684   -2.091216208
27   H     3.962615556    3.178448714   -2.997197281
28   H     2.026571472    4.523749717    0.382735296
29   H     3.190909821    3.430259876    1.190129601
30   H     1.684337479    3.855862546    2.010141809
31   H    -0.497506439    4.517837840    0.046667911
32   H    -0.841083196    3.926180166    1.709814861
33   H    -1.918671673    3.495169392    0.354768679
34   H     7.563257379    3.348676018   -1.297158292
35   H     6.123207149    4.391077378   -1.282165303
36   H     6.488138972    3.375300749   -2.720699298
---
H20C18,  RHF, CHARGE=0, MULT=1
HF=30.9
1    C    -0.000084407   -0.000013400    0.000094162
2    C     1.415670696   -0.000201550   -0.004895629
3    C     2.139979936    1.202904707    0.003205896
4    C     1.477316346    2.453947351    0.017645756
5    C     0.081787286    2.444207473   -0.213163230
6    C    -0.642815319    1.240084929   -0.222730442
7    C    -0.383686237    0.202432592    3.246378564
8    C     0.994976416    0.224769502    3.568064462
9    C     1.716611722    1.429056385    3.578847422
10   C     1.088621258    2.659326453    3.265030896
11   C    -0.322865261    2.655497436    3.177950159
12   C    -1.045617880    1.450202035    3.169185696
13   C    -1.110846101   -1.088952102    2.940168837
14   C    -0.796716072   -1.254488093    0.284625906
15   C     2.211372097    3.741407883    0.323135706
16   C     1.887201381    3.912157817    2.976914139
17   H     1.967758958   -0.941388677    0.021296048
18   H     3.230092699    1.155422807    0.033107755
19   H    -0.462497755    3.380667882   -0.348278497
```

```
20      H     -1.724292748     1.284549527    -0.364793430
21      H      1.527754485    -0.704189597     3.780616669
22      H      2.785251943     1.395451656     3.798553932
23      H     -0.872771283     3.592755520     3.076378483
24      H     -2.130795174     1.494567477     3.060719603
25      H     -2.203977880    -0.898282266     2.838635264
26      H     -1.012397387    -1.770648940     3.819419305
27      H     -0.507702371    -2.037371030    -0.457254320
28      H     -1.880951592    -1.068506444     0.105829663
29      H      1.543330529     4.615201728     0.145147678
30      H      3.047438238     3.857534448    -0.407842000
31      H      1.203339376     4.785941056     2.873199019
32      H      2.524141980     4.141258021     3.865234922
33      C      2.819003283     3.865588806     1.741605082
34      C     -0.629762216    -1.871599123     1.694476285
35      H      3.573973900     3.060069285     1.885118292
36      H      3.400815265     4.819378297     1.751257518
37      H      0.434040329    -2.167497408     1.838908601
38      H     -1.202788659    -2.831043246     1.686494342
---
H22C18, RHF, CHARGE=0, MULT=1
HF=13.7
1       C      0.000012464    -0.000259419     0.000011123
2       C      1.420346338    -0.000465273    -0.000106734
3       C      2.173190004     1.187165080     0.000095402
4       C      1.532988427     2.435398290     0.003908212
5       C      0.130690678     2.472558293     0.011469835
6       C     -0.614620780     1.280377041     0.010632555
7       C     -0.826398038    -1.313466639    -0.037481840
8       C     -0.053611775    -2.381951791    -0.882456418
9       C     -2.131669683    -1.074789997    -0.869413994
10      C     -1.146576701    -1.836943479     1.460142495
11      C      0.158504911    -2.078157092     2.291708198
12      C     -1.917091066    -0.767181347     2.305559887
13      C     -1.975147952    -3.148834406     1.422771281
14      C     -1.362382239    -4.430312262     1.408968808
15      C     -2.109061880    -5.621759336     1.408738109
16      C     -3.511258906    -5.582678346     1.419081109
17      C     -4.149754837    -4.333527102     1.425584943
18      C     -3.395664114    -3.146832074     1.426209106
19      H      1.989844752    -0.931273905     0.002526530
20      H      3.262817420     1.132820069    -0.002896180
21      H      2.113351951     3.357702640     0.001886268
22      H     -0.389247540     3.431456497     0.017854921
23      H     -1.700140933     1.391142745     0.021422624
24      H     -0.686212994    -3.260322406    -1.123907679
25      H      0.853506168    -2.778936410    -0.388540813
26      H      0.271576016    -1.954800355    -1.854944388
27      H     -2.652031803    -2.023899583    -1.111075887
28      H     -2.877531449    -0.431429957    -0.365205399
29      H     -1.896116583    -0.591718251    -1.841573331
30      H      0.680631804    -1.130029147     2.533458484
31      H      0.903073856    -2.722812868     1.787338276
32      H     -0.077792402    -2.560953328     3.263800368
33      H     -1.282704193     0.109848222     2.547230352
34      H     -2.823412861    -0.368194151     1.811772001
35      H     -2.243011653    -1.193969293     3.277971319
36      H     -0.276994334    -4.542355696     1.395254958
37      H     -1.590125511    -6.581201280     1.400630876
```

```
38      H     -4.093015237    -6.504120066     1.421177754
39      H     -5.239166157    -4.277812210     1.430438327
40      H     -3.963838518    -2.215272950     1.426862551
---
H36C18, RHF, CHARGE=0, MULT=1
HF=-90.4
1       C      0.190759470    -0.046427182    -0.010491041
2       C      1.724920346    -0.030985808     0.112453424
3       C      2.349132968     1.371815754     0.044937395
4       C      1.787137077     2.243815511    -1.088756375
5       C      0.252081087     2.219522240    -1.203020447
6       C     -0.364221826     0.796024764    -1.191060132
7       C     -1.919784588     0.800559136    -1.195576944
8       C     -2.577864096     1.143314853    -2.548077332
9       C     -4.121142364     1.123666507    -2.520751743
10      C     -4.764493192     1.494808065    -3.872541223
11      C     -6.306339613     1.523210181    -3.859791564
12      C     -6.928128949     1.895786234    -5.221784743
13      C     -8.469822892     1.931273374    -5.225119658
14      C     -9.083322995     2.293585269    -6.593496065
15      C    -10.625237425     2.328138746    -6.601027367
16      C    -11.239515417     2.686003061    -7.969870663
17      C    -12.780019273     2.725645070    -7.973881592
18      C    -13.409733453     3.071500591    -9.326065131
19      H     -0.253082697     0.302502189     0.949978274
20      H     -0.134440901    -1.105290094    -0.131198236
21      H      2.011008141    -0.508275884     1.077469059
22      H      2.167840280    -0.672586608    -0.683719941
23      H      3.451168342     1.274184745    -0.082989645
24      H      2.203403780     1.890898168     1.020124538
25      H      2.117043894     3.296480830    -0.933281559
26      H      2.236885948     1.927461900    -2.057988438
27      H     -0.181199155     2.830728257    -0.378465761
28      H     -0.026572726     2.739701553    -2.147174492
29      H     -0.033840761     0.284861058    -2.131120520
30      H     -2.290445800     1.504801404    -0.416153422
31      H     -2.277293640    -0.209661127    -0.887046359
32      H     -2.240735504     2.150125689    -2.883082651
33      H     -2.220961420     0.421449351    -3.318408195
34      H     -4.482502070     1.827684133    -1.736413581
35      H     -4.470289908     0.110648723    -2.215684238
36      H     -4.387956884     2.494081877    -4.191374006
37      H     -4.421575255     0.769546094    -4.645935109
38      H     -6.652600399     2.249154853    -3.088912077
39      H     -6.688374442     0.525683949    -3.543511504
40      H     -6.541835363     2.892498688    -5.536747349
41      H     -6.579967289     1.168208705    -5.990908567
42      H     -8.818887860     2.665772542    -4.463518517
43      H     -8.858256004     0.938098886    -4.902666292
44      H     -8.695742553     3.287896679    -6.913997671
45      H     -8.732056253     1.560552071    -7.355588103
46      H    -10.976276697     3.063670029    -5.841165364
47      H    -11.013165347     1.334974656    -6.277370591
48      H    -10.850145360     3.677209626    -8.297843387
49      H    -10.893739891     1.947858422    -8.729679551
50      H    -13.134003197     3.468085201    -7.221531176
51      H    -13.177637505     1.737835382    -7.644195826
52      H    -13.103225658     4.076804159    -9.679447628
53      H    -13.134633174     2.337382688   -10.110063551
```

```
54      H      -14.516394663    3.071287682    -9.243942458
---
H36C18, RHF, CHARGE=0, MULT=1
HF=-84.6
1    C    0.145447949    -0.163564126    -0.123302182
2    C    1.610728620    -0.226608261    0.351061412
3    C    2.155806249    1.213800556    0.378034783
4    H    -12.613785272   12.659079409   -2.491758692
5    C    1.030250737    2.151642509    -0.103525651
6    C    -0.117579679   1.269395109    -0.667397329
7    C    -1.553987699   1.778652413    -0.397729456
8    C    -1.956246548   3.047692605    -1.177829821
9    C    -3.410054206   3.504328478    -0.933934577
10   C    -3.790086629   4.790031164    -1.696953422
11   C    -5.235903642   5.269292027    -1.453811342
12   C    -5.587222257   6.579629370    -2.188073423
13   C    -7.033957241   7.063591512    -1.960203134
14   C    -7.372994873   8.378261792    -2.692444515
15   C    -8.823378055   8.860126789    -2.485497638
16   C    -9.156619349   10.174187639   -3.221704750
17   C    -10.608226861  10.654298505   -3.020649941
18   C    -10.942868141  11.972135494   -3.746067530
19   H    -0.535021939   -0.416727256    0.718108383
20   H    -0.049519685   -0.920219051    -0.912839728
21   H    1.680540266    -0.697859287    1.354419321
22   H    2.217378743    -0.862815404    -0.329004965
23   H    3.051466666    1.307169606    -0.273078461
24   H    2.490277468    1.492998983    1.399887763
25   H    -13.120009362  11.713676042   -3.927668323
26   H    -12.548053349  13.397715650   -4.119573502
27   H    0.685826678    2.789400584    0.738644197
28   H    1.405983792    2.848217041    -0.883117169
29   H    0.012748476    1.222732478    -1.777823267
30   H    -1.688196005   1.958706500    0.693340596
31   H    -2.265186502   0.961968995    -0.664564899
32   H    -1.267947514   3.880604198    -0.907823014
33   H    -1.813420488   2.864942018    -2.267764915
34   H    -3.563693353   3.668353819    0.157245190
35   H    -4.105477833   2.685152790    -1.228122553
36   H    -3.084926290   5.603653353    -1.409456000
37   H    -3.643927981   4.620583763    -2.788660566
38   H    -5.396140938   5.410301446    -0.360297495
39   H    -5.943508019   4.470115730    -1.773146210
40   H    -4.881456271   7.378366723    -1.862899284
41   H    -5.419410604   6.439714384    -3.280791156
42   H    -7.205692502   7.199854460    -0.867866024
43   H    -7.740704963   6.268197868    -2.290819016
44   H    -6.672891416   9.175063467    -2.350484980
45   H    -7.186881501   8.245239559    -3.782980834
46   H    -9.010236956   8.995626965    -1.395438106
47   H    -9.524425789   8.063907307    -2.826334625
48   H    -8.458200417   10.971386123   -2.877830358
49   H    -8.965690507   10.040145717   -4.311279855
50   H    -10.803188635  10.781420459   -1.930843135
51   H    -11.306564986  9.860483981    -3.372606594
52   H    -10.253662585  12.774864122   -3.395712145
53   H    -10.751088116  11.854893387   -4.837768950
54   C    -12.383619700  12.455474926   -3.557191201
---
```

```
H38C18, RHF, CHARGE=0, MULT=1
HF=-59.9
1    C    -0.043611597    0.087143453   -0.040325522
2    C     1.587249200    0.089460364   -0.065737578
3    C     2.435520184    1.466803810   -0.064272175
4    C     3.910280717    1.230255405    0.425214177
5    C     2.576047613    2.267552520   -1.390088741
6    C     1.885665426    2.470581755    1.002300020
7    C     2.231855045   -1.189424971   -0.865416560
8    C     3.326200095   -1.878276601    0.029783842
9    C     2.899661070   -0.893882446   -2.238183528
10   C     1.242865113   -2.371510829   -1.156563187
11   C    -0.913159570    0.928975467   -1.114225470
12   C    -2.376786459    0.367726667   -1.228606670
13   C    -1.084928472    2.461461799   -0.913119334
14   C    -0.365352343    0.733765833   -2.566414208
15   C    -0.673494771   -0.090568825    1.463918128
16   C    -1.383755329    1.149738940    2.076469742
17   C    -1.723425536   -1.261018407    1.461433000
18   C     0.334421041   -0.551001249    2.574168111
19   H    -0.264349690   -0.930845195   -0.434384679
20   H     1.819159614   -0.231275712    0.975205863
21   H     3.958693829    0.623225439    1.350885658
22   H     4.553673596    0.756903109   -0.339856217
23   H     4.410838210    2.193329915    0.667614533
24   H     2.946584351    1.668711634   -2.240310416
25   H    -0.149068273    3.033485845   -1.032988311
26   H     3.307016577    3.098653814   -1.271045693
27   H     1.968749959    2.074288781    2.033481752
28   H    -0.728499559    2.038599148    2.126965582
29   H     2.460109493    3.422652046    0.992442865
30   H     4.278428033   -1.318995152    0.079263497
31   H     2.982273951   -2.044764395    1.069785996
32   H     3.604576172   -2.877656261   -0.369551820
33   H     3.810401473   -0.270730549   -2.159174273
34   H     0.664576026    1.098544341   -2.702885625
35   H     3.237554771   -1.834272700   -2.727825930
36   H     0.453750951   -2.121738678   -1.891231164
37   H     0.763546174   -2.763761213   -0.237637313
38   H     1.779646029   -3.238227020   -1.602560079
39   H    -2.399928050   -0.733078914   -1.354450411
40   H    -3.017436348    0.636182540   -0.367878665
41   H    -2.898134117    0.785979960   -2.117365425
42   H     1.639507376    2.751371943   -1.717375401
43   H    -1.512685990    2.740858490    0.065886755
44   H    -1.783115273    2.880664138   -1.672164917
45   H     2.211816235   -0.410444488   -2.955686508
46   H    -0.394137554   -0.326374031   -2.887127646
47   H    -0.977717548    1.293903401   -3.306637735
48   H     0.835572903    2.758402803    0.838918742
49   H    -2.311502574    1.432987836    1.544406718
50   H    -1.704765947    0.950061764    3.123061357
51   H    -2.656469017   -1.026866160    0.918054553
52   H    -1.312526292   -2.194881014    1.028684833
53   H    -2.049298946   -1.511949011    2.494186071
54   H     0.848559214   -1.499303005    2.320164981
55   H     1.096683398    0.210262060    2.826763569
56   H    -0.194959747   -0.751309889    3.532105541
---
```

```
H38C18, RHF, CHARGE=0, MULT=1
HF=-99.1
1    C     0.000015726   -0.000044134    0.000024464
2    C     1.531232998    0.000029471   -0.000044039
3    C     2.174775539    1.400315646    0.000036647
4    C     3.716976801    1.384697764   -0.012156627
5    C     4.359025751    2.787193163    0.001365694
6    C     5.901295521    2.771482760   -0.023357872
7    C     6.543511210    4.173738967    0.003159980
8    C     8.085554780    4.158362520   -0.033447943
9    C     8.728044544    5.560206121    0.005400792
10   C    10.269794604    5.545278086   -0.042132913
11   C    10.912618626    6.946609245    0.008031617
12   C    12.454024002    6.932185633   -0.049150075
13   C    13.097231474    8.332953498    0.011042983
14   C    14.638331530    8.319048413   -0.054307570
15   C    15.281871247    9.719223222    0.014437089
16   C    16.822366676    9.706192923   -0.057403574
17   C    17.467859682   11.103574914    0.017854724
18   C    18.997282382   11.105244246   -0.056161644
19   H    -0.382375347   -1.041718131    0.005003103
20   H    -0.412904670    0.501407746   -0.898702707
21   H    -0.413084170    0.510181128    0.893645573
22   H     1.882598759   -0.570540428   -0.890788759
23   H     1.882934839   -0.570083015    0.890921300
24   H     1.822148304    1.960503039    0.896428733
25   H     1.808981748    1.967601244   -0.886596754
26   H     4.067551420    0.832059598   -0.913931914
27   H     4.082320887    0.808903332    0.869079554
28   H     4.014732329    3.335847970    0.907976158
29   H     3.987236030    3.366595564   -0.874793680
30   H     6.244742709    2.230651754   -0.934978739
31   H     6.273723833    2.184466382    0.847447246
32   H     6.206173066    4.710804115    0.919275872
33   H     6.165134385    4.764234367   -0.862705705
34   H     8.422067135    3.628790580   -0.954222036
35   H     8.464512943    3.560695673    0.827232181
36   H     8.397171604    6.086267909    0.930220853
37   H     8.343652900    6.161015296   -0.850662179
38   H    10.599882833    5.026153243   -0.971147783
39   H    10.654673773    4.937927358    0.809084721
40   H    10.587521050    7.462595464    0.940549290
41   H    10.522963395    7.556685826   -0.839049846
42   H    12.778423046    6.422398257   -0.985322758
43   H    12.844088091    6.316337689    0.793565766
44   H    12.777084862    8.840055145    0.950128112
45   H    12.703162195    8.951070950   -0.828127810
46   H    14.957860798    7.817278034   -0.996465441
47   H    15.032704793    7.696017702    0.781079541
48   H    14.965896311   10.218944601    0.958868475
49   H    14.884470857   10.344140723   -0.818084255
50   H    17.139317263    9.211292590   -1.004212428
51   H    17.221413584    9.077997898    0.772031210
52   H    17.160976546   11.604011271    0.965436192
53   H    17.075170370   11.739558700   -0.808904729
54   H    19.382344613   12.143868516    0.009147447
55   H    19.364906055   10.673611855   -1.009119105
56   H    19.451759589   10.529164635    0.775228395
---
```

```
H20C19, RHF, CHARGE=0, MULT=1
HF=61.9
1    C    -0.000238719    -0.000287733     0.000060030
2    C     1.405192897     0.000381226    -0.000297376
3    C     2.112008133     1.215196709     0.000041760
4    C     1.405111056     2.429601947     0.003530005
5    C    -0.000764439     2.430820909     0.005095002
6    C    -0.725455144     1.215634733    -0.000561295
7    C    -2.216327341     1.208781487     0.005001038
8    C    -2.911597182     1.060825220    -1.338100674
9    C    -4.113861881     2.231768978    -3.308778702
10   C    -4.603366319     3.536810456    -3.914190322
11   C    -6.044681150     3.870619863    -3.730527789
12   C    -6.465238725     4.732428518    -2.689910593
13   C    -7.025593374     3.340027743    -4.602670022
14   C    -8.384444514     3.656502398    -4.433114731
15   C    -8.788480417     4.509666595    -3.391948189
16   C    -7.825116974     5.047142576    -2.521740825
17   C    -2.897651697     1.297202069     1.166015607
18   C    -3.794305809     4.355607608    -4.617712255
19   H    -0.529091048    -0.954776386     0.003158080
20   H     1.947539496    -0.945774617     0.000130023
21   H     3.202208492     1.215195625    -0.001142831
22   H     1.947482683     3.375734379     0.006177952
23   H    -0.527879253     3.386092506     0.013665356
24   H    -2.224609638     0.563257286    -2.060309454
25   H    -3.773958147     0.362947522    -1.221397912
26   H    -3.419649704     1.739373523    -4.029766508
27   H    -4.971005309     1.528922017    -3.196608665
28   H    -5.734518704     5.165982232    -2.005287158
29   H    -6.734060779     2.678572690    -5.420134793
30   H    -9.127167960     3.238194979    -5.113555353
31   H    -9.842902730     4.753870069    -3.261037782
32   H    -8.131097211     5.711690288    -1.713009974
33   H    -2.414768603     1.384986873     2.138535074
34   H    -3.984787306     1.285743773     1.230951069
35   H    -4.132797762     5.285073838    -5.073952182
36   H    -2.737570162     4.157159650    -4.791826835
37   C    -3.407626390     2.391385301    -1.944890134
38   H    -4.105485092     2.880435980    -1.228491858
39   H    -2.543282369     3.083405984    -2.060984519
---
H38C19, RHF, CHARGE=0, MULT=1
HF=-89.5
1    C    -0.088359006    -0.192892967    -0.084226286
2    C     1.450022685    -0.232965023    -0.015871220
3    C     1.954126776     1.221259318     0.082375397
4    C     0.723389465     2.136693934     0.329912580
5    C    -0.518326049     1.287463779    -0.064435310
6    C     0.769441114     3.524463245    -0.353339781
7    C     1.797890859     4.510089993     0.240004842
8    C     1.813392843     5.894555362    -0.441235473
9    C     2.867809384     6.858602503     0.140563298
10   C     2.923517059     8.234040479    -0.555369035
11   C     4.005847938     9.174259329     0.013647437
12   C     4.075447879    10.551799241    -0.676829555
13   C     5.154935382    11.486155950    -0.092100163
14   C     5.230489851    12.867339116    -0.774119677
15   C     6.308597696    13.799755962    -0.183842399
```

```
16      C       6.394024618     15.173312784    -0.879624191
17      H      -0.460079511     -0.699943163    -0.999945930
18      H      -0.533336274     -0.743521940     0.772479169
19      H       1.874642713     -0.736824137    -0.910137869
20      H       1.789102746     -0.827324389     0.859720965
21      H       2.492785884      1.496520019    -0.849692315
22      H       2.691349502      1.326930145     0.906547525
23      H       0.656060700      2.314427701     1.432623720
24      H      -1.346424801      1.451394305     0.657395675
25      H      -0.917576118      1.578100172    -1.060004665
26      H      -0.243893277      3.983441035    -0.273883205
27      H       0.964652343      3.398308766    -1.442659581
28      H       1.587044796      4.644347155     1.325982135
29      H       2.816953440      4.066134777     0.172694011
30      H       0.804764589      6.358879360    -0.350452566
31      H       2.000153073      5.763090742    -1.531540356
32      H       2.661446725      7.009165290     1.225376583
33      H       3.871902072      6.379474414     0.077231221
34      H       1.928391322      8.727432560    -0.468646640
35      H       3.105359492      8.085958249    -1.644609652
36      H       3.821375174      9.322123241     1.102587570
37      H       4.999452502      8.676589551    -0.070346377
38      H       3.081475495     11.049304098    -0.600329150
39      H       4.268977685     10.405571100    -1.764346343
40      H       4.961864937     11.628593535     0.996122731
41      H       6.148305782     10.987455369    -0.171186193
42      H       4.237016037     13.366189795    -0.698140768
43      H       5.426987614     12.725526416    -1.861735221
44      H       6.104033478     13.952342926     0.900736608
45      H       7.300769670     13.296517667    -0.246745335
46      H       5.402613674     15.677214718    -0.810657145
47      H       6.591423395     15.018822941    -1.965276464
48      C       7.474547700     16.109676149    -0.301569585
49      H       8.467855809     15.609550352    -0.371870209
50      H       7.280347347     16.266908190     0.784272574
51      H       7.749356383     17.330552936    -2.088831645
52      C       7.552034610     17.480382355    -1.002007217
53      C       8.618251730     18.427084258    -0.444448298
54      H       6.561903578     17.987960397    -0.934686790
55      H       9.635976352     17.993194519    -0.518283871
56      H       8.432175072     18.676986050     0.619801400
57      H       8.620085531     19.378688391    -1.015431545
---
H38C19, RHF, CHARGE=0, MULT=1
HF=-95.3
1       C      -0.203841772     -0.157109998     0.005413475
2       C       1.331779502     -0.154830173     0.065537394
3       C       1.968961621      1.243607876    -0.025780043
4       C       1.415913877      2.124420034    -1.176790749
5       C      -0.137007388      2.137937649    -1.176069828
6       C       1.989686049      3.570174216    -1.174556891
7       C       3.452994768      3.704573083    -1.644657311
8       C       3.991479196      5.150937401    -1.624591445
9       C       5.466277423      5.265325308    -2.061621309
10      C       6.037320155      6.697061730    -2.008557088
11      C       7.506187364      6.794022014    -2.470307734
12      C       8.091608025      8.219870479    -2.419946439
13      C       9.552662856      8.314674094    -2.906665539
14      C      10.136723731      9.741123932    -2.855296837
```

```
15      C        11.592952522     9.846784944     -3.353144837
16      C        12.167593642    11.277025877     -3.294819854
17      H        -0.614772635     0.156816911      0.992376248
18      H        -0.559839352    -1.200626139     -0.151520169
19      H         1.735342161    -0.800153628     -0.748391540
20      H         1.656871432    -0.632587914      1.018101985
21      H         3.066528675     1.107209535     -0.154354101
22      H         1.837651566     1.764448299      0.950137529
23      H         1.731224361     1.647741667     -2.139751461
24      H        -0.500237483     2.634898881     -2.104662450
25      H        -0.513515223     2.757125805     -0.329913501
26      H         1.887658278     4.007310995     -0.155221826
27      H         1.357169642     4.201925705     -1.840991538
28      H         4.110526119     3.072323536     -1.006296906
29      H         3.539349452     3.300341152     -2.679479757
30      H         3.880741668     5.568451096     -0.597703222
31      H         3.362426578     5.785750003     -2.289641637
32      H         6.089113760     4.605067079     -1.414944119
33      H         5.566541476     4.871910758     -3.099446276
34      H         5.954773130     7.084698379     -0.967192583
35      H         5.410212225     7.365544592     -2.641972662
36      H         8.131209042     6.122963147     -1.837252371
37      H         7.585837084     6.405201382     -3.511572095
38      H         8.029825284     8.601728578     -1.375056589
39      H         7.457652879     8.896094551     -3.038131358
40      H        10.187422335     7.638505247     -2.289203587
41      H         9.613944618     7.933067924     -3.951669995
42      H        10.082710528    10.117932055     -1.808029609
43      H         9.495113776    10.418275029     -3.464764939
44      H        12.237026107     9.170165108     -2.745586668
45      H        11.648458891     9.472913293     -4.401253021
46      H        12.122495926    11.643535019     -2.243387037
47      H        11.515624376    11.956391899     -3.890930295
48      C        13.615972262    11.400451253     -3.806118776
49      H        13.668353341    11.050147920     -4.862938908
50      H        14.276410493    10.719156399     -3.221507161
51      H        14.199926861    13.211052870     -2.700177363
52      C        14.197466177    12.815455927     -3.736189855
53      H        13.629162691    13.526536489     -4.369379010
54      H        15.247478222    12.815779581     -4.094865395
55      C        -0.780616012     0.741944176     -1.099446555
56      H        -1.875783191     0.862530783     -0.933864082
57      H        -0.679080997     0.230562547     -2.084262296
---
H40C19, RHF, CHARGE=0, MULT=1
HF=-104
1       C         0.000005416    -0.000034199      0.000008729
2       C         1.531223935     0.000015205      0.000007115
3       C         2.174706746     1.400333921     -0.000017156
4       C         3.716954544     1.384792209      0.002663863
5       C         4.358801325     2.787441754      0.001474199
6       C         5.901256129     2.771665536      0.007391315
7       C         6.543041384     4.174364702      0.004692435
8       C         8.085486492     4.158633604      0.013953657
9       C         8.727241683     5.561343615      0.009729986
10      C        10.269662864     5.545673905      0.022281799
11      C        10.911400080     6.948382803      0.016508894
12      C        12.453791697     6.932772580      0.032327814
13      C        13.095519989     8.335477550      0.024872560
```

```
14    C       14.637875601    8.319931231    0.043937849
15    C       15.279612194    9.722616626    0.034621600
16    C       16.821898017    9.707184709    0.056945695
17    C       17.463910145   11.109382966    0.045626306
18    C       19.004740207   11.096426662    0.071174252
19    C       19.654989942   12.482641166    0.058800591
20    H       -0.382274703   -1.041830223    0.003864167
21    H       -0.412938204    0.502456996   -0.898103201
22    H       -0.413147371    0.509166667    0.894207278
23    H        1.882774213   -0.570281650   -0.890821933
24    H        1.882765344   -0.570320531    0.890824961
25    H        1.813949678    1.964803118    0.890447601
26    H        1.816908096    1.963316486   -0.892596048
27    H        4.076580413    0.819101188   -0.887393140
28    H        4.073400258    0.821946567    0.895800946
29    H        3.997211524    3.353945324    0.890224939
30    H        4.003922407    3.349184824   -0.892981327
31    H        6.262876701    2.204310160   -0.880795978
32    H        6.255975567    2.210726678    0.902402722
33    H        6.179547413    4.742688878    0.891497694
34    H        6.190083836    4.734247422   -0.891682657
35    H        8.449027068    3.589449742   -0.872277285
36    H        8.438294810    3.599563546    0.910886716
37    H        8.361860036    6.131477131    0.894595777
38    H        8.376168002    6.119373169   -0.888537498
39    H       10.635087356    4.974655343   -0.861994954
40    H       10.620580429    4.988462868    0.921118312
41    H       10.544152876    7.520335176    0.899426799
42    H       10.562198372    7.504545211   -0.883643541
43    H       12.821079927    6.359848809   -0.849943665
44    H       12.802827997    6.377515250    0.933101999
45    H       12.726436925    8.909317250    0.905791278
46    H       12.748176932    8.889694679   -0.877199520
47    H       15.007009645    7.745011245   -0.836246796
48    H       14.985037139    7.766729537    0.946701890
49    H       14.908680517   10.298473459    0.913447842
50    H       14.934150990   10.274772260   -0.869437066
51    H       17.193043979    9.130237725   -0.821076541
52    H       17.167328117    9.156218973    0.961741756
53    H       17.092111309   11.688260437    0.922255609
54    H       17.121300359   11.660302705   -0.860411117
55    H       19.386369621   10.521205120   -0.804011849
56    H       19.356667596   10.550718155    0.977219097
57    H       20.760241373   12.386960225    0.082054912
58    H       19.354933829   13.087569543    0.938366998
59    H       19.389772402   13.054876432   -0.853286990
---
H14C20, RHF, CHARGE=0, MULT=1
HF=76.9
1     C       -0.000271179    0.000004544    0.000067372
2     C        1.532930811    0.000588774   -0.000279914
3     C       -0.423428459    1.473791033    0.000366728
4     C       -0.422612036   -0.559759789   -1.363474160
5     C        2.087673421    0.739892628   -1.100796754
6     C        2.378293330   -0.619904179    0.923583556
7     C        0.131196077    2.212675777   -1.100590874
8     C       -1.252290044    2.113089354    0.926335390
9     C        0.132269482    0.178749081   -2.464588831
10    C       -1.250908634   -1.659313141   -1.603839522
```

```
11   C     1.013167910    1.348722648   -2.009896444
12   C     3.473827242    0.840605475   -1.248179025
13   C     3.779266056   -0.511910253    0.764314524
14   C    -0.157628762    3.572205449   -1.246989577
15   C    -1.537939021    3.489097962    0.768260309
16   C    -0.155816972   -0.201764742   -3.778152182
17   C    -1.535693289   -2.035352711   -2.937005818
18   C     4.319732779    0.209128002   -0.307272070
19   C    -0.997783158    4.209061985   -0.304053598
20   C    -0.995186623   -1.316181819   -4.009822905
21   H    -0.428429990   -0.568596735    0.847762220
22   H     1.976669182   -1.185666552    1.764556050
23   H    -1.682762954    1.569553702    1.767752668
24   H    -1.681265524   -2.231673953   -0.781720688
25   H     1.441629851    1.917289435   -2.857454536
26   H     3.913846657    1.397476524   -2.075824087
27   H     4.439663583   -0.995964130    1.484811682
28   H     0.252973784    4.150630025   -2.075080911
29   H    -2.185082945    3.988260487    1.490456219
30   H     0.255098254    0.345186943   -4.626986700
31   H    -2.182241826   -2.893289429   -3.125325723
32   H     5.401461122    0.287542081   -0.421857620
33   H    -1.223472624    5.270002170   -0.418085723
34   H    -1.220293040   -1.613648894   -5.034707516
 ---
H16C20, RHF, CHARGE=0, MULT=1
HF=45.1
1    C    -0.000746467   -0.000546200    0.000002805
2    C     1.374804712    0.000978285   -0.000012792
3    C     2.124677859    1.236498153    0.000027406
4    C     1.423440832    2.429965987    0.000085767
5    C    -0.021781480    2.469294482    0.000113384
6    C    -0.751633923    1.234446470    0.000063495
7    C    -2.175535666    1.268488829    0.000080012
8    C    -2.907707355    2.466931855    0.000147303
9    C    -2.159744556    3.709604547    0.000202848
10   C    -0.757831097    3.686676788    0.000185620
11   C    -4.383464358    2.522186542    0.000165634
12   C    -5.044132138    3.790773692    0.000222828
13   C    -4.248013634    5.012394375    0.000262180
14   C    -2.888482323    4.973640529    0.000266330
15   C    -6.466427120    3.857160686    0.000253612
16   C    -7.264836288    2.705717239    0.000245524
17   C    -6.602971090    1.447632645    0.000180965
18   C    -5.211469135    1.362033283    0.000136905
19   C     3.628837848    1.185142326    0.000043817
20   C    -8.768003182    2.774933532    0.000252074
21   H    -0.542807361   -0.947960346   -0.000033214
22   H     1.916839483   -0.946681667   -0.000055300
23   H     1.956167523    3.383592358    0.000117030
24   H    -2.689526601    0.305647266    0.000034186
25   H    -0.206447661    4.629900476    0.000228620
26   H    -4.769989953    5.971523536    0.000291431
27   H    -2.312740910    5.901455443    0.000312096
28   H    -6.940849348    4.841330632    0.000288751
29   H    -7.188464338    0.525981488    0.000169187
30   H    -4.773060595    0.363139154    0.000079632
31   H     4.006458332    0.652272116   -0.898071316
32   H     4.006386033    0.651400169    0.897697201
```

```
33      H         4.087917099     2.194656080     0.000543885
34      H        -9.187233973     2.274149562    -0.897931988
35      H        -9.187389965     2.272769211     0.897564994
36      H        -9.144918623     3.817838745     0.001046113
---
H16C20, RHF, CHARGE=0, MULT=1
HF=62.7
1       C        -0.000917672    -0.021918103     0.044547124
2       C         1.431018458    -0.039245847     0.013231098
3       C        -0.702122292     1.264615362    -0.030790493
4       C        -0.665052191    -1.278586782     0.241573242
5       C         2.137302605     1.220352282     0.058866819
6       C         2.131138201    -1.288349884     0.002057155
7       C         1.459373021     2.392689003     0.267672302
8       C         0.011972331     2.436342760     0.324543320
9       C        -0.727148744     3.627250767     0.731139239
10      C        -2.088925285     1.418074008    -0.474825532
11      C        -2.154408222     3.636743005     0.611847838
12      C        -0.080448733     4.765054632     1.319546667
13      C        -2.823651536     2.530513989    -0.075237591
14      C        -2.867554139     4.768270307     1.134709172
15      C        -4.303991542     2.656966586    -0.360656740
16      C        -2.648422856     0.435906085    -1.481589186
17      C        -0.799267471     5.851198145     1.796424001
18      C        -2.214389391     5.848751392     1.709416295
19      C         0.034079371    -2.478002980     0.266813911
20      C         1.446433961    -2.488919716     0.114276116
21      H        -1.741480859    -1.307610302     0.415690687
22      H         3.223989680     1.224238633    -0.041588659
23      H         3.219248352    -1.291591635    -0.080360643
24      H         2.032320491     3.317114145     0.352471500
25      H         1.005047513     4.791995239     1.422323626
26      H        -3.957171786     4.803932754     1.093548627
27      H        -4.506547740     3.500336223    -1.054584919
28      H        -4.753338773     1.753734724    -0.818364173
29      H        -4.875419215     2.825176616     0.576944511
30      H        -3.152409035     0.969212427    -2.316399081
31      H        -1.860951744    -0.176964892    -1.964959826
32      H        -3.385892103    -0.257539856    -1.026011380
33      H        -0.286642626     6.703026786     2.244393164
34      H        -2.780229074     6.696446546     2.096832206
35      H        -0.495098804    -3.420317097     0.414575107
36      H         1.980365415    -3.439442681     0.105480396
---
H16C20, RHF, CHARGE=0, MULT=1
HF=66.3
1       C        -0.017810916    -0.088446374     0.190514951
2       C         1.409718805    -0.093825018     0.171627975
3       C        -0.698558077     1.109592515     0.179848644
4       C         2.097290174     1.099764658     0.139632846
5       C         1.421643663     2.380412966     0.129533852
6       C        -0.016783861     2.386553114     0.146787069
7       C         2.151045818     3.627056972     0.142365227
8       C        -0.723684597     3.643425078     0.085193812
9       C         0.013647271     4.842143257     0.083665474
10      C        -2.234067112     3.653841698     0.020924056
11      C         1.444484217     4.828568023     0.351597810
12      C        -0.564482618     6.141510627    -0.277516609
13      C         2.057930921     6.097848086     0.810833350
```

```
14    C     3.632699466    3.618398830   -0.157560951
15    C     1.435366154    7.342056042    0.476193034
16    C     3.201673391    6.147315256    1.657004575
17    C     0.122691342    7.308427017   -0.156927909
18    C     2.054831984    8.569092548    0.835214983
19    C     3.783588539    7.362539120    2.032053829
20    C     3.227111187    8.582041966    1.594081917
21    H    -0.554378040   -1.037502241    0.214564057
22    H     1.940298046   -1.046366910    0.186841712
23    H    -1.788724782    1.072356217    0.193555834
24    H     3.187643450    1.059311787    0.140794252
25    H    -2.681018250    2.954829205    0.758628973
26    H    -2.589521475    3.367772214   -0.991918518
27    H    -2.677547024    4.640509596    0.262150510
28    H    -1.563449207    6.167962244   -0.716622900
29    H     4.232929612    3.257863928    0.703807462
30    H     4.017804793    4.617976324   -0.444020225
31    H     3.852960584    2.962076998   -1.027133450
32    H     3.634222441    5.230748067    2.060476022
33    H    -0.314056940    8.247171385   -0.502369842
34    H     1.603699695    9.514884054    0.530718226
35    H     4.667133211    7.368908352    2.671573950
36    H     3.699790600    9.526370766    1.864748487
---
H30C20, RHF, CHARGE=0, MULT=1
HF=3.4
1    C    -0.000178116    0.000189164    0.000090096
2    C     1.507474551   -0.000214031   -0.000027302
3    C    -0.429984386    1.418435329   -0.000323140
4    C     0.733677910    2.152412287   -0.000091730
5    C     1.964583203    1.289376598   -0.000009228
6    C     1.894332810   -1.421383734    0.000370216
7    C     0.761971465   -2.204448964    0.000856628
8    C    -0.430766989   -1.290315441    0.000697143
9    C    -1.872297120    1.910379204   -0.001144477
10    C    -2.593601231    1.392128274   -1.281154049
11    C    -2.594918485    1.392048735    1.278014775
12    C    -1.955203621    3.467226727   -0.001213958
13    C     3.373798763    1.871681902    0.000032565
14    C     3.564604582    2.741948990   -1.279601163
15    C     3.564490291    2.742167618    1.279587924
16    C     4.476161943    0.773007858    0.000006749
17    C     0.646999516   -3.724286285    0.001571007
18    C    -0.117638702   -4.180663064    1.280235364
19    C    -0.118541042   -4.181841144   -1.276157518
20    C     2.044144275   -4.413712257    0.001370180
21    H     0.809164031    3.232901527   -0.000167715
22    H     2.924521037   -1.750504781    0.000299228
23    H    -1.452461286   -1.647318637    0.001142877
24    H    -2.049733978    1.693494245   -2.199400316
25    H    -2.682002252    0.288134440   -1.297327456
26    H    -3.624159816    1.796351654   -1.357149886
27    H    -2.051763962    1.692889888    2.196823820
28    H    -2.683724893    0.288076313    1.293696085
29    H    -3.625397223    1.796484044    1.353527836
30    H    -1.472859805    3.907518861    0.894909306
31    H    -1.471175177    3.907450615   -0.896432232
32    H    -3.009289978    3.814367921   -0.002193150
33    H     2.891975305    3.622457417   -1.294676575
```

```
34      H        3.362468897      2.154999424     -2.198471603
35      H        4.601272613      3.130132129     -1.353631772
36      H        2.892282720      3.622999866      1.294047187
37      H        3.361696984      2.155601085      2.198540405
38      H        4.601292450      3.129854289      1.353949535
39      H        4.413256402      0.125398598     -0.897319154
40      H        4.413070675      0.125336628      0.897352358
41      H        5.491403884      1.221914467      0.000101347
42      H        0.377819422     -3.808849235      2.200034452
43      H       -1.163283273     -3.813521107      1.297291315
44      H       -0.165255231     -5.286783352      1.352133719
45      H        0.376822800     -3.811651809     -2.196670636
46      H       -1.163867574     -3.813901104     -1.293305057
47      H       -0.167087983     -5.288008775     -1.346412538
48      H        2.639595156     -4.142315670      0.896625391
49      H        2.639074506     -4.142720746     -0.894382588
50      H        1.948461617     -5.519389540      0.001646635
---
H36C20, RHF, CHARGE=0, MULT=1
HF=6.2
1       C        0.000230601     -0.000365032      0.000016908
2       C        1.533913115      0.000021638     -0.000258120
3       C        0.766838675      1.327912631      0.000249403
4       C        0.767276953      0.441996775      1.252147609
5       C        2.759299345     -0.707032749     -0.500879114
6       C        2.972256205     -2.041241639      0.276176933
7       C        4.020589904      0.187305021     -0.304589430
8       C        2.615246113     -1.034616344     -2.017718326
9       C        0.767623080      0.440887659      2.752842973
10      C        2.173235245      0.838690394      3.296156660
11      C        0.409868676     -0.975997872      3.294702507
12      C       -0.279977917      1.458569651      3.296827657
13      C        0.766458890      2.743062604     -0.499250951
14      C        1.122407665      2.783762091     -2.016071655
15      C        1.815299154      3.593665279      0.278914512
16      C       -0.638797344      3.387810546     -0.302539527
17      C       -1.224953770     -0.707943228     -0.500301263
18      C       -1.080326392     -2.247581470     -0.306065924
19      C       -2.486183519     -0.226501144      0.278605354
20      C       -1.438363218     -0.417727527     -2.016649187
21      H        3.826265329     -2.616347778     -0.138458422
22      H        2.082614869     -2.698501544      0.223415623
23      H        3.193120848     -1.864844422      1.347078016
24      H        3.977363846      1.104025063     -0.924650540
25      H        4.946046684     -0.353415777     -0.592998841
26      H        4.144955231      0.503579853      0.749335819
27      H        1.799896364     -1.759667147     -2.208042907
28      H        3.546286950     -1.483150676     -2.422671172
29      H        2.403176970     -0.129358571     -2.619214231
30      H        2.167783669      0.922533876      4.402953543
31      H        2.510217471      1.815746446      2.898610560
32      H        2.944775137      0.088414320      3.034390513
33      H       -0.626563372     -1.267860799      3.035132674
34      H        0.488048455     -1.015183961      4.401214781
35      H        1.085913783     -1.756187996      2.894065500
36      H       -0.351299120      1.409852376      4.403421323
37      H       -1.294197778      1.263074125      2.897599215
38      H       -0.015210360      2.502267796      3.037327375
39      H        1.044981851      3.814708042     -2.420011225
```

```
40      H       0.444803374     2.147798773    -2.618297246
41      H       2.158156425     2.440740525    -2.206564465
42      H       1.885800852     4.621605167    -0.133861606
43      H       2.829490814     3.152334896     0.225142486
44      H       1.552350109     3.694665510     1.350077580
45      H      -1.410805385     2.893054585    -0.923757479
46      H      -0.633172264     4.460107156    -0.589166384
47      H      -0.975290383     3.335643715     0.751169689
48      H      -2.011263587    -2.778716702    -0.594712390
49      H      -0.867974355    -2.514656422     0.747431039
50      H      -0.264939291    -2.667409568    -0.927041231
51      H      -2.443290008    -0.508458310     1.348834036
52      H      -3.411775316    -0.677074897    -0.136473411
53      H      -2.610006991     0.872784623     0.228496441
54      H      -2.291494686    -1.000872406    -2.421786333
55      H      -0.548313510    -0.684104317    -2.619243424
56      H      -1.660849259     0.650755589    -2.205281101
---
H38C20, RHF, CHARGE=0, MULT=1
HF=-71.6
1       C       0.000038309    -0.000056986    -0.000095703
2       C       1.540897419     0.000095275     0.000199372
3       C       2.173164334     1.424835838     0.000003343
4       C      -0.632417407     0.911841268    -1.063269639
5       C       0.031417159     2.294444902    -1.162459647
6       C       1.573225873     2.254191979    -1.165415066
7       C       3.746845278     1.384949763     0.176015984
8       C       6.664169234     0.030074853     0.180834401
9       C       7.767178464    -1.046436396     0.122917982
10      C       7.242165228    -2.467552686    -0.134493418
11      C       5.701206719    -0.001013021    -1.034945964
12      C       5.125099751    -1.439302776    -1.204473852
13      C       4.579571889     1.112118648    -1.146887023
14      C       6.199042672    -2.542615407    -1.260478423
15      C       5.029728705     2.356086247    -2.005416344
16      C       4.283520997     2.482220409     1.173858690
17      C       4.113437348     2.050584798     2.654763846
18      C       3.793013818     3.937200679     1.016126724
19      C       6.403764411     3.000143631    -1.722594832
20      C       4.903787969     2.084274597    -3.527915942
21      H      -0.355863487    -1.043522275    -0.160025076
22      H      -0.372163726     0.293030169     1.008570558
23      H       1.896041653    -0.586748135    -0.876005912
24      H       1.876154125    -0.559257079     0.903128809
25      H       1.788593972     1.905194151     0.936410537
26      H      -1.714684563     1.042349899    -0.834981655
27      H      -0.592114037     0.408219929    -2.056444889
28      H      -0.325241498     2.939511015    -0.327057590
29      H      -0.312586416     2.795014668    -2.096255663
30      H       1.934141977     3.304290371    -1.120684000
31      H       1.903223807     1.855050355    -2.149617761
32      H       3.897641152     0.435718983     0.754335277
33      H       7.166744425     1.018776560     0.251973958
34      H       6.106702093    -0.103189842     1.133752956
35      H       8.325247515    -1.035487873     1.087117625
36      H       8.513860861    -0.777490975    -0.659063079
37      H       6.805807501    -2.878648943     0.804783689
38      H       8.098788226    -3.133568781    -0.385522765
39      H       6.352207443     0.151146570    -1.934128558
```

```
40      H      4.534974875    -1.497810922    -2.147454268
41      H      4.415544207    -1.676932700    -0.381035825
42      H      3.820775922     0.619639219    -1.810567511
43      H      6.715802409    -2.509230481    -2.247194767
44      H      5.692986813    -3.534257217    -1.219848058
45      H      4.282524409     3.162911697    -1.804501583
46      H      5.390269422     2.529221319     1.023638098
47      H      3.052838683     2.014741810     2.973833192
48      H      4.552232601     1.051125789     2.848163389
49      H      4.636641429     2.760641384     3.328847676
50      H      3.882985448     4.313772156    -0.019818897
51      H      4.409499193     4.613948294     1.646399886
52      H      2.739595563     4.075891950     1.331687987
53      H      7.253080469     2.379489346    -2.071611660
54      H      6.567536101     3.216012584    -0.650356065
55      H      6.479216240     3.975477114    -2.250128112
56      H      5.632732506     1.333909315    -3.893279874
57      H      3.891631732     1.724974896    -3.802253579
58      H      5.074435104     3.017508262    -4.104673828
---
H40C20, RHF, CHARGE=0, MULT=1
HF=-100.2
1       C      0.226290638    -0.047387227     0.001789919
2       C      1.760647711    -0.030789443     0.121945449
3       C      2.383566474     1.372417024     0.052635440
4       C      1.818698022     2.242948338    -1.081016797
5       C      0.283399334     2.217920848    -1.190928267
6       C     -0.331546685     0.793866737    -1.178256326
7       C     -1.887165270     0.796294444    -1.181548027
8       C     -2.545559713     1.134708868    -2.535331315
9       C     -4.088525908     1.111752625    -2.509626119
10      C     -4.732703748     1.475846355    -3.863041848
11      C     -6.275049553     1.493992067    -3.850400107
12      C     -6.899252942     1.855285958    -5.214191781
13      C     -8.441291674     1.875041022    -5.220970115
14      C     -9.053171619     2.222574828    -6.593997227
15      C    -10.595239639     2.236776185    -6.610197719
16      C    -11.206496036     2.577472030    -7.985120343
17      C    -12.749005913     2.588408383    -7.996234118
18      C    -13.369236203     2.925810295    -9.367351962
19      C    -14.910495214     2.940380016    -9.370454714
20      C    -15.546451200     3.264338837   -10.725028389
21      H     -0.098602428    -1.106532349    -0.117289280
22      H     -0.215932976     0.301811459     0.962736876
23      H      2.202766933    -0.672335319    -0.674683724
24      H      2.049016118    -0.507310761     1.086706391
25      H      3.485517327     1.275613039    -0.077388492
26      H      2.239367789     1.891954141     1.027710192
27      H      2.265937954     1.924923490    -2.050936426
28      H      2.148770337     3.295951807    -0.927982653
29      H      0.001304614     2.738779504    -2.133610587
30      H     -0.147951604     2.827993578    -0.364542338
31      H     -0.001419860     0.282817865    -2.118470979
32      H     -2.258573358     1.501567608    -0.403588945
33      H     -2.243207862    -0.213615914    -0.870506925
34      H     -2.185993839     0.411745139    -3.303576310
35      H     -2.210180108     2.141543087    -2.872071208
36      H     -4.452178173     1.818711661    -1.729067987
37      H     -4.435358454     0.099108075    -2.200785518
```

```
38      H      -4.384204048     0.751527262    -4.634858354
39      H      -4.362605475     2.476975036    -4.183488111
40      H      -6.626386484     2.221538382    -3.083275647
41      H      -6.650242622     0.495276880    -3.529282305
42      H      -6.542182153     1.128801040    -5.979954616
43      H      -6.522639334     2.854982807    -5.531849360
44      H      -8.799734729     2.610761336    -4.464828640
45      H      -8.820855202     0.880070034    -4.893413934
46      H      -8.688204791     1.490541223    -7.350685694
47      H      -8.676961508     3.220387528    -6.917513297
48      H     -10.960104389     2.972848079    -5.857295320
49      H     -10.971614176     1.240869276    -6.281481126
50      H     -10.841316875     1.842404042    -8.738738367
51      H     -10.832105128     3.573985295    -8.313866433
52      H     -13.112164422     3.324049784    -7.242213262
53      H     -13.121295577     1.591499686    -7.665250872
54      H     -13.011711240     2.187883275   -10.121785772
55      H     -12.996567787     3.920904420    -9.702825343
56      H     -15.276041670     3.683552077    -8.624293824
57      H     -15.291569559     1.949080946    -9.031814401
58      H     -16.652969094     3.240892338   -10.642814546
59      H     -15.254557588     2.531872224   -11.504863960
60      H     -15.261630522     4.274228357   -11.083831379
---
H42C20, RHF, CHARGE=0, MULT=1
HF=-108.9
1       C      -0.000000664    -0.000065129     0.000022218
2       C       1.531201643    -0.000004182    -0.000058092
3       C       2.174675591     1.400321405     0.000038996
4       C       3.716855347     1.384852737    -0.015001329
5       C       4.358843546     2.787347336     0.001398287
6       C       5.900997987     2.771908601    -0.029879404
7       C       6.543252379     4.174016774     0.002708881
8       C       8.085030475     4.159140383    -0.043972847
9       C       8.727724599     5.560591482     0.004381566
10      C      10.269162334     5.546257501    -0.052599348
11      C      10.912356474     6.946887179     0.010632105
12      C      12.453433481     6.933181720    -0.055462464
13      C      13.097164396     8.332878459     0.021477877
14      C      14.637900813     8.319790501    -0.052226342
15      C      15.282173319     9.718516321     0.036874367
16      C      16.822619679     9.705973454    -0.042669781
17      C      17.467418639    11.103732610     0.056709869
18      C      19.007321965    11.092034778    -0.026824069
19      C      19.653950608    12.487106207     0.076373085
20      C      21.182742340    12.490001861    -0.009622263
21      H      -0.382432958    -1.041755159     0.005428104
22      H      -0.412930095     0.501067601    -0.898820552
23      H      -0.413100720     0.510540305     0.893455997
24      H       1.882543999    -0.570542786    -0.890826515
25      H       1.882931984    -0.570103371     0.890897487
26      H       1.823593502     1.959639371     0.897574145
27      H       1.807292490     1.968370574    -0.885406874
28      H       4.065698302     0.834820567    -0.919046490
29      H       4.083969378     0.806575244     0.863829858
30      H       4.018227499     3.332245016     0.911618789
31      H       3.983361781     3.370292024    -0.870857945
32      H       6.240540638     2.236754276    -0.946277316
33      H       6.277176935     2.179524967     0.835657844
```

```
34     H       6.211693274      4.704218838      0.924890804
35     H       6.159275936      4.770864164     -0.856322316
36     H       8.415590537      3.638607353     -0.972004350
37     H       8.469550552      3.553111716      0.808366186
38     H       8.402313575      6.077723512      0.936144551
39     H       8.338233476      6.169533065     -0.843594862
40     H      10.593555463      5.038436717     -0.989815373
41     H      10.659161213      4.928634422      0.788834657
42     H      10.592461112      7.451657402      0.951058316
43     H      10.518018387      7.567056818     -0.826886474
44     H      12.772361398      6.437085258     -1.000802097
45     H      12.848181287      6.305071526      0.775929519
46     H      12.781949922      8.826392023      0.969430057
47     H      12.698839102      8.963069493     -0.806618888
48     H      14.952240979      7.834049867     -1.004460669
49     H      15.036532976      7.682579008      0.770334058
50     H      14.970618870     10.202243287      0.991060598
51     H      14.880837969     10.357294405     -0.783146261
52     H      17.133410158      9.228899961     -1.000448875
53     H      17.224189225      9.061238519      0.772552455
54     H      17.158720368     11.579568475      1.015777070
55     H      17.064045992     11.749572649     -0.756763723
56     H      19.317019961     10.619467735     -0.987316148
57     H      19.412223374     10.444395652      0.784620268
58     H      19.354715485     12.965296865      1.037862152
59     H      19.255078319     13.142607275     -0.732120327
60     H      21.568645295     13.527092882      0.073672307
61     H      21.542900353     12.077070104     -0.973694573
62     H      21.643410444     11.897559022      0.806724363
---
N1, UHF, CHARGE=1, MULT=3
HF=448.3
1      N       0       0        0
---
N1, UHF, CHARGE=0, MULT=4
HF=113
1      N       0       0        0
---
H2N1, UHF, CHARGE=0, MULT=2
HF=45.5
1      H       0.000000000      0.000000000      0.000000000
2      N       1.001841068      0.000000000      0.000000000
3      H       1.252862827      0.958802407      0.000000000
EXPGEOM
1      N       0.00000          0.00000          0.14120
2      H       0.00000          0.80250         -0.49410
3      H       0.00000         -0.80250         -0.49410
---
H3N1, RHF, CHARGE=0, MULT=1
HF=-11, DIP=1.47, IE=10.85
1      N      -0.017192221     -0.006508967     -0.016394674
2      H       0.989510861     -0.016018273      0.017643365
3      H      -0.273556014      0.967374076      0.005083481
4      H      -0.316012991     -0.378812451      0.870655641
EXPGEOM
1      N       0.00000          0.00000          0.11420
2      H       0.00000          0.93750         -0.26640
3      H       0.81190         -0.46870         -0.26640
4      H      -0.81190         -0.46870         -0.26640
```

```
---
H4N1, RHF, CHARGE=1, MULT=1
HF=155
1    H     0.000000000    0.000000000    0.000000000
2    N     1.022819746    0.000000000    0.000000000
3    H     1.365871183    0.970289264    0.000000000
4    H     1.365865333   -0.485144780   -0.840297455
5    H     1.365865333   -0.485145666    0.840296943
EXPGEOM
1    N     0.00000    0.00000    0.00000
2    H     0.58890    0.58890    0.58890
3    H    -0.58890   -0.58890    0.58890
4    H    -0.58890    0.58890   -0.58890
5    H     0.58890   -0.58890   -0.58890
---
C1N1, UHF, CHARGE=0, MULT=2
HF=104
1    C     0.000000000    0.000000000    0.000000000
2    N     1.154575225    0.000000000    0.000000000
EXPGEOM
1    C     0.00000    0.00000   -0.63520
2    N     0.00000    0.00000    0.54450
---
H1C1N1, UHF, CHARGE=0, MULT=1
HF=32.3, DIP=2.98, IE=13.6
1    N     0.000000000    0.000000000    0.000000000
2    C     1.160401739    0.000000000    0.000000000
3    H     2.215523628    0.000036746    0.000000000
EXPGEOM
1    C     0.00000    0.00000   -0.50190
2    H     0.00000    0.00000   -1.56350
3    N     0.00000    0.00000    0.65360
---
H4C1N1, RHF, CHARGE=1, MULT=1
HF=178
1    C     0.000000000    0.000000000    0.000000000
2    N     1.309830279    0.000000000    0.000000000
3    H     1.858625094    0.849711657    0.000000000
4    H    -0.568566005    0.941398782   -0.004810094
5    H    -0.568568178   -0.941401785    0.003651049
6    H     1.858619953   -0.849707070    0.003593453
EXPGEOM
1    C     0.00000    0.00000   -0.67460
2    N     0.00000    0.00000    0.59970
3    H     0.00000    0.94120   -1.20570
4    H     0.00000   -0.94120   -1.20570
5    H     0.00000    0.86440    1.13030
6    H     0.00000   -0.86440    1.13030
---
H4C1N1, UHF, CHARGE=0, MULT=2
HF=37
1    N     0.000000000    0.000000000    0.000000000
2    C     1.436650675    0.000000000    0.000000000
3    H     1.915385345    1.006401249    0.000000000
4    H     1.797334753   -0.549872060   -0.901552259
5    H     1.797311723   -0.549881641    0.901528910
6    H    -0.349514999    0.940040030   -0.003491689
---
H4C1N1, RHF, CHARGE=-1, MULT=1
```

```
HF=30.5
1    N    -0.005367930     0.001031057    -0.006718149
2    C     1.381234471     0.000608911    -0.003546949
3    H     1.910777206     1.018605100     0.015020279
4    H     1.912778270    -0.487461477    -0.896657606
5    H     1.778093247    -0.544869768     0.895797784
6    H    -0.320093858     0.506252295    -0.821128406
---
H5C1N1, RHF, CHARGE=0, MULT=1
HF=-5.5, DIP=1.31, IE=9.6
1    N     0.000000000     0.000000000     0.000000000
2    C     1.459354897     0.000000000     0.000000000
3    H     1.912902049     1.021353928     0.000000000
4    H     1.830308430    -0.528083030    -0.907936370
5    H     1.824577718    -0.543936798     0.888073614
6    H    -0.341613505     0.533057736     0.784507672
7    H    -0.369814772     0.457120220    -0.817071066
EXPGEOM
1    C     0.05080     0.70550     0.00000
2    N     0.05080    -0.75850     0.00000
3    H    -0.94260     1.16460     0.00000
4    H     0.58940     1.06060     0.87730
5    H     0.58940     1.06060    -0.87730
6    H    -0.44830    -1.10450    -0.80890
7    H    -0.44830    -1.10450     0.80890
---
H3C2N1, RHF, CHARGE=0, MULT=1
HF=17.7, DIP=3.92, IE=12.21
1    C    -0.000238548    -0.006136078    -0.001136233
2    C     1.451431031     0.002271522    -0.002943699
3    H     1.836155374     1.043814996    -0.001404755
4    H     1.846319292    -0.516742707     0.895551852
5    H     1.844149631    -0.513225794    -0.904489395
6    N    -1.162402641    -0.012344012    -0.000011110
EXPGEOM
1    C     0.00000     0.00000    -1.17970
2    C     0.00000     0.00000     0.27900
3    N     0.00000     0.00000     1.43610
4    H     0.00000     1.02250    -1.54940
5    H     0.88550    -0.51120    -1.54940
6    H    -0.88550    -0.51120    -1.54940
---
H3C2N1, RHF, CHARGE=0, MULT=1
HF=39.1, DIP=3.85, IE=11.32
1    N     0.000000000     0.000000000     0.000000000
2    C     1.424250612     0.000000000     0.000000000
3    H     1.809792159     1.045740992     0.000000000
4    H     1.809788571    -0.522871257     0.905639582
5    H     1.809778790    -0.522873326    -0.905643167
6    C    -1.190875144     0.000000000    -0.000000000
EXPGEOM
1    C     0.00000     0.00000    -1.11910
2    N     0.00000     0.00000     0.31310
3    C     0.00000     0.00000     1.49480
4    H     0.00000     1.03210    -1.48190
5    H     0.89380    -0.51610    -1.48190
6    H    -0.89380    -0.51610    -1.48190
---
H5C2N1, RHF, CHARGE=0, MULT=1
```

```
HF=30.2, DIP=1.9, IE=9.9
 1    C    0.000000000     0.000000000     0.000000000
 2    C    1.515335936     0.000000000     0.000000000
 3    N    0.757632280     1.270378135     0.000000000
 4    H   -0.575339792    -0.209286729     0.913519127
 5    H   -0.554101111    -0.316867960    -0.896401727
 6    H    2.069470589    -0.317085590    -0.896274327
 7    H    2.090578696    -0.208628605     0.913747348
 8    H    0.757085615     1.741140112    -0.891532220
EXPGEOM
 1    N   -0.03860    0.87990    0.00000
 2    H    0.90560    1.26090    0.00000
 3    C   -0.03860   -0.39880    0.74250
 4    C   -0.03860   -0.39880   -0.74250
 5    H   -0.95410   -0.60900    1.28150
 6    H    0.86800   -0.70810    1.25140
 7    H   -0.95410   -0.60900   -1.28150
 8    H    0.86800   -0.70810   -1.25140
---
H6C2N1, RHF, CHARGE=-1, MULT=1
HF=24.7
 1    N    0.000000000     0.000000000     0.000000000
 2    C    1.400960508     0.000000000     0.000000000
 3    H    1.914441574     1.015406243     0.000000000
 4    H    1.778292435    -0.515818579     0.924091807
 5    H    1.915035816    -0.531526842    -0.864643551
 6    C   -0.655907664     0.599615535    -1.083014346
 7    H   -0.461063567     0.154313797    -2.111817709
 8    H   -1.767716846     0.523660803    -0.939839705
 9    H   -0.456289083     1.706231606    -1.256165171
---
H7C2N1, RHF, CHARGE=0, MULT=1
HF=-4.4, DIP=1.03, IE=8.93
 1    N    0.000000000     0.000000000     0.000000000
 2    C    1.461732040     0.000000000     0.000000000
 3    H    1.917767927     1.020045444     0.000000000
 4    H    1.823809776    -0.528647445     0.911055334
 5    H    1.846454361    -0.544281237    -0.891173520
 6    C   -0.676305418     0.641096552    -1.126171250
 7    H   -0.405307838     0.131936570    -2.077915261
 8    H   -1.777547439     0.551204669    -0.986593374
 9    H   -0.439513497     1.726947454    -1.241693042
10    H   -0.335920194     0.397777810     0.864216162
EXPGEOM
 1    N    0.02800    0.59700    0.00000
 2    H   -0.81380    1.16590    0.00000
 3    C    0.02800   -0.22570    1.20860
 4    C    0.02800   -0.22570   -1.20860
 5    H   -0.78950   -0.96290    1.24930
 6    H   -0.78950   -0.96290   -1.24930
 7    H    0.97200   -0.77300    1.26500
 8    H    0.97200   -0.77300   -1.26500
 9    H   -0.04150    0.41810    2.08670
10    H   -0.04150    0.41810   -2.08670
---
H7C2N1, RHF, CHARGE=0, MULT=1
HF=-11.4, DIP=1.22, IE=9.5
 1    C    0.000000000     0.000000000     0.000000000
 2    C    1.537349350     0.000000000     0.000000000
```

```
 3     N      2.203535611      1.306545865      0.000000000
 4     H      1.899832531     -0.562833348      0.896274552
 5     H      1.899274991     -0.563068957     -0.896340161
 6     H     -0.416121556      0.500996923     -0.897727961
 7     H     -0.376848091     -1.043118454      0.002349065
 8     H     -0.416162207      0.506326334      0.894749123
 9     H      1.917889214      1.845021211     -0.803407929
10     H      1.916800703      1.845646950      0.802597061
EXPGEOM
 1     N     -1.31270    -0.08680     0.00000
 2     C      0.00000     0.57940     0.00000
 3     C      1.21410    -0.35590     0.00000
 4     H      2.15730     0.20900     0.00000
 5     H      1.20610    -1.00290     0.88920
 6     H      1.20610    -1.00290    -0.88920
 7     H      0.03540     1.23690    -0.88060
 8     H      0.03540     1.23690     0.88060
 9     H     -1.36810    -0.70510     0.81420
10     H     -1.36810    -0.70510    -0.81420
---
H3C3N1, RHF, CHARGE=0, MULT=1
HF=44.1, DIP=3.87, IE=10.91
 1     C      0.000000000      0.000000000      0.000000000
 2     C      1.344003040      0.000000000      0.000000000
 3     H      1.904297667      0.941756547      0.000000000
 4     H     -0.577094903      0.923991974      0.000177889
 5     H     -0.611566075     -0.900984889      0.000001556
 6     C      2.158045208     -1.167370845     -0.000151164
 7     N      2.838950134     -2.110101618     -0.000972144
EXPGEOM
 1     C     -0.58720    -0.52910     0.00000
 2     N     -1.08200    -1.59530     0.00000
 3     C      0.00000     0.78960     0.00000
 4     H     -0.69590     1.62790     0.00000
 5     C      1.33450     0.97610     0.00000
 6     H      2.03120     0.13990     0.00000
 7     H      1.75460     1.97970     0.00000
---
H5C3N1, RHF, CHARGE=0, MULT=1
HF=12.1, IE=11.9
 1     C      0.000000000      0.000000000      0.000000000
 2     C      1.533604323      0.000000000      0.000000000
 3     C      2.122854882      1.332111474      0.000000000
 4     H      1.906594864     -0.554795446      0.893851145
 5     H      1.906635858     -0.554896596     -0.893757613
 6     H     -0.407702150      0.504899342     -0.898516066
 7     H     -0.377396908     -1.042839399      0.002993555
 8     H     -0.407919593      0.509856664      0.895564661
 9     N      2.607545311      2.388234921     -0.000499311
EXPGEOM
 1     C      1.51660     0.57270     0.00000
 2     C      0.00000     0.81520     0.00000
 3     C     -0.77820    -0.43250     0.00000
 4     N     -1.36300    -1.43950     0.00000
 5     H      2.05430     1.52640     0.00000
 6     H      1.82610     0.00760     0.88440
 7     H      1.82610     0.00760    -0.88440
 8     H     -0.29790     1.40160     0.87720
 9     H     -0.29790     1.40160    -0.87720
```

```
---
H7C3N1, RHF, CHARGE=0, MULT=1
HF=18.4
1    C     -0.005801926    -0.000806769     0.031880923
2    C      1.518660106     0.149460048     0.098673910
3    C      0.647720492     1.392565984    -0.040038344
4    N     -0.772395549    -0.502464784     1.150008264
5    H     -0.427962492    -0.396517626    -0.910932964
6    H      2.113904778    -0.189487405    -0.757863019
7    H      2.060110009    -0.029014734     1.035190948
8    H      0.627584279     1.933546204    -0.993292927
9    H      0.568664005     2.092236049     0.799947194
10   H     -0.394581955    -0.182098956     2.028542128
11   H     -1.705076390    -0.117716610     1.103990479
EXPGEOM
1    C     -0.31030     0.40710     0.00000
2    H     -1.25930     0.93130     0.00000
3    N      0.88590     1.19500     0.00000
4    C     -0.31030    -0.88580     0.74940
5    C     -0.31030    -0.88580    -0.74940
6    H      0.93400     1.78150     0.81790
7    H      0.93400     1.78150    -0.81790
8    H     -1.21000    -1.17220     1.27060
9    H     -1.21000    -1.17220    -1.27060
10   H      0.59770    -1.16370     1.25920
11   H      0.59770    -1.16370    -1.25920
---
H9C3N1, RHF, CHARGE=0, MULT=1
HF=-20
1    C     -0.029997285    -0.005979087    -0.003614076
2    C      1.516739404     0.017839075    -0.037945845
3    C      2.096953264     1.449738824     0.005379901
4    H     -0.482172631     0.471917761    -0.896329864
5    H     -0.414580134    -1.045397418     0.050979261
6    H     -0.410248244     0.528724036     0.890251317
7    H      1.866881689    -0.503361729     0.895944458
8    N      2.117789197    -0.716957149    -1.166319400
9    H      1.740225604    -0.393576515    -2.043968447
10   H      1.862109899    -1.691187190    -1.111353307
11   H      1.762928880     2.068760937    -0.851911306
12   H      1.776090706     1.964930320     0.934320091
13   H      3.205810827     1.436184038     0.002708638
EXPGEOM
1    C      0.29270     0.21980     0.00000
2    N     -0.90930     1.04710     0.00000
3    H      1.20920     0.82170     0.00000
4    C      0.29270    -0.63760     1.24890
5    C      0.29270    -0.63760    -1.24890
6    H     -0.92640     1.64670    -0.81270
7    H     -0.92640     1.64670     0.81270
8    H     -0.59490    -1.26870     1.26390
9    H     -0.59490    -1.26870    -1.26390
10   H      1.17830    -1.27070     1.27980
11   H      1.17830    -1.27070    -1.27980
12   H      0.28690    -0.01690     2.14530
13   H      0.28690    -0.01690    -2.14530
---
H9C3N1, RHF, CHARGE=0, MULT=1
HF=-16.8
```

```
    1    C      -0.000000491    -0.000001181    -0.000003038
    2    C       1.530384196    -0.000009464     0.000002448
    3    C       2.169553557     1.409701089     0.000002458
    4    H      -0.412224565     0.503504673    -0.897828191
    5    H      -0.382528628    -1.041595035     0.002657367
    6    H      -0.412488555     0.508636773     0.894790985
    7    H       1.881669522    -0.571003534     0.890835676
    8    H       1.881726632    -0.571206058    -0.890663451
    9    H       1.818414024     1.977742159    -0.897224203
   10    H       1.817693092     1.977741108     0.896989290
   11    N       3.635214905     1.459944066     0.000566168
   12    H       4.005030714     0.974079989    -0.801760488
   13    H       4.004304664     0.975112188     0.803918240
EXPGEOM
    1    C       1.44540     1.28590     0.00000
    2    C       0.00000     0.75690     0.00000
    3    C      -0.05470    -0.78240     0.00000
    4    N      -1.40490    -1.38050     0.00000
    5    H       1.47140     2.38970     0.00000
    6    H       1.99350     0.93350     0.89330
    7    H       1.99350     0.93350    -0.89330
    8    H      -0.54110     1.13640     0.88940
    9    H      -0.54110     1.13640    -0.88940
   10    H       0.48260    -1.16960    -0.88690
   11    H       0.48260    -1.16960     0.88690
   12    H      -1.92570    -1.04430    -0.81490
   13    H      -1.92570    -1.04430     0.81490
---
H9C3N1, RHF, CHARGE=0, MULT=1
HF=-5.7, DIP=0.61, IE=8.54
    1    C      -0.020790426    -0.015679613     0.007336321
    2    N       1.435384876     0.054251516    -0.128342842
    3    C       2.032895875     1.384382310     0.005840942
    4    H      -0.571200580     0.430528447    -0.857023317
    5    H      -0.347191487    -1.076109700     0.103944277
    6    H      -0.348825057     0.518707659     0.927689392
    7    H       1.644300351     1.893137146     0.916895985
    8    H       3.137783152     1.299725791     0.117390562
    9    H       1.830769989     2.052544980    -0.866906809
   10    C       2.028234978    -0.812182435    -1.149239610
   11    H       1.820357010    -0.474493912    -2.194161926
   12    H       3.133829510    -0.854920205    -1.022297946
   13    H       1.641181968    -1.851265093    -1.046941288
EXPGEOM
    1    N       0.00000     0.00000     0.38850
    2    C      -1.37050     0.00030    -0.06360
    3    C       0.68550     1.18670    -0.06350
    4    C       0.68500    -1.18700    -0.06350
    5    H      -1.45270     0.00030    -1.16110
    6    H       0.72660     1.25790    -1.16100
    7    H       0.72610    -1.25820    -1.16100
    8    H      -1.88040    -0.88220     0.31780
    9    H      -1.88000     0.88290     0.31780
   10    H       0.17630     2.06950     0.31780
   11    H       1.70460     1.18670     0.31780
   12    H       1.70420    -1.18730     0.31780
   13    H       0.17550    -2.06960     0.31780
---
H5C4N1, RHF, CHARGE=0, MULT=1
```

```
HF=33.6
1    C     0.000052650     0.000226559    -0.000166996
2    C     1.497904849    -0.000109326     0.000012500
3    C     3.700475649     1.079088678     0.000204365
4    N     4.863468275     1.080752111    -0.000767758
5    C     2.277027101     1.102926934     0.001267270
6    H    -0.402819799     0.511185274    -0.899504821
7    H    -0.404182232     0.517199509     0.895017519
8    H     1.952304748    -0.996635692    -0.001878965
9    H     1.848438744     2.111261527     0.003471815
10   H    -0.396989267    -1.035338100     0.002621970
EXPGEOM
1    H     2.64920     0.46030     0.00000
2    H     2.65860    -1.08400     0.88270
3    H     2.65860    -1.08400    -0.88270
4    C     2.26350    -0.56490     0.00000
5    H     0.27970    -1.56310     0.00000
6    C     0.76470    -0.58600     0.00000
7    H     0.44030     1.51120     0.00000
8    C     0.00000     0.51540     0.00000
9    N    -2.60350     0.41100     0.00000
10   C    -1.43850     0.44930     0.00000
---
H5C4N1, RHF, CHARGE=0, MULT=1
HF=32
1    C    -0.000015748    -0.000198650     0.000133279
2    C     1.496965182     0.000019823    -0.000111758
3    C     1.947272510     2.435299775    -0.000035224
4    N     1.661973732     3.562769846    -0.000034814
5    C     2.327529524     1.064273164     0.000074112
6    H    -0.406240555     0.512712138    -0.896771033
7    H    -0.405151035     0.508737986     0.899684586
8    H     1.948184542    -0.999163102     0.000169657
9    H     3.412675576     0.906730716     0.000257199
10   H    -0.393915061    -1.037075923    -0.002036017
EXPGEOM
1    C    -1.22120     0.22940     0.00000
2    N    -2.23470    -0.34640     0.00000
3    C     0.00000     0.99480     0.00000
4    H    -0.11020     2.07660     0.00000
5    C     1.21380     0.42110     0.00000
6    H     2.07740     1.08650     0.00000
7    C     1.49560    -1.05190     0.00000
8    H     0.57410    -1.64170     0.00000
9    H     2.08640    -1.32840     0.88270
10   H     2.08640    -1.32840    -0.88270
---
H5C4N1, RHF, CHARGE=0, MULT=1
HF=37.7
1    C     0.000044187    -0.000230952     0.000116141
2    C     1.341443448    -0.000146952    -0.000034812
3    C     3.636223046     0.952712252     0.000130435
4    N     4.780007669     0.748561615    -0.003718107
5    C     2.205624837     1.234577063     0.005156859
6    H    -0.614843088     0.898528285     0.003560555
7    H    -0.579863600    -0.922612173    -0.003696637
8    H     1.873875336    -0.956095177    -0.003557281
9    H     1.970574617     1.867086128    -0.885745450
10   H     1.974726219     1.856930647     0.904163945
```

```
---
H5C4N1, RHF, CHARGE=0, MULT=1
HF=25.9, DIP=1.74, IE=8.21
 1    C     0.000000000     0.000000000     0.000000000
 2    C     1.394543426     0.000000000     0.000000000
 3    N     1.812994438     1.333716785     0.000000000
 4    C     0.698656722     2.177678705     0.000053167
 5    C    -0.437476923     1.368983595     0.000024072
 6    H    -0.641808512    -0.868488962    -0.000020489
 7    H     2.106505549    -0.814579485    -0.000092602
 8    H     2.758710440     1.636104062     0.000020753
 9    H     0.806342313     3.254180543     0.000101513
10    H    -1.464107413     1.704086036     0.000062245
EXPGEOM
 1    N     0.00000     0.00000     1.12390
 2    H     0.00000     0.00000     2.13010
 3    C     0.00000     1.12590     0.33120
 4    C     0.00000    -1.12590     0.33120
 5    C     0.00000     0.71410    -0.98420
 6    C     0.00000    -0.71410    -0.98420
 7    H     0.00000     2.11210     0.76620
 8    H     0.00000    -2.11210     0.76620
 9    H     0.00000     1.36150    -1.84700
10    H     0.00000    -1.36150    -1.84700
---
H7C4N1, RHF, CHARGE=0, MULT=1
HF=7.5
 1    C    -0.000079921     0.000057475     0.000013334
 2    C     1.530791717    -0.000016331     0.000019173
 3    C     2.170123940     1.404320356     0.000009723
 4    C     3.627407561     1.370137457     0.001087171
 5    N     4.789370959     1.360733613     0.002264357
 6    H    -0.412072786     0.506469317     0.896200509
 7    H    -0.412121650     0.505798112    -0.896457517
 8    H    -0.380032199    -1.042096937     0.000415973
 9    H     1.887762680    -0.565223005     0.891504868
10    H     1.887933149    -0.564950135    -0.891583901
11    H     1.832178156     1.980460106     0.894195077
12    H     1.833233337     1.979532163    -0.895022417
EXPGEOM
 1    N    -2.60520     0.34650     0.00000
 2    C    -1.45410     0.48270     0.00000
 3    C     0.00000     0.62930     0.00000
 4    C     0.72210    -0.71820     0.00000
 5    C     2.23110    -0.53140     0.00000
 6    H     0.28450     1.20930     0.87680
 7    H     0.28450     1.20930    -0.87680
 8    H     0.41090    -1.28640     0.87450
 9    H     0.41090    -1.28640    -0.87450
10    H     2.74280    -1.49060     0.00000
11    H     2.55420     0.02240     0.88060
12    H     2.55420     0.02240    -0.88060
---
H7C4N1, RHF, CHARGE=0, MULT=1
HF=5.6
 1    C    -0.000001132    -0.000184289     0.000053686
 2    C     1.544803257     0.000142942    -0.000037569
 3    C     2.153792360     1.419795157    -0.000106750
 4    C     2.074052099    -0.800324903     1.104415879
```

```
5    N     2.501320122   -1.444477809    1.971806116
6    H    -0.393879528   -1.035298483   -0.049785778
7    H    -0.415761449    0.478601574    0.908800219
8    H    -0.389851896    0.546546327   -0.882296074
9    H     1.874573090   -0.500560105   -0.949149112
10   H     1.877625364    1.990661526    0.908707767
11   H     3.260448723    1.373850681   -0.050068761
12   H     1.805017784    1.993604906   -0.882280394
EXPGEOM
1    N     0.42560   -2.20130    0.00000
2    C     0.02440   -1.09320    0.00000
3    C    -0.45280    0.31660    0.00000
4    C     0.02440    1.03790    1.27930
5    C     0.02440    1.03790   -1.27930
6    H    -1.55800    0.27400    0.00000
7    H    -0.36610    2.07040    1.28570
8    H    -0.33220    0.52100    2.18570
9    H     1.12700    1.07840    1.30920
10   H    -0.36610    2.07040   -1.28570
11   H    -0.33220    0.52100   -2.18570
12   H     1.12700    1.07840   -1.30920
---
H9C4N1, RHF, CHARGE=0, MULT=1
HF=-0.8
1    C     0.000000000    0.000000000    0.000000000
2    C     1.540972205    0.000000000    0.000000000
3    C     1.983854132    1.479966383    0.000000000
4    C    -0.442967996    1.479918772    0.000297479
5    N     0.770402353    2.303063299   -0.143845645
6    H    -0.399474959   -0.534508711   -0.887496250
7    H    -0.401045659   -0.533792561    0.887679021
8    H     1.940604458   -0.534488552   -0.887445794
9    H     1.942053327   -0.533676410    0.887718853
10   H     2.662724926    1.704605143   -0.856901254
11   H     2.557666315    1.726261810    0.926183958
12   H    -1.122242890    1.704551913   -0.856285317
13   H    -1.016438621    1.726103187    0.926735944
14   H     0.770448370    3.058732729    0.522024336
EXPGEOM
1    N     0.55840   -1.09070    0.00000
2    H     0.41720   -2.10320    0.00000
3    C    -0.10640   -0.47320    1.15440
4    C    -0.10640   -0.47320   -1.15440
5    C    -0.10640    1.02540    0.77920
6    C    -0.10640    1.02540   -0.77920
7    H    -1.15560   -0.82660    1.28420
8    H    -1.15560   -0.82660   -1.28420
9    H     0.44140   -0.68550    2.08870
10   H     0.44140   -0.68550   -2.08870
11   H     0.80390    1.51460    1.16550
12   H     0.80390    1.51460   -1.16550
13   H    -0.97630    1.55340    1.20550
14   H    -0.97630    1.55340   -1.20550
---
H11C4N1, RHF, CHARGE=0, MULT=1
HF=-25.4
1    N     0.000064914    0.000038042   -0.000091821
2    C     1.474162069   -0.000040180    0.000083474
3    C     2.108103608    1.413416685    0.000007046
```

```
 4    C    2.006373407   -0.880505092    1.167290882
 5    C    1.636940780   -2.365742697    1.115971341
 6    H   -0.357623840    0.576678862    0.746362314
 7    H   -0.337973106    0.426357929   -0.849487052
 8    H    1.790620969   -0.487219734   -0.962836856
 9    H    1.890783295    1.977345172    0.929681727
10    H    3.210226502    1.349426644   -0.103044915
11    H    1.738482514    2.019341331   -0.853107153
12    H    3.119908017   -0.806712604    1.171453544
13    H    1.671608307   -0.462897234    2.144876807
14    H    0.550740328   -2.536271500    1.250525137
15    H    1.940559854   -2.834351952    0.157896003
16    H    2.160103187   -2.909569719    1.930992704
EXPGEOM
 1    N    0.55860    1.37750   -0.23930
 2    H   -0.21750    1.93980    0.10900
 3    H    1.40380    1.83800    0.10100
 4    C    1.77810   -0.71360   -0.01140
 5    H    1.90910   -0.74540   -1.09930
 6    H    2.64970   -0.20700    0.42340
 7    H    1.75840   -1.74000    0.37380
 8    C    0.48640    0.02290    0.33180
 9    H    0.38160    0.03970    1.43460
10    C   -0.73060   -0.70780   -0.23950
11    H   -0.60540   -0.77600   -1.32870
12    H   -0.73100   -1.73520    0.15220
13    C   -2.06860   -0.04120    0.08630
14    H   -2.90670   -0.65520   -0.26380
15    H   -2.15690    0.93740   -0.39960
16    H   -2.18760    0.09930    1.16900
---
H11C4N1, RHF, CHARGE=0, MULT=1
HF=-23.6
 1    C   -0.041069116   -0.002073602   -0.029009596
 2    C   -1.082323934    1.132399866    0.054834354
 3    C   -0.711029905   -1.392255027    0.015501292
 4    C    0.906786862    0.120836418   -1.258129841
 5    N    1.781233314    1.296936253   -1.172606770
 6    H    0.586683752    0.068075894    0.896683154
 7    H   -0.601930718    2.129898231    0.099465694
 8    H   -1.772829311    1.129597199   -0.813153532
 9    H   -1.697161849    1.030889552    0.973891037
10    H   -1.337970152   -1.586198280   -0.878101321
11    H    0.045708981   -2.201034319    0.077591750
12    H   -1.361426901   -1.490752800    0.908698559
13    H    1.502606087   -0.825364712   -1.358148291
14    H    0.306800488    0.202598462   -2.198067334
15    H    2.495265141    1.148362047   -0.475707008
16    H    2.280274917    1.397100852   -2.043787119
EXPGEOM
 1    N   -2.00540   -0.02410   -0.19170
 2    H   -2.80400   -0.60970    0.05180
 3    H   -2.12790    0.83790    0.33850
 4    C   -0.76700   -0.69150    0.23410
 5    H   -0.65520   -0.75720    1.33340
 6    H   -0.80270   -1.72320   -0.14070
 7    C    1.73530   -0.78070    0.02760
 8    H    1.87990   -0.79080    1.11650
 9    H    1.67920   -1.82170   -0.31510
```

```
10      H       2.62520    -0.32470    -0.42310
11      C       0.57290     1.45560     0.10840
12      H       1.47970     1.92590    -0.29140
13      H      -0.28420     2.04420    -0.23790
14      H       0.61950     1.51890     1.20500
15      C       0.47070    -0.00290    -0.34590
16      H       0.35650    -0.01370    -1.43940
---
H11C4N1, RHF, CHARGE=0, MULT=1
HF=-22.7
1     N     -0.000007031    -0.000102114    -0.000176486
2     C      1.468253074     0.000031827     0.000084357
3     C      2.030320697     1.439693672    -0.000009850
4     C      3.567988307     1.500184908     0.088195125
5     C      4.151836477     2.915718130     0.084474709
6     H     -0.347224631     0.307841331    -0.895716248
7     H     -0.327071727    -0.949413423     0.097061069
8     H      1.807592565    -0.540534111     0.918283119
9     H      1.889760006    -0.569480405    -0.869640000
10    H      1.700904628     1.965282905    -0.926187730
11    H      1.593620287     2.004998701     0.854593237
12    H      3.908277303     0.982124472     1.014650179
13    H      4.012815715     0.934602171    -0.763270499
14    H      3.895429486     3.465527027    -0.843772258
15    H      3.789589600     3.514104867     0.944923728
16    H      5.258651658     2.872749641     0.150773644
EXPGEOM
1     N      2.54360     0.44020     0.00000
2     C      1.34820    -0.41260     0.00000
3     C      0.00000     0.32440     0.00000
4     C     -1.21100    -0.61800     0.00000
5     C     -2.55080     0.12960     0.00000
6     H      2.48660     1.06760     0.80820
7     H      2.48660     1.06760    -0.80820
8     H      1.40230    -1.07620     0.88490
9     H      1.40230    -1.07620    -0.88490
10    H     -0.04900     0.98850    -0.88790
11    H     -0.04900     0.98850     0.88790
12    H     -1.15650    -1.28110     0.88610
13    H     -1.15650    -1.28110    -0.88610
14    H     -3.40450    -0.57120     0.00000
15    H     -2.64250     0.77570     0.89260
16    H     -2.64250     0.77570    -0.89260
---
H11C4N1, RHF, CHARGE=0, MULT=1
HF=-28.9
1     C     -0.025845324    -0.008468918    -0.005553385
2     C      1.533417760    -0.000043125     0.010858624
3     C      2.057656450     1.467704934    -0.001243994
4     H      3.165109868     1.510835209    -0.026722283
5     C      2.056559798    -0.733450498     1.282797932
6     N      2.114473049    -0.688166694    -1.167524100
7     H      1.723871093     2.014184372     0.904910335
8     H      1.683348407     2.034831568    -0.878287388
9     H     -0.433766167    -1.040170203    -0.008654250
10    H     -0.433023626     0.506424379    -0.899924012
11    H     -0.440567885     0.504857059     0.885083921
12    H      1.728234672    -0.210799715     2.204686782
13    H      3.163867187    -0.781796261     1.305700899
```

```
14    H     1.677633370   -1.774394714    1.346417650
15    H     1.779698677   -0.264215732   -2.019953015
16    H     1.788722440   -1.642305560   -1.210851256
EXPGEOM
1     C    -0.01170    0.01370    0.00000
2     N    -0.54360    1.38410    0.00000
3     C     1.52320   -0.02160    0.00000
4     C    -0.54360   -0.68410    1.25120
5     C    -0.54360   -0.68410   -1.25120
6     H    -0.20980    1.88820   -0.81270
7     H    -0.20980    1.88820    0.81270
8     H     1.90390   -1.04410    0.00000
9     H     1.91670    0.48460   -0.88250
10    H     1.91670    0.48460    0.88250
11    H    -0.18910   -0.18130    2.15260
12    H    -0.18910   -0.18130   -2.15260
13    H    -1.63190   -0.66560    1.25820
14    H    -0.20800   -1.71990    1.29250
15    H    -1.63190   -0.66560   -1.25820
16    H    -0.20800   -1.71990   -1.29250
---
H5C5N1, RHF, CHARGE=0, MULT=1
HF=34.6, DIP=2.22, IE=9.67
1     N     0.000293704   -0.000067690    0.000000000
2     C     1.353751814   -0.000420343    0.000000000
3     C     2.118744680    1.185775468    0.000000000
4     C     1.450547911    2.421382446    0.000000000
5     C     0.045716817    2.426895339    0.000000000
6     C    -0.639083481    1.192822288    0.000000000
7     H     2.010243235    3.356568237    0.000000000
8     H    -0.509454457    3.363636767    0.000000000
9     H    -1.733196401    1.149184398    0.000000000
10    H     3.206618788    1.138349622    0.000000000
11    H     1.831899116   -0.985446610    0.000000000
EXPGEOM
1     N     0.00000    0.00000    1.42330
2     C     0.00000    0.00000   -1.38660
3     C     0.00000    1.14190    0.72090
4     C     0.00000   -1.14190    0.72090
5     C     0.00000    1.19610   -0.67290
6     C     0.00000   -1.19610   -0.67290
7     H     0.00000    0.00000   -2.47510
8     H     0.00000    2.05870    1.30970
9     H     0.00000   -2.05870    1.30970
10    H     0.00000    2.15700   -1.18240
11    H     0.00000   -2.15700   -1.18240
---
H7C5N1, RHF, CHARGE=0, MULT=1
HF=24.6
1     C     0.000051034    0.000091719    0.000005089
2     C     1.390846928   -0.000217932   -0.000000823
3     C     1.828110541    1.369077425   -0.000005986
4     C     0.693863313    2.174116323   -0.000010441
5     N    -0.438903652    1.338175176    0.000286900
6     C    -1.831333353    1.768829900   -0.001303485
7     H    -0.700825503   -0.824815130   -0.000050037
8     H     2.033153726   -0.868250353    0.000021097
9     H     2.854813091    1.704661580    0.000029100
10    H     0.602279990    3.252528577   -0.000055690
```

```
11      H     -2.357168497    1.392279723   -0.909048495
12      H     -2.362948492    1.379851608    0.897754029
13      H     -1.902480577    2.878562918    0.006139782
EXPGEOM
1      C      2.07390     -0.00000      0.02770
2      H      2.46220     -0.88510     -0.47980
3      H      2.46210      0.88550     -0.47910
4      H      2.43190     -0.00040      1.06230
5      C     -1.49170      0.71330      0.01640
6      H     -2.35590      1.36240      0.02380
7      C     -1.49170     -0.71340      0.01640
8      H     -2.35590     -1.36240      0.02380
9      C     -0.17460     -1.11850     -0.01470
10      H      0.26090     -2.10780     -0.02530
11      C     -0.17460      1.11850     -0.01470
12      H      0.26090      2.10780     -0.02540
13      N      0.62660      0.00000     -0.04090
---
H9C5N1, RHF, CHARGE=0, MULT=1
HF=7.1
1      N      0.000019952    0.000035415   -0.000001072
2      C      1.470796253   -0.000226369    0.000009557
3      C      2.135633621    1.398751640   -0.000038376
4      C      1.334559476    2.438709399   -0.732089987
5      C      0.073007119    2.228660764   -1.150215407
6      C     -0.654125149    0.921982834   -0.943561499
7      H     -0.331646092    0.192852132    0.934500581
8      H      1.810948502   -0.571526074   -0.899203109
9      H      1.821736054   -0.577077855    0.889832366
10      H      3.147592593    1.317543982   -0.458535259
11      H      2.300529303    1.742871170    1.047904132
12      H      1.838377833    3.392519795   -0.901291599
13      H     -0.487788849    3.000353323   -1.680583056
14      H     -0.749148941    0.397072908   -1.928291446
15      H     -1.695614217    1.122120310   -0.589967234
EXPGEOM
1      H     -1.43780      2.08380      0.13510
2      C     -0.79270      1.21040      0.05700
3      H      0.97370      2.34840     -0.09450
4      C      0.53330      1.35450     -0.05190
5      H      2.34220      0.35300      0.53120
6      H      1.87030      0.06540     -1.13210
7      C      1.47230      0.17390     -0.11320
8      H      0.62850     -1.11740      1.40220
9      H      1.34470     -1.99600      0.05480
10      C      0.74970     -1.11430      0.31280
11      H     -2.32540     -0.15960     -0.57690
12      H     -1.85610     -0.33280      1.10130
13      C     -1.45860     -0.14790      0.09310
14      H     -0.49260     -1.28850     -1.28720
15      N     -0.58170     -1.25940     -0.27440
---
H9C5N1, RHF, CHARGE=0, MULT=1
HF=0.6
1      C     -0.000040292    0.000026887    0.000010735
2      C      1.531184427   -0.000043255   -0.000040274
3      C      2.216720531    1.396539520    0.000040756
4      C      2.066619551    2.187398834    1.319213226
5      C      3.632566057    1.262909822   -0.352958964
```

```
 6      N     4.753411558     1.167719645    -0.643926601
 7      H    -0.419928101     0.397228890     0.945621283
 8      H    -0.415004647     0.598717565    -0.835807840
 9      H    -0.374707130    -1.038202333    -0.115841245
10      H     1.890413989    -0.590213997     0.873728015
11      H     1.868824503    -0.556116575    -0.905943341
12      H     1.745064056     2.012144533    -0.810696203
13      H     2.472698394     1.630367119     2.186660606
14      H     2.602107822     3.156604346     1.257792158
15      H     1.002412984     2.415369909     1.527500587
EXPGEOM
 1      C    -1.52620    -0.28920     0.12050
 2      N    -2.59910    -0.69050    -0.09260
 3      C    -0.02670     1.64670    -0.15090
 4      H    -0.14080     1.67950    -1.23760
 5      H    -0.79070     2.28850     0.29320
 6      H     0.95310     2.05360     0.10780
 7      C     2.31130    -0.42670     0.07530
 8      H     2.46180    -0.39080     1.15890
 9      H     2.98440    -1.18710    -0.32920
10      H     2.61540     0.53720    -0.34050
11      C     0.86070    -0.76810    -0.26870
12      H     0.72090    -0.76620    -1.35500
13      H     0.63300    -1.78190     0.07520
14      C    -0.15800     0.20380     0.36160
15      H    -0.00970     0.20210     1.44830
---
H9C5N1, RHF, CHARGE=0, MULT=1
HF=2.7
 1      C    -0.000004367    -0.000015539     0.000012022
 2      C     1.531329133    -0.000030336    -0.000007669
 3      C     2.172109849     1.400961799    -0.000004714
 4      C     3.716384482     1.381229074    -0.000370236
 5      C     4.304818754     2.714955414     0.000243075
 6      N     4.791231597     3.770255566     0.000971793
 7      H    -0.412387227     0.505901227     0.896318535
 8      H    -0.412332378     0.505624030    -0.896357437
 9      H    -0.380947511    -1.042067689     0.000388703
10      H     1.884180793    -0.569099902     0.890932625
11      H     1.883943058    -0.569333687    -0.891059931
12      H     1.819218704     1.966992786    -0.892162738
13      H     1.819538330     1.966863299     0.892354062
14      H     4.094882372     0.831728632     0.894060214
15      H     4.094413424     0.832194685    -0.895218908
EXPGEOM
 1      C     0.00000     0.40540     0.00000
 2      C     1.42070     0.97490     0.00000
 3      C     1.43900     2.50480     0.00000
 4      C    -0.00320    -1.13260     0.00000
 5      N    -2.45100    -2.11020     0.00000
 6      C    -1.36540    -1.69060     0.00000
 7      H     2.46520     2.89010     0.00000
 8      H    -0.54750     0.76750     0.88050
 9      H    -0.54750     0.76750    -0.88050
10      H     1.96240     0.59920     0.88020
11      H     1.96240     0.59920    -0.88020
12      H     0.93000     2.90310    -0.88630
13      H     0.93000     2.90310     0.88630
14      H     0.52760    -1.51500    -0.88150
```

```
15      H        0.52760       -1.51500        0.88150
---
H9C5N1, RHF, CHARGE=0, MULT=1
HF=-0.8
1    C     -0.000110989    -0.000089156     0.000226119
2    C      1.557728333     0.000008547    -0.000062944
3    C     -0.540785803     1.460946386    -0.000153289
4    C     -0.540659583    -0.776974528    -1.237239844
5    C     -0.469635779    -0.674538799     1.219765141
6    N     -0.840577936    -1.207286058     2.183255994
7    H      1.962694907     0.533876097     0.883166848
8    H      1.962610308    -1.031809185     0.020224933
9    H      1.957778862     0.498915931    -0.905527684
10     H     -1.649049932     1.482548005     0.018651137
11     H     -0.181942450     2.025756179     0.883746915
12     H     -0.210416444     2.009731364    -0.904832514
13     H     -0.181809921    -1.825877863    -1.244864184
14     H     -1.648921737    -0.804089090    -1.245699570
15     H     -0.209930811    -0.303079463    -2.183229067
EXPGEOM
1    C      0.00000     0.00000    -0.28240
2    C      0.00000     0.00000     1.19820
3    C      0.00000     1.46380    -0.77320
4    C      1.26770    -0.73190    -0.77320
5    C     -1.26770    -0.73190    -0.77320
6    N      0.00000     0.00000     2.36250
7    H      0.00000     1.47760    -1.86890
8    H      1.27960    -0.73880    -1.86890
9    H     -1.27960    -0.73880    -1.86890
10     H     -0.88810     1.99790    -0.41790
11     H      0.88810     1.99790    -0.41790
12     H      2.17430    -0.22990    -0.41790
13     H      1.28620    -1.76810    -0.41790
14     H     -1.28620    -1.76810    -0.41790
15     H     -2.17430    -0.22990    -0.41790
---
H11C5N1, RHF, CHARGE=0, MULT=1
HF=-13.1
1    N     -0.020478688    -0.079592514    -0.007530258
2    C      1.248140823     0.496471673     0.458995853
3    C      2.333960295    -0.595091546     0.688337004
4    C      3.077452773    -0.239798050     1.991484657
5    C      2.303145343     0.896116252     2.687313954
6    C      1.084309841     1.249351799     1.811624117
7    H     -0.680723731     0.664042427    -0.181334505
8    H      0.121454422    -0.512788380    -0.908226414
9    H      1.652326953     1.229075588    -0.294111495
10     H      3.037030371    -0.625971688    -0.171753953
11     H      1.897704825    -1.613774318     0.766466806
12     H      4.121492615     0.073892969     1.774872469
13     H      3.155918044    -1.128447559     2.653285941
14     H      2.955057045     1.784982081     2.829383834
15     H      1.979382262     0.588659616     3.704431655
16     H      1.027126610     2.345467385     1.637583721
17     H      0.146100629     0.971973710     2.337951877
EXPGEOM
1    N      0.15520     2.14710     0.00000
2    C      0.45640     0.71740     0.00000
3    H      1.54060     0.50020     0.00000
```

```
 4      H      0.57100     2.59130     0.81490
 5      H      0.57100     2.59130    -0.81490
 6      C     -0.17450    -1.49500     0.77890
 7      C     -0.17450    -1.49500    -0.77890
 8      C     -0.17450    -0.00870    -1.19160
 9      C     -0.17450    -0.00870     1.19160
10      H     -1.03540    -2.02690    -1.19220
11      H     -1.03540    -2.02690     1.19220
12      H      0.71880    -1.99970     1.15910
13      H      0.71880    -1.99970    -1.15910
14      H     -1.19670     0.36800    -1.30540
15      H     -1.19670     0.36800     1.30540
16      H      0.35300     0.17220    -2.13410
17      H      0.35300     0.17220     2.13410
---
H11C5N1, RHF, CHARGE=0, MULT=1
HF=-11.3
 1    C    -0.000113666      0.000114732      0.000260029
 2    C     1.545386024     -0.000184604      0.000162190
 3    C     2.167466814      1.405935702     -0.000163980
 4    C     1.477781964      2.387553508     -0.961923693
 5    C    -0.065559762      2.313816472     -0.929712689
 6    N    -0.625606441      0.955264639     -0.924966712
 7    H    -0.375417794      0.224044584      1.029773145
 8    H    -0.367905048     -1.024990984     -0.250031812
 9    H     1.919176154     -0.571998043     -0.880169801
10    H     1.902169546     -0.557770012      0.895130123
11    H     3.245165091      1.327891716     -0.269712604
12    H     2.143985741      1.821668533      1.033283834
13    H     1.838961363      2.205643099     -2.000270707
14    H     1.790897776      3.425961859     -0.711441035
15    H    -0.478410083      2.862083475     -1.811687567
16    H    -0.448515297      2.846166864     -0.023659921
17    H    -0.601195647      0.579332390     -1.862260637
EXPGEOM
 1      C     -0.63770     1.32780     0.00000
 2      H     -0.56660     2.42390     0.00000
 3      H     -1.71010     1.07880     0.00000
 4      C      0.01340     0.74830    -1.26180
 5      C      0.01340     0.74830     1.26180
 6      C      0.01340    -0.77960     1.21190
 7      C      0.01340    -0.77960    -1.21190
 8      N      0.70040    -1.22490     0.00000
 9      H      0.77860    -2.24010     0.00000
10      H     -0.51480     1.09280    -2.16160
11      H     -0.51480     1.09280     2.16160
12      H      1.05340     1.09460    -1.33370
13      H      1.05340     1.09460     1.33370
14      H     -1.03420    -1.14070     1.26690
15      H     -1.03420    -1.14070    -1.26690
16      H      0.54570    -1.18640     2.08140
17      H      0.54570    -1.18640    -2.08140
---
H13C5N1, RHF, CHARGE=0, MULT=1
HF=-25.9
 1    C     0.000170423     -0.000003817     -0.000007942
 2    C     1.531389361     -0.000072741     -0.000042176
 3    C     2.173638885      1.401154095      0.000015528
 4    C     3.720417667      1.370168131     -0.004277602
```

```
 5      N      4.273861248     2.729905502     0.070730999
 6      C      5.716582631     2.847806752     0.281663000
 7      H     -0.412821566     0.506730991     0.895588817
 8      H     -0.412623005     0.504957239    -0.896891959
 9      H     -0.382131694    -1.041784170     0.000769950
10      H      1.882319672    -0.570625038     0.890839654
11      H      1.882722704    -0.571070885    -0.890748120
12      H      1.810611969     1.963480827    -0.891136090
13      H      1.819254301     1.964064910     0.893396983
14      H      4.083968617     0.788362084     0.878273754
15      H      4.081964383     0.820854991    -0.913219072
16      H      4.020755254     3.247340672    -0.758222905
17      H      6.000242576     2.406044333     1.262958921
18      H      6.336061716     2.354925767    -0.506625920
19      H      5.991567301     3.927114188     0.297392643
---
H7C6N1, RHF, CHARGE=0, MULT=1
HF=37.4
 1      C      0.000063821    -0.000334201    -0.000068720
 2      C      1.550230592     0.000148157     0.000065314
 3      C      1.938502331     1.460396155    -0.000025512
 4      C      0.855761843     2.281068475    -0.000609786
 5      C     -0.443166242     1.486132798    -0.001039608
 6      C      0.863913806     3.699037843    -0.000825783
 7      N      0.867363540     4.861765300    -0.001035427
 8      H     -0.397389720    -0.532344226     0.888975211
 9      H     -0.397347526    -0.533461256    -0.888331595
10      H      1.961161301    -0.521052742    -0.890602424
11      H      1.961012711    -0.520907017     0.890866312
12      H      2.981995608     1.757866395     0.000420986
13      H     -1.061377895     1.725601839     0.889788559
14      H     -1.060054085     1.724782682    -0.892968207
---
H7C6N1, RHF, CHARGE=0, MULT=1
HF=33.9
 1      C      0.000018830    -0.000163535    -0.000047836
 2      C      1.551179707     0.000350596     0.000028519
 3      C      1.939087887     1.460226386     0.000096973
 4      C      0.865694743     2.276750608     0.000425036
 5      C     -0.445705003     1.501954548     0.021036089
 6      C     -1.342109185     1.870935888    -1.067707978
 7      N     -2.064294828     2.174030512    -1.925801846
 8      H     -0.399110482    -0.527028535     0.891095066
 9      H     -0.393255164    -0.539897837    -0.885033679
10      H      1.958627362    -0.524053764    -0.890080558
11      H      1.958147927    -0.523559530     0.891090102
12      H      2.980873212     1.760600458     0.001362913
13      H      0.870324011     3.360642629     0.004167664
14      H     -0.980603653     1.715388943     0.980716923
---
H7C6N1, RHF, CHARGE=0, MULT=1
HF=23.7
 1      N     -0.000495989    -0.000564879    -0.000068396
 2      C      1.359154870     0.000352346     0.000175249
 3      C      2.093224773     1.213880869     0.000085880
 4      C      1.406381795     2.438773625    -0.000170638
 5      C      0.003167302     2.426950334    -0.000257899
 6      C     -0.657293701     1.180128604    -0.000621059
 7      C      2.032678391    -1.350004479     0.001216565
```

```
   8     H      3.183063139     1.202092984     0.000072229
   9     H      1.954817149     3.380841965    -0.000130439
  10     H     -0.566834933     3.354530724    -0.000328062
  11     H     -1.750955742     1.115751757    -0.000662963
  12     H      2.678951517    -1.459551321    -0.895430743
  13     H      2.674761386    -1.460076340     0.900764306
  14     H      1.313407975    -2.193046738    -0.000755859
EXPGEOM
   1     N      1.19010    0.25340    0.00000
   2     C      0.00000    0.87800    0.00000
   3     C     -1.20980    0.16940    0.00000
   4     C     -1.18120   -1.22560    0.00000
   5     C      0.05380   -1.87330    0.00000
   6     C      1.20700   -1.08360    0.00000
   7     C      0.03620    2.38640    0.00000
   8     H     -2.15460    0.70380    0.00000
   9     H     -2.10660   -1.79440    0.00000
  10     H      0.12670   -2.95570    0.00000
  11     H      2.19070   -1.55030    0.00000
  12     H     -0.96910    2.81460    0.00000
  13     H      0.57310    2.75070    0.88040
  14     H      0.57310    2.75070   -0.88040
---
H7C6N1, RHF, CHARGE=0, MULT=1
HF=24.8
   1     N      0.000036988    -0.000286600     0.000170014
   2     C      1.350707606     0.000119774     0.000201850
   3     C      2.137863072     1.183735026     0.000038913
   4     C      1.446407557     2.414500336     0.000002373
   5     C      0.041819495     2.419771185    -0.000103125
   6     C     -0.645167992     1.189511873    -0.000058452
   7     C      3.638556337     1.106084468    -0.000770865
   8     H      1.986457638     3.362577759    -0.000056625
   9     H     -0.510444316     3.358635443    -0.000298376
  10     H     -1.739103816     1.143082692    -0.000081655
  11     H      4.009070590     0.566920178     0.896650119
  12     H      4.007423675     0.566934148    -0.898764269
  13     H      4.112101742     2.108767952    -0.001028519
  14     H      1.815414406    -0.992776226     0.000184913
---
H7C6N1, RHF, CHARGE=0, MULT=1
HF=24.8
   1     N     -0.000035421    -0.000018470     0.000027382
   2     C      1.352077425     0.000001743    -0.000040657
   3     C      2.120915527     1.183179221    -0.000006712
   4     C      1.474295089     2.438423881     0.000012994
   5     C      0.060483210     2.427039202     0.000128749
   6     C     -0.629553364     1.197736231     0.000127782
   7     C      2.240652021     3.733426716     0.000475877
   8     H      3.208288695     1.108374832     0.000013971
   9     H     -0.508509903     3.356766434     0.000255151
  10     H     -1.724178894     1.162731075     0.000265090
  11     H      1.984913359     4.343031765    -0.891527840
  12     H      2.001345453     4.332029666     0.904414218
  13     H      3.337906295     3.575197645    -0.010697716
  14     H      1.830674592    -0.985149302    -0.000098559
---
H7C6N1, RHF, CHARGE=0, MULT=1
HF=20.8, DIP=1.53, IE=7.7
```

```
 1      C      0.000069298     0.000052190    -0.000110214
 2      C      2.830162060    -0.000089661     0.000209385
 3      C      0.706818144     1.214975133     0.000068894
 4      C      0.706951652    -1.214824741     0.000286876
 5      C      2.111001789     1.223276912     0.001200583
 6      C      2.111223901    -1.223579356     0.001550499
 7      N      4.248432775    -0.001985560    -0.117193237
 8      H     -1.089815061    -0.000302729    -0.001181149
 9      H      0.162599886     2.160435356    -0.001155399
10      H      0.162077660    -2.159984901    -0.000751570
11      H      2.632884314     2.181269766    -0.001204198
12      H      2.633499305    -2.181448761    -0.000981234
13      H      4.658984526     0.815387854     0.304720373
14      H      4.659250506    -0.797692566     0.344358789
EXPGEOM
 1      C     -0.00110     0.93500     0.00000
 2      C      0.00620     0.22190     1.20610
 3      C      0.00620    -1.17130     1.20210
 4      C      0.00740    -1.87910     0.00000
 5      C      0.00620    -1.17130    -1.20210
 6      C      0.00620     0.22190    -1.20610
 7      N      0.06350     2.34040     0.00000
 8      H      0.01070     0.76530     2.15100
 9      H      0.00710    -1.70670     2.15040
10      H      0.00780    -2.96700     0.00000
11      H      0.00710    -1.70670    -2.15040
12      H      0.01070     0.76530    -2.15100
13      H     -0.33780     2.76230    -0.83050
14      H     -0.33780     2.76230     0.83050
---
H9C6N1, RHF, CHARGE=0, MULT=1
HF=9.5
 1      C      0.000015686     0.000003981    -0.000013998
 2      C      1.490195965    -0.000045144     0.000042779
 3      C      2.413014744     1.055041970     0.000091354
 4      C      3.729803530     0.494777489     0.000039224
 5      C      3.609515219    -0.901769732    -0.000069986
 6      C      4.642782768    -1.975556084    -0.000441763
 7      N      2.237878507    -1.184140352    -0.000038100
 8      H     -0.410042521    -0.506392033    -0.900116871
 9      H     -0.410410809    -0.516065826     0.894409908
10      H     -0.392749261     1.037480493     0.005538020
11      H      2.181027358     2.110381180    -0.000001779
12      H      4.650964355     1.059557075     0.000221567
13      H      4.559131254    -2.625691409     0.896820791
14      H      4.558518818    -2.625605297    -0.897740027
15      H      5.662659849    -1.539219904    -0.000777669
16      H      1.848468873    -2.099300782    -0.000233298
---
H9C6N1, RHF, CHARGE=0, MULT=1
HF=11.7
 1      C     -0.000084069    -0.000172519    -0.000085850
 2      C      1.556226754    -0.000332310     0.000042049
 3      C      2.021055907     1.469671460    -0.000050681
 4      C      0.785782681     2.359868123    -0.237983215
 5      C     -0.445001633     1.441874739    -0.379295552
 6      C     -0.581930281    -1.031809965    -0.850226218
 7      N     -1.054525230    -1.862081064    -1.511629256
 8      H     -0.349432669    -0.210400064     1.043411754
```

```
 9     H     1.965184105    -0.532634881    -0.884652792
10     H     1.939278409    -0.541364903     0.890616622
11     H     2.788127332     1.637550281    -0.785945936
12     H     2.509317400     1.726179448     0.964783346
13     H     0.916057131     2.982042585    -1.148588632
14     H     0.648155086     3.071707289     0.603768160
15     H    -0.843058577     1.488621502    -1.415234221
16     H    -1.271117136     1.775181218     0.283512179
---
H13C6N1, RHF, CHARGE=0, MULT=1
HF=-20.2
 1     C    -0.000162095     0.000060969     0.000238716
 2     C     1.546154758    -0.000050356     0.000103615
 3     C     2.140852220     1.435239507    -0.000593437
 4     C     3.662720700     1.481910893    -0.217746018
 5     C     4.149012856     0.568012001    -1.352942423
 6     C     3.501230549    -0.829623898    -1.335099784
 7     N     2.043940924    -0.753427951    -1.167984950
 8     H    -0.427052868     0.434799365    -0.925340310
 9     H    -0.403090696    -1.028319935     0.107887443
10     H    -0.383748605     0.588985671     0.859054670
11     H     1.864730941    -0.494147225     0.962619758
12     H     1.905556219     1.916755008     0.976321483
13     H     1.643358881     2.059312583    -0.777236777
14     H     4.184023375     1.209664794     0.729019602
15     H     3.965829860     2.530913258    -0.436619280
16     H     5.254189056     0.448929739    -1.282358712
17     H     3.958295667     1.056672872    -2.335485374
18     H     3.991649838    -1.461915643    -0.548690499
19     H     3.700180916    -1.345333011    -2.307543931
20     H     1.682056066    -1.696078721    -1.136057273
EXPGEOM
 1     C    -1.86590    -0.06100     0.28660
 2     C    -1.21630     1.23220    -0.19590
 3     C     0.26560     1.25970     0.16790
 4     C     0.98220     0.00210    -0.32750
 5     N     0.33700    -1.22790     0.11850
 6     C    -1.07930    -1.27050    -0.21210
 7     C     2.44590    -0.02120     0.07470
 8     H    -2.90100    -0.12120    -0.04880
 9     H    -1.88810    -0.07280     1.37820
10     H    -1.32110     1.30400    -1.28000
11     H    -1.72730     2.09780     0.22220
12     H     0.74300     2.14840    -0.24530
13     H     0.37500     1.32160     1.25340
14     H     0.92730    -0.00540    -1.41870
15     H     0.44790    -1.31140     1.11570
16     H    -1.49560    -2.19390     0.18480
17     H    -1.16190    -1.33400    -1.29770
18     H     2.92730    -0.91700    -0.30550
19     H     2.54910    -0.01710     1.15950
20     H     2.97300     0.84880    -0.30930
---
H13C6N1, RHF, CHARGE=0, MULT=1
HF=-10.8
 1     N    -0.000153596    -0.000184396     0.000458582
 2     C     1.464034189    -0.000390974    -0.000110159
 3     C     2.206829654     1.357289612     0.000066008
 4     C     1.561851042     2.577100945    -0.679437001
```

```
 5      C       0.766348230     2.339916533    -1.973927292
 6      C      -0.634334051     1.731710028    -1.797909451
 7      C      -0.708273543     0.288928800    -1.249220221
 8      H      -0.355959053     0.572788970     0.749770744
 9      H       1.821472230    -0.609806336    -0.866909002
10      H       1.781449076    -0.560348602     0.917421613
11      H       3.205285806     1.181209468    -0.465419221
12      H       2.420030965     1.642650804     1.057494735
13      H       2.381760833     3.297167526    -0.913574748
14      H       0.913991207     3.104809019     0.059016758
15      H       1.363334418     1.720175504    -2.681294645
16      H       0.641622318     3.327459936    -2.479252226
17      H      -1.135990268     1.732433592    -2.794177560
18      H      -1.249937387     2.403735399    -1.155896501
19      H      -0.353484698    -0.424464652    -2.033748592
20      H      -1.787091622     0.043432974    -1.073568434
 ---
H13C6N1, RHF, CHARGE=0, MULT=1
HF=-25.1
 1      N       0.000003727     0.000028369    -0.000052895
 2      C       1.474847407    -0.000002731    -0.000056785
 3      C       2.000371127     1.462922990    -0.000028301
 4      C       3.522927213     1.568671521    -0.202109884
 5      C       4.064690724     0.735255180    -1.374770729
 6      C       3.525257575    -0.703647576    -1.412692522
 7      C       2.002680583    -0.814145308    -1.214486728
 8      H      -0.337544394    -0.948759211     0.066209185
 9      H      -0.336691304     0.449333735     0.838592834
10      H       1.866802198    -0.497425272     0.932642982
11      H       1.745198106     1.947914426     0.970466843
12      H       1.486590881     2.060142050    -0.787103201
13      H       4.044662118     1.270610214     0.736322970
14      H       3.787480244     2.638182331    -0.365920881
15      H       5.176494483     0.704657111    -1.315319030
16      H       3.827503530     1.247034157    -2.335658018
17      H       4.046785161    -1.315202341    -0.640805453
18      H       3.791847235    -1.163970779    -2.391314487
19      H       1.748699567    -1.890668548    -1.077245252
20      H       1.489795634    -0.493923899    -2.149549577
EXPGEOM
 1      C       1.87890    0.01390    0.29140
 2      C       1.18190   -1.25270   -0.21770
 3      C      -0.30350   -1.26660    0.16050
 4      C      -1.02960   -0.01290   -0.33130
 5      C      -0.32230    1.25190    0.17740
 6      C       1.16240    1.27410   -0.20580
 7      N      -2.45070   -0.10590    0.03330
 8      H       2.93160    0.02380   -0.02430
 9      H       1.87980    0.00920    1.39290
10      H       1.27590   -1.30060   -1.31410
11      H       1.68040   -2.14870    0.17760
12      H      -0.80840   -2.15340   -0.24340
13      H      -0.39950   -1.31870    1.25840
14      H      -0.98910   -0.00850   -1.43300
15      H      -0.41860    1.28520    1.27580
16      H      -0.83480    2.14500   -0.21140
17      H       1.64590    2.17540    0.19620
18      H       1.25230    1.33260   -1.30200
19      H      -2.94000    0.72520   -0.30600
```

```
   20    H      -2.52760     -0.07070     1.05290
---
H15C6N1, RHF, CHARGE=0, MULT=1
HF=-27.8
1     N     -0.005263364    0.098993950    0.263567861
2     C     -1.332371269    0.732080338    0.219926433
3     C     -2.451839409   -0.330457732    0.132602596
4     C     -3.865667411    0.254019500    0.065318128
5     C      1.152543130    1.005255266    0.221570973
6     C      2.477203561    0.213506743    0.132227689
7     C      3.726098454    1.096659681    0.060133333
8     H      0.058173228   -0.483342558    1.086343553
9     H     -1.513446095    1.392793566    1.108494058
10    H     -1.381578344    1.395587140   -0.678587219
11    H     -2.289030070   -0.969876491   -0.765368820
12    H     -2.399450658   -1.011279015    1.014223326
13    H     -4.112394372    0.845944751    0.970338616
14    H     -4.002009857    0.911016745   -0.817245295
15    H     -4.612068939   -0.563690289   -0.010679559
16    H      1.056283098    1.665114114   -0.676003513
17    H      1.185745078    1.687064499    1.111519272
18    H      2.578103753   -0.461932289    1.013573175
19    H      2.457332164   -0.447184276   -0.764908091
20    H      3.706164372    1.771983777   -0.819349895
21    H      3.840960015    1.724138887    0.967218443
22    H      4.634768789    0.465904716   -0.025215268
---
H15C6N1, RHF, CHARGE=0, MULT=1
HF=-32.6
1     C     -0.126457367    0.449003665    0.007393767
2     C     -1.490129937    0.895133257   -0.577808982
3     C     -0.230328423   -0.894051836    0.769138359
4     N      0.858556789    0.457823992   -1.092441155
5     C      2.220406253    0.981384118   -0.866136878
6     C      3.283419984   -0.099060658   -0.554510063
7     C      2.603571191    1.841368932   -2.096718458
8     H      0.161700505    1.225381095    0.766240100
9     H     -1.412063354    1.883531881   -1.073355356
10    H     -1.889253545    0.174736697   -1.319857736
11    H     -2.242610742    0.993609985    0.232437554
12    H     -0.496493776   -1.741524633    0.105508270
13    H      0.721312650   -1.150210666    1.276698435
14    H     -1.008424302   -0.832425248    1.557675529
15    H      0.895468056   -0.442552515   -1.542477661
16    H      2.206886694    1.677619060    0.015253255
17    H      3.049421058   -0.645105100    0.381572390
18    H      3.380997498   -0.848349654   -1.365914955
19    H      4.279384192    0.368588713   -0.412485150
20    H      2.671381641    1.243774613   -3.027917838
21    H      1.865395421    2.650813439   -2.266843315
22    H      3.589668893    2.325485570   -1.936320168
---
H15C6N1, RHF, CHARGE=0, MULT=1
HF=-22.1
1     C     -0.000084628   -0.000035728    0.000064236
2     C      1.537994855    0.000065595   -0.000101480
3     N      2.100649285    1.362672736    0.000010201
4     C      2.263318768    1.997886481   -1.320308327
5     C      1.716237328    3.435031029   -1.351970386
```

```
 6      H      1.777032631     3.827132333    -2.389270856
 7      H      2.295014458     4.124628841    -0.705801999
 8      H      0.654918192     3.484670529    -1.037348995
 9      H      1.725788075     1.408779912    -2.104863843
10      H      3.335882963     2.004622393    -1.647111733
11      C      3.166482509     1.626150051     0.983616704
12      H      1.881016165    -0.564326413     0.902402715
13      H      1.906323841    -0.599048499    -0.873453305
14      H     -0.366060941    -1.046093086     0.072076162
15      H     -0.426278515     0.434005004    -0.926164544
16      H     -0.416494912     0.559999513     0.860853608
17      C      2.610083748     2.024917432     2.360975970
18      H      3.828078304     2.455153740     0.628273212
19      H      3.847275393     0.743190200     1.102685581
20      H      3.454394622     2.270030347     3.039537530
21      H      2.030200194     1.210868172     2.839351310
22      H      1.957036299     2.918353559     2.302891649
EXPGEOM
 1    N      0.00000      0.00000      0.01740
 2    C     -0.33620      1.35000      0.44970
 3    C     -1.00110     -0.96620      0.44970
 4    C      1.33730     -0.38390      0.44970
 5    C      0.28570      2.41710     -0.44140
 6    C     -2.23610     -0.96110     -0.44140
 7    C      1.95040     -1.45600     -0.44140
 8    H     -1.42020      1.45710      0.41660
 9    H     -0.03980      1.51510      1.49730
10    H     -0.55180     -1.95850      0.41660
11    H     -1.29230     -0.79200      1.49730
12    H      1.97200      0.50140      0.41660
13    H      1.33200     -0.72310      1.49730
14    H      1.37430      2.36430     -0.43750
15    H     -0.05390      2.28820     -1.46820
16    H      0.00000      3.41260     -0.09900
17    H     -2.73470      0.00800     -0.43750
18    H     -1.95470     -1.19080     -1.46820
19    H     -2.95540     -1.70630     -0.09900
20    H      1.36050     -2.37230     -0.43750
21    H      2.00860     -1.09740     -1.46820
22    H      2.95540     -1.70630     -0.09900
---
H5C7N1, RHF, CHARGE=0, MULT=1
HF=51.5, IE=9.7
 1    C      0.000055501    -0.000038461     0.000018976
 2    C      1.405398727     0.000118558     0.000027141
 3    C      2.113613455     1.215127988     0.000011148
 4    C      1.412794848     2.434413454    -0.000005809
 5    C      0.007509170     2.442712659     0.000005860
 6    C     -0.714150511     1.223522784     0.000010569
 7    H      3.204111816     1.211741314     0.000009291
 8    H      1.960102776     3.377665903    -0.000018395
 9    H     -0.522765319     3.396259297    -0.000042571
10    C     -2.140968261     1.227935947    -0.000043633
11    N     -3.303562933     1.231545234    -0.000124935
12    H     -0.535767762    -0.950482510    -0.000012020
13    H      1.947334156    -0.946285527     0.000008382
EXPGEOM
 1    N      0.00000      0.00000      3.22770
 2    C      0.00000      0.00000      2.05210
```

```
3    C     0.00000     0.00000     0.60600
4    C     0.00000     1.22190    -0.09160
5    C     0.00000    -1.22190    -0.09160
6    C     0.00000     1.21560    -1.49030
7    C     0.00000    -1.21560    -1.49030
8    C     0.00000     0.00000    -2.18940
9    H     0.00000     2.16060     0.46250
10   H     0.00000    -2.16060     0.46250
11   H     0.00000     2.16030    -2.03440
12   H     0.00000    -2.16030    -2.03440
13   H     0.00000     0.00000    -3.27970
---
H9C7N1, RHF, CHARGE=0, MULT=1
HF=24.3
1    C    -0.000345275    -0.000044675     0.000601491
2    C     1.516524722    -0.000002845    -0.000210126
3    C     1.374534027     2.582633692     0.000753202
4    C    -0.136234749     2.509084337     0.286755349
5    C    -0.729660141     1.138226446     0.115738870
6    C    -0.636161738    -1.271282989    -0.148529037
7    N    -1.151401444    -2.306452018    -0.270141494
8    H     1.868673984    -0.250566744    -1.028359444
9    H     1.892093698    -0.811483950     0.663783595
10   H     1.534869400     2.748294479    -1.089053661
11   H     1.799615123     3.476582307     0.509527849
12   H    -0.670649463     3.223221724    -0.381254782
13   H    -0.341466497     2.850165289     1.329170969
14   H    -1.823057806     1.114071066     0.102577317
15   C     2.148863626     1.331519319     0.448328703
16   H     3.188140362     1.387364763     0.052997847
17   H     2.245455791     1.335699518     1.557874383
---
H9C7N1, RHF, CHARGE=0, MULT=1
HF=13.4
1    C     0.000065607     0.000092599    -0.000008332
2    C     1.509434327    -0.000180151     0.000094967
3    C     2.266852642     1.198179659    -0.000032526
4    C     3.667436282     1.120639838     0.000202452
5    C     4.278373358    -0.142100781     0.000331941
6    C     3.464718772    -1.302961714     0.000284930
7    C     4.047296310    -2.695336762    -0.000418113
8    N     2.110274081    -1.217073385     0.000384357
9    H    -0.385237354     0.528597063     0.897629510
10   H    -0.385175572     0.527006347    -0.898602594
11   H    -0.434085140    -1.019550050     0.000884919
12   H     1.772544630     2.169475679    -0.000196580
13   H     4.271951937     2.027913614     0.000192925
14   H     5.364969969    -0.223702207     0.000395815
15   H     4.682795053    -2.847478278     0.897631973
16   H     3.273995776    -3.489135307    -0.000839176
17   H     4.682613645    -2.846640233    -0.898830644
---
H9C7N1, RHF, CHARGE=0, MULT=1
HF=26.2
1    C     0.000000000     0.000000000     0.000000000
2    C     1.539167592     0.000000000     0.000000000
3    C     2.170277927     1.405488520     0.000000000
4    C     1.375552673     2.434870221    -0.754309118
5    C     0.128083882     2.240109187    -1.219057616
```

```
6    C     -0.644174272    0.945959358   -1.053697563
7    C     -0.824084161    0.294110488   -2.352094992
8    N     -0.980584382   -0.215745030   -3.384221285
9    H     -0.366451507    0.306793362    1.006663217
10   H     -0.364525833   -1.040475168   -0.148300117
11   H      1.909518487   -0.569752293   -0.882557875
12   H      1.896301427   -0.559975409    0.893622306
13   H      3.195861950    1.344556971   -0.431122153
14   H      2.298884306    1.759733756    1.050015283
15   H      1.878131805    3.394648828   -0.899381615
16   H     -0.404188058    3.034004675   -1.747940423
17   H     -1.668837703    1.208492779   -0.678410869
---
H9C7N1, RHF, CHARGE=0, MULT=1
HF=21
1    C     -0.009502856    0.021917791    0.025413033
2    C     -1.104878300   -0.833877918   -0.238530801
3    C     -0.999176102   -2.223321238   -0.054785738
4    C      0.207166076   -2.787011527    0.393883044
5    C      1.306171855   -1.951810337    0.656449992
6    C      1.198290279   -0.562089151    0.472872776
7    C     -0.124601321    1.524272801   -0.185550101
8    N     -0.610041762    2.289471282    0.969242054
9    H     -2.052384248   -0.423921124   -0.593102838
10   H     -1.856746711   -2.864005762   -0.263501891
11   H      0.290067966   -3.865008874    0.536287026
12   H      2.246940869   -2.381289997    1.003001812
13   H      2.069497037    0.061668626    0.681788531
14   H      0.869205782    1.945846521   -0.478477282
15   H     -0.808745057    1.736042191   -1.045032021
16   H     -1.511164646    1.946797510    1.264689301
17   H      0.000502808    2.154722797    1.760281627
---
H9C7N1, RHF, CHARGE=0, MULT=1
HF=14.6
1    N     -0.014630601    0.056489980   -0.122974917
2    C      0.056358786   -1.358602321    0.018504872
3    C     -1.127395071   -2.136970906    0.033334583
4    C     -1.037009899   -3.537311027    0.058009205
5    C      0.213388774   -4.174887891    0.068860879
6    C      1.409469942   -3.418651625    0.051774264
7    C      1.316884194   -2.010171636    0.029096787
8    C      2.745391449   -4.113969307    0.052962598
9    H      0.757200804    0.515631100    0.333717144
10   H     -0.852766006    0.430508447    0.292433986
11   H     -2.112458457   -1.668662164    0.021766362
12   H     -1.949409677   -4.135638280    0.068056265
13   H      0.251335776   -5.265097387    0.089462599
14   H      2.232395298   -1.415077780    0.014786158
15   H      2.852558759   -4.766197957    0.945220605
16   H      2.859273250   -4.750487817   -0.849948296
17   H      3.595756462   -3.402394326    0.063576598
---
H9C7N1, RHF, CHARGE=0, MULT=1
HF=20.1
1    C     -0.150474182   -0.166007813    0.297375610
2    N     -1.256986698    0.443532018   -0.448343454
3    C     -2.024579283   -0.409654138   -1.305865789
4    C     -1.490957495   -0.761589746   -2.571668095
```

```
5    C    -2.253493096   -1.532003279   -3.465993163
6    C    -3.548999586   -1.954642386   -3.120838209
7    C    -4.084543392   -1.604573281   -1.869881689
8    C    -3.331856732   -0.839235826   -0.963426793
9    H    -0.469975079   -0.981728965    0.989539659
10   H     0.606750918   -0.586144129   -0.400347560
11   H     0.343014144    0.623155210    0.909089319
12   H    -1.865481110    0.923736184    0.198762047
13   H    -0.493581579   -0.436586876   -2.871451167
14   H    -1.836446807   -1.799549945   -4.437887881
15   H    -4.135736707   -2.549623344   -3.820761436
16   H    -5.089506179   -1.929807902   -1.597964030
17   H    -3.773692418   -0.582861174    0.000786048
---
H9C7N1, RHF, CHARGE=0, MULT=1
HF=12.7
1    C    -0.045786296    0.055700741    0.001989935
2    C    -0.086486320   -1.449632166   -0.045415008
3    C    -0.727583895   -2.156801259    0.999028214
4    C    -0.804933739   -3.557969107    1.010486299
5    C    -0.233406693   -4.294607745   -0.039031649
6    C     0.412252400   -3.627436224   -1.089161188
7    C     0.492898122   -2.209816248   -1.107556728
8    N     1.056759547   -1.559428237   -2.243745238
9    H     0.998517237    0.427294797    0.083508100
10   H    -0.506103584    0.507152549   -0.901576751
11   H    -0.595642599    0.461246615    0.876575497
12   H    -1.182883727   -1.608461617    1.826765736
13   H    -1.306983554   -4.071065475    1.830910252
14   H    -0.288784760   -5.383832313   -0.038719692
15   H     0.849741025   -4.221694224   -1.893516157
16   H     1.692474246   -0.824024117   -1.978540886
17   H     1.592241637   -2.197174853   -2.810646757
---
H9C7N1, RHF, CHARGE=0, MULT=1
HF=10
1    N    -0.056308984    0.016479623    0.067792744
2    C     1.311557370   -0.343280627   -0.080122870
3    C     1.700769481   -1.705687076   -0.108032210
4    C     3.060673819   -2.054193000   -0.124569038
5    C     4.075368477   -1.071150783   -0.116828911
6    C     3.675186451    0.286113392   -0.099689827
7    C     2.321597377    0.652346736   -0.083246535
8    C     5.533785008   -1.437877503   -0.121862142
9    H    -0.256591983    0.908425601   -0.354800749
10   H    -0.670169577   -0.651323915   -0.370098712
11   H     0.955069851   -2.502216926   -0.110951541
12   H     3.322952129   -3.114055342   -0.140751114
13   H     4.428200721    1.077235363   -0.097495057
14   H     2.063731760    1.712372250   -0.068317680
15   H     6.047098502   -1.020933760   -1.014011263
16   H     6.043686586   -1.041579307    0.782176603
17   H     5.695529050   -2.534723730   -0.133825301
---
H11C7N1, RHF, CHARGE=0, MULT=1
HF=-0.9
1    C     0.000009918   -0.000009444    0.000028589
2    C     1.553212337    0.000069748   -0.000062375
3    C     2.168016403    1.410789124    0.000268422
```

```
  4    C      1.560365229    2.364421236   -1.042173129
  5    C      0.022978729    2.351074810   -1.078307047
  6    C     -0.596118764    0.942263911   -1.081154193
  7    C     -0.528088418   -1.358002664   -0.133639764
  8    N     -0.954958254   -2.434870023   -0.223458777
  9    H     -0.340511851    0.372964929    1.002120575
 10    H      1.933680129   -0.566547952   -0.879908304
 11    H      1.915339045   -0.547451932    0.899452086
 12    H      3.262536819    1.321982123   -0.185259116
 13    H      2.068596919    1.858706703    1.015498285
 14    H      1.957520046    2.112093873   -2.052099568
 15    H      1.908352365    3.400946012   -0.830519379
 16    H     -0.322543287    2.893701937   -1.987310620
 17    H     -0.373533413    2.928855294   -0.212046248
 18    H     -0.473071542    0.488830963   -2.090657975
 19    H     -1.693540503    1.034178452   -0.916141960
---
H13C7N1, RHF, CHARGE=0, MULT=1
HF=-0.9
  1    N      0.004910301    0.021778991   -0.248315761
  2    C      0.286032279    1.478948142   -0.216235649
  3    C      1.685519322    1.690701361   -0.848018581
  4    C      2.277523434    0.287428250   -1.078699648
  5    C      1.170443457   -0.735018338   -0.735196869
  6    C     -1.326441700   -0.253897359   -0.813187953
  7    C     -1.981817693    1.107680572   -1.136590397
  8    C     -0.925045953    2.194862745   -0.865732267
  9    H      0.337813172    1.784876540    0.861661693
 10    H      2.328818825    2.290897097   -0.170130546
 11    H      1.639154853    2.257735780   -1.801674077
 12    H      3.174754095    0.124700674   -0.444418301
 13    H      2.617083626    0.165496317   -2.129250729
 14    H      1.498670801   -1.428451523    0.076977867
 15    H      0.939987371   -1.379703738   -1.616467047
 16    H     -1.918217512   -0.819183677   -0.052909257
 17    H     -1.294806088   -0.900529967   -1.722906055
 18    H     -2.885835900    1.268364194   -0.511452180
 19    H     -2.327618962    1.139897754   -2.191625229
 20    H     -1.321374328    2.978328969   -0.185224452
 21    H     -0.653296427    2.719805490   -1.805991902
---
H13C7N1, RHF, CHARGE=0, MULT=1
HF=-7.4
  1    C     -0.000015343    0.000028695    0.000022389
  2    C      1.531267869   -0.000060030   -0.000013955
  3    C      2.174358030    1.400470461   -0.000014470
  4    C      3.716602640    1.383976687   -0.000097250
  5    C      4.356134952    2.787158535    0.000048386
  6    C      5.900601865    2.767353175    0.000009816
  7    C      6.488463807    4.101340758    0.000093186
  8    N      6.974430714    5.156845027    0.000244925
  9    H     -0.412723262    0.505726773    0.896343769
 10    H     -0.412785915    0.506213570   -0.896001394
 11    H     -0.381813803   -1.041822425   -0.000230312
 12    H      1.883115202   -0.570319568    0.890702622
 13    H      1.882991221   -0.570182803   -0.890893377
 14    H      1.816138693    1.964520599    0.891588732
 15    H      1.816030596    1.964348134   -0.891654085
 16    H      4.075750685    0.820629429    0.891556813
```

```
17      H        4.075713213     0.820997455    -0.892070376
18      H        4.004501823     3.353372731     0.892507184
19      H        4.004691083     3.353708693    -0.892154166
20      H        6.279175460     2.218015610     0.894502283
21      H        6.279236947     2.217994454    -0.894433155
---
H17C7N1, RHF, CHARGE=0, MULT=1
HF=-39.4
1       C        0.006937620     0.144915781     0.277546517
2       C        1.076153509    -0.885487371    -0.158824323
3       N        1.734309197    -0.582084969    -1.446626834
4       C        1.051633816    -0.973632505    -2.687649226
5       C        1.974583545    -0.751874585    -3.910297458
6       C        1.364127859    -1.234127085    -5.240706448
7       C        2.243352287    -0.987178995    -6.470042007
8       C        2.181032345    -1.080956174     0.908185719
9       H        0.422847529     1.166199776     0.394527644
10      H       -0.825059160     0.203267285    -0.453444190
11      H       -0.439319166    -0.146628786     1.250231156
12      H        0.549848310    -1.873383338    -0.255881933
13      H        1.974362748     0.397115897    -1.485117846
14      H        0.777887141    -2.055110457    -2.618857170
15      H        0.091224963    -0.415526037    -2.843287571
16      H        2.218276680     0.332156520    -3.998475163
17      H        2.941140119    -1.279535026    -3.743369766
18      H        1.147719221    -2.325533892    -5.176227228
19      H        0.381721394    -0.733745126    -5.405479903
20      H        2.420907411     0.094635507    -6.637974463
21      H        3.230731954    -1.483522159    -6.379453480
22      H        1.750527907    -1.387826358    -7.379844161
23      H        2.890757802    -1.878325105     0.608954929
24      H        2.765942517    -0.156930955     1.089991050
25      H        1.731383129    -1.387836120     1.875373741
---
H11C8N1, RHF, CHARGE=0, MULT=1
HF=18
1       C        0.000277408     0.000293684     0.000136746
2       C        1.559175825    -0.000438587    -0.000286649
3       C        1.920257128     1.512108196     0.000258450
4       C        0.515066489     2.225768649     0.012415422
5       C       -0.170692847     2.002748640    -1.389154963
6       C       -0.520628095     0.487714856    -1.385907866
7       C       -0.313713439     1.229826971     0.907461395
8       C        0.550810876     3.607817627     0.446244224
9       N        0.578637584     4.715629725     0.794935273
10      H       -0.440895723    -0.956138327     0.325256512
11      H        1.957839161    -0.514280782     0.898605227
12      H        1.987885479    -0.532906109    -0.872496656
13      H        2.518965050     1.783879708     0.893597414
14      H        2.528978654     1.798905013    -0.880133665
15      H       -1.081616630     2.626492769    -1.496596711
16      H        0.495488834     2.278439729    -2.230548966
17      H       -0.053197661    -0.051705769    -2.233677767
18      H       -1.613527454     0.322019968    -1.482260655
19      H        0.056605021     1.137434588     1.946134222
20      H       -1.389330142     1.477179437     0.985954292
---
H11C8N1, RHF, CHARGE=0, MULT=1
HF=39.6
```

```
   1    C      0.000584658     0.000563720     0.000615547
   2    C      1.560218963    -0.000714517    -0.000269768
   3    C      1.926657202     1.510680371     0.000232401
   4    C      0.515491338     2.215429220     0.010151927
   5    C     -0.173459782     2.004479876    -1.393019782
   6    C     -0.522859709     0.488907950    -1.384844689
   7    C     -0.316888593     1.226377613     0.912483410
   8    N      0.550359227     3.565379610     0.436485378
   9    C      0.578374190     4.699925126     0.795733765
  10    H     -0.438902758    -0.957798359     0.322270370
  11    H      1.957523438    -0.516056249     0.898575269
  12    H      1.986883356    -0.534575629    -0.872885835
  13    H      2.529405564     1.778034315     0.891953961
  14    H      2.535842340     1.793942661    -0.881083307
  15    H     -1.085061162     2.626708502    -1.504076886
  16    H      0.491519492     2.276019473    -2.236874207
  17    H     -0.057675986    -0.051913863    -2.233064841
  18    H     -1.616036169     0.322438811    -1.478132711
  19    H      0.051900611     1.128504793     1.951136192
  20    H     -1.393655444     1.469156340     0.990636848
---
H11C8N1, RHF, CHARGE=0, MULT=1
HF=8.3
   1    C     -0.000014662    -0.000045615    -0.000040633
   2    C      1.403438101    -0.000020464     0.000237954
   3    C      2.104449040     1.224874221    -0.000226758
   4    C      1.317641408     2.407278189    -0.003529571
   5    N     -0.031279886     2.412110006    -0.003445112
   6    C     -0.701532501     1.230055698    -0.001009059
   7    C     -2.208088141     1.306394844    -0.002309426
   8    C      3.614162268     1.283880074    -0.011532272
   9    C      4.258678286     1.307205800     1.379366567
  10    H     -0.540397415    -0.946672692     0.000633355
  11    H      1.936618674    -0.952160115    -0.000153761
  12    H      1.780083999     3.401376040    -0.007851769
  13    H     -2.617102653     0.822213497    -0.914448306
  14    H     -2.622512118     0.776039399     0.881136588
  15    H     -2.589136819     2.346709507     0.023733019
  16    H      3.950030912     2.183859303    -0.578048633
  17    H      4.017738179     0.413479235    -0.580079804
  18    H      4.002403822     0.402470800     1.966647584
  19    H      3.940803619     2.193365855     1.964738869
  20    H      5.363585445     1.344794842     1.287202750
---
H11C8N1, RHF, CHARGE=0, MULT=1
HF=24
   1    C     -0.059515992    -0.026340661     0.122304144
   2    C     -1.222201192    -0.812002770     0.097315578
   3    C     -1.145003999    -2.213535027     0.097785512
   4    C      0.119104602    -2.823067999     0.125100913
   5    C      1.291429233    -2.051387810     0.153685867
   6    C      1.233369299    -0.626896370     0.150920455
   7    N      2.409103417     0.160555764     0.156620720
   8    C      3.734723310    -0.457389570     0.209981471
   9    C      2.338297506     1.617838554     0.272933490
  10    H     -0.196167067     1.055025907     0.116988483
  11    H     -2.197214369    -0.322206922     0.076832258
  12    H     -2.050922696    -2.817993011     0.077333884
  13    H      0.195951053    -3.911671734     0.124517298
```

```
14      H         2.237255683      -2.592561010      0.177844427
15      H         3.880131787      -1.045158443      1.147342591
16      H         3.900745815      -1.130854516     -0.662953910
17      H         4.540844330       0.309736174      0.178815300
18      H         1.813023694       2.066414216     -0.602814573
19      H         1.812020224       1.933993997      1.204381799
20      H         3.353879032       2.072222872      0.305454312
---
H11C8N1, RHF, CHARGE=0, MULT=1
HF=13.4
1       C         0.058098671       0.140402159     -0.318980245
2       C        -0.698297006       1.340617720     -0.344595650
3       C        -2.018480679       1.351960224      0.135585458
4       C        -2.608191603       0.178187816      0.635508898
5       C        -1.869225006      -1.016638421      0.653071912
6       C        -0.545750697      -1.041775683      0.182786746
7       N         1.365869373       0.123159152     -0.895120479
8       C         2.455346080       0.876223806     -0.252208297
9       C         2.934622591       0.375940078      1.120661560
10      H        -0.276915599       2.264887472     -0.742170031
11      H        -2.590926358       2.280439274      0.114804933
12      H        -3.633403080       0.194269346      1.005304494
13      H        -2.322993524      -1.931622316      1.035685276
14      H         0.001356609      -1.985548962      0.209429982
15      H         1.665296657      -0.825597637     -1.063593508
16      H         3.318549298       0.850713088     -0.965461343
17      H         2.156374710       1.948752651     -0.163702080
18      H         2.154930777       0.483236829      1.901145506
19      H         3.245519671      -0.688196430      1.095352890
20      H         3.812838187       0.971398763      1.445237160
---
H15C8N1, RHF, CHARGE=0, MULT=1
HF=-10.4
1       C        -0.000331297       0.000833004     -0.000030555
2       C         1.550082214      -0.000342661      0.000294932
3       C         2.155691758       1.416968731     -0.000324797
4       C         1.082989043       2.536446863     -0.008814419
5       C         0.074594686       2.303101045     -1.163105926
6       C        -0.533820249       0.884131601     -1.157046847
7       C        -0.599730927       0.376397304      1.390139239
8       N        -0.683265619       1.798419987      1.740122075
9       C         0.399797304       2.720355998      1.381132828
10      H        -0.329628707      -1.048662683     -0.200881224
11      H         1.911074451      -0.557416821     -0.894500571
12      H         1.932977988      -0.575425863      0.872449591
13      H         2.812248230       1.539020525     -0.891955906
14      H         2.831416245       1.540469257      0.875030208
15      H         1.612863516       3.498514442     -0.217724947
16      H        -0.735049621       3.065589517     -1.125373604
17      H         0.590213567       2.479081143     -2.134563511
18      H        -1.644282549       0.945176275     -1.112865184
19      H        -0.311901136       0.383104166     -2.126546304
20      H        -1.632896290      -0.050467166      1.453175071
21      H        -0.013296984      -0.132781463      2.195538223
22      H        -1.562475811       2.172253586      1.414931618
23      H        -0.007036134       3.761674168      1.437005841
24      H         1.174847336       2.655497343      2.185823020
---
H15C8N1, RHF, CHARGE=0, MULT=1
```

```
HF=-12.1
1    C     0.000080004    -0.000046032    -0.000016873
2    C     1.531370911     0.000184185     0.000083157
3    C     2.174214267     1.400815519    -0.000055681
4    C     3.716493707     1.385260143    -0.000532162
5    C     4.357037067     2.788521271    -0.000217712
6    C     5.899012087     2.770990179    -0.000688688
7    C     6.537678829     4.177433535    -0.000333544
8    C     7.995070369     4.142240126    -0.000897018
9    N     9.157023905     4.132766874    -0.000943393
10   H    -0.412785409     0.506424828     0.895854033
11   H    -0.412832554     0.505463905    -0.896438400
12   H    -0.381707826    -1.041771039     0.000622302
13   H     1.883028156    -0.569676546     0.891181884
14   H     1.883432884    -0.570039148    -0.890647953
15   H     1.815527087     1.964354848     0.891582263
16   H     1.814990746     1.964611650    -0.891517424
17   H     4.075709109     0.821488127     0.890812237
18   H     4.075124315     0.822376107    -0.892661509
19   H     4.000232507     3.352731474     0.891858295
20   H     3.999642263     3.353257740    -0.891737773
21   H     6.262462076     2.211357442     0.891112777
22   H     6.262089389     2.212728472    -0.893699048
23   H     6.201816224     4.753296177     0.894397606
24   H     6.200783076     4.753685799    -0.894474416
---
H17C8N1, RHF, CHARGE=0, MULT=1
HF=-21.8
1    C     0.000228905    -0.000015754    -0.000025395
2    C     1.531484954    -0.000009634     0.000084330
3    C     2.174002137     1.401060270    -0.000115102
4    C     3.720181911     1.369452374     0.000986854
5    N     4.472762582     2.612565643     0.000942325
6    C     4.001779881     3.808013815     0.000546371
7    C     4.959239687     4.995589234     0.002101513
8    C     4.770874201     5.851110936    -1.267359984
9    C     4.817326139     5.808198989     1.305348082
10   H    -0.412645869     0.505154748     0.896506521
11   H    -0.412908951     0.506503279    -0.895807302
12   H    -0.381604881    -1.041898746    -0.000873106
13   H     1.882985780    -0.570817393    -0.890412294
14   H     1.882853027    -0.569236250     0.891828612
15   H     1.813217337     1.959408686    -0.894266032
16   H     1.811983244     1.959909054     0.893375293
17   H     4.075707213     0.802385419    -0.897857637
18   H     4.074836911     0.802382524     0.900194165
19   H     2.934440518     4.081630219    -0.000852256
20   H     6.014703397     4.620971144    -0.023263776
21   H     3.761339554     6.306502941    -1.323465487
22   H     4.916107870     5.240255511    -2.181561585
23   H     5.512603350     6.675014004    -1.300658704
24   H     5.000610294     5.168839093     2.192785514
25   H     3.808936456     6.256155449     1.415369632
26   H     5.556575935     6.634432491     1.336654418
---
H19C8N1, RHF, CHARGE=0, MULT=1
HF=-43.2
1    C     0.000059086     0.000058126     0.000088949
2    C     1.542109451     0.000153905    -0.000195035
```

```
 3     C     2.165713019     1.429833953    -0.000076366
 4     N     2.136018302     2.042960091     1.336592598
 5     C     2.917032278     3.277481879     1.508972484
 6     C     2.986865888     3.721935500     3.001387582
 7     C     3.740675602     2.724071909     3.903525468
 8     C     3.567218146     5.147826097     3.121195530
 9     C     2.102727972    -0.843211376    -1.166065398
10     H    -0.411414419     0.495534325     0.902479916
11     H    -0.419709433     0.517088986    -0.886786700
12     H    -0.395568378    -1.036752466    -0.000473465
13     H     1.865635135    -0.517310053     0.939109831
14     H     1.654287942     2.065750870    -0.769303429
15     H     3.233369274     1.358621586    -0.324444863
16     H     1.179939100     2.219546426     1.607130273
17     H     2.489951188     4.118000545     0.899902726
18     H     3.952019457     3.111671907     1.121436693
19     H     1.938675341     3.782105492     3.393324882
20     H     4.788864669     2.573427442     3.574330109
21     H     3.246628570     1.732510733     3.921063934
22     H     3.769395691     3.086819629     4.952377359
23     H     2.961174421     5.878448268     2.546911187
24     H     4.609910245     5.210238423     2.749504125
25     H     3.566778361     5.488615274     4.176880212
26     H     3.205073060    -0.944315789    -1.094600377
27     H     1.868253174    -0.401092783    -2.155679199
28     H     1.683828079    -1.870356198    -1.150379816
---
H19C8N1, RHF, CHARGE=0, MULT=1
HF=-43.9
 1     C     0.000099788    -0.000177704     0.000027660
 2     C     1.532218533     0.000198881    -0.000108903
 3     C     2.210973580     1.406671907     0.000095703
 4     N     3.602333139     1.215668121     0.460336278
 5     C     4.190419358     2.134688132     1.456994293
 6     C     4.956599107     1.253868341     2.494342235
 7     C     5.262248901     1.921113521     3.839165653
 8     C     5.066415019     3.249779844     0.835430018
 9     C     2.100985133     2.139112393    -1.359438250
10     H    -0.416095398     0.577780006     0.849901078
11     H    -0.424481607     0.418494505    -0.934480073
12     H    -0.375774536    -1.040591151     0.091720138
13     H     1.886858875    -0.590094541    -0.876307135
14     H     1.863049647    -0.565108561     0.902175691
15     H     1.663349242     2.039410497     0.748293881
16     H     4.224277116     1.088382506    -0.320902491
17     H     3.360561999     2.651423110     2.008623963
18     H     5.913808443     0.894608243     2.051419821
19     H     4.357965338     0.337910415     2.708194561
20     H     4.347023095     2.313301461     4.327319674
21     H     5.980726641     2.759805762     3.742573057
22     H     5.715452670     1.181218826     4.531594406
23     H     5.946307196     2.849342210     0.292367103
24     H     5.443632455     3.937305141     1.619144896
25     H     4.487616345     3.872145815     0.122973918
26     H     2.579994565     1.578065687    -2.187301870
27     H     2.570310912     3.143144055    -1.319691822
28     H     1.039418172     2.299091198    -1.636311835
---
H19C8N1, RHF, CHARGE=0, MULT=1
```

```
HF=-40.9
1    C    -0.065827247   -0.093508861   -0.051897256
2    C     1.265403228    0.663097661   -0.054153951
3    C     2.163515909    0.388203395    1.168049212
4    C     3.481853389    1.197154556    1.154893971
5    N     4.306360891    0.898642284    2.336547486
6    C     5.610492239    1.577693101    2.398360407
7    C     6.414554069    1.141214470    3.646242834
8    C     7.790194202    1.828429908    3.756690119
9    C     8.597115496    1.452231724    5.002542717
10   H     0.083402562   -1.192088584   -0.039635645
11   H    -0.691170416    0.174876243    0.823685779
12   H    -0.647298416    0.154425343   -0.964004374
13   H     1.817167900    0.402889301   -0.987019913
14   H     1.046528314    1.754432563   -0.118943183
15   H     1.589974265    0.621980637    2.094719398
16   H     2.399767322   -0.699624743    1.210157735
17   H     4.068876726    0.933827410    0.240525781
18   H     3.245911214    2.290864065    1.073230750
19   H     3.780240325    1.111870848    3.171893710
20   H     6.187696230    1.316433902    1.477281616
21   H     5.502535982    2.694291851    2.396738230
22   H     5.827476414    1.363012813    4.567310179
23   H     6.558038422    0.037111896    3.623121840
24   H     8.399815214    1.585100792    2.855881846
25   H     7.655570101    2.935124761    3.748937560
26   H     8.073558543    1.741903093    5.936375516
27   H     8.800477950    0.363315617    5.051020228
28   H     9.576561006    1.974354442    4.994242554
---
H19C8N1, RHF, CHARGE=0, MULT=1
HF=-41.8
1    C    -0.000010594    0.000008075   -0.000024759
2    C     1.531248136   -0.000029592    0.000010437
3    C     2.173204019    1.401347166   -0.000033755
4    C     3.720304644    1.370061581   -0.005699600
5    N     4.273283314    2.730931951    0.074712542
6    C     5.731121116    2.837431327    0.234559835
7    C     6.181237514    4.300857529    0.526544617
8    C     5.666005998    4.842086036    1.875455919
9    C     7.714621337    4.436703351    0.406752305
10   H    -0.412977586    0.508703046    0.894481921
11   H    -0.412786106    0.503107737   -0.897859268
12   H    -0.382182219   -1.041749396    0.003031511
13   H     1.882537706   -0.571386181   -0.890415420
14   H     1.882043984   -0.570646415    0.891032421
15   H     1.809619620    1.963495345   -0.891165685
16   H     1.818593561    1.964092729    0.893376929
17   H     4.080140209    0.823272487   -0.916770371
18   H     4.082766328    0.784941941    0.875051093
19   H     3.990447269    3.254494848   -0.741240219
20   H     6.269371823    2.467176573   -0.678456152
21   H     6.055297347    2.173171084    1.072759361
22   H     5.750842091    4.960301891   -0.270690281
23   H     6.022232063    4.234801042    2.732176122
24   H     4.558907828    4.862664322    1.913485650
25   H     6.011976440    5.884338573    2.038759151
26   H     8.067259700    4.138418834   -0.602002154
27   H     8.251723381    3.811945094    1.148699554
```

```
 28    H     8.032854351    5.488121414    0.560990103
---
H19C8N1, RHF, CHARGE=0, MULT=1
HF=-41.5
 1    C    -0.000086943   -0.000015669   -0.000042123
 2    C     1.531056813   -0.000106334    0.000044947
 3    C     2.174024023    1.400542785    0.000174950
 4    C     3.716269790    1.385741592    0.001274549
 5    C     4.357542903    2.788788576    0.001596163
 6    C     5.899930509    2.773497547    0.003342745
 7    C     6.540047891    4.177362314    0.001990078
 8    C     8.085548521    4.148708184    0.008049095
 9    N     8.646017979    5.502705151   -0.080683458
10    H    -0.413027489    0.505898017    0.896152057
11    H    -0.412951693    0.505999619   -0.896205754
12    H    -0.382402679   -1.041725720   -0.000160548
13    H     1.882652844   -0.570342457    0.891115880
14    H     1.882546932   -0.570796890   -0.890539360
15    H     1.815230969    1.963856334   -0.891699155
16    H     1.813825454    1.964170333    0.891507494
17    H     4.073921400    0.821841223    0.893455001
18    H     4.075003101    0.820946616   -0.889765078
19    H     4.000527703    3.352346397   -0.890773293
20    H     3.998474044    3.353359263    0.892751851
21    H     6.257006084    2.209906438    0.895973133
22    H     6.259626020    2.208231829   -0.887047086
23    H     6.186523656    4.739464863   -0.892395493
24    H     6.178539959    4.741921655    0.892482351
25    H     8.448380642    3.600970459    0.917244154
26    H     8.459208313    3.569465651   -0.872681536
27    H     8.482100740    6.004040747    0.778977121
28    H     9.649592020    5.437033926   -0.159375455
---
H7C9N1, RHF, CHARGE=0, MULT=1
HF=48.9
 1    C     0.000099828    0.000375140    0.000065873
 2    C     1.382632360   -0.000346746    0.000069847
 3    C     2.108137543    1.231627475    0.000072648
 4    C     1.433352865    2.438548200    0.000045063
 5    C    -0.002740757    2.472985089    0.000022882
 6    C    -0.731372456    1.239089856    0.000019683
 7    C    -2.164971287    1.316025251   -0.000020821
 8    C    -2.787871948    2.554640039    0.000036572
 9    N    -2.091085350    3.742724071    0.000099178
10    C    -0.763856635    3.700243860    0.000064640
11    H    -0.548488331   -0.942486457    0.000041292
12    H     1.937905611   -0.939457060    0.000084777
13    H     3.198341538    1.203149223    0.000004068
14    H     1.990271144    3.376805515   -0.000002327
15    H    -2.759747398    0.402600825   -0.000057408
16    H    -3.877276185    2.659266749    0.000040430
17    H    -0.249705446    4.668789698    0.000162702
---
H7C9N1, RHF, CHARGE=0, MULT=1
HF=47.9
 1    C     0.000123636    0.000165299   -0.000017524
 2    C     1.382165390   -0.000024970   -0.000009063
 3    C     2.111974265    1.230642429    0.000010913
 4    C     1.446784701    2.442175118    0.000037677
```

```
5    C     0.010149617    2.489916177    0.000050267
6    C    -0.716192590    1.251951444    0.000025318
7    N    -2.099402490    1.213649383    0.000031420
8    C    -2.768369027    2.360042421    0.000058395
9    C    -2.129502174    3.646368014    0.000101066
10   C    -0.750724467    3.709536762    0.000083048
11   H    -0.554946918   -0.939093292   -0.000032564
12   H     1.936475012   -0.939483062   -0.000017971
13   H     3.202100920    1.196095857   -0.000013539
14   H     2.010438223    3.376367493    0.000049534
15   H    -3.862291192    2.289790686    0.000057567
16   H    -2.742720042    4.546453762    0.000137735
17   H    -0.236163089    4.671181817    0.000109595
---
H9C9N1, RHF, CHARGE=0, MULT=1
HF=34.9
1    C     0.000249332    0.000367359    0.000038675
2    C     1.509211441   -0.000107251   -0.000165387
3    C     2.206628463    1.228587199   -0.000014915
4    C     3.608869200    1.263299258   -0.001972283
5    C     4.342183591    0.067677960   -0.004436480
6    C     3.689675065   -1.185471134   -0.003406403
7    C     2.262833336   -1.212075397   -0.000959542
8    C     4.510770709   -2.451471155   -0.004944018
9    C     1.579170412   -2.469752248    0.000863903
10   N     1.023798904   -3.491644868    0.002538486
11   H    -0.399881994   -0.509247074    0.901413723
12   H    -0.417997953    1.027605739   -0.002932179
13   H    -0.400198387   -0.514250796   -0.898325203
14   H     1.659700633    2.173318238    0.001572045
15   H     4.129977450    2.221748248   -0.001765195
16   H     5.432367651    0.121846077   -0.006854845
17   H     4.300262218   -3.066017943    0.895157309
18   H     4.297544762   -3.065705170   -0.904619558
19   H     5.600283685   -2.243316935   -0.006468765
---
H11C9N1, RHF, CHARGE=0, MULT=1
HF=36.2
1    C    -0.000088495   -0.000028180   -0.001256983
2    C     1.576334773   -0.001091170    0.004788043
3    C     2.074053387    1.483140161   -0.009935157
4    C     0.848804620    2.447097822   -0.076915292
5    C     0.027646985    2.143034323   -1.368521134
6    C    -0.486794781    0.682212462   -1.319300021
7    C    -0.481645969    0.859500831    1.159849235
8    C    -0.041880482    2.137840158    1.118954736
9    H    -0.390751980   -1.036085141    0.059609679
10   H     1.931847879   -0.503285339   -0.930498118
11   C     2.126752147   -0.777013939    1.112146019
12   H     2.684753239    1.720435501    0.886813329
13   H     2.743732895    1.644416317   -0.881626754
14   H     1.191632104    3.501947639   -0.074469619
15   H    -0.822197296    2.853927584   -1.452832887
16   H     0.651421082    2.309666567   -2.272518688
17   H    -1.596925511    0.657727932   -1.363425659
18   H    -0.138184164    0.111128418   -2.206277113
19   H    -1.123500667    0.429983446    1.921747990
20   H    -0.269790966    2.913074466    1.842811941
21   N     2.573346675   -1.410026730    1.978154998
```

```
---
H11C9N1, RHF, CHARGE=0, MULT=1
HF=36.6
1     C      0.000258374    -0.000114420     0.000136872
2     C      1.576600954    -0.000206577    -0.000060588
3     C      2.073621467     1.484000630    -0.000009408
4     C      0.848252613     2.446952396     0.073406581
5     C     -0.035598834     2.242054932    -1.196303044
6     C     -0.544815312     0.779240017    -1.237539495
7     C     -0.430667152     0.768948586     1.243517126
8     C      0.013177491     2.045990879     1.282173555
9     H     -0.387386385    -1.039094770     0.006754591
10    H      1.913293798    -0.482504170     0.953796939
11    C      2.155669462    -0.800559785    -1.074113889
12    H      2.740609410     1.652389844     0.872715175
13    H      2.687138249     1.714148683    -0.896328203
14    H      1.192575646     3.499117871     0.141051956
15    H     -0.889847967     2.952693948    -1.180448912
16    H      0.540911881     2.484253241    -2.114401388
17    H     -1.656017922     0.753697577    -1.230755344
18    H     -0.241472615     0.283154545    -2.183338885
19    H     -1.041810063     0.282675707     1.996252528
20    H     -0.178865410     2.763340388     2.072845101
21    N      2.631617060    -1.454600708    -1.908311628
---
H11C9N1, RHF, CHARGE=0, MULT=1
HF=19.6
1     C      0.000224336     0.000266259     0.000047032
2     C      1.401689810    -0.000340830     0.000377426
3     C      2.112304899     1.212706160    -0.000628709
4     C      1.415334077     2.430825285     0.003485907
5     C      0.003948376     2.465442918     0.008376650
6     C     -0.709834438     1.231198102    -0.005132392
7     N     -2.118777210     1.219305700    -0.121346602
8     C     -2.932842061     2.404109964     0.171112332
9     C     -2.146601465     3.641064641     0.644758081
10    C     -0.748749275     3.772253521     0.019230981
11    H     -0.528973123    -0.953569754    -0.001980258
12    H      1.942555640    -0.947731676     0.000132420
13    H      3.202248081     1.207284005    -0.002865411
14    H      1.983889768     3.362963542     0.004802817
15    H     -2.533380021     0.388703396     0.265965630
16    H     -3.709388267     2.152771708     0.937439895
17    H     -3.493231262     2.652189096    -0.767125831
18    H     -2.053320188     3.618545581     1.755149980
19    H     -2.742501182     4.551687628     0.409544608
20    H     -0.181254083     4.548328740     0.582065236
21    H     -0.836507753     4.154861285    -1.025064095
---
H11C9N1, RHF, CHARGE=0, MULT=1
HF=17
1     N     -0.000004249     0.000000214     0.000036489
2     C      1.349556259     0.000028114    -0.000038973
3     C      2.109011150     1.188076476     0.000092387
4     C      1.432677891     2.415639043    -0.005620993
5     C      0.019599010     2.436017578    -0.009456802
6     C     -0.663284049     1.185508565    -0.001873025
7     C     -2.176162124     1.079804303     0.033008135
8     C     -2.904328799     2.389435703    -0.318184788
```

```
9    C    -2.234604073    3.643155474    0.263033943
10   C    -0.731543560    3.746014317   -0.050807112
11   H     1.828059672   -0.985112185    0.000584395
12   H     3.197247447    1.148254496    0.003417912
13   H     2.004352179    3.344977025   -0.008337338
14   H    -2.481314875    0.751330336    1.055192668
15   H    -2.523681637    0.284027649   -0.664487530
16   H    -3.952669205    2.329991618    0.053036790
17   H    -2.980550523    2.487301186   -1.425394412
18   H    -2.384680744    3.667447706    1.366847672
19   H    -2.748442960    4.548686645   -0.132087409
20   H    -0.272726996    4.465698328    0.666415139
21   H    -0.598970525    4.195451897   -1.064064663
---
H13C9N1, RHF, CHARGE=0, MULT=1
HF=17.4
1    C    -0.026358752   -0.013655146    0.063522532
2    N     1.436756410   -0.070433783    0.089097401
3    C     2.048902257   -0.912400244    1.119577637
4    C     2.204622556    0.412544044   -1.000286284
5    C     3.616090687    0.235604609   -1.044004936
6    C     4.369483835    0.759336320   -2.105019856
7    C     3.755675834    1.465192452   -3.148785176
8    C     2.355593291    1.657706282   -3.142480717
9    C     1.599499070    1.128764436   -2.074545235
10   C     1.695607617    2.418419155   -4.262968613
11   H    -0.383954984    1.038357550   -0.024212744
12   H    -0.454163193   -0.608972395   -0.777993546
13   H    -0.461978098   -0.417536779    1.005104302
14   H     2.471185498   -1.854661136    0.696138810
15   H     2.861065800   -0.363983007    1.650970133
16   H     1.305808883   -1.208286644    1.893897528
17   H     4.161410736   -0.302491298   -0.268833031
18   H     5.451179799    0.613411917   -2.113774354
19   H     4.368070254    1.860695715   -3.959720926
20   H     0.521030945    1.289638392   -2.098837024
21   H     2.102047999    3.449658550   -4.333284450
22   H     1.870168917    1.916203159   -5.238019490
23   H     0.598156514    2.507863513   -4.131541164
---
H13C9N1, RHF, CHARGE=0, MULT=1
HF=16.5
1    C    -0.038295098   -0.026489299    0.061121959
2    N     1.419713065   -0.157922782    0.089932209
3    C     1.999834285   -0.981624976    1.152466162
4    C     2.212467763    0.302814128   -0.987346167
5    C     1.645115760    1.023623176   -2.077585172
6    C     2.443843459    1.505099449   -3.124395159
7    C     3.841575406    1.299675662   -3.152744090
8    C     4.404584637    0.581928842   -2.075953059
9    C     3.620384090    0.093042933   -1.020390381
10   C     4.679887316    1.829898411   -4.281605523
11   H    -0.342945632    1.043787775   -0.007668654
12   H    -0.489334772   -0.582367491   -0.794924236
13   H    -0.499320800   -0.429639245    0.990631740
14   H     2.417644906   -1.937970034    0.756916137
15   H     2.809134732   -0.432462244    1.687605785
16   H     1.237445593   -1.251451434    1.917309061
17   H     0.576560195    1.230026941   -2.141534745
```

```
18   H    1.956996495    2.053046374   -3.934497962
19   H    5.480009082    0.392293208   -2.046581771
20   H    4.143424315   -0.450355749   -0.233229244
21   H    4.330975472    1.433947123   -5.258763224
22   H    4.628755915    2.938593024   -4.327619390
23   H    5.749639971    1.555094650   -4.180423208
---
H13C9N1, RHF, CHARGE=0, MULT=1
HF=7.3
1    C   -0.053949908    0.178602707   -0.194950932
2    C    0.755945143    1.338259471   -0.296965188
3    C    2.070730251    1.358268625    0.220977913
4    C    2.571344066    0.190828249    0.843109104
5    C    1.780181803   -0.964666905    0.938445929
6    C    0.473273253   -0.980387629    0.427122800
7    N   -1.341974837    0.175900955   -0.817787891
8    C   -2.428641031    0.980566106   -0.234781477
9    C   -3.007805756    0.503609267    1.107463833
10   C    2.919717320    2.598243441    0.122421711
11   H    0.366822945    2.230728326   -0.791165020
12   H    3.581840133    0.174383795    1.254479710
13   H    2.183600641   -1.859108035    1.415407081
14   H   -0.116813951   -1.893828684    0.515766934
15   H   -1.660873816   -0.770825368   -0.960800903
16   H   -2.085459433    2.037894115   -0.126806490
17   H   -3.250861621    0.994667096   -0.995244984
18   H   -3.388232726   -0.536569180    1.052617717
19   H   -2.262761037    0.550926815    1.926646592
20   H   -3.859817176    1.153178861    1.395904955
21   H    2.630008073    3.236632165   -0.737198990
22   H    2.819991117    3.210006481    1.044546161
23   H    3.994281423    2.352696509   -0.004900083
---
H17C9N1, RHF, CHARGE=0, MULT=1
HF=-15.9
1    C   -0.000280930   -0.000075192   -0.000157611
2    N    1.483961503    0.000011102    0.001182073
3    C   -0.468950203    1.474969603    0.000076044
4    C    0.800724507    2.340776018   -0.057083259
5    C    2.024720886    1.380399666   -0.043738846
6    C    2.019061717   -0.947087214    1.009668908
7    C    0.796700158   -1.512134023    1.789695316
8    C   -0.470433707   -0.898892573    1.170056337
9    H   -0.330139280   -0.474318285   -0.961845490
10   H   -1.077775189    1.728945510    0.893390330
11   H   -1.124662518    1.670469405   -0.875492239
12   H    0.848465636    3.029940439    0.813901008
13   H    0.788699107    2.992456015   -0.956049357
14   H    2.630265253    1.612729076    0.872664549
15   C    2.970260228    1.546144298   -1.252534706
16   H    2.685282296   -0.425122408    1.746514884
17   C    2.881316952   -2.026485771    0.321004498
18   H    0.877289542   -1.247940036    2.866519655
19   H    0.752434632   -2.621051610    1.754939472
20   H   -1.047734127   -0.337082727    1.934039275
21   H   -1.155690510   -1.689724069    0.796705662
22   H    3.864532786    0.898080754   -1.157406307
23   H    3.333533922    2.594076599   -1.309864744
24   H    2.479346888    1.305146470   -2.216169981
```

```
25      H      3.254735758    -2.751203976     1.075002141
26      H      3.769950474    -1.581599931    -0.169669712
27      H      2.321539475    -2.596062493    -0.447299023
---
H17C9N1, RHF, CHARGE=0, MULT=1
HF=-27
1       C     -0.000759594    -0.035347095    -0.019644146
2       C      0.280829981     1.343008722    -0.710238500
3       C      1.672686072     1.911516941    -0.322645965
4       C      2.813837103     0.902308234    -0.548922782
5       C      2.535502101    -0.509207851    -0.005149338
6       C      1.128885114    -1.049376424    -0.323226944
7       C     -1.410344335    -0.561407703    -0.386228086
8       C     -2.515070964     0.495111054    -0.206541506
9       C     -2.147293557     1.891014100    -0.757498306
10      N     -0.778321621     2.339478475    -0.462234827
11      H      0.001990921     0.118571057     1.090481221
12      H      0.294539007     1.179784478    -1.822829987
13      H      1.685767874     2.240152987     0.741770873
14      H      1.881136242     2.823064149    -0.927475136
15      H      3.738659418     1.296842120    -0.069576028
16      H      3.043112383     0.839514114    -1.637694331
17      H      2.693791922    -0.518063499     1.097911566
18      H      3.291905268    -1.212058909    -0.423185892
19      H      0.963225814    -1.973814840     0.275594688
20      H      1.090588873    -1.366457416    -1.390174509
21      H     -1.652867573    -1.442304363     0.250706812
22      H     -1.427021697    -0.932271342    -1.436299252
23      H     -2.778066601     0.577700123     0.873186876
24      H     -3.440904624     0.142791322    -0.714808661
25      H     -2.859279847     2.648757150    -0.347764958
26      H     -2.283072162     1.904576151    -1.867570991
27      H     -0.738374174     2.663520220     0.493848356
---
H19C9N1, RHF, CHARGE=0, MULT=1
HF=-38.2
1       C      0.001342316     0.066833922    -0.059003545
2       C      1.317672540    -0.771375982    -0.186176246
3       C      2.464817442    -0.023472088     0.572055383
4       C      3.886995534    -0.411456681     0.147710143
5       C      4.079016326    -0.434109656    -1.374146271
6       C      2.967272597    -1.194887879    -2.171073286
7       C      2.988874526    -0.705381226    -3.658142976
8       C      3.235418096    -2.732204096    -2.178614223
9       C      1.044720164    -2.172068751     0.445521088
10      N      1.599393098    -0.956888839    -1.635602523
11      H      0.098292219     1.077630664    -0.507080090
12      H     -0.857317151    -0.431295438    -0.553153540
13      H     -0.271061565     0.213290856     1.006561923
14      H      2.360578696    -0.192694626     1.667771690
15      H      2.351272224     1.077466140     0.433722224
16      H      4.167943317    -1.394496601     0.588085691
17      H      4.604103551     0.320469038     0.587832897
18      H      5.073749799    -0.880029304    -1.601980475
19      H      4.142405606     0.620621551    -1.731096842
20      H      2.239706016    -1.236089580    -4.280292857
21      H      2.783109130     0.381461210    -3.748859795
22      H      3.983204616    -0.882886927    -4.117229672
23      H      3.431730627    -3.142182737    -1.168970024
```

```
24      H       2.383953609     -3.297265644    -2.608227967
25      H       4.130672352     -2.970435982    -2.791495147
26      H       0.344078403     -2.771560525    -0.169833187
27      H       1.965007626     -2.772307006     0.581445737
28      H       0.590421788     -2.065638935     1.453310247
29      H       1.213165596     -0.162703645    -2.126015029
---
N2, RHF, CHARGE=0, MULT=1
HF=0, IE=15.6
1       N       0.000000000      0.000000000     0.000000000
2       N       1.103816967      0.000000000     0.000000000
EXPGEOM
1       N       0.00000          0.00000         0.55000
2       N       0.00000          0.00000        -0.55000
---
H2N2, RHF, CHARGE=0, MULT=1
HF=36
1       N       0.000000000      0.000000000     0.000000000
2       N       1.220100431      0.000000000     0.000000000
3       H       1.591759007      0.955076514     0.000000000
4       H      -0.371652839     -0.955079619    -0.000010000
EXPGEOM
1       H      -0.99080         -0.90650         0.00000
2       H       0.99080          0.90650         0.00000
3       N       0.00000         -0.62520         0.00000
4       N       0.00000          0.62520         0.00000
---
H4N2, RHF, CHARGE=0, MULT=1
HF=22.8
1       N       0.000000000      0.000000000     0.000000000
2       N       1.397375652      0.000000000     0.000000000
3       H       1.698450781      0.975648398     0.000000000
4       H      -0.301439292     -0.975511801    -0.004573194
5       H      -0.301081591      0.331562417     0.917577992
6       H       1.698810811     -0.335807186    -0.915904767
EXPGEOM
1       N       0.00000          0.71670        -0.07590
2       N       0.00000         -0.71670        -0.07590
3       H      -0.22100          1.08800         0.83900
4       H       0.22100         -1.08800         0.83900
5       H       0.93550          1.01440        -0.30800
6       H      -0.93550         -1.01440        -0.30800
---
H2C1N2, RHF, CHARGE=0, MULT=1
HF=71, DIP=1.5, IE=9
1       C       0.000000000      0.000000000     0.000000000
2       H       1.086519936      0.000000000     0.000000000
3       H      -0.551667619      0.936083329     0.000000000
4       N      -0.649985227     -1.137475742     0.001160486
5       N      -1.216383450     -2.128782388     0.001970335
EXPGEOM
1       C       0.00000          0.00000        -1.14460
2       N       0.00000          0.00000         0.15520
3       N       0.00000          0.00000         1.29450
4       H       0.00000          0.95290        -1.64000
5       H       0.00000         -0.95290        -1.64000
---
H2C1N2, RHF, CHARGE=0, MULT=1
HF=79, DIP=1.59
```

```
1    N      0.000004410     0.000001322    -0.000001284
2    N      1.229398593    -0.000026170    -0.000000023
3    C      0.614914452     1.355219093     0.000000201
4    H      0.615139082     1.951276056    -0.922389643
5    H      0.615129424     1.951377342     0.922397408
EXPGEOM
1    C      0.00000      0.00000      0.80550
2    N      0.00000      0.61360     -0.53730
3    N      0.00000     -0.61360     -0.53730
4    H      0.93050      0.00000      1.34480
5    H     -0.93050      0.00000      1.34480
---
H6C1N2, RHF, CHARGE=0, MULT=1
HF=22.6, DIP=1.66, IE=9.3
1    N      0.034191542    -0.067458423    -0.008810794
2    N      1.390473450    -0.281510141    -0.273285775
3    H      1.791099209     0.622430380    -0.534353736
4    H     -0.414726855    -0.984799799     0.025931712
5    H     -0.373131566     0.351616190    -0.846730301
6    C      2.130610185    -0.738614368     0.924109204
7    H      2.042815664    -0.067349273     1.809364282
8    H      1.787468330    -1.753376730     1.224809522
9    H      3.208641806    -0.807510713     0.652840343
---
C2N2, RHF, CHARGE=0, MULT=1
HF=73.8, IE=13.36
1    N     -0.001285990     0.000000000     0.000000000
2    C      1.160285441     0.000000000     0.000000000
3    C      2.538091784     0.000000000     0.000000000
4    N      3.699697144     0.000000000     0.000000000
EXPGEOM
1    C      0.00000      0.00000      0.69140
2    C      0.00000      0.00000     -0.69140
3    N      0.00000      0.00000      1.85000
4    N      0.00000      0.00000     -1.85000
---
H8C2N2, RHF, CHARGE=0, MULT=1
HF=20
1    N      0.044676508    -0.023383023     0.051428599
2    N      1.444543147    -0.014092932     0.004302142
3    C      1.975471244     1.370440613     0.073053818
4    H     -0.252984310    -0.997770569     0.137106150
5    H     -0.231479693     0.364226904     0.956627267
6    C      1.953247497    -0.770708517    -1.167066449
7    H      3.066264686    -0.761296697    -1.146663526
8    H      1.631214357    -1.834865240    -1.101605112
9    H      1.623592088    -0.373134541    -2.155107562
10   H      3.087952768     1.330471323     0.073914864
11   H      1.654873689     2.031733271    -0.765144907
12   H      1.662888703     1.851816010     1.027356983
---
H8C2N2, RHF, CHARGE=0, MULT=1
HF=22
1    N      0.047699608    -0.060187691    -0.009553628
2    N      1.414922506    -0.309476937    -0.178667604
3    H      1.743125538     0.312713629    -0.922124746
4    H     -0.280534574    -0.683732627     0.732814952
5    C     -0.743436397    -0.407108078    -1.212420143
6    H     -1.821273299    -0.296773306    -0.954424193
```

```
7      H    -0.516688153     0.301437944    -2.039835878
8      H    -0.586822296    -1.443200278    -1.591744503
9      C     2.205567051     0.040035802     1.023823150
10     H     2.048747840     1.076979057     1.400747249
11     H     1.978352941    -0.666648275     1.852706800
12     H     3.283522853    -0.070906190     0.766621864
EXPGEOM
1      N    -0.16360     0.69980    -0.69880
2      N     0.16360    -0.69980    -0.69880
3      C     0.16360     1.43350     0.53370
4      C    -0.16360    -1.43350     0.53370
5      H    -1.17740     0.74100    -0.84720
6      H     1.17740    -0.74100    -0.84720
7      H    -0.10420     2.49470     0.37610
8      H    -0.34540     1.08130     1.45780
9      H     1.25690     1.38100     0.70280
10     H     0.10420    -2.49470     0.37610
11     H     0.34540    -1.08130     1.45780
12     H    -1.25690    -1.38100     0.70280
---
H4C3N2, RHF, CHARGE=0, MULT=1
HF=42.9
1      N    -0.000063318    -0.000058799     0.000002216
2      C     1.354898033     0.000019260    -0.000001094
3      C     1.851661561     1.351375541     0.000001233
4      C     0.716064034     2.160806559    -0.000002388
5      N    -0.369658826     1.280652346     0.000000066
6      H     1.924817720    -0.921281750    -0.000005996
7      H     2.885317318     1.657497014     0.000002835
8      H     0.587375950     3.234347442    -0.000016542
9      H    -1.341385284     1.527233620     0.000012881
EXPGEOM
1      H     2.10660     0.73130     0.00000
2      C     1.11180     0.30530     0.00000
3      H     1.26550    -1.89780     0.00000
4      C     0.66480    -0.99830     0.00000
5      H    -1.48230    -1.68350     0.00000
6      C    -0.74440    -0.89030     0.00000
7      N    -1.14750     0.38020     0.00000
8      H    -0.05070     2.09340     0.00000
9      N     0.00000     1.08500     0.00000
---
H4C3N2, RHF, CHARGE=0, MULT=1
HF=31.8
1      C     0.000000000     0.000000000     0.000000000
2      C     1.393626942     0.000000000     0.000000000
3      N     1.760497446     1.351067944     0.000000000
4      C     0.590243108     2.111763916    -0.000015074
5      N    -0.483226888     1.307265427    -0.000013883
6      H    -0.682932126    -0.839994760     0.000005177
7      H     2.122647360    -0.796797132     0.000006732
8      H     2.690787053     1.699284539     0.000061074
9      H     0.582863034     3.196274919    -0.000016370
EXPGEOM
1      N     0.00000     1.11480     0.00000
2      C    -1.09900     0.28620     0.00000
3      C     1.13060     0.30560     0.00000
4      N    -0.75180    -0.99760     0.00000
5      C     0.64220    -0.99340     0.00000
```

```
 6     H    -0.01390     2.12680     0.00000
 7     H    -2.11840     0.67050     0.00000
 8     H     2.13930     0.71410     0.00000
 9     H     1.21220    -1.92190     0.00000
---
H10C3N2, RHF, CHARGE=0, MULT=1
HF=-12.8
 1    N     0.015301762   -0.033258910    0.007001089
 2    C     1.467292685    0.142788942    0.182882454
 3    C     1.941594152    1.615275967    0.166658381
 4    C     1.934156580   -0.613170355    1.471411875
 5    N     1.639121158   -2.044129291    1.555065058
 6    H    -0.497618523    0.564615261    0.637240834
 7    H    -0.253389421    0.265014909   -0.918291644
 8    H     1.957833203   -0.354347722   -0.698983972
 9    H     1.528301604    2.205688396    1.009255269
10    H     3.047372005    1.672138041    0.228297435
11    H     1.642823135    2.120810485   -0.774983979
12    H     3.046376001   -0.506854549    1.545456577
13    H     1.507442980   -0.121709958    2.380514644
14    H     0.647641480   -2.201711223    1.643911566
15    H     1.914070677   -2.514514356    0.706960418
EXPGEOM
 1    N     0.47400     1.38530    -0.22260
 2    H    -0.28300     1.92930     0.19090
 3    H     1.33330     1.86250     0.05180
 4    N    -2.05420    -0.13260     0.02320
 5    H    -2.11290     0.75630    -0.47150
 6    H    -2.17970     0.06830     1.01580
 7    C    -0.73950    -0.73600    -0.20920
 8    H    -0.74900    -1.74760     0.21990
 9    H    -0.61200    -0.84210    -1.29410
10    C     1.76910    -0.67300    -0.03160
11    H     1.77580    -1.70780     0.33070
12    H     1.88540    -0.67960    -1.12170
13    H     2.63640    -0.15890     0.40220
14    C     0.46770     0.02790     0.34450
15    H     0.38310     0.03750     1.44960
---
C4N2, RHF, CHARGE=0, MULT=1
HF=126.5
 1    N     0.000000000    0.000000000    0.000000000
 2    C     1.163451925    0.000000000    0.000000000
 3    C     2.535264783    0.000000000    0.000000000
 4    C     3.735373452    0.000000000    0.000000000
 5    C     5.107186310    0.000000000    0.000000000
 6    N     6.270638219   -0.000203061    0.000000000
EXPGEOM
 1    C     0.00000     0.00000     0.60640
 2    C     0.00000     0.00000    -0.60640
 3    C     0.00000     0.00000     1.99440
 4    C     0.00000     0.00000    -1.99440
 5    N     0.00000     0.00000     3.16220
 6    N     0.00000     0.00000    -3.16220
---
H2C4N2, RHF, CHARGE=0, MULT=1
HF=81.3
 1    C     0.000024778    0.000029099    0.000012458
 2    C     1.352714309    0.000050073   -0.000012275
```

```
3    H      1.930642787     0.931173965     0.000013691
4    C     -0.792406890     1.182966866     0.000071089
5    N     -1.458240876     2.136272678     0.000137549
6    H     -0.578227429    -0.930907368    -0.000010019
7    C      2.145669800    -1.182566346    -0.000099952
8    N      2.811662265    -2.135748470    -0.000000725
EXPGEOM
1    C     -0.33660    0.57920    0.00000
2    C      0.33660   -0.57920    0.00000
3    C      0.33660    1.83440    0.00000
4    C     -0.33660   -1.83440    0.00000
5    N      0.86430    2.86710    0.00000
6    N     -0.86430   -2.86710    0.00000
7    H     -1.41530    0.59640    0.00000
8    H      1.41530   -0.59640    0.00000
---
H4C4N2, RHF, CHARGE=0, MULT=1
HF=47
1    C      0.000004664    -0.000011404     0.000008343
2    C      1.409309134     0.000144054    -0.000012813
3    N      2.112220454     1.155863113    -0.000011453
4    C      1.412054944     2.318534086     0.000014443
5    N      0.057687165     2.407388398     0.000010691
6    C     -0.646325516     1.252352058     0.000002824
7    H      1.983379800     3.256099850     0.000037265
8    H     -1.737078377     1.344738932    -0.000012758
9    H     -0.565778585    -0.928696284     0.000043709
10   H      1.991991395    -0.926549948    -0.000044864
EXPGEOM
1    C      0.00000     0.00000     1.35710
2    C      0.00000     0.00000    -1.30750
3    C      0.00000     1.18650     0.62240
4    C      0.00000    -1.18650     0.62240
5    N      0.00000     1.20230    -0.71770
6    N      0.00000    -1.20230    -0.71770
7    H      0.00000     0.00000     2.44210
8    H      0.00000     0.00000    -2.39520
9    H      0.00000     2.15620     1.11700
10   H      0.00000    -2.15620     1.11700
---
H4C4N2, RHF, CHARGE=0, MULT=1
HF=46.9
1    C      0.000008176     0.000013643    -0.000023215
2    C      1.416468976    -0.000050548     0.000012544
3    N      2.122827687     1.152983243    -0.000009532
4    C      1.416428465     2.306029198    -0.000047769
5    C     -0.000039426     2.306097292    -0.000125817
6    N     -0.706400191     1.153025043    -0.000129076
7    H     -0.576888653     3.235371790    -0.000139954
8    H      1.993313432     3.235268286    -0.000034301
9    H     -0.576839492    -0.929224478     0.000027884
10   H      1.993138405    -0.929306364     0.000044475
EXPGEOM
1    N      0.00000     0.00000     1.42520
2    N      0.00000     0.00000    -1.42520
3    C      0.00000     1.13220     0.70210
4    C      0.00000    -1.13220     0.70210
5    C      0.00000    -1.13220    -0.70210
6    C      0.00000     1.13220    -0.70210
```

```
  7    H     0.00000     2.07710     1.25840
  8    H     0.00000    -2.07710     1.25840
  9    H     0.00000    -2.07710    -1.25840
 10    H     0.00000     2.07710    -1.25840
---
H4C4N2, RHF, CHARGE=0, MULT=1
HF=66.5
  1    C     0.000138954     0.000012924     0.000035871
  2    C     1.410373408    -0.000034662    -0.000039335
  3    N     2.123826708     1.166383046     0.000045798
  4    N     1.538946913     2.299989377     0.000304162
  5    C     0.174864091     2.394768295     0.000538307
  6    C    -0.642486622     1.245492289     0.000395501
  7    H    -0.236864036     3.407027089     0.000910164
  8    H    -1.728068663     1.328412517     0.000563950
  9    H    -0.561192943    -0.932913523    -0.000169263
 10    H     1.996363779    -0.922393329    -0.000150799
EXPGEOM
  1    C     0.00000     0.69150     1.18570
  2    C     0.00000    -0.69150     1.18570
  3    C     0.00000    -1.32200    -0.07060
  4    C     0.00000     1.32200    -0.07060
  5    H     0.00000     1.27460     2.10180
  6    H     0.00000    -1.27460     2.10180
  7    H     0.00000    -2.40560    -0.15720
  8    H     0.00000     2.40560    -0.15720
  9    N     0.00000     0.66940    -1.23360
 10    N     0.00000    -0.66940    -1.23360
---
H4C4N2, RHF, CHARGE=0, MULT=1
HF=50.1
  1    C    -0.000009031    -0.000089577    -0.000140430
  2    C     1.546732683     0.000014772     0.000019893
  3    C     2.110815378     1.342848642    -0.000132578
  4    H     1.927786935    -0.546296564     0.895129739
  5    H     1.927504574    -0.546495967    -0.895062158
  6    H    -0.380588094     0.541269364    -0.898473277
  7    C    -0.564084419    -1.342901706     0.006718809
  8    H    -0.381210141     0.550976418     0.891946895
  9    N     2.580572202     2.405502795    -0.000284559
 10    N    -1.033795311    -2.405544256     0.011988888
EXPGEOM
  1    C     0.42780     0.64240     0.00000
  2    C    -0.42780    -0.64240     0.00000
  3    C    -0.42780     1.84200     0.00000
  4    C     0.42780    -1.84200     0.00000
  5    N     1.12170    -2.77270     0.00000
  6    N    -1.12170     2.77270     0.00000
  7    H     1.07270     0.66800     0.88420
  8    H     1.07270     0.66800    -0.88420
  9    H    -1.07270    -0.66800     0.88420
 10    H    -1.07270    -0.66800    -0.88420
---
H6C4N2, RHF, CHARGE=0, MULT=1
HF=21.5
  1    N    -0.000024467    -0.000031143     0.000030248
  2    C     1.397670279     0.000015216    -0.000004927
  3    C     1.762744180     1.345026259    -0.000003195
  4    N     0.629606562     2.151160735     0.000021231
```

```
5    C     -0.437385602    1.329469809    0.000037787
6    C     -1.872548736    1.741405142    0.000146469
7    H     -0.574292700   -0.811344743    0.000023636
8    H      1.975531343   -0.912620679   -0.000025478
9    H      2.752247045    1.784533776   -0.000004395
10   H     -2.399190123    1.352201611   -0.897343304
11   H     -2.398367310    1.354066798    0.898908841
12   H     -1.974486419    2.845365571   -0.000967855
EXPGEOM
1    C      2.10670   -0.02400    0.00000
2    H      2.49490    0.48610    0.88230
3    H      2.48100   -1.04290   -0.00110
4    H      2.49490    0.48800   -0.88110
5    N     -0.16520    1.04870    0.00010
6    H      0.15710    1.99940   -0.00050
7    C      0.61980   -0.06980   -0.00010
8    N     -0.11970   -1.15840   -0.00000
9    C     -1.42670   -0.72550    0.00000
10   H     -2.24720   -1.42000    0.00010
11   C     -1.48120    0.63940   -0.00010
12   H     -2.29770    1.33640    0.00000
---
H8C4N2, RHF, CHARGE=0, MULT=1
HF=27.3
1    N     -0.000003076    0.000017891   -0.000394234
2    C      1.161495812    0.000038143    0.000299850
3    C      2.628562588    0.010123907   -0.000088509
4    N      3.137510122   -1.297897557    0.421623119
5    C      3.690655921   -2.169173472   -0.615564520
6    C      3.727884665   -1.389749451    1.757203979
7    H      2.979273002    0.832617394    0.674677289
8    H      2.983889267    0.287571863   -1.025564952
9    H      3.012609620   -2.201725923   -1.498225893
10   H      4.698134517   -1.839911354   -0.969500079
11   H      3.788835408   -3.209252893   -0.230623814
12   H      3.071347779   -0.892806940    2.506914321
13   H      4.738736263   -0.917495954    1.820333883
14   H      3.831873951   -2.456435571    2.058484493
---
H8C4N2, RHF, CHARGE=0, MULT=1
HF=13.2
1    N      0.000034758   -0.000039042   -0.000047748
2    C      1.453781058    0.000019371    0.000028885
3    C      2.098550717    1.402489278    0.000002461
4    C      1.354127043    2.441066373   -0.865985437
5    N     -0.074371751    2.159833245   -1.024363516
6    C     -0.662435264    1.021198681   -0.439342627
7    H      1.816184374   -0.585665764   -0.882562527
8    H      1.785696441   -0.562321342    0.907386966
9    H      3.148394345    1.323748677   -0.360060645
10   H      2.165705622    1.778732430    1.046741798
11   H      1.805205818    2.475947285   -1.890425327
12   H      1.499782523    3.460293003   -0.427740498
13   H     -0.649343460    2.980986929   -0.994664689
14   H     -1.762040172    1.001547931   -0.427386434
---
H10C4N2, RHF, CHARGE=0, MULT=1
HF=6
1    N      0.000033394   -0.000044696    0.000116305
```

```
 2    C          1.468818159    -0.000180642     0.000139654
 3    C          2.117941247     1.406676004    -0.000013438
 4    N          1.470230202     2.292406295    -0.973119328
 5    C          0.006882569     2.372201040    -0.914005477
 6    C         -0.634278513     0.961631076    -0.911022354
 7    H         -0.337458601     0.152358778     0.940046138
 8    H          1.814068725    -0.577352111    -0.893107787
 9    H          1.827845757    -0.562763125     0.896590029
10    H          3.194892644     1.316272168    -0.285526206
11    H          2.112242258     1.830561000     1.038796110
12    H          1.859173460     3.219961008    -0.891720343
13    H         -0.340756600     2.933712529    -1.815972763
14    H         -0.356987851     2.959316625    -0.029765038
15    H         -1.713878497     1.056812686    -0.638143946
16    H         -0.606947899     0.530240312    -1.942141439
---
H6C5N2, RHF, CHARGE=0, MULT=1
HF=28.2
 1    N          0.000198494     0.000078341     0.000061826
 2    C          1.416352405    -0.000002313    -0.000118665
 3    C          2.162871150     1.210738110    -0.000243449
 4    C          3.556729576     1.134452210    -0.129613626
 5    C          4.166591917    -0.126082176    -0.252070335
 6    C          3.350568931    -1.275376610    -0.234584660
 7    N          2.006514382    -1.223928031    -0.102715478
 8    H         -0.397794961    -0.912327388     0.154495802
 9    H         -0.372432161     0.616770688     0.703725061
10    H          1.671461540     2.178539557     0.089249105
11    H          4.159396574     2.043143832    -0.137246714
12    H          5.246029680    -0.217766412    -0.357728778
13    H          3.778882977    -2.279802631    -0.322533848
---
H6C5N2, RHF, CHARGE=0, MULT=1
HF=34.5
 1    N         -0.022603568    -0.018843338     0.057523961
 2    C          2.443837660    -0.041659524    -0.116583859
 3    C          1.210436052     0.671886638    -0.096210530
 4    C          1.260829596     2.086961029    -0.104436398
 5    C          2.514312576     2.716555383    -0.132151463
 6    C          3.681967807     1.926841548    -0.150675629
 7    N          3.643853307     0.574134976    -0.144511684
 8    H         -0.783004263     0.481550841    -0.374454780
 9    H          0.004021580    -0.927394424    -0.377596305
10    H          0.353347063     2.691463758    -0.086481107
11    H          2.586850108     3.803538676    -0.138377308
12    H          4.678613936     2.378647299    -0.171543032
13    H          2.481665862    -1.136632957    -0.108764508
---
H6C5N2, RHF, CHARGE=0, MULT=1
HF=31.1
 1    N         -0.000066183    -0.000167286    -0.000032535
 2    C          3.736996674    -0.000037353    -0.000071279
 3    C          2.416703001     0.489841753    -0.000007044
 4    C          1.343810644    -0.430159343    -0.114916550
 5    C          1.665070907    -1.806373670    -0.232712785
 6    C          3.019433096    -2.192420626    -0.222505448
 7    N          4.042554675    -1.313809419    -0.110004687
 8    H         -0.136726934     0.935250558    -0.345010345
 9    H         -0.641034164    -0.598306672    -0.493962112
```

```
10      H     2.245377205    1.561426348    0.094494576
11      H     0.893395556   -2.569985434   -0.324170540
12      H     3.304117351   -3.246824882   -0.308461136
13      H     4.589727454    0.681759087    0.090441107
---
H6C5N2, RHF, CHARGE=0, MULT=1
HF=47.1
1       C    -0.000089859    0.000044974    0.000013101
2       C     1.561861947    0.000012859   -0.000021346
3       C     2.124472990    1.457069476   -0.000026741
4       C     2.051007079   -0.713611625   -1.190314306
5       N     2.435433958   -1.274082221   -2.131833288
6       C     2.051129697   -0.713325538    1.190383400
7       N     2.435406737   -1.273632157    2.132063164
8       H    -0.396182289    0.517223873    0.896269541
9       H    -0.400444596   -1.033213224   -0.000268142
10      H    -0.396150025    0.517714288   -0.895968876
11      H     1.784613948    2.012901013    0.896219896
12      H     3.232663431    1.458078552   -0.000047131
13      H     1.784478554    2.013009527   -0.896155701
---
H8C5N2, RHF, CHARGE=0, MULT=1
HF=16.3
1       C     0.025019309   -0.054505406   -0.169071972
2       N     1.270647036    0.218212092    0.264351287
3       C     1.622027178    1.456444671   -0.262538617
4       C     0.580399182    1.963805573   -1.036444323
5       N    -0.428416944    0.998079254   -0.971461286
6       C    -0.772240172   -1.282090054    0.159947330
7       C    -0.730186194   -2.373543112   -0.916530988
8       H     2.590472776    1.891398776   -0.050070138
9       H     0.469142023    2.877618979   -1.601270506
10      H    -1.314379166    1.063709441   -1.417483895
11      H    -0.398964456   -1.708994606    1.120196273
12      H    -1.834813258   -0.998075731    0.349241716
13      H    -1.235126447   -2.050022079   -1.849283676
14      H     0.308289127   -2.661867070   -1.172582637
15      H    -1.252226679   -3.282917373   -0.553862227
---
H10C5N2, RHF, CHARGE=0, MULT=1
HF=15.2
1       N    -0.000069651    0.000073618    0.000163266
2       C     1.482900333   -0.000059517   -0.000083057
3       C     2.074142795    1.418166111   -0.000068400
4       C    -0.708486374   -1.126637803    0.656172716
5       C    -1.986263032   -0.667292909    1.375481189
6       C    -0.581660523    0.446346127   -1.154720298
7       N    -1.099278323    0.885003411   -2.099526185
8       H     1.887390608   -0.579197302   -0.868795945
9       H     1.836176809   -0.527079673    0.919988295
10      H     1.889822334    1.954970428   -0.951650790
11      H     1.673310581    2.036416041    0.827677432
12      H     3.174271579    1.349198099    0.132993233
13      H    -0.955916370   -1.932673843   -0.080177305
14      H    -0.027221305   -1.593505866    1.408654732
15      H    -2.762474653   -0.306887868    0.671688096
16      H    -1.783436300    0.140067643    2.107058219
17      H    -2.415870421   -1.526571355    1.931995413
---
```

```
H12C5N2, RHF, CHARGE=0, MULT=1
HF=18.9
1    C    -0.069812828    -0.031998799    -0.003679077
2    C     1.453557620    -0.096299218     0.137213244
3    C     2.143495799     1.278804149     0.229618358
4    C     3.681983665     1.192974200     0.341348196
5    N     4.240858516     2.540073262     0.600721493
6    N     4.994830392     2.995890141    -0.245579170
7    C     5.559041183     4.331963402     0.016452059
8    H    -0.542943976     0.483555465     0.857108779
9    H    -0.376349652     0.498750228    -0.928051479
10   H    -0.492841125    -1.056748878    -0.052175327
11   H     1.863790720    -0.663820162    -0.729780855
12   H     1.698927002    -0.696580439     1.043719483
13   H     1.876824664     1.879359504    -0.670740710
14   H     1.735770055     1.828618113     1.108469945
15   H     3.977609248     0.571920351     1.223305077
16   H     4.105454685     0.690012669    -0.561515819
17   H     5.242493450     5.001765679    -0.814824530
18   H     5.265873273     4.805758313     0.979056612
19   H     6.668533039     4.241050828    -0.005630750
---
H14C4N2, RHF, CHARGE=0, MULT=1
HF=-8.3
1    N     1.538973936     0.368111912    -0.856555342
2    C     2.660906709     0.884423514    -0.072761567
3    C     1.833541757    -0.541851741    -1.974619484
4    C     1.752758932    -2.021241574    -1.527208801
5    C     2.023443663    -3.012017197    -2.683390459
6    N     2.053506848    -4.398311517    -2.201900083
7    H    -4.452986873    -0.238224379     1.598178009
8    H     1.000028510     1.147266436    -1.204108236
9    H    -4.088837065    -0.195328824     2.150929272
10   H     3.203805642     0.052173964     0.427349984
11   H     3.404852460     1.465817384    -0.670585534
12   H     2.264867265     1.561398486     0.718338637
13   H     2.833013594    -0.336562748    -2.436983496
14   H     1.081334553    -0.359032899    -2.783012627
15   H     2.484142728    -2.208167755    -0.709067713
16   H     0.742469721    -2.216669183    -1.099989777
17   H     1.262942293    -2.863621908    -3.494720353
18   H     3.017409879    -2.794610569    -3.148310314
19   H     1.126029864    -4.695783695    -1.939824415
20   H     2.319072632    -5.003628347    -2.963891628
---
H4C6N2, RHF, CHARGE=0, MULT=1
HF=67.1
1    N    -0.000478259    -0.000717567     0.000004208
2    C     1.348036570     0.000254361    -0.000000773
3    C     2.116153888     1.188906893     0.000016616
4    C     1.449471789     2.421442410     0.000103561
5    C     0.041520222     2.432631387     0.000129141
6    C    -0.649367750     1.198130234     0.000060042
7    C    -2.077712232     1.118812232     0.000008537
8    N    -3.237635779     1.055716141    -0.000041782
9    H     1.825871597    -0.985440035    -0.000033657
10   H     3.204161118     1.138395491    -0.000037598
11   H     2.007557095     3.357831359     0.000155750
12   H    -0.501126650     3.377488720     0.000201896
```

```
---
H4C6N2, RHF, CHARGE=0, MULT=1
HF=66.4
1    N     0.000176014     0.000138144    -0.000054909
2    C     1.353544213    -0.000235563     0.000002643
3    C     2.113713261     1.189122566     0.000110084
4    C     1.446187920     2.422895944     0.000005298
5    C     0.030404778     2.435399182    -0.000102968
6    C    -0.647759114     1.185436385    -0.000110835
7    C    -0.697169447     3.659055550    -0.000180712
8    N    -1.290221287     4.658896593    -0.000233394
9    H     1.832528771    -0.985179568    -0.000043104
10   H     3.202052370     1.145425434     0.000287437
11   H     2.014719889     3.353640042     0.000000813
12   H    -1.742170769     1.126388493    -0.000157094
---
H4C6N2, RHF, CHARGE=0, MULT=1
HF=67.8
1    N     0.000489782     0.000494201     0.000003945
2    C     1.353256114    -0.000735942    -0.000008933
3    C     2.122532842     1.182649045     0.000008042
4    C     1.459770577     2.431429111     0.000184965
5    C     0.045958316     2.428799677     0.000119848
6    C    -0.636901369     1.193445918     0.000066960
7    C     2.193912508     3.655344658     0.000673674
8    N     2.791773435     4.652073768     0.001216844
9    H     1.831508153    -0.986122051    -0.000042748
10   H     3.210492966     1.120582408    -0.000041066
11   H    -1.731409425     1.152255843     0.000108720
12   H    -0.520435731     3.359641210     0.000161742
---
H8C6N2, RHF, CHARGE=0, MULT=1
HF=30.1
1    N    -0.000036908    -0.000019851    -0.000044939
2    C     1.347509507     0.000071063     0.000047592
3    C     2.077721539     1.211616212     0.000087462
4    N     1.448579793     2.403274349    -0.000312551
5    C     0.090152198     2.419415554    -0.000797913
6    C    -0.648610199     1.193689401    -0.000460846
7    C    -2.155660304     1.108816278    -0.000347939
8    C    -0.538465983     3.791744944    -0.002200089
9    H     1.839980747    -0.976604986     0.000161690
10   H     3.171290961     1.232281739     0.000314990
11   H    -2.575159535     1.607380268    -0.899506807
12   H    -2.525012065     0.063402532    -0.000660286
13   H    -2.574823340     1.606762775     0.899333638
14   H     0.213555534     4.606474125     0.000359392
15   H    -1.172059402     3.930141217    -0.903584803
16   H    -1.177519313     3.929799382     0.895221613
---
H8C6N2, RHF, CHARGE=0, MULT=1
HF=35.7
1    C    -0.000011498     0.000012512    -0.000027209
2    C     1.457474496    -0.000043189     0.000003405
3    C     2.060248226     1.422060219    -0.000090518
4    C     3.601665604     1.441452024     0.000301554
5    C     4.204771779     2.863454466    -0.000062911
6    C     5.662268936     2.861924805     0.000115363
7    N     6.824098383     2.879560134     0.000261979
```

```
   8       N       -1.161780824       -0.019702818       -0.000012945
   9       H        1.808439772       -0.567230386       -0.894546320
  10       H        1.808318575       -0.567106898        0.894685343
  11       H        1.684853896        1.972435517       -0.892783941
  12       H        1.684003006        1.972622590        0.892132227
  13       H        3.977710334        0.890695249       -0.891820901
  14       H        3.977174982        0.891433278        0.893115774
  15       H        3.854354907        3.430205780       -0.894791406
  16       H        3.854153211        3.431058414        0.894343117
---
H8C6N2, RHF, CHARGE=0, MULT=1
HF=48.5
   1       N        0.302897884       -0.239968172        0.022459712
   2       N        1.032241247        0.370716101        1.051767615
   3       C        1.843961539        1.470225729        0.555652613
   4       C        1.432378500        2.776822317        0.910752007
   5       C        2.197462002        3.885595459        0.506640178
   6       C        3.369645295        3.704390643       -0.247220037
   7       C        3.780716102        2.406962509       -0.599632142
   8       C        3.024749833        1.290063324       -0.203547904
   9       H       -0.399012608        0.428841503       -0.298445285
  10       H       -0.262386514       -0.969741315        0.461279200
  11       H        1.678679461       -0.336896021        1.412821218
  12       H        0.528333141        2.940842305        1.499293530
  13       H        1.878023222        4.891504341        0.781844470
  14       H        3.959032315        4.567504286       -0.557830839
  15       H        4.690342858        2.264527127       -1.184192271
  16       H        3.363712956        0.293713831       -0.492094767
---
H12C6N2, RHF, CHARGE=0, MULT=1
HF=21.6
   1       N       -0.000241925        0.009043210       -0.003034071
   2       C        1.161828194        0.003974369       -0.001332033
   3       C        2.619432794        0.022820976        0.000104143
   4       N        5.325473022       -2.744784682        0.163709204
   5       C        5.898473468       -3.167639276        1.442454615
   6       C        5.980008212       -3.262793904       -1.038541333
   7       H        2.958868711        0.631814172        0.871369701
   8       H        2.962140198        0.560441861       -0.915838801
   9       H        6.895497543       -2.708697190        1.653130200
  10       H        5.218257145       -2.899338304        2.282213356
  11       H        6.025432069       -4.273997319        1.458446474
  12       H        6.986286949       -2.812706438       -1.222539124
  13       H        5.352667605       -3.068378386       -1.937761237
  14       H        6.112750487       -4.365678138       -0.958308653
  15       C        3.242955272       -1.389505731        0.056129777
  16       C        4.792580633       -1.375538697        0.093928059
  17       H        2.896331325       -1.969886636       -0.828548448
  18       H        2.852094659       -1.919641209        0.953784104
  19       H        5.137981581       -0.767468908        0.967519674
  20       H        5.180946421       -0.840847665       -0.809086773
---
H12C6N2, RHF, CHARGE=0, MULT=1
HF=35.9
   1       C        0.000031010       -0.000385846       -0.000060010
   2       C        1.621984537        0.000295158       -0.000202722
   3       N        1.425020046        1.501656724        0.000378958
   4       N        0.195977746        1.501252772        0.000380356
   5       C        2.349575686       -0.502817472       -1.261260138
```

```
 6    C    -0.726692152    -0.506234436     1.260420675
 7    C    -0.726902081    -0.504788785    -1.260990839
 8    C     2.349556580    -0.503948818     1.260711050
 9    H     2.230832482    -1.600851957    -1.364701178
10    H     1.983135523    -0.034498973    -2.196406809
11    H     3.436067308    -0.283367487    -1.207039453
12    H    -0.604174692    -1.604017526     1.363175509
13    H    -0.362529181    -0.037248767     2.196018243
14    H    -1.813962336    -0.290767804     1.205754818
15    H    -0.605920928    -1.602676561    -1.364251916
16    H    -0.361815886    -0.035470820    -2.196260197
17    H    -1.813898279    -0.287840755    -1.206555549
18    H     1.982899489    -0.036782825     2.196278433
19    H     2.231189963    -1.602206335     1.362856105
20    H     3.435981532    -0.284004323     1.206860504
 ---
H12C6N2, RHF, CHARGE=0, MULT=1
HF=21.6
 1    N     0.000186775    -0.000117980    -0.000126730
 2    C     1.491755944    -0.000360053     0.000122466
 3    C     2.011886991     1.470172369    -0.000210615
 4    N     0.851762964     2.407665537    -0.003816493
 5    C     0.024571758     2.170932887    -1.222253302
 6    C    -0.496415390     0.700882989    -1.219516157
 7    C    -0.497522649     0.705183939     1.216377059
 8    C     0.022064846     2.175823083     1.213723393
 9    H     1.863438822    -0.561913492    -0.888120605
10    H     2.651774854     1.674124548     0.889458408
11    H     2.656174920     1.672469093    -0.887129189
12    H    -0.818369368     2.899838248    -1.246467022
13    H     0.632919293     2.383802107    -2.131699986
14    H    -0.159987173     0.149391909    -2.127918129
15    H    -1.610299538     0.665092136    -1.241253606
16    H    -0.160855524     0.157768156     2.127075422
17    H     0.628225144     2.393199435     2.123579773
18    H    -0.821659018     2.904183192     1.233048908
19    H     1.863115305    -0.561798768     0.888601807
20    H    -1.611375697     0.668581940     1.237694097
EXPGEOM
 1    N     0.00000     0.00000     1.29510
 2    N     0.00000     0.00000    -1.29510
 3    C     0.09340     1.37540     0.77360
 4    C     1.14440    -0.76860     0.77360
 5    C    -1.23790    -0.60680     0.77360
 6    C    -0.09340     1.37540    -0.77360
 7    C    -1.14440    -0.76860    -0.77360
 8    C     1.23790    -0.60680    -0.77360
 9    H     1.07750     1.76600     1.05260
10    H    -0.66290     1.99160     1.27000
11    H     0.99070    -1.81620     1.05260
12    H     2.05620    -0.42170     1.27000
13    H    -2.06820     0.05010     1.05260
14    H    -1.39330    -1.56990     1.27000
15    H    -1.07750     1.76600    -1.05260
16    H     0.66290     1.99160    -1.27000
17    H    -0.99070    -1.81620    -1.05260
18    H    -2.05620    -0.42170    -1.27000
19    H     2.06820     0.05010    -1.05260
20    H     1.39330    -1.56990    -1.27000
```

```
---
H14C6N2, RHF, CHARGE=0, MULT=1
HF=8.6
1    C     0.000054330    -0.000039531    -0.000047948
2    C     1.543711597     0.000111005    -0.000136279
3    N     2.127042094     1.369936929     0.000043679
4    N     1.938422224     2.073555431     0.979522113
5    C     2.521133536     3.443581288     0.979293013
6    C     2.161333168    -0.920601000     1.074016383
7    C     4.064580645     3.448811916     0.955895726
8    C     1.886145039     4.367694052    -0.081777275
9    H    -0.430806296     0.324867145     0.967905064
10   H    -0.403050570     0.663587453    -0.792275728
11   H    -0.378400435    -1.022528226    -0.207644465
12   H     1.867524984    -0.405431862    -0.999345509
13   H     2.211233414     3.843511544     1.985073263
14   H     3.269616733    -0.901076255     1.033103088
15   H     1.853832621    -0.648147588     2.103211297
16   H     1.847220633    -1.970518318     0.898350118
17   H     4.482232405     2.781549662     1.737368224
18   H     4.481475428     3.132014360    -0.020985653
19   H     4.442266925     4.471348893     1.164425795
20   H     2.180461153     4.100681044    -1.116364330
21   H     0.778582305     4.344910007    -0.026072908
22   H     2.199784174     5.417596240     0.094237616
---
H14C6N2, RHF, CHARGE=0, MULT=1
HF=12.4
1    C    -0.082765975    -0.047990513     0.010389782
2    C     1.437645889     0.104867576    -0.093988846
3    C     1.913080683     1.567963148    -0.218995178
4    N     3.376745355     1.603382263    -0.445325324
5    N     4.033917864     2.278597425     0.332316067
6    C     5.497454584     2.316102260     0.105528451
7    C     5.969418081     3.780065097    -0.021864033
8    C     7.489364762     3.936642944    -0.127222243
9    H    -0.491710253     0.476861817     0.897553794
10   H    -0.600068806     0.347689247    -0.887039557
11   H    -0.352119408    -1.120204936     0.104425937
12   H     1.791550686    -0.484958480    -0.970803549
13   H     1.901803293    -0.362994328     0.804801358
14   H     1.454783381     2.050210948    -1.118162519
15   H     1.574968668     2.156770346     0.668286878
16   H     5.956964692     1.836028408     1.005365875
17   H     5.836405533     1.726217288    -0.780817969
18   H     5.503820218     4.245170679    -0.921768000
19   H     5.614571923     4.370121217     0.854228324
20   H     8.008173522     3.542686745     0.770123975
21   H     7.898857021     3.412519813    -1.014520407
22   H     7.755740166     5.009521627    -0.221852766
---
H6C7N2, RHF, CHARGE=0, MULT=1
HF=43.4
1    C     0.000179878     0.000348537    -0.000037387
2    C     1.425484963    -0.000199142     0.000026668
3    C    -0.735031404     1.186646525     0.000044716
4    C     0.004463507     2.391302580     0.000405148
5    C     1.443002639     2.404838949     0.000635000
6    C     2.163103197     1.184803591     0.000340454
```

```
 7    N    -0.365225308    3.739247804     0.000768211
 8    C     0.814436339    4.500062548     0.001253397
 9    N     1.903976209    3.729664348     0.001168503
10    H    -0.522599346   -0.957466021    -0.000171363
11    H     1.944802739   -0.959823647    -0.000205375
12    H    -1.823257408    1.180807464    -0.000156339
13    H     3.252158530    1.174265694     0.000343372
14    H    -1.293535597    4.091834208     0.000698151
15    H     0.803602942    5.586436257     0.001659730
---
H6C7N2, RHF, CHARGE=0, MULT=1
HF=58.1
 1    C     0.000347999   -0.000703218     0.012548119
 2    C     1.422235871   -0.000470836    -0.040632363
 3    C    -0.732523869    1.188097495     0.060993496
 4    C     0.006185908    2.392477768     0.047244215
 5    C     1.441552380    2.410152103    -0.011292687
 6    C     2.152079625    1.191526763    -0.050318835
 7    N    -0.391160176    3.741935382     0.179591074
 8    N     0.665091358    4.567833967     0.031829895
 9    C     1.779596494    3.827629246    -0.012041612
10    H    -0.524878581   -0.957385218     0.015267707
11    H     1.945437174   -0.957007826    -0.073460230
12    H    -1.819855590    1.181657237     0.107203894
13    H     3.240954143    1.182567832    -0.086281232
14    H    -1.274172689    4.069990670    -0.179206577
15    H     2.767972696    4.271834373    -0.050572145
---
H11C7N1, RHF, CHARGE=0, MULT=1
HF=50.4
 1    C    -0.038447697    0.001699050     0.854319040
 2    C     1.362728285   -0.082270399     0.917281119
 3    C    -0.659380685    1.154551533     0.345132048
 4    C     0.124689155    2.242603939    -0.115060690
 5    C     1.538291209    2.148553028    -0.058108438
 6    C     2.145440641    0.992213549     0.460531892
 7    N    -0.463868779    3.402752262    -0.715447619
 8    C    -0.519993309    4.626090397     0.092508734
 9    H     2.170441248    2.960953325    -0.419325980
10    H    -0.650080866   -0.831038909     1.203811125
11    H     1.840486888   -0.976373288     1.317762601
12    H    -1.749380475    1.191674685     0.309879833
13    H     3.233545367    0.927414281     0.503505042
14    H    -1.391250902    3.186473845    -1.051132377
15    H    -1.088850198    4.510954813     1.046315495
16    H     0.505727836    4.972502884     0.346572951
17    H    -1.016868294    5.424165776    -0.504489849
18    H    -4.495434139   11.883536472     3.381379040
19    H    -4.481752396   12.510637577     3.597088013
---
H10C7N2, RHF, CHARGE=0, MULT=1
HF=30.3
 1    C    -0.000010535    0.000003846     0.000062513
 2    C     1.558623145   -0.000141239    -0.000004690
 3    C     2.069938221    1.468841664    -0.000030465
 4    C     2.046182918   -0.712154072    -1.297832752
 5    C     2.083346236   -0.800320529     1.265800112
 6    C     3.542986728   -0.930892588     1.333330787
 7    N     4.694599191   -1.063207272     1.408396063
```

```
8    C     1.618376312   -0.266576152    2.550631724
9    N     1.256165919    0.123548068    3.583081058
10   H    -0.409918936   -1.029370988    0.047341383
11   H    -0.419588770    0.564626834    0.856266894
12   H    -0.399664316    0.469968179   -0.921698612
13   H     3.176512597    1.524138844   -0.027186389
14   H     1.732340799    2.026840162    0.896044961
15   H     1.695019712    2.020091855   -0.886758130
16   H     3.148917541   -0.683620091   -1.403101827
17   H     1.737245381   -1.777023206   -1.321933996
18   H     1.624261117   -0.227772615   -2.202042786
19   H     1.675565791   -1.846399697    1.192116820
---
H10C7N2, RHF, CHARGE=0, MULT=1
HF=17.8
1    C     0.000040620    0.000318106   -0.000024773
2    C     1.429958850    0.000235101   -0.000088263
3    N     2.109807946    1.175505407    0.000007463
4    C     1.412747420    2.326835590    0.000133080
5    C    -0.009048395    2.338997237    0.000344293
6    N    -0.688896947    1.169043178    0.000298726
7    C    -0.850750325   -1.247005083   -0.000363988
8    C     2.284973785   -1.243317329   -0.000322144
9    C    -0.818031956    3.610411784   -0.000196040
10   H     2.004772358    3.247702666    0.000030164
11   H    -0.640968738   -1.862845044    0.899509301
12   H    -1.936911237   -1.024437802   -0.001567981
13   H    -0.639235644   -1.863618983   -0.899284278
14   H     2.077156682   -1.860491117    0.899132873
15   H     3.370105177   -1.015927869   -0.000718676
16   H     2.076644863   -1.860609451   -0.899562479
17   H    -0.569403514    4.225479526    0.890383287
18   H    -0.589832399    4.210970102   -0.906036598
19   H    -1.910749055    3.426322748    0.013931986
---
H12C7N2, RHF, CHARGE=0, MULT=1
HF=19.8
1    C    -0.000004047   -0.000065912    0.000098989
2    C     1.535699448   -0.000072817    0.000113399
3    C     2.165224868    1.408636406   -0.000116806
4    N     1.544745684    2.286649626   -1.003940693
5    C     0.078393722    2.391206704   -0.978254891
6    C    -0.600222191    1.005313132   -0.993421692
7    C     2.283171566    3.510430733   -1.341181960
8    C     3.067364222    3.294886454   -2.561732768
9    N     3.688548417    3.134919932   -3.530059103
10   H    -0.367529695   -1.022281546   -0.244588473
11   H    -0.375418110    0.217654436    1.026670025
12   H     1.909081926   -0.574676076   -0.877910638
13   H     1.903224064   -0.545151344    0.899580203
14   H     3.255942010    1.308223477   -0.225020362
15   H     2.100884150    1.849505728    1.029810923
16   H    -0.263283536    2.958674503   -1.879242076
17   H    -0.278643398    2.975293947   -0.089051145
18   H    -1.681801380    1.145829729   -0.765768115
19   H    -0.558148751    0.582448465   -2.023061300
20   H     1.596594198    4.377878088   -1.512080474
21   H     2.975548352    3.831626378   -0.520438019
---
```

```
H14C7N2, RHF, CHARGE=0, MULT=1
HF=9.4
1    C      0.000128749     0.000033116    -0.000051672
2    C      1.569309228     0.000074614     0.000007029
3    N      1.905385291     1.468258358     0.000215168
4    N      0.946950125     2.211206501     0.000227184
5    C     -0.391154029     1.519653405    -0.000552565
6    C      2.177390900    -0.641011740    -1.273107821
7    C      2.177267945    -0.640854780     1.273252380
8    C     -1.162286963     1.948403476    -1.274576515
9    C     -1.164800210     1.948813527     1.271811212
10   H     -0.413644643    -0.533500474     0.879726731
11   H     -0.413572077    -0.533951695    -0.879654879
12   H      1.789040627    -0.177917544    -2.202944420
13   H      3.281644625    -0.543248336    -1.292522698
14   H      1.934769533    -1.722920629    -1.316633812
15   H      3.281441689    -0.542230455     1.293230886
16   H      1.788093604    -0.178256781     2.202994480
17   H      1.935458751    -1.722939926     1.316473013
18   H     -0.616230398     1.687233598    -2.203851368
19   H     -2.149632725     1.444003269    -1.318862150
20   H     -1.342635454     3.042275337    -1.294685141
21   H     -2.152047625     1.443981095     1.314510557
22   H     -0.620374777     1.688422481     2.202210942
23   H     -1.345725223     3.042551616     1.290943665
---
H4C8N2, RHF, CHARGE=0, MULT=1
HF=86.7
1    C     -0.000347505    -0.000409027    -0.000005002
2    C      1.404507801     0.000554879     0.000012947
3    C      2.114682533     1.226132612    -0.000005954
4    C      1.394872776     2.444918519    -0.000048826
5    C     -0.020523808     2.444771631    -0.000003291
6    C     -0.714096845     1.209645468    -0.000019145
7    C      3.541441858     1.232891946     0.000050607
8    N      4.703852159     1.236465802     0.000115564
9    C     -0.740694357     3.676523749     0.000140381
10   N     -1.328799251     4.679208773     0.000314523
11   H      1.942316739    -0.949106893     0.000044854
12   H      1.935953195     3.393294313    -0.000103516
13   H     -1.805179659     1.189320895    -0.000026619
14   H     -0.541166258    -0.947644992     0.000009533
---
H4C8N2, RHF, CHARGE=0, MULT=1
HF=87.8
1    C      0.000022784     0.000435803     0.000000085
2    C      1.405913983     0.000014554    -0.000005281
3    C      2.127975318     1.217183322     0.000012491
4    C      1.414078947     2.452914805     0.000038503
5    C     -0.000972003     2.434780390     0.000010181
6    C     -0.702659923     1.216487234     0.000001950
7    C      3.555728045     1.181791815     0.000024973
8    N      4.717428792     1.139263561     0.000046265
9    C      2.096523518     3.707484997     0.000073377
10   N      2.639892814     4.735150160     0.000104104
11   H      1.938438142    -0.952730149    -0.000010506
12   H     -0.542737039    -0.945612141     0.000001096
13   H     -0.560870066     3.371852988     0.000004165
14   H     -1.793158871     1.218611952     0.000010292
```

```
---
H4C8N2, RHF, CHARGE=0, MULT=1
HF=85.6
 1    C      -0.000026932    -0.000090552    -0.000042244
 2    C       1.404580659     0.000266159     0.000004218
 3    C       2.123031304     1.220240904    -0.000013147
 4    C       1.403734409     2.439861802     0.000030759
 5    C      -0.000846826     2.439661872     0.000100023
 6    C      -0.719291061     1.219585844     0.000051660
 7    C       3.550202063     1.221289544     0.000069686
 8    N       4.712645905     1.222582929     0.000160231
 9    C      -2.146421555     1.219138866     0.000568834
10    N      -3.308856167     1.219073347     0.001300934
11    H       1.936994003    -0.952511411     0.000085926
12    H       1.935496903     3.392837521     0.000099878
13    H      -0.531979810    -0.952893521     0.000004442
14    H      -0.533108888     3.392404109     0.000251109
---
H6C8N2, RHF, CHARGE=0, MULT=1
HF=78.9
 1    C      -0.000114950     0.000102076     0.000004822
 2    C       1.384945421     0.000328611     0.000004163
 3    C       2.109543252     1.231633247     0.000023882
 4    C       1.437593102     2.442889012     0.000075787
 5    C       0.004231315     2.471885945     0.000127860
 6    C      -0.720963577     1.239447084     0.000060815
 7    C      -2.159566624     1.348758678     0.000073731
 8    N      -2.788499070     2.530665170     0.000184487
 9    N      -2.128127230     3.652915722     0.000297001
10    C      -0.789520525     3.676701328     0.000273950
11    H      -0.550618426    -0.941664091    -0.000021304
12    H       1.939530729    -0.938996294    -0.000017442
13    H       3.200001514     1.202796464    -0.000011372
14    H       1.993995823     3.381180255     0.000066579
15    H      -2.796236810     0.458652955     0.000007344
16    H      -0.320168904     4.665330813     0.000382707
---
H6C8N2, RHF, CHARGE=0, MULT=1
HF=58.1
 1    C      -0.000099836     0.000000698    -0.000011194
 2    C       1.382783004     0.000417978    -0.000002601
 3    C       2.116946087     1.229316890    -0.000010187
 4    C       1.456475675     2.444316669    -0.000064923
 5    C       0.022472052     2.487977580    -0.000105379
 6    C      -0.708496457     1.254940025    -0.000061925
 7    N      -2.089859156     1.255653564    -0.000075514
 8    C      -2.715284537     2.430143569    -0.000137731
 9    N      -2.092824107     3.665486231    -0.000204590
10    C      -0.765743491     3.694403900    -0.000177765
11    H      -0.558728607    -0.936657589     0.000010162
12    H       1.935925554    -0.940197144     0.000002999
13    H       3.206915931     1.190446437     0.000031100
14    H       2.020234059     3.378411494    -0.000076583
15    H      -3.813416316     2.438783435    -0.000148651
16    H      -0.290479272     4.682061954    -0.000217785
---
H6C8N2, RHF, CHARGE=0, MULT=1
HF=57.4
 1    C       0.000001290    -0.000194485     0.000008643
```

```
2    C      1.382791964      0.000363874     0.000020321
3    C      2.119197048      1.228634701    -0.000016492
4    C      1.468196492      2.448691027    -0.000015670
5    C      0.028911416      2.483608277    -0.000040508
6    C     -0.708963572      1.252845498    -0.000012601
7    N     -2.092688521      1.256437107    -0.000004432
8    C     -2.695010595      2.437195643    -0.000030012
9    C     -1.951971122      3.676482066    -0.000091694
10   N     -0.626619864      3.702137545    -0.000096090
11   H     -0.557719502     -0.937646598     0.000021043
12   H      1.935237446     -0.940440831     0.000035611
13   H      3.209283985      1.184576431     0.000012314
14   H      2.032500964      3.382125215    -0.000010192
15   H     -3.789770773      2.445684764    -0.000000638
16   H     -2.475592262      4.637914254    -0.000136305
---
H12C8N2, RHF, CHARGE=0, MULT=1
HF=32.5
1    C      0.000065683     -0.000008639    -0.000013283
2    C      1.531396161      0.000126335     0.000031766
3    C      2.173029462      1.401371857    -0.000170677
4    C      3.715880289      1.384436633    -0.000478049
5    C      4.349054135      2.790884961    -0.000943215
6    C      5.908975416      2.807713997    -0.053878224
7    C      6.533249098      2.291942812     1.166344461
8    N      7.040109133      1.876160574     2.124983585
9    C      6.414733881      4.156736025    -0.324153745
10   N      6.823247807      5.219381984    -0.553358803
11   H     -0.412529094      0.506202887     0.896060459
12   H     -0.412544631      0.505623140    -0.896462967
13   H     -0.381299170     -1.041901441     0.000250071
14   H      1.883747297     -0.569588638    -0.890826936
15   H      1.883673566     -0.569245672     0.891054418
16   H      1.814530469      1.965169900    -0.891791371
17   H      1.815552437      1.965023686     0.891939503
18   H      4.074574206      0.824251390    -0.894341093
19   H      4.072006484      0.818015703     0.889465763
20   H      3.976257528      3.354391144    -0.886738714
21   H      4.011750051      3.355016883     0.897332576
22   H      6.243343016      2.155430824    -0.905705921
---
H12C8N2, RHF, CHARGE=0, MULT=1
HF=24.1
1    C      0.000056699      0.000154271    -0.000019783
2    C      1.564404785     -0.000046168     0.000019720
3    C     -0.478555645      1.489485113     0.000019526
4    C     -0.614177272     -0.842076955    -1.221524494
5    C     -2.178527789     -0.841131077    -1.222155390
6    C     -0.136117308     -2.331548286    -1.221202564
7    C     -0.181794631     -0.250482526    -2.499242893
8    N      0.141048257      0.191192769    -3.524554220
9    C     -0.431622275     -0.591773715     1.277771140
10   N     -0.753962554     -1.033726697     2.303117810
11   H      1.985049376     -1.015372630     0.136985194
12   H      1.977795875      0.414413490    -0.939904868
13   H      1.957471051      0.620229056     0.832048759
14   H     -1.573755356      1.579471797     0.138239390
15   H     -0.211496837      2.009429398    -0.940434300
16   H     -0.007234989      2.053735878     0.831270766
```

```
17    H    -2.592495626   -1.253600890   -0.281411515
18    H    -2.598578642    0.174034799   -1.361285135
19    H    -2.571591980   -1.463021180   -2.052948513
20    H    -0.405815015   -2.851911546   -0.281720140
21    H     0.959364263   -2.421952013   -1.356602556
22    H    -0.605422832   -2.895192193   -2.054031213
---
H12C8N2, RHF, CHARGE=0, MULT=1
HF=13.1
1     N     0.000178147    0.000083207    0.000607373
2     C     1.353983819   -0.000320151   -0.000138276
3     C     2.069395976    1.236929958    0.000066316
4     N     1.392947564    2.409582845    0.001273350
5     C     0.039151602    2.409937972    0.001970258
6     C    -0.676316780    1.172745852    0.001370309
7     C    -2.181098534    1.054211619    0.001835924
8     C    -0.608885531    3.773121905    0.002706198
9     C     2.002100666   -1.363446304   -0.001063971
10    C     3.574210989    1.355814646   -0.001029792
11    H    -2.611248683    1.545973184    0.899575875
12    H    -2.527584703    0.001016483    0.005169533
13    H    -2.610969017    1.540642458   -0.899114319
14    H    -1.241675384    3.904523935   -0.900263371
15    H     0.131062929    4.598772797    0.010169179
16    H    -1.253085955    3.898904400    0.898358820
17    H     2.638488601   -1.492213878   -0.901843623
18    H     2.642528239   -1.491658906    0.896863625
19    H     1.262218664   -2.189185052    0.000898989
20    H     4.005507267    0.866571783    0.897755513
21    H     4.003488517    0.866965083   -0.901126270
22    H     3.920260291    2.409175760   -0.001078807
---
H14C8N2, RHF, CHARGE=0, MULT=1
HF=22.1
1     C    -0.000309011   -0.000436138    0.000576928
2     C     1.569470951    0.000570466    0.000002260
3     C     2.080604657    1.461363555   -0.000176451
4     C     0.854443792    2.441701085    0.001037057
5     C     0.020851578    2.182801832   -1.303768145
6     C    -0.489238112    0.721726082   -1.304666507
7     N    -0.422632437    0.874012778    1.151302126
8     N    -0.020976228    2.021257990    1.151700129
9     C    -0.612933239   -1.400329450    0.192943526
10    C     1.249382069    3.917792642    0.193405862
11    H     1.954231918   -0.545129335   -0.888117579
12    H     1.965583835   -0.546046849    0.883240154
13    H     2.721044570    1.647787318   -0.888771123
14    H     2.731148065    1.641601940    0.882928003
15    H    -0.834634031    2.889240585   -1.372839226
16    H     0.643994858    2.376447353   -2.203007414
17    H    -1.598291248    0.701118956   -1.375764678
18    H    -0.120378568    0.182793066   -2.203594545
19    H    -0.272451780   -1.871707878    1.137152778
20    H    -1.720957953   -1.364117336    0.219426257
21    H    -0.320840934   -2.072816189   -0.640317516
22    H     0.361170528    4.581312238    0.220904840
23    H     1.810620318    4.073670395    1.137033805
24    H     1.896612738    4.261373593   -0.640251267
---
```

```
H16C8N2, RHF, CHARGE=0, MULT=1
HF=10
1    C    -0.000098399    -0.000232163    -0.000421101
2    C     1.558551419    -0.000958217    -0.000711825
3    C     2.184945403     1.403649794    -0.000173770
4    C     1.501802055     2.402640178    -0.982320088
5    N     0.034983958     2.140345960    -1.127595335
6    N    -0.567866589     1.174694375    -0.734480276
7    C    -0.549385716    -1.257632734    -0.735799773
8    C    -0.575356948     0.062947786     1.444647217
9    C     2.110808486     2.312007603    -2.412016170
10   C     1.614883794     3.860970778    -0.449568489
11   H     1.942725409    -0.564361584    -0.882018158
12   H     1.939534919    -0.558649791     0.884345238
13   H     3.267102922     1.309383964    -0.243971041
14   H     2.158601549     1.806718510     1.038124343
15   H    -0.221163146    -1.295648861    -1.794632242
16   H    -1.657730624    -1.287774145    -0.732362999
17   H    -0.189272347    -2.184977318    -0.243610205
18   H    -1.683413478     0.104610958     1.446417480
19   H    -0.210976857     0.949819024     2.001810362
20   H    -0.277405913    -0.835074950     2.024133827
21   H     2.042567570     1.288020853    -2.832028495
22   H     1.600750111     2.994146198    -3.122131630
23   H     3.184600335     2.591210358    -2.398461270
24   H     1.148369024     4.591536211    -1.140836290
25   H     1.124919443     3.981313414     0.538166616
26   H     2.679579147     4.149745747    -0.328774357
---
H18C8N2, RHF, CHARGE=0, MULT=1
HF=2.2
1    C    -0.096276468    -0.052798797    -0.005734688
2    C     1.431463181    -0.105971407     0.086239259
3    C     2.112211132     1.273126889     0.191928660
4    C     3.653187054     1.196100535     0.266598661
5    N     4.213879087     2.544197556     0.518361952
6    N     4.940156542     3.006600745    -0.348131141
7    C     5.499623544     4.355381952    -0.096755879
8    C     7.040607038     4.280758490    -0.018284315
9    C     7.718736156     5.661172548     0.087618927
10   C     9.246294518     5.610431553     0.183714298
11   H    -0.546732469     0.430813969     0.884532014
12   H    -0.435911608     0.503389962    -0.902741549
13   H    -0.511174800    -1.079384633    -0.073139784
14   H     1.710130044    -0.726910322     0.968618624
15   H     1.819752123    -0.646118038    -0.808188698
16   H     1.721495229     1.801076896     1.091845884
17   H     1.819598867     1.890377988    -0.688826593
18   H     3.973252132     0.572461974     1.138341315
19   H     4.057869634     0.699132086    -0.648387253
20   H     5.180928802     4.977318711    -0.970024565
21   H     5.092129153     4.853097538     0.816683696
22   H     7.331883547     3.665188144     0.863805425
23   H     7.434106944     3.752513089    -0.916573513
24   H     7.441561106     6.281140692    -0.796243389
25   H     7.327416152     6.202117281     0.979989818
26   H     9.584660141     5.052775783     1.080256402
27   H     9.699741479     5.129562875    -0.706530028
28   H     9.659252650     6.637627292     0.254057272
```

```
---
H18C8N2, RHF, CHARGE=0, MULT=1
HF=-8.7
1    C     0.000091840    -0.000018174    -0.000054389
2    C     1.555333131    -0.000136388     0.000037172
3    C    -0.530752066     1.465526799    -0.000199281
4    C    -0.531129269    -0.774861852    -1.243807537
5    N    -0.657856665    -0.651044323     1.175142439
6    N     0.041879653    -1.147002178     2.040597130
7    C    -0.615968900    -1.799187044     3.215159086
8    C    -2.170562057    -1.759630222     3.237208069
9    C    -0.122335743    -3.277373307     3.189232969
10   C    -0.048495184    -1.054038054     4.461112280
11   H     1.975986770    -1.025998845     0.001640388
12   H     1.975716128     0.537826624     0.873451792
13   H     1.942563604     0.506383728    -0.909019921
14   H    -1.638651368     1.509467496    -0.006904294
15   H    -0.180388084     2.029736318     0.888213590
16   H    -0.174720131     2.010652657    -0.898764695
17   H    -1.639067461    -0.791057901    -1.284971093
18   H    -0.181928592    -1.827614092    -1.252345900
19   H    -0.174321631    -0.301385005    -2.181660146
20   H    -2.564479900    -0.723344995     3.248909062
21   H    -2.617358003    -2.280268963     2.366210361
22   H    -2.557922327    -2.262836355     4.148083327
23   H    -0.497643717    -3.820682547     2.297986652
24   H     0.984088687    -3.349107541     3.181325269
25   H    -0.481037575    -3.825394311     4.084884403
26   H     1.059954998    -1.066169691     4.487190919
27   H    -0.370660056     0.007096256     4.487541997
28   H    -0.404782095    -1.530418488     5.397685569
---
H20C8N2, RHF, CHARGE=0, MULT=1
HF=-14.2
1    C    -0.181047273    -0.311273497    -0.782713368
2    C     0.414565390     0.957959420    -1.398537564
3    C     1.651550714     1.510943218    -0.664103049
4    C     2.252313113     2.771149605    -1.331534510
5    N     3.372091126     3.444792783    -0.622852474
6    N     4.544503051     2.696237193    -0.635561722
7    C     5.655868792     3.380634206    -1.347329337
8    C     6.259061790     4.635804907    -0.672586611
9    C     7.487285294     5.200522234    -1.412949186
10   C     8.097541319     6.452239566    -0.775881402
11   H     0.548638873    -1.146022573    -0.769724899
12   H    -0.523731125    -0.143503275     0.258464487
13   H    -1.059330048    -0.646258479    -1.372354442
14   H     0.677971678     0.742178533    -2.459904241
15   H    -0.379760900     1.739585253    -1.429131140
16   H     2.424330117     0.711171592    -0.609168391
17   H     1.368246135     1.745693720     0.388459955
18   H     2.581520146     2.525006930    -2.369702640
19   H     1.459380609     3.556424742    -1.426474445
20   H     6.450318699     2.598807708    -1.457463512
21   H     5.316460202     3.636155381    -2.379838554
22   H     5.484818243     5.433015876    -0.602318362
23   H     6.552812791     4.390956350     0.374683968
24   H     8.277445551     4.416505619    -1.474179386
25   H     7.208761997     5.442080592    -2.464782293
```

```
26      H        7.378954611      7.296484618     -0.755282757
27      H        8.431941452      6.265699921      0.264786106
28      H        8.983945587      6.783174828     -1.355603749
29      H        3.071476661      3.616645844      0.338437829
30      H        4.856750620      2.509766099      0.319346092
---
N3, UHF, CHARGE=0, MULT=2
HF=99
1       N        0.015081360      0.001298940     -0.010000000
2       N        1.182934816     -0.000574185     -0.010000000
3       N        2.350779420     -0.000916073     -0.010000000
EXPGEOM
1       N        0.00000          0.00000          0.00000
2       N        0.00000          0.00000          1.17990
3       N        0.00000          0.00000         -1.17990
---
H1N3, RHF, CHARGE=0, MULT=1
HF=70.3
1       N       -0.000004695     -0.000245739     -0.000175806
2       N        1.132118570      0.000087485      0.000259542
3       N        2.344717875      0.308530399     -0.000058142
4       H        2.955510741     -0.505566838      0.003244911
EXPGEOM
1       N        0.08550         -1.13070          0.00000
2       N        0.00000          0.11060          0.00000
3       N       -0.23830          1.21820          0.00000
4       H        1.06950         -1.38640          0.00000
---
H3C3N3, RHF, CHARGE=0, MULT=1
HF=54
1       N        0.000015971      0.000100143      0.000015803
2       C        1.356655742     -0.000018077     -0.000020979
3       N        2.102869836      1.132905060     -0.000011641
4       C        1.424751619      2.307853861      0.000021633
5       N        0.070462829      2.387581940      0.000059613
6       C       -0.608027027      1.212852010      0.000066446
7       H        1.877208889     -0.966449526     -0.000051850
8       H        2.001382168      3.241971024      0.000016194
9       H       -1.705301080      1.245204950      0.000111127
EXPGEOM
1       C        0.00000          1.29320          0.00000
2       C        1.11990         -0.64660          0.00000
3       C       -1.11990         -0.64660          0.00000
4       N        0.00000         -1.37800          0.00000
5       N        1.19340          0.68900          0.00000
6       N       -1.19340          0.68900          0.00000
7       H        0.00000          2.38080          0.00000
8       H        2.06190         -1.19040          0.00000
9       H       -2.06190         -1.19040          0.00000
---
H1C5N3, RHF, CHARGE=0, MULT=1
HF=124.4
1       C        0.000091154      0.000005996      0.000731182
2       C        1.363121276     -0.000150577     -0.000040319
3       C       -0.824317006      1.159865840      0.000504286
4       N       -1.522563130      2.089786592     -0.000296329
5       C        2.157556927      1.188788703      0.000584557
6       N        2.810049477      2.150384450      0.001380891
7       C        2.100008566     -1.227003039     -0.001067432
```

```
8    N     2.701926925    -2.221042136    -0.001946601
9    H    -0.557342479    -0.944698133     0.001293579
---
H3C5N3, RHF, CHARGE=0, MULT=1
HF=101
1    C    -0.000080809    -0.000034495    -0.000067264
2    C     1.566605208    -0.000001949     0.000062954
3    C     2.075296703     1.382066522     0.000128880
4    N     2.474363712     2.472033217     0.000212182
5    C     2.074755769    -0.691035107    -1.197135945
6    N     2.472411656    -1.236184685    -2.141531850
7    C     2.075368957    -0.691034454     1.196888833
8    N     2.474689121    -1.235799580     2.140842900
9    H    -0.391699368    -1.036098127     0.000216264
10   H    -0.392192846     0.518414235     0.896772889
11   H    -0.391756975     0.517594523    -0.897610608
---
H2C1N4, RHF, CHARGE=0, MULT=1
HF=76.6
1    N     0.000000000     0.000000000     0.000000000
2    N     1.274858947     0.000000000     0.000000000
3    N     1.675348132     1.280776264     0.000000000
4    C     0.538258588     2.090895661    -0.000058154
5    N    -0.503724156     1.229239048    -0.000091314
6    H     2.643984308     1.544371514     0.000018240
7    H     0.526493341     3.173089459     0.000155923
EXPGEOM
1    C     1.06250      0.23480     0.00000
2    N     0.00000      1.05300     0.00000
3    N    -1.11750      0.30240     0.00000
4    N    -0.72240     -0.92390     0.00000
5    N     0.64010     -1.00630     0.00000
6    H     2.08200      0.56130     0.00000
7    H    -0.05880      2.05400     0.00000
---
H2C1N4, RHF, CHARGE=0, MULT=1
HF=80
1    N     0.000021625     0.000037243    -0.000014374
2    N     1.338273431     0.000018832     0.000030999
3    N     1.726361319     1.207787395    -0.000023555
4    N    -0.517896399     1.227128850    -0.000087310
5    C     0.575392112     2.022525584    -0.000111428
6    H     0.591617715     3.105371517    -0.000276202
7    H    -0.571744253    -0.836744901     0.000016091
---
C6N4, RHF, CHARGE=0, MULT=1
HF=168.5
1    C     0.000000000     0.000000000     0.000000000
2    C     1.373479933     0.000000000     0.000000000
3    C    -0.771773782     1.204548606     0.000000000
4    N    -1.410629298     2.175467590     0.000032759
5    C     2.145327806     1.204506386    -0.000109360
6    N     2.784026218     2.175528573    -0.000228758
7    C     2.145112766    -1.204669551     0.000030078
8    N     2.784297713    -2.175370079     0.000087263
9    C    -0.771593509    -1.204694425     0.000071449
10   N    -1.410495135    -2.175581623     0.000051974
EXPGEOM
1    C     0.00000     0.00000     0.68860
```

```
 2    C      0.00000     0.00000    -0.68860
 3    C      0.00000     1.23810     1.43900
 4    C      0.00000    -1.23810     1.43900
 5    C      0.00000     1.23810    -1.43900
 6    C      0.00000    -1.23810    -1.43900
 7    N      0.00000     2.24060     2.05950
 8    N      0.00000    -2.24060     2.05950
 9    N      0.00000     2.24060    -2.05950
10    N      0.00000    -2.24060    -2.05950
---
H12C6N4, RHF, CHARGE=0, MULT=1
HF=47.6
 1    C      0.000144355    -0.001173161    -0.000044359
 2    N      1.493480580     0.000555003     0.001324661
 3    C      2.001861691     1.404640543    -0.000601624
 4    N      1.507224717     2.089084840     1.231016210
 5    C      0.013969236     2.105757673     1.240759610
 6    N     -0.475505596     0.695025519     1.232437341
 7    C      0.016649091    -0.022614810     2.446126354
 8    C      2.002923560    -0.723930989     1.203725010
 9    C      2.017162498     1.384242136     2.444718201
10    N      1.509741099    -0.020132687     2.425045656
11    H     -0.392677119    -1.047018295    -0.038204269
12    H     -0.393505201     0.505802721    -0.915286305
13    H      1.660140797     1.946309726    -0.916701429
14    H      3.118993167     1.419002884    -0.037568521
15    H     -0.368033044     2.649946207     2.139355123
16    H     -0.379535453     2.664393047     0.356086568
17    H     -0.375259257    -1.069186820     2.470864215
18    H     -0.364944357     0.467769620     3.375386968
19    H      3.120020856    -0.763621868     1.195928027
20    H      1.660593906    -1.788033610     1.197030460
21    H      1.685726243     1.909864272     3.373979821
22    H      3.134657317     1.399040027     2.468505022
---
H6C3N6, RHF, CHARGE=0, MULT=1
HF=12.4
 1    N      0.023617543     0.042316254     0.129359611
 2    C      1.387536368     0.010901961     0.068456667
 3    N      2.184625681     1.100077799    -0.139903458
 4    C      1.526068447     2.294465113    -0.209530099
 5    N      0.169177373     2.441222969    -0.162469513
 6    C     -0.534008277     1.280820046    -0.010019957
 7    N      2.014557037    -1.207571686     0.324116959
 8    N      2.297291829     3.455094574    -0.243886855
 9    N     -1.924050246     1.354151242    -0.103005270
10    H      2.941204786    -1.297624805    -0.052238270
11    H      1.466757934    -2.022164967     0.112047675
12    H      1.847165165     4.254601848    -0.652247328
13    H      3.222455213     3.341697505    -0.617562651
14    H     -2.418554370     0.606326133     0.350110657
15    H     -2.319186901     2.241030853     0.154767491
---
H9C10N1, RHF, CHARGE=0, MULT=1
HF=38
 1    C     -0.000154386    -0.000322761     0.000020815
 2    C      1.381754318     0.000498855    -0.000007557
 3    C      2.111054291     1.231576045    -0.000017921
 4    C      1.445102062     2.442594672    -0.000006131
```

```
 5    C     0.008628238    2.488137862    0.000007101
 6    C    -0.717721469    1.251595741    0.000001796
 7    N    -2.098448856    1.209072625    0.000024654
 8    C    -2.791107683    2.347960999    0.000101656
 9    C    -2.136186277    3.634609541    0.000117956
10    C    -0.758306389    3.701830083    0.000080292
11    C    -4.297311609    2.237959534    0.000364236
12    H    -0.554622392   -0.939771218    0.000088697
13    H     1.936724845   -0.938580683   -0.000023984
14    H     3.201093885    1.197568043   -0.000005477
15    H     2.007214120    3.377625848    0.000001868
16    H    -2.737034457    4.543989104    0.000183444
17    H    -0.249016349    4.666493762    0.000135846
18    H    -4.718681558    2.735220829    0.899494825
19    H    -4.719291162    2.737244649   -0.897346892
20    H    -4.655243703    1.189140388   -0.000703310
---
H9C10N1, RHF, CHARGE=0, MULT=1
HF=38.7
 1    C    -0.000131313   -0.000114446    0.000006988
 2    C     1.380861025    0.000811296    0.000002193
 3    C     2.098586157    1.236494183    0.000001019
 4    C     1.424743753    2.443542052    0.000007230
 5    C    -0.014115875    2.499727478    0.000023693
 6    C    -0.721824915    1.249290115    0.000024400
 7    N    -2.103101928    1.177121238    0.000082357
 8    C    -2.791895046    2.309307471    0.000117183
 9    C    -2.175643282    3.604078503    0.000096783
10    C    -0.792107127    3.723700610    0.000070984
11    H    -3.884628660    2.221835494    0.000177160
12    H    -0.553302420   -0.940702913    0.000004964
13    H     1.939639572   -0.935817212   -0.000002041
14    H     3.189244080    1.212525064   -0.000000021
15    H     2.004599408    3.367370916    0.000012661
16    H    -2.828505048    4.477691503    0.000106048
17    C    -0.145295542    5.085142973    0.000096876
18    H     0.488275054    5.227011217   -0.900208454
19    H     0.488310382    5.226914374    0.900418715
20    H    -0.894286833    5.903557289    0.000150299
---
H9C10N1, RHF, CHARGE=0, MULT=1
HF=38.5
 1    C     0.000265352    0.000063755   -0.000005568
 2    C     1.380904700   -0.000685325    0.000081906
 3    C     2.138123002    1.225365106    0.000026581
 4    C     1.448613933    2.431401270   -0.000260525
 5    C     0.012364194    2.480936984   -0.000493153
 6    C    -0.719461868    1.248235278   -0.000257642
 7    N    -2.102364716    1.213052254   -0.000257688
 8    C    -2.767217307    2.362298056   -0.000610757
 9    C    -2.122916884    3.645699204   -0.001028932
10    C    -0.743760485    3.703556716   -0.000956438
11    H    -3.861388109    2.296668430   -0.000571192
12    H    -0.552277795   -0.940910153    0.000103936
13    H     1.915683784   -0.952837312    0.000192000
14    C     3.642253621    1.163696821    0.000352563
15    H     1.991951632    3.378815226   -0.000314341
16    H    -2.732153614    4.548398234   -0.001400362
17    H    -0.225706896    4.663382802   -0.001249220
```

```
18      H      4.016103611     0.632123822      0.900777525
19      H      4.016206928     0.623713925     -0.894992934
20      H      4.108438250     2.169904588     -0.004415984
---
H9C10N1, RHF, CHARGE=0, MULT=1
HF=40.1
1       C      0.000083661     0.000004593      0.000055137
2       C      1.390560691     0.000191780      0.000034956
3       C      2.146320040     1.213080725      0.000026632
4       C      1.510581697     2.437789925      0.000084911
5       C      0.076115198     2.507712966      0.000005555
6       C     -0.678351083     1.285987125     -0.000075286
7       N     -2.062438539     1.301369808     -0.000263606
8       C     -2.700489146     2.464754710     -0.000306252
9       C     -2.027591881     3.732376829     -0.000134485
10      C     -0.647994005     3.749674979      0.000042989
11      H     -3.795946359     2.422656781     -0.000463301
12      C     -0.772629967    -1.294170610      0.000129576
13      H      1.949984428    -0.937974034      0.000045977
14      H      3.235471195     1.151871288     -0.000015811
15      H      2.092310463     3.360890695      0.000210238
16      H     -2.613243419     4.650457922     -0.000144670
17      H     -0.102931924     4.694411730      0.000177804
18      H     -1.417796342    -1.377494432      0.899280652
19      H     -1.416162069    -1.378462309     -0.899999955
20      H     -0.100009890    -2.177776149      0.001091824
---
H11C10N1, RHF, CHARGE=0, MULT=1
HF=25.4
1       C     -0.000095013    -0.000043212      0.000157775
2       C      1.424783469     0.000121237     -0.000319802
3       C      2.103717173     1.238898104      0.000362973
4       C      1.414668006     2.469239429      0.003512906
5       C      0.002874904     2.441069779      0.006450554
6       C     -0.721549107     1.229987997      0.003807078
7       C     -2.230466213     1.267292288      0.004307560
8       C      2.220884400    -1.282420697     -0.001772249
9       C      2.150782639     3.783525270      0.003045107
10      C     -0.712004947    -1.241709553     -0.003959126
11      N     -1.290689466    -2.250569397     -0.007921755
12      H      3.196390909     1.244307832     -0.002110888
13      H     -0.542298997     3.388137690      0.010788042
14      H     -2.643085347     0.766856187     -0.896509168
15      H     -2.642578160     0.763003813      0.903289158
16      H     -2.624430075     2.304016650      0.006723883
17      H      3.314617248    -1.097536296     -0.000649998
18      H      1.995751532    -1.893807201      0.896964177
19      H      1.997010571    -1.890839870     -0.902812451
20      H      1.896645068     4.377417536     -0.900079826
21      H      1.882840698     4.386586601      0.895972657
22      H      3.251711866     3.653104948      0.012411278
---
H11C10N1, RHF, CHARGE=0, MULT=1
HF=56.6
1       C      0.000027464     0.000041740      0.000062490
2       C      1.425804284    -0.000203986     -0.000076825
3       C      2.085373218     1.249133600      0.000126597
4       C      1.381603345     2.471200621      0.001412274
5       C     -0.030086111     2.428835766      0.001904483
```

```
  6    C    -0.748960586    1.214305682    0.000739388
  7    C    -2.256540660    1.240516161    0.000336096
  8    C     2.238191271   -1.270297824   -0.000318257
  9    C     2.102322424    3.793422601    0.001686593
 10    N    -0.678603726   -1.214712914   -0.000182474
 11    C    -1.261006330   -2.256974330   -0.000107092
 12    H     3.178130800    1.268805135   -0.000758137
 13    H    -0.582879818    3.371914745    0.003015009
 14    H    -2.669077402    0.736276943   -0.898719567
 15    H    -2.669150375    0.739411448    0.901116032
 16    H    -2.655730741    2.275781283   -0.001429265
 17    H     3.329101762   -1.067809851    0.000300632
 18    H     2.025914011   -1.885616224    0.899132395
 19    H     2.026691088   -1.884877404   -0.900518395
 20    H     1.835861376    4.389879959   -0.896261857
 21    H     1.834959808    4.389732645    0.899471527
 22    H     3.204829755    3.675571352    0.002245685
---
H15C10N1, RHF, CHARGE=0, MULT=1
HF=14.8
 1    N    -0.057631932    0.031360855   -0.496809166
 2    C     0.038286018    1.483006530   -0.265348454
 3    C    -0.101602240    2.340606280   -1.536305045
 4    C    -1.363455388   -0.584369334   -0.204097383
 5    C    -2.238991715   -0.837253471   -1.445018861
 6    C     1.080479584   -0.751053244   -0.812682676
 7    C     1.000502547   -2.169176674   -0.949550028
 8    C     2.130882992   -2.938425778   -1.266865613
 9    C     3.382283572   -2.335471561   -1.464307890
 10   C     3.483934278   -0.941439817   -1.342359097
 11   C     2.361625941   -0.159986544   -1.025926268
 12   H    -0.757616640    1.802436882    0.455342350
 13   H     0.993647859    1.735893341    0.257651909
 14   H     0.565463701    2.014585058   -2.357744082
 15   H    -1.139819129    2.334130698   -1.924424265
 16   H     0.154734806    3.394101940   -1.296614059
 17   H    -1.231171777   -1.538035143    0.364860127
 18   H    -1.940180761    0.070715101    0.497769811
 19   H    -2.617636190    0.108596307   -1.881431925
 20   H    -1.709702590   -1.387787260   -2.246723168
 21   H    -3.122154462   -1.442872617   -1.151606197
 22   H     0.067642409   -2.716944049   -0.818863852
 23   H     2.029279756   -4.020995764   -1.360945823
 24   H     4.256444700   -2.936917371   -1.708974643
 25   H     4.447492486   -0.452266713   -1.495585047
 26   H     2.527148954    0.914883724   -0.957470821
---
H19C10N1, RHF, CHARGE=0, MULT=1
HF=-21.9
 1    C     0.000003834   -0.000032216    0.000009709
 2    C     1.531281647    0.000102137    0.000189772
 3    C     2.174070254    1.400778296    0.000171694
 4    C     3.716320089    1.386003497    0.000323779
 5    C     4.357005455    2.789236891    0.001422319
 6    C     5.899514331    2.774017503    0.001883078
 7    C     6.539496348    4.177543988    0.003524617
 8    C     8.081510629    4.160566102    0.003451246
 9    C     8.719439394    5.567267233    0.006164136
 10   C    10.176862352    5.533096185    0.006613950
```

```
11    N     11.338794250    5.524532988    0.006851644
12    H     -0.412927263    0.506642956    0.895769196
13    H     -0.412968554    0.505282213   -0.896504056
14    H     -0.382249955   -1.041720787    0.000769866
15    H      1.882663446   -0.569870858    0.891263886
16    H      1.883245372   -0.570193792   -0.890427655
17    H      1.814718120    1.964501692   -0.891470052
18    H      1.814579573    1.964425257    0.891652760
19    H      4.074728708    0.821502837    0.891564069
20    H      4.075100046    0.822947751   -0.891745652
21    H      3.999193471    3.353696250   -0.890128717
22    H      3.998261748    3.352837234    0.893292289
23    H      6.258340095    2.210069323    0.893257149
24    H      6.259031022    2.211752913   -0.890312873
25    H      6.182059466    4.743088900   -0.887489572
26    H      6.182627424    4.740643704    0.896261113
27    H      8.444758378    3.599936947    0.894740913
28    H      8.445045570    3.603383982   -0.889930684
29    H      8.383256373    6.144785208   -0.887457601
30    H      8.382498161    6.141755751    0.901493703
---
H11C11N1, RHF, CHARGE=0, MULT=1
HF=28.9
1     C      0.000232777   -0.000184566   -0.000003106
2     C      1.380600206    0.000092478    0.000015334
3     C      2.137695537    1.226301285   -0.000024574
4     C      1.447842173    2.432276247   -0.000099679
5     C      0.011774965    2.479535689   -0.000177131
6     C     -0.720125394    1.248410007   -0.000100707
7     N     -2.100654768    1.208886694   -0.000125125
8     C     -2.789248753    2.350544936   -0.000211089
9     C     -2.129113150    3.634339312   -0.000291266
10    C     -0.750788601    3.696371969   -0.000273973
11    C     -4.296009018    2.246436417   -0.000124043
12    C      3.641688246    1.165048994   -0.000145663
13    H     -0.552018074   -0.941243381    0.000039619
14    H      1.915886324   -0.951724015    0.000070061
15    H      1.990247180    3.380300302   -0.000011017
16    H     -2.725790538    4.546512305   -0.000364540
17    H     -0.238936315    4.659696168   -0.000349914
18    H     -4.715526632    2.746293231    0.898474909
19    H     -4.715801455    2.746634584   -0.898337546
20    H     -4.658309730    1.199139590   -0.000315726
21    H      4.015822218    0.626182290    0.895696905
22    H      4.015761362    0.632697660   -0.900010286
23    H      4.107474807    2.171447868    0.003553662
---
H11C11N1, RHF, CHARGE=0, MULT=1
HF=29.1
1     C     -0.000032232   -0.000140165    0.000012318
2     C      1.389109649    0.000067080   -0.000047072
3     C      2.080740865    1.264348109   -0.000135340
4     C      1.395060411    2.462595626    0.000368570
5     C     -0.040692758    2.489070057    0.001066736
6     C     -0.739148041    1.238959676    0.000541095
7     N     -2.119272119    1.168162262    0.000990333
8     C     -2.834716710    2.292802915    0.002350771
9     C     -2.206500127    3.592694527    0.003020482
10    C     -0.830127111    3.687974184    0.002293949
```

```
11    C    -4.338461760    2.152100028     0.003143035
12    C     2.187461457   -1.276653968     0.000064339
13    H    -0.559237464   -0.938146577    -0.000202707
14    H     1.944125337    3.405582575     0.000127324
15    H    -2.825713306    4.489726130     0.004101646
16    H    -0.340088994    4.662486422     0.002650535
17    H    -4.769573064    2.641292785     0.902128274
18    H    -4.770658686    2.642021892    -0.894825780
19    H    -4.674974990    1.096217384     0.002805309
20    H     2.837551042   -1.336167428    -0.898227117
21    H     2.838657361   -1.335273761     0.897623370
22    H     1.543839012   -2.179556774     0.000918119
23    H     3.172511848    1.271393589    -0.000470223
---
H15C11N1, RHF, CHARGE=0, MULT=1
HF=-1.8
1     C    -0.000021181    0.000350617     0.000512594
2     C     1.553581792    0.000514144    -0.000257312
3     C     2.062204502    1.468369821     0.000164866
4     C     1.543956780    2.202570940    -1.267131102
5     C    -0.009589433    2.194793902    -1.261081186
6     C    -0.524868719    0.729147233    -1.267168035
7     C    -0.523632130    0.719708581     1.274142337
8     H    -0.368564042   -1.049740526     0.000871717
9     H     1.945167669   -0.549724749    -0.883434832
10    H     1.946121944   -0.551311546     0.881830679
11    H     3.175146995    1.472673055    -0.000477036
12    C     1.554802437    2.199550216     1.273366826
13    H     1.934044485    1.717184498    -2.188212050
14    H     1.930587047    3.244542771    -1.302411636
15    H    -0.384440610    2.720863449    -2.167340914
16    C    -0.533082497    2.931199746     0.002561481
17    H    -1.635822356    0.703532513    -1.302862530
18    H    -0.193606129    0.201637268    -2.188076839
19    C    -0.015651078    2.205905370     1.295844870
20    H    -1.634682045    0.686962458     1.304648176
21    H    -0.187073752    0.179732592     2.186030714
22    H     1.949462765    3.238632265     1.302609668
23    H     1.955547629    1.705989807     2.185449565
24    H    -1.644306630    2.966489540    -0.005970714
25    H    -0.203371074    3.993052309    -0.007294566
26    C    -0.503846145    2.892178428     2.493215645
27    N    -0.891226869    3.436589305     3.443625152
---
H15C11N1, RHF, CHARGE=0, MULT=1
HF=17.5
1     C     0.000668119    0.001271699     0.000587493
2     C     1.554217233   -0.000253826    -0.001146134
3     C     2.065807249    1.466770310     0.000319939
4     C     1.547131822    2.204708294    -1.264684346
5     C    -0.006518269    2.198757785    -1.257804619
6     C    -0.523919226    0.733851503    -1.265012399
7     C    -0.522277459    0.715069989     1.278050925
8     H    -0.368019744   -1.048805609    -0.003365705
9     H     1.944674632   -0.550600700    -0.884658537
10    H     1.946566740   -0.552722308     0.880438952
11    H     3.178704095    1.467641562    -0.003983755
12    C     1.565518941    2.197591277     1.277378988
13    H     1.936594670    1.721693160    -2.187290905
```

```
14         H       1.935091607    3.246213681    -1.298286407
15         H      -0.380279173    2.724794249    -2.164550315
16         C      -0.530246654    2.936673035     0.005694722
17         H      -1.634991684    0.710400551    -1.299741479
18         H      -0.194466535    0.207147230    -2.187097474
19         C      -0.006893717    2.200762803     1.293095273
20         H      -1.633247798    0.679554358     1.307757973
21         H      -0.185499760    0.168288535     2.185809950
22         H       1.965451392    3.234780655     1.305583887
23         H       1.971404723    1.700511466     2.185488748
24         H      -1.641533958    2.973792220    -0.006851555
25         H      -0.199678971    3.998310555    -0.007543761
26         N      -0.482430079    2.870803110     2.466982076
27         C      -0.877411807    3.427232704     3.442055545
---
H17C11N1, RHF, CHARGE=0, MULT=1
HF=-10.7
1     C       0.000127307    0.000067850    0.000184461
2     C       1.430955410   -0.000273220   -0.000414130
3     C       2.112503241    1.234506464    0.000685984
4     C       1.395139326    2.434747219    0.007059068
5     C      -0.006254204    2.417056290    0.011781988
6     C      -0.761356611    1.216007220    0.006406923
7     N      -0.719760921   -1.247989339   -0.006602820
8     C       2.253711106   -1.266012624   -0.001995622
9     C      -2.311946077    1.280722767    0.004736260
10    C      -2.862553005    2.745359160    0.023392554
11    C      -2.876178701    0.625859302   -1.293398468
12    C      -2.884639951    0.592321091    1.281600294
13    H       3.203367439    1.270983799   -0.003331851
14    H       1.925030692    3.388375640    0.008380291
15    H      -0.485545418    3.397019220    0.019792702
16    H      -0.456150687   -1.800083997    0.797335229
17    H      -0.452638129   -1.793508349   -0.814017199
18    H       2.050086212   -1.887571194    0.895497696
19    H       2.055121610   -1.882103749   -0.904529611
20    H       3.343107806   -1.055650169    0.001849450
21    H      -2.547623177    3.302396150    0.929262241
22    H      -2.548057313    3.325919500   -0.867701584
23    H      -3.973122395    2.755070834    0.023202033
24    H      -2.832608631   -0.479000903   -1.285865552
25    H      -2.324518005    0.977605154   -2.189494443
26    H      -3.945768524    0.885601392   -1.444300923
27    H      -2.328555329    0.908441667    2.188056224
28    H      -2.856660978   -0.512325055    1.238629673
29    H      -3.950564899    0.860935238    1.441501714
---
H21C11N1, RHF, CHARGE=0, MULT=1
HF=-27.1
1     C       0.000026747   -0.000327097   -0.000153114
2     C       1.542613192   -0.000067681    0.000010651
3     C       2.169478017    1.409388205   -0.000018183
4     C       3.712025962    1.408568220    0.000062793
5     C       4.337042517    2.818296391   -0.000387501
6     C       5.881595162    2.814597821   -0.000062328
7     C       6.455834459    4.154533183   -0.000321273
8     N       6.931322891    5.214773077   -0.000142351
9     H      -0.363930118    0.560836577    0.890929638
10    H      -0.363852843    0.559257607   -0.892444371
```

```
11        H         1.907080464     -0.560192089     0.891614671
12        H         1.906843396     -0.560038292    -0.891795015
13        H         1.806431317      1.969882077     0.891793436
14        H         1.806433785      1.969993854    -0.891790586
15        H         4.077017934      0.849649799     0.892198015
16        H         4.077338303      0.848870561    -0.891481350
17        H         3.979420840      3.381218250     0.891778884
18        H         3.979559187      3.380852412    -0.892842810
19        H         6.265752837      2.269599302     0.894684184
20        H         6.265982350      2.268808684    -0.894212630
21        C        -0.626602705     -1.409880554     0.000615975
22        C        -2.168937956     -1.410991392     0.001101470
23        C        -2.796991744     -2.818312399     0.001921522
24        C        -4.328118457     -2.834767688     0.002955434
25        H        -0.261968602     -1.969894608     0.892230796
26        H        -0.262619623     -1.970215482    -0.891092545
27        H        -2.533956508     -0.850721883     0.892460316
28        H        -2.534502530     -0.851724017    -0.890710696
29        H        -2.438536784     -3.384669324     0.892548072
30        H        -2.439890674     -3.384949634    -0.889071704
31        H        -4.745937483     -2.333000663     0.899195750
32        H        -4.746963711     -2.333540907    -0.893118621
33        H        -4.699223247     -3.880524864     0.003477474
---
H9C12N1, RHF, CHARGE=0, MULT=1
HF=50.1
1     C    -0.000353650     -0.000362200      0.000060013
2     C     1.405809864     -0.000289496     -0.000043701
3     C     2.136749695      1.206874912     -0.000257232
4     C     1.487466348      2.453563224     -0.006073924
5     C     0.081356816      2.454165085     -0.007491226
6     C    -0.679105891      1.229385196      0.003155679
7     N    -0.802098492      3.563459854     -0.107381063
8     C    -2.123604915      3.056060069      0.023845904
9     C    -2.090514991      1.614632931      0.023408157
10    C    -3.299514061      0.900096823      0.047776438
11    C    -4.510201702      1.614272553      0.084921215
12    C    -4.526556768      3.025332021      0.094670230
13    C    -3.334614797      3.769629131      0.062324888
14    H    -0.547029677     -0.943071012     -0.002180397
15    H     1.943192868     -0.949450752     -0.000196051
16    H     3.227297980      1.171312906      0.002688400
17    H     2.063227648      3.377731158     -0.012107867
18    H    -3.307404613     -0.189539208      0.038367138
19    H    -5.454964607      1.069662947      0.106433798
20    H    -5.483630308      3.548353810      0.126471402
21    H    -3.361234459      4.858220211      0.064447344
22    H    -0.567919393      4.395004769      0.400848798
---
H11C12N1, RHF, CHARGE=0, MULT=1
HF=44.1
1     C     0.000178445      0.000010297     -0.000011009
2     C     1.405708265     -0.000470416      0.000212642
3     C     2.113203378      1.213953338     -0.000207305
4     C     1.425817411      2.451542875      0.000756689
5     C     0.010341849      2.437577131     -0.002102940
6     C    -0.694981606      1.221290442     -0.001508818
7     C     2.173531381      3.737211141     -0.025928283
8     C     2.565237267      4.265756252     -1.279293620
```

```
 9    C     3.269799263    5.474823532   -1.378078738
10    C     3.598991339    6.185970193   -0.211931967
11    C     3.225415358    5.688349873    1.043672447
12    C     2.509556066    4.465626482    1.154452172
13    N     2.043827825    4.056428770    2.434971293
14    H    -0.547493042   -0.942723511    0.000192294
15    H     1.950158313   -0.945396070   -0.000719164
16    H     3.204159021    1.191585437   -0.003914147
17    H    -0.547773760    3.375257360   -0.004000714
18    H    -1.785553047    1.227570783   -0.002807510
19    H     2.315795533    3.726387313   -2.195475565
20    H     3.559289568    5.860282522   -2.355562940
21    H     4.146133151    7.126936399   -0.281934686
22    H     3.490796027    6.261205527    1.933934119
23    H     2.640410773    4.391680196    3.174251308
24    H     2.010556324    3.055428428    2.532859395
 ---
H11C12N1, RHF, CHARGE=0, MULT=1
HF=48.2
 1    C     0.03497        0.05385        0.25682
 2    C     1.43689        0.00071        0.19263
 3    C     2.18472        1.16780       -0.03643
 4    C     1.52024        2.39452       -0.20706
 5    C     0.11840        2.46360       -0.15184
 6    C    -0.64282        1.28932        0.08391
 7    C    -2.77093        2.30430        0.82505
 8    C    -2.70586        2.33823        2.23977
 9    C    -3.44597        3.29639        2.95273
10    C    -4.25542        4.22148        2.26983
11    C    -4.32536        4.18713        0.86682
12    C    -3.58917        3.23410        0.13960
13    N    -2.06954        1.33167        0.02552
14    H    -0.51569       -0.87054        0.43810
15    H     1.94627       -0.95493        0.32383
16    H     3.27260        1.12221       -0.08269
17    H     2.09653        3.30306       -0.38768
18    H    -0.36367        3.43053       -0.29923
19    H    -2.08770        1.62966        2.79318
20    H    -3.39175        3.32085        4.04172
21    H    -4.82761        4.96289        2.82810
22    H    -4.95398        4.90312        0.33606
23    H    -3.65880        3.22300       -0.94921
24    H    -2.46925        0.41955        0.19911
 ---
H23C12N1, RHF, CHARGE=0, MULT=1
HF=-31.8
 1    C     0.000082150    0.000018959    0.000004248
 2    C     1.531459404   -0.000008286   -0.000050072
 3    C     2.173443543    1.403863346    0.000163459
 4    C     3.716655657    1.372551959    0.037134447
 5    C     4.528669022    2.702311889   -0.156926538
 6    C     4.120051339    3.794019964    0.903638958
 7    C     2.893643973    4.548264482    0.670436452
 8    N     1.934508854    5.189662450    0.528993746
 9    C     4.355427481    3.257473586   -1.598462161
10    C     6.059051908    2.416622605    0.120130737
11    C     6.799826010    1.350011806   -0.715171184
12    C     8.306902690    1.259268388   -0.394069791
13    C     9.066207347    0.197474426   -1.194924310
```

```
14      H       -0.413057245     0.503033530     0.897550801
15      H       -0.412839168     0.507821589    -0.894963679
16      H       -0.380844047    -1.042308440    -0.002716924
17      H        1.883113751    -0.570597148     0.890510796
18      H        1.882297525    -0.569800604    -0.891454611
19      H        1.791478183     1.961182780     0.884722449
20      H        1.820646955     1.951802824    -0.901356311
21      H        4.016515364     0.924352961     1.013704947
22      H        4.051162391     0.655159423    -0.747545842
23      H        4.039376267     3.322160891     1.912206132
24      H        4.934894728     4.554984780     0.988006694
25      H        5.002235665     4.141979528    -1.772847211
26      H        4.615240936     2.502020220    -2.367011914
27      H        3.314550294     3.572216535    -1.810844970
28      H        6.610427431     3.378621479    -0.002139342
29      H        6.167276812     2.136320261     1.194355569
30      H        6.683384666     1.565057595    -1.800778332
31      H        6.343835463     0.349623698    -0.540818308
32      H        8.787440260     2.247818993    -0.578422073
33      H        8.444411115     1.045545896     0.691135218
34      H        9.014349799     0.387803018    -2.286071064
35      H        8.670857171    -0.821554700    -1.008441098
36      H       10.137962183     0.196633914    -0.907501185
---
H9C13N1, RHF, CHARGE=0, MULT=1
HF=58.2
1       C       -0.000246768     0.000012153    -0.000012656
2       C        1.391837212     0.000526145     0.000001241
3       C        2.109927035     1.222706356    -0.000009095
4       C        1.417142770     2.428545247    -0.000054399
5       C       -0.008790163     2.448509001    -0.000068310
6       C       -0.743559336     1.221542611    -0.000051896
7       C       -2.206307627     1.300616671    -0.000058646
8       C       -2.814123644     2.597140829    -0.000057099
9       N       -2.064328092     3.776462270    -0.000107683
10      C       -0.755799114     3.701142041    -0.000082126
11      C       -4.238491387     2.740652754     0.000007098
12      C       -5.047859344     1.610912004     0.000110282
13      C       -4.462090383     0.318582769     0.000048324
14      C       -3.078896757     0.168304179    -0.000031702
15      H       -0.514144629    -0.961520926     0.000017975
16      H        1.940059449    -0.942748211     0.000020649
17      H        3.200092469     1.208720046     0.000024892
18      H        1.970818133     3.369104847    -0.000071125
19      H       -0.212160742     4.654758719    -0.000092865
20      H       -6.134088565     1.707988860     0.000263838
21      H       -5.107920897    -0.560493844     0.000053575
22      H       -2.672022675    -0.843457674    -0.000048596
23      H       -4.686385708     3.735693559     0.000015713
---
H9C13N1, RHF, CHARGE=0, MULT=1
HF=59.9
1       C        0.000080526    -0.000064716     0.000003895
2       C        1.392910705     0.000210472     0.000021905
3       C        2.108502999     1.223098548    -0.000025902
4       C        1.417763466     2.430218010    -0.000219586
5       C       -0.010171848     2.458677591    -0.000287750
6       C       -0.736744515     1.224821412    -0.000097050
7       C       -2.204449518     1.278601930     0.000000443
```

```
 8    C       -2.887316291     2.536935966    -0.000238057
 9    C       -2.103950439     3.756849029    -0.000549616
10    C       -0.737899288     3.717264944    -0.000544121
11    C       -4.314637060     2.530417696    -0.000124053
12    C       -4.991381054     1.318333423     0.000252761
13    C       -4.232116662     0.112603822     0.000497252
14    N       -2.896438911     0.092400839     0.000364238
15    H       -0.520669134    -0.959165817     0.000075867
16    H        1.941642683    -0.942395939     0.000078975
17    H        3.198981700     1.209957780     0.000097845
18    H        1.975512197     3.368169097    -0.000313402
19    H       -2.628142533     4.714231514    -0.000775293
20    H       -0.165568559     4.647181312    -0.000745757
21    H       -4.867754182     3.470543714    -0.000324005
22    H       -6.079457537     1.274560096     0.000362821
23    H       -4.731375419    -0.862865592     0.000815515
---
H9C13N1, RHF, CHARGE=0, MULT=1
HF=65.5
 1    C        0.000104106     0.000560958    -0.000003639
 2    C        1.372522618    -0.000007538     0.000006784
 3    C        2.112051074     1.238250509    -0.000001736
 4    C        1.457693943     2.444950780    -0.000046404
 5    C        0.009005208     2.507303020    -0.000081757
 6    C       -0.729042271     1.261694775    -0.000040842
 7    N       -2.083705851     1.218735940    -0.000043309
 8    C       -2.783077978     2.379662691    -0.000089242
 9    C       -2.138075838     3.675898198    -0.000114414
10    C       -0.727623258     3.710556334    -0.000120711
11    C       -4.238130519     2.307348990    -0.000120289
12    C       -4.984042398     3.459323858    -0.000175902
13    C       -4.345916212     4.752798648    -0.000190052
14    C       -2.977305448     4.858416620    -0.000158176
15    H       -0.560273586    -0.935490530     0.000016371
16    H        1.931997117    -0.936558497     0.000021908
17    H        3.201722307     1.191525721     0.000044710
18    H        2.021582434     3.379150847    -0.000050167
19    H       -0.205774150     4.669441420    -0.000155906
20    H       -4.719676324     1.328381867    -0.000101613
21    H       -6.074324450     3.420158541    -0.000205377
22    H       -4.976804543     5.642493680    -0.000240831
23    H       -2.499098802     5.839225529    -0.000167130
---
H9C13N1, RHF, CHARGE=0, MULT=1
HF=55.9
 1    C       -0.000305412    -0.000320384    -0.000023739
 2    C        1.391129002     0.000623034    -0.000012502
 3    C        2.104673423     1.224671033     0.000004245
 4    C        1.406268914     2.426266580     0.000142140
 5    C       -0.022503701     2.449557268     0.000200446
 6    C       -0.752548210     1.216569914    -0.000013280
 7    C       -2.218359314     1.255047833    -0.000165117
 8    C       -2.880220818     2.526257231     0.000270130
 9    C       -2.107469301     3.759105771     0.000731594
10    C       -0.742831809     3.711951810     0.000598166
11    N       -4.247223474     2.656500154     0.000233030
12    C       -5.000769036     1.554260567    -0.000291371
13    C       -4.442874795     0.244328449    -0.000838011
14    C       -3.061267449     0.099999624    -0.000775967
```

```
15    H     -0.508237823    -0.965412211    -0.000022756
16    H      1.940616517    -0.941604890     0.000004104
17    H      3.194947517     1.215388200    -0.000056058
18    H      1.959346325     3.367525001     0.000145614
19    H     -2.634820665     4.714731536     0.001159598
20    H     -0.165550638     4.638899400     0.000844282
21    H     -6.085977583     1.703924307    -0.000298148
22    H     -5.100088284    -0.624100618    -0.001297408
23    H     -2.641514761    -0.906148532    -0.001226293
---
H9C13N1, RHF, CHARGE=0, MULT=1
HF=57.5
1     C      0.000073296     0.000094496     0.000002517
2     C      1.429974790     0.000073451    -0.000001179
3     C      2.153506186     1.189360052    -0.000000871
4     C      1.483218901     2.438494290     0.000001099
5     C      0.093044030     2.473532801     0.000005745
6     C     -0.665514053     1.265896510     0.000006924
7     C     -0.828329741    -1.208099434    -0.000003807
8     C     -2.251874124    -1.052858764    -0.000016234
9     N     -2.868924135     0.201083753    -0.000009188
10    C     -2.123883444     1.279329692     0.000001660
11    C     -0.315140320    -2.542405927    -0.000003994
12    C     -1.163022624    -3.645498658    -0.000000460
13    C     -2.571442474    -3.473480427    -0.000000899
14    C     -3.115441293    -2.194630688    -0.000017648
15    H      1.984114126    -0.938979342    -0.000005463
16    H      3.244390666     1.166955191    -0.000002880
17    H      2.062219627     3.362303920    -0.000000827
18    H     -0.422464773     3.435535222     0.000003800
19    H     -2.655397960     2.239693029     0.000015350
20    H      0.760656984    -2.720990835    -0.000004099
21    H     -0.748248492    -4.654436485    -0.000001212
22    H     -3.219427413    -4.350679308     0.000011392
23    H     -4.198150418    -2.059636342    -0.000031335
---
H11C13N1, RHF, CHARGE=0, MULT=1
HF=47.6
1     C      0.000159227     0.000340397    -0.000117495
2     C      1.402337370     0.000311134     0.000013996
3     C      2.128517508     1.212842597    -0.000127318
4     C      1.478966251     2.455521299     0.002251827
5     C      0.067002850     2.462878848     0.006040599
6     C     -0.682916791     1.230841791     0.000023376
7     N     -0.833708855     3.554821239     0.037987111
8     C     -2.150894863     3.037318823    -0.006172920
9     C     -2.093422188     1.596216317    -0.008146176
10    C     -3.287873871     0.851960254    -0.019912102
11    C     -4.513792491     1.532536638    -0.034759389
12    C     -4.559965350     2.945046264    -0.037956534
13    C     -3.388790787     3.716011389    -0.024215116
14    H     -0.547902902    -0.941685591    -0.000724374
15    H      1.944085803    -0.945991158    -0.000569818
16    H      3.219313514     1.179916795    -0.002467566
17    H      2.066641671     3.372015821     0.000342348
18    H     -3.265911847    -0.237714827    -0.018275514
19    H     -5.446679000     0.967919724    -0.044737729
20    H     -5.529550643     3.445729364    -0.051762713
21    H     -3.457916789     4.802361472    -0.028635459
```

| | | | | |
|---|---|---|---|---|
| 22 | C | -0.468691748 | 4.959202187 | -0.115881197 |
| 23 | H | -0.388198827 | 5.238726444 | -1.192900632 |
| 24 | H | -1.227189312 | 5.617947509 | 0.362100442 |
| 25 | H | 0.507598658 | 5.168449141 | 0.374926414 |

---

H15C13N1, RHF, CHARGE=0, MULT=1
HF=22.3

| | | | | |
|---|---|---|---|---|
| 1 | C | -0.000332324 | -0.000030878 | -0.000180002 |
| 2 | C | 1.393806504 | 0.000507764 | -0.000091372 |
| 3 | C | 2.127617304 | 1.218342524 | -0.000122467 |
| 4 | C | 1.484733242 | 2.455574414 | 0.002399638 |
| 5 | C | 0.065970939 | 2.463641449 | 0.006988163 |
| 6 | C | -0.690062264 | 1.237600993 | 0.001194967 |
| 7 | N | -0.840872958 | 3.543119473 | 0.022302053 |
| 8 | C | -2.149766298 | 3.010094308 | -0.010448319 |
| 9 | C | -2.088661140 | 1.614838479 | -0.007012884 |
| 10 | C | -3.299918900 | 0.738093732 | -0.030141083 |
| 11 | C | -4.603118593 | 1.511333373 | 0.264024377 |
| 12 | C | -4.653262226 | 2.933820533 | -0.323380419 |
| 13 | C | -3.418438017 | 3.811168795 | -0.019186040 |
| 14 | H | -0.551732449 | -0.940115965 | -0.002183677 |
| 15 | H | 1.939765128 | -0.943627908 | -0.001135212 |
| 16 | H | 3.217954518 | 1.176321726 | -0.003290326 |
| 17 | H | 2.067909693 | 3.374897700 | -0.001506897 |
| 18 | H | -3.197096494 | -0.084801476 | 0.713964942 |
| 19 | H | -4.755685274 | 1.565108164 | 1.366520315 |
| 20 | H | -4.788644152 | 2.869607071 | -1.427537785 |
| 21 | H | -3.553774367 | 4.308953952 | 0.970026713 |
| 22 | C | -0.470893199 | 4.954854567 | -0.029257754 |
| 23 | H | -0.271156103 | 5.272875137 | -1.079435908 |
| 24 | H | -1.279032186 | 5.597612930 | 0.383297664 |
| 25 | H | 0.443457990 | 5.144378257 | 0.576583865 |
| 26 | H | -3.377581147 | 0.243231105 | -1.027239609 |
| 27 | H | -5.466027852 | 0.930472098 | -0.133533094 |
| 28 | H | -5.559824964 | 3.451585849 | 0.064343151 |
| 29 | H | -3.374388933 | 4.626705000 | -0.777236443 |

---

H13C14N1, RHF, CHARGE=0, MULT=1
HF=40.6

| | | | | |
|---|---|---|---|---|
| 1 | C | -0.000059479 | 0.000193535 | 0.000260071 |
| 2 | C | 1.401766342 | -0.000177825 | -0.000278964 |
| 3 | C | 2.127989812 | 1.212463786 | 0.000220909 |
| 4 | C | 1.478327561 | 2.454635042 | 0.002482628 |
| 5 | C | 0.065631814 | 2.463822830 | 0.002881605 |
| 6 | C | -0.683318314 | 1.230988564 | 0.001529307 |
| 7 | C | -2.093708287 | 1.594007074 | 0.002057014 |
| 8 | C | -2.153585386 | 3.035105708 | 0.004472791 |
| 9 | N | -0.836799162 | 3.554005237 | -0.002234913 |
| 10 | C | -0.471378890 | 4.973255316 | 0.013200970 |
| 11 | C | -3.394760770 | 3.709703165 | 0.006180349 |
| 12 | C | -4.564120412 | 2.936816415 | 0.004053254 |
| 13 | C | -4.514864913 | 1.524145670 | 0.001059605 |
| 14 | C | -3.287253986 | 0.847272045 | 0.000138222 |
| 15 | H | -0.548401484 | -0.941554588 | -0.000542354 |
| 16 | H | 1.943472557 | -0.946401280 | -0.000637747 |
| 17 | H | 3.218773866 | 1.179288501 | -0.000864664 |
| 18 | H | 2.067711371 | 3.369812452 | 0.004797296 |
| 19 | C | -0.318917747 | 5.560926120 | 1.426457658 |
| 20 | H | -5.534867043 | 3.435536535 | 0.004736293 |

```
21   H    -5.446487005    0.957439186   -0.000507553
22   H    -3.262625092   -0.242407432   -0.002139241
23   H    -3.468168298    4.795723989    0.009671217
24   H     0.482823171    5.112297985   -0.552664569
25   H    -1.240207422    5.556309618   -0.551586327
26   H    -1.260780034    5.504483341    2.007430037
27   H     0.471296091    5.048891162    2.010498616
28   H    -0.037007801    6.631501888    1.347889949
---
H27C14N1, RHF, CHARGE=0, MULT=1
HF=-41.8
1    C    -0.000029600    0.000005906   -0.000005940
2    C     1.531213799    0.000197171    0.000176743
3    C     2.174029053    1.400690019    0.000023040
4    C     3.716287852    1.385452264   -0.001032662
5    C     4.358025444    2.788153665    0.000035515
6    C     5.900536980    2.772389556   -0.000986196
7    C     6.542277553    4.175138343    0.000180692
8    C     8.084798784    4.159116007   -0.000870047
9    C     8.726473891    5.561908339    0.000445737
10   C    10.268966429    5.545474465   -0.000615566
11   C    10.910551527    6.948277080    0.000541037
12   C    12.452449565    6.928901599   -0.000346723
13   C    13.092594223    8.334711789    0.001235009
14   C    14.549851073    8.297748466    0.000687522
15   N    15.711819455    8.286826045    0.000282053
16   H    -0.412868954    0.506967871    0.895638058
17   H    -0.413006937    0.505208827   -0.896563938
18   H    -0.382574399   -1.041596673    0.001012423
19   H     1.882713984   -0.569368001    0.891731431
20   H     1.883442860   -0.570616721   -0.890088294
21   H     1.815312083    1.963533014    0.892421292
22   H     1.813509127    1.964662147   -0.890824152
23   H     4.075035239    0.820280291    0.889752306
24   H     4.073894644    0.822302351   -0.893464768
25   H     4.000433121    3.351425187    0.892442556
26   H     3.999179364    3.353118392   -0.890773716
27   H     6.259411676    2.207272811    0.889722882
28   H     6.258266838    2.209480490   -0.893526488
29   H     6.184862556    4.738390503    0.892762125
30   H     6.183619468    4.740267077   -0.890550863
31   H     8.443768143    3.594007299    0.889785564
32   H     8.442564822    3.596453437   -0.893524580
33   H     8.369331791    6.125117592    0.893040391
34   H     8.368047455    6.127207009   -0.890301770
35   H    10.628142169    4.980916681    0.890231791
36   H    10.627050093    4.983181167   -0.893363640
37   H    10.554704080    7.512228834    0.893181585
38   H    10.553579226    7.514134167   -0.890405542
39   H    12.815434010    6.368284963    0.890860923
40   H    12.814530740    6.370704016   -0.893698151
41   H    12.757231899    8.910149112    0.896482157
42   H    12.756624343    8.912272688   -0.892580886
---
H8C9N2, RHF, CHARGE=0, MULT=1
HF=49.8
1    C     0.012165272   -0.022322830    0.017894134
2    C     1.403414284   -0.001160440    0.021702455
3    C     2.070285533    1.286862497    0.021829863
```

```
 4    N     1.421024525    2.439772077    0.012688331
 5    C     0.037364482    2.437710547    0.007939842
 6    C    -0.713201055    1.216414880    0.011810928
 7    C    -2.148879651    1.297403714    0.008090209
 8    C    -2.786568966    2.524012151   -0.000405946
 9    C    -2.030031904    3.737809582   -0.005628325
10    C    -0.647877423    3.705407893   -0.001239886
11    N     2.179705402   -1.185115798    0.150724264
12    H    -0.546155067   -0.959669584    0.021598383
13    H    -3.875789674    2.582384689   -0.003163674
14    H     3.165252326    1.353786419    0.031499333
15    H    -0.071205520    4.631407676   -0.004311620
16    H    -2.562566598    4.689691243   -0.012772286
17    H     3.045069826   -1.117090518   -0.361634385
18    H     1.695213292   -1.989385287   -0.215112897
19    H    -2.734358126    0.376863906    0.013137103
---
H8C9N2, RHF, CHARGE=0, MULT=1
HF=50.3
 1    C    -0.011374679   -0.003209164   -0.078591866
 2    C     1.369511113   -0.006148209   -0.049550596
 3    C     2.068841961    1.243261329    0.028451473
 4    N     1.448534661    2.415558888    0.062757406
 5    C     0.065262850    2.443144401    0.028186330
 6    C    -0.726579906    1.244496386   -0.026577554
 7    C    -2.172022875    1.383940827   -0.039287929
 8    C    -2.749614433    2.653932077   -0.025570616
 9    C    -1.941178681    3.828419088    0.009058728
10    C    -0.564042455    3.739062277    0.042745376
11    N    -2.980431702    0.214510354   -0.161959705
12    H    -0.543770255   -0.952251585   -0.151097895
13    H     1.937397632   -0.934792129   -0.087812572
14    H     3.164160318    1.263823113    0.060290683
15    H    -3.832875903    2.787971145   -0.041330105
16    H    -2.434367212    4.801960905    0.014110632
17    H     0.051054742    4.639144278    0.076735213
18    H    -2.808837930   -0.426399335    0.597953189
19    H    -3.962192609    0.437048174   -0.118622432
---
H8C9N2, RHF, CHARGE=0, MULT=1
HF=49.3
 1    C     0.017623902   -0.030947340   -0.003408499
 2    C     1.396629867    0.023889563   -0.003653975
 3    C     2.044402285    1.305724797   -0.004182657
 4    N     1.382145012    2.456189261   -0.006335904
 5    C    -0.000402592    2.424014034   -0.002997422
 6    C    -0.736844495    1.194061817   -0.001151010
 7    C    -2.170661028    1.242191460    0.006098588
 8    C    -2.845281120    2.462986331    0.015483441
 9    C    -2.092049971    3.694659224    0.011119457
10    C    -0.712404595    3.677602278    0.002046829
11    N    -4.260604853    2.521418973    0.147288909
12    H    -0.502088684   -0.989751487   -0.004090194
13    H     2.003927015   -0.880203107   -0.002657076
14    H     3.138555004    1.368770632   -0.003391209
15    H    -0.152299125    4.614426151    0.000273331
16    H    -2.616524721    4.652030564    0.021416824
17    H    -4.649348507    3.309326521   -0.345551860
18    H    -4.708441633    1.700108454   -0.226217092
```

```
19         H         -2.717272076      0.297464466      0.008619744
---
H8C9N2, RHF, CHARGE=0, MULT=1
HF=44.8
1    C    -0.000006742    -0.000396449    -0.000040013
2    C     1.379141131     0.000240809    -0.000134238
3    C     2.070255540     1.258965139     0.000040886
4    N     1.446087608     2.429138060    -0.011848202
5    C     0.063166618     2.458805652    -0.028505407
6    C    -0.711384871     1.250943514    -0.006951651
7    C    -2.142745990     1.340366524     0.013400800
8    C    -2.765029874     2.574903199     0.012703831
9    C    -2.006486915     3.780229548    -0.022832109
10   C    -0.608656318     3.749210841    -0.057154956
11   N     0.108922630     4.958067887    -0.237992698
12   H    -0.556878123    -0.938295979     0.005023280
13   H     1.953041133    -0.925493158     0.002280254
14   H     3.165810011     1.286675116     0.012576347
15   H    -3.854022103     2.643905685     0.034814581
16   H    -2.553126114     4.725080136    -0.030499884
17   H     1.085229579     4.890241476    -0.007113049
18   H    -0.272325866     5.705005875     0.320751029
19   H    -2.736786811     0.425873940     0.031579449
---
H18C9N2, RHF, CHARGE=0, MULT=1
HF=-1.1
1    C     0.000015271     0.000145021    -0.000026831
2    C     1.531908209     0.000023359     0.000003944
3    C     2.169012173     1.404400707     0.000422771
4    C     3.734728487     1.404093254     0.004704684
5    C     4.290345675     1.018845780    -1.303277710
6    N     4.747261952     0.694440734    -2.320790777
7    N     4.366691116     2.668392203     0.427574779
8    C     4.067632383     3.904343010    -0.317062851
9    C     5.329983828     4.698190861    -0.695556378
10   C     4.669039509     2.805418894     1.864210457
11   C     6.072839300     2.290485843     2.227824767
12   H    -0.412924394     0.510020022     0.893880292
13   H    -0.412576489     0.501359760    -0.898814224
14   H    -0.380204624    -1.042119537     0.004871281
15   H     1.882232033    -0.566934413     0.893500526
16   H     1.880946710    -0.574255354    -0.888336368
17   H     1.808115125     1.952583333     0.900301531
18   H     1.791942377     1.971560811    -0.879917790
19   H     4.064589972     0.602411320     0.723132352
20   H     3.376308254     4.569107510     0.263961686
21   H     3.522039080     3.667012623    -1.263338970
22   H     5.835659295     5.138184442     0.186987094
23   H     6.069085413     4.077825715    -1.239679426
24   H     5.042550136     5.539159276    -1.361461505
25   H     4.595165432     3.876642500     2.176874863
26   H     3.912729723     2.268809399     2.493073650
27   H     6.179607635     1.200125918     2.061267684
28   H     6.870021026     2.800653502     1.652260986
29   H     6.263115252     2.480790787     3.305107620
---
H8C10N2, RHF, CHARGE=0, MULT=1
HF=69.1
1    C    -0.000103518    -0.000105079     0.000048496
```

```
2    C     1.402767608    0.000208170   -0.000127414
3    C     2.080660006    1.231580192    0.000152184
4    C     1.327971286    2.431059176    0.000537372
5    N    -0.031761443    2.427384560    0.006851977
6    C    -0.675477911    1.239585077    0.003947998
7    C     1.958487124    3.782363589   -0.007972497
8    N     2.548017258    4.137536142   -1.180664123
9    C     3.131812426    5.352527701   -1.269959582
10   C     3.162635570    6.270409300   -0.197770642
11   C     2.560586047    5.904409384    1.015402962
12   C     1.948552786    4.642985594    1.116461671
13   H    -0.561656938   -0.932911026   -0.003477144
14   H     1.959873712   -0.936897023   -0.001968803
15   H     3.169869737    1.256038243   -0.003529039
16   H    -1.769471240    1.291404650    0.004867923
17   H     3.588311499    5.591057935   -2.236593171
18   H     2.564640915    6.585449963    1.866353391
19   H     1.472144039    4.339265156    2.048163743
20   H     3.643957344    7.240034231   -0.315966246
---
H8C10N2, RHF, CHARGE=0, MULT=1
HF=67.9
1    C    -0.000051500    0.000210314   -0.000058509
2    C     1.402850746   -0.000035659    0.000038948
3    C     2.080576516    1.231299894   -0.000290097
4    C     1.329543335    2.432042520   -0.001192550
5    N    -0.030183565    2.428333582   -0.001479373
6    C    -0.674612385    1.240666790   -0.000778111
7    C     1.968774300    3.776187841   -0.001553790
8    C     2.299372731    4.430252144    1.205515556
9    C     2.907052043    5.702603411    1.156761889
10   N     3.194795289    6.338476412   -0.001925407
11   C     2.877727599    5.716700605   -1.160561758
12   C     2.268307383    4.445045839   -1.208602726
13   H    -0.562091515   -0.932493518    0.000534144
14   H     1.959849442   -0.937095155    0.000450900
15   H     3.169848507    1.256954576    0.000211059
16   H    -1.768513610    1.293406428   -0.000720727
17   H     2.089964908    3.972381684    2.171897648
18   H     3.174581070    6.236718318    2.074693729
19   H     3.122178481    6.261467541   -2.078604727
20   H     2.033525579    3.999122109   -2.174698998
---
H8C10N2, RHF, CHARGE=0, MULT=1
HF=70.1
1    N     0.000249596    0.000444509   -0.000274704
2    C     1.352877754   -0.000384587   -0.000074798
3    C     2.122092561    1.182326801    0.000101373
4    C     1.467788686    2.434169579    0.000216473
5    C     0.055299974    2.428327237   -0.000379751
6    C    -0.632059986    1.196150224   -0.000410532
7    C     2.234541084    3.706079725    0.001121223
8    C     2.615429233    4.334479998   -1.205177570
9    C     3.342810529    5.542424953   -1.156649933
10   N     3.700882121    6.140206136    0.002683364
11   C     3.339056967    5.542956595    1.161202622
12   C     2.611371011    4.335156412    1.208436887
13   H     1.831194028   -0.985673823   -0.000028074
14   H     3.209705007    1.115336891    0.000103533
```

```
15    H    -1.726630257    1.159167959    -0.000698195
16    H     3.652194216    6.053551313    -2.074431416
17    H     3.645739319    6.054803615     2.079626588
18    H     2.349760861    3.903464760     2.174094008
19    H    -0.511409094    3.359132151    -0.000611630
20    H     2.356702821    3.902018350    -2.171428712
---
H10C10N2, RHF, CHARGE=0, MULT=1
HF=41.3
1     C     0.000008233   -0.000112508    -0.000013390
2     C     1.381959915    0.000020682     0.000004709
3     C     2.120374833    1.228963521     0.000011782
4     C     1.472086771    2.449377488    -0.000045236
5     C     0.031267340    2.483089365    -0.000133480
6     C    -0.706094122    1.256132111    -0.000109180
7     N    -2.083438222    1.271907111    -0.000184330
8     C    -2.713135082    2.446024284    -0.000353611
9     C    -1.961939995    3.698064903    -0.000208969
10    N    -0.628710933    3.691881135    -0.000155454
11    C    -4.222786959    2.373333294    -0.000808725
12    C    -2.586839936    5.075377435    -0.000447086
13    H    -0.559614797   -0.936063277     0.000032743
14    H     1.934593037   -0.940651468     0.000010526
15    H     3.210311920    1.182340639     0.000050779
16    H     2.036014414    3.382851547     0.000010253
17    H    -4.637989174    2.870973273     0.900601534
18    H    -4.599122170    1.330160585    -0.002266125
19    H    -4.637455775    2.873291224    -0.901181078
20    H    -2.270708210    5.651690632     0.894362603
21    H    -3.693874820    5.043776996     0.007036882
22    H    -2.282278803    5.645583671    -0.903170799
---
H12C10N2, RHF, CHARGE=0, MULT=1
HF=52.7
1     C    -0.039029217    0.010491065    -0.210824800
2     C     1.350681970    0.184032146    -0.476921953
3     C     1.980458808    1.420198839    -0.277076099
4     C     1.273845909    2.546554355     0.199443430
5     C    -0.102832123    2.378928888     0.465614090
6     C    -0.748152315    1.150251798     0.268919290
7     N    -0.681576036   -1.229981099    -0.408469920
8     C     0.040746948   -2.379644928    -0.957961650
9     C    -2.104676753   -1.401823255    -0.107295590
10    C     1.963034807    3.873810685     0.425218526
11    C     1.957758348    4.724004284    -0.761601999
12    N     1.958424695    5.407982457    -1.700887111
13    H     1.975196259   -0.630660658    -0.843968636
14    H     3.047180118    1.496645282    -0.500830830
15    H    -0.696691534    3.218589450     0.833610940
16    H    -1.813055828    1.113821675     0.498812214
17    H    -0.632012381   -3.256983288    -1.086858047
18    H     0.866271387   -2.700308202    -0.279844521
19    H     0.469157763   -2.150681595    -1.961930634
20    H    -2.437217523   -2.445767733    -0.303679114
21    H    -2.738633995   -0.734918505    -0.737314507
22    H    -2.319522090   -1.191360819     0.966629895
23    H     3.020875923    3.715873345     0.745325037
24    H     1.476944645    4.430332071     1.262256091
---
```

```
H16C10N2, RHF, CHARGE=0, MULT=1
HF=14.6
1    C    -0.000125516    -0.000102121    -0.000176649
2    C     1.578945311    -0.000124007     0.000156881
3    C    -0.461416803     1.402286798     0.000233930
4    C    -0.462399552    -0.578026514     1.277494154
5    C    -0.658158302    -0.804122067    -1.244695367
6    C    -2.214238633    -0.710186746    -1.155943783
7    C    -0.266071427    -2.312159811    -1.188695638
8    C    -0.164515834    -0.188252773    -2.604582503
9    H    -0.296601644     0.917635104    -2.577544708
10   C    -0.795557759    -0.668973720    -3.918688047
11   N    -0.815219843    -1.012379594     2.295886589
12   N    -0.814924271     2.508893526     0.022605577
13   H     1.940583966    -1.051118122    -0.049779485
14   C     2.294269217     0.669252629     1.182375098
15   H     1.938760869     0.503710776    -0.924592090
16   H    -2.584497611     0.318683583    -1.335525248
17   H    -2.602654337    -1.033951051    -0.169967952
18   H    -2.703126756    -1.364085876    -1.906793176
19   H     0.810072148    -2.477781374    -1.395225144
20   H    -0.488694098    -2.771842243    -0.205443567
21   H    -0.827125108    -2.902814888    -1.941600418
22   H     0.931755071    -0.363699536    -2.703272736
23   H    -0.708172147    -1.763627365    -4.063499255
24   H    -1.866092821    -0.396297609    -4.002262794
25   H    -0.270765809    -0.184895891    -4.769680882
26   H     2.103921317     0.149949589     2.142168088
27   H     2.012011518     1.732949324     1.306209162
28   H     3.390004779     0.638272547     1.005318525
---
H16C10N2, RHF, CHARGE=0, MULT=1
HF=15.3
1    C     0.000140528     0.000070559    -0.000064725
2    C     1.535151000     0.000043257    -0.000186596
3    C     2.290198090     1.384707532    -0.000217762
4    C     1.662060886     2.261208394     1.005209664
5    C     3.852696371     1.216381723     0.393819330
6    C     3.924839396     0.719416807     1.780044707
7    C     4.593256637     0.188881282    -0.540866552
8    C     6.085933306    -0.075172270    -0.297893464
9    C     4.604893008     2.588659071     0.369667496
10   C     2.100452207     2.072971835    -1.389752729
11   N     4.000078237     0.338755513     2.875574427
12   N     1.153894481     2.957146169     1.785193726
13   H    -0.423006991     0.405150626     0.940017316
14   H    -0.431530959     0.571841551    -0.844814323
15   H    -0.360589051    -1.045880935    -0.096228088
16   H     1.859096756    -0.578604296    -0.895598775
17   H     1.851660339    -0.595769377     0.885447645
18   H     4.487998420     0.521662758    -1.598096614
19   H     4.084044436    -0.799206344    -0.473301800
20   H     6.715695015     0.803444979    -0.541363472
21   H     6.305358208    -0.375949045     0.745072691
22   H     6.418639816    -0.905401363    -0.956076679
23   H     4.808942790     2.927941806    -0.665160799
24   H     4.040064893     3.392648262     0.880461711
25   H     5.582914210     2.518296360     0.888471843
26   H     2.602607590     3.058658137    -1.445060787
```

```
27         H     2.490320415    1.450314194   -2.218438898
28         H     1.027882226    2.260056714   -1.600886035
---
H16C10N2, RHF, CHARGE=0, MULT=1
HF=19
1     C    -0.000067788    0.000117719    0.000105390
2     C     1.567250931    0.000055722    0.000046981
3     C    -0.445492044    1.407856811   -0.000044205
4     C    -0.415932183   -0.495113412    1.332015135
5     C    -0.680023956   -0.872717411   -1.180277783
6     C    -2.227811578   -0.680640010   -1.130010125
7     C    -0.391771131   -2.385779707   -0.897184933
8     C    -0.109641244   -0.520741347   -2.625800882
9     C    -0.228795638    0.949111391   -3.093704172
10    C    -0.628077046   -1.426801241   -3.773570334
11    N    -0.718975710   -0.841004151    2.399181551
12    N    -0.781388714    2.518562497    0.062488492
13    H     1.986429650    0.512125808   -0.886677374
14    H     1.979589541   -1.026711099    0.029417489
15    H     1.956225293    0.534540709    0.892033492
16    H    -2.542022131    0.354293038   -1.366355613
17    H    -2.644261087   -0.925858632   -0.131797460
18    H    -2.745665984   -1.344436094   -1.851582436
19    H     0.668289684   -2.653462604   -1.078729614
20    H    -0.628092063   -2.680568749    0.144220672
21    H    -1.010230501   -3.049468884   -1.535000752
22    H     0.990182633   -0.731067365   -2.579875120
23    H     0.412830432    1.629968473   -2.502039101
24    H    -1.266705682    1.333040967   -3.062640191
25    H     0.121312879    1.052539500   -4.142984280
26    H    -0.447713423   -2.503192274   -3.586613014
27    H    -1.709610796   -1.292638809   -3.972606473
28    H    -0.089768523   -1.191914575   -4.716431907
---
H8C12N2, RHF, CHARGE=0, MULT=1
HF=80.9
1     C    -0.000006243    0.000349231    0.000011882
2     C     1.372316018   -0.000087412   -0.000009617
3     C     2.119845274    1.236091948    0.000007569
4     C     1.482098068    2.451141953    0.000076269
5     C     0.028356870    2.504174657    0.000116256
6     C    -0.722557360    1.263160661    0.000072887
7     N    -2.077080904    1.261405799    0.000097819
8     C    -2.703305854    2.462377312    0.000154384
9     C    -4.157179462    2.516687054    0.000177092
10    C    -4.794236969    3.732158069    0.000222990
11    C    -4.046106026    4.968006841    0.000244109
12    C    -2.673828250    4.966621443    0.000236108
13    C    -1.952220487    3.703308711    0.000197005
14    N    -0.597973367    3.705059011    0.000182315
15    H    -0.563483508   -0.933616850   -0.000038112
16    H     1.929944655   -0.937840477   -0.000036262
17    H     3.209408385    1.178182783   -0.000046781
18    H     2.047032519    3.384245660    0.000093930
19    H    -4.723006808    1.584159005    0.000152911
20    H    -5.883697493    3.790757573    0.000249601
21    H    -4.602893218    5.906272678    0.000260717
22    H    -2.109570241    5.900086151    0.000259116
---
```

```
H8C12N2, RHF, CHARGE=0, MULT=1
HF=89.9
1    C    -0.000046009    0.000071615    -0.000001560
2    C     1.393268021    0.000156084     0.000011182
3    C     2.118327149    1.219385092    -0.000011788
4    C     1.440714738    2.435735061    -0.000070420
5    C     0.014784567    2.442319285    -0.000092661
6    C    -0.732078143    1.225045880    -0.000042500
7    N    -0.634932539    3.703258505    -0.000153133
8    N    -1.862136338    3.808367889    -0.000145199
9    C    -2.716765687    2.676394147    -0.000086458
10   C    -2.187716752    1.349961399    -0.000043081
11   C    -3.117628094    0.267660329    -0.000000231
12   C    -4.490598158    0.505169195     0.000007818
13   C    -4.998148602    1.829734775    -0.000026291
14   C    -4.123100361    2.912908808    -0.000074328
15   H    -0.522781517   -0.956744812     0.000030931
16   H     1.938902365   -0.944684800     0.000026604
17   H     3.208626255    1.195096575     0.000023732
18   H     1.996319070    3.374973092    -0.000109925
19   H    -2.765806133   -0.764279836     0.000028173
20   H    -5.189018365   -0.333134552     0.000031993
21   H    -6.076788237    1.991245732    -0.000014234
22   H    -4.510278551    3.933038272    -0.000104599
---
H10C12N2, RHF, CHARGE=0, MULT=1
HF=107.7
1    C     0.000095429    0.000298626     0.000043716
2    C     1.405956457    0.000197763     0.000169964
3    C     2.104221884    1.229306087    -0.000387666
4    C     1.392147085    2.450187317    -0.022429948
5    C    -0.013718671    2.434066964    -0.022316203
6    C    -0.710947182    1.213197773    -0.008402665
7    N     3.543740550    1.236810774    -0.144011918
8    N     4.360820681    1.242814911     0.747852455
9    C     4.091576356    1.244322216     2.169314216
10   C     4.050695259    0.020546288     2.875826823
11   C     3.927649594    0.031249629     4.276350541
12   C     3.853675099    1.249431278     4.974393735
13   C     3.907173870    2.465108750     4.270115683
14   C     4.030274295    2.470804167     2.869532869
15   H    -0.540053784   -0.947331011     0.003939974
16   H     1.938730244   -0.951732237    -0.005242520
17   H     1.913848335    3.408019661    -0.045892192
18   H    -0.564584045    3.375248692    -0.035389132
19   H    -1.801221008    1.207205720    -0.007568161
20   H     4.118316600   -0.935182139     2.354229366
21   H     3.892781498   -0.912076898     4.822643977
22   H     3.757606262    1.251456356     6.060519996
23   H     3.856302492    3.410419851     4.811511120
24   H     4.082180353    3.424884561     2.343212505
---
H12C12N2, RHF, CHARGE=0, MULT=1
HF=50
1    C    -0.000421438   -0.000388889     0.000325633
2    C     1.409538975    0.000385812    -0.000091436
3    N     2.146763211    1.131991922    -0.000379513
4    C     1.493310630    2.322624259     0.004730294
5    C     0.080058337    2.406984053     0.001997194
```

```
6    C    -0.697700609    1.227316077   -0.000458626
7    C     2.372452837    3.528105498    0.016502761
8    C     3.115315218    3.943751031   -1.114739173
9    C     3.938950112    5.089304239   -1.036845411
10   C     3.972852423    5.772108937    0.198424474
11   C     3.199750330    5.296288661    1.277329821
12   N     2.413201506    4.201406237    1.195408365
13   H    -0.532976215   -0.951347525    0.001215535
14   H     1.976386008   -0.936976890   -0.000032481
15   H    -0.401257153    3.385488220    0.004618739
16   C    -2.202363623    1.264750175   -0.000602144
17   H     3.049270111    3.373552373   -2.041909886
18   C     4.747249658    5.569826921   -2.212042524
19   H     4.587333882    6.662098379    0.334369663
20   H     3.205845507    5.806269750    2.246720657
21   H    -2.609941091    0.735031424   -0.887211266
22   H    -2.607442935    0.772708967    0.908431593
23   H    -2.600346891    2.299187372   -0.022370161
24   H     5.823839714    5.638582428   -1.950276852
25   H     4.408270800    6.576668701   -2.534980313
26   H     4.663211672    4.895042113   -3.087601706
---
H20C12N2, RHF, CHARGE=0, MULT=1
HF=0.8
1    C    -0.000127185   -0.000150076   -0.000116971
2    C     1.533576830   -0.000306456    0.000042308
3    C     2.172921562    1.401485093   -0.000283607
4    C     1.621736444    2.412250462   -1.078819480
5    C     0.045520760    2.369081385   -1.034145449
6    C    -0.591768610    0.966263156   -1.033737948
7    N     2.109929799    3.793337770   -0.777382566
8    C     1.511841192    4.924677091   -1.522755968
9    C     1.725738004    6.267673352   -0.789052589
10   C     3.180763272    6.536441096   -0.383700214
11   C     3.868534127    5.290565360    0.189396876
12   C     3.565838461    3.989811672   -0.587647077
13   C     2.082524426    1.995616126   -2.423848718
14   N     2.445584501    1.673741829   -3.479687175
15   H    -0.377156799    0.257727587    1.016048481
16   H    -0.368154954   -1.030600696   -0.206854838
17   H     1.895792390   -0.544029217    0.902966577
18   H     1.906297909   -0.586622418   -0.870686089
19   H     2.041310929    1.845726944    1.012649389
20   H     3.265996984    1.241464932   -0.140402087
21   H    -0.306622248    2.911343514   -0.127139474
22   H    -0.382127477    2.902081743   -1.913464678
23   H    -1.682606100    1.083596686   -0.837584367
24   H    -0.518318561    0.516094579   -2.050012304
25   H     0.410834747    4.796776118   -1.638018989
26   H     1.917487518    4.997715943   -2.565713363
27   H     1.372124085    7.085808671   -1.457641545
28   H     1.072830148    6.307681721    0.112627218
29   H     3.211585224    7.356917035    0.368130377
30   H     3.753315686    6.911612081   -1.262977423
31   H     4.971982234    5.444819308    0.187413579
32   H     3.580883449    5.162309642    1.257831638
33   H     4.116147970    4.000384799   -1.564793189
34   H     4.014125270    3.154915464   -0.001292790
---
```

```
H16C13N2, RHF, CHARGE=0, MULT=1
HF=45.4
1    C    0.000182300    0.000087897   -0.000039538
2    C    1.405247944    0.000091773   -0.000025652
3    C    2.112818622    1.214280902    0.000027908
4    C    1.402752653    2.425624953   -0.003032605
5    C   -0.004083228    2.425218609   -0.006479095
6    C   -0.733460208    1.213442311   -0.001245837
7    C   -2.263594839    1.147378077    0.029541636
8    N   -2.802281932    0.723456572   -1.279229345
9    C   -2.665965735    1.642120858   -2.423358516
10   C   -2.872923355    0.913275398   -3.768816874
11   C   -4.123337381    0.023178014   -3.816985361
12   C   -4.310173210   -0.815646443   -2.544759460
13   C   -4.064978468   -0.037355667   -1.234617314
14   C   -2.906715004    2.369965955    0.549202470
15   N   -3.439462167    3.304884916    0.987766599
16   H   -0.517837659   -0.960912214    0.000714338
17   H    1.946537934   -0.946672817    0.000198742
18   H    3.202996822    1.215357435    0.001854152
19   H    1.941591894    3.373857133   -0.003890137
20   H   -0.512452903    3.390697429   -0.016603437
21   H   -2.523909545    0.350486506    0.785467724
22   H   -1.643444817    2.090215860   -2.429130487
23   H   -3.380496395    2.503830587   -2.353885639
24   H   -2.938898976    1.682381040   -4.572309885
25   H   -1.971453257    0.302753725   -4.004613501
26   H   -4.059244716   -0.654179170   -4.698351932
27   H   -5.026609519    0.653406023   -3.986434226
28   H   -5.348141651   -1.219638417   -2.519441492
29   H   -3.638863162   -1.703843316   -2.581251943
30   H   -4.944782803    0.624310149   -1.020868754
31   H   -4.025383818   -0.775947053   -0.396211928
---
H3C9N3, RHF, CHARGE=0, MULT=1
HF=121.8
1    C   -0.000007382    0.000035894   -0.000017033
2    C    1.415253628   -0.000037434    0.000012926
3    C    2.132345705    1.220105523   -0.000008289
4    C    1.424990973    2.445870291   -0.000110721
5    C    0.009785736    2.456582962   -0.000171964
6    C   -0.698175201    1.231073687   -0.000126185
7    C   -0.718253596   -1.232862816    0.000031716
8    N   -1.303341751   -2.237157308    0.000045809
9    C    3.559195558    1.214542748    0.000070125
10   N    4.721485158    1.210084985    0.000127626
11   C   -0.698749203    3.695002095   -0.000336992
12   N   -1.275900516    4.703879049   -0.000423761
13   H    1.957617845   -0.947937796    0.000029291
14   H    1.974782054    3.389555606   -0.000152402
15   H   -1.790311347    1.235444250   -0.000188384
---
H7C11N3, RHF, CHARGE=0, MULT=1
HF=129.2
1    C    0.000062035   -0.000005843    0.000033544
2    C    1.472435078    0.000172648   -0.000162314
3    N    2.633198479   -0.012006687   -0.000999612
4    C   -0.587693945    1.411677976    0.410048999
5    C   -0.465448173   -0.419733088   -1.332248895
```

```
 6     N    -0.828937841   -0.762080778   -2.380273712
 7     C    -0.454399247   -1.006503335    0.978681046
 8     N    -0.811689734   -1.798493676    1.748466929
 9     C    -0.232607842    2.574530219   -0.488771957
10     C    -1.068007937    2.940461497   -1.571939956
11     C    -0.756758534    4.042142783   -2.387351735
12     C     0.393629944    4.808224294   -2.133226831
13     C     1.225376113    4.469392651   -1.052340172
14     C     0.912747929    3.368076844   -0.236886041
15     H    -1.695790222    1.311011096    0.468676098
16     H    -0.250571615    1.627902962    1.449773987
17     H    -1.975592596    2.374022893   -1.788567216
18     H    -1.413246921    4.302062261   -3.218531112
19     H     0.637188553    5.660633092   -2.767975294
20     H     2.116310598    5.062348576   -0.842299911
21     H     1.571414724    3.140413284    0.603318349
---
H35C16N1, RHF, CHARGE=0, MULT=1
HF=-76.5
 1     C    -0.000016122    0.000065500   -0.000012782
 2     C     1.531199125    0.000035964    0.000040156
 3     C     2.174648399    1.400373945   -0.000010986
 4     C     3.716894872    1.384544173   -0.000528783
 5     C     4.358869628    2.787176300   -0.001728124
 6     C     5.901462978    2.770708335   -0.002565009
 7     C     6.542584495    4.174066284   -0.006714945
 8     C     8.089730187    4.141860140   -0.001819547
 9     N     8.643396216    5.501959363   -0.089351539
10     C    10.101691950    5.605007645   -0.253415032
11     C    10.526922273    7.071555031   -0.505135020
12     C    12.038191421    7.233705620   -0.770238097
13     C    12.477742081    8.691776194   -1.015993614
14     C    13.988121014    8.854859721   -1.283768461
15     C    14.428083580   10.312534311   -1.529589382
16     C    15.936658379   10.477612480   -1.797766206
17     C    16.388459237   11.919804180   -2.043961847
18     H    -0.413009919    0.506258527    0.895994847
19     H    -0.412991686    0.505716040   -0.896317131
20     H    -0.382455379   -1.041627676    0.000326809
21     H     1.882693809   -0.570212984    0.890963258
22     H     1.882794288   -0.570427577   -0.890635041
23     H     1.815708071    1.963923484    0.891773184
24     H     1.815087454    1.964254897   -0.891365090
25     H     4.075150190    0.820884196    0.891368155
26     H     4.074511491    0.819602040   -0.891824613
27     H     4.001608518    3.351789300    0.889917812
28     H     4.000578223    3.350946210   -0.893401832
29     H     6.259484092    2.208023033    0.890236360
30     H     6.258517050    2.204053972   -0.893072440
31     H     6.180697565    4.739456226    0.882974112
32     H     6.189242870    4.734833951   -0.901792179
33     H     8.449490408    3.599595356    0.911993119
34     H     8.451852720    3.551833263   -0.879364973
35     H     8.362626422    6.028223084    0.725551154
36     H    10.654064393    5.202597212    0.636253981
37     H    10.408034589    4.974750589   -1.124166445
38     H    10.245256088    7.695552799    0.374334069
39     H     9.958148044    7.475277229   -1.373463966
40     H    12.607162482    6.823404757    0.095652356
```

```
41   H       12.322147018     6.614942831    -1.652358882
42   H       12.195735119     9.310549690    -0.133306940
43   H       11.908125846     9.102562909    -1.880955054
44   H       14.557752774     8.443975475    -0.418729625
45   H       14.270320420     8.235878892    -2.166244057
46   H       14.147184728    10.932271862    -0.647123972
47   H       13.859071286    10.724578221    -2.394509158
48   H       16.513933945    10.070802100    -0.935348491
49   H       16.226399531     9.862833716    -2.681228030
50   H       15.889338648    12.363602782    -2.929131115
51   H       16.178435629    12.572818284    -1.172740931
52   H       17.482328170    11.951657155    -2.227781032
---
H15C18N1, RHF, CHARGE=0, MULT=1
HF=78.1
1    C       -0.000009259    -0.000006335     0.000119066
2    C        1.405028355     0.000178469     0.000168288
3    C        2.118929727     1.210758786     0.000314125
4    C        1.420608753     2.430153319     0.005858059
5    C        0.015444748     2.448552780     0.016429401
6    C       -0.709930949     1.228993554     0.010839612
7    N       -2.138896828     1.228519024     0.008625867
8    C       -2.848908475     1.974426187    -0.982221462
9    C       -3.773975014     2.982278796    -0.605153924
10   C       -2.637405146     1.708098352    -2.360152447
11   C       -3.337142179     2.442050396    -3.332963431
12   C       -4.256213454     3.435658925    -2.954340159
13   C       -4.472306452     3.699775255    -1.591152012
14   C       -2.855193729     0.555173503     1.045303963
15   C       -2.573383415     0.817554836     2.411163494
16   C       -3.278254968     0.135892539     3.417896921
17   C       -4.272633526    -0.799622546     3.085578261
18   C       -4.559801183    -1.058607764     1.734332859
19   C       -3.857242892    -0.394477707     0.714835462
20   H       -0.525537631    -0.955769029    -0.011145688
21   H        1.943697617    -0.948326123    -0.004243395
22   H        3.208945811     1.203765428    -0.004250114
23   H        1.971465821     3.371678633     0.006243828
24   H       -0.496050981     3.411828490     0.030992722
25   H       -3.957655560     3.218120755     0.443833359
26   H       -1.938323139     0.936646651    -2.685874836
27   H       -3.167183346     2.233963989    -4.390236638
28   H       -4.798701755     3.998535807    -3.714040049
29   H       -5.184170148     4.470610693    -1.292925610
30   H       -1.815030627     1.544941561     2.703463946
31   H       -3.052315339     0.339673874     4.465450635
32   H       -4.818215449    -1.322042238     3.871424713
33   H       -5.330601797    -1.784674516     1.472290182
34   H       -4.095572301    -0.626279191    -0.324069435
---
H12C18N2, RHF, CHARGE=0, MULT=1
HF=101.8
1    C        0.000026858    -0.000080260     0.000009171
2    C        1.381810609    -0.000121912    -0.000027206
3    C        2.098327369     1.252513240    -0.000359208
4    C        1.371390091     2.489278637     0.000020971
5    C       -0.065554110     2.442070950    -0.000559864
6    C       -0.730495228     1.230650310    -0.000517685
7    N        3.479354868     1.211381438    -0.000229215
```

```
 8     C     4.161598899    2.355810113    0.005384802
 9     C     3.514277915    3.643680578    0.004112059
10     C     2.135038690    3.705069476    0.001710375
11     C     5.650823376    2.226172351    0.010884484
12     C     6.383890433    1.871247945   -1.178467834
13     C     7.758650289    1.761636730   -1.115294743
14     C     8.432696354    2.003088509    0.129270192
15     C     7.624423856    2.353762390    1.261450395
16     N     6.249230540    2.460226123    1.177976513
17     C     8.248303236    2.606411706    2.537909679
18     C     9.621043633    2.508779633    2.661698159
19     C    10.432495109    2.159335242    1.535721633
20     C     9.856807806    1.912921859    0.303829342
21     H    -0.554286164   -0.939623366    0.000866478
22     H     1.936832398   -0.939252235    0.001471872
23     H    -0.628487018    3.376652094   -0.000360718
24     H    -1.820628862    1.195478888   -0.000238475
25     H     4.114986207    4.552745943    0.008502954
26     H     1.623063632    4.668195909    0.001938790
27     H     5.849759403    1.689777068   -2.110760923
28     H     8.334515351    1.492749354   -2.001856565
29     H     7.630815557    2.871563747    3.397373297
30     H    10.106242529    2.696178558    3.620615854
31     H    11.513019433    2.090197966    1.667256900
32     H    10.481533814    1.645973042   -0.549849032
---
H11C15N3, RHF, CHARGE=0, MULT=1
HF=94.3
 1     N    -0.000126227   -0.000380315   -0.000372197
 2     C     1.350895016    0.000147091   -0.000025291
 3     C     2.118481330    1.184996174   -0.000126104
 4     C     1.448759865    2.417797268   -0.001203658
 5     C     0.043106956    2.426844266    0.000529600
 6     C    -0.651648683    1.192953433    0.003736946
 7     C    -2.140409268    1.102896468    0.012925987
 8     N    -2.727969090    1.427367149    1.192470300
 9     C    -4.080046593    1.368691142    1.293523753
10     C    -4.899849512    0.976003389    0.207279772
11     C    -4.291583556    0.643599419   -1.013432794
12     C    -2.893324975    0.707208276   -1.119664855
13     C    -4.630155223    1.748588241    2.626875051
14     N    -5.262827806    2.951301925    2.670492205
15     C    -5.780302722    3.367686249    3.846994545
16     C    -5.698243558    2.608187950    5.034234370
17     C    -5.050715246    1.364410317    4.991055497
18     C    -4.507231744    0.924197526    3.771674441
19     H     1.826955789   -0.986191625    0.000200534
20     H     3.206302683    1.136701083    0.001436705
21     H     2.006683287    3.354294244   -0.001836576
22     H    -0.498276111    3.372439938    0.003058055
23     H    -5.983453375    0.933370486    0.311100425
24     H    -4.897204404    0.340064260   -1.867709623
25     H    -2.401131405    0.453280957   -2.057738731
26     H    -6.276092158    4.344007345    3.829589201
27     H    -6.129601251    2.985183090    5.960217998
28     H    -4.967755238    0.748505134    5.886614663
29     H    -3.997025380   -0.036859874    3.715434100
---
O1, UHF, CHARGE=0, MULT=3
```

```
HF=59.6
1    O    0.0000      0.0000      0.0000
---
H1O1, UHF, CHARGE=0, MULT=2
HF=9.5
1    H    0.000000000    0.000000000    0.000000000
2    O    0.936824079    0.000000000    0.000000000
EXPGEOM
1    O    0.00000    0.00000    0.10780
2    H    0.00000    0.00000   -0.86270
---
H1O1, UHF, CHARGE=-1, MULT=1
HF=-33.2
1    H    0.000000000    0.000000000    0.000000000
2    O    0.938524018    0.000000000    0.000000000
---
H2O1, RHF, CHARGE=0, MULT=1
HF=-57.8, DIP=1.85, IE=12.62
1    H    0.000000000    0.000000000    0.000000000
2    O    0.943000595    0.000000000    0.000000000
3    H    1.215558630    0.917366403    0.000000000
EXPGEOM
1    O    0.00000    0.00000    0.11770
2    H    0.00000    0.75750   -0.47070
3    H    0.00000   -0.75750   -0.47070
---
H3O1, UHF, CHARGE=1, MULT=1
HF=138.9
1    O    0.097678462    0.017621931   -0.001092390
2    H   -0.046655332   -0.935452493    0.016244250
3    H   -0.240343741    0.471494038   -0.781579012
4    H   -0.049484813    0.472611863    0.836016347
EXPGEOM
1    O    0.00000    0.00000    0.07770
2    H    0.00000    0.93260   -0.20710
3    H    0.80770   -0.46630   -0.20710
4    H   -0.80770   -0.46630   -0.20710
---
C1O1, RHF, CHARGE=0, MULT=1
HF=-26.4, DIP=0.11, IE=14.01
1    C    1.000000000    0.000000000    0.000000000
2    O    1.000000000    1.163228122    0.000000000
---
H1C1O1, UHF, CHARGE=1, MULT=1
HF=199
1    O    0.000000000    0.000000000    0.000000000
2    C    1.142434013    0.000000000    0.000000000
3    H    2.223033907   -0.000022381    0.000000000
---
H1C1O1, UHF, CHARGE=0, MULT=2
HF=10.4
1    O    0.000000000    0.000000000    0.000000000
2    C    1.185192635    0.000000000    0.000000000
3    H    2.044375390    0.645265575    0.000000000
EXPGEOM
1    C    0.06160    0.58870    0.00000
2    H   -0.86180    1.22240    0.00000
3    O    0.06160   -0.59430    0.00000
---
```

```
H2C1O1, RHF, CHARGE=0, MULT=1
HF=-26, DIP=2.33, IE=10.1
1    O     0.000000000     0.000000000     0.000000000
2    C     1.216508825     0.000000000     0.000000000
3    H     1.827249015     0.922233661     0.000000000
4    H     1.827246061    -0.922236511     0.000172606
EXPGEOM
1    O     0.00000     0.00000      0.67640
2    C     0.00000     0.00000     -0.53070
3    H     0.00000     0.93570     -1.11330
4    H     0.00000    -0.93570     -1.11330
---
H3C1O1, UHF, CHARGE=1, MULT=1
HF=168
1    O     0.000000000     0.000000000     0.000000000
2    C     1.275188127     0.000000000     0.000000000
3    H     1.884086962     0.920319130     0.000000000
4    H     1.774244226    -0.987530561    -0.000208789
5    H    -0.499708201     0.823225601     0.000082137
---
H3C1O1, UHF, CHARGE=-1, MULT=1
HF=-36
1    O    -0.003071173     0.004801218     0.002099372
2    C     1.284617682     0.008275470     0.012548238
3    H     1.791755317     1.045163595     0.001363808
4    H     1.788414175    -0.494947775     0.921186736
5    H     1.803374217    -0.521293441    -0.872405106
---
H4C1O1, RHF, CHARGE=0, MULT=1
HF=-48.1, DIP=1.7, IE=10.96
1    O     0.000000000     0.000000000     0.000000000
2    C     1.390675374     0.000000000     0.000000000
3    H     1.815342375     1.035436610     0.000000000
4    H    -0.348363299     0.428492263    -0.768781211
5    H     1.736847933    -0.520596945     0.923177155
6    H     1.815779413    -0.535588152    -0.885885387
EXPGEOM
1    C    -0.04670     0.66420      0.00000
2    O    -0.04670    -0.75630      0.00000
3    H    -1.08760     0.97730      0.00000
4    H     0.43810     1.07100      0.89020
5    H     0.43810     1.07100     -0.89020
6    H     0.86450    -1.05410      0.00000
---
H2C2O1, RHF, CHARGE=0, MULT=1
HF=-11.4, DIP=1.42, IE=9.64
1    O     0.000000000     0.000000000     0.000000000
2    C     1.184403668     0.000000000     0.000000000
3    C     2.503468481     0.000000000     0.000000000
4    H     3.074528386    -0.002835689    -0.922676102
5    H     3.074505228     0.002948061     0.922693094
EXPGEOM
1    C     0.00000     0.00000     -1.21130
2    C     0.00000     0.00000      0.10260
3    O     0.00000     0.00000      1.26490
4    H     0.00000     0.94000     -1.73370
5    H     0.00000    -0.94000     -1.73370
---
H4C2O1, RHF, CHARGE=0, MULT=1
```

```
HF=-39.7, DIP=2.69, IE=10.21
1    C    0.000000000     0.000000000     0.000000000
2    C    1.516805923     0.000000000     0.000000000
3    O    2.216673392     1.000972896     0.000000000
4    H    1.969235000    -1.015383456    -0.000404005
5    H   -0.435179218     1.018215964     0.011819021
6    H   -0.371315388    -0.523341608    -0.904576730
7    H   -0.371758994    -0.544211425     0.891932774
EXPGEOM
1    C    0.22970     0.40020     0.00000
2    C   -1.16480    -0.14910     0.00000
3    O    1.23100    -0.27660    -0.00000
4    H    0.30490     1.50260     0.00000
5    H   -1.14670    -1.23570     0.00040
6    H   -1.69790     0.21980     0.87830
7    H   -1.69760     0.21920    -0.87860
---
H4C2O1, RHF, CHARGE=0, MULT=1
HF=-12.6, DIP=1.89, IE=10.57
1    C    0.000000000     0.000000000     0.000000000
2    O    1.417371421     0.000000000     0.000000000
3    C    0.806984629     1.279320915     0.000000000
4    H   -0.494853746    -0.372316685     0.911921237
5    H   -0.494856912    -0.372265547    -0.911942483
6    H    0.929962314     1.886241360    -0.911918972
7    H    0.929997187     1.886259567     0.911900207
---
H5C2O1, UHF, CHARGE=-1, MULT=1
HF=-47.5
1    C    0.000000000     0.000000000     0.000000000
2    C    1.571796661     0.000000000     0.000000000
3    H    1.988084971     1.028565022     0.000000000
4    H    1.985481927    -0.527352377     0.884640357
5    H    1.974550696    -0.513991093    -0.900713746
6    H   -0.283774702     0.390404988    -1.052582740
7    H   -0.284076114    -1.117708013    -0.104131548
8    O   -0.573922300     0.617718162     0.981754489
---
H6C2O1, RHF, CHARGE=0, MULT=1
HF=-44, DIP=1.3, IE=10.04
1    O    0.001569385    -0.000856601    -0.015700203
2    C    1.397364450    -0.001143009    -0.005328138
3    C   -0.695970506     1.208298621    -0.005627851
4    H    1.728661109    -1.066447066    -0.004213248
5    H   -1.784127582     0.962725038    -0.008807962
6    H    1.833802189     0.499881533    -0.904836701
7    H    1.820002575     0.498193276     0.901537483
8    H   -0.475618950     1.838661250    -0.902759371
9    H   -0.478423140     1.822132472     0.903421752
EXPGEOM
1    O    0.00000     0.00000     0.60170
2    C    0.00000     1.16820    -0.20030
3    C    0.00000    -1.16820    -0.20030
4    H    0.00000     2.02090     0.47700
5    H    0.00000    -2.02090     0.47700
6    H    0.89020     1.21650    -0.84090
7    H   -0.89020     1.21650    -0.84090
8    H   -0.89020    -1.21650    -0.84090
9    H    0.89020    -1.21650    -0.84090
```

```
---
H6C2O1, RHF, CHARGE=0, MULT=1
HF=-56.2, DIP=1.69, IE=10.6
1    C     0.000000000     0.000000000     0.000000000
2    C     1.539191926     0.000000000     0.000000000
3    H     1.953056883     1.027945532     0.000000000
4    H     1.950938665    -0.536324560     0.878068128
5    H     1.904409100    -0.513383850    -0.913088595
6    H    -0.373449627     0.479922360    -0.945646843
7    H    -0.372858809    -1.060234984    -0.029847692
8    O    -0.486418363     0.669063556     1.124626069
9    H    -1.432995069     0.668925616     1.136231901
EXPGEOM
1    C     1.17150    -0.40490     0.00000
2    C     0.00000     0.56030     0.00000
3    O    -1.19450    -0.22360     0.00000
4    H    -1.94290     0.38350     0.00000
5    H     2.11800     0.13940     0.00000
6    H     1.13120    -1.04140     0.88470
7    H     1.13120    -1.04140    -0.88470
8    H     0.04490     1.20840     0.88530
9    H     0.04490     1.20840    -0.88530
---
H6C3O1, RHF, CHARGE=0, MULT=1
HF=-52, DIP=2.88, IE=9.72
1    O     0.000000000     0.000000000     0.000000000
2    C     1.227403441     0.000000000     0.000000000
3    C     2.017072419     1.307257610     0.000000000
4    C     2.019566296    -1.305292618     0.002397158
5    H     1.433726070     2.145543196    -0.429906225
6    H     2.292025454     1.576741107     1.040375843
7    H     2.950002801     1.211220072    -0.590212369
8    H     1.410375429    -2.164116902     0.347338189
9    H     2.376594650    -1.527991478    -1.023928347
10    H     2.902893114    -1.233440361     0.667908742
EXPGEOM
1    C     0.00000     0.00000     0.18720
2    O     0.00000     0.00000     1.40220
3    C     0.00000     1.28920    -0.61720
4    C     0.00000    -1.28920    -0.61720
5    H     0.00000     2.14250     0.06570
6    H     0.00000    -2.14250     0.06570
7    H     0.88390     1.33520    -1.26640
8    H    -0.88390     1.33520    -1.26640
9    H    -0.88390    -1.33520    -1.26640
10    H     0.88390    -1.33520    -1.26640
---
H6C3O1, RHF, CHARGE=0, MULT=1
HF=-45.5, IE=10
1    C     0.000000000     0.000000000     0.000000000
2    C     1.528975015     0.000000000     0.000000000
3    C    -0.643825585     1.384671414     0.000000000
4    O    -1.851769633     1.566368757     0.008609280
5    H    -0.362857794    -0.561398772     0.892181515
6    H    -0.363020018    -0.560431989    -0.892756862
7    H     1.910653411    -1.041466186     0.001375643
8    H     1.939831635     0.505447802    -0.897488658
9    H     1.939764186     0.507698046     0.896249928
10    H     0.058241314     2.245176738    -0.008857119
```

```
EXPGEOM
1    C      1.44500      0.44970      0.00000
2    C      0.00000      0.91420      0.00000
3    C     -0.99500     -0.21130      0.00000
4    O     -0.70280     -1.38450      0.00000
5    H      2.12130      1.30350      0.00000
6    H      1.65290     -0.15940      0.87830
7    H      1.65290     -0.15940     -0.87830
8    H     -0.22440      1.53830      0.87140
9    H     -0.22440      1.53830     -0.87140
10   H     -2.05620      0.09870      0.00000
---
H6C3O1, RHF, CHARGE=0, MULT=1
HF=-19.3
1    C      0.000000000     0.000000000     0.000000000
2    C      2.081751342     0.000000000     0.000000000
3    C      1.040869616     1.158715181     0.000000000
4    O      1.040895494    -0.971027022     0.008371186
5    H     -0.654549797    -0.067729704     0.896235813
6    H     -0.644905928    -0.075034197    -0.902836222
7    H      2.735688518    -0.067741040     0.896709154
8    H      2.726806102    -0.074586647    -0.902780781
9    H      1.040861373     1.805832661    -0.892834806
10   H      1.040728602     1.805136372     0.893355535
---
H8C3O1, RHF, CHARGE=0, MULT=1
HF=-65.1
1    C      0.000000000     0.000000000     0.000000000
2    C      1.552040577     0.000000000     0.000000000
3    C     -0.604090509     1.430001588     0.000000000
4    O     -0.533349220    -0.804207237    -1.018454096
5    H     -0.327770605    -0.493375933     0.959220912
6    H      1.950884781    -1.033887313     0.032923534
7    H      1.975294680     0.496213878    -0.896685174
8    H      1.936061092     0.532275750     0.893245050
9    H     -1.711552595     1.394548108     0.036567905
10   H     -0.313003333     2.012325862    -0.897474673
11   H     -0.260154948     1.992187536     0.891141751
12   H     -0.319932268    -0.480283091    -1.882265170
EXPGEOM
1    C      0.00050      0.04480      0.36410
2    C     -1.17490     -0.78950     -0.10270
3    C      1.31820     -0.53690     -0.08800
4    O     -0.06270      1.35750     -0.16820
5    H     -0.00670      0.08890      1.45750
6    H     -2.11630     -0.34400      0.22100
7    H     -1.17740     -0.84880     -1.19020
8    H     -1.11970     -1.79790      0.30550
9    H      2.13990      0.08920      0.25260
10   H      1.44850     -1.54040      0.31290
11   H      1.34650     -0.58580     -1.17560
12   H     -0.87580      1.76870      0.12170
---
H8C3O1, RHF, CHARGE=0, MULT=1
HF=-51.7
1    C      0.000000000     0.000000000     0.000000000
2    O      1.395782704     0.000000000     0.000000000
3    C      2.088054544     1.218610489     0.000000000
4    C      3.605085325     0.954022122     0.001536676
```

```
 5      H     -0.430774926     0.500027958      0.902809735
 6      H     -0.430634645     0.498880275     -0.903522765
 7      H     -0.330294186    -1.065670824      0.000731435
 8      H      1.820632438     1.838105088     -0.898257234
 9      H      1.818834485     1.839110472      0.896995343
10      H      3.925976810     0.389392010      0.899445778
11      H      3.927737702     0.387882055     -0.894762697
12      H      4.142650113     1.924663605      0.001140763
---
H8C3O1, RHF, CHARGE=0, MULT=1
HF=-61.2
 1      C      0.000000000     0.000000000      0.000000000
 2      C      1.529203144     0.000000000      0.000000000
 3      C     -0.691125583     1.390016641      0.000000000
 4      O     -0.479616235     2.151633180      1.150106146
 5      H     -0.362251418    -0.599011732      0.867255448
 6      H     -0.355135572    -0.538237665     -0.909704467
 7      H      1.908910189    -1.039854879     -0.081654584
 8      H      1.940645975     0.573619562     -0.855094583
 9      H      1.945795924     0.431906326      0.931859154
10      H     -1.789572667     1.228782078     -0.177513226
11      H     -0.319871813     2.001222333     -0.862317438
12      H     -0.938159282     1.795973469      1.898130062
EXPGEOM
 1      C     -1.45630    1.22570     0.00000
 2      C      0.00000    0.74340     0.00000
 3      C      0.10090   -0.77720     0.00000
 4      O      1.48270   -1.12570     0.00000
 5      H     -1.50820    2.32130     0.00000
 6      H     -1.99090    0.86250     0.88830
 7      H     -1.99090    0.86250    -0.88830
 8      H      0.52890    1.12370     0.88450
 9      H      0.52890    1.12370    -0.88450
10      H     -0.41090   -1.17780     0.89240
11      H     -0.41090   -1.17780    -0.89240
12      H      1.52500   -2.08420     0.00000
---
H4C4O1, RHF, CHARGE=0, MULT=1
HF=15.6
 1      C      0.000000000     0.000000000      0.000000000
 2      C      1.526560754     0.000000000      0.000000000
 3      C      2.166164847     1.295725588      0.000000000
 4      C      2.698220910     2.368274653     -0.000052444
 5      O      2.186383482    -1.035191523      0.000046127
 6      H     -0.380506231     0.525170805      0.899355730
 7      H     -0.380629095     0.526428951     -0.898559199
 8      H     -0.420509616    -1.024844238     -0.000648048
 9      H      3.164362053     3.311227210      0.000122051
EXPGEOM
 1      C      1.50160    0.73390     0.00000
 2      C      0.00000    0.50920     0.00000
 3      O     -0.82170    1.42090     0.00000
 4      C     -0.43350   -0.90380     0.00000
 5      C     -0.73250   -2.09900     0.00000
 6      H      1.71070    1.81390     0.00000
 7      H      1.94490    0.25410     0.89070
 8      H      1.94490    0.25410    -0.89070
 9      H     -1.04050   -3.13130     0.00000
---
```

```
H4C4O1, RHF, CHARGE=0, MULT=1
HF=-8.3, DIP=0.66, IE=8.88
1    O     0.000000000    0.000000000    0.000000000
2    C     1.366848838    0.000000000    0.000000000
3    C    -0.414425799    1.302663155    0.000000000
4    C     1.848169433    1.304035439    0.000000000
5    C     0.682783244    2.156105249    0.000000000
6    H     1.861007325   -0.964352699    0.000000000
7    H    -1.483036376    1.481770128   -0.000000000
8    H     2.878367762    1.623195909   -0.000000000
9    H     0.675053107    3.234726064    0.000000000
EXPGEOM
1    O     0.00000     0.00000     1.16370
2    C     0.00000     1.09060     0.34790
3    C     0.00000    -1.09060     0.34790
4    C     0.00000     0.72120    -0.96110
5    C     0.00000    -0.72120    -0.96110
6    H     0.00000     2.05130     0.84590
7    H     0.00000    -2.05130     0.84590
8    H     0.00000     1.37860    -1.82190
9    H     0.00000    -1.37860    -1.82190
---
H6C4O1, RHF, CHARGE=0, MULT=1
HF=-17.3
1    O     0.000000000    0.000000000    0.000000000
2    C     1.422712846    0.000000000    0.000000000
3    C     1.870302164    1.497453597    0.000000000
4    C     0.542534109    2.213726169    0.000194984
5    C    -0.457937551    1.292262251   -0.000258743
6    H     1.789771044   -0.551443569   -0.898680016
7    H     1.789099308   -0.550467706    0.899562042
8    H     2.483671884    1.744599826   -0.890770228
9    H     2.484006954    1.744415917    0.890604799
10   H     0.450271551    3.289351346    0.000785104
11   H    -1.539091391    1.409455454   -0.001386679
EXPGEOM
1    H     1.11930     2.02010     0.10660
2    C     0.61210     1.07260     0.05320
3    H     2.23370    -0.38440     0.03070
4    C     1.19030    -0.11710     0.01100
5    O     0.33550    -1.17460    -0.08750
6    H    -1.24150     1.21790    -1.05080
7    H    -1.45320     1.40950     0.68300
8    C    -0.87820     0.88200    -0.07730
9    H    -1.63930    -1.12400    -0.63310
10   H    -1.31860    -0.90830     1.09530
11   C    -0.98830    -0.64330     0.09110
---
H6C4O1, RHF, CHARGE=0, MULT=1
HF=-24
1    O     0.000000000    0.000000000    0.000000000
2    C     1.223284270    0.000000000    0.000000000
3    C     2.050183246    1.237876517    0.000000000
4    C     3.399107036    1.250770808   -0.006811298
5    C     4.256170971    2.479504592   -0.008575844
6    H     1.788825002   -0.956846597    0.000180720
7    H     1.486350098    2.176521043    0.006173480
8    H     3.963112557    0.311185702   -0.013459158
9    H     4.091216471    3.083314109   -0.925574458
```

```
10      H      4.040288362     3.125471523    0.868069345
11      H      5.332320764     2.213792798    0.028766278
EXPGEOM
1       C      2.38730        -0.48370       0.00000
2       C      0.89870        -0.66060       0.00000
3       C      0.00000         0.33380       0.00000
4       C     -1.44650         0.04800       0.00000
5       O     -2.30960         0.90660       0.00000
6       H     -1.72310        -1.02760       0.00000
7       H      0.29580         1.38240       0.00000
8       H      0.53300        -1.69120       0.00000
9       H      2.66340         0.57600       0.00000
10      H      2.83550        -0.95890       0.88240
11      H      2.83550        -0.95890      -0.88240
---
H6C4O1, RHF, CHARGE=0, MULT=1
HF=-3.3
1       C      0.000000000     0.000000000    0.000000000
2       C      1.349478582     0.000000000    0.000000000
3       O      2.021915331     1.190823292    0.000000000
4       C      3.377817255     1.312480509    0.007093471
5       C      4.331179416     0.360155091    0.040723759
6       H      5.383104726     0.639650494    0.043378708
7       H      4.170728271    -0.713358680    0.067599549
8       H      3.639325551     2.382387630   -0.016407843
9       H      1.927281966    -0.931093498   -0.001884141
10      H     -0.555139043    -0.935905670   -0.001404584
11      H     -0.622827790     0.890899330    0.001296221
EXPGEOM
1       O     -0.03690        -0.84650       0.05420
2       C      1.29720        -0.49040      -0.01840
3       C     -0.92380         0.17920       0.35500
4       C      1.79390         0.76180      -0.14850
5       C     -2.13660         0.24770      -0.23020
6       H      1.93050        -1.38100       0.02680
7       H     -0.57860         0.87100       1.13420
8       H      2.87820         0.89090      -0.18830
9       H      1.15900         1.64750      -0.23610
10      H     -2.84530         1.01500       0.09160
11      H     -2.43310        -0.46090      -1.00890
---
H8C4O1, RHF, CHARGE=0, MULT=1
HF=-48.9, IE=9.83
1       C     -0.043944300    -0.016142160   -0.005806364
2       C      1.483997379     0.008698217   -0.002692220
3       C      2.095064003     1.415047327    0.120047440
4       C      3.625892893     1.448275511    0.117009151
5       H      3.984541642     2.493942826    0.211261869
6       H      4.044668992     1.035215398   -0.822894231
7       H      4.052625692     0.871565874    0.962302861
8       H      1.727020822     2.052234859   -0.717385683
9       H      1.732710729     1.895053841    1.058680503
10      H      1.843749533    -0.474334708   -0.940436581
11      H      1.841296750    -0.628742258    0.838867722
12      H     -0.547744171     0.968557023    0.093877855
13      O     -0.698881162    -1.042333327   -0.109142035
EXPGEOM
1       C      2.27880        -0.36110       0.00000
2       H      2.80470        -1.32290       0.00000
```

```
3      H     2.60450    0.19840    0.88620
4      H     2.60450    0.19840   -0.88620
5      C     0.76240   -0.56160    0.00000
6      H     0.45830   -1.14910   -0.87520
7      H     0.45830   -1.14910    0.87520
8      C     0.00000    0.76110    0.00000
9      H     0.26900    1.37500    0.87500
10     H     0.26900    1.37500   -0.87500
11     C    -1.50130    0.60360    0.00000
12     H    -2.08600    1.54790    0.00000
13     O    -2.07770   -0.46570    0.00000
---
H8C4O1, RHF, CHARGE=0, MULT=1
HF=-51.6
1      O     0.012577378   -0.012333287   -0.025437691
2      C    -0.603211005   -0.721237222    0.754007954
3      C    -0.628147370   -2.256027809    0.687225285
4      C     0.016909261   -2.865116274    1.946847622
5      C    -2.057872573   -2.768387927    0.428165872
6      H    -1.196965082   -0.279585190    1.582583742
7      H    -0.006471792   -2.586705244   -0.183643216
8      H     1.054711862   -2.495916191    2.077803985
9      H    -0.548415209   -2.620220607    2.868624938
10     H     0.067510409   -3.969994688    1.867202983
11     H    -2.749606627   -2.521626049    1.258743153
12     H    -2.473615905   -2.328032619   -0.501021389
13     H    -2.063351883   -3.869945156    0.301587592
EXPGEOM
1      H    -0.21970    0.59170    2.17190
2      H    -0.21970    0.59170   -2.17190
3      H    -1.70520    0.12630   -1.31380
4      H    -1.70520    0.12630    1.31380
5      H    -1.15180    1.81010    1.27760
6      H    -1.15180    1.81010   -1.27760
7      C    -0.81480    0.76710   -1.26810
8      C    -0.81480    0.76710    1.26810
9      H     0.91490    1.08630    0.00000
10     C     0.00000    0.48070    0.00000
11     C     0.44050   -0.96640    0.00000
12     H    -0.39050   -1.70820    0.00000
13     O     1.59540   -1.34060    0.00000
---
H8C4O1, RHF, CHARGE=0, MULT=1
HF=-57.1
1      C     0.000000000    0.000000000    0.000000000
2      C     1.528964732    0.000000000    0.000000000
3      C     2.209846838    1.376923895    0.000000000
4      C     3.739356729    1.410316136   -0.018333807
5      O     2.159471201   -1.051612081    0.000929690
6      H    -0.385620886    0.527072379    0.895944945
7      H    -0.385454235    0.517082161   -0.901824865
8      H    -0.419869493   -1.025444954    0.005828358
9      H     1.833328374    1.940362137   -0.885397878
10     H     1.854335751    1.928657693    0.901385224
11     H     4.175299846    0.915536150    0.872425206
12     H     4.153262217    0.921954504   -0.923086723
13     H     4.089213182    2.463945945   -0.018857497
EXPGEOM
1      C     1.89650   -0.49620    0.00010
```

```
 2    C     0.52690    0.17080   -0.00020
 3    C    -0.68090   -0.76500   -0.00030
 4    C    -2.01740   -0.02100    0.00020
 5    O     0.40110    1.37780   -0.00010
 6    H     2.68480    0.27180   -0.00240
 7    H     2.00620   -1.13980    0.89180
 8    H     2.00500   -1.14490   -0.88790
 9    H    -2.86100   -0.73260   -0.00200
10    H    -2.10440    0.62490    0.89020
11    H    -2.10290    0.62840   -0.88730
12    H    -0.59360   -1.43110   -0.88160
13    H    -0.59310   -1.43110    0.88100
---
H8C4O1, RHF, CHARGE=0, MULT=1
HF=-44
 1    C     0.000000000    0.000000000    0.000000000
 2    C     1.539723207    0.000000000    0.000000000
 3    C     1.943557942    1.498968635    0.000000000
 4    C    -0.404675937    1.498400866   -0.022201356
 5    O     0.769326794    2.273921602   -0.015530318
 6    H    -0.408903616   -0.537441661   -0.880212043
 7    H    -0.408806471   -0.510905030    0.895971134
 8    H     1.948667421   -0.523874529   -0.888354914
 9    H     1.948810898   -0.524172232    0.888057530
10    H     2.568946256    1.756619516   -0.891190225
11    H     2.548569492    1.761950234    0.903697354
12    H    -1.007851790    1.747763690   -0.931070444
13    H    -1.032625456    1.769434873    0.863136361
---
H10C4O1, RHF, CHARGE=0, MULT=1
HF=-60.3, DIP=1.15, IE=9.6
 1    C     0.000000000    0.000000000    0.000000000
 2    C     1.540103775    0.000000000    0.000000000
 3    O     2.011123005    1.319870327    0.000000000
 4    C     3.392817434    1.554168330    0.000119629
 5    C     3.659865062    3.071123701    0.002841869
 6    H     4.755461718    3.246517926    0.001922512
 7    H     3.238867451    3.564278691    0.901469397
 8    H     3.236815964    3.567734006   -0.892934799
 9    H     3.884230835    1.094209455   -0.899386036
10    H     3.885836974    1.089859259    0.896316729
11    H     1.910486784   -0.565277876    0.897439207
12    H     1.910319491   -0.565137097   -0.897641918
13    H    -0.362740607   -1.048583243   -0.000079307
14    H    -0.414271575    0.501287367   -0.897258980
15    H    -0.414179933    0.501787382    0.897042013
EXPGEOM
 1    O     0.00000    0.00000    0.32890
 2    C     0.00000    1.19480   -0.52530
 3    C     0.00000   -1.19480   -0.52530
 4    C     0.00000    2.41410    0.38460
 5    C     0.00000   -2.41410    0.38460
 6    H     0.86900    1.18300   -1.13650
 7    H    -0.86900    1.18300   -1.13650
 8    H     0.86900   -1.18300   -1.13650
 9    H    -0.86900   -1.18300   -1.13650
10    H     0.00000    3.29830   -0.21000
11    H     0.00000   -3.29830   -0.21000
12    H    -0.86980    2.39870    1.00580
```

```
13    H      0.86980     2.39870     1.00580
14    H      0.86980    -2.39870     1.00580
15    H     -0.86980    -2.39870     1.00580
---
H10C4O1, RHF, CHARGE=0, MULT=1
HF=-74.7
1     O      0.000000000    0.000000000    0.000000000
2     C      1.409559764    0.000000000    0.000000000
3     C      1.833088307    1.505488088    0.000000000
4     C      1.957599246   -0.710044240    1.281814811
5     C      1.957694302   -0.710080239   -1.281762196
6     H      2.937970106    1.599809681    0.004273134
7     H      1.454919903    2.037337597   -0.895935338
8     H      1.448038004    2.039102097    0.891949480
9     H      3.062996542   -0.646314897    1.328071669
10    H      1.556649567   -0.250558199    2.207492436
11    H      1.687943870   -1.785840763    1.303979751
12    H      3.063057697   -0.645859625   -1.328153848
13    H      1.688454856   -1.785982581   -1.303707529
14    H      1.556443078   -0.250935087   -2.207476502
15    H     -0.357598724   -0.876766714    0.000099723
EXPGEOM
1     C     -0.00160     0.01810     0.00000
2     O     -0.48340     1.36470     0.00000
3     C      1.51730     0.14610     0.00000
4     C     -0.48340    -0.70670     1.25970
5     C     -0.48340    -0.70670    -1.25970
6     H     -1.45270     1.33380     0.00000
7     H      1.99090    -0.84300     0.00000
8     H      1.84510     0.69720    -0.88890
9     H      1.84510     0.69720     0.88890
10    H     -0.09080    -1.73010     1.30790
11    H     -0.09080    -1.73010    -1.30790
12    H     -0.15580    -0.15950     2.15110
13    H     -1.58080    -0.76450     1.27320
14    H     -0.15580    -0.15950    -2.15110
15    H     -1.58080    -0.76450    -1.27320
---
H8C5O1, RHF, CHARGE=0, MULT=1
HF=-31.1
1     O      0.000000000    0.000000000    0.000000000
2     C      1.377183930    0.000000000    0.000000000
3     C      2.047892442    1.341064650    0.000000000
4     C      1.891423857   -1.265122035   -0.000481497
5     C      0.765310807   -2.267249329   -0.001060050
6     C     -0.485909016   -1.335356776    0.000146274
7     H      1.768120439    1.930507653   -0.897775547
8     H      1.770386130    1.929424762    0.899168233
9     H      3.150797266    1.223671087   -0.001538181
10    H      2.931391698   -1.556045427   -0.000534857
11    H      0.789046757   -2.929635773    0.888814212
12    H      0.788451693   -2.927665790   -0.892455897
13    H     -1.129044260   -1.491507140    0.899651372
14    H     -1.130578560   -1.491064711   -0.898303159
EXPGEOM
1     C     -1.35770    -0.82360     0.05950
2     H     -1.87310    -1.37610    -0.72900
3     H     -1.69680    -1.20500     1.02830
4     O      0.05450    -1.09450    -0.05830
```

```
5    C    -1.54020     0.71340    -0.04690
6    H    -2.16480     1.10180     0.76410
7    H    -2.01790     1.00870    -0.98910
8    C    -0.10440     1.17750     0.03290
9    H     0.21240     2.21050     0.06480
10   C     2.19320     0.00540     0.00590
11   H     2.54150    -0.51130    -0.89360
12   H     2.64800     0.99630     0.04320
13   H     2.53210    -0.57520     0.86910
14   C     0.70620     0.11160     0.00010
---
H8C5O1, RHF, CHARGE=0, MULT=1
HF=-31.4
1    O    0.000000000    0.000000000    0.000000000
2    C    1.223328324    0.000000000    0.000000000
3    C    2.084923487    1.231292741    0.000000000
4    C    1.540431832    2.467310226    0.016910873
5    C    3.573784256    0.940451103   -0.018795493
6    C    4.556506007    2.114382335   -0.037798353
7    H    1.778792752   -0.962287751    0.000542003
8    H    2.116742282    3.391584607    0.018651015
9    H    0.466746160    2.656777468    0.030820326
10   H    3.814489595    0.314854533    0.875110482
11   H    3.791153062    0.306716909   -0.912911989
12   H    4.427556472    2.752201821   -0.935391040
13   H    4.459112032    2.755237131    0.861575483
14   H    5.597328776    1.728833208   -0.055107960
---
H8C5O1, RHF, CHARGE=0, MULT=1
HF=-27
1    O    0.000000000    0.000000000    0.000000000
2    C    1.360177661    0.000000000    0.000000000
3    C    2.116511909    1.123747682    0.000000000
4    C    1.522305939    2.500418799   -0.006201521
5    C    0.010021324    2.481940243    0.287055475
6    C   -0.705808371    1.201492587   -0.211816572
7    H    1.765158190   -1.018473124    0.018319674
8    H    3.203624606    1.058885523    0.001887619
9    H    1.712587532    2.981436174   -0.994249128
10   H    2.036803651    3.137676887    0.748463371
11   H   -0.475939746    3.361850442   -0.188615192
12   H   -0.157080183    2.593233125    1.382372598
13   H   -0.936359480    1.300394576   -1.306112058
14   H   -1.688237330    1.087913448    0.313555982
EXPGEOM
1    H     1.65630     1.84520    -0.18530
2    C     0.89060     1.08350    -0.07980
3    H    -0.72260     2.39900     0.09830
4    C    -0.41350     1.36000     0.06140
5    H    -2.30300     0.52530    -0.52520
6    H    -1.86600     0.18400     1.13840
7    C    -1.45680     0.27290     0.12360
8    H    -0.73900    -1.08180    -1.40940
9    H    -1.46240    -1.90770    -0.02580
10   C    -0.83590    -1.05920    -0.31920
11   H     1.03520    -2.13920    -0.01270
12   H     0.48100    -1.19940     1.39020
13   C     0.55410    -1.20830     0.29390
14   O     1.43610    -0.16480    -0.11850
```

```
---
H8C5O1, RHF, CHARGE=0, MULT=1
HF=-32.6
1    O     0.000000000      0.000000000      0.000000000
2    C     1.229451718      0.000000000      0.000000000
3    C     2.029822366      1.262569445      0.000000000
4    C     1.540646820      2.520791104     -0.029992192
5    C     2.373802340      3.767267661     -0.030814515
6    C     2.012980138     -1.311914761      0.000068405
7    H     3.113143043      1.106014528      0.027733841
8    H     0.461481615      2.706062938     -0.058841897
9    H     3.036498871      3.813132355      0.858657793
10   H     3.014153385      3.822580016     -0.936145779
11   H     1.734438234      4.673303633     -0.017913538
12   H     2.663243644     -1.371579233     -0.896234162
13   H     2.654437004     -1.376539768      0.902248572
14   H     1.349255275     -2.199113247     -0.006078162
---
H8C5O1, RHF, CHARGE=0, MULT=1
HF=-46
1    C     0.003149790     -0.073303300     -0.148502583
2    C     1.529867965     -0.011974400      0.080374627
3    C     1.995693847      1.438941012     -0.146723554
4    C    -0.485728126      1.336670575     -0.531393842
5    C     0.749275396      2.246199518     -0.530309505
6    O     0.741468661      3.438237554     -0.796389841
7    H    -0.241668760     -0.804110370     -0.948504360
8    H    -0.515879607     -0.432129092      0.765868473
9    H     2.055557913     -0.704769081     -0.611049349
10   H     1.785120469     -0.348613156      1.107712794
11   H     2.759514705      1.496788781     -0.951011387
12   H     2.462622792      1.859972905      0.769046632
13   H    -0.968496833      1.338596232     -1.531595614
14   H    -1.242358289      1.710941083      0.190673811
EXPGEOM
1    O     0.00000     0.00000     2.14520
2    C     0.00000     0.00000     0.92870
3    C     0.00000     1.24780     0.02440
4    C     0.00000    -1.24780     0.02440
5    C     0.32310     0.70380    -1.38020
6    C    -0.32310    -0.70380    -1.38020
7    H    -1.02330     1.67380     0.05490
8    H     1.02330    -1.67380     0.05490
9    H     0.69190     2.01530     0.41020
10   H    -0.69190    -2.01530     0.41020
11   H    -0.04670     1.34810    -2.19630
12   H     0.04670    -1.34810    -2.19630
13   H     1.42020     0.60500    -1.50100
14   H    -1.42020    -0.60500    -1.50100
---
H10C5O1, RHF, CHARGE=0, MULT=1
HF=-61.6
1    C     0.000159278      0.000142171     -0.000116502
2    C     1.531225540     -0.000127657     -0.000072441
3    C     2.185630438      1.390285266      0.000041689
4    C     3.162546032      1.740066650      1.133546102
5    C     2.506856730      2.419941525      2.338047093
6    O     1.966147052      2.191783154     -0.902411270
7    H    -0.410018154      0.565700274      0.860899895
```

```
 8     H     -0.415620967     0.439804624    -0.928298953
 9     H     -0.376314342    -1.041233592     0.074159997
10     H      1.882820604    -0.584550242     0.879684406
11     H      1.899733630    -0.550678206    -0.897093321
12     H      3.683484054     0.815672157     1.471115625
13     H      3.964338310     2.404716614     0.736053625
14     H      2.015971925     3.372916001     2.056166399
15     H      1.744786739     1.770750277     2.814659155
16     H      3.273939027     2.651579191     3.105388674
EXPGEOM
1     O    0.00000     0.00000     1.29480
2     C    0.00000     0.00000     0.06600
3     C    0.00000     1.29940    -0.74040
4     C    0.00000    -1.29940    -0.74040
5     C    0.00000     2.55910     0.13460
6     C    0.00000    -2.55910     0.13460
7     H    0.88270     1.27650    -1.40970
8     H   -0.88270     1.27650    -1.40970
9     H   -0.88270    -1.27650    -1.40970
10    H    0.88270    -1.27650    -1.40970
11    H    0.00000     3.46590    -0.49460
12    H   -0.88990     2.58120     0.78560
13    H    0.88990     2.58120     0.78560
14    H    0.00000    -3.46590    -0.49460
15    H    0.88990    -2.58120     0.78560
16    H   -0.88990    -2.58120     0.78560
---
H10C5O1, RHF, CHARGE=0, MULT=1
HF=-53.4
 1     O     -0.001515457    -0.028507780     0.026917646
 2     C      1.399801131    -0.026623526    -0.010489081
 3     C      2.031603937     1.386533996     0.068899420
 4     C      1.370790191     2.302464861     1.112143811
 5     C     -0.160756825     2.173783214     1.142046849
 6     C     -0.657269287     0.710850359     1.021151839
 7     H      1.688654431    -0.515380282    -0.975832521
 8     H      1.818009407    -0.671627580     0.808543319
 9     H      3.113381123     1.276325059     0.304393417
10     H      1.983813014     1.870027639    -0.932961308
11     H      1.648332648     3.359805791     0.902369102
12     H      1.783214441     2.076664455     2.122392630
13     H     -0.552182632     2.604237157     2.090263839
14     H     -0.602998139     2.784614281     0.322677878
15     H     -1.744925950     0.702508943     0.754735001
16     H     -0.570842260     0.199104525     2.017337093
EXPGEOM
1     O   -0.67410    -1.25610     0.00000
2     C    0.63050     1.32250     0.00000
3     H    1.70450     1.08050     0.00000
4     H    0.55380     2.41780     0.00000
5     C   -0.01900     0.72900     1.25600
6     C   -0.01900     0.72900    -1.25600
7     C   -0.01900    -0.79600    -1.17810
8     C   -0.01900    -0.79600     1.17810
9     H    0.50390     1.05660     2.16540
10    H    0.50390     1.05660    -2.16540
11    H   -1.05930     1.07360     1.33280
12    H   -1.05930     1.07360    -1.33280
13    H    1.01830    -1.17620    -1.19800
```

```
14      H       1.01830       -1.17620        1.19800
15      H      -0.55980       -1.24360       -2.01930
16      H      -0.55980       -1.24360        2.01930
---
H12C5O1, RHF, CHARGE=0, MULT=1
HF=-67.8
1    C    -0.004972355     0.000119682    -0.014410577
2    C     0.236353614     1.542500661    -0.106800126
3    C    -1.534539012    -0.310510212    -0.107953976
4    C     0.743642221    -0.714755575    -1.193212581
5    O     0.578282347    -0.555152465     1.148902733
6    C     0.213274719    -0.223249009     2.452724516
7    H    -0.325530948     2.103296970     0.666926269
8    H     1.308655853     1.799415846     0.006994355
9    H    -0.096526208     1.930787475    -1.090706253
10   H    -1.745222329    -1.389231336     0.036876272
11   H    -2.124028661     0.249469093     0.645788182
12   H    -1.929914974    -0.025918066    -1.104255499
13   H     1.834921093    -0.522933785    -1.163271148
14   H     0.597384248    -1.813121921    -1.166708887
15   H     0.367850504    -0.351506705    -2.171330926
16   H     0.381149177     0.853730342     2.699938228
17   H    -0.851819308    -0.463493152     2.690364781
18   H     0.858596337    -0.834849747     3.129159449
---
H5C6O1, UHF, CHARGE=-1, MULT=1
HF=-40.5
1    O     0.000000000     0.000000000     0.000000000
2    C     1.254273096     0.000000000     0.000000000
3    C     2.039904080     1.237668903     0.000000000
4    C     2.039922232    -1.236761594    -0.000031513
5    C     3.426444197     1.212128240     0.000033599
6    C     3.426105724    -1.214267347     0.000037601
7    C     4.145265920    -0.001258596     0.000052493
8    H     1.514933304     2.192797007     0.000099974
9    H     1.513716788    -2.191440128    -0.000082993
10   H     3.983517393     2.154552599     0.000061404
11   H     3.981454206    -2.157640806     0.000037844
12   H     5.233527013    -0.001198259     0.000036347
---
H6C6O1, RHF, CHARGE=0, MULT=1
HF=-23, DIP=1.45
1    O     0.000000000     0.000000000     0.000000000
2    C     1.358756036     0.000000000     0.000000000
3    C     2.134526669     1.189719403     0.000000000
4    C     2.005582685    -1.267362990     0.000010562
5    C     3.536608198     1.103593105     0.000009116
6    C     3.407265972    -1.325565652     0.000035105
7    C     4.176098546    -0.146952132    -0.000018543
8    H    -0.367990689     0.873441613    -0.000006479
9    H     1.661407407     2.172057217     0.000027590
10   H     1.424554887    -2.189765329     0.000039814
11   H     4.130869083     2.018493921     0.000069095
12   H     3.905114530    -2.296435143     0.000200792
13   H     5.264390430    -0.205043009    -0.000011066
EXPGEOM
1    C     0.00000      0.93700     0.00000
2    C    -1.20550      0.23510     0.00000
3    C    -1.19100     -1.16090     0.00000
```

```
4    C     0.01700    -1.85570    0.00000
5    C     1.21810    -1.14190    0.00000
6    C     1.21610     0.25020    0.00000
7    O     0.05890     2.30990    0.00000
8    H    -0.84540     2.66290    0.00000
9    H    -2.15340     0.77420    0.00000
10   H    -2.13450    -1.70380    0.00000
11   H     0.02440    -2.94350    0.00000
12   H     2.16780    -1.67400    0.00000
13   H     2.14150     0.82120    0.00000
---
H10C6O1, RHF, CHARGE=0, MULT=1
HF=-42.6
1    O    0.000000000    0.000000000    0.000000000
2    C    1.226707651    0.000000000    0.000000000
3    C    2.033535967    1.296977422    0.000000000
4    C    2.012949409   -1.276682816    0.056772482
5    C    2.253203882   -2.153695397   -0.945315608
6    C    1.722212988   -1.983636280   -2.348704926
7    C    3.081165411   -3.394028516   -0.699550187
8    H    2.541696643    1.424612172    0.977520851
9    H    2.808093104    1.273037635   -0.792441445
10   H    1.398344880    2.187925192   -0.173579504
11   H    2.412089007   -1.457687820    1.061451250
12   H    2.283588074   -2.594829629   -3.084781632
13   H    0.657314642   -2.293733631   -2.400761265
14   H    1.790182756   -0.931253752   -2.693510545
15   H    2.533928666   -4.304871346   -1.020688021
16   H    4.033638280   -3.346351255   -1.268579808
17   H    3.342713725   -3.531158900    0.369245954
---
H10C6O1, RHF, CHARGE=0, MULT=1
HF=-54
1    C    0.000000000    0.000000000    0.000000000
2    C    1.534212357    0.000000000    0.000000000
3    C    2.139310746    1.413627072    0.000000000
4    C    1.483495579    2.385720492    0.995227480
5    C   -0.053993269    2.382893511    0.950246939
6    C   -0.686879089    0.981478489    0.958413279
7    O   -0.643720738   -0.764390124   -0.708857387
8    H    1.927598065   -0.558686111   -0.878380408
9    H    1.873868064   -0.562413538    0.900298671
10   H    3.225945666    1.337145027    0.231725287
11   H    2.077316047    1.838492395   -1.028260767
12   H    1.848756546    3.416769187    0.786632014
13   H    1.825333170    2.146456383    2.028332960
14   H   -0.442586210    2.955562140    1.822761913
15   H   -0.399116357    2.935393904    0.046234269
16   H   -1.768952982    1.075290307    0.716048876
17   H   -0.637320651    0.545332068    1.982974144
EXPGEOM
1    C     0.42340    -1.07600    0.00000
2    C    -0.16820    -0.51100    1.28350
3    C    -0.16820    -0.51100   -1.28350
4    C    -0.16820     1.02860    1.26520
5    C    -0.16820     1.02860   -1.26520
6    C    -0.84100     1.57490    0.00000
7    O     1.30830    -1.89930    0.00000
8    H     0.39730    -0.91410    2.12970
```

```
9      H     -1.20480    -0.87200     1.36850
10     H      0.39730    -0.91410    -2.12970
11     H     -1.20480    -0.87200    -1.36850
12     H      0.86990     1.38860     1.30930
13     H     -0.66930     1.40920     2.16400
14     H      0.86990     1.38860    -1.30930
15     H     -0.66930     1.40920    -2.16400
16     H     -0.80610     2.67190     0.00000
17     H     -1.90510     1.29430     0.00000
---
H12C6O1, RHF, CHARGE=0, MULT=1
HF=-76.6
1    O    0.000000000    0.000000000    0.000000000
2    C    1.226910472    0.000000000    0.000000000
3    C    1.964122806    1.342549535    0.000000000
4    C    2.026962084   -1.344332833    0.014949388
5    C    3.245144984   -1.272443406   -0.948212489
6    C    2.505324172   -1.603218677    1.472281649
7    C    1.129617732   -2.535211330   -0.441534346
8    H    2.869705156    1.325311115    0.636178849
9    H    2.266443404    1.608225033   -1.033613526
10   H    1.316493848    2.159231428    0.379592356
11   H    2.933206202   -0.994539977   -1.975628556
12   H    4.005371231   -0.536218262   -0.618564788
13   H    3.763358899   -2.251226500   -1.014614901
14   H    3.205709362   -0.820489733    1.827111355
15   H    1.653640305   -1.630396221    2.181884694
16   H    3.036213130   -2.573538769    1.554753985
17   H    0.273519999   -2.701585907    0.242299587
18   H    0.722337513   -2.371622417   -1.459732519
19   H    1.707921715   -3.482142607   -0.468010462
---
H14C6O1, RHF, CHARGE=0, MULT=1
HF=-76.3
1    C    0.022398514    0.004854297    0.139952247
2    C    1.024788706   -1.035629695   -0.425928556
3    C    0.430018185   -1.864362750   -1.595343081
4    O    2.306126762   -0.515034927   -0.684569084
5    C    2.595733423    0.456090506   -1.664093767
6    C    3.173229802    1.724504313   -0.977442973
7    C    3.579758981   -0.143948708   -2.705924610
8    H    0.464647675    0.578433116    0.978920670
9    H   -0.325702238    0.729721026   -0.622590686
10   H   -0.873279882   -0.516114936    0.536007762
11   H    1.215819237   -1.771316737    0.411151203
12   H    0.101622381   -1.236933731   -2.447571130
13   H    1.159878031   -2.605049550   -1.978911714
14   H   -0.455214286   -2.430692847   -1.240021065
15   H    1.679983626    0.781843874   -2.230867767
16   H    2.444824734    2.171971190   -0.271846148
17   H    4.102307802    1.512270351   -0.412545958
18   H    3.406248361    2.494908902   -1.740692389
19   H    4.531338489   -0.473958767   -2.244550895
20   H    3.136664218   -1.016306773   -3.226559805
21   H    3.820269936    0.612179460   -3.481001846
---
H6C7O1, RHF, CHARGE=0, MULT=1
HF=-8.8, IE=9.7
1    C    0.000000000    0.000000000    0.000000000
```

```
 2     C     2.839950541    0.000000000    0.000000000
 3     C     0.706438585    1.216220994    0.000000000
 4     C     0.707139597   -1.213412686    0.000207465
 5     C     2.110506000    1.215316440   -0.000141522
 6     C     2.114221018   -1.215250102    0.000201328
 7     C     4.333765599    0.040074954   -0.000090068
 8     O     5.064534905   -0.940441695    0.002563772
 9     H    -1.090487108    0.000115884   -0.000070246
10     H     0.163782741    2.162018546   -0.000091071
11     H     0.165515736   -2.159976900    0.000485783
12     H     2.634520167    2.173231327   -0.000225897
13     H     2.633449902   -2.175557767    0.000363726
14     H     4.769775829    1.062476744   -0.003082053
EXPGEOM
 1     C     0.00000    0.57260    0.00000
 2     C    -1.04900   -0.36040    0.00000
 3     C    -0.76580   -1.72090    0.00000
 4     C     0.56320   -2.16080    0.00000
 5     C     1.61110   -1.23830    0.00000
 6     C     1.32840    0.12590    0.00000
 7     C    -0.27390    2.02600    0.00000
 8     O    -1.37960    2.53330    0.00000
 9     H     0.64340    2.66220    0.00000
10     H    -2.07110    0.00470    0.00000
11     H    -1.57510   -2.44490    0.00000
12     H     0.78000   -3.22510    0.00000
13     H     2.64050   -1.58350    0.00000
14     H     2.13520    0.85530    0.00000
---
H8C7O1, RHF, CHARGE=0, MULT=1
HF=-17.3, DIP=1.38, IE=8.4
 1     C     0.000000000    0.000000000    0.000000000
 2     C     2.827479487    0.000000000    0.000000000
 3     C     0.706796662    1.211593135    0.000000000
 4     C     0.706048412   -1.218053689   -0.000013176
 5     C     2.113657282    1.225525698   -0.000066241
 6     C     2.107252636   -1.231749384    0.000035388
 7     O     4.183555037   -0.126414089   -0.001237689
 8     C     5.069257992    0.957269765    0.009623425
 9     H    -1.089801637   -0.000070501    0.000007300
10     H     0.163777951    2.157951773   -0.000006489
11     H     0.158064320   -2.161477241   -0.000019724
12     H     2.619690838    2.190471127   -0.000183892
13     H     2.636591999   -2.185167216    0.000008279
14     H     6.098461410    0.524972948    0.008285551
15     H     4.960044729    1.593208493    0.921460777
16     H     4.964274447    1.608098470   -0.892152431
EXPGEOM
 1     C     0.00000    0.53100    0.00000
 2     C     0.92460   -0.52040    0.00000
 3     C     0.46420   -1.84080    0.00000
 4     C    -0.89890   -2.12400    0.00000
 5     C    -1.81510   -1.06680    0.00000
 6     C    -1.37470    0.25120    0.00000
 7     O     0.33580    1.85870    0.00000
 8     C     1.72090    2.19470    0.00000
 9     H     1.99060   -0.32820    0.00000
10     H     1.18820   -2.65080    0.00000
11     H    -1.24640   -3.15220    0.00000
```

```
12      H       -2.88200        -1.27130     0.00000
13      H       -2.07520         1.08020     0.00000
14      H        1.76240         3.28490     0.00000
15      H        2.22520         1.80870     0.89560
16      H        2.22520         1.80870    -0.89560
---
H8C7O1, RHF, CHARGE=0, MULT=1
HF=-31.9
1    C     0.000000000    0.000000000     0.000000000
2    C     1.418763048    0.000000000     0.000000000
3    C     2.139048689    1.216901582     0.000000000
4    C     1.417227786    2.430769191    -0.000040531
5    C     0.010566339    2.431570721    -0.000062368
6    C    -0.711557481    1.230975798    -0.000019260
7    C     3.645126320    1.200965238     0.000387426
8    O    -0.740430981   -1.139985267     0.000007442
9    H     1.964542221   -0.945261135     0.000047879
10   H     1.942317776    3.386989069    -0.000029571
11   H    -0.526279571    3.381725038    -0.000091290
12   H    -1.801404284    1.249017616    -0.000027454
13   H     4.035406155    0.683078806     0.901828358
14   H     4.036195139    0.671189233    -0.893788545
15   H     4.079642231    2.221051863    -0.006336418
16   H    -0.209013086   -1.924747086     0.000007966
---
H8C7O1, RHF, CHARGE=0, MULT=1
HF=-30.7
1    C     0.000000000    0.000000000     0.000000000
2    C     1.404968811    0.000000000     0.000000000
3    C     2.152846116    1.199041915     0.000000000
4    C     1.421746434    2.430518876     0.000027676
5    C     0.000367611    2.429605844    -0.000126274
6    C    -0.701017456    1.215725605     0.000065468
7    C     3.658400811    1.154191037    -0.000127765
8    O     2.132311375    3.589281158     0.000499444
9    H    -0.543669875   -0.944843123    -0.000055429
10   H     1.921684651   -0.962314608    -0.000035277
11   H    -0.560536271    3.365228679     0.000026304
12   H    -1.791728322    1.219171555     0.000053693
13   H     4.076806044    1.653854778    -0.898876161
14   H     4.076477906    1.653292734     0.899177401
15   H     4.043393906    0.113186570    -0.000311923
16   H     1.581261395    4.360450650     0.000903310
---
H8C7O1, RHF, CHARGE=0, MULT=1
HF=-29.9
1    O     0.000000000    0.000000000     0.000000000
2    C     1.358034311    0.000000000     0.000000000
3    C     2.137901098    1.184781299     0.000000000
4    C     3.539423172    1.098087066    -0.000040422
5    C     4.203362552   -0.148179901    -0.000089268
6    C     3.409133524   -1.321332110    -0.000186388
7    C     2.009150948   -1.264089547    -0.000223096
8    C     5.704428724   -0.242825173    -0.000532297
9    H    -0.367757538    0.873600158    -0.000064472
10   H     1.668584672    2.169185878    -0.000014293
11   H     4.113528160    2.027054993    -0.000035965
12   H     3.888607851   -2.302725250    -0.000196008
13   H     1.430860372   -2.188467225    -0.000218313
```

```
14      H        6.068571985    -0.793353222     0.892663638
15      H        6.068522247    -0.779192542    -0.902365408
16      H        6.191868404     0.753314424     0.007415690
---
H10C7O1, RHF, CHARGE=0, MULT=1
HF=11
1       C        0.000000000     0.000000000     0.000000000
2       C        1.567860231     0.000000000     0.000000000
3       C        2.050416065     1.390353802     0.000000000
4       C        2.444254922     2.523489005    -0.000104217
5       C        2.917413375     3.858427842     0.018342887
6       C        2.684570821     4.781887056    -0.931376443
7       O        2.069741967    -0.588951105     1.181364556
8       C        2.121502904    -0.735757777    -1.268952390
9       H       -0.404627456     0.505837183     0.899136538
10      H       -0.391408673     0.530807372    -0.890272724
11      H       -0.408554324    -1.030842788    -0.017645869
12      H        3.518781004     4.113686883     0.899760538
13      H        2.098090914     4.597806186    -1.829697456
14      H        3.075686380     5.795888830    -0.862706155
15      H        1.807716417    -1.495158544     1.263879969
16      H        1.790545827    -0.223695286    -2.194070816
17      H        3.229517640    -0.756000555    -1.273663331
18      H        1.766236556    -1.784982964    -1.320712340
---
H10C7O1, RHF, CHARGE=0, MULT=1
HF=-40.8
1       C        0.000000000     0.000000000     0.000000000
2       C        1.560317872     0.000000000     0.000000000
3       C        1.872337376     1.515838020     0.000000000
4       C        0.492534145     2.224914391    -0.034644198
5       C       -0.173027834     2.025336319     1.363037975
6       C       -0.523538368     0.508027504     1.380726276
7       C       -0.314377868     1.224189827    -0.920876042
8       H       -0.438664454    -0.960693952    -0.313972250
9       H        1.968721144    -0.497920866    -0.903021568
10      H        1.999365100    -0.517265087     0.875526539
11      O        2.970160248     2.041954746     0.012138685
12      H        0.513817533     3.270388496    -0.379480198
13      H       -1.079044472     2.657211196     1.462378369
14      H        0.497152980     2.311851567     2.197845474
15      H       -1.616255045     0.342195091     1.476096019
16      H       -0.056959760    -0.020582598     2.235719811
17      H        0.062890032     1.127356222    -1.956904469
18      H       -1.391011020     1.463140248    -1.009410169
---
H10C7O1, RHF, CHARGE=0, MULT=1
HF=-12.9
1       C        0.000000000     0.000000000     0.000000000
2       C        1.559726582     0.000000000     0.000000000
3       C        1.928362548     1.494702664     0.000000000
4       C        0.548792074     2.222102098    -0.000738682
5       C       -0.053291981     1.996630758     1.427813375
6       C       -0.427140102     0.481122244     1.428506707
7       C       -0.367355862     1.269203724    -0.843398115
8       H       -0.456518952    -0.947554700    -0.322244085
9       O        2.193300855     0.636557561    -1.100239183
10      H        2.090451813    -0.838716909     0.461404075
11      H        0.586230518     3.272674279    -0.324484172
```

```
12      H     -0.943956431     2.637526559     1.589035830
13      H      0.660021451     2.251409815     2.236738196
14      H     -1.513437331     0.327529816     1.590732377
15      H      0.086893614    -0.076034397     2.236944483
16      H     -0.124795120     1.208368114    -1.920254987
17      H      2.788826383     1.989174128     0.461779577
18      H     -1.442181929     1.534837302    -0.803343323
---
H10C7O1, RHF, CHARGE=0, MULT=1
HF=-32
1       C      0.000000000     0.000000000     0.000000000
2       C      1.559672311     0.000000000     0.000000000
3       C      1.934212187     1.513150038     0.000000000
4       C      0.553703136     2.239216047    -0.001951963
5       C     -0.169174400     2.034041008     1.364654829
6       C     -0.543754012     0.520881134     1.365893684
7       C     -0.259442226     1.251509276    -0.877632054
8       H     -0.448054095    -0.950914827    -0.327193611
9       H      1.954408676    -0.517096309    -0.898804627
10      H      1.982250350    -0.533229581     0.874603561
11      H      0.600207303     3.288855988    -0.330858968
12      H     -1.070930425     2.676642564     1.434924628
13      H      0.471741362     2.305017165     2.226944928
14      H     -1.641227691     0.373531836     1.435909983
15      H     -0.103132039    -0.016639666     2.228853260
16      O     -0.887484370     1.405778332    -1.903215520
17      H      2.525374776     1.786080796    -0.898331240
18      H      2.555445242     1.788252448     0.875271714
---
H12C7O1, RHF, CHARGE=0, MULT=1
HF=-38.5
1       O      0.000000000     0.000000000     0.000000000
2       C      1.399721765     0.000000000     0.000000000
3       C     -0.763406330     1.129136434     0.000000000
4       C     -2.249012725     0.783503238     0.022783256
5       C     -3.149065020     1.983904047     0.371603987
6       C     -2.690191133     3.314366426    -0.246266190
7       C     -1.196746709     3.615151061    -0.025916813
8       C     -0.303288897     2.408995726    -0.041947882
9       H      1.829588869     0.473973927    -0.915672797
10      H      1.830636002     0.507164129     0.897152191
11      H      1.717207226    -1.069988868     0.019215491
12      H     -2.448841430    -0.031849533     0.754764637
13      H     -2.535253274     0.382539689    -0.977869689
14      H     -3.207389855     2.090926046     1.479204396
15      H     -4.186908124     1.763093184     0.033480961
16      H     -3.298246902     4.143356455     0.181111949
17      H     -2.909412926     3.310074728    -1.338681899
18      H     -1.063193978     4.142024032     0.949171732
19      H     -0.854161084     4.332458753    -0.807879410
20      H      0.759353695     2.655035153    -0.074993828
---
H12C7O1, RHF, CHARGE=0, MULT=1
HF=-52
1       C      0.000000000     0.000000000     0.000000000
2       C      1.561469654     0.000000000     0.000000000
3       C      1.920060915     1.515730978     0.000000000
4       C      0.525290064     2.217368159    -0.000840334
5       C     -0.146936471     2.005678611     1.389706642
```

```
  6    C    -0.505134766     0.491611286     1.390648627
  7    C    -0.280580634     1.237035264    -0.950239153
  8    H    -0.440252338    -0.959280201    -0.316307495
  9    H     1.962587466    -0.517741488    -0.895583089
 10    H     1.987814043    -0.530767450     0.874510705
 11    H     0.562319124     3.272004069    -0.317639673
 12    H    -1.045312390     2.645821408     1.510010942
 13    H     0.524117988     2.275004211     2.230048317
 14    H    -1.594992591     0.322189089     1.511940035
 15    H    -0.025335285    -0.049226198     2.231079266
 16    O    -1.590216018     1.546392100    -1.286833276
 17    H     2.511695098     1.798255398    -0.895158384
 18    H     2.538977501     1.798854613     0.874755549
 19    H    -2.160942156     1.683852976    -0.545571533
 20    H     0.225430860     1.116650403    -1.943634732
---
H12C7O1, RHF, CHARGE=0, MULT=1
HF=-36.4
  1    C     0.000000000     0.000000000     0.000000000
  2    C     1.527598616     0.000000000     0.000000000
  3    C     2.161810103     1.402639603     0.000000000
  4    C     2.051924102     2.215075032     1.302457993
  5    C     0.652200247     2.692330256     1.730494598
  6    C    -0.354021287     1.606536303     2.153925862
  7    C    -0.876071396     0.747721027     1.002741860
  8    O    -0.693102163    -0.658755535     1.048108172
  9    H    -0.415380532    -0.184030073    -1.011226482
 10    H     1.864903490    -0.531652795    -0.920709023
 11    H     1.930354010    -0.596642911     0.849464584
 12    H     1.748363854     1.997742259    -0.846440027
 13    H     3.247953322     1.278108939    -0.224052287
 14    H     2.686955673     3.124821661     1.178322558
 15    H     2.516341721     1.639778598     2.136304544
 16    H     0.207620465     3.316838272     0.921920921
 17    H     0.786406421     3.379903910     2.599267424
 18    H    -1.231956653     2.111250247     2.621181720
 19    H     0.086270226     0.976466526     2.959284929
 20    H    -1.883206625     1.069059011     0.669629960
---
H12C7O1, RHF, CHARGE=0, MULT=1
HF=-59.3
  1    C    -0.163758044    -0.081341462     0.045735793
  2    C    -0.301379152     1.291245148     0.727902445
  3    C     0.671896282     2.407365965     0.305954548
  4    C     2.160255028     2.070236585     0.100439225
  5    C     2.687276221     0.880831574     0.912923458
  6    C     2.401170755    -0.523741162     0.365215127
  7    C     0.953528923    -0.997769111     0.575826914
  8    O     3.326197010     1.045174065     1.946068637
  9    H    -1.128843431    -0.624238414     0.185727918
 10    H    -0.053987593     0.048061085    -1.054998055
 11    H    -0.249108670     1.162526089     1.833883667
 12    H    -1.333194575     1.665520848     0.524707965
 13    H     0.599112801     3.214343359     1.072812198
 14    H     0.300609893     2.861213161    -0.643000830
 15    H     2.758635031     2.980389950     0.336092175
 16    H     2.348093317     1.869268958    -0.979345003
 17    H     3.083198743    -1.265457057     0.840866024
 18    H     2.654022106    -0.543400006    -0.719524673
```

```
19      H      0.788750726    -1.193465750     1.660530245
20      H      0.848939156    -1.988211040     0.073277916
---
H14C7O1, RHF, CHARGE=0, MULT=1
HF=-74.4
1       C      0.000000000     0.000000000     0.000000000
2       C      1.542486750     0.000000000     0.000000000
3       C      2.115850052     1.439040717     0.000000000
4       C      2.929482440     1.902710593    -1.234005228
5       C      2.248734644     3.081497710    -1.959465308
6       C      4.400805100     2.191655566    -0.872524541
7       C      2.141279939    -0.856155854     1.134802360
8       O      1.930796828     2.190550561     0.950520720
9       H     -0.425939160     0.405218182     0.938992184
10      H     -0.396364677     0.604344013    -0.841419370
11      H     -0.387812554    -1.031622279    -0.127493875
12      H      1.851294438    -0.495353857    -0.954947420
13      H      2.952296237     1.059845925    -1.970047847
14      H      1.199431480     2.834011172    -2.220091780
15      H      2.238131589     4.007344073    -1.351108234
16      H      2.776585575     3.309197411    -2.908442652
17      H      4.510264350     3.063686753    -0.197981360
18      H      4.865740540     1.316646657    -0.374047624
19      H      4.991739378     2.398709548    -1.788387337
20      H      3.249712262    -0.847730556     1.095791838
21      H      1.834953702    -0.505375785     2.140065059
22      H      1.818208671    -1.912824499     1.034721417
---
H14C7O1, RHF, CHARGE=0, MULT=1
HF=-71.9
1       H      0.000000000     0.000000000     0.000000000
2       C      1.121127511     0.000000000     0.000000000
3       C      1.553047591     1.516095646     0.000000000
4       C      3.055337733     1.729073432    -0.367513948
5       C      3.475104775     1.009332289    -1.659089954
6       C      3.010189502    -0.453024813    -1.747271253
7       C      1.545333491    -0.671456380    -1.333355498
8       O      1.181422947     2.199096202     1.163724269
9       C      1.531751202    -0.821209716     1.237466837
10      H      0.943312242     2.009675962    -0.811434394
11      H      3.716195331     1.399505252     0.466203405
12      H      3.246676389     2.819448454    -0.492429310
13      H      4.585464112     1.042128683    -1.742835906
14      H      3.090388924     1.573844024    -2.539568400
15      H      3.676807978    -1.093615472    -1.125880718
16      H      3.143078972    -0.810809310    -2.794021733
17      H      1.348728867    -1.766420400    -1.282646909
18      H      0.880091467    -0.294735255    -2.145456645
19      H      1.744079462     2.009559388     1.900916851
20      H      2.625089028    -0.815325965     1.418964094
21      H      1.221686501    -1.881131435     1.126777118
22      H      1.037074379    -0.444718043     2.156102819
---
H14C7O1, RHF, CHARGE=0, MULT=1
HF=-72.6
1       C      0.000000000     0.000000000     0.000000000
2       C      1.532563720     0.000000000     0.000000000
3       C      2.291314830     1.372801307     0.000000000
4       C      3.822315800     1.024108667     0.015963940
```

```
 5    C     4.570483450    0.719353800   -1.285359926
 6    C     1.942912366    2.226200227   -1.251499153
 7    C     1.924495875    2.212571972    1.261258600
 8    O     4.440587126    0.976988074    1.074903151
 9    H    -0.426043424    0.478053144    0.903998942
10    H    -0.425721708    0.507749413   -0.888074941
11    H    -0.365904989   -1.048551129   -0.017351438
12    H     1.864034561   -0.587332288    0.888455651
13    H     1.858100251   -0.589461764   -0.889186854
14    H     3.961400559    0.120645521   -1.989523461
15    H     4.860436081    1.664844313   -1.787455223
16    H     5.499985660    0.145479878   -1.092189991
17    H     2.617749210    3.102278425   -1.342514288
18    H     2.015009091    1.648313798   -2.194683312
19    H     0.908720877    2.623717149   -1.202129597
20    H     2.029297117    1.631749835    2.199166426
21    H     2.568110978    3.112063738    1.347697202
22    H     0.876309360    2.573349246    1.220150441
---
H14C7O1, RHF, CHARGE=0, MULT=1
HF=-71.3
 1    O     0.000000000    0.000000000    0.000000000
 2    C     1.224922458    0.000000000    0.000000000
 3    C     2.050955766    1.299589251    0.000000000
 4    C     1.268861942    2.625621452    0.012369907
 5    C     2.153065100    3.876947940    0.013540086
 6    C     2.050885321   -1.299629016   -0.005020021
 7    C     1.269641801   -2.625120780    0.048837372
 8    C     2.153835417   -3.876453410    0.048041048
 9    H     2.722605567    1.266780529    0.888921169
10    H     2.707915267    1.274657919   -0.900148626
11    H     0.597750249    2.671904082   -0.875754142
12    H     0.609026315    2.661906063    0.909357012
13    H     2.809368253    3.918775028    0.906432694
14    H     2.799892134    3.927329710   -0.885770443
15    H     1.521540322    4.789024325    0.021362507
16    H     2.681264384   -1.286943357   -0.924291203
17    H     2.749130542   -1.255686938    0.862604482
18    H     0.632872685   -2.646245466    0.962854161
19    H     0.576219739   -2.686176521   -0.821030679
20    H     2.777262044   -3.941691391   -0.866734856
21    H     2.832894584   -3.903719857    0.924307452
22    H     1.522918682   -4.788149728    0.086793483
---
H14C7O1, RHF, CHARGE=0, MULT=1
HF=-63.1
 1    C     0.000000000    0.000000000    0.000000000
 2    C     1.531260803    0.000000000    0.000000000
 3    C     2.174507014    1.400390690    0.000000000
 4    C     3.716779910    1.384257777   -0.001077074
 5    C     4.358065276    2.787035123   -0.000229235
 6    C     5.897883568    2.769984397   -0.003142066
 7    C     6.540562875    4.156669736   -0.000132623
 8    O     7.748285798    4.337382237    0.036626107
 9    H    -0.412769079    0.506072680    0.896170488
10    H    -0.412887146    0.506065228   -0.896100818
11    H    -0.382117194   -1.041708929    0.000157426
12    H     1.883111902   -0.570318720    0.890709445
13    H     1.882935213   -0.569971142   -0.891014025
```

```
14      H        1.814995642     1.964609986    -0.891132711
15      H        1.816334332     1.963909842     0.892100519
16      H        4.076040564     0.819927309     0.889897340
17      H        4.074736716     0.821580866    -0.893629960
18      H        4.000614701     3.353023378    -0.891164374
19      H        4.004091000     3.350588012     0.893635974
20      H        6.268435559     2.210032158     0.886252724
21      H        6.265498199     2.220350567    -0.900338877
22      H        5.841151498     5.018593648    -0.036442366
---
H14C7O1, RHF, CHARGE=0, MULT=1
HF=-75
1       C        0.000000000     0.000000000     0.000000000
2       C        1.541269100     0.000000000     0.000000000
3       C        2.187965560     1.389075021     0.000000000
4       C       -0.798994086    -0.720920650     1.136595689
5       C       -0.464256647    -2.239691120     1.106981161
6       C       -0.427286498    -0.102551511     2.513724440
7       C       -2.338062800    -0.558576380     0.936553894
8       O       -0.588128333     0.548228637    -0.926284868
9       H        1.943390559    -0.568778477     0.865924962
10      H        1.877493894    -0.557964767    -0.905832154
11      H        1.960368724     1.955633925    -0.924657108
12      H        1.856302863     1.998034719     0.864970101
13      H        3.291222831     1.288259005     0.065944837
14      H       -0.649694238    -2.675140531     0.103927858
15      H        0.593105802    -2.446534317     1.366901345
16      H       -1.085885072    -2.802393424     1.833274974
17      H        0.628940201    -0.288296414     2.793077041
18      H       -0.584523655     0.995290301     2.519190034
19      H       -1.048950893    -0.530202867     3.327019714
20      H       -2.646224941     0.506231150     0.946604558
21      H       -2.682071277    -1.001369180    -0.019652641
22      H       -2.899474343    -1.066305853     1.748400489
---
H16C7O1, RHF, CHARGE=0, MULT=1
HF=-81.2
1       O        0.000000000     0.000000000     0.000000000
2       C        1.396896383     0.000000000     0.000000000
3       C        1.917103735     1.460817059     0.000000000
4       C        3.456069901     1.558179686    -0.001475643
5       C        3.990722170     3.005092476    -0.000887650
6       C        5.529812495     3.105067179    -0.002264649
7       C        6.066343866     4.549711217    -0.001911017
8       C        7.593273741     4.664023248    -0.003095325
9       H       -0.338239630    -0.884151743    -0.000391200
10      H        1.792620234    -0.548933945     0.897404829
11      H        1.792800383    -0.549947259    -0.896589151
12      H        1.513093023     1.991869656    -0.891759411
13      H        1.515087886     1.991483206     0.892901467
14      H        3.856995293     1.021703346     0.889213672
15      H        3.855125326     1.023272015    -0.893936557
16      H        3.590133881     3.541250067    -0.891699385
17      H        3.592136510     3.539815776     0.891670481
18      H        5.931191501     2.569215842     0.888530055
19      H        5.929485960     2.570253732    -0.894462219
20      H        5.672086861     5.092490389    -0.892075446
21      H        5.673576878     5.091530064     0.889501450
22      H        8.043620153     4.189956509     0.892488939
```

```
23      H      8.042177388    4.190699614   -0.899796846
24      H      7.896807027    5.731352479   -0.002909743
---
H16C7O1, RHF, CHARGE=0, MULT=1
HF=-85.5
1    C     0.006642148   -0.019700711   -0.001849488
2    C    -0.946597448   -1.245055652    0.186876653
3    C     1.244258394   -0.154245084    0.944719053
4    C     0.501809861    0.023342608   -1.490961048
5    O    -0.667294071    1.212716708    0.164625159
6    C    -1.284316127    1.656221507    1.350748463
7    C    -0.605048842    2.973395855    1.818930225
8    C    -2.806902552    1.843624182    1.101541523
9    H    -1.286667762   -1.359370694    1.235565227
10   H    -1.850149305   -1.166668963   -0.450163879
11   H    -0.429599396   -2.186799790   -0.088505892
12   H     1.919988608    0.720947522    0.867051786
13   H     0.954042044   -0.258438813    2.009356858
14   H     1.838138359   -1.053880023    0.683737966
15   H    -0.342415418    0.122680490   -2.202332576
16   H     1.191414162    0.872335996   -1.670641586
17   H     1.046124918   -0.908321471   -1.747705468
18   H    -1.182766059    0.920730362    2.195585365
19   H     0.465143912    2.815349139    2.059850177
20   H    -0.663564726    3.772966874    1.054228301
21   H    -1.096492735    3.346625690    2.740691930
22   H    -3.015525248    2.564853304    0.286887411
23   H    -3.296846037    0.885338170    0.837374842
24   H    -3.298139340    2.218543205    2.022814975
---
H6C8O1, RHF, CHARGE=0, MULT=1
HF=3.3
1    C     0.000000000    0.000000000    0.000000000
2    C     1.398970128    0.000000000    0.000000000
3    C     2.146363958    1.208214971    0.000000000
4    C     1.517763368    2.457950903    0.000025667
5    C     0.107671332    2.501855110    0.000105099
6    C    -0.630396398    1.264737951   -0.000070354
7    O    -1.975104405    1.516326991   -0.000071424
8    C    -2.127853398    2.882462199   -0.000299802
9    H    -3.148354491    3.250851352   -0.000445482
10   H    -0.739489456    4.604113211   -0.000235520
11   C    -0.910735430    3.539019131   -0.000093863
12   H    -0.570093106   -0.926843422    0.000054331
13   H     1.935668835   -0.950135594   -0.000031322
14   H     3.235780322    1.151404741    0.000120509
15   H     2.101817130    3.377595066   -0.000043051
---
H8C8O1, RHF, CHARGE=0, MULT=1
HF=-7.2
1    C     0.000000000    0.000000000    0.000000000
2    C     1.428504025    0.000000000    0.000000000
3    C     2.142160378    1.197528881    0.000000000
4    C     1.416517457    2.413079808    0.000005007
5    C     0.014412733    2.414105864   -0.000029251
6    C    -0.711960680    1.198731337    0.000056479
7    C    -0.463633046   -1.446772923   -0.000030771
8    O     0.713755835   -2.230951168    0.000403985
9    H     3.231676951    1.212307083    0.000010411
```

```
10      H        1.959715528      3.359054420      0.000058204
11      H       -0.527877483      3.360495928      0.000003816
12      H       -1.801549243      1.214324367      0.000047495
13      C        1.891363123     -1.447232854      0.000552693
14      H        2.507337951     -1.694086151     -0.899857386
15      H        2.506734440     -1.694118205      0.901487313
16      H       -1.079595001     -1.693485982      0.900543214
17      H       -1.078931115     -1.693569887     -0.901011971
---
H8C8O1, RHF, CHARGE=0, MULT=1
HF=-11
1     C        0.000000000      0.000000000      0.000000000
2     C        1.405577614      0.000000000      0.000000000
3     C        2.112981833      1.215183629      0.000000000
4     C        1.420871799      2.448328138      0.003660919
5     C        0.006786633      2.437490946      0.002577072
6     C       -0.697144496      1.220489071      0.001383094
7     C        2.171039064      3.741260035      0.019452531
8     C        2.545430964      4.383987165      1.153011667
9     O        2.432797207      4.195491015     -1.239105274
10      H       -0.547358409     -0.942972034     -0.001716536
11      H        1.950699061     -0.944444894     -0.001926568
12      H        3.203880964      1.195348659     -0.003547020
13      H       -0.551714304      3.374788961      0.001021586
14      H       -1.787612904      1.225284990     -0.000194842
15      H        2.319176438      3.991862275      2.141041710
16      H        3.091018308      5.324761100      1.170870067
17      H        2.907885710      5.016041369     -1.243459817
---
H8C8O1, RHF, CHARGE=0, MULT=1
HF=-11.1
1     C        0.000000000      0.000000000      0.000000000
2     C        1.435102146      0.000000000      0.000000000
3     C        2.123920898      1.212412969      0.000000000
4     C        1.380416084      2.415725681      0.000139055
5     C       -0.023717458      2.400523485      0.000192048
6     C       -0.747643919      1.187416570      0.000121331
7     O       -0.521333866     -1.261897436     -0.000143655
8     C        0.534142342     -2.215637794     -0.000080768
9     H        3.213330240      1.243961631     -0.000034589
10      H        1.909507976      3.369210703      0.000170502
11      H       -0.571707664      3.344292922      0.000253914
12      H       -1.835820514      1.187279754      0.000203025
13      H        0.431548386     -2.870020861     -0.899358235
14      H        0.431198413     -2.869793375      0.899389690
15      C        1.890001834     -1.435961816     -0.000022677
16      H        2.501470599     -1.686250477     -0.891602431
17      H        2.501716948     -1.686644548      0.891182986
---
H8C8O1, RHF, CHARGE=0, MULT=1
HF=-20.7
1     C        0.000000000      0.000000000      0.000000000
2     C        1.412833758      0.000000000      0.000000000
3     C        2.125741386      1.212070554      0.000000000
4     C        1.435962046      2.436733161      0.003799467
5     C        0.030240712      2.446498602      0.009586657
6     C       -0.684518732      1.235902393      0.009181203
7     C       -0.765312425     -1.299154473      0.000900015
8     C       -1.121285987     -1.894275782     -1.359420442
```

```
 9      O      -1.081804507    -1.843584431     1.052578436
10      H       1.965620285    -0.940875013     0.003523658
11      H       3.216124028     1.199687507    -0.000599767
12      H       1.989409177     3.376284486     0.004487136
13      H      -0.508759668     3.394425174     0.016311203
14      H      -1.775364020     1.263414863     0.019968892
15      H      -1.748483909    -1.184889707    -1.935939229
16      H      -0.200208175    -2.096449736    -1.942303731
17      H      -1.681269202    -2.846053042    -1.268477404
EXPGEOM
1       C       1.72320      2.15350     0.00000
2       C       0.27700      1.68970     0.00000
3       C       0.00000      0.21570     0.00000
4       O      -0.63750      2.50140     0.00000
5       C       1.02360     -0.74410     0.00000
6       C      -1.33620     -0.21590     0.00000
7       C       0.71840     -2.10440     0.00000
8       C      -1.64040     -1.57240     0.00000
9       C      -0.61320     -2.52070     0.00000
10      H       1.74630      3.24330     0.00000
11      H       2.25270      1.77630     0.88190
12      H       2.25270      1.77630    -0.88190
13      H       2.06330     -0.43530     0.00000
14      H      -2.11920      0.53510     0.00000
15      H       1.51860     -2.83870     0.00000
16      H      -2.67730     -1.89570     0.00000
17      H      -0.85090     -3.58060     0.00000
---
H10C8O1, RHF, CHARGE=0, MULT=1
HF=-37.6
1       O       0.000000000     0.000000000     0.000000000
2       C       1.359959523     0.000000000     0.000000000
3       C       2.174376819     1.178748940     0.000000000
4       C       3.591962889     1.012026039     0.002252445
5       C       4.143655571    -0.288511134    -0.003783977
6       C       3.329508912    -1.431857237    -0.009719347
7       C       1.939005053    -1.300183544    -0.006607967
8       C       1.533062666     2.544723195    -0.002275105
9       C       4.549994034     2.179791975     0.011218810
10      H      -0.391930710     0.859252791     0.052222678
11      H       5.226542784    -0.427138984    -0.003990005
12      H       3.783869476    -2.423713377    -0.015568912
13      H       1.308994288    -2.190510695    -0.009306365
14      H       0.927862578     2.704593013     0.916688528
15      H       2.257821766     3.380261343    -0.042092635
16      H       0.868459674     2.676463240    -0.883783490
17      H       4.427925592     2.809423504    -0.895039060
18      H       4.401514446     2.818143398     0.907334064
19      H       5.610177546     1.852543370     0.028815095
---
H10C8O1, RHF, CHARGE=0, MULT=1
HF=-39
1       O       0.000000000     0.000000000     0.000000000
2       C       1.357942294     0.000000000     0.000000000
3       C       2.172642869     1.172146513     0.000000000
4       C       3.577461542     0.993115260     0.000805008
5       C       4.190556639    -0.275778704     0.000075820
6       C       3.352097843    -1.416756372    -0.002173594
7       C       1.960456136    -1.290918634    -0.001550729
```

```
 8    C      1.608953946    2.568711039    0.000030894
 9    H      4.214539722    1.881716737    0.001870763
10    H     -0.386792271    0.863656747    0.007799047
11    C      5.687020483   -0.430558786    0.000827650
12    H      3.788490529   -2.417832970   -0.003666183
13    H      1.340733130   -2.188803474   -0.002512500
14    H      2.403707572    3.342270278   -0.004219316
15    H      0.980797387    2.753878300   -0.897557317
16    H      0.987385213    2.756244697    0.901769731
17    H      6.027656095   -0.992951797    0.895918898
18    H      6.029222803   -0.984189490   -0.899142309
19    H      6.215401211    0.544464016    0.006108845
---
H10C8O1, RHF, CHARGE=0, MULT=1
HF=-38.7
 1    C      0.000000000    0.000000000    0.000000000
 2    C      1.427236354    0.000000000    0.000000000
 3    C      2.066163521    1.263134182    0.000000000
 4    C      1.347933067    2.464579298    0.000020754
 5    C     -0.068034428    2.464197540   -0.000518604
 6    C     -0.731309213    1.223836148   -0.000270199
 7    O     -0.769982930   -1.120257762   -0.000914232
 8    C      2.260782432   -1.253296043   -0.000226369
 9    C     -0.829570690    3.763334984   -0.000656467
10    H      3.157511401    1.316335979    0.000114517
11    H      1.897081545    3.407490341    0.000358359
12    H     -1.822697214    1.190375202   -0.000680376
13    H     -0.276349754   -1.927502951    0.005790611
14    H      3.347709197   -1.032930664   -0.001994298
15    H      2.063128537   -1.873707004    0.900191050
16    H      2.060450507   -1.875057633   -0.899032422
17    H     -0.575005201    4.370243826   -0.895022955
18    H     -0.584636673    4.364058506    0.900644436
19    H     -1.928014965    3.612847348   -0.007225049
---
H10C8O1, RHF, CHARGE=0, MULT=1
HF=-38.7
 1    C      0.000000000    0.000000000    0.000000000
 2    C      1.506725952    0.000000000    0.000000000
 3    C      2.189065668    1.238678448    0.000000000
 4    C      3.587837396    1.301311608   -0.000071538
 5    C      4.344907771    0.118911377    0.000757589
 6    C      3.728633818   -1.148429935    0.001952599
 7    C      2.293594544   -1.194990971    0.001513266
 8    C      4.576288095   -2.394858519    0.003442039
 9    O      1.748648259   -2.439802304    0.002779323
10    H     -0.417755659    1.027358683   -0.006872549
11    H     -0.408991245   -0.514421016   -0.895930788
12    H     -0.408233434   -0.501788199    0.903482016
13    H      1.625267354    2.173992795   -0.000222447
14    H      4.090596054    2.268807763   -0.000748071
15    H      5.434159974    0.199400834    0.001268771
16    H      4.381379559   -3.015394558    0.903176708
17    H      4.382866024   -3.017061976   -0.895444333
18    H      5.660919162   -2.158772095    0.004072312
19    H      0.802610369   -2.450648178    0.004695435
---
H10C8O1, RHF, CHARGE=0, MULT=1
HF=-34.7
```

```
   1    O      0.000000000     0.000000000     0.000000000
   2    C      1.359887897     0.000000000     0.000000000
   3    C      2.090829820     1.217983749     0.000000000
   4    C      3.492946381     1.197681108     0.009747012
   5    C      4.170299239    -0.029061522     0.022176035
   6    C      3.445518716    -1.234621501     0.020617502
   7    C      2.034019237    -1.269297666     0.003853797
   8    C      1.239585493    -2.560932812     0.008383924
   9    C      1.983381933    -3.888981120    -0.166532628
  10    H     -0.369484109     0.872669828    -0.016002843
  11    H      1.576720463     2.180195176    -0.006609432
  12    H      4.050177607     2.135177602     0.009152995
  13    H      5.260140693    -0.053238372     0.033330586
  14    H      4.023913107    -2.160229513     0.034273163
  15    H      0.675765079    -2.616256997     0.971331116
  16    H      0.474899453    -2.512918920    -0.804125784
  17    H      2.560814123    -3.925059052    -1.112295598
  18    H      2.676817890    -4.097469447     0.673018771
  19    H      1.250680431    -4.722780300    -0.197735133
---
H10C8O1, RHF, CHARGE=0, MULT=1
HF=-37.4
   1    C      0.000000000     0.000000000     0.000000000
   2    C      1.509339932     0.000000000     0.000000000
   3    C      2.184215835     1.242290395     0.000000000
   4    C      3.599956081     1.309459995     0.001274741
   5    C      4.355116113     0.108288938     0.001652361
   6    C      3.680483482    -1.117981966     0.000995754
   7    C      2.266949793    -1.205955702     0.000357213
   8    C      1.621352563    -2.567986195    -0.000043726
   9    O      4.290747826     2.479326491     0.001832583
  10    H     -0.401685221    -0.513703335     0.898495920
  11    H     -0.423063558     1.025623434     0.002062763
  12    H     -0.401657642    -0.510203355    -0.900498793
  13    H      1.603231122     2.166856962    -0.000677933
  14    H      5.445101584     0.129451461     0.002463355
  15    H      4.281746610    -2.030448091     0.001180264
  16    H      0.985988623    -2.713891770    -0.898964216
  17    H      2.368706750    -3.388064926    -0.000310618
  18    H      0.986345608    -2.714729557     0.899019552
  19    H      3.726654841     3.241004300     0.001966947
---
H10C8O1, RHF, CHARGE=0, MULT=1
HF=-38.6
   1    O      0.000000000     0.000000000     0.000000000
   2    C      1.359904548     0.000000000     0.000000000
   3    C      2.000544744     1.269218367     0.000000000
   4    C      3.406591138     1.347183610    -0.000018289
   5    C      4.160939079     0.148505019     0.000090529
   6    C      3.542662070    -1.119203715     0.000344777
   7    C      2.131001300    -1.188272273     0.000008010
   8    C      4.111132058     2.679293619    -0.000140648
   9    C      4.357113559    -2.386614649    -0.000708617
  10    H     -0.368221962    -0.873283606     0.000019805
  11    H      1.397188253     2.178315195    -0.000012761
  12    H      5.251193744     0.211531687    -0.000142910
  13    H      1.639233292    -2.162567584    -0.000162256
  14    H      4.756384461     2.785084331    -0.897650658
  15    H      4.755156111     2.786016508     0.898146114
```

```
16    H     3.404691826    3.533855907   -0.001498544
17    H     4.116585699   -3.013586190    0.883637700
18    H     4.151462216   -2.987800025   -0.911497697
19    H     5.448125083   -2.190398767    0.023776890
---
H10C8O1, RHF, CHARGE=0, MULT=1
HF=-34.9
1     C     0.000000000    0.000000000    0.000000000
2     C     1.532981843    0.000000000    0.000000000
3     C     2.158721544    1.379814583    0.000000000
4     C     2.465602895    2.030070483   -1.214729314
5     C     3.062941318    3.320885528   -1.210818771
6     C     3.355574581    3.965107063    0.018456326
7     C     3.048125402    3.310955293    1.221464141
8     C     2.458412001    2.037607538    1.217806641
9     O     3.326873234    3.879080610   -2.421935681
10    H    -0.411420771    0.506850178    0.895979556
11    H    -0.411322773    0.506941506   -0.895886200
12    H    -0.381471688   -1.041961222   -0.000011277
13    H     1.892192113   -0.574188523    0.885887366
14    H     1.891479619   -0.576068839   -0.884858861
15    H     2.246327324    1.544049821   -2.166901874
16    H     3.813257271    4.954105265    0.047192506
17    H     3.271331064    3.800638055    2.170781814
18    H     2.234372735    1.559715960    2.172621487
19    H     3.724680830    4.736484198   -2.352395197
---
H10C8O1, RHF, CHARGE=0, MULT=1
HF=-34.5
1     C     0.000000000    0.000000000    0.000000000
2     C     1.532884939    0.000000000    0.000000000
3     C     2.159519513    1.376824782    0.000000000
4     C     2.469984108    2.038617342   -1.209543984
5     C     3.063604084    3.310499725   -1.222127643
6     C     3.365213537    3.959800965    0.002828574
7     C     3.061053300    3.309733346    1.229381470
8     C     2.467971423    2.039593386    1.212653581
9     O     3.938556645    5.188815531    0.074678618
10    H    -0.411376292    0.508415334    0.895147777
11    H    -0.411535382    0.505939119   -0.896493812
12    H    -0.382627579   -1.041585817    0.001388745
13    H     1.891141714   -0.575518809    0.885648143
14    H     1.891009745   -0.576165943   -0.885239452
15    H     2.250558551    1.563210745   -2.168009601
16    H     3.285184944    3.782059890   -2.180149844
17    H     3.284181603    3.787755258    2.183669017
18    H     2.245442393    1.561808088    2.169298471
19    H     4.115300320    5.563093891   -0.778033630
---
H10C8O1, RHF, CHARGE=0, MULT=1
HF=-22.6
1     O     0.000000000    0.000000000    0.000000000
2     C     1.396501452    0.000000000    0.000000000
3     C     2.012262518    1.429640615    0.000000000
4     C     3.523100684    1.426708988   -0.077301950
5     C     4.189473119    1.460304258   -1.325627216
6     C     5.593202144    1.474287803   -1.397177524
7     C     6.361549018    1.455789627   -0.221051062
8     C     5.716801652    1.424005624    1.027029674
```

```
 9    C      4.313160270     1.409962854     1.097160954
10    H     -0.353929451     0.256621575     0.839736838
11    H      1.797899308    -0.579901371     0.873722254
12    H      1.712368026    -0.549862019    -0.923338674
13    H      1.594620201     2.012543220    -0.852805085
14    H      1.688427586     1.975339104     0.916731156
15    H      3.617738336     1.477247450    -2.255324444
16    H      6.085865317     1.500114701    -2.369793296
17    H      7.450287033     1.466679749    -0.276392057
18    H      6.305434424     1.410693590     1.944956525
19    H      3.839109739     1.387605652     2.080139195
---
H10C8O1, RHF, CHARGE=0, MULT=1
HF=-24.3
 1    C      0.000079374    -0.000339260     0.000018527
 2    C      1.402595374     0.000013495    -0.000021295
 3    C      2.123858642     1.208068117     0.000044131
 4    C      1.425786890     2.442467811    -0.000143054
 5    C     -0.001689247     2.440455680    -0.000015204
 6    C     -0.695859177     1.223586532     0.000176085
 7    O      1.995725463     3.679096176     0.003309304
 8    C      3.383433852     3.898507479    -0.026018476
 9    C      3.652545381     5.416032599    -0.031555621
10    H     -0.549076911    -0.941577008    -0.000021162
11    H      1.946556065    -0.945782650     0.000025883
12    H      3.212171092     1.157422590     0.000483313
13    H     -0.558876038     3.377846857     0.000225692
14    H     -1.786875827     1.225724587     0.000342620
15    H      3.887084677     3.438894287     0.865218159
16    H      3.850027523     3.438767305    -0.937059788
17    H      3.214101734     5.911599353    -0.920436763
18    H      3.250008441     5.912153786     0.873724841
19    H      4.748582190     5.586749857    -0.053507427
---
H10C8O1, RHF, CHARGE=0, MULT=1
HF=-26.3
 1    C     -0.096625344    -0.014559247     0.174249406
 2    C      2.688602164    -0.382430458    -0.161093152
 3    C      0.696444710     1.019692323    -0.344344406
 4    C      0.501364163    -1.239111040     0.526956586
 5    C      2.082642926     0.850501236    -0.513652086
 6    C      1.879377346    -1.434682177     0.363145735
 7    O      4.020152690    -0.662732913    -0.225065028
 8    C      4.969375957     0.065005537    -0.962846383
 9    C      4.992773868    -0.266097861    -2.468142312
10    H     -1.169227773     0.127953373     0.304375913
11    H      0.237598354     1.970513546    -0.619756389
12    H     -0.112455768    -2.044858852     0.932527830
13    H      2.656736097     1.685186850    -0.914360918
14    H      2.323342992    -2.390348416     0.643826503
15    H      5.959126546    -0.214901368    -0.512951004
16    H      4.868900761     1.171217997    -0.818952011
17    H      5.126345925    -1.350829569    -2.649984035
18    H      4.068058649     0.055670952    -2.986514738
19    H      5.844961920     0.262886845    -2.942137504
---
H12C8O1, RHF, CHARGE=0, MULT=1
HF=-48.9
 1    C      0.000000000     0.000000000     0.000000000
```

```
  2      C         1.565297000      0.000000000      0.000000000
  3      C         1.986854080      1.516842619      0.000000000
  4      C         1.587022576      2.023708925      1.415441977
  5      C         0.939363924      0.770469258      2.078072770
  6      C        -0.433664146      0.479862361      1.404202982
  7      O        -0.739053777     -0.310776462     -0.916241521
  8      C         2.248032016     -0.870134040     -1.053783530
  9      H         3.076694668      1.632013555     -0.171440734
 10      H         1.485178092      2.091093866     -0.804703275
 11      H         2.471396158      2.366396517      1.991284615
 12      H         0.897283373      2.890179156      1.370730767
 13      H         0.884217048      0.829166527      3.177238715
 14      H        -1.096846140      1.366108642      1.356511858
 15      H        -0.995725358     -0.309347269      1.944651718
 16      C         1.822260682     -0.380624717      1.503694763
 17      H         1.477224134     -1.395864277      1.779397186
 18      H         2.884851643     -0.326473405      1.808356554
 19      H         1.898049944     -1.921283801     -1.000100953
 20      H         2.048497680     -0.499485330     -2.079607515
 21      H         3.348545330     -0.881095637     -0.911433989
---
H12C8O1, RHF, CHARGE=0, MULT=1
HF=-52.2
  1      C         0.000000000      0.000000000      0.000000000
  2      C         1.554886633      0.000000000      0.000000000
  3      C         2.083494425      1.454454953      0.000000000
  4      C         0.889498293      2.450956027     -0.024203681
  5      C         0.039921724      2.203131340     -1.301824990
  6      C        -0.473179771      0.752325854     -1.267179349
  7      C        -0.535638143      0.771536500      1.239017985
  8      C         0.002805404      2.222616876      1.233180056
  9      O        -1.171520013      0.254103515     -2.136821834
 10      H        -0.375716818     -1.045385676      0.017589576
 11      H         1.935187312     -0.553020482      0.885895832
 12      H         1.936571407     -0.554011642     -0.885093229
 13      H         2.719016257      1.637429701      0.893414351
 14      H         2.746939843      1.626944004     -0.875233595
 15      H         1.272282752      3.494507234     -0.028204274
 16      H         0.643277532      2.377922212     -2.218774292
 17      H        -0.813409209      2.913360553     -1.354377026
 18      H        -0.234927442      0.247768644      2.172064388
 19      H        -1.647417242      0.767669271      1.239234155
 20      H         0.586363143      2.424874379      2.157317304
 21      H        -0.838163213      2.949475883      1.249027473
---
H12C8O1, RHF, CHARGE=0, MULT=1
HF=-52
  1      C         0.000000000      0.000000000      0.000000000
  2      C         1.548921740      0.000000000      0.000000000
  3      C         2.163064933      1.432233666      0.000000000
  4      C         2.308402611      1.822569696      1.492486656
  5      C         1.779020581      0.596171655      2.293840243
  6      C        -0.651741556      0.583121199      1.275171101
  7      C         2.053660997     -0.598152011      1.341680578
  8      H        -0.351881714     -1.047964332     -0.139715589
  9      H        -0.376416457      0.557727803     -0.886436819
 10      H         1.909809076     -0.580806246     -0.874539065
 11      H         1.543047689      2.163542607     -0.556916956
 12      H         3.151780327      1.426136272     -0.506080660
```

```
13         H       3.369935929     2.019320261     1.752364904
14         H       1.759159151     2.755825587     1.731342551
15         H       2.315436147     0.489079482     3.258974825
16         H       1.537523754    -1.532154617     1.640373292
17         H       3.133135146    -0.853919094     1.303581831
18         C       0.254417253     0.689788739     2.518310659
19         O      -0.220223947     0.848649105     3.635842186
20         H      -1.544428127    -0.032656066     1.528090790
21         H      -1.035612950     1.606509350     1.056645710
---
H12C8O1, RHF, CHARGE=0, MULT=1
HF=-52.9
1    C      0.000000000     0.000000000     0.000000000
2    C      1.547816599     0.000000000     0.000000000
3    C      2.154848609     1.435591806     0.000000000
4    C      2.296757008     1.823665008     1.493169715
5    C      1.764367265     0.596719297     2.293356998
6    C     -0.625079705     0.562347125     1.292995689
7    C      2.061568882    -0.592957722     1.340229021
8    O     -1.803406138     0.901012837     1.316234290
9    H     -0.366406162    -1.044412863    -0.130873688
10        H      -0.376429438     0.568837275    -0.878992368
11        H       1.907258790    -0.579392787    -0.876026165
12        H       1.530580297     2.163999840    -0.556077035
13        H       3.143218004     1.434695433    -0.506660085
14        H       3.358118011     2.021062184     1.754008017
15        H       1.748235526     2.758169365     1.728343004
16        H       2.297654843     0.496933486     3.261837042
17        H       1.561102446    -1.537082899     1.633285248
18        H       3.144751562    -0.832164047     1.300076087
19        C       0.242127181     0.667220655     2.564009964
20        H      -0.000036044     1.606570871     3.108936584
21        H      -0.047795663    -0.165728466     3.245449987
---
H12C8O1, RHF, CHARGE=0, MULT=1
HF=-46.2
1    C      0.000000000     0.000000000     0.000000000
2    C      1.549787447     0.000000000     0.000000000
3    C      2.185283363     1.420515412     0.000000000
4    C      2.298843926     1.837712487     1.490438054
5    C      1.724975387     0.649707597     2.315413615
6    C     -0.646837590     0.682647010     1.229094059
7    C      1.999715702    -0.540529139     1.373113057
8    H     -0.351762415    -1.055183868    -0.061798839
9    H     -0.370467929     0.493008130    -0.926220532
10        H       1.919346932    -0.599652193    -0.856743245
11        H       1.590348855     2.150956270    -0.584454141
12        H       3.187233392     1.390503508    -0.478583087
13        H       3.358665723     2.018814856     1.769630650
14        H       1.765093311     2.788714980     1.690128565
15        H       2.233135269     0.566353035     3.297633622
16        O       2.471219494    -1.628866581     1.643024309
17        H      -0.065707983     1.625678001     3.107941151
18        C       0.191171581     0.710162071     2.529969172
19        H      -0.107421541    -0.145756007     3.177489460
20        H      -1.614320905     0.175129074     1.444735162
21        H      -0.915462251     1.728879367     0.955761296
---
H12C8O1, RHF, CHARGE=0, MULT=1
```

```
HF=-55
1    C     0.000000000    0.000000000    0.000000000
2    C     1.562439661    0.000000000    0.000000000
3    C    -0.473647117    1.476627323    0.000000000
4    C     0.767244254    2.388064059   -0.037268781
5    C     1.996844479    1.473846927   -0.053427857
6    O     3.152498268    1.862120028   -0.126291394
7    C     2.035745028   -0.799390016    1.236883421
8    C    -0.461969139   -0.879855182    1.187354125
9    C     0.793021652   -1.443636508    1.882444983
10   H    -0.363092236   -0.473836026   -0.941696003
11   H     1.938296846   -0.501967852   -0.921584755
12   H    -1.099446743    1.717706771    0.884855639
13   H    -1.119821552    1.664466993   -0.884834580
14   H     0.799412332    3.059726039    0.847215163
15   H     0.758005105    3.042853166   -0.934695121
16   H     2.559981829   -0.154735846    1.974040830
17   H     2.771140331   -1.574629253    0.933227228
18   H    -1.108825711   -1.705176382    0.818752124
19   H    -1.083257503   -0.314610633    1.913757102
20   H     0.830649243   -2.549825830    1.781427895
21   H     0.765446215   -1.237007986    2.973633951
---
H12C8O1, RHF, CHARGE=0, MULT=1
HF=-49.4
1    C     0.000000000    0.000000000    0.000000000
2    C     1.563356773    0.000000000    0.000000000
3    C     1.879938592    1.494853088    0.000000000
4    C     0.742270889    2.172267343    0.803779930
5    C    -0.366610499    1.099142420    1.013987683
6    C    -0.346383216   -1.498694516    0.030694689
7    C     0.788659784   -2.149758733   -0.816419193
8    C     1.920239582   -1.095839126   -1.016557629
9    O     2.836572915    2.056415736   -0.504225982
10   H    -0.381370008    0.371983173   -0.988227426
11   H     1.961272335   -0.354965717    0.988351206
12   H     0.345162925    3.051285278    0.255177041
13   H     1.130394469    2.544673799    1.774365938
14   H    -1.375469864    1.517120811    0.821050839
15   H    -0.375268283    0.730731277    2.060466772
16   H    -1.346075572   -1.698782871   -0.405649900
17   H    -0.366545333   -1.911586377    1.060363380
18   H     0.398065887   -2.489358413   -1.798209747
19   H     1.182175533   -3.053961488   -0.307061558
20   H     1.939920946   -0.719964831   -2.060058998
21   H     2.921292427   -1.530076198   -0.819809567
---
H14C8O1, RHF, CHARGE=0, MULT=1
HF=-53.2
1    C     0.000000000    0.000000000    0.000000000
2    C     1.550389507    0.000000000    0.000000000
3    C     2.146483586    1.422097446    0.000000000
4    C     1.057800665    2.525472106    0.027768753
5    C    -0.539466581    0.887287907   -1.151992739
6    C     0.051551230    2.313978215   -1.133204429
7    C     0.364670452    2.646800452    1.423670350
8    O    -0.687924567    1.756349716    1.666303251
9    C    -0.596080429    0.386174078    1.393725708
10   H    -0.333894654   -1.048458650   -0.192647912
```

```
11      H        1.934571448     -0.573153660     0.872825713
12      H        1.913348619     -0.555926462    -0.894033842
13      H        2.838178342      1.543812645     0.862892595
14      H        2.781774993      1.560953056    -0.903899369
15      H        1.571361373      3.503382698    -0.140943418
16      H       -1.650104292      0.934531605    -1.110244510
17      H       -0.304456287      0.401055099    -2.125839661
18      H        0.564903645      2.517472634    -2.100261998
19      H       -0.766961308      3.064575456    -1.072353792
20      H       -0.081344028      3.671707770     1.519062930
21      H        1.136162258      2.569862871     2.235886795
22      H       -1.645471855     -0.004879437     1.459912364
23      H       -0.021099012     -0.135336507     2.205190526
---
H14C8O1, RHF, CHARGE=0, MULT=1
HF=-60.1
1       C        0.000000000      0.000000000     0.000000000
2       C        1.508743967      0.000000000     0.000000000
3       C        2.137956164      1.375661069     0.000000000
4       C        2.287815406     -1.109388600     0.001068811
5       C        1.847666829     -2.551023716     0.001994034
6       C        3.013163528     -3.560524447     0.010924577
7       C        2.643638579     -5.053024012     0.010126059
8       C        3.814459614     -6.036903665    -0.009198947
9       O        1.485362919     -5.454658599     0.024881107
10      H       -0.391526723      0.506190314     0.908040367
11      H       -0.393305251      0.540606889    -0.887045807
12      H       -0.444165630     -1.014435784    -0.019542805
13      H        1.816681387      1.958337071     0.888920750
14      H        1.843282979      1.945076435    -0.906611515
15      H        3.246451099      1.343145183     0.016997709
16      H        3.376386867     -0.975597259     0.000930628
17      H        1.205133363     -2.733738989     0.895105758
18      H        1.214957856     -2.737958143    -0.897262073
19      H        3.658633386     -3.379963977    -0.879722371
20      H        3.645882595     -3.379388300     0.910672885
21      H        4.419060211     -5.897739135    -0.928092148
22      H        4.471389021     -5.873813297     0.868837140
23      H        3.476447875     -7.091896803     0.014700717
---
H14C8O1, RHF, CHARGE=0, MULT=1
HF=6.4
1       C        0.000000000      0.000000000     0.000000000
2       C        1.568677874      0.000000000     0.000000000
3       C        2.019484628      1.544472650     0.000000000
4       C        1.031659505      2.741208617     0.425377402
5       C       -0.423270890      2.162716558     0.333884428
6       C       -1.948678583      2.115392556     0.694814758
7       C       -2.435399148      0.609217056     0.407728419
8       C       -1.412478479     -0.624951690     0.272722690
9       O       -0.111709380      0.970956527     1.006892974
10      H       -0.165139816      0.314862147    -1.083832812
11      H        1.970978827     -0.500242555    -0.903599597
12      H        1.998422582     -0.524189349     0.876009970
13      H        2.370873045      1.773107140    -1.029888921
14      H        2.907638223      1.615596435     0.662538041
15      H        1.169654550      3.595372018    -0.267440863
16      H        1.286294861      3.105857590     1.439834137
17      H       -0.521193229      2.139256488    -0.802258979
```

```
18    H    -2.532709577    2.816559248    0.065527672
19    H    -2.149101971    2.387304822    1.749758911
20    H    -3.024567739    0.639233481   -0.534795594
21    H    -3.146211207    0.345137703    1.218792963
22    H    -1.732365633   -1.276725710   -0.564907184
23    H    -1.437737623   -1.245335208    1.190063197
---
H14C8O1, RHF, CHARGE=0, MULT=1
HF=-68
1     C     0.000000000    0.000000000    0.000000000
2     C     1.556193256    0.000000000    0.000000000
3     C     2.079268282    1.455913281    0.000000000
4     C     0.878324573    2.443502147    0.007271054
5     C     0.007269161    2.220060180   -1.259825211
6     C    -0.484056878    0.730055252   -1.315139488
7     C    -0.499348814    0.735970274    1.274527873
8     C     0.016474016    2.193838786    1.276533257
9     O    -1.845540891    0.579298775   -1.586208880
10    H    -0.833104041    2.910042918    1.314881546
11    H     0.612203839    2.395572650    2.193216696
12    H    -0.369159304   -1.048486827    0.007029996
13    H     1.946303603   -0.555738872   -0.880168535
14    H     1.938945908   -0.550853952    0.886776201
15    H     2.724109890    1.638955020   -0.886919996
16    H     2.731773678    1.636339088    0.881771835
17    H     1.257004488    3.489249028    0.014234974
18    H    -0.837575359    2.942266834   -1.269308514
19    H     0.595578733    2.448401702   -2.175069355
20    H     0.021070648    0.229705084   -2.187907477
21    H    -1.608650139    0.713288719    1.338781288
22    H    -0.149146023    0.199816077    2.183970785
23    H    -2.402157357    1.031413545   -0.968743966
---
H14C8O1, RHF, CHARGE=0, MULT=1
HF=-39.5
1     C    -0.000664111    0.000087106    0.000711607
2     C     1.526850112    0.000024517   -0.000726798
3     C     2.126354481    1.420191341    0.000995643
4     C     2.684127648    1.969887842    1.325752649
5     C     1.909932628    1.784572831    2.643476788
6     C     0.392079764    2.050835818    2.665985031
7     C    -0.522087376    0.812427986    2.571610681
8     C    -0.918569113    0.382422439    1.159867560
9     H    -1.915011023    0.787807071    0.885330040
10    O    -0.707168414   -0.964445802    0.762503979
11    H     1.940575172   -0.608674143    0.833218313
12    H     1.857806902   -0.522328658   -0.929231700
13    H     1.383693871    2.144924608   -0.404435413
14    H     2.969412617    1.432334596   -0.731130594
15    H     2.859405165    3.062410062    1.174391372
16    H     3.696108922    1.525484045    1.485011520
17    H     2.377550835    2.483791799    3.379392308
18    H     2.121222754    0.766888602    3.044070571
19    H     0.111849365    2.795143837    1.888439276
20    H     0.160978524    2.549214975    3.638406383
21    H    -0.074541653   -0.037739877    3.134060800
22    H    -1.469819867    1.049676050    3.112319291
23    H    -0.400894967    0.149924365   -1.024748253
---
```

```
H14C8O1, RHF, CHARGE=0, MULT=1
HF=-65.1
1    C    0.000000000    0.000000000    0.000000000
2    C    1.538899581    0.000000000    0.000000000
3    C   -0.645415070    1.392549487    0.000000000
4    C    2.251788452    0.342007112    1.321188553
5    C   -1.086806709    1.997222732    1.339874204
6    C    2.301409708    1.820194033    1.753826268
7    C   -0.070379361    2.957373082    1.982503292
8    C    1.166447567    2.337765985    2.658587206
9    O   -0.831778400    1.996600624   -1.051609000
10   H   -0.343719266   -0.551962845   -0.906558096
11   H   -0.381119086   -0.591200696    0.862222414
12   H    1.915308251    0.661243775   -0.813048705
13   H    1.860018985   -1.031206609   -0.284927065
14   H    3.309932773    0.002269401    1.205999160
15   H    1.834102900   -0.283677064    2.142309074
16   H   -2.037918156    2.558400716    1.181360202
17   H   -1.345684220    1.182970731    2.052781936
18   H    3.251996691    1.953759219    2.325697247
19   H    2.408182627    2.476433759    0.861180579
20   H    0.250283601    3.720128077    1.237284824
21   H   -0.620282078    3.526824541    2.770559882
22   H    1.605933869    3.136029171    3.305125281
23   H    0.843796858    1.538311244    3.363530943
---
H16C8O1, RHF, CHARGE=0, MULT=1
HF=-80.9
1    O   -0.045359913    0.010316510   -0.100019564
2    C    1.105034436    0.050054071   -0.521934151
3    C    1.730469191    1.410577491   -0.985213198
4    C    2.093437525    1.330481972   -2.495140858
5    C    2.993967152    1.729204991   -0.136831030
6    C    0.720218907    2.585810534   -0.796237718
7    C    1.925212720   -1.267214166   -0.591130672
8    C    2.160631789   -1.867357723    0.811838234
9    C    1.293508933   -2.286130002   -1.564236904
10   H    2.895870752    0.594626277   -2.701773561
11   H    1.214379225    1.045696390   -3.108335888
12   H    2.456380012    2.308634186   -2.872553325
13   H    3.818833140    1.011068167   -0.315055136
14   H    2.767719432    1.715224730    0.948741862
15   H    3.394636505    2.735665150   -0.377255362
16   H   -0.203059350    2.439339948   -1.391773151
17   H    0.423623732    2.710690471    0.264497127
18   H    1.165853770    3.548479459   -1.123311608
19   H    2.938987130   -1.039865921   -1.002188934
20   H    2.617745097   -1.123235424    1.495246304
21   H    1.225083627   -2.228366592    1.283085623
22   H    2.859674584   -2.726997984    0.751394720
23   H    0.310745809   -2.658936682   -1.214099508
24   H    1.147119871   -1.842292174   -2.569719858
25   H    1.960350733   -3.164235453   -1.689521933
---
H16C8O1, RHF, CHARGE=0, MULT=1
HF=-82.5
1    O   -0.000003967    0.000033699   -0.000012711
2    C    1.226003951   -0.000002247    0.000006469
3    C    2.009435940    1.313628273   -0.000004874
```

```
4    C    2.064204610   -1.289190314   -0.000612771
5    C    1.294948953   -2.624124502    0.019190455
6    C    2.212868864   -3.865027117    0.020375462
7    C    1.451758180   -5.206200920    0.039839561
8    C    2.365197089   -6.447410338    0.040982516
9    C    1.623271164   -7.786772029    0.060449539
10   H    2.645878982    1.382970730    0.905162484
11   H    2.664598334    1.372880373   -0.892438551
12   H    1.343761460    2.199411221   -0.012201020
13   H    2.716750750   -1.255833484   -0.903745993
14   H    2.739651743   -1.243007277    0.884877642
15   H    0.640804288   -2.655720688    0.920058828
16   H    0.621658507   -2.671559625   -0.866668007
17   H    2.870755931   -3.832760156   -0.878314339
18   H    2.889387059   -3.817436051    0.904569761
19   H    0.794033385   -5.239138796    0.938647598
20   H    0.775585697   -5.254480341   -0.844492575
21   H    3.025076117   -6.423902388   -0.857080364
22   H    3.043998590   -6.408391492    0.924249176
23   H    0.990113420   -7.894948749    0.964291498
24   H    0.971319909   -7.910601908   -0.827893403
25   H    2.349398650   -8.625858199    0.060212689
---
H16C8O1, RHF, CHARGE=0, MULT=1
HF=-78.5
1    C    0.090152439   -0.182976337   -0.282675705
2    C    1.334475362    0.109915005    0.560030714
3    C    2.126878277    1.461054428    0.417646767
4    C    3.660067056    1.138654590    0.164999436
5    C    3.938269495    0.306067566   -1.108463232
6    C    1.550232302    2.366838285   -0.706262340
7    C    1.922593424    2.209769214    1.773968801
8    C    4.611412483    2.357744516    0.202215419
9    O    1.690471273   -0.724644132    1.387736948
10   H   -0.709312478    0.555252912   -0.071596411
11   H    0.322347989   -0.160402207   -1.365452658
12   H   -0.325410578   -1.187365490   -0.060455782
13   H    3.990846775    0.497826206    1.023680533
14   H    3.288228173   -0.587770182   -1.183018765
15   H    3.806184032    0.891631096   -2.040544304
16   H    4.984004753   -0.066979516   -1.100249292
17   H    2.101346916    3.326113259   -0.782578152
18   H    1.594288912    1.891134555   -1.705841509
19   H    0.490546549    2.636036074   -0.515698754
20   H    2.469663336    1.728259073    2.608532136
21   H    2.262138338    3.263845414    1.722519604
22   H    0.848996705    2.237617920    2.055692015
23   H    4.426597499    3.079971846   -0.617233842
24   H    4.546160131    2.910379387    1.160267025
25   H    5.665943051    2.022069304    0.108348655
---
H16C8O1, RHF, CHARGE=0, MULT=1
HF=-80.9
1    C   -0.11637       -0.15300        0.26899
2    C    1.38267        0.07001        0.05012
3    C    2.12508        0.67574        1.25713
4    C    3.63387        0.89339        1.02155
5    C    4.37070        1.50475        2.23027
6    C    5.88063        1.68278        1.99776
```

```
7    C      6.83751         0.60482         2.52997
8    C      7.47060         0.94518         3.88180
9    O      6.31251         2.66576         1.40434
10   H     -0.31168        -0.85196         1.10738
11   H     -0.64739         0.79569         0.48685
12   H     -0.57459        -0.58966        -0.64233
13   H      1.51429         0.73176        -0.83698
14   H      1.84786        -0.90622        -0.22005
15   H      1.65342         1.64973         1.52213
16   H      1.98597         0.01016         2.13994
17   H      4.10683        -0.08141         0.76177
18   H      3.77392         1.55590         0.13727
19   H      3.92786         2.49662         2.47853
20   H      4.21206         0.86989         3.13047
21   H      6.29825        -0.36541         2.61448
22   H      7.64582         0.43371         1.78189
23   H      8.09355         1.86020         3.82677
24   H      6.70438         1.10033         4.66812
25   H      8.12528         0.11294         4.21288
---
H16C8O1, RHF, CHARGE=0, MULT=1
HF=-83.5
1    O   0.00000         0.00000         0.00000
2    C   1.22677         0.00000         0.00000
3    C   2.01231         1.32223         0.00000
4    C   2.17002         1.94164        -1.40168
5    C   2.97867         3.24187        -1.43325
6    C   2.01260        -1.32229        -0.00065
7    C   2.16685        -1.94298         1.40224
8    C   2.98594        -3.24843         1.42050
9    C   3.14427        -3.88236         2.80539
10   H   1.49659         2.04512         0.67317
11   H   3.01547         1.15315         0.45133
12   H   2.65954         1.20534        -2.08029
13   H   1.16307         2.14148        -1.83466
14   H   2.51061         4.03371        -0.81433
15   H   4.01491         3.09229        -1.06799
16   H   3.04339         3.62459        -2.47258
17   H   1.49780        -2.04338        -0.67637
18   H   3.01614        -1.15151        -0.45036
19   H   2.65065        -1.20114         2.07840
20   H   1.15616        -2.14144         1.82621
21   H   2.50962        -3.99473         0.74357
22   H   4.00145        -3.05553         1.00340
23   H   3.66486        -3.20487         3.51218
24   H   2.16512        -4.15120         3.25066
25   H   3.74433        -4.81281         2.73256
---
H16C8O1, RHF, CHARGE=0, MULT=1
HF=-69.8
1    C     -0.000051299    -0.000019530    -0.000015001
2    C      1.531200289     0.000181644     0.000170723
3    C      2.173466581     1.400988850    -0.000025104
4    C      3.715693191     1.386497330     0.000137494
5    C      4.355699174     2.789964031    -0.000601669
6    C      5.898071639     2.775193966     0.000229504
7    C      6.535590720     4.176942728    -0.000728986
8    C      8.063974481     4.171307770    -0.001589936
9    O      8.739823486     5.189019775    -0.013650497
```

```
10      H     -0.412730290     0.506311010     0.896049326
11      H     -0.412966776     0.505708544    -0.896205601
12      H     -0.382296503    -1.041676494     0.000448894
13      H      1.882676105    -0.569310292     0.891685318
14      H      1.883528440    -0.570484657    -0.890330470
15      H      1.813838099     1.964491368    -0.891487718
16      H      1.814147278     1.964363799     0.891699917
17      H      4.074391100     0.823498898     0.892204808
18      H      4.074594258     0.822229327    -0.891042604
19      H      3.998345995     3.353197519    -0.893003819
20      H      3.997456105     3.354650279     0.890542368
21      H      6.258134937     2.214458589     0.893490497
22      H      6.259140037     2.212220153    -0.891243883
23      H      6.188091894     4.746429652    -0.893449420
24      H      6.190899310     4.746431624     0.893198968
25      H      8.548053829     3.171864119     0.010577924
---
H18C8O1, RHF, CHARGE=0, MULT=1
HF=-85.3
1       O     -0.000032497    -0.000069610    -0.000171493
2       C      1.396881148     0.000265309     0.000085638
3       C      1.916583740     1.461014237    -0.000008298
4       C      3.455559464     1.558983471    -0.001533132
5       C      3.989633654     3.006122877    -0.001211151
6       C      5.528926671     3.106464279    -0.002582457
7       C      6.063427710     4.553192726    -0.003100490
8       C      7.601056822     4.655751332    -0.004247746
9       C      8.146736922     6.086440572    -0.005114529
10      H     -0.338458113    -0.884102710    -0.000107077
11      H      1.792449457    -0.549141657     0.897362067
12      H      1.793164135    -0.549507280    -0.896645354
13      H      1.512176550     1.991759948    -0.891759322
14      H      1.514159712     1.990916200     0.893132425
15      H      3.856510764     1.023057335     0.889334044
16      H      3.854859113     1.024017413    -0.893799534
17      H      3.588934413     3.541872037    -0.892154082
18      H      3.590571082     3.540807821     0.891107111
19      H      5.929505498     2.570977277     0.888619654
20      H      5.927913770     2.570830430    -0.894456931
21      H      5.663778636     5.089583992    -0.894274356
22      H      5.665368232     5.089588385     0.888777925
23      H      8.009325473     4.124361959     0.886441765
24      H      8.008130025     4.123615487    -0.895045590
25      H      7.820318721     6.652641284    -0.900978960
26      H      7.822432203     6.652733431     0.891375427
27      H      9.256310534     6.072253952    -0.006463333
---
H18C8O1, RHF, CHARGE=0, MULT=1
HF=-86.3
1       C     -0.000029594     0.000019365     0.000008603
2       C      1.531261111     0.000023342     0.000159705
3       C      2.171554638     1.401523417     0.000015768
4       C      3.724827721     1.364711629    -0.000205839
5       O      4.195738093     2.683841939     0.003086861
6       C      5.563797123     3.042813752    -0.032480663
7       C      6.350454462     2.493916531     1.202479987
8       C      6.254701804     2.581362156    -1.357478620
9       C      5.543768749     4.610819023     0.021851659
10      H     -0.412702853     0.505709398     0.896337449
```

```
11      H       -0.412824382     0.505962824    -0.896137609
12      H       -0.381982916    -1.041866860    -0.000120953
13      H        1.882212886    -0.570724672     0.890967949
14      H        1.882383490    -0.570843575    -0.890698559
15      H        1.817307823     1.965889178    -0.892406236
16      H        1.817459709     1.965855271     0.892466305
17      H        4.084626178     0.795368683     0.897312815
18      H        4.084060910     0.795627766    -0.897831934
19      H        5.840940020     2.736843577     2.156563439
20      H        6.481326159     1.393731903     1.164402308
21      H        7.366129587     2.937126530     1.245783355
22      H        6.388986271     1.481902778    -1.402574619
23      H        5.673528478     2.883628304    -2.251782556
24      H        7.263634725     3.032577734    -1.448478514
25      H        4.995733514     5.044599864    -0.838513465
26      H        5.061127892     4.983662065     0.947552754
27      H        6.576021035     5.015732875    -0.002161868
---
H18C8O1, RHF, CHARGE=0, MULT=1
HF=-90.8
1       C       -0.000046728     0.000078118    -0.000028282
2       C        1.529699535    -0.000079069    -0.000067026
3       C        2.249546531     1.391631186     0.000079429
4       O        2.042932815     2.012518904     1.245963122
5       C        1.687930211     3.365004632     1.462682827
6       C        2.753409092     4.369038679     0.914079508
7       C        0.283875311     3.718486570     0.871432066
8       C        1.630331600     3.484454438     3.027393614
9       C        3.771701515     1.213325968    -0.274088346
10      H       -0.417864449     0.632099914    -0.809704346
11      H       -0.420519164     0.350605916     0.963018604
12      H       -0.374173801    -1.032804720    -0.163142316
13      H        1.863253397    -0.545213354    -0.914603737
14      H        1.882429713    -0.605171429     0.866054796
15      H        1.834090132     1.989790759    -0.855906710
16      H        3.769678354     4.143394857     1.294310694
17      H        2.799744121     4.372196747    -0.193370463
18      H        2.509085523     5.404681086     1.227679423
19      H        0.268660622     3.682877217    -0.236107530
20      H       -0.505553916     3.033730876     1.239509907
21      H       -0.012242868     4.746746076     1.163920204
22      H        0.880618015     2.795475006     3.465431445
23      H        2.609011527     3.252630431     3.493361272
24      H        1.352645111     4.514743713     3.330014473
25      H        4.287954606     2.187412982    -0.381668055
26      H        4.281665313     0.650178438     0.532321026
27      H        3.927518490     0.663810076    -1.225062029
---
H18C8O1, RHF, CHARGE=0, MULT=1
HF=-79.8
1       C        0.000056231    -0.000006491     0.000010172
2       C        1.531365422     0.000240477     0.000095391
3       C        2.171337930     1.401710211    -0.000023186
4       C        3.722272645     1.365478754    -0.001161850
5       O        4.207324150     2.681189930    -0.000794621
6       C        5.592512862     2.899389843     0.000463478
7       C        5.860942010     4.427367787    -0.007286598
8       C        7.360731331     4.780215829    -0.003939857
9       C        7.661420602     6.281736145    -0.012077627
```

```
10      H       -0.412617000     0.506582617     0.895855681
11      H       -0.412749370     0.505016819    -0.896625113
12      H       -0.381756785    -1.041912549     0.000842542
13      H        1.882475583    -0.569990276     0.891399006
14      H        1.882894705    -0.570845100    -0.890512052
15      H        1.817014366     1.966793878    -0.891947966
16      H        1.818186137     1.966021717     0.892915544
17      H        4.091655894     0.800580637     0.896484404
18      H        4.090250753     0.801922034    -0.900284727
19      H        6.071677965     2.431999671     0.902271792
20      H        6.075488072     2.422843404    -0.894684236
21      H        5.379797004     4.879785349    -0.904148297
22      H        5.373316833     4.890357964     0.880686548
23      H        7.848188208     4.329308801     0.891428898
24      H        7.854627387     4.318371930    -0.890206462
25      H        7.249991611     6.780172850    -0.913011701
26      H        7.243115748     6.791317610     0.879411857
27      H        8.758024704     6.451546376    -0.008840824
---
H18C8O1, RHF, CHARGE=0, MULT=1
HF=-86.3
1       C       -0.103198964    -0.010607835     0.003120797
2       C        1.328436067     0.527721327     0.084768028
3       C        2.441172902    -0.427743595    -0.459270034
4       C        2.433658770    -0.581382491    -2.003958205
5       O        3.716279749    -0.146307528     0.069499848
6       C        4.447749896     1.035984100    -0.168205795
7       C        4.742106588     1.716532844     1.198610042
8       C        5.752832366     0.682576117    -0.959940350
9       C        6.388494996     1.847402826    -1.724723606
10      H       -0.456064130    -0.113538335    -1.042656960
11      H       -0.205617233    -0.998344102     0.496101455
12      H       -0.796617551     0.689079655     0.514769151
13      H        1.364615607     1.504576593    -0.449483422
14      H        1.535619063     0.752763157     1.156786274
15      H        2.223185088    -1.454725719    -0.041381183
16      H        2.538627651     0.384694809    -2.536314834
17      H        3.250127465    -1.249417347    -2.344815968
18      H        1.482530718    -1.042919583    -2.338173025
19      H        3.863901747     1.774790021    -0.782258210
20      H        5.321256749     1.061700926     1.879476363
21      H        5.320896658     2.649992295     1.049624331
22      H        3.804760287     1.997543012     1.719351359
23      H        6.508973159     0.255652227    -0.262303877
24      H        5.540748839    -0.124974109    -1.697571295
25      H        5.696163626     2.273635973    -2.479158240
26      H        6.707753051     2.670796898    -1.055069615
27      H        7.291235540     1.495320441    -2.265704778
---
H18C8O1, RHF, CHARGE=0, MULT=1
HF=-86.3
1       C        0.035026130     0.035981017    -0.076146021
2       C       -1.363440355    -0.635529630    -0.287826172
3       O       -1.895191322    -1.100779801     0.934894870
4       C       -2.588671773    -0.470915358     1.991710575
5       C       -4.112689493    -0.333795561     1.659630675
6       C       -2.421214965    -1.492091434     3.178757189
7       C       -2.020615595     0.908598113     2.448611851
8       C       -2.288502079     0.314733531    -1.109982068
```

```
9    C    -1.163958871   -1.963737064   -1.109914680
10   H    -0.042781870    1.067643769    0.319060893
11   H     0.670603794   -0.543441260    0.622865764
12   H     0.582812587    0.107368294   -1.038383086
13   H    -4.324695104    0.515291124    0.980521341
14   H    -4.517251864   -1.250817572    1.186642895
15   H    -4.698711335   -0.152982076    2.584160885
16   H    -2.841617861   -2.486842706    2.929500185
17   H    -1.355286869   -1.637317746    3.446289596
18   H    -2.944409153   -1.128629316    4.086804224
19   H    -0.948571818    0.848125646    2.722787329
20   H    -2.129516378    1.697761974    1.679584499
21   H    -2.565048207    1.267632895    3.346789552
22   H    -3.282522402   -0.135513066   -1.303046896
23   H    -2.449618173    1.293709551   -0.618356848
24   H    -1.832931140    0.530316588   -2.098951480
25   H    -0.481153976   -2.668523507   -0.594651105
26   H    -2.124774959   -2.490956109   -1.276207222
27   H    -0.727528880   -1.746492154   -2.106205784
---
H18C8O1, RHF, CHARGE=0, MULT=1
HF=-88
1    H     0.000002313    0.000195919   -0.000883215
2    C     1.108885785   -0.000043793    0.000156785
3    C     1.695678617    1.427418542   -0.000034281
4    C     1.298156918    2.184344892   -1.311164792
5    O     2.085022858    3.335015817   -1.440742055
6    C     1.897532755    4.322214183   -2.437125351
7    C     1.944334659    3.731849689   -3.884198276
8    C     0.562714052    5.108785723   -2.221626499
9    C     3.111325049    5.296259504   -2.237560667
10   C     1.331601163    2.193671823    1.287233455
11   H     1.445559531   -0.574744305   -0.887012138
12   H     1.443174679   -0.564133000    0.894915599
13   H     2.809350993    1.308804010   -0.001955525
14   H     1.455934126    1.497227622   -2.185535999
15   H     0.201289801    2.415972849   -1.294082552
16   H     2.846201707    3.108172301   -4.046650594
17   H     1.058756095    3.105515790   -4.113332623
18   H     1.964862861    4.546749502   -4.635931338
19   H    -0.331121873    4.492084002   -2.444141977
20   H     0.464023279    5.473944079   -1.179732343
21   H     0.518772061    5.992760237   -2.889725033
22   H     3.116869623    5.743855828   -1.223534555
23   H     4.080880782    4.778027618   -2.379243020
24   H     3.068048430    6.128794232   -2.969062791
25   H     1.862500721    3.164461418    1.349554035
26   H     0.243933605    2.398520117    1.358699773
27   H     1.620801627    1.611481606    2.186982499
---
H10C9O1, RHF, CHARGE=0, MULT=1
HF=-15.1
1    C     0.020836202   -0.028674275   -0.039446154
2    C     1.425068677   -0.053231021   -0.024735922
3    C     2.146010094    1.152135483   -0.045083574
4    C     1.460487257    2.377629669   -0.078367105
5    C     0.048245428    2.416540783   -0.092279153
6    C    -0.679877941    1.196827281   -0.075743888
7    C    -0.689810262    3.733612216   -0.144006272
```

```
 8    C    -2.207724932     3.608566164     0.144595193
 9    O    -2.781753134     2.442977947    -0.377345254
10    C    -2.202188108     1.202330333    -0.079345896
11    H    -0.524139424    -0.974301291    -0.023124495
12    H     1.953077462    -1.006992381     0.001943231
13    H     3.236373010     1.137820525    -0.035473014
14    H     2.035955664     3.304925569    -0.093779491
15    H    -0.540480258     4.192345574    -1.149172601
16    H    -0.256326140     4.448104503     0.592157039
17    H    -2.754279912     4.466902454    -0.322834326
18    H    -2.387695440     3.681466998     1.250825215
19    H    -2.590767587     0.487289245    -0.850822053
20    H    -2.569422993     0.834721484     0.918294075
---
H10C9O1, RHF, CHARGE=0, MULT=1
HF=-19.7
 1    C     0.022648541    -0.026959204     0.037904283
 2    C     1.423861780    -0.051737598     0.052045658
 3    C     2.157373462     1.146394484     0.129179985
 4    C     1.488848253     2.378510225     0.192471279
 5    C     0.078160905     2.446506120     0.178475797
 6    C    -0.651421540     1.223478431     0.100556523
 7    C    -0.646710266     3.767315245     0.260299487
 8    C    -2.133748381     3.659465791    -0.116880101
 9    C    -2.778555813     2.328670554     0.332387608
10    O    -2.011772743     1.182379713     0.052452965
11    H    -0.536929683    -0.960723791    -0.021555400
12    H     1.948249466    -1.007179046     0.003657224
13    H     3.247077478     1.117476026     0.139981671
14    H     2.077394051     3.296093950     0.252591682
15    H    -0.162448122     4.516878903    -0.405762926
16    H    -0.552784680     4.170307847     1.296630809
17    H    -2.250610128     3.777696704    -1.217863472
18    H    -2.699796256     4.499400281     0.343266890
19    H    -3.758295351     2.183947614    -0.190817964
20    H    -3.003772424     2.365898234     1.431574006
---
H10C9O1, RHF, CHARGE=0, MULT=1
HF=-22.6
 1    O     0.000020713     0.000090253    -0.000109907
 2    C     1.226593914    -0.000009058    -0.000037272
 3    C     2.018109895     1.322310989    -0.000392993
 4    C     2.140032862     1.945034105    -1.373616609
 5    C     3.286522852     1.726272102    -2.174154292
 6    C     3.408531604     2.318976118    -3.442676358
 7    C     2.386312654     3.146371117    -3.937554441
 8    C     1.243327458     3.378015155    -3.153858193
 9    C     1.121646874     2.785111526    -1.884887878
10    C     2.016196538    -1.306629575     0.026759921
11    H     1.529533099     2.038405265     0.700599347
12    H     3.028971195     1.145690231     0.431859789
13    H     4.099310899     1.093051885    -1.814415856
14    H     4.300564325     2.135441902    -4.042732175
15    H     2.479927746     3.605739062    -4.921882806
16    H     0.445719706     4.019930174    -3.529694141
17    H     0.223343402     2.984814202    -1.298028851
18    H     1.356724373    -2.196181701    -0.011310090
19    H     2.705872141    -1.362418942    -0.838935311
20    H     2.616519312    -1.370918165     0.957039977
```

```
---
H12C9O1, RHF, CHARGE=0, MULT=1
HF=-41.9
1    C     0.016604911   -0.028278892    0.030956580
2    C     1.346177102   -0.082257335   -0.406128459
3    C     1.950453300    1.066897167   -0.937282744
4    C     1.221985515    2.264020572   -1.029874892
5    C    -0.123438007    2.370452602   -0.602396785
6    C    -0.721324428    1.184947615   -0.061788136
7    O    -2.011578946    1.244993809    0.363027881
8    C    -1.064936260    4.125778576   -2.194435179
9    C    -0.240997234    4.804214143    0.149834156
10   H     2.985115551    1.033617572   -1.278708053
11   H    -0.440119292   -0.929878845    0.442150070
12   H     1.906703478   -1.015063019   -0.331587446
13   C    -0.871147845    3.696681256   -0.723698206
14   H     1.733450028    3.132248654   -1.451224026
15   H    -2.330007256    0.415839110    0.693815778
16   H    -1.543497587    3.316808983   -2.783092181
17   H    -0.112058342    4.386942064   -2.696835088
18   H    -1.725950669    5.014733571   -2.258146571
19   H     0.744111737    5.143894158   -0.227414494
20   H    -0.099478926    4.453203592    1.191838393
21   H    -0.904734717    5.693015534    0.190177826
22   H    -1.909473834    3.573524825   -0.323851469
---
H12C9O1, RHF, CHARGE=0, MULT=1
HF=-42.3
1    C     0.000117912   -0.000079202   -0.000206463
2    C     1.407842334   -0.000727266   -0.000335389
3    C     2.168047596    1.190996798   -0.000014899
4    C     1.475129686    2.416361085    0.000833804
5    C     0.063836902    2.492118377    0.001409654
6    C    -0.672142038    1.267194413    0.000835563
7    O    -2.029450698    1.214760929    0.001056124
8    C     3.671938671    1.134189716   -0.000921934
9    C    -0.749341682   -1.308481237   -0.001857282
10   C    -0.591715601    3.849106804    0.002475548
11   H     4.048948781    0.601572496   -0.899636331
12   H     1.937166975   -0.957453502   -0.000806908
13   H     4.050094806    0.598650617    0.895374048
14   H     2.050448963    3.346023453    0.000977919
15   H    -2.451978106    2.061385977    0.002308148
16   H     4.134881083    2.141941205    0.000468716
17   H    -1.390928430   -1.405379234   -0.902695792
18   H    -1.395219524   -1.405874086    0.895798335
19   H    -0.062957846   -2.180784090   -0.000478909
20   H    -1.227736351    3.995354175   -0.896835178
21   H     0.150612820    4.672932654    0.001925529
22   H    -1.225883183    3.995059100    0.903218706
---
H12C9O1, RHF, CHARGE=0, MULT=1
HF=-41.9
1    C    -0.000266200    0.000088687    0.000084406
2    C     1.402659418    0.000181607    0.000025137
3    C     2.121737215    1.205254704   -0.000241011
4    C     1.455470303    2.456108886   -0.001027180
5    C     0.043443512    2.461925946   -0.000437989
6    C    -0.686553256    1.240908184    0.000090408
```

```
 7     O    -2.042657713    1.337591128    0.000540753
 8     C     2.070370410    4.571381220    1.286951894
 9     C     2.063522143    4.575367385   -1.287232297
10     H     3.211922683    1.158489882   -0.000029512
11     H    -0.537780424   -0.948245911    0.000030119
12     H     1.939972157   -0.949560874    0.000285159
13     C     2.271173375    3.746531192   -0.001939649
14     H    -0.506173235    3.404570705   -0.000222625
15     H    -2.471161490    0.492159140    0.000124328
16     H     2.233406541    3.944355737    2.187084066
17     H     1.055485443    5.009799302    1.361039600
18     H     2.798101571    5.407801888    1.332531760
19     H     1.050096716    5.018546046   -1.351948170
20     H     2.216702938    3.950186847   -2.190368181
21     H     2.794573085    5.408664207   -1.336217077
22     H     3.356369991    3.460786718   -0.005509299
---
H12C9O1, RHF, CHARGE=0, MULT=1
HF=-41.9
 1     C    -0.004323520   -0.004283448   -0.002967183
 2     C     1.394378104    0.001709770    0.090539838
 3     C     2.102065041    1.067075459    0.700446444
 4     C     1.341581642    2.139793070    1.220450678
 5     C    -0.059864255    2.160392605    1.141288495
 6     C    -0.748734605    1.084479682    0.525401449
 7     O    -2.100678581    1.031062384    0.408970143
 8     C     4.138276464    0.954230073    2.231060753
 9     C     4.284872424    2.157330762   -0.040553220
10     H    -0.600267796    3.010877183    1.558234744
11     H    -0.508871200   -0.845068256   -0.480484564
12     H     1.936475889   -0.849948953   -0.326071521
13     C     3.623787229    1.027978687    0.777870870
14     H     1.833297436    2.987715975    1.701625433
15     H    -2.532199316    1.790442304    0.777604837
16     H     3.654105238    0.120814062    2.779711282
17     H     3.949086529    1.886267627    2.799968847
18     H     5.232205109    0.769469774    2.250809159
19     H     4.086255444    3.163550607    0.379315573
20     H     3.919895038    2.150923085   -1.087980406
21     H     5.385836811    2.023490748   -0.074981333
22     H     3.964566058    0.074164513    0.294638702
---
H14C9O1, RHF, CHARGE=0, MULT=1
HF=-55.6
 1     C     0.000034081   -0.000131952    0.000145575
 2     C     1.563200109    0.000085954    0.000045651
 3     C     2.236711059    1.383038280   -0.000102522
 4     C     1.482927338    2.445129818   -0.745069530
 5     C     0.225606428    2.311357870   -1.230715249
 6     C    -0.502922030    0.995238894   -1.093607409
 7     C    -0.479512370   -1.455704316   -0.282116909
 8     C    -0.578103411    0.453041465    1.371249682
 9     O    -1.456120368    0.722955557   -1.814440068
10     C    -0.480054786    3.448780404   -1.926186208
11     H     1.937738128   -0.573101778    0.878052469
12     H     1.923473659   -0.554968997   -0.897585992
13     H     2.392116241    1.733627819    1.047788886
14     H     3.257150648    1.285006012   -0.439139906
15     H     2.035227750    3.383126040   -0.858165928
```

```
16      H     -1.582770829    -1.548560764    -0.230146767
17      H     -0.161826527    -1.807050379    -1.284613443
18      H     -0.059038696    -2.163007175     0.462723632
19      H     -1.687008198     0.428847360     1.369200091
20      H     -0.276936573     1.485755464     1.638093639
21      H     -0.233728107    -0.209021463     2.191786371
22      H     -0.605813226     3.237217173    -3.008507223
23      H     -1.483864737     3.627223543    -1.487908144
24      H      0.083686849     4.401041935    -1.841171883
---
H14C9O1, RHF, CHARGE=0, MULT=1
HF=-57.3
1       C      0.000279550     0.000098839    -0.000060605
2       C      1.551638970    -0.000719676    -0.000323217
3       C      2.206896566     1.394190825     0.000331200
4       C      1.503245732     2.448144947     0.877494748
5       C     -0.046987396     2.393193054     0.860328202
6       C     -0.705919459     2.897218121    -0.452184164
7       C     -1.340359068     1.808045082    -1.340629361
8       C     -0.654923054     0.426466393    -1.341745140
9       C     -0.477594010     0.938276008     1.123663643
10      O     -1.112714346     0.574377942     2.101209810
11      H      1.925607346    -0.577043672    -0.876049003
12      H      1.897969761    -0.565664313     0.896172242
13      H      2.276845472     1.770375232    -1.045774833
14      H      3.260515322     1.290559276     0.347633074
15      H      1.845011735     3.461860310     0.570600875
16      H      1.842654805     2.327656907     1.932621886
17      H     -1.502081166     3.629174054    -0.183534107
18      H      0.029391913     3.474203834    -1.056081881
19      H     -1.419834745    -0.336895859    -1.613894657
20      H      0.094463303     0.395034387    -2.163854031
21      H     -0.322996692    -1.047485889     0.209819220
22      H     -0.402371057     3.054975985     1.686076720
23      H     -2.405303912     1.676585926    -1.036587303
24      H     -1.375343580     2.184517481    -2.388465444
---
H14C9O1, RHF, CHARGE=0, MULT=1
HF=-59.7
1       O     -0.000684632     0.000555485    -0.002606981
2       C      1.221315335     0.000383901    -0.000505025
3       C      2.096815030     1.255753285     0.000956099
4       C      3.573498940     0.782953683     0.005544762
5       C      2.095992256    -1.255373527     0.004698138
6       C      3.573165845    -0.784205430     0.023737823
7       C      4.433957962     1.433085877     1.111725608
8       C      5.810973602     0.774798004     1.301705260
9       C      5.795960833    -0.764292801     1.347904602
10      C      4.411494089    -1.408990712     1.161091671
11      H      1.888564510     1.869389719    -0.902149579
12      H      1.845703364     1.893079099     0.875040191
13      H      4.022435354     1.095999243    -0.970314635
14      H      1.895412070    -1.865399333    -0.902790149
15      H      1.837312572    -1.896306832     0.873901631
16      H      4.039722013    -1.121031542    -0.935581816
17      H      4.585158561     2.508270318     0.862011632
18      H      3.892083918     1.422179664     2.085200551
19      H      6.492270053     1.107413436     0.484657696
20      H      6.260127153     1.166759536     2.242743422
```

```
21     H     6.491117055    -1.157316804     0.571047704
22     H     6.212595571    -1.107276673     2.322481322
23     H     4.551174721    -2.493863587     0.948656340
24     H     3.857655885    -1.358822888     2.126622124
---
H14C9O1, RHF, CHARGE=0, MULT=1
HF=-59.6
1      O    -0.000069730     0.000397550     0.001588917
2      C     1.220701638    -0.000117701    -0.000564240
3      C     2.105997880     1.260797734    -0.000555183
4      C     3.528142004     0.724181442     0.282010009
5      C     2.105218499    -1.261619931    -0.004298242
6      C     3.525755350    -0.726403556    -0.298834416
7      C     4.758130012     1.508276861    -0.207089045
8      C     6.067173205     0.758996055     0.128135197
9      C     6.066297942    -0.758723027    -0.152675149
10     C     4.758636117    -1.509689754     0.183974219
11     H     2.028076982     1.779758085    -0.979135125
12     H     1.783222150     1.985547100     0.774896230
13     H     3.615438189     0.668343445     1.398218356
14     H     1.775635635    -1.988183126    -0.775000698
15     H     2.034655464    -1.778017185     0.976338444
16     H     3.605462695    -0.671828899    -1.415866619
17     H     4.701247136     1.698024182    -1.301965924
18     H     4.781858388     2.511136394     0.274687666
19     H     6.898959135     1.224129876    -0.448301719
20     H     6.313663292     0.927436004     1.202073615
21     H     6.310551886    -0.926211781    -1.227184142
22     H     6.899346489    -1.224001107     0.421831349
23     H     4.782179031    -2.511725171    -0.299802509
24     H     4.704822392    -1.702307724     1.278697961
---
H16C9O1, RHF, CHARGE=0, MULT=1
HF=-66.9
1      C    -0.000917076     0.000191146     0.000132781
2      C     1.535279210     0.001824849    -0.000227836
3      C     2.228924670     1.376279763     0.001023334
4      C     1.896390194     2.331117956     1.164779354
5      C     1.086457845     3.600150030     0.832307214
6      C    -0.436837738     3.461656471     0.636980823
7      C    -0.945467932     3.135288406    -0.783159362
8      C    -1.637537642     1.774753229    -0.998588485
9      C    -0.743360302     0.542311285    -1.229898840
10     O    -0.619159169    -0.486839240     0.941826197
11     H     1.870238113    -0.557337658    -0.905616990
12     H     1.911087363    -0.587660937     0.868153150
13     H     2.060498841     1.877870791    -0.978664091
14     H     3.326706174     1.175121188     0.030349893
15     H     2.867678751     2.677701858     1.594056524
16     H     1.397390868     1.782446380     1.994652565
17     H     1.538826559     4.113918663    -0.045814012
18     H     1.236775872     4.302508025     1.688356689
19     H    -0.843580571     2.739099111     1.378434054
20     H    -0.884687954     4.444221573     0.924763693
21     H    -0.132458224     3.257990707    -1.533201188
22     H    -1.697286887     3.917290584    -1.050430052
23     H    -2.273867462     1.876584898    -1.911553744
24     H    -2.349797864     1.575640628    -0.166729485
25     H    -0.021169054     0.756492256    -2.048622486
```

```
26      H       -1.389399433     -0.281368753    -1.618423001
---
H18C9O1, RHF, CHARGE=0, MULT=1
HF=-85.5
1       C       0.267512155      0.371689953    -0.371443043
2       C       1.271148615     -0.284493900     0.598392405
3       C       1.990992212      0.732930691     1.524546328
4       C       1.096079317      1.512082489     2.505405477
5       C       1.122420958      3.050664447     2.454065069
6       C       1.510396704      3.773002777     3.772641297
7       C       2.950462115      3.471555247     4.234378025
8       C       1.254040397      5.292424501     3.668227857
9       C       2.291784439     -1.167549542    -0.152331999
10      O       0.379857291      0.922413028     3.308612507
11      H       0.751835935      1.113643396    -1.038300923
12      H      -0.545180001      0.890546850     0.175682306
13      H      -0.216624835     -0.391323795    -1.014716420
14      H       0.682221575     -0.979377557     1.249900566
15      H       2.743494787      0.191622930     2.145574366
16      H       2.573683824      1.445393381     0.899547099
17      H       1.800622478      3.388846137     1.639648095
18      H       0.099680455      3.374826760     2.147825221
19      H       0.828003307      3.400114566     4.578530605
20      H       3.087115893      2.394405590     4.458803025
21      H       3.703774985      3.755365324     3.471776524
22      H       3.191885162      4.025290606     5.164883967
23      H       1.897510833      5.775806221     2.905582718
24      H       0.198531235      5.503980498     3.400946572
25      H       1.448964414      5.792901026     4.638615621
26      H       2.929997223     -0.578668751    -0.841726227
27      H       2.960999337     -1.698136858     0.555413791
28      H       1.775454788     -1.943376987    -0.753294642
---
H18C9O1, RHF, CHARGE=0, MULT=1
HF=-81.5
1       O       0.000010120     -0.000022171    -0.000022483
2       C       4.623741590     -0.000066629    -0.000135116
3       C       5.580184930      1.210214292    -0.000062800
4       C       7.031596066      0.861232731     0.387865802
5       C       7.989841297      2.068122761     0.389808445
6       C       9.435025053      1.735551701     0.771528635
7       C       3.171867050      0.348827500    -0.388184782
8       C       2.218999238     -0.863220297    -0.383015777
9       C       0.779077761     -0.514422968    -0.796126586
10      C       0.342996321     -0.856936281    -2.219012407
11      H       5.576001779      1.676978243    -1.011851309
12      H       5.190634787      1.982134501     0.702539285
13      H       7.036498308      0.392652217     1.398847384
14      H       7.421711294      0.090068381    -0.315748349
15      H       7.992682454      2.541534971    -0.619289047
16      H       7.609484006      2.843718241     1.094091825
17      H       9.504817894      1.309849204     1.793027961
18      H       9.890646034      1.008718812     0.068838571
19      H      10.057252638      2.654028745     0.750254268
20      H       4.626919942     -0.467641973     1.011130833
21      H       3.165388112      0.811211541    -1.401941158
22      H       2.610356228     -1.652370750    -1.063697931
23      H       2.196534050     -1.321927858     0.632067191
24      H       1.061613829     -0.449955073    -2.958226608
```

```
25      H       0.298023824     -1.957927178    -2.345579329
26      H      -0.657325623     -0.446742747    -2.462169955
27      H       5.012864795     -0.771844016    -0.703496120
28      H       2.782208278      1.121915319     0.312322697
---
H18C9O1, RHF, CHARGE=0, MULT=1
HF=-83.1
1       C      -0.055549706     -0.077236471    -0.007523136
2       C       1.488618525     -0.293146039     0.063131345
3       C       2.291762668      1.098621350    -0.025248319
4       C       1.812515430      1.941424663    -1.265935469
5       C       0.873121897      3.145010116    -1.119658198
6       C       1.856899738     -1.281261434    -1.089829187
7       C       1.783447799     -1.047790246     1.399108674
8       C       3.835649496      0.890448441    -0.181617728
9       C       2.130834801      1.944515050     1.277329754
10      O       2.222206064      1.675474914    -2.392525040
11      H      -0.380836847      0.334446725    -0.983836087
12      H      -0.430326483      0.601482106     0.783723596
13      H      -0.599318658     -1.036919063     0.122032191
14      H       0.068047385      2.976361097    -0.380162766
15      H       1.451695974      4.039745137    -0.810856378
16      H       0.377813854      3.386374522    -2.083020399
17      H       1.657481150     -0.868770969    -2.096704391
18      H       2.919313120     -1.593761277    -1.057503106
19      H       1.257630061     -2.214175795    -1.014216329
20      H       1.394514315     -0.511316523     2.287564481
21      H       2.865369554     -1.221875634     1.560703410
22      H       1.299493394     -2.047491998     1.407362100
23      H       4.274319771      0.362207795     0.688458155
24      H       4.117693959      0.315239058    -1.084087381
25      H       4.356328283      1.869157982    -0.254533302
26      H       1.076649080      2.092960461     1.581090046
27      H       2.648977448      1.480741755     2.141187844
28      H       2.580637395      2.953065479     1.160298281
---
H18C9O1, RHF, CHARGE=0, MULT=1
HF=-82.4
1       O       0.000002441     -0.000010832    -0.000072136
2       C       1.224975937      0.000002886     0.000018969
3       C       2.051391145      1.299551395     0.000009187
4       C       1.267654076      2.625773180    -0.031513217
5       C       2.169048777      3.876925015    -0.034241419
6       C       1.412033075      5.207590835    -0.064875995
7       C       2.051235854     -1.299637336    -0.004707079
8       C       1.270377149     -2.623932091     0.098413024
9       C       2.171861432     -3.875062447     0.097889479
10      C       1.417684498     -5.203894779     0.198442634
11      H       2.697387862      1.280263395     0.908157695
12      H       2.733462790      1.259145215    -0.880533973
13      H       0.617840580      2.646620747    -0.935887530
14      H       0.589391443      2.672620889     0.850694068
15      H       2.824215103      3.865706872     0.867536089
16      H       2.853095871      3.841052163    -0.913657580
17      H       0.782962720      5.303602649    -0.972854834
18      H       0.754034797      5.328438295     0.819305278
19      H       2.129074124      6.054486404    -0.065075416
20      H       2.652913096     -1.302768873    -0.943090762
21      H       2.774992270     -1.238577637     0.840547918
```

```
22         H         0.661162904      -2.618603756      1.030850977
23         H         0.553350790      -2.695902589     -0.750787042
24         H         2.892724031      -3.814954133      0.946031761
25         H         2.788095757      -3.889475850     -0.830879771
26         H         0.827697906      -5.273699264      1.134621559
27         H         0.722888202      -5.349460942     -0.653267677
28         H         2.134605179      -6.050868462      0.192359171
---
H18C9O1, RHF, CHARGE=0, MULT=1
HF=-82.7
1     O     0.000048892      0.000055401      0.000012160
2     C     1.228098982      0.000009236     -0.000041870
3     C     1.929032035      1.412101353     -0.000058431
4     C     3.166939811      1.503849735      0.931774719
5     C     2.322451262      1.777562927     -1.460675658
6     C     0.924247907      2.501696377      0.510667812
7     C     1.928916718     -1.412166816      0.000134745
8     C     3.167685709     -1.504299056     -0.930590708
9     C     2.321222053     -1.777983437      1.460937393
10    C     0.924262248     -2.501355527     -0.511568241
11    H     2.910024130      1.235183457      1.976345563
12    H     4.007020244      0.859221934      0.612206719
13    H     3.570543234      2.538435357      0.953858699
14    H     3.196744199      1.202357454     -1.822024948
15    H     1.488077846      1.596289428     -2.168245487
16    H     2.597165272      2.849462715     -1.545356975
17    H     0.051059307      2.622102183     -0.160955834
18    H     0.544168531      2.264551946      1.524793715
19    H     1.413995136      3.496107828      0.571777337
20    H     2.911619611     -1.236446919     -1.975553680
21    H     4.007497761     -0.859390107     -0.610864249
22    H     3.571362631     -2.538901360     -0.951632373
23    H     3.196376760     -1.204183583      1.822485444
24    H     1.487110742     -1.595232156      2.168172130
25    H     2.594143878     -2.850327078      1.545936145
26    H     0.050485700     -2.621437089      0.159204337
27    H     0.545394712     -2.264078921     -1.526198230
28    H     1.413611833     -3.496102056     -0.572079260
---
H20C9O1, RHF, CHARGE=0, MULT=1
HF=-89.8
1     C    -0.000051667      0.000000699     -0.000005720
2     C     1.531206024      0.000102211      0.000045606
3     C     2.174206450      1.400657074      0.000002283
4     C     3.716411743      1.385759620      0.000455285
5     C     4.357605182      2.788660631      0.000332366
6     C     5.900064385      2.773345763      0.001183016
7     C     6.540976494      4.176413101      0.000592705
8     C     8.082930248      4.158931993      0.002013751
9     C     8.710220027      5.576892214     -0.000134457
10    O    10.103275065      5.473636802      0.003118195
11    H    -0.412999916      0.506376341      0.895888480
12    H    -0.413036952      0.505552819     -0.896360182
13    H    -0.382574721     -1.041594843      0.000525072
14    H     1.882544384     -0.569896324      0.891155105
15    H     1.882926712     -0.570473613     -0.890609954
16    H     1.814948431      1.964085169     -0.891580610
17    H     1.814390759      1.964464288      0.891372548
18    H     4.074408136      0.821748447      0.892239678
```

```
19    H     4.074920304    0.821593404   -0.891034555
20    H     3.999446961    3.352405806   -0.891575517
21    H     3.998759930    3.352922435    0.891661244
22    H     6.258099865    2.210167207    0.893416900
23    H     6.259099436    2.209161154   -0.890020321
24    H     6.182615203    4.739611595   -0.891728766
25    H     6.181322329    4.741039920    0.891491950
26    H     8.444685700    3.601296141    0.895554624
27    H     8.446344082    3.598407373   -0.889335035
28    H     8.358099080    6.151637999   -0.899654339
29    H     8.354506095    6.157354915    0.894136133
30    H    10.505930307    6.330448677    0.001156836
---
H20C9O1, RHF, CHARGE=0, MULT=1
HF=-91
1     O    -0.000005423    0.000019257   -0.000013118
2     C     1.400700664   -0.000035526   -0.000007801
3     C     1.890603910    1.474712800   -0.000014538
4     C     3.427223923    1.602587586   -0.017361971
5     C     3.934612569    3.057814001   -0.015393728
6     C    -0.797482581   -1.154027003    0.184356818
7     C    -0.464466354   -2.285482427   -0.841582603
8     C    -0.680396414   -1.702136533    1.644838520
9     C    -2.261150464   -0.646674664   -0.063416715
10    H    -2.388196538   -0.245223290   -1.088835870
11    H     1.803706735   -0.527142983   -0.905625359
12    H     1.825885410   -0.534345772    0.889481923
13    H     1.468020946    2.001345214   -0.885623953
14    H     1.488413635    1.994391510    0.899139189
15    H     3.828545648    1.080782068   -0.916540079
16    H     3.849508275    1.071193702    0.866288131
17    H     3.519358524    3.597054275   -0.898074061
18    H     3.542183410    3.587389556    0.883371854
19    C     5.459005215    3.202104905   -0.034017957
20    H     0.528464874   -2.742784992   -0.657906627
21    H    -0.472691300   -1.911708914   -1.885205183
22    H    -1.211894715   -3.102195495   -0.776409718
23    H     0.306282650   -2.166487242    1.843613855
24    H    -0.829635847   -0.902227260    2.397563130
25    H    -1.446641838   -2.481740116    1.831683195
26    H    -2.544106042    0.155841046    0.646853115
27    H    -2.988244139   -1.474677010    0.062925363
28    H     5.929131014    2.732775419    0.853881698
29    H     5.906394663    2.741808614   -0.938194394
30    H     5.741760695    4.275190103   -0.032093837
---
H20C9O1, RHF, CHARGE=0, MULT=1
HF=-91.4
1     O    -0.000018394   -0.000018064   -0.000757133
2     C     1.400939240    0.000017937    0.000020506
3     C     1.890936414    1.474495496   -0.000015008
4     C     3.426187523    1.603882145   -0.014703094
5     C     3.943615500    3.045120362   -0.017589283
6     C    -0.798434699   -1.147303537    0.210522102
7     C    -0.420813538   -2.333657266   -0.734838314
8     C    -0.704999498   -1.623454613    1.701948839
9     C    -2.280636851   -0.665165167   -0.057176630
10    H    -2.550017479    0.098990188    0.708966516
11    H     1.802572515   -0.526986534   -0.906371851
```

```
12      H       1.827089477     -0.534316214     0.888749802
13      H       1.470580922      2.001685423    -0.886372986
14      H       1.488441363      1.995339204     0.898295444
15      H       3.836300663      1.081893636    -0.910250727
16      H       3.853465568      1.079351423     0.871337752
17      H       3.599611530      3.605268613    -0.910533170
18      H       3.614533489      3.603776711     0.881891409
19      H       5.053239936      3.052476450    -0.026781094
20      H       0.542026044     -2.804609513    -0.449422412
21      H      -0.329117758     -2.017968027    -1.792960292
22      H      -1.192312188     -3.128861722    -0.690319559
23      H       0.285617117     -2.058774758     1.943096010
24      H      -0.881570506     -0.791468550     2.412789491
25      H      -1.459945317     -2.407997096     1.911801042
26      H      -2.956978654     -1.532524003     0.127201947
27      C      -2.628382525     -0.083258865    -1.430388955
28      H      -2.039582713      0.823068972    -1.672279908
29      H      -3.700441811      0.208273578    -1.444228530
30      H      -2.480909897     -0.814690632    -2.250143251
---
N1O1, RHF, CHARGE=1, MULT=1
HF=237
1    N    0.000000000    0.000000000    0.000000000
2    O    1.080146610    0.000000000    0.000000000
EXPGEOM
1    N    0.00000    0.00000    -0.56730
2    O    0.00000    0.00000     0.49640
---
N1O1,  UHF, CHARGE=0, MULT=2
HF=21.6
1    N    0.000000000    0.000000000    0.000000000
2    O    1.122532799    0.000000000    0.000000000
EXPGEOM
1    O    0.00000    0.00000     0.53610
2    N    0.00000    0.00000    -0.61270
---
C1N1O1,  UHF, CHARGE=0, MULT=2
HF=38.1
1    N    0.000085341    0.000005844    0.000003044
2    C    1.231298107   -0.000005826   -0.000001717
3    O    2.436084631   -0.000026409    0.000000811
EXPGEOM
1    N    0.00000    0.00000    -1.26570
2    C    0.00000    0.00000    -0.03820
3    O    0.00000    0.00000     1.13610
---
H1C1N1O1, RHF, CHARGE=0, MULT=1
HF=-24.3, IE=11.6
1    H    0.000000000    0.000000000    0.000000000
2    N    0.997737655    0.000000000    0.000000000
3    C    1.625794206    1.079163565    0.000000000
4    O    2.428737926    1.950809928    0.000213041
EXPGEOM
1    H    1.57170   -1.14830    0.00000
2    N    0.56620   -1.03640    0.00000
3    C    0.00000    0.05250    0.00000
4    O   -0.69190    1.01100    0.00000
---
H3C1N1O1, RHF, CHARGE=0, MULT=1
```

```
HF=-44.5
1    C    -0.000035773    -0.000032997    -0.000048694
2    N     1.388919897    -0.000078819    -0.000014538
3    H     1.903784002     0.848064256    -0.000010499
4    H    -0.459869309     1.008197200     0.000057393
5    O    -0.630156571    -1.052909330    -0.000204486
6    H     1.931565395    -0.833474001    -0.000307907
EXPGEOM
1    C     0.00000     0.41920     0.00000
2    O     1.19720     0.23000     0.00000
3    N    -0.93880    -0.55980     0.00000
4    H    -0.44320     1.42290     0.00000
5    H    -0.64730    -1.52010     0.00000
6    H    -1.91550    -0.33980     0.00000
---
H5C2N1O1, RHF, CHARGE=0, MULT=1
HF=-5.4
1    C    -0.000015100     0.000041280    -0.000024396
2    C     1.501737238     0.000023982     0.000094878
3    N     2.147652162     1.139181174    -0.000021374
4    O     3.448171472     1.072489781    -0.000061905
5    H    -0.374451193    -0.528329453    -0.902371415
6    H    -0.448011159     1.013614871     0.008493253
7    H    -0.374655152    -0.543337384     0.893246925
8    H     1.975190599    -0.991696290     0.000132173
9    H     3.790841498     1.968356332    -0.000024589
EXPGEOM
1    C    -1.44210     0.48020     0.00000
2    C     0.00000     0.86330     0.00000
3    N     1.01840     0.09420     0.00000
4    O     0.64860    -1.26640     0.00000
5    H     1.50480    -1.70130     0.00000
6    H    -1.56770    -0.59750     0.00000
7    H    -1.93360     0.89900     0.87930
8    H    -1.93360     0.89900    -0.87930
9    H     0.26500     1.91110     0.00000
---
H5C2N1O1, RHF, CHARGE=0, MULT=1
HF=-57
1    C     0.000114755    -0.000005998    -0.000018632
2    C     1.524510161     0.000025471     0.000009258
3    O     2.222008972     1.015608727    -0.000038352
4    N     2.161434064    -1.239520790     0.000555519
5    H     1.659855138    -2.095452583     0.000654441
6    H     3.152689115    -1.324527378     0.000959015
7    H    -0.397589812     1.034561177    -0.008950748
8    H    -0.396589782    -0.520869388    -0.895212770
9    H    -0.395612902    -0.505369270     0.904528031
EXPGEOM
1    C    -1.35960    -0.29940    -0.00010
2    C     0.07790     0.15440     0.00180
3    N     0.98900    -0.85780    -0.03150
4    O     0.40340     1.31700     0.00400
5    H    -2.00050     0.57190    -0.05940
6    H    -1.55180    -0.95400    -0.84820
7    H    -1.58040    -0.85120     0.91210
8    H     1.95620    -0.61910     0.06920
9    H     0.71660    -1.80950     0.10450
---
```

```
H3C3N1O1, RHF, CHARGE=0, MULT=1
HF=19.6
1    C     0.000033099     0.000128447    -0.000002547
2    N     1.350720816    -0.000056099     0.000011123
3    O     1.778435991     1.228277268    -0.000010132
4    C     0.722558352     2.106028366    -0.000032268
5    C    -0.456743769     1.375225792    -0.000038071
6    H    -0.584862361    -0.910729878     0.000042731
7    H     0.947075209     3.165837648    -0.000011637
8    H    -1.474799684     1.727065801    -0.000091884
EXPGEOM
1    H    -0.19110     2.19310     0.00000
2    C     0.00000     1.12830     0.00000
3    H     2.15320     0.70340     0.00000
4    C     1.12440     0.37250     0.00000
5    O    -1.09230     0.33750     0.00000
6    H     1.16370    -1.89620     0.00000
7    C     0.61930    -0.95900     0.00000
8    N    -0.69280    -0.99300     0.00000
---
H3C3N1O1, RHF, CHARGE=0, MULT=1
HF=-3.7
1    N     0.000010557    -0.000205071    -0.000002365
2    C     1.336884438     0.000067587     0.000008619
3    O     1.856320510     1.261468376    -0.000016675
4    C     0.789593335     2.117706595    -0.000069797
5    C    -0.369692979     1.349767226    -0.000052322
6    H     0.989219145     3.180904797    -0.000145177
7    H    -1.408385071     1.649941865    -0.000036558
8    H     2.009032133    -0.853103793     0.000027490
EXPGEOM
1    O    -1.10180     0.31060     0.00000
2    C     0.00000     1.09530     0.00000
3    N     1.12570     0.46450     0.00000
4    C     0.75170    -0.87760     0.00000
5    C    -0.59660    -0.96090     0.00000
6    H    -0.16420     2.15420     0.00000
7    H     1.47270    -1.66960     0.00000
8    H    -1.30430    -1.76260     0.00000
---
H5C3N1O1, RHF, CHARGE=0, MULT=1
HF=-31.1
1    O     0.000041240    -0.000372795     0.000247000
2    C     1.234177570    -0.000030222    -0.000125228
3    C     2.091638721     1.221568596     0.000094213
4    C     1.641114446     2.487296163    -0.015552886
5    N     1.887689822    -1.231409807    -0.000237642
6    H     3.172955393     1.051020907     0.014525027
7    H     0.591284120     2.776542195    -0.030291641
8    H     2.324910321     3.336855203    -0.014178698
9    H     2.876217926    -1.312388750    -0.000641587
10   H     1.388075100    -2.091557485    -0.000366597
EXPGEOM
1    C     0.80060    -0.65420    -0.03040
2    C    -0.47710     0.12760    -0.00840
3    N    -1.61600    -0.63850    -0.03280
4    O    -0.51230     1.34800    -0.00900
5    C     1.97690    -0.02490     0.02590
6    H     0.74150    -1.74050    -0.09650
```

```
7    H    -2.48430    -0.14800     0.13790
8    H    -1.58340    -1.60970     0.24200
9    H     2.01380     1.05980     0.08450
10   H     2.91900    -0.56680     0.01150
---
H5C3N1O1, RHF, CHARGE=0, MULT=1
HF=-8.5
1    N    -0.000203593   0.000058221   0.000010702
2    C     1.160754498   0.000145890  -0.000140733
3    C     2.627589106  -0.031009286   0.000331726
4    O     3.096783170   1.288797156   0.004017219
5    C     4.474058248   1.532181043   0.017323997
6    H     2.980302804  -0.598348084  -0.903856630
7    H     2.979137029  -0.604247824   0.900874497
8    H     4.990595648   1.112810062  -0.880855110
9    H     4.972464751   1.115662190   0.926879503
10   H     4.609096104   2.639186110   0.016908235
EXPGEOM
1    C     1.54700    -0.79760     0.13890
2    O     1.10170     0.43870    -0.44150
3    C    -0.04600     0.94750     0.21300
4    C    -1.24550     0.06840     0.03430
5    N    -2.17610    -0.64110    -0.09890
6    H     2.46260    -1.08070    -0.39970
7    H     0.78480    -1.59080     0.01940
8    H     1.77390    -0.66320     1.21500
9    H    -0.26170     1.93250    -0.22960
10   H     0.12580     1.07130     1.30140
---
H7C3N1O1, RHF, CHARGE=0, MULT=1
HF=-45.8, DIP=3.82
1    N    -0.000101462   0.000228278   0.000180827
2    C     1.463768397  -0.000195374   0.000129746
3    H     1.862543759   1.041554854   0.000222685
4    H     1.878002117  -0.515953218   0.894714704
5    H     1.861528100  -0.522856237  -0.901275453
6    C    -0.649656511   0.655139429  -1.139778415
7    H    -0.338643769   0.170764052  -2.095645912
8    H    -1.757582809   0.608045902  -1.086175226
9    H    -0.359749968   1.731577960  -1.187996163
10   C    -0.703011070  -0.596544312   1.067377321
11   H    -0.061043318  -1.047057402   1.851398357
12   O    -1.926377822  -0.612323453   1.115575512
EXPGEOM
1    C     0.69380    -0.83170     0.00000
2    O     0.20440    -1.94310     0.00000
3    N     0.00000     0.34730     0.00000
4    C    -1.45290     0.34200     0.00000
5    C     0.67250     1.63020     0.00000
6    H     1.79670    -0.65700     0.00000
7    H    -1.79090    -0.70410     0.00000
8    H    -1.84650     0.85520     0.89790
9    H    -1.84650     0.85520    -0.89790
10   H     1.76410     1.47620     0.00000
11   H     0.40410     2.22220    -0.89630
12   H     0.40410     2.22220     0.89630
---
H7C3N1O1, RHF, CHARGE=0, MULT=1
HF=-59.3
```

```
 1   C    0.000056973   -0.000091906   -0.000017068
 2   C    1.523602296    0.000150451   -0.000096095
 3   O    2.201319052    1.029352453    0.000081598
 4   N    2.220005238   -1.220785539    0.000013953
 5   C    1.676265175   -2.565051919   -0.000802546
 6   H    3.221107358   -1.171890356   -0.000789107
 7   H   -0.392626233    1.037115624   -0.002572141
 8   H   -0.400703453   -0.509042824   -0.899815587
 9   H   -0.400567607   -0.504671716    0.902299180
10   H    1.056963533   -2.760198687   -0.907855150
11   H    1.042943108   -2.755637534    0.897416113
12   H    2.508734256   -3.304845213    0.007750259
---
H7C3N1O1, RHF, CHARGE=0, MULT=1
HF=-61.9
 1   O   -0.000014709   -0.000077567   -0.000004248
 2   C    1.231415534    0.000061653    0.000007791
 3   C    2.128094785    1.243560008    0.000002500
 4   C    1.406415341    2.594413478   -0.003887409
 5   N    1.884641570   -1.232141802    0.000197393
 6   H    2.794551398    1.198548809   -0.893109196
 7   H    2.790295002    1.201289014    0.896435829
 8   H    0.772932953    2.722646569   -0.903982425
 9   H    0.767390041    2.725081050    0.891923060
10   H    2.155329775    3.413960721   -0.002625945
11   H    2.873403380   -1.312914433    0.000259970
12   H    1.386001216   -2.093162809    0.000308153
EXPGEOM
 1   C   -1.97740    0.03420    0.00020
 2   C   -0.66830   -0.79200   -0.00020
 3   C    0.56240    0.13940   -0.00010
 4   O    0.47080    1.38360   -0.00010
 5   N    1.77060   -0.52660    0.00030
 6   H   -2.85210   -0.63070   -0.00010
 7   H   -2.01480    0.67950    0.88660
 8   H   -2.01480    0.68020   -0.88580
 9   H   -0.61920   -1.44110   -0.88890
10   H   -0.61910   -1.44170    0.88810
11   H    1.83280   -1.53660   -0.00040
12   H    2.62570    0.01800   -0.00000
---
H9C3N1O1, RHF, CHARGE=0, MULT=1
HF=-48.6
 1   N    0.061388206    0.086352896   -0.210444257
 2   C    1.499335227    0.015647936    0.038792267
 3   C   -0.553107566    1.407845157   -0.077796137
 4   C   -0.533156027   -0.862974501   -1.165875568
 5   O   -0.334529122   -0.492574449   -2.496978062
 6   H    1.844035256   -1.043212699    0.067976636
 7   H    2.111185910    0.546463116   -0.731183503
 8   H    1.740133260    0.468552655    1.027890743
 9   H   -0.217715060    2.131707950   -0.859404047
10   H   -1.662234840    1.334516006   -0.137240489
11   H   -0.301608191    1.845440019    0.916275812
12   H   -1.643370147   -0.906180592   -0.997758403
13   H   -0.129353626   -1.892353610   -0.954154719
14   H    0.515805074   -0.741764722   -2.829275199
---
H5C4N1O1, RHF, CHARGE=0, MULT=1
```

```
HF=8.5
1    C    -0.000041089    0.000058181   -0.000011577
2    C     1.385530919    0.000046170    0.000012286
3    O     1.819366880    1.302251265   -0.000005122
4    N     0.810908463    2.120330699   -0.000025951
5    C    -0.345653086    1.414069250   -0.000058621
6    C    -1.695345045    2.048260685   -0.000625526
7    H    -0.676251114   -0.838745514    0.000010996
8    H     2.134950229   -0.782316592    0.000037383
9    H    -2.263112015    1.742421594   -0.904719432
10   H    -2.272485274    1.724601238    0.891236358
11   H    -1.646605098    3.155716543    0.010626998
EXPGEOM
1    C     2.13400    0.02140    0.00010
2    H     2.52230    0.53560    0.88350
3    H     2.50440   -1.00450    0.00020
4    H     2.52230    0.53540   -0.88350
5    C    -1.48240    0.57600    0.00010
6    O    -1.38170   -0.76370    0.00000
7    N    -0.03540   -1.10630   -0.00010
8    C     0.63690    0.02190    0.00000
9    C    -0.24970    1.14970   -0.00010
10   H     0.00510    2.19780   -0.00020
11   H    -2.48610    0.97500    0.00020
---
H5C4N1O1, RHF, CHARGE=0, MULT=1
HF=8.1
1    C     0.000000000    0.000000000    0.000000000
2    C     1.392968678    0.000000000    0.000000000
3    C    -0.345935988    1.404791211    0.000000000
4    N     0.797814614    2.122518462    0.000488711
5    O     1.812850707    1.313176808    0.000355745
6    C     2.422200656   -1.082377785   -0.000039490
7    H    -1.325841041    1.864817042   -0.000118617
8    H    -0.678623347   -0.836924482   -0.000205222
9    H     3.076926528   -1.015754627   -0.893766704
10   H     3.064245022   -1.026878625    0.903637432
11   H     1.931669963   -2.076983084   -0.009626854
EXPGEOM
1    C     2.10680    0.01730   -0.00000
2    H     2.46140   -0.52110   -0.88310
3    H     2.54350    1.01650   -0.00060
4    H     2.46140   -0.52000    0.88370
5    C     0.61830    0.11160   -0.00000
6    O    -0.06210   -1.05300    0.00000
7    N    -1.42700   -0.79110    0.00000
8    C    -1.53470    0.51530   -0.00000
9    H    -2.51770    0.96670   -0.00010
10   C    -0.26050    1.15680    0.00000
11   H    -0.04290    2.21300    0.00020
---
H7C4N1O1, RHF, CHARGE=0, MULT=1
HF=-47.2
1    C     0.000122660    0.000006125   -0.000026704
2    C     1.545067568    0.000114099   -0.000000731
3    C     1.994335387    1.484921673    0.000025894
4    N     0.755932630    2.249349981    0.002327821
5    C    -0.412959358    1.475359702    0.001533844
6    O    -1.530585416    1.975047411    0.002259351
```

```
  7    H    -0.409187250    -0.519243566     0.891743148
  8    H    -0.408798672    -0.517297149    -0.893142024
  9    H     1.944361964    -0.532474557    -0.888408487
 10    H     1.944361864    -0.532480895     0.888399777
 11    H     2.614695424     1.730436868    -0.895689117
 12    H     2.617513223     1.729616518     0.893944174
 13    H     0.768019399     3.244632250     0.002594976
EXPGEOM
  1    N     0.08220    -1.09810    -0.12320
  2    C     0.90540     0.00360    -0.01310
  3    C    -0.01350     1.22260     0.14570
  4    C    -1.41100     0.68520    -0.20960
  5    C    -1.31660    -0.81010     0.16720
  6    O     2.11760    -0.00850    -0.02860
  7    H     0.48720    -2.01880     0.00670
  8    H     0.04500     1.54190     1.20350
  9    H     0.33450     2.06040    -0.47810
 10    H    -2.23440     1.19980     0.31150
 11    H    -1.58090     0.77280    -1.29740
 12    H    -1.56190    -0.96800     1.23810
 13    H    -1.99150    -1.44160    -0.43490
---
H7C4N1O1, RHF, CHARGE=0, MULT=1
HF=-31.2
  1    N    -0.000093155    -0.000036893     0.000018815
  2    C     1.309533088    -0.000017399     0.000003757
  3    C    -0.468057266     1.390697667    -0.000004873
  4    C     0.851142511     2.227237941    -0.000174197
  5    O     1.880129055     1.250436412    -0.000122767
  6    C     2.214639291    -1.200348193     0.000188764
  7    H    -1.098258406     1.589559430     0.895681217
  8    H    -1.098340800     1.589562098    -0.895620912
  9    H     0.953029970     2.879762769     0.899081227
 10    H     0.952874885     2.879641582    -0.899650359
 11    H     2.864441506    -1.199143922    -0.899334765
 12    H     2.864190505    -1.199059804     0.899888861
 13    H     1.628019788    -2.140813278     0.000163559
EXPGEOM
  1    H    -1.89620    -1.21600     0.90200
  2    H    -1.89820    -1.21730    -0.89910
  3    C    -1.44580    -0.78710     0.00060
  4    H    -1.95080     1.20170     0.89470
  5    H    -1.94960     1.20060    -0.89650
  6    C    -1.46980     0.79160    -0.00040
  7    N     0.00280     1.20300     0.00020
  8    O     0.01660    -1.13030    -0.00070
  9    C     0.69650     0.12210    -0.00000
 10    H     2.52350    -0.55380     0.89070
 11    H     2.63220     1.00350     0.00030
 12    H     2.52370    -0.55350    -0.89050
 13    C     2.19620    -0.00060     0.00020
---
H7C4N1O1, RHF, CHARGE=0, MULT=1
HF=-37.7
  1    O     0.270420305    -0.112152814     0.301556068
  2    C     1.080572710     0.124266426    -0.594255557
  3    C     1.838147412     1.419633699    -0.737639819
  4    C     1.101411850     2.541275361    -1.423547336
  5    C     3.080818740     1.546474041    -0.232628415
```

```
  6     N     1.347623674    -0.858909098   -1.543913485
  7     H     0.789816793     2.247199917   -2.447787011
  8     H     0.190200901     2.822949815   -0.856073545
  9     H     1.726984687     3.452414279   -1.519513955
 10     H     3.610373774     0.741899954    0.276921872
 11     H     3.659024146     2.468222725   -0.289278952
 12     H     1.956449222    -0.700803485   -2.311651089
 13     H     0.871397462    -1.732578022   -1.547818670
EXPGEOM
 1     N    -1.67090     0.75110     0.25220
 2     H    -2.62980     0.44130     0.32890
 3     H    -1.40500     1.55840     0.79510
 4     C     1.67420    -0.90390     0.26300
 5     H     1.49220    -1.36100     1.24040
 6     H     1.55270    -1.69300    -0.48270
 7     H     2.70140    -0.53420     0.23360
 8     C     1.05710     1.46880    -0.27510
 9     H     0.33210     2.23690    -0.52340
 10     H     2.10080     1.76790    -0.29480
 11     C    -0.74750    -0.22750    -0.03470
 12     O    -1.06590    -1.36740    -0.34590
 13     C     0.69600     0.20690    -0.00230
 ---
H9C4N1O1, RHF, CHARGE=0, MULT=1
HF=-67.5
 1     C     0.000087966     0.000261440   -0.000152285
 2     C     1.543205469    -0.000210534    0.000336777
 3     C     2.149300179     1.418587419    0.000000574
 4     C     2.101326102    -0.845475554    1.164325318
 5     O     1.927265097    -0.593089340    2.357466514
 6     N     2.850153663    -1.973845993    0.837818549
 7     H    -0.425118830     0.512273073    0.885323266
 8     H    -0.395750445    -1.035848423   -0.017204937
 9     H    -0.386170915     0.513848345   -0.904532401
 10     H     1.851242233    -0.467366465   -0.970322535
 11     H     3.257522369     1.375541590   -0.019701683
 12     H     1.847532788     2.010113134    0.886751042
 13     H     1.826461956     1.976894270   -0.902964047
 14     H     3.048418549    -2.235402071   -0.098335628
 15     H     3.249140901    -2.559440144    1.537017103
EXPGEOM
 1     C    -1.35270    -1.26480    -0.02480
 2     H    -0.83980    -2.16900    -0.36610
 3     H    -1.40610    -1.29500     1.06580
 4     H    -2.36980    -1.27630    -0.42760
 5     C    -1.35280     1.26480    -0.02460
 6     H    -0.83980     2.16910    -0.36570
 7     H    -1.40610     1.29480     1.06600
 8     C    -0.61220     0.00000    -0.47310
 9     H    -0.54220     0.00010    -1.56830
 10     N     1.82150     0.00000    -0.75940
 11     H     2.76640     0.00020    -0.40350
 12     H     1.67890     0.00040    -1.75630
 13     C     0.78970     0.00000     0.13930
 14     O     0.96820    -0.00010     1.34980
 15     H    -2.36980     1.27640    -0.42740
 ---
H9C4N1O1, RHF, CHARGE=0, MULT=1
HF=-66.7
```

```
 1    O    -0.000043080     0.000169986    -0.000011535
 2    C     1.231308723     0.000127606     0.000046643
 3    C     2.134296035     1.240847144    -0.000083920
 4    C     1.403556476     2.597174361     0.021861011
 5    C     2.337451991     3.811937620     0.022867474
 6    N     1.882024648    -1.233849581    -0.001356536
 7    H     2.784126989     1.194033527    -0.905029133
 8    H     2.808462009     1.178805985     0.885899238
 9    H     0.730088514     2.672675786    -0.862226299
10    H     0.750845660     2.655032839     0.922790665
11    H     2.998520694     3.824845797     0.913134199
12    H     2.982148052     3.839131306    -0.878956101
13    H     1.742781299     4.748373487     0.035728366
14    H     2.870542447    -1.317634555    -0.001613618
15    H     1.380850628    -2.093504769    -0.002280119
EXPGEOM
 1    C    -2.73160    -0.19530     0.03350
 2    H    -2.88140    -0.31000     1.11210
 3    H    -3.51860     0.46480    -0.34190
 4    H    -2.87720    -1.17850    -0.42600
 5    C    -1.34560     0.36700    -0.27410
 6    H    -1.23220     0.50830    -1.35490
 7    H    -1.22220     1.35570     0.17450
 8    C    -0.21770    -0.53130     0.23390
 9    H    -0.25380    -1.51180    -0.25600
10    H    -0.35280    -0.71890     1.30770
11    N     2.18100    -0.77700    -0.15540
12    H     3.12310    -0.41640    -0.18810
13    H     2.04780    -1.77500    -0.13620
14    C     1.15770     0.10790     0.05500
15    O     1.34030     1.31640     0.11340
---
H9C4N1O1, RHF, CHARGE=0, MULT=1
HF=-67.5
 1    O    -0.034763990    -0.048422487    -0.015648030
 2    C     1.181690328     0.018065877     0.167503641
 3    C     2.000332958     1.325844252     0.147138795
 4    C     1.930849266     2.008444531    -1.234992104
 5    H     2.248917356     1.311225475    -2.037029241
 6    N     1.871005221    -1.167554659     0.411627878
 7    H     3.080687501     1.082104147     0.315929693
 8    C     1.588382460     2.258753179     1.305196851
 9    H     0.912413328     2.367796764    -1.481076887
10    H     2.614033434     2.882042638    -1.269352895
11    H     1.675489965     1.739141536     2.281216207
12    H     0.547538770     2.624948664     1.206464428
13    H     2.255513483     3.144400682     1.343676164
14    H     2.849864889    -1.197783993     0.569454441
15    H     1.407793825    -2.047997990     0.440514059
---
H11C4N1O1, RHF, CHARGE=0, MULT=1
HF=-29.1
 1    N     0.012011121     0.037462971    -0.233090992
 2    C     1.497290949    -0.008178886    -0.001988581
 3    C    -0.563180710     1.402889461     0.024736106
 4    O    -0.289917103    -0.380601009    -1.459660592
 5    H    -0.826862255    -1.172074721    -1.381231505
 6    C     2.041664012    -1.440868061     0.083627363
 7    H     2.068521769     0.559553442    -0.778321570
```

```
  8    H     1.699730379     0.496115200     0.974611478
  9    H    -0.251793454     2.151263774    -0.745836162
 10    C    -2.094131208     1.391745475     0.134066221
 11    H    -0.151222032     1.760694833     1.000187921
 12    H     1.468888601    -2.067441294     0.796525174
 13    H     2.045901291    -1.954569604    -0.898547685
 14    H     3.093196922    -1.407495416     0.439735470
 15    H    -2.460726471     0.616483936     0.836404345
 16    H    -2.590137690     1.230912512    -0.844023830
 17    H    -2.436629110     2.376681702     0.516266531
EXPGEOM
 1    N    -0.15110    -0.14960    0.00000
 2    O     1.10900    -0.79690    0.00000
 3    H     1.78540    -0.08620    0.00000
 4    C    -0.27430    -0.22850    2.46320
 5    C    -0.27430    -0.22850   -2.46320
 6    C    -0.27430     0.64560    1.21620
 7    C    -0.27430     0.64560   -1.21620
 8    H     0.66120    -0.78360    2.54610
 9    H     0.66120    -0.78360   -2.54610
 10   H     0.53870     1.39670   -1.27580
 11   H     0.53870     1.39670    1.27580
 12   H    -1.21600     1.19860   -1.14420
 13   H    -1.21600     1.19860    1.14420
 14   H    -1.09530    -0.94750    2.41590
 15   H    -1.09530    -0.94750   -2.41590
 16   H    -0.39630     0.38730    3.35900
 17   H    -0.39630     0.38730   -3.35900
---
H5C5N1O1, RHF, CHARGE=0, MULT=1
HF=-19
  1    C    -0.000180078    -0.000484843    -0.000009398
  2    C     1.408284132     0.000627213     0.000002910
  3    C     2.087410309     1.233781012     0.000000627
  4    C     1.351387126     2.425319416    -0.000021270
  5    C    -0.069844936     2.322932562    -0.000024971
  6    N    -0.737462398     1.133734340    -0.000011729
  7    H    -1.761402149     3.272216055    -0.000044298
  8    H    -0.571627505    -0.934994797    -0.000017190
  9    H     1.958595926    -0.938454847     0.000018649
 10    H     3.177501947     1.261876990     0.000017303
 11    H     1.847393566     3.394070357    -0.000037709
 12    O    -0.825868181     3.439774357    -0.000040007
EXPGEOM
 1    N    -1.18160     0.30170    0.00000
 2    C     0.00000     0.90720    0.00000
 3    C     1.22970     0.22420    0.00000
 4    C     1.19080    -1.16150    0.00000
 5    C    -0.04820    -1.82120    0.00000
 6    C    -1.19910    -1.04320    0.00000
 7    O     0.02030     2.25840    0.00000
 8    H     2.15800     0.78280    0.00000
 9    H     2.11560    -1.73060    0.00000
 10   H    -0.11120    -2.90310    0.00000
 11   H    -2.18400    -1.50450    0.00000
 12   H    -0.90830     2.54340    0.00000
---
H5C5N1O1, RHF, CHARGE=0, MULT=1
HF=-10.4
```

```
    1    C       0.000004964      0.000011951      0.000031591
    2    C       1.408301759      0.000005783      0.000007694
    3    C       2.097056758      1.223688612     -0.000004247
    4    C       1.337655366      2.418757814      0.000016209
    5    C      -0.089946041      2.326554887      0.000033359
    6    N      -0.730026100      1.141898336     -0.000020197
    7    H       2.824211313      3.654851049      0.000015123
    8    H      -0.572354548     -0.932739291      0.000206666
    9    H       1.957818265     -0.940828595      0.000125916
   10    H       3.186903752      1.234236000      0.000023894
   11    O       1.876322536      3.664506674      0.000065204
   12    H      -0.725872796      3.217904013      0.000207183
EXPGEOM
    1    N      -1.20730     -1.15870     0.00000
    2    C      -1.18440      0.17710     0.00000
    3    C       0.00000      0.92490     0.00000
    4    C       1.21940      0.24220     0.00000
    5    C       1.19490     -1.14910     0.00000
    6    C      -0.03950     -1.80830     0.00000
    7    O       0.02080      2.28660     0.00000
    8    H      -2.15010      0.68520     0.00000
    9    H       2.14770      0.80360     0.00000
   10    H       2.11950     -1.71720     0.00000
   11    H      -0.08670     -2.89450     0.00000
   12    H      -0.88800      2.62030     0.00000
---
H5C5N1O1, RHF, CHARGE=0, MULT=1
HF=-7.2
    1    C       0.000008058      0.000031930      0.000174651
    2    C       1.407214040     -0.000038019      0.000192652
    3    C       2.080910049      1.251239296     -0.000019760
    4    C       1.301729207      2.437706913      0.000024175
    5    C      -0.101931471      2.318195623     -0.000065308
    6    N      -0.747418608      1.130097124     -0.000154028
    7    H       3.814098475      2.108784871      0.000068818
    8    H      -0.561465049     -0.940808001      0.000431621
    9    H       1.954288134     -0.941483192      0.000497858
   10    O       3.434317017      1.239798045      0.000123407
   11    H       1.757155243      3.426770584      0.000197644
   12    H      -0.740135873      3.208593712      0.000094042
EXPGEOM
    1    N       0.01900     -1.87600     0.00000
    2    C      -1.12470     -1.18380     0.00000
    3    C      -1.19630      0.21060     0.00000
    4    C       0.00000      0.93550     0.00000
    5    C       1.20780      0.22920     0.00000
    6    C       1.15530     -1.16190     0.00000
    7    O       0.05380      2.29060     0.00000
    8    H      -2.04390     -1.76650     0.00000
    9    H      -2.15950      0.71380     0.00000
   10    H       2.15030      0.76430     0.00000
   11    H       2.08110     -1.73340     0.00000
   12    H      -0.84440      2.65180     0.00000
---
H5C5N1O1, RHF, CHARGE=0, MULT=1
HF=21
    1    C       0.000029403     -0.000195773     -0.000004096
    2    C       1.399747493      0.000027893      0.000009804
    3    C       2.122180475      1.203598542      0.000015727
```

```
4    C     1.409693442    2.413177009   -0.000087858
5    C     0.010176977    2.425719236   -0.000106480
6    N    -0.710460636    1.215856324   -0.000038042
7    O    -1.938247139    1.220988317   -0.000018247
8    H    -0.586607364   -0.921233363    0.000026857
9    H     1.923136597   -0.957582728    0.000023565
10   H     3.211084629    1.199216780    0.000148857
11   H     1.941513669    3.366094269   -0.000137645
12   H    -0.568204058    3.352114035   -0.000150300
---
H7C5N1O1, RHF, CHARGE=0, MULT=1
HF=-4.3
1    O    -0.000014202   -0.000019862   -0.000035436
2    N     1.296068426   -0.000018973    0.000046284
3    C     1.752059438    1.275580657    0.000013632
4    C     0.595061588    2.155099489   -0.000070375
5    C    -0.492266866    1.287185951   -0.000109948
6    C     3.200415630    1.630761236    0.000069034
7    C    -1.972158691    1.489712562   -0.000458362
8    H     0.598109964    3.232944409   -0.000130229
9    H     3.447903724    2.236444603   -0.897298453
10   H     3.448487438    2.234175074    0.898779933
11   H     3.858419366    0.738525616   -0.001245590
12   H    -2.441448107    1.032707609    0.895518684
13   H    -2.439375561    1.041378600   -0.901951587
14   H    -2.210290809    2.572865183    0.004454453
---
H9C5N1O1, RHF, CHARGE=0, MULT=1
HF=-50.4
1    C    -0.004772065    0.017096425    0.089742508
2    C     1.542592843    0.008746689    0.050424136
3    C     1.987332824    1.476856776   -0.085734656
4    N    -0.407651381    1.423955268   -0.027561268
5    C    -1.819442758    1.790129317   -0.030773852
6    C     0.696655568    2.299735702   -0.132583462
7    O     0.578829834    3.512849932   -0.241452392
8    H    -0.387593240   -0.429017476    1.039549248
9    H    -0.433727279   -0.590453977   -0.743617156
10   H     1.912713088   -0.603024725   -0.799046495
11   H     1.958783599   -0.453260832    0.970144264
12   H     2.618498278    1.795618718    0.770500137
13   H     2.584964556    1.641213028   -1.006978285
14   H    -2.309699100    1.467682470    0.918462493
15   H    -1.964467075    2.886450807   -0.133956807
16   H    -2.350605220    1.293544278   -0.877353009
---
H9C5N1O1, RHF, CHARGE=0, MULT=1
HF=-35.6
1    O    -0.000015166    0.000093456   -0.000240906
2    C     1.418423803   -0.000061487    0.000075041
3    C     1.798080750    1.514623478    0.000015503
4    C    -0.450143314    1.300181452   -0.004327233
5    C    -1.945616602    1.514593608   -0.007554955
6    C    -2.413209226    2.973519772   -0.006295626
7    N     0.500978765    2.200070848   -0.005052632
8    H     1.794178807   -0.543360895   -0.899270633
9    H     1.793913862   -0.543271003    0.899424268
10   H     2.388277550    1.805782135    0.898029366
11   H     2.395367540    1.804468615   -0.893439193
```

```
12      H       -2.374077643     0.998951356     0.884556445
13      H       -2.369468932     1.001790683    -0.903529672
14      H       -2.065969025     3.519997273     0.892945548
15      H       -2.063066678     3.522246056    -0.902883124
16      H       -3.523049255     3.004161484    -0.008022938
---
H9C5N1O1, RHF, CHARGE=0, MULT=1
HF=-24.9
1       N        0.020053701     0.041081349     0.241658879
2       C        1.471504248     0.082993508     0.442546149
3       C       -0.737602503     1.270127104     0.500612489
4       C       -0.684871282    -1.170307876     0.188343725
5       C       -0.165576884    -2.424219324     0.292788058
6       C       -0.986901846    -3.646140960     0.154113013
7       O       -0.547070208    -4.784967704     0.253569452
8       H        1.760881785    -0.254320957     1.465695539
9       H        1.988223312    -0.560924244    -0.304461383
10      H        1.856279266     1.118734397     0.307328535
11      H       -1.740271842     1.233666345     0.018784629
12      H       -0.886565029     1.439156754     1.593785326
13      H       -0.203077672     2.153300084     0.083827853
14      H       -1.764274232    -1.042686417     0.012656842
15      H        0.893822304    -2.603588006     0.487071943
16      H       -2.068555620    -3.482155876    -0.051089370
---
H11C5N1O1, RHF, CHARGE=0, MULT=1
HF=-43
1       N        0.354701277     0.013501340    -0.018429064
2       C        1.482656848     0.019645465    -0.958743883
3       C       -0.428241096     1.253405318     0.027329053
4       C        0.592572451    -0.647437726     1.272350691
5       C        0.239824687    -2.152511524     1.172090139
6       C       -1.231701066    -2.555053852     1.143238386
7       O        1.130783247    -2.995141431     1.139325125
8       H        2.317704449     0.690772115    -0.642309995
9       H        1.897065816    -1.005367324    -1.079347796
10      H        1.137149246     0.358611628    -1.961660572
11      H       -1.376565257     1.091758667     0.588013280
12      H        0.112303294     2.103531041     0.509616134
13      H       -0.703972121     1.568317829    -1.004453059
14      H        1.648277161    -0.521892417     1.622094480
15      H       -0.035124238    -0.198062239     2.080111436
16      H       -1.821085610    -1.965954352     1.873381766
17      H       -1.653502767    -2.379741880     0.133122829
18      H       -1.372935868    -3.626804675     1.389650118
---
H11C5N1O1, RHF, CHARGE=0, MULT=1
HF=-74.8
1       C       -0.000118497    -0.000106336     0.000064855
2       C        1.555901043     0.000042405    -0.000064247
3       C       -0.550400835     1.455396603     0.000027422
4       C       -0.488065885    -0.705623682    -1.304215124
5       C       -0.527319040    -0.765791333     1.251147429
6       O       -1.220357725    -1.783414466     1.202322330
7       N       -0.205997191    -0.287359879     2.521102382
8       H        1.981392081     0.600868921     0.828817589
9       H        1.961211312    -1.027887564     0.093173698
10      H        1.955920704     0.430769708    -0.940818610
11      H       -1.654659862     1.471015108     0.099656802
```

```
12      H       -0.133917416     2.067760750      0.825014683
13      H       -0.294049561     1.980198876     -0.943227035
14      H       -0.119821164    -1.748665467     -1.372676545
15      H       -1.594127184    -0.731235963     -1.370876150
16      H       -0.119783840    -0.170480259     -2.204124473
17      H        0.344461813     0.521323404      2.681200554
18      H       -0.524352016    -0.748735210      3.344160890
---
H11C5N1O1, RHF, CHARGE=0, MULT=1
HF=-59.8
1       C        0.000037257     0.000028481     -0.000008357
2       C        1.533346808    -0.000125719     -0.000094168
3       C        2.192870766     1.386459875      0.000120267
4       N        3.500891008     1.519984090      0.546954731
5       C        4.462008838     0.416494272      0.516203808
6       C        4.018260682     2.841228117      0.927074861
7       O        1.638153774     2.394968984     -0.431682609
8       H       -0.411838247     0.547178822      0.871422964
9       H       -0.418515189     0.450041163     -0.921835036
10      H       -0.368492149    -1.045679983      0.056779380
11      H        1.866948151    -0.583605634      0.887910250
12      H        1.888290860    -0.550425932     -0.902757509
13      H        4.322900018    -0.258301008      1.393833745
14      H        4.368042990    -0.192315570     -0.412326326
15      H        5.510142823     0.791340760      0.546351257
16      H        4.427573311     3.393909827      0.049106038
17      H        3.226884053     3.464727754      1.399440170
18      H        4.835700194     2.733928366      1.677065791
---
H7C6N1O1, RHF, CHARGE=0, MULT=1
HF=-28.8
1       O        0.000038031    -0.000062433      0.000111653
2       C        1.349306032    -0.000009338     -0.000025431
3       C        2.060145962     1.233935916     -0.000020684
4       C        3.457832549     1.167462424     -0.000229500
5       C        4.094498491    -0.087432308     -0.000458567
6       C        3.313483357    -1.267845177     -0.000375221
7       N        1.955872547    -1.219100506     -0.000163547
8       H        5.182665863    -0.142511093     -0.000690161
9       H       -0.385558765    -0.868732135      0.000134403
10      H        1.538355144     2.188868858      0.000137339
11      H        4.051160829     2.082638566     -0.000251169
12      C        3.930185736    -2.645597983     -0.001013889
13      H        4.563868155    -2.782614099     -0.902684544
14      H        4.574221474    -2.779107839      0.893802919
15      H        3.177368316    -3.458847336      0.004933612
---
H7C6N1O1, RHF, CHARGE=0, MULT=1
HF=-20.2
1       H        0.000111316    -0.000414912      0.000024031
2       C        1.094949840     0.001247285      0.000028572
3       C        1.829906287     1.202572913     -0.000006352
4       C        3.229611616     1.139083491     -0.000087745
5       C        3.851396648    -0.135180212     -0.000165957
6       C        3.034576684    -1.317453887     -0.000135230
7       N        1.683906262    -1.214795586     -0.000003784
8       O        5.199811205    -0.295518339     -0.000375138
9       H        1.318016100     2.164130207      0.000053805
10      H        3.813428269     2.059881872     -0.000051876
```

```
11      C       3.583892834     -2.724066264    -0.000449656
12      H       5.668241953      0.528668856    -0.000095688
13      H       4.209196194     -2.899724031    -0.901071575
14      H       4.213076363     -2.898756925     0.897596896
15      H       2.784457116     -3.492674318     0.001687825
---
H7C6N1O1, RHF, CHARGE=0, MULT=1
HF=-16.7
1       H       0.000033858      0.000073653    -0.000039775
2       C       1.095470842     -0.000023379     0.000031866
3       C       1.854517972      1.205670597     0.000000750
4       C       3.270063276      1.109296948     0.000157679
5       C       3.846660704     -0.167285047     0.000300952
6       C       3.024034595     -1.322910160     0.000270094
7       N       1.670241037     -1.218803621     0.000121206
8       H       4.932858748     -0.264690709     0.000438421
9       H       0.353265892      2.424144645    -0.000243976
10      O       1.301161861      2.444919369    -0.000124689
11      H       3.899676885      1.998998993     0.000184152
12      C       3.593188715     -2.719486342     0.000390736
13      H       4.227300053     -2.878388435    -0.897595697
14      H       4.227303851     -2.878190020     0.898415448
15      H       2.812029028     -3.505597982     0.000466938
---
H7C6N1O1, RHF, CHARGE=0, MULT=1
HF=-17.1
1       H       0.000032029     -0.000353674    -0.000100458
2       C       1.096088178     -0.000058133     0.000038933
3       C       1.816384733      1.208346869    -0.000026191
4       C       3.234106381      1.137380228     0.000145220
5       C       3.849635666     -0.141308567     0.000537519
6       C       3.035261174     -1.299104724     0.000743648
7       N       1.679408347     -1.220242994     0.000308898
8       H       4.934899681     -0.241571334     0.000817114
9       H       4.864323800      2.178513761     0.000241098
10      H       1.290663629      2.161771712    -0.000152254
11      O       3.924153779      2.302429084     0.000080463
12      C       3.624499058     -2.689354966     0.001501949
13      H       4.259195214     -2.839401422    -0.897433021
14      H       4.261743365     -2.837410552     0.899028761
15      H       2.854127712     -3.486010961     0.003522946
---
H7C6N1O1, RHF, CHARGE=0, MULT=1
HF=-28.8
1       N      -0.000288378     -0.000250071     0.000154107
2       C       1.396780276      0.000251580    -0.000037564
3       C       2.052450821      1.215195604    -0.000083982
4       C       1.312453023      2.454084004    -0.000218086
5       C      -0.051865191      2.447917796    -0.000253997
6       C      -0.804029698      1.180756706    -0.000172701
7       O      -2.024299987      1.029354977    -0.000224012
8       C       2.115459469     -1.327022751     0.001715960
9       H       3.140675563      1.265009253     0.000129392
10      H       1.872666481      3.390635825    -0.000101801
11      H      -0.633372594      3.369473103    -0.000528324
12      H       2.792124793     -1.399721360    -0.875638468
13      H       2.731755868     -1.428259019     0.919824589
14      H       1.434264357     -2.201202591    -0.035967204
15      H      -0.467307978     -0.888260975     0.000372673
```

```
---
H7C6N1O1, RHF, CHARGE=0, MULT=1
HF=-23.6
1     O     0.000257072     0.000032666     0.001112388
2     C     1.358748447    -0.000061460    -0.000203976
3     C     2.129857593     1.192165076    -0.000007323
4     C     3.530420880     1.102620441    -0.007804681
5     C     4.177841458    -0.141023861    -0.016660058
6     C     3.412619681    -1.337351653    -0.012479167
7     C     1.998272008    -1.267970829    -0.007987991
8     H     5.267843092    -0.170821467    -0.020578529
9     H    -0.368102266     0.872748703     0.031553792
10    H     1.656429647     2.173769077     0.006735623
11    H     4.125318858     2.017727557    -0.007277819
12    N     4.063131567    -2.595960565     0.099059363
13    H     1.395403098    -2.176913859    -0.004115427
14    H     4.960582726    -2.598898193    -0.357407970
15    H     3.524987748    -3.339752141    -0.314471179
---
H7C6N1O1, RHF, CHARGE=0, MULT=1
HF=-25
1     O     0.000031684     0.000033644     0.000005606
2     C     1.353005083    -0.000084656     0.000004869
3     C     1.980945392     1.281410578    -0.000038101
4     C     3.376851431     1.376259179    -0.000306149
5     C     4.174274608     0.215021293    -0.000452481
6     C     3.570935505    -1.048870558    -0.000223208
7     C     2.160949112    -1.178155454     0.000044066
8     H     5.260906926     0.301136404    -0.000736358
9     H    -0.390123729    -0.863952825    -0.000812475
10    H     1.378021490     2.190243116     0.000060951
11    H     3.851993795     2.358412765    -0.000524766
12    H     4.206134448    -1.937262184    -0.000280351
13    N     1.489932570    -2.448988562     0.000168087
14    H     1.787462181    -2.985558726    -0.802868642
15    H     1.785489972    -2.984310967     0.804774198
---
H7C6N1O1, RHF, CHARGE=0, MULT=1
HF=-21.6
1     O    -0.039284241     0.027314198     0.006250705
2     C     1.320357508    -0.001291395     0.007530366
3     C     2.124668899     1.166709454     0.007276213
4     C     3.523640827     1.058230509     0.004463063
5     C     4.150492106    -0.211773749     0.001701673
6     C     3.340269841    -1.376888168    -0.001037200
7     C     1.942576115    -1.279420793     0.002644499
8     N     5.566753243    -0.317312811     0.123251487
9     H    -0.386945175     0.907684428     0.053382723
10    H     1.674514762     2.160378186     0.011683010
11    H     4.115639182     1.975266737     0.007651205
12    H     3.791870699    -2.370835218    -0.001708966
13    H     1.342418097    -2.190048575     0.002643873
14    H     5.911577296    -1.155765789    -0.317139665
15    H     6.032586275     0.450445541    -0.334492366
---
H9C6N1O1, RHF, CHARGE=0, MULT=1
HF=-4.8
1     N    -0.000001021    -0.000000602     0.000013495
2     C     1.352230905     0.000001402     0.000001017
```

```
3    C     2.150047169    1.261928509   -0.000002464
4    C     1.802012211   -1.392367205   -0.000055678
5    C     3.210400282   -1.871798032   -0.000363707
6    C     0.606522970   -2.117498991   -0.000129161
7    C     0.257566789   -3.572070388   -0.001112307
8    O    -0.436325628   -1.219041136    0.000074702
9    H     2.799638349    1.310299188    0.899244025
10   H     2.801225241    1.309242178   -0.898378947
11   H     1.509842652    2.167456791   -0.001444011
12   H     3.751658653   -1.511441165   -0.900890947
13   H     3.754727844   -1.503817955    0.895183119
14   H     3.279306045   -2.978368038    0.004166426
15   H    -0.336287739   -3.840515474    0.897706470
16   H    -0.336457256   -3.839209100   -0.900194454
17   H     1.170903858   -4.199485999   -0.001690011
---
H11C6N1O1, RHF, CHARGE=0, MULT=1
HF=-57.3
1    C     0.000000000    0.000000000    0.000000000
2    C     1.547808242    0.000000000    0.000000000
3    C     2.283604884    1.349532967    0.000000000
4    C     1.852500207    2.401190868    1.036462492
5    C     0.357597217    2.758234798    1.145621436
6    C    -0.492912829    2.505311463   -0.100937752
7    N    -0.620291452    1.186907185   -0.565724112
8    O    -1.091463157    3.388691769   -0.714855637
9    H    -0.370533871   -0.158888217    1.046053803
10   H    -0.354448866   -0.887452137   -0.582334105
11   H     1.898148879   -0.568211608   -0.893635193
12   H     1.880566039   -0.594319533    0.882860472
13   H     2.232788080    1.797177818   -1.019816413
14   H     3.366468274    1.138509611    0.171094841
15   H     2.418496186    3.335749159    0.811533187
16   H     2.197919606    2.076246065    2.045645137
17   H     0.280074135    3.836868536    1.416972089
18   H    -0.099455323    2.210066687    2.002267759
19   H    -1.186558599    1.030454380   -1.377037689
---
H11C6N1O1, RHF, CHARGE=0, MULT=1
HF=-17.9
1    C    -0.000256530   -0.000305800    0.000133229
2    C     1.516844460    0.001270835   -0.000413172
3    C     2.124128447    1.417438417    0.000239486
4    C     1.459896132    2.398524777   -0.979241410
5    C    -0.076249906    2.382210533   -0.925281897
6    C    -0.692733597    0.971448565   -0.946354950
7    N    -0.814063276   -0.765101959    0.689882199
8    O    -0.311324655   -1.647594539    1.501493596
9    H     1.861600362   -0.556839595   -0.903409203
10   H     1.938719587   -0.547434951    0.870897143
11   H     3.208043745    1.339979034   -0.243795704
12   H     2.073947265    1.835028481    1.032118108
13   H     1.819689724    3.429183573   -0.759058197
14   H     1.798349415    2.175379486   -2.017133641
15   H    -0.476182316    2.959928847   -1.789471393
16   H    -0.421913200    2.921430215   -0.013671976
17   H    -1.779427492    1.061421959   -0.719910700
18   H    -0.632106934    0.549998237   -1.978052843
19   H    -1.040653535   -2.104198201    1.925888853
```

```
---
H13C6N1O1, RHF, CHARGE=0, MULT=1
HF=-68.6
1    N     -0.102140145    0.007516016    0.229111631
2    C      1.138907311   -0.037357115    0.919000071
3    C      1.950923714    1.234672865    1.157359795
4    C     -0.735478429    1.271435137   -0.178743532
5    C     -1.656562998    1.902028866    0.881013579
6    C     -0.767766332   -1.243407911   -0.195456424
7    C     -1.643582395   -1.912767469    0.877145091
8    O      1.551996990   -1.129952504    1.309633754
9    H      2.236631611    1.715609640    0.200157427
10   H      1.399902683    1.966865121    1.780021908
11   H      2.892242214    0.996732193    1.695097680
12   H      0.040734274    2.019032490   -0.477375476
13   H     -1.334323980    1.107111420   -1.110459702
14   H     -2.548451385    1.276998760    1.084301116
15   H     -1.142372891    2.080565098    1.845791814
16   H     -2.018655837    2.883797664    0.510060572
17   H     -1.407986048   -1.033190311   -1.089682402
18   H     -0.002783787   -1.974799991   -0.557725682
19   H     -1.091787262   -2.139872446    1.809487633
20   H     -2.521579375   -1.292031528    1.146651779
21   H     -2.029623771   -2.874275407    0.477335640
---
H13C6N1O1, RHF, CHARGE=0, MULT=1
HF=-64.7
1    C     -0.085407431   -0.054273082   -0.101079216
2    C      1.435448139   -0.020626588    0.084198462
3    C      2.030660771    1.402008639    0.131463906
4    C      3.557220478    1.474979916    0.293430675
5    N      4.204840638    2.728057806    0.099822428
6    C      3.482534956    3.995378297    0.221275127
7    C      5.670893030    2.796778839    0.015625082
8    O      4.257799811    0.499928144    0.559468326
9    H     -0.615147766    0.434372218    0.741539700
10   H     -0.400592622    0.448496888   -1.037798377
11   H     -0.438523629   -1.104626420   -0.151966218
12   H      1.903567558   -0.594298557   -0.748143255
13   H      1.684534484   -0.564947988    1.023475204
14   H      1.744723066    1.931297105   -0.806244956
15   H      1.564314149    1.953994061    0.980065871
16   H      2.789944485    4.147460225   -0.639521801
17   H      2.888681273    4.044728489    1.164075561
18   H      4.180041426    4.862425864    0.235711395
19   H      6.145261490    2.704711880    1.020710794
20   H      6.074254318    1.995590311   -0.643662334
21   H      5.993043750    3.767430158   -0.426035597
---
H5C7N1O1, RHF, CHARGE=0, MULT=1
HF=10.8
1    C      0.000023852   -0.000356058   -0.000059496
2    C      1.400620396    0.000217203    0.000006281
3    C      2.154780402    1.205516233    0.000024915
4    C      1.543472019    2.466094622   -0.000017060
5    C      0.134197809    2.466445627   -0.000120150
6    C     -0.641854075    1.255574116   -0.000190255
7    N     -1.999733670    1.624436653   -0.000412399
8    C     -1.993864015    2.954840375   -0.000403759
```

```
 9      O       -0.740619991    3.517547075   -0.000222702
10      H       -0.568029120   -0.929116972   -0.000011524
11      H        1.936957425   -0.950144765   -0.000003056
12      H        3.244084707    1.140906625    0.000048962
13      H        2.123706609    3.386142610   -0.000012795
14      H       -2.855123180    3.619361895   -0.000499532
---
H5C7N1O1, RHF, CHARGE=0, MULT=1
HF=-3.5
 1      C       -0.000007293   -0.000008948   -0.000012240
 2      C        1.417547618   -0.000036502    0.000003522
 3      C        2.114094126    1.220729575    0.000000792
 4      C        1.415913626    2.439897661   -0.000021332
 5      C        0.008699161    2.440502702   -0.000037244
 6      C       -0.706420419    1.232247417   -0.000025503
 7      N       -0.789007641   -1.162596743    0.000020186
 8      O       -0.434000167   -3.551848174    0.000612595
 9      H       -0.532927804    3.387459570   -0.000069703
10      H       -1.796900044    1.252220378   -0.000035638
11      C       -0.463611746   -2.368114226    0.000326259
12      H        1.962364472    3.383044441   -0.000027031
13      H        3.204864643    1.217571398    0.000007647
14      H        1.985398967   -0.930965960    0.000000818
---
H7C7N1O1, RHF, CHARGE=0, MULT=1
HF=-24.1
 1      C        0.000006938    0.000004623    0.000001697
 2      C        1.413283302   -0.000006969   -0.000004795
 3      C        2.124010416    1.213190906    0.000008275
 4      C        1.433014048    2.437428605    0.002998217
 5      C        0.027166282    2.445728765    0.007055933
 6      C       -0.687467037    1.234742897    0.007313323
 7      C       -0.760774122   -1.295406394    0.028256376
 8      N       -1.087920119   -1.844318306   -1.208565345
 9      O       -1.100872032   -1.882351203    1.054956840
10      H        1.966921588   -0.940379784    0.004343502
11      H        3.214454831    1.202272401    0.000398439
12      H        1.985480229    3.377556395    0.004239904
13      H       -0.512562098    3.393164695    0.013158214
14      H       -1.778377798    1.261014373    0.017471441
15      H       -0.836577213   -1.407473675   -2.063451644
16      H       -1.590540850   -2.698676766   -1.293168057
---
H11C7N1O1, RHF, CHARGE=0, MULT=1
HF=-7
 1      C       -0.060996785    0.115325168    0.178105549
 2      N        1.389934994    0.083168236   -0.021756566
 3      C        2.141804874    1.311893229    0.252129902
 4      C        2.102750610   -1.126751896   -0.065718867
 5      C        1.586797614   -2.384070613    0.042272775
 6      C        2.389062198   -3.595085346   -0.095456301
 7      C        1.907540324   -4.859573396   -0.015492997
 8      C        2.738550078   -6.079804562   -0.161780430
 9      O        2.295678472   -7.220280533   -0.110816435
10      H       -0.352744801   -0.227883195    1.199017801
11      H       -0.572436830   -0.528923004   -0.572624221
12      H       -0.452622431    1.148950769    0.045745595
13      H        3.131412242    1.297253772   -0.257276622
14      H        2.318506861    1.455320087    1.345048099
```

```
15      H        1.589127010     2.200797563    -0.127091390
16      H        3.181145622    -0.994364103    -0.240349502
17      H        0.525191357    -2.551483684     0.238850670
18      H        3.459906850    -3.438885343    -0.278249559
19      H        0.843182062    -5.037392879     0.165807631
20      H        3.824977222    -5.907131897    -0.326563744
---
H13C7N1O1, RHF, CHARGE=0, MULT=1
HF=-39.4
1       C       -0.000003862    -0.000003369     0.000001340
2       C        1.555186947     0.000023099     0.000013164
3       C        2.153337776     1.478262305    -0.000007199
4       C        1.787226352     2.264686782    -1.199991300
5       N        1.525679946     2.899618782    -2.136781832
6       O        1.733899566     2.164073211     1.151246927
7       C        2.512752289     3.188451490     1.702573338
8       C        2.064968833    -0.769807892     1.255280690
9       C        2.066396030    -0.746964580    -1.268837631
10      H       -0.420445469     0.479426720     0.906032579
11      H       -0.415857805     0.531179350    -0.879451087
12      H       -0.395634953    -1.036731820    -0.031213534
13      H        3.278043975     1.380542484    -0.027147792
14      H        3.529038266     2.836712294     2.007393952
15      H        2.641740087     4.054542315     1.009032538
16      H        1.976633667     3.545362979     2.613442840
17      H        3.172009950    -0.752241224     1.325649245
18      H        1.665219125    -0.351636664     2.200194710
19      H        1.757627544    -1.835582432     1.221708903
20      H        1.673325050    -0.304858786    -2.205962755
21      H        3.173611473    -0.735429110    -1.335714327
22      H        1.751553298    -1.810653989    -1.260977753
---
H15C7N1O1, RHF, CHARGE=0, MULT=1
HF=-55.8
1       C        0.436692730     0.259250646    -0.146241979
2       C        1.189690044    -0.695884375    -1.067922278
3       O        1.336889436    -0.437572236    -2.258623358
4       C        1.738817663    -2.026402716    -0.492721918
5       N        2.849468967    -1.772584107     0.437151623
6       C        2.623496783    -2.173036918     1.839594875
7       C        3.173587722    -1.152966646     2.850486005
8       C        4.209255519    -2.008844453    -0.093131614
9       C        4.866351525    -0.727634864    -0.632188135
10      H        1.155904195     0.889092755     0.415377022
11      H       -0.189376695    -0.292358207     0.582740440
12      H       -0.234526309     0.936059250    -0.712463542
13      H        0.888313035    -2.559503821    -0.000031101
14      H        2.043483612    -2.700934682    -1.331875563
15      H        1.527572769    -2.280767470     2.036310040
16      H        3.062500060    -3.181683167     2.056516115
17      H        4.279840611    -1.091089131     2.837798988
18      H        2.773750163    -0.134425857     2.676762591
19      H        2.874188550    -1.459815754     3.875108706
20      H        4.871279409    -2.431489861     0.703650265
21      H        4.206072234    -2.783310433    -0.902254206
22      H        4.373449754    -0.355023276    -1.551260316
23      H        4.865366173     0.092199527     0.113148718
24      H        5.925035219    -0.943184608    -0.889966586
---
```

```
H15C7N1O1, RHF, CHARGE=0, MULT=1
HF=-68.4
1    C     0.015021888     0.031071101     0.033915060
2    C     1.569150910    -0.015120677     0.066119697
3    C    -0.514817889     1.493231064     0.024657893
4    C    -0.483646315    -0.622994569     1.370401099
5    C    -0.553201992    -0.785272900    -1.177930949
6    O    -1.267786418    -1.766136030    -0.964448372
7    N    -0.317363109    -0.482967937    -2.548456120
8    C     0.523385125     0.602414965    -3.056576799
9    C    -0.902098932    -1.311963055    -3.619665905
10   H     1.949845944    -1.030199889    -0.165946462
11   H     2.039287643     0.692083160    -0.643624711
12   H     1.957664627     0.260623708     1.068792053
13   H    -1.589452476     1.534326916    -0.245186469
14   H     0.034900610     2.149287201    -0.677292790
15   H    -0.408357563     1.964486552     1.024031890
16   H    -1.589180301    -0.621029403     1.447858910
17   H    -0.133519609    -1.669039616     1.478025638
18   H    -0.099801837    -0.063885858     2.249552352
19   H     0.153515863     1.601344329    -2.735023485
20   H     1.586234826     0.488026027    -2.746097582
21   H     0.525988766     0.618630996    -4.171341866
22   H    -0.092669337    -1.792676525    -4.219868224
23   H    -1.565825961    -2.123826823    -3.258734458
24   H    -1.514552120    -0.675578283    -4.302499638
---
H5C8N1O1, RHF, CHARGE=0, MULT=1
HF=28.1
1    C    -0.000020948     0.000035494     0.000020824
2    C     1.405807945     0.000603825     0.000184078
3    C     2.111923213     1.216692307    -0.000100828
4    C     1.408372034     2.434214263     0.000424863
5    C     0.002528062     2.438504979     0.000212625
6    C    -0.713030063     1.220073314    -0.001997835
7    C    -2.214995317     1.221435066    -0.001829875
8    O    -2.889629425     1.222123540     1.017623985
9    C    -2.846687818     1.221749879    -1.316355050
10   N    -3.338273616     1.222032918    -2.368264846
11   H    -0.532599195    -0.952480966     0.003116388
12   H     1.947909356    -0.945739074     0.001292757
13   H     3.202471893     1.215376429     0.000267199
14   H     1.953107396     3.378986920     0.001853544
15   H    -0.527626979     3.392345458     0.003273953
---
H9C8N1O1, RHF, CHARGE=0, MULT=1
HF=33.4
1    C    -0.000063375    -0.000127078    -0.000009187
2    C     1.414225456     0.000870960    -0.000315839
3    C     2.137589593     1.202310627    -0.000054520
4    C     1.462443155     2.431378645     0.005773794
5    C     0.049348940     2.488492898     0.007534481
6    C    -0.661624480     1.257816322    -0.004923615
7    N    -2.132041403     1.286034020     0.126015355
8    O    -2.776290427     1.305388518    -0.840605429
9    C    -0.737562647    -1.316393493     0.004746003
10   H     1.965203152    -0.941547747    -0.000707591
11   H     3.228223392     1.180804909    -0.002965809
12   H     2.050147739     3.351462343     0.009427121
```

```
13    C     -0.635398230    3.832857656    0.023508552
14    H     -1.440265648   -1.387989592    0.860884467
15    H     -1.313521305   -1.457732606   -0.933840464
16    H     -0.045119157   -2.179247807    0.088791352
17    H     -1.211305328    4.002472964   -0.910405629
18    H      0.091136691    4.667232217    0.107651487
19    H     -1.329237310    3.927054041    0.884758771
---
H9C8N1O1, RHF, CHARGE=0, MULT=1
HF=-18.1
1     C     -0.000000583   -0.000095501    0.000023089
2     C      1.405878136    0.000045858    0.000081297
3     C      2.117781935    1.212477422   -0.000122498
4     C      1.413167188    2.441530596   -0.000500932
5     C     -0.003576251    2.439768616   -0.000073555
6     C     -0.700495624    1.218799305    0.000173800
7     N      2.131390741    3.683051627   -0.004704346
8     C      2.474723620    4.279762648   -1.301573707
9     C      2.471861041    4.273381546    1.232396711
10    O      3.083053300    5.332895141    1.288799481
11    H     -0.545995766   -0.943842387   -0.000228617
12    H      1.949714530   -0.945400703   -0.000089807
13    H      3.208596352    1.188888252   -0.000825583
14    H     -0.567490030    3.373739173   -0.000369400
15    H     -1.791212818    1.219446060    0.000010526
16    H      2.111436723    3.643876928   -2.139331451
17    H      2.008705023    5.285960000   -1.411618632
18    H      3.578615830    4.386470604   -1.409497427
19    H      2.143118701    3.703123891    2.123681328
---
H17C8N1O1, RHF, CHARGE=0, MULT=1
HF=-35.7
1     C     -0.000005553    0.000004427    0.000005984
2     C      1.531218301   -0.000008007   -0.000003587
3     C      2.174759063    1.400207325    0.000001546
4     C      3.716943992    1.384190781   -0.003214125
5     C      4.358844501    2.786783040    0.000239929
6     C      5.901150887    2.769688519   -0.006982414
7     C      6.541847650    4.173475362    0.000888110
8     C      8.054219365    4.149726451   -0.009673801
9     N      8.691349747    5.294862505   -0.031951477
10    O      9.992333114    5.237310289   -0.038500930
11    H     -0.412865115    0.506196926    0.896064742
12    H     -0.413006454    0.505887833   -0.896144307
13    H     -0.382383434   -1.041657651    0.000300020
14    H      1.882892606   -0.570325442    0.890779120
15    H      1.882610558   -0.570519083   -0.890825848
16    H      1.814008903    1.964988301   -0.890279031
17    H      1.817144362    1.963022365    0.892811879
18    H      4.077053371    0.817218239    0.885805238
19    H      4.072927868    0.822693007   -0.897432422
20    H      3.997230818    3.355245276   -0.887214900
21    H      4.005267673    3.346767925    0.896283277
22    H      6.264214748    2.200121348    0.879281953
23    H      6.255905052    2.214086888   -0.905560746
24    H      6.178532282    4.747560790   -0.883742579
25    H      6.193985188    4.730977397    0.902287098
26    H      8.547193579    3.168343314    0.003854968
27    H     10.327988020    6.135758222   -0.055220083
```

```
---
H17C8N1O1, RHF, CHARGE=0, MULT=1
HF=-43.2
1    H    -0.000010888    0.000057263    0.000039293
2    C     1.108864811   -0.000036478   -0.000048057
3    C     1.679247801    1.420971149    0.000080856
4    C     1.280258881    2.274156010    1.219911549
5    C     1.861309803    3.702748496    1.206060510
6    C     1.458671576    4.554509495    2.429774526
7    C     2.045900922    5.980738773    2.403110008
8    C     1.721964472    6.924605288    3.558250596
9    N     0.952266211    6.450055417    4.509820209
10   O     0.648417889    7.217976085    5.514083284
11   H     1.448472767   -0.581421332    0.880984467
12   H     1.441607206   -0.545639954   -0.907284083
13   H     2.790336069    1.353845205   -0.059055608
14   H     1.350370985    1.931938128   -0.934405043
15   H     0.168784762    2.332789042    1.272964915
16   H     1.610521549    1.754813090    2.148789605
17   H     2.972664300    3.642328401    1.152269536
18   H     1.531329428    4.220035843    0.275875100
19   H     0.347587692    4.612563623    2.476104254
20   H     1.793619581    4.033505217    3.355108056
21   H     3.158252059    5.905725635    2.343695922
22   H     1.711398848    6.485180044    1.464904247
23   C     2.331795302    8.306662587    3.465915908
24   H     0.086104656    6.715200593    6.107179825
25   H     3.439066340    8.236456676    3.415681531
26   H     1.979234400    8.822006152    2.547393771
27   H     2.089235101    8.967968096    4.320580632
---
H17C8N1O1, RHF, CHARGE=0, MULT=1
HF=-41.3
1    H    -0.000001129    0.000209216    0.000317072
2    C     1.108898597   -0.000114364   -0.000022536
3    C     1.679467577    1.420870806   -0.000337550
4    C     1.281769954    2.274263903    1.219826302
5    C     1.862471774    3.703522644    1.202723560
6    C     1.461555873    4.552818379    2.428816783
7    C     1.985053212    5.979848830    2.391015839
8    N     3.158640174    6.411740031    2.795510084
9    O     4.011798098    5.558878223    3.280043238
10   H     1.448635759   -0.581524459    0.880926772
11   H     1.441347252   -0.545398705   -0.907478671
12   H     2.790426806    1.353777729   -0.060302808
13   H     1.349228325    1.932008390   -0.934226852
14   H     0.170218937    2.331833608    1.274681763
15   H     1.613832052    1.756267810    2.148810839
16   H     2.972384437    3.643437519    1.141627482
17   H     1.526440497    4.221938414    0.275599079
18   H     0.349465731    4.581614682    2.506024318
19   H     1.800367785    4.047334494    3.362477595
20   C     1.079709239    7.076855927    1.839649875
21   H     4.801794945    6.043510642    3.528693406
22   C     0.257853845    7.815361251    2.902499675
23   H     1.681435205    7.822842544    1.269201311
24   H     0.382059562    6.633807596    1.091148049
25   H     0.902682687    8.363271889    3.618199784
26   H    -0.384381900    7.123658915    3.484078535
```

```
27    H       -0.408004005     8.558074625     2.415973535
---
H17C8N1O1, RHF, CHARGE=0, MULT=1
HF=-43.3
1     H       -0.000245105     0.000280015     0.000807535
2     C        1.108478998    -0.000288622    -0.000199866
3     C        1.679115408     1.420868086     0.000156655
4     C        1.285189393     2.270632453     1.224450508
5     C        1.884894188     3.693210128     1.216125032
6     C        1.469643869     4.553706807     2.399621097
7     N        0.411409183     5.328936656     2.484673492
8     O       -0.419590016     5.375400469     1.486203343
9     H        1.449022434    -0.583234719     0.879397069
10    H        1.440521175    -0.544370034    -0.908456454
11    H        2.789857970     1.353638377    -0.064456768
12    H        1.345211061     1.934171603    -0.931249066
13    H        0.175049169     2.335383411     1.277331272
14    H        1.612583037     1.744131360     2.150052417
15    H        2.997586503     3.619629502     1.198059848
16    H        1.620238079     4.205700101     0.262246183
17    C        2.378452383     4.607624396     3.626190215
18    H       -1.119748460     5.990029763     1.716146149
19    C        1.792704244     3.959821678     4.897878786
20    H        3.346954573     4.112699060     3.383277493
21    H        2.637347768     5.672035307     3.837725014
22    C        2.752491293     3.928565449     6.091336491
23    H        1.477831606     2.914999594     4.673062047
24    H        0.869986392     4.502361168     5.206091429
25    H        3.069029186     4.947298258     6.393580102
26    H        3.666465578     3.339896136     5.872344617
27    H        2.256840148     3.460066705     6.966568568
---
N2O1, RHF, CHARGE=0, MULT=1
HF=19.6, DIP=0.17
1    N     0.0000    0.0000    0.0000
2    N     1.1278    0.0000    0.0000
3    O     2.3092    0.0018    0.0000
EXPGEOM
1    N     0.00000    0.00000   -1.20180
2    N     0.00000    0.00000   -0.07220
3    O     0.00000    0.00000    1.11470
---
H4C1N2O1, RHF, CHARGE=0, MULT=1
HF=-58.7
1    O     0.019206721   -0.027097528   -0.095618909
2    C     1.244188076   -0.007568761    0.029610190
3    N     1.981564724    1.178865616   -0.110179618
4    N     1.996050462   -1.178771743    0.334525495
5    H     1.474520529    2.038944119   -0.058238448
6    H     2.889236389    1.233600069    0.301703642
7    H     1.470736984   -2.027175776    0.204521879
8    H     2.853961562   -1.262699024   -0.182633058
EXPGEOM
1    C     0.00000    0.00000    0.14450
2    O     0.00000    0.00000    1.35780
3    N     0.00000    1.15670   -0.61310
4    N     0.00000   -1.15670   -0.61310
5    H    -0.22680    1.97210   -0.07000
6    H    -0.46660    1.11070   -1.50280
```

```
7    H     0.22680    -1.97210    -0.07000
8    H     0.46660    -1.11070    -1.50280
---
H6C2N2O1, RHF, CHARGE=0, MULT=1
HF=-56.3
1    N    -0.033717359    0.022650596    0.041809749
2    C     1.364649051   -0.041325405   -0.110475407
3    N     2.022158040    1.202627687   -0.285770974
4    C     3.474739446    1.333310234   -0.164054966
5    O     1.958749097   -1.119265505   -0.086293383
6    H    -0.539902537   -0.829227781   -0.093037353
7    H    -0.521937960    0.821637242   -0.303875600
8    H     1.545659518    1.987481215    0.124162856
9    H     3.753433615    2.385805387   -0.401855868
10   H     3.990964898    0.666286354   -0.888604475
11   H     3.849485404    1.103157234    0.860781291
EXPGEOM
1    C    -1.86350    0.05510    0.12690
2    H    -2.07090    0.87610   -0.57000
3    H    -2.69370   -0.65190    0.04810
4    H    -1.83500    0.45960    1.14710
5    H    -0.63140   -1.64480   -0.08690
6    N    -0.63790   -0.64410   -0.23910
7    C     0.62010   -0.11430   -0.02860
8    H     0.00340    1.78290   -0.53440
9    H     1.61070    1.61860   -0.03180
10   N     0.66110    1.27330    0.04300
11   O     1.61440   -0.81120    0.10140
---
H6C4N2O1, RHF, CHARGE=0, MULT=1
HF=25.6
1    O     0.000020918    0.000061162    0.000003000
2    N     1.300938159   -0.000072526    0.000004876
3    C     1.729053166    1.279619978   -0.000017614
4    C     0.502862458    2.091331338    0.000015241
5    N    -0.508116439    1.197506106    0.000008668
6    C     3.162708113    1.686533438   -0.000352717
7    C     0.317772923    3.570062201   -0.000126830
8    H     3.390805049    2.294361883   -0.901131540
9    H     3.847578990    0.814654054    0.003144128
10   H     3.389619691    2.300487462    0.896546913
11   H    -0.752317876    3.860147433    0.004769578
12   H     0.787130721    4.017157328   -0.901605020
13   H     0.795653522    4.018736899    0.896060857
---
H10C4N2O1, RHF, CHARGE=0, MULT=1
HF=-69.3
1    C     0.000047165    0.000066877    0.000024977
2    C     1.548415106   -0.000280306   -0.000103130
3    N     2.093644681    1.365472111    0.000159635
4    C     2.484163362    2.091806311    1.132596064
5    N     2.773050469    3.471614469    0.894527570
6    C     2.152616113   -0.794932103   -1.183859545
7    O     2.544298249    1.606372989    2.262454185
8    H    -0.404660888    0.540870630    0.878914268
9    H    -0.424950640    0.469599542   -0.910232384
10   H    -0.382114702   -1.040073968    0.050154428
11   H     1.878590785   -0.524253045    0.935684690
12   H     1.925858871    1.886521158   -0.838753862
```

```
13      H      3.369945122     3.632163396     0.101399117
14      H      3.195685089     3.920873351     1.689984328
15      H      3.260525374    -0.803907277    -1.139620240
16      H      1.856583520    -0.386009774    -2.171132408
17      H      1.813465628    -1.850509324    -1.144205003
---
H8C5N2O1, RHF, CHARGE=0, MULT=1
HF=1.2
1       O     -0.000001510    -0.000047812     0.000020872
2       C      1.374129823     0.000040220     0.000038157
3       C      1.834314675     1.323825620    -0.000045043
4       C      0.590219391     2.089903616    -0.005664595
5       N     -0.447240006     1.221581250    -0.006672556
6       C      0.380702888     3.568818719    -0.013283226
7       C      3.233373432     1.823747918    -0.014750228
8       N      2.016048813    -1.241372180    -0.122933651
9       H      0.868595432     4.024844664    -0.900286095
10      H      0.821307502     4.027147069     0.897168811
11      H     -0.693113257     3.844470833    -0.041879370
12      H      3.704519099     1.659410969    -1.007627871
13      H      3.290131312     2.908917065     0.205546188
14      H      3.856895381     1.304901136     0.744759020
15      H      1.436102651    -2.017735514     0.154658733
16      H      2.855831342    -1.270622100     0.433821352
---
H12C5N2O1, RHF, CHARGE=0, MULT=1
HF=-73.4
1       C      0.000022935     0.000027492     0.000007834
2       N      1.400834766     0.000015413     0.000309522
3       H      1.855718799     0.892492165    -0.000087524
4       C      2.263138545    -1.134109439    -0.361792154
5       N     -0.603594051     1.284030680     0.176484387
6       H     -0.213038747     1.818715611     0.933173375
7       H     -1.597743343     1.226959576     0.323903566
8       O     -0.678366277    -1.012034793    -0.177215864
9       H      1.678786366    -2.086288101    -0.263050289
10      C      2.718816550    -1.040295376    -1.840567702
11      C      3.468667989    -1.232086242     0.622738424
12      H      4.148950604    -2.029911638     0.241791138
13      H      4.064730106    -0.290793218     0.599932643
14      C      3.131538635    -1.563822495     2.079165032
15      H      3.355051423    -0.153254059    -2.035158160
16      H      1.849930958    -0.988871473    -2.527252914
17      H      3.303238730    -1.940446856    -2.120870666
18      H      2.576563999    -0.745910849     2.580361187
19      H      4.068045857    -1.724730646     2.653582514
20      H      2.526429639    -2.488700410     2.164702146
---
H12C5N2O1, RHF, CHARGE=0, MULT=1
HF=-65.1
1       C      0.000052605     0.000164437    -0.000011165
2       N      1.434249163    -0.000314209     0.000103898
3       C      2.169354417     1.272656942     0.000064572
4       C      2.129620191    -1.251037295    -0.341728244
5       N     -0.594471914     0.463061651     1.170528644
6       H     -0.074027366     0.608763528     2.004746776
7       H     -1.579975556     0.383413089     1.292835581
8       O     -0.678948874    -0.369285037    -0.956009553
9       H      1.987496999     1.794280021     0.975345778
```

```
10     H     3.266585594     1.057437015    -0.019731910
11     C     1.842981235     2.247286440    -1.144771544
12     H     3.058324810    -1.307859954     0.283318131
13     H     1.495062078    -2.112377062    -0.014035897
14     C     2.514871259    -1.456422695    -1.816113265
15     H     1.971793695     1.788048161    -2.144381105
16     H     0.809405988     2.642250967    -1.082001895
17     H     2.530898758     3.116915038    -1.086814995
18     H     1.646848676    -1.391332021    -2.500263955
19     H     3.274441471    -0.724188626    -2.157352490
20     H     2.958226462    -2.467524635    -1.935188257
---
H12C5N2O1, RHF, CHARGE=0, MULT=1
HF=-49.1
1      C     0.000197611    -0.000293037     0.000210907
2      N     1.428598147    -0.000127099    -0.001056480
3      C     2.148751721     1.279336915    -0.000017843
4      C     2.125787414    -1.133816614     0.624145244
5      N    -0.614730006     0.696317921    -1.087487932
6      C    -0.347735411     0.247023448    -2.459794361
7      C    -1.925645269     1.321028621    -0.855883845
8      O    -0.654411750    -0.560740837     0.873639001
9      H     1.605995265     2.070538211    -0.559225343
10     H     2.319167904     1.657592715     1.037162531
11     H     3.143191435     1.149792812    -0.486389200
12     H     2.163641682    -1.060569957     1.736293284
13     H     1.635475501    -2.095039027     0.352160353
14     H     3.174083628    -1.178017752     0.248541192
15     H     0.691718418    -0.124332404    -2.585256427
16     H    -1.038314224    -0.573726453    -2.771252581
17     H    -0.485108508     1.098292093    -3.165807218
18     H    -2.768103157     0.590525548    -0.882059313
19     H    -1.942594315     1.841177216     0.127752124
20     H    -2.114612615     2.088621488    -1.641260890
---
H5C4N3O1, RHF, CHARGE=0, MULT=1
HF=-14.2
1      N     0.000038396    -0.000087174    -0.000051936
2      C     1.353367504     0.000036427    -0.000236745
3      C     2.115681902     1.181119823     0.000405707
4      C     1.405318829     2.412604205    -0.001929077
5      N     0.043886790     2.441998196     0.016935668
6      C    -0.591292462     1.232353522     0.004107005
7      O    -1.932224636     1.222341241     0.007325342
8      N     2.040537965     3.658822651    -0.126044550
9      H     1.827450420    -0.987997711    -0.001173510
10     H     3.201998092     1.139620094    -0.008672954
11     H    -2.319151281     2.091648197    -0.002614517
12     H     1.465526685     4.449351805     0.109946023
13     H     2.904405108     3.717520352     0.383872599
---
O2, RHF, CHARGE=0, MULT=1
HF=22
1      O     0.000000000     0.000000000     0.000000000
2      O     1.134699292     0.000000000     0.000000000
EXPGEOM
1      O     0.00000     0.00000     0.60410
2      O     0.00000     0.00000    -0.60410
---
```

```
O2, UHF, CHARGE=0, MULT=3
HF=0
1    O     0.000000000     0.000000000     0.000000000
2    O     1.133669707     0.000000000     0.000000000
EXPGEOM
1    O     0.00000     0.00000     0.60410
2    O     0.00000     0.00000    -0.60410
---
H2O2, RHF, CHARGE=0, MULT=1
HF=-32.5
1    O     0.022130011    -0.059899876     0.017475845
2    O     1.315644818     0.007867468    -0.032988291
3    H     1.554788308     0.930915936     0.086334111
4    H    -0.211694317    -0.991997736     0.019011784
EXPGEOM
1    O     0.00000     0.72720    -0.05930
2    O     0.00000    -0.72720    -0.05930
3    H     0.78470     0.89420     0.47470
4    H    -0.78470    -0.89420     0.47470
---
C1O2, RHF, CHARGE=0, MULT=1
HF=-94.1
1    O     0.000000000     0.000000000     0.000000000
2    C     1.186179375     0.000000000     0.000000000
3    O     2.372359739     0.000230891     0.000000000
EXPGEOM
1    C     0.00000     0.00000     0.00000
2    O     0.00000     0.00000     1.16260
3    O     0.00000     0.00000    -1.16260
---
H1C1O2, RHF, CHARGE=-1, MULT=1
HF=-106.6
1    O    -0.000051809    -0.000460116     0.000000000
2    C     1.243939388     0.000075248     0.000000000
3    O     2.005822869     0.983470929     0.000000000
4    H     1.732248070    -0.994897496     0.000000000
EXPGEOM
1    C     0.00000     0.00000     0.30820
2    H     0.00000     0.00000     1.44900
3    O     0.00000     1.13210    -0.20610
4    O     0.00000    -1.13210    -0.20610
---
H2C1O2, RHF, CHARGE=0, MULT=1
HF=-90.5, DIP=1.41, IE=11.51
1    O     0.000000000     0.000000000     0.000000000
2    C     1.226851956     0.000000000     0.000000000
3    O     1.915319957     1.165935588     0.000000000
4    H     1.395476875     1.960022422     0.000000000
5    H     1.888522403    -0.884921927     0.000000000
EXPGEOM
1    C     0.00000     0.38430     0.00000
2    O    -0.89680    -0.62500     0.00000
3    O     1.17760     0.19580     0.00000
4    H    -0.46140     1.38110     0.00000
5    H    -1.78460    -0.25290     0.00000
---
H2C2O2, RHF, CHARGE=0, MULT=1
HF=-50.7, IE=10.59
1    O    -0.002219841    -0.007674491     0.000000000
```

```
2    C    1.217646784    0.001269962    0.000000000
3    C    2.017500212    1.304193561    0.000000000
4    O    3.237332030    1.314616740    0.000000000
5    H    1.421672482    2.238707522    0.000000000
6    H    1.811934430   -0.934164940    0.000000000
EXPGEOM
1    C   -0.33260    0.68310    0.00000
2    C    0.33260   -0.68310    0.00000
3    H   -1.44110    0.67280    0.00000
4    H    1.44110   -0.67280    0.00000
5    O    0.33260    1.70820    0.00000
6    O   -0.33260   -1.70820    0.00000
---
H3C2O2, RHF, CHARGE=-1, MULT=1
HF=-122.5
1    C    0.000019327   -0.000011723   -0.000013350
2    C    1.562680232    0.000070150    0.000035253
3    H    1.910298865    1.029112710   -0.000023176
4    H    1.913689901   -0.496388213    0.900120412
5    H    1.932025840   -0.517349582   -0.880846399
6    O   -0.540758276   -0.506618288   -0.991504624
7    O   -0.533919120    0.509462279    0.994285678
EXPGEOM
1    C    0.04630   -1.34590    0.00000
2    C    0.00000    0.21390    0.00000
3    H    1.07770   -1.72030    0.00000
4    H   -0.48340   -1.73170    0.88290
5    H   -0.48340   -1.73170   -0.88290
6    O   -1.16280    0.69860    0.00000
7    O    1.11420    0.79830    0.00000
---
H4C2O2, RHF, CHARGE=0, MULT=1
HF=-103.3, DIP=1.74, IE=10.8
1    C    0.000000000    0.000000000    0.000000000
2    C    1.522084975    0.000000000    0.000000000
3    H    1.913262039    1.037235427    0.000000000
4    H    1.910749514   -0.523924609    0.896385774
5    H    1.913059132   -0.516576086   -0.899040243
6    O   -0.736617589   -0.511027341   -0.843373667
7    O   -0.565954868    0.639170685    1.056826530
8    H   -1.515145884    0.647117607    1.071117325
EXPGEOM
1    C    0.98430   -0.98000    0.00000
2    C    0.00000    0.14120    0.00000
3    O    0.29680    1.30920    0.00000
4    H    1.97180   -0.54080    0.00000
5    H    0.84810   -1.60460    0.88040
6    H    0.84810   -1.60460   -0.88040
7    O   -1.26990   -0.27780    0.00000
8    H   -1.78850    0.53110    0.00000
---
H4C2O2, RHF, CHARGE=0, MULT=1
HF=-83.6, DIP=1.77, IE=11.02
1    O   -0.000003326    0.000647510    0.001935751
2    C    1.224507610   -0.000059322   -0.002414249
3    O    1.943964876    1.150807107   -0.010338446
4    C    1.411056351    2.450014022   -0.005051401
5    H    1.883326402   -0.888457025   -0.001517355
6    H    2.278771284    3.151445082   -0.015048750
```

```
7    H     0.782799239    2.652378648   -0.904767618
8    H     0.803338364    2.652718094    0.908908614
EXPGEOM
1    C     1.34690    0.40240    0.00000
2    O     0.00000    0.86800    0.00000
3    C    -0.92070   -0.08640    0.00000
4    O    -0.71150   -1.26300    0.00000
5    H     1.96840    1.28940    0.00000
6    H     1.53910   -0.19700    0.88510
7    H     1.53910   -0.19700   -0.88510
8    H    -1.91200    0.36820    0.00000
---
H6C2O2, RHF, CHARGE=0, MULT=1
HF=-30.1, IE=10.6
1    C    -0.224762728   -0.039744487    0.055057639
2    O     1.184705238    0.042421032    0.169327904
3    O     1.622091122    1.266799888    0.159127993
4    C     3.024527933    1.347657181   -0.023489627
5    H     3.269285591    2.435119370    0.009530558
6    H     3.353847279    0.937635962   -1.007401287
7    H     3.594519152    0.828712082    0.782909171
8    H    -0.606224482    0.399416645   -0.896929167
9    H    -0.462967513   -1.129157810    0.067017267
10   H    -0.753982259    0.451404687    0.905433419
EXPGEOM
1    O    -0.44450    0.58240    0.00000
2    O     0.44450   -0.58240    0.00000
3    C     0.44450    1.68090    0.00000
4    C    -0.44450   -1.68090    0.00000
5    H    -0.19570    2.56330    0.00000
6    H     1.07230    1.67760    0.89430
7    H     1.07230    1.67760   -0.89430
8    H     0.19570   -2.56330    0.00000
9    H    -1.07230   -1.67760    0.89430
10   H    -1.07230   -1.67760   -0.89430
---
H6C2O2, RHF, CHARGE=0, MULT=1
HF=-93.9
1    C     0.043572035   -0.129972310    0.068154746
2    C     1.425038664    0.283207867   -0.541998267
3    H     1.277021478    1.103280257   -1.291411364
4    H     2.040299692    0.721334302    0.289482208
5    O     2.113018397   -0.734112091   -1.198597036
6    H    -0.503046776    0.816318034    0.328134739
7    O    -0.736391891   -0.961515753   -0.731516094
8    H     0.205280110   -0.676630269    1.033126274
9    H    -0.976660434   -0.549504221   -1.549687897
10   H     2.348868196   -1.441060521   -0.614234909
EXPGEOM
1    C     0.00000    0.75760    0.00000
2    C     0.00000   -0.75760    0.00000
3    O     1.36590   -1.16370    0.00000
4    O    -1.36590    1.16370    0.00000
5    H     1.38960   -2.13240    0.00000
6    H    -1.38960    2.13240    0.00000
7    H    -0.53470   -1.11870    0.89100
8    H    -0.53470   -1.11870   -0.89100
9    H     0.53470    1.11870    0.89100
10   H     0.53470    1.11870   -0.89100
```

```
---
C3O2, RHF, CHARGE=0, MULT=1
HF=-22.4, IE=10.6
1    O    0.000000000     0.000000000     0.000000000
2    C    1.180263116     0.000000000     0.000000000
3    C    2.451028243    -0.000260812     0.000000000
4    C    3.721794410     0.000066889    -0.000005878
5    O    4.902058046     0.000220378    -0.000008322
EXPGEOM
1    C    0.00000    0.00000     0.00000
2    C    0.00000    0.00000     1.27450
3    C    0.00000    0.00000    -1.27450
4    O    0.00000    0.00000     2.43520
5    O    0.00000    0.00000    -2.43520
---
H4C3O2, RHF, CHARGE=0, MULT=1
HF=-64.8
1    O   -0.000009116    -0.000248388    -0.000039196
2    C    1.218006171     0.000097079    -0.000069938
3    C    2.056518645     1.281319658    -0.000000709
4    C    2.373045087     1.900181425    -1.354686324
5    O    2.455197971     1.746781837     1.059793297
6    H    1.816863509    -0.932889855    -0.005540555
7    H    2.739483353     1.125514842    -2.058495044
8    H    1.457345128     2.356850514    -1.782374091
9    H    3.148762709     2.687958446    -1.283647621
EXPGEOM
1    C   -0.83960    -0.71930     0.00000
2    C    0.00000     0.54970     0.00000
3    C    1.48650     0.38090     0.00000
4    O   -0.35660    -1.82610     0.00000
5    O   -0.58120     1.61370     0.00000
6    H   -1.92150    -0.53360     0.00000
7    H    1.96450     1.35350     0.00000
8    H    1.78870    -0.19450     0.87290
9    H    1.78870    -0.19450    -0.87290
---
H4C3O2,   RHF, CHARGE=0, MULT=1
HF=-79
1    O   -0.000068561     0.000078413     0.000037884
2    C    1.233006083    -0.000067078    -0.000007256
3    C    2.136043586     1.183188218    -0.000479230
4    C    1.729081234     2.463975844    -0.015647909
5    O    1.877406359    -1.197357868     0.000240975
6    H    3.207169810     0.958151953     0.012879786
7    H    0.689890510     2.789198501    -0.031051824
8    H    2.442754651     3.288544419    -0.014423428
9    H    1.314424766    -1.961670632     0.000483272
EXPGEOM
1    O    1.32610     0.34200     0.00000
2    H    1.74790     1.21090     0.00000
3    O   -0.46800     1.67270     0.00000
4    C    0.00000     0.56390     0.00000
5    C   -0.80940    -0.67270     0.00000
6    H   -1.87460    -0.49970     0.00000
7    C   -0.28940    -1.89240     0.00000
8    H    0.77700    -2.05480     0.00000
9    H   -0.92250    -2.76690     0.00000
---
```

```
H4C3O2,  RHF, CHARGE=0, MULT=1
HF=-67.6, IE=10.6
1    C    -0.000010034    -0.000013195    -0.000004792
2    O     1.431689655     0.000027060    -0.000003754
3    C    -0.018489269     1.564056630    -0.000004304
4    C     1.509111039     1.384057543    -0.000489096
5    O     2.537396464     2.023503188    -0.001512905
6    H    -0.421757346    -0.491861266     0.901806572
7    H    -0.421740338    -0.491687761    -0.902022697
8    H    -0.440089840     2.047569334     0.896761553
9    H    -0.440149921     2.047673809    -0.896710313
EXPGEOM
1    O    -1.03680    -0.27840     0.00000
2    C     0.00000     0.62530     0.00000
3    O    -0.06590     1.81690     0.00000
4    C     1.08540    -0.44490     0.00000
5    H     1.71360    -0.44230     0.89510
6    H     1.71360    -0.44230    -0.89510
7    C    -0.10630    -1.41160     0.00000
8    H    -0.24000    -2.01800    -0.89970
9    H    -0.24000    -2.01800     0.89970
---
H6C3O2,  RHF, CHARGE=0, MULT=1
HF=-72.1
1    O     0.028645445     0.027353796    -0.208346732
2    C     1.362464459     0.068805824     0.247431043
3    O     1.680959281     1.422924084     0.478319262
4    C     0.598637420     2.271770402     0.187301856
5    C    -0.537865615     1.311822759    -0.285677758
6    H     2.054167471    -0.370889233    -0.523377756
7    H     1.471556888    -0.527268526     1.195173546
8    H     0.312403002     2.855088205     1.095793715
9    H     0.888520900     3.009046891    -0.600191184
10   H    -1.444370864     1.372082916     0.364067419
11   H    -0.867168010     1.524900574    -1.331447786
EXPGEOM
1    C     1.15970    -0.08860     0.16430
2    C    -0.84200     0.85460     0.13880
3    C    -1.02900    -0.67080    -0.03460
4    O     0.49820     1.06590    -0.27990
5    O     0.30290    -1.17970    -0.09540
6    H     2.09480    -0.21260    -0.40310
7    H     1.37810    -0.01870     1.25470
8    H    -0.97260     1.15520     1.19860
9    H    -1.54710    -0.92360    -0.97640
10   H    -1.58310    -1.12010     0.81000
11   H    -1.51070     1.45930    -0.49250
---
H6C3O2,  RHF, CHARGE=0, MULT=1
HF=-95.2
1    O     0.000164279    -0.000081837     0.000288294
2    O     2.262842971    -0.000204773     0.000063389
3    C     1.206663958     0.619391948    -0.000074046
4    C    -0.237238278    -1.389698448    -0.013095535
5    C    -0.317219758    -2.004660349     1.397271028
6    H     1.091287845     1.719971437    -0.002457492
7    H    -1.224809426    -1.517855864    -0.530612946
8    H     0.515518318    -1.944152323    -0.628867313
9    H     0.654620583    -1.966857541     1.927869351
```

```
10      H     -1.072509326    -1.495000480     2.028403774
11      H     -0.612040976    -3.070493178     1.310642061
EXPGEOM
1    C     -2.18960    -0.26150    0.00000
2    C     -0.69770    -0.56540    0.00000
3    O      0.00000     0.71900    0.00000
4    C      1.34920     0.62140    0.00000
5    O      1.98810    -0.41450    0.00000
6    H     -2.75660    -1.20870    0.00000
7    H     -2.47020     0.31640    0.89660
8    H     -2.47020     0.31640   -0.89660
9    H     -0.38570    -1.13160   -0.89360
10   H     -0.38570    -1.13160    0.89360
11   H      1.79140     1.63550    0.00000
---
H6C3O2,   RHF, CHARGE=0, MULT=1
HF=-97.9, DIP=1.72, IE=10.6
1    C     -0.000132788    -0.000037037     0.000000902
2    C      1.523664226     0.000083377     0.000076316
3    O      2.066164494     1.249282662    -0.000081359
4    C      3.440314747     1.538631491    -0.002584447
5    H      3.529971276     2.651214797    -0.001096278
6    H      3.954039782     1.144629935    -0.911463403
7    H      3.958282016     1.142032077     0.902744780
8    O      2.254353422    -0.987338961    -0.000089556
9    H     -0.391652146     0.523641150     0.895137343
10   H     -0.390630642     0.507220462    -0.905010258
11   H     -0.389478648    -1.037561737     0.009296942
EXPGEOM
1    C      1.12040     1.49970     0.00000
2    C      0.00000     0.49170     0.00000
3    O     -1.18560     0.74870     0.00000
4    O      0.48910    -0.77060     0.00000
5    C     -0.51600    -1.80050     0.00000
6    H      0.69840     2.50650     0.00000
7    H      1.74970     1.35460     0.88460
8    H      1.74970     1.35460    -0.88460
9    H      0.03710    -2.74120     0.00000
10   H     -1.14480    -1.72180     0.89130
11   H     -1.14480    -1.72180    -0.89130
---
H6C3O2,   RHF, CHARGE=0, MULT=1
HF=-108.4, DIP=1.75, IE=10.5
1    C     -0.000049846    -0.000087419     0.000043010
2    C      1.531417194     0.000062111    -0.000009266
3    C     -0.663793324     1.378609696    -0.000131096
4    O     -0.120158734     2.482402942    -0.010581192
5    O     -2.022489894     1.330550029     0.012439801
6    H     -0.360124168    -0.559928788     0.894329673
7    H     -0.360561604    -0.559092296    -0.894679523
8    H      1.900457383    -1.046684250     0.006533584
9    H      1.945425875     0.494937507    -0.900933707
10   H      1.945636085     0.506066141     0.894671716
11   H     -2.461872173     2.172134524     0.012205066
EXPGEOM
1    C      0.00000     0.57150     0.00000
2    C     -0.60800    -0.81900     0.00000
3    C      0.45580    -1.92350     0.00000
4    O     -0.96860     1.54050     0.00000
```

```
 5    O      1.19000     0.83670     0.00000
 6    H     -1.26850    -0.88780     0.88350
 7    H     -1.26850    -0.88780    -0.88350
 8    H     -0.02680    -2.91560     0.00000
 9    H      1.10070    -1.84540    -0.89090
10    H      1.10070    -1.84540     0.89090
11    H     -0.49620     2.39070     0.00000
---
H8C3O2,  RHF, CHARGE=0, MULT=1
HF=-97.6
 1    O     -0.000115973    0.000008548    0.000012584
 2    C      1.396307603    0.000010984   -0.000010674
 3    C      1.914227422    1.461264891    0.000008781
 4    C      3.462535349    1.539798987    0.004331699
 5    O      3.861990505    2.877658435    0.017096063
 6    H     -0.338755565   -0.883932886    0.006009963
 7    H      1.792249061   -0.549926812   -0.896964083
 8    H      1.792666592   -0.550205646    0.896566571
 9    H      1.513870750    1.995148782    0.891324071
10    H      1.519946732    1.993459448   -0.894979817
11    H      3.878236816    1.010716793   -0.896380999
12    H      3.874624826    0.994891961    0.897044630
13    H      4.805930829    2.949192026    0.021811486
EXPGEOM
 1    C     -0.02420     1.03970    -0.35140
 2    C     -1.25720     0.40990     0.30420
 3    O     -1.49820    -0.92210    -0.13650
 4    C      1.28710     0.47370     0.17720
 5    O      1.30070    -0.94060    -0.09260
 6    H     -0.08120     0.87850    -1.43740
 7    H     -0.02270     2.12710    -0.17270
 8    H      2.08940    -1.32280     0.32990
 9    H     -0.64060    -1.38220    -0.06640
10    H      2.14540     0.96470    -0.31050
11    H      1.36080     0.65640     1.26350
12    H     -1.14990     0.45630     1.40560
13    H     -2.15580     0.98340     0.04040
---
H8C3O2,  RHF, CHARGE=0, MULT=1
HF=-90.1, IE=9.8
 1    C     -0.000088577    0.000046116   -0.000013044
 2    O      1.396039559   -0.000252072    0.000132336
 3    C      2.088465924    1.219322854    0.000033204
 4    C      3.616829305    0.901453530    0.008088757
 5    O      4.318903361    2.109337711    0.016275647
 6    H     -0.329191213   -1.065834190    0.005409451
 7    H     -0.429798427    0.494607383   -0.906102777
 8    H     -0.430317143    0.503920235    0.900658746
 9    H      1.819000659    1.837415966    0.896876233
10    H      1.828615675    1.831425583   -0.903871389
11    H      3.889431089    0.287488139   -0.891599959
12    H      3.879829048    0.280733756    0.905967967
13    H      5.253175308    1.958225743    0.021571484
---
H8C3O2,  RHF, CHARGE=0, MULT=1
HF=-83.2
 1    C     -0.203344314    0.040934447    0.028273082
 2    O      0.646506798    1.103574662   -0.288005130
 3    C      1.571307379    1.613318804    0.631917908
```

```
  4    O     2.695185556    0.779183661    0.690213856
  5    C     3.699871856    0.819612268   -0.279938015
  6    H    -0.882701780   -0.106394917   -0.844941489
  7    H    -0.832884077    0.250972975    0.928289997
  8    H     0.344560621   -0.915793980    0.207679304
  9    H     1.163174943    1.675996613    1.680693885
 10    H     1.814918234    2.661022437    0.295423503
 11    H     4.526757999    0.160624360    0.077982439
 12    H     4.114059922    1.847781134   -0.427261101
 13    H     3.356599499    0.441782631   -1.273188110
EXPGEOM
 1    C     0.00000    0.00000    0.94910
 2    H    -0.71130    0.55580    1.57650
 3    H     0.71130   -0.55580    1.57650
 4    O     0.78470    0.87110    0.18010
 5    O    -0.78470   -0.87110    0.18010
 6    C     0.00000    1.77880   -0.58540
 7    C     0.00000   -1.77880   -0.58540
 8    H     0.70220    2.44580   -1.09150
 9    H    -0.70220   -2.44580   -1.09150
10    H    -0.61260    1.25390   -1.32740
11    H    -0.65940    2.37150    0.06620
12    H     0.61260   -1.25390   -1.32740
13    H     0.65940   -2.37150    0.06620
---
H8C3O2,  RHF, CHARGE=0, MULT=1
HF=-102.7
 1    C    -0.052876788   -0.032054671   -0.066344469
 2    C     1.494603433    0.040136879   -0.009936958
 3    C     2.070243219    1.510457050    0.054123523
 4    O     1.895968705    2.102689253    1.305202095
 5    O     2.000105104   -0.618932051   -1.140257963
 6    H    -0.398095577   -1.085089457   -0.108698158
 7    H    -0.469287583    0.496057452   -0.947262563
 8    H    -0.495250595    0.420611204    0.844106898
 9    H     1.818257171   -0.495476858    0.927847478
10    H     1.627161852    2.131884310   -0.766664737
11    H     3.176196665    1.483120932   -0.131752861
12    H     1.081953137    2.581982717    1.362201919
13    H     2.871280284   -0.955384600   -0.984365548
EXPGEOM
 1    O    -0.43330    1.37380   -0.15880
 2    H     0.50370    1.63760   -0.15530
 3    O     1.94230   -0.04640    0.00910
 4    H     2.15720   -0.16570    0.94850
 5    C     0.71190   -0.73290   -0.24490
 6    H     0.61290   -0.76800   -1.33510
 7    H     0.74110   -1.76440    0.13750
 8    C    -0.45910    0.04390    0.34270
 9    H    -0.34640    0.05610    1.44410
10    C    -1.80450   -0.56090   -0.01680
11    H    -1.91890   -0.59290   -1.10620
12    H    -1.90090   -1.57740    0.38370
13    H    -2.61050    0.05530    0.39400
---
H6C4O2,  RHF, CHARGE=0, MULT=1
HF=-88.1
 1    C     0.000000000    0.000000000    0.000000000
 2    C     1.499308880    0.000000000    0.000000000
```

```
3    C     2.270096066    1.108837262    0.000000000
4    C     3.759201692    1.102074355   -0.000804362
5    O     4.506605384    0.121366656   -0.012312452
6    O     4.323290426    2.339659692    0.012723524
7    H    -0.404649394    0.533736743    0.885244655
8    H    -0.402605473    0.492719059   -0.909721452
9    H    -0.396387611   -1.035456266    0.023388893
10   H     1.946695100   -0.999855453   -0.000633529
11   H     1.811256267    2.102549984    0.000633314
12   H     5.272390670    2.353103814    0.012152106
---
H6C4O2,  RHF, CHARGE=0, MULT=1
HF=-87.8
1    C    -0.000079219    0.000023766    0.000269114
2    C     1.507080122   -0.000022937    0.000014228
3    C    -0.751429591    1.118260482   -0.000075828
4    C    -0.652173587   -1.355272053    0.026878262
5    O    -0.896145234   -2.029632468    1.026096846
6    O    -0.987684125   -1.840528390   -1.196450086
7    H     1.903453563   -0.472260790    0.923058067
8    H     1.906288116   -0.559278618   -0.871533351
9    H     1.920206196    1.027988789   -0.050362010
10   H    -1.840949978    1.114603906    0.002250903
11   H    -0.327974392    2.122017870   -0.000178978
12   H    -1.392056328   -2.699766286   -1.197625594
---
H6C4O2,  RHF, CHARGE=0, MULT=1
HF=-78.2
1    C    -0.096604945   -0.067536714   -0.085559515
2    C     1.422748880    0.010859274   -0.157200819
3    C     2.064449743    1.408233304   -0.062783113
4    C     2.624881100    1.828925694    1.289417716
5    H     2.812175125    2.919954414    1.340279462
6    H     3.585154740    1.306002299    1.475219278
7    H     1.922279437    1.567733553    2.105837018
8    O     2.136207930   -0.974302180   -0.292453400
9    O     2.113875847    2.120324051   -1.056928089
10   H    -0.483719310    0.561658987    0.740848114
11   H    -0.536037284    0.296505314   -1.036683535
12   H    -0.451623440   -1.103616127    0.082741905
EXPGEOM
1    C    -0.08230    0.76880    0.00000
2    C     0.08230   -0.76880    0.00000
3    C     1.19670    1.57930    0.00000
4    C    -1.19670   -1.57930    0.00000
5    O    -1.19670    1.24980    0.00000
6    O     1.19670   -1.24980    0.00000
7    H     0.95150    2.65170    0.00000
8    H    -0.95150   -2.65170    0.00000
9    H     1.80340    1.32360    0.88550
10   H     1.80340    1.32360   -0.88550
11   H    -1.80340   -1.32360    0.88550
12   H    -1.80340   -1.32360   -0.88550
---
H6C4O2, RHF, CHARGE=0, MULT=1
HF=-87
1    O    -0.000109860   -0.000094711    0.000016208
2    C     1.412362856   -0.000068973    0.000010725
3    C     1.891670900    1.481457481    0.000015838
```

```
4      C        0.600403929     2.321304450     0.003389830
5      C       -0.518271515     1.271410331     0.002583287
6      O       -1.731329870     1.402700257     0.003750067
7      H        1.778755996    -0.552975757    -0.899263986
8      H        1.778625480    -0.553003672     0.899380365
9      H        2.517310464     1.696497232    -0.890109386
10     H        2.521425579     1.695251857     0.887497016
11     H        0.523076851     2.977124643    -0.887887051
12     H        0.525404114     2.973316253     0.897640534
---
H6C4O2, RHF, CHARGE=0, MULT=1
HF=-79.6
1      C       -0.000038793     0.000021735     0.000011788
2      C        1.343814998    -0.000029608    -0.000035737
3      C        2.197079202     1.221994965    -0.000044593
4      O        3.531810185     0.946220199     0.016618346
5      C        4.556754062     1.905485100     0.020953045
6      O        1.816668940     2.392198891    -0.013366594
7      H       -0.625211876     0.891682186     0.000309954
8      H       -0.569701548    -0.929854362    -0.000182678
9      H        1.881979900    -0.953277226    -0.000161183
10     H        4.541424384     2.546852212    -0.892269322
11     H        4.518868032     2.563120059     0.921862670
12     H        5.519050933     1.340077268     0.038158888
EXPGEOM
1      H       -3.02700        -0.90130        -0.00330
2      H       -2.42430         0.53550        -0.88560
3      H       -2.42710         0.52990         0.88970
4      C       -2.30100        -0.09040         0.00030
5      O       -1.01620        -0.71950         0.00020
6      O       -0.05150         1.33070        -0.00030
7      C        0.03820         0.11910        -0.00050
8      H        1.23470        -1.73330        -0.00120
9      C        1.31280        -0.65120        -0.00040
10     H        3.42830        -0.55630         0.00050
11     H        2.52870         1.06790         0.00120
12     C        2.48800        -0.01630         0.00050
---
H8C4O2,  RHF, CHARGE=0, MULT=1
HF=-67.1
1      C       -0.173909416     0.033689904    -0.023936596
2      C        1.179159002    -0.014959061    -0.050158229
3      C        2.561094532     1.949129721     0.130753494
4      C        2.404845464    -2.074681425    -0.279815016
5      O        1.986710190     0.800462023     0.697501161
6      O        1.895032340    -0.886727981    -0.827095292
7      H       -0.798142655    -0.678295051    -0.558430787
8      H       -0.724731153     0.787646835     0.533132157
9      H        3.281451626     1.712150888    -0.688378299
10     H        1.792247543     2.651054772    -0.275090783
11     H        3.115451696     2.467912703     0.948287894
12     H        3.193618910    -1.890100561     0.488239600
13     H        1.608025773    -2.705471589     0.184418066
14     H        2.861058917    -2.646675765    -1.122079957
EXPGEOM
1      C        0.00000         0.00000         1.41880
2      C        0.00000         0.00000         0.07430
3      H        0.00000         0.92060         1.98020
4      H        0.00000        -0.92060         1.98020
```

```
5      O      0.00000     1.08110    -0.73900
6      O      0.00000    -1.08110    -0.73900
7      C      0.00000     2.35350    -0.10860
8      C      0.00000    -2.35350    -0.10860
9      H      0.00000     3.08250    -0.91710
10     H      0.00000    -3.08250    -0.91710
11     H     -0.89290     2.48630     0.51050
12     H      0.89290     2.48630     0.51050
13     H      0.89290    -2.48630     0.51050
14     H     -0.89290    -2.48630     0.51050
---
H8C4O2,   RHF, CHARGE=0, MULT=1
HF=-80.9
1      O      0.000462299    0.000166870   -0.000536469
2      C      1.403017405   -0.000631822    0.000351840
3      C      2.007061208    1.425799863    0.000164834
4      C      1.156818817    2.437302060   -0.808075375
5      O     -0.219878450    2.181039930   -0.730529554
6      C     -0.684355808    0.864550645   -0.867304778
7      H      1.795321278   -0.580579984   -0.877750778
8      H      1.724339653   -0.555821885    0.917320404
9      H      3.034022776    1.389104987   -0.423021721
10     H      2.116163731    1.784231113    1.047678686
11     H      1.493231706    2.447567522   -1.879453993
12     H      1.314024835    3.473347046   -0.415830957
13     H     -1.769137818    0.854011145   -0.571925131
14     H     -0.610881299    0.513398140   -1.939470073
EXPGEOM
1      C     -0.62750    -1.20290     0.00000
2      O      0.01600    -0.76480     1.17290
3      O      0.01600    -0.76480    -1.17290
4      C      0.01600     0.66240     1.23980
5      C      0.01600     0.66240    -1.23980
6      C      0.69260     1.24160     0.00000
7      H     -0.58500    -2.29370     0.00000
8      H     -1.67880    -0.85500     0.00000
9      H      0.54330     0.92680     2.16170
10     H     -1.02290     1.03010     1.31650
11     H      0.54330     0.92680    -2.16170
12     H     -1.02290     1.03010    -1.31650
13     H      1.75020     0.95110     0.00000
14     H      0.63370     2.33910     0.00000
---
H8C4O2,   RHF, CHARGE=0, MULT=1
HF=-75.5
1      C     -0.000087932   -0.000022452   -0.000104102
2      C      1.560094950   -0.000079658    0.000041421
3      C     -0.000287742    2.407703941    0.000072497
4      C      1.560081779    2.407421767    0.007415194
5      O      2.109103056    1.202208695    0.467414614
6      O     -0.549431031    1.203815463   -0.463222646
7      H     -0.383042065   -0.241121204    1.026710695
8      H     -0.372146977   -0.807717184   -0.679147439
9      H      1.943017871   -0.237588031   -1.027543123
10     H      1.932138007   -0.810177331    0.676233218
11     H     -0.369131325    3.214770103   -0.682036847
12     H     -0.387610152    2.650922775    1.024671630
13     H      1.928730952    3.212088869    0.691992385
14     H      1.947483474    2.653017016   -1.016631073
```

```
EXPGEOM
1    C     0.00000     0.76220     1.16110
2    C     0.00000     0.76220    -1.16110
3    C     0.00000    -0.76220     1.16110
4    C     0.00000    -0.76220    -1.16110
5    O    -0.64960    -1.25950     0.00000
6    O     0.64960     1.25950     0.00000
7    H    -1.04080     1.12480     1.19540
8    H    -1.04080     1.12480    -1.19540
9    H     1.04080    -1.12480    -1.19540
10   H     1.04080    -1.12480     1.19540
11   H     0.54700     1.15510    -2.02440
12   H     0.54700     1.15510     2.02440
13   H    -0.54700    -1.15510    -2.02440
14   H    -0.54700    -1.15510     2.02440
---
H8C4O2,  RHF, CHARGE=0, MULT=1
HF=-106.5
1    O    -0.000079064    0.000081603   -0.000084329
2    O     2.246534986    0.000196951   -0.000019294
3    C     1.046490843    0.643228134    0.000102918
4    C     1.174827298    2.162109090    0.000398564
5    C     2.417975045   -1.399271683    0.000207495
6    C     3.924682688   -1.721704551    0.002323928
7    H     1.722189649    2.508536951    0.900001756
8    H     1.720373759    2.508676699   -0.900190326
9    H     0.173590962    2.637331182    0.001537059
10   H     1.936864838   -1.862031853    0.900911632
11   H     1.939272106   -1.861779421   -0.901943200
12   H     4.437070436   -1.318252630   -0.893635543
13   H     4.434203600   -1.319704328    0.900574422
14   H     4.056782679   -2.823204566    0.001523635
EXPGEOM
1    C    -2.31840     0.04490     0.00000
2    C    -0.91060    -0.51470     0.00000
3    O     0.00000     0.49450     0.00000
4    O    -0.61250    -1.69150     0.00000
5    C     1.37730     0.05250     0.00000
6    C     2.24480     1.30110     0.00000
7    H    -3.04350    -0.78190     0.00000
8    H    -2.46460     0.68010     0.89050
9    H    -2.46460     0.68010    -0.89050
10   H     1.55710    -0.57600     0.89120
11   H     1.55710    -0.57600    -0.89120
12   H     3.31250     1.01730     0.00000
13   H     2.04390     1.91440    -0.89590
14   H     2.04390     1.91440     0.89590
---
H10C4O2, RHF, CHARGE=0, MULT=1
HF=-81.9
1    O    -0.000010751    0.000027305   -0.000038753
2    O     3.602762550   -0.000040217    0.000049725
3    C     4.731692904    0.821406902   -0.000013874
4    C     2.325464419    0.579218487   -0.006711238
5    C     1.277230442   -0.579388562    0.001756108
6    C    -1.129053091   -0.821154101    0.012356320
7    H     4.782774139    1.481909712    0.900782693
8    H     4.788435510    1.474110609   -0.906117963
9    H     5.625222419    0.153309735    0.005673483
```

```
10      H      2.176680124     1.240001082     0.887649333
11      H      2.179280438     1.224143492    -0.913106418
12      H      1.423266267    -1.237803944    -0.894941292
13      H      1.425973600    -1.226936486     0.905819298
14      H     -1.179446412    -1.469198385     0.922191439
15      H     -1.186693137    -1.486266944    -0.884657216
16      H     -2.022481184    -0.152904645     0.009109608
EXPGEOM
1       O      1.41320         1.15560         0.00000
2       O     -1.41320        -1.15560         0.00000
3       C     -1.56990        -2.61320         0.00000
4       C      1.56990         2.61320         0.00000
5       C      0.00000         0.76640         0.00000
6       C      0.00000        -0.76640         0.00000
7       H      2.64930         2.80300         0.00000
8       H     -2.64930        -2.80300         0.00000
9       H     -1.11590        -3.07120         0.89550
10      H     -1.11590        -3.07120        -0.89550
11      H      1.11590         3.07120         0.89550
12      H      1.11590         3.07120        -0.89550
13      H     -0.52710         1.13690         0.89410
14      H     -0.52710         1.13690        -0.89410
15      H      0.52710        -1.13690         0.89410
16      H      0.52710        -1.13690        -0.89410
---
H10C4O2,   RHF, CHARGE=0, MULT=1
HF=-101.8
1       C      0.000064971     0.000018666     0.000083099
2       C      1.550619002    -0.000110510     0.000181200
3       C      2.155883751     1.417777285     0.000038243
4       C      3.706440253     1.417516861    -0.000241408
5       O      4.174960240     2.733581092     0.002705997
6       O     -0.468047833    -1.316193349    -0.001322277
7       H     -0.384252783     0.555706873     0.898228770
8       H     -0.384872000     0.558084749    -0.896178797
9       H      1.914664726    -0.557626660     0.893029008
10      H      1.915372043    -0.558518697    -0.891817988
11      H      1.791578296     1.975986182    -0.892368920
12      H      1.791586979     1.975640825     0.892667838
13      H      4.090845968     0.862646918    -0.898754755
14      H      4.091454330     0.858504831     0.895421454
15      H      5.121394650     2.755670513     0.002579183
16      H     -1.414452266    -1.338899435    -0.001165656
EXPGEOM
1       O      1.26500         2.81730         0.00000
2       O     -1.26500        -2.81730         0.00000
3       C      1.42690         1.35250         0.00000
4       C     -1.42690        -1.35250         0.00000
5       C      0.00000         0.77470         0.00000
6       C      0.00000        -0.77470         0.00000
7       H      2.17240         3.22290         0.00000
8       H     -2.17240        -3.22290         0.00000
9       H     -0.52250         1.15950         0.88830
10      H     -0.52250         1.15950        -0.88830
11      H      0.52250        -1.15950         0.88830
12      H      0.52250        -1.15950        -0.88830
13      H     -1.97010        -0.99800        -0.89350
14      H     -1.97010        -0.99800         0.89350
15      H      1.97010         0.99800        -0.89350
```

```
16      H      1.97010      0.99800      0.89350
---
H10C4O2,  RHF, CHARGE=0, MULT=1
HF=-46.1
1     C     -0.159262246    -0.378223518    -0.013759127
2     C      1.363410663    -0.303937216    -0.230495023
3     O      1.883711074     1.019946414    -0.230680804
4     O      1.788053457     1.610086396    -1.385773227
5     C      2.260790233     2.951797647    -1.370662748
6     C      1.193849033     3.981276360    -0.954206104
7     H      1.609607690     5.003373611    -1.070954853
8     H      0.886152356     3.862955321     0.103718464
9     H      0.284837724     3.912808815    -1.585070495
10      H      2.577312464     3.151107270    -2.427946523
11      H      3.173270002     3.075737935    -0.734344526
12      H      1.885930901    -0.808344422     0.623759154
13      H      1.641264245    -0.868725499    -1.156291061
14      H     -0.725606481     0.054648147    -0.861945983
15      H     -0.461955558    -1.441037686     0.083891906
16      H     -0.471534861     0.147218744     0.911148593
---
H10C4O2,  RHF, CHARGE=0, MULT=1
HF=-93.1
1     C     -0.000401989     0.001070940    -0.000020394
2     C      1.559606359    -0.001625423     0.000680381
3     O      2.033846354     1.324090722     0.001938822
4     C      3.021673225     1.756230657    -0.883773935
5     O      2.004970336    -0.601600068     1.190908387
6     C      3.210151279    -1.301025256     1.267434301
7     H     -0.378649082    -1.041126847     0.015618468
8     H     -0.430020018     0.538292549     0.867153109
9     H     -0.375903718     0.481276615    -0.926545014
10      H      1.908843056    -0.574872379    -0.911801848
11      H      4.005011044     1.250574680    -0.722686955
12      H      2.735778570     1.605893294    -1.955304110
13      H      3.161308654     2.849731832    -0.711485218
14      H      3.297009179    -2.100695175     0.489573721
15      H      4.105063916    -0.638899974     1.175693436
16      H      3.242609333    -1.787847972     2.271145713
EXPGEOM
1     O     -0.48540    -0.46630     1.11290
2     O     -0.48540    -0.46630    -1.11290
3     C     -0.48540     0.33070     2.27420
4     C     -0.48540     0.33070    -2.27420
5     C      0.13450     0.11970     0.00000
6     C      1.63510    -0.15690     0.00000
7     H      0.52610     0.52390     2.65830
8     H      0.52610     0.52390    -2.65830
9     H     -1.04170    -0.22170     3.03240
10      H     -1.04170    -0.22170    -3.03240
11      H     -0.98130     1.29790     2.10590
12      H     -0.98130     1.29790    -2.10590
13      H     -0.05130     1.21150     0.00000
14      H      1.78730    -1.23810     0.00000
15      H      2.11520     0.27080    -0.88490
16      H      2.11520     0.27080     0.88490
---
H8C5O2,  RHF, CHARGE=0, MULT=1
HF=-91.9, IE=9.15
```

```
  1    C     -0.112316922    -0.145540885     0.220183737
  2    C      1.404323946    -0.120145093     0.385827939
  3    C      2.124267126     1.239404544     0.395300292
  4    C      2.523338960     1.699049323     1.808117021
  5    C      4.010385338     1.822094924     2.124473766
  6    H      4.190265325     2.079410203     3.187223063
  7    H      4.465212916     2.618982754     1.501104994
  8    H      4.534772889     0.868553986     1.914429878
  9    O      1.663548260     1.976505976     2.637414519
 10    H      1.480889388     2.023191600    -0.063410463
 11    H      3.023127203     1.175677119    -0.257974056
 12    O      2.048219316    -1.156881375     0.510366272
 13    H     -0.489316201     0.762701108    -0.288803085
 14    H     -0.595574970    -0.206956918     1.216824130
 15    H     -0.438135478    -1.021491850    -0.376310633
EXPGEOM
  1    C      0.00000    0.00000    1.02770
  2    C      0.00000    1.22760    0.11050
  3    C      0.00000   -1.22760    0.11050
  4    C     -1.35760    1.74100   -0.31610
  5    C      1.35760   -1.74100   -0.31610
  6    O      1.05590    1.70850   -0.28460
  7    O     -1.05590   -1.70850   -0.28460
  8    H     -0.90210   -0.03000    1.65080
  9    H      0.90210    0.03000    1.65080
 10    H     -1.23920    2.49910   -1.09620
 11    H     -1.97070    0.90390   -0.67750
 12    H     -1.87570    2.17790    0.54980
 13    H      1.23920   -2.49910   -1.09620
 14    H      1.97070   -0.90390   -0.67750
 15    H      1.87570   -2.17790    0.54980
---
H10C5O2,  RHF, CHARGE=0, MULT=1
HF=-111.5
  1    C     -0.000029007    -0.000004086     0.000067973
  2    C      1.531491228    -0.000063521    -0.000085582
  3    C      2.195097158     1.381412373     0.000083440
  4    C      4.431250485     2.409089218    -0.025489976
  5    C      5.885047308     1.899225123    -0.051524507
  6    O      3.555375368     1.304534613    -0.021198862
  7    O      1.644304393     2.478727716     0.017381586
  8    H     -0.414767050     0.497034636     0.899510387
  9    H     -0.414561491     0.503074677    -0.896260610
 10    H     -0.368567934    -1.047007189    -0.003471750
 11    H      1.891547066    -0.559240114     0.894378072
 12    H      1.890017180    -0.560575040    -0.894540183
 13    H      4.277028576     3.046764930     0.883705060
 14    H      4.249398498     3.061280003    -0.919076890
 15    H      6.124589614     1.281403469     0.836699483
 16    H      6.098114707     1.297382475    -0.957247230
 17    H      6.571504303     2.770752818    -0.053886231
---
H10C5O2,  RHF, CHARGE=0, MULT=1
HF=-115.1
  1    C     -0.077174496     0.014144565    -0.000712325
  2    C      1.273547407    -0.645688871     0.256530702
  3    O      1.948893884    -0.964556935    -0.879775358
  4    C      3.236998596    -1.547395063    -0.964038906
  5    C      3.132784538    -3.090352775    -1.082102662
```

```
6    C    3.964596545   -0.893947133   -2.170197121
7    O    1.749152699   -0.898895805    1.360706084
8    H   -0.254263103    0.817091571    0.742998034
9    H   -0.147339119    0.463876687   -1.010430331
10   H   -0.881751086   -0.742652904    0.100251113
11   H    3.855429393   -1.318503586   -0.052795481
12   H    2.627960163   -3.533372776   -0.200868948
13   H    2.575015969   -3.406886092   -1.986071322
14   H    4.148666406   -3.531744590   -1.136807035
15   H    3.463969783   -1.107461444   -3.135178198
16   H    4.021818600    0.207429685   -2.055718180
17   H    5.003403738   -1.277751289   -2.231281338
---
H12C5O2, RHF, CHARGE=0, MULT=1
HF=-105.6
1    C   -0.000002163   -0.000001986   -0.000000219
2    C    1.550625753    0.000001584    0.000000417
3    C    2.156743745    1.418092385   -0.000103085
4    C    3.698650349    1.435160935   -0.002121989
5    C    4.293253268    2.867215395   -0.000833178
6    O   -0.468673130   -1.315832795    0.004536464
7    O    5.688117674    2.795779183   -0.004606567
8    H   -0.385311822    0.560600163    0.894754447
9    H   -0.385112808    0.554425841   -0.898705829
10   H    1.916175192   -0.556838740    0.892528295
11   H    1.915546942   -0.557401654   -0.892430465
12   H    1.784373162    1.974480012   -0.891001374
13   H    1.785772072    1.973509954    0.892206962
14   H    4.074575704    0.882654425    0.888875404
15   H    4.072080285    0.886181456   -0.896332057
16   H    3.923698581    3.438268968   -0.895622385
17   H    3.928459764    3.435179210    0.897865791
18   H   -1.415176210   -1.338242869    0.001632044
19   H    6.071490253    3.661445599   -0.001091476
---
H4C6O2, RHF, CHARGE=0, MULT=1
HF=-29.3
1    C   -0.003190502   -0.001644205    0.000000000
2    C    1.498062273   -0.003431844    0.000000000
3    C    2.217481608    1.138078500    0.000000000
4    C    1.568519074    2.491855041    0.000000000
5    C    0.067260028    2.493593941    0.000000000
6    C   -0.652322017    1.352076427    0.000000000
7    O   -0.657472237   -1.038761109    0.000000000
8    H   -0.425945695    3.468239150    0.000000000
9    H    1.990988903   -0.978452740    0.000000000
10   H   -1.744471968    1.376257475    0.000000000
11   H    3.309593198    1.113949406    0.000000000
12   O    2.222571265    3.529161395    0.000000000
---
H6C6O2, RHF, CHARGE=0, MULT=1
HF=-65.7
1    O    0.000029404   -0.000040793   -0.000081228
2    C    1.358385270    0.000030594    0.000037834
3    C    2.129916542    1.187436189   -0.000004044
4    C    3.534121826    1.122649451    0.000015393
5    C    4.186462879   -0.118186328    0.000561530
6    C    3.443514498   -1.311568169    0.000934278
7    C    2.027986402   -1.273659417    0.000170210
```

```
8      H      5.276000543   -0.163086681    0.000583275
9      H     -0.362976281    0.876024816   -0.000220339
10     H      1.647570764    2.166094487   -0.000057922
11     H      4.114866835    2.045559616   -0.000280833
12     H      3.976075937   -2.263872478    0.001669294
13     O      1.257587316   -2.392405466   -0.000699307
14     H      1.773244529   -3.188241340   -0.000642443
---
H6C6O2, RHF, CHARGE=0, MULT=1
HF=-66.2
1      O      0.000016244    0.000049267   -0.000003805
2      C      1.360377233   -0.000010628    0.000002177
3      C      2.139074316    1.183488408   -0.000002500
4      C      3.541168054    1.106044927   -0.000005600
5      C      4.184748801   -0.156021896    0.000009475
6      C      3.403178674   -1.345859840    0.000028588
7      C      2.006219660   -1.268685740    0.000019850
8      O      5.536858241   -0.305520348    0.000003038
9      H     -0.365941845    0.874184217   -0.000014322
10     H      1.668984045    2.167975043   -0.000004622
11     H      4.116815567    2.032776527   -0.000022643
12     H      3.884771331   -2.324739321    0.000057992
13     H      1.419763940   -2.188576307    0.000011482
14     H      5.996576888    0.523219148   -0.000008870
---
H6C6O2, RHF, CHARGE=0, MULT=1
HF=-68
1      C      0.000000000    0.000000000    0.000000000
2      C      1.419862642    0.000000000    0.000000000
3      C      2.107170217    1.224622184    0.000000000
4      C      1.423237274    2.447944436   -0.000056562
5      C      0.000265114    2.446990627   -0.000131406
6      C     -0.718806703    1.224356046    0.000032162
7      O     -0.737542595   -1.140213331   -0.000024439
8      O     -0.614342199    3.657198134   -0.000102331
9      H      1.986454369   -0.930838358   -0.000077972
10     H      3.198964925    1.222394929    0.000012971
11     H      1.981752047    3.383548554   -0.000071303
12     H     -1.809038673    1.215871792   -0.000125775
13     H     -0.205005566   -1.924391920   -0.000406183
14     H     -1.560201030    3.593304940   -0.000134288
---
H8C6O2, RHF, CHARGE=0, MULT=1
HF=-80.2
1      C      0.000000000    0.000000000    0.000000000
2      C      1.535832473    0.000000000    0.000000000
3      C      2.160354387    1.402030098    0.000000000
4      C      1.595045387    2.337672227   -1.074413935
5      C      0.264012285    1.967891939   -1.740461201
6      C     -0.643720148    1.056890348   -0.905886779
7      O     -1.863045669    1.159974979   -0.976666748
8      H      3.261699331    1.303078591   -0.132305453
9      H     -0.360375011   -1.006723656   -0.312854326
10     H     -0.369925308    0.155001293    1.039939248
11     H      1.910178830   -0.573950969   -0.878757866
12     H      1.895011232   -0.553741245    0.896941728
13     O      2.192465386    3.356533208   -1.400940385
14     H      0.491020916    1.439786327   -2.696937238
15     H     -0.284329659    2.896843496   -2.017489413
```

```
16     H     2.007772011    1.885720804    0.992737147
---
H8C6O2, RHF, CHARGE=0, MULT=1
HF=-79.5
1    C     0.000000000    0.000000000    0.000000000
2    C     1.536422240    0.000000000    0.000000000
3    C     2.149511993    1.406243866    0.000000000
4    C     1.536475428    2.419269384   -0.975260027
5    C     0.000070440    2.419255748   -0.976195188
6    C    -0.613752555    1.013429536   -0.974520593
7    O    -1.562574837    0.723962030   -1.692358523
8    O     3.097354710    1.696197746    0.718841579
9    H    -0.355005439   -1.029729078   -0.228661912
10     H    -0.379378608    0.237901081    1.020902955
11     H     1.916480357   -0.536608103   -0.900074215
12     H     1.890794475   -0.583770310    0.878900184
13     H     1.916781123    2.180643297   -1.995589242
14     H     1.891285772    3.449378251   -0.747844589
15     H    -0.354253409    3.002209892   -1.855577598
16     H    -0.380115015    2.956374758   -0.076402736
---
H10C6O2, RHF, CHARGE=0, MULT=1
HF=-105.1
1    C     0.094258710   -0.124832668   -0.034228604
2    C     1.339509999    0.308189704   -0.801653093
3    C     2.102868542    1.542680614   -0.291397293
4    C     3.522944462    1.215671575    0.207489107
5    C     4.696071827    1.774255027   -0.607778382
6    C     6.079390291    1.177357942   -0.337470581
7    O     1.718334125   -0.285755951   -1.805381112
8    O     3.674577817    0.565109581    1.234902235
9    H    -0.689540837    0.655993874   -0.111745907
10     H     0.331010132   -0.278708688    1.037473108
11     H    -0.329728482   -1.070866453   -0.425705609
12     H     4.467306541    1.646054285   -1.690783938
13     H     4.732613381    2.873323125   -0.418177225
14     H     6.822179232    1.633374117   -1.025127864
15     H     2.138229053    2.297472224   -1.108400436
16     H     1.541894051    2.029621668    0.537610002
17     H     6.095499299    0.082062319   -0.507157270
18     H     6.421037249    1.372863840    0.698692398
---
H10C6O2, RHF, CHARGE=0, MULT=1
HF=-94.7
1    C     0.000000000    0.000000000    0.000000000
2    C     1.538863413    0.000000000    0.000000000
3    C     2.232809357    1.374407950    0.000000000
4    C     2.012577083    2.246265569   -1.247663766
5    C     0.629960722    2.928911685   -1.421681330
6    O    -0.490816691    2.223723318   -0.961429260
7    C    -0.629851800    0.859768323   -1.097984354
8    O    -1.322575768    0.464981330   -2.025260530
9    H    -0.383171851    0.349649008    0.986108965
10     H    -0.352978341   -1.050512641   -0.111523563
11     H     1.902504772   -0.598206069   -0.867164721
12     H     1.873354302   -0.552552244    0.908997233
13     H     3.329822594    1.192458974    0.092898362
14     H     1.946228359    1.941054877    0.914941371
15     H     2.239882701    1.649405737   -2.160980530
```

```
16         H         2.770273193     3.064189393    -1.231712360
17         H         0.496726595     3.180288857    -2.507905044
18         H         0.621463905     3.901997481    -0.863945999
---
H10C6O2, RHF, CHARGE=0, MULT=1
HF=-102.5
1     C         0.170670268    -0.136348503     0.287461666
2     C         1.333598552     0.346194821    -0.576265339
3     C         2.219304009     1.541074843    -0.132189246
4     C         3.412159599     1.040606915     0.721703967
5     C         4.812387916     1.218004669     0.139565395
6     O         3.239464815     0.524597865     1.819830027
7     O         1.583234166    -0.201649766    -1.645561586
8     C         1.458755470     2.700831213     0.538278617
9     H        -0.107785397    -1.181717492     0.042656397
10        H         0.411477495    -0.104456633     1.367144908
11        H        -0.719593166     0.501298679     0.108380981
12        H         2.645988764     1.983586217    -1.069357515
13        H         5.011651987     2.288714587    -0.068097734
14        H         5.599311967     0.856535154     0.831362021
15        H         4.909591809     0.652108351    -0.808860554
16        H         1.014994777     2.429461895     1.515656921
17        H         2.142778085     3.557471423     0.710221654
18        H         0.640674185     3.059997647    -0.119069538
---
H10C6O2, RHF, CHARGE=0, MULT=1
HF=-89.8
1     C         0.000000000     0.000000000     0.000000000
2     C         1.540735902     0.000000000     0.000000000
3     O         2.000962361     1.332151596     0.000000000
4     C         3.311369209     1.707077613     0.009047192
5     O         4.193581972     0.849250241     0.022117370
6     C         3.494544628     3.187558147     0.002714207
7     C         4.693082515     3.808284258    -0.007612183
8     C         4.906214227     5.291394894    -0.013765779
9     H        -0.415878450     0.501738292    -0.896274742
10        H        -0.415563140     0.499195575     0.897894853
11        H        -0.359407317    -1.049621428    -0.001336994
12        H         1.912978849    -0.552289195    -0.901901566
13        H         1.912367643    -0.553707980     0.901180054
14        H         2.568333986     3.770419547     0.007777024
15        H         5.628130942     3.236440627    -0.012267557
16        H         5.482143805     5.589559613    -0.915115698
17        H         3.961297497     5.870838311    -0.011434922
18        H         5.490433874     5.595659621     0.880190775
---
H12C6O2, RHF, CHARGE=0, MULT=1
HF=-72.4
1     C        -0.005602881    -0.006894125     0.001643001
2     O         1.391251473     0.010916216     0.040601506
3     C         2.127890290     1.204812100     0.015873161
4     O         1.972069121     1.751691091    -1.276119625
5     C         1.889727561     3.130690723    -1.467969711
6     C         3.593805620     0.838391030     0.278628812
7     C         4.245901236     1.083931104     1.429487265
8     C         5.673268952     0.737087005     1.720875799
9     H        -0.471513587     0.690708887     0.741601453
10        H        -0.413832920     0.240443432    -1.007897718
11        H        -0.322622300    -1.046242472     0.258009052
```

```
12     H      1.755158669     1.923120627     0.810814692
13     H      0.993318596     3.580483173    -0.972068541
14     H      2.792902582     3.675031374    -1.096588701
15     H      1.803175517     3.305909056    -2.566669481
16     H      4.093785695     0.335108814    -0.554526319
17     H      3.741028919     1.578134059     2.267412019
18     H      6.184205680     0.240705032     0.871399650
19     H      5.731290610     0.053075700     2.593859237
20     H      6.249183431     1.653722221     1.969215492
---
H12C6O2, RHF, CHARGE=0, MULT=1
HF=-116.2
1    C     0.000000000     0.000000000     0.000000000
2    C     1.528299293     0.000000000     0.000000000
3    O     2.157618513     1.053297768     0.000000000
4    C     2.229135851    -1.367298362    -0.053619086
5    C     3.379286936    -1.676773408     0.983475805
6    O     4.526768571    -0.915069132     0.708448794
7    C     2.892582891    -1.465422073     2.454954003
8    C     3.823257243    -3.169895402     0.796085964
9    H    -0.385148590    -0.511436637     0.905189558
10    H    -0.384489831    -0.531952645    -0.893661942
11    H    -0.419841550     1.025470205    -0.011846766
12    H     1.456787339    -2.161353701     0.055515170
13    H     2.632610647    -1.473779252    -1.088215004
14    H     4.436338281    -0.002689680     0.942808921
15    H     2.627678053    -0.410209959     2.667100230
16    H     1.995467135    -2.080761081     2.667514485
17    H     3.675762891    -1.755101221     3.184607072
18    H     2.983773694    -3.861155907     1.013254990
19    H     4.164864986    -3.371125208    -0.238927781
20    H     4.656642037    -3.433667886     1.477881904
---
H12C6O2, RHF, CHARGE=0, MULT=1
HF=-100.7
1    C     0.000000000     0.000000000     0.000000000
2    C     1.576092571     0.000000000     0.000000000
3    O     2.123865092     1.289846803     0.000000000
4    C     1.660359499     2.233289288    -0.927636758
5    O     0.260101354     2.299982796    -0.907094697
6    C    -0.487169069     1.117960811    -0.998555752
7    H     2.056078207     3.239768521    -0.620374142
8    H    -1.550965095     1.399750148    -0.796359096
9    C    -0.543799591     0.274180083     1.429782782
10    C    -0.518394840    -1.382691348    -0.477323969
11    H     1.960951090    -0.580074880    -0.881656299
12    H     1.976768963    -0.513991419     0.909480735
13    H     2.045282681     2.010207116    -1.967038590
14    H    -0.462271458     0.729689096    -2.052428379
15    H    -1.653241198     0.276539201     1.445315827
16    H    -0.208662037     1.251685974     1.829430333
17    H    -0.208211078    -0.505020621     2.144842370
18    H    -1.626851953    -1.412707574    -0.504201059
19    H    -0.159949727    -1.633007581    -1.496887608
20    H    -0.180951416    -2.196592580     0.196483873
---
H12C6O2, RHF, CHARGE=0, MULT=1
HF=-102.3
1    C     0.000000000     0.000000000     0.000000000
```

```
2    C     1.549126807    0.000000000    0.000000000
3    O     2.094638557    1.291305273    0.000000000
4    C     1.578019972    2.253013856   -0.888723743
5    O     0.178888663    2.324883972   -0.755230998
6    C    -0.587247052    1.149001694   -0.874197494
7    C    -2.059312214    1.495683659   -0.540289234
8    C     2.197712489    3.642239606   -0.562614496
9    H    -0.365853686    0.076476015    1.047483147
10   H    -0.362750048   -0.977130010   -0.386081111
11   H     1.931962144   -0.590432731   -0.875355891
12   H     1.933469035   -0.510843617    0.918846106
13   H     1.854323082    1.981372048   -1.956955434
14   H    -0.570518078    0.797759972   -1.947663649
15   H    -2.691483714    0.591358084   -0.654703698
16   H    -2.183517508    1.872221305    0.494125276
17   H    -2.452115326    2.267315253   -1.232696852
18   H     1.810280497    4.403453803   -1.269397461
19   H     1.972162562    3.979661765    0.467663776
20   H     3.299339935    3.600625336   -0.680094616
---
H12C6O2, RHF, CHARGE=0, MULT=1
HF=-115.9
1    C     0.000000000    0.000000000    0.000000000
2    C     1.540854144    0.000000000    0.000000000
3    O     2.001571081    1.332276442    0.000000000
4    C     3.311240003    1.709992047    0.001809096
5    C     3.476181655    3.235170579    0.007218801
6    C     4.933274666    3.735856071    0.013459966
7    C     5.069485001    5.261960317    0.021883752
8    O     4.192661698    0.855217558   -0.001887037
9    H    -0.359480122   -1.049428535    0.016264387
10   H    -0.414959704    0.487149128   -0.904696632
11   H    -0.415894259    0.514039186    0.889300315
12   H     1.912712396   -0.552995779   -0.901580322
13   H     1.913141960   -0.552972589    0.901370403
14   H     2.944553299    3.633301730    0.902257975
15   H     2.949737021    3.638789461   -0.888503361
16   H     5.468554506    3.339952403   -0.879792249
17   H     5.464191602    3.330792200    0.905182704
18   H     6.141556029    5.547029343    0.028499598
19   H     4.598601036    5.714177371    0.918168100
20   H     4.607154797    5.723120972   -0.874292793
---
H12C6O2, RHF, CHARGE=0, MULT=1
HF=-122.5
1    C     0.000000000    0.000000000    0.000000000
2    C     1.531342633    0.000000000    0.000000000
3    C     2.173410059    1.400826005    0.000000000
4    C     3.716437596    1.383376384    0.000651313
5    C     4.345133749    2.791508862    0.000737929
6    C     5.876813084    2.819197401    0.000338199
7    O     6.396002282    4.076173623   -0.010650087
8    O     6.657987897    1.868686923    0.009192352
9    H    -0.412942611    0.505872796   -0.896175570
10   H    -0.412790512    0.505874838    0.896254684
11   H    -0.381740614   -1.041819102    0.000039541
12   H     1.883448003   -0.569715442   -0.890854610
13   H     1.883217301   -0.569699831    0.891013051
14   H     1.814437817    1.964453329   -0.891770041
```

```
15      H       1.814027771     1.964275556     0.891745760
16      H       4.074379611     0.822643981    -0.892516618
17      H       4.073536923     0.822670797     0.894181946
18      H       3.994687097     3.355954808    -0.894184280
19      H       3.996024565     3.355391588     0.896613514
20      H       7.343889936     4.127491755    -0.010764481
EXPGEOM
1       C       4.07840        -0.21720         0.00010
2       C       2.73660         0.52630        -0.00010
3       C       1.52950        -0.42170         0.00010
4       C       0.18630         0.31860        -0.00010
5       C      -1.00910        -0.63640         0.00010
6       C      -2.34090         0.08680        -0.00010
7       O      -3.37670        -0.78570         0.00010
8       O      -2.49440         1.28300        -0.00020
9       H       4.92970         0.48630        -0.00000
10      H       4.17250        -0.86280         0.89290
11      H       4.17260        -0.86320        -0.89240
12      H       2.67780         1.18840        -0.88630
13      H       2.67770         1.18870         0.88580
14      H       1.58590        -1.08540         0.88700
15      H       1.58590        -1.08580        -0.88650
16      H       0.12100         0.98120        -0.88330
17      H       0.12100         0.98160         0.88290
18      H      -0.99180        -1.30430         0.88210
19      H      -0.99180        -1.30470        -0.88160
20      H      -4.17660        -0.23680        -0.00000
---
H12C6O2, RHF, CHARGE=0, MULT=1
HF=-117.7
1       C      -0.065358427    -0.034659093    -0.052157929
2       C       1.412708234    -0.040623888     0.350871095
3       C       1.990392084     1.321602356     0.826072638
4       C       3.534432291     1.283917401     0.802790894
5       O       4.087656619     2.246157449     0.013923522
6       C       5.464235676     2.464485874    -0.159077170
7       O       4.261294783     0.502975505     1.411330992
8       C       1.496924773     1.776851726     2.216417773
9       H      -0.280257395     0.722222573    -0.833275432
10      H      -0.734849313     0.164630564     0.808202905
11      H      -0.344856818    -1.026732382    -0.463734048
12      H       1.999143289    -0.385527432    -0.532967560
13      H       1.561897094    -0.812398821     1.139853743
14      H       1.648643457     2.098434254     0.096623772
15      H       5.976020125     2.721728406     0.798721812
16      H       5.978586394     1.586246921    -0.616298550
17      H       5.561894435     3.331841629    -0.854884008
18      H       1.693718510     1.022997537     3.004146661
19      H       1.990900352     2.722311280     2.521053647
20      H       0.405895033     1.972951739     2.200324317
---
H12C6O2, RHF, CHARGE=0, MULT=1
HF=-118.2
1       C       0.014182251     0.028424717    -0.025562962
2       C       1.570635288     0.011070463     0.011387054
3       C      -0.513618676     1.492390651     0.018490832
4       C      -0.463114058    -0.639141958    -1.351721697
5       C      -0.541906295    -0.762031745     1.194480620
6       O      -0.188844672    -0.247364843     2.405654867
```

```
 7    C     -0.555555633    -0.775615138     3.654173073
 8    O     -1.241964456    -1.770566798     1.159708358
 9    H      1.979006816     0.569148530     0.877124015
10    H      1.962199793    -1.024915808     0.067138736
11    H      1.993587568     0.477131882    -0.902129126
12    H     -0.122945634     2.055608597     0.889227775
13    H     -1.621003403     1.521185732     0.071022318
14    H     -0.209632437     2.050943014    -0.890452275
15    H     -0.104745607    -1.684298491    -1.440228314
16    H     -1.568411224    -0.651849041    -1.433343674
17    H     -0.076733484    -0.086387636    -2.232943137
18    H     -1.661245309    -0.784763830     3.803319764
19    H     -0.163777776    -1.808925640     3.808785518
20    H     -0.101775858    -0.107551126     4.424837963
---
H12C6O2, RHF, CHARGE=0, MULT=1
HF=-119
 1    C     -0.077914444     0.080652719     0.222145120
 2    O      1.296666738     0.368895398     0.197806676
 3    C      1.905781634     1.217895903    -0.674759778
 4    C      3.430511429     1.284073141    -0.521802507
 5    C      3.979366884     2.649130145    -0.023101834
 6    C      3.612546495     2.956539597     1.443222579
 7    C      5.501514349     2.748608519    -0.264222223
 8    O      1.233522137     1.821275620    -1.507745492
 9    H     -0.421622202    -0.412317464    -0.718490735
10    H     -0.700533797     0.989596105     0.396379269
11    H     -0.237897397    -0.625171803     1.071305723
12    H      3.851198495     1.061384916    -1.530820874
13    H      3.788055297     0.467119694     0.143698916
14    H      3.514313770     3.453321451    -0.648623554
15    H      2.514126144     3.015778451     1.582022141
16    H      4.000280262     2.188118644     2.142306666
17    H      4.028650071     3.936102033     1.757282659
18    H      6.063975880     1.960658424     0.276172155
19    H      5.740779680     2.655866500    -1.343377485
20    H      5.891464618     3.731389625     0.070675524
---
H12C6O2, RHF, CHARGE=0, MULT=1
HF=-112.7
 1    C      0.000000000     0.000000000     0.000000000
 2    C      1.531383796     0.000000000     0.000000000
 3    C      2.171668498     1.402672294     0.000000000
 4    C      3.713529428     1.374447676    -0.002052106
 5    C      4.387939765     2.751936112     0.002700951
 6    O      5.747746291     2.663276989     0.047526177
 7    C      6.634271823     3.751924674     0.061209045
 8    O      3.854189148     3.857169626    -0.030880934
 9    H     -0.412487545     0.505533703    -0.896430407
10    H     -0.412492748     0.505914688     0.896218760
11    H     -0.380954142    -1.042140873     0.000145968
12    H      1.883595098    -0.569825143     0.890874575
13    H      1.883441618    -0.569835421    -0.890959587
14    H      1.812919865     1.964619411    -0.892138921
15    H      1.816564064     1.963095220     0.894588884
16    H      4.075796753     0.812316007     0.889444801
17    H      4.074466172     0.823693027    -0.901360794
18    H      6.551228115     4.377830815    -0.858885301
19    H      6.489867464     4.405146643     0.954390572
```

```
20     H     7.663372785    3.321449403    0.102820810
---
H12C6O2, RHF, CHARGE=0, MULT=1
HF=-123.4
1      C     0.000000000    0.000000000    0.000000000
2      C     1.562465098    0.000000000    0.000000000
3      C    -0.559542629    1.458631877    0.000000000
4      C    -0.518415639   -0.754455318   -1.276216850
5      O    -0.527039202   -0.768736633    1.074207256
6      C    -0.420815937   -0.605434176    2.418308771
7      O     0.213804200    0.325478402    2.905929543
8      C    -1.162556106   -1.686397189    3.200707181
9      H     1.991595547    0.624989148    0.806902039
10     H     1.971882446   -1.023965798    0.115653892
11     H     1.945392380    0.404913721   -0.959145217
12     H    -1.660622923    1.475278591    0.129788780
13     H    -0.119050083    2.086997568    0.798193311
14     H    -0.330682915    1.956632799   -0.964463802
15     H    -0.153528497   -1.800513673   -1.312780091
16     H    -1.625817664   -0.785325016   -1.314413883
17     H    -0.167036616   -0.246188385   -2.196939446
18     H    -2.243842965   -1.674854825    2.957055387
19     H    -0.757367066   -2.690417024    2.963050803
20     H    -1.053800852   -1.520292862    4.291129295
---
H12C6O2, RHF, CHARGE=0, MULT=1
HF=-98.2
1      C     0.000000000    0.000000000    0.000000000
2      C     1.573551159    0.000000000    0.000000000
3      O     2.075939867    1.316326731    0.000000000
4      C     1.585569572    2.249725190   -0.922245431
5      O     0.186646488    2.312008179   -0.861509253
6      C    -0.550041897    1.122944956   -0.938777979
7      H     1.986669320    3.259067312   -0.631003919
8      H    -1.599433324    1.381018652   -0.646045344
9      H    -0.342318030    0.250447570    1.035539265
10     C    -0.617202299   -1.366573435   -0.347904623
11     H     1.923042726   -0.529773078   -0.933376740
12     C     2.219532447   -0.732224221    1.205013344
13     H     1.941996837    2.022603818   -1.970832328
14     H    -0.592781675    0.763912753   -2.002084589
15     H     1.965979161   -1.811042484    1.180009280
16     H     1.884024404   -0.323318415    2.178462069
17     H     3.325293750   -0.655633673    1.167775458
18     H    -0.273027374   -1.750114945   -1.329640290
19     H    -0.367302941   -2.128645345    0.417397375
20     H    -1.724354051   -1.299842998   -0.383964007
---
H14C6O2, RHF, CHARGE=0, MULT=1
HF=-108.4
1      C    -0.154265887    0.285031572   -0.339341545
2      C     1.287241668   -0.205852011   -0.004923344
3      O     2.016028805    0.868792213    0.528958836
4      C     3.415417948    0.943938353    0.440736197
5      C     3.851993445    2.337514799   -0.054264045
6      O     1.359939460   -1.309403426    0.861495836
7      C     0.906258263   -1.320135275    2.188295653
8      C     0.810081446   -2.777960762    2.681359390
9      H    -0.782548905   -0.570781645   -0.658360329
```

```
10      H      -0.658687816     0.787383475     0.508504503
11      H      -0.115929004     1.005538454    -1.181395594
12      H       1.744252082    -0.586524132    -0.971443050
13      H       3.849381553     0.759197102     1.459575843
14      H       3.857553509     0.163655667    -0.231355221
15      H       3.537009906     2.518987186    -1.101611522
16      H       3.438767103     3.151677172     0.573437013
17      H       4.958281022     2.407870416    -0.017735347
18      H       1.606776228    -0.752388267     2.856158638
19      H      -0.097003502    -0.832543552     2.302716810
20      H       0.048115632    -3.354248173     2.119431797
21      H       1.777281455    -3.311931416     2.595995873
22      H       0.514932909    -2.779660694     3.751004910
---
H14C6O2, RHF, CHARGE=0, MULT=1
HF=-101.7
1       C      -0.071432022    -0.101205954     0.099465453
2       O       1.315637962     0.043249303     0.174715202
3       C       1.951733506     1.287369528     0.051039787
4       O       1.931925826     1.619574175    -1.319845687
5       C       1.385791729     2.821826196    -1.769998710
6       C       3.439963191     1.150320757     0.537916826
7       C       3.602637428     0.933183291     2.054374753
8       C       5.055414774     0.956130900     2.539908374
9       H      -0.623583197     0.637685272     0.732145422
10      H      -0.457806160    -0.014958732    -0.945181039
11      H      -0.312588088    -1.124976829     0.473782658
12      H       1.433945579     2.075776645     0.679144676
13      H       0.294455555     2.914506153    -1.543938127
14      H       1.898364412     3.717515281    -1.337970708
15      H       1.514389033     2.847628174    -2.878007187
16      H       3.969758185     2.087949371     0.249737396
17      H       3.942514265     0.323480917    -0.011645730
18      H       3.036744725     1.717316139     2.609195053
19      H       3.151388656    -0.041661808     2.348384028
20      H       5.668766713     0.176545662     2.044969888
21      H       5.537519917     1.937315342     2.353261290
22      H       5.095458509     0.768233387     3.632788836
---
H14C6O2, RHF, CHARGE=0, MULT=1
HF=-97.6
1       H       0.000000000     0.000000000     0.000000000
2       C       1.108163872     0.000000000     0.000000000
3       C       1.683120424     1.428979983     0.000000000
4       O       1.259874732     2.106523204    -1.151668421
5       C       1.669447507     3.432266828    -1.355351846
6       C       1.052234140     3.916218307    -2.707539236
7       O       1.462103467     5.241635024    -2.912563948
8       C       1.040239962     5.918168454    -4.065341897
9       C       1.615408243     7.346857739    -4.066330335
10      H       1.452343011    -0.584242507    -0.876498217
11      H       1.448114851    -0.527614172     0.914830199
12      H       1.350665282     1.962226508     0.931073401
13      H       2.803958064     1.377436182     0.053289236
14      H       1.326128437     4.095210499    -0.517770638
15      H       2.787922969     3.510290032    -1.393281145
16      H       1.393098657     3.252749910    -3.545672283
17      H      -0.066457941     3.839479274    -2.669079111
18      H       1.372843974     5.383853488    -4.995844359
```

```
19      H       -0.080594457    5.970716420    -4.119419025
20      H        1.270508788    7.932778567    -3.191207139
21      H        2.723565750    7.347497104    -4.065620194
22      H        1.276302417    7.873701813    -4.981996343
---
H14C6O2, RHF, CHARGE=0, MULT=1
HF=-109.8
1       O        0.000000000    0.000000000     0.000000000
2       C        1.396863220    0.000000000     0.000000000
3       C        1.916847565    1.460648056     0.000000000
4       C        3.455854071    1.558385414    -0.000111857
5       C        3.989789008    3.005599002    -0.000095513
6       C        5.528706453    3.103167256    -0.000077978
7       C        6.048289034    4.564078640    -0.000232633
8       O        7.445193979    4.564614454    -0.000840051
9       H       -0.338846730   -0.883994709     0.000339317
10      H        1.793031124   -0.549365904     0.896876287
11      H        1.792728452   -0.549733102    -0.896735359
12      H        1.513612690    1.991091879     0.892552052
13      H        1.513231292    1.991251394    -0.892284200
14      H        3.856586611    1.023479570     0.891501628
15      H        3.856145614    1.022989227    -0.891672501
16      H        3.589301967    3.541044119     0.891316274
17      H        3.590063578    3.540618934    -0.892120411
18      H        5.932004403    2.572887348     0.892479721
19      H        5.931902879    2.572450865    -0.892456472
20      H        5.653506986    5.111840967     0.898353107
21      H        5.651421400    5.114629797    -0.895921844
22      H        7.782456107    5.449116102    -0.011702340
---
H14C6O2, RHF, CHARGE=0, MULT=1
HF=-129.2
1       C        0.000000000    0.000000000     0.000000000
2       C        1.569636012    0.000000000     0.000000000
3       C        2.189023005    1.500271316     0.000000000
4       C        3.758860331    1.501444131     0.002919287
5       C        1.707742570    2.349259287     1.228991867
6       C        2.049900896   -0.847353708    -1.230596208
7       O        2.039700913   -0.589101179     1.191841124
8       O        1.720499292    2.088494498    -1.192872673
9       H       -0.414775673    0.527604426    -0.879982640
10      H       -0.425344183    0.469938413     0.908594008
11      H       -0.395958416   -1.037800355    -0.029982426
12      H        4.171538086    0.978166695     0.886487770
13      H        4.186302339    1.027127885    -0.902354399
14      H        4.154509424    2.539429185     0.028636250
15      H        0.611273055    2.507771879     1.233562270
16      H        1.985546204    1.878689563     2.191434264
17      H        2.169682446    3.359765869     1.218752927
18      H        1.768567106   -0.377741490    -2.192435357
19      H        3.146638072   -1.003615227    -1.237721125
20      H        1.589935624   -1.858808727    -1.219082262
21      H        1.745768782   -1.483328034     1.292250175
22      H        2.018952529    2.980898321    -1.296120654
---
H6C7O2, RHF, CHARGE=0, MULT=1
HF=-25.3
1       O        0.000000000    0.000000000     0.000000000
2       C        1.223406387    0.000000000     0.000000000
```

```
 3    C     2.047776793    1.239065116    0.000000000
 4    C     3.401625296    1.242781379    0.000503415
 5    C     4.238857844    2.438088421    0.000398783
 6    C     3.970050784    3.813696669    0.000513339
 7    C     5.245147930    4.479966017    0.000253222
 8    C     6.201821200    3.469688734    0.000002080
 9    O     5.602202029    2.247473167    0.000098033
10    H     1.789376288   -0.956351014   -0.000013803
11    H     1.480013202    2.174025176   -0.000532569
12    H     3.949184248    0.292026626    0.000934350
13    H     3.004137667    4.295647810    0.000761073
14    H     5.415605761    5.545177448    0.000230901
15    H     7.286163850    3.488871221   -0.000247778
---
H6C7O2, RHF, CHARGE=0, MULT=1
HF=-70.1, IE=9.8
 1    C    -0.089357672   -0.025368582   -0.010893633
 2    C     2.754198829    0.014358701    0.032433266
 3    C     0.601689198    1.196175739    0.052104789
 4    C     0.635534415   -1.228416300   -0.051867560
 5    C     2.007593059    1.218466422    0.073175816
 6    C     2.040728705   -1.211074679   -0.030532606
 7    C     4.250810177   -0.010818550    0.051597164
 8    O     4.967150320   -1.013257689   -0.005655093
 9    O     4.870592525    1.196724639    0.141604836
10    H    -1.179577380   -0.040011714   -0.028148879
11    H     0.047047699    2.134647915    0.084812466
12    H     0.107900880   -2.181571878   -0.100745329
13    H     2.500927430    2.190765751    0.121193343
14    H     2.568182344   -2.166493120   -0.064000185
15    H     5.819392545    1.164703048    0.151591762
EXPGEOM
 1    C     0.00000    0.21950    0.00000
 2    C     1.28000   -0.34800    0.00000
 3    C     1.42350   -1.73430    0.00000
 4    C     0.29080   -2.55370    0.00000
 5    C    -0.98580   -1.98600    0.00000
 6    C    -1.13630   -0.59880    0.00000
 7    C    -0.10020    1.70540    0.00000
 8    O     0.84710    2.46670    0.00000
 9    O    -1.38270    2.14980    0.00000
10    H     2.14690    0.30830    0.00000
11    H     2.41800   -2.17660    0.00000
12    H     0.40350   -3.63680    0.00000
13    H    -1.86700   -2.62510    0.00000
14    H    -2.12660   -0.15160    0.00000
15    H    -1.32180    3.12510    0.00000
---
H6C7O2, RHF, CHARGE=0, MULT=1
HF=-51.6
 1    C     0.000000000    0.000000000    0.000000000
 2    C     1.405313148    0.000000000    0.000000000
 3    C     2.119805068    1.210925730    0.000000000
 4    C     1.429613244    2.436271012   -0.001691242
 5    C     0.024617535    2.462510013   -0.000851111
 6    C    -0.692522467    1.238297930    0.003992456
 7    O    -2.062302543    1.251798425   -0.095410377
 8    C    -2.845070031    1.262895854    1.029431461
 9    O    -4.051898662    1.276392866    0.842630758
```

```
10      H      -0.539829006    -0.947459361    -0.010293345
11      H       1.943833047    -0.948609855    -0.003126565
12      H       3.210167523     1.200003051    -0.001055505
13      H       1.986517016     3.374095900    -0.006153884
14      H      -0.496575522     3.420421790    -0.011248596
15      H      -2.344918171     1.259343573     2.017385732
---
H6C7O2, RHF, CHARGE=0, MULT=1
HF=-37.2
1       C      -0.030010819    -0.031300380     0.049554096
2       C       1.446768722    -0.127668748     0.387686005
3       C       2.340754059     0.905009013     0.285413407
4       C       2.041090821     2.284432105    -0.055507122
5       C       0.863889470     2.945240608     0.095854737
6       C      -0.382795532     2.409733681     0.623550771
7       C      -0.777701202     1.115538630     0.637497536
8       O      -0.568417610    -0.858036499    -0.675720349
9       O       1.783392128    -1.376324913     0.791280213
10      H       3.404433367     0.720460298     0.472714817
11      H       2.899768804     2.847459760    -0.440035204
12      H       0.824437398     4.006667706    -0.172058929
13      H      -1.060105582     3.164159785     1.041389340
14      H      -1.741578426     0.848765492     1.081881555
15      H       2.705953436    -1.475997333     0.988108863
---
H8C7O2, RHF, CHARGE=0, MULT=1
HF=-71.5
1       C       0.000272662    -0.000035685     0.000092418
2       C       1.430127559     0.000027916     0.000192925
3       C       2.095906240     1.247228677    -0.001656311
4       C       1.393779234     2.460935057    -0.007256208
5       C      -0.008243967     2.470286843    -0.006606955
6       C      -0.715633935     1.246244190    -0.002033291
7       C       2.231691965    -1.276847768     0.001254214
8       O      -0.617245423    -1.209009349     0.002279111
9       O      -2.078326533     1.186008495     0.000307786
10      H       3.187334038     1.280678271     0.000277281
11      H       1.939853956     3.404937846    -0.012050438
12      H      -0.535938634     3.425002171    -0.010341046
13      H       2.017310517    -1.887194528    -0.900934317
14      H       2.009141059    -1.892655085     0.897589254
15      H       3.324179814    -1.081526446     0.006908389
16      H      -1.563204239    -1.168032182    -0.006342414
17      H      -2.482641606     2.042449854    -0.001932830
---
H8C7O2, RHF, CHARGE=0, MULT=1
HF=-71.3
1       C       0.000000000     0.000000000     0.000000000
2       C       1.414349673     0.000000000     0.000000000
3       C       2.129504813     1.209386983     0.000000000
4       C       1.466399258     2.453181073     0.000612101
5       C       0.051360334     2.461716787     0.001456947
6       C      -0.693548035     1.259380488     0.000607406
7       H       1.971973530    -0.938069750     0.000008579
8       O      -0.743467716    -1.136041395    -0.000398220
9       O      -2.052162741     1.230655475     0.000359855
10      H       3.220132487     1.164375738    -0.000362913
11      C       2.228774687     3.751230528     0.000817025
12      H      -0.472517771     3.420179661     0.002894457
```

```
13         H          -0.209568303    -1.919786928     0.000573559
14         H          -2.434400232     2.098451008     0.000766811
15         H           1.982949822     4.355903159     0.899363945
16         H           1.980290854     4.357892380    -0.895598812
17         H           3.326884740     3.597502659    -0.000687747
---
H10C7O2, RHF, CHARGE=0, MULT=1
HF=-69.2
1     C     0.536470272    -0.158268416     0.584058317
2     C     1.163624841     1.025646048     0.637610146
3     C     1.799643602     1.725692397    -0.501091375
4     C     1.158167571     2.638696580    -1.255438450
5     C     3.237235824     1.358398309    -0.761237678
6     O     3.629396494     0.428014396    -1.459040178
7     O     4.120696656     2.178183122    -0.129689558
8     C     5.523169051     2.036666293    -0.166976286
9     C     6.166495774     3.181127491     0.639305698
10        H     0.425726905    -0.752591422    -0.321246231
11        H     0.087442168    -0.616398979     1.464792401
12        H     1.237138311     1.568869323     1.586449063
13        H     1.624252339     3.178064192    -2.079250684
14        H     0.114002857     2.906903706    -1.098229072
15        H     5.835865923     1.051673103     0.267904721
16        H     5.903918445     2.065463654    -1.220961274
17        H     5.927359950     4.174981111     0.211084473
18        H     5.843856095     3.176034648     1.699438658
19        H     7.269078578     3.059350095     0.619505260
---
H10C7O2, RHF, CHARGE=0, MULT=1
HF=-59.8
1     C     0.000000000     0.000000000     0.000000000
2     C     1.534132156     0.000000000     0.000000000
3     C     2.129740614     1.323627228     0.000000000
4     C     2.636209029     2.409751959     0.000138418
5     C     3.224044886     3.720608928     0.000090498
6     O     4.585859714     3.686516759    -0.013621698
7     C     5.416291217     4.826166984    -0.018036062
8     C     6.889305319     4.374144931    -0.038370125
9     O     2.614271387     4.787253025     0.010950715
10        H    -0.406821117     0.508485866     0.896861319
11        H    -0.406846327     0.508407636    -0.896886449
12        H    -0.379532139    -1.042032127     0.000123459
13        H     1.903879480    -0.560186920     0.892795643
14        H     1.904093722    -0.558102580    -0.893929558
15        H     5.212417881     5.466584210    -0.915476725
16        H     5.233323873     5.459631953     0.888439874
17        H     7.150849393     3.769936938     0.852925167
18        H     7.128229440     3.777638682    -0.941152214
19        H     7.540187159     5.272535855    -0.042802296
---
H10C7O2, RHF, CHARGE=0, MULT=1
HF=-56.8
1     C     0.000000000     0.000000000     0.000000000
2     C     1.444906377     0.000000000     0.000000000
3     C     2.643830274    -0.003329598     0.000000000
4     C     4.090761236     0.023622505     0.003390886
5     C     4.743195606    -1.368022294    -0.001706261
6     O     6.104092851    -1.298431561    -0.004600481
7     C     6.973384385    -2.408353817    -0.008767466
```

```
8    C     8.430300880   -1.906385661   -0.011250371
9    O     4.182432613   -2.459177889   -0.002958289
10   H    -0.396770535   -0.505730809   -0.906369741
11   H    -0.396604670   -0.531930970    0.891377757
12   H    -0.396857018    1.037979785    0.015171489
13   H     4.448823979    0.590228322   -0.890188445
14   H     4.444891575    0.580922824    0.904447440
15   H     6.804256301   -3.052798739    0.892965473
16   H     6.799528912   -3.050962424   -0.910647453
17   H     8.658325874   -1.296630127   -0.907991541
18   H     8.661874570   -1.299037272    0.886267942
19   H     9.111803491   -2.781772613   -0.013700489
---
H10C7O2, RHF, CHARGE=0, MULT=1
HF=-55.7
1    C     0.000000000    0.000000000    0.000000000
2    C     1.196913820    0.000000000    0.000000000
3    C     2.647928457    0.034158534    0.000000000
4    C     3.274195803   -1.376191554   -0.078703861
5    C     4.807365332   -1.405752288   -0.079833948
6    O     5.299732622   -2.675359839   -0.124172133
7    O     5.579395052   -0.451357841   -0.048081066
8    C     6.667630994   -3.016306297   -0.137634759
9    H    -1.050942900   -0.001074206   -0.000078408
10   H     2.991464348    0.653782214   -0.862568122
11   H     2.993951587    0.554881385    0.924763881
12   H     2.916010789   -1.984479630    0.783434999
13   H     2.919030235   -1.883901751   -1.004950010
14   H     7.184113749   -2.566251717   -1.025133588
15   H     7.186902198   -2.627373857    0.776666375
16   C     6.803725859   -4.550185333   -0.189906585
17   H     7.880954665   -4.815139101   -0.200761035
18   H     6.343334639   -5.038383066    0.691972051
19   H     6.340250320   -4.977496888   -1.101297058
---
H12C7O2, RHF, CHARGE=0, MULT=1
HF=-104.9
1    C    -0.202475005    0.058848680   -0.348664882
2    C     1.145545504    0.344754990    0.317838720
3    C     2.171425782    1.143227696   -0.495901841
4    C     3.649927074    1.013181953   -0.085336777
5    C     4.291916831    2.344316968    0.348399393
6    C     5.475688850    2.847145907   -0.486727452
7    C     6.242370577    4.061697054    0.041582604
8    O     1.858158553    1.864005480   -1.436309140
9    O     3.860427538    2.940891314    1.327799472
10   H    -0.081522007   -0.523528410   -1.284151573
11   H    -0.756817402    0.987742301   -0.587803736
12   H    -0.836674769   -0.540403919    0.337321650
13   H     0.971101775    0.908501612    1.264194596
14   H     1.600500321   -0.630158287    0.608024941
15   H     3.766036656    0.290440389    0.753036521
16   H     4.203598012    0.569268470   -0.942726597
17   H     6.194683693    2.003392767   -0.605966514
18   H     5.089049418    3.084240848   -1.505434172
19   H     5.589231103    4.949875188    0.152328712
20   H     6.715882957    3.854747645    1.022173390
21   H     7.051603894    4.328345511   -0.669808323
---
```

```
H12C7O2, RHF, CHARGE=0, MULT=1
HF=-105.1
1    C     0.309333670     0.456865849     0.072850759
2    C     1.565228937    -0.230739867    -0.460431294
3    O     2.356554048     0.367794149    -1.179979945
4    C     1.794774067    -1.707502066    -0.048733055
5    C     2.857767474    -1.771646086     1.084050363
6    C     4.351943866    -1.747005338     0.767977025
7    O     2.473583917    -1.852949573     2.246862835
8    C     2.101066864    -2.671220685    -1.224935727
9    C     0.951349787    -2.903220636    -2.211533304
10   H    -0.598903723    -0.048562365    -0.313196078
11   H     0.285713996     0.422614882     1.180381251
12   H     0.257741784     1.521093534    -0.232483040
13   H     0.833366391    -2.074705087     0.394273891
14   H     4.711216086    -2.780420844     0.581780427
15   H     4.586743275    -1.132987216    -0.121883501
16   H     4.935679725    -1.334078900     1.616058150
17   H     2.989069077    -2.324178361    -1.799090083
18   H     2.379209489    -3.661885775    -0.793397094
19   H     0.661788321    -1.976326754    -2.745180798
20   H     0.047189019    -3.303425968    -1.709689080
21   H     1.261045820    -3.643697174    -2.977906603
---
H12C7O2, RHF, CHARGE=0, MULT=1
HF=-108.2
1    C     0.000000000     0.000000000     0.000000000
2    C     1.542588612     0.000000000     0.000000000
3    C     2.116529986     1.435424633     0.000000000
4    C     2.968792264     1.877256809    -1.203058246
5    C     4.300017779     2.547270528    -0.820322797
6    C     4.500099048     4.007735732    -1.212439849
7    O     5.171646675     1.908584890    -0.241140582
8    O     1.886269304     2.222163116     0.911197895
9    C     2.145707768    -0.851370083     1.136591434
10   H    -0.394607663     0.610611054    -0.838096666
11   H    -0.426808006     0.398644381     0.941207962
12   H    -0.386789411    -1.030945053    -0.134843217
13   H     1.851325122    -0.493117198    -0.956208448
14   H     3.201608191     1.013529885    -1.864772104
15   H     2.347022677     2.568407978    -1.816117404
16   H     4.436264525     4.119437927    -2.314068512
17   H     3.718223295     4.642575928    -0.749488890
18   H     5.486703417     4.395622491    -0.890122094
19   H     3.253452104    -0.848014616     1.089422979
20   H     1.846378864    -0.491090420     2.140592095
21   H     1.815879062    -1.906684776     1.045653300
---
H12C7O2, RHF, CHARGE=0, MULT=1
HF=-89.7
1    C     0.000000000     0.000000000     0.000000000
2    C     1.343936563     0.000000000     0.000000000
3    C     2.197406218     1.222367713     0.000000000
4    O     3.530910998     0.943967135    -0.028107478
5    C     4.558954827     1.909401031    -0.032897396
6    C     5.923109071     1.169007288    -0.064943206
7    C     7.129503595     2.127293298    -0.074081291
8    C     8.493574173     1.432098253    -0.104948329
9    O     1.814348726     2.391550375     0.022483768
```

```
10      H        -0.625378233     0.891531077    -0.000513988
11      H        -0.569706275    -0.929855435     0.000375941
12      H         1.881790728    -0.953292855     0.000272558
13      H         4.497450073     2.557397095     0.879589046
14      H         4.466930271     2.581420670    -0.925044183
15      H         5.996735061     0.494590505     0.818412299
16      H         5.966619684     0.515763045    -0.966035189
17      H         7.090243484     2.784630482     0.825331572
18      H         7.060254346     2.805273269    -0.956200773
19      H         8.646572309     0.783828664     0.781514151
20      H         8.616387373     0.804196553    -1.010549139
21      H         9.306651685     2.187125245    -0.110096107
---
H12C7O2, RHF, CHARGE=0, MULT=1
HF=-94.3
1       C         0.000000000     0.000000000     0.000000000
2       C         1.531400683     0.000000000     0.000000000
3       C         2.174288511     1.363557518     0.000000000
4       C         3.509642101     1.568374771    -0.019116410
5       C         4.156774561     2.912253935    -0.017470178
6       O         3.596822533     4.007528158     0.014465746
7       O         5.517939833     2.847492715    -0.056473014
8       C         6.379727635     3.962436256    -0.063306850
9       C         7.839143759     3.470221173    -0.112133036
10      H        -0.410184158     0.507362847    -0.896468017
11      H        -0.410529014     0.505411282     0.897365920
12      H        -0.380715622    -1.041949312    -0.001298982
13      H         1.886913623    -0.568104863     0.892571783
14      H         1.886450691    -0.570254722    -0.891302042
15      H         1.480164334     2.210154736     0.018125825
16      H         4.200517148     0.719752189    -0.037191158
17      H         6.230265608     4.591567332     0.852628677
18      H         6.179527907     4.619871254    -0.949175006
19      H         8.097740371     2.849552820     0.768766929
20      H         8.047199465     2.876445091    -1.024346058
21      H         8.515390800     4.349656271    -0.117754300
---
H12C7O2, RHF, CHARGE=0, MULT=1
HF=-92.6
1       C         0.000000000     0.000000000     0.000000000
2       C         1.497116576     0.000000000     0.000000000
3       C         2.290730474     1.087657283     0.000000000
4       C         3.795724355     1.078215945    -0.021226690
5       C         4.393290218     1.472298342     1.336985561
6       O         4.690003193     2.609641774     1.692905194
7       O         4.594514201     0.417111874     2.172183657
8       C         5.124657634     0.503848677     3.475715966
9       C         5.184152013    -0.908108972     4.089641165
10      H        -0.383946614    -0.532917775    -0.895556633
11      H        -0.437679978     1.018583425    -0.002886766
12      H        -0.383319160    -0.527560058     0.898960968
13      H         1.931724050    -1.006186165     0.002126801
14      H         1.857714694     2.093726704    -0.001481616
15      H         4.196847431     0.090213629    -0.339677390
16      H         4.150963104     1.809428422    -0.784951968
17      H         4.490606040     1.163979113     4.122916478
18      H         6.154074093     0.947263745     3.461381803
19      H         4.180427769    -1.371634688     4.164169070
20      H         5.834867883    -1.588366687     3.504972647
```

```
21     H       5.601878696     -0.836819035     5.114985456
---
H12C7O2, RHF, CHARGE=0, MULT=1
HF=-92.1
1      C       0.000000000      0.000000000     0.000000000
2      C       1.341612068      0.000000000     0.000000000
3      C       2.207453385      1.232958361     0.000000000
4      C       3.720175391      0.934223318     0.021891358
5      C       4.627261099      2.171625390     0.022190396
6      O       5.949578535      1.841429251     0.040741581
7      C       7.016444052      2.762578924     0.044763727
8      C       8.349197336      1.989009119     0.061281084
9      O       4.294076881      3.353315541     0.008322100
10     H      -0.614304207      0.899066007     0.001167214
11     H      -0.580754015     -0.921821116    -0.001147279
12     H       1.866914335     -0.960491514    -0.001776741
13     H       1.964799892      1.841400297    -0.902871475
14     H       1.942365944      1.858332035     0.884687315
15     H       3.987147494      0.314801926    -0.865354025
16     H       3.964348781      0.331204349     0.926871527
17     H       6.981528910      3.421052481    -0.861979571
18     H       6.964537767      3.433081234     0.941589656
19     H       8.449117965      1.353143688     0.963368366
20     H       8.464678717      1.341525889    -0.830567707
21     H       9.186871086      2.716323455     0.064098071
---
H12C7O2, RHF, CHARGE=0, MULT=1
HF=-94.2
1      C       0.000000000      0.000000000     0.000000000
2      C       1.540800622      0.000000000     0.000000000
3      O       2.000924435      1.332121704     0.000000000
4      C       3.311537340      1.706692254    -0.008784831
5      O       4.193230632      0.848229915    -0.014492538
6      C       3.494563803      3.186926419    -0.009034057
7      C       4.693877123      3.808741760    -0.033042347
8      C       4.866484989      5.306395008    -0.030853258
9      C       6.316082655      5.796690540    -0.089266513
10     H      -0.415646982      0.499442418    -0.897680417
11     H      -0.415659037      0.501402510     0.896580771
12     H      -0.359416467     -1.049632625     0.001175891
13     H       1.913286515     -0.552501624     0.901689147
14     H       1.912528335     -0.554143893    -0.900941701
15     H       2.568460319      3.769773402     0.012087273
16     H       5.622285302      3.228702949    -0.055033236
17     H       4.386312821      5.725349332     0.885525472
18     H       4.311043710      5.738890583    -0.896830640
19     H       6.904222263      5.443843986     0.781841893
20     H       6.828557443      5.453315046    -1.010538968
21     H       6.343175698      6.905704841    -0.084572214
---
H12C7O2, RHF, CHARGE=0, MULT=1
HF=-98.3
1      O       0.000000000      0.000000000     0.000000000
2      C       1.400632973      0.000000000     0.000000000
3      C       2.138870106      1.364334157     0.000000000
4      C       1.514802967      2.576517045    -0.707967833
5      C       0.132569075      3.101487661    -0.275708986
6      C      -0.279533174      3.025328679     1.205048930
7      C      -0.429255214      1.649036723     1.885247105
```

```
 8    C    -0.828204466    0.482831082    0.976159493
 9    O    -1.921912404   -0.072137349    1.008075560
10    H     1.788143229   -0.605593231    0.863073030
11    H     1.681192048   -0.559904525   -0.931927492
12    H     2.394960947    1.642312986    1.048122743
13    H     3.124872785    1.174459552   -0.489654932
14    H     2.246884954    3.413448676   -0.602313560
15    H     1.462119377    2.360886909   -1.801061142
16    H     0.096128675    4.181445925   -0.560346115
17    H    -0.653511081    2.617843987   -0.900186258
18    H     0.437106591    3.624110049    1.814024157
19    H    -1.260001899    3.551946262    1.291347658
20    H     0.502400179    1.383770557    2.433893819
21    H    -1.205577262    1.747953304    2.680438666
---
H12C7O2, RHF, CHARGE=0, MULT=1
HF=-98.2
 1    C    -0.143053491    0.022793908    0.010816448
 2    C     1.354070590    0.039862854   -0.048263259
 3    C     2.126014491    1.146494019   -0.039132801
 4    C     3.617296923    1.137387178   -0.111417124
 5    O     4.338572879    0.190871025   -0.423024569
 6    O     4.163326365    2.343565922    0.207980314
 7    C     5.537241246    2.684182226    0.182948567
 8    C     6.174624317    2.488267004    1.583521910
 9    C     5.641603021    4.145909080   -0.330365669
10    H    -0.554174253   -0.428548640   -0.916868830
11    H    -0.590005795    1.031115308    0.121487865
12    H    -0.484391595   -0.592521958    0.869645283
13    H     1.797725428   -0.960760626   -0.104080049
14    H     1.672016069    2.140513700    0.021214392
15    H     6.111589227    2.045115999   -0.542719967
16    H     5.681488270    3.105308998    2.360919563
17    H     6.124156975    1.430008460    1.908789010
18    H     7.246044871    2.772653625    1.552056424
19    H     5.129149646    4.866449003    0.337290693
20    H     5.201334792    4.247813059   -1.342783069
21    H     6.707726938    4.443260217   -0.400418372
---
H12C7O2, RHF, CHARGE=0, MULT=1
HF=-94.4
 1    C    -0.171024753    0.069664375   -0.008532945
 2    C     1.326709926    0.061470382   -0.052598466
 3    C     2.118034778    1.154825537   -0.032096015
 4    C     3.608888430    1.117112848   -0.078618801
 5    O     4.323776723    0.119243931   -0.165408922
 6    O     4.175922779    2.354862631   -0.014908245
 7    C     5.562104268    2.612257160   -0.043052143
 8    C     5.777289152    4.145999189    0.034535619
 9    C     7.251442643    4.559106905    0.013220967
10    H    -0.582307235   -0.440888265   -0.904740772
11    H    -0.602787272    1.089941630    0.025311976
12    H    -0.528194510   -0.478022887    0.889079063
13    H     1.752968238   -0.946974436   -0.104417206
14    H     1.679222318    2.155840347    0.021432580
15    H     6.075676781    2.110731493    0.817837837
16    H     6.021873345    2.210115678   -0.982784621
17    H     5.308829047    4.543613366    0.963875618
18    H     5.255579912    4.641597116   -0.815976875
```

```
19      H       7.811237301     4.139690765     0.873536811
20      H       7.760635351     4.230958639    -0.915598964
21      H       7.336906607     5.664045108     0.067218151
---
H14C7O2, RHF, CHARGE=0, MULT=1
HF=-106.8
1       C       0.000000000     0.000000000     0.000000000
2       C       1.549286089     0.000000000     0.000000000
3       C       2.174483535     1.429078923     0.000000000
4       C       3.668313230     1.410159715    -0.448601895
5       O       3.872049726     0.508169548    -1.510464853
6       C       3.364945739    -0.799593155    -1.395829640
7       O       1.988655656    -0.750887931    -1.106964515
8       C       4.203086475     2.797593984    -0.883665026
9       C       3.584010560    -1.556482086    -2.737262773
10       H     -0.395953723    -1.032654635     0.076800866
11       H     -0.425294032     0.460406752    -0.913388399
12       H     -0.374474061     0.567557856     0.876394129
13       H      1.874339570    -0.510964020     0.953548965
14       H      2.111136223     1.851295947     1.026362823
15       H      1.585245508     2.106237672    -0.655951736
16       H      4.284788187     1.088881352     0.441731567
17       H      3.916406237    -1.362369259    -0.576934958
18       H      3.677489216     3.194630056    -1.774531258
19       H      5.285024241     2.749348852    -1.120253632
20       H      4.073674406     3.525284922    -0.056641073
21       H      3.200182996    -2.593147582    -2.653071327
22       H      3.076278126    -1.068229677    -3.591666847
23       H      4.667669609    -1.613880374    -2.964283847
---
H14C7O2, RHF, CHARGE=0, MULT=1
HF=-95
1       C       0.000000000     0.000000000     0.000000000
2       C       1.587982707     0.000000000     0.000000000
3       C       2.063717430     1.463828686     0.000000000
4       C       0.831433481     2.383667271    -0.067021759
5       C      -0.420353882     1.507620686    -0.252249210
6       O      -0.539853311    -0.744337949    -1.062052654
7       C      -0.581683738    -2.140984977    -1.128121833
8       O      -0.526275997    -0.599841299     1.161941057
9       C      -0.608663776    -0.008924012     2.420909335
10       H      1.984712439    -0.520506999    -0.897672699
11       H      2.004424101    -0.548226216     0.870716874
12       H      2.738374558     1.652076394    -0.863016916
13       H      2.665526527     1.681346035     0.908801085
14       H      0.926690883     3.109663050    -0.902687513
15       H      0.748808901     2.995415725     0.857676713
16       H     -1.240659334     1.836116352     0.417843936
17       H     -0.817388122     1.637355157    -1.282607057
18       H     -1.449781499    -2.562824648    -0.566041325
19       H      0.345946777    -2.642122569    -0.761078741
20       H     -0.710397902    -2.404636774    -2.206779588
21       H      0.323783290     0.519721177     2.734474907
22       H     -1.456815041     0.715244189     2.496229832
23       H     -0.797719038    -0.832475776     3.151979959
---
H14C7O2, RHF, CHARGE=0, MULT=1
HF=-128.8
1       C      -0.041512155    -0.162952299     0.114710235
```

```
 2    C     1.476150199     0.041881277     0.079773680
 3    C     2.044969323     1.520912140     0.110728679
 4    C     1.563467278     2.329795165    -1.135958414
 5    C     1.644129728     2.246769668     1.434809666
 6    O     3.451481490     1.290359432     0.061124344
 7    C     4.505285870     2.144489666     0.059428293
 8    O     4.350915411     3.362695936     0.084166256
 9    C     5.856155710     1.432939453     0.023478047
10    H    -0.552020327     0.273550445    -0.766488334
11    H    -0.511429074     0.258518563     1.025233721
12    H    -0.262065679    -1.251687666     0.110428414
13    H     1.857746119    -0.465224923    -0.837639685
14    H     1.906705761    -0.521031987     0.941169941
15    H     1.942600357     3.370252431    -1.143363308
16    H     1.884216710     1.851051052    -2.083327003
17    H     0.457482455     2.400586338    -1.154799512
18    H     2.040038442     3.279261658     1.493949657
19    H     2.008101399     1.698675042     2.327552645
20    H     0.542085141     2.329577116     1.519295019
21    H     5.766500724     0.330378470    -0.021110909
22    H     6.435270498     1.694163233     0.932711185
23    H     6.427320119     1.766858869    -0.866643619
---
H14C7O2, RHF, CHARGE=0, MULT=1
HF=-123.4
 1    C     0.000000000     0.000000000     0.000000000
 2    C     1.540624418     0.000000000     0.000000000
 3    O     2.097435058     1.295478994     0.000000000
 4    C     2.406405831     2.045667635     1.092469380
 5    C     3.029092216     3.414493052     0.735402276
 6    C     4.497167604     3.476562273     1.241056263
 7    H     4.997333768     2.510985961     0.992275349
 8    C     5.358876467     4.602176179     0.659813645
 9    O     2.183346720     1.613367086     2.220139779
10    H    -0.413553167     0.595374003    -0.838423062
11    H    -0.420647183     0.399381725     0.943806237
12    H    -0.361599946    -1.042347431    -0.114769083
13    H     1.910956379    -0.479300911    -0.945135620
14    H     1.922461512    -0.631274747     0.841884272
15    H     3.058124816     3.505869051    -0.379654023
16    C     2.125975868     4.558974744     1.242946705
17    H     4.513208736     3.554842876     2.351888161
18    H     1.079462062     4.415657003     0.903791350
19    H     2.117069300     4.635849801     2.348133597
20    H     2.465352772     5.534143130     0.838993149
21    H     5.014663489     5.604503653     0.983989274
22    H     5.370480748     4.587083259    -0.448706406
23    H     6.407257555     4.486755258     1.005260519
---
H14C7O2, RHF, CHARGE=0, MULT=1
HF=-126
 1    C     0.000000000     0.000000000     0.000000000
 2    C     1.540622384     0.000000000     0.000000000
 3    O     2.097982602     1.295311293     0.000000000
 4    C     2.416553595     2.041751976     1.091883122
 5    C     3.098821156     3.373309636     0.749317317
 6    C     2.304362089     4.642504538     1.163171440
 7    C     0.996266860     4.832230320     0.368374227
 8    C     3.199935472     5.898810237     1.093084356
```

```
 9    O     2.185592346    1.615487825    2.220758800
10    H    -0.413338820    0.600344202   -0.834981836
11    H    -0.421266814    0.392853502    0.946157501
12    H    -0.361070355   -1.041829341   -0.121569091
13    H     1.910184109   -0.478595554   -0.945911778
14    H     1.922701126   -0.631181136    0.841714180
15    H     4.080103913    3.360818589    1.280074389
16    H     3.339512519    3.418047151   -0.336238344
17    H     2.015099199    4.527096768    2.238862218
18    H     1.176525850    4.912711168   -0.722491959
19    H     0.296067485    3.988995728    0.534009001
20    H     0.469548756    5.753938372    0.690477828
21    H     3.551838330    6.106722245    0.062497243
22    H     4.094455273    5.789857602    1.739757354
23    H     2.651032168    6.794764779    1.448275252
---
H14C7O2, RHF, CHARGE=0, MULT=1
HF=-121.2
 1    C    -0.000079285   -0.000080744    0.000006947
 2    C     1.540768977   -0.000048810    0.000028231
 3    O     2.001034137    1.332282794    0.000437822
 4    C     3.310321860    1.711144209    0.007548480
 5    C     3.473616250    3.236719399    0.007645372
 6    C     4.931896762    3.737246259    0.039706349
 7    C     5.053758680    5.274385915    0.041887679
 8    C     6.492377478    5.798273624    0.072178206
 9    O     4.192410650    0.857029336    0.011766178
10    H    -0.415648361    0.501187641   -0.896674877
11    H    -0.415404749    0.500003285    0.897530562
12    H    -0.359567034   -1.049605554   -0.000672199
13    H     1.913207132   -0.552679349   -0.901443531
14    H     1.912702723   -0.553504336    0.901254934
15    H     2.964316539    3.633265342   -0.900917943
16    H     2.924019285    3.639246218    0.889768264
17    H     5.479094360    3.332018574   -0.841521931
18    H     5.439288022    3.332280218    0.944643900
19    H     4.546909823    5.688998803   -0.860061752
20    H     4.509334785    5.689700972    0.921389342
21    H     7.033877381    5.464334587    0.980285020
22    H     7.071883670    5.463954610   -0.811967454
23    H     6.493879226    6.907807701    0.071954424
---
H14C7O2, RHF, CHARGE=0, MULT=1
HF=-122.3
 1    C     0.000000000    0.000000000    0.000000000
 2    C     1.555019087    0.000000000    0.000000000
 3    C     2.071908371    1.467354541    0.000000000
 4    C     2.054081428   -0.702732536   -1.300382050
 5    C     2.126510397   -0.807549758    1.216412118
 6    C     1.758248193   -0.326992654    2.627140298
 7    O     2.328259159    0.552371759    3.268653941
 8    O     0.730870751   -1.015663717    3.197334418
 9    C     0.230754822   -0.806153888    4.493089187
10    H    -0.416635316    0.570293473    0.854155509
11    H    -0.410717096   -1.028516161    0.052953563
12    H    -0.401866441    0.466469344   -0.923107833
13    H     1.657878789    2.056958648    0.841659536
14    H     3.177391004    1.512311545    0.072118572
15    H     1.782322923    1.994177538   -0.932768447
```

```
16      H      1.699893290     -1.751891920    -1.363000068
17      H      3.161533030     -0.723898614    -1.356600024
18      H      1.690661560     -0.181038002    -2.209054862
19      H      1.826023310     -1.875672686     1.115011684
20      H      3.240912854     -0.804231520     1.165524007
21      H      0.999671846     -0.995501094     5.279361123
22      H     -0.174579570      0.224495722     4.629517099
23      H     -0.602859241     -1.535224543     4.631892887
---
H14C7O2, RHF, CHARGE=0, MULT=1
HF=-118
1       H      0.000000000      0.000000000     0.000000000
2       C      1.115910197      0.000000000     0.000000000
3       O      1.649535849      1.298693837     0.000000000
4       C      1.455812376      2.231190879    -0.975722123
5       C      2.175395270      3.551475151    -0.672438336
6       C      2.017687298      4.643708589    -1.750295815
7       C      2.758271897      5.953570598    -1.407819730
8       C      2.608477271      7.054517627    -2.475895997
9       O      0.764303948      1.952119459    -1.951160739
10      H      2.993203017      8.806104453    -1.201105505
11      H      4.438050194      8.209242025    -2.074607916
12      H      3.166008655      9.103828843    -2.957464192
13      H      1.474251803     -0.605358539    -0.866394087
14      H      1.468870541     -0.487417925     0.940129569
15      H      1.790138904      3.932257747     0.301635917
16      H      3.257126817      3.326160371    -0.528176695
17      H      0.935529667      4.864436109    -1.892795462
18      H      2.395825104      4.255566895    -2.723112629
19      H      2.383172413      6.340601432    -0.432515000
20      H      3.840837694      5.733690097    -1.261812060
21      H      2.979344516      6.674741448    -3.455865687
22      H      1.527789022      7.284968736    -2.621443783
23      C      3.342417916      8.359384037    -2.153923356
---
H16C7O2, RHF, CHARGE=0, MULT=1
HF=-104.3
1       C      0.000000000      0.000000000     0.000000000
2       C      1.540122799      0.000000000     0.000000000
3       O      2.010435842      1.320535243     0.000000000
4       C      3.392335632      1.555065538     0.000224834
5       C      3.639126510      3.086821041    -0.000015920
6       C      5.150483717      3.438165750     0.000411688
7       O      5.288777523      4.832966256    -0.001356577
8       C      6.572578328      5.395181288     0.001865908
9       C      6.464285404      6.931444385    -0.002208509
10      H     -0.414422222      0.501341467    -0.897083970
11      H      1.910678463     -0.565052921     0.897534873
12      H      1.911116537     -0.564681874    -0.897529452
13      H      3.879287423      1.088347169     0.898497436
14      H      3.879398179      1.089308657    -0.898485519
15      H      3.153557281      3.540365692     0.893300926
16      H      3.151812689      3.540126618    -0.892552030
17      H      5.649159420      2.986607872     0.899977257
18      H      5.650318066      2.982911297    -0.896610303
19      H      7.158937031      5.067147668     0.902277223
20      H      7.165515116      5.062845940    -0.892658008
21      H      7.484952723      7.366636457     0.000332093
22      H     -0.414163393      0.501230815     0.897310782
```

```
23      H       -0.362143448    -1.048832142     0.000185046
24      H        5.937526217    7.307172167     -0.901811332
25      H        5.932519836    7.311745934      0.892496296
---
H16C7O2, RHF, CHARGE=0, MULT=1
HF=-114.1
1       O        0.000000000    0.000000000     0.000000000
2       C        1.396832082    0.000000000     0.000000000
3       C        1.916960577    1.460752456     0.000000000
4       C        3.455893973    1.558375782    -0.001162315
5       C        3.990246823    3.005465845     0.000103293
6       C        5.529583364    3.105114305    -0.001921597
7       C        6.061744987    4.552343251     0.000127449
8       C        7.610164254    4.633944622    -0.002215527
9       O        8.009391716    5.972476691    -0.000555688
10      H       -0.338484712   -0.884098871    -0.000178026
11      H        1.792722446   -0.549302933     0.897072665
12      H        1.792849220   -0.549669969    -0.896799792
13      H        1.514332205    1.990832351     0.893016430
14      H        1.513040854    1.991662226    -0.891919692
15      H        3.857090876    1.021722004     0.889294868
16      H        3.855270482    1.023987603    -0.893838172
17      H        3.592430007    3.539746196     0.893173951
18      H        3.589410856    3.542170052    -0.890209296
19      H        5.930389014    2.566928107     0.887785141
20      H        5.927504519    2.571028001    -0.895427958
21      H        5.669737239    5.088277617     0.894317695
22      H        5.666770713    5.091803031    -0.890658937
23      H        8.024094592    4.097062185     0.894402278
24      H        8.023378098    4.098453012    -0.899730784
25      H        8.953306304    6.043518169    -0.012843037
---
H8C8O2, RHF, CHARGE=0, MULT=1
HF=-78.4
1       C        0.000000000    0.000000000     0.000000000
2       C        1.411488737    0.000000000     0.000000000
3       C        2.109772613    1.218758485     0.000000000
4       C        1.409640010    2.436251085    -0.005400333
5       C       -0.005235740    2.461864356    -0.010802573
6       C       -0.696734982    1.228635425    -0.007256162
7       C       -0.755220020   -1.294973065     0.027811777
8       O       -1.042211491   -1.802059602    -1.198910234
9       O       -1.118600067   -1.908081487     1.030053080
10      C       -0.744072773    3.773816041    -0.023251682
11      H        1.969176635   -0.937700878     0.002297393
12      H        3.200353553    1.218939183     0.001991765
13      H        1.975524619    3.369858345    -0.006196116
14      H       -1.788806893    1.223100756    -0.009355447
15      H       -1.519758973   -2.622821548    -1.199596116
16      H       -0.466135516    4.393628823     0.855016117
17      H       -1.844511341    3.641394955     0.002988241
18      H       -0.503174116    4.352519295    -0.939916206
---
H8C8O2, RHF, CHARGE=0, MULT=1
HF=-66.8
1       C        0.000000000    0.000000000     0.000000000
2       C        1.412554365    0.000000000     0.000000000
3       C        2.122178619    1.214259108     0.000000000
4       C        1.427340483    2.436398269    -0.002363807
```

```
5    C      0.021373339    2.443672433   -0.002792921
6    C     -0.692080490    1.232131202   -0.000641869
7    C     -0.758229652   -1.294904502   -0.023390943
8    O     -1.041686778   -1.773306854    1.219203383
9    O     -1.114604837   -1.908070857   -1.025178797
10   C     -1.738116629   -2.962726369    1.488743353
11   H      1.966602629   -0.940080939    0.000104280
12   H      3.212445174    1.205173385    0.001069774
13   H      1.978054920    3.377690403   -0.003455794
14   H     -0.519028395    3.390727952   -0.004794950
15   H     -1.782925369    1.255670185   -0.000587894
16   H     -1.212283902   -3.861761834    1.087986639
17   H     -1.793591054   -3.053046250    2.599817207
18   H     -2.778003157   -2.945716648    1.084020261
---
H8C8O2, RHF, CHARGE=0, MULT=1
HF=-76.6
1    C      0.000000000    0.000000000    0.000000000
2    C      1.413737933    0.000000000    0.000000000
3    C      2.132167002    1.206004507    0.000000000
4    C      1.437531577    2.426032889   -0.000482763
5    C      0.033268430    2.432729302    0.001473051
6    C     -0.714441805    1.230133159    0.002313768
7    C     -0.697999696   -1.330484692    0.019025433
8    O     -0.938913236   -1.855056639   -1.210572559
9    O     -1.033562707   -1.966856655    1.016696034
10   C     -2.221738007    1.292827273    0.004606037
11   H      1.964904920   -0.942375581    0.000875125
12   H      3.222294290    1.192683150    0.000072883
13   H      1.986138428    3.368713887   -0.001566118
14   H     -0.480764956    3.395944032    0.002412311
15   H     -1.368745512   -2.701864838   -1.215423644
16   H     -2.596186110    2.334843203    0.071418172
17   H     -2.648730268    0.742566755    0.868353724
18   H     -2.639570838    0.858207637   -0.927510529
---
H8C8O2, RHF, CHARGE=0, MULT=1
HF=-79
1    C      0.000000000    0.000000000    0.000000000
2    C      1.411335982    0.000000000    0.000000000
3    C      2.122525496    1.212013646    0.000000000
4    C      1.450355039    2.456104298   -0.001843103
5    C      0.035704781    2.443688266   -0.000081965
6    C     -0.681969202    1.236032948   -0.000156908
7    C     -0.761998115   -1.290432574    0.022859137
8    O     -1.058246613   -1.787411009   -1.205784212
9    O     -1.124391175   -1.907947860    1.022813430
10   C      2.211902150    3.754640258    0.013088388
11   H      1.965879186   -0.939922738    0.001327155
12   H      3.213470072    1.175860224    0.002032244
13   H     -0.521925164    3.382076003    0.001830595
14   H     -1.772824839    1.267431593    0.001111784
15   H     -1.541153052   -2.605041261   -1.209435563
16   H      3.234898578    3.643292622   -0.400638654
17   H      2.308316570    4.132700366    1.053192172
18   H      1.702042860    4.536731053   -0.586316347
---
N1O2, RHF, CHARGE=1, MULT=1
HF=233
```

```
1    O     0.000000000     0.000000000     0.000000000
2    N     1.132283947     0.000000000     0.000000000
3    O     2.264567807    -0.000766173     0.000000000
---
N1O2, UHF, CHARGE=0, MULT=2
HF=7.9
1    O     0.000000000     0.000000000     0.000000000
2    N     1.174078912     0.000000000     0.000000000
3    O     1.977691088     0.856024891     0.000000000
EXPGEOM
1    N     0.00000     0.00000     0.32210
2    O     0.00000     1.09920    -0.14090
3    O     0.00000    -1.09920    -0.14090
---
H1N1O2, RHF, CHARGE=0, MULT=1
HF=-18.8, DIP=1.86
1    O     0.000000000     0.000000000     0.000000000
2    N     1.312132873     0.000000000     0.000000000
3    O     1.772869346     1.071248419     0.000000000
4    H    -0.327978271    -0.899691125    -0.000122520
EXPGEOM
1    H    -0.55770    -1.25190     0.00000
2    O    -1.02560    -0.39080     0.00000
3    N     0.00000     0.55870     0.00000
4    O     1.09540     0.05840     0.00000
---
H3C1N1O2, RHF, CHARGE=0, MULT=1
HF=-15.8, IE=11
1    C     0.000000000     0.000000000     0.000000000
2    O     1.413451935     0.000000000     0.000000000
3    N     1.998830977     1.178814608     0.000000000
4    O     3.161610641     1.107417589     0.000262753
5    H    -0.313544849    -1.070218646    -0.000989585
6    H    -0.422750228     0.495806124    -0.905809515
7    H    -0.422467302     0.494374541     0.906707523
EXPGEOM
1    C     1.32240     0.31520     0.00000
2    O     0.00000     0.90120     0.00000
3    H     1.99230     1.17730     0.00000
4    H     1.47280    -0.29920     0.89460
5    H     1.47280    -0.29920    -0.89460
6    N    -1.05090    -0.03270     0.00000
7    O    -0.68950    -1.18130     0.00000
---
H3C1N1O2, RHF, CHARGE=0, MULT=1
HF=-17.9
1    C     0.000000000     0.000000000     0.000000000
2    N     1.545628772     0.000000000     0.000000000
3    O     2.135006732     1.056702758     0.000000000
4    O     2.140141119    -1.053416540     0.008648045
5    H    -0.373221057    -1.017430926    -0.232442942
6    H    -0.357876769     0.717493725    -0.766058102
7    H    -0.343328597     0.310335790     1.008422113
EXPGEOM
1    C     0.00200    -1.32590     0.00000
2    N    -0.01400     0.16910     0.00000
3    H     1.05220    -1.63350     0.00000
4    H    -0.49960    -1.66170     0.90880
5    H    -0.49960    -1.66170    -0.90880
```

```
6    O    0.00200    0.73300    -1.10230
7    O    0.00200    0.73300     1.10230
---
H5C2N1O2, RHF, CHARGE=0, MULT=1
HF=-25.9, IE=11.3
1    C    0.000000000    0.000000000     0.000000000
2    C    1.540146320    0.000000000     0.000000000
3    O    2.099499509    1.304238231     0.000000000
4    N    2.292370219    1.867673924    -1.172556068
5    O    2.751788530    2.935602671    -1.091974304
6    H   -0.414861673    0.603387115     0.831684945
7    H   -0.417747224    0.390012449    -0.949487868
8    H   -0.361235881   -1.041331697     0.123836338
9    H    1.916178034   -0.480868658     0.940553288
10   H    1.919635554   -0.621070349    -0.851571024
---
H5C2N1O2, RHF, CHARGE=0, MULT=1
HF=-93.7
1    C    0.031440705    0.001390095    -0.001871426
2    C    1.573500430    0.012441229    -0.002737891
3    N    2.206961520    1.325557442     0.003230531
4    H    1.930630379   -0.550420639     0.895215533
5    H    1.927750775   -0.540709975    -0.908000045
6    O   -0.717327323    0.979003506    -0.001854635
7    O   -0.515049017   -1.239473106    -0.001995979
8    H   -1.464200199   -1.270922122    -0.003209533
9    H    1.931830318    1.869303353    -0.799254487
10   H    1.937295459    1.859263876     0.814178217
EXPGEOM
1    C    0.00000     0.55060    0.00000
2    O    1.17760     0.81190    0.00000
3    O   -0.96940     1.49270    0.00000
4    C   -0.58480    -0.85640    0.00000
5    N    0.40070    -1.92780    0.00000
6    H   -0.50710     2.34580    0.00000
7    H   -1.24570    -0.94560    0.88130
8    H   -1.24570    -0.94560   -0.88130
9    H    1.01850    -1.78120    0.80320
10   H    1.01850    -1.78120   -0.80320
---
H5C2N1O2, RHF, CHARGE=0, MULT=1
HF=-101.6
1    O    0.000000000    0.000000000     0.000000000
2    C    1.232580194    0.000000000     0.000000000
3    N    2.006777766    1.146714476     0.000000000
4    O    1.990243065   -1.130967951     0.000131027
5    C    1.448954506   -2.428474247    -0.000516713
6    H    3.000206942    1.126231266     0.000159660
7    H    1.575709600    2.042980307     0.000222805
8    H    0.832208795   -2.629605068     0.907220566
9    H    0.831351389   -2.628772137    -0.907793112
10   H    2.317510433   -3.129504910    -0.001027526
---
H5C2N1O2, RHF, CHARGE=0, MULT=1
HF=-23.5
1    C    0.000000000    0.000000000     0.000000000
2    C    1.529471718    0.000000000     0.000000000
3    H    1.953167942    1.023819523     0.000000000
4    H    1.953170894   -0.555532028     0.860029302
```

```
5     H     1.874265825    -0.504935439   -0.927364964
6     H    -0.389307050     0.542372298   -0.893141893
7     H    -0.389598191    -1.044622038   -0.020923883
8     N    -0.658764136     0.679104848    1.236964926
9     O     0.012428230     1.151621520    2.124229524
10    O    -1.867652280     0.718280282    1.282720437
---
H7C3N1O2, RHF, CHARGE=0, MULT=1
HF=-99.1, IE=8.9
1     C     0.024641017     0.033091499    0.091026737
2     C     1.571124867    -0.009235472   -0.024048026
3     N     2.201877415     1.311203011    0.068189398
4     H     1.963245726    -0.603124148    0.846034863
5     C     2.014393938    -0.730357462   -1.319182595
6     O    -0.716299220     0.933123010   -0.305184575
7     O    -0.528661935    -1.043718901    0.701216757
8     H    -1.475776595    -1.034735991    0.771509220
9     H     1.891624434     1.930086525   -0.664572148
10    H     1.954168097     1.759725913    0.936266538
11    H     1.638220087    -0.230571563   -2.234391035
12    H     1.640998361    -1.774612972   -1.326403894
13    H     3.120357386    -0.773175557   -1.384052108
EXPGEOM
1     N     1.02490     1.38860    -0.08640
2     C     1.51040    -1.10870    -0.14650
3     C     0.64340     0.03990     0.37410
4     C    -0.84740    -0.18990     0.04890
5     O    -1.55370     0.94340    -0.08520
6     O    -1.34030    -1.28420    -0.03870
7     H     1.29800     1.33780    -1.07290
8     H     1.86100     1.69990     0.41110
9     H     1.13860    -2.07340     0.23330
10    H     1.47790    -1.14590    -1.25110
11    H     0.69110     0.05870     1.48090
12    H     2.55990    -0.97480     0.17000
13    H    -0.88680     1.65590    -0.03460
---
H7C3N1O2, RHF, CHARGE=0, MULT=1
HF=-101
1     N    -0.071431767    -0.015142579   -0.025761522
2     C     1.396981716     0.005492439   -0.015752303
3     C     1.903581548     1.466851161   -0.024437155
4     C     3.425278980     1.614896660    0.070149670
5     O     3.847591388     2.906415417    0.034502352
6     O     4.272941673     0.728261737    0.172625261
7     H    -0.390226999    -0.934770648    0.237501449
8     H    -0.414169340     0.126843920   -0.963664144
9     H     1.834219003    -0.554115018   -0.882047138
10    H     1.747365541    -0.514546709    0.909426603
11    H     1.452927233     2.023478928    0.829674747
12    H     1.567989526     1.976043428   -0.957516048
13    H     4.786545521     3.032043966    0.096765020
EXPGEOM
1     N     1.92270     0.57640     0.13760
2     C     1.35840    -0.68610    -0.38230
3     C     0.02310    -0.98200     0.31650
4     C    -1.09000     0.03500     0.02940
5     O    -0.70270     1.31990    -0.04170
6     O    -2.24080    -0.29240    -0.10510
```

```
7      H      2.72070      0.86410     -0.43290
8      H      2.29810      0.41570      1.07670
9      H      2.04060     -1.55050     -0.26040
10     H      1.18780     -0.55640     -1.46660
11     H     -0.36190     -1.97160      0.02410
12     H      0.18270     -1.00720      1.41330
13     H      0.27260      1.34950      0.07560
---
H7C3N1O2, RHF, CHARGE=0, MULT=1
HF=-31.9
1    C     0.000000000     0.000000000     0.000000000
2    C     1.550801336     0.000000000     0.000000000
3    C     2.167358368     1.422983883     0.000000000
4    O     2.007500543    -0.699811861     1.156160228
5    N     2.742400520    -1.768400004     0.946528054
6    O     3.080106142    -2.287184247     1.934400680
7    H    -0.400306844    -1.033263839    -0.038250397
8    H    -0.422830220     0.495030431     0.896326373
9    H    -0.375113235     0.534801938    -0.895972372
10   H     1.883474550    -0.506437265    -0.948678667
11   H     1.885191705     2.006094551     0.898643173
12   H     3.274346191     1.379909045    -0.044186040
13   H     1.821400461     1.981707855    -0.893186645
EXPGEOM
1    H     0.54370     -0.00000     -1.35610
2    C     0.81050      0.00000     -0.29880
3    H     2.50920      1.30380     -0.49150
4    H     1.78040      1.30010      1.11860
5    H     0.98120      2.14710     -0.21060
6    C     1.56550      1.26760      0.05150
7    H     2.51070     -1.30180     -0.49140
8    H     0.98360     -2.14690     -0.21130
9    H     1.78140     -1.29950      1.11850
10   C     1.56680     -1.26680      0.05130
11   O    -0.43190     -0.00080      0.44630
12   O    -2.56280     -0.00010      0.14910
13   N    -1.54140      0.00000     -0.43760
---
H7C3N1O2, RHF, CHARGE=0, MULT=1
HF=-87.8
1    C    -0.034305211     0.068499484    -0.135465443
2    N     1.420476497     0.011224627     0.013778824
3    C     2.137215166     1.280289637     0.151890652
4    C     3.585653293     1.060957326    -0.322581769
5    O     3.875945682     1.604127219    -1.530382119
6    O     4.470987852     0.470751259     0.295105180
7    H    -0.431575309    -0.971052461    -0.171546544
8    H    -0.301395376     0.575000082    -1.089742516
9    H    -0.555661158     0.602398766     0.695174087
10   H     1.657962239    -0.603250833     0.778896180
11   H     2.168870669     1.656490478     1.206817787
12   H     1.644591471     2.080962395    -0.449103261
13   H     4.759480147     1.455039494    -1.844882903
---
H7C3N1O2, RHF, CHARGE=0, MULT=1
HF=-28.4
1    C     0.000000000     0.000000000     0.000000000
2    C     1.549497647     0.000000000     0.000000000
3    H     1.935051370     1.052383690     0.000000000
```

```
   4    O      1.996018667    -0.695505232     1.155059271
   5    N      3.299263379    -0.792855613     1.304267775
   6    O      3.602128673    -1.373417233     2.268162992
   7    H     -0.374323755     0.479231543     0.933492090
   8    C     -0.612447756     0.723872575    -1.202436855
   9    H     -0.374525115    -1.049130256     0.013749759
  10    H      1.935367637    -0.491557090    -0.930368531
  11    H     -0.314789471     1.791462414    -1.238038721
  12    H     -0.315213048     0.255467601    -2.162563970
  13    H     -1.719503567     0.687984650    -1.142162691
EXPGEOM
   1    C     -1.90900    2.22810    0.00000
   2    C     -1.51410    0.74490    0.00000
   3    C      0.00000    0.56910    0.00000
   4    O      0.26890   -0.84520    0.00000
   5    N      1.68950   -1.04170    0.00000
   6    O      1.94440   -2.19740    0.00000
   7    H     -3.00700    2.34550    0.00000
   8    H     -1.51310    2.74600    0.89390
   9    H     -1.51310    2.74600   -0.89390
  10    H     -1.93040    0.23410   -0.88870
  11    H     -1.93040    0.23410    0.88870
  12    H      0.45010    1.03710    0.89760
  13    H      0.45010    1.03710   -0.89760
---
H7C3N1O2, RHF, CHARGE=0, MULT=1
HF=-106.7
   1    O      0.000000000     0.000000000     0.000000000
   2    C      1.232495218     0.000000000     0.000000000
   3    N      2.003946155     1.149389986     0.000000000
   4    O      1.995482665    -1.126950908     0.000805579
   5    C      1.452976447    -2.429919205    -0.006130118
   6    C      2.606343789    -3.451390325    -0.006774158
   7    H      2.997346316     1.131898325     0.000735326
   8    H      1.570429784     2.044407246     0.001077829
   9    H      0.805107208    -2.598024356     0.893024441
  10    H      0.810762730    -2.591278432    -0.910542806
  11    H      3.251590762    -3.349447994    -0.901939540
  12    H      3.247058914    -3.355719686     0.892332826
  13    H      2.178542809    -4.474975126    -0.011466295
EXPGEOM
   1    C      2.54520   -0.24860    0.05090
   2    H      2.68980   -0.85350   -0.84130
   3    H      3.35770    0.47310    0.11890
   4    H      2.58330   -0.90100    0.92010
   5    C      1.21930    0.47250   -0.00780
   6    H      1.05680    1.07300    0.88790
   7    H      1.18360    1.12960   -0.88180
   8    O      0.19780   -0.52630   -0.11110
   9    N     -1.37210    1.16430    0.08060
  10    H     -0.73780    1.82360   -0.33250
  11    H     -2.34740    1.38090   -0.02810
  12    C     -1.11040   -0.18490   -0.01000
  13    O     -1.96100   -1.03740    0.03550
---
H5C4N1O2, RHF, CHARGE=0, MULT=1
HF=-58.2
   1    C      0.000000000     0.000000000     0.000000000
   2    O      1.406230664     0.000000000     0.000000000
```

```
3    C      2.191339998     1.110601787     0.000000000
4    O      1.688687605     2.228933620     0.002512394
5    C      3.687215848     0.763178100     0.005372296
6    C      4.558642155     1.922212626    -0.113291992
7    N      5.278279529     2.829021127    -0.207427134
8    H     -0.421142303     0.491607516     0.908673147
9    H     -0.421610687     0.495421318    -0.906323624
10   H     -0.314300241    -1.070739690    -0.002211616
11   H      3.928211401     0.225017039     0.953577262
12   H      3.906607173     0.061998213    -0.835199629
---
H5C4N1O2, RHF, CHARGE=0, MULT=1
HF=-89.8
1    N      0.000000000     0.000000000     0.000000000
2    C      1.416785049     0.000000000     0.000000000
3    C      1.887237988     1.458902149     0.000000000
4    C      0.595471775     2.309552440     0.000356347
5    C     -0.559363203     1.301436098     0.000305199
6    O     -1.765239678     1.489444919     0.000514035
7    O      2.065300642    -1.033837951    -0.000125343
8    H     -0.549833099    -0.834739279    -0.000071205
9    H      2.513589912     1.666184331    -0.892726876
10   H      2.514007902     1.665917384     0.892499306
11   H      0.537823874     2.967464663    -0.891864747
12   H      0.538452100     2.966587498     0.893319777
---
H9C4N1O2, RHF, CHARGE=0, MULT=1
HF=-39.1
1    C      0.000000000     0.000000000     0.000000000
2    C      1.532557515     0.000000000     0.000000000
3    C      2.226395838     1.387003558     0.000000000
4    C      3.751302273     1.325905749     0.227782229
5    N      1.932798056     2.168184818    -1.326869355
6    O      1.360975045     3.231912291    -1.259655363
7    O      2.283407816     1.717581284    -2.392902782
8    H     -0.368753421    -1.035637148     0.151696162
9    H     -0.414320147     0.624338151     0.817046970
10   H      1.871821296    -0.525471574     0.924684534
11   H      1.893964359    -0.620953537    -0.850589546
12   H      1.784164487     2.008030752     0.820857048
13   H      4.273040190     0.680271525    -0.505668174
14   H      4.208804814     2.335121186     0.185427659
15   H      3.952314712     0.918135057     1.240284094
16   H     -0.422589992     0.362570423    -0.957590123
---
H9C4N1O2, RHF, CHARGE=0, MULT=1
HF=-42.2
1    C     -0.052432182    -0.001528503     0.016698573
2    C      1.502405751     0.015871503     0.029631363
3    C      2.023664965     1.481506244     0.037685941
4    H      3.130570631     1.537079524     0.039125014
5    C      2.080719069    -0.767813828     1.241731280
6    N      1.993199049    -0.725298557    -1.284496867
7    H      1.670883547     1.994331165     0.957177758
8    H      1.652030337     2.067839152    -0.826222089
9    H     -0.469918182    -1.027203348     0.056339863
10   H     -0.462853532     0.497332333    -0.884398724
11   H     -0.441760284     0.541024418     0.903108271
12   H      1.823825081    -0.242215931     2.184711715
```

```
13      H        3.185815805     -0.844877097     1.195471004
14      H        1.671917562     -1.794735633     1.320374201
15      O        2.677263124     -0.148978298    -2.097256531
16      O        1.676218872     -1.878594351    -1.466338044
---
H9C4N1O2, RHF, CHARGE=0, MULT=1
HF=-105
1       N       -0.010002309     -0.039351043     0.127297787
2       C        1.352434196      0.029852458    -0.413369194
3       C        1.988034027      1.412757338    -0.139297796
4       C        3.444046379      1.519333655    -0.639008719
5       C        4.049842050      2.908912143    -0.421230878
6       O        4.692373090      3.076920112     0.762833514
7       O        4.015642312      3.851546601    -1.212085062
8       H       -0.637288263      0.510466396    -0.439489232
9       H       -0.345872368     -0.987423720     0.050651618
10      H        1.383703468     -0.189022084    -1.512799166
11      H        1.958248320     -0.769905507     0.080751024
12      H        1.377695608      2.206127342    -0.628429585
13      H        1.960149247      1.619218202     0.954655004
14      H        3.492732699      1.296419700    -1.729678861
15      H        4.078096614      0.754118886    -0.137318100
16      H        5.073613882      3.935054546     0.904029234
---
H9C4N1O2, RHF, CHARGE=0, MULT=1
HF=-36.1
1       C        0.000000000      0.000000000     0.000000000
2       O        1.420068935      0.000000000     0.000000000
3       N        1.999351058      1.181050561     0.000000000
4       O        3.162611813      1.113988328     0.006509561
5       C       -0.515169216     -1.472031662    -0.079399606
6       H        0.015757865     -2.062355061     0.710537443
7       C       -0.211780031     -2.155852478    -1.428325120
8       C       -2.017894440     -1.534580545     0.269086171
9       H       -0.401271834      0.601052273    -0.856218994
10      H       -2.213520865     -1.123532449     1.280670184
11      H       -2.640503789     -0.966898152    -0.451577567
12      H       -2.377706431     -2.583601746     0.271172449
13      H       -0.379415302      0.481437518     0.939072370
14      H        0.878704198     -2.226020853    -1.613413550
15      H       -0.665363084     -1.613577664    -2.282641965
16      H       -0.607183401     -3.192392469    -1.440379094
---
H9C4N1O2, RHF, CHARGE=0, MULT=1
HF=-34.8
1       C        0.000000000      0.000000000     0.000000000
2       C        1.531440315      0.000000000     0.000000000
3       C        2.171973111      1.400889033     0.000000000
4       C        3.722918659      1.361575249     0.020069969
5       O        4.297111988      2.649336464    -0.146807040
6       N        4.520692985      3.348043208     0.944471823
7       O        4.989760697      4.392233888     0.725647131
8       H       -0.412479030      0.504289772     0.897100730
9       H       -0.412213086      0.506936083    -0.895761187
10      H       -0.380523113     -1.042214483    -0.001749948
11      H        1.882957910     -0.570511834     0.890906524
12      H        1.883756215     -0.569627672    -0.891000752
13      H        1.828816946      1.960065726    -0.900049364
14      H        1.805491424      1.970117342     0.884380096
```

```
15      H      4.087071378     0.860103902     0.952800637
16      H      4.108106177     0.759875644    -0.843621808
---
H9C4N1O2, RHF, CHARGE=0, MULT=1
HF=-36.5
1       C      0.000000000     0.000000000     0.000000000
2       C      1.551634022     0.000000000     0.000000000
3       O      1.989720678     1.359410328     0.000000000
4       N      2.729808848     1.741131929     1.015854830
5       O      3.050243224     2.860015546     0.948435081
6       H     -0.422639526     0.455262580    -0.917248027
7       H     -0.382415582    -1.037475164     0.075379585
8       C      2.204472840    -0.716881122    -1.225062962
9       H     -0.398746084     0.558892057     0.871026184
10      H      3.292095470    -0.470916709    -1.248852241
11      H      1.779463732    -0.310474872    -2.170741707
12      C      2.075768954    -2.242710819    -1.238709486
13      H      1.884736003    -0.535113637     0.932011138
14      H      2.471338123    -2.701703611    -0.310067253
15      H      1.026014463    -2.575632848    -1.363127129
16      H      2.655650622    -2.659480570    -2.088178259
---
H9C4N1O2, RHF, CHARGE=0, MULT=1
HF=-41
1       C      0.000000000     0.000000000     0.000000000
2       C      1.562365510     0.000000000     0.000000000
3       O      2.080276826     1.335992940     0.000000000
4       N      1.807395689     2.119750710     1.017681191
5       O      2.280682676     3.179936433     0.907794344
6       H     -0.411163161     0.634744329    -0.810678473
7       H     -0.384032872    -1.028617930    -0.155294641
8       C      2.113824544    -0.820986675     1.209520121
9       C      2.094618102    -0.608030826    -1.342206468
10      H     -0.422650564     0.364388028     0.957564668
11      H      3.219806706    -0.772750847     1.269327567
12      H      1.710471664    -0.467226787     2.179338417
13      H      1.831718655    -1.888722404     1.109576203
14      H      3.202321354    -0.615161094    -1.380217841
15      H      1.732432436    -0.040929774    -2.223004643
16      H      1.749058679    -1.655728453    -1.453805881
---
H11C4N1O2, RHF, CHARGE=0, MULT=1
HF=-94.9
1       O      0.000000000     0.000000000     0.000000000
2       C      1.397152093     0.000000000     0.000000000
3       C      1.986489111     1.445707302     0.000000000
4       N      3.455466521     1.389450487    -0.020884229
5       C      4.156202600     2.659961686    -0.258484355
6       C      5.675318649     2.393394412    -0.500104613
7       O      6.384009955     3.570871719    -0.751561252
8       H     -0.353062416     0.154009985    -0.864744421
9       H      1.713149624    -0.545965142     0.926415200
10      H      1.802750234    -0.577916733    -0.871849747
11      H      1.594231333     2.012077468     0.883369063
12      H      1.629960537     1.992163286    -0.907595651
13      H      3.781501529     0.993011508     0.849604604
14      H      4.029748163     3.381211239     0.589306906
15      H      3.717042673     3.153915784    -1.159946775
16      H      6.137344694     1.935446962     0.412600015
```

```
17      H       5.805372759     1.655749585    -1.335390158
18      H       6.281290221     3.859811208    -1.647161348
---
H5C5N1O2, RHF, CHARGE=0, MULT=1
HF=-61.2
1       N       0.000000000     0.000000000     0.000000000
2       C       1.434800717     0.000000000     0.000000000
3       C       1.853368484     1.447467515     0.000000000
4       C       0.751592863     2.230190999    -0.000781671
5       C      -0.474846415     1.355119269    -0.001601355
6       C      -0.868315486    -1.179578337    -0.002067829
7       O       2.114417450    -1.011346273    -0.000001922
8       O      -1.655547175     1.656362933    -0.003202009
9       H       2.894830821     1.744323501     0.000604737
10      H       0.687492973     3.311129313    -0.000959781
11      H      -1.506675856    -1.189466067    -0.915431096
12      H      -1.529325073    -1.175885301     0.895012486
13      H      -0.275889917    -2.119340537     0.012749502
---
H7C5N1O2, RHF, CHARGE=0, MULT=1
HF=-94.1
1       C       0.000000000     0.000000000     0.000000000
2       C       1.527909399     0.000000000     0.000000000
3       C       2.161271480     1.398908119     0.000000000
4       C       1.488342596     2.376683612     0.974071098
5       C      -0.038668160     2.342884932     0.944062094
6       N      -0.660244195     1.155093052     0.486249197
7       O      -0.693575971    -0.939506931    -0.375533084
8       O      -0.761923670     3.265193272     1.305576229
9       H       1.868276130    -0.573556407     0.893622684
10      H       1.892215024    -0.562609278    -0.889419843
11      H       3.239115910     1.308888416     0.263697123
12      H       2.133653895     1.820960404    -1.030748303
13      H       1.813513330     2.160699182     2.018599600
14      H       1.835048739     3.410358271     0.746857223
15      H      -1.667253679     1.137522945     0.488083322
---
H7C5N1O2, RHF, CHARGE=0, MULT=1
HF=-93.1
1       N       0.000000000     0.000000000     0.000000000
2       C       1.427910841     0.000000000     0.000000000
3       C       1.927145587     1.448257156     0.000000000
4       C       0.657347597     2.320756248    -0.001717507
5       C      -0.513182989     1.333372925    -0.002815424
6       C      -0.853662624    -1.194582429    -0.002383236
7       O       2.089916599    -1.025250516    -0.000120439
8       O      -1.709110396     1.577217778    -0.005513174
9       H       2.560040150     1.640776960    -0.891813304
10      H       2.558119300     1.641331538     0.893063779
11      H       0.611524571     2.979568720    -0.894318612
12      H       0.609395961     2.979952560     0.890492770
13      H      -1.520913528    -1.195097787     0.889874492
14      H      -1.483772010    -1.217659328    -0.921081051
15      H      -0.251190620    -2.127507875     0.022431423
---
H9C5N1O2, RHF, CHARGE=0, MULT=1
HF=-87.5
1       C      -0.044556715    -0.069346376    -0.024053365
2       C       1.496046070    -0.169635093    -0.031693237
```

```
   3    C     -0.392109090    1.421470895    0.151701780
   4    C      0.953129994    2.165411783    0.393460035
   5    N      2.034689092    1.167182608    0.282553065
   6    C      1.089315524    2.805769631    1.791639973
   7    O      1.540655424    4.087284804    1.769999068
   8    O      0.840392466    2.286877828    2.877388346
   9    H     -0.476276461   -0.676923881    0.799058751
  10    H     -0.475344478   -0.472752673   -0.964790759
  11    H      1.869898867   -0.547823080   -1.013336295
  12    H      1.864713864   -0.882460925    0.743447093
  13    H     -1.122062219    1.569220717    0.973040548
  14    H     -0.880045314    1.824664339   -0.762336278
  15    H      1.059886164    2.965259727   -0.383592972
  16    H      2.706881183    1.453348270   -0.411845008
  17    H      1.668061094    4.490473890    2.620132174
EXPGEOM
  1     C     -0.04810    0.15030    0.80590
  2     C      0.91530    1.24280    0.33190
  3     C      1.98310    0.48760   -0.48300
  4     C      1.44340   -0.93540   -0.57310
  5     N      0.72320   -1.06750    0.68380
  6     C     -1.33660    0.17550    0.00070
  7     O     -1.87140    1.16280   -0.41710
  8     O     -1.84720   -1.05630   -0.15930
  9     H     -0.35600    0.30570    1.83820
 10     H      1.35740    1.74580    1.18340
 11     H      0.38410    1.98090   -0.25560
 12     H      2.93010    0.48700    0.04530
 13     H      2.14190    0.92620   -1.46160
 14     H      2.22750   -1.68080   -0.64640
 15     H      0.79390   -1.03730   -1.44980
 16     H      0.14010   -1.88680    0.71200
 17     H     -2.67490   -0.94470   -0.63530
---
H11C5N1O2, RHF, CHARGE=0, MULT=1
HF=-110
   1    N      0.000000000    0.000000000    0.000000000
   2    C      1.468069081    0.000000000    0.000000000
   3    C      2.024329057    1.442466664    0.000000000
   4    C      3.564677236    1.503754395   -0.079186792
   5    C      4.119736156    2.942466113   -0.088027488
   6    C      5.646211597    3.048359755   -0.164536994
   7    O      6.098517905    4.329521794   -0.220839640
   8    O      6.474925926    2.138888528   -0.176219171
   9    H     -0.347134218    0.276052942    0.905964522
  10    H     -0.327970674   -0.944767214   -0.131592196
  11    H      1.809722621   -0.538699858   -0.918507126
  12    H      1.890667902   -0.567618924    0.869961772
  13    H      1.590048779    2.002485952   -0.859304638
  14    H      1.687146291    1.968571912    0.922818441
  15    H      3.904337344    0.975673436   -0.999052790
  16    H      3.996590297    0.949866680    0.785208752
  17    H      3.694316607    3.499618215   -0.954479180
  18    H      3.790399155    3.475542218    0.833800696
  19    H      7.041406924    4.429167473   -0.268632086
---
H11C5N1O2, RHF, CHARGE=0, MULT=1
HF=-88.5
   1    C      0.000000000    0.000000000    0.000000000
```

```
 2      O      1.404235051     0.000000000     0.000000000
 3      C      2.190117213     1.110137751     0.000000000
 4      C      3.703922049     0.804362713     0.066734066
 5      N      4.119489164     0.789713800     1.469925204
 6      C      4.842875519     1.965681205     1.963233984
 7      O      1.682927852     2.225948866    -0.070306306
 8      C      4.522999396    -0.502613528     2.031113626
 9      H     -0.423192895     0.467622137    -0.920753217
10      H     -0.424572331     0.517351890     0.892828216
11      H     -0.317182700    -1.069854108     0.028735404
12      H      3.922641448    -0.162123545    -0.448568087
13      H      4.241828548     1.584717859    -0.527193562
14      H      5.902292958     2.005524004     1.610447136
15      H      4.338667080     2.902556867     1.638028875
16      H      4.855493239     1.962516993     3.076784173
17      H      5.522922543    -0.843781787     1.667413486
18      H      3.779941614    -1.290270772     1.774694689
19      H      4.570009484    -0.436883273     3.141685602
---
H11C5N1O2, RHF, CHARGE=0, MULT=1
HF=-45.8
 1      C      0.000000000     0.000000000     0.000000000
 2      C      1.575505350     0.000000000     0.000000000
 3      O      2.047884609     1.356091466     0.000000000
 4      N      1.826629043     2.111916062     1.050503334
 5      O      2.255568979     3.189357553     0.926726713
 6      H     -0.356586604     0.687296446    -0.802954691
 7      C     -0.719959651    -1.339217743    -0.189143002
 8      C      2.162526010    -0.792813155     1.212347036
 9      C      2.155996660    -0.574563361    -1.337813047
10      H     -0.366164966     0.434731627     0.959303800
11      H      3.256452171    -0.638655156     1.310929521
12      H      1.695476549    -0.501361840     2.174214063
13      H      1.998279973    -1.881548823     1.085340237
14      H      3.257037098    -0.453004326    -1.390310788
15      H      1.721552715    -0.071649472    -2.224727627
16      H      1.943036146    -1.658437516    -1.426473671
17      H     -0.453143296    -2.080196494     0.590219403
18      H     -1.816562820    -1.176963743    -0.123913564
19      H     -0.517439459    -1.795193491    -1.178446418
---
H11C5N1O2, RHF, CHARGE=0, MULT=1
HF=-108.2
 1      H      0.000000000     0.000000000     0.000000000
 2      N      1.007974670     0.000000000     0.000000000
 3      C      1.517731782     1.375717642     0.000000000
 4      C      0.949733600     2.217088130    -1.179073295
 5      O      0.302324900     3.257409457    -1.068994418
 6      C      3.085877161     1.401987199     0.087572243
 7      C      3.653163661     2.839355778     0.113755669
 8      C      3.624308003     0.587148882     1.286332823
 9      H      1.288194472    -0.510953951    -0.821927761
10      H      3.495023014     0.916454492    -0.834535346
11      H      4.755769659     2.823801786     0.237111429
12      H      3.453563980     3.377574922    -0.834698987
13      H      3.233307877     3.443291040     0.942932046
14      H      4.730573810     0.663520439     1.342302342
15      H      3.215674294     0.945635891     2.252440827
16      H      3.388374141    -0.491358332     1.195417239
```

```
17      H      1.117860286      1.854287448      0.935960877
18      O      1.169336095      1.721151468     -2.423687078
19      H      0.825368163      2.241350200     -3.140031660
---
H13C5N1O2, RHF, CHARGE=0, MULT=1
HF=-85
1       C      0.000000000      0.000000000      0.000000000
2       O      1.397136107      0.000000000      0.000000000
3       C      2.153179569      1.181936572      0.000000000
4       O      1.939916713      1.796092350      1.245500417
5       C      2.307955496      3.123304128      1.477431001
6       N      3.579322412      0.836292733     -0.163380655
7       C      4.066356480      0.780404785     -1.546323177
8       C      4.217180435     -0.120745739      0.751785315
9       H     -0.434571343      0.686499678     -0.768742597
10      H     -0.431384978      0.275912919      0.992171349
11      H     -0.324371575     -1.040676796     -0.241874227
12      H      1.838162985      1.869230757     -0.849508943
13      H      1.838815225      3.831115259      0.748631590
14      H      3.412194681      3.284303224      1.448565185
15      H      1.944319646      3.387386671      2.499398837
16      H      3.758554803      1.693071338     -2.106605951
17      H      3.698867558     -0.106984905     -2.115083294
18      H      5.179778671      0.750665667     -1.556725623
19      H      3.802482934     -0.052871260      1.780209061
20      H      4.119719996     -1.181025195      0.416940964
21      H      5.306068343      0.114127633      0.818063151
---
H5C6N1O2, RHF, CHARGE=0, MULT=1
HF=-52.9
1       C      0.000000000      0.000000000      0.000000000
2       C      1.418450480      0.000000000      0.000000000
3       N      2.150692707      1.136047755      0.000000000
4       C      1.503568641      2.324392865      0.007001351
5       C     -0.667455542      1.244198974      0.009704545
6       C      0.095881166      2.422156560      0.013197910
7       C     -0.765334603     -1.285554036     -0.027828627
8       O     -1.195773901     -1.856013403     -1.028217910
9       O     -0.979552059     -1.833589575      1.195497384
10      H      1.995996151     -0.931327976     -0.004180926
11      H      2.138956241      3.216421779      0.007636114
12      H     -1.756762156      1.300332294      0.013112452
13      H     -0.387876313      3.397872322      0.020228605
14      H     -1.461777139     -2.651808421      1.197555018
---
H5C6N1O2, RHF, CHARGE=0, MULT=1
HF=15.4, IE=9.9
1       C      0.000000000      0.000000000      0.000000000
2       C      2.814068219      0.000000000      0.000000000
3       C      0.703801116      1.216935076      0.000000000
4       C      0.704534879     -1.216822600      0.000389587
5       C      2.110254940      1.228609210     -0.000067766
6       C      2.110995886     -1.228816416      0.000241093
7       N      4.312826639      0.000490241     -0.000185590
8       O      4.918028762      1.050693113     -0.000077946
9       O      4.918216900     -1.049630996     -0.000430601
10      H     -1.090813304     -0.000508517     -0.000186174
11      H      0.159183709      2.161919924     -0.000085644
12      H      0.159877105     -2.161864556      0.000714077
```

```
13      H       2.619345019     2.194284365     -0.000942766
14      H       2.620071930     -2.194483836    0.000452867
EXPGEOM
1       C       0.00000         0.00000         0.24180
2       C       0.00000         1.22160         -0.42480
3       C       0.00000         1.20990         -1.81810
4       C       0.00000         0.00000         -2.51450
5       C       0.00000         -1.20990        -1.81810
6       C       0.00000         -1.22160        -0.42480
7       N       0.00000         0.00000         1.71270
8       O       0.00000         -1.08290        2.28800
9       O       0.00000         1.08290         2.28800
10      H       0.00000         2.14780         0.13950
11      H       0.00000         2.15270         -2.36090
12      H       0.00000         0.00000         -3.60260
13      H       0.00000         -2.15270        -2.36090
14      H       0.00000         -2.14780        0.13950
---
H9C6N1O2, RHF, CHARGE=0, MULT=1
HF=-74.6
1       C       0.000000000     0.000000000     0.000000000
2       C       1.546931609     0.000000000     0.000000000
3       C       2.124893796     1.435484922     0.000000000
4       O       2.845965536     1.714247872     -1.115455522
5       C       3.531868277     2.917257066     -1.392361190
6       C       2.661497949     3.958950609     -2.120440134
7       C       2.099860797     -0.750796859    1.125270108
8       N       2.544380390     -1.364568517    2.005396607
9       O       1.976434301     2.273741352     0.882914364
10      H       -0.422096001    0.447085645     0.921117793
11      H       -0.387364530    0.568391803     -0.869236349
12      H       -0.380496697    -1.038327971    -0.080935915
13      H       1.873003369     -0.529623937    -0.933639377
14      H       3.985389272     3.369790630     -0.474683676
15      H       4.387332678     2.619172520     -2.054891224
16      H       2.218040730     3.549559674     -3.050071391
17      H       1.836752555     4.336927901     -1.484760146
18      H       3.294890831     4.825310954     -2.400476949
---
H13C6N1O2, RHF, CHARGE=0, MULT=1
HF=-97.3
1       C       0.000000000     0.000000000     0.000000000
2       C       1.540732601     0.000000000     0.000000000
3       O       2.094219574     1.296705295     0.000000000
4       C       2.443535337     2.025480012     -1.092738896
5       C       3.125247682     3.370111677     -0.748789483
6       N       4.573780222     3.165271157     -0.699102589
7       C       5.352486715     3.627578860     -1.852111580
8       O       2.196058341     1.615368362     -2.223051922
9       C       5.218401806     3.289320729     0.610754707
10      H       -0.412885163    0.585492350     0.845765709
11      H       -0.421140858    0.410730473     -0.938642189
12      H       -0.361940324    -1.043468804    0.102946253
13      H       1.910528327     -0.480492048    0.944707972
14      H       1.922217177     -0.630662390    -0.842300519
15      H       2.828762790     4.117538037     -1.526316585
16      H       2.722272013     3.769025577     0.213375613
17      H       4.868085202     3.316981477     -2.804276421
18      H       5.476601174     4.737772401     -1.877344405
```

```
19      H       6.368459035     3.172348787     -1.829844892
20      H       5.285593622     4.346568231     0.966339889
21      H       4.661267449     2.706341089     1.377784326
22      H       6.252972529     2.879888379     0.566926614
---
H13C6N1O2, RHF, CHARGE=0, MULT=1
HF=-115.3
1       O       0.000000000     0.000000000     0.000000000
2       C       1.230366999     0.000000000     0.000000000
3       C       2.180116196     1.201937424     0.000000000
4       C       1.489997395     2.579622783     0.062860846
5       C       2.481271452     3.762774893     0.066590097
6       C       1.802054169     5.145104547     0.128237785
7       C       2.801516765     6.329561053     0.131648071
8       N       2.215325189     7.672266706     0.189098011
9       O       1.872920208     -1.198318051    -0.003465347
10      H       2.871148145     1.092019114     0.867499023
11      H       2.804212351     1.138833024     -0.921468086
12      H       0.801340205     2.684830295     -0.805997909
13      H       0.858704617     2.631189261     0.978917880
14      H       3.172202747     3.657191456     0.934361257
15      H       3.115873565     3.709981617     -0.847763012
16      H       1.111376742     5.249743012     -0.740064541
17      H       1.166446611     5.198099253     1.042009909
18      H       3.492766747     6.232416860     1.005531446
19      H       3.437342231     6.284196268     -0.787398374
20      H       1.593694613     7.820503218     -0.590685153
21      H       1.644211447     7.773794607     1.013717765
22      H       1.311646622     -1.963913247    -0.005732982
---
H13C6N1O2, RHF, CHARGE=0, MULT=1
HF=-110
1       H       -0.126826906    0.243846627     -0.360503199
2       N       0.880647887     0.196893874     -0.356099504
3       C       1.439579289     1.552460450     -0.258386400
4       C       2.943334931     1.468255695     -0.612720631
5       O       3.501768933     2.141043762     -1.478735362
6       C       1.093921505     2.366734627     1.046389672
7       C       1.419118575     3.879586186     0.872709857
8       C       0.636237243     4.844390560     1.770670864
9       C       1.661807932     1.800146260     2.363843643
10      H       1.121449702     -0.378515458    0.434845552
11      H       -0.021951078    2.290338740     1.135807847
12      H       1.217852095     4.182154494     -0.181314844
13      H       0.879869790     5.891693390     1.495821000
14      H       -0.459762719    4.718557073     1.659012605
15      H       0.885320835     4.720974800     2.843447975
16      H       2.507431780     4.051951605     1.035642591
17      H       1.173241587     2.282493089     3.235926997
18      H       1.473217530     0.712948054     2.467925914
19      H       2.752694916     1.963847165     2.465588168
20      H       0.959255404     2.119988084     -1.105512315
21      O       3.683995949     0.571694058     0.085261983
22      H       4.589113457     0.488951294     -0.190487934
---
H13C6N1O2, RHF, CHARGE=0, MULT=1
HF=-113.1
1       H       0.001171499     -0.007612800    0.094051512
2       N       1.004956312     -0.004760220    0.003172139
```

```
 3    C     1.509194422    1.369452846   -0.044437138
 4    C     0.861778693    2.260657554   -1.142719468
 5    O     0.482699820    3.420862023   -0.990003773
 6    C     3.063424763    1.409110533   -0.169389631
 7    C     3.880071676    0.834500077    1.022581389
 8    C     5.321866197    0.492367760    0.584712065
 9    C     3.874614806    1.744807724    2.268119548
10    H     1.211742907   -0.517612953   -0.838941651
11    H     3.365599004    2.470287551   -0.328981672
12    H     3.415990269   -0.134758765    1.333925048
13    H     5.895854824    0.047016656    1.422894653
14    H     5.325100489   -0.251136484   -0.238536677
15    H     5.878498778    1.385144405    0.234502385
16    H     4.476758641    1.297578430    3.085852879
17    H     4.293271002    2.749497073    2.056549386
18    H     2.850127379    1.884859707    2.667354738
19    H     3.358373171    0.869830767   -1.099185624
20    H     1.212555088    1.841738554    0.931035396
21    O     0.708107113    1.685611386   -2.362882007
22    H     0.326397513    2.238747864   -3.033971653
---
H13C6N1O2, RHF, CHARGE=0, MULT=1
HF=-94.3
 1    N     0.000000000    0.000000000    0.000000000
 2    C     1.470464351    0.000000000    0.000000000
 3    C     1.919416959    1.489006685    0.000000000
 4    O     1.997590294    2.239649950    0.968025601
 5    O     2.256077372    1.943303061   -1.237913699
 6    C     2.627633560    3.262156761   -1.545161352
 7    C    -0.650399791   -0.746455076   -1.081273956
 8    C    -0.745532947    0.061898395    1.260036998
 9    C     2.169738690   -0.821691468    1.110086048
10    H     1.828664773   -0.461030923   -0.961129449
11    H     2.802978830    3.293694483   -2.647099738
12    H     1.827733046    3.998627725   -1.294674929
13    H     3.569844494    3.566815603   -1.030231384
14    H    -0.210628983   -0.474281113   -2.067271600
15    H    -0.563944253   -1.854027607   -0.961454023
16    H    -1.734500426   -0.495591566   -1.123951724
17    H    -0.855764755   -0.935449248    1.751948683
18    H    -0.264820159    0.749360703    1.988665846
19    H    -1.770408536    0.454962217    1.065387653
20    H     3.267414919   -0.810184475    0.945761727
21    H     1.984840603   -0.444937734    2.134422466
22    H     1.843357989   -1.881334168    1.074723138
---
H15C6N1O2, RHF, CHARGE=0, MULT=1
HF=-92.5
 1    C     0.000000000    0.000000000    0.000000000
 2    O     1.396944952    0.000000000    0.000000000
 3    C     2.236953796    1.132458044    0.000000000
 4    C     2.529263204    1.571808587   -1.485466053
 5    O     1.558329072    2.141559905    0.712594155
 6    N     3.458673685    0.756975318    0.766711649
 7    C     3.311193501    0.275367224    2.151851442
 8    C     4.561550295    0.096455594    0.054886881
 9    C     2.043657461    3.433865713    0.925094290
10    H    -0.429661927    0.251672647    0.999618453
11    H    -0.446488664    0.689993101   -0.756448797
```

```
12      H      -0.316238929     -1.041002123     -0.257037041
13      H       3.329510295      2.335566372     -1.544685431
14      H       2.827684030      0.712475274     -2.117963242
15      H       1.621563944      2.010503893     -1.949331615
16      H       2.440792692      0.724255419      2.675572914
17      H       3.203361915     -0.833475746      2.223113325
18      H       4.220509254      0.565828604      2.730579457
19      H       4.267142899     -0.851242480     -0.454596052
20      H       5.001682377      0.778961902     -0.707439362
21      H       5.386104216     -0.152557982      0.761923341
22      H       3.115419256      3.481659132      1.228996925
23      H       1.915659154      4.077491377      0.018726169
24      H       1.430447297      3.877027929      1.747263955
---
H7C7N1O2, RHF, CHARGE=0, MULT=1
HF=-69.2
1       C       0.000000000      0.000000000      0.000000000
2       C       1.418630585      0.000000000      0.000000000
3       C       2.130362457      1.218264196      0.000000000
4       C       1.429004642      2.442826608      0.002056738
5       C       0.023350720      2.443330313      0.002598015
6       C      -0.693979935      1.237643282      0.000618586
7       N      -0.715415844     -1.223393695      0.118228420
8       C       3.630494045      1.207665992     -0.027036826
9       O       4.214491667      1.289666002      1.195685538
10      O       4.341445375      1.134276151     -1.028181949
11      H      -0.515632398      3.391890765      0.004828625
12      H      -1.784534920      1.270710971      0.003197523
13      H      -1.619704190     -1.173313192     -0.322508670
14      H      -0.226273523     -1.989511336     -0.315743691
15      H       1.967886987     -0.943418790      0.002224215
16      H       5.164100771      1.284622550      1.195409658
17      H       1.964278352      3.393047768      0.001852641
---
H7C7N1O2, RHF, CHARGE=0, MULT=1
HF=-70.8
1       C       0.000000000      0.000000000      0.000000000
2       C       1.425114344      0.000000000      0.000000000
3       C       2.115243447      1.235012887      0.000000000
4       C       1.419536305      2.453006918     -0.007711522
5       C       0.013278062      2.451169201     -0.018459277
6       C      -0.693070638      1.241380340     -0.011069803
7       N      -0.721736531     -1.219590709     -0.103384987
8       C       2.224990387     -1.269276666      0.010032220
9       O       2.730614697     -1.625783808     -1.197153746
10      O       2.463729626     -1.972546251      0.992030371
11      H      -0.531519344      3.396509235     -0.028079866
12      H      -1.783908428      1.270215648     -0.015604758
13      H      -1.707258607     -1.093206317      0.057996078
14      H      -0.411336592     -1.910727079      0.560278427
15      H       3.229500536     -2.433903195     -1.207381178
16      H       1.966911826      3.395434392     -0.007333548
17      H       3.207029001      1.251635453      0.006548485
---
H7C7N1O2, RHF, CHARGE=0, MULT=1
HF=-70.2
1       C       0.000000000      0.000000000      0.000000000
2       C       1.419315449      0.000000000      0.000000000
3       C       2.131044753      1.209053121      0.000000000
```

```
    4    C         1.447801880       2.445378698      -0.005916288
    5    C         0.035506579       2.447385435      -0.005610176
    6    C        -0.684556544       1.243436702      -0.004788534
    7    N        -0.719396630      -1.216210367       0.119677242
    8    C         2.209538136       3.734419054      -0.034917304
    9    O         2.514318744       4.234691930       1.190416260
   10    O         2.566520144       4.350645238      -1.038043502
   11    H         3.222173801       1.179526508       0.002121489
   12    H        -0.515115323       3.390106270      -0.008128108
   13    H        -1.774773340       1.283952602      -0.003333230
   14    H        -1.628319341      -1.168077135      -0.310174099
   15    H        -0.234703746      -1.991722167      -0.300856837
   16    H         1.978994321      -0.936679885       0.004544827
   17    H         2.997560961       5.052108603       1.188421191
 ---
H7C7N1O2, RHF, CHARGE=0, MULT=1
HF=7.4
    1    C         0.000000000       0.000000000       0.000000000
    2    C         1.414623087       0.000000000       0.000000000
    3    C         2.147046853       1.200981257       0.000000000
    4    C         1.446067707       2.421615223      -0.006353221
    5    C         0.038124671       2.461190370      -0.010577480
    6    C        -0.670761226       1.247546049      -0.005612447
    7    C        -0.784659180      -1.285748302       0.002001432
    8    N         2.216607826       3.711532252      -0.008265139
    9    O         2.516155410       4.219609183       1.048066755
   10    O         2.521104640       4.214504692      -1.065582381
   11    H         3.237393324       1.163591444       0.003946742
   12    H        -0.510922476       3.403889345      -0.015562519
   13    H        -1.761859235       1.283206180      -0.006002551
   14    H        -1.500533057      -1.311729462       0.850022821
   15    H        -1.364604967      -1.390546115      -0.939112518
   16    H        -0.133936133      -2.178497285       0.092862137
   17    H         1.966097426      -0.942027551       0.001791934
 ---
H7C7N1O2, RHF, CHARGE=0, MULT=1
HF=7.3
    1    C         0.000000000       0.000000000       0.000000000
    2    C         1.416417329       0.000000000       0.000000000
    3    C         2.133534885       1.208582480       0.000000000
    4    C         1.452881915       2.438745726       0.000171717
    5    C         0.046943698       2.452137446       0.002854137
    6    C        -0.673730450       1.245795805       0.002892329
    7    C        -0.770660345      -1.293029309       0.024556423
    8    O        -0.264374026      -2.502669122      -1.986520728
    9    O        -2.075996463      -1.426610713      -1.986289460
   10    H        -0.488740583       3.402135957       0.003202117
   11    H        -1.764629736       1.286226484       0.003975542
   12    H        -0.213437965      -2.109942709       0.537438865
   13    N        -1.058823179      -1.777284526      -1.434229485
   14    H         2.011128650       3.375342840      -0.002111100
   15    H         3.224042343       1.189678515      -0.002059333
   16    H         1.971963442      -0.939682998      -0.001679433
   17    H        -1.754404086      -1.191997805       0.537096002
 ---
H15C7N1O2, RHF, CHARGE=0, MULT=1
HF=-108.1
    1    C         0.000000000       0.000000000       0.000000000
    2    C         1.570594320       0.000000000       0.000000000
```

```
 3     O    2.287304558    0.995998254    0.000000000
 4     O    2.151510844   -1.232098756    0.045968413
 5     N   -0.390019293    0.947694726   -1.069012374
 6     C   -0.095480899    0.612712146   -2.466816039
 7     C   -1.544647735    1.832529975   -0.895802955
 8     C    3.533681894   -1.478912827    0.076550415
 9     C   -0.637677443   -1.409760750   -0.227607236
10     C   -0.407631183    0.486832240    1.430887354
11     H    0.798187101   -0.040324948   -2.568965313
12     H   -0.943950011    0.093007173   -2.976061885
13     H    0.119305112    1.547849143   -3.035130444
14     H   -2.496967645    1.281150095   -0.704173293
15     H   -1.379953886    2.545273459   -0.056535553
16     H   -1.704164749    2.450936333   -1.808601621
17     H    4.054854389   -1.089860286   -0.830202111
18     H    4.021318175   -1.048297547    0.983258855
19     H    3.656168424   -2.588152976    0.104346723
20     H   -1.742417826   -1.335141829   -0.303665516
21     H   -0.279540949   -1.904801896   -1.151564504
22     H   -0.418371101   -2.095802126    0.615575971
23     H   -0.059603814    1.514875079    1.653377872
24     H   -1.505486420    0.459960808    1.583472323
25     H    0.036473424   -0.178988952    2.200692439
---
H4C2N2O2, RHF, CHARGE=0, MULT=1
HF=-92.5
 1     N   -0.024780356   -0.033304241    0.005218420
 2     C    1.357414601    0.005007398    0.005071887
 3     C    2.018631510    1.389423009   -0.006180474
 4     N    3.400866435    1.427559831    0.000666560
 5     O    1.386489190    2.446834710   -0.020580618
 6     O    1.989825407   -1.052357673    0.014870201
 7     H   -0.593790761    0.781996994   -0.005339808
 8     H   -0.517860486   -0.899142323    0.010957907
 9     H    3.969614944    0.612090342    0.012360111
10     H    3.894333762    2.293078289   -0.005166170
EXPGEOM
 1     C    0.00000    0.76980    0.00000
 2     C    0.00000   -0.76980    0.00000
 3     O    1.04350    1.40960    0.00000
 4     O   -1.04350   -1.40960    0.00000
 5     N   -1.24660    1.27640    0.00000
 6     N    1.24660   -1.27640    0.00000
 7     H   -1.38240    2.27650    0.00000
 8     H   -2.03450    0.64380    0.00000
 9     H    1.38240   -2.27650    0.00000
10     H    2.03450   -0.64380    0.00000
---
H6C2N2O2, RHF, CHARGE=0, MULT=1
HF=-3.2
 1     N    0.000000000    0.000000000    0.000000000
 2     C    1.486284010    0.000000000    0.000000000
 3     H    1.841871917    1.054268641    0.000000000
 4     H    1.883815929   -0.481730533    0.920077201
 5     H    1.911863545   -0.510166972   -0.894336606
 6     C   -0.666666851    1.105544733   -0.735578099
 7     H   -0.710799347    0.920926930   -1.833281484
 8     H   -1.695135828    1.282215638   -0.352080767
 9     H   -0.088104072    2.040267908   -0.560770618
```

```
10      N        -0.649373095    -1.233964330     0.060087784
11      O        -1.817488202    -1.293265993    -0.252677771
12      O        -0.041499742    -2.185588563     0.496258065
EXPGEOM
1       O         1.35370         1.09440          0.03300
2       O         1.35370        -1.09450          0.03310
3       H        -1.19920         1.50970          1.11450
4       H        -2.26890         1.14800         -0.25910
5       H        -0.76990         2.05650         -0.51290
6       C        -1.23470         1.25820          0.05210
7       H        -1.19930        -1.50960          1.11450
8       H        -0.77010        -2.05650         -0.51290
9       H        -2.26900        -1.14780         -0.25910
10      C        -1.23490        -1.25810          0.05210
11      N         0.79910        -0.00000         -0.02890
12      N        -0.56560        -0.00000         -0.23380
---
H6C3N2O2, RHF, CHARGE=0, MULT=1
HF=-105.6
1       C        -0.098820723     0.125818714    -0.006993136
2       C         1.415836556     0.267978436    -0.011531247
3       O         1.993475176     1.353023069    -0.030017630
4       N         2.282238196    -0.867004507    -0.000692169
5       C         1.986119172    -2.250169684     0.023648151
6       O         0.847962536    -2.701033376     0.148597059
7       N         3.078861814    -3.100857758    -0.128644862
8       H        -0.453058063    -0.369408583     0.918688220
9       H        -0.452190385    -0.446961246    -0.887213152
10      H        -0.575440601     1.127613502    -0.050121543
11      H         3.260973258    -0.632707912    -0.026903555
12      H         4.021159438    -2.787127644    -0.129505631
13      H         2.956669141    -4.085837010    -0.042143483
---
H6C3N2O2, RHF, CHARGE=0, MULT=1
HF=-99.5
1       O        -0.586971903     0.077759238     0.529329900
2       C        -1.166144856     0.598057722    -0.425914448
3       C        -0.530791285     1.682679877    -1.306383800
4       C        -0.935457291     3.096108900    -0.866079667
5       O        -1.861384799     3.725621014    -1.380743360
6       N        -0.193327154     3.693151715     0.144089169
7       N        -2.446740870     0.171620118    -0.751226645
8       H         0.577255366     1.569411606    -1.285496593
9       H        -0.814836307     1.535306015    -2.373460605
10      H         0.517982462     3.208327577     0.640404144
11      H        -0.424724661     4.594017374     0.499730970
12      H        -2.986639653     0.602018429    -1.465842324
13      H        -2.928005704    -0.503486523    -0.199598082
---
H4C4N2O2, RHF, CHARGE=0, MULT=1
HF=36.3
1       N         0.000000000     0.000000000     0.000000000
2       C         1.403782341     0.000000000     0.000000000
3       C         2.126228440     1.192357647     0.000000000
4       N         1.476753771     2.437045730     0.000124721
5       C         0.072853283     2.437001400     0.000214776
6       C        -0.649562467     1.244647962     0.000183635
7       O        -0.636906050    -1.050496320    -0.000074457
8       O         2.113507887     3.487606408     0.000137665
```

```
 9    H    1.912476293   -0.968235725   -0.000050838
10    H    3.219962823    1.189738000   -0.000050230
11    H   -0.435711349    3.405309477    0.000308058
12    H   -1.743320150    1.247196424    0.000282269
---
H4C4N2O2, RHF, CHARGE=0, MULT=1
HF=-72.4
 1    C   -0.004966585   -0.008984104   -0.000833602
 2    C    1.471233225   -0.006763081   -0.002446187
 3    C    2.176338255    1.156282921   -0.001635919
 4    N    1.510146847    2.383641829    0.000513619
 5    C    0.099730749    2.503658727    0.001661676
 6    N   -0.603988377    1.288986179    0.001119723
 7    O   -0.751543472   -0.980377974   -0.001120710
 8    O   -0.448717432    3.601584370    0.003206024
 9    H    1.974280612   -0.973066451   -0.004642417
10    H    3.269598276    1.204699580   -0.002741951
11    H    2.058140746    3.222678470    0.001158880
12    H   -1.607845187    1.354383025    0.001905225
EXPGEOM
 1    C    1.26490    0.38850    0.00000
 2    C    1.23210   -1.06020    0.00000
 3    N    0.00000    0.99410    0.00000
 4    C    0.05340   -1.70200    0.00000
 5    O    2.28040    1.06750    0.00000
 6    N   -1.13850   -1.02100    0.00000
 7    C   -1.22210    0.36090    0.00000
 8    O   -2.29190    0.94720    0.00000
 9    H    2.17450   -1.57760    0.00000
10    H   -0.03860    2.00070    0.00000
11    H   -0.02190   -2.77910    0.00000
12    H   -2.02200   -1.49740    0.00000
---
H6C5N2O2, RHF, CHARGE=0, MULT=1
HF=-78.6
 1    C    0.002415295   -0.010706291   -0.001508996
 2    C    1.489459981   -0.019895153   -0.001142312
 3    C    2.168382396    1.167667655    0.001602932
 4    N    1.494158239    2.388168182    0.002717831
 5    C    0.084895408    2.504420517    0.000663401
 6    N   -0.604135834    1.283547381   -0.001149915
 7    O   -0.754954276   -0.974278490   -0.002054364
 8    C    2.202738883   -1.344667505   -0.004819798
 9    O   -0.470782468    3.598602172    0.000752074
10    H    3.261381085    1.244590418    0.003099715
11    H   -1.609162795    1.338266653   -0.001609111
12    H    3.305686265   -1.219665611    0.000014111
13    H    1.943614149   -1.935451168   -0.908048800
14    H    1.937012371   -1.945029896    0.890041589
15    H    2.039394787    3.229306091    0.004460992
---
H6C6N2O2, RHF, CHARGE=0, MULT=1
HF=14.9
 1    C   -0.000567016   -0.000474564    0.000221106
 2    C    1.420796924    0.000640287   -0.000467768
 3    C    2.116120789    1.238585512    0.000690457
 4    C    1.369644231    2.432483275    0.005442605
 5    C   -0.037283000    2.445231876    0.010023579
 6    C   -0.710914521    1.209157026    0.007006521
```

```
7    N     2.136169683   -1.218367337    0.120237121
8    N     2.107464351    3.743860672    0.001900309
9    O     2.348728745    4.288427765    1.054585833
10   O     2.442421387    4.226289800   -1.056059008
11   H     3.207141272    1.262724246    0.002133969
12   H     3.054469496   -1.168194857   -0.289171033
13   H    -1.801951447    1.194191543    0.010533452
14   H    -0.609453727    3.373602482    0.015106154
15   H     1.658167920   -1.987799067   -0.318836259
16   H    -0.558445616   -0.938475852    0.000197028
---
H6C6N2O2, RHF, CHARGE=0, MULT=1
HF=13.2
1    C     0.000000000    0.000000000    0.000000000
2    C     1.421743669    0.000000000    0.000000000
3    C     2.134187795    1.208625368    0.000000000
4    C     1.422147321    2.425056485   -0.002792485
5    C     0.013238074    2.453350989   -0.006428520
6    C    -0.693307261    1.241050869   -0.006311907
7    N    -0.713112274   -1.214726702    0.123754471
8    N     2.181457933    3.719625294   -0.006772594
9    O     2.438315596    4.248163338   -1.064726293
10   O     2.521592405    4.208621536    1.046184769
11   H     1.982040340   -0.936182106    0.005464137
12   H     3.225322174    1.184299260    0.002306398
13   H    -0.230597514   -1.997846490   -0.283317359
14   H    -1.783988727    1.273175135   -0.006271387
15   H    -0.540515961    3.393810439   -0.008040106
16   H    -1.630950353   -1.175791442   -0.285975531
---
H14C6N2O2, RHF, CHARGE=0, MULT=1
HF=-102.5
1    H    -0.137473956    0.047948325    0.018974035
2    N    -0.012104413    0.275662538    0.992862377
3    C     0.636127685    1.586205105    1.131153853
4    C     2.009497964    1.639503054    0.408854413
5    O     2.938290056    0.842485876    0.542929939
6    C     0.742356488    1.961543891    2.641354991
7    C     1.008455474    3.455725776    2.922388261
8    C     1.060099286    3.802431268    4.424259253
9    C     1.224374584    5.317443964    4.707086063
10   N     1.313900111    5.708447614    6.117168501
11   H     0.547412077   -0.466206381    1.383455238
12   H    -0.053682256    2.321902376    0.637069785
13   H     0.499509144    5.397442149    6.623586829
14   H     2.098319440    5.256085937    6.560415220
15   H     2.146232082    5.695153034    4.198992032
16   H     0.362102090    5.874607572    4.263058379
17   H     0.130438769    3.436877531    4.917936231
18   H     1.902950668    3.244919083    4.894102075
19   H     1.971954907    3.759430014    2.454309746
20   H     0.214441256    4.065134349    2.433546632
21   H    -0.217420810    1.678672054    3.132843138
22   H     1.533395672    1.347614648    3.128580504
23   O     2.154972806    2.663826024   -0.467226983
24   H     2.992964635    2.708533845   -0.912412773
---
H5C2N3O2, RHF, CHARGE=0, MULT=1
HF=-104.4
```

```
   1    O       -0.148301710     -0.023228562      0.605785934
   2    C       -1.252486483     -0.104525154      0.068607247
   3    N       -1.991066033      1.032811625     -0.304035148
   4    C       -1.658462601      2.405110519     -0.059402870
   5    O       -2.508458344      3.156499959      0.414481831
   6    N       -0.413791883      2.850423298     -0.476508835
   7    N       -1.820149911     -1.351793572     -0.308410082
   8    H       -2.913830823      0.893560301     -0.675246202
   9    H        0.324562260      2.227312741     -0.716670606
  10    H       -0.114851287      3.761133916     -0.197289513
  11    H       -2.807245012     -1.437207070     -0.143346232
  12    H       -1.358052383     -2.140364984      0.111927085
---
O3, RHF, CHARGE=0, MULT=1
HF=34.1, DIP=0.53, IE=12.75
   1    O        0.000000000      0.000000000      0.000000000
   2    O        1.189359703      0.000000000      0.000000000
   3    O        1.740366500      1.063437049      0.000000000
EXPGEOM
   1    O        0.00000          0.00000          0.44320
   2    O        0.00000          1.08550         -0.22160
   3    O        0.00000         -1.08550         -0.22160
---
H6C3O3, RHF, CHARGE=0, MULT=1
HF=-111.3
   1    O        0.006907411     -0.013623858     -0.063404408
   2    C        1.407106144     -0.055994824      0.034607109
   3    O        1.949095985      1.233150547     -0.071484333
   4    C        1.506314250      2.110239671     -1.075669485
   5    O        0.244587599      1.728370579     -1.559755899
   6    C       -0.647323143      1.038119077     -0.726171357
   7    H        1.831088319     -0.748040260     -0.750666722
   8    H        1.684029196     -0.463732928      1.045836782
   9    H        2.218760966      2.115575599     -1.948387717
  10    H        1.478228894      3.148682384     -0.638225033
  11    H       -1.124783004      1.726813711      0.030805884
  12    H       -1.462189588      0.610319988     -1.373277274
EXPGEOM
   1    C        0.00000          1.33460          0.18680
   2    C       -1.15580         -0.66730          0.18680
   3    C        1.15580         -0.66730          0.18680
   4    O       -1.17390          0.67780         -0.27350
   5    O        1.17390          0.67780         -0.27350
   6    O        0.00000         -1.35560         -0.27350
   7    H        0.00000          2.34860         -0.23270
   8    H        0.00000          1.35470          1.29990
   9    H       -2.03400         -1.17430         -0.23270
  10    H       -1.17320         -0.67730          1.29990
  11    H        2.03400         -1.17430         -0.23270
  12    H        1.17320         -0.67730          1.29990
---
H6C3O3, RHF, CHARGE=0, MULT=1
HF=-133.1
   1    C        0.000000000      0.000000000      0.000000000
   2    O        1.405593599      0.000000000      0.000000000
   3    C        2.190517774      1.107703857      0.000000000
   4    C        3.707886294      0.787683733     -0.013473596
   5    O        4.417849008      1.650921004     -0.837500730
   6    O        1.689345861      2.228123155      0.040169921
```

```
7    H    -0.315852063   -1.069085819   -0.052194207
8    H    -0.425091135    0.537655654   -0.880164869
9    H    -0.419098012    0.447714269    0.932334452
10   H     3.891751869   -0.248921134   -0.394045629
11   H     4.062234580    0.821530573    1.051370198
12   H     4.500441436    2.522044173   -0.473363169
---
H8C3O3, RHF, CHARGE=0, MULT=1
HF=-138.1
1    C    -0.181729902    0.024555433   -0.124636537
2    C     1.392737846    0.047559882   -0.033222258
3    C     2.031329829    1.491998905    0.007815165
4    O    -0.797509318    0.565783016    1.004004632
5    O     1.858233194   -0.649742287   -1.157025099
6    O     1.971889286    2.076348313    1.271025558
7    H    -0.520452294   -1.030938040   -0.299865674
8    H    -0.527661145    0.625124205   -1.004563930
9    H     1.697039934   -0.494716650    0.906578116
10   H     1.561725319    2.140015088   -0.776772619
11   H     3.120146860    1.419475057   -0.254184964
12   H    -0.866254525   -0.064585173    1.707094352
13   H     2.727729430   -0.996434132   -1.015037131
14   H     1.172800640    2.566393479    1.404030793
EXPGEOM
1    H    -1.59740   -0.53910    1.42260
2    H    -0.17800   -1.52980    0.97450
3    C    -0.83700   -0.71640    0.65020
4    H     0.34270    0.88280    1.43730
5    C    -0.03050    0.55910    0.46090
6    H     1.65790    1.32370   -0.59800
7    H     0.80720    0.00720   -1.44000
8    C     1.16810    0.35070   -0.46120
9    H     2.77180   -0.77090   -0.43730
10   O     2.02590   -0.60260    0.15240
11   H    -2.02130   -1.75220   -0.52730
12   O    -1.44080   -0.98590   -0.61390
13   H    -1.40840    1.17980   -0.72210
14   O    -0.85740    1.59330   -0.03970
---
H2C4O3, RHF, CHARGE=0, MULT=1
HF=-95.2, IE=10.84
1    C     0.000000000    0.000000000    0.000000000
2    C     1.352632485    0.000000000    0.000000000
3    C     1.807638920    1.440963194    0.000000000
4    C    -0.455119844    1.440813453   -0.000072393
5    O     0.676218227    2.249815117    0.000044622
6    O     2.907649845    1.957761425   -0.000050091
7    O    -1.555157013    1.957668393   -0.000273810
8    H    -0.693589716   -0.829715087    0.000024813
9    H     2.046268934   -0.829662550    0.000074849
EXPGEOM
1    O     0.00000    0.00000    0.97700
2    C     0.00000    1.13600    0.16020
3    C     0.00000   -1.13600    0.16020
4    O     0.00000    2.25200    0.60590
5    O     0.00000   -2.25200    0.60590
6    C     0.00000    0.67410   -1.26850
7    C     0.00000   -0.67410   -1.26850
8    H     0.00000    1.37310   -2.10520
```

```
 9    H     0.00000    -1.37310    -2.10520
---
H6C4O3, RHF, CHARGE=0, MULT=1
HF=-137.1
1     C    -0.048667489   -0.059945683   -0.040032529
2     C     1.454576143    0.015916889   -0.266487392
3     O     1.982193453    1.273125287   -0.081003007
4     C     3.239914513    1.618401534    0.356196593
5     C     3.736384604    2.949373699   -0.190715370
6     H     2.997414097    3.755499990   -0.010319788
7     H     3.915678841    2.871544836   -1.282039541
8     H     4.687769656    3.238439044    0.298571493
9     O     3.799165682    0.888823202    1.163267691
10    O     2.196251320   -0.877517511   -0.650560979
11    H    -0.402372799   -1.105617127   -0.136627649
12    H    -0.584464175    0.558540874   -0.787980485
13    H    -0.313203395    0.301016571    0.973998856
---
H10C4O3, RHF, CHARGE=0, MULT=1
HF=-127.1
1     C     0.000000000    0.000000000    0.000000000
2     O     1.397883541    0.000000000    0.000000000
3     C     2.247033085    1.110990984    0.000000000
4     O    -0.482650233    1.302051669    0.212167821
5     C    -1.656453142    1.539691460    0.932116836
6     O    -0.495072006   -0.485856829   -1.216122635
7     C    -0.167188750   -0.027480405   -2.494498481
8     H    -0.338325414   -0.744217398    0.789752880
9     H     2.041580915    1.831835634   -0.826779139
10    H     2.212898837    1.668002413    0.967988859
11    H     3.283197461    0.715385237   -0.134956966
12    H    -1.584747495    1.181673655    1.988802661
13    H    -2.553347549    1.065011132    0.465666641
14    H    -1.812561989    2.644513604    0.947315767
15    H    -0.277720795    1.077019009   -2.614771033
16    H     0.871525606   -0.308019916   -2.795019546
17    H    -0.875006561   -0.523146272   -3.202609104
---
H14C6O3, RHF, CHARGE=0, MULT=1
HF=-124.6
1     C     0.000000000    0.000000000    0.000000000
2     O     1.396171420    0.000000000    0.000000000
3     C     2.088886404    1.219286325    0.000000000
4     C     3.618118909    0.897172043    0.001172138
5     O     4.309888665    2.117396253    0.001382229
6     C     5.712549845    2.108822696   -0.009792729
7     C     6.195395960    3.595139115   -0.003811678
8     O     7.597606152    3.588432611   -0.015398957
9     C     8.292689738    4.799336622   -0.013102695
10    H    -0.429780475    0.499618634    0.903333675
11    H    -0.429832248    0.499273070   -0.903467923
12    H    -0.329410044   -1.065919536    0.000363996
13    H     1.824962703    1.833885554   -0.900843650
14    H     1.823795477    1.834716644    0.899918522
15    H     3.881672281    0.283186473    0.902401319
16    H     3.882484854    0.281772954   -0.898819044
17    H     6.121004309    1.567726629    0.884039014
18    H     6.106543355    1.579523716   -0.917086445
19    H     5.786544083    4.137113516   -0.897115461
```

```
20      H       5.801308352     4.124693100     0.903503164
21      H       8.065835315     5.426762603    -0.910421342
22      H       8.081198319     5.414523298     0.896340164
23      H       9.381075337     4.554448081    -0.024116331
---
H6C7O3, RHF, CHARGE=0, MULT=1
HF=-112.1
1       C       0.000000000     0.000000000     0.000000000
2       C       1.423549918     0.000000000     0.000000000
3       C       2.133411050     1.228903445     0.000000000
4       C       1.423664532     2.449334500     0.000706148
5       C      -0.685110061     1.222010747     0.000770006
6       C       0.014097130     2.444123923     0.000711976
7       H      -0.557782972    -0.936946731    -0.000223118
8       O       2.037964228    -1.210970092    -0.001232672
9       H      -1.776270648     1.228074630     0.001476801
10      H       3.224457762     1.236590892     0.000666426
11      H      -0.547288561     3.379156301     0.002308219
12      C       2.174290840     3.748740289     0.024832251
13      O       2.553623602     4.354578798     1.025500765
14      O       2.436826833     4.265868259    -1.202861841
15      H       2.905921829     5.091551502    -1.206072352
16      H       2.983765064    -1.147570080     0.008860763
---
H6C7O3, RHF, CHARGE=0, MULT=1
HF=-118.5, IE=9.8
1       C       0.000000000     0.000000000     0.000000000
2       C       1.398084242     0.000000000     0.000000000
3       C       2.109889395     1.238042612     0.000000000
4       C       1.392432586     2.469819064    -0.006612633
5       C      -0.716283563     1.213692274     0.000724070
6       C      -0.024427252     2.430544883    -0.004012358
7       H      -0.541263290    -0.947585483     0.000939741
8       H       1.940024346    -0.946464629     0.000256750
9       H      -1.806118017     1.204100599     0.004104637
10      O       3.460407132     1.130023261    -0.015184507
11      H      -0.593976740     3.362301574    -0.003884126
12      C       2.067017706     3.807365418     0.002923010
13      O       2.558033252     4.372663341     0.979731997
14      O       2.112625689     4.422884721    -1.206167432
15      H       2.503447920     5.288550467    -1.210015373
16      H       3.923099054     1.942979606     0.130876757
---
H6C7O3, RHF, CHARGE=0, MULT=1
HF=-117.7
1       C      -0.030017756    -0.081008336     0.006555731
2       C       1.391288397    -0.056036295    -0.034878399
3       C       2.068382570     1.170983163    -0.045173305
4       C       1.369717866     2.405557477    -0.015040202
5       C      -0.744777128     1.145965853     0.036440787
6       C      -0.047460140     2.362447017     0.025831679
7       O      -0.631076921    -1.294185193     0.015387325
8       H       1.962735759    -0.984532731    -0.058864400
9       H      -1.834739831     1.162252105     0.067564510
10      H       3.159703655     1.142020763    -0.077382215
11      H      -0.633954139     3.284012703     0.050407472
12      C       2.063096351     3.728757211    -0.024972692
13      O       1.524430907     4.838635380    -0.007581727
14      O       3.422861081     3.694850167    -0.055857668
```

```
15      H        3.847678809       4.543750233      -0.061614169
16      H       -1.577578293      -1.242355844       0.043794194
---
H14C7O3, RHF, CHARGE=0, MULT=1
HF=-161.7
1       C        0.000000000       0.000000000       0.000000000
2       C        1.569409243       0.000000000       0.000000000
3       C        2.051220484       1.493665288       0.000000000
4       O        2.038513176      -0.648530300       1.160872993
5       C        2.196059084      -0.859916746      -1.234780225
6       C        1.704561915      -2.348985441      -1.222874539
7       O        1.710749580      -0.190521161      -2.387751067
8       C        3.764079102      -0.852111204      -1.220838813
9       C        1.952772757      -0.523823259      -3.685620571
10      O        1.430690742       0.197814784      -4.524080102
11      H       -0.415503376       0.438910582      -0.927042157
12      H       -0.425028714      -1.016737873       0.114004387
13      H       -0.394642435       0.601050031       0.846905118
14      H        3.149192740       1.586082846       0.115723873
15      H        1.762333799       2.023285190      -0.927693515
16      H        1.598705066       2.054634192       0.845491960
17      H        1.757892909      -0.216347332       1.955021507
18      H        1.922530246      -2.843800948      -0.256980885
19      H        0.616692273      -2.439145391      -1.412342690
20      H        2.212950952      -2.950447750      -2.005814180
21      H        4.185634089       0.155664923      -1.405804665
22      H        4.165859678      -1.215027855      -0.255326783
23      H        4.180683804      -1.518024552      -2.005722561
24      H        2.587799020      -1.400695879      -3.914984982
---
N1O3, UHF, CHARGE=-1, MULT=1
HF=-74.7
1       N        0.000000000       0.000000000       0.000000000
2       O        1.235194016       0.000000000       0.000000000
3       O       -0.617633334       1.069686546       0.000000000
4       O       -0.617542508      -1.069747241      -0.000053607
---
H1N1O3, RHF, CHARGE=0, MULT=1
HF=-32.1, DIP=2.17
1       O        0.000000000       0.000000000       0.000000000
2       N        1.210961921       0.000000000       0.000000000
3       O        1.926159169       0.966425664       0.000000000
4       O        1.848268025      -1.178237510       0.000000000
5       H        1.260986959      -1.940318027       0.000000000
EXPGEOM
1       N        0.00000           0.15190           0.00000
2       O       -0.27160          -1.23610           0.00000
3       O        1.18690           0.46160           0.00000
4       O       -0.99440           0.84440           0.00000
5       H        0.63270          -1.62240           0.00000
---
H3C1N1O3, RHF, CHARGE=0, MULT=1
HF=-29.1
1       C        0.000000000       0.000000000       0.000000000
2       H        1.114574943       0.000000000       0.000000000
3       H       -0.360454383       1.056457035       0.000000000
4       H       -0.370771817      -0.506551497      -0.920860660
5       O       -0.542261470      -0.560902673       1.187057955
6       N       -0.344080730      -1.839482737       1.545930072
```

```
7    O    -0.877642079   -2.147956932    2.578955678
8    O     0.323592847   -2.566041859    0.847032744
EXPGEOM
1    N     0.00000    0.61290    0.00000
2    O    -1.21200    0.58240    0.00000
3    O     0.75300    1.55580    0.00000
4    O     0.67260   -0.56980    0.00000
5    C    -0.13270   -1.77000    0.00000
6    H     0.59550   -2.57870    0.00000
7    H    -0.75400   -1.81950    0.89220
8    H    -0.75400   -1.81950   -0.89220
---
H3C2N1O3, RHF, CHARGE=0, MULT=1
HF=-132
1    N     0.000000000    0.000000000    0.000000000
2    C     1.383123415    0.000000000    0.000000000
3    C     2.049989410    1.375879532    0.000000000
4    O     2.387052983    1.811460777    1.234181368
5    O     2.024311479   -1.044836957   -0.009879182
6    O     2.288111999    2.055147909   -0.993469822
7    H    -0.535110985    0.836879842   -0.018625230
8    H    -0.521902405   -0.847937144   -0.029106582
9    H     2.863340262    2.633122228    1.265662017
EXPGEOM
1    C     0.00560   -0.79610    0.00000
2    C     0.00000    0.74950    0.00000
3    O    -1.06590    1.34670    0.00000
4    O     1.01490   -1.45200    0.00000
5    O    -1.22960   -1.29210    0.00000
6    N     1.23400    1.28650    0.00000
7    H     1.34860    2.29160    0.00000
8    H     2.03590    0.66900    0.00000
9    H    -1.81170   -0.50810    0.00000
---
H5C2N1O3, RHF, CHARGE=0, MULT=1
HF=-36.8
1    C     0.000000000    0.000000000    0.000000000
2    H     1.119714223    0.000000000    0.000000000
3    C    -0.518542674    1.451833777    0.000000000
4    H    -0.341530217   -0.521974254   -0.929626371
5    O    -0.510851978   -0.638848800    1.169053834
6    N    -0.263402267   -1.924760314    1.460030360
7    O    -0.770485532   -2.300701629    2.484641372
8    O     0.416650734   -2.596233578    0.719170171
9    H    -0.159337974    1.950946669   -0.923286842
10   H    -1.625299298    1.500024344    0.006244396
11   H    -0.140361473    2.026529788    0.868400225
EXPGEOM
1    N     1.10810   -0.31620    0.00000
2    O     0.00000    0.57520    0.00000
3    O     2.16560    0.26380    0.00000
4    O     0.86880   -1.50760    0.00000
5    C    -1.27420   -0.11290    0.00000
6    C    -2.32080    0.99200    0.00000
7    H    -1.34930   -0.75370    0.89610
8    H    -1.34930   -0.75370   -0.89610
9    H    -3.32880    0.54080    0.00000
10   H    -2.21700    1.62680    0.89660
11   H    -2.21700    1.62680   -0.89660
```

```
---
H7C3N1O3, RHF, CHARGE=0, MULT=1
HF=-133.5
1    H     0.001859415    -0.084901304    -0.327085622
2    N     0.990867308    -0.122023335    -0.134232748
3    C     1.535535722     1.239169693    -0.061096289
4    C     3.065139921     1.101959623    -0.238553910
5    O     3.849407848     0.543714166     0.526601600
6    C     1.199026059     2.062333068     1.237351195
7    O    -0.174555786     2.236470211     1.405542852
8    H     1.079797107    -0.603508603     0.747658250
9    H     1.737736213     3.044619606     1.184515802
10   H    -0.485517089     3.033020257     0.998532684
11   H     1.578292478     1.536187359     2.151068677
12   H     1.120324899     1.807623014    -0.936045595
13   O     3.549201500     1.678626519    -1.367137109
14   H     4.478762956     1.559864038    -1.520873040
EXPGEOM
1    C     0.80010    -0.53270     0.01870
2    O     2.07260    -0.33160    -0.39360
3    O     0.43850    -1.54730     0.58240
4    C    -0.09450     0.67000    -0.30920
5    C    -1.52490     0.42300     0.19190
6    O    -2.12430    -0.71560    -0.40130
7    N     0.40770     1.93530     0.23310
8    H     2.58170    -1.12290    -0.16560
9    H    -0.12680     0.75200    -1.40950
10   H    -1.51620     0.35450     1.29760
11   H    -2.13100     1.29740    -0.08730
12   H    -1.67560    -1.49450    -0.05210
13   H     1.31060     2.17810    -0.16790
14   H     0.52490     1.88170     1.24440
---
H3C4N1O3, RHF, CHARGE=0, MULT=1
HF=-6.9
1    O     0.000000000     0.000000000     0.000000000
2    C     1.361014544     0.000000000     0.000000000
3    C     1.853658980     1.306636200     0.000000000
4    C     0.705315200     2.173738141    -0.000032288
5    C    -0.392797518     1.309551234     0.000046100
6    N    -1.841361712     1.552215283     0.000265443
7    O    -2.617783536     0.624686547    -0.007036885
8    O    -2.222836162     2.703499880     0.007632694
9    H     1.854672589    -0.966845211     0.000036678
10   H     2.888601611     1.612595129     0.000019660
11   H     0.707194847     3.254111290    -0.000028197
---
H9C4N1O3, RHF, CHARGE=0, MULT=1
HF=-140.5
1    H     0.216921126    -0.166432558    -0.089085625
2    N     1.201824850    -0.075682363     0.105820737
3    C     1.543569772     1.338393920     0.285869269
4    C     3.083102889     1.454768192     0.178495497
5    O     3.688386200     2.050219175    -0.710722894
6    C     0.956347084     2.090529648     1.553038842
7    O     1.002854400     1.304139802     2.707957948
8    C    -0.508820856     2.564153435     1.345428479
9    H     1.365513063    -0.619378631     0.938442205
10   H     1.571771776     3.026288514     1.681138985
```

```
11       H        1.809712447     1.423525433     3.189031248
12       H       -0.576243967     3.228067647     0.460013049
13       H       -1.215317053     1.724010810     1.198838075
14       H       -0.856389399     3.146897045     2.222650504
15       H        1.134921233     1.880251668    -0.610829826
16       O        3.798728149     0.853817190     1.162863014
17       H        4.741573199     0.865621411     1.048678526
---
H5C6N1O3, RHF, CHARGE=0, MULT=1
HF=-25.2
1        C        0.000000000     0.000000000     0.000000000
2        C        1.421368203     0.000000000     0.000000000
3        C        2.121533895     1.216686562     0.000000000
4        C        1.436669047     2.444993565     0.001175883
5        C        0.027749866     2.425687761     0.003744384
6        C       -0.715194910     1.231924502     0.002244225
7        O       -0.744446602    -1.133136774    -0.001458347
8        N       -0.713204972     3.735620219     0.007187805
9        O       -0.998617443     4.246874989     1.065660312
10       O       -1.003511973     4.249913009    -1.048467280
11       H        1.983090397    -0.935092964    -0.000426025
12       H        3.212637891     1.210051886    -0.000942553
13       H        2.000980236     3.377908743     0.000862341
14       H       -0.221117599    -1.923843407    -0.005448333
15       H       -1.806172746     1.235459286     0.003177711
---
H5C6N1O3, RHF, CHARGE=0, MULT=1
HF=-30.8
1        C        0.000000000     0.000000000     0.000000000
2        C        1.430909577     0.000000000     0.000000000
3        C        2.135995995     1.206550503     0.000000000
4        C        1.454186298     2.443063887    -0.000138804
5        C        0.055527989     2.472159831     0.004635316
6        C       -0.668091845     1.255477297     0.006620285
7        O       -0.583685663    -1.217570607     0.020737127
8        N       -2.165282878     1.344069216     0.014880787
9        O       -2.772887239     1.170165673    -1.018380402
10       O       -2.737824055     1.589580292     1.051706403
11       H        1.975582081    -0.945303240    -0.000137233
12       H        3.227389835     1.193959315    -0.001175265
13       H        2.018846328     3.375455525    -0.003266891
14       H       -0.460544377     3.434183263     0.005191506
15       H       -1.520461104    -1.227475087    -0.117421877
---
H5C6N1O3, RHF, CHARGE=0, MULT=1
HF=-27.4
1        C        0.000000000     0.000000000     0.000000000
2        C        1.422382988     0.000000000     0.000000000
3        C        2.124265866     1.215634155     0.000000000
4        C        1.403754020     2.425677969    -0.004219435
5        C       -0.007657227     2.447011538    -0.004565279
6        C       -0.711601743     1.235179726    -0.001891514
7        O       -0.739895225    -1.132981385    -0.000300543
8        N        2.154035675     3.725258424    -0.006108360
9        O        2.522150439     4.195204732     1.046331138
10       O        2.376116345     4.276422092    -1.060089006
11       H        1.985460271    -0.933853880     0.000811956
12       H        3.215671540     1.198483057     0.001328158
13       H       -0.217387619    -1.924551360     0.004323080
```

```
14      H      -1.802308494     1.248475317    -0.001367792
15      H      -0.564419994     3.385913886    -0.006931976
---
N2O3, RHF, CHARGE=0, MULT=1
HF=19.8
1    N      0.004559217     0.008790381     0.000000000
2    N      1.527987791     0.001547784     0.000000000
3    O      2.033844613     1.021710130     0.000000000
4    O     -0.610116914     1.045324051     0.000000000
5    O     -0.499485622    -1.080966369     0.000000000
EXPGEOM
1    N     -0.70380        -1.06750         0.00000
2    N      0.00000         0.63140         0.00000
3    O      0.20200        -1.77320         0.00000
4    O      1.20630         0.61400         0.00000
5    O     -0.79240         1.54080         0.00000
---
H2C2O4, RHF, CHARGE=0, MULT=1
HF=-175, IE=11.2
1    O      0.000000000     0.000000000     0.000000000
2    C      1.226563180     0.000000000     0.000000000
3    C      2.070394593     1.289794055     0.000000000
4    O      3.297022362     1.289782348    -0.000144200
5    O      1.469726816     2.497305386     0.000168947
6    O      1.828285874    -1.207019584    -0.000207041
7    H      0.521154151     2.491976237     0.000282781
8    H      2.776847801    -1.201275171    -0.000235823
EXPGEOM
1    C     -0.05340         0.76290         0.00000
2    C      0.05340        -0.76290         0.00000
3    O      1.12280         1.36670         0.00000
4    O     -1.12280        -1.36670         0.00000
5    O     -1.12280         1.31410         0.00000
6    O      1.12280        -1.31410         0.00000
7    H      1.78510         0.65510         0.00000
8    H     -1.78510        -0.65510         0.00000
---
H6C2O4, RHF, CHARGE=0, MULT=1
HF=-124.5
1    O      0.199604423    -0.564074725    -0.103201064
2    C      1.463748926    -0.048917877    -0.364507857
3    O      1.556895123     1.286433514     0.134170239
4    O      1.741236960     1.366493296     1.412082676
5    C      1.063082355     2.455909703     2.039259308
6    O     -0.284727511     2.152560852     2.190296391
7    H     -0.488745055    -0.138474184    -0.595932905
8    H      1.680109488     0.043970155    -1.466947036
9    H      2.219475368    -0.745391443     0.089389640
10     H     1.136550807     3.416202394     1.460603287
11     H     1.594675902     2.590987160     3.024320806
12     H    -0.444369019     1.437579065     2.791338951
---
H4C4O4, RHF, CHARGE=0, MULT=1
HF=-146.3
1    C     -0.025015755    -0.003323759    -0.093008048
2    O      1.339731664     0.008889979     0.010774779
3    C      2.090476657     1.144237176    -0.325964609
4    C      1.346698391     2.471385314    -0.078213127
5    O     -0.019254306     2.459232536     0.006671638
```

```
 6    C    -0.766112532    1.325166303   -0.342407310
 7    O    -0.552240799   -1.098391291    0.035344851
 8    O     1.872679281    3.565038886    0.066026771
 9    H     2.395425614    1.099723274   -1.404776104
10    H     3.025074603    1.110546828    0.290210294
11    H    -1.707381764    1.356322710    0.263679124
12    H    -1.059610037    1.374237252   -1.424242986
---
H6C4O4, RHF, CHARGE=0, MULT=1
HF=-169.5
 1    C     0.000000000    0.000000000    0.000000000
 2    O     1.408367215    0.000000000    0.000000000
 3    C     2.184697195    1.106699445    0.000000000
 4    C     3.673307449    0.759759344   -0.039803259
 5    O     4.200290043    0.653820692    1.200640316
 6    C     5.550483149    0.367928614    1.482095429
 7    O     1.729621243    2.241631088    0.013053212
 8    O     4.331084369    0.610438187   -1.059845814
 9    H    -0.313171793   -1.070638089    0.001067629
10    H    -0.418354006    0.493978531   -0.908437373
11    H    -0.418822810    0.495821099    0.907255325
12    H     5.642332511    0.353212924    2.593708642
13    H     6.236833785    1.148376273    1.077029600
14    H     5.859207139   -0.627675959    1.084845887
---
H8C4O4, RHF, CHARGE=0, MULT=1
HF=-148.2
 1    C     0.000000000    0.000000000    0.000000000
 2    O     1.398744585    0.000000000    0.000000000
 3    C     2.184020900    1.157939030    0.000000000
 4    O     2.378097927    1.613995889   -1.307892002
 5    C     1.497710979    2.465399172   -1.983674623
 6    O     0.493097486    1.718703991   -2.608124937
 7    C    -0.688731832    1.310799757   -1.981016688
 8    O    -0.486957086    0.101637291   -1.307473170
 9    H    -0.431621546    0.790004283    0.684044765
10    H    -0.323571773   -1.006529665    0.388595990
11    H     3.197084454    0.861717502    0.393439035
12    H     1.769677777    1.960093014    0.680520146
13    H     2.090005842    2.962380479   -2.802544973
14    H     1.081534374    3.276755807   -1.315457843
15    H    -1.435013573    1.106124571   -2.799202060
16    H    -1.121226919    2.111935445   -1.310608840
---
H8C5O4, RHF, CHARGE=0, MULT=1
HF=-176.4
 1    C    -0.095166442    0.009761307   -0.087025001
 2    O     1.288444436    0.237931203   -0.196522401
 3    C     1.990111144    1.161933979    0.509206265
 4    O     1.433832397    1.865189241    1.347089220
 5    C     3.486784152    1.176286212    0.176141828
 6    C     3.989815334    2.574442211   -0.202119288
 7    O     3.482145981    3.322253201   -1.032279216
 8    O     5.127207092    2.930512133    0.449136970
 9    C     5.802277835    4.153923126    0.289246362
10    H    -0.392268163   -0.298982560    0.943342216
11    H    -0.693344770    0.903446795   -0.382942140
12    H    -0.331716624   -0.824681112   -0.789257138
13    H     4.027428711    0.773595905    1.063204021
```

```
14      H       3.725040468     0.497278649    -0.674505344
15      H       5.172538907     5.023812287     0.591108739
16      H       6.153162491     4.308389120    -0.758683406
17      H       6.694943377     4.111053158     0.957610260
---
H8C5O4, RHF, CHARGE=0, MULT=1
HF=-203.1
1       C      -0.169603554    -0.076413446    -0.007331301
2       O       3.404444854     1.574030260     0.381451668
3       C       2.556467083     0.973067570     1.252557413
4       O       2.923198750    -0.039950785     1.843817550
5       C       1.183654608     1.657833983     1.409091940
6       C       1.364884578     3.163231669     1.703234594
7       O       1.018258477     4.114455263     1.007924599
8       O       1.969949723     3.410494843     2.893884990
9       H      -0.573559030    -0.543425371     0.913121242
10      H       0.681880244    -0.691613224    -0.359757909
11      H      -0.960253915    -0.132387985    -0.783864923
12      H       4.256562690     1.165113552     0.287798731
13      C       0.225778805     1.388054583     0.213871382
14      H       0.699054905     1.217802286     2.319239365
15      H       2.102543796     4.326875169     3.104759727
16      H      -0.711309539     1.966186977     0.390655590
17      H       0.664277255     1.784724189    -0.729216511
---
H8C5O4, RHF, CHARGE=0, MULT=1
HF=-186.6
1       C      -0.191138178     0.173845356     0.051467235
2       C       0.972513581     0.217196524    -0.931672454
3       O       0.887579635     0.279060144    -2.154218788
4       O       2.198961500     0.185344957    -0.333664502
5       C       3.454505951     0.190853670    -0.965142791
6       O       3.741849273    -1.119210992    -1.384815778
7       C       4.262655396    -2.132427677    -0.633056432
8       O       4.564431014    -1.932580307     0.539335773
9       C       4.413386937    -3.445418989    -1.392165424
10      H      -0.668937564    -0.826358551     0.010037245
11      H       0.122210804     0.365961160     1.095884054
12      H      -0.946331423     0.936942539    -0.226084255
13      H       3.496351316     0.838516658    -1.884402464
14      H       4.188499681     0.600039547    -0.216606452
15      H       4.413739063    -3.306166348    -2.490691772
16      H       3.575192685    -4.121637588    -1.126588671
17      H       5.365649339    -3.938412263    -1.109681556
---
H12C5O4, RHF, CHARGE=0, MULT=1
HF=-173.8
1       C       0.005510127     0.002735494    -0.003383841
2       O       1.313173446     0.317524288     0.382353450
3       C       2.412874442     0.401824221    -0.482061111
4       O      -0.860589909     0.260857631     1.064356396
5       C      -0.817242921     1.388363426     1.894982596
6       O      -0.337808396     0.786549360    -1.109840106
7       C      -1.416781782     0.548308578    -1.971051019
8       O      -0.096110717    -1.348899792    -0.351268810
9       C      -0.093043835    -2.417295570     0.555198048
10      H       2.512559432    -0.479001275    -1.160293225
11      H       2.391481846     1.327229385    -1.106250117
12      H       3.320376738     0.443177148     0.168660426
```

```
13      H       -0.004647777     1.315426392     2.657739079
14      H       -0.695418022     2.348228985     1.339561537
15      H       -1.796069624     1.419267886     2.433372126
16      H       -2.367096833     0.300767924    -1.441417818
17      H       -1.575108265     1.494319275    -2.544423626
18      H       -1.196714755    -0.267807411    -2.701011334
19      H        0.718880014    -2.352160838     1.317984830
20      H       -1.068807706    -2.521031186     1.087823864
21      H        0.072234162    -3.341720980    -0.050178318
---
H10C6O4, RHF, CHARGE=0, MULT=1
HF=-194.2
1       O       -0.287353349     0.435328271     0.311600337
2       C       -0.157739083     0.955859466    -0.792564556
3       C        0.555991003     2.268204840    -1.094442395
4       O       -0.677347339     0.349992628    -1.898722948
5       C       -1.340374532    -0.895400568    -1.947979922
6       O       -0.322667226    -1.853520282    -1.740378742
7       C       -0.401596996    -2.988210220    -0.986953270
8       O       -1.490396422    -3.374525958    -0.574057078
9       C        0.945354203    -3.669587118    -0.777258132
10      C       -1.981295517    -1.046131264    -3.361547656
11      H        0.109187677     2.799507674    -1.957491050
12      H        1.622080697     2.063875753    -1.322647591
13      H        0.508388099     2.941799141    -0.214996846
14      H       -2.158064271    -0.946944737    -1.170813156
15      H        1.554290093    -3.089825710    -0.054328556
16      H        1.509186547    -3.753058842    -1.727633484
17      H        0.802713655    -4.691467132    -0.371897828
18      H       -2.494637979    -2.025528320    -3.439610240
19      H       -1.238496545    -0.979884206    -4.180192053
20      H       -2.741274086    -0.254311217    -3.517623159
---
H10C6O4, RHF, CHARGE=0, MULT=1
HF=-183.7
1       C        0.375988978     0.864465382    -0.134009990
2       O        1.545955658     1.158210447     0.588974734
3       C        2.525221762     0.267285705     0.892413519
4       O        2.427979361    -0.907045846     0.550616648
5       C        3.703749858     0.862921609     1.690824680
6       C        4.148581742     2.227486855     1.124301220
7       O        4.036929535     3.325692369     1.657649325
8       O        4.739162023     2.115008820    -0.096477279
9       C        5.230411349     3.186117124    -0.862853972
10      C        3.397188036     0.892022077     3.205143802
11      H        0.598551344     0.539544435    -1.178055713
12      H       -0.243332652     0.080791966     0.363017874
13      H       -0.214962208     1.810146973    -0.174200972
14      H        4.570066293     0.162830483     1.559563557
15      H        6.068424457     3.718031853    -0.353024940
16      H        4.435019764     3.926411092    -1.115859996
17      H        5.616842206     2.743465147    -1.811602832
18      H        2.548666632     1.556895568     3.458355685
19      H        4.285583559     1.235372981     3.772865266
20      H        3.149756300    -0.127850754     3.564594691
---
H7C4N1O4, RHF, CHARGE=0, MULT=1
HF=-185.9
1       H       -0.459915954     0.288963228    -0.096910593
```

```
 2    N    -0.006996736    0.396355211    -0.990315009
 3    C     0.739450743    1.657950367    -1.039240939
 4    C     0.906069128    1.977453908    -2.546614596
 5    O     1.599261054    1.362669464    -3.354727907
 6    C     2.120518182    1.700127912    -0.313679950
 7    H     2.772838225    0.858312769    -0.641856339
 8    C     2.018926353    1.635660152     1.213369000
 9    H     0.596513114   -0.403821366    -1.097719508
10    H     2.649528713    2.632984012    -0.613848911
11    O     2.251242213    2.808486083     1.852229904
12    O     1.769516362    0.633423041     1.883391895
13    H     0.093851541    2.448958263    -0.572199402
14    O     0.194482023    3.051010433    -2.968288299
15    H     0.236715529    3.229609254    -3.900174785
16    H     2.218767181    2.773935088     2.800914719
---
N2O4, RHF, CHARGE=0, MULT=1
HF=2.2, IE=11.4
 1    N     0.000000000    0.000000000    0.000000000
 2    N     1.615006977    0.000000000    0.000000000
 3    O     2.122233017    1.075382763    0.000000000
 4    O    -0.507321616    0.003386315   -1.075394557
 5    O    -0.507317481   -0.003351927    1.075396055
 6    O     2.122275602   -1.075460238   -0.000060389
EXPGEOM
 1    N     0.00000    0.00000    0.90320
 2    N     0.00000    0.00000   -0.90320
 3    O     0.00000    1.10400    1.36590
 4    O     0.00000   -1.10400    1.36590
 5    O     0.00000    1.10400   -1.36590
 6    O     0.00000   -1.10400   -1.36590
---
H2C1N2O4, RHF, CHARGE=0, MULT=1
HF=-13.3
 1    C     0.000000000    0.000000000    0.000000000
 2    N     1.550283994    0.000000000    0.000000000
 3    N    -0.487607674    1.471476220    0.000000000
 4    H    -0.354042282   -0.527168606    0.917022721
 5    H    -0.388694399   -0.501138375   -0.917751075
 6    O     2.121851333   -0.601559081   -0.876116994
 7    O     2.133536419    0.565825308    0.892530256
 8    O    -0.124665719    2.206835219   -0.885724941
 9    O    -1.247973156    1.821378797    0.869301382
---
H3C1N3O4, RHF, CHARGE=0, MULT=1
HF=10.3
 1    C     0.205373359   -0.276140618    0.394130202
 2    N     1.614971312   -0.087799456   -0.087752429
 3    N     2.105736928    1.240114958   -0.274905336
 4    N     2.010008558   -0.993016702   -1.147408925
 5    O     1.671157929   -0.745792274   -2.276453971
 6    O     2.651694449   -1.941169397   -0.795799165
 7    O     1.431095296    2.158174993    0.117301618
 8    O     3.216035768    1.348222238   -0.726737115
 9    H     0.023975533   -1.365888091    0.527466652
10    H    -0.559361889    0.135490095   -0.300629200
11    H     0.107095576    0.209527088    1.389920454
---
H4C3O5, RHF, CHARGE=0, MULT=1
```

```
    HF=-227.5
    1    C    -0.043530280     0.034079063     0.016464701
    2    C     1.511528214     0.065686898    -0.024266702
    3    C     2.025716056     1.535413554    -0.059985579
    4    O    -0.562799462     0.361146996     1.222731317
    5    O    -0.793231084    -0.257067307    -0.911855697
    6    O     2.039193842    -0.700057927    -1.053843129
    7    O     2.669960526     2.085852485     0.826950619
    8    O     1.710932181     2.220461954    -1.185747380
    9    H     1.896131726    -0.404360076     0.925513974
    10   H    -1.510291839     0.332049523     1.285292018
    11   H     1.718874859    -0.474246589    -1.917836706
    12   H     2.029841639     3.114149064    -1.229746294
    ---
    H10C5O5, RHF, CHARGE=0, MULT=1
    HF=-186.4
    1    O     0.000000000     0.000000000     0.000000000
    2    C     1.396618783     0.000000000     0.000000000
    3    O     1.865927977     1.324135386     0.000000000
    4    C     2.666832359     1.872300010     1.003325755
    5    O     1.820990065     2.364248780     2.011793765
    6    C     1.849518411     1.925580357     3.339262256
    7    O     1.152272063     0.718962806     3.458945504
    8    C    -0.245232826     0.603702518     3.381691206
    9    O    -0.562193243    -0.093806546     2.213493953
    10   C    -0.850027371     0.515428834     0.981576478
    11   H     1.700970618    -0.469936270    -0.980470483
    12   H     1.832499984    -0.632272553     0.823680128
    13   H     3.202706795     2.751437656     0.543233515
    14   H     3.448420703     1.154691852     1.384343415
    15   H     1.402117885     2.752601354     3.961491215
    16   H     2.891673498     1.732147660     3.722064876
    17   H    -0.781955849     1.590957255     3.468382860
    18   H    -0.575683141    -0.043920848     4.243917085
    19   H    -1.892460528     0.206538012     0.675673485
    20   H    -0.832703486     1.640923065     1.017300495
    ---
    N2O5, RHF, CHARGE=0, MULT=1
    HF=2.7, IE=12.3
    1    O     0.171720052    -0.128499075     0.381092110
    2    N     1.254884217     0.067678198    -0.087376045
    3    O     1.638243497     1.391035247    -0.067484278
    4    N     2.908011074     1.753936637     0.316789429
    5    O     3.408048452     1.153778871     1.229004021
    6    O     3.340545161     2.692178438    -0.286442911
    7    O     1.994515064    -0.712440640    -0.621372486
    EXPGEOM
    1    O     0.00000     0.00000     0.84400
    2    N     0.00000     1.21320     0.01870
    3    N     0.00000    -1.21320     0.01870
    4    O     0.62990     2.08510     0.53980
    5    O    -0.62990    -2.08510     0.53980
    6    O    -0.66260     1.17190    -0.97810
    7    O     0.66260    -1.17190    -0.97810
    ---
    H10C10O1, RHF, CHARGE=0, MULT=1
    HF=-11.5
    1    O    -0.000048373    -0.000043755     0.000205461
    2    C     1.227116676    -0.000065817     0.000215668
```

```
3    C     2.032872078    1.296979173   -0.000060539
4    C     2.014752378   -1.275474824    0.027789194
5    C     1.968488198   -2.239136226   -0.910994435
6    C     2.760069469   -3.489677522   -0.888168039
7    C     3.961932826   -3.592550211   -1.627601958
8    C     4.700520245   -4.788374963   -1.630913479
9    C     4.249220844   -5.899960106   -0.898213121
10   C     3.055366614   -5.810357622   -0.161668039
11   C     2.315252894   -4.615120076   -0.155729825
12   H     2.514937950    1.443101705    0.987888312
13   H     2.826708596    1.259486577   -0.772537715
14   H     1.401856098    2.184057916   -0.206069783
15   H     2.663806733   -1.368196595    0.903889330
16   H     1.317853002   -2.146285473   -1.788524082
17   H     4.326561554   -2.740281033   -2.203242561
18   H     5.625847585   -4.852628892   -2.204304105
19   H     4.822182945   -6.827566508   -0.901608304
20   H     2.700331224   -6.670052824    0.407584024
21   H     1.389765994   -4.564641693    0.420086503
---
H14C10O1, RHF, CHARGE=0, MULT=1
HF=-55.1
1    C    -0.000178133    0.000127587   -0.000918218
2    C     1.553761394   -0.000424277    0.000066645
3    C     2.038147461    1.469695710   -0.000068742
4    C     1.530549596    2.249282967    1.237079549
5    C    -0.023138734    2.225721601    1.222162150
6    C    -0.519603347    0.752855066    1.256260953
7    C     2.092524187   -0.686120180    1.286286489
8    C     0.008084404    0.051775310    2.539782800
9    C     2.068649870    1.539597599    2.510421315
10   C     1.562732092    0.069765402    2.537316205
11   H    -0.386611489   -1.042046837   -0.020922108
12   H    -0.387157461    0.473281645   -0.929698388
13   H     1.924939086   -0.545883739   -0.894714960
14   O     2.739608320    1.956435832   -0.872796979
15   H     3.204310300   -0.703949139    1.279584549
16   H     1.885573549    3.302656562    1.221441285
17   H    -0.410230913    2.749904095    0.321110901
18   H    -0.426563960    2.795189591    2.087426085
19   H    -1.631955933    0.739747590    1.259363997
20   H     1.783321408   -1.753575532    1.313569666
21   H    -0.370072866   -0.992258292    2.597827953
22   H    -0.387106944    0.554054009    3.449690700
23   H     1.741265532    2.083572913    3.423021898
24   H     3.179839943    1.573143806    2.532752604
25   H     1.939084743   -0.431485745    3.456247331
---
H14C10O1, RHF, CHARGE=0, MULT=1
HF=-47.4
1    C     0.025355963    0.035602714    0.052533374
2    C     1.373664783   -0.024317942   -0.392742665
3    C     1.957465966    1.200833550   -0.834886204
4    C     1.195392005    2.408264943   -0.822814630
5    C    -0.126127024    2.410380020   -0.375662811
6    C    -0.740298421    1.216751136    0.073769467
7    C     2.051716848   -1.393227120   -0.346010616
8    C     2.473665644   -1.935966126   -1.728840386
9    C     3.164300348   -1.489277469    0.720566706
```

```
10    H     1.273014091    -2.124716127    0.003733048
11    O     3.222806511     1.362826378   -1.299755959
12    H    -0.449958689    -0.886204686    0.400294495
13    C    -2.164436990     1.228627702    0.558722409
14    H     1.644355092     3.342480273   -1.164855899
15    H     3.430312605    -1.514319297   -2.095246032
16    H     1.699531270    -1.727144614   -2.494335767
17    H     2.604376701    -3.037360715   -1.683885290
18    H     2.824196950    -1.065120143    1.686911303
19    H     4.095650576    -0.962354951    0.434245534
20    H     3.434342963    -2.550306971    0.902365194
21    H    -2.265833835     1.854367891    1.470829328
22    H    -2.845503447     1.642136662   -0.214818615
23    H    -2.536176695     0.215112375    0.812593307
24    H     3.764715413     0.590944328   -1.245986473
25    H    -0.682533460     3.349884246   -0.376255604
---
H14C10O1, RHF, CHARGE=0, MULT=1
HF=-44.1
1     C    -0.042037460    -0.049099718    0.025428614
2     C     1.314732706    -0.005546705    0.406999833
3     C     2.067282184     1.197478914    0.325185276
4     C     1.469454182     2.412995814   -0.141514730
5     C     0.114936649     2.328970118   -0.538300324
6     C    -0.626430810     1.139574331   -0.458776385
7     C     2.489730515     4.174184004   -1.672195019
8     C     1.553815693     4.847217364    0.628630871
9     H    -1.669796728     1.152814705   -0.778755322
10    H     2.993000729     3.366488176   -2.241261648
11    O     3.377616039     1.219247113    0.689129355
12    C    -0.827272229    -1.328920360    0.133077670
13    H     1.791593424    -0.917345730    0.773350250
14    C     2.223106658     3.738887882   -0.214824162
15    H    -0.398435534     3.212564002   -0.924519342
16    H     1.562201540     4.441039103   -2.217041622
17    H     3.157657268     5.060097343   -1.700051412
18    H     0.609383444     5.218979628    0.183993686
19    H     1.325366169     4.484731118    1.651441385
20    H     2.233280661     5.718279126    0.733127513
21    H    -0.746151186    -1.763754303    1.151604638
22    H    -0.450819661    -2.082316326   -0.590741315
23    H    -1.906237056    -1.180445688   -0.074086483
24    H     3.236729559     3.611212641    0.243119338
25    H     3.694716812     0.375266153    0.981090184
---
H14C10O1, RHF, CHARGE=0, MULT=1
HF=-45.4
1     C     0.049273791     0.019649296    0.008690795
2     C     1.388270908     0.005518417   -0.459659283
3     C     1.987864171     1.273775696   -0.742739808
4     C     1.267868377     2.506809995   -0.561006748
5     C    -0.056875225     2.440637482   -0.090712184
6     C    -0.664787497     1.207889359    0.192831128
7     C     2.041096707    -1.369955422   -0.603476929
8     C     2.409040013    -1.747417690   -2.055047324
9     C     3.178740347    -1.622295028    0.409996642
10    H     1.255561465    -2.123988896   -0.322458263
11    O     3.257788440     1.434491780   -1.196941550
12    H    -0.460280250    -0.918962842    0.240390012
```

```
13      H       -1.692222241     1.176454990     0.556426151
14      C        1.884475305     3.851040606    -0.856847513
15      H        3.363915278    -1.303403115    -2.399174950
16      H        1.616616785    -1.432505772    -2.763509826
17      H        2.516261054    -2.848412049    -2.148381786
18      H        2.878629935    -1.305755303     1.429395250
19      H        4.119422134    -1.093742621     0.159860101
20      H        3.418662411    -2.704905383     0.457657705
21      H        2.787377842     4.029515129    -0.236065827
22      H        2.174649268     3.936613250    -1.925056397
23      H        1.181232236     4.684035103    -0.647208521
24      H        3.740258704     0.632051741    -1.318437580
25      H       -0.634517326     3.354881596     0.061260308
---
H14C10O1, RHF, CHARGE=0, MULT=1
HF=-46.4
1       C       -0.000587786    -0.012845839     0.005666424
2       C        1.347567908    -0.140321597     0.414458381
3       C        1.965186792     1.025413043     0.958666652
4       C        1.244320425     2.249511273     1.076545272
5       C       -0.097258147     2.350191113     0.659935898
6       C       -0.706865328     1.190655627     0.121168959
7       C       -1.098539676     4.092766186     2.233506317
8       C       -0.318639717     4.767333942    -0.124865074
9       H       -1.744692189     1.216426776    -0.215679422
10      H       -0.521526267    -0.875396943    -0.416768247
11      O        3.249846234     1.080050267     1.400132922
12      H       -1.521582580     3.267483346     2.841720803
13      C        2.050668175    -1.462278340     0.261287340
14      H        1.754854195     3.116974842     1.499375331
15      C       -0.892471852     3.648190039     0.769921801
16      H       -0.157569277     4.420355003     2.718267730
17      H       -1.812570800     4.940403353     2.286859970
18      H        0.663657992     5.137376326     0.229940641
19      H       -0.188850109     4.412629495    -1.167600282
20      H       -1.009530816     5.635033852    -0.156246173
21      H        1.395003268    -2.234990416    -0.189388834
22      H        2.385762113    -1.860868172     1.242689074
23      H        2.940709774    -1.376082982    -0.398152991
24      H       -1.922697038     3.448534239     0.372656332
25      H        3.721781310     0.264376188     1.314714885
---
H14C10O1, RHF, CHARGE=0, MULT=1
HF=-43.6
1       C       -0.011808742     0.009135638    -0.005878702
2       C        1.347017674     0.023050268    -0.452638669
3       C        2.055828707     1.267650558    -0.412383724
4       C        1.407360095     2.480346433    -0.051464640
5       C        0.057779251     2.452256853     0.306527094
6       C       -0.633135868     1.231649397     0.341659131
7       C       -1.314830578    -1.525326772     1.573637019
8       C       -1.965842237    -1.365062047    -0.913178323
9       C        2.021240890    -1.227006883    -0.966715933
10      O        3.369040228     1.414943396    -0.743160939
11      H       -0.481690365    -1.414457087     2.297043422
12      H       -1.679765884     1.255911754     0.651101514
13      H        1.950862121     3.425982017    -0.052739108
14      C       -0.827585729    -1.280914376     0.127186813
15      H       -2.123325845    -0.833689211     1.882272855
```

```
16    H    -1.707100372   -2.558116200    1.679453370
17    H    -1.581243424   -1.188600398   -1.938046231
18    H    -2.773036782   -0.629248753   -0.726841682
19    H    -2.428074878   -2.373844450   -0.904269650
20    H     1.338858557   -1.846208680   -1.584592159
21    H     2.884891144   -1.018040684   -1.631746124
22    H     2.394789639   -1.852573125   -0.127603829
23    H    -0.166793951   -2.158314006   -0.087490037
24    H     3.884312801    0.630742031   -0.616322066
25    H    -0.457269283    3.377171727    0.570601531
---
H14C10O1, RHF, CHARGE=0, MULT=1
HF=-41
1     C    -0.004392712   -0.003661366    0.005848731
2     C     1.349165284   -0.034229447   -0.467015746
3     C     1.942247938    1.219985418   -0.824195611
4     C     1.223304825    2.447278203   -0.709967959
5     C    -0.089551450    2.440982189   -0.241771503
6     C    -0.693392593    1.224192119    0.110039431
7     C     2.482025291   -1.780958682   -1.999399988
8     C     3.237493975   -1.513510764    0.463371609
9     C    -0.773610396   -1.239542508    0.415875875
10    O     3.201593923    1.402901907   -1.302347825
11    H     1.705319749   -1.479891149   -2.730371576
12    H    -1.723296013    1.249254867    0.471745880
13    H     1.693988060    3.392569495   -0.985590989
14    C     2.087984104   -1.372980829   -0.560810943
15    H     3.446488588   -1.351100158   -2.335738268
16    H     2.585986861   -2.884426534   -2.065459782
17    H     2.917278670   -1.169059503    1.467683868
18    H     4.149372163   -0.945815192    0.193125131
19    H     3.537562320   -2.577681753    0.558693619
20    H    -0.283614583   -1.764114659    1.262751170
21    H    -1.803090957   -0.998888093    0.754008293
22    H    -0.882618669   -1.951755303   -0.429227753
23    H     1.372179550   -2.180564748   -0.257613788
24    H     3.731958750    0.623037606   -1.340690222
25    H    -0.645573323    3.374692691   -0.148867085
---
H14C10O1, RHF, CHARGE=0, MULT=1
HF=-50.2
1     C     0.001012851   -0.029830168   -0.003619094
2     C     1.388561594   -0.074677923    0.256170623
3     C     2.080650451    1.107505886    0.615620726
4     C     1.385804861    2.343564354    0.714307923
5     C     0.000782215    2.402026208    0.455691173
6     C    -0.671383758    1.206185737    0.100099781
7     C    -0.818580429    4.278244830    1.981400173
8     C    -0.325547526    4.744297342   -0.501581597
9     H    -1.744483914    1.238675906   -0.101384007
10    C    -0.733748913   -1.289443418   -0.382424282
11    O     3.414532235    1.132828072    0.880340271
12    H    -1.159619921    3.509324440    2.704266012
13    H     1.922232275   -1.023403396    0.178586458
14    H     1.939084355    3.242093819    0.991312582
15    C    -0.783887873    3.708909529    0.547278072
16    H     0.170587041    4.644168607    2.320933499
17    H    -1.528567604    5.128489847    2.046186797
18    H     0.690092870    5.138027788   -0.298746159
```

```
19         H         -0.318704114     4.301077346    -1.518144189
20         H         -1.019304532     5.609964548    -0.524925740
21         H         -0.661532632    -2.047207827     0.426437306
22         H         -0.307479751    -1.734570606    -1.306052840
23         H         -1.811766965    -1.112318277    -0.571451751
24         H         -1.853005668     3.482279842     0.292827515
25         H          3.820584214     0.279703321     0.805908182
---
H14C10O1, RHF, CHARGE=0, MULT=1
HF=-49.4
1     C         -0.000121958    -0.000583175     0.000087468
2     C          1.415778393     0.000877061     0.000088791
3     C          2.074636985     1.267182087     0.000091439
4     C          1.310569608     2.468115816     0.000522703
5     C         -0.086489473     2.414496848     0.000153134
6     C         -0.776481491     1.177622244    -0.000493443
7     C         -2.883912839     0.516939279    -1.279533026
8     C         -2.882086230     0.548933753     1.294136052
9     H         -0.508975090    -0.967936724     0.000755817
10    O          3.421537287     1.441101438    -0.000193336
11    H         -2.442919450     0.976152821    -2.187586646
12    C          2.153271104    -1.312435237     0.000188960
13    H          1.808331017     3.439040637     0.001098682
14    C         -2.301051198     1.155376649    -0.000851820
15    H         -2.703890033    -0.575039990    -1.333539668
16    H         -3.981063189     0.675187886    -1.330544117
17    H         -2.690120453    -0.539028267     1.380957541
18    H         -2.448970002     1.039592150     2.189465503
19    H         -3.980985901     0.696739677     1.337248939
20    H          2.797421471    -1.417129084    -0.899148478
21    H          1.464604806    -2.181815534     0.000111731
22    H          2.796868622    -1.417303676     0.899964386
23    H         -2.655028520     2.220236571    -0.014539007
24    H          3.915356218     0.633868225    -0.000337099
25    H         -0.639019814     3.356755460     0.000615230
---
H14C10O1, RHF, CHARGE=0, MULT=1
HF=-44
1     C         -0.001846941    -0.068600334    -0.023973740
2     C          1.406419674    -0.026189749     0.158227760
3     C          2.068184147     1.140162304     0.602879192
4     C          1.301781999     2.303232339     0.876220802
5     C         -0.082505040     2.254511251     0.698723563
6     C         -0.773971489     1.092417118     0.252256440
7     C         -2.805557797     1.142549316    -1.329274105
8     C         -3.084880081     0.368753867     1.124012533
9     C         -0.534810746    -1.395114696    -0.511984236
10    O          3.411740200     1.218303242     0.787098843
11    H         -2.173954771     1.747215283    -2.011763924
12    H          1.992116993    -0.923927016    -0.053119639
13    H          1.776275718     3.222884955     1.219185488
14    C         -2.292080668     1.235807737     0.122974402
15    H         -2.831596363     0.109502253    -1.725477133
16    H         -3.839324696     1.541825217    -1.397382366
17    H         -3.132034129    -0.700908483     0.843811244
18    H         -2.641505869     0.430030465     2.138755882
19    H         -4.131758120     0.730625382     1.200025995
20    H         -0.282217667    -2.203810162     0.207252334
21    H         -0.080617220    -1.657422573    -1.492128736
```

```
22      H      -1.629451647    -1.432589990    -0.653425236
23      H      -2.540957212     2.290201227     0.426893997
24      H       3.859306855     0.407237657     0.586234256
25      H      -0.641292251     3.168028241     0.920581132
---
H14C10O1, RHF, CHARGE=0, MULT=1
HF=-44.2
1       C      -0.052879209     0.056078617    -0.026556942
2       C       1.316099226     0.022401449     0.354453189
3       C       2.083099124     1.199572150     0.513810008
4       C       1.471965447     2.461512051     0.292050712
5       C       0.127212385     2.494136770    -0.079178336
6       C      -0.661013687     1.321518898    -0.249020511
7       C      -1.121521719    -1.666650538    -1.610533260
8       C      -1.848207100    -1.560298398     0.871814518
9       C      -2.092313278     1.568293785    -0.656336781
10      O       3.393137875     1.196476505     0.873609565
11      H      -0.302880978    -1.415361412    -2.315279331
12      H      -2.034927959    -1.149473905    -1.960611579
13      H       2.032992201     3.388917763     0.406900017
14      C      -0.727991098    -1.313314506    -0.160565780
15      H       0.049377471    -2.083025346     0.101470396
16      H      -1.310820713    -2.756895513    -1.701778985
17      H      -2.791067557    -1.029536709     0.639826101
18      H      -1.527311115    -1.246391611     1.886032881
19      H      -2.089666276    -2.642698230     0.923889528
20      H      -2.132937074     2.146517430    -1.604943363
21      H      -2.693177371     0.656575951    -0.822435355
22      H      -2.620924886     2.158645366     0.123652643
23      H       3.737955129     0.323501827     1.006142934
24      H      -0.328963824     3.473954215    -0.245123282
25      H       1.798802646    -0.941821532     0.531102180
---
H14C10O1, RHF, CHARGE=0, MULT=1
HF=-44.3
1       C      -0.000276069    -0.000299816    -0.000199202
2       C       1.402446111     0.000254644    -0.000073022
3       C       2.076715473     1.244720309     0.000000139
4       C       1.338570203     2.463008726     0.000104135
5       C      -0.073757682     2.467014215     0.000129292
6       C      -0.722508617     1.201253113     0.000106535
7       C      -0.910576884     3.765142953     0.000890872
8       C      -1.793624120     3.802163030     1.285373582
9       C      -1.804573397     3.798000314    -1.276119233
10      C      -0.046398672     5.062357534    -0.005399325
11      O       3.432277527     1.352813504    -0.000078774
12      H      -0.536481648    -0.950712151    -0.000197957
13      H       1.943459516    -0.945823030    -0.000074863
14      H       1.908414947     3.392782681     0.000242406
15      H      -1.811146644     1.127467567     0.000409679
16      H      -1.175121596     3.701755235     2.200415494
17      H      -2.547931612     2.990627211     1.306034563
18      H      -2.353689930     4.756624874     1.365750589
19      H      -1.194326098     3.690647977    -2.195904350
20      H      -2.561913240     2.989091318    -1.285781218
21      H      -2.362179270     4.753891770    -1.356857021
22      H       0.604154816     5.135106349     0.889526750
23      H       0.598554849     5.129859265    -0.904770181
24      H      -0.687008723     5.969267964    -0.005884628
```

```
25     H      3.867403705     0.510845364     0.000173050
---
H14C10O1, RHF, CHARGE=0, MULT=1
HF=-45.8
1      C      0.076909479    -0.044401030     0.019077675
2      C      1.280025363     0.041492266     0.771310241
3      C      1.671041821     1.347609336     1.200716388
4      C      0.878522689     2.491460819     0.875173571
5      C     -0.295870343     2.351877519     0.131521779
6      C     -0.700209424     1.075781582    -0.298853455
7      C      2.020862487    -1.271149379     1.035805232
8      C      3.388666334    -1.342728409     0.297599213
9      C      3.926676030    -2.750589576     0.019952204
10     C      2.077194951    -1.648338513     2.533223497
11     O      2.777703170     1.649866153     1.927976715
12     H     -0.273431313    -1.017004678    -0.335719369
13     H     -1.615856670     0.958377411    -0.878712143
14     H      1.187552501     3.484008241     1.207005754
15     H      1.393729266    -2.080027709     0.573923139
16     H      4.170189119    -0.786465206     0.863010273
17     H      3.290858906    -0.819464088    -0.682003020
18     H      3.212784675    -3.360296249    -0.569642706
19     H      4.156395697    -3.302382134     0.953427548
20     H      4.868079797    -2.684955933    -0.564389468
21     H      1.147295337    -1.348588903     3.057556448
22     H      2.928234008    -1.187165908     3.071738496
23     H      2.176450866    -2.747200386     2.652732639
24     H      3.313415995     0.907049874     2.158787282
25     H     -0.895676720     3.229084724    -0.113373608
---
H14C10O1, RHF, CHARGE=0, MULT=1
HF=-47.6
1      C      0.068092108     0.096952611    -0.671211148
2      C      1.332048573    -0.039351119    -0.044587292
3      C      1.831808459     1.157158152     0.581704488
4      C      1.097957164     2.376520388     0.552260900
5      C     -0.143729881     2.447599298    -0.089215483
6      C     -0.659403023     1.299088812    -0.701746652
7      C      2.086085361    -1.390555118    -0.058706392
8      C      2.283356658    -1.910392856     1.398459256
9      C      1.311867080    -2.512365554    -0.824403798
10     C      3.455810520    -1.232580730    -0.788356003
11     O      3.032394011     1.124650796     1.219025166
12     H     -0.408327735    -0.747858990    -1.171783186
13     H     -1.626037657     1.331917428    -1.204930875
14     H      1.492267322     3.275253728     1.029909010
15     H      1.333141531    -1.866881813     1.969175452
16     H      3.040901003    -1.341033138     1.969341623
17     H      2.624717981    -2.967080282     1.407387310
18     H      1.146451490    -2.255488093    -1.890477780
19     H      0.326943578    -2.733360913    -0.364995807
20     H      1.880037070    -3.466553782    -0.817301831
21     H      4.205007687    -0.669739669    -0.200739498
22     H      3.332475322    -0.709850836    -1.759142675
23     H      3.915062067    -2.220439495    -1.002585756
24     H      3.281412543     1.962917209     1.584578171
25     H     -0.698609271     3.385824474    -0.109512877
---
H14C10O1, RHF, CHARGE=0, MULT=1
```

```
HF=-45.6
1    C    0.082032657     0.004858331    -0.091332618
2    C    1.472955510     0.019733119     0.098680679
3    C    2.071473753     1.055571503     0.860394506
4    C    1.248318583     2.068558249     1.421171273
5    C   -0.137611922     2.029603260     1.214020000
6    C   -0.755818713     1.002721982     0.457694263
7    C   -2.931003102    -0.224609328     0.951724399
8    C   -4.423162756    -0.096096980     1.273591180
9    C   -2.269994200     1.003453973     0.263483524
10   H   -0.338921374    -0.810994421    -0.682609968
11   O    3.407416660     1.138313835     1.089863312
12   C   -2.655973282     1.179859321    -1.221072203
13   H    2.075382845    -0.772685714    -0.346725354
14   H    1.682104916     2.876509534     2.011019800
15   H   -0.741504747     2.824566553     1.657570291
16   H   -2.398157264    -0.422457530     1.911317749
17   H   -2.783741022    -1.135964559     0.328268025
18   H   -4.637222330     0.790383591     1.904314128
19   H   -5.046046201    -0.021575712     0.359930173
20   H   -4.766248405    -0.991322591     1.833288068
21   H   -2.670893292     1.911815196     0.784236500
22   H   -2.105489318     2.031050814    -1.671914630
23   H   -2.439485789     0.279934795    -1.830511236
24   H   -3.737460844     1.401355124    -1.327921131
25   H    3.894169947     0.430901397     0.688390780
---
H14C10O1, RHF, CHARGE=0, MULT=1
HF=-44.5
1    C   -0.000211275    -0.000142584     0.000285792
2    C    1.405417672     0.000490959    -0.000178391
3    C    2.114275673     1.225771726     0.000030850
4    C    1.379113179     2.440977647     0.000332365
5    C   -0.021154555     2.406432363     0.001281039
6    C   -0.761101402     1.191866201     0.002077684
7    C   -2.302883451     1.216236502     0.003229905
8    C   -2.804396474     1.939574533    -1.283702374
9    C   -2.804413995     1.963114516     1.276761182
10   C   -2.943271890    -0.204592803     0.017400290
11   O    3.469408314     1.311547906    -0.000292286
12   H   -0.477491063    -0.981478802    -0.001000483
13   H    1.930852649    -0.955410265    -0.000727263
14   H    1.892754538     3.402961198    -0.000214939
15   H   -0.533633214     3.371140589     0.001120867
16   H   -2.395737217     1.462080043    -2.197393167
17   H   -2.512175893     3.008141222    -1.311001581
18   H   -3.910967434     1.912052297    -1.360939038
19   H   -2.390775981     1.506320587     2.198790632
20   H   -2.518101975     3.033659114     1.281673754
21   H   -3.910623003     1.931565229     1.357790366
22   H   -2.661996823    -0.798545369    -0.875794087
23   H   -2.654410787    -0.783606860     0.917965508
24   H   -4.051840524    -0.144562548     0.021214820
25   H    3.891457803     0.462738672    -0.000581271
---
H16C10O1, RHF, CHARGE=0, MULT=1
HF=-74.3
1    C   -0.000238837     0.000158855    -0.000003925
2    C    1.577689294    -0.000424385    -0.000024329
```

```
 3    C     2.077512409    1.496211482    0.000885948
 4    C     1.529371955    2.232576309    1.252520348
 5    C    -0.023943317    2.224383630    1.229176223
 6    C    -0.526261787    0.754445857    1.250652650
 7    C     2.078122517   -0.699747728    1.320449273
 8    C    -0.024558500    0.049168888    2.540487549
 9    C     2.037944062    1.530261256    2.541484924
10    C     1.528537205    0.063022990    2.557253731
11    H    -0.394166697   -1.039433291   -0.014730426
12    H    -0.396592350    0.475722750   -0.924425689
13    O     2.101088678   -0.726027348   -1.078292078
14    H     3.188394813    1.540158453   -0.014672663
15    H     1.750987169    2.023365952   -0.922874680
16    H     1.891058754    3.285257789    1.241333699
17    H    -0.405578020    2.756232332    0.330220893
18    H    -0.431993886    2.790751530    2.094724206
19    H    -1.639149647    0.746706722    1.237071819
20    H     1.751959807   -1.762259389    1.351072119
21    H     3.189051375   -0.730653088    1.353302584
22    H    -0.408375707   -0.993004176    2.592271432
23    H    -0.431053549    0.552212649    3.445198476
24    H     1.691902981    2.075505882    3.446701661
25    H     3.148277412    1.560789446    2.592421941
26    H     1.889660033   -0.440766785    3.481794818
27    H     1.839928699   -0.363012239   -1.913034586
---
H16C10O1, RHF, CHARGE=0, MULT=1
HF=-71.5
 1    C     0.000312262    0.000175318   -0.000303906
 2    C     1.554071731    0.000520595    0.000834245
 3    C     2.061680021    1.471420636   -0.001186767
 4    C     1.537053793    2.205839128    1.264545042
 5    C    -0.016931408    2.204342542    1.258583238
 6    C    -0.523347015    0.734850244    1.264648450
 7    C     2.136983898   -0.717580899    1.280391170
 8    C    -0.007741215    0.012667040    2.540122523
 9    C     2.052240341    1.484055018    2.541431331
10    C     1.545879583    0.012594046    2.548737721
11    H    -0.402568407   -1.034893353   -0.046236087
12    H    -0.383534300    0.490647234   -0.921982210
13    H     1.921709477   -0.511884616   -0.915565824
14    H     3.172415921    1.505085973   -0.040861823
15    H     1.727136630    1.995667003   -0.923021047
16    H     1.905789271    3.255660713    1.260763158
17    H    -0.406719125    2.749699006    0.372119583
18    H    -0.413593515    2.759129289    2.136722347
19    H    -1.635995593    0.727604842    1.261276941
20    O     2.002717545   -2.106619678    1.286781765
21    H     3.253689449   -0.568500150    1.283115425
22    H    -0.411076769   -1.022000480    2.593765832
23    H    -0.397690732    0.512745316    3.454354031
24    H     1.708614064    2.016376230    3.455706108
25    H     3.162516908    1.519197038    2.590162164
26    H     1.908118049   -0.491331002    3.472050676
27    H     1.105454991   -2.404995087    1.283377967
---
H16C10O1, RHF, CHARGE=0, MULT=1
HF=-63.9
 1    C    -0.013972418   -0.021623415   -0.002647975
```

```
 2    C    1.530155186   -0.020134211    0.002979569
 3    C    1.834006344    1.507181664    0.004989405
 4    C    1.466302609    2.048479939    1.420454970
 5    C   -0.085243770    2.001832711    1.421241563
 6    C   -0.457558610    1.482758094   -0.020398529
 7    C   -1.910770624    1.742157989   -0.423170022
 8    O   -0.734597772   -1.003073739    0.020764562
 9    C    0.699937359    2.111763331   -0.930539519
10    C    0.737432761    1.601787546   -2.397202914
11    C    0.726999293    3.659162863   -1.037793892
12    H    1.955433540   -0.538978016   -0.879538966
13    H    1.940489907   -0.547579936    0.888214117
14    H    2.866107619    1.746205370   -0.300658294
15    H    1.857895258    3.072062883    1.587988151
16    H    1.901922935    1.434429781    2.235454673
17    H   -0.526126546    2.997139341    1.631035383
18    H   -0.476692074    1.331820638    2.214670795
19    H   -2.138138029    2.827299110   -0.455093188
20    H   -2.615570348    1.282059003    0.299799234
21    H   -2.142944544    1.325592688   -1.424011250
22    H   -0.038168037    2.097299635   -3.017190885
23    H    0.563401393    0.513934102   -2.508506753
24    H    1.722801955    1.818254518   -2.859959821
25    H   -0.086932998    4.033736090   -1.692679720
26    H    1.689512075    3.999241088   -1.474754026
27    H    0.612671240    4.190519350   -0.073852559
---
H16C10O1, RHF, CHARGE=0, MULT=1
HF=-68.6
 1    C   -0.107377267   -0.018963348    0.039618650
 2    C   -0.924316402    1.168681835    0.574414868
 3    C   -0.083263538    2.205695325    1.331105271
 4    C    0.853673902    1.587757147    2.378903867
 5    C    1.649176275    0.342635814    1.929737904
 6    C    0.853967849   -0.711660356    1.060003970
 7    C    0.061021971   -1.699730325    1.954919695
 8    C    1.979482300   -1.475646450    0.278995574
 9    C    3.190549196   -0.539128987    0.224198546
10    C    2.968645847    0.643595630    1.170515436
11    O    4.182117138   -0.711857898   -0.468120703
12    H    0.476374691    0.342596122   -0.839098512
13    H   -0.815808514   -0.773870485   -0.372068418
14    H   -1.425632445    1.667968440   -0.287002298
15    H   -1.749647977    0.804879864    1.227469486
16    H   -0.763529411    2.927589600    1.838173581
17    H    0.506303064    2.810876486    0.605327182
18    H    0.252474455    1.312219980    3.277183969
19    H    1.564065277    2.370192441    2.731204123
20    H    1.953718417   -0.165976911    2.880364645
21    H    0.718735203   -2.215435727    2.684634143
22    H   -0.731368402   -1.190019088    2.538549616
23    H   -0.431033160   -2.487209993    1.347353094
24    H    2.278244611   -2.411717162    0.800463892
25    H    1.656561690   -1.783512288   -0.736910585
26    H    2.924794806    1.587344921    0.586736582
27    H    3.816167969    0.750203125    1.880661430
---
H16C10O1, RHF, CHARGE=0, MULT=1
HF=-65.8
```

```
 1    C      0.045274419    -0.047825318   -0.056722746
 2    C      1.578047399     0.017074285    0.136525027
 3    C      2.256398539     1.388322447   -0.073575833
 4    C      1.632531287     2.315936775   -1.140155151
 5    C      0.099241822     2.244115391   -1.093591967
 6    C     -0.461822673     0.779641900   -1.272258457
 7    C     -0.053339145     0.067710787   -2.594626660
 8    C     -2.002626271     1.051184609   -1.329252046
 9    C     -2.134195288     2.432315387   -2.001937531
10    C     -0.765248595     3.138139574   -2.009731249
11    O     -3.167901915     2.896008612   -2.456410891
12    H     -0.449629522     0.303842472    0.877291243
13    H     -0.256435516    -1.114113911   -0.161999598
14    H      1.807724296    -0.328017268    1.171342491
15    H      2.061762219    -0.729328174   -0.534574607
16    H      3.324713853     1.215217403   -0.337062640
17    H      2.271203938     1.929740130    0.900957475
18    H      2.027534864     2.066811270   -2.149018796
19    H      1.962967880     3.361420552   -0.944023466
20    H     -0.168791719     2.578133214   -0.052783405
21    H     -0.365080859     0.635266188   -3.494981660
22    H      1.038976459    -0.089935638   -2.685051729
23    H     -0.529064846    -0.933127149   -2.661301990
24    H     -2.558210330     0.269139352   -1.885233143
25    H     -2.458021137     1.090603926   -0.316306505
26    H     -0.839250643     4.173218305   -1.616291001
27    H     -0.370986997     3.218867241   -3.043823602
---
H18C10O1, RHF, CHARGE=0, MULT=1
HF=-61.9
 1    C      0.000424865     0.000589894   -0.001014330
 2    C      1.583915630     0.000178385    0.001271488
 3    C      2.231191764     1.443352596   -0.001701013
 4    C      1.295087246     2.656075515    0.181596369
 5    C      0.033133640     2.540463495   -0.656724727
 6    C     -0.620585254     1.162367752   -0.758857167
 7    C     -1.404078235     2.338598784   -0.110700782
 8    C     -1.736168503     2.393713027    1.376636825
 9    C     -2.544918893     2.938526756   -0.940442707
10    C      2.107332776    -0.799529496   -1.242294159
11    O      2.042537037    -0.609812946    1.187104782
12    H     -0.373172602    -0.026609942    1.046989607
13    H     -0.371062068    -0.948006248   -0.453060726
14    H      2.771069660     1.587256930   -0.966256831
15    H      3.018239644     1.508967217    0.783434017
16    H      1.849342139     3.574577607   -0.121777872
17    H      1.071894234     2.797538377    1.261473426
18    H      0.113309058     3.142643324   -1.576277924
19    H     -0.983198080     0.838560596   -1.747886778
20    H     -0.944524686     1.979858252    2.031389146
21    H     -1.901937713     3.442572106    1.701707675
22    H     -2.661950663     1.820368602    1.593483970
23    H     -2.722862107     3.997632412   -0.661241866
24    H     -2.342842485     2.922390185   -2.030714858
25    H     -3.488965383     2.377974690   -0.778879664
26    H      1.777599075    -0.326602513   -2.188499175
27    H      3.214634578    -0.850318846   -1.258597897
28    H      1.734183669    -1.844687076   -1.249486827
29    H      1.761709395    -1.511329129    1.260211866
```

```
---
H18C10O1, RHF, CHARGE=0, MULT=1
HF=-72.9
1    C     0.060139618    0.295528131   -0.031974624
2    C     1.508796377   -0.138096814   -0.330808974
3    C     2.550412233    0.961470987   -0.619128856
4    C     2.393999253    1.765958677   -1.926226719
5    C     1.847057997    3.200188014   -1.764822763
6    C     1.031670753    3.747600239   -2.954165525
7    C    -0.499666099    3.821813169   -2.791127599
8    C    -1.303992157    2.505678754   -2.817942053
9    C    -1.571203309    1.815652871   -1.462132594
10   C    -0.869753759    0.463915557   -1.245673799
11   O    -1.108898902   -0.486715405   -1.984388065
12   H    -0.403484704   -0.481478040    0.622801682
13   H     0.066571129    1.223561397    0.580134262
14   H     1.526431387   -0.884676533   -1.156280317
15   H     1.870267789   -0.689959430    0.570878709
16   H     3.544052411    0.452965274   -0.655993710
17   H     2.610368556    1.646788035    0.256808229
18   H     3.399850828    1.839648304   -2.403740227
19   H     1.779757635    1.187043789   -2.650272977
20   H     2.719774614    3.878313944   -1.608513529
21   H     1.249308334    3.291179428   -0.831131658
22   H     1.380683223    4.791832053   -3.143144837
23   H     1.275863846    3.189858242   -3.886429709
24   H    -0.876162110    4.444816834   -3.638928954
25   H    -0.745833465    4.403519360   -1.873850130
26   H    -2.300956645    2.745350079   -3.260324180
27   H    -0.834801514    1.799809451   -3.538020560
28   H    -2.668484541    1.628364593   -1.373510291
29   H    -1.332446705    2.502614303   -0.621626500
---
H20C10O1, RHF, CHARGE=0, MULT=1
HF=-94.1
1    O     0.000015809    0.000357975    0.001411477
2    C     1.225526799   -0.000222445   -0.000216975
3    C     2.006032536    1.361804900   -0.000179617
4    C     2.906217109    1.445136928   -1.265836387
5    C     2.863898736    1.478461983    1.290935367
6    C     1.022286040    2.573581240   -0.032431229
7    C     2.002301581   -1.331973046   -0.088507961
8    C     1.583809529   -2.558386700    0.797285833
9    C     0.220125764   -3.160875212    0.351163017
10   C     1.515982888   -2.177664903    2.302202396
11   C     2.680119377   -3.655243683    0.612168122
12   H     2.319431655    1.273841665   -2.191296674
13   H     3.729722522    0.703538105   -1.252847536
14   H     3.381787204    2.443351265   -1.355747167
15   H     2.243338103    1.354916958    2.201602145
16   H     3.673508778    0.723204845    1.334507967
17   H     3.354539767    2.471364242    1.358054068
18   H     0.354628434    2.590077925    0.852257769
19   H     0.384099016    2.565275094   -0.938786321
20   H     1.577013551    3.535070810   -0.036388708
21   H     1.950567928   -1.634795311   -1.162137435
22   H     3.077473847   -1.132541604    0.115159055
23   H     0.182832025   -3.319889579   -0.745646078
24   H    -0.639776488   -2.520300120    0.626005810
```

```
25   H    0.043676469   -4.147218205    0.829558838
26   H    2.474807718   -1.750185424    2.660468844
27   H    0.722439509   -1.433967267    2.513686634
28   H    1.300152016   -3.064971418    2.932639093
29   H    2.776196000   -3.963719726   -0.448868911
30   H    3.678165463   -3.303722598    0.945067757
31   H    2.447046107   -4.567664717    1.197996064
---
H22C10O1, RHF, CHARGE=0, MULT=1
HF=-94.5
1    C   -0.000038552   -0.000002771   -0.000018882
2    C    1.531191348    0.000121421    0.000061901
3    C    2.174223371    1.400646829   -0.000084149
4    C    3.716453022    1.385693013   -0.003174838
5    C    4.357715287    2.788640143   -0.000912876
6    C    5.900207558    2.773249181   -0.007422620
7    C    6.541487446    4.176233275   -0.003511021
8    C    8.083934604    4.160504316   -0.013201895
9    C    8.723121204    5.563883200   -0.008048087
10   C   10.273226207    5.529300688   -0.019579415
11   O   10.771767266    6.834199112   -0.014940269
12   H   -0.413011949    0.506683336    0.895715245
13   H   -0.412997277    0.505232338   -0.896572462
14   H   -0.382493879   -1.041669617    0.000820985
15   H    1.882763613   -0.569893019    0.891158443
16   H    1.882973189   -0.570555044   -0.890451031
17   H    1.816412098    1.963287334    0.892725246
18   H    1.813065924    1.965113300   -0.890357844
19   H    4.076689826    0.819375679    0.886244899
20   H    4.072785239    0.823834783   -0.896969065
21   H    4.003097787    3.349489149    0.894188758
22   H    3.995718756    3.355738260   -0.889075004
23   H    6.262443601    2.205084294    0.880035082
24   H    6.254757622    2.213362918   -0.903181677
25   H    6.188977299    4.735212115    0.893529451
26   H    6.178046173    4.745315585   -0.889908384
27   H    8.447527567    3.590116814    0.872409973
28   H    8.436335497    3.601828527   -0.910801585
29   H    8.377560026    6.123266784    0.890940594
30   H    8.364520869    6.135785845   -0.893942382
31   H   10.651947756    4.957793980    0.871024612
32   H   10.639270382    4.969137421   -0.922670067
33   H   11.718377249    6.834652301   -0.022232004
---
H22C10O1, RHF, CHARGE=0, MULT=1
HF=-93.1
1    C   -0.263153742   -0.296329870    0.300754912
2    C    1.144480088   -0.898912715    0.317542718
3    C    2.286330577    0.134875061    0.368970626
4    C    3.696083971   -0.487728652    0.400952835
5    C    4.832947810    0.569699720    0.468750428
6    O    6.103148629    0.000350171    0.642604244
7    C    6.886152409   -0.356569964   -0.462541590
8    C    8.297752688   -0.806937829   -0.001900114
9    C    9.224877855    0.311677479    0.511030160
10   C   10.653920388   -0.163810083    0.840321926
11   C   11.597637365    0.937706291    1.330756009
12   H   -0.460562021    0.318119891    1.202371720
13   H   -0.426894194    0.344345604   -0.589304849
```

```
14      H       -1.023963059    -1.103588399     0.274522461
15      H        1.223506873    -1.583480617     1.193579724
16      H        1.264076486    -1.539077392    -0.587094083
17      H        2.150339191     0.780176113     1.267173042
18      H        2.204489627     0.808667141    -0.515167643
19      H        3.778840928    -1.169106474     1.277690470
20      H        3.833277296    -1.121897215    -0.504282375
21      H        4.675718434     1.234499233     1.358855204
22      H        4.787096822     1.233621551    -0.433998534
23      H       11.255271546     1.377473660     2.289173898
24      H       11.692514429     1.760806761     0.593457828
25      H        6.974538373     0.486690379    -1.198464514
26      H        6.418202393    -1.207726995    -1.027837837
27      H        8.199052697    -1.598046646     0.775262178
28      H        8.769073289    -1.298071035    -0.885623084
29      H        8.779220679     0.774057694     1.420601656
30      H        9.280729892     1.118761014    -0.255751959
31      H       10.609907843    -0.961415085     1.617727308
32      H       11.106007213    -0.636661429    -0.062238039
33      H       12.613785603     0.524040873     1.498000501
---
H14C11O1, RHF, CHARGE=0, MULT=1
HF=-45.2
1       C        0.173644999    -0.089270141    -0.020392157
2       C        1.575285275     0.051747930    -0.186549005
3       C        2.121043340     1.352941959    -0.109583748
4       C        1.339726822     2.508648954     0.114292198
5       C       -0.067149869     2.359675523     0.269942586
6       C       -0.615330524     1.059793075     0.197624757
7       C       -0.522864657    -1.429230402    -0.077491141
8       C       -0.790978087    -2.135014970     1.250568599
9       O       -0.881343500    -1.909111840    -1.147328648
10      C        2.485264910    -1.120610156    -0.455894623
11      C       -0.985833011     3.531637772     0.509519039
12      C        2.021728061     3.852991530     0.173934536
13      H        3.200968279     1.470357259    -0.233423330
14      H       -1.695932961     0.936935955     0.310506258
15      H        0.077068380    -2.046047935     1.933086013
16      H       -1.003253981    -3.213837720     1.112571699
17      H       -1.668638261    -1.674805903     1.748690233
18      H        3.557103552    -0.837083906    -0.418228893
19      H        2.345025213    -1.928821286     0.291554594
20      H        2.294441006    -1.544635310    -1.464040638
21      H       -0.760791813     4.022415306     1.479886270
22      H       -0.891457031     4.291637343    -0.293951466
23      H       -2.053243137     3.231015741     0.540208905
24      H        1.661271576     4.520592293    -0.636764409
25      H        1.836359681     4.356022876     1.145938100
26      H        3.122236326     3.773557057     0.059475212
---
H14C11O1, RHF, CHARGE=0, MULT=1
HF=-49
1       C       -0.005701435    -0.002702117     0.004169209
2       C        1.401642565     0.073728353    -0.185357933
3       C        2.038448781     1.331936249    -0.085530078
4       C        1.320581184     2.514502057     0.182194992
5       C       -0.077174070     2.414429256     0.346956003
6       C       -0.756433561     1.179143477     0.260197059
7       C       -0.709359109    -1.341106023    -0.072537740
```

```
8    C     -0.835674734    -2.156884464     1.214568535
9    O     -1.160506650    -1.756481496    -1.134871641
10   C      2.244535891    -1.138036263    -0.504308798
11   C     -2.257269594     1.166054620     0.426061149
12   C      2.008849327     3.850181964     0.284297632
13   H      3.121039536     1.391273459    -0.222557423
14   H     -0.487738720    -1.589864906     2.099700952
15   H     -0.229915924    -3.083244897     1.143611414
16   H     -1.889971881    -2.450613767     1.392826436
17   H      3.318391147    -0.883605601    -0.618951607
18   H      2.183803336    -1.898105417     0.302431897
19   H      1.924720496    -1.611492902    -1.455605047
20   H     -2.759701052     0.861837447    -0.515743771
21   H     -2.574173989     0.470299376     1.230520190
22   H     -2.657599648     2.164410357     0.697628316
23   H      1.637645758     4.546016708    -0.497509211
24   H      1.822419716     4.317522702     1.274157554
25   H      3.107840245     3.771860991     0.160381735
26   H     -0.648743577     3.324702306     0.545181233
---
H16C11O1, RHF, CHARGE=0, MULT=1
HF=-49.5
1    C     -0.000127899    -0.000399661    -0.000073659
2    C      1.437828117     0.000026808     0.000504456
3    C      2.039135896     1.283485964    -0.000697266
4    C      1.326738328     2.504470004    -0.003395808
5    C     -0.081824921     2.451009254    -0.004151587
6    C     -0.736705481     1.216531706    -0.002058762
7    O     -0.672237809    -1.181582588     0.001301087
8    C      2.038420515     3.830686758    -0.009536443
9    C      2.302391973    -1.284586932     0.002277856
10   C      2.018538170    -2.123031172    -1.282129891
11   C      3.838071900    -0.991249276     0.000742570
12   C      2.019985371    -2.118007419     1.290273883
13   H      3.127680295     1.371991239     0.000453332
14   H     -0.675718663     3.366923812    -0.006412039
15   H     -1.828484315     1.205832992    -0.001941020
16   H     -1.613904565    -1.072163750     0.000685254
17   H      1.816642038     4.392777842    -0.941479245
18   H      1.719210367     4.457558732     0.849756089
19   H      3.140271314     3.724898573     0.055586743
20   H      1.034660515    -2.628592069    -1.268658710
21   H      2.054414139    -1.486214157    -2.189781929
22   H      2.772773832    -2.927248342    -1.414343352
23   H      4.158053049    -0.427001647    -0.898758336
24   H      4.159179675    -0.423812610     0.897900819
25   H      4.425109934    -1.934007617     0.002292985
26   H      2.061014472    -1.478432600     2.195780037
27   H      1.034346929    -2.620223894     1.281706161
28   H      2.771969604    -2.924362240     1.422293342
---
H16C11O1, RHF, CHARGE=0, MULT=1
HF=-50.5
1    C     -0.057658594    -0.066182937     0.050494826
2    C      0.964395516     0.124940818     0.994419631
3    C      1.699795153     1.322289393     1.012165342
4    C      1.436288112     2.375094962     0.095904563
5    C      0.410934373     2.150044535    -0.854194873
6    C     -0.327218350     0.952279164    -0.876047573
```

```
   7    C     2.259467310    3.695580211    0.161207610
   8    C     3.727837360    3.446573367   -0.416593858
   9    C     3.760584712    2.904623760   -1.862380751
  10    C     4.693327974    4.645860149   -0.275401011
  11    C     1.533273140    4.891251556   -0.548659329
  12    O     2.439485358    4.050659066    1.517401805
  13    H    -0.631017933   -0.993125832    0.036405512
  14    H     1.191812323   -0.658673178    1.718153764
  15    H     2.489816347    1.412289298    1.760876393
  16    H     0.159726083    2.895728170   -1.609648275
  17    H    -1.113029886    0.818456756   -1.620525046
  18    H     4.174086018    2.634748480    0.215322950
  19    H     3.359880172    3.623583189   -2.604024017
  20    H     3.192080258    1.959436352   -1.967657200
  21    H     4.804799356    2.674341526   -2.162685808
  22    H     4.403726857    5.506735011   -0.910960748
  23    H     4.769438887    5.006662107    0.769193574
  24    H     5.719041900    4.346882968   -0.579759984
  25    H     2.027970052    5.858888867   -0.327970751
  26    H     1.514773253    4.781593272   -1.649648532
  27    H     0.482080349    4.989163587   -0.205442393
  28    H     1.614687204    4.182618494    1.963860179
---
H20C11O1, RHF, CHARGE=0, MULT=1
HF=-77
   1    C    -0.151150855    0.131598747   -0.024148337
   2    C     1.388865988    0.100337881   -0.130804354
   3    C     2.135092999    1.449861702   -0.103097940
   4    C     1.851557799    2.490036812   -1.205025143
   5    C     2.228379187    2.097395659   -2.649165039
   6    C     1.335927074    2.709942181   -3.752695059
   7    C     0.892158879    1.757954891   -4.883084953
   8    C    -0.468732487    1.043249307   -4.745432904
   9    C    -0.589107050   -0.098908655   -3.713513223
  10    C    -1.314625370    0.262133601   -2.399798375
  11    C    -0.879427585   -0.573322196   -1.183144333
  12    O    -1.141041365   -1.770084191   -1.116780345
  13    H    -0.447570999   -0.357216226    0.933464254
  14    H    -0.534409482    1.171331120    0.063841627
  15    H     1.705286458   -0.473321627   -1.030213108
  16    H     1.771711399   -0.494223784    0.733276584
  17    H     3.227116557    1.221183139   -0.102419220
  18    H     1.932107760    1.943233186    0.877016508
  19    H     2.423193997    3.411822746   -0.940247863
  20    H     0.781641204    2.789833584   -1.144900697
  21    H     3.279999877    2.421308119   -2.833308061
  22    H     2.255062738    0.991350885   -2.751990569
  23    H     1.909959953    3.543069569   -4.224330051
  24    H     0.437943881    3.199771330   -3.314523536
  25    H     0.830498026    2.369698618   -5.816115365
  26    H     1.686399179    1.003709326   -5.083756224
  27    H    -0.695205120    0.603687273   -5.747143266
  28    H    -1.265989066    1.802246100   -4.577046816
  29    H    -1.152190144   -0.933614404   -4.193998993
  30    H     0.418178817   -0.518356029   -3.498638835
  31    H    -2.412116178    0.121881302   -2.541340544
  32    H    -1.188363539    1.340642120   -2.165427659
---
H22C11O1, RHF, CHARGE=0, MULT=1
```

```
HF=-100.7
1    C    -0.000050419     0.000007161     0.000192712
2    O     1.226707518     0.000016811    -0.000816485
3    C    -0.785737535     1.326889642     0.000200273
4    C    -0.842526805     2.168556886    -1.321764305
5    C    -1.387932926     1.326111269    -2.508431051
6    C     0.555759431     2.738962715    -1.695601357
7    C    -1.812207195     3.365696756    -1.073494469
8    C    -0.786569124    -1.326308403     0.001645442
9    C    -0.825943235    -2.177923535     1.317794636
10   C    -1.348599883    -1.341662830     2.518966266
11   C     0.575507924    -2.757252424     1.665231553
12   C    -1.804396824    -3.369226977     1.075974995
13   H    -1.829952409     1.111853850     0.322256996
14   H    -0.354606665     1.960622458     0.810915234
15   H    -2.381542500     0.889085115    -2.280568936
16   H    -0.709857522     0.489819005    -2.772563195
17   H    -1.505886312     1.944162899    -3.422270598
18   H     1.023403130     3.270017309    -0.842111356
19   H     1.258518508     1.948136173    -2.023452764
20   H     0.481582842     3.465285530    -2.531774489
21   H    -2.836646618     3.021719631    -0.823363041
22   H    -1.466240692     4.007098039    -0.237501952
23   H    -1.897074019     4.013680510    -1.969637343
24   H    -0.367246447    -1.953889070    -0.820000181
25   H    -1.835069434    -1.108252158    -0.303893711
26   H    -2.344420713    -0.899956729     2.310551113
27   H    -0.663545290    -0.509085233     2.777042193
28   H    -1.453393474    -1.965309666     3.430406410
29   H     1.286232443    -1.972593164     1.990549446
30   H     1.028457246    -3.281658810     0.799710717
31   H     0.510791596    -3.491617124     2.495057540
32   H    -2.831243358    -3.018569961     0.844621027
33   H    -1.474218964    -4.005660748     0.229858680
34   H    -1.877840543    -4.023601693     1.968469395
---
H22C11O1, RHF, CHARGE=0, MULT=1
HF=-92.6
1    C   0.00000      0.00000      0.00000
2    C   1.53126      0.00000      0.00000
3    C   2.17373      1.40078      0.00000
4    C   3.71623      1.38410      0.00024
5    C   4.35355      2.78777     -0.00186
6    C   5.89206      2.76361      0.01313
7    C   6.63967      2.81916     -1.33031
8    C   7.60842      1.64450     -1.57387
9    C   8.29084      1.67922     -2.95711
10   C   9.26873      0.51391     -3.20344
11   C   9.95285      0.53327     -4.57325
12   O   6.50253      2.74855      1.07718
13   H  -0.41280      0.50605     -0.89616
14   H  -0.41284      0.50578      0.89621
15   H  -0.38189     -1.04178     -0.00026
16   H   1.88306     -0.56987     -0.89102
17   H   1.88329     -0.56990      0.89077
18   H   1.81528      1.96443     -0.89191
19   H   1.81520      1.96464      0.89153
20   H   4.07400      0.81990     -0.89123
21   H   4.07561      0.82289      0.89270
```

```
22     H   4.00798      3.36064     -0.89143
23     H   3.99513      3.36301      0.88269
24     H   5.90307      2.86463     -2.16304
25     H   7.19782      3.78380     -1.36233
26     H   7.04994      0.68650     -1.46648
27     H   8.39169      1.64356     -0.78245
28     H   7.50883      1.66998     -3.75093
29     H   8.83873      2.64293     -3.06914
30     H   8.72740     -0.45429     -3.09419
31     H  10.05747      0.51824     -2.41604
32     H   9.21886      0.47291     -5.40225
33     H  10.55656      1.45164     -4.72033
34     H  10.63734     -0.33455     -4.67070
---
H24C11O1, RHF, CHARGE=0, MULT=1
HF=-91.1
1      C   0.000027687    0.000059106   -0.000062068
2      C   1.531293042    0.000283375   -0.000239211
3      C   2.174373177    1.400752513   -0.000018517
4      C   3.716592581    1.385573916    0.006787863
5      C   4.358201849    2.788389500    0.004400501
6      C   5.900645390    2.772745278    0.012497317
7      C   6.542132726    4.175637173    0.009128219
8      C   8.084573810    4.159671592    0.018175290
9      C   8.723619902    5.563269628    0.013802824
10     C  10.274690557    5.526491635    0.024511737
11     O  10.761355462    6.841773213    0.018405841
12     C  12.138862562    7.066551236    0.030790893
13     H  -0.413008225    0.507629668   -0.895298687
14     H  -0.412946923    0.504381460    0.896959172
15     H  -0.382319584   -1.041723628   -0.001941558
16     H   1.882840598   -0.569510684   -0.891384608
17     H   1.883239230   -0.570415431    0.890059061
18     H   1.818784956    1.962541714   -0.894246953
19     H   1.811202426    1.966203045    0.888856942
20     H   4.078817321    0.818123714   -0.881106698
21     H   4.070886457    0.824739296    0.902051730
22     H   4.004623371    3.348806880   -0.891385183
23     H   3.995383137    3.356143331    0.891857014
24     H   6.263594792    2.204467122   -0.874560830
25     H   6.254301421    2.212893771    0.908625949
26     H   6.189466206    4.735208543   -0.887489766
27     H   6.179149752    4.744256273    0.895906240
28     H   8.447709998    3.590113582   -0.868114258
29     H   8.437247396    3.600269834    0.915137498
30     H   8.377141253    6.123146279   -0.884420158
31     H   8.365332088    6.134161390    0.900433558
32     H  10.650108860    4.956457100   -0.867415893
33     H  10.637721885    4.969124737    0.929455871
34     H  12.653712723    6.636016378   -0.863755241
35     H  12.634982295    6.650242688    0.942516560
36     H  12.293081414    8.171493809    0.023637637
---
H8C12O1, RHF, CHARGE=0, MULT=1
HF=11.3
1      C  -0.000027043    0.000217019    0.000000698
2      C   1.406004712   -0.000219398    0.000005635
3      C   2.147276837    1.201996186    0.000008578
4      C   1.511020921    2.455969311   -0.000009260
```

```
5    C     0.108219719    2.497141802   -0.000013869
6    C    -0.632160991    1.257366095   -0.000004633
7    O    -1.988477925    1.475826607   -0.000000309
8    C    -2.168907638    2.837708202   -0.000017197
9    C    -3.408095515    3.504676243    0.000041035
10   C    -3.371298663    4.910113430   -0.000052273
11   C    -2.148950995    5.617775244   -0.000040318
12   C    -0.912858058    4.947082014    0.000035604
13   C    -0.909384563    3.543584114   -0.000016918
14   H    -0.564052475   -0.930629564    0.000005579
15   H     1.938234014   -0.952655250    0.000007068
16   H     3.236989924    1.153846855    0.000024093
17   H     2.099650508    3.372817328   -0.000010375
18   H    -4.353619691    2.965967525    0.000186568
19   H    -4.309066415    5.467983355   -0.000119765
20   H    -2.166617191    6.708395525   -0.000096076
21   H     0.019572717    5.510714765    0.000079618
---
H10C12O1, RHF, CHARGE=0, MULT=1
HF=5.1
1    C     0.000212530   -0.000062743    0.000122423
2    C     1.405768911   -0.000595832   -0.000099043
3    C     2.113724968    1.213777935    0.000193626
4    C     1.425025262    2.449627871    0.001344289
5    C     0.010250068    2.436850888    0.001107968
6    C    -0.695426050    1.221155597    0.000521626
7    C     2.172529041    3.735574091    0.001601377
8    C     2.530989618    4.359718078   -1.214239878
9    C     3.238075065    5.575167710   -1.210067733
10   C     3.602539400    6.193830958   -0.006119652
11   C     3.247560311    5.575711438    1.223812181
12   C     2.533809447    4.348745437    1.222320031
13   O     3.622286492    6.212452088    2.364836299
14   H    -0.547745162   -0.942562740   -0.000255162
15   H     1.950200945   -0.945530519   -0.000657087
16   H     3.204709472    1.193055934   -0.000798997
17   H    -0.547049596    3.374928812    0.001026492
18   H    -1.785957255    1.227530940    0.000171972
19   H     2.262172323    3.903860695   -2.167917187
20   H     3.507286983    6.044368683   -2.157793224
21   H     2.259187633    3.870959396    2.164084194
22   H     3.356878786    5.750645852    3.148788518
23   H     4.150855064    7.135846933   -0.021701358
---
H10C12O1, RHF, CHARGE=0, MULT=1
HF=4
1    C     0.000513024   -0.000214637   -0.000014084
2    C     1.405823831   -0.000768804    0.000410660
3    C     2.114176897    1.213454563   -0.000999853
4    C     1.426516211    2.450365883   -0.004428798
5    C     0.011389647    2.436858507   -0.004085000
6    C    -0.694470315    1.221294387   -0.001651831
7    C     2.173484298    3.735679906   -0.020036267
8    C     2.525617717    4.324554438   -1.255551040
9    C     3.231228616    5.537936410   -1.318944189
10   C     3.603085647    6.194796578   -0.135179430
11   C     3.272346336    5.643287525    1.110897485
12   C     2.559074595    4.415535057    1.178763031
13   O     2.219482847    3.846298358    2.364598410
```

```
14   H   -0.547746240   -0.942605281    0.001724075
15   H    1.950102440   -0.945846562    0.002745880
16   H    3.204983059    1.191438593    0.002387317
17   H   -0.546706242    3.374444792   -0.003878513
18   H   -1.784941940    1.228145805   -0.000050560
19   H    2.246153820    3.830573478   -2.188470795
20   H    3.488813293    5.967166137   -2.287245667
21   H    4.151146636    7.136811663   -0.181351161
22   H    3.570237764    6.168736924    2.019102331
23   H    2.505024368    4.354071259    3.112507163
---
H10C12O1, RHF, CHARGE=0, MULT=1
HF=0
1    C    0.000028859    0.000049432   -0.000030242
2    C    1.405471137   -0.000224967    0.000097347
3    C    2.113429269    1.214277233    0.000106792
4    C    1.425822420    2.451109266   -0.000011122
5    C    0.010745844    2.437152525   -0.000485588
6    C   -0.695123523    1.221554686   -0.000395319
7    C    2.172783748    3.735351906   -0.000177234
8    C    2.529675397    4.366361394    1.212705073
9    C    3.235186050    5.580017589    1.223536768
10   C    3.602229262    6.193256971   -0.002501173
11   C    3.249873775    5.569423318   -1.230517041
12   C    2.545069237    4.357639281   -1.215964774
13   O    4.284613544    7.365298913   -0.074486581
14   H   -0.547997709   -0.942433619   -0.000131548
15   H    1.950262128   -0.944972288   -0.000204949
16   H    3.204316388    1.192958763    0.000179033
17   H   -0.547049170    3.374957722   -0.001150312
18   H   -1.785645010    1.227859616   -0.000688179
19   H    2.257080975    3.911557421    2.167104646
20   H    3.491526029    6.034868334    2.181141807
21   H    2.282603283    3.893145023   -2.168623752
22   H    3.521943412    6.023412657   -2.183854049
23   H    4.487329433    7.726527247    0.778121914
---
H16C12O1, RHF, CHARGE=0, MULT=1
HF=-38.4
1    C   -0.028002624    0.030753818    0.035950855
2    C    1.259523598    0.304382700   -0.482796506
3    C    1.909495755    1.517601164   -0.208959832
4    C    1.272219737    2.479203462    0.591162930
5    C   -0.010313814    2.223081533    1.100012149
6    C   -0.687400432    1.007982750    0.831437383
7    C   -0.648644643   -1.307763422   -0.303210134
8    C   -0.219974145   -2.493976978    0.578236631
9    O   -1.387408598   -1.421671894   -1.273792456
10   H    1.766039028   -0.430193495   -1.112111527
11   C   -2.074329486    0.802441534    1.386663629
12   H    1.769432522    3.423739219    0.814568170
13   H    2.901870574    1.710695428   -0.616941507
14   C   -1.147722302   -3.736420781    0.647508557
15   H   -0.035374237   -2.118005951    1.609753218
16   H    0.774139485   -2.811326119    0.181866492
17   C   -2.466401966   -3.478935644    1.404238952
18   H   -1.424816866   -4.018939906   -0.399788662
19   H   -2.292582874   -3.162910621    2.453025750
20   H   -3.073756379   -2.694041313    0.911527611
```

```
21    H    -2.813313580    0.645000773    0.573954178
22    H    -2.109510916   -0.074806266    2.065638897
23    H    -2.424097675    1.675358623    1.974725213
24    H    -3.090936141   -4.395611374    1.431491603
25    H     0.502280470   -5.197583546    0.622403016
26    H    -0.058311980   -4.781468942    2.271337806
27    H    -0.486885700    2.989753090    1.714371724
28    C    -0.393972919   -4.953516486    1.228730733
29    H    -1.039310015   -5.855790319    1.228288964
---
H18C12O1, RHF, CHARGE=0, MULT=1
HF=-60.7
1     C     0.016058417   -0.075073726    0.133163072
2     C     0.830203705    0.058090218   -1.173307690
3     C     1.462585387    1.442167169   -1.330884872
4     C     1.184639904    2.306544194   -2.447319181
5     C     1.784814155    3.602132107   -2.569804943
6     C     1.579170586    4.598627298   -3.712640856
7     C     0.174264366    5.239327299   -3.732900503
8     C     2.042620421    4.081243445   -5.091683473
9     C     1.867312418   -1.072273725   -1.354236012
10    C     2.368180080    1.909867877   -0.354707876
11    C     2.961601457    3.176509271   -0.450765352
12    C     2.667607297    4.005409118   -1.538444912
13    O     0.365689535    1.787874550   -3.403173417
14    H    -0.686033814    0.775344093    0.251160306
15    H     0.652017111   -0.114117790    1.039594188
16    H    -0.592186760   -1.003117896    0.118145448
17    H     0.077873910   -0.105999590   -1.985665923
18    H     2.268263681    5.462850570   -3.505324986
19    H    -0.605873483    4.580745939   -4.163013633
20    H    -0.151196451    5.519804315   -2.710490238
21    H     0.184332299    6.167054127   -4.342225846
22    H     1.318754555    3.395121894   -5.572580480
23    H     3.011109352    3.547525387   -5.011606998
24    H     2.192037896    4.932808520   -5.788216460
25    H     2.621927982   -1.097322070   -0.542907166
26    H     2.409925864   -0.960723605   -2.314734820
27    H     1.367335086   -2.062686361   -1.372048797
28    H     2.628763443    1.290785861    0.506271332
29    H     3.652965469    3.514037952    0.321932362
30    H     3.148765504    4.985713690   -1.575352411
31    H    -0.167160103    2.432925811   -3.843935652
---
H10C13O1, RHF, CHARGE=0, MULT=1
HF=11.9
1     C     0.000648693   -0.001096082   -0.000599871
2     C     1.406427881    0.000495097    0.000276530
3     C     2.112166900    1.216090820    0.000196668
4     C     1.405826964    2.431292542    0.002135695
5     C    -0.000260518    2.432091190   -0.000799424
6     C    -0.718723587    1.215542109   -0.008242987
7     C    -2.226590798    1.211864008   -0.011679515
8     O    -2.867070649    1.036470650    1.017436759
9     C    -2.911740358    1.433931604   -1.336507839
10    C    -3.430118533    2.705227507   -1.671795201
11    C    -3.056277905    0.373869679   -2.259400080
12    C    -3.693706522    0.583806463   -3.494926857
13    C    -4.199375734    1.853714425   -3.822197150
```

```
14    C    -4.066867844    2.913101728   -2.907920355
15    H    -0.527259826   -0.956187268    0.008587595
16    H     1.949221981   -0.945253777    0.003980711
17    H     3.202400134    1.216381423    0.001170525
18    H     1.948191373    3.377366216    0.007673017
19    H    -0.527501689    3.387260434    0.008206878
20    H    -3.345644622    3.539618474   -0.973699747
21    H    -2.679718256   -0.622368909   -2.022453501
22    H    -3.796810494   -0.242273351   -4.199337127
23    H    -4.693986351    2.016035056   -4.780302577
24    H    -4.460968551    3.899388086   -3.155147743
---
H10C14O1, RHF, CHARGE=0, MULT=1
HF=5.6
1     C    -0.000036798    0.000160081   -0.000073171
2     C     1.504139605   -0.000353523    0.000032186
3     C     2.226113030    1.222674156   -0.000049000
4     C     1.529757422    2.562836863   -0.000188583
5     C     0.019562005    2.538749188    0.001706887
6     C    -0.715121231    1.323295231    0.001805075
7     C    -2.131799357    1.381544183    0.003681281
8     C     2.227369424   -1.219894360    0.000142909
9     C     3.641362041    1.193466205   -0.000019188
10    C    -0.677231929    3.770860717    0.003390131
11    C     3.631347994   -1.232961041    0.000153713
12    C     4.342036783   -0.022106879    0.000061722
13    C    -2.809000123    2.611608547    0.005331931
14    C    -2.079668982    3.811374786    0.005163980
15    O    -0.630165817   -1.056613942   -0.001695366
16    H     1.871239209    3.136777496   -0.896425315
17    H     1.873498805    3.138007292    0.894396937
18    H    -2.730032815    0.467889121    0.003916867
19    H     1.706503356   -2.179733029    0.000198868
20    H     4.208849780    2.126037631   -0.000033383
21    H    -0.124921888    4.712531955    0.003362482
22    H     4.166802041   -2.182856009    0.000219305
23    H     5.432660054   -0.024559706    0.000054995
24    H    -3.899223980    2.633092311    0.006746515
25    H    -2.599163673    4.770440981    0.006440262
---
H7C9N1O1, RHF, CHARGE=0, MULT=1
HF=-6.1
1     C     0.000014490    0.000347076   -0.000011088
2     C     1.395063331    0.000220271   -0.000021856
3     C     2.115671893    1.217957704   -0.000012667
4     C     1.433457327    2.435339663   -0.000065180
5     C     0.012605551    2.477696799   -0.000015235
6     C    -0.705536992    1.240586643    0.000046097
7     N    -2.103751618    1.270207732    0.000184798
8     C    -2.880017283    2.463302915    0.000349923
9     C    -2.104488772    3.722645985    0.000309563
10    C    -0.745966201    3.723509826    0.000063548
11    H    -0.537349688   -0.948563098   -0.000030942
12    H     1.938813391   -0.945982390   -0.000049828
13    H     3.205548035    1.199046882   -0.000005793
14    H     2.000999657    3.367716665   -0.000143301
15    O    -4.103252150    2.345307416    0.000474825
16    H    -2.678563631    4.650054961    0.000481730
17    H    -0.194080402    4.666075227   -0.000048081
```

```
18      H     -2.583941475     0.389546369      0.000186858
---
H7C9N1O1, RHF, CHARGE=0, MULT=1
HF=33.3
1       C     -0.000228130    -0.000044542      0.000363622
2       C      1.385413617     0.000301524     -0.000424634
3       C      1.727279818     1.415338243     -0.000041668
4       N      0.571153359     2.122572858      0.001580170
5       O     -0.435636578     1.301924257      0.001568118
6       C      3.053085789     2.051444777     -0.001772235
7       C      3.799559130     2.156755123      1.195731834
8       C      5.072929091     2.751959709      1.192397799
9       C      5.618195509     3.250104973     -0.003744896
10      C      4.883925776     3.152319798     -1.198278491
11      C      3.609566471     2.558449450     -1.199724931
12      H     -0.748992412    -0.783022654     -0.000190991
13      H      2.064119104    -0.836341005     -0.002320890
14      H      3.391521828     1.779035256      2.134488341
15      H      5.637704619     2.827976302      2.122148888
16      H      6.606218156     3.711328800     -0.004495824
17      H      5.301868903     3.539325687     -2.128235873
18      H      3.054261352     2.493578782     -2.136619523
---
H7C9N1O1, RHF, CHARGE=0, MULT=1
HF=5
1       C     -0.000306492    -0.000030679     -0.000108278
2       C      1.381718414     0.000760431     -0.000040819
3       C      2.104456597     1.233168602      0.000006175
4       C      1.434655623     2.443755759      0.000015863
5       C     -0.002224830     2.491786242      0.000002834
6       C     -0.721735086     1.248615148     -0.000073374
7       N     -2.103617411     1.183048133     -0.000127889
8       C     -2.794671188     2.315909999     -0.000178138
9       C     -2.192862038     3.612609294     -0.000111192
10      C     -0.798607604     3.708474880     -0.000004163
11      H     -0.553354422    -0.940478722     -0.000119106
12      H      1.937639711    -0.937903779      0.000011283
13      H      3.194723824     1.205071630      0.000003326
14      H      2.011349109     3.369984906      0.000029906
15      H     -3.887306428     2.223061453     -0.000287427
16      H     -2.840830049     4.487948484     -0.000150231
17      O     -0.140119230     4.890723397      0.000079486
18      H     -0.719934786     5.641443217      0.000053927
---
H7C9N1O1, RHF, CHARGE=0, MULT=1
HF=38.3
1       C      0.000200084    -0.000058870      0.000347980
2       C      1.393557713     0.000080296      0.000302070
3       C      1.736500830     1.405353694     -0.000074327
4       N      0.591740532     2.121967958     -0.000724745
5       O     -0.423133910     1.312491262     -0.000962424
6       C     -1.012314089    -1.068217226      0.002666068
7       C     -1.801281600    -1.320740728      1.150030268
8       C     -2.750074192    -2.358219033      1.147739737
9       C     -2.925044481    -3.154743922      0.002982847
10      C     -2.147139545    -2.910900528     -1.142642659
11      C     -1.196990851    -1.875273725     -1.145370708
12      H      2.715405549     1.867582508      0.000589329
13      H      2.071157314    -0.837567637      0.003303964
```

```
14      H     -1.679656555    -0.713795001     2.048583789
15      H     -3.351956177    -2.543157012     2.038095270
16      H     -3.662014273    -3.958420502     0.002470551
17      H     -2.281152095    -3.525339720    -2.033270194
18      H     -0.605702307    -1.697878996    -2.045112199
---
H7C9N1O1, RHF, CHARGE=0, MULT=1
HF=1.6
1       C     -0.000056718    -0.000018735    -0.000029317
2       C      1.401761471     0.000280302     0.000000975
3       C      2.117552463     1.230596531     0.000002934
4       C      1.457419047     2.447531699    -0.000042690
5       C      0.025274642     2.497026153    -0.000079782
6       C     -0.713775025     1.268760723    -0.000083910
7       N     -2.096841942     1.247535967    -0.000121914
8       C     -2.758460967     2.396505405    -0.000111827
9       C     -2.106135781     3.677184249    -0.000076334
10      C     -0.728057378     3.724788270    -0.000081760
11      O     -0.619014341    -1.203198197    -0.000011007
12      H      1.966220735    -0.932932968     0.000022262
13      H      3.208760903     1.197907548     0.000042376
14      H      2.025496638     3.378380724    -0.000049589
15      H     -3.852625731     2.334949086    -0.000124879
16      H     -2.710666814     4.583203600    -0.000045812
17      H     -0.202762232     4.680705055    -0.000073885
18      H     -1.565795683    -1.155552149    -0.000066733
---
H7C9N1O1, RHF, CHARGE=0, MULT=1
HF=16.8
1       C      0.000096587     0.000051216    -0.000045426
2       C      1.406301613    -0.000047023    -0.000058003
3       C      2.113998131     1.214483367     0.000013891
4       C      1.410575545     2.431947620    -0.001706381
5       C      0.004968736     2.435861129    -0.000906816
6       C     -0.714253428     1.219085260     0.004738011
7       C     -2.218533663     1.224661591    -0.001486673
8       O     -2.857875673     1.206762384    -1.044473295
9       C     -2.931018043     1.276251189     1.367879524
10      C     -3.374080811    -0.035707201     1.810987579
11      N     -3.735225032    -1.075801899     2.181849054
12      H     -0.529974120    -0.953778029    -0.008097115
13      H      1.947501360    -0.946935289    -0.002589710
14      H      3.204513099     1.212701623    -0.000946645
15      H      1.955179883     3.376677442    -0.006044798
16      H     -0.523484329     3.390629373    -0.008887276
17      H     -2.249137866     1.708520837     2.136424975
18      H     -3.810176039     1.961164440     1.309630272
---
H11C9N1O1, RHF, CHARGE=0, MULT=1
HF=25.7
1       C      0.000000968     0.000269641     0.000151974
2       C      1.409909071    -0.000089622     0.000288419
3       C      2.162317206     1.197064784     0.000028480
4       C      1.454456947     2.427092224    -0.008713922
5       C      0.034927195     2.472586711     0.001652697
6       C     -0.665311966     1.245650708     0.003345539
7       N      2.211025403     3.687999920     0.121578522
8       O      2.538986410     4.243179330    -0.844896019
9       H     -1.409404535    -1.361728716    -0.920454489
```

```
10      H       -1.464293793    -1.337159309     0.874296552
11      H       -0.135950539    -2.180726500     0.029226487
12      C       -0.787198886    -1.283941529    -0.004060303
13      H        1.938327824    -0.956548013     0.000599055
14      C        3.669424514     1.118506490     0.008488675
15      C       -0.743334751     3.765502490     0.010424444
16      H       -1.758114048     1.257403058     0.007259273
17      H        4.102951133     1.680924597     0.861602140
18      H        4.030916790     0.073712770     0.100894257
19      H        4.097724469     1.523899903    -0.932177243
20      H       -0.589490652     4.332907511    -0.931426353
21      H       -0.448844459     4.413786560     0.861825547
22      H       -1.835269333     3.592877399     0.105872721
---
H11C9N1O1, RHF, CHARGE=0, MULT=1
HF=-20.6
1       C       -0.083351491    -0.061151327     0.041916895
2       C        0.918375900    -0.031612439     1.027723548
3       C        1.688210850     1.129207042     1.219771872
4       C        1.467172656     2.274424981     0.420857654
5       C        0.450259432     2.239890817    -0.560729547
6       C       -0.316709285     1.076613872    -0.749919616
7       C        2.243442828     3.539687149     0.670988787
8       O        1.847252707     4.385695626     1.470342793
9       N        3.453691831     3.753059833    -0.032926221
10      C        3.956573276     2.778889372    -1.004974912
11      C        4.283947941     4.947093089     0.169901993
12      H       -0.679654418    -0.962140249    -0.105517434
13      H        1.098136181    -0.909484739     1.649136502
14      H        2.454246382     1.135150520     1.997089954
15      H        0.246320207     3.116075937    -1.178506951
16      H       -1.096962991     1.060523618    -1.511498750
17      H        4.157465445     1.791240315    -0.528695886
18      H        3.234660655     2.627805363    -1.840726998
19      H        4.913398468     3.121783386    -1.459180043
20      H        4.437443259     5.481279795    -0.798052514
21      H        3.840035805     5.675730548     0.880100006
22      H        5.285458826     4.658728488     0.570324826
---
H13C9N1O1, RHF, CHARGE=0, MULT=1
HF=5.2
1       C       -0.081101819     0.140662011     0.154051092
2       N        1.364991176     0.005988507    -0.035493756
3       C        2.195182639     1.183873261     0.234593396
4       C        1.996058178    -1.251144825    -0.039947249
5       C        1.397856773    -2.468253719     0.094817695
6       C        2.121022296    -3.732057711    -0.001043869
7       C        1.549169612    -4.961461747     0.081414602
8       C        2.272850690    -6.228223029    -0.014906600
9       C        1.694461064    -7.453046247    -0.003811944
10      C        2.442390256    -8.732241452    -0.103469143
11      O        1.908749959    -9.833178181    -0.143725693
12      H       -0.631436364    -0.487694346    -0.582463712
13      H       -0.401863818    -0.153855932     1.181673728
14      H       -0.401547476     1.194122547    -0.008711053
15      H        2.352561341     1.340439827     1.328751807
16      H        3.194810811     1.087564226    -0.245638350
17      H        1.717674051     2.100680708    -0.179167735
18      H        3.082000942    -1.194785028    -0.208480625
```

```
19    H     0.325390479   -2.560191459    0.282948775
20    H     3.204465342   -3.652733571   -0.151898270
21    H     0.467533318   -5.047359267    0.228931106
22    H     3.363259085   -6.145391880   -0.101958784
23    H     0.607799627   -7.552869144    0.080535215
24    H     3.550196727   -8.642392054   -0.140731764
---
H17C9N1O1, RHF, CHARGE=0, MULT=1
HF=-65.4
1     C    -0.000157369   -0.000072158   -0.000272789
2     C     1.528771796    0.000171947    0.000059799
3     C     2.152692757    1.435914393   -0.000072803
4     N     1.420647814    2.388990181   -0.875438180
5     C    -0.055130582    2.362858274   -1.054146138
6     C    -0.650930367    0.915189335   -1.037823689
7     C    -0.381355210    3.024085897   -2.435002212
8     C    -0.708906585    3.239739555    0.059444200
9     C     3.626031841    1.342743596   -0.520983758
10    C     2.196201530    2.017884020    1.447674573
11    O    -0.656068150   -0.714861689    0.747365337
12    H     1.848950706   -0.565659144   -0.906453010
13    H     1.921996089   -0.578414089    0.865112790
14    H     1.844478082    2.356715022   -1.791653009
15    H    -1.751370833    0.965580090   -0.882640881
16    H    -0.512504699    0.423830637   -2.029756165
17    H     0.063025046    2.466271992   -3.285333121
18    H    -1.476633550    3.055120931   -2.608096521
19    H    -0.008432183    4.066836936   -2.491636282
20    H    -1.771566599    3.452079730   -0.183205199
21    H    -0.707606844    2.746989131    1.050881461
22    H    -0.195018475    4.215775493    0.168405239
23    H     3.682926777    0.949321303   -1.557201516
24    H     4.129893918    2.330507276   -0.513474331
25    H     4.226522435    0.660739201    0.115313111
26    H     2.491571237    3.086184774    1.452195004
27    H     1.225463680    1.935016512    1.973738234
28    H     2.935923006    1.468792656    2.067769317
---
H9C10N1O1, RHF, CHARGE=0, MULT=1
HF=-5.6
1     C     0.000162940    0.000553685   -0.000039390
2     C     1.383369940   -0.000013614   -0.000001861
3     C     2.112142193    1.229171404   -0.000020856
4     C     1.444883213    2.439144167   -0.000093631
5     C     0.002077349    2.466486860   -0.000200441
6     C    -0.735718347    1.235575783   -0.000158990
7     C    -2.183482614    1.351416215   -0.000273393
8     C    -2.771107710    2.618905606   -0.000393886
9     C    -1.935496043    3.787754914   -0.000299663
10    N    -0.605265280    3.706801405   -0.000314970
11    C    -2.537302323    5.174058246   -0.000390552
12    O    -2.904740780    0.205692633   -0.000241095
13    H    -0.532539492   -0.951576814    0.000000735
14    H     1.935789309   -0.940401151    0.000037869
15    H     3.202569067    1.196253535    0.000053465
16    H     2.000581238    3.378004524   -0.000071469
17    H    -3.853632349    2.745885020   -0.000492636
18    H    -3.173697921    5.316636392   -0.899204521
19    H    -3.174545162    5.316657337    0.897801163
```

```
20      H     -1.773806108     5.977338766      0.000022565
21      H     -3.841692939     0.353431547     -0.000330141
---
H9C10N1O1, RHF, CHARGE=0, MULT=1
HF=-9.4
1       C     -0.000089601     0.000248685     -0.000012679
2       C      1.401593010    -0.000185039     -0.000068962
3       C      2.118607307     1.229543217      0.000005053
4       C      1.459299386     2.446866590      0.000155876
5       C      0.027263616     2.495903662      0.000315999
6       C     -0.713556919     1.270067495      0.000264797
7       N     -2.094247646     1.247082637      0.000609548
8       C     -2.777950270     2.389513027      0.000998054
9       C     -2.108097724     3.669411529      0.000989445
10      C     -0.730832682     3.718853744      0.000614660
11      O     -0.619422409    -1.202802052     -0.000219389
12      C     -4.284940780     2.293904461      0.001511461
13      H      1.965530748    -0.933711226     -0.000167579
14      H      3.209564739     1.195923584     -0.000062227
15      H      2.027104050     3.377835739      0.000185001
16      H     -2.698530709     4.585780346      0.001287945
17      H     -0.209485444     4.677064061      0.000570445
18      H     -1.566149323    -1.155045488      0.000006511
19      H     -4.702136864     2.797106718     -0.896171056
20      H     -4.701679424     2.794704499      0.900857928
21      H     -4.652918927     1.248488566      0.000258888
---
H9C10N1O1, RHF, CHARGE=0, MULT=1
HF=24.5
1       C     -0.000096684    -0.000217340      0.000173709
2       C      1.391604215     0.000103590     -0.000299480
3       C      1.744134795     1.409622345     -0.000011961
4       N      0.590902841     2.121056540      0.001019969
5       O     -0.421853011     1.312293803      0.001483562
6       C     -1.012680028    -1.068289168     -0.001490880
7       C     -1.825011156    -1.300810724     -1.136652874
8       C     -2.773829420    -2.338467707     -1.132967436
9       C     -2.925856265    -3.155233392      0.000699988
10      C     -2.124576458    -2.931349465      1.134073874
11      C     -1.174483136    -1.895786608      1.135501308
12      C      3.096528258     2.037984973     -0.000886251
13      H      2.065358773    -0.841092336     -0.003871331
14      H     -1.721905294    -0.678773444     -2.027123032
15      H     -3.393655838    -2.507736930     -2.014080779
16      H     -3.662892156    -3.958922069      0.001765811
17      H     -2.240245555    -3.561315740      2.016627601
18      H     -0.565077196    -1.734288616      2.026058783
19      H      3.668235866     1.721378113      0.897012383
20      H      3.667446054     1.720708785     -0.899044666
21      H      3.052393578     3.145849924     -0.001277788
---
H9C10N1O1, RHF, CHARGE=0, MULT=1
HF=-14.6
1       C     -0.000098771     0.000165148     -0.000014204
2       C      1.382895761    -0.000082951     -0.000005152
3       C      2.109649843     1.229461551      0.000002744
4       C      1.436236046     2.435604736      0.000021320
5       C     -0.006481751     2.455762633      0.000008067
6       C     -0.751644744     1.227192704      0.000022712
```

```
7    C    -2.201674648    1.304976191   -0.000033256
8    C    -2.792193890    2.559676316   -0.000050721
9    C    -1.950527375    3.734068539   -0.000045732
10   N    -0.613208040    3.694902707   -0.000034257
11   O    -2.532343102    4.951425071   -0.000119776
12   C    -3.066128517    0.068708143   -0.000352744
13   H    -0.522501966   -0.957500582   -0.000056423
14   H     1.935803775   -0.940234237   -0.000000747
15   H     3.200022099    1.200271824   -0.000002504
16   H     1.987828158    3.376924700    0.000056902
17   H    -3.874941292    2.688379220   -0.000138505
18   H    -1.915562620    5.674737189   -0.000122315
19   H    -4.148006838    0.314054689   -0.000082065
20   H    -2.875337518   -0.550759004   -0.901356090
21   H    -2.875148899   -0.551681830    0.899940544
---
H9C10N1O1, RHF, CHARGE=0, MULT=1
HF=23.6
1    C     0.088023164   -0.034556494   -0.311764330
2    C     1.382364679    0.110912206    0.332781970
3    C     1.852752766    1.336472124   -0.128319767
4    O     0.901951782    1.858085032   -0.978796794
5    N    -0.115931826    1.061730873   -1.081783938
6    C    -0.889471723   -1.126914026   -0.198773033
7    C    -2.070012256   -0.962059883    0.563902703
8    C    -2.994629018   -2.014334068    0.683159091
9    C    -2.755884331   -3.242117981    0.042082629
10   C    -1.587347461   -3.415608288   -0.719945549
11   C    -0.659391703   -2.366794427   -0.840899416
12   C     3.096198860    2.120671972    0.136138421
13   H     1.851993866   -0.582393884    1.011236742
14   H    -2.274227942   -0.015908117    1.067138285
15   H    -3.899817700   -1.874810555    1.274983532
16   H    -3.474924313   -4.056485866    0.134427101
17   H    -1.399420109   -4.365938638   -1.220828201
18   H     0.240258630   -2.518786327   -1.439450378
19   H     3.916576890    1.441936176    0.447088215
20   H     3.438717160    2.680630998   -0.757811975
21   H     2.925657110    2.850611214    0.955626641
---
H9C10N1O1, RHF, CHARGE=0, MULT=1
HF=7.2
1    C    -0.000147718    0.000049749    0.000214186
2    C     1.405735348    0.000019005   -0.000154803
3    C     2.112891355    1.215133269    0.000011915
4    C     1.408668743    2.431820855   -0.001178591
5    C     0.002696679    2.435098751   -0.000672912
6    C    -0.715825728    1.218635380    0.004494633
7    C    -2.222099448    1.218131308    0.000385842
8    C    -2.926353000    1.147191860    1.366805396
9    C    -3.910384953    2.304947865    1.636481836
10   C    -4.495550965    2.247647781    2.970424210
11   N    -4.972928245    2.218647662    4.029358319
12   O    -2.851162378    1.244370516   -1.050523000
13   H    -0.531213618   -0.953202675   -0.006006880
14   H     1.947424184   -0.946369585   -0.002434897
15   H     3.203223543    1.213758826   -0.000828231
16   H     1.952650138    3.376980585   -0.004801916
17   H    -0.525453988    3.389945241   -0.008062599
```

```
18   H    -3.467185746    0.174175025    1.421460061
19   H    -2.160801494    1.123515161    2.173864307
20   H    -3.389387653    3.285223941    1.523325365
21   H    -4.738761379    2.293830101    0.889350997
---
H11C10N1O1, RHF, CHARGE=0, MULT=1
HF=32.7
1    C     0.000328639    0.000053788    0.000127572
2    C     1.411281265   -0.000406951   -0.000109237
3    C     2.162977417    1.194791345    0.000086188
4    C     1.462988002    2.419257500    0.004086267
5    C     0.052033533    2.476067010    0.004664434
6    C    -0.683002082    1.253401893    0.001889689
7    N    -3.273949531    1.307791246    0.000024117
8    O    -4.463629404    1.332733370   -0.000619285
9    C    -0.743600444   -1.312625310   -0.001465867
10   H     1.942456088   -0.955304272   -0.000189244
11   C     3.668536790    1.151936232    0.009460143
12   H     2.030841863    3.352763439    0.007150783
13   C    -0.636278804    3.818818434    0.008226954
14   H    -1.387013508   -1.409367953    0.898030799
15   H    -1.386761931   -1.407201149   -0.901477268
16   H    -0.057701667   -2.184290767   -0.002409487
17   H     0.085368490    4.661028934    0.012085725
18   H    -1.276151239    3.938207955    0.907567595
19   H    -1.273713006    3.944424793   -0.892045747
20   C    -2.102252598    1.282934063    0.001107046
21   H     4.041861216    0.858613456    1.013605246
22   H     4.056649858    0.417628108   -0.726718156
23   H     4.120019058    2.133324212   -0.240689274
---
H13C10N1O1, RHF, CHARGE=0, MULT=1
HF=-11.1
1    C     0.039667872    0.005572052   -0.125215892
2    C     1.272005803    0.633960288    0.165986904
3    C     1.626450424    1.837788236   -0.466489178
4    C     0.753592902    2.431765893   -1.394918990
5    C    -0.477113089    1.818845101   -1.684774189
6    C    -0.835382750    0.614849599   -1.052137330
7    C    -0.339689151   -1.284214528    0.554284306
8    O    -1.074087646   -1.288055594    1.534824591
9    N    -0.512561533   -3.021963467   -1.187103586
10   C     0.247900224   -3.170404207   -2.432014874
11   C    -1.565675208   -4.019090206   -0.963218969
12   C     0.251130103   -2.602380770   -0.005320228
13   H     1.960191402    0.193702173    0.889237909
14   H     2.580327962    2.312382950   -0.233926814
15   H    -1.159741647    2.278014165   -2.400597556
16   H    -1.798534040    0.158681352   -1.286671061
17   H     0.826498022   -2.243164883   -2.642830707
18   H     0.967319797   -4.024759949   -2.412729489
19   H    -0.445213958   -3.330919763   -3.288044313
20   H    -1.167607810   -5.045956523   -0.775290851
21   H    -2.196574724   -3.734433894   -0.092451126
22   H    -2.231742223   -4.071268389   -1.854365024
23   H     1.030038024    3.365260780   -1.885827717
24   H     0.255949838   -3.386647004    0.792913274
25   H     1.327872857   -2.433205393   -0.250674558
---
```

```
H13C10N1O1, RHF, CHARGE=0, MULT=1
HF=-26.9
1    C     0.034082966    0.025973069    0.001715971
2    C     1.114042020    0.398419278    0.833075378
3    C     1.742916816    1.643824415    0.672165390
4    C     1.311434228    2.561980648   -0.314465119
5    C     0.222317825    2.186659306   -1.133060145
6    C    -0.409614463    0.940783459   -0.977911968
7    C    -0.684372262   -1.279689299    0.213015705
8    O    -1.642262508   -1.366205006    0.978814042
9    N    -0.242546718   -2.427148447   -0.489842281
10   C     0.886362363   -2.368539044   -1.422452420
11   C    -0.869467140   -3.744606818   -0.321790592
12   C     1.995370805    3.893727792   -0.471505346
13   H     1.469421571   -0.275491800    1.614763891
14   H     2.576886993    1.897577734    1.329343108
15   H    -0.149617599    2.866302423   -1.901935798
16   H    -1.253319289    0.693241449   -1.624726390
17   H     0.704452382   -1.631917018   -2.238810664
18   H     1.832386720   -2.091696834   -0.901580654
19   H     1.057453576   -3.355513267   -1.908337974
20   H    -1.716170124   -3.739904627    0.395604863
21   H    -1.264679311   -4.112554831   -1.298637902
22   H    -0.121699863   -4.484221238    0.052333026
23   H     1.567519738    4.494601390   -1.299009050
24   H     1.902880781    4.496148855    0.456668268
25   H     3.076679265    3.760701019   -0.686170992
---
H13C11N1O1, RHF, CHARGE=0, MULT=1
HF=-12.8
1    C     0.000366168    0.000068236   -0.000125222
2    C     1.405950771   -0.000460400    0.000071103
3    C     2.113626464    1.214441234   -0.000006756
4    C     1.422239334    2.446234395   -0.000889823
5    C     0.009382579    2.437020457    0.002621989
6    C    -0.696234344    1.220772574    0.001219699
7    C     2.183450109    3.750979746    0.001035546
8    C     2.626886094    4.246667968   -1.333626429
9    C     3.848583533    4.748474626   -1.679068397
10   N     4.097461946    5.304968303   -2.956626373
11   C     3.244998674    5.109668795   -4.125806438
12   C     5.023859083    4.849950282   -0.723991545
13   O     2.376444080    4.358411223    1.049935970
14   H    -0.547246271   -0.942777067    0.000444801
15   H     1.950793130   -0.945029332    0.001982745
16   H     3.204422147    1.192678097    0.004354111
17   H    -0.549113642    3.374442691    0.008717647
18   H    -1.786739759    1.226749109    0.003707216
19   H     1.821614361    4.170532333   -2.069821007
20   H     5.068938081    5.268209312   -3.215018632
21   H     3.080512068    4.035514313   -4.380908905
22   H     2.254009972    5.592025995   -3.972437627
23   H     3.727664152    5.595375518   -5.004854156
24   H     6.000268244    4.967419057   -1.238935067
25   H     4.898978018    5.724355690   -0.051896197
26   H     5.112095724    3.937493101   -0.099440307
---
H15C11N1O1, RHF, CHARGE=0, MULT=1
HF=-16.7
```

```
 1   C    0.000046915   -0.000056549   -0.000107264
 2   N    1.466961284    0.000058221    0.000339670
 3   C    2.204649125    1.277234957   -0.000021475
 4   C    2.327235745    1.754491669   -1.479872073
 5   C    3.626012985    1.456732057   -2.186270249
 6   C    3.771064295    0.284340691   -2.961001220
 7   C    4.981500433    0.015297429   -3.625117330
 8   C    6.057207818    0.914394482   -3.527793075
 9   C    5.918885671    2.087720747   -2.765551352
10   C    4.710971029    2.359023618   -2.100232354
11   C    2.078674963   -1.093480333    0.763588660
12   C    1.685214398    2.374980570    0.958598851
13   O    1.429505386    2.353984082   -2.056725000
14   H   -0.441385384    0.128605327    1.018317403
15   H   -0.413078264    0.799654732   -0.650723193
16   H   -0.366615429   -0.972731041   -0.403593204
17   H    3.254228157    1.071106855    0.347063370
18   H    2.945736722   -0.423294706   -3.055535435
19   H    5.082561572   -0.893533173   -4.219269365
20   H    6.994608080    0.704581670   -4.043420708
21   H    6.748649055    2.791190418   -2.691384619
22   H    4.620698076    3.278685698   -1.519815742
23   H    3.148852876   -1.217516680    0.481758319
24   H    2.033429061   -0.942295624    1.869536820
25   H    1.565895919   -2.055180449    0.534340009
26   H    2.371955029    3.246752584    0.935988502
27   H    0.672650757    2.746212247    0.708315832
28   H    1.660115784    2.006402783    2.004600881
 ---
H15C11N1O1, RHF, CHARGE=0, MULT=1
HF=7.4
 1   C    0.000163799    0.000123155   -0.000014780
 2   C    1.557139507    0.000190721    0.000234388
 3   C   -0.502938056    1.478541633   -0.000016766
 4   C   -0.579044259   -0.727649786   -1.248602248
 5   N   -0.491818034   -0.742363450    1.272708711
 6   C   -0.167569908   -2.036570352    1.478381432
 7   C   -0.594664343   -2.812760675    2.658516250
 8   C    0.199013183   -2.843573816    3.829851715
 9   C   -0.185804364   -3.623356029    4.934439457
10   C   -1.367592579   -4.382616148    4.887364948
11   C   -2.164521699   -4.359466836    3.729824157
12   C   -1.783021830   -3.580627810    2.623584295
13   O   -1.178038951   -0.114959206    2.072655808
14   H    1.969607093    0.427412816    0.936408148
15   H    1.988404096   -1.013082457   -0.124039828
16   H    1.940431668    0.612885539   -0.841965117
17   H   -1.608620069    1.549838600   -0.003266335
18   H   -0.123393365    2.054892717    0.866964728
19   H   -0.142142321    1.994465648   -0.915269983
20   H   -0.169037162   -1.748683656   -1.381612344
21   H   -1.683313446   -0.813759428   -1.200562270
22   H   -0.329462909   -0.163485023   -2.171007424
23   H    0.444491853   -2.570762739    0.748338142
24   H    1.118515311   -2.259115679    3.887236807
25   H    0.434681666   -3.636091447    5.831135463
26   H   -1.665987159   -4.985259191    5.745709220
27   H   -3.083573703   -4.945008327    3.689168647
28   H   -2.416948152   -3.573212692    1.735662604
```

```
---
H17C11N1O1, RHF, CHARGE=0, MULT=1
HF=-76.2
1    C    0.000200611    0.000225396   -0.000117929
2    C    1.559831680   -0.000415864    0.000082461
3    C    2.081799359    1.464070009   -0.000206766
4    C    1.560506748    2.202975129   -1.263269372
5    C    0.002775486    2.192788880   -1.264165914
6    C   -0.487551661    0.703801053   -1.315808466
7    C   -0.491158261    0.727843672    1.283363496
8    H   -0.359342794   -1.052395074    0.013585538
9    H    1.956030259   -0.556520627   -0.877671780
10   H    1.945694726   -0.552110310    0.885015873
11   H    3.194204291    1.459048495   -0.007742362
12   C    1.575512715    2.200484672    1.271197064
13   H    1.958458253    1.729839740   -2.187662058
14   H    1.939822566    3.247996386   -1.291033995
15   H   -0.351879398    2.745848381   -2.162173754
16   C   -0.513048140    2.918282746    0.010086281
17   C   -1.980871762    0.535760061   -1.662983264
18   H    0.049079779    0.193750819   -2.157978784
19   H   -1.596346633    0.702817554    1.377245924
20   C    0.021438672    2.193803590    1.275575026
21   H   -0.121676072    0.191615777    2.186233471
22   H    1.964889566    3.241584644    1.302487405
23   H    1.972937266    1.715878067    2.189492886
24   O   -2.877546749    0.199717295   -0.889082305
25   H   -1.623201592    2.961051432    0.022408783
26   H   -0.185953301    3.981880954    0.002308532
27   H   -0.346865563    2.720155787    2.183716157
28   N   -2.339075160    0.754060881   -2.994573178
29   H   -1.687060162    1.016910601   -3.694273464
30   H   -3.277817051    0.648731189   -3.307892740
---
H23C11N1O1, RHF, CHARGE=0, MULT=1
HF=-89.4
1    C   -0.000023451   -0.000003335    0.000055473
2    C    1.531209632    0.000047023    0.000030398
3    C    2.174446747    1.400490574    0.000011512
4    C    3.716578157    1.384795529   -0.002606380
5    C    4.358148418    2.787691365    0.000527842
6    C    5.900621828    2.771282319   -0.006687648
7    C    6.540869897    4.175207483    0.000435532
8    C    8.084192662    4.152017589   -0.011723499
9    C    8.702756782    5.558378269   -0.059334960
10   N    8.919531571    6.286238005    1.145280944
11   C    9.254105784    5.604336017    2.397007227
12   C    9.138798433    7.738791373    1.101474456
13   O    8.980475638    6.099808718   -1.128246246
14   H   -0.412892302    0.505778915    0.896318397
15   H   -0.413037960    0.506063296   -0.896007295
16   H   -0.382279751   -1.041732005   -0.000080844
17   H    1.882863659   -0.570148350    0.890944266
18   H    1.882902369   -0.570359864   -0.890677296
19   H    1.813837997    1.964928839   -0.890500425
20   H    1.816743784    1.963446779    0.892680393
21   H    4.076579212    0.818226363    0.886847517
22   H    4.073394901    0.823211957   -0.896402319
23   H    3.996572763    3.355917299   -0.886987212
```

```
24    H       4.004337734     3.347757502     0.896463158
25    H       6.263459401     2.201240486     0.879356432
26    H       6.254928864     2.214332648    -0.904218171
27    H       6.179483413     4.745817372    -0.885326045
28    H       6.188005586     4.732038907     0.897999398
29    H       8.462907501     3.572919269     0.857955452
30    H       8.441187040     3.593988562    -0.909092492
31    H       8.350411520     5.132398041     2.848687336
32    H      10.026974576     4.814616958     2.245876208
33    H       9.663918671     6.314523590     3.149796470
34    H      10.196681510     7.993114160     0.856599312
35    H       8.475803258     8.224179726     0.351434303
36    H       8.895635109     8.194355637     2.089145054
---
H9C12N1O1, RHF, CHARGE=0, MULT=1
HF=22.5
1     C       0.000138567     0.000032401     0.000149811
2     C       1.407863425    -0.000128720    -0.000182782
3     C       2.123475833     1.207018064    -0.000112640
4     C       1.447035735     2.441089487    -0.003437963
5     C       0.033806573     2.450591684    -0.001640938
6     C      -0.701068421     1.223891116     0.003007066
7     N      -2.124261457     1.261011031     0.107492046
8     C      -2.745127364     2.515793152    -0.171007030
9     C      -1.952311795     3.705822029    -0.171487474
10    C      -2.556747897     4.971343430    -0.346838504
11    C      -3.951344864     5.046099973    -0.519932354
12    C      -4.734233488     3.881361941    -0.524140585
13    C      -4.140481527     2.616860695    -0.351833989
14    O      -0.594451347     3.668205471     0.007936044
15    H      -0.531004982    -0.952669523     0.002193169
16    H       1.943711246    -0.950038079     0.001517235
17    H       3.213975801     1.193321935     0.001835284
18    H       2.015251863     3.371483447    -0.006655306
19    H      -2.556583578     0.507429478    -0.407106965
20    H      -1.959716945     5.883506013    -0.349002370
21    H      -4.423866166     6.020110498    -0.652327879
22    H      -5.813984455     3.952353275    -0.660112199
23    H      -4.772730978     1.727881930    -0.356873582
---
H17C12N1O1, RHF, CHARGE=0, MULT=1
HF=-23
1     C       0.000020024     0.000005034     0.000059845
2     C       1.549901062     0.000007135     0.000207040
3     O       2.187503234     1.046248996    -0.000140757
4     C       2.259003323    -1.328672692    -0.051286906
5     C       2.424452994    -2.004557253    -1.281979243
6     C       3.097657812    -3.236990061    -1.335294395
7     C       3.618100381    -3.809072447    -0.161071861
8     C       3.464957234    -3.141592594     1.065973806
9     C       2.792800976    -1.907025552     1.121933146
10    N      -0.522193912     0.035064773     1.374517269
11    C      -0.628232926     1.374506624     1.993585165
12    C      -0.048540447     1.401778914     3.417404142
13    C      -1.651732314    -0.878427590     1.648860705
14    C      -1.190569401    -2.283813424     2.067669566
15    H      -0.370332497     0.855644693    -0.621150666
16    H      -0.349335094    -0.915862613    -0.534631310
17    H       2.035708468    -1.573402796    -2.206096707
```

```
18   H     3.218026573   -3.747883222   -2.291013483
19   H     4.140460985   -4.765268324   -0.203254567
20   H     3.870410558   -3.578761658    1.979044301
21   H     2.691853642   -1.401768402    2.083542319
22   H    -0.086533283    2.137356920    1.383278692
23   H    -1.690992667    1.730402588    2.020775736
24   H    -0.615945292    0.759549466    4.120523023
25   H     1.011379242    1.080269471    3.443035521
26   H    -0.093616060    2.439676720    3.809571101
27   H    -2.288279185   -0.469421336    2.472502104
28   H    -2.338032132   -0.963374889    0.766646784
29   H    -0.683974021   -2.827597850    1.245795344
30   H    -0.502279173   -2.256611835    2.935489188
31   H    -2.076474220   -2.885443654    2.361993289
---
H9C13N1O1, RHF, CHARGE=0, MULT=1
HF=8.3
1    C    -0.000011792    0.000272572    0.000002997
2    C     1.496907281   -0.000028424    0.000030026
3    C     2.211162326    1.234161395   -0.000123415
4    N     1.503512924    2.441336054    0.003046529
5    C     0.107115119    2.530156775   -0.010242283
6    C    -0.673578350    1.336915022   -0.010430729
7    C    -2.086783245    1.471216244   -0.020587263
8    C    -2.698009306    2.731260590   -0.029786704
9    C    -1.910982682    3.900726782   -0.029038131
10   C    -0.516396141    3.810925938   -0.019511212
11   C     2.252502086   -1.201880328    0.000346167
12   C     3.652824550   -1.181050908    0.000761178
13   C     4.342915577    0.048068413    0.000912106
14   C     3.635450102    1.253317632    0.000391221
15   O    -0.645541653   -1.047804183    0.008379682
16   H     2.028129005    3.293006857   -0.009200563
17   H    -2.725200367    0.584985335   -0.021456951
18   H    -3.785227300    2.809801910   -0.037514740
19   H    -2.392817901    4.879665056   -0.036069669
20   H     0.077835148    4.725232232   -0.019111267
21   H     1.748010521   -2.170509053    0.000217847
22   H     4.212175508   -2.116669096    0.001016683
23   H     5.433992516    0.058277669    0.001364783
24   H     4.184784340    2.195230368    0.000457789
---
H11C13N1O1, RHF, CHARGE=0, MULT=1
HF=62.9
1    C     0.000332515   -0.000136248   -0.000046197
2    C     1.406125954   -0.001324116   -0.000013684
3    C     2.115042060    1.212552141    0.000363764
4    C     1.427024579    2.449732285   -0.006254926
5    C     0.011256186    2.436651283    0.004672582
6    C    -0.695025185    1.221382050    0.004374569
7    C     2.166202671    3.723068014    0.067352708
8    N     2.584067686    4.438044308   -0.995792359
9    C     2.327047581    3.984775372   -2.381254372
10   C     3.284756633    3.181351707   -3.037950661
11   C     3.047983925    2.784572362   -4.366610834
12   C     1.878839894    3.191613075   -5.033672076
13   C     0.939056865    4.004175358   -4.374600504
14   C     1.157273635    4.412448659   -3.046210327
15   O     3.197568949    5.497812220   -0.871957252
```

```
16      H       -0.547864044    -0.942508919    -0.000269334
17      H       1.950006128     -0.946522939    0.002676079
18      H       3.206063069     1.189598935     0.011881756
19      H       -0.547034799    3.374142817     0.019374416
20      H       -1.785549207    1.228613968     0.010490903
21      H       2.386482692     4.109316352     1.069159736
22      H       4.203741658     2.867734524     -2.541111409
23      H       3.779871717     2.160988281     -4.881288590
24      H       1.702921944     2.880735692     -6.064014990
25      H       0.036446654     4.325287703     -4.895638995
26      H       0.424503216     5.054951147     -2.556171186
---
H19C13N1O1, RHF, CHARGE=0, MULT=1
HF=-29.7
1       C       0.000004698     -0.000004061    0.000029205
2       C       1.538737920     0.000028693     -0.000016522
3       N       2.102797338     1.365383094     -0.000010320
4       C       3.207866135     1.687350343     0.925777374
5       C       2.626354675     1.895943969     2.358333823
6       C       2.461009364     3.318539185     2.832761982
7       C       1.204107748     3.959447707     2.759104278
8       C       1.049642792     5.281062050     3.215451508
9       C       2.145158967     5.975195620     3.756184073
10      C       3.397860927     5.342559134     3.841730318
11      C       3.555261298     4.022635954     3.385454308
12      O       2.331500346     0.963561609     3.094913565
13      C       4.427862256     0.733033922     0.905298538
14      C       2.177897643     2.016267314     -1.325913047
15      C       1.451197768     3.371244612     -1.365446028
16      H       -0.419491260    0.584196031     0.842640465
17      H       -0.427268289    0.402784666     -0.939981939
18      H       -0.362706300    -1.044514045    0.103354631
19      H       1.875926287     -0.574998738    0.894684561
20      H       1.904230055     -0.592551277    -0.879536293
21      H       3.622784656     2.686613735     0.619661210
22      H       0.338550519     3.435715553     2.350479297
23      H       0.074663940     5.765427545     3.151157606
24      H       2.024212824     6.999211110     4.110636996
25      H       4.249900702     5.875070042     4.265403614
26      H       4.534038900     3.547393861     3.469322497
27      H       5.237895056     1.147878742     1.540728362
28      H       4.207854501     -0.286197606    1.277028252
29      H       4.837730122     0.632831926     -0.120505903
30      H       3.238367119     2.164755528     -1.655346367
31      H       1.725933502     1.359962958     -2.110842099
32      H       0.393498720     3.290987505     -1.045915643
33      H       1.942244973     4.134236925     -0.729472881
34      H       1.457171870     3.757889985     -2.406591099
---
H21C13N1O1, RHF, CHARGE=0, MULT=1
HF=-68.4
1       C       -0.000108618    -0.000585185    0.000153206
2       C       1.559389109     0.001059949     -0.001844951
3       C       2.079060035     1.465582217     0.001242827
4       C       1.552045033     2.207266712     -1.257239208
5       C       -0.005981767    2.195940624     -1.254617779
6       C       -0.497122215    0.706799650     -1.312847222
7       C       -0.491051541    0.721411167     1.287072237
8       H       -0.352817668    -1.055508387    0.013316527
```

```
9    H     1.954911674   -0.552413283   -0.881128541
10   H     1.945888214   -0.552001548    0.882086661
11   H     3.191761565    1.462586611   -0.009630812
12   C     1.575009317    2.196565649    1.276767462
13   H     1.948071949    1.737654641   -2.184415330
14   H     1.929498539    3.253079718   -1.283400600
15   H    -0.359723752    2.758506828   -2.145751627
16   C    -0.518931300    2.914773389    0.024802434
17   C    -1.995595033    0.533763295   -1.645080436
18   H     0.033918136    0.197163069   -2.156295219
19   H    -1.596723189    0.696266480    1.379531602
20   C     0.021181351    2.187230321    1.285859888
21   H    -0.123608306    0.180963444    2.187929674
22   H     1.962542293    3.238021482    1.310797127
23   H     1.975703150    1.708903218    2.192229119
24   O    -2.813242715   -0.004144705   -0.906432706
25   H    -1.629407976    2.953667862    0.041783574
26   H    -0.196049662    3.979382118    0.020247057
27   H    -0.344661947    2.708762241    2.197943703
28   N    -2.463427490    1.056232396   -2.902768338
29   C    -1.838434824    0.593367785   -4.145581585
30   C    -3.847641583    1.536662953   -3.009100761
31   H    -0.833435396    1.055512195   -4.279824389
32   H    -1.716324271   -0.515643630   -4.168626277
33   H    -2.446730555    0.876228533   -5.033709613
34   H    -3.912249760    2.306095390   -3.813819905
35   H    -4.175282145    2.024155382   -2.064683661
36   H    -4.571019469    0.722768507   -3.251423545
---
H10C9O2, RHF, CHARGE=0, MULT=1
HF=-49.1
1    C     0.000082009    0.000369958    0.000193707
2    C     1.404399898   -0.000038364   -0.000139288
3    C     2.115728797    1.213464074    0.000066552
4    C     1.438411974    2.456257412   -0.008300014
5    C     0.022791180    2.438948657    0.009472860
6    C    -0.687245353    1.225108183    0.008546694
7    C     2.210721626    3.785260458    0.051046373
8    O     3.494035458    3.710817229   -0.542332254
9    C     3.639484684    4.597912350   -1.622150110
10   C     2.291317933    5.377962239   -1.669011306
11   O     1.539216952    4.842123185   -0.610369511
12   H    -0.550887654   -0.940394380   -0.002893181
13   H     1.949503419   -0.944622668    0.001147974
14   H     3.206081624    1.164877729    0.011307340
15   H    -0.552020559    3.366622410    0.028500907
16   H    -1.777702062    1.237908473    0.016193482
17   H     2.333674796    4.061363589    1.142174336
18   H     3.838999130    4.033086417   -2.565308341
19   H     4.515906464    5.268289068   -1.448576186
20   H     1.752150517    5.240393464   -2.637832198
21   H     2.427016108    6.476803654   -1.521001331
---
H10C9O2, RHF, CHARGE=0, MULT=1
HF=-82.7
1    C     0.111512841   -0.142062173    0.226311033
2    C     1.528537180   -0.131938931    0.105553212
3    C     2.195855264    1.125162285   -0.030665974
4    C     1.436172564    2.317202876   -0.031387836
```

```
  5    C     0.039663431    2.293734510    0.095273648
  6    C    -0.622118192    1.066122185    0.222500823
  7    C    -0.687361289   -1.411266053    0.337053598
  8    O    -0.929628715   -1.813772236    1.611834866
  9    O    -1.150807580   -2.066757104   -0.595183106
 10    C     2.329905439   -1.413434443    0.119839662
 11    C     3.694154053    1.233120407   -0.180616811
 12    H     1.931290384    3.285512113   -0.132347674
 13    H    -0.527428840    3.225118221    0.093472907
 14    H    -1.709632161    1.055819652    0.316712884
 15    H    -1.451520912   -2.602498957    1.697958442
 16    H     2.903604774   -1.533622109   -0.823383462
 17    H     3.046733303   -1.420946413    0.968158342
 18    H     1.712067179   -2.325939993    0.226755142
 19    H     4.223130626    0.780265136    0.683711393
 20    H     4.033745032    2.287385571   -0.244002178
 21    H     4.042418702    0.728558066   -1.106218424
---
H10C9O2, RHF, CHARGE=0, MULT=1
HF=-84.9
  1    C     0.000086075    0.000012113    0.000475709
  2    C     1.411965941   -0.001430282   -0.000946295
  3    C     2.135517540    1.200993870    0.000244343
  4    C     1.464344589    2.442540459    0.003590977
  5    C     0.050741772    2.432096774    0.006446404
  6    C    -0.703345102    1.235303402    0.005176906
  7    C    -0.705340662   -1.326008315   -0.027989786
  8    O    -0.908762616   -1.879899503    1.195515523
  9    O    -1.077495225   -1.935690121   -1.029396715
 10    C    -2.210272965    1.308061211    0.001682671
 11    C     2.217700098    3.746482193    0.003310544
 12    H     1.962447903   -0.944404852   -0.004558797
 13    H     3.225927591    1.157732401   -0.001942608
 14    H    -0.477838658    3.388694983    0.009015909
 15    H    -1.343455457   -2.724288144    1.193764533
 16    H    -2.626982221    0.882796772   -0.935242224
 17    H    -2.579046623    2.351514617    0.076200243
 18    H    -2.644501137    0.753571546    0.859335745
 19    H     1.967232410    4.347083065   -0.896375383
 20    H     3.316699008    3.600100188    0.006152894
 21    H     1.963106985    4.350133419    0.899743255
---
H10C9O2, RHF, CHARGE=0, MULT=1
HF=-83.9
  1    C     0.000004862   -0.000272394    0.000525233
  2    C     1.413726443    0.000326743   -0.002505261
  3    C     2.155663609    1.200867138    0.001364055
  4    C     1.435302279    2.416775813    0.014976384
  5    C     0.032900231    2.424327892    0.020159071
  6    C    -0.718584080    1.225277148    0.011187318
  7    C    -0.695075295   -1.332170330   -0.032416583
  8    O    -0.914824201   -1.881048988    1.190551028
  9    O    -1.046647938   -1.949693462   -1.036467924
 10    C    -2.225292414    1.287802528    0.008770058
 11    C     3.660814508    1.200080264   -0.011347777
 12    H     1.944423885   -0.954814639   -0.009408914
 13    H     1.966377401    3.370752665    0.020993848
 14    H    -0.478647984    3.389057220    0.030549363
 15    H    -1.342895709   -2.728722929    1.185904466
```

```
16      H     -2.643487009    0.823620318   -0.908627014
17      H     -2.600043967    2.331290318    0.042782449
18      H     -2.651550313    0.764404725    0.889752843
19      H      4.047815777    1.721141509   -0.912389457
20      H      4.084643303    0.175649464   -0.016045127
21      H      4.063925348    1.719310409    0.883435073
---
H10C9O2, RHF, CHARGE=0, MULT=1
HF=-81.6
1       C      0.000528221    0.008115100    0.000423900
2       C      1.422380560    0.032633511   -0.079567628
3       C      2.076375357    1.286347414   -0.092486998
4       C      1.347777210    2.482196293   -0.030079767
5       C     -0.051592145    2.446505830    0.044118540
6       C     -0.750042867    1.216986207    0.059906776
7       C     -0.719631561   -1.314502916    0.000592754
8       O     -0.903451237   -1.855801970    1.233271239
9       O     -1.137790933   -1.921869492   -0.984272465
10      C      2.257025274   -1.223494827   -0.149756245
11      C     -2.257773958    1.237763842    0.135159307
12      H      3.165270287    1.339630197   -0.151142461
13      H      1.869690837    3.440025150   -0.039385257
14      H     -0.596954958    3.391082791    0.089528762
15      H     -1.351416510   -2.693171054    1.247839440
16      H      1.960190663   -1.863118952   -1.006416768
17      H      3.335973779   -0.999604708   -0.278775766
18      H      2.162026819   -1.819600506    0.781781801
19      H     -2.711728126    0.769698062   -0.762832920
20      H     -2.625975700    0.701219599    1.034003729
21      H     -2.657560942    2.270904858    0.195028081
---
H10C9O2, RHF, CHARGE=0, MULT=1
HF=-86.6
1       C      0.000663251    0.000134221   -0.000138014
2       C      1.410044601   -0.001011957    0.000168438
3       C      2.108481347    1.215399891   -0.000259954
4       C      1.437892361    2.461275686   -0.000951389
5       C      0.012455226    2.469103246   -0.001077178
6       C     -0.680019160    1.234701014   -0.000258410
7       C     -0.762234276   -1.290232753   -0.022561776
8       O     -1.053907200   -1.789690983    1.206154019
9       O     -1.127639718   -1.906271435   -1.022329771
10      C     -0.785982617    3.749218061   -0.002035944
11      C      2.253718129    3.730168942   -0.000286912
12      H      1.968402405   -0.938440884   -0.000514545
13      H      3.200094706    1.181116967    0.000015256
14      H     -1.772722871    1.234146761   -0.000891904
15      H     -1.536121927   -2.607739015    1.209795843
16      H     -0.565206221    4.359671689   -0.902638007
17      H     -0.563052940    4.362322078    0.896269335
18      H     -1.879701195    3.563948145   -0.000463871
19      H      2.029075373    4.353675592   -0.890858347
20      H      3.344443017    3.528824808   -0.016243275
21      H      2.051329503    4.335710617    0.908028841
---
H10C9O2, RHF, CHARGE=0, MULT=1
HF=-87.1
1       C      0.000264675   -0.001133768    0.000092845
2       C      1.412230882   -0.000535458   -0.000017062
```

```
 3    C     2.133296537    1.212812490    0.000156071
 4    C     1.407905423    2.427945003    0.001347812
 5    C    -0.003590704    2.454119775    0.002069897
 6    C    -0.699367392    1.222441252    0.001850740
 7    C    -0.750104425   -1.299360932   -0.023560817
 8    O    -1.073369159   -1.781572128    1.204188166
 9    O    -1.081106738   -1.934230002   -1.023464491
10    C    -0.764420570    3.753822980    0.003570475
11    C     3.639380435    1.229065649   -0.001097290
12    H     1.949152776   -0.951422150   -0.001233789
13    H     1.958978843    3.371419221    0.001448988
14    H    -1.791471806    1.220298061    0.001757270
15    H    -1.547875908   -2.604094907    1.207384177
16    H    -1.423585274    3.828115323   -0.886948192
17    H    -0.094068266    4.636866024   -0.009383669
18    H    -1.402055127    3.837722711    0.908842209
19    H     4.028746624    1.759394665   -0.895619142
20    H     4.073725425    0.208991106   -0.008561761
21    H     4.030030972    1.747024806    0.900104424
---
H10C9O2, RHF, CHARGE=0, MULT=1
HF=-82.9
 1    C     0.000745688    0.000405971    0.000103628
 2    C     1.412919618   -0.000015016    0.000053982
 3    C     2.147415441    1.209104263    0.000212298
 4    C     1.428638274    2.427539711    0.001050192
 5    C     0.023751506    2.435601481    0.000746186
 6    C    -0.694623386    1.228773140   -0.000254877
 7    C    -0.754262039   -1.294829647    0.022281690
 8    O    -1.039335794   -1.797710403   -1.206630713
 9    O    -1.121245380   -1.910608001    1.021619054
10    C     3.661544631    1.197297675    0.012970865
11    C     4.306540455    1.195084386   -1.377865008
12    H     1.945538443   -0.953621777    0.001794757
13    H     1.958905385    3.381718410    0.003041342
14    H    -0.513331290    3.384875230    0.001627814
15    H    -1.785171396    1.253107597    0.000210333
16    H    -1.518310010   -2.617649757   -1.210732840
17    H     4.037754192    2.079226542    0.582505424
18    H     4.025683752    0.308696163    0.579641968
19    H     5.412154333    1.188699400   -1.285911791
20    H     4.024438678    2.093357596   -1.963093406
21    H     4.014181722    0.301494472   -1.965109445
---
H10C9O2, RHF, CHARGE=0, MULT=1
HF=-85
 1    C     0.000019210   -0.000080644    0.000053169
 2    C     1.411453372    0.000023591    0.000082406
 3    C     2.122531975    1.211896367   -0.000083700
 4    C     1.451196833    2.457112503   -0.001994896
 5    C     0.036671165    2.443894869   -0.001918903
 6    C    -0.681243187    1.235942203   -0.001054123
 7    C    -0.761919900   -1.290535506    0.024138410
 8    O    -1.047114249   -1.794736568   -1.204198749
 9    O    -1.132908493   -1.901932373    1.024648654
10    C     2.221161985    3.760696282    0.010335931
11    C     2.550617169    4.314112467   -1.380899137
12    H     1.966277286   -0.939736897    0.001655297
13    H     3.213345513    1.173858640    0.002465636
```

```
14      H      -0.523371371    3.380908901    -0.001148049
15      H      -1.772095552    1.267607000     0.000473205
16      H      -1.529777153   -2.612599580    -1.207472310
17      H       1.645983570    4.530134079     0.576504238
18      H       3.170997307    3.629452070     0.579501589
19      H       3.112331211    5.266391976    -1.288791461
20      H       1.633792304    4.521378059    -1.968513093
21      H       3.177216095    3.610794922    -1.965289731
---
H10C9O2, RHF, CHARGE=0, MULT=1
HF=-74.7
1       C       0.006901141    0.010876145    0.035178850
2       C       1.420898916    0.056976858    0.006563194
3       C       2.107550261    1.284076812    0.000623494
4       C       1.415518200    2.514855380    0.022414698
5       C       0.002020338    2.468993680    0.051582749
6       C      -0.688792697    1.247044531    0.057899695
7       C      -0.772538111   -1.269689235    0.040496486
8       O      -0.020730650   -2.406987608    0.030177776
9       O      -1.999093381   -1.370738884    0.052323197
10      C       2.142692407    3.832863903    0.014342707
11      C      -0.532552027   -3.714515640    0.022925930
12      H       2.018489515   -0.856376197   -0.012041771
13      H       3.198838069    1.266296822   -0.021507777
14      H      -0.576043711    3.394977257    0.069662140
15      H      -1.780214530    1.276208419    0.080564396
16      H       1.864869041    4.430738715   -0.879169227
17      H       3.244400807    3.708698381   -0.001272557
18      H       1.890213958    4.428628692    0.916617805
19      H      -1.152268723   -3.922792629   -0.881576807
20      H       0.349078023   -4.399212876    0.009623726
21      H      -1.138345672   -3.938986265    0.932793267
---
H8C10O2, RHF, CHARGE=0, MULT=1
HF=-47.9
1       C       0.076896379   -0.009894465    0.043296367
2       C       1.459178937    0.025122486    0.015639246
3       C       2.148708564    1.271993886   -0.057919719
4       C       1.431217482    2.452420391   -0.090515823
5       C      -0.006937982    2.450104775   -0.056389458
6       C      -0.706302728    1.198687574   -0.002539569
7       C      -2.158194207    1.218327828    0.005467895
8       C      -2.862116962    2.440857154    0.007818554
9       C      -2.133146877    3.686040047   -0.027105694
10      C      -0.756511541    3.678317195   -0.066259670
11      O      -4.212081681    2.561231935    0.035320313
12      H      -2.679341923    4.630752578   -0.025916823
13      H      -0.416279356   -0.980540347    0.113499582
14      H       2.037455963   -0.899070432    0.052788577
15      H       3.238663961    1.278129426   -0.084203418
16      H       1.959377331    3.406382112   -0.141914504
17      O      -2.888175235    0.064195023    0.086571160
18      H      -0.217046169    4.627137426   -0.100781972
19      H      -2.692098613   -0.526588865   -0.630266712
20      H      -4.675575515    1.743522453    0.153583211
---
H8C10O2, RHF, CHARGE=0, MULT=1
HF=-50.5
1       C      -0.000138554   -0.000011279    0.000007381
```

```
2    C     1.383499770    0.000054672    0.000005775
3    C     2.102358479    1.232172972   -0.000003394
4    C     1.418265363    2.432606082   -0.000046854
5    C    -0.021766003    2.470374531   -0.000067009
6    C    -0.745406374    1.230776621   -0.000006255
7    C    -2.200675129    1.297961854    0.000035957
8    C    -2.875784962    2.521088987   -0.000018664
9    C    -2.122781626    3.746280743   -0.000121445
10   C    -0.726999845    3.719556143   -0.000139303
11   H    -3.965388350    2.569490091    0.000024019
12   O    -2.852533708    4.891293350   -0.000188652
13   H    -0.524228243   -0.957199197    0.000018034
14   H     1.938527340   -0.938538043    0.000014425
15   H     3.192924355    1.212101485    0.000025696
16   H     1.973497753    3.371917353   -0.000061879
17   O    -2.868082806    0.116066533    0.000147552
18   H    -0.148666452    4.644254861   -0.000207985
19   H    -3.810656193    0.219578286    0.000155826
20   H    -2.315315657    5.672442530   -0.000277900
---
H8C10O2, RHF, CHARGE=0, MULT=1
HF=-47.1
1    C     0.000021253   -0.000014830   -0.000005205
2    C     1.383591305    0.000006397   -0.000000457
3    C     2.097531100    1.232092618    0.000000197
4    C     1.409728210    2.432629805   -0.000014198
5    C    -0.029055495    2.471999174   -0.000031330
6    C    -0.749686942    1.228713578   -0.000002930
7    C    -2.206710205    1.264935341   -0.000003265
8    C    -2.891679238    2.479508963   -0.000050143
9    C    -2.179265031    3.708826028   -0.000084244
10   C    -0.784786255    3.718274551   -0.000048695
11   H    -3.982047704    2.524020408   -0.000056754
12   H    -2.759775150    4.632736856   -0.000135081
13   H    -0.521114635   -0.958963854   -0.000021762
14   H     1.938816412   -0.938684799   -0.000009680
15   H     3.188029977    1.217063726    0.000006669
16   H     1.982619409    3.361395783   -0.000024549
17   O    -2.854080945    0.067921876    0.000012250
18   O    -0.068273106    4.875214807   -0.000035167
19   H    -3.797348308    0.161246717    0.000353699
20   H    -0.618284104    5.647152064   -0.000075251
---
H8C10O2, RHF, CHARGE=0, MULT=1
HF=-46.1
1    C    -0.000289118    0.000378965   -0.000011313
2    C     1.382736527   -0.000166297    0.000000253
3    C     2.101388866    1.233351086   -0.000010000
4    C     1.419188997    2.436227417   -0.000027037
5    C    -0.018835962    2.471694660   -0.000035478
6    C    -0.739904624    1.234137384   -0.000032550
7    C    -2.176513769    1.261788026   -0.000054356
8    C    -2.878938081    2.464970876   -0.000083000
9    C    -2.143212400    3.727375230   -0.000090184
10   C    -0.750260647    3.708533322   -0.000056924
11   O    -4.233854371    2.544677068   -0.000111698
12   O    -2.880091105    4.867227875   -0.000120846
13   H    -0.542750697   -0.946526932   -0.000001121
14   H     1.939102966   -0.938082893    0.000020184
```

```
15      H     3.191702114    1.212090846   -0.000001174
16      H     1.975372817    3.375064186   -0.000035855
17      H    -2.709896651    0.309038068   -0.000053535
18      H    -0.183578070    4.641733270   -0.000041646
19      H    -4.650298841    1.692502252   -0.000099929
20      H    -2.344010786    5.649850298   -0.000179828
---
H10C10O2, RHF, CHARGE=0, MULT=1
HF=-58.3
1       O     0.387077996    0.699868748    0.800973067
2       C     1.339627771    0.407019760    0.086994989
3       C     2.041608064    1.450703260   -0.777081827
4       C     1.888899716   -1.029272875    0.072859412
5       C     1.180758367   -1.997174269   -0.887299015
6       C     1.614035230   -3.438182877   -0.812800567
7       C     2.621701318   -3.925296182   -1.675558345
8       C     3.025215060   -5.270090033   -1.611336315
9       C     2.425018324   -6.144369601   -0.688183581
10      C     1.417678117   -5.669564023    0.169797160
11      C     1.011637011   -4.324589168    0.107582620
12      H     2.438452899    1.005713379   -1.710268683
13      H     1.357369626    2.273918843   -1.063747492
14      H     2.891118703    1.888518808   -0.213310395
15      H     2.970006823   -1.011173528   -0.192307185
16      O     0.324010282   -1.633456026   -1.682526220
17      H     3.095389527   -3.263174920   -2.402289806
18      H     3.803811095   -5.635348255   -2.281852383
19      H     2.738287275   -7.187563162   -0.639011947
20      H     0.946937025   -6.344735534    0.884946547
21      H     0.223363638   -3.976193827    0.776882083
22      H     1.832443046   -1.450619919    1.102495648
---
H12C10O2, RHF, CHARGE=0, MULT=1
HF=-90.2
1       C    -0.042866003   -0.030051228    0.197160576
2       C     1.368286466   -0.051493723    0.035763004
3       C     2.089017371    1.185417700    0.015652160
4       C     1.382311497    2.422384898    0.071676857
5       C    -0.023339889    2.404741375    0.214413833
6       C    -0.727310179    1.199571359    0.290329126
7       C    -0.882941388   -1.275556154    0.263641993
8       O    -0.996756060   -1.807224749    1.508706664
9       O    -1.480513098   -1.807757430   -0.670738829
10      C     2.097066800   -1.369200770   -0.113508943
11      C     2.065822640    3.766544699   -0.015820487
12      C     3.599342027    1.187120988   -0.077090612
13      H    -0.586073724    3.339008106    0.269309914
14      H    -1.811318406    1.225873222    0.417549886
15      H    -1.542151322   -2.581884951    1.572860797
16      H     1.425633197   -2.218626190   -0.347773972
17      H     2.632469861   -1.629394255    0.824212361
18      H     2.834623737   -1.337319454   -0.942355753
19      H     2.670259575    3.967275392    0.893861956
20      H     1.338503452    4.599370596   -0.108213783
21      H     2.731241021    3.830591005   -0.901658127
22      H     3.923705147    1.166267000   -1.139436419
23      H     4.062148956    0.321057941    0.438277854
24      H     4.056783171    2.079944558    0.394659686
---
```

```
H12C10O2, RHF, CHARGE=0, MULT=1
HF=-91.4
1    C     0.095738706    -0.088685951    -0.176940938
2    C     1.510663790    -0.050598986    -0.044545500
3    C     2.133978617     1.215425489     0.170608209
4    C     1.340475882     2.383593507     0.212778332
5    C    -0.062436908     2.352757613     0.062901424
6    C    -0.670222477     1.096808360    -0.129997871
7    C    -0.666468969    -1.376208113    -0.331957061
8    O    -0.976375581    -1.699049888    -1.614509683
9    O    -1.044970477    -2.112184018     0.578203100
10   C     2.333870862    -1.316570935    -0.106453745
11   C    -0.873599559     3.620579656     0.097702777
12   C     3.625198562     1.362617220     0.358983622
13   H     1.826885653     3.350297528     0.368005540
14   H    -1.755483026     1.038144305    -0.242415231
15   H    -1.476430596    -2.498970990    -1.723852352
16   H     1.816453344    -2.147635995    -0.624857552
17   H     2.579837101    -1.665899398     0.918959376
18   H     3.285160297    -1.168481344    -0.657733371
19   H    -0.501160949     4.322697753     0.872098205
20   H    -1.942561531     3.428223020     0.321246952
21   H    -0.823733262     4.138710142    -0.883438461
22   H     3.910311184     2.392951463     0.656125252
23   H     4.170469132     1.136030194    -0.581531476
24   H     4.005685591     0.689103821     1.154727849
---
H12C10O2, RHF, CHARGE=0, MULT=1
HF=-88.7
1    C    -0.000413641    -0.000843874     0.000170080
2    C     1.427234124     0.000485484    -0.000101538
3    C     2.119093064     1.246006647    -0.000283834
4    C     1.365334884     2.441131289     0.014036174
5    C    -0.033401582     2.425971071     0.023560342
6    C    -0.747890641     1.208611800     0.015070347
7    C    -0.747999461    -1.309968484     0.003583222
8    O    -1.017485360    -1.808117047    -1.231759793
9    O    -1.127542671    -1.935655749     0.992687660
10   C     2.176958646    -1.314076393    -0.006244644
11   C    -2.257075780     1.246402450     0.024237530
12   C     3.624442375     1.358055232    -0.016614950
13   H     1.870312619     3.409904754     0.017943478
14   H    -0.565500458     3.379315438     0.036686118
15   H    -1.492952331    -2.630109638    -1.244721756
16   H     2.026369083    -1.848324818    -0.968014559
17   H     1.842419859    -1.975361444     0.820029993
18   H     3.271096090    -1.202291333     0.122596021
19   H    -2.669109144     0.732584080     0.917361252
20   H    -2.678877911     0.765740847    -0.882852509
21   H    -2.646541221     2.284958097     0.045131237
22   H     3.964379855     2.413683367    -0.056315368
23   H     4.059009146     0.852459353    -0.904137524
24   H     4.071854464     0.914346347     0.897339645
---
H12C10O2, RHF, CHARGE=0, MULT=1
HF=-92.3
1    C    -0.036037241    -0.053560237     0.025168570
2    C     1.366404317     0.001098838     0.166217219
3    C     2.082763721     1.218794816     0.142461606
```

```
 4    C     1.354055894    2.428515057   -0.034830350
 5    C    -0.050879137    2.359895866   -0.174585919
 6    C    -0.773318603    1.146103519   -0.149412435
 7    C    -0.700236638   -1.399736326    0.091632134
 8    O    -0.772363080   -2.049718974   -1.098856857
 9    O    -1.147160448   -1.946171512    1.098946006
10    C     3.581809924    1.190145178    0.307262183
11    C    -2.274352246    1.164558197   -0.297002549
12    C     2.027196157    3.778448053   -0.074155222
13    H     1.916305742   -0.934142392    0.300816960
14    H    -0.607526570    3.291491370   -0.307434035
15    H    -1.180804067   -2.906600075   -1.071376290
16    H     4.092430336    1.617122078   -0.581289680
17    H     3.971777819    0.159746190    0.437073491
18    H     3.898319326    1.768398662    1.200608723
19    H    -2.663322583    2.183800075   -0.497493566
20    H    -2.768369997    0.804801657    0.629818144
21    H    -2.606354469    0.523943678   -1.140182306
22    H     1.309950554    4.602081534   -0.268751536
23    H     2.792643566    3.824768974   -0.876587170
24    H     2.523804207    4.004112945    0.892832676
---
H12C10O2, RHF, CHARGE=0, MULT=1
HF=-89.4
 1    C    -0.049345678   -0.035004418   -0.069520390
 2    C     1.368479804   -0.032000746    0.040608574
 3    C     2.043175758    1.209442054    0.089687532
 4    C     1.354803254    2.438006730    0.035773510
 5    C    -0.051308958    2.403828703   -0.079608930
 6    C    -0.771125812    1.189752502   -0.134638430
 7    C    -0.795020640   -1.343268582   -0.085985562
 8    O    -0.977140674   -1.876731006   -1.322105726
 9    O    -1.235864187   -1.944716368    0.892650448
10    C     2.181582184   -1.302697887    0.110351292
11    C    -2.276011632    1.242561789   -0.250002523
12    C     2.084104888    3.754325765    0.101034499
13    H     3.132572741    1.217419602    0.174083851
14    H    -0.597487535    3.349256965   -0.125820781
15    H    -1.446578150   -2.702059361   -1.344574985
16    H     2.064998624   -1.905779996   -0.814078174
17    H     1.881726335   -1.927397344    0.977119655
18    H     3.266205353   -1.099160194    0.224299747
19    H    -2.760927967    0.819477338    0.654447528
20    H    -2.637085050    0.681075706   -1.136251219
21    H    -2.649922960    2.281341615   -0.359039112
22    H     1.918329881    4.343316278   -0.825587013
23    H     3.179088891    3.627083781    0.219469489
24    H     1.728544796    4.360519549    0.960691115
---
H12C10O2, RHF, CHARGE=0, MULT=1
HF=-85.9
 1    C    -0.018257437    0.039420095    0.162836704
 2    C     1.358346627    0.115461743    0.485119455
 3    C     2.095856670    1.280584805    0.230442698
 4    C     1.455442966    2.387201507   -0.349677589
 5    C     0.091718776    2.317142471   -0.673082941
 6    C    -0.678418890    1.149556549   -0.434744379
 7    C    -0.719762363   -1.243043479    0.514604567
 8    O    -1.277286974   -1.501299525    1.580203386
```

```
 9    O    -0.675129531   -2.195160473   -0.453520050
10    C    -2.159833581    1.124198401   -0.811168273
11    C    -2.394566543    1.360921495   -2.319687492
12    C    -2.999957645    2.061214428    0.085091236
13    H     1.864244035   -0.737934995    0.941620583
14    H     3.155243188    1.325597573    0.483061859
15    H     2.015176395    3.301955184   -0.549453552
16    H    -0.368917936    3.200760800   -1.118986598
17    H    -1.100445023   -3.016182820   -0.236622149
18    H    -2.563886586    0.097362875   -0.619934334
19    H    -1.749514106    0.697732831   -2.930851952
20    H    -2.192763385    2.405843814   -2.628056572
21    H    -3.447187764    1.135475480   -2.587134285
22    H    -2.783525868    3.133052031   -0.093929080
23    H    -2.813540201    1.855181603    1.158332214
24    H    -4.082975105    1.905403715   -0.097880311
---
H12C10O2, RHF, CHARGE=0, MULT=1
HF=-62.6
 1    C    -0.088268509    0.014423991    0.128553872
 2    C     1.229109507    0.504449967    0.798900723
 3    O     1.456615143    1.772140140    0.238765006
 4    C     0.507646180    2.108186767   -0.765437288
 5    O    -0.461009000    1.068430922   -0.721395705
 6    C    -0.173312195    3.475376705   -0.411732675
 7    C     1.179874184    2.183901832   -2.160785408
 8    C     0.468609287    1.840230961   -3.336000071
 9    C     1.070764559    1.934492104   -4.603478527
10    C     2.395263190    2.383559160   -4.728695157
11    C     3.110656705    2.743553338   -3.575143421
12    C     2.509810930    2.648554696   -2.307153176
13    H     0.055205437   -0.927189357   -0.455826730
14    H    -0.901455362   -0.180263043    0.869068442
15    H     1.139002758    0.590702675    1.909118888
16    H     2.091925810   -0.174336432    0.591347105
17    H    -0.960789499    3.721399924   -1.151209803
18    H     0.571466958    4.295491907   -0.418243537
19    H    -0.644835545    3.455987899    0.591143405
20    H    -0.566593428    1.497034700   -3.290952642
21    H     0.502505259    1.658725289   -5.492485789
22    H     2.862202603    2.454542592   -5.711325296
23    H     4.137647728    3.100707389   -3.660021514
24    H     3.101901799    2.948412353   -1.440638886
---
H12C10O2, RHF, CHARGE=0, MULT=1
HF=-93.2
 1    C     0.009049611   -0.020852355   -0.024963114
 2    C     1.418274480    0.035295169    0.032304776
 3    C     2.119613409    1.262492847    0.046974109
 4    C     1.396620176    2.491490433    0.012004589
 5    C    -0.027067520    2.452105808   -0.059038099
 6    C    -0.691871173    1.205524237   -0.070241661
 7    C    -0.756010831   -1.307995383   -0.045297808
 8    O    -0.017574020   -2.448433635    0.019539607
 9    O    -1.981362441   -1.430249459   -0.115510705
10    C     2.131832695    3.812800797    0.021725101
11    C    -0.875163410    3.700484072   -0.126672448
12    C     3.628520871    1.207796915    0.100539882
13    H     1.996863221   -0.891038601    0.065160063
```

```
14    H     -1.784087079    1.197147173   -0.117296834
15    H     -0.517624823   -3.255346764    0.006012177
16    H      2.401976181    4.106150991   -1.015537598
17    H      3.064586628    3.777529142    0.619946421
18    H      1.535382300    4.638111171    0.459725650
19    H     -0.800050767    4.287417634    0.812819541
20    H     -1.950686045    3.470775714   -0.273836704
21    H     -0.573018187    4.351724541   -0.973193356
22    H      4.088628317    1.799616324   -0.717793068
23    H      4.016372277    0.173423074   -0.003924612
24    H      4.007908156    1.595334263    1.069751393
---
H12C10O2, RHF, CHARGE=0, MULT=1
HF=-89.8
1     C     -0.037916421   -0.124095846   -0.199909238
2     C      1.273124314   -0.086562763   -0.720575853
3     C      1.903077742    1.151736418   -0.922995431
4     C      1.237037653    2.348798943   -0.611425403
5     C     -0.078771936    2.339505812   -0.088563426
6     C     -0.700403251    1.083838242    0.111678135
7     C     -0.734858668   -1.436721365    0.000103547
8     O     -1.497513719   -1.987561358   -0.792030375
9     O     -0.465224345   -2.039305275    1.186872596
10    C     -0.778488036    3.655802052    0.236219564
11    C     -1.066010660    3.814360702    1.744437901
12    C     -2.023238509    3.894635690   -0.644973396
13    H      1.803086928   -1.006984208   -0.969390903
14    H      2.916415857    1.184618593   -1.325179980
15    H      1.758941826    3.292364596   -0.782711733
16    H     -1.714372545    1.036745638    0.514467354
17    H     -0.900248120   -2.871248096    1.329980336
18    H     -0.069808464    4.487268092   -0.019818745
19    H     -0.147183687    3.644626349    2.341815734
20    H     -1.839770291    3.110005777    2.109073117
21    H     -1.419457274    4.841889617    1.968455827
22    H     -2.855040524    3.202358821   -0.407281938
23    H     -1.777204685    3.770301378   -1.719149620
24    H     -2.402626575    4.928688422   -0.511596924
---
H12C10O2, RHF, CHARGE=0, MULT=1
HF=-91.5
1     C     -0.058288512   -0.121149440   -0.102620255
2     C      1.133038672    0.022523518   -0.845173803
3     C      1.816855858    1.249947543   -0.867110099
4     C      1.336631246    2.374982230   -0.155146506
5     C      0.141209485    2.215388666    0.586221176
6     C     -0.546512009    0.990374853    0.616034841
7     C     -0.803068954   -1.421743580   -0.096892647
8     O     -1.688762464   -1.760271412   -0.880202053
9     O     -0.428320289   -2.277951819    0.888489807
10    C      2.054679080    3.720836896   -0.168223846
11    C      2.068386873    4.364010058   -1.571499529
12    C      3.454438105    3.648498730    0.478365685
13    H      1.537095938   -0.817614341   -1.412691198
14    H      2.735499353    1.312273299   -1.453293002
15    H     -0.273244771    3.050619966    1.154080927
16    H     -1.464324260    0.912669501    1.201304000
17    H     -0.891234493   -3.106916997    0.906324691
18    H      1.465804671    4.425596715    0.476357240
```

```
19    H     1.044720594     4.407318847    -1.996054997
20    H     2.707428000     3.811974414    -2.288734873
21    H     2.448102876     5.405157427    -1.520070944
22    H     4.176232937     3.065980579    -0.127440584
23    H     3.402104716     3.180930796     1.482660328
24    H     3.876059756     4.666149425     0.610444756
---
H14C10O2, RHF, CHARGE=0, MULT=1
HF=-90.6
1     C    -0.034874056    -0.011621018    -0.103302169
2     C     1.323936862     0.067821998    -0.485357599
3     C     2.057335113     1.258582566    -0.391830442
4     C     1.483995498     2.457292400     0.079465361
5     C     0.113662107     2.409253172     0.479128691
6     C    -0.629386923     1.180669274     0.404996735
7     C    -0.796931924    -1.328911748    -0.243706656
8     C    -0.948350872    -1.759521281    -1.719642069
9     C    -0.213718506    -2.443162327     0.653185930
10    C     2.305930334     3.718438947     0.145546606
11    O    -1.948282453     1.242959168     0.768457228
12    O    -0.428307372     3.560871288     0.952485029
13    H     1.841504485    -0.812514805    -0.871656879
14    H     3.104660751     1.241956689    -0.702442331
15    H    -1.847284911    -1.188635663     0.118784265
16    H    -1.379410061    -0.939061746    -2.328135657
17    H     0.013999689    -2.054502346    -2.182871857
18    H    -1.635460256    -2.626966433    -1.802206085
19    H     0.776152442    -2.800662829     0.307196066
20    H    -0.097835305    -2.090445556     1.698370202
21    H    -0.893354918    -3.320110503     0.675645144
22    H     2.355560931     4.116799875     1.180355831
23    H     1.881492732     4.507943605    -0.509628147
24    H     3.351363670     3.549776318    -0.186790562
25    H    -2.124439980     0.711324489     1.535480067
26    H    -1.375537706     3.573488273     0.961971584
---
H18C10O2, RHF, CHARGE=0, MULT=1
HF=-130.4
1     C    -0.193525652    -0.054052354    -0.283251105
2     C     1.299973761    -0.206537612     0.054632642
3     C     2.023356366     1.160726945     0.257499393
4     C     1.746132399     2.158632174    -0.911315169
5     C     0.249181790     2.298839563    -1.239252100
6     C    -0.490759439     0.960541022    -1.399919342
7     O     3.425305056     0.980347951     0.328155399
8     C     4.168480771     0.792397646     1.453913644
9     O     3.621585063     0.713716246     2.550259730
10    C     5.670643428     0.717790361     1.147216728
11    C     6.560685288     0.301996222     2.334342634
12    C     8.056947605     0.259549559     2.007010404
13    H    -0.593109092    -1.049273390    -0.584438622
14    H    -0.755467498     0.230866694     0.635822247
15    H     1.389882896    -0.817009994     0.981041012
16    H     1.799622757    -0.792561795    -0.749811898
17    H     1.626657792     1.630490478     1.198712412
18    H     2.145653915     3.161507108    -0.637938069
19    H     2.294396831     1.847995369    -1.829278415
20    H     0.145154118     2.884641965    -2.180978164
21    H    -0.249937188     2.906613324    -0.449899190
```

```
22     H      -1.587026837     1.154411693    -1.430516334
23     H      -0.235416280     0.511474299    -2.387159106
24     H       5.984875753     1.722135161     0.779354724
25     H       5.823440939     0.002488925     0.306672106
26     H       6.408291452     1.005738184     3.184306519
27     H       6.249518415    -0.704234566     2.697649190
28     H       8.439439790     1.252978931     1.696460369
29     H       8.282846413    -0.460475829     1.194830507
30     H       8.632481199    -0.055089294     2.901635034
---
H20C10O2, RHF, CHARGE=0, MULT=1
HF=-136.2
1      C      -0.000038415    -0.000006391    -0.000003399
2      C       1.540817719     0.000068353     0.000035140
3      O       2.001477371     1.332314675    -0.000164403
4      C       3.310819215     1.710912273    -0.008953405
5      C       3.474471045     3.236439770    -0.013322723
6      C       4.933632105     3.736122840    -0.004727553
7      C       5.054359168     5.274799614    -0.008852120
8      C       6.509536826     5.786413648     0.002490393
9      C       6.633031726     7.323835840    -0.003022206
10     C       8.085849205     7.837650767     0.010650280
11     C       8.224615405     9.362571875     0.004331987
12     H      -0.415675430     0.501170986     0.896670966
13     H      -0.415400489     0.499943928    -0.897532239
14     O       4.192693551     0.856615295    -0.013304499
15     H      -0.359466388    -1.049574093     0.000646787
16     H       1.913185392    -0.552559694     0.901615526
17     H       1.912586929    -0.553442881    -0.901221971
18     H       2.938465815     3.640088265     0.876568047
19     H       2.951316614     3.632047554    -0.914460322
20     H       5.453538562     3.332641746     0.893692503
21     H       5.466249828     3.327098174    -0.893017845
22     H       4.516908825     5.685074125     0.876872255
23     H       4.532496006     5.678980492    -0.906595255
24     H       7.029913761     5.382462736     0.901059726
25     H       7.047347364     5.374649391    -0.882420962
26     H       6.094321889     7.736409970     0.880913832
27     H       6.113761744     7.728431388    -0.902133777
28     H       8.611750759     7.440519810     0.909666405
29     H       8.631888171     7.431351223    -0.871926087
30     H       7.768637685     9.815041100    -0.899513610
31     H       7.747770422     9.824356319     0.892597640
32     H       9.296703126     9.648935556     0.015334851
---
H22C10O2, RHF, CHARGE=0, MULT=1
HF=-125
1      O      -0.000028536    -0.000013015    -0.000048268
2      C       1.396834241    -0.000028174     0.000024118
3      C       1.917333915     1.460508835    -0.000010675
4      C       3.456371207     1.557405078    -0.000666263
5      C       3.991393567     3.004198616    -0.000210497
6      C       5.530786411     3.103433454    -0.001385650
7      C       6.065998252     4.550138382    -0.000594541
8      C       7.605387758     4.649322648    -0.002255572
9      C       8.140452531     6.096086579    -0.000780866
10     C       9.679450798     6.192926784    -0.004221030
11     C      10.199890006     7.653467504     0.004006229
12     O      11.596757760     7.653356450     0.000621417
```

```
13      H       -0.338477815    -0.884096119    -0.000114892
14      H        1.792655511    -0.550021843    -0.896643147
15      H        1.792604910    -0.549527432     0.897095231
16      H        1.513696386     1.991325733    -0.892108176
17      H        1.514833817     1.990973339     0.892803748
18      H        3.856021661     1.022109539    -0.892591759
19      H        3.856188001     1.021434843     0.890579651
20      H        3.591193468     3.540389330    -0.891117672
21      H        3.592792204     3.539085211     0.892224820
22      H        5.929490125     2.568295580    -0.893636398
23      H        5.930857772     2.567043067     0.889558690
24      H        5.665581146     5.086924300    -0.891090855
25      H        5.667824546     5.085034306     0.892102781
26      H        8.003890770     4.115014251    -0.895033553
27      H        8.005867704     4.112697892     0.888302071
28      H        7.738630015     6.633577984    -0.890406666
29      H        7.742605701     6.629957107     0.892774971
30      H       10.079914198     5.668097818    -0.901302632
31      H       10.084892648     5.656445115     0.883610853
32      H        9.802005785     8.208491704    -0.888558264
33      H        9.806331830     8.198149940     0.904897306
34      H       11.935034363     8.537491457     0.005467009
---
H8C11O2, RHF, CHARGE=0, MULT=1
HF=-53.3
1       C       -0.000162857     0.000065935     0.000077320
2       C        1.382311390     0.000142470    -0.000051532
3       C        2.104771818     1.231694681    -0.000048779
4       C        1.420039186     2.431087792    -0.000924227
5       C       -0.019088269     2.466082260    -0.001280748
6       C       -0.751918264     1.229916606     0.000493307
7       C       -2.198624930     1.286353452     0.001324716
8       C       -2.853070184     2.512663863    -0.001143007
9       C       -2.116870989     3.734925279    -0.004521474
10      C       -0.736660164     3.712830145    -0.003897662
11      C       -3.029256499     0.034675572    -0.016504232
12      O       -3.422512419    -0.569209864    -1.013383073
13      O       -3.380144944    -0.422816484     1.213630362
14      H       -0.522245433    -0.957759751    -0.001654429
15      H        1.936743841    -0.939287638    -0.000760353
16      H        3.194958408     1.209625803     0.000131711
17      H        1.973020853     3.372023444    -0.001712997
18      H       -3.943551187     2.571754218    -0.001749475
19      H       -2.662050029     4.679433800    -0.007392597
20      H       -0.177367498     4.650232982    -0.005770230
21      H       -3.915822185    -1.206884200     1.219455701
---
H8C11O2, RHF, CHARGE=0, MULT=1
HF=-55.6
1       C       -0.000042151    -0.000485716     0.000117784
2       C        1.382004635     0.000463809    -0.000126060
3       C        2.101406372     1.235294987     0.000068034
4       C        1.420817249     2.438416618     0.000632363
5       C       -0.017826082     2.472996767     0.001093110
6       C       -0.740094204     1.234277351     0.000880791
7       C       -2.178715755     1.265404181     0.001559591
8       C       -2.871934866     2.468062582     0.001340486
9       C       -2.143448248     3.705405885     0.001803524
10      C       -0.762117736     3.704007149     0.001578882
```

```
11      C       -4.370950051    2.497756905    -0.021591483
12      O       -5.086397454    2.538688957    -1.021138268
13      O       -4.950351418    2.477690789     1.206512220
14      H       -0.543335047   -0.946969980    -0.000446126
15      H        1.939478721   -0.936912098    -0.000582708
16      H        3.191868152    1.212437562    -0.000247762
17      H        1.975700188    3.377990721     0.000625261
18      H       -2.718538040    0.315965201     0.000785255
19      H       -2.687759800    4.651165678     0.001355073
20      H       -0.221463910    4.652122105     0.001486522
21      H       -5.899637708    2.502066366     1.210146902
---
H14C11O2, RHF, CHARGE=0, MULT=1
HF=-95.3
1       C       -0.000132843    0.000152004    -0.000157529
2       C        1.418242284   -0.000305173     0.000066727
3       C        2.103943497    1.253104429    -0.000032670
4       C        1.378241150    2.476812329     0.132221045
5       C       -0.045308502    2.459684664     0.098190845
6       C       -0.706535026    1.217480658     0.026625056
7       C       -0.823761379   -1.258363935    -0.019672928
8       O       -1.076564654   -1.744124722    -1.263055125
9       O       -1.296999920   -1.838840606     0.956239179
10      C        2.110228634    3.789856787     0.309057336
11      C       -0.890440653    3.711044655     0.128768877
12      C        3.610916582    1.284181821    -0.148863135
13      C        2.180393559   -1.307192684     0.001674078
14      H       -1.799277517    1.194669475    -0.001193234
15      H       -1.611961652   -2.527941149    -1.293151589
16      H        2.393949557    4.216901245    -0.676204186
17      H        3.031033282    3.680117781     0.918274798
18      H        1.503625976    4.551224560     0.839752670
19      H       -0.861066414    4.188942446     1.130578004
20      H       -1.956525642    3.500263300    -0.096068827
21      H       -0.552477561    4.452653829    -0.624528165
22      H        3.964537184    2.185493058    -0.690372568
23      H        4.000663950    0.429691114    -0.738508439
24      H        4.106600032    1.263990795     0.844957576
25      H        2.558634315   -1.541875626    -1.016015398
26      H        1.565475542   -2.169699552     0.326793514
27      H        3.042681364   -1.280706863     0.699850736
---
H14C11O2, RHF, CHARGE=0, MULT=1
HF=-95.2
1       C        0.009028751    0.012425032    -0.038423661
2       C        1.431280513    0.042400717    -0.050181374
3       C        2.115803759    1.294365929     0.041952971
4       C        1.364795978    2.502996283     0.038245112
5       C       -0.044418406    2.434592484     0.032737506
6       C       -0.746723329    1.213999498     0.016160165
7       C       -0.725867618   -1.303845106    -0.047355580
8       O       -1.107152597   -1.733797117    -1.278851473
9       O       -1.009172922   -1.987482081     0.935906824
10      C        2.003341522    3.872459418     0.041093688
11      C       -2.256252309    1.238410102     0.050362554
12      C        3.625288607    1.340262002     0.135511696
13      C        2.223907597   -1.244287543    -0.151238085
14      H       -0.615764196    3.366765206     0.046125853
15      H       -1.576760403   -2.559233984    -1.294300165
```

```
16      H        2.798085940     3.958266953    -0.728395970
17      H        2.447817761     4.101599627     1.032523024
18      H        1.269402163     4.675558900    -0.177209201
19      H       -2.656135362     0.622964127     0.882452858
20      H       -2.683838177     0.861012991    -0.902035518
21      H       -2.653159289     2.264224260     0.196752413
22      H        4.080272208     1.306426304    -0.877524405
23      H        4.039633441     0.498749565     0.727649934
24      H        3.997437497     2.255702497     0.638095914
25      H        1.634158972    -2.100352001    -0.533522905
26      H        2.616891720    -1.539452726     0.844914259
27      H        3.079598761    -1.146319557    -0.851229656
---
H14C11O2, RHF, CHARGE=0, MULT=1
HF=-95.6
1       C        0.038072789     0.038171804    -0.015907411
2       C        1.462990535     0.042727871     0.013825147
3       C        2.139012261     1.292841291     0.078316347
4       C        1.376115138     2.478344701     0.035060669
5       C       -0.031748043     2.485484080    -0.033171130
6       C       -0.723330255     1.240992509    -0.021782320
7       C       -0.682658634    -1.287209222    -0.016875553
8       O       -1.067724745    -1.720930436    -1.246293365
9       O       -0.939393215    -1.981933065     0.965884229
10      C       -0.737520655     3.818118613    -0.112424970
11      C       -2.236170239     1.216817819    -0.033527299
12      C        3.640149660     1.414455547     0.189864239
13      C        2.243596885    -1.253521384     0.000346497
14      H        1.902013350     3.437338456     0.056100983
15      H       -1.524742717    -2.553472592    -1.259706594
16      H       -1.462343869     3.848271275    -0.952360666
17      H       -0.029816996     4.656411256    -0.278324746
18      H       -1.282679971     4.038443471     0.829498039
19      H       -2.664926493     0.289573560     0.395187086
20      H       -2.613985129     1.311970894    -1.073880637
21      H       -2.671613584     2.042474247     0.565895135
22      H        3.961018471     2.457359949     0.391993281
23      H        4.137657350     1.100751189    -0.751999508
24      H        4.039864084     0.796927508     1.020832663
25      H        1.761083884    -2.037495236    -0.617606731
26      H        2.353981550    -1.647607981     1.032856493
27      H        3.260635821    -1.136186897    -0.424815310
---
H14C11O2, RHF, CHARGE=0, MULT=1
HF=-95.2
1       C        0.000335431     0.000266205     0.000015277
2       C        1.408824310    -0.001504934    -0.000356925
3       C        2.129620733     1.206830786    -0.000439401
4       C        1.485205956     2.467155514    -0.001453037
5       C       -0.665183110     1.244118513     0.000480757
6       C        0.064543614     2.443337792     0.000720113
7       H       -1.755764637     1.289241556     0.002122254
8       H       -0.511917590     3.370444147     0.003411064
9       C       -0.770327645    -1.284413390     0.022541495
10      O       -1.125918172    -1.906398449     1.022144863
11      O       -1.083470526    -1.771160492    -1.206077266
12      C        2.245129093     3.810162937    -0.000623471
13      C        1.888693169     4.594458950     1.299046846
14      H        1.962273113    -0.942362153     0.000751893
```

```
15      H        3.216990932     1.120125565     0.001141108
16      C        1.838135873     4.631970316    -1.261795312
17      C        3.794366644     3.643310286    -0.036344296
18      H       -1.572316666    -2.585395542    -1.209665693
19      H        2.107030124     3.991644122     2.203890150
20      H        0.818518110     4.878958561     1.341896543
21      H        2.471010218     5.535478543     1.377326946
22      H        2.009151392     4.051387902    -2.191084343
23      H        0.770625896     4.928958010    -1.248964711
24      H        2.426135751     5.569299524    -1.342068273
25      H        4.134277591     3.104779148    -0.944050476
26      H        4.176093553     3.096526962     0.849580106
27      H        4.302736107     4.630170283    -0.042600999
---
H20C11O2, RHF, CHARGE=0, MULT=1
HF=-123.6
1       C        0.033493720    -0.472085824    -0.040128596
2       C        1.305789649    -0.881113796     0.710844228
3       C        2.593667441    -0.148623938     0.279045051
4       C        2.701454280     1.358010087     0.604343007
5       C        2.398773955     2.307720942    -0.571182929
6       C        2.405876311     3.817961773    -0.265239555
7       C        1.314674370     4.420459866     0.641623967
8       C       -0.139359058     4.447758919     0.120174135
9       C       -1.105930422     3.391280625     0.703470290
10      C       -1.676819021     2.370734425    -0.303121848
11      C       -1.973770554     0.952236846     0.255221988
12      O       -0.825510173     0.273030203     0.708227548
13      O       -0.259527808    -0.787076300    -1.190467631
14      H        1.163122964    -0.770394931     1.809095307
15      H        1.453688774    -1.973019867     0.534331792
16      H        3.433537410    -0.664735213     0.803496687
17      H        2.772116998    -0.326825335    -0.805071556
18      H        2.053881236     1.599123005     1.475430554
19      H        3.744630998     1.552168695     0.949510115
20      H        3.170356355     2.137597859    -1.360658877
21      H        1.430550967     2.036384133    -1.043581601
22      H        3.390623472     4.081210015     0.189621388
23      H        2.381478672     4.350123791    -1.246117415
24      H        1.358808720     3.942505988     1.645466514
25      H        1.614770131     5.482406493     0.819934094
26      H       -0.558416218     5.451342867     0.373694564
27      H       -0.143560566     4.414719489    -0.991745275
28      H       -0.621203450     2.867252473     1.555960543
29      H       -1.968236506     3.933134168     1.160404382
30      H       -2.644187456     2.765816773    -0.693052494
31      H       -1.018881176     2.276806274    -1.194705091
32      H       -2.649572201     1.023175634     1.147623763
33      H       -2.537523431     0.370204776    -0.517481531
---
H22C11O2, RHF, CHARGE=0, MULT=1
HF=-141.1
1       C       -0.000081224    -0.000063099     0.000022665
2       C        1.540779150     0.000084413     0.000032671
3       O        2.001061440     1.332446261    -0.000374297
4       C        3.310151302     1.711837882    -0.008645019
5       O        4.192572706     0.858106495    -0.011771174
6       C        3.472624703     3.237513991    -0.012194055
7       C        4.931324839     3.738259521    -0.029667185
```

```
8    C     5.050671926    5.277036918   -0.034730091
9    C     6.505332665    5.789956929   -0.052490136
10   C     6.626943414    7.327724758   -0.059030161
11   C     8.081032263    7.841564257   -0.077059748
12   C     8.204971705    9.377628182   -0.084025688
13   C     9.643147251    9.902971196   -0.102237496
14   H    -0.415535705    0.499225405   -0.897828167
15   H    -0.415609456    0.501692855    0.896447333
16   H    -0.359444809   -1.049581445    0.001461779
17   H     1.912907028   -0.552411921    0.901755033
18   H     1.912810627   -0.553512939   -0.901094046
19   H     2.953138691    3.637640797    0.889026274
20   H     2.932426110    3.636008957   -0.901994943
21   H     5.467601215    3.334647113    0.859007095
22   H     5.447946281    3.330045513   -0.927820322
23   H     4.530864316    5.685791614    0.862149977
24   H     4.510504531    5.681714262   -0.921302891
25   H     7.044261599    5.385539662    0.834818166
26   H     7.024725838    5.379388614   -0.948641616
27   H     6.107164716    7.737748619    0.837219461
28   H     6.086893783    7.731876284   -0.945947926
29   H     8.621357712    7.438391336    0.810160776
30   H     8.601751003    7.431799198   -0.972966330
31   H     7.689236115    9.795986001    0.811203163
32   H     7.668883676    9.789573544   -0.970235585
33   H    10.214716109    9.572790718    0.788671436
34   H    10.194465500    9.566168150   -1.003436594
35   H     9.644279627   11.012623869   -0.106360160
---
H24C11O2, RHF, CHARGE=0, MULT=1
HF=-132.1
1    C    -0.000024774   -0.000001284    0.000019905
2    C     1.531269721   -0.000009933    0.000016852
3    C     2.171411676    1.401530247    0.000011672
4    C     3.723031830    1.364798634   -0.001226221
5    O     4.204984722    2.681254036    0.009244451
6    C     5.582489993    2.965332762   -0.022600284
7    O     5.857603816    3.476709636   -1.304394615
8    C     6.351092627    2.667352820   -2.336937353
9    C     6.755973970    3.573825037   -3.530230309
10   C     7.309128828    2.784025396   -4.732019383
11   C     7.717146976    3.648539647   -5.928285357
12   C     5.923054408    4.086667730    1.023850893
13   C     5.808573632    3.665014234    2.490926611
14   H    -0.412666767    0.506272840   -0.895962824
15   H    -0.412554958    0.505337997    0.896610594
16   H    -0.381731728   -1.041951259   -0.000464971
17   H     1.882493685   -0.570856642   -0.890784030
18   H     1.882553433   -0.570735809    0.890885762
19   H     1.817958882    1.966751223   -0.892098207
20   H     1.818070305    1.965874313    0.892766227
21   H     4.086870403    0.797742826   -0.898686411
22   H     4.092553371    0.797907806    0.895544841
23   H     6.213363479    2.054152791    0.208356975
24   H     5.580868152    1.923805138   -2.672562888
25   H     7.238825915    2.069047655   -1.996244404
26   H     5.871839952    4.167234561   -3.856277319
27   H     7.521916679    4.307490939   -3.190172923
28   H     6.548065106    2.045582175   -5.075737837
```

```
29      H        8.194926713     2.188031239    -4.411055451
30      H        6.860428682     4.224889097    -6.332207318
31      H        8.517101786     4.368991479    -5.662929291
32      H        8.104073244     3.008120763    -6.747776176
33      H        6.972598456     4.414155243     0.838294120
34      H        5.281120246     4.979903041     0.853730252
35      H        6.419844574     2.766860882     2.715155662
36      H        4.763082354     3.448121473     2.787078222
37      H        6.173389993     4.484408011     3.144890681
---
H16C12O2, RHF, CHARGE=0, MULT=1
HF=-74.4
1       C       -0.048638717    -0.079291055    -0.095286613
2       C        1.352730535    -0.129377434    -0.005474534
3       C        2.059520409     0.766138704     0.832113616
4       C        1.320395566     1.714555743     1.575765030
5       C       -0.083043633     1.763155378     1.485014248
6       C       -0.771428001     0.867671855     0.650456092
7       C        3.591408371     0.677954152     0.900037598
8       O        3.990855465     0.425820596     2.226888762
9       C        5.355118380     0.464057708     2.545411805
10      C        6.075921761     1.769108106     2.034467884
11      C        5.529892892     2.095991125     0.592781071
12      O        4.148691093     1.896688958     0.466577667
13      C        7.606838428     1.523389666     1.962905742
14      C        5.804192339     2.953375779     3.003224140
15      H       -0.575017839    -0.777961856    -0.746361960
16      H        1.885405142    -0.875921916    -0.597445688
17      H        1.822855705     2.426409055     2.232858951
18      H       -0.636529056     2.501069460     2.066734193
19      H       -1.858802388     0.906135548     0.581186003
20      H        3.951754305    -0.165198506     0.229246514
21      H        5.420502778     0.380355051     3.659418419
22      H        5.871212209    -0.445578889     2.135045423
23      H        5.719037340     3.166301119     0.326015123
24      H        6.077496279     1.483132924    -0.173100012
25      H        8.144090418     2.420572314     1.593416024
26      H        7.863715979     0.686119867     1.281766144
27      H        8.024609512     1.274237928     2.959725048
28      H        6.313566279     3.878405347     2.662595705
29      H        4.725203969     3.187861414     3.092283896
30      H        6.176759997     2.731689498     4.024688649
---
H16C12O2, RHF, CHARGE=0, MULT=1
HF=-101.1
1       C       -0.000249416    -0.000122237    -0.000214811
2       C        1.420364446     0.000344795     0.000487301
3       C        2.111571993     1.246341708    -0.000109199
4       C        1.376078873     2.463685735     0.087672506
5       C       -0.041407975     2.451372704    -0.057956909
6       C       -0.741851624     1.210925173    -0.044213276
7       C       -0.738422997    -1.315181293     0.045126245
8       O       -1.092289002    -1.925590184     1.053433091
9       O       -1.020671645    -1.841980779    -1.175933093
10      C        2.199639085    -1.298617643     0.004576771
11      C        3.622157937     1.278967951    -0.106100426
12      C        2.097816826     3.768126314     0.358602536
13      C       -0.798196133     3.750370672    -0.241502311
14      C       -2.256175277     1.193977624    -0.075564456
```

```
15    H    -1.485344512   -2.670231391   -1.164754167
16    H     1.598275226   -2.182568191    0.292533447
17    H     3.039342789   -1.269620979    0.730109891
18    H     2.613093458   -1.507228524   -1.005320292
19    H     3.993642298    2.197924795   -0.604148649
20    H     4.091586906    1.220479028    0.898797836
21    H     4.023378408    0.445353450   -0.717918961
22    H     3.006640031    3.625362292    0.978640165
23    H     2.402712816    4.262566917   -0.587835371
24    H     1.473581060    4.483281974    0.932773694
25    H    -1.684462753    3.634024090   -0.898225469
26    H    -1.146348541    4.149637217    0.734658923
27    H    -0.185337057    4.534449974   -0.731568886
28    H    -2.690925310    1.979873425    0.576446531
29    H    -2.624767515    1.353785325   -1.111353853
30    H    -2.699373707    0.244892642    0.283784754
 ---
H22C12O2, RHF, CHARGE=0, MULT=1
HF=-126.3
1     C    -0.000121574   -0.000133680   -0.000146341
2     C     1.567051702    0.000121251    0.000069250
3     C     2.312738910    1.347342044   -0.000023147
4     O     2.373171828    2.048784689    1.002710139
5     C     3.095953242    1.723650939   -1.272237235
6     C     3.289991059    3.235575576   -1.513223435
7     O     2.290534034    3.939357135   -1.599302198
8     C     4.721783710    3.835265972   -1.698587514
9     C     5.573809895    3.572320793   -0.424459939
10    C     5.386253805    3.188625139   -2.947554970
11    C     4.659435934    5.378265942   -1.924838052
12    C    -0.564641724    0.781398890   -1.219408453
13    C    -0.572851318    0.608376979    1.312044947
14    C    -0.466495887   -1.485922252   -0.099669784
15    H     1.919290759   -0.586897662   -0.879208991
16    H     1.928098757   -0.565761911    0.891606679
17    H     4.067779846    1.186848143   -1.210645719
18    H     2.579456501    1.317660902   -2.171970276
19    H     5.080308540    3.974246006    0.483075457
20    H     5.759629271    2.493201452   -0.253281615
21    H     6.569409088    4.055680721   -0.503049971
22    H     4.751537017    3.304293718   -3.849751625
23    H     5.576245468    2.104769300   -2.813787569
24    H     6.366541309    3.657741216   -3.169406380
25    H     4.076764828    5.641778692   -2.830296690
26    H     4.202448100    5.903949435   -1.062662993
27    H     5.676920485    5.800205482   -2.061177416
28    H    -0.167296042    0.388818179   -2.177727947
29    H    -0.318707418    1.860631581   -1.173261626
30    H    -1.670291105    0.704604467   -1.269936981
31    H    -0.400496536    1.700352312    1.379442692
32    H    -0.122283467    0.141206250    2.210969396
33    H    -1.670044187    0.454364992    1.378657570
34    H    -0.104497391   -1.970138095   -1.029692390
35    H    -0.095748522   -2.090589688    0.752887667
36    H    -1.572656323   -1.565986964   -0.100596200
 ---
H24C12O2, RHF, CHARGE=0, MULT=1
HF=-146.3
1     C     0.000012078    0.000000138    0.000047957
```

```
 2    C     1.540839440    -0.000005301    -0.000067866
 3    O     2.001353764     1.332140210    -0.000194513
 4    C     3.310804245     1.710552980    -0.002238624
 5    C     3.474804975     3.236065877    -0.002460690
 6    C     4.933885672     3.735973884    -0.000746126
 7    C     5.054148809     5.274739936    -0.003109162
 8    C     6.509411312     5.786484753    -0.002514362
 9    C     6.632289004     7.324198589    -0.004088713
10    C     8.087223970     7.836694077    -0.003707333
11    C     8.210741479     9.374082781    -0.004571891
12    C     9.663480856     9.888378637    -0.004430830
13    O     4.192456034     0.856005669    -0.003118858
14    C     9.801961665    11.413316305    -0.007212796
15    H    -0.359464360    -1.049433812     0.015610853
16    H    -0.414950053     0.487015566    -0.904772783
17    H    -0.415919859     0.514170422     0.889185015
18    H     1.912735731    -0.553362076    -0.901415487
19    H     1.913207540    -0.552907118     0.901432857
20    H     2.944016392     3.636358962     0.892050433
21    H     2.946486547     3.635453543    -0.898856858
22    H     5.461848124     3.328403353    -0.892536802
23    H     5.458668476     3.331290239     0.894291409
24    H     4.523664335     5.683258379     0.887478185
25    H     4.525378770     5.680407740    -0.896019377
26    H     7.038961575     5.377663596    -0.893431273
27    H     7.037562443     5.379581032     0.890113272
28    H     6.102074558     7.732580535     0.886760384
29    H     6.103229192     7.730630930    -0.896525278
30    H     8.617023631     7.428590550    -0.894917439
31    H     8.616279772     7.429567845     0.888390248
32    H     7.681532052     9.783206951     0.886679926
33    H     7.681891823     9.782222184    -0.896480438
34    H    10.200376608     9.485089326    -0.894124345
35    H    10.198945820     9.488244720     0.887519667
36    H    10.873975587    11.699878499    -0.003463855
37    H     9.331731264    11.872501354     0.885887175
38    H     9.339164269    11.868245792    -0.906368867
---
H8C13O2, RHF, CHARGE=0, MULT=1
HF=-23.5
 1    C    -0.000089720    -0.000226058     0.000067612
 2    C     1.496265266     0.000374192     0.000010864
 3    C     2.194648227     1.242808960     0.000073110
 4    O     1.550769616     2.444784494    -0.000093745
 5    C     0.189078966     2.515358774    -0.000156442
 6    C    -0.636974700     1.353913062    -0.000078695
 7    C    -2.041853261     1.544059281    -0.000186857
 8    C     2.267598203    -1.189160888    -0.000079863
 9    C     3.616837295     1.294583009     0.000414418
10    C    -0.370717508     3.823827384    -0.000298846
11    C     3.668544670    -1.142640868     0.000071135
12    C     4.340191655     0.096272381     0.000468634
13    C    -2.597381501     2.830887930    -0.000560681
14    C    -1.763220816     3.966911700    -0.000468746
15    O    -0.658750479    -1.037586444     0.000316638
16    H    -2.714957150     0.684124455    -0.000078498
17    H     1.777016517    -2.164656137    -0.000197714
18    H     4.142090495     2.249948515     0.000602027
19    H     0.269311346     4.706320942    -0.000259166
```

```
20    H     4.242768852   -2.069528027   -0.000117280
21    H     5.430974520    0.121906319    0.000879392
22    H    -3.680663143    2.955026187   -0.000552196
23    H    -2.205400022    4.964367036   -0.000626631
---
H10C13O2, RHF, CHARGE=0, MULT=1
HF=-34
1     O     0.027609303    0.002655435   -0.302871892
2     C     1.214395759   -0.036028113    0.000437514
3     C     2.143158374    1.134827978    0.132512648
4     O     1.828960367   -1.230805626    0.260274897
5     C     1.192132655   -2.450105753    0.245688333
6     C     1.131995536   -3.203447007   -0.954402116
7     C     0.579945210   -4.495533354   -0.926195272
8     C     0.093961539   -5.038992261    0.276131819
9     C     0.156971951   -4.288243512    1.463450342
10    C     0.702561291   -2.993121822    1.460902349
11    C     2.220491698    1.846262941    1.351269847
12    C     3.081014572    2.951486705    1.471442614
13    C     3.868623024    3.356967094    0.379662074
14    C     3.792612425    2.655304497   -0.836218024
15    C     2.935265423    1.547960072   -0.962064593
16    H     1.506681020   -2.798413736   -1.894736334
17    H     0.529357571   -5.079718993   -1.845930292
18    H    -0.331678360   -6.042600193    0.287869456
19    H    -0.220227104   -4.712205190    2.394997532
20    H     0.745237962   -2.424842024    2.390398852
21    H     1.614941323    1.546486097    2.207968437
22    H     3.136410176    3.494635137    2.415391432
23    H     4.536403102    4.213767602    0.475316870
24    H     4.400300474    2.968607916   -1.685837670
25    H     2.889492235    1.014742790   -1.912832040
---
H9C8N1O2, RHF, CHARGE=0, MULT=1
HF=2.1
1     C    -0.000169291   -0.000346082    0.000172932
2     C     1.402562152    0.000175958   -0.000205211
3     C     2.135876264    1.210868382    0.000012125
4     C     1.383047741    2.415553194    0.001273636
5     C    -0.037007705    2.455199862    0.002225387
6     C    -0.709617404    1.209877994    0.001647182
7     H    -0.542492030   -0.947095490   -0.000769345
8     N     2.133553894    3.725640317    0.001810352
9     O     2.427351720    4.238610053    1.057873709
10    O     2.427977315    4.239432190   -1.053719599
11    H     1.925790399   -0.958080473   -0.001221777
12    C     3.644654850    1.154469839   -0.001231899
13    C    -0.849698278    3.727667019    0.004880078
14    H    -1.800840329    1.175938726    0.002458684
15    H     4.067962387    1.645360449   -0.901715282
16    H     4.015855885    0.108956806   -0.003811639
17    H     4.069159631    1.641506519    0.900890844
18    H    -0.636896163    4.347211566   -0.890613570
19    H    -0.648425277    4.334299585    0.911881801
20    H    -1.939161196    3.518471320   -0.003963285
---
H9C8N1O2, RHF, CHARGE=0, MULT=1
HF=-67
1     C     0.000347355   -0.000495166   -0.000100492
```

```
 2      C      1.405352234     -0.000787661    -0.000112056
 3      C      2.112565378      1.213701963     0.000080232
 4      C      1.434765908      2.457432500    -0.000523665
 5      C      0.019590019      2.438089991     0.003208391
 6      C     -0.689017885      1.223392891     0.002208815
 7      C      2.248842786      3.752560628    -0.008230496
 8      N      1.984739505      4.677181820    -1.120308489
 9      C      2.078528141      4.532248418     1.317913986
10      O      3.109044726      4.372694400     2.186875651
11      O      1.132957919      5.245216724     1.648724630
12      H     -0.549638062     -0.941826326    -0.000550378
13      H      1.950894100     -0.945161858     0.000513864
14      H      3.203910509      1.176879769     0.002475800
15      H     -0.556948017      3.364818189     0.013078945
16      H     -1.779592722      1.234933052     0.005404008
17      H      3.335957003      3.482144271    -0.109702043
18      H      2.207899470      4.221650730    -1.992175694
19      H      1.009893000      4.924382197    -1.186970305
20      H      3.034354621      4.854892608     3.001467109
---
H9C8N1O2, RHF, CHARGE=0, MULT=1
HF=2.1
 1      C      0.000078364      0.000359811    -0.000082820
 2      C      1.420579233     -0.001047424    -0.000323272
 3      C      2.060432280      1.261476555     0.000453266
 4      C      1.319475730      2.452511435     0.001596922
 5      C     -0.082725826      2.414107413     0.001808326
 6      C     -0.784265324      1.184793691     0.001317343
 7      N     -0.714980210     -1.329503611    -0.002027918
 8      O     -0.991637590     -1.854373754     1.052786824
 9      O     -0.998689092     -1.846863318    -1.058690196
10      C      2.266650934     -1.251468892    -0.001030847
11      H      3.150285337      1.323918647     0.000080843
12      H     -0.631167586      3.358055914     0.002266011
13      C     -2.294036889      1.202108724     0.001406541
14      H      1.836368388      3.413334787     0.002293061
15      H      3.350214287     -1.013435952    -0.004668835
16      H      2.073282243     -1.871494929    -0.900642454
17      H      2.078268058     -1.868135927     0.901902437
18      H     -2.706091246      0.690319777     0.895435998
19      H     -2.704293898      0.714488966    -0.906958748
20      H     -2.692103082      2.237711759     0.015478181
---
H9C8N1O2, RHF, CHARGE=0, MULT=1
HF=-64.2
 1      C      0.000836247      0.000135512    -0.000024166
 2      C      1.406403315     -0.000719981    -0.000022836
 3      C      2.121700135      1.208939248     0.000061403
 4      C      1.425257518      2.443972361     0.012737739
 5      C      0.006993178      2.439318040     0.007630611
 6      C     -0.694001330      1.221677471    -0.000399691
 7      N      2.171634711      3.663430458    -0.074321031
 8      C      2.966092056      4.102671327     1.072764623
 9      C      2.152877050      4.557659132     2.303950776
10      O      2.615859137      4.095466253     3.491334657
11      O      1.171811545      5.301436015     2.299753921
12      H     -0.547464605     -0.941979668    -0.002228716
13      H      1.948936044     -0.947034404    -0.005521263
14      H      3.211869714      1.175116926    -0.018661461
```

```
15         H         -0.560801172      3.371231489      0.006686424
16         H         -1.784756773      1.227486883     -0.005213240
17         H          1.595452915      4.416856102     -0.418634928
18         H          3.683913421      3.297284238      1.356868614
19         H          3.578892570      4.981104322      0.745266362
20         H          2.136084632      4.381398761      4.259427256
---
H17C8N1O2, RHF, CHARGE=0, MULT=1
HF=-125
1          C          0.000011582      0.000090785      0.000019004
2          C          1.545923864     -0.000021749     -0.000015607
3          C          2.159840265      1.415297104      0.000004432
4          C          3.702381402      1.428042952      0.009728654
5          C          4.316290285      2.843215532      0.002585798
6          C          5.859638009      2.854123209      0.015595107
7          C          6.461834409      4.273919557      0.004702330
8          C          7.992602429      4.330237377      0.015264879
9          N         -0.533481910     -1.364580911     -0.090758501
10         O          8.790798982      3.395723502      0.073852147
11         O          8.488850941      5.594942691     -0.045958569
12         H         -0.376612002      0.540853771      0.907816811
13         H         -0.380702505      0.572711072     -0.881996833
14         H          1.914301965     -0.555258984      0.893342972
15         H          1.912741164     -0.558081542     -0.891375684
16         H          1.797104148      1.970197123     -0.895578315
17         H          1.786180470      1.975957049      0.887692918
18         H          4.065448762      0.878657258      0.908434712
19         H          4.077147707      0.864406453     -0.875132293
20         H          3.955327287      3.392536528     -0.897131829
21         H          3.940461404      3.408406323      0.886273637
22         H          6.219770674      2.310197294      0.918147108
23         H          6.235522347      2.290750394     -0.868472336
24         H          6.105955579      4.821926260     -0.898165732
25         H          6.097325851      4.840704364      0.892635319
26         H         -0.370217346     -1.859916358      0.772624402
27         H         -1.536914671     -1.318363892     -0.182190288
28         H          9.435592592      5.664207537     -0.039597446
---
H7C9N1O2, RHF, CHARGE=0, MULT=1
HF=-55.9
1          C         -0.000413011     -0.000780557      0.000018335
2          C          1.399167631     -0.000836975     -0.000005304
3          C          2.129626746      1.216551441     -0.000011305
4          C          1.415280803      2.412483927     -0.000007662
5          C         -0.017746295      2.412993416      0.000031108
6          C         -0.731357371      1.216525499      0.000046772
7          C         -0.482461710      3.837063336     -0.000008583
8          N          0.698085325      4.652298598      0.000847952
9          C          1.878920691      3.835825349     -0.000069565
10         O          3.008394767      4.296854150     -0.000741778
11         O         -1.613663150      4.293290492     -0.000560057
12         C          0.725969230      6.117748467     -0.000825513
13         H         -0.544083433     -0.946573366     -0.000003429
14         H          1.943405567     -0.946421922     -0.000023745
15         H          3.219514862      1.197802279     -0.000012611
16         H         -1.821148467      1.196634463      0.000092306
17         H          1.264839335      6.496865674      0.897896559
18         H          1.243692273      6.495261760     -0.912663480
19         H         -0.299392254      6.545474814      0.011182329
```

```
---
H11C9N1O2, RHF, CHARGE=0, MULT=1
HF=-6.4
1    C     0.000282020    0.000098446    0.000403550
2    C     1.410620235   -0.000163998   -0.000538880
3    C     2.082178630    1.242277526   -0.000681125
4    C     1.389714043    2.474832159    0.001506765
5    C    -0.028573231    2.415486660    0.006990920
6    C    -0.754752708    1.197083755    0.005271501
7    C     2.195114445   -1.286875661   -0.001628014
8    C    -2.262592117    1.110833977    0.010027022
9    C     2.181998555    3.760572377   -0.001622434
10   N    -0.801266719    3.712431923    0.017523979
11   O    -1.109698692    4.226359537   -1.033878787
12   O    -1.098464104    4.215001121    1.077636945
13   H    -0.528541886   -0.955982124   -0.002514677
14   H     3.175090052    1.248693766   -0.001505457
15   H     2.830861382   -1.358561545   -0.908995873
16   H     2.857442351   -1.345310442    0.887441137
17   H     1.541273906   -2.181856766    0.014907684
18   H    -2.614906604    0.058737281    0.000992663
19   H    -2.691919918    1.582923931    0.917979261
20   H    -2.699416046    1.600773932   -0.884498314
21   H     1.955920470    4.374030995   -0.898126844
22   H     1.973424033    4.366060993    0.904473848
23   H     3.275081570    3.570405605   -0.013389298
---
H11C9N1O2, RHF, CHARGE=0, MULT=1
HF=-69.3
1    C    -0.070104152    0.021833641   -0.055879884
2    C     1.074222701    0.701730848    0.426917416
3    C     1.003514722    2.109544253    0.555855655
4    C    -0.169990052    2.810153516    0.226559692
5    C    -1.299308893    2.117876162   -0.241919735
6    C    -1.244599541    0.720688012   -0.383418702
7    C     2.349417093   -0.047559145    0.751830944
8    C     2.319536937   -0.846837717    2.093274956
9    C     3.622837975   -1.662296762    2.276895864
10   O     4.758869124   -1.215604695    2.426409164
11   O     3.446066357   -3.007486531    2.260129207
12   H    -0.054776468   -1.061952105   -0.185562707
13   H     1.866403855    2.676857234    0.909883951
14   H    -0.200971409    3.895030949    0.334093831
15   H    -2.210200123    2.660374267   -0.495966479
16   H    -2.114886138    0.175184385   -0.750131533
17   H     2.556859767   -0.768381637   -0.073961284
18   H     3.210824228    0.658679563    0.747689245
19   H     4.228188900   -3.524136701    2.411641011
20   N     2.152355465   -0.072594536    3.330011615
21   H     1.455203428   -1.561206767    2.039655217
22   H     1.201385343    0.252714525    3.405138038
23   H     2.727280449    0.755587152    3.331051945
---
H7C10N1O2, RHF, CHARGE=0, MULT=1
HF=8.6
1    C     0.000299717    0.000064237    0.000033645
2    C     1.386499792   -0.000289992    0.000007789
3    C     2.112549696    1.224210174    0.000006421
4    C     1.424965560    2.424530024    0.000043521
```

```
5    C     -0.010353299     2.455033338     0.000152340
6    C     -0.752401868     1.227788005     0.000149222
7    C     -2.211417340     1.275758745     0.000296366
8    C     -2.870689363     2.525855476     0.000348029
9    C     -2.095660160     3.752719601     0.000367124
10   C     -0.723002940     3.707328499     0.000285374
11   N     -2.812776299    -0.038177968     0.000452700
12   O     -3.974767422    -0.162346572     0.000603861
13   O     -4.198402652     2.748930359     0.000414099
14   H     -0.508438769    -0.965189512    -0.000068773
15   H      1.935874674    -0.942985126    -0.000012705
16   H      3.202622023     1.202737157    -0.000031885
17   H      1.978249399     3.365499673     0.000011530
18   H     -2.613603927     4.713402847     0.000448669
19   H     -0.154955250     4.640778061     0.000329705
20   H     -4.765570854     1.992424819     0.000351898
---
H13C10N1O2, RHF, CHARGE=0, MULT=1
HF=-53.7
1    C     -0.005909181     0.016290509    -0.052440271
2    C      1.184273131     0.338308412     0.632582221
3    C      1.841418854     1.561703716     0.417003125
4    C      1.300870260     2.502985755    -0.495827386
5    C      0.089556379     2.190163533    -1.181234671
6    C     -0.547855202     0.964551484    -0.952842894
7    C     -0.726242716    -1.271895652     0.231804544
8    O     -1.488878909    -1.389810352     1.189215152
9    N     -0.529918387    -2.375821824    -0.636713061
10   C      0.495304938    -2.355432932    -1.681489117
11   C     -1.275380853    -3.625228069    -0.430777875
12   H      1.615949358    -0.363773427     1.349211335
13   H      2.760530212     1.754814555     0.969553979
14   H     -0.349771414     2.899312580    -1.883876724
15   H     -1.479887642     0.751796956    -1.480757435
16   H      0.242157738    -1.620009119    -2.480392377
17   H      1.498190864    -2.096262752    -1.269334083
18   H      0.596612168    -3.349657612    -2.171232339
19   H     -2.362566143    -3.429707989    -0.291085766
20   H     -1.178938511    -4.297034236    -1.313517684
21   H     -0.895714155    -4.179633276     0.459244681
22   O      1.847248673     3.708062742    -0.805527045
23   H      2.784213263     4.368318349     0.956707917
24   H      3.885794206     3.677018561    -0.308611778
25   C      2.949822779     4.263550522    -0.142669258
26   H      3.096117826     5.282842460    -0.573038607
---
H15C10N1O2, RHF, CHARGE=0, MULT=1
HF=-45.7
1    C     -0.000044687     0.000176731    -0.000107886
2    C      1.567867995    -0.000195014     0.000019636
3    C      2.074030182     1.470983242    -0.000012345
4    C      1.554664588     2.206585377    -1.266495215
5    C      0.000972195     2.212647156    -1.251168472
6    C     -0.532729233     0.751323003    -1.268797775
7    C     -0.510693212     0.739439323     1.285785453
8    H      1.981358261    -0.531872949    -0.884129737
9    H      1.963506115    -0.540943515     0.887486803
10   H      3.186682248     1.464666640    -0.007819769
11   C      1.572086965     2.201829792     1.274969020
```

```
12      H     1.936627614    1.717667468   -2.189076442
13      H     1.949116775    3.245382893   -1.303916768
14      H    -0.374472908    2.739598762   -2.156482926
15      C    -0.507951296    2.946599496    0.019443889
16      N    -0.513340104   -1.485659246   -0.040980384
17      H    -0.212564207    0.251097385   -2.208417320
18      H    -1.621333589    0.748860217    1.335648561
19      C     0.019119677    2.202664111    1.276753460
20      H    -0.168796373    0.225544472    2.210517702
21      H     1.965102912    3.240887021    1.311679099
22      H     1.966077995    1.710081772    2.190914274
23      O    -1.259041820   -1.902525396    0.814656774
24      H    -1.618694479    2.994644316    0.027561384
25      H    -0.171814065    4.006187595    0.020979050
26      H    -0.348031820    2.719036598    2.191550314
27      O    -0.159639087   -2.211680996   -0.942388279
28      H    -1.644538323    0.750613454   -1.292705212
---
H15C10N1O2, RHF, CHARGE=0, MULT=1
HF=-43
1       C    -0.000013041   -0.000452404   -0.000019211
2       C     1.559984916   -0.000286128   -0.000140441
3       C     2.082767074    1.462453630   -0.000319174
4       C     1.566560114    2.195228111   -1.269010378
5       C     0.006527226    2.200404855   -1.273752857
6       C    -0.471204711    0.709175696   -1.314456304
7       C    -0.507923904    0.739897344    1.270067455
8       H    -0.344896000   -1.057689896    0.025693647
9       H     1.953416272   -0.558981255   -0.877322100
10      H     1.939862608   -0.552826195    0.886623417
11      H     3.195060503    1.456756782   -0.004783532
12      C     1.572698394    2.198386515    1.270055202
13      H     1.961928054    1.713785979   -2.190163307
14      H     1.949727426    3.238326047   -1.302957595
15      H    -0.333530500    2.750275317   -2.178817955
16      C    -0.502088355    2.933869751   -0.000042757
17      N    -2.002472674    0.544144813   -1.607385066
18      H    -0.021343197    0.194213706   -2.202125409
19      H    -1.616690982    0.724827905    1.334021156
20      C     0.018550738    2.201186017    1.266723449
21      H    -0.164503418    0.203143481    2.181877946
22      H     1.966915805    3.237413650    1.305511440
23      H     1.962285742    1.708735422    2.188992127
24      O    -2.584286289   -0.447184221   -1.229872451
25      H    -1.610612032    3.000345065    0.015770546
26      H    -0.155493218    3.990762900   -0.009950603
27      H    -0.350894078    2.727795803    2.174300869
28      O    -2.580186660    1.369050660   -2.277908549
---
H8C8N2O2, RHF, CHARGE=0, MULT=1
HF=-70.3
1       C     0.000084610    0.000094726   -0.000089819
2       C     1.405320191   -0.000271370    0.000154163
3       C     2.116757015    1.211631171   -0.000163020
4       C     1.422000918    2.441556169   -0.004705792
5       C     0.009440455    2.438129573   -0.000832001
6       C    -0.708880271    1.221801912    0.003336788
7       C    -2.212128895    1.227988779   -0.026260398
8       O    -2.883723376    1.212820830   -1.057444993
```

```
 9    N    -2.853236128    1.241621240     1.207674444
10    C     2.178125282    3.740872159     0.024173521
11    O     2.529927079    4.313741447     1.055043087
12    N     2.491479441    4.299998833    -1.209930153
13    H     1.947123872   -0.946818905     0.000423495
14    H     3.207824572    1.189989157     0.005522084
15    H    -0.533056879    3.386154209    -0.001136097
16    H    -2.351677638    1.283065101     2.063400070
17    H    -3.844823377    1.265844689     1.287925380
18    H     2.208032977    3.884737128    -2.065933504
19    H     2.976213427    5.165313611    -1.290432806
20    H    -0.533805470   -0.951775081    -0.005239044
---
H8C8N2O2, RHF, CHARGE=0, MULT=1
HF=-69.7
 1    C     0.000418136    0.000408326    -0.000095415
 2    C     1.412568657   -0.001033392     0.000177728
 3    C     2.128511724    1.208967199    -0.000083011
 4    C     1.447329761    2.445904981     0.003252909
 5    C     0.035183039    2.447329627     0.002895847
 6    C    -0.680860292    1.237330045     0.003280523
 7    C    -0.764879336   -1.293821160     0.030817211
 8    O    -1.106111291   -1.871697612     1.062284201
 9    N    -1.092111706   -1.846090147    -1.202854077
10    C     2.213110423    3.739766872    -0.027784517
11    O     2.559724550    4.314192374    -1.059343329
12    N     2.534189609    4.295586416     1.205860175
13    H     1.965939954   -0.941863217     0.003406534
14    H     3.219492890    1.177207753    -0.005657992
15    H    -0.517708723    3.388353573    -0.000465987
16    H    -1.771827868    1.269249944     0.009059618
17    H    -0.836657567   -1.415362443    -2.059945618
18    H    -1.596051713   -2.700386851    -1.283126647
19    H     2.274557378    3.867313489     2.062915953
20    H     3.038744334    5.149564645     1.285925894
---
H10C8N2O2, RHF, CHARGE=0, MULT=1
HF=17.4
 1    C     0.000057669   -0.000187799    -0.000314454
 2    C     1.417022401    0.000135744     0.000100879
 3    C     2.126225249    1.214484463    -0.000020181
 4    C     1.447600646    2.445238955     0.000456326
 5    C     0.038472767    2.430766094     0.001770483
 6    C    -0.702881096    1.233061236     0.001281007
 7    N    -0.790255621   -1.198301545    -0.003431232
 8    N    -0.699490568    3.741043666     0.003144550
 9    O    -0.949848789    4.276717158    -1.052321606
10    O    -1.024186091    4.232762550     1.059713465
11    H     1.981210557   -0.934237829     0.000189249
12    H     3.217151295    1.202747723    -0.000370653
13    C    -0.796219000   -1.969057560    -1.251842101
14    H    -1.794650492    1.245774655     0.001313107
15    C    -0.823402636   -1.955556961     1.252784700
16    H    -0.888024565   -1.286552014    -2.126042973
17    H     0.121722508   -2.588362894    -1.396481294
18    H    -1.673680501   -2.654507518    -1.260592550
19    H    -0.931754915   -1.263703991     2.117730772
20    H    -1.702321376   -2.639022918     1.251190607
21    H     0.090375015   -2.574495630     1.422122875
```

```
22         H          2.015314244        3.376282517        0.000015243
 ---
H10C8N2O2, RHF, CHARGE=0, MULT=1
HF=16.1
1          C         -0.000026601        0.000421299        0.000354315
2          C          1.417822915        0.000229991       -0.000489374
3          C          2.132253621        1.211200631        0.000305915
4          C          1.416904666        2.424340365        0.002016903
5          C          0.008546059        2.452926266        0.003135825
6          C         -0.696540811        1.236155857        0.002101293
7          N         -0.788078308       -1.199747228       -0.000715378
8          N          2.175956815        3.721403583        0.002448631
9          O          2.460905401        4.236819288        1.059236396
10         O          2.486226023        4.222177293       -1.054225794
11         H          1.982802572       -0.933403350       -0.001701536
12         H          3.223155924        1.187917516       -0.000841179
13         C         -0.798716709       -1.970421364        1.247783397
14         H         -1.787914440        1.261400327        0.002549423
15         H         -0.546292380        3.392474699        0.004810923
16         C         -0.812933930       -1.957394746       -1.257045484
17         H          0.120727604       -2.585821623        1.398487169
18         H         -0.899044956       -1.287868177        2.121150894
19         H         -1.673617032       -2.659286586        1.251250880
20         H         -0.916839917       -1.265568452       -2.122532490
21         H          0.102240215       -2.575622786       -1.421413597
22         H         -1.691477984       -2.641611406       -1.260409485
 ---
H12C11N2O2, RHF, CHARGE=0, MULT=1
HF=-51.6
1          C          0.031886818        0.006020189       -0.572792230
2          C          1.427887488       -0.029767260       -0.328175588
3          C          2.159974160        1.210192894       -0.380437923
4          C          1.538719922        2.453562914       -0.650866052
5          C          0.163017436        2.442794446       -0.877099986
6          C         -0.580903290        1.230727909       -0.840532627
7          N          3.509319004        0.908460524       -0.163436047
8          C          2.391757060       -1.080319462       -0.016093034
9          C          3.635557237       -0.460783761        0.100531294
10         C          2.124828391       -2.548050186        0.104185097
11         C          2.199013320       -3.110639299        1.560944208
12         C          1.082991174       -2.558661288        2.475994216
13         O         -0.128013363       -2.732474514        2.348300841
14         O          1.530378169       -1.817527817        3.520766020
15         H         -0.561453036       -0.907743642       -0.552615053
16         H          2.108388739        3.380761752       -0.682219883
17         H         -0.359889753        3.376952928       -1.087356982
18         H         -1.654965274        1.269284074       -1.027650928
19         H          0.856895967       -1.486299399        4.102580648
20         N          2.125361348       -4.576430062        1.671293349
21         H          3.196479688       -2.814694761        1.983166076
22         H          2.962417824       -4.987123890        1.287046317
23         H          1.364737839       -4.951946016        1.125602438
24         H          4.211981274        1.581949486        0.041157555
25         H          4.609759167       -0.881168585        0.325978952
26         H          2.880444359       -3.092702957       -0.513523804
27         H          1.140472904       -2.796265367       -0.352447731
 ---
H9C6N3O2, RHF, CHARGE=0, MULT=1
HF=-63.5
```

```
 1    H     0.240150439   -0.266837627    0.147009849
 2    N     1.218957140   -0.162319565   -0.070171723
 3    C     1.639432731    1.232904249    0.135928413
 4    C     0.884149559    2.201251935   -0.812327861
 5    O     1.092545708    2.369941320   -2.013797316
 6    C     3.189720538    1.321888976    0.003987527
 7    C     3.836699019    2.586500366    0.484023069
 8    C     5.176466628    2.771722657    0.846400783
 9    N     5.290247100    4.120906640    1.190495535
10    C     4.036291887    4.709986778    1.029588265
11    N     3.151584204    3.798782938    0.603595506
12    H     3.499295018    1.154946354   -1.055300104
13    H     6.018792932    2.095494893    0.887115966
14    H     6.122182867    4.574130213    1.490794975
15    H     3.835259729    5.757305485    1.228964087
16    H     3.629419416    0.477377468    0.588066450
17    H     1.374909785    1.487710558    1.196868984
18    H     1.321048403   -0.445840870   -1.032043277
19    O    -0.142589653    2.870888364   -0.231813994
20    H    -0.564413805    3.522876065   -0.778012904
---
H14C6N4O2, RHF, CHARGE=0, MULT=1
HF=-84.3
 1    H     0.026959158    0.068995581   -0.060645172
 2    N     0.995792985    0.063691834    0.217701359
 3    C     1.513439633    1.433490963    0.292496999
 4    C     1.402197855    2.199721733   -1.054168760
 5    O     1.491638255    1.716505014   -2.183014527
 6    C     2.985739566    1.453642873    0.809979530
 7    C     3.159858793    0.971379019    2.262867387
 8    C     4.556171402    1.183916847    2.902605753
 9    N     5.629063403    0.277146077    2.469890032
10    C     6.598872017    0.618563291    1.492533182
11    N     7.508994681   -0.437693172    1.178670103
12    N     6.596380396    1.797238718    0.945147656
13    H     0.868022245    1.972578116    1.037096696
14    H     5.290393265   -0.665852080    2.358503538
15    H     8.421039041   -0.093774517    0.925939531
16    H     7.645014158   -1.094917459    1.928881767
17    H     7.369380465    1.985841837    0.334035955
18    H     4.867819972    2.249600877    2.809011596
19    H     4.442438885    0.994914415    4.004167204
20    H     2.438843555    1.520586581    2.914053076
21    H     2.880009389   -0.103478726    2.340780442
22    H     3.623417382    0.844066628    0.131834510
23    H     3.357659673    2.501022747    0.735238747
24    H     1.480914072   -0.487018363   -0.473367052
25    O     1.167186672    3.529401492   -0.932844752
26    H     1.119477245    4.014417060   -1.748077828
---
H14C10O3, RHF, CHARGE=0, MULT=1
HF=-103.6
 1    C    -0.121394974    0.167463223   -0.077447779
 2    C     1.269344096    0.322218780    0.063816836
 3    C     1.798501115    1.485543238    0.645625123
 4    C     0.923373968    2.488494734    1.092678560
 5    C    -0.467812113    2.333773817    0.952945689
 6    C    -1.018865012    1.176896831    0.352124708
 7    C    -2.556428486    0.970090987    0.194670503
```

```
 8    O    -3.048277154    0.044297133    1.132676195
 9    O    -2.807451501    0.450279031   -1.085337050
10    O    -3.319354101    2.144277044    0.365465832
11    C    -3.516968360    3.133699762   -0.601298988
12    C    -3.863072113   -0.395629390   -1.448611841
13    C    -3.205270734    0.283530994    2.500880947
14    H    -0.485821599   -0.759031767   -0.524962417
15    H     1.937749126   -0.468271152   -0.279791130
16    H     2.876649742    1.607293066    0.752288738
17    H     1.320966461    3.393698881    1.553140249
18    H    -1.101151503    3.136459016    1.334185871
19    H    -2.622482618    3.335065607   -1.237427548
20    H    -4.373514845    2.886716292   -1.275253753
21    H    -3.772737847    4.074964596   -0.055490036
22    H    -3.668310501   -1.455421714   -1.155061590
23    H    -4.852803202   -0.091542406   -1.032923965
24    H    -3.930376956   -0.354355824   -2.563720901
25    H    -4.059973696    0.965806508    2.725017968
26    H    -2.289569093    0.701673331    2.984063447
27    H    -3.422214753   -0.706751490    2.970952851
---
H22C10O3, RHF, CHARGE=0, MULT=1
HF=-169.2
 1    C     0.000005850    0.000002132    0.000090202
 2    C     1.530367434   -0.000002647   -0.000220170
 3    C     2.161110958    1.418264773    0.000051426
 4    O     3.561085781    1.371264523    0.086901548
 5    C     4.353611717    1.588857621   -1.048965808
 6    C     5.867856522    1.453966153   -0.637355672
 7    C     6.854097752    1.683643372   -1.847640766
 8    O     6.708588367    0.618814132   -2.745325309
 9    C     7.635515876    0.304270421   -3.767725024
10    C     8.965218739   -0.265782577   -3.172130533
11    C     7.940592570    1.522775388   -4.698207824
12    C     6.917974730   -0.815490350   -4.599966504
13    O     6.227961650    2.392102139    0.337726848
14    H    -0.413845277    0.508272077   -0.894585057
15    H    -0.413626361    0.502709095    0.897801961
16    H    -0.381416762   -1.041886799   -0.003173735
17    H     1.887140821   -0.564590270    0.891428627
18    H     1.884079406   -0.569193183   -0.890215877
19    H     1.807451781    1.987752885    0.900197018
20    H     1.804580208    1.991536288   -0.894724033
21    H     4.158874037    2.602602362   -1.488691010
22    H     4.121358200    0.841587192   -1.850801468
23    H     6.027610260    0.407871622   -0.251196082
24    H     7.898616460    1.765638241   -1.450334456
25    H     6.626539263    2.662182399   -2.346785618
26    H     9.562956890    0.510703308   -2.654163675
27    H     8.773717036   -1.079043963   -2.443661225
28    H     9.605178038   -0.682455984   -3.976216539
29    H     7.012310666    1.985894425   -5.089080136
30    H     8.517063114    2.315298839   -4.180101884
31    H     8.544779465    1.203077643   -5.571392846
32    H     5.966207855   -0.454433398   -5.038872592
33    H     6.683808552   -1.703416324   -3.979223802
34    H     7.563467834   -1.153999129   -5.435861324
35    H     5.907817464    2.150782865    1.195320938
---
```

```
H24C11O3, RHF, CHARGE=0, MULT=1
HF=-172.8
1    C    0.000010233    -0.000000964   -0.000006577
2    C    1.531351826    -0.000006091    0.000008128
3    C    2.171836125     1.401278329    0.000005427
4    C    3.722971477     1.364513505   -0.003215183
5    O    4.208400520     2.680545066    0.000499779
6    C    5.593326548     2.900919849   -0.001989664
7    C    5.855374468     4.453531733    0.017114357
8    O    5.306306258     5.087822014   -1.104024629
9    C    7.391575698     4.814840949    0.013060674
10   O    7.941561965     4.423245364    1.239701022
11   C    9.193549655     4.864762928    1.730959530
12   C   10.357231575     4.627251405    0.714567455
13   C    9.139066831     6.372362883    2.145425489
14   C    9.433271401     3.983944283    3.006957998
15   H   -0.412643301     0.505833122   -0.896273062
16   H   -0.412715136     0.505783112    0.896258543
17   H   -0.381858480    -1.041881523   -0.000097672
18   H    1.882675724    -0.570710270   -0.890796537
19   H    1.882659207    -0.570394664    0.891033901
20   H    1.816264815     1.966813910   -0.891244469
21   H    1.819652141     1.965298464    0.893565177
22   H    4.089525240     0.803959791   -0.904650511
23   H    4.093486516     0.797946063    0.892706899
24   H    6.076634016     2.434981800   -0.900874971
25   H    6.072050012     2.434300251    0.897678705
26   H    5.402114435     4.867384746    0.961667261
27   H    4.367994808     5.188290161   -1.028777351
28   H    7.502333842     5.914577726   -0.171429870
29   H    7.904435911     4.296408094   -0.839386263
30   H   10.384515903     3.578520747    0.356269072
31   H   10.280166374     5.282898820   -0.175973829
32   H   11.336683436     4.841127274    1.188570834
33   H    8.273352484     6.586268387    2.803852360
34   H    9.066233772     7.048416479    1.270114374
35   H   10.056888794     6.657901249    2.698329997
36   H    8.631024010     4.122406566    3.759166827
37   H    9.475263390     2.904438205    2.758492917
38   H   10.394262831     4.252273048    3.491146538
---
H11C9N1O3, RHF, CHARGE=0, MULT=1
HF=-111.6
1    C   -0.082388357     0.029203673    0.027980409
2    C    1.147934108     0.591466542    0.450227701
3    C    1.219129792     1.997541734    0.575734095
4    C    0.114297939     2.819093751    0.302826330
5    C   -1.111298723     2.238179249   -0.113888887
6    C   -1.201357580     0.825408753   -0.249574953
7    C    2.359541631    -0.277960144    0.705462623
8    C    2.329643698    -1.080071455    2.045155143
9    C    3.566385692    -2.004981856    2.155500570
10   O    4.738705379    -1.657940766    2.287932008
11   O    3.281312446    -3.329772329    2.086275201
12   H   -0.177452507    -1.052074408   -0.094594682
13   H    2.149768016     2.477251592    0.887194874
14   H    0.218001662     3.899080572    0.413771456
15   O   -2.222985495     2.962241689   -0.396655525
16   H   -2.131467270     0.355076495   -0.570496487
```

```
17      H       2.453147858     -1.010128207    -0.131352333
18      H       3.283231754      0.342771809     0.654813741
19      H       4.023550532     -3.913324301     2.188637863
20      N       2.293020306     -0.302320220     3.290361950
21      H       1.406123998     -1.718055617     2.036457353
22      H       1.378852276      0.103955610     3.413386941
23      H       2.936938797      0.473102653     3.263626832
24      H      -2.090559229      3.896373516    -0.303708251
---
H4C4N2O3, RHF, CHARGE=0, MULT=1
HF=-121.8
1       C       0.000133557     -0.000026596     0.000052949
2       N       1.408929219      0.000432532     0.000208185
3       C       2.241407389      1.143768972    -0.000126663
4       C       1.569330259      2.512771663    -0.002646469
5       C       0.044348745      2.516234519    -0.001150636
6       N      -0.617393284      1.266296680    -0.000326832
7       O      -0.649566149     -1.039539508     0.000359097
8       O       3.454319011      0.971889692     0.001315034
9       O      -0.642675546      3.530491571    -0.000918483
10      H       1.852378625     -0.904318159     0.000933076
11      H       1.920851250      3.074900516    -0.900697883
12      H       1.922793824      3.079402061     0.891783403
13      H      -1.624960412      1.267666113     0.000173053
---
H8C4N2O3, RHF, CHARGE=0, MULT=1
HF=-137.8
1       H       0.080554984     -0.125885903    -0.379409737
2       N       0.989420644      0.151305291    -0.044286751
3       C       1.191182975      1.591279342    -0.237686294
4       C       2.722645594      1.752492983    -0.425198408
5       O       3.585981784      1.644785894     0.444027697
6       C       0.676153955      2.547076546     0.882364874
7       C      -0.849009323      2.681132745     0.989680603
8       O      -1.653924442      1.778578563     0.751108653
9       N      -1.345192087      3.908463217     1.415224229
10      H       0.679415826      1.889576600    -1.191005860
11      H      -2.319978532      4.063279692     1.546582041
12      H      -0.756558900      4.671159102     1.654080606
13      H       1.047603675      2.222078709     1.882102724
14      H       1.123574078      3.549939951     0.692834244
15      H       1.020083827     -0.104584575     0.929776201
16      O       3.100776088      2.038666030    -1.695380686
17      H       4.037896660      2.072219494    -1.845443545
---
H8C4N2O3, RHF, CHARGE=0, MULT=1
HF=-136.4
1       C      -0.029146298      0.428473802    -0.098838641
2       C       1.487858344      0.675260340    -0.254339204
3       O      -0.804220454      1.196437195     0.467445069
4       N       1.972202369      2.000922978     0.117365773
5       H       2.029265310     -0.077411522     0.372087830
6       H       1.782921945      0.498444050    -1.318946948
7       H       1.503120092      2.719606048    -0.411043438
8       C      -2.226387049     -2.051442738     0.647787263
9       C      -1.889382012     -1.251438608    -0.628859817
10      O      -3.273593729     -1.973548941     1.287003459
11      N      -0.524844539     -0.758808635    -0.667267416
12      H      -2.053088148     -1.929237334    -1.505234608
```

```
13      H       -2.617278707    -0.410251265    -0.730612922
14      H        0.140231219   -1.395351361    -1.059434558
15      H        1.786263653    2.197943675     1.088176641
16      O       -1.280537890   -2.940684639     1.044937574
17      H       -1.489824760   -3.441099389     1.824516002
---
H10C5N2O3, RHF, CHARGE=0, MULT=1
HF=-142.4
1       H       -0.015436122    0.110939483    -0.283734491
2       N        0.990463258    0.040052364    -0.297078276
3       C        1.581939305    1.383170274    -0.202353629
4       C        3.026210970    1.269116444    -0.745014672
5       O        3.452309799    1.837365675    -1.749395209
6       C        1.472991705    2.055120395     1.201782115
7       C        1.896497975    3.537506129     1.241140119
8       C        1.723969202    4.227826403     2.602481831
9       N        2.220616649    5.525646821     2.701526083
10      O        1.179799656    3.738347782     3.593077550
11      H        1.233403848   -0.530786304     0.497893906
12      H        2.067846484    1.482762360     1.948238193
13      H        1.301790978    4.110964286     0.492445120
14      H        2.158954569    6.048811642     3.546180045
15      H        2.684348404    5.980692630     1.951450048
16      H        2.965777422    3.623411467     0.940822596
17      H        0.408260080    1.981950963     1.525086127
18      H        1.017249477    2.031156842    -0.927454372
19      O        3.867445248    0.476795457    -0.035634633
20      H        4.732335082    0.360116673    -0.410417793
---
H12C6N2O3, RHF, CHARGE=0, MULT=1
HF=-152.4
1       C        0.063886285   -0.003743753     0.073669890
2       C        1.601315545    0.019606996    -0.139667768
3       N       -0.536163669    0.523028087    -1.162936457
4       C       -0.462752601   -1.382204011     0.540900143
5       O        2.245783639   -0.928887584    -0.579618093
6       H       -0.202710235    0.715583421     0.895540328
7       H       -1.520023790    0.691511607    -1.016949890
8       H        0.062602417   -1.707437146     1.461138901
9       H       -0.337235335   -2.175956590    -0.221476531
10      H       -1.543918426   -1.318457543     0.783991801
11      C        3.678452406    1.511517281     0.096654417
12      C        4.335739942    1.531901746     1.504037054
13      N        2.255923367    1.210827191     0.214657464
14      C        3.934185663    2.829751775    -0.675804694
15      O        5.377531311    0.955014307     1.810939652
16      H        4.193095616    0.697344134    -0.479806642
17      H        1.680863860    1.977538215     0.504317106
18      H        3.458906951    3.711800319    -0.202441369
19      H        3.547495012    2.751749228    -1.712401524
20      H        5.023566655    3.027528614    -0.741271502
21      H       -0.487467850   -0.156365402    -1.907268035
22      O        3.696039384    2.265415178     2.450261414
23      H        4.094301576    2.256425089     3.312394108
---
H6C9N2O3, RHF, CHARGE=0, MULT=1
HF=-0.4
1       C       -0.000054626    0.000313454     0.000152141
2       C        1.379729607    0.000057869    -0.000308363
```

```
 3    C      2.081845904    1.252867739    0.000047862
 4    N      1.456716097    2.421878556    0.000979001
 5    C      0.075093806    2.450637342    0.001848101
 6    C     -0.715539412    1.253068885    0.001451320
 7    C     -2.143767924    1.412955762    0.002664723
 8    C     -2.755446961    2.662269938    0.004578124
 9    C     -1.967602886    3.843014500    0.004773035
10    C     -0.566137012    3.760602223    0.003253115
11    N     -3.033140282    0.201204768    0.001514212
12    O     -3.388374856   -0.271939414    1.057174636
13    O     -3.389271531   -0.268907815   -1.055239447
14    O      0.116792667    4.921757859    0.003299092
15    H     -0.533523214   -0.950689989   -0.000497613
16    H      1.948099064   -0.929675016   -0.000863819
17    H      3.177376888    1.277738885   -0.000585936
18    H     -3.842397741    2.773270155    0.006132183
19    H     -2.476400485    4.807875568    0.005922203
20    H      1.061414555    4.831356977    0.002313566
---
H16C10N2O3, RHF, CHARGE=0, MULT=1
HF=-131.4
 1    C      0.120473448    0.071179111   -0.157883680
 2    C      1.465982845    0.712952891   -0.549584094
 3    C      1.334971697    2.224819421   -0.272190367
 4    C     -0.768419698    1.210000350    0.428108374
 5    C     -1.074489870    1.010529206    1.933520155
 6    N     -0.064311842    2.468728308    0.117195138
 7    O     -0.318703984    1.422331032    2.813733574
 8    H      1.701655689    0.521395811   -1.617747281
 9    H      2.302151019    0.273594050    0.035184650
10    H      0.272399432   -0.763226647    0.556981766
11    H     -0.377693628   -0.373873456   -1.046054357
12    H     -1.728399775    1.266048780   -0.145654238
13    H      2.063127799    2.557655959    0.506072028
14    H      1.550954502    2.831262994   -1.184364069
15    C     -3.972436265   -0.558684294    3.793328891
16    C     -4.456388480   -0.660833562    2.337769752
17    C     -3.335023621   -0.089935938    1.445536203
18    C     -2.513694718   -0.017966137    3.752941699
19    C     -1.495970164   -1.047113348    4.299898256
20    N     -2.237420824    0.310404277    2.341021407
21    O     -0.935161001   -1.944324001    3.673367296
22    H     -4.679249544   -1.714146646    2.064167682
23    H     -5.402603122   -0.097287187    2.191859516
24    H     -4.612974013    0.136565321    4.377781242
25    H     -4.046236098   -1.541093648    4.304604227
26    H     -2.455006076    0.912635264    4.375213497
27    H     -2.988721352   -0.853496826    0.708245918
28    H     -3.712984124    0.785838414    0.864116594
29    O     -1.247280571   -0.919495182    5.627620719
30    H     -0.127565606    3.156498479    0.848595729
31    H     -0.591072144   -1.508727176    5.979338437
---
H20C10N2O3, RHF, CHARGE=0, MULT=1
HF=-160.6
 1    C      0.000084455    0.000047671    0.000058913
 2    C      1.556521751   -0.000101440    0.000050938
 3    N      1.978827332    1.400192546    0.000008407
 4    C      2.427813095    2.162147381   -1.094173206
```

```
5    C     2.832180315    3.640852817   -0.807873632
6    N     4.250450961    3.866076759   -1.124137939
7    H     4.450173062    3.698599895   -2.098066149
8    H     2.733177286    3.840330184    0.292868109
9    H     2.030502263    4.517688463   -2.652304376
10   H    -0.188647928    5.274574593   -1.735556440
11   H     0.208712675    4.561168368   -0.144768688
12   H     0.039409183    3.519984619   -1.594444617
13   H     1.653056262    6.852264782   -1.776576994
14   H     3.358011811    6.361235632   -1.685948759
15   H     2.357190885    6.384954367   -0.199364071
16   C     1.919830721    4.683184588   -1.550422610
17   O    -0.732607628    0.564895310    0.811974059
18   C     2.349946356    6.143872078   -1.281434217
19   C     2.164128075   -0.798967800    1.213197099
20   C     3.709364755   -0.767805223    1.228966178
21   C     1.654754139   -2.257159600    1.271558697
22   C     0.419645469    4.495572719   -1.231331182
23   H     1.862923879    1.879568345    0.871685302
24   H     1.915088184   -0.499593501   -0.936564881
25   H     1.825722951   -0.309128442    2.161430825
26   H     2.129925476   -2.803484054    2.112192854
27   H     1.875894875   -2.817035320    0.340838137
28   H     0.561315640   -2.304163539    1.446513536
29   H     4.104413442   -1.383660210    2.063377624
30   H     4.101062101    0.257183640    1.385306627
31   H     4.146475055   -1.158881121    0.288403992
32   H     4.828933626    3.221592293   -0.608060920
33   O    -0.564504440   -0.679640998   -1.027256593
34   H    -1.513463648   -0.657950702   -1.062158587
35   O     2.533196299    1.633436466   -2.198856264
---
H3C3N3O3, RHF, CHARGE=0, MULT=1
HF=-134.8
1    C     0.000069850   -0.000073145   -0.000004659
2    N     1.409618283    0.000181012    0.000003938
3    C     2.221669292    1.152350768    0.000004662
4    N     1.516322739    2.372788442    0.000042717
5    C     0.112522977    2.500186695    0.000068649
6    N    -0.591798658    1.279223845    0.000041351
7    O    -0.659067517   -1.031347802   -0.000041372
8    O     3.444361703    1.097381492   -0.000028049
9    O    -0.451094828    3.586613927    0.000109138
10   H     2.058514439    3.221281712    0.000063622
11   H    -1.597711754    1.324427118    0.000048323
12   H     1.873238027   -0.893689973   -0.000016452
---
H9C6N3O3, RHF, CHARGE=0, MULT=1
HF=-141.1
1    C     0.000004062   -0.000005285   -0.000002986
2    N     1.423530121    0.000006938   -0.000993591
3    C     2.136627500    1.226666244    0.000336752
4    N     1.332664257    2.393957757    0.121550218
5    C     0.073119671    2.336034592    0.781531766
6    N    -0.587944593    1.082406824    0.712141847
7    O    -0.661499296   -0.870429613   -0.547311110
8    C    -1.955314452    0.937976775    1.266476220
9    O     3.354844866    1.279184095   -0.109771238
10   C     2.153512265   -1.279301088   -0.166723673
```

```
11      O      -0.411338922     3.304439411     1.352633052
12      C       1.907192852     3.714310684    -0.231954221
13      H      -2.197353893    -0.132582455     1.443528168
14      H      -2.706767992     1.365384949     0.565003152
15      H      -2.035345817     1.457059443     2.247107101
16      H       2.405603052    -1.449455949    -1.237694359
17      H       3.090968344    -1.273241888     0.432001444
18      H       1.538453017    -2.130946112     0.197678965
19      H       2.474160774     4.144755968     0.623696875
20      H       1.099171160     4.420817826    -0.523848462
21      H       2.588093909     3.616167499    -1.106093587
---
H9C6N3O3, RHF, CHARGE=0, MULT=1
HF=-70.1
1       C       0.000082258     0.000005540    -0.000002811
2       N       1.365760038     0.000068470     0.000096886
3       C       1.928792423     1.246616126    -0.000007874
4       N       1.245860399     2.429299793    -0.000344334
5       C      -0.115163840     2.293626096    -0.000675903
6       N      -0.797951883     1.110903661    -0.000450928
7       O      -0.672713211    -1.154620591     0.000355598
8       C      -0.053151293    -2.423996817     0.000864526
9       O       3.265127794     1.241285666     0.000277482
10      C       4.055040655     2.412256281     0.000350579
11      O      -0.778642225     3.453580889    -0.001294604
12      C      -2.187664241     3.552270919    -0.001521695
13      H       0.571673861    -2.589576235    -0.907482740
14      H       0.570006728    -2.589531795     0.910377994
15      H      -0.886582124    -3.166484232     0.000093476
16      H       3.886275595     3.035741050     0.909046444
17      H       3.887093130     3.035254111    -0.908831801
18      H       5.114652955     2.061300683     0.000915069
19      H      -2.643385390     3.095387543     0.907592882
20      H      -2.643204575     3.094494425    -0.910278934
21      H      -2.413425849     4.645423531    -0.002016073
---
H12C7O4, RHF, CHARGE=0, MULT=1
HF=-199.3
1       C      -0.000051629    -0.000014223    -0.000038817
2       C       1.569739427    -0.000069894     0.000052059
3       C       2.182391742     1.445345272    -0.000008822
4       O       2.091970095    -0.790798170     1.064169038
5       C       2.010493085    -0.653319928     2.412196399
6       O       1.396424014     0.279843207     2.920609169
7       C       2.747502254    -1.761175668     3.160124879
8       O       2.092958359    -0.790249106    -1.063956334
9       C       2.009206040    -0.653249313    -2.411905522
10      O       1.389020144     0.276281492    -2.919583149
11      C       2.751524435    -1.756892841    -3.160780705
12      H      -0.403702483     0.528588007    -0.885261513
13      H      -0.409187012    -1.030371405    -0.004351430
14      H      -0.403617786     0.520852248     0.889838470
15      H       3.290742314     1.419836695     0.001929649
16      H       1.858519382     2.021388811    -0.888434106
17      H       1.855094891     2.022648673     0.886352049
18      H       2.323848678    -2.753695205     2.907268147
19      H       3.824953636    -1.759547963     2.900071333
20      H       2.655942018    -1.615028362     4.254980883
21      H       3.829202692    -1.749499667    -2.901764116
```

```
22      H      2.333488877     -2.751737276    -2.907720614
23      H      2.658156711     -1.611006335    -4.255527481
---
H12C7O4, RHF, CHARGE=0, MULT=1
HF=-190.1
1       C      -0.000012481    -0.000011368     0.000082241
2       C       1.533983859    -0.000052328    -0.000056739
3       C       2.106351858     1.423277210    -0.000008739
4       O       1.924311681     2.265471649     0.874209510
5       O      -0.714917714     0.530212075    -0.845218024
6       O      -0.538316138    -0.693929536     1.034761344
7       C      -1.920876603    -0.831444892     1.285968980
8       C      -2.117218724    -1.661267027     2.569015984
9       O       2.896305858     1.687135093    -1.071454642
10      C       3.521574818     2.926225883    -1.331201909
11      C       4.296060741     2.828679098    -2.659403891
12      H       1.945856255    -0.527811888     0.890124914
13      H       1.869510549    -0.579066646    -0.890471257
14      H      -2.405783729     0.171196041     1.411087709
15      H      -2.432141880    -1.341459750     0.428553421
16      H      -1.688249393    -2.678741063     2.475314896
17      H      -1.659506093    -1.174092759     3.452764896
18      H      -3.204086046    -1.767593442     2.764127463
19      H       4.229116948     3.195305029    -0.504487694
20      H       2.764235202     3.749582473    -1.399140092
21      H       3.628234464     2.601386952    -3.513878097
22      H       5.086297110     2.052611035    -2.623301052
23      H       4.786274355     3.803295736    -2.860655272
---
H12C7O4, RHF, CHARGE=0, MULT=1
HF=-191.9
1       C       0.000171202    -0.000079460    -0.000286823
2       C       1.559849056     0.000211931     0.000194181
3       C       1.999727861     1.496524171     0.000112959
4       O       1.555212993     2.353721312     0.758780133
5       O      -0.697881262     0.688492919    -0.739543761
6       O      -0.582492425    -0.868522159     0.864824884
7       C      -1.968680807    -1.010251582     1.058604112
8       H      -2.446436607    -0.063498907     1.404425137
9       O       2.972723660     1.819181392    -0.889517976
10      C       3.490836741     3.111979151    -1.088208958
11      H       3.999819914     3.505914429    -0.176755057
12      C       2.198079060    -0.649915476     1.266145693
13      C       2.005023777    -0.797515297    -1.263914830
14      H      -2.488723003    -1.361524979     0.136075980
15      H      -2.096573852    -1.783461887     1.853052110
16      H       3.300693512    -0.525002655     1.253142268
17      H       1.823051769    -0.194852954     2.204193322
18      H       1.993762205    -1.738044625     1.314011708
19      H       2.706103495     3.838046629    -1.406247417
20      H       4.246065167     3.024770888    -1.905223837
21      H       3.107312896    -0.899954145    -1.314833602
22      H       1.666989616    -0.316180034    -2.202874043
23      H       1.582107360    -1.823318902    -1.245332938
---
H16C7O4, RHF, CHARGE=0, MULT=1
HF=-177.1
1       C      -1.54747         2.49061         0.70635
2       C      -0.24746         1.66556         0.75221
```

```
 3    O     -0.11119          0.93928         -0.44082
 4    C      0.97647          0.08508         -0.65910
 5    O      2.07018          0.82275         -1.12545
 6    C      2.14116          1.31720         -2.44178
 7    O      3.48911          1.58585         -2.69307
 8    C      4.46313          0.59879         -2.90335
 9    O      4.34635          0.10668         -4.21059
10    C      4.93049          0.76772         -5.30105
11    C      4.67133         -0.03638         -6.58929
12    H     -1.65428          3.05114          1.65785
13    H     -2.44308          1.84857          0.58938
14    H     -1.54151          3.22692         -0.12170
15    H     -0.27778          0.97637          1.63896
16    H      0.62197          2.35417          0.91936
17    H      0.63008         -0.69785         -1.39328
18    H      1.32236         -0.44782          0.27150
19    H      1.60058          2.30225         -2.53376
20    H      1.68772          0.60493         -3.18981
21    H      4.36696         -0.28833         -2.21711
22    H      5.45790          1.08584         -2.69615
23    H      6.03829          0.88385         -5.15453
24    H      4.52046          1.80467         -5.42449
25    H      5.14328          0.48832         -7.44541
26    H      3.58899         -0.13294         -6.80660
27    H      5.10383         -1.05532         -6.53679
---
H16C8O4, RHF, CHARGE=0, MULT=1
HF=-150.8
 1    C      0.000074624      0.000019329     -0.000165868
 2    C      1.563325894      0.000001750      0.000005883
 3    O      2.040818449      1.315781381     -0.000652847
 4    C      3.388456358      1.562002012     -0.288474905
 5    C      3.599201246      3.080425336     -0.595113912
 6    O      3.123445134      3.442165693     -1.860178496
 7    C      1.968463560      4.214010554     -2.019861681
 8    C      1.112836787      3.695294113     -3.223063633
 9    O      0.595761464      2.415496187     -2.995346965
10    C     -0.777010568      2.185766928     -2.850070167
11    C     -1.134398028      1.615497950     -1.438132991
12    O     -0.539771463      0.365995060     -1.237878096
13    H     -0.389132311      0.610577165      0.854027662
14    H     -0.329629363     -1.055926795      0.195245774
15    H      1.944813357     -0.597072135     -0.868044247
16    H      1.904008811     -0.538651986      0.926241387
17    H      3.766820556      0.943555363     -1.143053795
18    H      4.030054514      1.300505783      0.597082714
19    H      4.707298929      3.265189256     -0.601056748
20    H      3.183976821      3.707058553      0.234662011
21    H      2.272299717      5.271475770     -2.253039670
22    H      1.335333103      4.274017256     -1.098561319
23    H      1.742353015      3.646438558     -4.149539019
24    H      0.323003177      4.466576500     -3.422762883
25    H     -1.090009468      1.461284782     -3.647030773
26    H     -1.402275935      3.105007365     -3.000452807
27    H     -2.251975051      1.493686077     -1.397864343
28    H     -0.880312365      2.368023992     -0.649493164
---
H6C10O4, RHF, CHARGE=0, MULT=1
HF=-119.3
```

```
   1    C     0.000006261     0.000046075    -0.000168140
   2    C     1.440288316    -0.000050705    -0.000015346
   3    C     2.139511715     1.190148690     0.000115349
   4    C     1.438334164     2.448077307     0.000080045
   5    C     0.024806235     2.470994080    -0.008970420
   6    C    -0.708058943     1.223820546    -0.007881784
   7    C    -2.211623654     1.311894029    -0.043135640
   8    C    -2.769180146     2.492438224    -0.781350952
   9    C    -2.086939401     3.652335420    -0.784537297
  10    C    -0.783460651     3.742016871    -0.048022396
  11    O    -2.950597370     0.503407823     0.511267828
  12    O    -0.436274969     4.782680017     0.502835642
  13    O    -0.536954920    -1.241392458    -0.043514512
  14    O     2.260333567     3.522701821    -0.047346854
  15    H     1.983145365    -0.947628298     0.000915285
  16    H     3.231537501     1.177249717     0.000936554
  17    H    -3.727598483     2.368877385    -1.286436343
  18    H    -2.445409437     4.547579675    -1.293624023
  19    H    -1.455886218    -1.303089536     0.171347412
  20    H     1.876509850     4.347669055     0.210962170
---
H10C10O4, RHF, CHARGE=0, MULT=1
HF=-150.4
   1    C    -0.000016012     0.000078241    -0.000019923
   2    C     1.412528266     0.000241915    -0.000068258
   3    C     2.108785838     1.229224869     0.000001813
   4    C     1.397189901     2.441098737    -0.001254671
   5    C    -0.008114480     2.441021179    -0.001316993
   6    C    -0.715173681     1.218190703    -0.001706810
   7    C    -2.216571500     1.214246904    -0.025586497
   8    O    -2.923537381     1.253503456    -1.027643409
   9    O    -2.770353528     1.164629934     1.216162326
  10    C    -4.148936148     1.147411668     1.487487686
  11    C     2.169289750    -1.296554044    -0.021525378
  12    O     2.554595537    -1.892101045    -1.022720340
  13    O     2.415245891    -1.791987727     1.221473442
  14    C     3.099616494    -2.988243632     1.495296637
  15    H    -0.542420548    -0.947403428    -0.001349262
  16    H     3.199841706     1.250428088    -0.000090033
  17    H     1.938990375     3.387672767    -0.002438766
  18    H    -0.542599765     3.392497204    -0.002113728
  19    H    -4.655365357     0.255205455     1.048611441
  20    H    -4.663137120     2.067959052     1.122242297
  21    H    -4.252092648     1.101909427     2.597803392
  22    H     4.150913739    -2.969073584     1.121505798
  23    H     2.582519055    -3.879124599     1.066493081
  24    H     3.121492697    -3.092846101     2.606181219
---
H10C10O4, RHF, CHARGE=0, MULT=1
HF=-144.9
   1    C     0.026613739    -0.101876274    -0.156921804
   2    C     1.375873043     0.212831506     0.152223081
   3    C     2.088001757     1.105063872    -0.681600747
   4    C     1.476786091     1.675377675    -1.811455035
   5    C     0.144805939     1.358099530    -2.121879325
   6    C    -0.574972532     0.474804435    -1.299135386
   7    C    -0.811779409    -1.009323637     0.702530344
   8    O    -1.654043280    -0.650561429     1.518956534
   9    O    -0.576379302    -2.330273718     0.479113173
```

```
10    C     -1.229491744   -3.390130439    1.131199301
11    C      2.075672290   -0.352661431    1.358615733
12    O      2.003058258    0.080255853    2.504265445
13    O      2.859170953   -1.425342067    1.067162648
14    C      3.634918811   -2.136169554    1.998623205
15    H     -1.610074738    0.240782926   -1.554370025
16    H      3.124233174    1.362871704   -0.455340795
17    H      2.038897759    2.362520472   -2.444747329
18    H     -0.333659470    1.796100718   -2.998508458
19    H     -2.322109022   -3.412130706    0.905204802
20    H     -1.092340986   -3.359203799    2.237991364
21    H     -0.769131523   -4.330304276    0.744176279
22    H      3.016218359   -2.591567029    2.807775407
23    H      4.422380034   -1.497668042    2.465126016
24    H      4.135502902   -2.956959860    1.432073339
---
H22C10O4, RHF, CHARGE=0, MULT=1
HF=-157
1     C      0.000006897   -0.000000084    0.000008166
2     C      1.561746504   -0.000044825    0.000014998
3     C      2.107062260    1.462878426    0.000006964
4     C      2.092069305   -0.762915198   -1.261079542
5     O      2.090521487   -0.770972791    1.094295660
6     O      1.769905741   -0.361499471    2.280038047
7     C      2.833499090   -0.418987032    3.226952592
8     C      2.697326133   -1.673028490    4.170037512
9     O      2.849094013   -2.872957718    3.464422160
10    C      3.782244211   -1.698790743    5.312639984
11    O      3.532285881   -0.641018659    6.197243821
12    C      4.276908026   -0.504864807    7.379069754
13    C      5.454113780    0.495703945    7.228700846
14    C      6.207918502    0.768406113    8.532394757
15    H     -0.410742693   -1.027352543    0.069975074
16    H     -0.422162893    0.585971163    0.840444818
17    H     -0.386260746    0.449950408   -0.937520618
18    H      1.711803423    2.061472486    0.844730366
19    H      3.213857426    1.489994208    0.060891342
20    H      1.813473850    1.984888662   -0.933551788
21    H      1.730344384   -1.810312835   -1.293212183
22    H      3.199932593   -0.790458154   -1.291936568
23    H      1.745537922   -0.262155792   -2.187823791
24    H      3.844884887   -0.390174575    2.753119079
25    H      2.728827442    0.513860995    3.837912360
26    H      1.682679129   -1.628590348    4.657018713
27    H      2.043497899   -3.140835747    3.045793855
28    H      3.729573436   -2.686060511    5.844514136
29    H      4.811900623   -1.629119129    4.873739433
30    H      3.558804637   -0.127056978    8.154199574
31    H      4.661782208   -1.487682457    7.757057146
32    H      5.071246599    1.462587282    6.829663302
33    H      6.179874497    0.113397280    6.474652006
34    H      6.656954430   -0.155096232    8.951336963
35    H      5.547900354    1.207321168    9.307880014
36    H      7.033403178    1.487486855    8.352101117
---
H12C11O4, RHF, CHARGE=0, MULT=1
HF=-167.3
1     C      0.000636924    0.000297599    0.000109133
2     C      1.406145638   -0.000927037   -0.000091430
```

```
3    C     2.114640223    1.212949014   -0.000085564
4    C     1.430644014    2.451949243   -0.002602233
5    C     0.016522518    2.438887243    0.001273698
6    C    -0.690989101    1.222930215    0.001362393
7    C     2.246129540    3.757209480   -0.004709373
8    O     1.850565638    4.625489649   -1.049353947
9    C     2.387669549    4.695046375   -2.300416444
10   C     1.844545849    5.856083554   -3.124970531
11   O     1.939669071    4.459137772    1.181387303
12   C     2.571072278    5.570685991    1.658865700
13   C     1.959253494    6.100778941    2.950224495
14   O     3.209280644    3.858977712   -2.663483988
15   O     3.536781830    6.036400896    1.061448965
16   H    -0.548681025   -0.941519572   -0.000520626
17   H     1.950752914   -0.945591783   -0.000844327
18   H     3.205741359    1.178913290   -0.001376991
19   H    -0.552361676    3.370227708    0.005755323
20   H    -1.781496595    1.232434833    0.002995079
21   H     3.351053139    3.525328509   -0.050228422
22   H     2.110674155    6.824089731   -2.654794756
23   H     2.272302234    5.838566874   -4.147288479
24   H     0.741158471    5.795917602   -3.211659602
25   H     2.510485861    6.995588981    3.302174788
26   H     0.900288233    6.387404831    2.791476755
27   H     2.002020629    5.332683554    3.748305161
---
H24C11O4, RHF, CHARGE=0, MULT=1
HF=-161.3
1    C    -0.000032363   -0.000181437   -0.000028679
2    C     1.561735073    0.000142707   -0.000002804
3    C    -0.548578494    1.462274654    0.000025835
4    C    -0.529775346   -0.763189056   -1.261724015
5    O    -0.532529668   -0.767227210    1.092393998
6    O    -0.224914686   -0.329377759    2.276040051
7    C    -0.858057181   -1.061735536    3.325163295
8    O     0.036743375   -2.053373909    3.742565408
9    C    -0.023618796   -3.367672556    3.249518562
10   C     0.934232008   -4.256005990    4.088276983
11   C     0.958833342   -5.725562280    3.626005451
12   C     1.892455346   -6.631135694    4.434191583
13   C    -1.137216867   -0.048904106    4.531115765
14   C     0.086675899    0.727292989    5.082733572
15   O    -1.853666644   -0.716308849    5.530686672
16   H     1.972711323   -1.027439326    0.065951181
17   H     1.984368269    0.583091480    0.842043805
18   H     1.947098602    0.453348692   -0.936337904
19   H    -0.142563510    2.066045146    0.835746998
20   H    -1.654348316    1.487130666    0.077179549
21   H    -0.269246011    1.979530904   -0.940673635
22   H    -0.167512909   -1.810417302   -1.294394277
23   H    -1.637631014   -0.791431733   -1.293260140
24   H    -0.183070270   -0.262355446   -2.188279337
25   H    -1.853775180   -1.484889197    2.999211686
26   H    -1.068224596   -3.774611231    3.309038025
27   H     0.273356593   -3.407742601    2.169315698
28   H     0.631041649   -4.214522895    5.159436993
29   H     1.964870660   -3.837362404    4.035103262
30   H    -0.071011020   -6.149489426    3.675169831
31   H     1.262543735   -5.774400187    2.554577560
```

```
32      H        2.946849518     -6.292914001     4.377061422
33      H        1.605301375     -6.669827247     5.504518714
34      H        1.853261126     -7.668039304     4.041081437
35      H       -1.854565217      0.715538660     4.116413212
36      H        0.531221644      1.372040683     4.299487162
37      H        0.884177441      0.062425422     5.468675093
38      H       -0.225703075      1.393967077     5.913900620
39      H       -1.301124093     -1.231567758     6.101888736
---
H5C6N1O4, RHF, CHARGE=0, MULT=1
HF=-69.3
1       C       -0.055591105     -0.032891343    -0.039601736
2       C        1.309104853     -0.029719180     0.298762825
3       C        1.991042786      1.196313629     0.371506393
4       C        1.317759163      2.418648743     0.106887301
5       C       -0.764748579      1.153439122    -0.312678449
6       C       -0.079651765      2.388356050    -0.241668277
7       O       -0.683138612      3.582484819    -0.489798565
8       O        2.058274843      3.545093212     0.205613431
9       N       -0.788495097     -1.342409549    -0.112881946
10      O       -0.717378230     -2.001975871    -1.124580943
11      O       -1.435289718     -1.709227832     0.841520634
12      H        1.852055702     -0.953073797     0.507156887
13      H        3.050189379      1.197576150     0.635058974
14      H       -1.823822306      1.123292403    -0.575055548
15      H       -1.597184365      3.497890850    -0.725095627
16      H        1.589934900      4.347860961     0.019889218
---
H5C7N1O4, RHF, CHARGE=0, MULT=1
HF=-68.7
1       C       -0.001305206     -0.033075267    -0.020817629
2       C        1.405673165     -0.057494977    -0.080888123
3       C        2.105026240      1.163861215    -0.065623367
4       C        1.408299871      2.390644724     0.009681290
5       C       -0.003469306      2.387983419     0.069720223
6       C       -0.719333147      1.176320564     0.055153090
7       N       -0.756796114     -1.333436885    -0.041504216
8       O       -1.150924507     -1.765068926    -1.100369752
9       O       -0.951402294     -1.916715449     0.999778709
10      C        2.163332127      3.688360025     0.049555892
11      O        2.582114411      4.251667624     1.058082267
12      O        2.372443025      4.245411256    -1.169010096
13      H        1.964012337     -0.993075026    -0.138360187
14      H        3.195353872      1.146061710    -0.111695179
15      H       -0.558439811      3.325869180     0.128711000
16      H       -1.809016810      1.197659304     0.103054714
17      H        2.845267421      5.069405090    -1.167144045
---
H4C6N2O4, RHF, CHARGE=0, MULT=1
HF=11.3
1       C        0.237078863      1.372353724    -1.192727199
2       C        2.710354332      0.382473179    -0.306583792
3       C        0.663404283      1.569041672     0.137313020
4       C        1.075213771      0.665090725    -2.075032099
5       C        1.900438391      1.084236435     0.610239007
6       C        2.317331789      0.163372723    -1.643408743
7       H       -0.721450920      1.752771694    -1.550929528
8       N       -0.233157852      2.317950287     1.085086594
9       H        0.757949417      0.503739511    -3.106335817
```

```
10      H       2.217872441     1.245476701     1.642462978
11      H       2.949777863    -0.382467662    -2.345911591
12      N       4.038378643    -0.150347254     0.157794313
13      O       5.025767799     0.530115201     0.004588197
14      O       4.082272851    -1.245667450     0.667493395
15      O      -0.988088108     1.690849703     1.791111674
16      O      -0.178193953     3.525289013     1.111440420
---
H4C6N2O4, RHF, CHARGE=0, MULT=1
HF=20.2
1       C       0.121123738    -0.035090754     0.072564776
2       C       1.530377718    -0.004971895     0.122680907
3       C       2.260654368     1.198634912     0.038923908
4       C       1.521864091     2.392211559    -0.099565423
5       C      -0.577655003     1.178556484    -0.067644233
6       C       0.112513842     2.402053740    -0.155076917
7       H      -0.435964582    -0.971365538     0.139131681
8       N       2.280305802    -1.300877195     0.268819791
9       H      -1.668053223     1.170849850    -0.108515735
10      H       3.351810513     1.205521160     0.080257510
11      H      -0.451688287     3.330109547    -0.264073763
12      N       2.261937447     3.698585666    -0.190909469
13      O       2.733055369     4.030569986    -1.253572583
14      O       2.362549780     4.382904386     0.800719722
15      O       2.643308883    -1.880223893    -0.728155942
16      O       2.495188462    -1.729216847     1.378756900
---
H4C6N2O4, RHF, CHARGE=0, MULT=1
HF=13.3
1       C       0.000006487     0.000184797    -0.000121028
2       C       1.409524578    -0.000332443     0.000079515
3       C       2.158882248     1.193466105    -0.000222917
4       C       1.477341354     2.426200374    -0.002272459
5       C      -0.681571447     1.233126823    -0.000896162
6       C       0.067810259     2.426780394    -0.002787430
7       H      -0.576478987    -0.926562506     0.000529745
8       N       2.137557315    -1.317583163     0.000662397
9       H      -1.772949327     1.237867486     0.000121093
10      H       3.250256970     1.188756062     0.000884131
11      H       2.053948630     3.352820460    -0.003211145
12      N      -0.659845152     3.744152011    -0.005312900
13      O      -0.946000322     4.254714972     1.052492046
14      O      -0.937979636     4.255276572    -1.065056398
15      O       2.410389460    -1.833982836    -1.057839037
16      O       2.429372491    -1.822653797     1.059568285
---
H6C7N2O4, RHF, CHARGE=0, MULT=1
HF=7.9
1       C      -0.000043560    -0.000379338    -0.000283267
2       C       1.419346234     0.000652894     0.000006803
3       C       2.188357704     1.184701915     0.000095278
4       C       1.500294534     2.412496191    -0.000545648
5       C       0.092794870     2.470339274    -0.002157043
6       C      -0.634647842     1.268151721    -0.001703464
7       C      -0.840711629    -1.254322190     0.001004499
8       N       2.186059599    -1.297451802     0.000587420
9       N       2.287112208     3.693974777     0.000868290
10      O       2.489064898    -1.797557460     1.058935347
11      O       2.492992157    -1.796299710    -1.057249014
```

```
12      O        2.593240898     4.189951617     1.060047237
13      O        2.587779873     4.195611598    -1.057205172
14      H        3.280403495     1.150586284     0.000698716
15      H       -0.443726077     3.421067892    -0.003852234
16      H       -1.725190069     1.329666048    -0.002177518
17      H       -1.925198264    -1.022129857     0.001528913
18      H       -0.642406145    -1.870585766    -0.900113891
19      H       -0.641387876    -1.869413289     0.902672925
---
H6C7N2O4, RHF, CHARGE=0, MULT=1
HF=8.3
1       C        0.000025595    -0.000073135     0.000323793
2       C        1.404295175     0.000040600     0.000095513
3       C        2.117648624     1.213146947    -0.000002284
4       C        1.433727940     2.453705510     0.002679429
5       C        0.012496790     2.435803797     0.000242268
6       C       -0.694401648     1.224302394    -0.000843400
7       C        2.121295772     3.795491333     0.012219643
8       N        3.247875909     3.880391854     1.094352556
9       O        2.901293596     4.043493517     2.239786792
10      O        4.409496327     3.788766210     0.782074264
11      N        2.771420619     4.171691129    -1.365888677
12      O        3.183216424     3.316568946    -2.108083100
13      O        2.810236188     5.344635044    -1.654668395
14      H       -0.549703584    -0.942005876     0.000462563
15      H        1.950361529    -0.944291331    -0.001228562
16      H        3.208023438     1.161604231    -0.007702892
17      H       -0.555102229     3.368630632    -0.000318275
18      H       -1.784916116     1.233297756    -0.002224771
19      H        1.405148680     4.626489274     0.251261973
---
H18C8O5, RHF, CHARGE=0, MULT=1
HF=-216.5
1       C        0.60903    -0.32622     1.97256
2       C        1.30938     0.41588     0.81841
3       O        2.05732    -0.49702     0.06026
4       C        2.85059    -0.08226    -1.01740
5       O        2.03720     0.08187    -2.14757
6       C        1.65333    -0.97537    -2.98499
7       O        2.68202    -1.19758    -3.91178
8       C        2.82495    -0.46775    -5.10068
9       O        1.99132    -1.04154    -6.07137
10      C        2.35934    -2.14633    -6.85226
11      O        3.12709    -1.70365    -7.93724
12      C        2.53316    -1.17387    -9.09231
13      C        3.62493    -0.85084   -10.13012
14      H        0.02303     0.40278     2.56944
15      H        1.33476    -0.81494     2.65267
16      H       -0.09122    -1.10118     1.60229
17      H        1.96692     1.22207     1.24263
18      H        0.53769     0.93783     0.19333
19      H        3.65094    -0.86359    -1.14752
20      H        3.36098     0.90684    -0.84001
21      H        0.67598    -0.67672    -3.45760
22      H        1.49108    -1.94758    -2.44073
23      H        3.91799    -0.49580    -5.36931
24      H        2.51940     0.61197    -5.00770
25      H        1.40312    -2.65631    -7.16183
26      H        2.98735    -2.90003    -6.29958
```

```
27   H     1.95077    -0.24177    -8.86721
28   H     1.80192    -1.90025    -9.53933
29   H     3.14717    -0.43635   -11.04172
30   H     4.19500    -1.75348   -10.42742
31   H     4.34518    -0.09781    -9.75311
---
H20C10O5, RHF, CHARGE=0, MULT=1
HF=-191.1
1    C    -0.023282158    0.088688257   -0.168151456
2    C     1.506258218    0.018182732    0.137265302
3    O     1.857704476    1.082771039    0.971976147
4    C     3.142466094    1.147510802    1.518491777
5    C     3.265058229    2.439329450    2.386819060
6    O     3.251572099    3.570541327    1.562801792
7    C     2.880059275    4.823968300    2.060204996
8    C     1.513035903    5.315203030    1.482081760
9    O     0.461766717    4.631602846    2.103542627
10   C    -0.854701702    4.883332423    1.700751338
11   C    -1.808564561    3.818414947    2.328247884
12   O    -1.740938681    2.628379710    1.594628831
13   C    -2.208570879    1.439848724    2.166557410
14   C    -2.071547152    0.266007360    1.142719933
15   O    -0.752548659   -0.175200459    0.997165204
16   H    -0.250411507   -0.695091376   -0.942615152
17   H    -0.274944460    1.073185831   -0.637358442
18   H     2.052019056    0.076506871   -0.843600338
19   H     1.769411530   -0.977489427    0.581429768
20   H     3.937643940    1.128606567    0.727525336
21   H     3.347707543    0.268976048    2.188898430
22   H     4.244040708    2.384506679    2.938284504
23   H     2.465944560    2.451144501    3.170857032
24   H     3.662714970    5.555404396    1.720605696
25   H     2.870662072    4.876657970    3.178428044
26   H     1.495788002    5.190997307    0.368231480
27   H     1.442687576    6.420629995    1.675846632
28   H    -0.967059727    4.903155276    0.585914703
29   H    -1.189992120    5.893041331    2.065885826
30   H    -2.852490952    4.235634444    2.289533100
31   H    -1.572384729    3.673651178    3.414028321
32   H    -3.296687250    1.522276864    2.437956445
33   H    -1.666463597    1.192228404    3.115980652
34   H    -2.560727145    0.537692877    0.172841568
35   H    -2.657811140   -0.598650608    1.556900724
---
H12C6N2O5, RHF, CHARGE=0, MULT=1
HF=-211.6
1    C     0.038787582   -0.036470383   -0.070167434
2    C     1.591201807   -0.075222130   -0.132586956
3    C     2.206699588    1.360604720   -0.278015283
4    N     1.922725731   -1.034390059   -1.194920452
5    O    -0.655330973    0.765889295   -0.687751721
6    O     3.599947033    1.262987423   -0.251514473
7    H     4.003613791    2.118991934   -0.273300789
8    H     1.864720335    1.857454412   -1.224172623
9    H     1.834847890    1.999416137    0.567109320
10   H     1.983580345   -0.490076707    0.834614334
11   H     2.875179299   -1.348572623   -1.091663685
12   C    -2.301247045   -1.076155259    2.453171975
13   C    -1.993489577   -1.241853844    0.941338863
```

```
14      C       -2.391392574    -2.661795393     0.385418052
15      N       -0.564940390    -0.993827724     0.765119226
16      O       -2.001239131    -1.865585232     3.349455492
17      O       -3.729086280    -2.940864735     0.665788830
18      H       -3.827679467    -3.737366413     1.167547335
19      H       -2.256716519    -2.648773877    -0.728838313
20      H       -1.705190100    -3.455225929     0.778694701
21      H       -2.593172408    -0.489230051     0.367679293
22      H        0.037881434    -1.683837590     1.170664341
23      H        1.873975199    -0.602778481    -2.105254579
24      O       -2.953710178     0.064923802     2.774522757
25      H       -3.156714454     0.176392666     3.695699707
---
H16C8N2O5, RHF, CHARGE=0, MULT=1
HF=-230.2
1       C        0.032033638     0.098068146    -0.111667610
2       C        1.574542955    -0.217767883    -0.314117906
3       C       -0.273183227     1.619753963    -0.071882868
4       C        2.467797323     0.305863122     0.842673593
5       N       -0.874506455    -0.441853014    -1.135238985
6       O        1.755812902    -1.586994195    -0.555139015
7       O        0.137474443     2.424755900    -0.903984760
8       H       -0.511846231    -0.293976526    -2.064912349
9       H       -0.264890866    -0.371375227     0.864134683
10      H        2.449488248     1.412018182     0.887756980
11      H        1.922316209     0.292223499    -1.256116236
12      H        3.524193904     0.005380935     0.687017192
13      H        2.150221921    -0.079219336     1.832806391
14      H        1.801149313    -2.099705301     0.239935980
15      C       -1.512581778     3.427213834     1.264156016
16      C       -2.986778259     3.685413955     0.739150005
17      C       -1.318326835     3.736288758     2.779048955
18      C       -4.121568621     2.842781778     1.380657233
19      N       -1.066525771     2.063435397     1.002092123
20      O       -3.220621201     5.060505172     0.899421738
21      O       -1.450583742     2.921673464     3.688866034
22      H       -1.451099984     1.357065063     1.599079272
23      H       -0.847717803     4.129699098     0.693227838
24      H       -3.936625885     1.758848113     1.251219215
25      H       -2.977391583     3.425846085    -0.358661851
26      H       -5.089310468     3.064731920     0.883719579
27      H       -4.249923094     3.044558859     2.461737410
28      H       -3.900721559     5.371893064     0.319606256
29      H       -0.959910021    -1.441172191    -1.032510267
30      O       -0.941662177     4.977737317     3.178047617
31      H       -0.914947963     5.631619644     2.493711887
---
H10C7O6, RHF, CHARGE=0, MULT=1
HF=-252.1
1       C       -0.084330656    -0.025529215     0.146369599
2       C        1.456516090     0.096192693     0.156878105
3       C        1.902171965     1.572379358     0.050934054
4       C        2.058646734    -0.581605849     1.408930017
5       O        1.462387006    -0.852928395     2.443997341
6       O       -0.884324802     0.865011035     0.405275436
7       O        2.367167327     2.256500938     0.954557216
8       O        3.371071905    -0.887312992     1.251047138
9       C        4.194343531    -1.453166351     2.242834608
10      O        1.748958017     2.070649843    -1.201650516
```

```
11      C        2.014365071     3.400133880    -1.580807914
12      O       -0.501991418    -1.265870625    -0.212084780
13      C       -1.844365026    -1.688201166    -0.247248256
14      H        1.848694212    -0.442178101    -0.744732117
15      H        5.216847541    -1.517875330     1.800730802
16      H        3.862059757    -2.480680069     2.522870372
17      H        4.242956895    -0.824510546     3.162627652
18      H        1.759333851     3.473071112    -2.664644836
19      H        3.090091379     3.665629669    -1.449927515
20      H        1.389527086     4.132283237    -1.017565947
21      H       -1.826254810    -2.763303800    -0.545287867
22      H       -2.441624205    -1.120819357    -0.999665904
23      H       -2.340130381    -1.602382049     0.748170070
---
H3C2N3O6, RHF, CHARGE=0, MULT=1
HF=-12.4
1       N        0.005761290     0.015307958     0.006610638
2       C        1.576543165    -0.037063394    -0.062744852
3       N        2.084706519     1.450184196     0.010605212
4       O        1.309380044     2.318675327     0.313882809
5       N        2.066098208    -0.750641418     1.251253180
6       C        2.079978449    -0.761983251    -1.340706762
7       O        3.243752880     1.657516041    -0.243896282
8       O       -0.588454316     0.308347641    -0.999485530
9       O       -0.542026960    -0.236431133     1.047786580
10      O        1.976881549    -1.950930513     1.301435542
11      O        2.507917689    -0.082811644     2.149108632
12      H        1.883241105    -0.155520815    -2.249137225
13      H        1.573673827    -1.740911374    -1.471869845
14      H        3.172484059    -0.950721307    -1.290262147
---
C1N4O8, RHF, CHARGE=0, MULT=1
HF=18.5
1       N        0.006388146    -0.001270894    -0.001887138
2       C        1.575677323    -0.004729209     0.001384272
3       N        2.073774649     1.483259088    -0.003512509
4       O        2.970074218     1.777561041    -0.746780713
5       N        2.109082256    -0.745538313     1.277538202
6       N        2.113740854    -0.756641188    -1.266469946
7       O        1.524962478     2.249338313     0.741434479
8       O       -0.562037084    -0.746075107     0.749749686
9       O       -0.541301851     0.753033617    -0.759533629
10      O        2.045667555    -1.945054467     1.284923545
11      O        2.552670510    -0.079793881     2.173420396
12      O        3.272911574    -1.071391565    -1.269954593
13      O        1.342606809    -0.980095291    -2.159987979
---
H5C3N3O9, RHF, CHARGE=0, MULT=1
HF=-64.7
1       C        0.000191971     0.000260205    -0.000186628
2       C        1.583200577    -0.000048759     0.000106133
3       C        2.194529875     1.452451804     0.000093044
4       O       -0.518727314     0.602694490     1.174133050
5       O        3.601380384     1.439165323    -0.175552443
6       O        1.928183563    -0.669584343    -1.208119333
7       H        1.927969790    -0.547114297     0.916257981
8       H       -0.381487620    -1.036059660    -0.174923905
9       H       -0.395620737     0.635177775    -0.836287029
10      H        1.866656092     1.995280097     0.919962649
```

```
 11      H       1.813896228     2.039995756    -0.876979962
 12      N      -0.917064335    -0.079469590     2.270131615
 13      N       4.491488311     1.478396842     0.840183587
 14      N       2.971716820    -1.516314717    -1.345724360
 15      O      -0.712449784    -1.266882409     2.344531160
 16      O      -1.447452180     0.610184256     3.097382550
 17      O       4.108149399     1.337953565     1.976731485
 18      O       5.624338895     1.646360363     0.481596747
 19      O       3.670679447    -1.771440532    -0.396231576
 20      O       3.086664489    -1.948934776    -2.459755791
---
H21C14N1O1, RHF, CHARGE=0, MULT=1
HF=-12.2
 1       C       0.000113363     0.000117528     0.000032418
 2       C       1.561762938     0.000077392     0.000007967
 3       C      -0.502942489     1.473522460    -0.000394501
 4       C      -0.501146576    -0.784070055    -1.248110488
 5       N      -0.543938296    -0.702249772     1.274170734
 6       C      -1.873441156    -0.827043411     1.471594324
 7       C      -2.469821453    -1.477572652     2.653599932
 8       C      -2.705840524    -2.870431102     2.680142275
 9       C      -3.320644027    -3.476413861     3.788912622
 10      C      -3.723190547    -2.720018218     4.914719717
 11      C      -3.481552147    -1.325723258     4.880583587
 12      C      -2.867694466    -0.714596179     3.774850076
 13      O       0.270915828    -1.131731966     2.084084762
 14      H       1.985644158     0.547628472     0.865017804
 15      H       1.984821222    -1.024075189     0.000147041
 16      H       1.934018068     0.504705461    -0.916889219
 17      H      -1.602140951     1.553848619    -0.116910161
 18      H      -0.225209255     2.004305161     0.932444811
 19      H      -0.052881953     2.031866678    -0.847191372
 20      H      -0.231296356    -1.858145837    -1.196208302
 21      H      -1.599058655    -0.720709183    -1.385565448
 22      H      -0.042589967    -0.370431207    -2.170102205
 23      H      -2.576951190    -0.433206122     0.734695366
 24      H      -2.409548167    -3.495534868     1.835901953
 25      H      -3.478271853    -4.555988078     3.758637646
 26      C      -4.390413323    -3.352253201     6.131491287
 27      H      -3.769094119    -0.694072640     5.723509478
 28      H      -2.698998754     0.363476367     3.794543351
 29      C      -3.454872245    -4.329814460     6.874150935
 30      C      -5.764362705    -3.970101907     5.795967501
 31      H      -4.605645366    -2.531325773     6.865608095
 32      H      -2.482353200    -3.849078230     7.103714555
 33      H      -3.247334203    -5.249149983     6.291420037
 34      H      -3.903672901    -4.642648458     7.839602531
 35      H      -5.683932095    -4.871400455     5.156343264
 36      H      -6.408884468    -3.238473425     5.267352931
 37      H      -6.296585786    -4.267212849     6.723190088
---
H8C14O2, RHF, CHARGE=0, MULT=1
HF=-18.1
 1       C       0.000008428    -0.000558415     0.000811786
 2       C       1.504572629     0.000273105    -0.000467971
 3       C       2.231445521     1.225619157    -0.000983815
 4       C       1.510670846     2.546246830    -0.003478931
 5       C       0.006193975     2.545242897    -0.024192176
 6       C      -0.720736929     1.319948497    -0.023296781
```

```
7    C    -2.134290847    1.360079523    -0.043848403
8    C    -0.707210210    3.766233754    -0.043854434
9    C     3.645091992    1.185239938     0.001226236
10   C     2.218079544   -1.220841044     0.000416788
11   C     3.625246630   -1.243207095     0.002121615
12   C     4.339585203   -0.038773240     0.002863246
13   C    -2.828527677    2.584093751    -0.063673301
14   C    -2.114249252    3.788440771    -0.063450738
15   O    -0.627521636   -1.058056655     0.022209009
16   O     2.137805798    3.604127140     0.010571080
17   H    -2.723806323    0.440880300    -0.044709719
18   H    -0.183327468    4.724435353    -0.044434039
19   H     4.234599144    2.104473580     0.001618633
20   H     1.694214881   -2.179008228    -0.000170629
21   H     4.152995551   -2.197656184     0.003147561
22   H     5.430178310   -0.044372293     0.004738194
23   H    -3.919059296    2.589315294    -0.078862977
24   H    -2.641820912    4.742900105    -0.078098009
---
H8C14O2, RHF, CHARGE=0, MULT=1
HF=-11.1
1    C    -0.000087633   -0.000293892     0.000040855
2    C     1.487278229    0.000643891    -0.002497096
3    C     2.189794513    1.248784172     0.002282200
4    C     1.389544102    2.510587665     0.111605617
5    C     0.100693261    2.333468142     0.927738278
6    C    -0.702565094    1.143217592     0.499981250
7    C    -2.110374771    1.149985860     0.543084663
8    C    -0.762073299   -1.083217110    -0.491639804
9    C    -2.845338354    0.045800945     0.071081297
10   C    -2.169501779   -1.063698130    -0.453367372
11   C     3.597212567    1.271679413    -0.049565490
12   C     2.249212087   -1.188926219     0.005526891
13   C     3.656483783   -1.155814368    -0.031668220
14   C     4.331942954    0.070873780    -0.071715990
15   O     1.685402663    3.575412475    -0.410712207
16   O    -0.192096337    3.081012222     1.849157442
17   H    -2.651813673    2.013944331     0.933278629
18   H    -0.278045616   -1.959099668    -0.926217095
19   H    -3.934705667    0.060651944     0.108749619
20   H    -2.731279839   -1.916803101    -0.836267190
21   H     4.139051123    2.219191734    -0.063682074
22   H     1.765282369   -2.165501022     0.052668237
23   H     4.218199605   -2.091045904    -0.028553636
24   H     5.421194451    0.099840484    -0.112401835
---
H11C13N1O2, RHF, CHARGE=0, MULT=1
HF=12.9
1    C    -0.000024052   -0.000040246     0.000003054
2    C     1.406449586    0.000010096     0.000070345
3    C     2.118638112    1.213235488     0.000006128
4    C     1.399862371    2.428339934     0.006994254
5    C    -0.011735056    2.441301940    -0.002866454
6    C    -0.704862384    1.217182708    -0.002470278
7    N     2.142683152    3.709935613    -0.012564322
8    C     2.490762070    4.253219021     1.172761164
9    C     3.286196730    5.484553274     1.316024601
10   C     2.700311219    6.773949067     1.500772270
11   C     3.527915016    7.917513888     1.701203898
```

```
12   C       4.920396532    7.779671214    1.706520287
13   C       5.510013974    6.516140844    1.514152626
14   C       4.700420344    5.388953379    1.321756333
15   O       2.382989611    4.204160416   -1.109047342
16   O       1.363335687    7.005997203    1.533986569
17   H      -0.544558158   -0.944960221   -0.001207367
18   H       1.951313956   -0.944723683   -0.003454034
19   H       3.209276081    1.195450836   -0.009247294
20   H      -0.574006244    3.375827080   -0.015015313
21   H      -1.795408397    1.215260714   -0.007883211
22   H       2.208516131    3.723672211    2.088188433
23   H       3.082034651    8.901844861    1.848651107
24   H       5.551571373    8.656696017    1.857457927
25   H       6.595527394    6.415204039    1.513233780
26   H       5.176703302    4.417259434    1.176089044
27   H       0.843067674    6.291782133    1.194581901
---
H13C14N1O2, RHF, CHARGE=0, MULT=1
HF=26.2
1    C      -0.000018007   -0.000087337    0.000211671
2    C       1.406417681    0.000120207   -0.000124144
3    C       2.118430746    1.213397979    0.000047421
4    C       1.399783786    2.428523494    0.007933602
5    C      -0.011971130    2.441191676   -0.000406672
6    C      -0.704977384    1.216933027   -0.000771926
7    N       2.139323001    3.711590782   -0.012248130
8    C       2.468360561    4.264549770    1.171966390
9    C       3.201573239    5.534273709    1.311625250
10   C       2.518160218    6.769936946    1.336789293
11   C       3.206452311    7.982066518    1.513095300
12   C       4.616550866    7.978643175    1.661945364
13   C       5.315794974    6.738010707    1.638716471
14   C       4.609249857    5.540450712    1.466339954
15   O       2.403957262    4.190378915   -1.111393906
16   C       5.020655568   10.365944237    1.451681968
17   H      -0.544569830   -0.944970315   -0.001796200
18   H       1.951209330   -0.944622513   -0.004191033
19   H       3.209075673    1.195945489   -0.009631844
20   H      -0.573773372    3.376165642   -0.010210773
21   H      -1.795525467    1.214737339   -0.005581939
22   H       2.178970670    3.743428475    2.089341678
23   H       2.628677911    8.906019473    1.531058432
24   O       5.379027007    9.084448493    1.889774114
25   H       6.399866607    6.708180039    1.754687854
26   H       5.164966458    4.601051667    1.449160984
27   H       1.433061549    6.799294631    1.219962081
28   H       4.197344643   10.806029612    2.065507285
29   H       4.714157961   10.391444800    0.378490036
30   H       5.924545531   11.009297212    1.573683003
---
H17C15N1O2, RHF, CHARGE=0, MULT=1
HF=-44.1
1    C       0.000750480    0.000544525   -0.000040461
2    C       1.405772513   -0.000717928   -0.000161752
3    C       2.118244659    1.226153787    0.000101133
4    N       3.545146017    1.206249959    0.095060385
5    C       4.345200929    0.702794413   -1.029220432
6    C       4.417209131    1.609545180   -2.313040818
7    C       5.247007885    0.956787000   -3.485637254
```

```
8     O     4.535184313    -0.144332380   -3.990842907
9     C     5.008617105    -0.789012530   -5.105652216
10    C     4.531861607    -0.418300473   -6.390090341
11    C     4.936410666    -1.154624193   -7.516254700
12    C    -0.713402893     1.210980644    0.012575045
13    C    -0.013844662     2.429512480    0.025132254
14    C     1.390832778     2.444430920    0.016907301
15    C     5.881045450    -1.899735522   -4.964756294
16    C     6.272157400    -2.621382106   -6.105087777
17    C     5.804692484    -2.251848294   -7.378483414
18    O     4.904797493     2.894069827   -2.035500311
19    H    -0.538495611    -0.947759203   -0.004653954
20    H     1.929302946    -0.957551778    0.008291572
21    H     3.892200309     2.100843118    0.405183017
22    H     5.378179591     0.537321208   -0.628825257
23    H     3.958030172    -0.300678039   -1.323415319
24    H     3.369868405     1.772024467   -2.691354849
25    H     5.422373355     1.724219358   -4.285032592
26    H     6.262266938     0.645768021   -3.121699689
27    H     3.851270760     0.424561830   -6.515556782
28    H     4.571125603    -0.872077703   -8.504518669
29    H    -1.803358613     1.204606530    0.014600109
30    H    -0.562877423     3.371893341    0.038987248
31    H     1.904260200     3.407332044    0.024742994
32    H     6.246876216    -2.206788333   -3.984399732
33    H     6.941818700    -3.475957392   -5.998968775
34    H     6.113209950    -2.817124754   -8.258225740
35    H     5.804201131     2.888741453   -1.737802160
---
H10C14O3, RHF, CHARGE=0, MULT=1
HF=-76.3
1     C    -0.000017126     0.000003442   -0.000110087
2     C     1.406019296    -0.000024882    0.000129331
3     C     2.114071041     1.214783543    0.000149399
4     C     1.412702692     2.433344951    0.000597168
5     C     0.006646328     2.440302243    0.000548717
6     C    -0.709708559     1.222296342    0.000828383
7     C    -2.208162189     1.221526205    0.023816676
8     O    -2.776273491     1.291050885   -1.227181137
9     C    -3.972482607     0.763933981   -1.652539568
10    C    -4.595540503     1.529518873   -2.780353505
11    C    -4.296320017     1.199087638   -4.121382619
12    C    -4.901883573     1.905896754   -5.175347504
13    C    -5.807949070     2.945597877   -4.901501081
14    C    -6.108744170     3.278455824   -3.568948576
15    C    -5.506489241     2.575837843   -2.510346271
16    O    -2.933131094     1.233292010    1.007866132
17    O    -4.380279854    -0.272405466   -1.148290594
18    H    -0.533357733    -0.952098479    0.001862641
19    H     1.947179267    -0.946763507    0.000702787
20    H     3.204562823     1.211777407    0.000157142
21    H     1.958769878     3.377218075    0.001309734
22    H    -0.521934644     3.394837502    0.001521504
23    H    -3.595911578     0.394743581   -4.351652331
24    H    -4.666960436     1.644803619   -6.207712916
25    H    -6.276603684     3.492160777   -5.720536864
26    H    -6.811640637     4.083759640   -3.353230062
27    H    -5.752993902     2.846053503   -1.482257151
---
```

```
H24C12N2O3, RHF, CHARGE=0, MULT=1
HF=-176.5
1    C    -0.287664271    0.106158441    0.414499243
2    C     1.171990119    0.256860307    0.985612770
3    C     1.517708837    1.716391621    1.358960259
4    C     1.415730707   -0.701498695    2.190881346
5    C     2.874025073   -1.106744035    2.433841720
6    C    -0.657076507    1.168054794   -0.666190246
7    O    -0.151191583    1.199000317   -1.786355707
8    N    -0.501416274   -1.253052610   -0.106956035
9    H     0.091540900   -1.450471184   -0.898053968
10   H    -0.993949598    0.214221150    1.280914231
11   H     1.881057887   -0.032694929    0.170018938
12   H     1.554172229    2.370054452    0.464500678
13   H     0.790901652    2.152034828    2.073715330
14   H     2.520325225    1.782694628    1.827893581
15   H     1.016578040   -0.233423482    3.121028376
16   H     0.837381396   -1.644166964    2.065669301
17   H     3.313756018   -1.614174351    1.551508488
18   H     3.520490593   -0.242669432    2.685848229
19   H     2.928642131   -1.815619261    3.286373235
20   C    -2.251080585    3.113955749   -1.202175717
21   C    -3.810040494    3.250440553   -0.990980249
22   C    -4.239740461    3.961251073    0.310261433
23   C    -4.499029706    3.864625122   -2.246139020
24   C    -5.991945203    3.555960934   -2.409683701
25   C    -1.452892762    4.428176005   -0.981930199
26   O    -0.943059941    4.815791260    0.068943343
27   N    -1.694969669    2.064760557   -0.347588117
28   H    -1.994386825    2.078877527    0.607051502
29   H    -2.098257488    2.801042912   -2.269193837
30   H    -4.187495876    2.197736675   -0.911887926
31   H    -3.701948748    3.574029946    1.198640733
32   H    -4.074962608    5.056508075    0.271452841
33   H    -5.319249584    3.795055667    0.505249487
34   H    -4.367156864    4.970390197   -2.248655744
35   H    -3.989420017    3.494109193   -3.165679014
36   H    -6.194592531    2.465897415   -2.399884170
37   H    -6.609814744    4.024467893   -1.618029646
38   H    -6.350567773    3.951643787   -3.382689869
39   H    -1.448261929   -1.355064761   -0.437748261
40   O    -1.305274416    5.190285019   -2.093659607
41   H    -0.787421053    5.979089757   -1.984307614
---
H24C12N2O3, RHF, CHARGE=0, MULT=1
HF=-177.1
1    O    -0.043921331   -0.132784802   -0.252073946
2    C     0.993886669    0.036914614    0.381692191
3    C     1.809040514    1.362195160    0.364223703
4    C     0.887039131    2.596168353    0.619743481
5    N     2.559262429    1.338884820   -0.900674866
6    C     1.587235970    3.946774291    0.943567459
7    C     2.145210951    4.021774145    2.380062197
8    H     2.569203743    1.336885401    1.188337800
9    H     0.229502462    2.740577593   -0.268145753
10   H     0.203696248    2.344024325    1.463131395
11   C     0.653559674    5.140807857    0.642293108
12   H     0.360865324    5.158720550   -0.427592757
13   H    -0.275095500    5.110002329    1.246509627
```

```
14      H       2.463095061      4.065087677     0.256327088
15      H       1.350143792      3.912974188     3.144898164
16      H       2.900635992      3.232325156     2.566672529
17      H       2.649983874      4.993783375     2.557019339
18      H       1.161149831      6.104248343     0.853707198
19      H       3.292073447      2.030647901    -0.881552714
20      O      -0.489499161     -2.767220040     3.320592850
21      C       0.630039074     -2.554047060     2.858883218
22      C       0.955647552     -2.342845287     1.354056879
23      C       1.916901360     -3.412747590     0.745014268
24      N       1.494798362     -0.990946787     1.201931861
25      C       1.405314720     -4.881083995     0.716900555
26      C       2.575998403     -5.868837116     0.512271774
27      H      -0.022896916     -2.412907667     0.812109413
28      H       2.153818521     -3.100453166    -0.298625645
29      H       2.885776953     -3.385119680     1.294129318
30      C       0.281939513     -5.127206161    -0.311905184
31      H      -0.622391775     -4.529896861    -0.080836479
32      H       0.598039130     -4.875974692    -1.344568338
33      H       0.976368512     -5.120771520     1.722579645
34      H       3.081157365     -5.723989169    -0.463770712
35      H       3.340966362     -5.756114738     1.307531348
36      H       2.219429228     -6.918290518     0.554196423
37      H      -0.029918325     -6.192050459    -0.306918544
38      H       2.406446611     -0.850356335     1.592017733
39      H       1.972196118      1.586431554    -1.682477124
40      O       1.680571830     -2.479869071     3.715696635
41      H       1.468131054     -2.597244357     4.633926990
---
H14C12O4, RHF, CHARGE=0, MULT=1
HF=-167.3
1       C       0.061366649     -0.046576041     0.101969134
2       C       1.410411311     -0.184207802    -0.307295876
3       C       2.267706946      0.927808499    -0.361128243
4       C       1.794197397      2.203468888    -0.009576831
5       C       0.457349217      2.358671017     0.393096076
6       C      -0.400312602      1.245575061     0.447154761
7       C      -0.861985305     -1.244922919     0.134430250
8       C      -0.736933495     -2.097678541     1.452654468
9       O      -1.431854781     -1.429318817     2.482776016
10      O      -1.410757967     -3.333133214     1.312858783
11      C      -0.845568032     -4.553365474     1.082560098
12      C      -1.823036114     -5.712167282     1.236988538
13      O       0.336032230     -4.630276817     0.761093200
14      C      -1.424506212     -1.742703534     3.810127857
15      O      -0.744618608     -2.680668071     4.217183133
16      C      -2.296244388     -0.830725095     4.665372509
17      H       1.804658956     -1.162164480    -0.590284305
18      H       3.303437097      0.798409649    -0.677327551
19      H       2.459162498      3.066401043    -0.050615542
20      H       0.080835097      3.345459996     0.664737031
21      H      -1.434753183      1.397249020     0.760644551
22      H      -1.917000234     -0.925305468    -0.019894399
23      H      -0.626145012     -1.906828553    -0.730944128
24      H       0.351668174     -2.227384540     1.720119913
25      H      -2.080979501     -5.853059735     2.306051339
26      H      -2.756468888     -5.528455935     0.668539396
27      H      -1.370981035     -6.651954507     0.861906163
28      H      -2.943154416     -1.444160420     5.324991921
```

```
29      H     -1.648768024    -0.192861691     5.301259714
30      H     -2.949853269    -0.169720309     4.064036215
---
H14C12O4, RHF, CHARGE=0, MULT=1
HF=-161.8
1       C      0.000757255     0.000217200    -0.000136970
2       C      1.406254524    -0.000691340    -0.000161577
3       C      2.114814845     1.212799311    -0.000055856
4       C      1.434013944     2.454697261    -0.002682322
5       C      0.019052326     2.437115972     0.002415066
6       C     -0.690366880     1.223187907     0.002237219
7       C      2.199978752     3.760543262     0.026926393
8       C      2.740122934     4.217525245    -1.378738452
9       O      1.643918150     4.586548822    -2.185836103
10      C      1.670644755     4.870750954    -3.520122130
11      C      0.303341693     5.235530446    -4.085837799
12      O      3.503042796     5.401504563    -1.253118284
13      C      4.863011417     5.509298147    -1.272673785
14      C      5.353274771     6.948545387    -1.375743801
15      O      2.727577748     4.811993009    -4.141339352
16      O      5.558295697     4.502431509    -1.181081974
17      H     -0.548801554    -0.941340915    -0.000582403
18      H      1.950781200    -0.945585877     0.000755786
19      H      3.205942180     1.180889285     0.005470279
20      H     -0.544885784     3.371636559     0.007381909
21      H     -1.780731126     1.233433598     0.004895936
22      H      1.569866671     4.563323125     0.474556988
23      H      3.073208272     3.652282667     0.709283473
24      H      3.322593391     3.374251418    -1.851094888
25      H     -0.108671775     6.126244576    -3.570536469
26      H      0.379908761     5.466508284    -5.167168914
27      H     -0.407997713     4.394360615    -3.961973207
28      H      5.044655292     7.394396852    -2.342623122
29      H      4.940832361     7.566574067    -0.553201135
30      H      6.459280489     6.986210245    -1.314340786
---
H6C14O4, RHF, CHARGE=0, MULT=1
HF=-50
1       C      0.000396349     0.000096350     0.000085726
2       C      1.411935929    -0.000492898    -0.000001011
3       C      2.116500678     1.209770969    -0.000209272
4       C      1.418790847     2.436797541     0.001137169
5       C      0.012738822     2.450733009    -0.000337615
6       C     -0.706112387     1.215939952    -0.002464337
7       C     -2.208656064     1.280021226    -0.017333389
8       C     -2.766787149     2.456215572    -0.787386710
9       C     -2.081374154     3.635540989    -0.781994229
10      C     -0.784332810     3.725993017    -0.009160463
11      C     -4.063882545     2.332663990    -1.553215155
12      C     -4.771963385     3.634587843    -1.790154732
13      C     -4.095054673     4.800786337    -1.783438063
14      C     -2.614672951     4.830216463    -1.538934012
15      O     -2.927050831     0.490988459     0.579640830
16      O     -0.452746951     4.736419259     0.594268826
17      O     -4.481559054     1.265638762    -1.980964915
18      O     -1.894131590     5.727181728    -1.953952433
19      H     -0.527247585    -0.955283738    -0.001868340
20      H      1.950198217    -0.949096950    -0.000209056
21      H      3.207251747     1.210001155    -0.000891853
```

```
22      H       1.988650171     3.367609957    -0.000437818
23      H      -5.846868145     3.591281793    -1.973641164
24      H      -4.589796280     5.757087950    -1.961428823
---
H8C14O4, RHF, CHARGE=0, MULT=1
HF=-112.6
1       C      -0.000231485    -0.000170077    -0.000148159
2       C       1.409955092     0.000016357    -0.000384314
3       C       2.115152130     1.211751131     0.000421774
4       C       1.418520667     2.437860996     0.002747810
5       C       0.011641335     2.448469274    -0.001524374
6       C      -0.704793192     1.217659185    -0.003627534
7       C      -2.205978271     1.303746408    -0.018099285
8       C      -2.769215883     2.411670094    -0.870037090
9       C      -2.045330937     3.655450618    -0.866194244
10      C      -0.804568961     3.711206795    -0.011907284
11      C      -3.927827553     2.279559890    -1.671702056
12      C      -4.364827983     3.380068268    -2.485119593
13      C      -3.667353361     4.576110292    -2.480554997
14      C      -2.498589003     4.733185410    -1.661941748
15      O      -2.909697166     0.550697285     0.647472113
16      O      -0.499383344     4.692468486     0.657827206
17      O      -4.674168776     1.157195666    -1.810236619
18      O      -1.873769646     5.930204874    -1.794935137
19      H      -0.529581674    -0.954255312    -0.000456896
20      H       1.949954533    -0.947706275    -0.000154851
21      H       3.205896109     1.210611305     0.000544711
22      H       1.986189628     3.369715257     0.004428515
23      H      -5.253569284     3.275239394    -3.110705558
24      H      -4.010765939     5.404116584    -3.104203466
25      H      -4.619673886     0.543566218    -1.092115439
26      H      -1.437997961     6.239629775    -1.013218285
---
H10C14O4, RHF, CHARGE=0, MULT=1
HF=-104.5
1       C      -0.000237699    -0.000278794     0.000017536
2       C       1.406088287     0.000672747     0.000228102
3       C       2.109715683     1.218334500     0.000226754
4       C       1.419308451     2.442548435    -0.001077637
5       C       0.001745058     2.433074047    -0.003182817
6       C      -0.715050325     1.209899642    -0.001550583
7       O      -0.687103878     3.623498524     0.104182290
8       C      -1.099873941     4.355342193    -0.962988952
9       C      -1.863998691     5.605667744    -0.526833878
10      O      -1.037539486     6.674769965    -0.390162388
11      C      -1.468449228     7.935471595    -0.032679522
12      C      -1.803582294     8.871433639    -1.043445721
13      C      -2.135846866    10.185892142    -0.673245801
14      C      -2.135560344    10.568608479     0.680053335
15      C      -1.799794387     9.634885970     1.676636265
16      C      -1.463752206     8.314022910     1.333439377
17      O      -3.069549466     5.662407518    -0.338658294
18      O      -0.892583532     4.026726926    -2.121010554
19      H      -0.543325060    -0.946165733     0.000744865
20      H       1.951611410    -0.943484802     0.000197010
21      H       3.200350638     1.215925421     0.000913224
22      H       1.978909928     3.378435557     0.001363881
23      H      -1.805445852     1.193203041     0.000733832
24      H      -1.805710632     8.589992913    -2.096966345
```

```
25      H       -2.395770293    10.912868794    -1.443719409
26      H       -2.396105430    11.590555708     0.957025434
27      H       -1.799655508     9.935478739     2.725106361
28      H       -1.203449608     7.601851722     2.117084612
---
H12C14O4, RHF, CHARGE=0, MULT=1
HF=-132.7
1       C       -0.000556479     0.000094958     0.000093806
2       C        1.380973519     0.000224821    -0.000127651
3       C        2.108109155     1.238525946     0.000082787
4       C        1.414028975     2.440661081     0.002045164
5       C       -0.024968433     2.469294236     0.001297822
6       C       -0.745233730     1.231113964     0.000913959
7       C       -2.184173709     1.259590018     0.001653178
8       C       -2.878561256     2.461661606     0.000811472
9       C       -2.151615242     3.699931916     0.000811428
10      C       -0.770099902     3.700069374     0.000858596
11      C       -4.380085958     2.488436836    -0.023086006
12      O       -5.087128623     2.524552957    -1.025423844
13      O       -4.935106099     2.469207282     1.219229192
14      C       -6.313272999     2.495142185     1.490513989
15      C        3.609642159     1.211195357    -0.025004285
16      O        4.315804000     1.171051023    -1.027781373
17      O        4.165808612     1.234760273     1.216710015
18      C        5.544268660     1.208832061     1.486894912
19      H       -0.540226980    -0.948611225    -0.000811265
20      H        1.925718892    -0.945369790    -0.001598890
21      H        1.952016508     3.391076112     0.002741126
22      H       -2.722162211     0.309196053     0.001398687
23      H       -2.696448661     4.645467550    -0.000434633
24      H       -0.230581968     4.648869814     0.000224060
25      H       -6.802198184     3.423173794     1.109780579
26      H       -6.844498607     1.610014279     1.066875683
27      H       -6.417599632     2.471263423     2.601440183
28      H        6.032029948     0.278852681     1.109502639
29      H        6.075956476     2.091797019     1.059193103
30      H        5.649581939     1.237165289     2.597609898
---
H16C10O6, RHF, CHARGE=0, MULT=1
HF=-280.4
1       C       -0.168451161     0.028560114    -0.201852096
2       C        1.338249310     0.148661543     0.095386042
3       O        1.707009183     1.512514368     0.057938234
4       C        2.929182738     1.993476628     0.396848046
5       C        2.987871371     3.537868610     0.346353484
6       C        2.219476561     4.166481799     1.531447362
7       O        1.918581891     5.471876168     1.317895340
8       C        1.286462596     6.317317617     2.257735570
9       C        1.223496239     7.745538904     1.684246010
10      O        3.855241755     1.243555484     0.681725572
11      O        1.878993152     3.591209498     2.558014872
12      C        4.449455230     4.039773157     0.297984177
13      O        5.064205366     4.564818947     1.218483265
14      O        5.016865250     3.857765993    -0.920471987
15      C        6.360941098     4.156553434    -1.238702877
16      C        6.654062451     3.677962313    -2.673445505
17      H       -0.783400269     0.559011456     0.552019524
18      H       -0.428354300     0.428215092    -1.202163056
19      H       -0.452997717    -1.043419495    -0.179583344
```

```
20      H        1.557778647     -0.302074842     1.097212601
21      H        1.920452645     -0.433880740    -0.665193184
22      H        2.482212544      3.865774467    -0.598767974
23      H        0.249016542      5.953726986     2.476936856
24      H        1.845853529      6.328742202     3.228431459
25      H        2.232480833      8.156316922     1.481235918
26      H        0.637220254      7.790580008     0.745113061
27      H        0.729522932      8.406997190     2.425342219
28      H        6.543991851      5.260064935    -1.164622585
29      H        7.065485254      3.654987700    -0.526595255
30      H        6.531344936      2.581631427    -2.777297974
31      H        5.998908587      4.170853481    -3.418833408
32      H        7.704969535      3.928330091    -2.926154514
---
H18C11O6, RHF, CHARGE=0, MULT=1
HF=-286.2
1       C       -0.000148299     -0.000077777     0.000070223
2       C        1.562423758      0.000097017    -0.000153651
3       C        2.107785359      1.459895248     0.000156710
4       O        1.626532785      2.212599285     1.022782027
5       C        1.975661855      3.559574669     1.267485351
6       O        2.898687117      1.932715162    -0.807435908
7       C        2.104574369     -0.759770722     1.253886708
8       O        1.492256124     -1.605032352     1.895463041
9       O        3.367750869     -0.402609763     1.590838013
10      C        2.080291877     -0.754898009    -1.266142949
11      O        2.810489572     -1.739947332    -1.253068621
12      O        1.622123319     -0.243530382    -2.434370622
13      C        4.189149347     -1.035058079     2.551447994
14      C        4.067423631     -0.412973697     3.955369987
15      C        2.024433459     -0.677480759    -3.718183380
16      C        1.330327874      0.194864824    -4.781361424
17      C        1.214074684      4.056313681     2.511197433
18      H       -0.401621846      0.687783358    -0.770767617
19      H       -0.398725569     -1.012781983    -0.211190665
20      H       -0.410362023      0.322984561     0.977457483
21      H        3.078608886      3.660665548     1.438013512
22      H        1.716248742      4.204352269     0.388121197
23      H        4.010300312     -2.138423658     2.603661709
24      H        5.236911691     -0.899592289     2.173114853
25      H        4.247629036      0.680516978     3.941079663
26      H        3.073060102     -0.590698666     4.410041699
27      H        4.828144536     -0.872251479     4.619342921
28      H        3.135825026     -0.599538006    -3.836336791
29      H        1.748221166     -1.752320366    -3.876622762
30      H        0.226143240      0.119315904    -4.724544631
31      H        1.608594443      1.263260099    -4.688585649
32      H        1.642797240     -0.152187148    -5.787712582
33      H        0.115703445      4.009217508     2.372269210
34      H        1.472355602      3.472960570     3.417119682
35      H        1.487701492      5.114870717     2.699047201
---
H5C7N3O6, RHF, CHARGE=0, MULT=1
HF=12.9
1       C        0.000065363      0.000122033    -0.000054800
2       C        1.423670227      0.000083632     0.000272333
3       C        2.021787374      1.291658508    -0.000428987
4       C        1.286869589      2.496642174     0.025080898
5       C       -0.117815699      2.412944392     0.043240483
```

```
 6    C    -0.784117285    1.173386891     0.026642347
 7    C     2.236525698   -1.276381775    -0.025044258
 8    N     3.523153178    1.446560449    -0.041824416
 9    H     1.790939340    3.467098646     0.028074070
10    N    -0.928010377    3.681020463     0.077126114
11    H    -1.876636481    1.123988063     0.032779082
12    N    -0.770989802   -1.297541529    -0.041567455
13    O    -1.173484289   -1.768719319     0.995972588
14    O    -0.984218304   -1.811091100    -1.114413733
15    O     4.120872230    1.606328495     0.996207452
16    O     4.077476721    1.432110028    -1.115530117
17    O    -1.348198941    4.128087741    -0.963618624
18    O    -1.131255133    4.209904436     1.144113174
19    H     3.224010720   -1.167062261     0.464802838
20    H     1.737630383   -2.109378674     0.508768059
21    H     2.406555715   -1.590249594    -1.077190984
---
H9C9N1O7, RHF, CHARGE=0, MULT=1
HF=-184.4
 1    O     0.000046494   -0.000012178    -0.000207332
 2    C     1.364025425   -0.000072565    -0.000007934
 3    C     1.880583352    1.298122048     0.000006443
 4    C     0.722850703    2.150445595     0.001413230
 5    C    -0.394265905    1.304796187     0.001677263
 6    C    -1.891881455    1.622816864    -0.006491171
 7    O    -2.602520102    0.865482806     0.944293748
 8    C    -2.765317262    1.162122941     2.269526236
 9    C    -3.714744817    0.212409617     2.988239992
10    O    -2.155690636    2.110903679     2.751612529
11    O    -2.389957350    1.211048910    -1.255734852
12    C    -3.651196042    1.426605912    -1.744898265
13    O    -4.427839155    2.142280984    -1.122493145
14    C    -3.906891822    0.724744664    -3.072427702
15    N     2.012166246   -1.320783072     0.003272837
16    O     1.351728303   -2.326138370     0.127655213
17    O     3.218050035   -1.359125497    -0.117241839
18    H     2.915427651    1.608571380     0.000613794
19    H     0.727312094    3.230386885     0.007753375
20    H    -2.017390202    2.735536927     0.143339677
21    H    -3.721386008    0.425796932     4.075813464
22    H    -3.406065872   -0.842171015     2.843729985
23    H    -4.746675582    0.336541727     2.602573675
24    H    -3.175862714    1.053638295    -3.837980474
25    H    -3.824849518   -0.374157584    -2.952531468
26    H    -4.924939045    0.959021281    -3.442657769
---
H5C7N3O7, RHF, CHARGE=0, MULT=1
HF=-5.8
 1    C    -0.221125414    0.036858878    -0.205843833
 2    C     1.203536355    0.007854284    -0.125883990
 3    C     1.890896136    1.239876106    -0.347788983
 4    C     1.208323162    2.446213936    -0.598178177
 5    C    -0.200933412    2.416332989    -0.637470714
 6    C    -0.934494600    1.226472781    -0.454747221
 7    O     1.887087981   -1.148131480     0.006882962
 8    N     3.395067093    1.270257700    -0.334045350
 9    H     1.755858416    3.379207364    -0.758597174
10    N    -0.945692092    3.699800858    -0.887455450
11    H    -2.026967061    1.225831981    -0.504706396
```

```
12      N     -1.010419008    -1.233094255    -0.038788839
13      O     -1.271423488    -1.889398858    -1.017105725
14      O     -1.371527111    -1.545490889     1.072454170
15      O      3.963134576     1.313238936     0.732471601
16      O      3.985165229     1.267390156    -1.386862076
17      O     -1.383626650     3.912687073    -1.992921255
18      O     -1.080719319     4.477279434     0.027865650
19      C      2.217481966    -1.725960427     1.259444287
20      H      2.073047737    -1.031485803     2.118076081
21      H      1.592189709    -2.635183254     1.419890926
22      H      3.288935191    -2.030900776     1.211480743
---
H7C8N3O7, RHF, CHARGE=0, MULT=1
HF=-20.1
1       C     -0.113573995     0.007022051    -0.034192452
2       C      1.305366640     0.006300387     0.122796598
3       C      1.950092623     1.287692629     0.118928771
4       C      1.237789457     2.496108916     0.029887003
5       C     -0.168506634     2.431397153    -0.069085344
6       C     -0.857494249     1.204461774    -0.113251000
7       O      2.095138650    -1.081544886     0.134264238
8       N      3.451881647     1.362023865     0.194978354
9       H      1.757989648     3.458184412     0.039763776
10      N     -0.953910556     3.711562024    -0.137036066
11      H     -1.946568060     1.178517132    -0.215381651
12      N     -0.887970631    -1.274930675    -0.166158436
13      O     -1.504682131    -1.685109883     0.790329957
14      O     -0.893455419    -1.842099033    -1.232250487
15      O      4.085817481     1.371808585    -0.831823991
16      O      3.973697280     1.430129989     1.283442994
17      O     -1.618826727     4.033628227     0.819504317
18      O     -0.899515760     4.376885661    -1.144154178
19      C      2.105488638    -2.142283822     1.079167436
20      H      1.228532098    -2.817563709     0.921010026
21      H      3.017687458    -2.733844489     0.802432181
22      C      2.191257694    -1.727727324     2.559107464
23      H      3.132981902    -1.190765905     2.786349611
24      H      1.342458576    -1.089060662     2.873917157
25      H      2.167927576    -2.645058643     3.182292248
---
F1, UHF, CHARGE=0, MULT=2
HF=18.9
1    F    0.000000000    0.000000000    0.000000000
---
F1, RHF, CHARGE=-1, MULT=1
HF=-61.0
1    F    0.000000000    0.000000000    0.000000000
---
H1F1, RHF, CHARGE=0, MULT=1
HF=-65.1, DIP=1.83, IE=16.06
1    H    0.000000000    0.000000000    0.000000000
2    F    0.956345846    0.000000000    0.000000000
EXPGEOM
1    F    0.00000    0.00000     0.09260
2    H    0.00000    0.00000    -0.83340
---
C1F1, UHF, CHARGE=0, MULT=2
HF=61.0
1    C    0.000000000    0.000000000    0.000000000
```

```
2    F    1.263205874     0.000000000     0.000000000
EXPGEOM
1    C    0.00000    0.00000    -0.77130
2    F    0.00000    0.00000     0.51420
---
H2C1F1, RHF, CHARGE=1, MULT=1
HF=200.3
1    H    0.000000000     0.000000000     0.000000000
2    C    1.107465390     0.000000000     0.000000000
3    H    1.693877290     0.939501913     0.000000000
4    F    1.721512584    -1.107144723     0.000087998
---
H3C1F1, RHF, CHARGE=0, MULT=1
HF=-56.8, DIP=1.86, IE=13.31
1    F   -0.007062878     0.010965470     0.000866855
2    C    1.340023648     0.000373645     0.001024994
3    H    1.741673919     1.043402027     0.001065765
4    H    1.729900309    -0.525354898    -0.904950679
5    H    1.729603291    -0.525696443     0.906960767
EXPGEOM
1    C    0.00000    0.00000    -0.64050
2    F    0.00000    0.00000     0.75760
3    H    0.00000    1.03720    -0.99170
4    H    0.89820   -0.51860    -0.99170
5    H   -0.89820   -0.51860    -0.99170
---
H1C2F1, RHF, CHARGE=0, MULT=1
HF=30.0, DIP=0.70, IE=11.3
1    H    0.000000000     0.000000000     0.000000000
2    C    1.049045890     0.000000000     0.000000000
3    C    2.241731460     0.000000000     0.000000000
4    F    3.519183140     0.000000000     0.000000000
EXPGEOM
1    C    0.00000    0.00000    -0.09420
2    C    0.00000    0.00000    -1.29470
3    F    0.00000    0.00000     1.18790
4    H    0.00000    0.00000    -2.35830
---
H3C2F1, RHF, CHARGE=0, MULT=1
HF=-32.5, DIP=1.43, IE=10.58
1    C    0.000000000     0.000000000     0.000000000
2    C    1.350552409     0.000000000     0.000000000
3    H    1.949110199     0.921750225     0.000000000
4    H   -0.558248400     0.933251415     0.000000000
5    H   -0.616412235    -0.895586524    -0.000000000
6    F    2.078069964    -1.106582227    -0.000000000
EXPGEOM
1    C    0.00000    0.43580    0.00000
2    C    1.19290   -0.14110    0.00000
3    F   -1.15000   -0.28190    0.00000
4    H   -0.19500    1.50460    0.00000
5    H    1.30910   -1.21920    0.00000
6    H    2.07820    0.48390    0.00000
---
H4C2F1, RHF, CHARGE=1, MULT=1
HF=166.0
1    C    0.000000000     0.000000000     0.000000000
2    C    1.498922060     0.000000000     0.000000000
3    F    2.178342476     1.079266000     0.000000000
```

```
4    H     2.077483792    -0.949372044    -0.000600384
5    H    -0.455868650     1.011004590     0.014929622
6    H    -0.356961004    -0.545227742    -0.906801065
7    H    -0.358136387    -0.573474552     0.888366039
---
H5C2F1, RHF, CHARGE=0, MULT=1
HF=-62.9, DIP=1.96, IE=12.43
1    C     0.000000000     0.000000000     0.000000000
2    C     1.546249237     0.000000000     0.000000000
3    F     2.076219323     1.244388743     0.000000000
4    H     1.917912015    -0.560163768     0.899597936
5    H     1.917961132    -0.560247441    -0.899515292
6    H    -0.407914131     0.502978996    -0.898750727
7    H    -0.364969740    -1.046284267     0.003646804
8    H    -0.408223731     0.509257154     0.895065139
EXPGEOM
1    C     0.00000     0.55760     0.00000
2    C     1.13790    -0.44130     0.00000
3    F    -1.23920    -0.11420     0.00000
4    H     0.01760     1.19470     0.89250
5    H     0.01760     1.19470    -0.89250
6    H     2.09450     0.09390     0.00000
7    H     1.09790    -1.07640     0.88960
8    H     1.09790    -1.07640    -0.88960
---
H7C3F1, RHF, CHARGE=0, MULT=1
HF=-69.4, IE=11.08
1    C     0.000000000     0.000000000     0.000000000
2    C     1.556587520     0.000000000     0.000000000
3    C    -0.607950102     1.433115695     0.000000000
4    F    -0.485417741    -0.733212448     1.036022658
5    H    -0.328884310    -0.496839373    -0.959223212
6    H     1.949796547    -1.035428883    -0.038929876
7    H    -1.714657408     1.390734171    -0.039917008
8    H     1.935835434     0.538462925    -0.890770816
9    H     1.972795025     0.488702705     0.902609996
10   H    -0.259776046     1.992787902    -0.890359080
11   H    -0.321469694     2.006731149     0.903110504
EXPGEOM
1    C     0.28190     0.23820     0.00000
2    F    -0.87980     1.05600     0.00000
3    H     1.12950     0.93630     0.00000
4    C     0.28190    -0.58930     1.27290
5    C     0.28190    -0.58930    -1.27290
6    H     1.19830    -1.18560     1.33020
7    H     1.19830    -1.18560    -1.33020
8    H     0.23430     0.05610     2.15380
9    H     0.23430     0.05610    -2.15380
10   H    -0.57530    -1.26940     1.28730
11   H    -0.57530    -1.26940    -1.28730
---
H5C6F1, RHF, CHARGE=0, MULT=1
HF=-27.8, DIP=1.66, IE=9.19
1    C     0.000000000     0.000000000     0.000000000
2    C     1.422388981     0.000000000     0.000000000
3    C     2.137708801     1.229375025     0.000000000
4    C     1.424504888     2.439082447     0.000000000
5    C     0.018207313     2.441355692     0.000000000
6    C    -0.686949748     1.224720656     0.000000000
```

```
  7      H    -0.555762588   -0.937740707    0.000000000
  8      H    -0.525463219    3.386327121   -0.000000000
  9      F     2.083323491   -1.149347808    0.000000000
 10      H    -1.777829030    1.229359823   -0.000000000
 11      H     3.227728173    1.238155687    0.000000000
 12      H     1.968830016    3.384437665    0.000000000
EXPGEOM
 1    F    0.00000    0.00000    2.28790
 2    C    0.00000    0.00000    0.92890
 3    C    0.00000    1.21800    0.25940
 4    C    0.00000   -1.21800    0.25940
 5    C    0.00000    1.20800   -1.13640
 6    C    0.00000   -1.20800   -1.13640
 7    C    0.00000    0.00000   -1.83600
 8    H    0.00000    2.14380    0.82440
 9    H    0.00000   -2.14380    0.82440
10    H    0.00000    2.15030   -1.67590
11    H    0.00000   -2.15030   -1.67590
12    H    0.00000    0.00000   -2.92130
---
H11C6F1, RHF, CHARGE=0, MULT=1
HF=-80.5
 1    C    0.000000000    0.000000000    0.000000000
 2    C    1.565481124    0.000000000    0.000000000
 3    C    2.156450895    1.450006505    0.000000000
 4    C    1.533519019    2.359342465   -1.072061342
 5    C   -0.003519449    2.332729856   -1.109177578
 6    C   -0.605971099    0.918240916   -1.073294165
 7    H   -0.359877373   -1.039422308   -0.170546453
 8    H   -0.380031763    0.297015930    1.003198654
 9    H    1.907722477   -0.507661945   -0.947765615
10    F    2.047804192   -0.717947747    1.046492186
11    H    3.254827750    1.391187626   -0.171175481
12    H    2.025064868    1.913835098    1.003242331
13    H    1.872405723    3.404976501   -0.892020424
14    H    1.933577181    2.080814359   -2.074099187
15    H   -0.351500521    2.850376384   -2.031808394
16    H   -0.405512554    2.928655677   -0.257913562
17    H   -1.702761956    0.996928619   -0.895732222
18    H   -0.495954891    0.442323515   -2.074912818
---
H19C9F1, RHF, CHARGE=0, MULT=1
HF=-101.2
 1    C    0.000000000    0.000000000    0.000000000
 2    C    1.531238364    0.000000000    0.000000000
 3    C    2.174735215    1.400255940    0.000000000
 4    C    3.717010283    1.384216655    0.000973062
 5    C    4.358850265    2.786917779    0.000566660
 6    C    5.901384724    2.770165069    0.001986211
 7    C    6.542872985    4.173053638    0.001485150
 8    C    8.083076652    4.155081905    0.002996809
 9    C    8.716385802    5.578390845    0.002433354
10    F   10.068652752    5.535400112    0.002978540
11    H   -0.412988416    0.505409180    0.896414658
12    H   -0.412800702    0.506301758   -0.896021436
13    H   -0.382042926   -1.041814291   -0.000707479
14    H    1.882422968   -0.570228058    0.890993501
15    H    1.882376390   -0.571117063   -0.890545699
16    H    1.815117830    1.964418218    0.891095309
```

```
17      H      1.816487691      1.963703840     -0.892097102
18      H      4.074408274      0.820793715      0.893313078
19      H      4.075611576      0.819495937     -0.890072870
20      H      4.000323823      3.351728997      0.891587504
21      H      4.001834775      3.350545080     -0.891820389
22      H      6.259114518      2.207786548      0.894777890
23      H      6.260720794      2.205279183     -0.888582614
24      H      6.183921146      4.738167256      0.892352274
25      H      6.186151563      4.736537622     -0.891327805
26      H      8.449563571      3.599930474      0.895867475
27      H      8.451087143      3.598019487     -0.888085758
28      H      8.369657116      6.152382993      0.902919348
29      H      8.370174216      6.152317210     -0.898151121
---
C1N1F1, RHF, CHARGE=0, MULT=1
HF=8.6, DIP=2.17
1       N      0.000000000      0.000000000      0.000000000
2       C      1.160281250      0.000000000      0.000000000
3       F      2.433363920      0.000000358      0.000000000
EXPGEOM
1       C      0.00000      0.00000     -0.15070
2       F      0.00000      0.00000      1.12120
3       N      0.00000      0.00000     -1.31240
---
O1F1, UHF, CHARGE=0, MULT=2
HF=26.1
1       F      0.000000000      0.000000000      0.000000000
2       O      1.225002931      0.000000000      0.000000000
EXPGEOM
1       O      0.00000      0.00000     -0.71960
2       F      0.00000      0.00000      0.63970
---
H1O1F1, RHF, CHARGE=0, MULT=1
HF=-20.9, DIP=2.23
1       H      0.000000000      0.000000000      0.000000000
2       O      0.963965100      0.000000000      0.000000000
3       F      1.355501396      1.215434371      0.000000000
EXPGEOM
1       O      0.05370      0.71460      0.00000
2       H     -0.91270      0.85110      0.00000
3       F      0.05370     -0.72980      0.00000
---
H1C1O1F1, RHF, CHARGE=0, MULT=1
HF=-90, DIP=2.02
1       O      0.000000000      0.000000000      0.000000000
2       C      1.221663739      0.000000000      0.000000000
3       F      1.842670060      1.173595222      0.000000000
4       H      1.889649177     -0.881114882      0.000000400
EXPGEOM
1       C      0.00000      0.40260      0.00000
2       O      1.15440      0.14420      0.00000
3       F     -0.97320     -0.55130      0.00000
4       H     -0.47600      1.39300      0.00000
---
H3C2O1F1, RHF, CHARGE=0, MULT=1
HF=-106.4
1       C      0.000000000      0.000000000      0.000000000
2       C      1.526089594      0.000000000      0.000000000
3       H      1.910895715      1.039433562      0.000000000
```

```
  4    H     1.918143967    -0.517930706     0.896807300
  5    H     1.910156722    -0.519990855    -0.900366424
  6    F    -0.603516682     0.601874016    -1.023791163
  7    O    -0.738527921    -0.495885354     0.843035812
EXPGEOM
  1    C     0.00000        0.19030       0.00000
  2    C     1.08410       -0.84220       0.00000
  3    O     0.08070        1.37550       0.00000
  4    F    -1.24090       -0.41910       0.00000
  5    H     2.05860       -0.35520       0.00000
  6    H     0.97960       -1.48260       0.88100
  7    H     0.97960       -1.48260      -0.88100
 ---
N1O1F1, RHF, CHARGE=0, MULT=1
HF=-15.7, DIP=1.81, IE=12.94
  1    F     0.000000000    0.000000000    0.000000000
  2    N     1.304897240    0.000000000    0.000000000
  3    O     1.774190822    1.061716326    0.000000000
EXPGEOM
  1    F    -0.96190       -0.66560       0.00000
  2    N     0.00000        0.57610       0.00000
  3    O     1.08210        0.24480       0.00000
 ---
O2F1, UHF, CHARGE=0, MULT=2
HF=3.0
  1    F     0.000000000    0.000000000    0.000000000
  2    O     1.311894590    0.000000000    0.000000000
  3    O     1.771223822    1.101924284    0.000000000
EXPGEOM
  1    O     1.14710        0.25240       0.00000
  2    O     0.00000        0.57470       0.00000
  3    F    -1.01970       -0.73520       0.00000
 ---
H5C7O2F1, RHF, CHARGE=0, MULT=1
HF=-118.4, IE=9.9
  1    C    -0.059493126     0.104884920    -0.036937888
  2    C     2.773756606     0.009650280     0.033712875
  3    C     0.676069225     1.103037318     0.659775857
  4    C     0.635008521    -0.925493903    -0.688219425
  5    C     2.096824324     1.055234458     0.696108173
  6    C     2.039999314    -0.978821428    -0.658035373
  7    H    -1.148959631     0.135599596    -0.067670817
  8    F     0.035559345     2.084384809     1.278699758
  9    H     0.077915281    -1.694582787    -1.225211505
 10    H     2.652780223     1.823961449     1.234866068
 11    H     2.552917076    -1.792131869    -1.173709345
 12    C     4.272252338    -0.062011310     0.081957742
 13    O     4.946584363    -0.657726090     0.919692280
 14    O     4.894035441     0.609597204    -0.920162633
 15    H     5.843166669     0.576324664    -0.907388424
 ---
N1O2F1, RHF, CHARGE=0, MULT=1
HF=-26.0, DIP=0.47, IE=13.51
  1    O     0.000000000    0.000000000    0.000000000
  2    N     1.201165270    0.000000000    0.000000000
  3    O     1.951957789    0.937586544    0.000000000
  4    F     1.776500874   -1.197837424    0.000010275
 ---
N1O3F1, RHF, CHARGE=0, MULT=1
```

```
HF=2.5
1    O      0.000000000     0.000000000      0.000000000
2    N      1.393231440     0.000000000      0.000000000
3    O      1.848414893     1.108766330      0.000000000
4    F     -0.534300439    -1.152181409     -0.000000000
5    O      1.984553537    -1.043595331      0.000000000
EXPGEOM
1    N      0.00000         0.65830          0.00000
2    O     -0.69070        -0.76010          0.00000
3    O      1.18180         0.66590          0.00000
4    O     -0.86660         1.47360          0.00000
5    F      0.33370        -1.73820          0.00000
---
H1C1N2O4F1, RHF, CHARGE=0, MULT=1
HF=-56.1
1    F      0.000000000     0.000000000      0.000000000
2    C      1.322125819     0.000000000      0.000000000
3    N      1.807386387     1.507930240      0.000000000
4    N      1.861319823    -0.653370703     -1.334140337
5    O      1.088910392    -0.988301882     -2.192672677
6    O      3.054571245    -0.814633293     -1.413168886
7    O      2.397063540     1.891335215      0.977952797
8    O      1.566683389     2.204898254     -0.950725888
9    H      1.769342072    -0.548394892      0.876381080
---
F2, RHF, CHARGE=0, MULT=1
HF=0
1    F      0.000000000     0.000000000      0.000000000
2    F      1.265522769     0.000000000      0.000000000
EXPGEOM
1    F      0.00000         0.00000          0.70670
2    F      0.00000         0.00000         -0.70670
---
C1F2, RHF, CHARGE=0, MULT=1
HF=-45
1    F      0.000000000     0.000000000      0.000000000
2    C      1.304197620     0.000000000      0.000000000
3    F      1.714775703     1.233460838      0.000000000
EXPGEOM
1    C      0.00000         0.00000          0.60410
2    F      0.00000         1.04430         -0.20140
3    F      0.00000        -1.04430         -0.20140
---
H1C1F2, RHF, CHARGE=1, MULT=1
HF=142.4
1    H      0.000000000     0.000000000      0.000000000
2    C      1.118049140     0.000000000      0.000000000
3    F      1.816881452     1.072915395      0.000000000
4    F      1.815920878    -1.073589897      0.000652281
---
H2C1F2, UHF, CHARGE=1, MULT=2
HF=185.2
1    H      0.000000000     0.000000000      0.000000000
2    C      1.209004484     0.000000000      0.000000000
3    F      1.760497897     1.185535557      0.000000000
4    F      1.759347566    -0.771539975     -0.900890070
5    H      1.079327687    -0.502569406      1.091438786
---
H2C1F2, RHF, CHARGE=0, MULT=1
```

```
HF=-108.1, DIP=1.96, IE=13.17
1    H    0.000000000    0.000000000    0.000000000
2    C    1.129761260    0.000000000    0.000000000
3    F    1.600355520    1.268135300    0.000000000
4    F    1.601351394   -0.590116329   -1.116595243
5    H    1.480012741   -0.548600794    0.915216351
EXPGEOM
1    C    0.00000    0.00000    0.50700
2    H   -0.91760    0.00000    1.10590
3    H    0.91760    0.00000    1.10590
4    F    0.00000    1.11470   -0.29190
5    F    0.00000   -1.11470   -0.29190
---
C2F2, RHF, CHARGE=0, MULT=1
HF=5.0, IE=11.2
1    C    0.000029773    0.008918935    0.000000000
2    C    1.186308461    0.002805277    0.000000000
3    F    2.460803770   -0.003379026    0.000000000
4    F   -1.274463859    0.015310465    0.000000000
---
H2C2F2, RHF, CHARGE=0, MULT=1
HF=-80.5, IE=10.72
1    C    0.020777621   -0.025632999   -0.010000000
2    C    1.387614627   -0.030175401   -0.010000000
3    H    1.960566374    0.905045371   -0.010000000
4    H   -0.545795410    0.913466896   -0.010000000
5    F   -0.718796685   -1.118660885   -0.010000000
6    F    2.120040602   -1.128026841   -0.010000000
EXPGEOM
1    C    0.00000    0.00000    1.39030
2    C    0.00000    0.00000    0.06680
3    H    0.00000    0.93850    1.92670
4    H    0.00000   -0.93850    1.92670
5    F    0.00000    1.08780   -0.69980
6    F    0.00000   -1.08780   -0.69980
---
H3C2F2, RHF, CHARGE=1, MULT=1
HF=107.0
1    C    0.000000000    0.000000000    0.000000000
2    C    1.532971310    0.000000000    0.000000000
3    F    2.250158508    1.069590551    0.000000000
4    F    2.257843553   -1.063681343    0.000985619
5    H   -0.362400956    0.562260362   -0.889875570
6    H   -0.395021420   -1.035845542   -0.025696438
7    H   -0.359642076    0.515763466    0.919028335
---
H4C2F2, RHF, CHARGE=0, MULT=1
HF=-118.8, DIP=2.3, IE=12.8
1    C    0.000000000    0.000000000    0.000000000
2    C    1.568590010    0.000000000    0.000000000
3    F    2.085294644    1.254643739    0.000000000
4    F    2.085322062   -0.605779459    1.098654993
5    H    1.928234482   -0.547664085   -0.926988529
6    H   -0.372314367    0.503144493   -0.913174834
7    H   -0.372054177   -1.042717986   -0.000588483
8    H   -0.416754956    0.521166771    0.882897037
EXPGEOM
1    C    0.32170    0.16640    0.00000
2    C   -0.90020    1.04920    0.00000
```

```
3    H     1.27510    0.70910    0.00000
4    F     0.32170   -0.65560    1.11090
5    F     0.32170   -0.65560   -1.11090
6    H    -1.80100    0.43000    0.00000
7    H    -0.89700    1.68390    0.88990
8    H    -0.89700    1.68390   -0.88990
---
H4C6F2, RHF, CHARGE=0, MULT=1
HF=-67.7
1    C     0.000000000    0.000000000    0.000000000
2    C     1.403225398    0.000000000    0.000000000
3    C     2.120582466    1.210067739    0.000000000
4    C     1.447030513    2.441206354    0.000246892
5    C     0.026977641    2.469411293    0.000441656
6    C    -0.706156209    1.232383366    0.000278807
7    F    -2.029101103    1.243808961    0.000343246
8    F    -0.617982285    3.624532866    0.000799394
9    H    -0.544832864   -0.944813719   -0.000179561
10   H     1.940961389   -0.949060484   -0.000101996
11   H     3.211269703    1.193722114   -0.000137360
12   H     2.014464311    3.372633927    0.000340075
EXPGEOM
1    C     0.00000    0.69830   -0.53500
2    C     0.00000   -0.69830   -0.53500
3    C     0.00000   -1.40160    0.66240
4    C     0.00000   -0.69760    1.86780
5    C     0.00000    0.69760    1.86780
6    C     0.00000    1.40160    0.66240
7    F     0.00000    1.35900   -1.71220
8    F     0.00000   -1.35900   -1.71220
9    H     0.00000   -2.48610    0.63330
10   H     0.00000   -1.24370    2.80530
11   H     0.00000    1.24370    2.80530
12   H     0.00000    2.48610    0.63330
---
H4C6F2, RHF, CHARGE=0, MULT=1
HF=-73.9
1    C     0.000000000    0.000000000    0.000000000
2    C     1.420129661    0.000000000    0.000000000
3    C     2.146904897    1.222694334    0.000000000
4    C     1.445267501    2.438541844    0.000031078
5    C     0.041779285    2.470403579    0.000032224
6    C    -0.681721628    1.246015307   -0.000070295
7    F    -2.006310792    1.264801031   -0.000111581
8    F     2.072573046   -1.152969113    0.000013790
9    H    -0.555688053   -0.938040945   -0.000017212
10   H     3.236910703    1.221778513    0.000006123
11   H     2.001898750    3.377457136    0.000086313
12   H    -0.482722951    3.425901662    0.000175278
EXPGEOM
1    C     0.00000    0.00000    1.77250
2    C     0.00000    1.21640    1.08640
3    C     0.00000   -1.21640    1.08640
4    C     0.00000    1.18680   -0.30470
5    C     0.00000   -1.18680   -0.30470
6    C     0.00000    0.00000   -1.03120
7    F     0.00000    2.35890   -0.98820
8    F     0.00000   -2.35890   -0.98820
9    H     0.00000    2.16800    1.60780
```

```
10      H       0.00000        -2.16800       1.60780
11      H       0.00000         0.00000       2.85910
12      H       0.00000         0.00000      -2.11570
---
H4C6F2, RHF, CHARGE=0, MULT=1
HF=-73.3
1     C      0.000000000     0.000000000     0.000000000
2     C      1.422174355     0.000000000     0.000000000
3     C      2.137062468     1.229491882     0.000000000
4     C      1.437817269     2.444243705     0.000036667
5     C      0.015628460     2.444410995     0.000022528
6     C     -0.699042035     1.214846880    -0.000008725
7     F      2.082851963    -1.148718269    -0.000013201
8     F     -0.645181241     3.593019430    -0.000031567
9     H     -0.553466743    -0.939874410    -0.000028656
10     H      3.227774065     1.235919825    -0.000009071
11     H      1.991442626     3.384050848     0.000058622
12     H     -1.789752999     1.208535219     0.000006615
EXPGEOM
1     C      0.00000         0.00000        1.36890
2     C      0.00000         0.00000       -1.36890
3     C      0.00000         1.21710        0.69790
4     C      0.00000        -1.21710        0.69790
5     C      0.00000        -1.21710       -0.69790
6     C      0.00000         1.21710       -0.69790
7     F      0.00000         0.00000        2.72750
8     F      0.00000         0.00000       -2.72750
9     H      0.00000         2.14470        1.25940
10     H      0.00000        -2.14470        1.25940
11     H      0.00000        -2.14470       -1.25940
12     H      0.00000         2.14470       -1.25940
---
H9C4N1F2, RHF, CHARGE=0, MULT=1
HF=-46
1     C     -0.000154198     0.000255377    -0.000216220
2     C      1.557764047    -0.000224737     0.000536756
3     C      2.108691591     1.452018627    -0.000044834
4     C      2.071291075    -0.785556982    -1.243067678
5     N      1.990456315    -0.841703568     1.257412084
6     F      1.598988413    -0.289454856     2.379094461
7     F      3.292732466    -0.939834222     1.362308380
8     H     -0.417012160    -1.022602945     0.095994872
9     H     -0.379758220     0.424064038    -0.952846008
10     H     -0.425198638     0.615881700     0.817632299
11     H      1.782213983     1.985404736    -0.916555069
12     H      3.216648259     1.482152175     0.017466312
13     H      1.745509596     2.040203898     0.866482867
14     H      1.672595488    -0.328324675    -2.172334840
15     H      1.745499514    -1.845147243    -1.229231079
16     H      3.176316411    -0.773407753    -1.327666037
---
H7C7N1F2, RHF, CHARGE=0, MULT=1
HF=1.8
1     C      0.077000933    -0.050748901     0.022238946
2     C      1.485936651     0.048644422     0.114834265
3     C      2.122436732     1.301666168     0.101688354
4     C      1.363363637     2.479545188    -0.007317148
5     C     -0.036491042     2.397036950    -0.101171843
6     C     -0.674330786     1.144471113    -0.085738529
```

```
7    C    -0.608653490    -1.394422449     0.070430962
8    N    -1.148212647    -1.741950707    -1.345074298
9    F    -1.976414033    -2.755981178    -1.273464958
10   F    -0.179742145    -2.130304196    -2.136845917
11   H     2.100368923    -0.849823819     0.197249452
12   H     3.209188604     1.357754398     0.174426301
13   H     1.857789360     3.451314426    -0.019784739
14   H    -0.631795128     3.306725588    -0.187348004
15   H    -1.762733940     1.109619200    -0.159797152
16   H    -1.495859246    -1.356348187     0.745344917
17   H     0.061688580    -2.194866084     0.460008446
---
N2F2, RHF, CHARGE=0, MULT=1
HF=16.4, DIP=0.16
1    N    -0.013583377     0.029708793    -0.010000000
2    N     1.226144114     0.004601143    -0.010000000
3    F     1.916998897     1.083974929    -0.010000000
4    F    -0.660339662     1.136039828    -0.010000000
EXPGEOM
1    F     0.00000    1.18200   -0.55120
2    N     0.00000    0.61250    0.70870
3    N     0.00000   -0.61250    0.70870
4    F     0.00000   -1.18200   -0.55120
---
N2F2, RHF, CHARGE=0, MULT=1
HF=19.4, IE=13.4
1    N     0.000601939    -0.041242323    -0.010000000
2    N     1.261958288    -0.010415454    -0.010000000
3    F     1.712353532     1.185049568    -0.010000000
4    F    -0.449941907    -1.236626108    -0.010000000
EXPGEOM
1    F     0.59630    1.53840    0.00000
2    N     0.59630    0.15760    0.00000
3    N    -0.59630   -0.15760    0.00000
4    F    -0.59630   -1.53840    0.00000
---
O1F2, RHF, CHARGE=0, MULT=1
HF=5.9, DIP=0.3, IE=13.26
1    F     0.000000000     0.000000000     0.000000000
2    O     1.281225399     0.000000000     0.000000000
3    F     1.699770779     1.210933916     0.000000000
EXPGEOM
1    O     0.00000    0.00000    0.60650
2    F     0.00000    1.10440   -0.26960
3    F     0.00000   -1.10440   -0.26960
---
C1O1F2, RHF, CHARGE=0, MULT=1
HF=-152.7, DIP=0.95
1    O     0.000000000     0.000000000     0.000000000
2    C     1.219359327     0.000000000     0.000000000
3    F     1.957048666     1.090121755     0.000000000
4    F     1.957027613    -1.090132814     0.000000000
---
C1F3, UHF, CHARGE=1, MULT=1
HF=99.3
1    C     0.000004157    -0.000979405     0.001473300
2    F    -0.000010506    -1.281081436     0.001409267
3    F     0.000001170     0.639169024    -1.107069966
4    F    -0.000004321     0.638976467     1.110125847
```

```
---
C1F3, UHF, CHARGE=0, MULT=2
HF=-112.4
1    C    0.001238412     0.011546772    -0.005518750
2    F   -0.161986524    -1.290637174    -0.000524996
3    F   -0.346124454     0.615947389    -1.117447031
4    F   -0.351929126     0.624821288     1.099773906
EXPGEOM
1    C    0.00000    0.00000    0.33110
2    F    0.00000    1.26660   -0.07360
3    F    1.09690   -0.63330   -0.07360
4    F   -1.09690   -0.63330   -0.07360
---
C1F3, UHF, CHARGE=-1, MULT=1
HF=-163.4
1    C   -0.004078221     0.013655299    -0.003249375
2    F   -0.468301784    -1.274949549    -0.001365023
3    F   -0.604928087     0.591982851    -1.089744759
4    F   -0.613118064     0.598095463     1.075235916
---
H1C1F3, RHF, CHARGE=0, MULT=1
HF=-166.3, DIP=1.65, IE=14.8
1    H    0.000000000     0.000000000     0.000000000
2    C    1.136202281     0.000000000     0.000000000
3    F    1.631726581     1.259076696     0.000000000
4    F    1.631671477    -0.629587912    -1.090444708
5    F    1.631695611    -0.629598675     1.090426367
EXPGEOM
1    C    0.00000    0.00000    0.34190
2    H    0.00000    0.00000    1.43460
3    F    0.00000    1.25960   -0.12910
4    F    1.09090   -0.62980   -0.12910
5    F   -1.09090   -0.62980   -0.12910
---
H1C2F3, RHF, CHARGE=0, MULT=1
HF=-117.3, DIP=1.3, IE=10.54
1    C   -0.000021926    -0.000074248    -0.000537173
2    C    1.376087879     0.000085296    -0.002230322
3    H    1.967823347     0.921716926     0.000923799
4    F   -0.728538946     1.103074055     0.001959211
5    F   -0.779366210    -1.066143203    -0.000183567
6    F    2.064299443    -1.123395576    -0.007158408
EXPGEOM
1    C    0.00000    0.43630    0.00000
2    C   -0.70240   -0.69760    0.00000
3    F    1.31500    0.51140    0.00000
4    F   -0.57100    1.62890    0.00000
5    F   -0.07760   -1.88640    0.00000
6    H   -1.78380   -0.71840    0.00000
---
H2C2F3, UHF, CHARGE=1, MULT=1
HF=114
1    C   -0.000042451    -0.000007004     0.000021556
2    C    1.610172570     0.000005672     0.000033700
3    F    2.079436247     1.258879923     0.000055967
4    F    2.078202897    -0.641412376     1.083809087
5    F    2.028011797    -0.633805155    -1.111246047
6    H   -0.550666824     0.828561055    -0.461300336
7    H   -0.549907881    -0.835384563     0.449819613
```

```
---
H2C2F3, UHF, CHARGE=1, MULT=1
HF=81
1    C     0.000015109    -0.000039152     0.000140663
2    C     1.604407836     0.000028002    -0.000150608
3    F     2.338799217     1.051093212     0.000220031
4    F     2.290674229    -1.085980293    -0.000607394
5    H    -0.306388586    -0.559676745    -0.925866385
6    H    -0.305772745    -0.571313415     0.919091533
7    F    -0.452360755     1.248950904     0.008144427
---
H2C2F3, UHF, CHARGE=0, MULT=2
HF=-123.6
1    C    -0.000037339     0.000049340     0.000040085
2    C     1.539703632     0.000000080    -0.000073987
3    F     2.067036371     1.251745352     0.000085391
4    F     2.067532685    -0.630700704     1.080921848
5    F     2.051139606    -0.626932051    -1.092066625
6    H    -0.539268471     0.798208185    -0.489301350
7    H    -0.540076907    -0.806675558     0.474185784
---
H3C2F3, RHF, CHARGE=0, MULT=1
HF=-178.9, DIP=2.32, IE=13.8
1    C     0.002423380    -0.014169882    -0.002657019
2    C     1.588786923     0.008385781     0.003453616
3    F     2.093728463     1.268919294     0.004452266
4    F     2.115771409    -0.609639171     1.091836136
5    F     2.125287940    -0.611341866    -1.079195025
6    H    -0.380314061     0.490670209    -0.909838368
7    H    -0.362624363    -1.058558659     0.006973531
8    H    -0.388015641     0.509747718     0.890227003
EXPGEOM
1    C     0.00000     0.00000     1.47570
2    C     0.00000     0.00000    -0.02670
3    H     0.00000    -1.03360     1.83510
4    H     0.89510     0.51680     1.83510
5    H    -0.89510     0.51680     1.83510
6    F     0.00000     1.25850    -0.52590
7    F    -1.08990    -0.62930    -0.52590
8    F     1.08990    -0.62930    -0.52590
---
H5C7F3, RHF, CHARGE=0, MULT=1
HF=-143.2, IE=9.68
1    C    -0.055053307    -0.004516902     0.006453928
2    C     2.779586723     0.001720589    -0.015471628
3    C     0.649577843     1.211580315     0.009614996
4    C     0.655200828    -1.217275737     0.003805256
5    C     2.055553403     1.217154912     0.002015828
6    C     2.061341371    -1.216640654    -0.002148929
7    H    -1.145511166    -0.007185468     0.008140607
8    H     0.106614405     2.157353119     0.019234110
9    H     0.116617450    -2.165584022     0.008428668
10   H     2.568711817     2.180541016     0.011556821
11   H     2.578459506    -2.177898626     0.006311108
12   C     4.341287321     0.006678099    -0.009904783
13   F     4.842146437     0.030343681     1.253662684
14   F     4.894659283    -1.079500813    -0.609825748
15   F     4.884377614     1.077164892    -0.646920348
---
```

```
N1F3, RHF, CHARGE=0, MULT=1
HF=-31.6, DIP=0.24, IE=13.73
1    N    -0.008844447     0.026637938     0.036352218
2    F     1.303274016     0.014582873    -0.053604187
3    F    -0.366972075     1.292179355     0.018321007
4    F    -0.299140017    -0.420392315     1.238800480
EXPGEOM
1    N     0.00000     0.00000     0.49230
2    F     0.00000     1.24390    -0.12760
3    F     1.07720    -0.62190    -0.12760
4    F    -1.07720    -0.62190    -0.12760
---
C2N1F3, RHF, CHARGE=0, MULT=1
HF=-118.4, DIP=1.26
1    C    -0.000520785     0.005312238    -0.003172033
2    C     1.497077785     0.003301292    -0.002264432
3    F     2.004603585     1.259503822    -0.001421589
4    F     2.000017412    -0.626420543     1.086731939
5    F     2.002367196    -0.625468811    -1.090888850
6    N    -1.159244318     0.006070137    -0.003352059
EXPGEOM
1    C     0.00000     0.00000    -0.29450
2    C     0.00000     0.00000     1.13070
3    N     0.00000     0.00000     2.34170
4    F     0.00000     1.33560    -0.79290
5    F     1.15670    -0.66780    -0.79290
6    F    -1.15670    -0.66780    -0.79290
---
C2N1F3, UHF, CHARGE=0, MULT=1
HF=-99.7
1    C     0.000028416     0.000007604     0.000190177
2    N     1.434203501    -0.000026365    -0.000208159
3    F    -0.472058009     1.263435010    -0.000057230
4    F    -0.471487449    -0.631388667     1.094666191
5    F    -0.472697656    -0.631957919    -1.093590869
6    C     2.633262774     0.000002549     0.000088603
---
N1O1F3, RHF, CHARGE=0, MULT=1
HF=-39
1    F     0.000000000     0.000000000     0.000000000
2    N     1.354574694     0.000000000     0.000000000
3    F     1.690727882     1.312193045     0.000000000
4    F     1.690889265    -0.433291099     1.238601730
5    O     1.860107563    -0.651300405    -0.917924363
EXPGEOM
1    N     0.00000     0.00000     0.20090
2    O     0.00000     0.00000     1.37830
3    F     0.00000     1.27600    -0.46050
4    F     1.10500    -0.63800    -0.46050
5    F    -1.10500    -0.63800    -0.46050
---
H1C2O2F3, RHF, CHARGE=0, MULT=1
HF=-255, DIP=2.28, IE=12
1    C     0.040374429     0.002805722     0.005158845
2    C     1.633613533    -0.016020645    -0.026248942
3    F     2.145231879     1.234726600    -0.069441654
4    F     2.147544940    -0.629893426     1.063184308
5    F     2.105850116    -0.673693034    -1.108653987
6    O    -0.680969704    -0.489410588    -0.848116727
```

```
7    O    -0.439478388    0.631997955    1.092609821
8    H    -1.387207422    0.673630528    1.162881365
EXPGEOM
1    C    0.08850    0.59240    0.00000
2    C    -0.29610   -0.90080    0.00000
3    O    0.81770   -1.65520    0.00000
4    O    -1.43360   -1.30170    0.00000
5    F    -1.00610    1.35120    0.00000
6    F    0.81770    0.88470    1.08790
7    F    0.81770    0.88470   -1.08790
8    H    0.51030   -2.57960    0.00000
---
C1F4, RHF, CHARGE=0, MULT=1
HF=-223.3, IE=16.23
1    F    0.000113305    0.000033182   -0.000159214
2    C    1.347031322   -0.000108399   -0.000031727
3    F    1.795989471    1.269850933   -0.000207851
4    F    1.796195847   -0.635337406   -1.099485826
5    F    1.795915313   -0.635123759    1.099710313
EXPGEOM
1    C    0.00000    0.00000    0.00000
2    F    0.76820    0.76820    0.76820
3    F   -0.76820   -0.76820    0.76820
4    F   -0.76820    0.76820   -0.76820
5    F    0.76820   -0.76820   -0.76820
---
C2F4, RHF, CHARGE=0, MULT=1
HF=-157.9, IE=10.5
1    C   -0.000071472    0.000024767   -0.000074651
2    C    1.380520651   -0.000024343    0.000056373
3    F    2.121771501    1.089701898   -0.000116993
4    F   -0.741486761    1.089655368   -0.000772930
5    F   -0.741319551   -1.089705815    0.000278973
6    F    2.121929411   -1.089666649    0.000735399
---
H2C6F4, RHF, CHARGE=0, MULT=1
HF=-154.6
1    C   -0.000026786    0.000019828   -0.000100582
2    C    1.417176888    0.000008556    0.000051222
3    C    2.139007013    1.244808079   -0.000052387
4    C    1.435006385    2.474778055   -0.000181341
5    C    0.017862768    2.474751321   -0.000130113
6    C   -0.704025616    1.229995605   -0.000418388
7    F   -2.025525437    1.233292310   -0.001031483
8    F   -0.641399296    3.620160822    0.000197746
9    H   -0.547671524   -0.944362947    0.000185736
10   F    2.076362864   -1.145294822    0.000238181
11   F    3.460497069    1.241514905    0.000113902
12   H    1.982629644    3.419180096   -0.000357208
---
N2F4, RHF, CHARGE=0, MULT=1
HF=-2, IE=12
1    N    0.024542445    0.018606250   -0.017042323
2    N    1.525162806    0.014931183   -0.019081423
3    F    2.007080436    1.225296721   -0.043565220
4    F   -0.450916186    1.231625423   -0.034331022
5    F   -0.457455251   -0.580818556   -1.068858705
6    F    2.000703162   -0.592207086   -1.069272343
---
```

```
C1O1F4, RHF, CHARGE=0, MULT=1
HF=-182.8
1    C    -0.000007081    -0.000097251     0.000089906
2    O     1.443131771     0.000026005    -0.000337507
3    F    -0.367329754     1.289366005    -0.000190416
4    F    -0.473877556    -0.611168097     1.098153458
5    F    -0.474770539    -0.612008233    -1.096981545
6    F     1.976229867    -1.157289273    -0.001023505
---
C1O2F4, RHF, CHARGE=0, MULT=1
HF=-134.9
1    C     0.000513232    -0.013654356    -0.011084798
2    F     1.333724507    -0.107149742    -0.030644211
3    F    -0.376402614     1.268585773    -0.000624981
4    O    -0.428198110    -0.700206050     1.177895335
5    F    -1.671586268    -0.597286077     1.437596837
6    O    -0.584971835    -0.589977566    -1.191714125
7    F    -0.242434732    -1.794773044    -1.426493249
---
C1N1F5, RHF, CHARGE=0, MULT=1
HF=-169
1    C     0.000127498     0.000073053     0.000224469
2    N     1.590596182    -0.000060557    -0.000036938
3    F    -0.404513230     1.276619800    -0.000329909
4    F    -0.513507800    -0.636018984     1.064319264
5    F    -0.406937203    -0.602291397    -1.124519140
6    F     2.033963249    -1.225009661     0.004175543
7    F     2.033854317     0.567385940     1.085737916
---
C1N3F5, RHF, CHARGE=0, MULT=1
HF=22.9
1    N     0.000076770    -0.000000516     0.000000203
2    C     1.502534384     0.000000862    -0.000000940
3    N     2.145342269     1.355822898    -0.000000491
4    N     2.171544091    -1.122550410    -0.000166885
5    F    -0.454960155    -0.633266498     1.049848508
6    F    -0.455226990    -0.633138307    -1.049817905
7    F     1.770339204     2.037749230     1.050455560
8    F     1.769230112     2.038212543    -1.049837737
9    F     3.445041955    -1.104209630    -0.000113471
---
C2F6, RHF, CHARGE=0, MULT=1
HF=-321.2, IE=14.6
1    C    -0.000005962    -0.000049157    -0.000076415
2    C     1.674310080    -0.000045854     0.000058087
3    F     2.150811537     1.259080707    -0.000208160
4    F     2.150768862    -0.629327562     1.090672351
5    F    -0.476314561    -1.259203946     0.013092500
6    F    -0.476292351     0.617764302    -1.097247383
7    F    -0.477048005     0.640931242     1.083493180
8    F     2.150972663    -0.629865006    -1.090153643
EXPGEOM
1    C     0.00000     0.00000     0.76660
2    C     0.00000     0.00000    -0.76660
3    F     0.00000     1.26150     1.21880
4    F    -1.09250    -0.63080     1.21880
5    F     1.09250    -0.63080     1.21880
6    F     0.00000    -1.26150    -1.21880
7    F    -1.09250     0.63080    -1.21880
```

```
     8    F      1.09250     0.63080    -1.21880
---
C4F6, RHF, CHARGE=0, MULT=1
HF=-253.4
1     C    -0.037161268     0.088285957    -0.045913285
2     C     1.457800220     0.007562899    -0.091969140
3     C    -0.747017275     1.027608925     0.680221122
4     C     2.206261174    -0.964570762     0.547065761
5     F    -0.690657581    -0.808305188    -0.765745467
6     F     2.069696958     0.935563228    -0.808471797
7     F    -2.061084170     1.115155033     0.720832883
8     F    -0.200568517     1.957740416     1.437741744
9     F     3.519847110    -1.056789701     0.503810121
10    F     1.702074885    -1.925835811     1.295055986
EXPGEOM
1     C     0.15390     1.87640     0.00000
2     C     0.46880     0.56250     0.00000
3     C    -0.46880    -0.56250     0.00000
4     C    -0.15390    -1.87640     0.00000
5     F     1.07890     2.83320     0.00000
6     F    -1.07890     2.37130     0.00000
7     F     1.78690     0.23440     0.00000
8     F    -1.78690    -0.23440     0.00000
9     F     1.07890    -2.37130     0.00000
10    F    -1.07890    -2.83320     0.00000
---
C1N2F6, RHF, CHARGE=0, MULT=1
HF=-108.8
1     C     0.000065741     0.000254547     0.000233169
2     N     1.580912798     0.000017761    -0.001145481
3     N    -0.483677333     1.502924630     0.000185167
4     F    -0.495829181    -0.583375901    -1.089202227
5     F     2.054327598    -0.700560037    -0.991603775
6     F     2.052386694    -0.527945903     1.092233146
7     F    -0.027466721     2.127352290    -1.047762205
8     F    -0.025349162     2.128482322     1.046478486
9     F    -0.505856318    -0.589408849     1.082250259
---
C2O1F6, RHF, CHARGE=0, MULT=1
HF=-369
1     O    -0.000027640    -0.000050047     0.000067050
2     C     1.405852908    -0.000019238    -0.000035226
3     C    -0.915874757     1.066575089    -0.000023712
4     F     1.780916197    -1.288063773     0.001251546
5     F    -2.137404046     0.512031714    -0.006035660
6     F     1.888021264     0.610092716    -1.097749694
7     F     1.888330931     0.612413379     1.096260405
8     F    -0.762498152     1.834213437    -1.094110708
9     F    -0.770120663     1.827247651     1.099898727
---
C3O1F6, RHF, CHARGE=0, MULT=1
HF=-342.6, IE=12.1
1     O    -0.012241193     0.011882301    -0.012609432
2     C     1.195591655     0.040018242    -0.000010605
3     C     1.980045882     1.441728441     0.011663272
4     C     2.044662958    -1.323582222     0.004183961
5     F     1.126429635     2.488969054     0.010550358
6     F     2.759504534     1.565942120     1.108011084
7     F     2.773665743     1.574951214    -1.073448828
```

```
   8     F      1.240415186    -2.409201533     0.003546330
   9     F      2.839028123    -1.416857784    -1.084439028
  10     F      2.833366675    -1.414283840     1.097118462
 ---
H5C7N2O4F1, RHF, CHARGE=0, MULT=1
HF=-16.9
   1     C      0.057969203    -0.003707037     0.023275573
   2     C      1.470749739     0.015379066    -0.108360294
   3     C      2.176495017     1.231381589    -0.113889188
   4     C      1.493323319     2.452333907     0.015611017
   5     C      0.093885562     2.449867983     0.154169715
   6     C     -0.616832971     1.238978842     0.161515441
   7     C     -0.730001988    -1.311060445     0.028145007
   8     N     -0.269191186    -2.266632146    -1.170477900
   9     F     -2.047985021    -1.199817426    -0.106954524
  10     N     -0.523853427    -2.156356420     1.378288954
  11     O     -1.298220131    -3.047466996     1.614856784
  12     O      0.377201810    -1.863332299     2.121055908
  13     O      0.412556813    -3.231667365    -0.936300593
  14     O     -0.612453083    -1.962991785    -2.284246324
  15     H      2.052875874    -0.903482056    -0.200855023
  16     H      3.262425242     1.221650675    -0.216636413
  17     H      2.043325085     3.394187582     0.010460536
  18     H     -0.446089099     3.391903113     0.257670787
  19     H     -1.701526904     1.286006858     0.277242900
 ---
H4C7F4, RHF, CHARGE=0, MULT=1
HF=-189.4
   1     C     -0.000159765    -0.000009759    -0.000134192
   2     C      1.411380791     0.000708039    -0.000008592
   3     C      2.125024201     1.214216815     0.000046694
   4     C      1.456392976     2.445741748     0.000112659
   5     C      0.033658365     2.454463562     0.000219462
   6     C     -0.692601676     1.233125694     0.000158807
   7     C     -0.823507229    -1.330396218    -0.000941331
   8     H      1.982350192    -0.929380174    -0.000003306
   9     H      3.216068828     1.193308961    -0.000076348
  10     H      2.020688848     3.378941290     0.000091179
  11     F     -0.619897701     3.606823664     0.000356492
  12     H     -1.784064970     1.264115063     0.000061172
  13     F     -1.631327496    -1.434561369    -1.088543078
  14     F     -1.631934119    -1.435767074     1.086081502
  15     F     -0.067099037    -2.457707794    -0.001413909
 ---
H10C6N2F4, RHF, CHARGE=0, MULT=1
HF=-41.6
   1     C      0.000031805     0.000094118     0.000018321
   2     C      1.541799135    -0.000239157    -0.000023731
   3     C      2.171536889     1.402068221    -0.000028089
   4     C      1.528782619     2.379492163    -0.997479191
   5     C     -0.012589237     2.357153833    -1.010530022
   6     C     -0.637253130     0.925637661    -1.094961967
   7     N     -0.321630283     0.281832560    -2.482849648
   8     N     -2.197609742     0.868021227    -0.942237890
   9     F     -2.799313550     1.649561387    -1.801716587
  10     F     -2.596231809     1.272915275     0.235369503
  11     F     -0.930802311    -0.863416634    -2.651483340
  12     F     -0.701039097     1.028445639    -3.487870049
  13     H     -0.340178154     0.340434088     1.005598908
```

```
14    H    -0.358871728   -1.048771578   -0.093816047
15    H     1.928153328   -0.586249719   -0.863596189
16    H     1.882765353   -0.548036867    0.908558114
17    H     3.256564421    1.311082564   -0.232672846
18    H     2.117166756    1.832379316    1.026527058
19    H     1.921970528    2.183332032   -2.020216156
20    H     1.852905667    3.415658118   -0.746066633
21    H    -0.365149563    3.012565028   -1.837440551
22    H    -0.377353457    2.841873964   -0.074708804
---
H12C6N2F4, RHF, CHARGE=0, MULT=1
HF=-52.8
1     C    -0.064644170    0.106376436   -0.223119354
2     C     1.477901372    0.134843277   -0.010153075
3     C     2.112184337    1.553432229    0.160046040
4     N     3.620626387    1.422544078   -0.245811475
5     C     1.993148192    2.051825699    1.631136034
6     N     1.736388843    3.551337202    1.910177326
7     F     2.699405678    4.317260628    1.468899594
8     F     0.642375779    3.971874897    1.324377497
9     F     4.294753066    2.511078831    0.025066789
10    F     3.732323146    1.269234172   -1.542465324
11    C    -0.488593429    0.459025552   -1.664690164
12    C    -0.658837889   -1.248915551    0.221718595
13    H    -0.529580647    0.877773192    0.441093981
14    H     1.946320125   -0.405764138   -0.862918628
15    H     1.738574495   -0.469763854    0.889339195
16    H     1.611079020    2.271087650   -0.533738283
17    H     2.895496471    1.800302192    2.236549920
18    H     1.131454121    1.552686622    2.135327131
19    H    -1.593918998    0.460837595   -1.758621653
20    H    -0.088950458   -0.260588436   -2.406962962
21    H    -0.141102655    1.470437880   -1.956575329
22    H    -1.759839299   -1.259887285    0.090680803
23    H    -0.458829158   -1.439095218    1.295977181
24    H    -0.242709840   -2.098782203   -0.355842445
---
H14C7N2F4, RHF, CHARGE=0, MULT=1
HF=-52.8
1     C    -0.127517163   -0.076584096    0.067390170
2     C     1.394292561    0.036682580    0.197995581
3     C     1.959546376    1.439372942   -0.099298660
4     C     3.493802618    1.541441542    0.024704452
5     C     4.042639101    2.961247466   -0.233628381
6     C     5.582661680    3.051639124   -0.147162326
7     C     6.146634689    4.488448426   -0.336364071
8     N     7.657656790    4.398449133   -0.674741318
9     N     5.746753391    5.404297699    0.845389814
10    F     8.391446164    4.077679391    0.357930065
11    F     8.114184060    5.548468581   -1.100246798
12    F     6.055600773    6.651810931    0.599118116
13    F     6.354846117    5.095598123    1.960424483
14    H    -0.456038401   -1.107921216    0.312624939
15    H    -0.470212366    0.145678372   -0.963761189
16    H    -0.654027003    0.614392017    0.756777884
17    H     1.863460382   -0.706376207   -0.487776065
18    H     1.683319655   -0.268419021    1.230303374
19    H     1.487514293    2.173543073    0.593605902
20    H     1.658202634    1.741387386   -1.128578620
```

```
21     H     3.965917654    0.829115919   -0.689425993
22     H     3.801380037    1.214718396    1.044451261
23     H     3.578768949    3.660068050    0.497639078
24     H     3.719583668    3.301199707   -1.244717582
25     H     6.016059960    2.414967570   -0.954835816
26     H     5.930523867    2.612937432    0.814617250
27     H     5.674761941    4.953557882   -1.243031558
---
H1C6F5, RHF, CHARGE=0, MULT=1
HF=-192.5, IE=9.75
1      C    -0.000051478   -0.000045489    0.000000000
2      C     1.417052785    0.000168643    0.000000000
3      C     2.148446453    1.236538410    0.000000000
4      C     1.433412705    2.479176286    0.000000000
5      C    -0.000324241    2.478694165    0.000000000
6      C    -0.707068646    1.227950702    0.000000000
7      H    -0.546547741   -0.945341323    0.000000000
8      F    -0.662509902    3.619537647    0.000000000
9      F     2.073884839   -1.145358933    0.000000000
10     F    -2.027521339    1.225674955    0.000000000
11     F     3.467496150    1.231848756    0.000000000
12     F     2.093173353    3.620619267    0.000000000
---
H3C7F5, RHF, CHARGE=0, MULT=1
HF=-201.6
1      C    -0.000049485   -0.000112915    0.000005578
2      C     1.435845189    0.000097400   -0.000016476
3      C     2.147116833    1.242089576    0.000006274
4      C     1.410657313    2.467937523   -0.000073500
5      C    -0.027013088    2.441573964   -0.000033167
6      C    -0.755439636    1.214326905    0.000389257
7      C    -2.266753105    1.171703775    0.000631112
8      F    -0.616500611   -1.169553546   -0.000361119
9      F     2.099310241   -1.140792463   -0.000101774
10     F     3.465195866    1.255550223    0.000027241
11     F     2.050640641    3.622231898   -0.000221153
12     F    -0.659932931    3.601856328   -0.000424264
13     H    -2.640685030    0.632604540    0.896317655
14     H    -2.640702494    0.648152598   -0.904273075
15     H    -2.731975217    2.176347102    0.009755143
---
H1C6O1F5, RHF, CHARGE=0, MULT=1
HF=-228.8
1      C    -0.000337400   -0.000168728   -0.000106220
2      C     1.431968375   -0.000098317   -0.000011173
3      C     2.152794463    1.240011432   -0.000051551
4      C     1.435144748    2.478224422   -0.000642409
5      C     0.002053750    2.475150606   -0.001418237
6      C    -0.723196446    1.240948994   -0.000762996
7      O    -2.068888075    1.149577109   -0.001100791
8      F    -0.649811814   -1.148676458    0.000424911
9      F     2.090962187   -1.142023162    0.000323343
10     F     3.472404844    1.238162678    0.000493571
11     F     2.090313562    3.622169810   -0.000540721
12     F    -0.651680650    3.626126722   -0.002450845
13     H    -2.524819629    1.982020319    0.002072391
---
C6F6, RHF, CHARGE=0, MULT=1
HF=-242.5, IE=10.9
```

```
1   C    -0.000298040    0.000087378    0.000031969
2   C     1.433609931    0.000421043    0.000220304
3   C     2.149949302    1.242420901   -0.000316670
4   C     1.432523523    2.483951426   -0.002180889
5   C    -0.001415238    2.483746710   -0.002486872
6   C    -0.717752708    1.241592957   -0.001115855
7   F    -0.659021191   -1.141570427    0.001146917
8   F    -0.660839682    3.624993715   -0.003717981
9   F     2.093112657   -1.140810614    0.000813182
10   F   -2.035781853    1.240894045   -0.000968532
11   F    3.468049217    1.242961026    0.000923689
12   F    2.091230182    3.625559867   -0.003645640
EXPGEOM
1   C     0.00000    1.38280    0.00000
2   C     1.19750    0.69140    0.00000
3   C     1.19750   -0.69140    0.00000
4   C     0.00000   -1.38280    0.00000
5   C    -1.19750   -0.69140    0.00000
6   C    -1.19750    0.69140    0.00000
7   F     0.00000    2.72680    0.00000
8   F     2.36150    1.36340    0.00000
9   F     2.36150   -1.36340    0.00000
10   F    0.00000   -2.72680    0.00000
11   F   -2.36150   -1.36340    0.00000
12   F   -2.36150    1.36340    0.00000
---
C3F8, RHF, CHARGE=0, MULT=1
HF=-426.2
1   C     0.000039523    0.000021996    0.000051435
2   C     1.671482820   -0.000366216   -0.000149687
3   C     2.351945927    1.526329952    0.000101355
4   F    -0.488343540    0.595955658   -1.104394458
5   F     1.961190602    2.235818715   -1.075706944
6   F    -0.488802405    0.647733148    1.074717863
7   F     2.004328803    2.215690007    1.103456517
8   F    -0.467704913   -1.263111479    0.029605583
9   F     3.696128452    1.439493511   -0.027003155
10   F    2.103655304   -0.667352110   -1.082255180
11   F    2.105151080   -0.667424036    1.081212448
EXPGEOM
1   C     0.00000    0.00000    0.59040
2   C     0.00000    1.30450   -0.23480
3   C     0.00000   -1.30450   -0.23480
4   F     1.09780    0.00000    1.36810
5   F    -1.09780    0.00000    1.36810
6   F     0.00000    2.34030    0.59980
7   F     0.00000   -2.34030    0.59980
8   F     1.08390    1.36830   -1.00400
9   F    -1.08390    1.36830   -1.00400
10   F   -1.08390   -1.36830   -1.00400
11   F    1.08390   -1.36830   -1.00400
---
C4F8, RHF, CHARGE=0, MULT=1
HF=-389.9
1   C     0.000080764   -0.000044262   -0.000049094
2   C     1.588003926   -0.000023428    0.000026754
3   C     2.380038812    1.129628448    0.000153058
4   C     3.967848922    1.129711578   -0.000825077
5   F    -0.492478096    0.629943370    1.090459623
```

```
6       F     -0.492054730    0.629758167   -1.090834034
7       F     -0.513616251   -1.249452539    0.000077852
8       F      4.459349201    0.504007404   -1.094279984
9       F      2.125627408   -1.199817699   -0.000301969
10      F      1.842352233    2.329398691    0.000219460
11      F      4.461310296    0.495683572    1.086877664
12      F      4.481587424    2.379123621    0.003596178
---
C4F8, RHF, CHARGE=0, MULT=1
HF=-369.5
1       C     -0.000037563   -0.000387766    0.000024440
2       C      2.325367531    0.000392463    0.000023501
3       C      1.162268635    1.162606541   -0.000064542
4       C      1.163059609   -1.162598006    0.001613348
5       F     -0.793320275    0.000470485    1.075969810
6       F     -0.792110552   -0.001709599   -1.076791166
7       F      3.118655297    0.001708633    1.075963116
8       F      3.117434434   -0.000441472   -1.076796891
9       F      1.162053179    1.955297206   -1.076385466
10      F      1.162019233    1.955079867    1.076535219
11      F      1.163293040   -1.956768896   -1.073666121
12      F      1.163288673   -1.953593945    1.079250613
---
C7F8, RHF, CHARGE=0, MULT=1
HF=-356.8
1       C      0.000108517   -0.000033529    0.000028296
2       C      1.435853293   -0.000036622    0.000073948
3       C      2.149259566    1.242245852   -0.000060959
4       C      1.413513537    2.468901690   -0.000406562
5       C     -0.026296327    2.448883512   -0.000377088
6       C     -0.749715763    1.219623144   -0.000413754
7       C     -2.320390069    1.150615240   -0.000117718
8       F     -0.618622463   -1.163576658    0.000558483
9       F      2.096942742   -1.141238636    0.000337688
10      F      3.464706374    1.254000210    0.000264734
11      F      2.053945649    3.622122910   -0.000693330
12      F     -0.649281938    3.609054102   -0.000152622
13      F     -2.787111593    0.498779927    1.093570152
14      F     -2.787590387    0.489491782   -1.087971135
15      F     -2.937178412    2.355064974   -0.005317708
---
C1N4F8, RHF, CHARGE=0, MULT=1
HF=0.4
1       C      0.000000467    0.000055792    0.000023254
2       N      1.553216478    0.000009255    0.000045461
3       N     -0.400190233    1.500886743   -0.000039060
4       N     -0.750812970   -0.617406715    1.211417509
5       N     -0.402755805   -0.884422807   -1.211520658
6       F      2.018176354    0.643600006    1.035677925
7       F      2.019027987   -1.215713755    0.089024791
8       F     -0.432152245   -0.004198569    2.318427614
9       F     -0.414723263   -1.867133948    1.379093171
10      F      0.104517311   -0.416233558   -2.318877708
11      F     -1.697190622   -0.885025273   -1.377109477
12      F     -1.695034475    1.637498954   -0.087432961
13      F      0.100369814    2.115775358   -1.036493145
---
C3O2F8, RHF, CHARGE=0, MULT=1
HF=-520.6
```

```
  1    C     0.000024453    -0.000011060    -0.000035940
  2    O     1.406559361    -0.000028673     0.000764872
  3    C     2.319217093     1.064746276    -0.000039307
  4    O     2.519953457     1.501206852     1.317605385
  5    C     3.416243887     1.053112186     2.304533598
  6    F     3.009667833     1.601884516     3.459599636
  7    F     4.663090083     1.469517784     2.020410930
  8    F     3.404627969    -0.287331035     2.413830373
  9    F     3.445220176     0.604942438    -0.563411468
 10    F     1.868952629     2.121408616    -0.691168394
 11    F    -0.376996328    -1.242841354     0.337674877
 12    F    -0.459422374     0.293805380    -1.229375036
 13    F    -0.499733470     0.878656294     0.887141651
---
C4F10, RHF, CHARGE=0, MULT=1
HF=-515.3
  1    C    -0.000001405    -0.000164341     0.000052458
  2    C     1.676747148     0.000089333    -0.000177824
  3    C     2.352077753     1.523000000     0.000131026
  4    C     4.026510033     1.524404339    -0.085807788
  5    F     4.570756558     0.827803008     0.929438565
  6    F     4.491938018     2.786894733    -0.018401711
  7    F     4.457760771     0.986243537    -1.241881163
  8    F     1.986503628     2.171154734     1.118151703
  9    F     1.877685285     2.226225778    -1.041006847
 10    F     2.096982928    -0.677432489    -1.080719284
 11    F    -0.490296625     0.645245750     1.074938570
 12    F    -0.490889770     0.592462062    -1.104516504
 13    F     2.098156625    -0.674538505     1.081910906
 14    F    -0.461862797    -1.265401701     0.030199918
---
H6C7O4F6, RHF, CHARGE=0, MULT=1
HF=-460.5
  1    C     3.226804608     0.407848945     0.570210536
  2    O     3.383688274     0.791055875    -0.779990224
  3    C     2.406495731     0.766922881    -1.705315019
  4    C     2.889620077     1.297238451    -3.123515424
  5    C     2.971934880     0.089906310    -4.277040034
  6    C     4.099745014    -1.112681556    -3.998782019
  7    C     5.593251308    -0.569532371    -3.978185228
  8    O     6.164216165    -0.659727894    -2.762729202
  9    C     7.489172904    -0.274301704    -2.461978954
 10    O     1.266768145     0.394912412    -1.476095187
 11    O     6.165588417    -0.134666176    -4.964665914
 12    F     4.082733015     1.921336004    -3.060632230
 13    F     2.014136237     2.230444590    -3.554967256
 14    F     3.233703219     0.659337508    -5.468390869
 15    F     1.759039909    -0.483593010    -4.389421852
 16    F     3.988892758    -2.027315724    -4.986127720
 17    F     3.769851117    -1.761233357    -2.864228391
 18    H     4.214920968     0.582281685     1.056419337
 19    H     2.961310883    -0.670398411     0.667319051
 20    H     2.456169983     1.023172355     1.090854015
 21    H     7.633769710    -0.489777351    -1.377243169
 22    H     7.657130939     0.813464876    -2.639646905
 23    H     8.234661401    -0.860385241    -3.048585602
---
C6F10, RHF, CHARGE=0, MULT=1
HF=-461.9
```

```
  1    C      0.000243787    0.000202275   -0.000061213
  2    C      1.570349783   -0.000540621    0.000298345
  3    C      2.366858789    1.121792776    0.001730180
  4    C      1.852827655    2.604569569   -0.059429414
  5    C      0.232218719    2.700062840    0.279603372
  6    C     -0.597086097    1.476344698   -0.461510154
  7    F     -0.481434607   -0.944231130   -0.832119864
  8    F     -0.461796052   -0.307984716    1.229407811
  9    F      2.105771772   -1.202531680    0.029162672
 10    F      3.678374443    1.012419490    0.027407726
 11    F      2.526271226    3.380703576    0.813059298
 12    F      2.085554863    3.121664800   -1.283369345
 13    F      0.039609417    2.638738854    1.609869873
 14    F     -0.251943007    3.887046424   -0.123204364
 15    F     -1.902909206    1.560300256   -0.156190650
 16    F     -0.505847627    1.608502282   -1.797513340
---
C5N1F11, RHF, CHARGE=0, MULT=1
HF=-478.9
  1    N      0.468294191   -0.073305654   -0.683071912
  2    C      1.408499935    0.287006923    0.461003855
  3    C      2.362044312    1.516483843   -0.112545742
  4    C      1.644692541    2.336531186   -1.366389866
  5    C      0.001696997    2.413038429   -1.173157901
  6    C     -0.655511792    0.899245428   -1.040387818
  7    F      0.052196995   -1.298745669   -0.629484346
  8    F      2.141914891   -0.787688824    0.774395234
  9    F      0.733010041    0.639653946    1.567159130
 10    F      3.540413210    1.021381299   -0.524583391
 11    F      2.633801086    2.367574512    0.892452450
 12    F      1.937018192    1.748083290   -2.537880887
 13    F      2.145394342    3.582618790   -1.431166568
 14    F     -0.549181467    3.054272708   -2.216946316
 15    F     -0.290623115    3.140727454   -0.080285093
 16    F     -1.649761951    0.859990202   -0.140546217
 17    F     -1.160219816    0.488925973   -2.211381539
---
C2N5F11, RHF, CHARGE=0, MULT=1
HF=-84.4
  1    N      0.538699147    0.195324515   -0.887840279
  2    C      1.674615395   -0.284696879    0.109597077
  3    N      2.324904551    1.023613647    0.570223158
  4    C      3.510605489    1.646333419   -0.103210316
  5    N      4.815099280    1.073983118    0.528364943
  6    N      3.337590246    3.173955168    0.193305456
  7    F     -0.380909633    0.853874272   -0.242907500
  8    F     -0.046352258   -0.847218061   -1.410764070
  9    F      2.500934724   -1.094772828   -0.552281552
 10    F      1.170067973   -0.954875151    1.149367459
 11    F      2.445574740    1.042437513    1.863771902
 12    N      3.362397651    1.367268386   -1.623404541
 13    F      5.859030881    1.790121430    0.216259679
 14    F      5.062769899   -0.138820830    0.120364954
 15    F      4.190832907    3.872888146   -0.507222160
 16    F      3.571578253    3.462693620    1.441355722
 17    F      4.492322635    1.570558362   -2.244353680
 18    F      2.489876039    2.158704607   -2.181875494
---
C6F12, RHF, CHARGE=0, MULT=1
```

```
HF=-590.5
1   C   -0.000243169   -0.000086543   -0.000255598
2   C    1.657552658    0.000496541   -0.000035669
3   C    2.278401954    1.537488598    0.000004299
4   C    1.654802669    2.463424049   -1.225554575
5   C   -0.002888295    2.460962860   -1.229859265
6   C   -0.624064922    0.923892521   -1.227344288
7   F   -0.442519479   -1.262700980   -0.128595558
8   F   -0.454468175    0.448838853    1.182241504
9   F    2.100620511   -0.656345619    1.085593313
10  F    2.112406809   -0.673615444   -1.069997345
11  F    3.614741491    1.475032720   -0.128722098
12  F    2.032670764    2.125737474    1.183206555
13  F    2.094949103    3.726439128   -1.093481366
14  F    2.112015240    2.017879211   -2.408213131
15  F   -0.460925961    3.137375706   -0.162712402
16  F   -0.443745392    3.114700312   -2.318386880
17  F   -1.960387255    0.986792418   -1.098364468
18  F   -0.378348503    0.333560355   -2.409475661
---
```